\documentclass[openany,11pt]{book}

\newcount\twosided \twosided=0

\usepackage{Bbook}
\usepackage{axodraw}  

\renewcommand{\d}{{\rm d}}
\def\OMIT#1{{}}

\newcommand{\be}{\begin{equation}}
\newcommand{\ee}{\end{equation}}
\newcommand{\bea}{\begin{eqnarray}}
\newcommand{\eea}{\end{eqnarray}}
\newcommand{\beq}{\begin{equation}}
\newcommand{\eeq}{\end{equation}} 
\newcommand{\beqa}{\begin{eqnarray}}
\newcommand{\eeqa}{\end{eqnarray}}
\newcommand{\bey}{\begin{eqnarray}}
\newcommand{\eey}{\end{eqnarray}}
\newcommand{\nn}{\nonumber \\}
\newcommand{\no}{\nonumber }

\def\ifmath#1{\relax\ifmmode#1\else$#1$\fi}
\def\etal{{\it et~al.,}}

\def\ie{{\it i.e.}}

\def\D0{D\O}

\newcommand{\lt}{\left}
\newcommand{\rt}{\right}
\newcommand{\ov}{\overline}

\renewcommand{\Im}{{\rm Im}\,}
\renewcommand{\Re}{{\rm Re}\,}
\newcommand{\imag}{\Im}
\newcommand{\real}{\Re}
\newcommand{\diag}{{\rm diag}\,}

\def\BR{{\rm BR}}

\def\gsim{\gtrsim}
\def\lsim{\lesssim}

\def\lesssim{\mathrel{\mathpalette\vereq<}}
\def\gtrsim{\mathrel{\mathpalette\vereq>}}
\def\vereq#1#2{\lower4pt\vbox{\baselineskip1pt\lineskip1pt
    \ialign{\\$#1\hfill##\hfil\\$\crcr#2\crcr\sim\crcr}}}

\def\openone{\leavevmode\hbox{\small1\kern-3.8pt\normalsize1}}%

\newcommand{\gev}{\,\mbox{GeV}}

\newcommand{\keV}{\ensuremath{\mathrm{ke\kern -0.1em V}}}
\newcommand{\eV}{\ensuremath{\mathrm{e\kern -0.1em V}}}
\newcommand{\MeV}{\ensuremath{\mathrm{Me\kern -0.1em V}}}
\newcommand{\GeV}{\ensuremath{\mathrm{Ge\kern -0.1em V}}}
\newcommand{\TeV}{\ensuremath{\mathrm{Te\kern -0.1em V}}}

\def\vek#1{\mbox{\protect\boldmath $#1$}}

\newcommand{\ds}{\displaystyle}

\renewcommand{\C}{\ensuremath{C}}
\newcommand{\p}{\ensuremath{P}}
\newcommand{\CP}{\ensuremath{CP}}
\newcommand{\T}{\ensuremath{T}}
\newcommand{\CPT}{\ensuremath{C\!PT}}

\newcommand{\Bbar}{\,\overline{\!B}}

\newcommand{\Kbar}{\,\overline{\!K}}
\def\B0bar{\Bbar{}^0}

\newcommand{\adg}[1]{\ensuremath{{\cal A}^{#1}_{\rm \Delta\Gamma}}}

\newcommand{\gunt}[1]{\ensuremath{ \Gamma \lt[#1,t\rt] }}
\newcommand{\guntf}{\ensuremath{\Gamma  [f,t] }}
\newcommand{\guntfb}{\ensuremath{\Gamma  [\ov{f},t] }}
\newcommand{\guntfcpp}{\ensuremath{\Gamma [ f_{\rm CP+} ,t] }}

\def\lqcd{\Lambda_{\rm QCD}}
\newcommand{\as}{\ensuremath{\alpha_s}}
\newcommand{\e}{\epsilon}
\newcommand{\bra}[1]{\ensuremath{\langle #1 |}}
\newcommand{\ket}[1]{\ensuremath{| #1 \rangle }}
\newcommand{\braket}[2]{\ensuremath{\langle #1 | #2 \rangle }}
\newcommand{\epm}[2]{
 \raisebox{-0.5ex}{\shortstack[l]{$\scriptstyle+#1$\\$\scriptstyle-#2$}}}

\newcommand{\eq}[1]{(\ref{#1})}
\newcommand{\fig}[1]{Fig.~\ref{#1}}
\newcommand{\tab}[1]{Table~\ref{#1}}

\newcommand{\prl}{Phys.\ Rev. Lett.}
\newcommand{\prd}{Phys.\ Rev.~D}
\newcommand{\prep}{Phys.\ Rep.}
\newcommand{\plb}{Phys.\ Lett.~B}

\newcommand{\rmp}{Rev.\ Mod.\ Phys.}

\newcommand{\PRL}[3]{{\em Phys. Rev. Lett.} {\bf #1},\ #2 (#3)}
\newcommand{\PRD}[3]{{\em Phys. Rev. } {\bf #1},\ #2 (#3)}
\newcommand{\PLB}[3]{{\em Phys. Lett.} {\bf #1},\ #2 (#3)}
\newcommand{\NPB}[3]{{\em Nucl. Phys.} {\bf #1},\ #2 (#3)}
\newcommand{\RMP}[3]{{\em Rev. Mod. Phys.} {\bf #1},\ #2 (#3)}
\newcommand{\ZPC}[3]{{\em Z. Phys.} {\bf #1},\ #2 (#3)}
\newcommand{\EPJ}[3]{{\em Eur. Phys. J.} {\bf #1},\ #2 (#3)}
\newcommand{\NPproc}[3]{{\em Nucl. Phys. B} (Proc. Suppl.){\bf #1},\ #2 (#3)}

\def\bsg{\ifmmode B\to X_s\gamma\else $B\to X_s\gamma$\fi}

\newcommand{\bbm}{$B^0 - \Bbar{}^0$ mixing}
\newcommand{\bbmd}{$B_d^0\!-\!\Bbar{}_d^0\,$ mixing}
\newcommand{\bbms}{$B_s^0\!-\!\Bbar{}_s^0\,$ mixing}
\newcommand{\kkm}{$K^0\!-\!\ov{K}{}^0\,$ mixing}
\newcommand{\ddm}{$D^0\!-\!\ov{D}{}^0\,$ mixing}
\newcommand{\dbt}{\ensuremath{|\Delta B| \! = \!2}}
\newcommand{\dbo}{\ensuremath{|\Delta B| \! = \!1}}
\newcommand{\dst}{\ensuremath{|\Delta S|  = 2}}
\newcommand{\dso}{\ensuremath{|\Delta S|  = 1}}
\newcommand{\dft}{\ensuremath{|\Delta F|  = 2}}
\newcommand{\dfo}{\ensuremath{|\Delta F|  = 1}}
\newcommand{\hbo}{\ensuremath{H^\mathrm{|\Delta B| = 1}}}
\newcommand{\dm}{\ensuremath{\Delta m}}
\newcommand{\dms}{\ensuremath{\Delta m_{B_s}}}
\newcommand{\dmd}{\ensuremath{\Delta m_{B_d}}}
\newcommand{\dg}{\ensuremath{\Delta \Gamma}}
\newcommand{\dgs}{\ensuremath{\Delta \Gamma_{B_s}}}

\newcommand{\gtf}{\ensuremath{\Gamma (B^0(t) \rightarrow f )}}
\newcommand{\gbtf}{\ensuremath{\Gamma (\Bbar{}^0(t) \rightarrow f )}}
\newcommand{\gtfb}{\ensuremath{\Gamma (B^0(t) \rightarrow \ov{f} )}}
\newcommand{\gbtfb}{\ensuremath{\Gamma (\Bbar{}^0(t) \rightarrow \ov{f} )}}
\newcommand{\Bun}{\raisebox{7.5pt}{$\scriptscriptstyle(\hspace*{6.5pt})$}
  \hspace*{-9.7pt}\!\Bbar}
\newcommand{\Bdun}{\raisebox{7.5pt}{$\scriptscriptstyle(\hspace*{6.5pt})$}
  \hspace*{-9.7pt}\!\Bbar_{d}}
\newcommand{\Bsun}{\raisebox{7.7pt}{$\scriptscriptstyle(\hspace*{6.5pt})$}
  \hspace*{-9.7pt}\!\Bbar_{s}}
\newcommand{\ega}{a}

\begin{document}

\frontmatter

\begin{titlepage}

\vspace*{-1.6cm}
\centerline{\raisebox{11.2pt}{\epsfxsize 1cm\epsffile{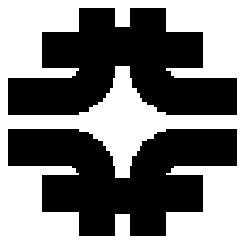}} \hfill 
  \vbox{\hbox{FERMILAB-Pub-01/197}\hbox{hep-ph/0201071}\hbox{December, 2001}}}

\begin{center}

\vspace*{1cm}
{\bf\Huge \boldmath $B$ Physics at the Tevatron: \\[12pt]
  Run II and Beyond}
\vspace*{1.75cm}

\includegraphics*[scale=0.8]{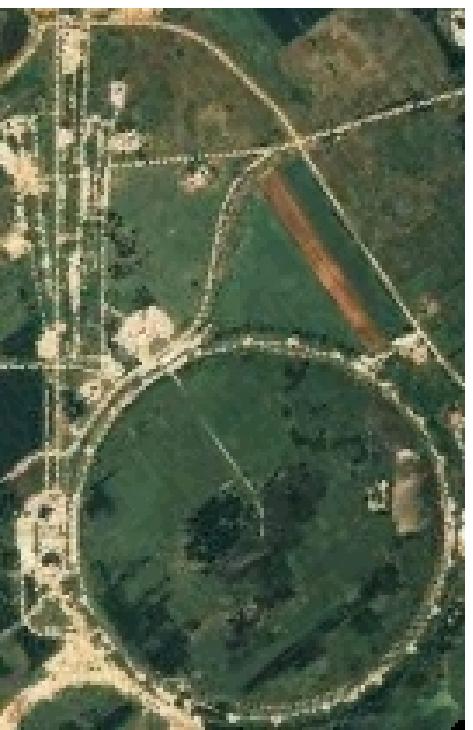}

\vspace*{1.25cm}

{\bf Authors} \vspace*{-2mm}

\addtolength{\hsize}{-.4cm}
\parbox{\hsize}{\author{%
K.~Anikeev$^{\,23}$,	
D.~Atwood$^{\,18}$,	
F.~Azfar$^{\,31}$,	
S.~Bailey$^{\,11}$,	
C.W.~Bauer$^{\,5}$,	
W.~Bell$^{\,10}$,	
G.~Bodwin$^{\,1}$,	
E.~Braaten$^{\,30}$,	
G.~Burdman$^{\,21}$,	
J.N.~Butler$^{\,9}$,	
K.~Byrum$^{\,1}$,	
N.~Cason$^{\,29}$,	
A.~Cerri$^{\,33}$,	
H.W.K.~Cheung$^{\,9}$,	
A.~Dighe$^{\,34}$,	
S.~Donati$^{\,33}$,	
R.K.~Ellis$^{\,9}$,	
A.~Falk$^{\,19}$,	
G.~Feild$^{\,45}$,	
S.~Fleming$^{\,6}$,	
I.~Furic$^{\,23}$,	
S.~Gardner$^{\,20}$,	
Y.~Grossman$^{\,38}$,	
G.~Gutierrez$^{\,9}$,	
W.~Hao$^{\,39}$,	
B.W.~Harris$^{\,1,26}$,	
J.~Hewett$^{\,35}$,	
G.~Hiller$^{\,35}$,	
R.~Jesik$^{\,14,13}$,	
M.~Jones$^{\,32}$,	
P.A.~Kasper$^{\,9}$,	
A.~El-Khadra$^{\,12}$,	
M.~Kirk$^{\,2}$,	
V.V.~Kiselev$^{\,17}$,	
J.~Kroll$^{\,32}$,	
A.S.~Kronfeld$^{\,9}$,	
R.~Kutschke$^{\,9}$,	
V.E.~Kuznetsov$^{\,4}$,	
E.~Laenen$^{\,27}$,	
J.~Lee$^{\,8,1}$,	
A.K.~Leibovich$^{\,6,9}$,	
J.D.~Lewis$^{\,9}$,	
Z.~Ligeti$^{\,9,21}$,	
A.K.~Likhoded$^{\,17}$,	
H.E.~Logan$^{\,9}$,	
M.~Luke$^{\,40}$,	
A.~Maciel$^{\,28}$,	
G.~Majumder$^{\,37}$,	
P.~Maksimovi\'c$^{\,11,21}$,	
M.~Martin$^{\,31}$,	
S.~Menary$^{\,46}$,	
P.~Nason$^{\,24}$,	
U.~Nierste$^{\,9,7}$,	
Y.~Nir$^{\,43}$,	
L.~Nogach$^{\,17}$,	
E.~Norrbin$^{\,22}$,	
C.~Oleari$^{\,44}$,	
V.~Papadimitriou$^{\,39}$,	
M.~Paulini$^{\,6}$,	
C.~Paus$^{\,23}$,	
M.~Petteni$^{\,13}$,	
R.~Poling$^{\,25}$,	
M.~Procario$^{\,6}$,	
G.~Punzi$^{\,33}$,	
H.~Quinn$^{\,35}$,	
A.~Rakitine$^{\,23}$,	
G.~Ridolfi$^{\,15}$,	
K.~Shestermanov$^{\,17}$,	
G.~Signorelli$^{\,33}$,	
J.P.~Silva$^{\,35}$,	
T.~Skwarnicki$^{\,37}$,	
A.~Smith$^{\,25}$,	
B.~Speakman$^{\,25}$,	
K.~Stenson$^{\,42}$,	
F.~Stichelbaut$^{\,36}$, 
S.~Stone$^{\,37}$,	
K.~Sumorok$^{\,23}$,	
M.~Tanaka$^{\,41,1}$,	
W.~Taylor$^{\,36}$,	
W.~Trischuk$^{\,40}$,	
J.~Tseng$^{\,23}$,	
R.~Van~Kooten$^{\,14}$,	
A.~Vasiliev$^{\,17}$,	
M.~Voloshin$^{\,25,16}$,	
J.C.~Wang$^{\,37}$,	
A.B.~Wicklund$^{\,1}$,	
F.~W\"urthwein$^{\,23}$,	
N.~Xuan$^{\,29}$,	
J.~Yarba$^{\,9}$,	
K.~Yip$^{\,9,3}$,	
A.~Zieminski$^{\,14}$	
}
}
\addtolength{\hsize}{.4cm}

\vspace*{-1.25cm}
\end{center}

\end{titlepage}

\thispagestyle{empty}
\begin{center}
\sl\vspace*{-26pt}
${}^{1}$Argonne National Laboratory, Argonne, Illinois \\
${}^{2}$Brandeis Universtiy, Waltham, Massachusetts \\
${}^{3}$Brookhaven National Laboratory, Upton, New York \\
${}^{4}$University of California, Riverside, California \\
${}^{5}$University of California, San Diego, California \\
${}^{6}$Carnegie Mellon University, Pittsburgh, Pennsylvania \\
${}^{7}$CERN, Geneva, Switzerland \\
${}^{8}$Deutsches Elektronen-Synchrotron (DESY), Hamburg, Germany \\
${}^{9}$Fermi National Accelerator Laboratory, Batavia, Illinois \\
${}^{10}$Glasgow University, Glasgow, United Kingdom \\
${}^{11}$Harvard University, Cambridge, Massachusetts \\
${}^{12}$University of Illinois, Urbana-Champaign, Illinois \\
${}^{13}$Imperial College, London, United Kingdom \\
${}^{14}$Indiana University, Bloomington, Indiana \\
${}^{15}$INFN, Genoa, Italy \\
${}^{16}$Institute for Theoretical and Experimental Physics, Moscow, Russia \\
${}^{17}$Institute for High Energy Physics, Protvino, Russia \\
${}^{18}$Iowa State University, Ames, Iowa \\
${}^{19}$The Johns Hopkins University, Baltimore, Maryland \\
${}^{20}$University of Kentucky, Lexington, Kentucky \\
${}^{21}$Lawrence Berkeley National Laboratory, Berkeley, California \\
${}^{22}$Lund University, Lund, Sweden \\
${}^{23}$Massachusetts Institute of Technology, Cambridge, Massachusetts \\
${}^{24}$INFN \& Universit\`a degli Studi di Milano, Milan, Italy \\
${}^{25}$University of Minnesota, Minneapolis, Minnesota \\
${}^{26}$Robert Morris College, Moon Township, Pennsylvania \\
${}^{27}$NIKHEF, Amsterdam, The Netherlands \\
${}^{28}$Northern Illinois University, DeKalb, Illinois \\
${}^{29}$University of Notre Dame, Notre Dame, Indiana \\
${}^{30}$The Ohio State University, Columbus, Ohio \\
${}^{31}$Oxford University, Oxford, United Kingdom \\
${}^{32}$University of Pennsylvania, Philadelphia, Pennsylvania \\
${}^{33}$INFN \& Universit\`a e Scuola Normale Superiore di Pisa, Pisa, Italy \\
${}^{34}$Max-Planck-Institut f\"ur Physik, Munich, Germany \\
${}^{35}$Stanford Linear Accelerator Center, Stanford, California \\
${}^{36}$State University of New York, Stony Brook, New York \\
${}^{37}$Syracuse University, Syracuse, New York \\
${}^{38}$Technion--Israel Institute of Technology, Haifa, Israel \\
${}^{39}$Texas Tech University, Lubbock, Texas \\
${}^{40}$University of Toronto, Toronto, Canada \\
${}^{41}$University of Tsukuba, Tsukuba, Japan  \\
${}^{42}$Vanderbilt University, Nashville, Tennessee \\
${}^{43}$Weizmann Institute of Science, Rehovot, Israel \\
${}^{44}$University of Wisconsin, Madison, Wisconsin \\
${}^{45}$Yale University, New Haven, Connecticut \\
${}^{46}$York University, Toronto, Canada
\end{center}

\chapter*{Preface}

This report presents the results of the workshop devoted to study {\it $B$
Physics at the Tevatron: Run~II and Beyond}.  Like other workshops on the
physics potential of Run~II of the Tevatron, held at Fermilab from 1998--2000,
this workshop brought together theorists from around the world and
experimenters from the CDF, D\O, and BTeV collaborations, and elsewhere.

There were two general meetings held at Fermilab: during September 23--25,
1999, and February 24--26, 2000.  The working groups also held additional
interim meetings to report on their progress and plan further work. The
resulting physics studies can be found in Chapters~6--9.  The other chapters in
this report provide theoretical background on $B$ decays (Chapter~1), common
experimental issues (Chapter~2), and brief descriptions of the CDF, D\O, and
BTeV detectors (Chapters~3--5).

Flavor physics and \CP~violation will be a major focus of experimental
high-energy physics in the coming decade. The preceding decade verified at a
high precision the gauge sector of the Standard Model through a wide range of
experimental tests.  While many extensions of the Standard Model contain new
sources of flavor and \CP~violation, these sectors of the theory are poorly
tested at present.  Precise tests of the flavor sector of the Standard Model
and the origin of \CP~violation will come from sometimes competitive and
sometimes complementary measurements at the Tevatron and at the $e^+e^-$ $B$
factories.  Chapter~10 summarizes the results of the workshops for the most
interesting processes.

This report represents the status of the field around the Summer of 2001.  Both
the state of the theory and the experimental possibilities continue to
advance.  The results presented here are thus not a final view of what the
experiments can achieve.  This report concentrates on aspects of $B$ physics
accessible mainly to hadron colliders, and it is hoped that it will prove a
ready and complete reference and aid new collaboration members and maybe others
interested in the field as well.

These workshops could not have been organized without the help of many people.
The organizers would like to especially thank the support of Mike Witherell and
the Fermilab Directorate; the help of Cynthia Sazama and Patti Poole of the
Fermilab Conference Office with the general meetings' organization; and Lois
Deringer and Laura Sedlacek in the Fermilab Theory Group for taking care of so
many things.


\chapter*{Workshop Structure}

The workshop formed four semi-autonomous working groups.  Each was led by two
theory conveners and a convener from each experiment. 

\begin{itemize}

\item
{\bf CP Violation} \\ 
Yossi Nir, Helen Quinn, Manfred Paulini (CDF), 
Rick Jesik (D\O), Tomasz Skwarnicki (BTeV) 

\item
{\bf Rare and Semileptonic Decays} \\ 
Aida El-Khadra, Mike Luke, Jonathan Lewis (CDF), 
Andrzej Zieminski (D\O), Ron Poling (BTeV) 

\item
{\bf Mixing and Lifetimes} \\ 
Ulrich Nierste, Mikhail Voloshin, Christoph Paus (CDF), 
Neal Cason (D\O), Harry Cheung (BTeV) 

\item
{\bf Production, Fragmentation, Spectroscopy} \\  
Eric Braaten, Keith Ellis, Eric Laenen, 
William Trischuk (CDF), Rick Van Kooten (D\O), Scott Menary (BTeV) 

\end{itemize} 
Chapters \mbox{6 -- 9} of this report present the work done in these groups.

The programs of the general meetings, including transparencies of all talks,
working group information, and other documentation for this workshop can be
found on the WorldWideWeb at \verb"http://www-theory.lbl.gov/Brun2/".

The workshop was organized by
Rick Jesik, 
Andreas Kronfeld,
Rob Kutschke,
Zoltan Ligeti,
Manfred Paulini, and
Barry Wicklund.

\OMIT{
In addition, Chapter 1 provides a theoretical introduction, Chapters 2 -- 5
give introductions to the detectors and simulation tools, and Chapter 10
highlights the main results.  Chapter 1 was convened by U.~Nierste, Z.~Ligeti,
and A.~Kronfeld; Chapter 2 was convened by R.~Kutschke, R.~Jesik, and
M.~Paulini; Chapter 3 was convened by M.~Paulini and B.~Wicklund; Chapter 4 was
convened by R.~Jesik; and Chapter 5 was convened by R.~Kutschke.
}

\tableofcontents
\listoffigures
\addcontentsline{toc}{chapter}{List of Figures}
\listoftables
\addcontentsline{toc}{chapter}{List of Tables}

\clearpage{\pagestyle{empty}\cleardoublepage}

\mainmatter

\chapter{Common Theoretical Issues}

\authors{U.~Nierste, Z.~Ligeti, A.~S.~Kronfeld}

\section{Introduction}
\label{ch1:sec:intro}

This chapter provides some of the theoretical background needed to
interpret measurements studied by Working Groups~1--3
(and reported in Chapters~6--8, respectively).
These three working groups deal with the decay of $b$-flavored hadrons,
and they are independent of the production mechanism.
The theory of $b$ decays requires
some elementary concepts on symmetries and mixing,
some knowledge of the standard electroweak theory, and
some information on how the $b$ quark is bound into hadrons.
On the other hand, the theory of production, fragmentation, and
spectroscopy---the subjects of Working Group~4 (and Chapter~9)---is
separate, dealing entirely with aspects of the strong interactions.
Hence, the theoretical background needed for Working Group~4 is
entirely in Chapter~9, and theoretical issues common to the
working groups studying decays are collected together in this chapter.%
\footnote{There is, unavoidably, some overlap with theoretical material
in Chapters~6--8.  We have attempted to use consistent conventions and
notation throughout.}
Although the experimental study of $CP$ violation in the $B$ system is just
beginning~\cite{CDFs2b,Bas2b,Bes2b}, there are several theoretical
reviews~\cite{babook,BLS,book,bsbook,cern} that the reader may want to consult
for details not covered here.

We start in Sec.~\ref{ch1:sec:cpvsm} by reviewing how flavor mixing
and \CP\ violation arise in the Standard Model.
Experiments of the past decade have verified the
$SU(3)\times SU(2)\times U(1)$ gauge structure of elementary particle
interactions, in a comprehensive and very precise way.
By comparison, tests of the flavor interactions are not yet nearly as
broad or detailed.
The Standard Model, in which quark masses and mixing arise from
Yukawa interactions with the Higgs field, still serves as the current
foundation for discussing flavor physics.
Sec.~\ref{ch1:sec:cpvsm} discusses the standard flavor sector,
leading to the Cabibbo-Kobayashi-Maskawa (CKM) matrix, which contains
a \CP\ violating parameter for three generations.
By construction, the CKM matrix is unitary, which implies several
relations among its entries and, hence, between \CP\ conserving and \CP\
violating observables.
Furthermore, the same construction shows how, in the Standard Model,
neutral currents conserve flavor at the tree level, which is known as
the Glashow-Iliopoulos-Maiani (GIM) effect.

We emphasize that in extensions of the Standard Model the CKM mechanisms
can exist side-by-side with other sources of \CP\ and flavor violation.
Many measurements are therefore needed to test whether the standard
patterns prevail.
Quark interactions are obscured by confinement, however.
Therefore, Sec.~\ref{ch1:sec:cpvsm} concludes with a brief summary of
the ways of avoiding or reducing uncertainties from nonperturbative
QCD, and the much of the rest of the chapter revisits various aspects
in greater detail.

Sec.~\ref{ch1:sect:gen} covers aspects of $B$ mesons that can be
discussed without reference to their underlying dynamics.
The strong interactions conserve the quantum numbers \p, \T, and \C\
of parity, time reversal, and charge conjugation.
We therefore start by discussing the transformation properties of
currents and hadrons under these discrete symmetry transformations.
Once these concepts---and associated phase conventions---are fixed,
one can discuss the mixing between neutral mesons.
(In the Standard Model, neutral meson mixing is induced through
one-loop effects.)
Although a flavored neutral meson and its anti-particle form a two-state
quantum mechanical system, the particles are not stable, so the two mass
eigenstates can have different decay widths in general.
Consequently, the physical description of decay during mixing contains
formulae that are not always simple and phase conventions that are not
always transparent.
Sec.~\ref{ch1:sect:gen} provides such a general set of formulae,
derived with a self-consistent set of conventions.

The general discussion of Sec.~\ref{ch1:sect:gen}, leads to a useful
classification of \CP\ violation.
There can be \CP\ violation in mixing, in decay, and in the interference
of decays with and without mixing.
Sec.~\ref{ch1:sec:aspects} gives the concrete mathematical definition
of these three types of \CP\ violation, illustrated with examples.
\CP\ violation in many $B$ decays is principally of one type or
another, although in general two or more of these types may be present,
as is the case with some kaon decays.
We also work through two important examples of \CP\ in the interference
of amplitudes: $B\to J/\psi K_S$ and $B_s\to D_s^\pm K^\mp$, where it is
possible to use the \CP\ invariance of QCD to show that the \CP\
asymmetry of these decays is independent of the hadronic transition
amplitude.

To gain a comprehensive view of flavor physics, and decide whether the
standard model correctly describes flavor-changing interactions, one
has to consider QCD.
Sec.~\ref{1:tools} discusses several theoretical tools to separate
scales, so that the nonperturbative hadronic physics can
be treated separately from physics at higher scales, where perturbation theory
is accurate.
Indeed, in $B$ decays several length scales are involved:
the scale of QCD, $\lqcd$, 
the mass of the $b$ quark, $m_b$,
the higher masses of the $W$ and $Z$ bosons and the top quark, and,
possibly, higher scales of new physics.
The first step is to separate out weak (and higher) scales with an
operator product expansion, leading to an effective weak Hamiltonian
for flavor-changing processes.
This formalism applies for all flavor physics.
For $b$ quarks, one finds further simplifications, because
$m_b\gg\lqcd$.
Two tools are used to separate these scales, heavy quark effective
theory for hadronic matrix elements, and the heavy quark expansion for
inclusive decay rates.
Sec.~\ref{1:tools} also includes a brief overview of lattice QCD,
which is a promising numerical method to compute hadronic matrix
elements of the electroweak Hamiltonian, when there is at most one hadron
in the final state.
In particular, we discuss how heavy quark effective theory can be used
to control uncertainties in the numerical calculation.

A further predictive aspect of the CKM mechanism, with only one
parameter to describe \CP\ violation, is that observables in the $B$
and $B_s$ systems are connected to those in the kaon system.
Sec.~\ref{1:sect:kaon} gives an overview of \kkm\ using the same
formalism as in our treatment of \bbm\ in Sec.~\ref{ch1:sect:gen}.
This also gives us the opportunity to introduce the most important
constraints from kaon physics: not only those currently available but
also those that could be measured in the coming decade.
Finally, Sec.~\ref{ch1:SM-expect} gives a summary of expectations for
measurements of the unitarity triangle, based on global fits of kaon
mixing and \CP\ conserving observables in $B$~physics.

\boldmath
\section{$CP$ Violation in the Standard Model}
\label{ch1:sec:cpvsm}
\unboldmath

As mentioned in the introduction, in the Standard Model quark masses,
flavor violation, and $CP$ violation all arise from Yukawa interactions
among the quark fields and the Higgs field.
In this section we review how these phenomena appear, leading to the
Cabibbo-Kobayashi-Maskawa (CKM) matrix.
  From a theoretical point of view, the CKM mechanism could, and probably
does, exist along with other sources of $CP$ violation.
We therefore also discuss some of the important features of the CKM
model, to provide a framework for testing it.

\subsection{Yukawa interactions and the CKM matrix}
\label{1:subsec:yukawa}

Let us begin by recalling some of the most elementary aspects of
particle physics.
Experiments have demonstrated that there are several species, or
flavors, of quarks and leptons.
They are the down-type quarks ($d$, $s$, $b$), up-type quarks
($u$, $c$, $t$), charged leptons ($e$, $\mu$, $\tau$),
and neutrinos ($\nu_e$, $\nu_\mu$, $\nu_\tau$).
They interact through the exchange of gauge bosons:
the weak bosons $W^\pm$ and $Z^0$, the photon, and the gluons.
These interactions are dictated by local gauge invariance, with gauge
group $SU(3)\times SU(2)\times U(1)_Y$.
With this gauge symmetry, and the observed quantum numbers of the
fermions, at least one scalar field is needed to accommodate quark
masses, and, in turns out, the couplings to this field can generate
flavor and $CP$ violation.

One of the most striking features of the charged-current weak 
interactions is that they do not couple solely to a vector current 
(as in QED and QCD) but to the linear combination of vector and axial 
vector currents $V-A$.
As a consequence, the electroweak theory is a \emph{chiral} gauge theory,
which means that left- and right-handed fermions transform differently
under the electroweak gauge group $SU(2)\times U(1)_Y$.
The right-handed fermions do not couple to $W^\pm$, and they are
singlets under $SU(2)$:
\begin{eqnarray}
    E_R &=& (e_R, \mu_R, \tau_R)\,,\qquad Y_E = -1\,; \nonumber\\
    U_R &=& (u_R,  c_R,  t_R)\,,   \qquad Y_U = \frac{2}{3}\,; \\
    D_R &=& (d_R,  s_R,  b_R)\,,   \qquad Y_D = -\frac{1}{3}\,;\nonumber
\end{eqnarray}
where the hypercharge~$Y$ is given.
For convenience below, the three generations are grouped together.
The gauge and kinetic interactions for $G$ generations of these
fields are
\begin{equation}
    {\cal L}_R = \sum_{i=1}^G
        \bar{E}^i_R(i\kern+0.1em  /\kern-0.55em \partial
           -  g_1Y_E \kern+0.1em  /\kern-0.65em B)E^i_R +
        \bar{D}^i_R(i\kern+0.1em  /\kern-0.65em D
           -  g_1Y_D \kern+0.1em  /\kern-0.65em B)D^i_R +
        \bar{U}^i_R(i\kern+0.15em /\kern-0.65em D
           -  g_1Y_U \kern+0.1em  /\kern-0.65em B)U^i_R \,,
\end{equation}
where $B$ is the gauge boson of $U(1)_Y$, with coupling~$g_1$, and
$D^\mu$ is the covariant derivative of QCD:
quarks are triplets under color $SU(3)$.
On the other hand, the left-handed fermions do couple to~$W^\pm$, so
they are doublets under $SU(2)$:
\begin{eqnarray}
L_L &=& \left(
    \pmatrix{\nu_e \cr e}_L,
    \pmatrix{\nu_\mu \cr \mu}_L,
    \pmatrix{\nu_\tau \cr \tau}_L\,
    \right), \qquad Y_L = -\frac{1}{2} \,; \nonumber\\
Q_L &=& \left(
    \pmatrix{u \cr d}_L,
    \pmatrix{c \cr s}_L,
    \pmatrix{t \cr b}_L\,
    \right), \qquad\quad\  Y_Q = \frac{1}{6}\,.
\end{eqnarray}
The $SU(2)$ quantum number is called weak isospin, and the third
component~$I_3$ distinguishes upper and lower entries.
The gauge and kinetic interactions for $G$ generations of these
fields are
\begin{equation}
    {\cal L}_L = \sum_{i=1}^G
        \bar{L}^i_L(i\kern+0.1em /\kern-0.55em \partial
           -  g_1Y_L\kern+0.1em /\kern-0.65em B
           -  g_2   \kern+0.1em /\kern-0.65em W)L^i_L +
        \bar{Q}^i_L(i\kern+0.1em /\kern-0.65em D
           -  g_1Y_Q\kern+0.1em /\kern-0.65em B
           -  g_2   \kern+0.1em /\kern-0.65em W)Q^i_L \,,
\end{equation}
where $W=W^a\sigma_a/2$ are the gauge bosons of $SU(2)$, with gauge
coupling~$g_2$.
Note that as far as gauge interactions are concerned, the generations 
are simply copies of each other, and ${\cal L}_R+{\cal L}_L$ possesses 
a large global flavor symmetry.
For $G$ generations, the symmetry group is $U(G)^5$, that is, a $U(G)$ 
symmetry for each of~$E_R$, $U_R$, $D_R$, $L_L$, and~$Q_L$.

The assignments of $SU(2)$ and $U(1)_Y$ quantum numbers follow from simple,
experimentally determined properties of weak decays.
For example, by the mid-1980s measurements of decays of
$b$-flavored hadrons had shown the weak isospin of~$b_L$
to be~$-\frac{1}{2}$~\cite{Langacker:1988ce}.
Consequently, its isopartner~$t_L$ had to exist, for symmetry reasons,
although several years passed before the top quark was observed at the
Tevatron.
In contrast, gauge symmetry does not motivate the inclusion of
right-handed neutrinos, which would be neutral under all three gauge
groups.
For this reason, they are usually omitted from the ``standard'' model.

With only gauge fields and fermions, the model is incomplete.
In particular, it does not accommodate the observed non-zero masses of
the quarks, charged leptons, and weak gauge bosons $W^\pm$ and~$Z^0$.
For example, masses for the charged fermions%
\footnote{Because it is completely neutral, a right-handed neutrino
may have a so-called Majorana mass term, coupling neutrino to neutrino,
instead of---or in addition to---a Dirac mass term, coupling neutrino
to anti-neutrino.  For this reason neutrino masses are even more
perplexing than quark and charged lepton masses.}
normally would come from interactions that couple the
left- and right-handed components of the field, such as
\begin{equation}
    {\cal L}_m = -m\left( \bar{\psi}_R \psi_L +
        \bar{\psi}_L \psi_R \right),
\end{equation}
where, in the case at hand, $\psi\in\{e,\mu,\tau,d,s,b,u,c,t\}$.
With the fields introduced above, one would have to combine a component
of a doublet with a singlet, which would violate $SU(2)$.
Any pairing of left- and right-handed fields with the listed
hypercharges would violate $U(1)_Y$ as well.

To construct gauge invariant interactions coupling left- and
right-handed fermions, at least one additional field is necessary.
For simplicity, let us begin with only the first generation leptons.
Consider
\begin{equation}
    {\cal L}_Y = - ye^{i\delta}\, \bar{l}_L\phi\, e_R -
        ye^{-i\delta}\, \bar{e}_R\,\phi^\dagger l_L,
    \label{1:eq:Yuk}
\end{equation}
where $\bar{l}_L=\left(\bar{\nu}_L,\ \bar{e}_L\right)$,
and $y$ is real.
If the quantum numbers of~$\phi$ are chosen suitably, then
the interaction~${\cal L}_I$ would be gauge (and Lorentz) invariant.
To preserve Lorentz invariance, $\phi$ must have spin~0.
To preserve invariance under $U(1)_Y$, the hypercharge
of~$\phi$ must be $Y_\phi=Y_L-Y_E=+\frac{1}{2}$.
To preserve invariance under $SU(2)$ $\phi$, must be a
doublet,
\begin{equation}
    \phi = \pmatrix{ \phi^+ \cr \phi^0 }.
\end{equation}
The superscripts denote the electric charge $Q=Y+I_3$.
An interaction similar to~${\cal L}_Y$ was first introduced by Yukawa
to describe the decay $\pi^+\to\mu^+\nu_\mu$, so it is called a Yukawa
interaction, and the coupling~$y$ is called a Yukawa coupling.

At first glance, the interaction in Eq.~(\ref{1:eq:Yuk}) appears to 
violate $CP$, with a strength proportional to $y\sin\delta$.
One may, however, remove $\delta$, by exploiting the invariance of 
${\cal L}_R+{\cal L}_L$ under independent changes in the phases 
of~$e_R$ and~$l_L$.
Thus, the one-generation Yukawa interaction has only one real 
parameter, $y$, and it conserves~$CP$.

Since it is charged under $SU(2) \times U(1)_Y$, the field~$\phi$
has gauge interactions, which are dictated by symmetry.
The scalar field may also have self-interactions, which are not dictated
by symmetry.
If one limits one's attention to renormalizable interactions
\begin{equation}
    V(\phi) = - \lambda v^2\phi^\dagger\phi +
        \lambda(\phi^\dagger\phi)^2,
\end{equation}
with two new parameters, $v$ and~$\lambda$.
The state with no propagating particles, called the vacuum, is realized 
when $\phi$ minimizes~$V(\phi)$.
The quartic coupling~$\lambda$ must be positive; otherwise the
potential energy would be unbounded from below, and the vacuum would
be unstable.
If $v^2<0$, then there is a single minimum of the potential, with 
vacuum expectation value $\langle\phi\rangle=0$; this possibility does 
not interest us here.
If $v^2>0$, then~$V(\phi)$ takes the shape of a sombrero with a 
three-dimensional family of minima:
\begin{equation}
    \langle\phi\rangle = e^{i\langle\xi^a\rangle\sigma_a/2v}
        \pmatrix{ 0 \cr v/\sqrt{2}},
    \label{1:eq:vev}
\end{equation}
parametrized by~$\langle\xi^a\rangle$.
Through an $x$-independent $SU(2)$ transformation, one can
set~$\langle\xi^a\rangle=0$.
Although the Lagrangian fully respects local $SU(2) \times U(1)_Y$
gauge symmetry, the vacuum solution of the equations of motion given
in Eq.~(\ref{1:eq:vev}) does not:
this is called spontaneous (as opposed to explicit) symmetry breaking.

Physical particles arise from fluctuations around the solution of
the equations of motion, so one writes
\begin{equation}
    \phi(x) = e^{i\xi^a(x)\sigma_a/2v}
        \pmatrix{ 0 \cr [v+h(x)]/\sqrt{2} }.
    \label{1:eq:phi}
\end{equation}
The vacuum expectation values of the fluctuation fields are
$\langle\xi^a\rangle=\langle h\rangle=0$.
Masses of the physical particles are found by inserting
Eq.~(\ref{1:eq:phi}) into the expressions for the interactions in
the Lagrangian and examining the quadratic terms.
By comparing the $\bar{e}_Re_L$ terms in ${\cal L}_Y$ and~${\cal L}_m$,
one sees that the electron mass in this model is $m_e=yv/\sqrt{2}$.
Similarly, from $V(\phi)$ the field $h$ is seen to have a (squared)
mass $m_h^2=2\lambda v^2$, and from the kinetic energy of the
scalar field non-zero masses for three of the gauge bosons arise:
$m_{W^\pm}^2 = \frac{1}{4}g_2^2v^2$ and
$m_{Z^0}^2   = \frac{1}{4}(g_1^2+g_2^2)v^2$,
where $Z^0$ is the massive linear combination of $W^3$ and $B$.
(The orthogonal combination is the massless photon~$\gamma$.)
The amplitude for muon decay is, to excellent approximation,
proportional to $g_2^2/m_W^2$.
Therefore, one can obtain the vacuum expectation value from the Fermi
decay constant, finding $v=246$~GeV.

Repeating this construction with $\left(\bar{u}_L,\ \bar{d}_L\right)$
and $d_R$ requires a doublet with hypercharge $Y_Q-Y_D=+\frac{1}{2}$.
The Standard Model uses the same doublet as for leptons.
Repeating it with $\left(\bar{u}_L,\ \bar{d}_L\right)$
and $u_R$ requires a doublet with hypercharge~$Y_Q-Y_U=-\frac{1}{2}$.
The Standard Model uses the charge-conjugate 
\begin{equation}
    \tilde{\phi} \equiv i\sigma_2\phi^* =
        \pmatrix{ \overline{\phi^0} \cr  -\phi^- }
\end{equation}
of the doublet used for leptons.
In the one-generation case, three real Yukawa couplings are 
introduced, leading to masses for the electron, down quark, and 
up~quark.

With $G$ generations the full set of Yukawa interactions is complicated.
It is instructive to leave $G$ arbitrary for now, and to compare the 
physics for $G=2$, 3, 4, later on.
The generations may interact with each other as in
\begin{equation}
    {\cal L}_Y = - \sum_{i,j=1}^G \left[
        \hat{y}^e_{ij} \bar{L}^i_L\phi\, E^j_R +
        \hat{y}^d_{ij} \bar{Q}^i_L\phi\, D^j_R +
        \hat{y}^u_{ij} \bar{Q}^i_L\tilde{\phi}\, U^j_R +
		{\rm h.c.}\right],
\end{equation}
because no symmetry would enforce a simpler structure.
For $G$ generations, the Yukawa matrices are complex $G\times G$
matrices.
At first glance, each matrix~$\hat{y}^a$ seems to introduce $2G^2$ 
parameters: $G^2$ that are real and $CP$-conserving, and another $G^2$ 
that are imaginary and $CP$ violating.
But, as in the one-generation case, one must think carefully about 
physically equivalent matrices before understanding how many physical
parameters there really are.

Let us consider the leptons first.
As mentioned above, the non-Yukawa part of the Lagrangian is invariant 
under the following transformation of generations
\begin{eqnarray}
    E_R \mapsto R E_R\,, & \qquad & \bar{E}_R \mapsto \bar{E}_R R^\dagger, 
    \nonumber\\
    L_L \mapsto S L_L\,, & \qquad & \bar{L}_L \mapsto \bar{L}_L S^\dagger,
    \label{1:eq:UGlepton}
\end{eqnarray}
where $R\in U(G)_{E_R}$ and $S\in U(G)_{L_L}$.
That means that the Yukawa matrix $\hat{y}^e$ is equivalent to 
${y^e=S\hat{y}^eR^\dagger}$.
By suitable choice, $y^e$ can be made diagonal, real, and
non-negative.
The leptons' Yukawa interactions now read
\begin{equation}
    {\cal L}_{Yl} = -\sum_{i=1}^G \left[
        y^e_{i} \bar{L}^i_L\phi\, E^i_R + {\rm h.c.} \right].
    \label{1:eq:Yl}
\end{equation}
Note that if $S$ and $R$ achieve this structure, so do $S'=DS$ and
$R'=DR$, where $D=\diag(e^{i\varphi_1},\ldots,e^{i\varphi_G})$.
Thus, part of the transformation from $\hat{y}^e$ to $y^e$ is redundant
and must not be counted twice.
(The freedom to choose these phases leads to global conservation of
lepton flavor.)
Hence, the transformation removes $2G^2-G$ parameters, leaving $G$
independent entries in~$y^e$.
Since all are real, there is no $CP$ violation.

For the quarks the reasoning is the same but the algebra is trickier.
There are now three distinct $U(G)$ symmetries, and
the non-Yukawa Lagrangian is invariant under 
\begin{eqnarray}
    D_R \mapsto R_d D_R\,, &\qquad& \bar{D}_R \mapsto \bar{D}_R R_d^\dagger\,, 
    \nonumber\\
    U_R \mapsto R_u U_R\,, &\qquad& \bar{U}_R \mapsto \bar{U}_R R_u^\dagger\,, 
     \\
    Q_L \mapsto S_u Q_L\,, &\qquad& \bar{Q}_L \mapsto \bar{Q}_L S_u^\dagger\,.
    \nonumber
    \label{1:eq:UGquark}
\end{eqnarray}
One may again exploit $S_u$ and $R_u$ to transform $\hat{y}^u$ into the
diagonal, real, non-negative form~$y^u$.
Then the transform of $\hat{y}^d$ is, in general, neither real nor diagonal.
Instead
\begin{equation}
    S_u \hat{y}^d R_d^\dagger = V y^d \,,
    \label{1:eq:VydR}
\end{equation}
where $y^d=S_d\hat{y}^dR_d^\dagger$ \emph{is} diagonal, real, and
non-negative, and
\begin{equation}
    V = S_uS_d^\dagger
    \label{1:eq:CKM}
\end{equation}
is the Cabibbo-Kobayashi-Maskawa (CKM)
matrix~\cite{Cabibbo:1963yz,Kobayashi:1973fv}.
By construction, $V$ is a $G\times G$ unitary matrix.
The quarks' Yukawa interactions now read
\begin{equation}
    {\cal L}_{Yq} = 
        - \sum_{i,j=1}^G \left[y^d_{j}\bar{Q}^i_L\phi\,  V_{ij}D^j_R +
        {\rm h.c.} \right]
        - \sum_{i=1}^G   \left[y^u_{i}\bar{Q}^i_L\tilde{\phi}\,U^i_R +
        {\rm h.c.} \right].
    \label{1:eq:Yq}
\end{equation}
If $S_u$, $R_u$, and $R_d$ achieve this structure, so do
$e^{i\varphi}S_u$, $e^{i\varphi}R_u$, and $e^{i\varphi}R_d$.
(The freedom to choose this phase leads to global conservation of
total baryon number.)
Thus, the manipulations remove $3G^2-1$ parameters from the $4G^2$ in 
two arbitrary $G\times G$ matrices, leaving $G^2+1$.
Of these, $2G$ are in $y^u$ and $y^d$, and the other 
$(G-1)^2$ are in the CKM matrix~$V$.
One can also count separately the real and imaginary parameters.
Since a $G\times G$ unitary matrix has $\frac{1}{2}G(G-1)$ real and
$\frac{1}{2}G(G+1)$ imaginary components, one finds that the CKM matrix
has $\frac{1}{2}G(G-1)$ real, $CP$-conserving parameters,
and $\frac{1}{2}(G-1)(G-2)$ imaginary, $CP$ violating parameters.
For example, the case $G=2$ has no $CP$ violation from this mechanism, 
$G=3$ has a single $CP$ violating parameter,%
\index{CP violation@\CP\ violation!Kobayashi-Maskawa mechanism}
and $G=4$ has three.

The CKM matrix $V$ arises from the misalignment of the matrices~$S_u$
and $S_d$.
Under circumstances that preserve some of the flavor symmetry, they can
be partially aligned, and then $V$ contains even fewer physical
parameters.
In an example with three generations, if two entries either in~$y^u$
or in~$y^d$ are equal, partial re-alignment removes one real angle
and one phase.
Therefore, the CKM mechanism leads to $CP$ violation only if
like-charged quarks all have distinct masses.

Substituting Eq.~(\ref{1:eq:phi}) into ${\cal L}_Y$ and keeping
quadratic terms shows that the masses are
\begin{equation}
    m^a_{k} = \frac{v}{\sqrt{2}}\, y^a_{k} \,,
\end{equation}
for $k=1, 2, 3$, and $a=e, d, u$.
For quarks this is easiest to see if one sets
\begin{equation}
    Q_L = \pmatrix{ U_L \cr VD_L },
    \label{1:eq:D}
\end{equation}
which diagonalizes the mass terms for the down-like quarks in 
Eq.~(\ref{1:eq:Yq}).
In this basis the CKM matrix migrates to the charged-current vertex:
\begin{equation}
    {\cal L}_{\bar{U}WD} = - \frac{g_2}{\sqrt{2}} \left[ 
        \bar{U}_L            \kern+0.1em /\kern-0.65em W^+ V D_L +
        \bar{D}_L V^\dagger \kern+0.1em /\kern-0.65em W^- U_L
        \right],
    \label{1:eq:UWD}
\end{equation}
where $W^\pm=(W^1\mp iW^2)/\sqrt{2}$.
The basis in \eq{1:eq:D}, with diagonal mass matrices and the CKM matrix
in the charged currents of quarks, is usually adopted in phenomenology.

Note that the neutral current interactions are unaffected by writing
$V$ in \eq{1:eq:D}.
Thus, there are no flavor-changing neutral currents (FCNC) at the tree
level in the Standard Model.
This is known as the Glashow-Iliopoulus-Maiani (GIM) effect~\cite{gim}.
Even at the loop level, where the FCNCs do arise, the GIM mechanism
can suppress processes by a factor $m_q^2/m_W^2$, which is very small,
except in the case of the top quark.
GIM suppression and Cabibbo suppression (i.e., factors of $\lambda$)
both imply near-null predictions for several processes.
Observation of any of these would constitute a clear signal of
non-standard physics.

Note that quark and lepton masses arise from the same microscopic
interactions as CKM flavor violation.
In Nature, the quark and lepton masses vary over orders of magnitude.
Thus, the large flavor symmetry that would arise in the absence of
Yukawa interactions is severely broken.
In the Standard Model, the Yukawa couplings are simply chosen to
contrive the observed masses.
This is unsatisfactory, but we lack the detailed experimental
information needed to develop a deeper theory of flavor.

\subsection{General models}
\label{1:subsec:general}

The foregoing discussion makes clear that the unitary CKM matrix
arises in an algebraic way.
Therefore, the mechanism can survive in models with a more complicated
Higgs sector.
The Standard Model is a model of economy: a single doublet generates
mass for the gauge bosons, charged leptons, down-like quarks, and
up-like quarks.
In models with two doublets (and three or more generations), the CKM
source of $CP$ violations remains, but there can be additional $CP$
violation in the Higgs sector~\cite{Gunion:1989we}.

To emphasize this point, let us consider an extreme example with
\begin{equation}
    {\cal L}_Y =
        \bar{L}^i_L\Phi^e_{ij} E^j_R +
        \bar{Q}^i_L\Phi^d_{ij} D^j_R +
        \bar{Q}^i_L\tilde{\Phi}^u_{ij} U^j_R + {\rm h.c.},
    \label{1:eq:ugly}
\end{equation}
where $i,j$ run over generations, and we take the basis in which
gauge interactions do not change generations.
The tilde on $\tilde{\Phi}_u$ is introduced so that, with
$\tilde{\Phi}^u=i\sigma_2{\Phi^u}^*$, all $\Phi^a$ are (matrix) 
fields with hypercharge~$+\frac{1}{2}$.
Here the Yukawa couplings are absorbed into the
fields---Eq.~(\ref{1:eq:ugly}) is hideous enough as it is.
Since all~$\Phi^a$ are doublets under $SU(2)$, they all would participate 
in electroweak symmetry breaking.

Suppose the Higgs potential, now a complicated function of all the 
scalar fields, leads to vacuum expectation values of the form
\begin{equation}
    \langle\Phi^a_{ij}\rangle = \pmatrix{ 0 \cr \hat{m}^a_{ij} }.
\end{equation}
Then the $\hat{m}^a_{ij}$ are mass matrices, and the algebra leading to
the real, physical masses~$m^a_k$ and the CKM matrix is just as above.
The CKM matrix, $V$, survives and should lead to $CP$ violation, because
there is no good reason for the phase in~$V$ to be small.
There would, however, almost certainly be new sources of $CP$ 
violation from the Higgs sector.

\boldmath
\subsection{$CP$ violation from a unitary CKM matrix}
\label{ch1:sect:ut}
\unboldmath

In the standard, one-doublet, model, we see that flavor and $CP$
violation arise solely through the CKM matrix.
Furthermore, in more general settings, the CKM matrix can still
arise, but there may be other sources of $CP$ violation as well.
If the CKM matrix is the only source of $CP$ violation, there are
many relations between $CP$-conserving and $CP$ violating observables
that arise from the fact that $V$ is a unitary matrix.
This section outlines a framework for testing whether these
constraints are, in fact, realized.

A useful way~\cite{Jarlskog:1985ht} of gauging the size of $CP$
violation starts with the commutator of the mass matrices,
$C=[\hat{m}^u\hat{m}^{u\dagger}, \hat{m}^d\hat{m}^{d\dagger}]$, which
can be re-written
\begin{equation}
    C = S_u^\dagger \left[(m^u)^2,\, V(m^d)^2\, V^\dagger\right] S_u\,,
\end{equation}
to show that $\det C$ depends on the physical masses and~$V$.
After some algebra one finds
\begin{equation}
    \det C = - 2i F_u F_d J\,,
    \label{1:eq:CJ}
\end{equation}
where
\begin{eqnarray}
    F_u & = & (m_u^2-m_c^2) (m_c^2-m_t^2) (m_t^2-m_u^2)\,, \\
    F_d & = & (m_d^2-m_s^2) (m_s^2-m_b^2) (m_b^2-m_d^2)\,, \\
     J  & = & \imag[V_{11}V^*_{21}\,V_{22}V^*_{12}]\,.
\end{eqnarray}
To arrive at Eq.~(\ref{1:eq:CJ}) one makes repeated use of the
property $VV^\dagger=\openone$, especially that
\begin{equation}
     J = \imag[V_{ij}V^*_{kj}\,V_{kl}V^*_{il}]\,
        \sum_m \varepsilon_{ikm} \sum_n \varepsilon_{jln}\,,
\end{equation}
for all combinations of $i$, $j$, $k$, and~$l$.
The determinant $\det C$ captures several essential features of $CP$
violation from the CKM mechanism.
It is imaginary, reminding us that $CP$ violation stems from a complex
coupling.
More significantly, there is no $CP$ violation unless $F_u$, $F_d$,
and~$J$ are all different from zero.
Non-vanishing $F_u$ and $F_d$ codify the requirements on the quark
masses given above.
Non-vanishing~$J$ codifies requirements on~$V$, which are
clearest after choosing a specific parameterization.
The key point, however, is that the value taken by $J$ is independent of
the parameterization, by construction of $\det C$.

To emphasize the physical transitions associated with the CKM matrix,
it is usually written
\begin{equation}
    V = \left( \begin{array}{ccc}
                V_{ud} & V_{us} & V_{ub} \\
                V_{cd} & V_{cs} & V_{cb} \\
                V_{td} & V_{ts} & V_{tb} 
        \end{array} \right),
\end{equation}
so that the entries are labeled by the quark flavors.
  From Eq.~(\ref{1:eq:UWD}),
the vertex at which a $b$ quark decays to a $W^-$ and $c$ quark is 
proportional to $V_{cb}$; similarly, the vertex at which a $c$ quark 
decays to a $W^+$ and $s$ quark is proportional to $V_{cs}^*$.
Because $V$ is unitary, $|V_{ud}|^2 + |V_{us}|^2 + |V_{ub}|^2 =1$, and
similarly for all other rows and columns.
These constraints give information on unmeasured (or poorly measured)
elements of~$V$.
For example, because $|V_{cb}|$ and $|V_{ub}|$ are known to be small,
$|V_{tb}|$ should be very close to 1---if, indeed, there are only
three generations.
Furthermore, $|V_{ts}|$ and $|V_{td}|$ must also be small.

\begin{figure}[bt]
\centerline{
    \epsfysize 4cm
	\epsffile{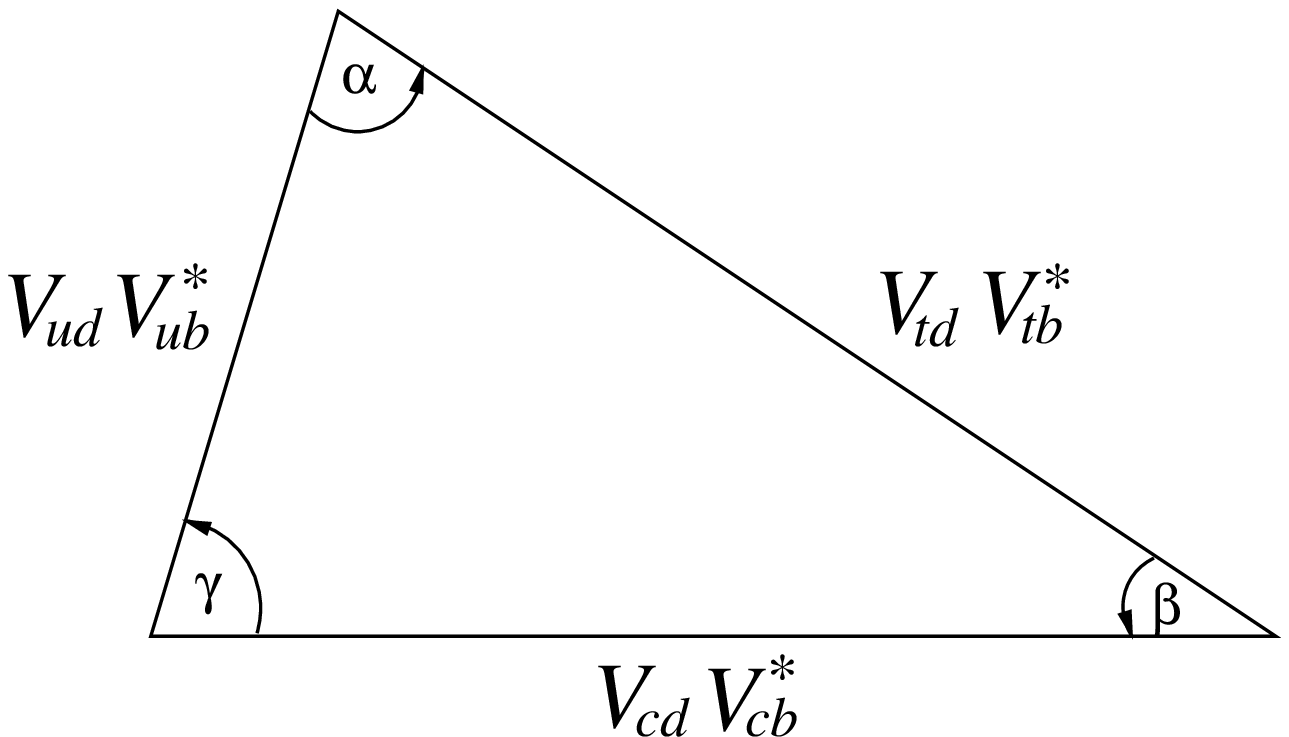}
	\hfill
    \epsfysize 4cm
	\epsffile{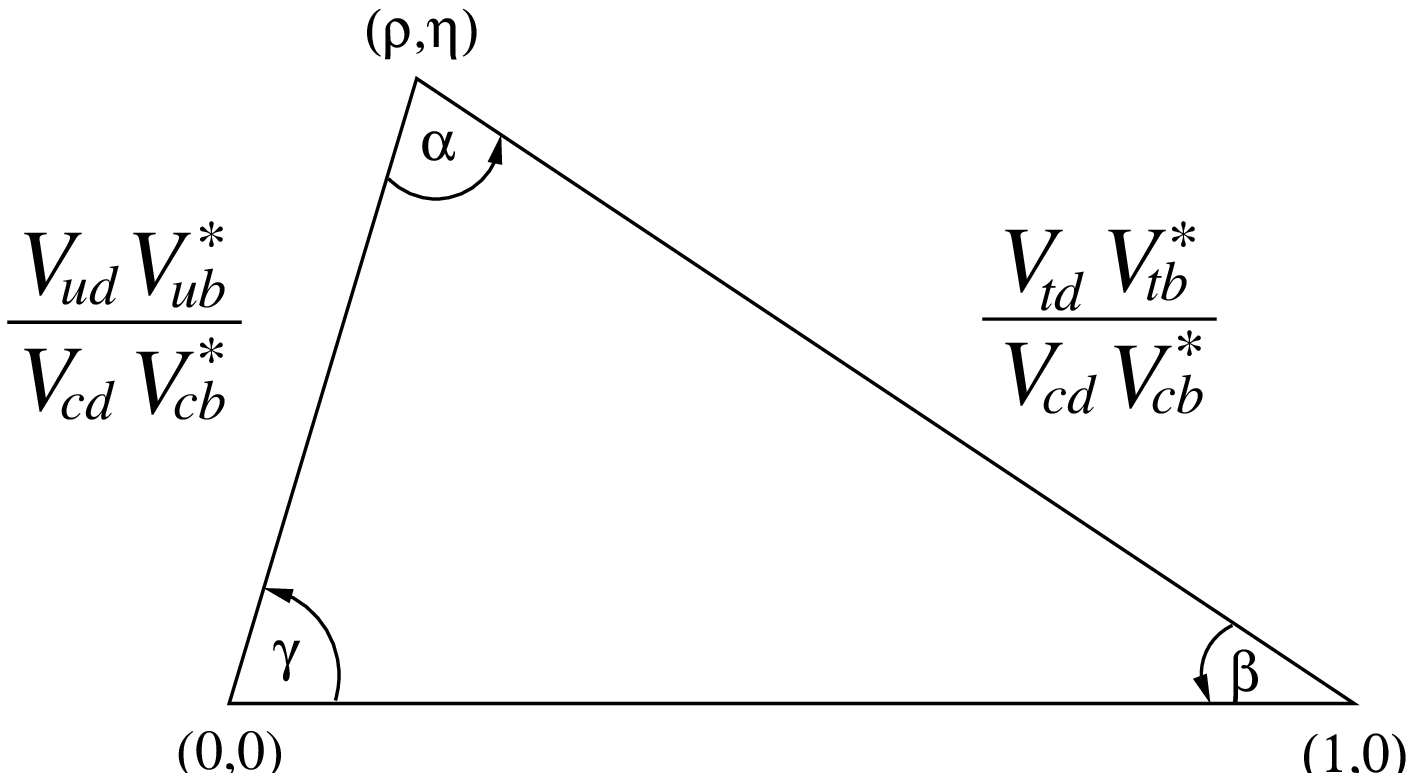}
	}
\vskip 6pt
\caption[The unitarity triangle]{The unitarity triangle.
The version on the left directly expresses Eq.~(\ref{1:eq:UT}).
The rescaled version shows the definition of $(\bar{\rho},\bar{\eta})$.}
\label{1:fig:UT}
\end{figure}

Even more interesting constraints come from the orthogonality of
columns (or rows) of a unitary matrix.
Taking the first and third columns of~$V$, one has
\begin{equation}
     V_{ud}V^*_{ub} + V_{cd}V^*_{cb} + V_{td}V^*_{tb} = 0\,.
     \label{1:eq:UT}
\end{equation}
Equation~(\ref{1:eq:UT}) says that the three terms in the sum
trace out a triangle on the complex plane.
Because it is a consequence of the unitarity property of~$V$, this
triangle is called the ``unitarity triangle,''
shown in Fig.~\ref{1:fig:UT}.
The lengths of the sides are simply $|V_{ud}V^*_{ub}|$, etc.,
and the angles are
\begin{equation}
    \alpha = \arg\left[ - \frac{V_{td}V_{tb}^*}{V_{ud}V_{ub}^*}\right],
    \qquad
    \beta  = \arg\left[ - \frac{V_{cd}V_{cb}^*}{V_{td}V_{tb}^*}\right],
    \qquad
    \gamma = \arg\left[ - \frac{V_{ud}V_{ub}^*}{V_{cd}V_{cb}^*}\right].
    \label{1:eq:beta}
\end{equation}
The notation $\beta \equiv \phi_1$, $\alpha \equiv \phi_2$,
$\gamma \equiv \phi_3$ is also used.
By construction $\alpha+\beta+\gamma=\pi$.
The area of the triangle is $|J|/2$ and the terms trace out the triangle
in a counter-clockwise (clockwise) sense if $J$ is positive (negative).
In fact, there are five more unitarity triangles, all with area $|J|/2$
and orientation linked to the sign of~$J$.

The unitarity triangle(s) are useful because they provide a simple,
vivid summary of the CKM mechanism.
Separate measurements of lengths, through decay and mixing rates,
and angles, through $CP$ asymmetries, should fit together.
Furthermore, when one combines measurements---from the $B$, $B_s$, $K$,
and $D$ systems, as well as from hadronic $W$ decays---all triangles
should have the same area and orientation.
If there are non-CKM contributions to flavor or $CP$ violation,
however, the interpretation of rates and asymmetries as measurements
of the sides and angles no longer holds.
The triangle built from experimentally defined sides and angles will
not fit with the CKM picture.

In the parameterization favored by the Particle Data Book~\cite{pdg}
\begin{equation}
    V = \left( \begin{array}{ccc}
                c_{12}c_{13} & s_{12}c_{13} & s_{13}e^{-i\delta_{13}} \\
                -s_{12}c_{23}-c_{12}s_{23}s_{13}e^{i\delta_{13}} &
                 c_{12}c_{23}-s_{12}s_{23}s_{13}e^{i\delta_{13}} &
                s_{23}c_{13} \\
                 s_{12}s_{23}-c_{12}c_{23}s_{13}e^{i\delta_{13}} &
                -c_{12}s_{23}-s_{12}c_{23}s_{13}e^{i\delta_{13}} &
                c_{23}c_{13}
        \end{array} \right),
    \label{1:eq:VPDG}
\end{equation}
where $c_{ij}=\cos\theta_{ij}$ and $s_{ij}=\sin\theta_{ij}$.
The real angles~$\theta_{ij}$ may be chosen so that
$0\le\theta_{ij}\le\pi/2$, and the phase $\delta_{13}$ so that
$0\le\delta_{13}<2\pi$.
With Eq.~(\ref{1:eq:VPDG}) the Jarlskog invariant becomes
\begin{equation}
    J = c_{12} c_{23} c_{13}^2 s_{12} s_{23} s_{13} \sin\delta_{13}\,.
\end{equation}
The parameters must satisfy
\begin{equation}
    \delta_{13}\neq 0,\;\pi\,; \qquad \theta_{ij}\neq 0,\;\pi/2\,;
\end{equation}
otherwise $J$ vanishes.
Since $CP$ violation is proportional to $J$, 
the CKM matrix must not only have complex entries, but also 
non-trivial mixing;
otherwise the KM phase~$\delta_{13}$ can be removed.

A convenient parameterization of the CKM matrix is due to
Wolfenstein~\cite{wolf}.
It stems from the observation that the measured matrix obeys a
hierarchy, with diagonal elements close to~1, and progressively
smaller elements away from the diagonal.
This hierarchy can be formalized by defining
$\lambda$, $A$, $\rho$, and $\eta$ via
\begin{equation}
    \lambda \equiv s_{12}\,, \qquad 
    A \equiv s_{23}/\lambda^2\,, \qquad
    \rho+i\eta \equiv s_{13}e^{i\delta_{13}}/A\lambda^3\,.
\end{equation}
  From experiment $\lambda\approx 0.22$, $A\approx 0.8$,
and $\sqrt{\rho^2+\eta^2}\approx 0.4$, so it is phenomenologically
useful to expand $V$ in powers of~$\lambda$:
\begin{equation}
    V = \left( \begin{array}{ccc}
                1-\frac{1}{2}\lambda^2 & \lambda & A\lambda^3(\rho-i\eta) \\
                -\lambda & 1-\frac{1}{2}\lambda^2 & A\lambda^2 \\
                 A\lambda^3(1-\rho-i\eta) & -A\lambda^2 & 1
        \end{array} \right) + {\cal O}(\lambda^4)\,.
\end{equation}
The most interesting correction at ${\cal O}(\lambda^4)$ for our purposes is
$\imag V_{ts}=-A\lambda^4\eta$.
The Jarlskog invariant can now be expressed $J = A^2\lambda^6\eta \approx
(7\times10^{-5})\eta$. One sees that CKM $CP$ violation is small not because
$\delta_{13}$ is small but because flavor violation must also occur, and flavor
violation is suppressed, empirically, by powers of~$\lambda$.

The unitarity triangle in Eq.~(\ref{1:eq:UT}) is special, because its
three sides are all of order $A\lambda^3$.
The triangle formed from the orthogonality of the first and third
rows also has this property, but it is not accessible, because
the top quark decays before the mesons needed to measure the angles are
bound.
The other triangles are all long and thin, with sides
$(\lambda,  \lambda,  A\lambda^5)$ (e.g., for the kaon) or
$(\lambda^2,\lambda^2,A\lambda^4)$ (e.g., for the $B_s$ meson).

It is customary to rescale Eq.~(\ref{1:eq:UT}) by the common factor
$A\lambda^3$, to focus on the less well-determined parameters
$(\rho,\eta)$.
In the context of the Wolfenstein parameterization, there are many ways
to do this.
Since we anticipate precision in experimental measurements, and also in
theoretical calculations of some important hadronic transition
amplitudes, it is useful to choose an exact rescaling.
We choose to divide all three terms in Eq.~(\ref{1:eq:UT}) by
$V_{cd}V_{cb}^*$ and define%
\footnote{This definition differs at ${\cal O}(\lambda^4)$ from the original
one in Ref.~\cite{blo}.}
\begin{equation}
	\bar{\rho} + i \bar{\eta} \equiv
		- \frac{ V_{ud} V_{ub}^* }{ V_{cd} V_{cb}^* }.
\end{equation}
Then the rescaled triangle, also shown in Fig.~\ref{1:fig:UT}, has its
apex in the complex plane at $(\bar{\rho},\bar{\eta})$.
The angles of the triangle are easily expressed
\begin{equation}
    \alpha = \tan^{-1}\left(
		\frac{\bar{\eta}}{\bar{\eta}^2+\bar{\rho}(\bar{\rho}-1)}
		\right)\,, \quad
    \beta  = \tan^{-1}\left(\frac{\bar{\eta}}{1-\bar{\rho}}\right)\,, \quad
    \gamma = \tan^{-1}\left(\frac{\bar{\eta}}{\bar{\rho}}\right)\,,
    \label{1:eq:betarhoeta}
\end{equation}
Since $\bar{\eta}$, $\bar{\rho}$, and $1-\bar{\rho}$ could easily be of
comparable size, the angles and, thus, the corresponding $CP$
asymmetries could be large.

At the Tevatron there is also copious production of $B_s$~mesons.
The corresponding unitarity triangle is
\begin{equation}
     V_{us}V^*_{ub} + V_{cs}V^*_{cb} + V_{ts}V^*_{tb} = 0 \,,
     \label{1:eq:UTBs}
\end{equation}
replacing the $d$~quark with~$s$.
In Eq.~(\ref{1:eq:UTBs}) the first side is much shorter than the other
two.
Therefore, the opposing angle
\begin{equation}
    \beta_s  = \arg\left[ - \frac{V_{ts}V_{tb}^*}{V_{cs}V_{cb}^*}\right]
        = \lambda^2 \eta + {\cal O}(\lambda^4)
    \label{1:eq:betasdef}
\end{equation}
is small, of order one degree.
Therefore, the asymmetries in $B_s\to \psi\eta^{(\prime)}$
and $B_s\to \psi\phi$ are much smaller than in the corresponding $B$
decays.
On the other hand, this asymmetry is sensitive to new physics
in $B_s^0 - \B0bar_s$ mixing.
In the standard model, as discussed below, mixing is induced by
loop processes.
When, as here, there is also Cabibbo suppression, it is easy for the
non-standard phenomena to compete.
Thus, in the short term a measurement of $\beta_s$ represents a search
for new physics, whereas in the long term it would be a verification of
the CKM picture.

The unitarity triangle for the $D$ system comes from the 
orthogonality of the top two rows of the CKM matrix.
It is even longer and thinner than the one for the $B_s$ system.
Consequently, a non-zero measurement of the $CP$ asymmetry associated with
the small angle is a clear sign of new physics.
It seems that experiments to measure the $D$-system unitarity triangle 
are not yet feasible.

\subsection{Hadronic uncertainties and clean measurements}
\label{1:subsec:clean}

At a superficial level, the way to test the CKM picture is to measure
rates and asymmetries that are sensitive to the sides and angles
(of all triangles), in as many ways as possible.
A~serious obstacle, however, is that the quarks are confined in
hadrons.
Consequently, most relations between experimental observables and
Lagrangian-level couplings, like the CKM matrix, involve hadronic matrix
elements.
In this subsection, we summarize briefly how to treat the matrix
elements, with an eye toward identifying processes that are relatively
free of hadronic uncertainty.

There are several approaches for treating the hadronic matrix elements,
none of which is universally useful.
In Sec.~\ref{1:tools} we introduce some of the essential tools in 
greater detail.
Here we illustrate the possibilities with examples.
\begin{itemize}
    \item Perfect (or essentially perfect) symmetries of QCD, such as $C$
        or $CP$: When a single CKM factor dominates a process, the QCD
        part of the amplitude cancels in ratios such as the $CP$
        asymmetry in interference between decays with and without
        mixing. The two best examples are in the asymmetries for%
\footnote{Here, and in the rest of this chapter, $\psi$ stands for any
charmonium state, $J/\psi$, $\psi'$, $\chi_c$, etc.}
	$B\to \psi K_S$ and $B_s\to D_s^\pm K^\mp$.
	Assuming that $CP$ violation comes only from the CKM matrix,
	the first class of modes cleanly yields~$\sin 2\beta$, and
	the other pair of modes cleanly yields~$\sin(\gamma-2\beta_s)$.
    \item Approximate symmetries, such as isospin, flavor $SU(3)$, chiral
	symmetry, or heavy quark symmetry: The best-known examples are when the
	symmetry restricts a form factor for semileptonic decays. Isospin and
	$n\to pe\bar{\nu}_e$ give $|V_{ud}|$; flavor $SU(3)$ and $K\to\pi
	e\bar{\nu}_e$ give $|V_{us}|$; and heavy quark symmetry and $B\to
	D^*\ell\bar\nu_\ell$ give $|V_{cb}|$. The hadronic uncertainty is now
	in the deviation from the symmetry limit. An even more intriguing use
	of isospin is to relate the form factor of $K^0\to\pi^+e\bar{\nu}_e$ to
	that of $K^{0,\pm}\to\pi^{0,\pm} \nu\bar\nu$. The rare $\nu\bar\nu$
	decays are, thus, essentially free of hadronic uncertainties.%
\footnote{In the charged mode there is an uncertainty stemming from the 
uncertainty in the charmed quark mass. It has been estimated to be around 5\%
in $|V_{ts}V_{td}|$~\cite{Buchalla:1996fp}.  Power suppressed corrections to
$K\to \pi\nu\bar\nu$ have also been estimated and found to be
small~\cite{mcsup}.}
    \item Lattice QCD:
        This computational method is sound, in principle, for hadronic
        matrix elements with at most one final-state hadron.
        Limitations in computer power have led to an approximation,
        called the quenched approximation, whose error is difficult
        to quantify.
        With increases in computer resources, lattice results should, in
        the future, play a more important role in determining the sides 
        of the unitarity triangles.
        For more details, see Sec.\ref{1:lattice}.
    \item Perturbative QCD for exclusive processes:
        It may be possible to calculate the strong phases of
		certain\index{strong phase shifts!from perturbative QCD}
        nonleptonic $B$ decays using perturbative QCD.
        This is in some ways analogous to computing cross sections
        in hadronic collisions, and the nonperturbative information
        is captured in light-cone distribution
        amplitudes~\cite{Lepage:1980fj}.
        There are, at present, two different
        approaches~\cite{Chang:1997dw,Beneke:1999br},
        whose practical relevance remains an open question.
    \item QCD sum rules:
        Like lattice and perturbative QCD, sum rules are based on QCD
        and field theory.
        Uncertainty estimates are usually semi-quantitative and it is
        difficult to reduce them in a controlled manner.
    \item Models of QCD, such as quark models, naive factorization, etc:
        These techniques can be applied for back-of-the-envelope
        estimates.
        There is no prospect for providing a quotable error and, thus,
        should not be used in quantitative work.
\end{itemize}
In summary, at the present time the cleanest observables are the
$CP$ asymmetries in $\B0bar_d$ decays to charmonium+kaon and 
in $\B0bar_s$ decays to $D_s^\pm K^\mp$.
The rare decay $K_L\to\pi^0\nu\bar\nu$ is free of theoretical 
uncertainties at a similar level, but presents a big experimental 
challenge.
Semileptonic decays restricted by symmetries as well as 
$K^\pm\to\pi^\pm\nu\bar\nu$ are a step down, but still good.
With enough computer power to overcome the quenched approximation,
lattice QCD could yield, during the course of Run~II, controlled
uncertainty estimates for neutral-meson mixing and leptonic and
semileptonic decays of a few percent.

\boldmath
\section{General Formalism for Mixing and $CP$ Violation}
\label{ch1:sect:gen} 
\unboldmath

This section is devoted to the general formalism for meson mixing and 
$CP$ violation. Much of the material can also be found in other
review articles and reports \cite{babook,BLS,cern}, but some
topics require a different viewpoint in the light of the $B$ physics
program at a hadron collider: unlike the $B$ factories the Tevatron
will be able to study $B_s^0$ mesons. The two mass eigenstates in the
$B_s^0$ system may involve a sizable width difference \dg, which must
be included in the formulae for the $B_s^0$ time evolution. We
consequently present these formulae including all effects from a
non-vanishing \dg.  

Many details of the formalism depend on conventions, particularly in 
the choice of the complex
phases that unavoidably appear in any \CP\ violating physical system.
We would like to discourage the reader from combining formulae from
different sources, so we try to give a comprehensive and self-contained
presentation of the subject. We start by introducing the discrete
transformation \C, \p\ and \T\ in Sec.~\ref{subs:disc}. 
Experimentally we know that \C, \p\ and \T\ are
symmetries of the electromagnetic and strong interactions, 
so the corresponding quantum numbers can be used to classify the
hadron states.
The description in Sec.~\ref{subs:mix} of the time evolution of the 
neutral $B$ meson system is
applicable to both the $B_d^0$ and the $B_s^0$ meson systems.
Sec.~\ref{subs:unt} deals with untagged $B^0$ decays and
Sec.~\ref{subs:cp} presents the formulae for \CP\ asymmetries.
Finally, in Sec.~\ref{subs:cp} we discuss phase conventions and
rephasing invariant quantities.

\subsection{Discrete transformations}
\label{subs:disc}

\index{parity} \index{time reversal} \index{charge conjugation}
In this section we introduce the parity, \p, time reversal, \T, 
and charge conjugation, \C, transformations. \p\ and
\T\ are defined through their action on coordinate vectors
$x=(x^0, x^1, x^2, x^3)$: \p\ flips the sign of the spatial
coordinates $x^1$,$x^2$,$x^3$ and \T\ changes the time component
$t=x^0$ into $-t$. Adopting the convention $g_{\mu \nu} = \mbox{diag}
(1,-1,-1,-1) $ for the Lorentz metric, one can compactly express the
transformations in terms of $x_{\mu} = g_{\mu \nu} x^{\nu}$:
\bey
\p: && \quad x^{\mu} \to x_{\mu} \,, \nn 
\T: && \quad x^{\mu} \to -x_{\mu} \,. \label{defpt} 
\eey
The definition \eq{defpt} implies that the derivative operator
$\partial^\mu=\partial/\partial x_\mu$
and the momentum $p^\mu$ transform under $P$ and $T$ 
in the same way as $x^{\mu}$. 
Finally, \C\ interchanges particles and anti-particles.
Apart from the weak interactions, these transformations are symmetries
of the Standard Model.
It is therefore convenient to classify hadronic states by their \C, 
\p\ and \T\ quantum numbers, which are multiplicative and take the 
values~$\pm1$.


The Lagrangian of the Standard Model and its possible extensions
contain bilinear currents of the quark fields, to which gauge bosons and
scalar fields couple. 
For example, as discussed in Sec.~\ref{ch1:sec:cpvsm}, 
the $W$ boson field $W^{\mu}$ couples
to the chiral vector current $\ov{b}_L \gamma_{\mu} c_L$.
Quark bilinears also appear in composite operators which represent the 
Standard Model interactions in low energy effective Hamiltonians, 
cf., Sec.~\ref{1:Heff}.
To understand how these interactions work, it is helpful to list the
transformation of the quark bilinears under \C, \p, \T\ and the
combined transformations \CP\ and \CPT. For illustration we specify to
currents involving a $b$ and a $d$ field.  The generic transformation
under some discrete symmetry $\mathit{X}$ is
\beq
\mathit{X}: \qquad \ov{b}\, \Gamma d \;\; \to \;\;  
    \mathit{X} \, \ov{b}\, \Gamma d \, \mathit{X}^{-1}, 
\eeq
and \tab{cpt} lists the transformation for the chiral scalar, vector
and magnetic
currents.\index{CP transformation@\CP\ transformation!quark currents}
\begin{table}[tp]
\begin{center}
$ \begin{array}{r@{\quad}|@{\quad}l@{\quad}l@{\quad}l}
\hline\hline
\mbox{~~current} &\ds  \ov{b}_{R}\, d_{L} \,(x^{\rho}) &\ds  
    \phantom{-}  \ov{b}_{L}\, \gamma_{\mu} \, d_{L}\,  (x^{\rho}) &\ds   
    \phantom{-} \ov{b}_{R}\, \sigma_{\mu \nu} d_{L}\,  (x^{\rho}) \\ \hline 
\C\ &\ds   \ov{d}_{R}\, b_{L} \, (x^{\rho}) \, \eta_C &\ds  
      - \ov{d}_{R} \, \gamma_{\mu}\,  b_{R}  (x^{\rho}) \, \eta_C &\ds  
     - \ov{d}_{R}\, \sigma_{\mu \nu}\, b_{L}  (x^{\rho}) \\
\p\ &\ds    \ov{b}_{L}\, d_{R} \, (x_{\rho}) \, \eta_P &\ds  
   \phantom{-}\ov{b}_{R}\, \gamma^{\mu} \, d_{R}\,  (x_{\rho}) \, \eta_P &\ds  
   \phantom{-}\ov{b}_{L}\, \sigma^{\mu \nu}\, d_{R}\,  (x_{\rho})\, \eta_P \\
\CP\ &\ds   \ov{d}_{L}\, b_{R}\, (x_{\rho})\, \eta_C  \eta_P &\ds  
      - \ov{d}_{L}\, \gamma^{\mu}\,  b_{L}\,  (x_{\rho})\,  \eta_C \eta_P&\ds  
     - \ov{d}_{L}\, \sigma^{\mu \nu}\, b_{R}\,  (x_{\rho})\, \eta_C \eta_P \\
\T\ &\ds    \ov{b}_{R}\, d_{L}\, (-x_{\rho})\, \eta_T &\ds  
  \phantom{-}\ov{b}_{L}\, \gamma^{\mu}\, d_{L}\,  (-x_{\rho}) \,  \eta_T &\ds  
  \phantom{-}\ov{b}_{R}\, \sigma^{\mu \nu}\, d_{L}\,  (-x_{\rho})\, \eta_T \\
\CPT\ &\ds  
\ov{d}_{L}\, b_{R}\, (-x^{\rho}) \, \eta_C  \eta_P \eta_T &\ds   
 - \ov{d}_{L}\, \gamma_{\mu}\,b_{L}\,(-x^{\rho})\, \eta_C  \eta_P \eta_T 
&\ds 
\phantom{-}\ov{d}_{L}\, \sigma_{\mu \nu}\, b_{R}\,  (-x^{\rho}) 
    \, \eta_C  \eta_P \eta_T \\ \hline\hline
\end{array} $ \\[6pt]
\caption[\C, \p\ and \T\ transformation properties of quark bilinears]%
{\C, \p\ and \T\ transformation properties of the chiral scalar, vector
and magnetic currents.
The coordinate $x$ in parentheses is the argument of both quark fields.}
\label{cpt}
\end{center}
\end{table}
Here $\sigma_{\mu \nu}=(i/2)[ \gamma_{\mu}, \gamma_{\nu} ] $. 
The transformation laws for the currents with opposite chirality are
obtained by interchanging $L\leftrightarrow R$ in \tab{cpt}.
The phase factors
\beq
\eta_X = \eta_X^{bd} = e^{i \lt( \phi_X^d
  - \phi_X^b \rt)}, \qquad  X=C,P,T \,. \label{phs} 
\eeq
depend on the quark flavors, but for simplicity the flavor indices of 
the $\eta_X$s have been omitted in \tab{cpt}.
One can absorb these
arbitrary phase factors $\exp(i \phi_X^q)$ into the definitions 
of the discrete transformations for every quark field in the theory. 
This feature originates from the freedom to redefine any quark field 
by a phase transformation\index{phase transformation!quark field}
\beq
q \to q \, e^{i \phi^q } . \label{pht} 
\eeq 
In the absence of flavor-changing couplings the change in \eq{pht}
is a $U(1)$ symmetry transformation leaving the Lagrangian
invariant. The corresponding conserved quantum number is the flavor of
the quark $q$. After including the flavor-changing interactions,
the phase transformations in \eq{pht} change the phases of the
flavor-changing couplings. The flavor symmetry is broken and every
phase transformation \eq{pht} leads to a different, but physically
equivalent Lagrangian. In the case of the Standard Model these
transform the Yukawa couplings and, hence, the CKM matrix from one phase 
convention into another.\index{phase convention!CKM matrix}

The currents in \tab{cpt} create and destroy the meson states with the
appropriate quantum numbers. Since the QCD interaction, which binds
the quarks into mesons, conserves \C, \p\ and \T, the  
meson states transform like the corresponding currents in \tab{cpt}.%
\index{CP transformation@\CP\ transformation!B meson@$B$ meson}%
\index{B meson@$B$ meson!CP transformation@\CP\ transformation}%
\index{B meson@$B$ meson!decay constant $f_B$}
For example, the $B_d$ meson is pseudoscalar and transforms under \CP\ as
\beqa
\CP\, \ket{\ov{B}{}_d^0 \, (P^{\rho})} &=&
  - \eta_P^{bd}\, \eta_C^{bd}\, \ket{B_d^0 \, (P_{\rho})} \,, 
  \nonumber\\
\CP\, \ket{B_d^0 \, (P^{\rho})} &=&
  - \eta_P^{bd}{}^* \eta_C^{bd}{}^* \, \ket{\ov{B}{}_d^0 \, (P_{\rho}) } \,. 
\label{mcp}
\eeqa
The vacuum state $\ket{0}$ is invariant under \C, \p\ and \T.  
Hence one finds, for example,
\beq
\bra{0}\, \ov{b} \gamma_{\mu} \gamma_5 d (x)\, \ket{B_d^0 (P)} 
\stackrel{CP}{=}
\bra{0}\, \ov{d} \gamma_{\mu} \gamma_5 b (x)\, \ket{\ov{B}{}_d^0 (P)}
    = i f_{B_d} P_{\mu} e^{-i P\cdot x} \,, 
\label{deffb}
\eeq
which is the definition of the $B$ meson decay constant $f_{B_d}$.
The phases $\eta_P^{bd} \eta_C^{bd}$ from the \CP\ transformation
of the pseudovector current $ \ov{b} \gamma_{\mu} \gamma_5 d =
\ov{b}_R \gamma_{\mu} d_R-\ov{b}_L \gamma_{\mu} d_L$
and $\eta_P^{bd}{}^* \eta_C^{bd}{}^* $ from \eq{mcp} cancel in the
first relation in \eq{deffb}. We can further multiply $\ket{B_d^0
(P)}$ and $\ket{\ov{B}{}_d^0 (P)}$ by another common phase factor
(unrelated to \CP) to choose $f_{B_d}$ positive.
\index{parity}\index{charge conjugation}%

Although \C\ and \p\ are unitary transformations, \T\ is
anti-unitary \index{time reversal}\index{anti-unitary transformation}%
(i.e., $T^\dagger T=1$ and
$\langle T\phi | T\psi\rangle =\langle \psi | \phi \rangle$).  
Thus, for example,
\beq
\T\, \ket{B_d^0 (P^{\rho})} = 
	\bra{B_d^0 (-P_{\rho})} \,. 
\eeq
The anti-unitary property of \T\ means also that $c$-numbers, such as 
the CKM matrix, are transformed into their complex conjugates.

\tab{cptg} lists the transformation properties of the vector 
bosons\index{CP transformation@\CP\ transformation!vector boson}
and the scalar Higgs field
$H$\index{CP transformation@\CP\ transformation!Higgs boson}
appearing in the Standard Model.
\begin{table}[tp]
\begin{center}
$ \begin{array}{r@{\quad}|@{\quad}c@{\quad}l@{\quad}l}
\hline\hline
& \mbox{photon, gluon, $Z$ boson} & \mbox{$W$ boson} & \mbox{Higgs} \\
\mbox{field}
&\ds  V^{\mu} \,(x^{\rho}) = A^{\mu} (x^{\rho}) \,, 
    A^{\mu,\, a} (x^{\rho}) \,, Z^{\mu} (x^{\rho}) 
  &\ds  W^{\pm, \mu}  \,(x^{\rho})  &\ds  H  \,(x^{\rho}) \\ \hline
\C\ &\ds   - V^{\mu} \,(x^{\rho}) 
    &\ds - W^{\mp, \mu} \,(x^{\rho})  
    &\ds  H \,(x^{\rho}) \\ 
\p\ &\ds  \phantom{-} V_{\mu} \,(x_{\rho}) 
    &\ds  \phantom{-} W^{\pm}_{\mu} \,(x_{\rho}) 
    &\ds  H \,(x_{\rho}) \\ 
\CP\ &\ds   - V_{\mu} \,(x_{\rho}) 
    &\ds   - W^{\mp}_{\mu}  \,(x_{\rho}) 
    &\ds  H \,(x_{\rho}) \\ 
\T\ &\ds  \phantom{-} V_{\mu} \,(-x_{\rho}) 
    &\ds  \phantom{-} W^{\pm}_{\mu}  \,( -x_{\rho}) 
    &\ds  H (- x_{\rho}) \\ 
\CPT\ &\ds -  V^{\mu} \,( - x^{\rho})
 &\ds  - W^{\mp, \mu}  \,( - x^{\rho})   
    &\ds   H  \,(- x^{\rho}) \\ \hline\hline
\end{array} $ \\[6pt]
\caption[\C, \p\ and \T\ transformation properties of bosons]%
{\C, \p and \T\ transformation properties of bosons in the 
Standard Model.}
\label{cptg}
\end{center}
\end{table}
The transformation properties of the photon and gluon field are
deduced from the experimental observation that QED and QCD conserve \C, 
\p\ and \T quantum numbers. 
For the weak gauge bosons the absence of \CP\ and \T\
violation in the gauge sector fixes the transformation properties of
$W^{\pm, \mu}$ and $Z^{\mu}$ under \CP\ and \T. The assignment of the
\C\ and \p\ transformations to the weak gauge bosons and the Higgs in
\tab{cptg} is chosen such that the Standard Model conserves \C\ and \p\
in the absence of fermion fields. These assignments do not impose
additional selection rules on the Standard Model interactions and
therefore have no observable consequences.

\index{CP violation@\CP\ violation!Standard Model}
\index{CP violation@\CP\ violation!complex couplings}
  From \tab{cpt} one can see why the weak interaction in the
Standard Model violates \C\ and \p. These transformations flip the
chirality of the quark fields, but left- and right-handed fields
belong to different representations of the $SU(2)$ gauge group. The
combined transformation \CP, however, maps the quark fields onto
fields with the same chirality. 
Still, the currents and their $CP$ conjugates (i.e., the first and 
fourth rows of \tab{cpt}) are not identical: instead they are 
Hermitian conjugates of each other.
Since the Lagrangian of any quantm field theory is Hermitian, it 
contains for each coupling of a quark current to a vector field its 
Hermitian conjugate coupling as well.
For example, the coupling of the $W$ to $b$ and $u$ quarks in the 
Standard Model is
\beq\label{1:Lagsample}
{\cal L} = - \frac{g_2}{\sqrt{2}} \lt[
    V_{ub} \, \ov{u}_L\, \gamma^{\mu} b_L \, W^{+}_{\mu} +
    V_{ub}^* \, \ov{b}_L\, \gamma^{\mu} u_L \, W^{-}_{\mu} \rt] . 
 \label{wex}  
\eeq
  From Tables~\ref{cpt} and \ref{cptg} one derives the \CP\ transformation
\beq
\CP \, {\cal L} \, (\CP)^{-1}
= - \frac{g_2}{\sqrt 2} \, 
    \lt[ V_{ub} \, \ov{b}_L\, \gamma^{\mu} u_L \, W^{-}_{\mu} 
     + V_{ub}^* \, \ov{u}_L\, \gamma^{\mu} b_L \, W^{+}_{\mu} \rt] ,
\eeq
which is the same only if $V_{ub}=V_{ub}^*$.
This illuminates why \CP\ violation is related to complex phases in
couplings. Yet complex couplings alone are not sufficient for a theory
to violate \CP.  A~phase rotation \eq{pht} of the quark fields in the
\CP\ transformed Lagrangian changes the phases of the couplings. If we
can in this way rotate the phases in $\CP \, {\cal L} \, (\CP)^{-1}$
back into those in ${\cal L}$, then \CP\ is conserved. In our example
\eq{wex} the choice $\phi^b-\phi^u=2 \arg V_{ub}$ would transform 
$\CP \, {\cal L} \, (\CP)^{-1}$ back into ${\cal L}$. 
As outlined in Sec.~\ref{ch1:sec:cpvsm}, Kobayashi and Maskawa%
\index{CP violation@\CP\ violation!Kobayashi-Maskawa mechanism}
realized that it is not possible to remove all the phases, once there 
are more than two quark generations~\cite{Kobayashi:1973fv}.

It is also illustrative to apply
the time reversal transformation to \eq{wex}.  It does not modify the 
currents, but, due to its anti-unitary\index{anti-unitary transformation}
character, it flips the phases
of the couplings and thereby leads to the same result as the \CP\
transformation. In our example we have disregarded the changes in the 
arguments $x^{\rho}$ of the fields shown in \eq{defpt}. Since physical 
observables depend on the action, $S=\int \!\d^4 x\, {\cal L} (x)$, rather 
than on ${\cal L}$, the sign of $x^{\rho}$ can be absorbed into a
change of the integration variables.

\index{CPT theorem@\CPT\ theorem|}%
From \tab{cpt} and \tab{cptg} one
can verify that the action of the
Standard Model is  invariant under the combined transformation~\CPT.
The \CPT\ transformation\index{CPT transformation@\CPT\ transformation}
simply turns the currents and the vector fields into their Hermitian
conjugates.
Due to $ {\cal L}= {\cal L}^\dagger$ one has
\beq 
S = \int\! \d^4 x \, {\cal L} (x) =  
    \int\! \d^4 x^\prime\,  {\cal L} (-x^\prime ) =
    \int\! \d^4 x^\prime\, \CPT \, {\cal L}( x^\prime)\, (\CPT)^{-1} 
     =  \CPT \, S \, (\CPT)^{-1}\, . 
\eeq
This \CPT\ {\it theorem} holds in any local Poincar\'e invariant
quantum field theory \cite{lp}.
It implies that particles and antiparticles have the same masses and
total decay widths.
In certain string
theories\index{CPT violation@\CPT\ violation!string theory} $CPT$
violation may be possible, and at low energies manifests itself in the
violation of Poincar\'e invariance or of quantum mechanics \cite{ks}.
In the standard framework of quantum field theory, however, the $CPT$
theorem is built in from the very beginning.
For example, the Feynman diagram for
any decay or scattering process and its $CPT$ conjugate diagram are
simply related by complex conjugation and give the same result for the
decay rate or cross section.  Unless stated otherwise it is always
assumed that $CPT$ invariance holds in all the formulae in this
report. In this context it is meaningless to distinguish $CP$ violation
and $T$ violation.%
\index{CP violation@\CP\ violation!T violation@\T\ violation}%

\subsection{Time evolution and mixing}
\label{subs:mix} 
\index{B mixing@$B$ mixing|(}

In this section we list the necessary formulae to describe \bbmd\ and 
\bbms.
The formulae are general and apply to both $B_d^0$ and to $B_s^0$ 
mesons, although with different values of the parameters.
Eqs.~(\ref{mgmat})--(\ref{defphi}) are even correct for 
\kkm\index{K mixing@$K$ mixing} and \ddm.\index{D mixing@$D$ mixing}
In the following, the notation $B^0$ represents either of the two 
neutral $B$ meson species with the standard convention that $B^0$ 
($\ov{B}{}^0$) contains a $\ov{b}$~antiquark (a $b$~quark).

\begin{figure}[bt]
\centerline{\epsfysize=3cm \epsffile{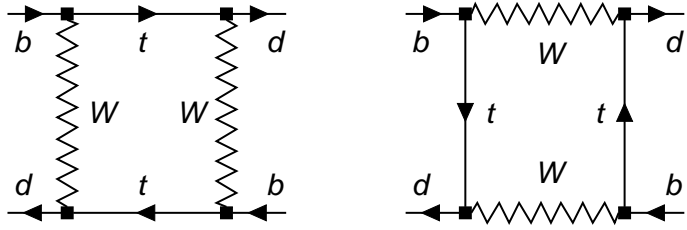}}
\caption{Standard Model box diagrams inducing
  $B_d^0-\ov{B}{}_d^0$ mixing.}
\label{ch1:fig:box}
\end{figure}

\bbm\ refers to transitions between the two flavor eigenstates
\ket{B^0} and \ket{\ov{B}{}^0}. In the Standard Model \bbm\ is caused
by the fourth order flavor-changing weak interaction described by the
box diagrams in \fig{ch1:fig:box}. 
Such transitions are called \dbt\
transitions, because they change the bottom quantum number by two
units.  In the Standard Model \dbt \index{$|\Delta B|=2$ transition} 
amplitudes are small, so measurements of \bbm\ could easily be 
sensitive to new physics.\index{B mixing@$B$ mixing!new physics}

\bbm\ induces oscillations between $B^0$ and $\ov{B}{}^0$.  
An initially produced $B^0$ or $\ov{B}{}^0$ evolves in time into a
superposition of $B^0$ and $\ov{B}{}^0$. Let \ket{B^0 (t)} denote the
state vector of a $B$ meson which is tagged as a $B^0$ at time $t=0$,
i.e., $\ket{B^0 (t=0)}=\ket{B^0}$.  Likewise \ket{\ov{B}{}^0 (t)}
represents a $B$ meson initially tagged as a $\ov{B}{}^0$.  The time
evolution of these states is governed by a Schr\"odinger equation:
\index{Schr\"odinger equation}
\beq
i\, \frac{\d}{\d\, t} \pmatrix{ \ds \ket{B(t)} \cr
  \ds \ket{\ov{B}(t)} \cr } 
= \left( M - i\, \frac{\Gamma}{2} \right) \pmatrix{ \ds \ket{B(t)} \cr
  \ds \ket{\ov{B}(t)} \cr } . 
\label{mgmat}
\eeq
The \textit{mass matrix}\ $M$ and the \textit{decay matrix}\ $\Gamma$
are $t$-independent, Hermitian $2\times 2$ matrices. 
\index{mass matrix $M$}\index{decay matrix $\Gamma$}
\CPT\ invariance implies that
\beq
M_{11} = M_{22} \,, \qquad \qquad  \Gamma_{11} = \Gamma_{22} \,. 
\eeq
\dbt\ transitions induce non-zero off-diagonal elements in \eq{mgmat}, 
so that the 
mass eigenstates of the neutral $B$ meson are different from the
flavor eigenstates \ket{B^0} and \ket{\ov{B}{}^0}. The mass
eigenstates are defined as the eigenvectors of $M - i\,\Gamma/2$. We
express them in terms of the flavor eigenstates as
\index{mass eigenstates}
\bey
\mbox{Lighter eigenstate:} \quad \ket{B_L} &=& 
    p \ket{B^0} + q \ket{\ov{B}{}^0} \,,  \nn
\mbox{Heavier eigenstate:} \quad \ket{B_H} &=& 
    p \ket{B^0} - q \ket{\ov{B}{}^0} \,,
\label{defpq}
\eey
with $\lt|p\rt|^2+\lt|q\rt|^2 = 1$.
Note that, in general, $\ket{B_L}$ and $\ket{B_H}$ are not orthogonal 
to each other.

The time evolution of the mass eigenstates is governed by
the two eigenvalues $M_H -i\,\Gamma_H/2$ and $M_L -i\,\Gamma_L/2$:
\beq
\ket{B_{H,L} (t) } = e^{-(i M_{H,L} + \Gamma_{H,L}/2)t} \, 
  \ket{B_{H,L} } \,, \label{thl}
\eeq
where $\ket{B_{H,L}}$ (without the time argument) denotes the mass 
eigenstates at time $t=0$: $\ket{B_{H,L}}=\ket{B_{H,L} (t=0)}$.  
We adopt the following definitions for the average mass and width and
the mass and width differences of the $B$ meson eigenstates: 
\index{B meson@$B$ meson!mass difference \dm}
\index{B meson@$B$ meson!width difference $\dg$}
\index{B meson@$B$ meson!width difference $\dg$!sign convention}
\begin{equation}
\begin{array}{rclrcl}
m & = & \displaystyle \frac{ M_H + M_L}{2} =  M_{11} \,, \qquad & 
  \Gamma & = & \displaystyle \frac{\Gamma_L + \Gamma_H}2 = \Gamma_{11} \,,
  \\[6pt] 
\dm & = & M_H - M_L \,, & \dg & = & \Gamma_L - \Gamma_H \,.
\end{array}\label{mg} 
\end{equation}
\dm\ is positive by definition. Note that the sign convention for 
\dg\ is opposite to the one used in Refs.~\cite{babook,BLS,book,cern}. 
In our convention the Standard Model prediction
for \dg\ is positive.

We can find the time evolution of \ket{B (t)} and \ket{\ov{B}(t)} as 
follows.
We first invert \eq{defpq} to express \ket{B^0} and \ket{\ov{B}{}^0} in
terms of the mass eigenstates and using their time evolution in
\eq{thl}:
\bey
\ket{B^0 (t)} &=& \frac{1}{2 p} 
    \lt[ \, e^{-i M_L t - \Gamma_L t/2} \, \ket{B_L} \, + \,  
     e^{-i M_H t - \Gamma_H t/2} \, \ket{B_H} \, \rt] , \nn
\ket{\ov{B}{}^0 (t)} &=& \frac{1}{2 q}
    \lt[ \, e^{-i M_L t - \Gamma_L t/2} \, \ket{B_L} \, - \,  
     e^{-i M_H t - \Gamma_H t/2} \, \ket{B_H} \, \rt] .
\label{tes}
\eey
These expressions will be very useful in the discussion of $B_s$ mixing.%
\footnote{The Schr\"odinger equation,\index{Schr\"odinger equation} 
\eq{mgmat}, is not exactly valid, but the result of the so-called 
\textit{Wigner-Weisskopf approximation}~\cite{wwa}
\index{Wigner-Weisskopf approximation} to the decay problem. 
In general, there are tiny corrections to the exponential decay laws in
\eq{tes} at very short and very large times \cite{kha}. 
These corrections are irrelevant for the mixing and
\CP\ studies at Run~II, but they must be taken into account in high
precision searches for \CPT\ violation \cite{azi}.  
\index{CPT violation@\CPT\ violation}}
With \eq{defpq} we next eliminate the mass eigenstates in \eq{tes} in
favor of the flavor eigenstates:
\bey 
\ket{B^0 (t)} &=& \phantom{\frac{p}{q}\,} 
  g_+ (t)\, \ket{B^0} + \frac{q}{p}\, g_- (t)\, \ket{\ov{B}{}^0} \,, \nn 
\ket{\ov{B}{}^0 (t)} &=& \frac{p}{q}\, g_- (t)\, \ket{B^0} 
  + \phantom{\frac{q}{p}\,} g_+(t)\, \ket{\ov{B}{}^0} \,,
  \label{tgg}    
\eey 
where
\bey
g_+ (t) &=& e^{-i m t} \, e^{-\Gamma t/2} 
   \lt[ \phantom{-} \cosh\frac{\dg \, t}{4}\, \cos\frac{\dm\, t}{2} -   
      i \sinh\frac{\dg \, t}{4}\, \sin \frac{\dm \, t}{2} \, \rt] , \nn
g_- (t) &=& e^{-i m t} \, e^{-\Gamma t/2} 
   \lt[- \sinh\frac{\dg \, t}{4}\, \cos \frac{\dm \, t}{2} +
      i \cosh\frac{\dg \, t}{4}\, \sin \frac{\dm \, t}{2} \, \rt] . 
\label{gpgm} 
\eey
Note that---owing to $\dg \neq 0$---the coefficient $g_+ (t)$ 
has no zeros, and $g_- (t)$ vanishes only at $t=0$. Hence an initially 
produced $B^0$ will never turn into a pure $\ov{B}{}^0$ or back into a
pure $B^0$. The coefficients in \eq{gpgm} will enter the formulae for
the decay asymmetries in the combinations
\bey
| g_\pm (t) |^2 & = & \frac{e^{- \Gamma t}}{2} 
  \lt[ \cosh \frac{\Delta \Gamma \, t}{2} \pm  
    \cos\lt( \Delta m\, t \rt) \rt] , \nn
g_+^* (t)\, g_- (t) & = & \frac{e^{- \Gamma t}}{2} 
  \lt[ - \sinh \frac{\Delta \Gamma \, t}{2} +
    i \sin \lt( \Delta m\, t \rt) \rt] .
\label{gpgms}
\eey 

In a given theory, such as the Standard Model, one can calculate the
off-diagonal elements $M_{12}$ and $\Gamma_{12}$ entering \eq{mgmat}
from \dbt\ diagrams. In order to exploit the formulae
(\ref{tes})--(\ref{gpgm}) for the time evolution we still need to express
$\dm$, $\dg$ and $q/p$ in terms of $M_{12}$ and $\Gamma_{12}$.  By
solving for the eigenvalues and eigenvectors of $M -i\,\Gamma/2$ one
finds%
\index{mass matrix $M$!eigenvalues}\index{mass matrix $M$!eigenvectors}%
\index{decay matrix $\Gamma$!eigenvalues}%
\index{decay matrix $\Gamma$!eigenvectors}%
\index{$q/p$}
\begin{mathletters}
\label{mgqp}
\bey
\lt( \dm \rt)^2 - \frac{1}{4} \lt( \dg \rt)^2 
  &=& 4  \lt| M_{12} \rt|^2 -  \lt| \Gamma_{12} \rt|^2 \,,
  \label{mgqp:a} \\[2mm]
\dm\, \dg &=& -4\, \real ( M_{12} \Gamma_{12}^* ) \,,
  \label{mgqp:b} \\[2mm]
\frac{q}{p} &=& - \frac{\dm + i \, \dg/2}{2 M_{12} -i\, \Gamma_{12} } 
  = - \frac{2 M_{12}^* -i\, \Gamma_{12}^*}{\dm + i \, \dg/2} \,.
\label{mgqp:c}
\eey
\end{mathletters}%
\index{phase!in mass matrix, $\phi_M$}\index{phase!in decay matrix}%
The relative phase between $M_{12}$ and $\Gamma_{12}$ appears in 
many observables related to $B$ mixing.
We introduce
\index{phase!of \CP\ violation in mixing, $\phi$}
\beq
	\phi = \arg \left( -\frac{M_{12}}{\Gamma_{12}} \right).
	\label{defphi}
\eeq
Now one can solve \eq{mgqp} for \dm\ and \dg\ in terms of $|M_{12}|$, 
$|\Gamma_{12}|$ and $\phi$.

The general solution is not illuminating, but a simple,
approximate solution may be derived when
\beq
	\lt| \Gamma_{12} \rt| \ll \lt| M_{12} \rt | \,,
		\qquad \mbox{and} \qquad 
	\dg \ll \dm \,.
	\label{dgsmall} 
\eeq
These inequalities hold (empirically) for both $B^0$ systems.
We first note that $|\Gamma_{12}| \leq \Gamma$ always,
because $\Gamma_{12}$ stems from the decays into final states common
to $B^0$ and $\ov{B}{}^0$.  
For the $B_s^0$ meson the lower bound on $\dms$ establishes experimentally
that $\Gamma _{B_s}\ll \dms$.  
Hence $\Gamma_{12}^s \ll \dms$, and Eqs.~(\ref{mgqp:a}) and (\ref{mgqp:b}) 
imply $\dms \approx 2 |M_{12}^s|$ and $|\dgs| \leq 2|\Gamma_{12}^s|$, 
so that \eq{dgsmall} holds.  
For the $B_d^0$ meson the experiments give 
$\dmd \approx 0.75 \,\Gamma_{B_d}$.
The Standard Model predicts 
$|\Gamma_{12}^{d}|/\Gamma_{B_d} = {\cal O}(1\% )$, but $\Gamma_{12}^{d}$ 
stems solely from CKM-suppressed
decay channels (common to $B_d^0$ and $\ov{B}{}_d^0$) and could
therefore be affected by new physics. 
New decay channels would, however,
also increase $\Gamma_{B_d}$ and potentially  conflict with the precisely
measured semileptonic branching ratio.  A~conservative estimate is
$|\Gamma_{12}^{d}|/\Gamma_{B_d} < 10\% $. Hence for both the $B_s^0$ and
$B_d^0$ system an expansion in $\Gamma_{12}/M_{12}$ and
$\dg/\dm$ is a good approximation, and we easily find
\index{B meson@$B$ meson!mass difference \dm}
\index{B meson@$B$ meson!width difference $\dg$}
\begin{mathletters}
\label{mgsol}
\bey
\dm &=& 2\, |M_{12}| \lt[ 1 + 
  {\cal O} \lt( \lt| \frac{\Gamma_{12}}{M_{12}} \rt|^2 \rt) \rt] ,
  \label{mgsol:a} \\
\dg &=& 2\, |\Gamma_{12}| \cos \phi \lt[ 1+  
  {\cal O} \lt( \lt| \frac{\Gamma_{12}}{M_{12}} \rt|^2 \rt) \rt] .
  \label{mgsol:b}
\eey
\end{mathletters}%
We also need an approximate expression for $q/p$ in \eq{mgqp}.
It is convenient to define a small parameter \index{$A_{AA}$@$\ega$!definition}
\beq
\ega = \imag \frac{\Gamma_{12}}{M_{12}} 
  =  \lt| \frac{\Gamma_{12}}{M_{12}} \rt| \sin \phi \,, \label{defepsg} 
\eeq
because occasionally we need to keep terms of order $\ega$.
Then $q/p$ becomes
\beq
\frac{q}{p} = - e^{- i \phi_M} \lt[ 1 - \frac{\ega}{2} \rt]  
  + {\cal O} \lt( \lt| \frac{\Gamma_{12}}{M_{12}} \rt|^2 \rt) ,
\label{qpsol}  
\eeq
where $\phi_M$ is the phase of 
$M_{12}$,\index{phase!in mass matrix, $\phi_M$}
\beq
  M_{12} = |M_{12}|\, e^{i \phi_M} \,. \label{defphm}
\eeq 
Note that \eq{qpsol} and the normalization condition $|p|^2+|q|^2=1$
imply 
\beq
\lt| p \rt| = \frac{1}{\sqrt{2}} \lt( 1 + \frac{\ega}{4} \rt) 
  + {\cal O} \lt( \lt| \frac{\Gamma_{12}}{M_{12}} \rt|^2 \rt), \qquad
\lt| q \rt| = \frac{1}{\sqrt2} \lt( 1 - \frac{\ega}{4} \rt) 
  + {\cal O} \lt( \lt| \frac{\Gamma_{12}}{M_{12}} \rt|^2 \rt) .
\eeq

We are now prepared to exhibit the time-dependent decay rate 
\gtf\ of an initially tagged $B^0$ into some final state $f$. It is
defined as\index{time evolution!decay rate} 
\beq
\gtf = \frac{1}{N_B}\, \frac{\d N(B^0 (t) \to f)}{\d t} \,, 
\label{defgtf}
\eeq
where $\d N(B^0(t) \to f)$ denotes the number of decays of a
$B$ meson tagged as a $B^0$ at $t=0$ into the final state $f$ occurring
within the time interval between $t$ and $t+\d t$. $N_B$ is the total
number of $B^0$'s produced at time $t=0$. An analogous
definition holds for \gbtf .  
One has 
\beq
\gtf = {\cal N}_f \lt| \langle f \ket{B^0 (t)}  \rt|^2 , \qquad 
  \gbtf = {\cal N}_f \lt| \langle f \ket{\ov{B}{}^0 (t)}  \rt|^2 . 
\label{gtfaf}
\eeq
Here $ {\cal N}_f$ is a time-independent normalization factor. To calculate
\gtf\ we introduce the two decay amplitudes \index{decay amplitude}
\beq 
A_f = \langle f \ket{B^0} \,, \qquad \qquad
\ov{A}_f = \langle f \ket{\ov{B}{}^0} \,, \label{defaf}
\eeq
and the quantity \index{$\lambda_f$!definition}
\beq
\lambda_f = \frac{q}{p}\, \frac{\ov{A}_f}{A_f}
  \simeq - e^{- i \phi_M}\, \frac{\ov{A}_f}{A_f}\,
  \lt[ 1 - \frac{\ega}{2} \rt] . \label{deflaf}
\eeq
We will see in the following sections that $\lambda_f$ plays the
pivotal role in \CP\ asymmetries and other observables in $B$ mixing.
Finally with \eq{tgg}, \eq{gpgms} and $|p/q|^2 = (1+\ega)$
we find the desired formulae for the decay rates:
\index{time evolution!decay rate} 
\bey 
\gtf &=&  {\cal N}_f \, | A_f |^2 \, e^{-\Gamma t}\,
  \Bigg\{ \frac{1 + \lt| \lambda_f \rt|^2}2\, \cosh \frac{\dg \, t}{2} + 
  \frac{ 1 - \lt| \lambda_f \rt|^2}2\, \cos ( \dm \, t )  \no \\*
&& \qquad \qquad \qquad
  - \real \lambda_f \, \sinh \frac{\dg \, t}{2} 
  - \imag \lambda_f \, \sin \lt( \dm \, t \rt) \Bigg\} \,, 
\label{gtfres} \\
\gbtf &=& {\cal N}_f \, | A_f |^2 \, ( 1+ \ega ) \,
  e^{-\Gamma t}\, \Bigg\{ \frac{1 + \lt| \lambda_f \rt|^2}2\,
    \cosh \frac{\dg \, t}{2}  
  - \frac{1 - \lt| \lambda_f \rt|^2}2\, \cos ( \dm \, t ) \no \\*
&&  \qquad \qquad \qquad        
    - \real \lambda_f \, \sinh \frac{\dg \, t}{2} 
    + \imag \lambda_f \, \sin ( \dm \, t ) \Bigg\} \,.
   \label{gbtfres}
\eey
Next we consider the decay into $\ov{f}$, which denotes the 
\CP\ conjugate state to
$f$,\index{CP conjugate state@\CP\ conjugate state}
\beq
\ket{\ov{f}} = \CP\, \ket{f} \,. 
\eeq
For example, for $f=D_s^- \pi^+$ the \CP\ conjugate state is 
$\ov{f}=D_s^+\pi^-$. 
The decay rate into $\ov{f}$ can be obtained from \eq{gtfres} and
\eq{gbtfres} by simply replacing $f$ with $\ov{f}$. Yet 
$| A_{\ov{f}} |$ and $| A_f |$ are unrelated, unless $f$
is a \CP\ eigenstate, fulfilling $\ket{\ov{f}} = \pm \ket{f}$.  On the
other hand the \CP\ transformation relates $| \ov{A}_{\ov{f}} |$
to $| A_f |$, so it is more useful to factor out $| \ov{A}_{\ov{f}} |$,
\bey 
\gtfb &=&  {\cal N}_f \lt| \ov{A}_{\ov{f}} \rt|^2 e^{-\Gamma t}\,
  ( 1 - \ega) \, \Bigg\{ \frac{1 + 
  | \lambda_{\ov{f}} |^{-2}}{2}\, \cosh \frac{\dg \, t}{2} 
  - \frac{ 1 - | \lambda_{\ov{f}} |^{-2}}{2}\, \cos ( \dm \, t ) \no \\*
&& \qquad \qquad \qquad 
  - \real \frac{1}{\lambda_{\ov{f}}}\, \sinh \frac{\dg \, t}{2} \, 
  + \imag \frac{1}{\lambda_{\ov{f}}}\, \sin (\dm \, t) \Bigg\} \,, 
\label{gtfbres} \\[2pt]
\gbtfb &=& {\cal N}_f \lt| \ov{A}_{\ov{f}} \rt|^2 e^{-\Gamma t}\,
  \Bigg\{ \frac{1 + | \lambda_{\ov{f}} |^{-2}}2\,
    \cosh \frac{\dg \, t}{2}  
  + \frac{1 - | \lambda_{\ov{f}} |^{-2}}2\, \cos ( \dm \, t ) \no\\*    
&& \qquad \qquad \qquad 
  - \real \frac{1}{\lambda_{\ov{f}}}\, \sinh \frac{\dg \, t}{2} 
  - \imag \frac{1}{\lambda_{\ov{f}}}\, \sin ( \dm \, t ) \Bigg\} \,.
   \label{gbtfbres}
\eey
Here we set ${\cal N}_{\ov{f}}={\cal N}_f$, because these
normalization factors arise from kinematics.  In
Eqs.~\eq{gtfres}--\eq{gbtfbres} we consistently keep terms of order
$\ega$, which appear explicitly in the prefactor in
\eq{gbtfres}, \eq{gtfbres} and are implicit in $\lambda_f$ through 
\eq{deflaf}.%
\footnote{We have omitted terms of order $|\Gamma_{12}/M_{12}|^2$ in
Eqs.~(\ref{deflaf})--(\ref{gbtfbres}) and will do this throughout the
report. In most applications one can set $\ega$ to zero and
often also \dg\ can be neglected, so that the expressions in
Eqs.~(\ref{gtfres})--(\ref{gbtfbres}) simplify considerably.}

We now apply the derived formalism to the decay rate into a 
\textit{flavor-specific}\ final state $f$
\index{flavor-specific} meaning that a $B^0$ can
decay into $f$, while $\ov{B}{}^0$ cannot. 
Examples are $f=D_s^- \pi^+$ (from $B_s^0$) and $f=X \ell^+ \nu_\ell$.
\index{decay!$B_s \to D_s^- \pi^+$}
\index{decay!$B_s \to X \ell^+ \nu_\ell$}
\index{decay!$B_d \to X \ell^+ \nu_\ell$}
In such decays $\ov{A}_f=A_{\ov{f}}=0$ by definition and, hence,
$\lambda_f=1/\lambda_{\ov{f}}=0$.
Therefore,
\bey
\gtf &=&  {\cal N}_f \, \lt| A_f \rt|^2 \, e^{-\Gamma t} \,
  \frac{1}{2} \lt[ \cosh \frac{\dg \, t}{2} 
  + \cos ( \dm \, t ) \rt] \qquad\qquad \! \mbox{ for }  \ov{A}_f=0 \,,\\
\label{gtfs}
\gtfb &=& {\cal N}_f \, \lt| \ov{A}_{\ov{f}} \rt|^2 \,  
    ( 1- \ega ) \,
    e^{-\Gamma t} \,
    \frac{1}{2}  
    \lt[ \cosh \frac{\dg \, t}{2}  
    - \cos ( \dm \, t ) \rt] \quad  \mbox{for }  A_{\ov{f}}=0 \,.
    \label{gtfbs}
\eey
Flavor-specific decays can be used to measure \dm\ via the
asymmetry in decays from mixed and unmixed $B$s:
\index{mixing asymmetry}
\beq
{\cal A}_0 (t) = \frac{\gtf - \gtfb}{\gtf + \gtfb } \,. 
\label{defa0}
\eeq
The amplitudes $A_f$ and $\ov{A}_{\ov{f}}$ are related to each other by
\CP\ conjugation. If there is no \CP\ violation in the decay amplitude 
(i.e., no \textit{direct \CP\ violation}), $|A_f|$ and
$|\ov{A}_{\ov{f}}|$ are equal. This is the case for decays like 
$B_s \to D_s^- \pi^+$ and $B \to X \ell^+ \nu_\ell $
conventionally used to measure \dm. Then the 
mixing asymmetry in \eq{defa0} reads
\index{mixing asymmetry}
\index{decay!$B_s \to D_s^- \pi^+$}
\index{decay!$B_s \to X \ell^+ \nu_\ell$}
\index{decay!$B_d \to X \ell^+ \nu_\ell$}
\beq
{\cal A}_0 (t) = \frac{\cos ( \dm\, t ) }{\cosh (\dg \,t/2)} 
  + \frac{\ega }{2} 
  \lt[1- \frac{\cos^2 ( \dm\, t )}{\cosh^2 (\dg \, t/2)}  \rt] ,
\label{resa0} 
\eeq
where we have allowed for a non-zero width difference.%
\index{B mixing@$B$ mixing|)}

\boldmath
\subsection{Time evolution of untagged $B^0$ mesons}
\label{subs:unt} 
\unboldmath
\index{time evolution!untagged $B$}

Since $B^0$'s and $\ov{B}{}^0$'s are produced in equal numbers at the
Tevatron, the untagged decay rate for the decay $\Bun \to f$ reads
\bey
\guntf &=& \gtf + \gbtf         \label{guntf}  \\*
&=& {\cal N}_f \lt| A_f \rt|^2  
    \lt( 1 + |\lambda_f|^2  \rt) e^{- \Gamma t}  \lt[
    \cosh \frac{\dg  \, t}{2} +  \, 
    \sinh \frac{\dg  t}{2} \, A_{\Delta \Gamma} \rt]
    + {\cal O} ( \ega ) \, \nonumber
\eey 
with\index{$A_{\Delta \Gamma}$!definition}
\beq
A_{\Delta \Gamma} = - \frac{2\, \real \lambda_f}{1+ \lt| \lambda_f \rt|^2} \,.
    \label{adeg}
\eeq
  From this equation one realizes that untagged samples are interesting
for the determination of \dg. The fit of an untagged decay distribution to 
\eq{guntf} involves the overall normalization factor 
${\cal N}_f\, |A_f|^2\, (1 + |\lambda_f|^2)$.  From
\eq{defgtf} one realizes that by integrating \guntf\ over all times
one obtains the branching ratio for the decay of an untagged $B^0$
into the final state $f$:
\index{untagged $B$!branching ratio}
\bey
{\cal B} \big( \!\Bun \to f \, \big) &=& \frac{1}{2}   
    \int_0^\infty \! \d t \, \guntf\
    = \frac{{\cal N}_f}{2} \, | A_f |^2 \, 
    \frac{\Gamma \lt( 1+ | \lambda_f |^2 \rt) - 
    \dg \, \real \lambda_f }{\Gamma^2-(\dg/2)^2}
    + {\cal O} ( \ega ) \no \\[2mm]
 &=&    \frac{{\cal N}_f}{2} \, | A_f |^2 \lt( 1 + |\lambda_f|^2 \rt)
    \frac{1}{\Gamma} \lt[ 1 + 
    \frac{\dg}{2 \Gamma} \, A_{\Delta \Gamma} \, + 
    {\cal O} \lt( \frac{(\dg)^2}{\Gamma^2} \rt) \rt] . 
    \label{untnorm} 
\eey
Relation \eq{untnorm} allows to eliminate $ {\cal N}_f \, | A_f|^2\,[
1 + |\lambda_f|^2]$ from \eq{guntf}, if the branching ratio is
known. If both ${\cal B}\, \big( \!\Bun \to f \big)$ and \dg\ are
known, a one-parameter fit to the measured untagged time evolution
\eq{guntf} allows to determine $ A_{\Delta \Gamma}$, which is of key
interest for \CP\ studies.

Finally we write down a more intuitive expression for \guntf. From 
\eq{gtfaf} and \eq{tes} one immediately finds
\beq
\guntf = {\cal N}_f
    \lt[ \,  e^{-\Gamma_L t} \lt| \langle f \ket{B_L} \rt|^2 + 
    e^{-\Gamma_H t} \lt| \langle f \ket{B_H} \rt|^2  \rt] +  
    {\cal O} ( \ega ) \,.
    \label{guntfeig}
\eeq
With \eq{defpq} one recovers \eq{guntf} from \eq{guntfeig}. Now
\eq{guntfeig} nicely shows that the decay of the untagged sample into
some final state $f$ is governed by two exponentials. If $B_s$ mixing
is correctly described by the Standard Model, the mass eigenstates 
$\ket{B_L}$ and $\ket{B_H}$ are to a high precision also
\CP\ eigenstates and \eq{guntfeig} proves useful for the description of 
decays into \CP\ eigenstates.

\boldmath
\subsection{Time-dependent and time-integrated $CP$ asymmetries}
\label{subs:cp}
\unboldmath
\index{CP asymmetry@\CP\ asymmetry!time-dependent}

The \CP\ asymmetry for the decay of a charged $B$ into the final state
$f$ reads
\index{CP asymmetry@\CP\ asymmetry!in charged $B$ decay}
\beq
a_f = \frac{\Gamma (B^- \to f )-\Gamma (B^+ \to \ov{f} ) }{
      \Gamma (B^- \to f )+\Gamma (B^+ \to
      \ov{f} ) }
    \qquad \quad \mbox{with~~} \ket{\ov{f}} = \CP\, \ket{f} 
    \label{defacpch}.
\eeq 
Defining 
\beq
A_f = \langle f \ket{B^+}  \qquad \mbox{and} \qquad
  \ov{A}_{\ov{f}} = \langle \ov{f} \ket{B^-}  \label{defafch}
\eeq
in analogy to \eq{defaf} one finds 
\beq
a_f = - \frac{1- \lt| \ov{A}_{\ov{f}}/A_f \rt|^2}{1+ \lt|
    \ov{A}_{\ov{f}}/A_f \rt|^2} \,. \label{acpdirch}
\eeq
Since charged $B$ mesons cannot mix, a non-zero $a_f$ can only occur
through \CP\ violation in the \dbo\ matrix elements $A_f$ and
$\ov{A}_{\ov{f}}$.  This is called \textit{direct}\ \CP\ violation and
stems from $|A_f| \neq |\ov{A}_{\ov{f}}|$. 
\index{CP violation@\CP\ violation!direct}
\index{CP violation@\CP\ violation!in decay}

Next we consider the decay of a  neutral $B$ meson into a \CP\
eigenstate $f=f_{CP}=\eta_f \ov{f}$. Here $\eta_f=\pm 1$ is the \CP\
quantum number of $f$. An example for this situation is the decay
$B_s^0 \to D_s^+ D_s^-$\index{decay!$B_s \to D_s^+ D_s^-$},
where $\eta_f=+1$.    
We define the time-dependent \CP\ asymmetry as  
\beq
a_f(t) = \frac{ \gbtf - \gtf }{ \gbtf + \gtf } \,. \label{defacp}
\eeq
Using \eq{gtfres} and \eq{gbtfres} one finds 
\beq
a_f(t) = - \frac{A_{CP}^{\rm dir} \cos ( \Delta m  \, t ) + 
       A_{CP}^{\rm mix} \sin ( \Delta m \, t )}{
       \cosh (\Delta \Gamma \, t/ 2) + 
       A_{\Delta \Gamma} \sinh  (\Delta \Gamma \, t / 2) }
    + {\cal O} ( \ega ) \,, 
    \label{acp} 
\eeq
where $A_{\Delta \Gamma}$ is defined in \eq{adeg}, and
the \textit{direct} and \textit{mixing-induced} (or
\textit{interference type}) \CP\ asymmetries are
\index{CP asymmetry@\CP\ asymmetry!mixing-induced}%
\index{CP asymmetry@\CP\ asymmetry!interference type}%
\index{CP violation@\CP\ violation!interference type}%
\index{$A_{CP}^{\rm dir}$ and $A_{CP}^{\rm mix}$ definitions}%
\beq
A_{CP}^{\rm dir} =
    \frac{1- \lt| \lambda_f \rt|^2}{1+ \lt| \lambda_f 
    \rt|^2} \,, \qquad
A_{CP}^{\rm mix} = - \frac{2\, \imag \lambda_f}{1+ \lt| \lambda_f \rt|^2} \,,
    \label{dirmix} 
\eeq
$A_{CP}^{\rm mix}$ stems from the interference of the decay amplitudes of
the unmixed and the mixed $B$, i.e., of $\ov{B}{}^0 \to f$ and
$B^0 \to f $. It is discussed in more detail in Sec.~\ref{ch1:sect:ty}.
Note that the quantities in \eq{dirmix} and \eq{adeg} are not
independent, 
$|A_{CP}^{\rm dir}|^2 + |A_{CP}^{\rm mix}|^2 + |A_{\Delta\Gamma}|^2 = 1$.

\index{CP asymmetry@\CP\ asymmetry!time-integrated}
The time integrated asymmetry reads%
\footnote{
Our sign conventions for the \CP\ asymmetries in \eq{defacpch} and 
\eq{defacp} are opposite to those in \cite{cern}. 
Our definitions of $A_{CP}^{\rm dir}$, $A_{CP}^{\rm mix}$ and 
$A_{\Delta \Gamma}$ are the same as in \cite{cern}, taking into 
account that the quantity $\xi_f^{(q)}$ of \cite{cern} equals 
$-\lambda_f$.}
\beq 
a_f^{\rm int} = \frac{ \int_0^\infty \d t \lt[ \gbtf - \gtf \rt] }{ 
             \int_0^\infty \d t \lt[ \gbtf + \gtf \rt] }
= - \frac{1 + y^2}{1 + x^2} \,
    \frac{ A_{CP}^{\rm dir} + A_{CP}^{\rm mix} \, x }{
    1 + A_{\Delta \Gamma} \, y } \,.
    \label{acpint}
\eeq
Here the quantities $x$ and $y$ are defined as 
\index{mixing parameter!$x$}
\index{mixing parameter!$y$}
\beq
x = \frac{\dm}{\Gamma} \,, \qquad 
  y = \frac{\dg}{2\, \Gamma} \,. \label{defxy}
\eeq
Thus, even without following the time evolution, a measurement of
$a_f^{\rm int}$ puts constraints on $\dm$ and $\dg$.

\subsection{Phase conventions}
\label{subs:phc}
\index{phase convention}

In Sec.~\ref{subs:disc} we learned that there is no unique way to define the
\CP\ transformation, because it involves an arbitrary phase factor
$\eta_{CP}\equiv\eta_{C}\eta_{P}$ (see \tab{cpt} and Eq.~\eq{phs}). This
arbitrariness stems from the fact that phases of  quark fields are unobservable
and phase redefinitions as in \eq{pht} transform the Lagrangian into a
physically equivalent one. This feature implies that the phases of the
flavor-changing couplings in our Lagrangian are not fixed and the phase
rotation \eq{pht} transforms one phase convention for these couplings into
another one. 
Of course, physical
observables are independent of these phase conventions. Hence it is
worth noting which of the quantities defined in the previous sections are
invariant, when $\eta_{CP}$ or the phases of the quark fields are changed. 
It is also important to identify the quantities that do depend on phase
conventions to avoid mistakes when combining convention dependent quantities
into an invariant observable.

The phases of 
\beq
M_{12} \,,\  \Gamma_{12} \,,\  \frac{q}{p} \,, 
  \mbox{ and } \frac{\ov{A}_f}{A_f} \,. 
\label{notinv}
\eeq
depend on the phase convention of the \CP\
transformation or the phase convention of the \CP\ violating couplings.
In particular, the phase $\phi_M$ of the mixing amplitude $M_{12}$ 
(defined in \eq{defphm}) is convention dependent.
When speaking informally, one often says that a given process, such as
$B_d^0\to\psi K_S$, measures the phase of the \dbt\ amplitude, 
i.e.,~$\phi_M$.
Such statements refer to a specific phase convention, in which the decay 
amplitude of the process has a vanishing (or negligible) phase.  
The following quantities are independent of phase conventions:
\beq
\lt|\frac{q}{p} \rt| \,,\ \lt| \frac{\ov{A}_f}{A_f} \rt| \,,\ 
  \ega \,,\  \phi \,,\  \dm \,,\  \dg \,,\ 
  \mbox{and } \lambda_f \,. 
\label{inv}
\eeq
The only complex quantity here is $\lambda_f$.
Its phase is a physical observable. 

We have shown that the arbitrary phases accompanying the 
\CP\ transformation stem from the freedom to rephase the quark fields,
see \eq{pht}. The corresponding phase factors $\eta_{CP}$ in the \CP\
transformed quark bilinears are sufficient to parameterize this
arbitrariness and likewise appear in the \CP\ transformations of the
mesons and the quantities in \eq{notinv}. In some discussions of this
issue authors allow for phases different from $\eta_{CP}$ accompanying
the \CP\ transformation \eq{mcp} of the meson states. This is simply
equivalent to using our transformation
\eq{mcp} followed by a multiplication of $\ket{B_d^0}$ and 
$\ket{\ov{B}{}_d^0}$ with extra phase factors (unrelated to \CP ),
which do not affect observables.  This would further introduce an
extra inconvenient phase into \eq{deffb}. The quantities in \eq{inv} are
still invariant under such an extra rephasing and no new information
is gained from this generalization.
Unless stated otherwise, we will use the phase convention
\index{phase convention}
$\eta_{CP}=1$, i.e.,
\beq
\CP\, \ket{\ov{B}{}^0 (P^{\rho})} =
    - \ket{B^0 (P_{\rho})} \,, \qquad 
\CP\, \ket{B^0 (P^{\rho})} =
    - \ket{\ov{B}{}^0 (P_{\rho}) } \,.  \label{mcpconv}
\eeq
For the phases of the CKM elements we use the convention of the 
Particle Data Group, \eq{1:eq:VPDG}.

\boldmath
\section{Aspects of $CP$ Violation }
\label{ch1:sec:aspects}
\unboldmath

\OMIT{
Thoughts:
{\bf This section needs some work: it should have an intro tying it 
into the larger picture.  I also don't get the choice of subsections,
because $B_s\to D_sK$ is $CP$ violation in interference between decays 
with and without mixing.  Finally, is there overlap with WG1?  
There is also Uli's point about whether $\psi K$ gets inference 
through $K - \Kbar$ mixing or through $\pi\pi$ final states.}
}

\boldmath
\subsection{The three types of $CP$ violation}
\label{ch1:sect:ty}
\unboldmath
As discussed in Sec.~\ref{subs:phc}, there are three phase convention
independent physical $CP$ violating observables
\begin{equation}
\bigg| \frac qp \bigg|\,, \qquad 
  \lt| \frac{\ov{A}_{\ov{f}}}{A_f} \rt|\,, \qquad  
  \lambda_f = \frac qp\, \frac{\ov{A}_f}{A_f} \,.
\end{equation}
If any one of these quantities is not equal to 1 (or $-1$ for $\lambda_f$),
then $CP$ is violated in the particular decay.  In fact, there are decays 
where only one of these types of $CP$ violations occur (to a very good
approximation).

\subsubsection*{\boldmath $CP$ violation in mixing ($|q/p| \neq 1$) }
\index{CP violation@\CP\ violation!in mixing}

It follows from Eq.~(\ref{mgqp:c}) that 
\begin{equation}
\bigg| \frac qp \bigg|^2 = \bigg| { 2M_{12}^* - i\,\Gamma_{12}^* \over
  2M_{12} - i\,\Gamma_{12} } \bigg| \,.
\end{equation}
If $CP$ were conserved, then the relative phase between $M_{12}$ and
$\Gamma_{12}$ would vanish, and so $|q/p| = 1$.  If $|q/p| \neq 1$, then
$CP$ is violated.  This is called $CP$ violation in mixing, because it
results from the mass eigenstates being different from the $CP$
eigenstates.  It follows from Eq.~(\ref{defpq}) that
$\braket{B_H}{B_L} = |p|^2 - |q|^2$, and so the two physical states
are orthogonal if and only if $CP$ is conserved in $|\Delta B| = 2$
amplitudes.

The simplest example of this type of $CP$ violation is the neutral meson 
semileptonic decay asymmetry to ``wrong sign" leptons
\index{CP asymmetry@\CP\ asymmetry!in flavor-specific decay}
\index{CP asymmetry@\CP\ asymmetry!semileptonic}
\index{decay!$B_d \to X \ell^+ \nu$ }
\begin{eqnarray}
a_{\rm sl}(t) &=&
{\Gamma(\B0bar(t) \to \ell^+\nu X) - \Gamma(B^0(t) \to \ell^-\bar\nu X) \over
  \Gamma(\B0bar(t) \to \ell^+\nu X) + \Gamma(B^0(t) \to \ell^-\bar\nu X) }
  \nonumber\\*
&=& { |p/q|^2 - |q/p|^2 \over |p/q|^2 + |q/p|^2 }
  = { 1 - |q/p|^4 \over 1 + |q/p|^4 } 
  = \ega + {\cal O}(\ega^2) \,.
  \label{1:eq:asl(t)}
\end{eqnarray}
The second line follows from Eq.~(\ref{tgg}).  In $B$ meson decay such an
asymmetry is expected to be ${\cal O}(10^{-2})$.  The calculation of $|q/p|-1$
involves $\imag (\Gamma_{12} / M_{12})$, which suffers from hadronic
uncertainties.  Thus, it would be difficult to relate the observation of such
an asymmetry to CKM parameters.  This type of $CP$ violation can also be
observed in any decay for which $A_f \gg A_{\ov{f}}\,$, such as decays to 
flavor specific final states (for which $A_{\ov{f}} = 0$), e.g., $B_{(s)}\to
D_{(s)}^-\pi^+$.  In kaon decays this asymmetry was recently measured by
CPLEAR~\cite{cplear} in agreement with the expectation that it should be equal
to $4\,\real \e_K$.
\index{decay!$B_d \to D^- \pi^+$ }
\index{decay!$B_s \to D_s^- \pi^+$ }

\boldmath
\subsubsection*{$CP$ violation in decay ($|\ov{A}_{\ov{f}}/A_f| \neq 1$) }
\unboldmath
\index{CP violation@\CP\ violation!in decay}
\index{CP violation@\CP\ violation!direct}
\index{$\lambda_f$!$|\lambda_f|\neq 1$}

For any final state $f$, the quantity $|\ov{A}_{\ov{f}} / A_f|$ is a phase
convention independent physical observable.  There are two types of complex
phases which can appear in $\ov{A}_{\ov{f}}$ and $A_f$ defined in
Eq.~(\ref{defaf}).  Complex parameters in the Lagrangian which enter a decay
amplitude also enter the $CP$ conjugate amplitude but in complex conjugate
form.  In the Standard Model such parameters only occur in the CKM matrix. 
These so-called weak phases enter $\ov{A}_{\ov{f}}$ and $A_f$ with opposite
signs.  Another type of phase can arise even when the Lagrangian is real, from
absorptive parts of decay amplitudes.\index{absorptive part}  
These correspond to on-shell intermediate
states rescattering into the desired final state.  Such rescattering is usually
dominated by strong interactions, and give rise to $CP$ conserving
strong phases, which enter $\ov{A}_{\ov{f}}$ and $A_f$ with the same signs. 
Thus one can write $\ov{A}_{\ov{f}}$ and $A_f$ as
\begin{equation}
A_f = \sum_k A_k\, e^{i(\delta_k + \phi_k)} \,, \qquad
\ov{A}_{\ov{f}} = \sum_k A_k\, e^{i(\delta_k - \phi_k)} \,,
\end{equation}
where $k$ label the separate contributions to the amplitudes, $A_k$ are the
magnitudes of each term, $\delta_k$ are the strong phases, and $\phi_k$ are the
weak phases.  The individual phases $\delta_k$ and $\phi_k$ are convention
dependent, but the phase differences between different terms, $\delta_i -
\delta_j$ and $\phi_i - \phi_j$, are physical.

Clearly, if $|\ov{A}_{\ov{f}} / A_f| \neq 1$ then $CP$ is violated.  This is
called $CP$ violation in decay, or direct $CP$ violation.  It occurs due to
interference between various terms in the decay amplitude, and requires that at
least two terms differ both in their strong and in their weak phases.  The
simplest example of this is direct $CP$ violation in charged $B$ decays
\begin{equation}
{\Gamma(B^- \to f) - \Gamma(B^+ \to \ov{f}) \over
  \Gamma(B^- \to f) + \Gamma(B^+ \to \ov{f}) }
  = - { 1 - |\ov{A}_{\ov{f}}/A_f|^2 \over 1 + |\ov{A}_{\ov{f}}/A_f|^2 } \,.
\end{equation}
To extract the interesting weak phases from such $CP$ violating observables,
one needs to know the amplitudes $A_k$ and their strong phases $\delta_k$.  
The problem is that theorists do not know how to compute these from first
principles, and most estimates are unreliable.  The only experimental
observation of direct $CP$ violation so far is $\real \e_K'$ in kaon decays.

This type of $CP$ violation can also occur in neutral $B$ decays in conjunction
with the others.  In such cases direct $CP$ violation is rarely beneficial, and
is typically a source of hadronic uncertainties that are hard to control.

\boldmath
\subsubsection*{$CP$ violation in the interference between decay and mixing 
($\lambda_f \neq \pm1)$ }
\unboldmath
\index{CP violation@\CP\ violation!interference type}
\index{$\lambda_f$!$\imag \lambda_f\neq 0$}

Another type of $CP$ violation is possible in neutral $B$ decay into a $CP$
eigenstate final state, $f_{CP}$.  If $CP$ is conserved, then not only
$|q/p| = 1$ and $|\ov{A}_f/A_f| = 1$, but the relative phase between $q/p$
and $\ov{A}_f/A_f$ also vanishes.  In this case it is convenient to rewrite 
\beq
\lambda_{f_{CP}} = \frac qp\, \frac{\ov{A}_{f_{CP}}}{A_{f_{CP}}} 
  = \eta_{f_{CP}}\, \frac qp\, \frac{\ov{A}_{\ov{f}_{CP}}}{A_{f_{CP}}} \,,
\eeq
where $\eta_{f_{CP}} = \pm 1$ is the $CP$ eigenvalue of $f_{CP}$.  This form of
$\lambda_{f_{CP}}$ is useful for calculations, because $A_{f_{CP}}$ and
$\ov{A}_{\ov{f}_{CP}}$ are related by $CP$ as discussed in the previous
subsection.  If $\lambda_{f_{CP}} \neq \pm 1$ then $CP$ is violated.  This is
called $CP$ violation in the interference between decays with and without
mixing, because it results from the $CP$ violating interference between $B^0
\to f_{CP}$ and $B^0 \to \ov{B}{}^0 \to f_{CP}$.

As derived in Eq.~(\ref{acp}), the time dependent asymmetry is
\index{CP asymmetry@\CP\ asymmetry!time-dependent}
\begin{eqnarray}
a_f(t) &=&
{\Gamma(\B0bar(t) \to f) - \Gamma(B^0(t) \to f) \over
  \Gamma(\B0bar(t) \to f) + \Gamma(B^0(t) \to f) }
  \nonumber\\*
&=& - {(1-|\lambda_f|^2) \cos(\Delta m\, t) 
  - 2\, \imag \lambda_f \sin(\Delta m\, t) 
  \over (1+|\lambda_f|^2) \cosh(\Delta\Gamma\, t/2) 
  - 2\, \real \lambda_f \sinh(\Delta\Gamma\, t/2) }
  + {\cal O}(\ega) \,.
\label{1:cpmix}
\end{eqnarray}
This asymmetry is non-zero if any type of $CP$ violation occurs.  In
particular, it is possible that $\imag \lambda_f \neq 0$, but
$|\lambda_f| = 1$ to a good approximation, because $|q/p| \simeq 1$
and $|\ov{A}_f/A_f| \simeq 1$.  In both the $B_d$ and $B_s$ systems
$|q/p| - 1 \lesssim {\cal O}(10^{-2})$.
Furthermore, if only one amplitude contributes to a decay,
then $|\ov{A}_f/A_f| = 1$ automatically.
These modes are ``clean'', because in such cases $A_f$ drops out and
\index{CP violation@\CP\ violation!clean modes}
\begin{equation}
a_f(t) = { \imag \lambda_f \sin(\Delta m\, t) \over \cosh(\Delta\Gamma\, t/2) 
  - \real \lambda_f \sinh(\Delta\Gamma\, t/2) } \,,
\end{equation}
measures $\imag \lambda_f$, which is given by a weak phase.  In addition, if
$\Delta\Gamma$ can be neglected then $a_f(t)$ further simplifies to
$a_f(t) = \imag \lambda_f \sin(\Delta m\, t)$.

The best known example of this type of $CP$ violation (and also the one where
$|\lambda_f| = 1$ holds to a very good accuracy) is the asymmetry in $B \to
\psi K_S$,\index{decay!$B_d \to \psi K_S$} where $\psi$ denotes any
charmonium state.
The decay is dominated by the tree level $b\to c\bar cs$ transition
and its $CP$ conjugate.  In the phase convention (\ref{mcpconv}) one finds
\begin{equation}
{\ov{A}_{\psi K_S} \over A_{\psi K_S} } = 
  \bigg( {V_{cb} V_{cs}^* \over V_{cb}^* V_{cs}} \bigg)\,
  \bigg( {V_{cs} V_{cd}^* \over V_{cs}^* V_{cd}} \bigg) \,.
\end{equation}
The overall plus sign arises from \eq{mcpconv} and because $\psi K_S$
is $CP$ odd, $\eta_{\psi K_S} = -1$, and the last factor is $(q/p)^*$ in
$K^0 - \Kbar^0$ mixing.  This is crucial, because in the absence of
$K^0 - \Kbar^0$ mixing there could be no interference between 
$\B0bar\to \psi \Kbar^0$ and $B^0\to \psi K^0$.  There are also penguin
contributions to this decay, which have different weak and
strong phases.  These are discussed in detail in Chapter~6, where they
are shown to give rise to hadronic uncertainties suppressed by 
$\lambda^2$.  Then one finds
\begin{equation}
\lambda_{\psi K_S} = - \bigg( { V_{tb}^* V_{td} \over V_{tb} V_{td}^*} \bigg)\,
  \bigg( {V_{cb} V_{cs}^* \over V_{cb}^* V_{cs}} \bigg)\,
  \bigg( {V_{cs} V_{cd}^* \over V_{cs}^* V_{cd}} \bigg) 
= - e^{-2i\beta} \,,
\end{equation}
where the first factor is the Standard Model value of $q/p$ in $B_d$
mixing.  Thus, $a_{\psi K_S}(t)$ measures $\imag \lambda_{\psi K_S} =
\sin2\beta$ cleanly.
\index{CP asymmetry@\CP\ asymmetry!$\sin2\beta$}

Of significant interest are some final states which are not pure $CP$
eigenstates, but have $CP$ self conjugate particle content and can be
decomposed in $CP$ even and odd partial waves.  In some cases an angular
analysis can separate the various components, and may provide theoretically
clean information.  An example is $B_s\to \psi\,\phi$ discussed in Chapters 6
and 8.  There are many cases when $CP$ violation in decay occurs in addition to
$CP$ violation in the interference between mixing and decay.  Then the
asymmetry in Eq.~(\ref{1:cpmix}) depends on the ratio of different decay
amplitudes and their strong phases, which introduce hadronic uncertainties.  In
some cases it is possible to remove (or reduce) these by measuring several
rates related by isospin symmetry.  An example is $B_d\to \rho\,\pi$ (or
$\pi\,\pi$) discussed in Chapter~6.
\index{decay!$B_d \to \pi \pi $ }
\index{decay!$B_d \to \rho \pi $ }

\boldmath
\subsection{Decays to non-$CP$ eigenstates}
\unboldmath
\index{CP violation@\CP\ violation!clean modes}
\index{CP violation@\CP\ violation!in non-$CP$ eigenstates}

In certain decays to final states which are not $CP$ eigenstates, it is still
possible to extract weak phases model independently from the interference
between mixing and decay.  This occurs if both $B^0$ and $\B0bar$ can decay
into a particular final state and its $CP$ conjugate, but there is only one
contribution to each of these decay amplitudes.  In this case no assumptions
about hadronic physics are needed, even though $|\ov{A}_f/A_f| \neq 1$ and
$|\ov{A}_{\ov{f}}/A_{\ov{f}}| \neq 1$.

\index{decay!$B_s \to D_s^\pm K^\mp$ }
The most important example is $B_s\to D_s^\pm K^\mp$, which allows a model
independent determination of $\gamma$~\cite{ADK}.
Both $\B0bar_s$ and $B_s^0$
can decay to $D_s^+ K^-$ and $D_s^- K^+$, but the only decay processes are the
tree level $b\to c\bar u s$ and $b\to u \bar c s$ transitions, and their $CP$
conjugates.  One can easily see that
\begin{equation}
{\ov{A}_{D_s^+ K^-} \over A_{D_s^+ K^-} } = \frac{A_1}{A_2}
  \bigg( {V_{cb} V_{us}^* \over V_{ub}^* V_{cs}} \bigg) \,, \qquad
{\ov{A}_{D_s^- K^+} \over A_{D_s^- K^+} } = \frac{A_2}{A_1}
  \bigg( {V_{ub} V_{cs}^* \over V_{cb}^* V_{us}} \bigg) \,, 
\label{1:A12ratio}
\end{equation}
where the ratio of amplitudes, $A_1/A_2$, includes the strong phases, and is an
unknown complex number of order unity.  It is important for the utility of this
method that $|V_{cb} V_{us}|$ and $|V_{ub} V_{cs}|$ are comparable in
magnitude, since both are of order $\lambda^3$ in the Wolfenstein
parameterization.  Eqs.~(\ref{gtfres}) and (\ref{gbtfres}) show that measuring
the four time dependent decay rates determine both $\lambda_{D_s^+K^-}$ and
$\lambda_{D_s^-K^+}$.  The unknown $A_1/A_2$ ratio drops out from their product
\begin{equation}
\lambda_{D_s^+ K^-}\, \lambda_{D_s^- K^+} = 
  \bigg( {V_{tb}^* V_{ts} \over V_{tb} V_{ts}^*} \bigg)^{\!2}
  \bigg( {V_{cb} V_{us}^* \over V_{ub}^* V_{cs}} \bigg)
  \bigg( {V_{ub} V_{cs}^* \over V_{cb}^* V_{us}} \bigg)
= e^{-2i(\gamma-2\beta_s-\beta_K)}\,.
\end{equation}
The first factor is the Standard Model value of $q/p$ in $B_s$ mixing.  The
angles $\beta_s$ and $\beta_K$ occur in ``squashed" unitarity triangles;
$\beta_s$ defined in Eq.~(\ref{1:eq:betasdef}) is of order $\lambda^2$ and
$\beta_K = \arg(-V_{cs}V_{cd}^*/V_{us}V_{ud}^*)$ is of order $\lambda^4$.
Thus, this mode can provide a precise determination of $\gamma$ (or
$\gamma-2\beta_s)$; the determination of $\beta_s$ is discussed in Chapter 6,
e.g., from $B_s \to \psi\, \eta^{(\prime)}$.

\index{decay!$B_d \to D^\pm \pi^\mp$ }
\index{decay!$B_d \to D^{*\pm} \pi^\mp$ }
In exact analogy to the above, the $B_d\to D^{(*)}{}^\pm \pi^\mp$ decays can
determine $\gamma + 2\beta$, since $\lambda_{D^+\pi^-}\, \lambda_{D^-\pi^+} =
\exp\, [-2i(\gamma + 2\beta)]$.  In this case, however, the two decay
amplitudes differ in magnitude by order $\lambda^2$, and therefore the $CP$
asymmetries are expected to be much smaller, at the percent level.

\boldmath
\subsection{$\Delta F=2$ vs.~$\Delta F=1$ \CP\ violation}
\unboldmath
\index{CP violation@\CP\ violation!direct}
\index{CP violation@\CP\ violation!indirect}
\index{CP violation@\CP\ violation!in $\Delta F=1$ transitions}

At low energies flavor-changing transitions are described by effective
Hamiltonians, which are discussed in detail in Sec.~\ref{1:Heff}.
Decays are mediated by the $\Delta F=1$ Hamiltonian $H^{\dfo}$, whereas
mixing is induced by the $\Delta F=2$ Hamiltonian.
The changing flavor is $F=B$ for $B$ decays and $F=S$ for $K$ decays.
In kaon physics it is customary to distinguish $\Delta F=1$ \CP\
violation, which is often called \emph{direct} \CP\ violation, from
$\Delta F=2$ \CP\ violation, called \emph{indirect} \CP\ violation.
Here we compare this classification with the three types of \CP\
violation in $B$ decays discussed in Sec.~\ref{ch1:sect:ty}.

If we can find phase transformations of the quark fields in
\eq{pht} which leave the Hamiltonian invariant, $\CP\, H^{\dfo}
(\CP)^{-1} = H^{\dfo}$, then we conclude that the \dfo\ interaction
conserves \CP. Analogously we could define \CP\ violation and \CP\ 
conservation in $H^{\dft}$, but a $B$ physics experiment probes only
one matrix element of $H^{\dft}$, namely $M_{12}$. One can always find
a phase transformation which renders $M_{12}$ real and thereby shifts
the \CP\ violation from $H^{\dft}$ completely into $H^{\dfo}$. The
converse is not true, since one can explore the different couplings in
$H^{\dfo}$ by studying different decay modes. This leaves three
scenarios to be experimentally distinguished:%
\index{CP violation@\CP\ violation!superweak}
\begin{itemize}
\item[i)] With rephasing of the quark fields one can achieve
     $\CP\, H (\CP)^{-1} =  H$ for both $H^{\dft}$ and 
     $H^{\dfo}$: The theory conserves \CP.     
\item[ii)] One can rephase the quark fields such that 
     $\CP\, H^{\dfo} (\CP)^{-1} = H^{\dfo}$, but for this phase 
     transformation  $\CP\,H^{\dft} (\CP)^{-1} \neq H^{\dft}$.
     This scenario is called \emph{superweak} \cite{w62}.            
\item[iii)]  $\CP\,H^{\dfo} (\CP)^{-1} \neq H^{\dfo}$ for any phase 
    convention of the quark fields. This scenario is realized 
    in the CKM mechanism of the Standard Model.
\end{itemize}
Historically, after the discovery of \CP\ violation in 1964 \cite{ccft},
it was of prime interest to distinguish the second from the third 
scenario in kaon physics. The recent establishment of $\e_K^\prime
\neq 0$ has shown that possibility iii) is realized in kaon physics. 
\index{CP violation@\CP\ violation!kaon physics!$\e_K^\prime$}
\index{CP violation@\CP\ violation!kaon physics!$\e_K$}

It is difficult (but possible) to build a viable theory with
$\e_K^\prime \neq 0$ in which \CP\ violation in the $B$ system is of
the superweak type. Still we can play the rules of the kaon game and
ask, what must be measured to rule out the superweak scenario. 
Clearly, \CP\ violation in decay unambiguously proves $\dfo$ \CP\ violation. 
\CP\ violation in mixing purely
measures \CP\ violation in the \dft\ transition. It measures the
relative phase between $M_{12}$ and the decay matrix
$\Gamma_{12}$. $\Gamma_{12}$ arises at second order in the \dfo\
interaction, from $\sum_f A_f^* \ov{A}_f$, where $A_f$ and $\ov{A}_f$
are the \dfo\ decay amplitudes introduced in
\eq{defafch}. $M_{12}$ receives contributions at first order in
$H^{\dft}$ and at second order in $H^{\dfo}$. Interference type \CP\
violation measures the difference between the mixing phase
$\phi_M=\arg M_{12}$ and twice the weak phase $\phi_f$ of some decay
amplitude $\ov{A}_f$. Both types of \CP\ violations are therefore
sensitive to relative phases between $H^{\dft}$ and $H^{\dfo}$. Yet
the measurement of a single \CP\ violating observable of either type
is not sufficient to rule out the superweak scenario, because we can
always rephase $\Gamma_{12}$ or $A_f$ to be real. However, the measurement 
of interference type \CP\ violation in two different decay modes with
different results would prove that two weak
phases in $H^{\dfo}$ are different. Since $\phi_{f_1}-\phi_{f_2}$ is
a rephasing invariant observable, no field transformation in
\eq{pht} can render $H^{\dfo}$ real and \dfo\ \CP\ violation is
established. Hence for example the measurement of different \CP\
asymmetries in $B_d \to J/\psi K_S$ and $B_d \to \pi^+ \pi^-$ is
sufficient to rule out the superweak scenario. Interestingly,
$\e_K^\prime$ contains both of the discussed types of $\Delta S=1$
\CP\ violation: \CP\ violation in decay and the difference of two
interference type \CP\ violating phases.  Since in both $K$-
and $B$ physics the dominant decay modes have the same weak phases,
essentially no new information is gained by comparing \CP\ violation
in mixing with interference type
\CP\ violation in a dominant decay mode. We will see this in
Sec.~\ref{1:sect:kaon} when comparing $\e_K$ with the semileptonic
\CP\ asymmetry in $K_L$ decays.

\section{Theoretical Tools}
\label{1:tools}

This section provides a brief review of the tools used to
derive theoretical predictions for $B$ mixing and decays.
The theory of $b$ production and fragmentation is discussed in
Chapter~9.

The principal aim of $B$ physics is to learn about the short distance
dynamics of nature.  Short distance physics couples to $b$ quarks,
while experiments detect $b$-flavored hadrons.  One therefore needs
to connect the properties of these hadrons in terms of the
underlying $b$ quark dynamics.  Except for a few special cases, this
requires an understanding of the long distance, nonperturbative
properties of QCD.  It is then useful to separate long distance
physics from short distance using an operator product expansion (OPE)
or an effective field theory.
The basic idea is that interactions at higher scales give rise to
local operators at lower scales. This allows us to think about the
short distance phenomena responsible for the flavor structure in
nature independent of the complications due to hadronic physics, which
can then be attacked separately.  This strategy can lead to very
practical results: the hadronic part of an interesting process may be
related by exact or approximate symmetries to the hadronic part of a
less interesting or more easily measured process.

In the description of $B$ decays several short distances arise.  
$CP$ and flavor violation stem from the weak scale and, probably, even shorter
distances.  These scales are separated from the scale $m_B$ with an OPE,
leading to an effective Hamiltonian for flavor changing
processes.\index{operator product expansion} This is reviewed in
Sec.~\ref{1:Heff}.  Furthermore, the $b$ and (to a lesser extent) the $c$
quark masses are much larger than $\Lambda_{\rm QCD}$.  In the limit
$\Lambda_{\rm QCD}/m_Q\to 0$, the bound state dynamics simplify.  Implications
for exclusive processes are discussed in Sec.~\ref{1:HQET}.  For inclusive
decays one can apply an OPE again, the so-called heavy quark expansion,
reviewed in Sec.~\ref{1:HQE}.
Despite the simplifications, these expansions still require hadronic matrix
elements, so we briefly review lattice QCD in Sec.~\ref{1:lattice}.

\subsection{Effective Hamiltonians}
\label{1:Heff}

\begin{figure}[bt]
\centerline{\epsfysize=4cm \epsffile{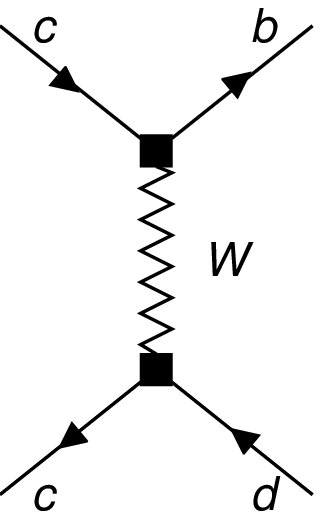} \hspace{2cm}
  \epsfysize=3.5cm \epsffile{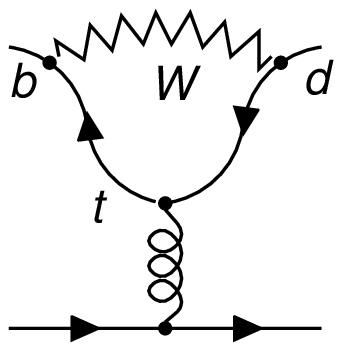} }\vspace{3ex}
\caption[Standard Model weak decay diagrams]%
{Standard Model $W$ exchange diagram and penguin diagram with 
internal top quark for the decay $b\to c \ov{c} d$.}
\label{ch1:fig:wpeng}
\end{figure}

To predict the decay rate of a $B$ meson into some final
state $f$, one must calculate the transition amplitude ${\cal M}$
for $B\to f$. In general there are many contributions to
${\cal M}$, each of which is, at the quark level, pictorially
represented by Feynman diagrams such as those in
Fig.~\ref{ch1:fig:wpeng}.

Quark diagrams are a poor description for the decay amplitude of a
$B$ meson. The quarks feel the strong interaction, whose nature
changes drastically over the distances at which it is probed: At short
distances much smaller than $1/\Lambda_{\rm QCD}$ the strong
interaction can be described perturbatively by dressing the lowest
order diagrams in Fig.~\ref{ch1:fig:wpeng} with gluons.  When traveling
over a distance of order $1/\Lambda_{\rm QCD}$, however, quarks and
gluons hadronize and QCD becomes nonperturbative.  Therefore the
physics from different length scales, or, equivalently, from different
energy scales must be treated differently.  One theoretical tool for
this is the \emph{operator product expansion} (OPE)~\cite{wil}.
\index{operator product expansion}  
Schematically the decay amplitude ${\cal M}$ is expressed as
\beq
{\cal M} = - \frac{4\,G_F}{\sqrt{2}}\, V_{\rm CKM} 
    \sum_{j} C_j (\mu)\, \langle f | O_j (\mu) | B \rangle 
    \bigg[ 1+{\cal O} \bigg( \frac{m_b^2}{M_W^2} \bigg)\bigg] ,
\label{ope}
\eeq
where $\mu$ is a renormalization scale.
Physics from distances shorter than $\mu^{-1}$ is contained in the
Wilson coefficients\index{Wilson coefficient} $C_j$, and
physics from distances longer than $\mu^{-1}$ is accounted for by the
hadronic matrix elements $\bra{f}O_j\ket{B}$ of the
local operators~$O_j$.
In principle, there are infinitely many terms in the OPE, but higher
dimension operators yield contributions suppressed by powers of
$m_b^2/m_W^2$.
From a practical point of view, therefore, the sum in \eq{ope} ranges
over operators of dimension five and six.

All dependence on heavy masses $M\gg\mu$ such as $m_t$, $M_W$ or the 
masses of new undiscovered heavy
particles is contained in $C_j$. By convention one factors out
$4\,G_F/\sqrt{2}$ and the CKM factors, which are denoted by
$V_{\rm CKM}$ in \eq{ope}. 
On the other hand, the matrix element $\langle f | O_j | B\rangle$ 
of the $B\to f$ transition contains information from scales, such as 
$\Lambda_{\rm QCD}$, that are below~$\mu$. 
Therefore, they can only be evaluated using nonperturbative methods 
such as lattice calculations (cf., Sec.~\ref{1:lattice}), QCD sum 
rules, or by using related processes to obtain them from experiment.

An important feature of the OPE in (\ref{ope}) is the
universality of the coefficients $C_j$; they are independent of the
external states, i.e., their numerical value is the same for all final
states $f$ in (\ref{ope}). Therefore one can view the
$C_j$'s as effective coupling constants and the $O_j$'s as the
corresponding interaction vertices. Thus one can introduce
the \emph{effective Hamiltonian} 
\index{Hamiltonian!$|\Delta B|=1$}
\beq
\hbo = \frac{4\, G_F}{\sqrt{2}}\, V_{\rm CKM} 
  \sum_{j} C_j \, O_j + \mathrm{h.c.} \label{heff}
\eeq
An amplitude calculated from \hbo\, defined at a scale of order $m_b$,
reproduces the corresponding Standard Model result up to corrections 
of order $m_b^2/M_W^2$ as indicated in (\ref{ope}). 
Hard QCD effects can be included
perturbatively in the Wilson coefficients, i.e., by calculating
Feynman diagrams with quarks and gluons.

\begin{figure}[bt]
\centerline{\epsfxsize=0.98\textwidth \epsffile{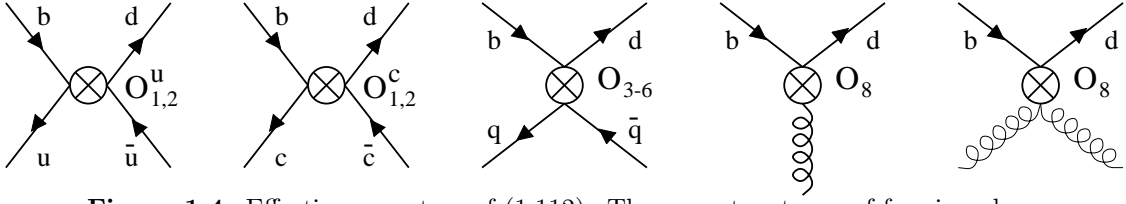}}
\caption[Effective Hamiltonian weak decay diagrams]%
{Effective operators of \eq{basis}.
There are two types of fermion-gluon couplings associated
with the chromomagnetic operator $O_8$.}
\label{ch1:fig:ops}
\end{figure}

The set of
operators $O_j$ needed in (\ref{heff}) depends on the flavor
structure of the physical process under consideration. 
Pictorially the operators are
obtained by contracting the lines corresponding to heavy particles in
the Feynman diagrams to a point. The tree level diagram involving the
$W$ boson in Fig.~\ref{ch1:fig:wpeng} generates the operator $O_2^c$
shown in Fig.~\ref{ch1:fig:ops}.  
In the Standard Model only two operators occur for
$b\to c \ov{u} d$ transitions,
\index{operator!current-current}
\index{operator!$O_{1-2}$}
\beq \label{1:O12def}
O_1=\bar{b}_L^{\alpha} \gamma_{\mu} c_L^{\beta} \, 
  \bar{u}_L^{\beta} \gamma^{\mu} d_L^{\alpha} \,, \qquad 
O_2=\bar{b}_L^{\alpha} \gamma_{\mu} c_L^{\alpha} \, 
  \bar{u}_L^{\beta} \gamma^{\mu} d_L^{\beta} \,,
\eeq
where $\alpha$ and $\beta$ are color indices.  These arise from $W$
exchange shown in Fig.~\ref{ch1:fig:wpeng}, and QCD corrections to it.
Operators and Wilson coefficients at different scales $\mu_1$ and $\mu_2$
are related by a renormalization group transformation.
\index{renormalization group} $C_{1,2} (\mu_1)$ is not just a function 
of $C_{1,2} (\mu_2)$, but a linear combination of both $C_1 (\mu_2)$
and $C_2 (\mu_2)$. This feature is called operator mixing. It is
convenient to introduce the linear combinations $O_\pm = (O_2 \pm
O_1)/2$, which do not mix with each other.  Their coefficients can be
more easily calculated and are related to $C_1$ and $C_2$ by
$C_{\pm}=C_2\pm C_1$.
\index{operator!$O_\pm$}

The Hamiltonian for $\Delta B=1$ and $\Delta C=\Delta S=0$ transitions
consists of more operators, because it must also accommodate for
the so-called penguin diagram with an internal top quark, shown in
Fig.~\ref{ch1:fig:wpeng}.
The corresponding operator basis reads
\index{operator!$O_{3-6}$}
\index{operator!$O_{8}$}
\index{operator!QCD penguin}
\index{operator!chromomagnetic}
\begin{equation}
  \begin{array}{rclrcl}
O_1^{c} &=& 
  \bar{d}{}_L^{\alpha} \gamma_{\mu}  c_L^{\beta}  \, 
  \bar{c}{}_L^{\beta}  \gamma^{\mu}  b_L^{\alpha} \,, \qquad &
O_1^{u} &=& 
  \bar{d}{}_L^{\alpha} \gamma_{\mu}  u_L^{\beta}  \, 
  \bar{u}{}_L^{\beta}  \gamma^{\mu}  b_L^{\alpha} \,, \\[3mm]
O_2^{c} &=& 
  \bar{d}{}_L^{\alpha} \gamma_{\mu}  c_L^{\alpha}  \, 
  \bar{c}{}_L^{\beta}  \gamma^{\mu}  b_L^{\beta} \,, \qquad &
O_2^{u} &=& 
  \bar{d}{}_L^{\alpha} \gamma_{\mu}  u_L^{\alpha}  \, 
  \bar{u}{}_L^{\beta}  \gamma^{\mu}  b_L^{\beta} \,, \\[3mm]
O_3 &=& \displaystyle \sum_{q=u,d,s,c,b} \!\!
  \bar{d}{}_L^{\alpha} \gamma_{\mu}  b_L^{\alpha}  \, 
  \bar{q}{}_L^{\beta}  \gamma^{\mu}  q_L^{\beta} \,, \qquad & 
O_4 &=& \displaystyle \sum_{q=u,d,s,c,b} \!\!
  \bar{d}{}_L^{\alpha} \gamma_{\mu}  b_L^{\beta}  \, 
  \bar{q}{}_L^{\beta}  \gamma^{\mu}  q_L^{\alpha} \,, \\[5mm]
O_5 &=& \displaystyle \sum_{q=u,d,s,c,b} \!\!
  \bar{d}{}_L^{\alpha} \gamma_{\mu}  b_L^{\alpha}  \, 
  \bar{q}{}_R^{\beta}  \gamma^{\mu}  q_R^{\beta} \,, \qquad &
O_6 &=& \displaystyle \sum_{q=u,d,s,c,b} \!\!
  \bar{d}{}_L^{\alpha} \gamma_{\mu}  b_L^{\beta}  \, 
  \bar{q}{}_R^{\beta}  \gamma^{\mu}  q_R^{\alpha} \,, \\[5mm]
O_8 & = & \displaystyle - \frac{g}{16 \pi^2} \, m_b \, 
    \bar{d}_L \sigma^{\mu \nu} G^a_{\mu \nu} T^a b_R \,. 
  \end{array}
\label{basis}
\end{equation}
These operators are also depicted in Fig.~\ref{ch1:fig:ops}.
In $O_8$, $G_{\mu\nu}^{a}$ is the chromomagnetic field strength tensor.
The operators are grouped into classes, based on their origin:
$O_1$ and $O_2$ are called \emph{current-current operators},
$O_3$ through $O_6$ are called \emph{four-quark penguin operators},
and $O_8$ is called the \emph{chromomagnetic penguin operator}.%
\footnote{In the literature one also finds $O_7$ and $O_8$ with the
opposite signs. In QCD and QED the sign of the gauge coupling is
convention dependent, and \eq{basis} is consistent with the values
for $C_8$ in Table~\ref{tab}, if the Feynman rule for the quark-gluon
coupling is chosen as $+ig$.}

The operators in \eq{basis} arise from the lowest order in the
electroweak interaction, i.e., diagrams involving a single $W$ bosons
plus QCD corrections to it. In some cases, especially when isospin breaking
plays a role, one also needs to consider penguin diagrams which are of
higher order in the electroweak fine structure constant $\alpha_{ew}$.
They give rise to the \emph{electroweak penguin operators}:
\index{operator!electroweak penguin}
\index{operator!magnetic}
\index{operator!$O_{7}$}
\index{operator!$O_{7-10}^{\rm ew}$}
\begin{equation}
  \begin{array}{rclrcl}
O_7 &=& \displaystyle - {e \over 16\pi^2}\, m_b\, 
  \bar d{}_L^{\alpha} \, \sigma^{\mu\nu} F_{\mu\nu}\, b_R^{\alpha} \,, \\[3mm]
O_7^{\rm ew} &=& \displaystyle \frac{3}{2} 
\sum_{q=u,d,s,c,b} \!\! e_q \, 
  \bar{d}{}_L^{\alpha} \gamma_{\mu}  b_L^{\alpha}  \, 
  \bar{q}{}_R^{\beta}  \gamma^{\mu}  q_R^{\beta} \,,  \qquad &
O_8^{\rm ew} &=& \displaystyle \frac{3}{2} 
\sum_{q=u,d,s,c,b} \!\!  e_q \, 
  \bar{d}{}_L^{\alpha} \gamma_{\mu}  b_L^{\beta}  \, 
  \bar{q}{}_R^{\beta}  \gamma^{\mu}  q_R^{\alpha} \,, \\[5mm]
O_9^{\rm ew} &=& \displaystyle \frac{3}{2} 
   \sum_{q=u,d,s,c,b} \!\! e_q \, 
  \bar{d}{}_L^{\alpha} \gamma_{\mu}  b_L^{\alpha}  \, 
  \bar{q}{}_L^{\beta}  \gamma^{\mu}  q_L^{\beta} \,,  \qquad &
O_{10}^{\rm ew} &=& \displaystyle \frac{3}{2} 
   \sum_{q=u,d,s,c,b} \!\! e_q \, 
  \bar{d}{}_L^{\alpha} \gamma_{\mu}  b_L^{\beta}  \, 
  \bar{q}{}_L^{\beta}  \gamma^{\mu}  q_L^{\alpha} \,.
  \end{array}
\label{basis2}
\end{equation}
Here $F^{\mu\nu}$ is the electromagnetic field strength tensor, and
$e_q$ denotes the charge of quark $q$.  
The magnetic (penguin) operator $O_7$ is also of key importance
for the radiative decay $b \to d \gamma$.
Eqs.~(\ref{basis}) and (\ref{basis2}) reveal that
there is no consensus yet on how to number the operators
consecutively.

For semileptonic decays the following additional operators occur
\index{operator!semileptonic}
\index{operator!$O_{9-11}$}
\begin{equation}
  \begin{array}{rclrcl}
O_9 &=& \displaystyle {e^2 \over 16\pi^2}\, \bar d_L\, \gamma_\mu\, b_L \,
  \bar\ell\, \gamma^\mu\, \ell \,,  \qquad &
O_{10} &=& \displaystyle {e^2 \over 16\pi^2}\, \bar d_L\, \gamma_\mu\, b_L \,
  \bar\ell\, \gamma^\mu\gamma_5\, \ell \,, \\[3mm]
O_{11} &=& \displaystyle {e^2 \over 32\pi^2\sin^2\theta_W}\, 
  \bar d_L\, \gamma_\mu\, b_L \, \bar\nu_L \, \gamma^\mu \, \nu_L \,,
  \end{array}
\label{basis3}
\end{equation}
and the counterparts of these with $\bar d_L$ replaced by $\bar s_L$.

Hence the $\Delta B=1$ and $\Delta C=\Delta S=0$ Hamiltonian reads:
\beq
\hbo = \frac{4\, G_F}{\sqrt{2}}\, \bigg\{ 
    \sum_{j=1}^2 C_j \left( \xi_c O_j^c + \xi_u O_j^u \right) 
    - \xi_t \sum_{j=3}^{11} C_j \, O_j  
    - \xi_t \sum_{j =7}^{10} C_j^{\rm ew} \, O_j \bigg\} 
    + \mathrm{h.c.} \,, 
\label{hd}
\eeq
where
\beq
\xi_{q} = V_{q b}^* V_{q d} \,.
\eeq
Note that $  \xi_u + \xi_c + \xi_t = 0$ by unitarity of the CKM matrix.
The corresponding operator basis for $b\to s$
transitions is obtained by simply exchanging $d$ with $s$ in
\eq{basis}, \eq{basis2} and \eq{basis3} and changing $\xi_i$
accordingly.  

The operators introduced above are sufficient to describe nonleptonic
transitions in the Standard Model to order~$G_F$.
In extensions of the Standard Model, on the other hand, the
short distance structure can be very different.
Additional operators with new Dirac structures, whose standard Wilson 
coefficients vanish, could enter the effective Hamiltonian. 
A~list of these operators, including their RG evolution, can
be found in \cite{bmu}.

\index{renormalization group}
\index{Wilson coefficient}
In general the QCD corrections to the transition amplitude ${\cal M}
(B \to f)$ contain large logarithms such as $\ln (m_b/M_W)$
which need to be resummed to all orders in $\alpha_s$.  The OPE splits
these logarithms as $\ln (m_b/M_W) = \ln (\mu/M_W) - \ln (\mu/m_b)$. 
The former term resides in the Wilson coefficients, the latter
logarithm is contained in the matrix element.  Such large
logarithms can be summed to all orders by solving
\emph{renormalization group} (RG) equations for the
$C_j$'s. These RG-improved perturbation series are
well-behaved. The minimal way to include QCD corrections is the
\emph{leading logarithmic approximation}. The corresponding 
\emph{leading order} (LO) Wilson coefficients comprise $[\alpha_s \ln
(m_b/M_W)]^n$ to all orders $n=0,1,2,\ldots$ in perturbation
theory. This approximation has certain conceptual deficits and is too
crude for the precision of the experiments and the accuracy of present
day lattice calculations of the hadronic matrix elements. The
\emph{next-to-leading order}\ (NLO) results for the $C_j$'s comprises
in addition terms of order $\alpha_s \, [\alpha_s
\ln (m_b/M_W)]^n, n=0,1,2,\ldots$. The Wilson coefficients depend on
the unphysical scale $\mu$ at which the OPE is performed. Starting
from the NLO the $C_j$'s further depend on the
\emph{renormalization scheme}, which is related to the way one treats
divergent loops in Feynman diagrams.  In an exact calculation both the
scale and scheme dependence cancels between the coefficients and the
matrix elements, but in practice the calculation of matrix elements
with the correct scale and scheme dependence can be a non-trivial
task.  The appearance of the scale and scheme dependence in the
coefficients is inevitable.  The OPE enforces the short distance
physics involving heavy masses like $M_W$ and $m_t$ to belong to the
$C_j$'s, while the long distance physics is contained in the matrix
elements. But a constant number can be attributed to either of them.
Switching from one scheme to another or changing the scale $\mu$ just
shuffles constant terms between the Wilson coefficients and the matrix
elements.  There is no unique definition of ``scheme independent''
Wilson coefficients.

The numerical values for the renormalization group improved Wilson
coefficients can be found in Table~\ref{tab}.
\begin{table}[p]
\[
\begin{array}{clc|*{7}{@{\hspace{1em}}r}}
\hline\hline
\alpha_s (M_Z) &  \mathrm{scheme} &  
  \mu\,(\mathrm{GeV}) & 
	\multicolumn{1}{c}{C_1} & 
	\multicolumn{1}{c}{C_2~} & 
	\multicolumn{1}{c}{C_3~} & 
	\multicolumn{1}{c}{C_4} & 
	\multicolumn{1}{c}{C_5~} & 
	\multicolumn{1}{c}{C_6} & 
	\multicolumn{1}{c}{C_8}
\\ \hline
0.112 &
   {\rm LO} & 4.8 
&       -0.229    &        1.097    &         0.010   &       -0.024 
&        0.007    &       -0.029    &        -0.146 \\
&     & 2.4 
&        -0.325   &        1.149    &        0.015    &        -0.033
&        0.009    &        -0.043   &         -0.161 \\ 
&     & 9.6
&        -0.155   &        1.062    &        0.007    &        -0.016
&        0.005    &        -0.019   &        -0.133   \\ 
\cline{2-10}
& {\rm NDR} & 4.8 
&        -0.160   &        1.066    &        0.011    &        -0.031
&        0.008    &        -0.035   &        \\ 
&     & 2.4
&        -0.245   &        1.110    &        0.017    &        -0.043
&        0.009    &        -0.052   &        \\
&     & 9.6
&        -0.093   &        1.036    &        0.008    &        -0.021
&        0.006    &        -0.023   &        \\
\cline{2-10}
& {\rm HV}  & 4.8 
&        -0.177   &        0.993    &        0.009    &        -0.024
&        0.007    &        -0.026   &         \\
&     & 2.4 
&        -0.260    &        1.020     &        0.014    &        -0.033
&        0.010     &        -0.038   &        \\
&     & 9.6 
&        -0.111   &        0.975    &        0.006    &        -0.015
&        0.005    &        -0.017   &         \\
\hline
0.118 & 
LO & 4.8 &  -0.249 & 1.108 & 0.011 & -0.026 & 0.008 & -0.031 & -0.149 \\ 
&  & 2.4 & -0.361 &  1.169 & 0.017 & -0.036 & 0.010 & -0.048 & -0.166 \\
&  & 9.6 & -0.167 &  1.067 & 0.007 & -0.018 & 0.005 & -0.020 & -0.135 \\
\cline{2-10}
& NDR & 4.8  &  -0.174 & 1.073 & 0.013 & -0.034 & 0.009 & -0.038 &  \\
&     & 2.4  &  -0.272 & 1.124 & 0.020 & -0.047 & 0.010 & -0.060 &  \\
&     & 9.6  &  -0.100 & 1.039 & 0.008 & -0.024 & 0.006 & -0.025 &  \\
\cline{2-10}
& HV & 4.8 & -0.192 & 0.993 & 0.010 & -0.026 & 0.008 & -0.028 & \\ 
&    & 2.4 & -0.286 & 1.022 & 0.016 & -0.036 & 0.011 & -0.042 & \\
&    & 9.6 & -0.120 & 0.972 & 0.006 & -0.017 & 0.005 & -0.018 & \\
\hline
0.124 & 
  LO  & 4.8 
&        -0.272   &        1.120    &        0.012    &        -0.028
&        0.008    &        -0.035   &        -0.153   \\ 
&     & 2.4 
&        -0.403   &        1.194    &        0.019    &        -0.040
&        0.011    &        -0.055   &        -0.172    \\
&     & 9.6 
&        -0.180   &        1.073    &        0.008    &        -0.019
&        0.006    &        -0.022   &        -0.138    \\ 
\cline{2-10}
& NDR & 4.8 
&        -0.190   &        1.082    &        0.014    &        -0.037
&        0.009    &        -0.043   &        \\
&     & 2.4  
&        -0.303   &        1.142    &        0.022    &        -0.054
&        0.011    &        -0.069   &        \\
&     & 9.6
&        -0.108   &        1.042    &        0.009    &        -0.025
&        0.007    &        -0.028   &        \\
\cline{2-10}
& HV  & 4.8 
&        -0.208   &        0.993    &        0.011    &        -0.028
&        0.008    &        -0.031   &        \\
&     & 2.4 
&        -0.316   &        1.025    &        0.018    &        -0.040
&        0.012    &        -0.048   &        \\
&     & 9.6
&        -0.129   &        0.970    &        0.007    &        -0.018
&        0.006    &        -0.019   &        \\
\hline\hline
\end{array}
\] \vspace*{-2pt}
\caption[Wilson coefficients in the leading and next-to-leading
order]{QCD Wilson coefficients in the leading and next-to-leading
order. The NLO running of $\alpha_s$ has been used in both the LO
and NLO coefficients. $\alpha_s(M_Z) =0.112$, $0.118$, $0.124$ implies 
$\alpha_s(4.8\, \mathrm{GeV}) = 0.196$, $0.216$, $0.238$. The 
corresponding values of the five-flavor QCD scale parameter 
$\Lambda_{\overline{\mathrm{MS}}}$ are $159$, $226$ and $312$ MeV.
The dependence on $m_t (m_t)$, here taken as $168$ GeV, is negligible.   
The NLO coefficients are listed for the NDR and HV scheme. 
There are two different conventions for the HV scheme, here we use the
one adopted in \cite{nlo}. The HV coefficients tabulated in
\cite{bbl} are related to our $C_j^{HV}\!$'s by 
$C_j^{HV}(\!\!\mbox{\cite{bbl}}) = 
  [1+ 16/3 \cdot \alpha_s (\mu)/(4\pi) ] C_j^{HV} $. 
Small QED corrections have been omitted.\index{Wilson coefficient!table} }
\label{tab}
\end{table}
The NLO coefficients are
listed for two popular schemes, the \emph{naive dimensional
regularization}\ (NDR) scheme and the \emph{'t Hooft-Veltman}\ (HV)
scheme. These results have been independently obtained by the Rome and
Munich groups \cite{nlo}. The situation with $C_8$ is special: To
obtain the LO values for $C_{1-6}$ in Table~\ref{tab} one must
calculate one-loop diagrams. The calculation of $C_8$, however,
already involves two-loop diagrams in the leading order. This implies
that even the LO expression for $C_8$ is scheme dependent. The
tabulated value corresponds to the commonly used ``effective''
coefficient $C_8$ introduced in
\cite{cfmrs}, which is defined in a scheme independent way.
To know the NLO value for $C_8$ one must calculate three-loop
diagrams.  The operator basis in \eq{basis} is badly suited for this
calculation and hence a different one has been used \cite{cmm}. For
the basis in \eq{basis} the NLO value for $C_8$ is not known, we
therefore leave the corresponding rows open. In Table~\ref{tab} small
corrections proportional to $\alpha_{ew}$ have been omitted. For the 
Wilson coefficients of the electroweak penguin operators in
\eq{basis2} and the semileptonic operators in \eq{basis3}  we refer
the reader to \cite{bbl}.

\index{Hamiltonian!$|\Delta B|=2$}
We can derive an effective Hamiltonian for the \dbt\ transition, which
induces \bbmd, just in the same way as discussed above for \dbo. 
In the Standard Model only a single operator $Q$ arises:%
\footnote{Once again in \eq{ch1:uli:h2}, new short distance 
physics can generate Wilson coefficients for additional operators.}
\beq
H^{|\Delta B|=2} = \frac{G_F^2}{4 \pi^2}\, ( V_{tb} V_{td}^* )^2 \,
    C^{|\Delta B|=2}( m_t, M_W, \mu )\, Q (\mu) + h.c.
\label{ch1:uli:h2}
\eeq
with
\beq 
  Q = \ov{d}_L \gamma_{\nu} b_L \, \ov{d}_L \gamma^{\nu} b_L .
\label{ch1:uli:defq}
\eeq
The Wilson coefficient\index{Wilson coefficient!$|\Delta B|=2$} is 
\beq
C^{|\Delta B|=2} ( m_t, M_W, \mu ) = M_W^2\, 
    S \bigg( \frac{m_t^2}{M_W^2} \bigg)\, \eta_B\, b_B(\mu) \,.
\label{ch1:uli:c2} 
\eeq
It contains the \textit{Inami-Lim function}\ \cite{il}
\beq
S(x) = x \left[ \frac{1}{4} + \frac{9}{4}
    \frac{1}{1-x} - \frac{3}{2} \frac{1}{(1-x)^2} \right]
    - \frac{3}{2} \left[ \frac{x}{1-x} \right]^3 \ln x \,,
\label{ch1:uli:sxt}
\eeq
which is calculated from the box diagram in Fig.~\ref{ch1:fig:box}.
The coefficients $\eta_B$ and $b_B$ in (\ref{ch1:uli:c2}) account for
short distance QCD corrections. In the next-to-leading order of QCD
one finds $\eta_B=0.55$ \cite{bjw}. $b_B$ depends on the
renormalization scale $\mu={\cal O}(m_b)$, at which the matrix element
$\bra{B^0_d} Q \ket{\ov{B}{}^0_d}$ is calculated.  $b_B(\mu)$ equals
$[ \alpha_s (\mu) ]^{-6/23}$ in the LO. The $\mu$-dependence of
$b_B(\mu)$ cancels the $\mu$-dependence of the matrix element to the
calculated order. The same remark applies to the dependence of
$b_B(\mu)$ on the renormalization scheme in which the calculation is
carried out.  One parameterizes the hadronic matrix elements as
\bey
\bra{B^0} Q (\mu) \ket{\ov{B}{}^0} &=& 
    \frac{2}{3} f_B^2 m_B^2 \frac{\widehat{B}_B}{b_B (\mu)} 
 , \label{defbb}
\eey
so that $\widehat{B}_B$ \index{hadronic parameter!$\widehat{B}_B$}
is scale and scheme independent.  The
effective Hamiltonian for \bbms\ is obtained as usual by replacing $d$
with $s$ in \eq{ch1:uli:h2} and
\eq{ch1:uli:defq}.  The Wilson coefficient in \eq{ch1:uli:c2} does not
depend on the light quark flavor.

\subsection{Heavy quark effective theory}
\label{1:HQET}
\index{heavy quark effective theory (HQET)|(}

In hadrons composed of a heavy quark and light degrees of freedom (light 
quarks, antiquarks, and gluons), the binding energy, which is of 
order $\lqcd$, is small compared to the heavy quark mass~$m_Q$.
In the limit $m_Q\gg\lqcd$, the heavy quark acts approximately as a 
static color-triplet source,%
\footnote{For the same reason, heavy quark symmetries also apply to 
hadrons composed of two heavy and a light quark, because the color 
quantum numbers of the two heavy quarks combine to an antitriplet.}
and its spin and flavor do not affect the light degrees of freedom.
This is analogous to atomic physics, where isotopes with different 
nuclei have nearly the same properties.
Thus, the properties of heavy-light hadrons are related by a 
symmetry, called heavy quark symmetry\index{heavy quark symmetry} 
(HQS)~\cite{HQS,Shur,Nuss,VS,PoWi,Eich,Grin,Geor}.
In practice, only the $b$ and $c$ quarks have masses large enough for 
HQS to be useful.%
\footnote{The top quark also satisfies $m_t\gg\lqcd$, but it 
decays before it hadronizes.}  
This results in an $SU(2N_h)$ spin-flavor symmetry, where 
$N_h=1$ or 2, depending on the problem at hand.

The heavy quark spin-flavor symmetries are helpful for understanding 
many aspects of the spectroscopy and decays of heavy hadrons from 
first principles.
For example, in the infinite mass limit, mass splittings between 
$b$-flavored hadrons can be related to those between charmed hadrons, and 
many semileptonic and radiative decay form factors can be related to one 
another.
There are corrections to the HQS limit from long distances and from 
short distances.
The former are suppressed by powers of~$\lqcd/m_Q$.
They must be calculated by nonperturbative methods, but HQS again
imposes relations among these terms.
The latter arise from the exchange of hard virtual gluons, so they can 
be calculated accurately in a perturbation series in~$\alpha_s(m_Q)$.
The heavy quark effective theory (HQET) provides a convenient 
framework for treating these effects~\cite{Eich,Grin,Geor,FGL,FGGW,Bjor,Luke}.
In leading order the effective theory reproduces the model independent 
predictions of HQS, and both series of symmetry breaking corrections are
developed in a systematic, consistent way.

To see how the heavy quark symmetries arise, it is instructive to 
look at the infinite-mass limit of the Feynman rules.
For momentum $p=m_Qv+k$, with $v^2=1$ and $k\ll m_Q$, the propagator 
of a heavy quark becomes 
\begin{equation}
{i\over p\!\!\!\slash - m_Q} = {i(p\!\!\!\slash + m_Q)\over p^2-m_Q^2}
= {i(m_Q v\!\!\!\slash + k\!\!\!\slash + m_Q)\over 2m_Q\,v\cdot k +k^2}
= {i\over v\cdot k}\,{1+ v\!\!\!\slash \over2} + \ldots \,.
\end{equation}
As $m_Q\to\infty$ it is independent of its mass, and in this way 
heavy quark \emph{flavor} symmetry emerges.
In a Feynman diagram, the quark-gluon vertex appears between two 
propagators and, hence, for $m_Q\to\infty$, sandwiched between the 
projection operator
\begin{equation}
	P_+(v) = \frac{1+ v\!\!\!\slash}{2}\,.
\end{equation}
Consequently the gamma matrix at the vertex becomes
\begin{equation}
	P_+ \gamma^\mu P_+ = v^\mu\,P_+\,.
\end{equation}
Thus, both the vertex and the propagator depend on gamma matrices
only through~$P_+$.
Since $P_+^2=P_+$, all these factors reduce to a single one, and in 
this way heavy quark \emph{spin} symmetry emerges.

The construction of HQET \cite{HQS} starts by removing the mass-dependent piece
of the momentum operator by a field redefinition. One introduces a field
$h_v(x)$, which annihilates a heavy quark with velocity $v$ \cite{Geor},
\begin{equation}
h_v(x) = e^{i m_Q v\cdot x}\, P_+(v)\, Q(x) \,,
\label{field}
\end{equation}
where $Q(x)$ denotes the quark field in full QCD.
Here the physical interpretation of the projection operator~$P_+$ is 
that $h_v$ represents just the heavy quark (rather than antiquark) 
components of~$Q$.  
If $p$ is the total momentum of the heavy quark, then the field $h_v$ 
carries the residual momentum $k=p-m_Qv$.
Inside a hadron, the residual momentum~$k\sim{\cal O}(\lqcd)$.
Since the phase factor in Eq.~(\ref{field}) effectively removes the
mass of the heavy quark from the states, it is the mass difference
\begin{equation}
	\bar\Lambda = m_H - m_Q \,,
	\label{Ldef}
\end{equation}
where $m_H$ is the hadron mass, that determines the $x$-dependence of 
hadronic matrix elements in HQET~\cite{Luke}.
It is also this parameter that sets the characteristic scale of the
$1/m_Q$ expansion.
Because of heavy quark flavor symmetry $\bar\Lambda=m_B-m_b=m_D-m_c$, 
and because of heavy quark spin symmetry $\bar\Lambda=m_{B^*}-m_b$,
in both cases up to ${\cal O}(\lqcd^2/m_Q)$ corrections.
Other heavy hadrons, for example heavy-flavored baryons, have a 
distinct ``$\bar{\Lambda}$'', but the flavor symmetry implies
$m_{\Lambda_b}-m_b=m_{\Lambda_c}-m_c$, 
up to ${\cal O}(\lqcd^2/m_Q)$.
 
The HQET Lagrangian is constructed from the field~$h_v$.
Including the leading $1/m_Q$ corrections, it is~\cite{Eich,Geor,FGL}
\begin{equation}
{\cal L}_{\rm HQET} = \bar h_v\,i v\!\cdot\!D\,h_v
+ {1\over 2 m_Q}\,\Big[ O_{\rm kin}
+ C_{\rm mag}(\mu)\,O_{\rm mag}(\mu) \Big] + {\cal O}(1/m_Q^2) \,,
\label{Lag}
\end{equation}
where $D^\mu = \partial^\mu - i g_s T_a A_a^\mu$ is the color $SU(3)$ covariant
derivative. 
The leading term respects both the spin and flavor symmetries, and 
reproduces the heavy quark propagator derived above. 
The symmetry breaking operators appearing at order $1/m_Q$ are
\begin{equation}
O_{\rm kin} = \bar h_v\,(i D)^2\, h_v \,, \qquad
O_{\rm mag} = {g_s\over 2}\,\bar h_v\,\sigma_{\mu\nu}\, G^{\mu\nu}\, h_v \,.
\label{Lag1}
\end{equation}
Here $G^{\mu\nu}$ is the gluon field strength tensor defined by
$[iD^\mu,iD^\nu] = i g_s G^{\mu\nu}$.  In the rest frame of the hadron, 
$O_{\rm kin}$ describes the kinetic energy resulting from the residual motion
of the heavy quark, whereas $O_{\rm mag}$ corresponds to the chromomagnetic
coupling of the heavy quark spin to the gluon field. 
While $O_{\rm kin}$ violates only the heavy quark flavor symmetry, 
$O_{\rm mag}$ violates the spin symmetry as well.

In the operators of the electroweak Hamiltonian, the QCD field~$Q$
must also be replaced with~$h_v$ and a series of higher-dimension
operators to describe $1/m_Q$ effects.
The short distance behavior can be matched using perturbation theory.
The matrix elements of the HQET operators still cannot be calculated in
perturbatively, but HQS restricts their form.
The best known example is in exclusive semileptonic $b\to c$
corrections.
In $B\to D^{(*)}\ell\nu$ and $\Lambda_b\to \Lambda_c\ell\nu$, let $v$
($v'$) be the velocity of the initial (final) heavy-light hadron.
HQS requires that the mesonic decays are described by a set of
heavy quark spin- and mass-independent functions of the kinematic
variable $w=v\cdot v'$.
The baryonic decay is described by another function of~$w$.
When $v=v'$ the symmetry becomes larger---from $SU(2)_v\times SU(2)_{v'}$
to~$SU(4)$---so there are further restrictions.
One is that symmetry limit of the form factor is completely determined
by symmetry (at $w=1$).
Furthermore, HQS also requires that the $1/m_Q$ corrections to
$B\to D^*\ell\nu$ and $\Lambda_b\to \Lambda_c\ell\nu$ vanish for $w=1$.

The utility of HQET is not limited to exclusive decays.
Matrix elements of the effective Lagrangian play an important
role in inclusive semileptonic and radiative decays.
One defines
\begin{eqnarray}
\lambda_1 &=& {1\over 2\,m_M}\, 
  \langle M(v)|\, O_{\rm kin}\, |M(v) \rangle \,, \nonumber\\
d_M\,\lambda_2 &=& {1\over 2\, m_M}\, 
  \langle M(v)|\, O_{\rm mag}\, |M(v)\rangle \,,
\label{l12def}
\end{eqnarray}
where $M$ denotes a $B$ or $B^*$ meson, and $d_M=3,\,-1$ for $B$ and $B^*$,
\index{heavy quark effective theory (HQET)!$\lambda_1$ and $\lambda_2$}
respectively.  
Strictly speaking, both $\lambda_1$ and $\lambda_2$ depend on the 
renormalization scale~$\mu$.
For $\lambda_1$, however, there is no $\mu$ dependence if 
$O_{\rm kin}$ is renormalized in the $\ov{\rm MS}$ scheme.
For $\lambda_2$, the $\mu$ dependence is canceled by the coefficient
$C_{\rm mag}(\mu)$ in \eq{Lag}.

HQET provides an expansion of the heavy meson masses
in terms of the  heavy quark masses,
\begin{equation}
m_B = m_b + \bar\Lambda - {\lambda_1 + 3\lambda_2 \over 2m_b} +\ldots \,, \qquad
m_{B^*} = m_b + \bar\Lambda - {\lambda_1-\lambda_2 \over 2m_b} +\ldots\,.
\label{quarkmass}
\end{equation}
Consequently, the value of $\lambda_2$ is related to the mass splitting 
between the vector and the pseudoscalar mesons, 
\begin{equation}
\lambda_2 = {m_{B^*}^2 - m_B^2 \over 4} 
  + {\cal O} \Big( \lqcd^3/ m_b \Big) \,,
\end{equation}
taking $\mu=m_b$ and $C_{\rm mag}(m_b)=1$.
From the measured $B$ and $B^*$ masses one finds
$\lambda_2(m_b) \simeq 0.12~{\rm GeV}^2$.
These formulae will play an important role in the description of both
inclusive and exclusive heavy meson decays in the following chapters.  

It was only recognized recently that HQS also yields important
simplifications in the description of heavy-to-light radiative
and semileptonic decays in the region of large recoil
(small~$q^2$)~\cite{Charles}.  In the infinite mass limit, the
three form factors which parameterize the vector and tensor current matrix
elements in $B\to K\ell^+\ell^-$ are related to a single function of $q^2$, and
the seven form factors which occur in $B\to K^*\ell^+\ell^-$ are related to
only two functions of $q^2$.  In contrast to the predictions of HQS in the
region of small recoil, in this case it is not known yet how to formulate the
subleading corrections suppressed by powers of $\lqcd/m_Q$.  Nevertheless,
these relations play a very important role in Chapter~7, where they will be
discussed in detail.%
\index{heavy quark effective theory (HQET)|)}

\subsection{Heavy quark expansion}
\label{1:HQE}
\index{heavy quark expansion|(}

In inclusive $B$ decays, when many final states are summed over,
certain model independent formulae can be derived.
In this section we examine how the large $b$ quark mass, $m_b \gg \lqcd$, 
allows one to extract reliable information about such decays. In most of the
phase space the energy release, which can be as large as  ${\cal O}(m_b)$, is
much larger than the typical scale of hadronic interactions. The large energy
release implies a short distance, and we can use the same tools as before---an
operator product expansion~\cite{CGG,incl,MaWi}
(though not the same OPE as in Sec.~\ref{1:Heff}) and HQET---to
separate short and long distances.
In this way, inclusive decay rates can be described with a double
series in $\lqcd/m_b$ and $\alpha_s(m_b)$.

Inclusive decay widths are given by the sum over all final states.
Schematically, the width is given by
\begin{equation}
	\Gamma \sim \sum_X
		\langle B |O^\dagger | X\rangle\, 
		\langle X |O         | B\rangle\,.
\end{equation}
where $X$ is any final state.
One can also limit $X$ to $X_c$ or $X_u$, i.e., to final states with or
without a charmed quark, respectively.
  From Sec.~\ref{1:Heff}, we see that inclusive semileptonic 
$B$ decays are mediated by operators of the form
\begin{equation}
O_\ell \sim \bar{q}_L\gamma^\mu  b_L\,
  \bar{\ell}_{L1}\gamma_\mu  \ell_{L2}\,, 
\end{equation}
and nonleptonic decays are mediated by four-quark operators of the 
form
\begin{equation}
O_h \sim \bar{q}_L\gamma^\mu b_L\,
  \bar{q}_{L1}\gamma_\mu q_{L2}\,.
\end{equation}
Although these operators are superficially similar, we shall see that 
they have to be treated differently, because in $O_h$ hard gluons 
can be exchanged among all four quark fields.
We start by showing in detail how the OPE and HQET are used to 
describe inclusive semileptonic $B$ decays.
We then explain what restrictions arise for nonleptonic decay rates and 
lifetimes.
Finally, we treat the width difference in the $B_s$ system, which is 
of special interest to Tevatron experiments.

\boldmath
\subsubsection{Inclusive semileptonic $B$ decays} 
\unboldmath
\label{1:subsubsec:incsemi}
\index{B decay@$B$ decay!semileptonic}

In semileptonic decays, one may factorize the matrix 
element of the four-fermion operator
\begin{equation}
\langle X\ell\, \bar\nu_\ell|\, O_\ell\, |B\rangle = 
  \langle X|\, \bar q\, \gamma^\mu P_L\, b\, |B\rangle\, 
  \langle \ell\, \bar\nu_\ell|\, \bar\ell\, \gamma_\mu P_L\, \nu_\ell\, 
  |0\rangle\,,
\end{equation}
neglecting electroweak loop corrections.
Then the decay rate can be written in the form
\begin{equation}
\frac{\d^2\Gamma}{\d y\,\d q^2} \sim \int \d(q\cdot v)\, 
  L_{\mu\nu}(p_\ell,p_{\bar\nu})\, 
  W^{\mu\nu}(q\cdot v, q^2)\,,
  \label{1:eq:dGamma}
\end{equation}
where $L_{\mu\nu}$ is the lepton tensor and $W^{\mu\nu}$ is the hadron 
tensor.
The momentum of the decaying $b$ quark is written as $p_b^\mu=m_bv^\mu$,
$q^\mu = p_\ell^\mu + p_{\bar\nu}^\mu$, and we have introduced the
dimensionless variable $y=2E_\ell/m_b$.
Since the antineutrino is not detected, its energy or, equivalently,
$q\cdot v=E_\ell+E_{\bar\nu}$ is integrated over.
The lepton tensor
$L^{\mu\rho}=2(p^\mu_\ell p^\rho_{\bar\nu}+p^\rho_\ell p^\mu_{\bar\nu}%
- g^{\mu\rho} p_\ell \cdot p_{\bar\nu}%
-i\varepsilon^{\mu\rho\alpha\beta}p_{\ell\alpha}p_{\bar\nu\beta})$.
The hadron tensor $W^{\mu\nu}$ contains all strong interaction physics
relevant for the semileptonic decay, and it can be expressed as
\begin{equation}
W^{\mu\nu} = \sum_X\, (2\pi)^3\, \delta^4(p_B-q-p_X)\,
  { \langle \Bbar |\, {J^\mu}^\dagger\, | X \rangle\, 
  \langle X|\, J^\nu\, |\Bbar\rangle \over 2m_B } \,,
\end{equation}
where $J^\mu = \bar q\, \gamma^\mu P_L\, b$.  

\begin{figure}[bt] 
\centerline{\epsfysize 2.5cm \epsffile{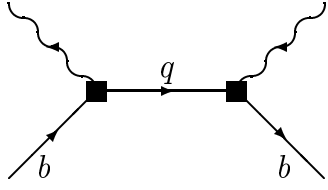}}
\caption{OPE diagram for semileptonic and radiative $B$ decays.}
\label{1:ope1}
\end{figure}

\begin{figure}[bt]
\centerline{\epsfysize 4cm \epsffile{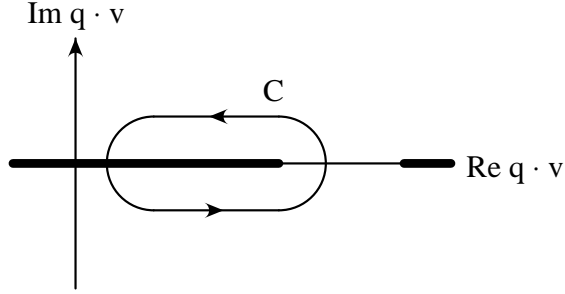}}
\caption[Analytic structure of $T^{\mu\nu}$ in the $q\cdot v$ plane.]{The
analytic structure of $T^{\mu\nu}$ in the $q\cdot v$ plane, with $q^2$
fixed.  The cuts corresponding to $B$ decay (left) and to an unphysical process 
(right) are both shown, together with the integration contour for computing 
the decay rate. }
\label{1:cuts}
\end{figure}

The optical theorem can be used to relate $W^{\mu\nu}$ to the 
discontinuity across a cut of the forward scattering matrix element of 
a time ordered product
\begin{equation}
T^{\mu\nu} = - i \int \d^4 x\, e^{-iq\cdot x}\, { \langle \Bbar |\, 
  T \{ {J^\mu}^\dagger(x)\, J^\nu(0) \} |\Bbar\rangle \over 2 m_B }\,.
\label{1:top}
\end{equation}
To show that
\begin{equation}
	W^{\mu\nu} = - \frac1\pi\, \imag  T^{\mu\nu} \,,
	\label{1:Wmunu}
\end{equation}
one inserts a complete set of states between the currents
in the two possible time orderings in~$T^{\mu\nu}$. 
Using $\langle A| J(x) | B\rangle =
\langle A| J(0) | B\rangle\, e^{i(p_A-p_B)\cdot x}$ and the identity
$\theta(x^0) = i/(2\pi)\, \int_{-\infty}^{+\infty} \d\omega\, [ e^{-i\omega
x^0} \! / (\omega + i\varepsilon) ]$, the $\d^4x$ integration gives 
(in the $B$ rest frame, so $q\cdot v=q^0$)
\begin{eqnarray}
T^{\mu\nu} &=& \sum_{X_q}\, 
  { \langle \Bbar |\, J^\mu{}^\dagger\, | X_q \rangle\, 
  \langle X_q|\, J^\nu\, |\Bbar\rangle \over 
  2m_B\, (m_B - E_X - q^0 + i\varepsilon)}\,
  (2\pi)^3\, \delta^3(\vek{q} + \vek{p}_X) \nonumber\\*
&-& \sum_{X_{\bar qbb}}\, 
  { \langle \Bbar |\, J^\nu\, | X_{\bar qbb} \rangle\, 
  \langle X_{\bar qbb}|\, {J^\mu}^\dagger\, |\Bbar\rangle \over 
  2m_B\, (E_X - m_B - q^0 - i\varepsilon)}\, 
  (2\pi)^3\, \delta^3(\vek{q} - \vek{p}_X) \,.
\label{1:twoterms}
\end{eqnarray}
This form shows that, for fixed $q^2$,
$T^{\mu\nu}$ has cuts in the complex $q^0$ plane
corresponding to physical processes.
The first sum in Eq.~(\ref{1:twoterms}) corresponds to $B$ decay shown
in Fig.~\ref{1:ope1}, with intermediate states containing a $q$ quark
(and arbitrary number of gluons and light quark-antiquark pairs).
It leads to a cut for
$q^0=q\cdot v < (m_B^2 + q^2 - m_{X_q^{\rm min}}^2)/2m_B$,  
towards the left in Fig.~\ref{1:cuts}.
For charmed final states $m_{X_q^{\rm min}}^2=m_D^2$ and for charmless
final states $m_{X_q^{\rm min}}^2=m_\pi^2$.
The second sum in Eq.~(\ref{1:twoterms}) corresponds to an unphysical
process with a $\bar q$ and two $b$ quarks in the intermediate state.
It leads to another cut for
$q^0=q\cdot v > (m_{X_{\bar qbb}^{\rm min}}^2 - m_B^2 -q^2) / 2m_B$,
towards the right in Fig.~\ref{1:cuts}.
The imaginary part can be read off using 
$\imag(A+i\varepsilon)^{-1}=-\pi\delta(A)$, and \eq{1:Wmunu} follows 
immediately, because the kinematics of the decay process allow only the
first sum to contribute.

Because $W^{\mu\nu}$ is the discontinuity across the left cut in
Fig.~\ref{1:cuts}, the integral in \eq{1:eq:dGamma} can be replaced
with a contour integral of $L_{\mu\nu}T^{\mu\nu}$.
The two cuts are well separated (unless $m_q \to 0$ and $q^2 \to
m_b^2$), so one may deform the contour away from the cuts~\cite{PQW},
as shown in Fig.~\ref{1:cuts}.
The equivalence of the sum over hadronic states with a contour ranging
far from the physical region is called ``global duality".
This procedure is advantageous, because $T^{\mu\nu}$ can be
reliably described by an operator product expansion (OPE) far (compared
to $\lqcd$) from its singularities in the complex $q\cdot v$ plane
\cite{CGG,incl,MaWi}.
One simply replaces the time ordered product
\begin{equation}
- i \int \d^4 x\, e^{-iq\cdot x}\, T \{ J^\mu{}^\dagger(x)\, J^\nu(0) \} \,,
\label{1:topb}
\end{equation}
appearing in Eq.~(\ref{1:top}), with a series of local operators
multiplied with Wilson coefficients.
The Wilson coefficients of this OPE can again be evaluated in a
perturbation series in $\alpha_s(m_b)$.
Higher dimension operators in the OPE incorporate higher powers of
$\lqcd/m_b$.

Unfortunately, the contour $C$ must still approach the cut near the
low $q\cdot v$ endpoint of the integration.
Using the OPE directly in the physical region is an assumption called
``local duality".
It introduces an uncertainty to the calculation, which can be argued
to be small.
First, in semileptonic and radiative decays the fraction of the contour
which has to be within order $\lqcd$ from the cut scales as $\lqcd/m_b$.
Second, since the energy release to the hadronic final state is large
compared to $\lqcd$, the imaginary part of $T^{\mu\nu}$ is dominated
by multiparticle states, so it is expected to be a smooth function.
In the end, the violation of local duality is believed to be
exponentially suppressed in the $m_Q\to\infty$ limit, but it is not
well understood how well it works at the scale of the $b$ quark mass.
In semileptonic decay the agreement between the inclusive and exclusive
determinations of $|V_{cb}|$ suggests that duality violation is at
most a few percent.
But there is no known relation between the size of duality violation
in semileptonic and nonleptonic $B$ decays~\cite{FWD}, or between
these processes and others, such as $e^+e^-\to{\rm hadrons}$.

At lowest order in $\lqcd/m_b$ the OPE leads to operators of the form
$\bar b\Gamma b$ occur, where $\Gamma$ is any Dirac matrix.
For $\Gamma=\gamma^\mu$ or $\gamma^\mu\gamma_5$ their matrix elements
are known to all orders in $\lqcd/m_B$
\begin{eqnarray}
\langle \Bbar(p_B) |\, \bar b\, \gamma^\mu b \, |\Bbar(p_B)\rangle &=&
  2p_B^\mu = 2m_B\, v^\mu \,, \nonumber\\*
\langle \Bbar(p_B) |\, \bar b\, \gamma^\mu\gamma_5\, b \, |\Bbar(p_B)\rangle 
  &=& 0 \,.
  \label{1:eq:hqe3}
\end{eqnarray}
by conservation of the $b$ quark number current and parity invariance
of strong interactions, respectively.
The matrix elements for other gamma matrices can be related by heavy quark
symmetry to these plus order $\lqcd^2/m_b^2$ corrections.
Consequently, at the leading order in $\lqcd/m_b$ inclusive decay
rates are given by the rate for $b$ quark decay, multiplied with a
Wilson coefficient that does not depend on the decaying hadron.

To compute subleading corrections in $\lqcd/m_b$, it is convenient to
use HQET.
\index{heavy quark effective theory (HQET)!in inclusive decays}
There are no order $\lqcd/m_b$ corrections because the matrix element 
of any gauge invariant dimension-4 two-quark operator vanishes,
\begin{equation}
\langle \Bbar(v) |\, \bar h_v^{(b)}\, iD_\alpha \Gamma\, h_v^{(b)}\, 
  |\Bbar(v)\rangle = 0 \,,
  \label{1:eq:hqe4}
\end{equation}
because contracting the left-hand side by $v^\alpha$ gives zero due
to the equation of motion following from (\ref{Lag}).
Thus, the leading nonperturbative corrections to $b$ quark decay occur
at order $\lqcd^2/m_b^2$.
The operators that appear are again $O_{\rm kin}$ and $O_{\rm mag}$
so the same hadronic elements $\lambda_1$ and $\lambda_2$,%
\index{heavy quark effective theory (HQET)!$\lambda_1$ and $\lambda_2$}
defined in Eq.~(\ref{l12def}), appear again.

Combining the matrix elements from Eqs.~(\ref{1:eq:hqe3}),
(\ref{1:eq:hqe4}) and~(\ref{l12def}) with the Wilson coefficients leads
to expressions of the form
\begin{eqnarray}
\frac{\d^2\Gamma}{\d y\,\d q^2} =
	\pmatrix{ b{\rm ~quark} \cr {\rm decay}\cr } \times \bigg\{ 1 &+&
  \frac{\alpha_s}\pi\, A_1 + \frac{\alpha_s^2}{\pi^2}\, A_2 + \ldots 
  + \frac{f(\lambda_1,\lambda_2)}{m_B^2}\, 
  \Big[ 1 + {\cal O}(\alpha_s) + \ldots \Big] \nonumber\\*
  &+& {\cal O}(\lqcd^3/m_B^3) + \ldots \bigg\} \,.
\label{1:blahblah}
\end{eqnarray}
The differential rate may be integrated to obtain the full rate.
The description in \eq{1:blahblah} is model independent, although
$\lambda_1$ must be determined either from data~\cite{Gremm:1996yn}
or from lattice QCD~\cite{Crisafulli:1995pg}.
For most quantities of interest the functions $f$, $A_1$, and the
part of $A_2$ proportional to $\beta_0$, the first coefficient of
the $\beta$-function, are known.
Corrections to the $m_b\to\infty$ limit are expected to be under
control in parts of the $\Bbar\to X_q\, \ell\, \bar\nu$ phase space
where several hadronic final states are allowed (but not required)
to contribute with invariant mass and energy satisfying
$m_X^2 \gg m_q^2 + \lqcd E_X$.

\boldmath
\subsubsection{Inclusive nonleptonic $B$ decays}
\unboldmath
\index{B decay@$B$ decay!nonleptonic}

\begin{figure}[bt]
\centerline{\epsfysize 2.5cm \epsffile{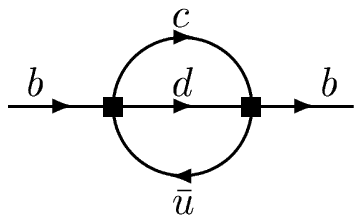} \hspace*{1cm}
  \epsfysize 2.5cm \epsffile{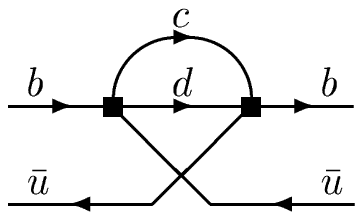}}
\caption[OPE diagrams for nonleptonic $B$ decays.]{OPE diagrams for 
nonleptonic $B$ decays.  The left one is the leading contribution, while
the ``Pauli interference" diagram on the right corresponds to a dimension-6 
contribution of order $16\pi^2\, (\lqcd^3/m_B^3)$.}
\label{1:ope2}
\end{figure}

Inclusive nonleptonic decays can also be studied using the OPE,
and much of the discussion in Sec.~\ref{1:subsubsec:incsemi}
applies here also.
In this case, however, there are no ``external" variables, such as
$q^2$ and $q\cdot v$, since all particles in the final state interact
strongly.
For this reason, only the fully integrated inclusive width can be treated
with the OPE, term-by-term in the weak Hamiltonian.
For example, the $B$ decay width corresponding to the $b\to c\bar u d$
effective Hamiltonian in \eq{heff}--\eq{1:O12def} is given by
\begin{eqnarray}
\Gamma &=& \frac1{2m_B}\, \sum_X (2\pi)^4\, \delta^4(p_B-p_X)
  \left| \langle X(p_X)|\, H^{|\Delta B|=1}(0)\, |\Bbar(p_B)\rangle \right|^2 
  \nonumber\\*
&=& \frac1{2m_B}\, \imag  \langle \Bbar |\, i \int \d^4 x\,
  T \lt\{ H^{|\Delta B|=1}(x)\, H^{|\Delta B|=1}(0) \rt\} |\Bbar\rangle \,.
\label{1:nonlept}
\end{eqnarray}
Because one has to use the OPE directly in the physical region,
the results are more sensitive to violations of local duality than
in the case of semileptonic and radiative decays.
The leading term in the OPE corresponds to the left diagram in
Fig.~\ref{1:ope2},
whose imaginary part gives the total nonleptonic width.

The result is again of the form shown in Eq.~(\ref{1:blahblah}).  
An important new ingredient at order $\lqcd^3/m_B^3$ are certain
contributions due to four-quark operators involving the spectator quark.
They are usually called ``weak annihilation", ``$W$~exchange", and
``Pauli interference" contributions.
(The last is sketched on the right in Fig.~\ref{1:ope2}).
They contain one less loop than the diagram on the left, so they
are enhanced by a relative factor of~$16\pi^2$.
They are expected to be
more important than the dimension-5 contributions proportional to
$\lambda_1$ and $\lambda_2$.  The matrix elements of the resulting
four-quark operators are poorly known.  Such contributions are expected
to explain the $D^\pm - D^0$ lifetime difference.

\boldmath
\subsubsection{$B_s$ width difference, $\Delta\Gamma$}
\unboldmath
\index{width difference!$B_s$}

\begin{figure}[bt]
\centerline{\epsfysize 2.5cm \epsffile{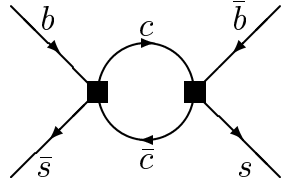}}
\caption{OPE diagram for the $B_s$ width difference.}
\label{1:ope3}
\end{figure}

Another important application, especially for the Tevatron, 
is for the $B_s$ width difference.  
The off-diagonal element of the width matrix (cf., Sec.~\ref{subs:mix}) 
is given by
\begin{eqnarray}
\Gamma_{12} &=& \frac1{2m_{B_s}}\, \sum_X (2\pi)^4\, \delta^4(p_{B_s}-p_X)\, 
  \langle B_s |\, H^{|\Delta B|=1}\, | X \rangle\, 
  \langle X|\, H^{|\Delta B|=1}\, | \Bbar_s\rangle  \nonumber\\*
&=& \frac1{2m_{B_s}}\, \imag  \langle B_s |\, i \int \d^4 x\, 
  T \left\{ H^{|\Delta B|=1}(x)\, H^{|\Delta B|=1}(0) \right\} | 
  \Bbar_s\rangle \,.
\label{1:delgam}
\end{eqnarray}
The first line defines $\Gamma_{12}$, and the second
line can be verified by inserting a complete set of intermediate states. 
The corresponding diagram is shown in Fig.~\ref{1:ope3}.
$\Gamma_{12}$ arises from final states~$X$ which are common to both
$B_s$ and $\Bbar_s$ decay.  Therefore, the spectator quark is involved, and
Eq.~(\ref{1:delgam}) is dominated by the $b \to c \bar c s$ part of the weak
Hamiltonian, $O_1$ and $O_2$ in Eq.~(\ref{basis}), with the others,
$O_3$ through $O_6$, making very small contributions.

Thus, the naive estimate of the $B_s$ width difference is $\Delta
\Gamma_{B_s} / \Gamma_{B_s} = 2\, |\Gamma_{12}| \cos\phi / \Gamma_{B_s} \sim
16\pi^2 (\lqcd^3/m_B^3) \sim 0.1$.  In the $B_d$ system the common decay
modes of $B^0$ and $\B0bar$ are suppressed relative to the leading ones by
the Cabibbo angle, and therefore the naive estimate is $\Delta\Gamma_{B_d} /
\Gamma_{B_d} \lesssim 1\%$.
See the discussion following \eq{dgsmall} and Chapter~8 for more details.
\index{heavy quark expansion|)}


\subsection{Lattice QCD}
\label{1:lattice}

If one considers the long term goal of ``measuring'' the Wilson
coefficients of the electroweak Hamiltonian, as outlined elsewhere,
then it is clear that it will be important to gain theoretical control
over hadronic matrix elements.
Since QCD is a completely well-defined quantum field theory, the
calculation of hadronic matrix elements should be, in principle,
possible.
The main difficulty is that hadronic wavefunctions are sensitive mostly
to the long distances where QCD becomes nonperturbative.

The difficulties of the bound-state problem in QCD led
Wilson~\cite{Wilson:1974sk} to formulate gauge field theory on a
discrete spacetime, or lattice.
The basic idea starts with the functional integral for correlation
functions in QCD
\begin{equation}
    \langle O_1\cdots O_n \rangle = \frac{1}{Z}
        \int \prod_{x,\mu} dA_\mu(x) \prod_x d\psi(x) d\bar{\psi}(x)
        \, O_1\cdots O_n \, e^{-S_{\rm QCD}}
    \label{1:eq:funcint}
\end{equation}
where $Z$ is defined so that $\langle 1\rangle=1$.
For QCD $A_\mu$ is the gluon field, $\psi$ and $\bar{\psi}$ are the
quark and antiquark fields, and $S_{\rm QCD}$ is the QCD action.
The $O_i$ are operators for creating and annihilating the hadrons of 
interest and also terms in the electroweak Hamiltonian.
The continuous spacetime is then replaced with a discrete grid of
points, or lattice.
Then the quark variables live on sites;
the gluons on links connecting the sites.
With quarks on sites and gluons on links, it is possible to devise
lattice actions that respect gauge symmetry.
As in discrete approximations to partial differential equations,
derivatives in the Lagrangian are replaced with difference operators.

The breakthrough of the lattice formulation is that it turns quantum
field theory into a mathematically well-defined problem in statistical
mechanics.
Condensed matter theorists and mathematical physicists have devised a
variety of methods for tackling such problems, only one of which is
weak-coupling perturbation theory.
In the years immediately following Wilson's work, many of these tools were
tried, for example analytical strong coupling expansions.
The strong coupling limit is especially appealing, because confinement
emerges immediately~\cite{Kogut:1983ds}.

Strong coupling is, however, not the whole story.
Owing to asymptotic freedom, the continuum limit of lattice QCD is
controlled by weak coupling.
Unfortunately, strong coupling expansions do not converge quickly
enough to reach into the weak-coupling regime, at least with the simple
discretizations that have been used till now.
Consequently, results from strong coupling expansions for hadron masses
and matrix elements are not close enough to continuum QCD to apply to
particle phenomenology.

Since such analytical methods have not borne out, the tool of choice
now is to compute (the discrete version of) Eq.~(\ref{1:eq:funcint})
numerically via Monte Carlo integration.
This numerical method has, over the years, developed several
specialized features, and corresponding jargon, that often make its
results impenetrable to non-experts.
Moreover, as with any numerical method, there are several 
sources of systematic uncertainty.
Most of the systematic effects can, however, be controlled with
effective field theories, i.e., with techniques like those explained
in the previous sections.
After reviewing the basic elements of the Monte Carlo method,
we cover the systematic effects.
First is the so-called quenched approximation, which is difficult to
control, but also not a fundamental limitation.
Other uncertainties, which can be controlled, are reviewed next,
emphasizing the role of effective field theories.
It is hoped that in this way non-experts can learn to make simple
estimates of size systematic uncertainties, without repeating all the
steps of the numerical analysis.
We end with a comment on the (unsatisfactory) status of computing
strong phase shifts for $B$ decays.

\subsubsection{Monte Carlo integration}

This part of the method is well understood and, these days, rarely
leads to controversy.
For completeness, however, we include a short explanation, focusing on
the points that limit the range of applicability of the method.
A~more thorough treatment aimed at experimenters can be found in
Ref.~\cite{DiPierro:2000nt}.

The first salient observation is that there are very many variables.
Continuum field theory has uncountably many degrees of freedom.
Field theory on an infinite lattice still has an infinite number of
degrees of freedom, but at least countably infinite.
(This makes the products over $x$ in Eq.~(\ref{1:eq:funcint})
well-defined.)
To keep the number finite (for a computer with finite memory),
one must also introduce a finite spacetime volume.
This may seem alarming, but what one has done is simply to introduce an
ultraviolet cutoff (the lattice) and an infrared cutoff (the finite
volume).
This is usual in quantum field theory, and field theoretic techniques
can be used to understand how to extract cutoff-free quantities from
numerically calculable cutoff quantities.

Even with a finite lattice, the number of integration variables is
large.
If one only demands a volume a few times the size of a hadron and also
several grid points within a hadron's diameter, one already requires
at least, say, 10 points along each direction.
In four-dimensional spacetime this leads to $\sim 32\times 10^4$ gluonic
variables.
With so many variables, the only feasible methods are based on Monte
Carlo integration.
The basic idea of Monte Carlo integration is simple: generate an
ensemble of random variables and approximate the integrals in
Eq.~(\ref{1:eq:funcint}) by ensemble averages.

Quarks pose special problems, principally because, to implement Fermi
statistics, fermi\-onic variables are Grassmann numbers.
In all cases of interest, the quark action can be written
\begin{equation}
    S_F = \sum_{\alpha\beta}
        \bar{\psi}_\alpha M_{\alpha\beta} \psi_\beta,
\end{equation}
where $\alpha$ and $\beta$ are multi-indices for (discrete) spacetime,
spin and internal quantum numbers.
The matrix $M_{\alpha\beta}$ is some discretization of the Dirac
operator $\kern+0.1em /\kern-0.65em D+m$.
Note that it depends on the gauge field, but one may integrate over the
gauge fields after integrating over the quark fields.
Then, because the quark action is a quadratic form, the integral can be
carried out exactly:
\begin{equation}
    \int \prod_{\alpha\beta} d\bar{\psi}_\alpha d\psi_\beta\,
        e^{- \bar{\psi} M \psi} = \det M .
    \label{1:eq:quarkdet}
\end{equation}
Similarly, products $\psi_\alpha\bar{\psi}_\beta$ in the integrand are
replaced with quark propagators~$[M^{-1}]_{\alpha\beta}$.
The computation of $M^{-1}$ is demanding, and the computation of $\det M$
(or, more precisely, changes in $\det M$ as the gauge field is changed)
is very demanding.

With the quarks integrated analytically, it is the gluons that are
subject to the Monte Carlo method.
The factor with the action is now $\det M e^{-S}$, where $S$ is now just
the gluons' action.
Both $\det M$ and $e^{-S}$ are the exponential of a number that scales
with the spacetime volume.
In Minkowski spacetime the exponent is an imaginary number, so
there are wild fluctuations for moderate changes in the gauge field.
On the other hand, in Euclidean spacetime, with an imaginary time
variable, $S$ is real.
In that case (assuming $\det M$ is positive definite) one can devise a
Monte Carlo with \emph{importance sampling}, which means that the
random number generator creates gauges field weighted according to
$\det M e^{-S}$.
Because importance sampling is essential, only in Euclidean spacetime is 
lattice QCD numerically tractable.

Importance sampling works well if $\det M$ is positive.
For pairs of equal-mass quarks, this is easy to achieve.
As a result, most calculations of $\det M$ are for 2 or 4 flavors.
Note that a physically desirable situation with three flavors, with the
strange quark's mass different from that of two lighter quarks, must
either cope with (occasional) non-positive weights, or find a (new)
discretization with $\det M$ positive flavor by flavor.

The choice of imaginary time has an important practical advantage.
Consider the two-point correlation function
\begin{equation}
    C_2(t) = \bra{0} \Phi_B(t) \Phi_B^\dagger(0) \ket{0},
\end{equation}
where $\Phi_B$ is an operator with the quantum numbers of the $B$ meson at
rest.
Inserting a complete set of states between $B$ and $B^\dagger$
\begin{equation}
    C_2(t) = \sum_n \frac{1}{2m_n} \bra{0} \Phi_B \ket{B_n}
        \bra{B_n} \Phi_B^\dagger \ket{0} e^{im_nt},
\end{equation}
where $m_n$ is the mass of $\ket{B_n}$, the $n$th radial excitation
of the $B$~meson.
For real $t$ it would be difficult to disentangle all these
contributions.
If, however, $t=ix_4$, with $x_4$ real and positive, then one has a sum
of damped exponentials.
For large $x_4$ the lowest-lying state dominates and
\begin{equation}
    C_2(x_4) = (2m_B)^{-1} |\bra{0} \Phi_B \ket{B}|^2 e^{-m_Bx_4} + \cdots,
    \label{1:eq:C2exp}
\end{equation}
where $\ket{B}$ is the lowest-lying state and $m_B$ its mass.
The omitted terms are exponentially suppressed.
It is straightforward to test when the first term dominates a
numerically computed correlation function, and then fit the exponential
form to obtain the mass.

This technique for isolating the lowest-lying state is essential also
for obtaining hadronic matrix elements.
For $B^0 - \B0bar$ mixing, for example, one must compute the matrix
element $\bra{B^0}Q\ket{\B0bar}$, given in Eq.~(\ref{ch1:uli:defq}).
One uses a three-point correlation function
\begin{equation}
    C_Q(x_4,y_4) = \bra{0} \Phi_B(x_4+y_4) Q(y_4) \Phi_B(0) \ket{0},
\end{equation}
where only the Euclidean times of the operators have been written out.
Inserting complete sets of states and taking $x_4$ and $y_4$ large
enough,
\begin{equation}
    C_Q(x_4,y_4) = (2m_B)^{-1} (2m_{\Bbar})^{-1}
    	\bra{0} \Phi_B \ket{B} \bra{B} Q \ket{\Bbar}
        \bra{\Bbar} \Phi_B\ket{0} e^{-m_Bx_4-m_{\Bbar}y_4}.
    \label{1:eq:CQexp}
\end{equation}
The amplitude ($\bra{0}\Phi_B\ket{B}=\bra{\Bbar}\Phi_B\ket{0}$)
and mass ($m_B=m_{\Bbar}$) are obtained from $C_2$, leaving
$\bra{B} Q \ket{\Bbar}$ to be determined from~$C_Q$.
Similarly, to obtain amplitudes for $B$ decays to a single hadron (plus
leptons or photons), simply replace one of the $\Phi_B$ operators with
one for the desired hadron and $Q$ with the desired operator.
To compute the purely leptonic decay, simply replace $\Phi_B$ in $C_2$
with the charged current.

These methods are conceptually clean and technically feasible for
the calculation of masses and hadronic matrix elements
with at most one hadron in the final state.
The procedure for computing correlation functions is as follows.
First generate an ensemble of lattice gluon fields with the appropriate
weight.
Next form the desired product $O_1\cdots O_n$, with quark variables
exactly integrated out to form propagators~$M^{-1}$.
Then take the average over the ensemble.
Finally, fit the Euclidean time dependence of 
Eqs.~(\ref{1:eq:C2exp}) and~(\ref{1:eq:CQexp}).
Note that since the same ensemble is used for many similar correlation
functions, the statistical fluctuations within the ensemble are
correlated.
This is not a concern, as long as the correlations are propagated
sensibly through the analysis.

\subsubsection{Quenched approximation}

Any perusal of the literature on lattice QCD quickly comes across
something called the\index{lattice QCD!quenched approximation}
``quenched approximation.''
As mentioned above, the factor $\det M$ in Eq.~(\ref{1:eq:quarkdet}) is
difficult to incorporate.
The determinant generates sea quarks inside a hadron.
The quenched approximation replaces $\det M$ with $1$ \emph{and}
compensates the corresponding omission of the sea quarks with shifts in
the bare couplings.
This is analogous to a dielectric approximation in electromagnetism,
and it fails under similar circumstances.
In particular, if one is interested in comparing two quantities that are
sensitive to somewhat different energy scales, one cannot expect the
same dielectric shift to suffice.
Another name for the quenched approximation is the ``valence''
approximation, which makes clearer that the valence quarks (and gluons)
in hadrons are treated fully, and the sea quarks merely modeled.

It is not easy to estimate quantitatively the effect of quenching.
For $\alpha_s$~\cite{El-Khadra:1992vn}
and the quark masses~\cite{Mackenzie:1994zw}
one can compute the short distance
contribution to the quenching shift, but that is only a start.
The quenched approximation can be cast as the first term in a
systematic expansion~\cite{Sexton:1997ud}, but it is about as difficult
to compute the next term as to restore the fermion determinant.
In the context of heavy quark physics one should note that the
CP-PACS~\cite{AliKhan:2000eg} and MILC~\cite{Bernard:2000nv} groups now
have unquenched calculations\index{lattice QCD!unquenched $f_B$}
of the heavy-light decay constants $f_B$,
$f_{B_s}$, $f_D$, and~$f_{D_s}$.
Both have results at several lattice spacings, so they can study the
$a$~dependence.
Their results are about 10--15\% higher than the most mature estimates
from the quenched approximation.

\subsubsection{Controllable systematic uncertainties}

By a controllable systematic uncertainty we mean an uncertainty that can
be incrementally improved in a well-defined way.
In lattice QCD they arise from the ultraviolet and infrared cutoffs,
and also from the fact that quark masses are freely adjustable and, for
technical reasons, not always adjusted to their physical values.
Because these effects are subject to theoretical control, the errors
they introduce can largely be reduced to a level that is essentially
statistical, given enough computing.

One of the least troublesome systematic effects comes from the finite
volume.
Finite-volume effects can be understood separately from lattice-spacing
effects with an effective massive quantum field
theory~\cite{Luscher:1986dn}.
In some cases adjusting the volume at will is, at least in principle,
a boon, yielding valuable information, such as scattering lengths and
resonance widths.

The computer algorithms for computing the quark propagator $M^{-1}$ 
converge more quickly at masses near that of the strange quark than 
for lighter masses.
Consequently, the Monte Carlo is run at a sequence of light quark masses
typically in the range $0.2m_s\lesssim m_q\lesssim m_s$.
(The up and down quark masses are far smaller still and not reached.)
The dependence on $m_q$ can be understood and controlled via the chiral
Lagrangian~\cite{Gasser:1984yg}, another effective field theory.
A~recent development is to show in detail how to extract physical
information from results at practical values of the light quark
masses~\cite{Sharpe:2000bc}.

A special difficulty with heavy quarks is the effect of non-zero lattice
spacing.
The bottom and charmed quark masses are large in lattice units.
For this reason it is frequently (but incorrectly) stated that heavy
quarks cannot be directly accommodated by a lattice.
  From the inception of HQET and NRQCD, these effective field theories
have been used to treat heavy quarks, and more recently it has been
shown how to use these tools to understand the discretization effects
of heavy quarks discretized with the original Wilson
formulation~\cite{Wilson:1977nj}.

\index{lattice QCD!lattice spacing effects!light quarks and gluons|(}
Let us first recall how lattice-spacing effects are controlled for
systems of light quarks.
Long ago, Symanzik introduced a local effective Lagrangian
(LE${\cal L}$) to describe cutoff effects~\cite{Symanzik:1979ph}.
One writes
\begin{equation}
    {\cal L}_{\rm lat} \doteq {\cal L}_{\rm cont} +
        \sum_{i} a^{s_{{\cal O}_i}} K_i(a;\mu) {\cal O}_i(\mu),
\end{equation}
where $s_{{\cal O}_i}={\rm dim\,}{\cal O}_i-4$.
The symbol $\doteq$ means
``has the same (on-shell) matrix elements as''.
For operators such as $Q$,
needed for mixing,\index{hadronic parameter!in lattice QCD}
\begin{equation}
    Q_{\rm lat} \doteq Z^{-1}_Q(a;\mu) Q_{\rm cont}(\mu) +
        \sum_{i} a^{s_{{\cal Q}_i}} C_i(a;\mu) {\cal Q}_i(\mu),
\end{equation}
where now $s_{{\cal Q}_i}={\rm dim\,}{\cal Q}_i-{\rm dim\,}Q$.
The continuum operators ${\cal O}_i$, $Q_{\rm cont}$, and
${\cal Q}_i$ are defined in a mass-independent scheme
at scale~$\mu$.
They do not depend on the lattice spacing~$a$.
The coefficients $K_i$, $Z_Q$, and $C_i$ account for short distance
effects, so they do depend on~$a$.

If $a$ is small enough the higher terms can be treated as
perturbations.
So, the $a$ dependence of $\bra{B} Q_{\rm lat} \ket{\Bbar}$ is
\begin{equation}
    \bra{B} Q_{\rm lat} \ket{\Bbar} =
         Z_Q ^{-1} \bra{B} Q_{\rm cont} \ket{\Bbar} +
        a K_{\sigma F} \bra{B}T\, Q_{\rm cont} \int\!d^4x\,
                \bar{\psi} \sigma\cdot F\psi \ket{\Bbar} +
        a C_1 \bra{B} {\cal Q}_1 \ket{\Bbar},
    \label{1:eq:Qlat}
\end{equation}
keeping only contributions of order~$a$.
To reduce the unwanted terms one might try to reduce $a$ greatly, but
CPU time goes as $a^{-\rm (5~or~6)}$.
It is more effective to use a sequence of lattice spacings and
extrapolate, with Eq.~(\ref{1:eq:Qlat}) as a guide.
It is even better to adjust things so $K_{\sigma F}$ and $K_1$ are
${\cal O}(\alpha_s^\ell)$~\cite{Sheikholeslami:1985ij} or
${\cal O}(a)$~\cite{Jansen:1996ck}, which is called Symanzik improvement.
For light hadrons, a combination of improvement and extrapolation is 
best.
Note that one still has to adjust $Q_{\rm lat}$ so that $Z_Q=1$.
In some cases the needed adjustment can be made nonpertubatively, even
though it is a short distance quantity.
When that is possible, lattice QCD can provide results with no
perturbative uncertainty, although perturbative uncertainty
may reenter through the electroweak Hamiltonian.
\index{lattice QCD!lattice spacing effects!light quarks and gluons|)}

\index{lattice QCD!lattice spacing effects!heavy quarks|(}
The Symanzik theory, as usually applied, assumes $m_qa\ll1$.
The bottom and charmed quarks' masses in lattice units are at present
large: $m_ba\sim1$--2 and $m_ca$ about a third of that.
It will not be possible to reduce $a$ enough to make $m_ba\ll1$
for many, many years.
So, other methods are needed to control the lattice spacing
effects of heavy quarks.
There are several alternatives:
\begin{enumerate}
    \item static approximation~\cite{Eichten:1990zv}
    \item lattice NRQCD~\cite{Lepage:1987gg}
    \item extrapolation from $m_Q\lesssim m_c$ up to $m_b$
    \item[$3'\!.$] combine 3 with 1
    \item normalize systematically to HQET~\cite{El-Khadra:1997mp}
    \item anisotropic lattices with temporal lattice spacing
        $a_t\ll a$~\cite{Kla99}
\end{enumerate}
All but the last use the heavy quark expansion in some way.
The first two discretize continuum HQET;
method~1 stops at the leading term, and method~2 carries the
heavy quark expansion out to the desired order.
Methods~3 and~$3'$ keep the heavy quark mass artificially small and
appeal to the $1/m_Q$ expansion to extrapolate back up to $m_b$.
Method~4 uses the same lattice action as method~3, but uses the
heavy quark expansion to normalize and improve it.
Methods~2 and~4 are able to calculate matrix elements directly at
the $b$-quark mass.
Method~5 has only recently been applied to heavy-light
mesons~\cite{Collins:2001pe}, and, like the other methods, it requires
that spatial momenta are much less than~$m_Q$.

The methods can be compared and contrasted by \emph{describing} the 
lattice theories with HQET~\cite{Kronfeld:2000ck}.%
\index{heavy quark effective theory (HQET)!for lattice QCD}
This is, in a sense, the opposite of \emph{discretizing} HQET.
One writes down a (continuum) effective Lagrangian
\begin{equation}
    {\cal L}_{\rm lat} \doteq
        \sum_{n} {\cal C}^{(n)}_{\rm lat}(m_Qa; \mu) 
                O^{(n)}_{\rm HQET}(\mu),
    \label{eq:hqet}
\end{equation}
with the operators $O^{(n)}_{\rm HQET}$ defined exactly as in
Sec.~\ref{1:HQET}, so they do not depend on $m_Q$ or~$a$.
As long as $m_Q\gg\Lambda_{\rm QCD}$ this description makes
sense.
There are two short distances, $1/m_Q$ and the lattice spacing~$a$,
so the short distance coefficients ${\cal C}^{(n)}_{\rm lat}$ depend 
on~$m_Qa$.
Since all dependence on $m_Qa$ is isolated into the coefficients,
this description shows that heavy quark lattice artifacts arise only
from the mismatch of the ${\cal C}^{(n)}_{\rm lat}$ and their continuum 
analogs~${\cal C}^{(n)}_{\rm cont}$.

For methods~1 and~2, Eq.~(\ref{eq:hqet}) is just a
Symanzik~LE${\cal L}$.
For lattice NRQCD we recover the result that some of the 
coefficients ${\cal C}^{(n)}_{\rm lat}$ have power-law divergences
as $a\to0$~\cite{Lepage:1987gg}.
So, to obtain continuum (NR)QCD, one must add more and more terms to
the action.
(This is just a generic feature of effective field theories, namely,
that accuracy is improved by adding more terms, rather than taking the 
cutoff too high.)
The truncation leaves a systematic error, which, in practice, is
usually accounted for conservatively.

Eq.~(\ref{eq:hqet}) is more illuminating for methods~3--5, which 
use the same actions, but with different normalization conditions.
The lattice quarks are Wilson fermions~\cite{Wilson:1977nj},
which have the same degrees of freedom and heavy quark symmetries as
continuum~quarks.
Thus, the HQET description is admissible for all~$m_Qa$.
Method~4 matches the coefficients of Eq.~(\ref{eq:hqet}) term by term
to Eq.~(\ref{Lag}), by adjusting the lattice action and operators.
In practice, this is possible only to finite order, so there are
errors $({\cal C}^{(n)}_{\rm lat}-{\cal C}^{(n)}_{\rm cont})%
\langle O^{(n)}_{\rm HQET}\rangle$.
The rough size of matrix element here is
$\Lambda_{\rm QCD}^{{\rm dim\,}O-4}$.
The coefficients balance the dimensions with $a$ and $1/m_Q$.
If ${\cal C}^{(n)}_{\rm lat}$ is matched to ${\cal C}^{(n)}_{\rm cont}$
in perturbation theory, the difference is of order~$\alpha_s^\ell$.
Method~3 artificially reduces $m_Qa$ until the mismatch is of 
order~$(m_Qa)^2$.
This would be fine if $m_Qa$ were small enough, but with currently
available lattices, $m_Qa$ is small only if $m_Q$ is reduced until the
heavy quark expansion falls apart.
In method~5 the temporal lattice spacing~$a_t$ is smaller than the
spatial lattice spacing.
The behavior of the mismatch
${\cal C}^{(n)}_{\rm lat}-{\cal C}^{(n)}_{\rm cont}$ for practical
values of~$m_Q$ and~$a_t$ is still an open
question~\cite{Harada:2001ei}.

The non-expert can get a feel for which methods are most appropriate by
asking himself what order in $\Lambda_{\rm QCD}/m_b$ is needed.
For zeroth order, method~1 will do.
Perhaps the only quantity where this is sufficiently accurate is the
mass of the $b$ quark, where the most advanced
calculation~\cite{Gimenez:2000cj} neglects the subleading
term~$\lambda_1/m_b$ in Eq.~(\ref{quarkmass}).
For matrix elements, the first non-trivial terms are those of
Eq.~(\ref{Lag1}), so the other methods must be used.
With method~3 one should check that $m_Q/\Lambda_{\rm QCD}$ is large
enough; so far, all work with this method is worrisome in this
respect.
\index{lattice QCD!lattice spacing effects!heavy quarks|)}

Most of the matrix elements that are of interest to $B$ physics will
soon be recalculated, like $f_B$~\cite{AliKhan:2000eg,Bernard:2000nv}
and $m_b$~\cite{Gimenez:2000cj}, with two flavors of sea quarks.
It seems, therefore, not useful tabulate quenched results.
One can consult recent reviews focusing on the status of matrix
elements instead~\cite{Hashimoto:2000bk}.

\subsubsection{Strong phases of nonleptonic decays}
\label{1:subsubsec:phases}

In considering $CP$ asymmetries one encounters strong
phase\index{strong phase shifts!from lattice QCD} shifts.
It is therefore interesting to consider computing them in lattice QCD.

A short summary is that this is still an unsolved problem, at least for
inelastic decays, such as $B$ decays.
This does not mean that it is an unsolvable problem, but at this time
numerical lattice calculations are not helpful for computing scattering
phases above the inelastic threshold.

Often an even bleaker picture is painted, based on a superficial
understanding a theorem of Maiani and Testa~\cite{Maiani:1990ca}.
The theorem assumes an infinite volume and is, thus, relevant only to
extremely large volumes.
In volumes of $(\mbox{2--6 fm})^3$ it is possible to disentangle phase
information, because the scattering phase shift enters into the
finite-volume boundary conditions of the final-state two-body wave
function~\cite{Luscher:1991ux}.
This works, however, only in the kinematic region with two-body final
states.
This has been worked out explicitly for kaon
decays~\cite{Lellouch:2000pv}, giving also references to earlier work.

\section{Constraints from Kaon Physics}
\label{1:sect:kaon}

There are two strong reasons for the discussion of the neutral kaon
system in a report on $B$ physics.
First, for more than 30 years the only observation of \CP\ 
violation was in the neutral kaon system. Over these years
the formalism used to describe \CP\ violation has changed, partly
because our theoretical understanding of the subject has improved.
One example of this development is the present classification of three,
rather than two, types of \CP\ violation, as explained in
Sec.~\ref{ch1:sect:ty}. 
Second, the Standard Model expresses all \CP\ violating quantities in
terms of the same CKM phase. The consistency of the experiments in $B$
physics with those in the kaon system therefore provides a stringent
test of the Standard Model. In practice both $B$ and $K$ data are used
to overconstrain the unitarity triangle: the indirect constraints on
$\sin 2\beta$, in particular its sign, rely largely on $\epsilon_K$. Any
future inconsistency in the overdetermined unitarity triangle
indicates new physics either in the $B$ or $K$ system or in both. 

Sec.\ref{subs:nks} describes the neutral kaon system with the modern
formalism and makes contact with the formalism traditionally used for
kaon physics.  To show how kaon measurements shape our expectations
for $B$ physics, Sec.~\ref{sect:epsk} discusses the already measured
\CP\ violating quantities $\e_K$ and $\e_K^\prime$.  In a similar
vein, Sec.~\ref{subs:kpnn} deals with the rare decays
$K^+\to\pi^+\nu\bar\nu$ and $K_L\to\pi^0\nu\bar\nu$, which are the
target of new high-precision kaon experiments.

\subsection{The neutral kaon system}
\label{subs:nks}
\index{K mixing@$K$ mixing}%
\index{CP violation@\CP\ violation!kaon physics} 

\CP\ violation in \kkm\ was discovered in 1964 \cite{ccft}. The quantity
$\epsilon_K$, which is discussed in Sec.~\ref{sect:epsk}, is of key
importance to test the CKM mechanism of \CP\ violation, because new
physics enters $K$ and $B$ physics in different ways. We introduce
the neutral kaon system using the same formalism as for the $B$-meson
system as derived in Sec.~\ref{ch1:sect:gen} and translate it to the
traditional notation.

The lighter mass eigenstate of the neutral kaon is $\ket{K_S}$
and the heavier one is $\ket{K_L}$, where the subscripts refer to their
short and long lifetimes.
They are\index{kaon!mass eigenstates}
\begin{eqnarray}
\ket{K_S} &=& p\, \ket{K^0} + q\, \ket{\Kbar^0} 
  = \frac{\lt( 1+ \ov{\e} \rt) \ket{K^0} -        
    \lt( 1- \ov{\e} \rt) \ket{\ov{K}{}^0}}{
    \sqrt{2 \lt( 1+|\ov{\e}|^2 \rt)}} \,,
    \nn 
\ket{K_L} &=& p\, \ket{K^0} - q\, \ket{\Kbar^0}
  = \frac{\lt( 1+ \ov{\e} \rt) \ket{K^0} +        
    \lt( 1- \ov{\e} \rt) \ket{\ov{K}{}^0}}{
    \sqrt{ 2\lt(  1+|\ov{\e}|^2 \rt)}} \,. 
\end{eqnarray}
The quantity 
\begin{equation}
  \ov{\e} = \frac{1+q/p}{1-q/p}
\end{equation}
depends on phase conventions.
(The parameter $\ov{\e}$ is not to be confused with the well-known
parameter~$\e_K$, defined in \eq{ek}.)

\CP\ conservation in \dst\ transitions
corresponds to $\ov{\e}=0$, in which case $\ket{K_S}$ and $\ket{K_L}$
become the \CP\ even and \CP\ odd eigenstates.  \CP\ violation in mixing
is well-established from the semileptonic \CP\ asymmetry
\index{decay!$K_L \to \ell^{\pm}\nu\,\pi^{\mp}$}
\bey
\delta (\ell) &=&   
  {\Gamma(K_L \to \ell^+\nu\,\pi^-) - \Gamma(K_L\to \ell^-\bar\nu\,\pi^+) \over
   \Gamma(K_L \to \ell^+\nu\,\pi^-) + \Gamma(K_L\to \ell^-\bar\nu\,\pi^+)} \nn
&=& \frac{1 - |q/p|^2}{1 + |q/p|^2} 
  = \frac{2\, \real \ov{\e} }{1 + |\ov{\e}|^2} \; = \; 
    \lt( 3.27 \pm 0.12 \rt) \times 10^{-3} \,.
   \label{ksl}
\eey 
\index{CP violation@\CP\ violation!kaon physics!semileptonic asymmetry}%
The quoted numerical value is the average for $\ell=e$ and
$\mu$ \cite{pdg}.  From \eq{ksl} it is clear that in the kaon
system $|q/p|$ is close to one.
In the $B$ systems $|q/p|$ is close to one because the width difference
is smaller than the mass difference.
Here, however, they are comparable~\cite{pdg}%
\index{kaon!mass difference $\dm_K$}%
\index{kaon!width difference $\dg_K$}%
\index{kaon!$q/p$}
\beq
  \Delta m_K = \lt( 0.5301 \pm 0.0014 \rt) \times 10^{10}\, s^{-1}, \qquad
  \Delta \Gamma_K = \lt( 1.1174 \pm 0.0010 \rt) \times 10^{10}\, s^{-1} \,.
\label{dmdgk}
\eeq
\index{phase!of \CP\ violation in mixing, $\phi$!kaon}%
Hence one concludes that $|q/p|-1$ is so small, because the relative
phase $\phi$ between $M_{12}$ and $-\Gamma_{12}$ (cf., \eq{defphi}) is
close to zero. Expanding in $\phi$ one easily finds from \eq{mgqp} that
\begin{mathletters}
\label{mgsolk}
\bey
\lt| M_{12} \rt| &=& \frac{\dm_K}{2} + 
                {\cal O} \lt( \phi^2 \rt), \qquad \qquad
\lt| \Gamma_{12} \rt| = \frac{\dg_K}{2} + 
                {\cal O} \lt( \phi^2 \rt), \label{mgsolk:a}\\
\frac{q}{p} &=& - e^{- i \phi_M} \lt[ 1 - \phi \, 
        \frac{\dg_K/2}{\dm_K+i \, \dg_K/2} + 
        {\cal O} \lt( \phi^2 \rt)  \rt] . 
    \label{mgsolk:b}
\eey
\end{mathletters}%
Hence \eq{mgsolk:b} and \eq{ksl} allow us to solve for the 
\CP\ violating phase $\phi$:
\bey
\phi &=& 
    \frac{ \lt( \dm_K \rt)^2+ \lt( \dg_K/2 \rt)^2}{\dm_K \dg_K/2} 
    \, \delta (\ell) \, + \, {\cal O} \lt( \phi^2 \rt) 
    = \lt( 6.6 \pm 0.2 \rt) \cdot 10^{-3} \,. \label{phknum1}
\eey 

\index{decay!$K \to \pi \pi$}
In the literature on $K\to\pi\pi$ decays the following 
amplitude ratios are introduced:
\begin{eqnarray}
\eta_{+-} &=& \frac{\bra{\pi^+\pi^-} K_L \rangle}{ 
        \bra{\pi^+\pi^-} K_S \rangle} \,,
 \qquad \qquad
\eta_{00} = \frac{\bra{\pi^0\pi^0} K_L \rangle}{ 
    \bra{\pi^0\pi^0} K_S \rangle}  \,.
\end{eqnarray}
If \CP\ were conserved, both would vanish. 
The moduli and phases of $\eta_{+-}$ and $\eta_{00}$ have been 
measured to be
\bey
\lt|\eta_{+-} \rt|
  = \lt( 2.285 \pm 0.019 \rt) \cdot 10^{-3} \,, && \qquad
\phi_{+-} 
  = 43.5^\circ \pm 0.6^\circ \,,  \nn 
\lt|\eta_{00} \rt|
  = \lt( 2.275 \pm 0.019 \rt) \cdot 10^{-3} \,, && \qquad\;\, 
\phi_{00} 
  = 43.4^\circ \pm 1.0^\circ \,,
    \label{epmnum}
\eey
according to the PDG fit \cite{pdg}. 
All three types of \CP\ violation lead to non-zero 
$\eta_{+-}$ and $\eta_{00}$.
To separate \dst\ from \dso\ \CP\ violation one introduces isospin 
states
\bey
\ket{\pi^0 \pi^0 } & = & 
      \sqrt{\frac{1}{3}} \, \ket{\lt(\pi \pi \rt)_{I=0}}  
    - \sqrt{\frac{2}{3}} \, \ket{\lt(\pi \pi \rt)_{I=2}} \,, \nn
\ket{\pi^+ \pi^- } & = & 
      \sqrt{\frac{2}{3}} \, \ket{\lt(\pi \pi \rt)_{I=0}}  
    + \sqrt{\frac{1}{3}} \, \ket{\lt(\pi \pi \rt)_{I=2}}  \,, \no
\eey
and isospin amplitudes
\beq
A_I = \bra{ \lt( \pi \pi  \rt)_I} K^0 \rangle \,, \qquad 
  \ov{A}_I = \bra{ \lt( \pi \pi  \rt)_I} \ov{K}{}^0 \rangle \,, 
  \qquad \qquad I=0,2 \,.
\eeq
\index{kaon!isospin}%
The strong final state interaction of the two-pion final states is
highly constrained by kinematics and conservation laws: the \CP\
invariance of the strong interaction forbids a two-pion state to
scatter into a three-pion state and the rescattering into a state with
four or more pions is kinematically forbidden. Furthermore, isospin is an
almost exact symmetry of QCD and forbids the rescattering between the
two isospin eigenstates. Hence the final state interaction of the
$I=0$ and $I=2$ states is only elastic and, thus, fully described by two
scattering phases.
This feature is known as \emph{Watson's theorem}~\cite{wat}.
Hence we can write
\bey
A_I & = & |A_I| e^{i \Phi_I} e^{i \delta_I} \,, \qquad \qquad
  \ov{A}_I = -|A_I| e^{-i \Phi_I} e^{i \delta_I} \,, \nn 
\lambda_I & = & \frac{q}{p} \frac{\ov{A}_I}{A_I} 
    = e^{-i \lt( 2   \Phi_I + \phi_M   \rt) } 
    \, \lt[ 1 - \phi \frac{\dg_K/2}{\dm_K + i\, \dg/2} \rt] + 
     {\cal O} (\phi^2) \,, \label{ai}
\eey
where the two scattering phases~$\delta_I$ are empirically determined
to be $\delta_0\approx 37^\circ$ and $\delta_2\approx -7^\circ$.
Several weak amplitudes (with different \CP\ violating
phases) contribute to $|A_I| e^{i \Phi_I}$, but the presence of a single
strong phase allows to write $A_I$ as in \eq{ai}, ensuring
$|\ov{A}_I/A_I|=1$.  Therefore, there is no direct \CP\ violation in $K
\to (\pi \pi)_I$. Note that our definition of $A_0$ and $A_2$ includes
both the weak and strong phases, in accordance with the formalism used
in $B$ physics. In the kaon literature the $A_I$'s are commonly defined
without the factors $e^{i\delta_I}$.   

A simplification arises from the experimental observation that 
$|A_0|\simeq 22 \, |A_2|$, which is called $\Delta I =1/2$ rule. This
enhancement of $|A_0|$ allows to expand in $|A_2/A_0|$. The \CP\
violating quantity $\e_K$ reads
\index{CP violation@\CP\ violation!kaon physics!$\e_K$}
\beq
\e_K = \frac{\eta_{00}+2\, \eta_{+-}}{3} =
    \frac{  \bra{(\pi\pi)_{I=0}} K_L \rangle }{ 
        \bra{(\pi\pi)_{I=0}} K_S \rangle  } 
    \lt[ 1+ {\cal O} \lt( \frac{A_2^2}{A_0^2}  \rt) \rt]
    = \frac{1-\lambda_0}{1+\lambda_0} \,. 
\label{ek} 
\eeq
Hence $\e_K$ is defined in a way that to zeroth and first order in
$A_2/A_0$ only a single strong amplitude contributes and therefore
\CP\ violation in decay is absent. The $I=0$ two-pion state 
dominates the $K_S$ width $\Gamma_S$. Thus 
$\bra{K^0}(\pi\pi)_{I=0} \rangle \bra{(\pi\pi)_{I=0}} \ov{K}{}^0
\rangle$ almost saturates $\Gamma_{12}$, so that the phase 
$\phi_M-\phi$ of $-\Gamma_{12}$ equals $-2 \Phi_0$ up to tiny
corrections of order $A_2^2/A_0^2$ and $\Gamma_L/\Gamma_S$. This
implies that $\e_K $ does not provide any additional information
compared to the semileptonic asymmetry in \eq{ksl}. We find from
\eq{ai}
\beq
\lambda_0 = 1 - i \, \phi \frac{\dm_K}{\dm_K + i \, \dg_K } + 
     {\cal O} \lt( \phi^2,\frac{A_2^2}{A_0^2},
               \frac{\Gamma_L}{\Gamma_S}  \rt) ,
\eeq
and \eq{ek} evaluates to
\index{CP violation@\CP\ violation!kaon physics!$\e_K$}
\beq
\e_K = \frac{\phi}{2} \frac{\dm_K}{\sqrt{(\dm_K)^2 + (\dg_K/2)^2  }} \, 
    e^{i \phi_{\e}} 
       +  {\cal O} \lt( \phi^2,\frac{A_2^2}{A_0^2},
               \frac{\Gamma_L}{\Gamma_S}  \rt)  
    \quad \mbox{ with~~}
    \phi_{\e} = \arctan \frac{\dm_K}{\dg_K/2} \,. 
   \label{ekres}
\eeq 
  From \eq{epmnum} one finds the experimental value:
\beq
\e_K = e^{i\, (0.97 \pm 0.02)\, \pi/4}\, (2.28 \pm 0.02) \times 10^{-3} \,.  
\eeq
Therefore \eq{ekres} yields
\index{phase!of \CP\ violation in mixing, $\phi$!kaon}
\beq
\phi = \lt( 6.63 \pm 0.06 \rt) \times 10^{-3} \,, \label{phknum}
\eeq
in perfect agreement with \eq{phknum1}.  A~numerical accident leads to
$\dm_K \approx \dg_K/2$, which explains why the phase $\phi_{\e}$ in
\eq{ekres} is so close to $\pi/4$.

To first order in $\phi$ one finds from \eq{ek} 
\beq
\e_K \simeq \frac{1}{2} \lt[ 1 -\lambda_0  \rt] \; \simeq \;
    \frac{1}{2} 
    \lt( 1 -\lt|\frac{q}{p}\rt| - i \, \imag \lambda_0 \rt) .
\eeq
Therefore $\real \e_K$ measures \CP\ violation in mixing and $\imag \e_K$ 
measures interference type \CP\ violation.
\index{CP violation@\CP\ violation!in mixing!kaon}
\index{CP violation@\CP\ violation!interference type!kaon}

\CP\ violation in \dso\ transitions is characterized by 
\index{CP violation@\CP\ violation!kaon physics!$\e_K^\prime$}
\bey
\e_K^{\prime} & = & 
 \frac{\eta_{+-} - \, \eta_{00}}{3} = 
    \frac{\e_K}{\sqrt{2}} \lt[ 
    \frac{  \bra{(\pi\pi)_{I=2}} K_L \rangle }{ 
        \bra{(\pi\pi)_{I=0}} K_L \rangle  } 
    - 
    \frac{  \bra{(\pi\pi)_{I=2}} K_S \rangle }{ 
        \bra{(\pi\pi)_{I=0}} K_S \rangle  } \rt] 
    \lt[ 1+ {\cal O} \lt( \frac{A_2 }{A_0 }  \rt) \rt] \nn 
&=& \frac{A_2}{A_0}\, \frac{1}{\sqrt{2}} \lt[ 
    \frac{1-\lambda_2}{1+ \lambda_0}  
    - \frac{(1-\lambda_0)(1+\lambda_2)}{(1+ \lambda_0)^2} \rt] 
    . \label{ekp} 
\eey
Next we use  
\beq
\lambda_2 = \lambda_0\, e^{2 \, i\, \lt(\Phi_0-\Phi_2 \rt)} \,,
\eeq
and expand to first order in the small phases:
\beq
\e_K^{\prime} = \frac{1}{2\, \sqrt{2}} \,  
     \frac{A_2}{A_0} \, 
    ( \lambda_0-\lambda_2 ) + 
    {\cal O} \lt( \frac{A_2^2 }{A_0^2 }\,, \phi^2\,, 
    (\Phi_0-\Phi_2)^2 \rt)  
   = \frac{1}{\sqrt{2}}\, \frac{A_2}{A_0}\, i\, 
    ( \Phi_2-\Phi_0 ) \,.   \label{ekpres}
\eeq
A non-vanishing value of $\e_K^{\prime}$ implies different \CP\
violating phases in the two isospin amplitudes and therefore 
\dso\ \CP\ violation.
\index{CP violation@\CP\ violation!in $\Delta F=1$ transitions}
Since experimentally $\real \e_K^\prime >0$,
one finds $ \Phi_2 > \Phi_0$. The phase of $\e_K^{\prime}$ is
$90^\circ +\delta_2-\delta_0 \simeq 46^\circ $ and
$\e_K^{\prime}/\e_K$ is almost real and positive.

Since \eq{ekpres} does not depend on $q/p$, there is no contribution
from \CP\ violation in mixing to $\e_K^{\prime}$.  The strong phases
drop out in the combination
\beq
\imag \frac{A_0}{A_2}\, \e_K^\prime \simeq   \frac{1}{2\, \sqrt{2}} \,  
    ( \imag \lambda_0 - \imag \lambda_2 ) \,. \label{ekpcpm}
\eeq
Since we work to first order in $\phi$, we can set $|\lambda_I|=1$,
and therefore \eq{ekpcpm} purely measures interference type \CP\
violation. From the definition in \eq{ekp} one further finds that
\index{CP violation@\CP\ violation!interference type!kaon}
\beq
\real  \e_K^\prime \simeq \frac{1}{6} 
    \lt( 1- \lt| \frac{A_{\pi^0\pi^0}\, \ov{A}_{\pi^+\pi^-} }{
                  \ov{A}_{\pi^0\pi^0}\, A_{\pi^+\pi^-} }
    \rt|\, \rt)
 \simeq \frac{1}{\sqrt{2}}\, \frac{|A_2|}{|A_0|}
    \, \sin \lt( \delta_0-\delta_2 \rt)
    \lt( \Phi_2 - \Phi_0 \rt)
\eeq
originates solely from $|\ov{A}_f/A_f|\neq 1$.  Hence $\real
\e_K^\prime $ measures \CP\ violation in decay.
\index{CP violation@\CP\ violation!in decay!kaon}

Experimentally the quantity 
$|\eta_{00}/\eta_{+-}|^2= 1- 6\, \real \e_K^\prime/\e_K $ has been
determined. Recent results are 
\bey
\real \frac{\e_K'}{\e_K}  &=& (20.7 \pm 2.8) \times 10^{-4} \qquad 
  {\rm (KTeV)}~\mbox{\cite{KTeV}} \,, \nonumber\\*
\real  \frac{\e_K'}{\e_K} &=& (15.3 \pm 2.6) \times 10^{-4} \qquad 
  {\rm (NA48)}~\mbox{\cite{NA48}} .  
\label{epnum}
\eey 
We therefore find from \eq{ekpres} that the difference of the \CP\
violating phases is tiny:
\beq
\Phi_2-\Phi_0 =  (1.5 \pm 0.2) \cdot 10^{-4} \quad  {\rm (KTeV)}\,, \qquad 
  \Phi_2-\Phi_0 = (1.1 \pm 0.2) \cdot 10^{-4} \quad  {\rm (NA48)}\,. 
\label{ph20num}
\eeq

\boldmath
\subsection{Phenomenology of $\e_K$ and $\e_K^\prime$}
\label{sect:epsk}
\unboldmath
\index{phase!in mass matrix, $\phi_M$!kaon}

In order to exploit the precise measurement of $\phi=-\arg
M_{12}/\Gamma_{12}$ from $\e_K$ in \eq{phknum} one must calculate
the phases of 
\beq
M_{12} = \frac{1}{2 m_K}\, \bra{K^0} H^{\dst} \ket{\ov{K}{}^0} 
    - \mbox{Disp}\, \frac{i}{4 m_K} \int \! \d^4 x \, 
    \bra{K^0} H^{\dso} (x)\, H^{\dso} (0) \ket{\ov{K}{}^0} \,.    
\label{m12k} 
\eeq
and\index{phase!in decay matrix!kaon}
\begin{eqnarray}
\Gamma_{12} & = & \mbox{Abs}\,  
    \frac{i}{2 m_K} \int \! \d^4 x \, 
    \bra{K^0} H^{\dso} (x)\, H^{\dso} (0) \ket{\ov{K}{}^0} 
\label{ga12k} \\
&=& \frac{1}{2 m_K} \sum_f (2\pi)^4 \delta^4 ( p_K-p_f )  
  \bra{K^0} H^{\dso}\ket{f}\, \bra{f} H^{\dso} \ket{\ov{K}{}^0}
  \simeq \frac{1}{2 m_K}\, A_0^*\, \ov{A}_0 \,. \no
\end{eqnarray}
Here Abs\index{absorptive part} denotes the absorptive part of the 
amplitude.  It is calculated by retaining only the imaginary part of 
the loop integration while keeping both real and imaginary parts of 
complex coupling constants.  Analogously, the dispersive part 
Disp\index{dispersive part} is obtained from the real part of the loop 
integral.

The second term in \eq{m12k} shows that, at second order, also the
\dso\ Hamiltonian contributes to $M_{12}$. In the $B$ system the
corresponding contribution is negligibly small.  The Standard Model
\dst\ Hamiltonian reads
\bey 
H^{\dst} = \frac{G_{F}^2}{4 \pi^2}\, M_W\, \Big[ &&
          \lambda_c^{*2} \, \eta_1  \, 
          S (x_c)   \:  + \:
          \lambda_t^{*2} \, \eta_2  \, 
          S (x_t)  \nn 
&& {}       + 2\, \lambda_{c}^* \, \lambda_t^* \, \eta_3 \, 
          S (x_c,x_t) \Big]  
    b_K (\mu)\, Q_K (\mu) +  \mathrm{h.c.} \label{hs2}
\eey
It involves the \dst\ operator 
\beq
Q_K (\mu) = \ov{d}_L \gamma_{\nu} s_L \, \ov{d}_L \gamma^{\nu} s_L \,.
\eeq
\index{Cabibbo-Kobayashi-Maskawa matrix!$V_{td}$}%
\index{Cabibbo-Kobayashi-Maskawa matrix!$V_{ts}$}%
In \eq{hs2} $\lambda_q=V_{qd}V_{qs}^*$, $x_q=m_q^2/M_W^2$ and $S(x)$ is
the Inami-Lim function introduced in \eq{ch1:uli:sxt}. The third
function $S(x_c,x_t)$ comes from the box diagram with one charmed and
one top quark.  One finds $S(x_c) \simeq x_c$, $S(x_c,x_t) \simeq x_c
(0.6-\ln x_c)$ and $S (x_t)\simeq 2.4$ for $m_t\simeq 167\,$GeV in
the ${\ov{\rm MS}}$ scheme. Short distance QCD corrections are
contained in the $\eta_i$'s. In the ${\ov{\rm MS}}$ scheme the
next-to-leading order results are $\eta_1=1.4 \pm 0.3$, $\eta_2=0.57
\pm 0.01$ and $\eta_3=0.47 \pm 0.04 $ \cite{hn1}. $\eta_1$ strongly
depends on $m_c$ and $\alpha_s$, the quoted range corresponds to
$m_c=1.3\,$GeV. A~common factor of the QCD coefficients is 
$b_K(\mu)$, the kaon analogue of $b_B(\mu)$ encountered in
\eq{ch1:uli:c2}. The matrix element of $Q_K$ is parameterized as
\beq
\bra{K^0} Q_K (\mu) \ket{\ov{K}{}^0} =
    \frac{2}{3}\, f_K^2\, m_K^2\, \frac{\widehat{B}_K}{b_K (\mu)} \,,
 \label{defbk} 
\eeq
where $f_K$ is the kaon decay constant.%
\index{hadronic parameter!$\widehat{B}_K$}

\CP\ violation in the  kaon system  is related to the squashed 
unitarity triangle with sides $|\lambda_u|$, $|\lambda_c|$ and
$|\lambda_t|$. In the limit $\lambda_t=0$ all \CP\ violation vanishes,
thus \CP\ violation is governed by the small parameter $\imag
(\lambda_t/\lambda_u)$. This explains the smallness of the measured
phases in \eq{phknum} and \eq{ph20num}. This pattern is a feature of
the CKM mechanism of \CP\ violation and need not hold in extensions of
the Standard Model. Hence kaon physics provides a fertile testing
ground for non-standard \CP\ violation related to the first two
quark generations.  

The presence of the second term in \eq{m12k} impedes the clean
calculation of the mixing phase $\phi_M =\arg M_{12}$ in terms of the
CKM phases. It constitutes a long distance contribution, which is not
proportional to $\widehat{B}_K$. Since both terms in \eq{m12k} have
different weak phases, $\phi_M$ involves the ratio of the two hadronic
matrix elements. This is different from the case in \bbm\, where only
one hadronic matrix element contributes in the Standard Model, which
therefore cancels from $\phi_M$.  The long distance \dso\ piece is
hard to calculate and is usually eliminated with the help of the
experimental value of $\dm_K=2\, |M_{12}|$ in \eq{dmdgk}. Then,
however, our expression for the mixing phase $\phi_M$ still depends on
the hadronic parameter $\widehat{B}_K$. 

The phases of both $\phi_M$ and $\arg \Gamma_{12}$ are close to $\arg
\lambda_u$, which vanishes in the CKM phase convention. The dominant 
corrections to $\phi_M$ stems from the term proportional to
$\lambda_t^{*2}$ in $H^{\dst}$. For $\arg \Gamma_{12}$ we need the 
\dso\ Hamiltonian, which is obtained from the \dbo\ Hamiltonian in 
\eq{hd} by replacing $\xi_{u,c,t}$ with $\lambda_{u,c,t}$ and
replacing the $b$ quark field in the operators by an $s$ field.  The
leading contribution to $\arg (-\Gamma_{12})\approx - 2\, \Phi_0$, is
proportional to $\imag \lambda_t \, \bra{(\pi\pi)_{I=0}} Q_6
\ket{\ov{K}{}^0}/|A_0|$.  The $\Delta I =1/2$ enhancement of $|A_0|$
suppresses $\Phi_0$, which is calculated to $\Phi_0={\cal O} (2 \cdot
10^{-4})$ \cite{ep1}.  Hence in the CKM phase convention $\arg
(-\Gamma_{12})$ contributes roughly 6\% to the measured phase $\phi$
in \eq{phknum} and one can approximate $\phi\approx \phi_M$.  After
expressing the CKM elements in \eq{hs2} in terms of the improved
Wolfenstein parameters the constraint from the measured value in
\eq{phknum}  can be cast in the form \cite{n1,hn1}
\beq
5.3 \times 10^{-4} = \widehat{B}_K A^2 \, \ov{\eta} \,
    \lt\{ \lt[ 1 - \ov{\rho} + \Delta  \lt(\ov{\rho},\ov{\eta} \rt) \rt] 
     A^2 \lambda^4 \eta_2\, S(x_t)
       + \eta_3 S(x_c,x_t) - \eta_1 x_c \right\} . 
\label{cons2}  
\eeq
\index{CP violation@\CP\ violation!kaon physics!$\e_K$ constraint on $\ov{\rho},\ov{\eta}$}%
\index{unitarity triangle!kaon physics}%
In the absence of the small term $\Delta \lt( \ov{\rho},\ov{\eta}
\rt) = \lambda^2 \, \lt( \ov{\rho} -\ov{\rho}^2 -\ov{\eta}^2 \rt)$
this equation defines a hyperbola in the $(\ov{\rho},\ov{\eta})$ plane.
The largest uncertainties in \eq{cons2} stem from $\widehat{B}_K$ and 
$A=|V_{cb}|/\lambda^2$, which enters the largest term in \eq{cons2} raised
to the fourth power. Hence, reducing the error of $|V_{cb}|$ improves the 
$\e_K$ constraint.

It is more difficult to analyze $\e_K^\prime$, because the weak phases
$\Phi_0$ and $\Phi_2$ are much harder to compute than $\phi_M$.
$\Phi_2$ is essentially proportional to $\imag \lambda_t \,
\bra{(\pi\pi)_{I=2}}\, Q_8\, \ket{\ov{K}{}^0}/|A_2|$.  The two matrix
elements entering $\Phi_2-\Phi_0$ are difficult to calculate and
numerically tend to cancel each other. Especially there is a
controversy about $\bra{(\pi\pi)_{I=0}} Q_6 \ket{\ov{K}{}^0}$ and the
different theoretical estimates can accommodate for both the KTeV and
the NA48 result in \eq{ph20num} \cite{ep1,kns}. Even after the
experimental discrepancy in \eq{epnum} is resolved, $\e_K^\prime$ will
not immediately be useful to determine $\imag \lambda_t \simeq A^2
\lambda ^5 \ov{\eta}$.  Nevertheless, $\e_K^\prime$ can be useful to
constrain new physics contributions \cite{bs}. For recent overviews on
$\e_K^\prime$ we refer to \cite{eprev}.

\boldmath
\subsection{$K \to \pi \nu\bar\nu$}
\label{subs:kpnn}
\unboldmath
\index{kaon!rare decay}
\index{decay!$K^+ \to \pi^+ \nu\bar\nu$}
\index{Cabibbo-Kobayashi-Maskawa matrix!$V_{td}$}
\index{Cabibbo-Kobayashi-Maskawa matrix!$V_{ts}$}

Rare kaon decays triggered by loop-induced $s\to d$ transitions can
provide information on $\lambda_t$ and thereby on the shape of the
unitarity triangle. Final states with charged leptons are poorly
suited for a clean extraction of this information, because they
involve diagrams with photon-meson couplings. Such diagrams are
affected by long distance hadronic effects and are hard to evaluate.
The decays $K^+ \to \pi^+\nu\ov{\nu}$ and $K_L \to \pi^0\nu\ov{\nu}$,
however, are theoretically very clean, with negligible hadronic
uncertainties. The $K \to \pi$ form factors can be extracted from the
well-measured $K_{\ell3}$ decays. So far two 
$K^+ \to \pi^+\nu\ov{\nu}$ events have been observed \cite{bnl},
corresponding to a branching ratio
\beq
{\cal B} (K^+ \to \pi^+\nu\ov{\nu}) = \Big(1.57 \epm{1.75}{0.82}\Big) \times 10^{-10} 
    \,. \label{kppexp}
\eeq
Experimental proposals at BNL and Fermilab aim at a measurement  
of $K^+\to \pi^+\nu\ov{\nu}$ and $K_L \to \pi^0\nu\ov{\nu}$
at the 10\% level. The constraint on the improved
Wolfenstein parameters $(\ov{\rho},\ov{\eta})$ can be cast in the form
\cite{bbl}
\beq
\frac{{\cal B}(K^+ \to \pi^+ \nu \ov{\nu}) }{4.57 \cdot 10^{-11}} 
= A^4 X^2(x_t) \lt( 1- \lambda^2 \rt)
    \bigg[ \bigg( \frac{\ov{\eta}}{ 1 - \lambda^2 } \bigg)^2 
     + \lt( \rho_0 - \ov{\rho} \rt)^2 \bigg]  
    \label{kpp} .
\eeq
Here $X (x_t)\simeq 1.50$ comprises the dependence on $m_t$ and the
NLO short distance QCD corrections \cite{kpnncalc}.  $\rho_0 \approx
1+0.27/A^2$ contains the contribution from the internal charm loop
\cite{kpnncalc}.  The quoted numerical value corresponds to a $\ov{{\rm
MS}}$ mass of $m_c=1.3\,$GeV.  The largest theoretical uncertainty in
\eq{kpp}, of order 5\%, stems from the charm contribution. Further the
fourth power of $A$ introduces a sizable parametric uncertainty.  The
equation in \eq{kpp} describes an ellipse in the
$(\ov{\rho},\ov{\eta})$ plane centered at $(\rho_0,0)$. By
inserting typical values for the Wolfenstein parameters (e.g.,
$\lambda=0.22$, $A=0.8$, $\ov{\rho}=0.2$ and $\ov{\eta}=0.4$ ) into
\eq{kpp} one finds that \eq{kppexp} is compatible with the Standard
Model.

\index{CP violation@\CP\ violation!interference type!kaon}
\index{decay!$K_L \to \pi^0 \nu\bar\nu$}
In the Standard Model the decay $K_L \to \pi^0 \nu \ov{\nu}$ is \CP\
violating. It measures interference type \CP\ violation, the
associated phase $\arg \lambda_{\pi^0 \nu \ov{\nu}}$ is large, of
order $\ov{\eta}/(1-\ov{\rho})$. This is in sharp contrast to the
small phases we found in \eq{phknum} and \eq{ph20num}. A~measurement
of ${\cal B}(K_L \to \pi^0 \nu \ov{\nu})$ establishes $\arg \lambda_{\pi^0
\nu \ov{\nu}} \neq \phi $ and therefore implies \CP\ violation in the
\dso\ Hamiltonian. This is the same situation as with $\imag
(\e_K^\prime\, A_0/A_2)$ in \eq{ekpcpm}, which also proves \dso\ 
\CP\ violation from the difference of two interference type \CP\ 
violating phases. $K_L \to \pi^0 \nu \ov{\nu}$ is even cleaner than
$K^+\to \pi^+\nu\ov{\nu}$, because the charm contribution is
negligible.  ${\cal B}(K_L \to \pi^0 \nu \ov{\nu})$ is proportional to
$(\imag \lambda_t)^2 \propto \ov{\eta}^2$ and therefore determines
the height of the unitarity triangle
\beq
\frac{{\cal B}(K_L \to \pi^0 \nu \ov{\nu}) }{1.91 \cdot 10^{-10}} 
= A^4 X^2(x_t)\, \ov{\eta}^2 \lt( 1+ \lambda^2 \rt)
    \label{kpp0} .
\eeq
Hence the two discussed branching ratios allow for a precise
construction of the unitarity triangle from kaon physics alone.
Moreover the ratio of the two branching ratios is almost independent
of $A$ and $m_t$. It allows for a determination of $\sin 2\beta$ with a
similar precision as from $a_{CP} (B\to \psi K_S)$ \cite{kpnnphen}.
New physics may enter $s\to d$ transitions in a different way than $b \to
s$ and $b \to d$ transitions. Hence comparing of the unitarity
triangles from $K$ physics and from $B$ physics provides an excellent test
of the Standard Model.
\index{unitarity triangle!kaon physics}
\index{decay!$B_d \to \psi K_S$}

\section{Standard Model Expectations }
\label{ch1:SM-expect}

\OMIT{
{\bf I believe that it has to be reconsidered now [after 1.5 years] whether
this is the right way to finish this section.  ---Z}
{\bf I think it's fine. -- A}
}

This section outlines what is known about the CKM matrix at the present
time, and what the pattern of expectations is for some of the most
interesting processes in the Standard Model.


Since most of the existing data apart from $\sin2\beta$ come from $CP$
conserving measurements, it is convenient to present the constraints on the CKM
matrix using the Wolfenstein parameterization.  Magnitudes of CKM matrix
elements are simply related to $\lambda$, $A$, $\bar\rho$, and $\bar\eta$. The
best known of these is $\lambda$, the Cabibbo angle, which is known at the 1\%
level.  The parameter $A$ is determined by $|V_{cb}|$, which is known with a
5\% error.  The uncertainty in $\bar\rho$ and $\bar\eta$ is significantly
larger.  The most important constraints come
from\index{unitarity triangle!kaon physics}%
\index{CP violation@\CP\ violation!kaon physics!$\e_K$ constraint on $\ov{\rho},\ov{\eta}$}%
\index{unitarity triangle!and Vub@and $V_{ub}$}%
\index{unitarity triangle!and \bbmd}%
\index{unitarity triangle!and \bbms}%
\begin{itemize} \vspace*{-8pt} \itemsep -2pt

\item $CP$ violation in $K^0\!-\!\Kbar^0$ mixing described by the $\epsilon_K$
parameter;

\item $|V_{ub}/V_{cb}|$ measured from semileptonic $B$ decays;

\item $B^0\!-\!\B0bar$ mixing;

\item The lower limit on $B_s\!-\!\Bbar_s$ mixing.

\end{itemize} \vspace*{-8pt}

\begin{figure}[t]
\centerline{\epsfxsize 0.85\textwidth \epsffile{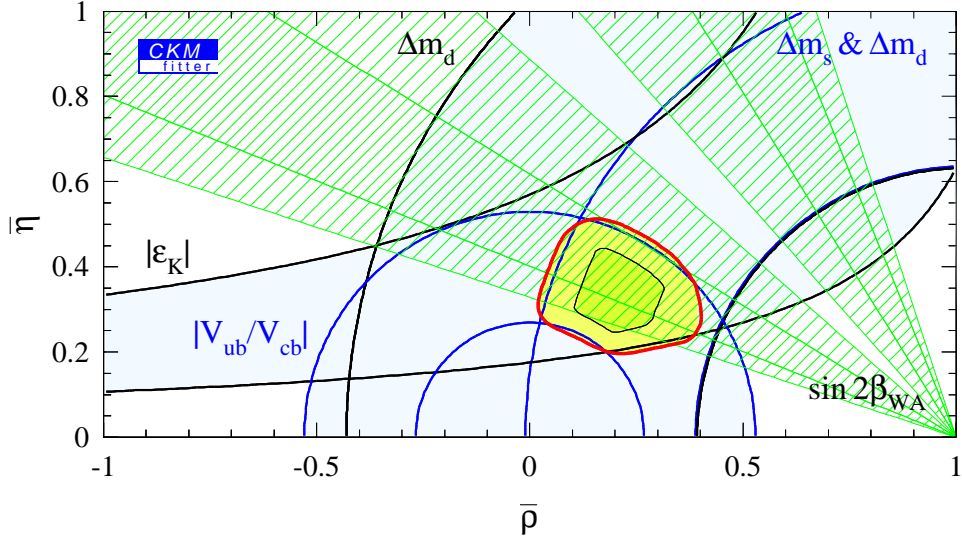}}
\caption[The allowed region in the $\bar\rho-\bar\eta$ plane.]{The allowed
region in the $\bar\rho-\bar\eta$ plane. Also shown are the individual
constraints, and the world average $\sin2\beta$. (From Ref.~\cite{CKMfitter}.)
}
\label{1:refit}
\end{figure}

\begin{figure}[t]
\centerline{\epsfxsize 0.85\textwidth \epsffile{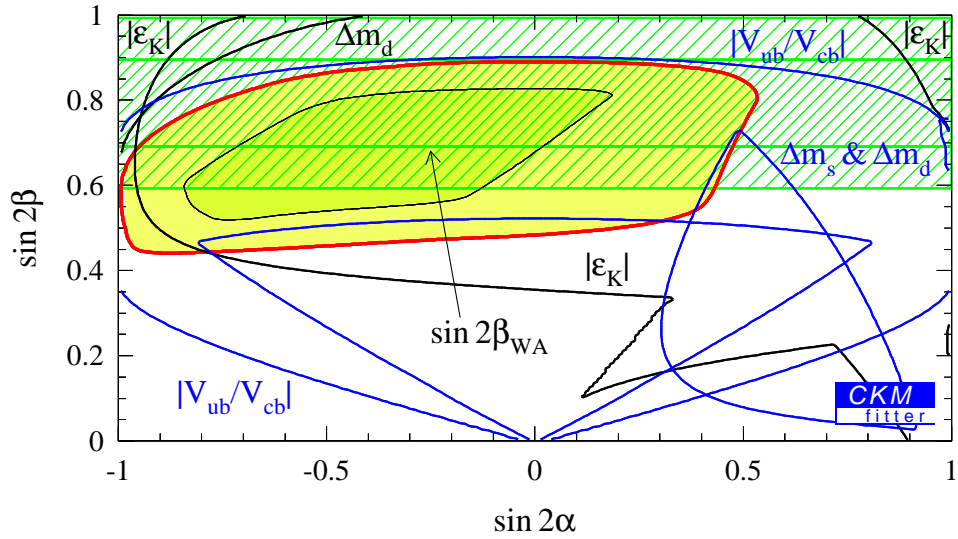}}
\caption[The allowed region in the $\sin2\alpha-\sin2\beta$ plane.]{The
allowed region in the $\sin2\alpha-\sin2\beta$ plane~\cite{CKMfitter}. }
\label{1:abfit}
\end{figure}

\begin{figure}[t]
\centerline{\epsfxsize 0.85\textwidth \epsffile{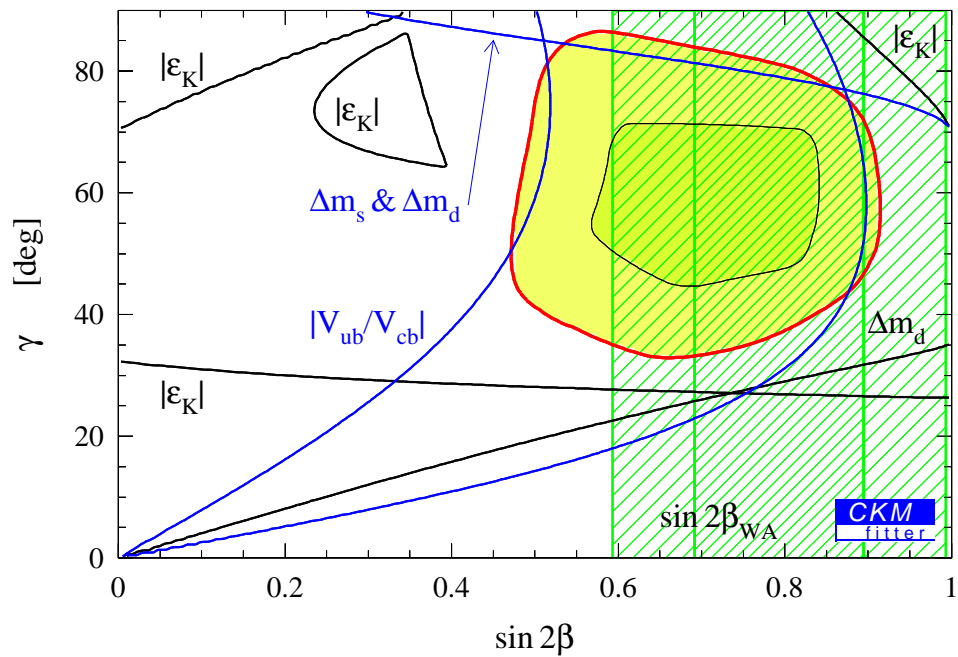}}
\caption[The allowed region in the $\sin2\beta-\gamma$ plane.]{The
allowed region in the $\sin2\beta-\gamma$ plane~\cite{CKMfitter}. }
\label{1:bgfit}
\end{figure}

A problem in translating these to constraints on the CKM matrix is related to
theoretical uncertainties.  We follow the point of view adopted in the BaBar
book~\cite{babook} that no confidence level can be attached to model dependent
theory errors.  Fig.~\ref{1:refit} shows the result of such an analysis from
Ref.~\cite{CKMfitter}. Fig.~\ref{1:abfit} shows the same fit on the
$\sin2\alpha-\sin2\beta$ plane.  Note that any value for $\sin2\alpha$ would
still be allowed if $|V_{ub}|$ were slightly larger, or if $\Delta m_{B_s}$
were slightly smaller than their allowed ranges.  Fig.~\ref{1:bgfit} shows the
allowed range in the $\sin2\beta - \gamma$ plane, and that $\gamma$ is already
constrained.

Some of the uncertainties entering these constraints will be significantly
reduced during Run~II.
The hadronic matrix elements $B_K$, $f_{B_d}^2 B_{B_d}$, and
$f_{B_s}^2 B_{B_s}$ need to be determined by unquenched lattice QCD
calculations.
The theoretical uncertainties in $|V_{ub}|$ and $|V_{cb}|$ will also be reduced
to the few percent level by unquenched lattice calculations of the exclusive
$\B0bar_d \to \pi\, \ell\, \bar\nu$%
\index{decay!$B_d \to \pi \ell \nu$}
and $\B0bar_d\to D^{(*)} \ell\, \bar\nu$%
\index{decay!$B_d \to  D^{(*)} \ell \nu$}
form factors in the region of phase space where the momentum of the final
hadron is small.
As discussed in Sec.~\ref{1:lattice}, these lattice calculations are
straightforward in principle, but a variety of uncertainties must be
brought under control.
The uncertainties in these two CKM matrix elements may be
reduced in the next few years, even without recourse to lattice QCD, 
using inclusive semileptonic decays.
The error in $|V_{cb}|$ may be reduced to 2--3\% with precise
determinations of a short distance $b$ quark mass and by gaining more
confidence about the smallness of quark-hadron duality violation.
On a similar timescale the error in $|V_{ub}|$ may be reduced to the 5--10\%
level \cite{Vub} by pursuing several model independent determinations.


\clearpage{\pagestyle{empty}\cleardoublepage}
\chapter{Common Experimental Issues}

\authors{R.~Kutschke, M. Paulini}   

\section{Introduction}

  This chapter will discuss the experimental issues
which underlie $B$ physics at CDF, D\O\ and BTeV.
Many of these issues also apply to charm physics, which will 
also be discussed.
The chapter will be painted in fairly broad strokes and 
the reader is referred to the subsequent
chapters and to the experiments' own Technical Design
Reports (TDR)~\cite{cdf_tdr}\cite{cdf_tof_l00}\cite{d0_tdr}\cite{btev_tdr}
for more details on specific experiments.

  During Run~II, the Fermilab Tevatron will collide counter-rotating
proton $p$ and anti-proton $\bar{p}$ beams at a 
center-of-mass energy of 2~TeV.  Some other design parameters of the
Tevatron for Run~II are summarized in Table~\ref{2:tab:tevatron}.
In rough terms there
are three processes which take place at this energy and which
are important to the design
of a $B$ physics experiment.  These are the production of
$b\bar{b}$ pairs, the production of $c\bar{c}$ pairs
and all of the light quark and gluon processes which
contribute to the background; the cross-sections
for these processes are summarized in 
Table~\ref{2:tab:xsecs}.  There are no known
processes which produce a single $b$ or a single $\bar{b}$
at a significant rate, only processes which produce pairs.
Despite this, one usually talks about $b$ production,
not $b\bar{b}$ production.
Similarly, there are important sources of $c\bar{c}$ production
but not of single $c$ or $\bar{c}$ production.  The
theory behind the production of heavy quark pairs
in $p\bar{p}$ collisions is discussed in 
chapter~9.
There are, of course, many
other interesting processes
which occur, including top quark production, Higgs boson
production and perhaps even the production of supersymmetric
particles.  The cross-sections for these processes, however, are
small enough that they do not have any impact on
how one designs a $B$ physics experiment for the Tevatron.
\index{b-quark production@$b$-quark production! impact on experiments} 

  After a $b\bar{b}$ pair is produced, it hadronizes to
form pairs of $b$ hadrons including $B$ mesons, such
as $B_d$, $B_u$, $B_s$, $B_c$,
and $b$ baryons such as 
$\Lambda_b$, $\Xi_b$, $\Omega_b$, $\Xi_{bc}$, $\Omega_{cc}$ etc.
All of these states decay weakly, with a significant
lifetime and, therefore, with a significant decay length.
Excited states of these $b$ hadrons are also produced,
all of which decay strongly or electromagnetically to one
of the weakly decaying $b$ hadrons.  A similar
picture exists for the hadronization of 
$c\bar{c}$ pairs into hadrons.
Therefore the route to
all of $b$ and $c$ physics goes through the weakly decaying states.

  One shorthand which will be used in the following is,
\begin{equation}
\sigma_{BG} = \sigma_{tot}-\sigma_{c\bar{c}}-\sigma_{b\bar{b}}.
\end{equation}
This stands for the ``background'' cross-section; that is for the
total hadronic cross-section with the
$c\bar{c}$ and $b\bar{b}$ pieces excluded.  The 
$c\bar{c}$ piece is treated separately because it is interesting
to study in its own right and because it has a few critical
properties which are more like $b$'s than background.
At our current level of precision, 
$\sigma_{BG}\simeq\sigma_{tot}\simeq 75$~mb\cite{bgxsec}.
This includes elastic $p\bar{p}$ scattering, diffractive scattering
and inelastic scattering.
Because $\sigma_{BG}\simeq\sigma_{tot}$ many authors are 
are careless about distinguishing between the two.

\begin{table}[tbf]
\begin{center}
\begin{tabular}{cc} \hline
Quantity & Value \\ \hline \hline
Center of Mass Energy         & 2~TeV \\
Peak Instantaneous Luminosity & 
                        $2\times 10^{32}\; {\rm cm}^{-2}\; {\rm s}^{-1}$ 
                        \\ 
Yearly Integrated Luminosity & 2~fb$^{-1}$/year \\
Time between bunch crossings &
          \multicolumn{1}{l}{\quad\quad\quad 396~ns for $\simeq$ 2 years} \\
       &  \multicolumn{1}{l}{\quad\quad\quad 132~ns afterwards} \\
Luminous region & $(\sigma_x,\sigma_y,\sigma_z) = 
                   ( 0.003, 0.003, 30.)$~cm \\ \hline
\end{tabular}
\medskip
\caption[Tevatron parameters for Run~II]
{Tevatron parameters for Run~II. The conversion
from peak instantaneous luminosity to yearly integrated
luminosity assumes that a year consists of $10^7$ useful
seconds, as discussed in Section~\protect{\ref{2:sec:lumi}}.
\index{Tevatron parameters}
}
\label{2:tab:tevatron}
\end{center}
\end{table}

\begin{table}[tbf]
\begin{center}
\begin{tabular}{ccc} \hline
Quantity & ~~Value (mb)~~ & Comment \\ \hline \hline
$\sigma_{tot}$ & $\approx 75$ & Total hadronic cross-section including \\
                               && elastic, diffractive and inelastic 
                                  processes. \\
$\sigma_{c\bar{c}}$ & $\approx 1$  & Charm pair production cross-section. 
                                         \\
$\sigma_{b\bar{b}}$ & $\approx 0.1$ & Beauty pair production cross-section.
                                         \\
$\sigma_{BG}$ & $\approx 75$& The chapter's short-hand for
                      $\sigma_{tot}-\sigma_{c\bar{c}}-\sigma_{b\bar{b}}$.
\\ \hline
\end{tabular}
\medskip
\caption[Approximate values of cross-sections]
{Approximate values of the cross-sections which are of interest to a $B$
physics experiment using $p\bar{p}$ collisions at a center-of-mass energy of
2~TeV.  The estimate for the total cross-section is from
Ref.~\protect{\cite{bgxsec}}. The estimate of $\sigma_{b\bar{b}}$ is
discussed in Section~\ref{2:ssec:xsec}.%
\index{cross-sections!table of}%
}
\label{2:tab:xsecs}
\end{center}
\end{table}

  Throughout this chapter, the $z$ axis is defined to lie
along the beam direction and quantities such as $p_T$ are measured
with respect to this axis.  The variable $\varphi$ is the azimuth
around the $z$ axis and $\theta$ is the polar angle relative
to the $z$ axis.

\section{Separating \lowercase{$b$} and \lowercase{$c$} Hadrons from the
Backgrounds}

Inspection of Table~\ref{2:tab:xsecs} shows that
the cross-section for $b$ production is about 1.5 parts in 1000 of the 
total cross-section.  Moreover, many of the $B$ physics
processes of interest have product branching 
fractions of $10^{-6}$ or smaller.\footnote{
The product branching fraction is defined
as the product of all of the branching fractions in a decay
chain.  An example of such a decay chain is
$B_d\to J/\psi K^0$, $J/\psi\to \mu^+\mu^-$, 
$K^0\to K^0_S$, $K^0_S\to \pi^+\pi^-$.  
The branching fraction for the first decay is
about $1\times 10^{-3}$, but the product of all branching
fractions in the chain is much smaller, about $2\times 10^{-5}$.
}
Therefore one is often looking for signals
of a few parts per billion of the total cross-section!
%

  The signature which allows one to see this needle in
a haystack is the lifetime of the $b$ quark.  The $B_d$, $B_u$ and
$B_s$ mesons each have a lifetime $\tau$ of approximately
1.5~ps, or $c\tau=450$~$\mu$m.  When the momentum spectrum
of the $B$ mesons is folded in, the mean decay length of
all produced $B$ mesons is on the order of a few mm. 
Therefore almost all $B$ mesons decay inside the beam pipe.
The resolution on the decay length varies from one decay
mode to another and from one experiment to another but typical
values fall in the range of 50~$\mu$m to 100~$\mu$m;
therefore the $B$ decay vertices will be well resolved
and will be readily separated from the $p\bar{p}$ interaction vertex.
The $B_c$, the weakly decaying $b$ baryons and the weakly decaying 
charmed hadrons have somewhat shorter lifetimes\cite{PDG}, but
most of them have a long enough lifetime that their decay vertices too
will also be well separated from  $p\bar{p}$ interaction vertex.

The myriad background processes, with their much larger cross-sections,
do not produce particles which have this type of decay length
signature.  This brings us to the magic bullet: it is the presence 
of distinct secondary vertices which allows
the experiments to extract the $b$ and $c$ signals from the background.

About 85\% of all weakly decaying
$b$ hadrons decay into one charmed hadron plus long lived particles.
Long lived particles include pions, kaons, protons, photons,
charged leptons and neutrinos, all of which are stable enough to
escape the interaction region and leave tracks in the detector.   
Some of these particles, such as the $K^0_S$ and $\Lambda$ do decay
but their lifetimes are very long compared to the those of the $b$ 
and $c$ hadrons.\footnote{
The one exception is the $\tau$ lepton but the branching
ratio of $b\to c \tau\nu$ is small, $(2.6\pm0.4)\%$\cite{PDG}.}
About 
15\% of weakly decaying $b$ hadrons decay into 2 charmed hadrons, plus
long lived particles; the decay  $B^0\to D^{*+} D^{*-}$ is an example.
And about 1\% of all weakly decaying $b$ hadrons 
decay into only long lived particles; 
the decay $B^0\to \pi^+\pi^-$ is an example.
Therefore a typical $b\bar{b}$ event has 5
distinct vertices, all inside the beam pipe: the primary $p\bar{p}$ interaction
vertex, the two secondary $B$ decay vertices and the two
tertiary $c$ decay vertices.  On the other hand, many of the
most interesting decays involve charmless decays of the $b$ and
a typical event containing one of these
these decays has 4 vertices inside the beam pipe:
the primary $p\bar{p}$ vertex, 
the vertex from the signal charmless $b$, and the $b$ and $c$ vertices
from the other $b$ (or $\bar{b}$) produced in the $p\bar{p}$ 
interaction.

\begin{figure}
\centerline{\epsfig{file=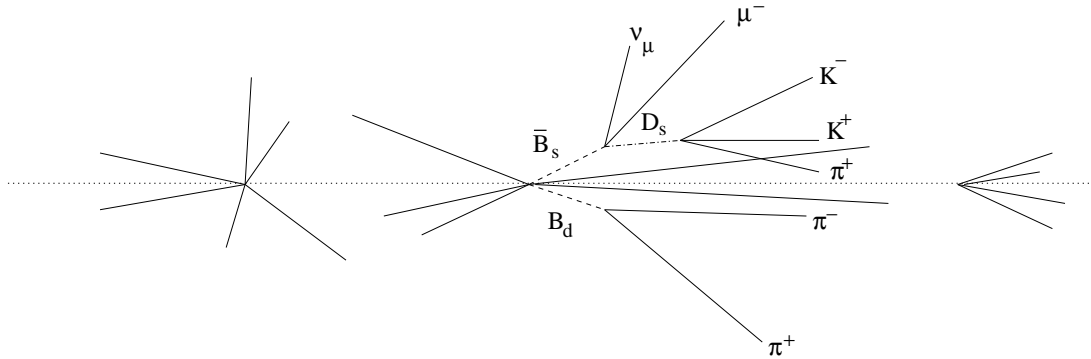, width=0.95\hsize}} 
\caption[A cartoon of a $b\bar{b}$ event]
{A cartoon of an interesting
$b\bar{b}$ event at the Tevatron.  In this
beam-crossing the bunches undergo three independent primary
interactions.  The one in the middle produces a $b\bar{b}$
pair plus some other hadrons while the ones to either side are 
background interactions.  In this event, one $b$ undergoes a charmless
decay while the other decays semileptonically to charm, which decays
hadronically to light hadrons.  The cartoon is meant to emphasize
the topological properties of an event: it is not to scale and does
not correctly represent the number of tracks
in a vertex or the distribution of track directions.
}
\label{2:fig:cartoon}
\index{event! cartoon of a typical}
\end{figure}

  To be complete, one more detail must be added to the description of typical
$b\bar{b}$ events.  For the running conditions anticipated  for Run~II, each
beam crossing which contains a $b\bar{b}$ interaction will also contain several
background interactions which contain no $b\bar{b}$ or $c\bar{c}$ pairs. This
is discussed in more detail in Section~\ref{2:sec:multi}.
Fig.~\ref{2:fig:cartoon} shows a cartoon of a $b\bar{b}$  event with one
charmless $b$ decay. Throughout this chapter the word ``event'' should be
understood to include all of the interactions within one beam crossing, both
the signal and background interactions.

Similarly, a typical $c\bar{c}$ interaction has
three distinct vertices inside the beam pipe: the primary vertex
plus two secondary vertices, which come from the decay of the
two charmed hadrons.  A typical beam crossing which contains
a $c\bar{c}$ interaction will also contain a few background interactions.

In an event containing a $b\bar{b}$ or
a $c\bar{c}$ pair, the stable daughters
of the $b$ and $c$ hadrons usually have a large impact parameter with
respect to the primary vertex.  
Because the beam spot is very narrow, roughly 30 $\mu$m in diameter, 
these tracks will also have a large $r\varphi$ impact parameter
with respect to the beam line.  A track is said to be detached
if the impact parameter, divided by its error, is large; this definition
is used both for 2D and 3D impact parameters.

  While the reconstruction of the full vertex topology
of an event is a very powerful tool to reduce backgrounds, it is often
too inefficient or too slow to be useful.  In particular present
computing technologies are too slow to allow full exploitation
of the topology at trigger time.  However a powerful
trigger can be made by looking for the presence of a few
detached tracks.   All of CDF, D\O\ and BTeV
have design triggers which make some sort of detachment requirement,
with the sophistication of that requirement changing from one
experiment to the next.   BTeV exploits detachment at all trigger
levels, including level 1, while the other experiments introduce
detachment cuts only at higher levels.  
The reader is referred to chapters~3 to~5
for further details about the triggers of each experiment.

In addition to detachment, there are other properties which can be 
used to identify events which contain $b$ quarks.  For example,
selecting events with one or two 
leptons of moderate to high $p_T$ is an excellent way to select 
events containing $b\bar{b}$ pairs while rejecting background events.  
CDF and D\O\ have successfully used single lepton and di-lepton triggers,
without any detachment requirement, to select events for their Run~I $B$ 
physics program.  
All of the Run~II detectors plan some sort of lepton triggers,
including single high $p_T$ leptons, di-leptons and 
$\psi\to\mu^+\mu^-$ triggers.
The experiments envisage some, but not all, of these triggers also
to include detachment information. For example,
when evidence of detachment is present, one can lower $p_T$
thresholds and still have an acceptable background suppression.
But when detachment information is ignored, or unavailable,
the triggers require higher $p_T$ thresholds.
Those triggers 
which do not include detachment information will provide
a useful sample for calibrating the detachment based triggers.

One of the limitations of Run~I was that the experiments
could only trigger on $b$ events with leptons in the final state.
For Run~II and beyond, both CDF and BTeV have triggers which rely only on
detachment and which are capable of triggering all hadronic final states.
The development of triggers which rely only on detachment 
is  one of the major advances since Run~I.

  Because of their topological similarities to $b\bar{b}$ events,
some $c\bar{c}$ events will also pass
these triggers.  Charm events, however, have properties which
are intermediate to the $b$ events and the background events: 
their decay lengths are shorter,
their impact parameters smaller and their stable daughters have 
both a softer momentum spectrum and a softer $p_T$ spectrum.
Therefore the cuts which reduce the background
to an acceptable level are much less efficient for $c\bar{c}$ 
events than they are for $b\bar{b}$ events.  The CDF and
D\O\ experiments do not expect that significant $c\bar{c}$ samples will
pass their trigger and have not discussed a charm physics
program.  They can, of course, do some charm physics with the
charm which is produced via $B$ decay.  BTeV, on the other hand,
expects that a significant fraction of the events which pass their
trigger, will contain $c\bar{c}$ events and they plan a
charm physics program to exploit that data.

  In summary, the long lifetimes of the weakly decaying $b$
and $c$ hadrons are the magic bullet which allow the $b$ and $c$
physics to be extracted from the background.  At trigger time
minimal cuts will be made on detachment and the offline analyses
will make more complete use of the topological information.
Various lepton based triggers, some with detachment requirements
and some without, will form a second set of triggers.

\section{Sources of Backgrounds } 
\label{2:sec:backgrounds}

The most pernicious backgrounds are those which peak in the
signal region and which can fake signals.  One example of
this is a true $B^{0}\to K^+ \pi^-$ being misreconstructed as
a $B^{0}\to \pi^+ \pi^-$ decay; this results in a peak which is
almost at the correct mass, with almost the correct width.
This sort of problem is very mode specific and will be discussed,
as needed, in the working group chapters.

  A second class of backgrounds is combinatoric background
within true $b\bar{b}$ and $c\bar{c}$ events, events which have
the correct topological properties to pass the trigger.
Suppose that one is looking for the decay, 
$B^0\to D^- \pi^+$ followed by $D^-\to K^+\pi^-\pi^-$.
All $b\bar{b}$ and $c\bar{c}$ events which produce a reconstructed 
$D^-\to K^+\pi^-\pi^-$ candidate have potential to produce background.
If another track, perhaps from the main vertex, forms a good vertex 
with the $D^-$ candidate this will be considered, incorrectly, as
a $B$ candidate.  This sort of background will not peak near the 
$B$ mass but it will produce background entries throughout the 
$D^-\pi^+$ mass plot, thereby diluting the signal.  This sort
of background can be reduced by demanding that tracks which 
participate in a $B$ candidate be inconsistent with the primary vertex.
In addition, improved vertexing precision will reduce the number of
random $D^-\pi^+$ combinations which form a vertex with an
acceptable $\chi^2$.

   There are many other background sources
of secondary vertices and detached
tracks: strange particles, interactions
of particles with the detector material, misreconstructed tracks,
multiple interactions per beam crossing, and
mis-reconstructed vertices.  While none of these backgrounds
can create fake mass peaks at the $B$ mass, they can dilute signals 
and they can overwhelm a poorly designed trigger.

At first thought one might summarily dismiss the strange hadrons as a 
source of background.  After all, they typically have lifetimes 
100 to 1000 times
longer than those of the $b$ hadrons; so only a small fraction will
decay inside the beam pipe with a decay length typical
of that for $B$ decay.  However they are produced about
a few thousand times more frequently, a few per background
interaction.  Moreover the most probable decay time of
an exponential distribution is zero, so some of the strange
hadrons will have decay lengths of a few~mm.
There is a powerful countermeasure against most of the
strange particle
background: the trigger must ignore tracks with an
impact parameter which is too large.  
One might worry that the contradictory requirements of a 
large detachment but a small impact parameter
might leave no window to accept the physics.   The answer is
clear if one recalls the definition of detachment, an impact
parameter divided by its error:  make the error
small.  In practice the detectors have sufficiently good
resolution that this background is reduced to  acceptable level.

The strange hadrons have masses much less than the those of
the $b$ hadrons;  therefore, an isolated decay of a strange particle
is unlikely to be confused for a $b$ decay.  If however,
another track, or tracks, pass close enough to the strange 
particle decay vertex that the reconstruction code incorrectly 
assigns that track to that strange particle's decay vertex,
the combination can contribute combinatorial background beneath
a $B$ signal.  This sort of background can be reduced by 
building a vertex detector with sufficiently high precision.

Another source of background comes from the
interaction of the tracks with the detector and support materials.
Photons can pair convert and hadrons can undergo
inelastic collisions.  There may be several of these secondary interactions
for each primary $p\bar{p}$ interaction.  Again, this sort of background
can be suppressed, at the trigger level, by excluding
tracks with too much detachment.  And, in the offline analysis, one can
exclude vertices which occur in the detector
material.  Having excellent resolution on vertex position
is again the secret to background reduction.

Other sorts of interaction with the detector material includes
Rutherford scattering and the tails of multiple 
Coulomb scattering.  At the trigger level, the way to deal with these 
tracks is to make sure that the detachment cuts are large enough.
At the offline reconstruction level, these tracks can often be
rejected by cutting on the confidence level of the
track fit.

Mis-reconstructed tracks are tracks which have incorrect
hit assignments.  The most direct way to deal with this problem
is to ensure a sufficiently small occupancy in the detectors.
For example, the upper limit for the long 
dimension of the BTeV pixels is set by such a study --- if the pixels 
are too long then the two track separation degrades and errors
in pattern recognition result.  This, in turn, creates false detached
tracks.  In the offline analysis, one can also reject mis-reconstructed
tracks on the basis of a bad track fit confidence level.

Multiple interactions in one beam crossing are another source of
background.   Consider the case that 
two background interactions occur in the same beam 
crossing.  In this case there are two chances that a background
interaction might trigger the detector.  But there is the
additional complication that the trigger might fire
based on some information from one vertex and some information 
from the other vertex.  This last problem can be reduced by
doing 3D vertexing in the trigger.

At the Tevatron the luminous region has a length of
$\sigma_z\approx30$~cm so it is reasonable to expect the triggers
to behave acceptably with a few background interactions per beam
crossing; most of the time the interactions will be well separated.
When testing trigger algorithms it is important to measure how
the trigger degrades with an increasing number of interactions per
crossing.  The trigger performance should degrade smoothly, without
sudden drops.  

The last background class is misreconstructed vertices, which includes
both errors made when all of the tracks are well measured but also
errors made when one of the tracks suffers from one of the diseases
mentioned above.   The solution is to ensure sufficient tracking 
precision that fake track rates are small and sufficient vertex
precision that the rates for accidental vertices are small.

The above discussion has presented a number of factors  which
bound the detachment required at the trigger level from below
and which bound impact parameter cuts from above.
Using detailed simulations of their detector response,
the experiments have shown that their proposed triggers will
reduce these backgrounds to an acceptable level and that their
detectors have enough rejection power to obtain an acceptable
signal-to-background ratio during offline analysis.  The common
thread running through the discussion is that improved
vertex resolution reduces every one of these backgrounds.

\section{Basics of $b$ Production Physics}
\label{2:ssec:xsec}

\begin{figure}
\centerline{\epsfig{file=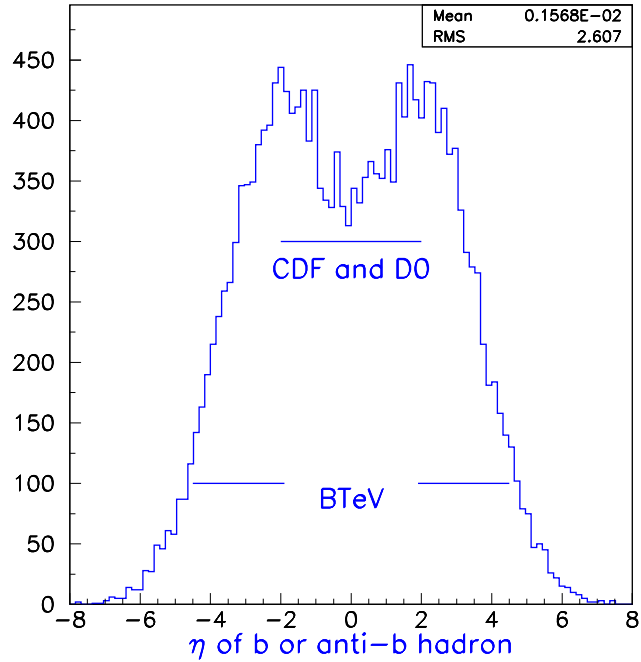, height=3.6in}} 
\caption[Production cross-section vs $\eta$ for $B$ mesons (from {\tt PYTHIA})]
{The production cross-section of $B$ mesons
vs $\eta$.  The plot is from the {\tt PYTHIA}
event generator and does not contain detector effects. 
The horizontal lines show the regions of
$\eta$ which are covered by the three detectors.
CDF and D\O\ do not cover all regions of $\eta$ with
equally quality;  the barrel region, in which they make their
best measurements covers approximately $|\eta|<1.0$.
}
\label{2:fig:vseta}
\index{psuedo-rapidity ($\eta$)!coverage of detectors}
\index{$\eta$ (psuedo-rapidity)!coverage of detectors}
\end{figure}

 At a $p\bar{p}$ collider, it is usually most convenient
to describe particle production
in terms of three variables,
$p_T$, $y$ and $\varphi$, where $p_T$ is the transverse
momentum of the particle with respect to the beam line,
$\varphi$ is the azimuth around the beam line and where
the rapidity, $y$, is a measure of the polar angle, $\theta$, relative
to the beam line,
\begin{equation}
y = \frac{1}{2}\ln\left( \frac{E+P_{\parallel}}{E-P_{\parallel}}
       \right).
\end{equation}
\index{rapidity ($y$)!definition of}
\index{$y$ (rapidity)!definition of}
For historical reasons people sometimes work in units of pseudo-rapidity,
$\eta$, instead of $y$,
\begin{equation}
\eta = -\ln\left( \tan\theta/2\right).
\end{equation}
\index{psuedo-rapidity ($\eta$)!definition of}
\index{$\eta$ (psuedo-rapidity)!definition of}
For massless particles $\eta=y$ and for
highly relativistic particles $\eta$ approaches $y$.
The utility of the variable $\eta$ can be seen
in Fig.~\ref{2:fig:vseta}, which shows the prediction of
the {\tt PYTHIA} Monte Carlo event generator for
the production of $b$ flavored hadrons as a function of $\eta$.
The production is approximately flat in the central $\eta$
region, falling off at large $|\eta|$,  a general
feature of particle production in hadronic collisions.
The figure also shows the regions of $\eta$ which are covered by 
the three detectors.

\begin{figure}[thb]
\centerline{\epsfig{file=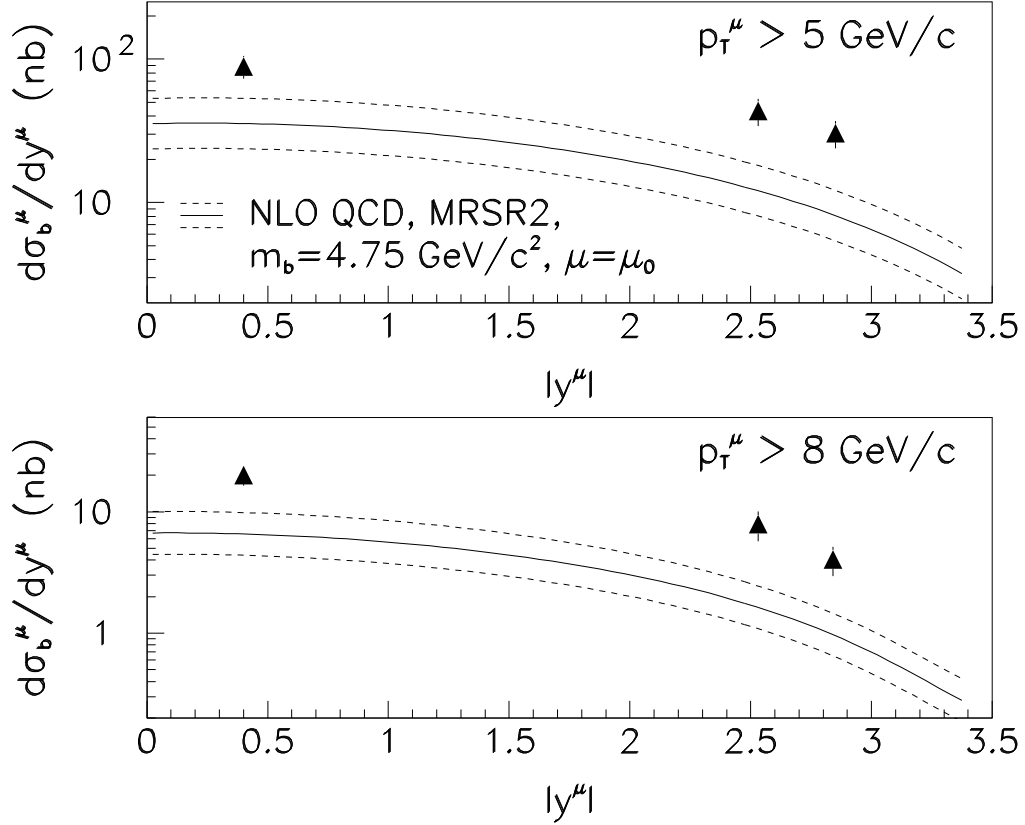, width=\hsize}} 
\caption[$\sigma$($\mu$ from B decay) vs $y_{\mu}$]
{The cross-section for muons from $b$-decay as a function of the rapidity the
of muon, $y^{\mu}$, measured by D\O\ .  The solid curve is the prediction of
the next-to-leading order QCD calculation for a $b$-quark mass of 4.75 GeV. The
dashed curves represent the estimated theoretical 1$\sigma$ error band.}
\label{2:fig:d0xsec}
\index{cross-sections!D\O\ measurement}
\end{figure}

During Run~I, both CDF and D\O\ studied the production of $b$ quarks 
in $p\bar{p}$ collisions at a center-of-mass energy of 1.8~TeV.
Both CDF and D\O\ have studied the central rapidity region $|\eta|<1$
and D\O\ has also studied the forward region, $2.4< y < 3.2$.
The data of D\O\ are shown in Fig.~\ref{2:fig:d0xsec}.
Both CDF and D\O\ find that the
$b\bar{b}$ production cross-section in the central region is
underestimated by the Mangano, Nason and Ridolfi (MNR) next-to-leading order
QCD calculation~\cite{MNR} by
a factor of more than two.
The D\O\ data in the higher $y^{\mu}$ region is $3.6\pm 0.8$ times 
higher than the QCD calculation.

  When predicting their sensitivities for physics at Run~II,
CDF and D\O\ normalize their predictions to the
cross-sections which they measured in Run~I.
Not only does BTeV not 
have previous data, there are no experimental data at all 
over much of the range of the BTeV acceptance, $1.9\le|\eta|\le4.5$.
Instead BTeV uses the following procedure.  When integrated
over $\eta$ and $p_T$, the QCD predictions shown in 
Fig.~\ref{2:fig:d0xsec} predict a total $b\bar{b}$ production
cross-section of 50~$\mu$b.  Since all of the experimental data
is more than a factor of two above the theoretical calculations,
BTeV estimates the total cross-section
to be 100$\mu$b.  BTeV then uses the predictions of {\tt PYTHIA} to describe
how the cross-section is distributed over $p_T,\eta,\varphi$.
Within regions of phase space covered by CDF and D\O\, {\tt PYTHIA} has
done a good job of describing the most important experimental correlations.
 
\begin{figure}
\centerline{\epsfig{file=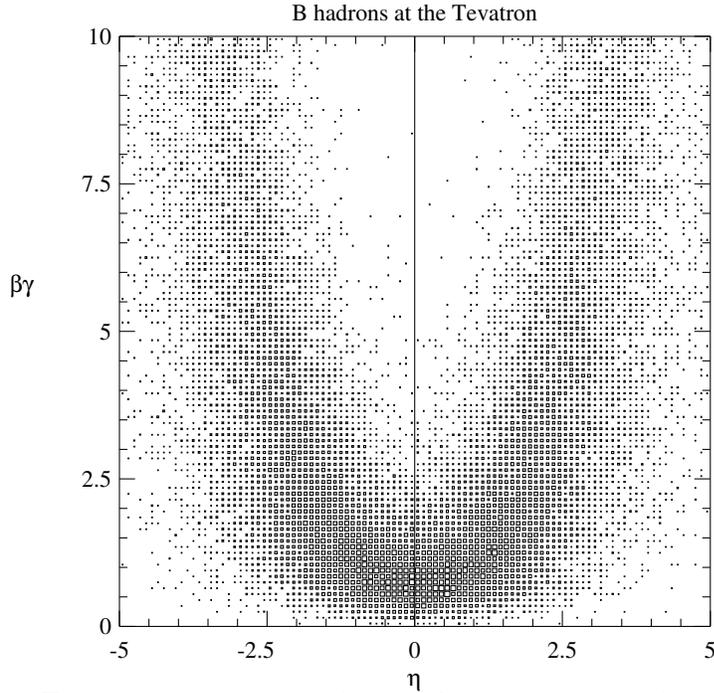, height=3.6in}} 
\caption[$\sigma_{b\overline{b}}$ as a function of $\beta\gamma$ vs $\eta$]
{The production cross-section for  $B$ mesons as a function
of $\beta\gamma$ and $\eta$ plane.  The plot is from the {\tt PYTHIA}
event generator and does not contain detector effects. }
\label{2:fig:bgvseta}
\index{psuedo-rapidity ($\eta$)!$\beta$ vs $\eta$}
\index{$\eta$ (psuedo-rapidity)!$\beta$ vs $\eta$}
\end{figure}

\begin{figure}
\centerline{\epsfig{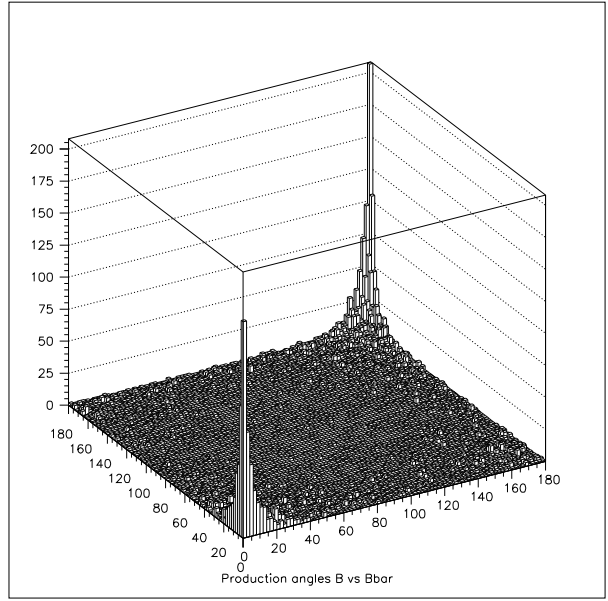}}
\vspace*{4pt}
\caption[Production angle correlations for $B\overline{B}$ pairs]
{The production angle (in degrees) for the hadron
containing a $b$ quark plotted versus the production angle for a hadron
containing $\bar{b}$ quark.  The plot is from the {\tt PYTHIA}
event generator and does not contain detector effects. 
One must be careful interpreting this plot since the natural
axes are $\eta$, not $\theta$.
}
\label{2:fig:thetavstheta}
\end{figure}

\begin{figure}
\centerline{\epsfig{file=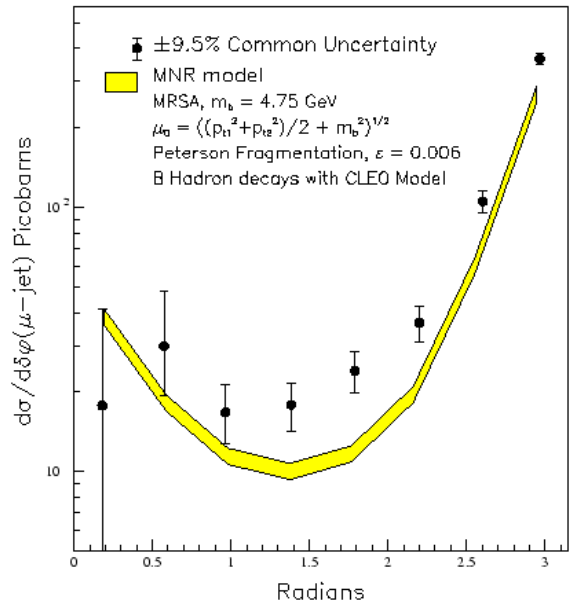, height=3.2in}} 
\caption[Azimuthal angle correlation between $B\overline{B}$ pairs]
{The differential $\delta\varphi$ cross-sections for
$p^{\mu}_T> 9 $ GeV/c, $|\eta^{\mu}|<$0.6, E$^{\bar{b}}_T>$10 GeV,
$\big|\eta^{\bar{b}}\big|<1.5$ compared with theoretical predictions. The
data points have a common systematic uncertainty of $\pm$9.5\%. The uncertainty
in the theory curve arises from the error on the muonic branching fraction
and the uncertainty in the fragmentation model.
}
\label{2:fig:vsphi}
\end{figure}

  Other properties of $b\bar{b}$ production are illustrated
in Fig.~\ref{2:fig:bgvseta} through~\ref{2:fig:vsphi}.
Fig.~\ref{2:fig:bgvseta} shows, for $B$ mesons, the prediction
of the {\tt PYTHIA} event generator for the cross-section as
a function of $\beta\gamma$ vs $\eta$.  The figure shows that the
bulk of the cross-section is concentrated in the central region and
that forward going $B$ mesons have a much higher momentum than
do $B$ mesons produced in the central region.  This implies that
in the forward region a greater fraction of the cross-section
has long decay lengths, while in the central region there are
more events to start with.  The implications of this
tradeoff will be discussed further in Section~\ref{2:sec:centfor}.

   Fig.~\ref{2:fig:thetavstheta} illustrates another of the
the properties of $b\bar{b}$ production, that $b$~hadron
and the $\bar{b}$~hadron have an RMS separation of about one unit
of $\eta$.  The figure was  made using generator level tracks from the
{\tt PYTHIA} event generator and shows
the cross-section as a function of the polar angle of one 
$B$ vs the polar angle of the other $B$.  
In a two-arm forward detector, such as BTeV,
if one $B$ is produced in a particular arm, then the other $B$
is highly likely to be produced in the same arm.
This is important for measurements which make use 
of opposite side tagging (see Section~\ref{2:sec:exp_common_tagging}).
The choice of axes for this figure, $\theta$ rather than $\eta$,
exaggerates this effect.  Some other consequences of this distribution
are discussed in Section~\ref{2:sec:centfor}.

  Fig.~\ref{2:fig:vsphi} shows the azimuthal correlation
between a $b$ and its $\bar{b}$ partner.  The data are for D\O\
events in which two muons are reconstructed, both consistent
with coming from the decay of a $b$ hadron.  
The horizontal axis is $\delta\varphi$,
the difference in azimuth between the two muons.  Since the selection
criteria imply that the two $B$'s have a significant momentum,
the muons tend to follow the $B$ direction.  Therefore $\delta\varphi$
is a measure of the difference in azimuth between the two $b$ hadrons
in the event.  The band shows the prediction of MNR\cite{MNR}.
The $b$ hadrons are preferentially produced back to back in azimuth
and the gross shape is reproduced well by the model. 
It has already been noted that
the MNR prediction underestimates the cross-section.

  While the production of $b\bar{b}$ pairs is well described
by perturbative QCD and knowledge of the structure functions 
of the proton, the hadronization, or fragmentation, of these
quarks into the final state hadrons is described by models.
These models are usually realized as computer codes for event
generators, the most commonly used being {\tt PYTHIA}\cite{pythia},
{\tt ISAJET}\cite{isajet} and {\tt HERWIG}\cite{herwig}.  
One of the properties which must be input to the event generators is
the fraction of time that the $b$ quark fragments into each of the
allowed hadrons, $B^{-}$, $\bar{B}^0$, $\bar{B}_s$, $B_c^-$ 
or one of the $b$ baryons.  A recent measurement from
CDF\cite{cdfprodfrac} gives,
$f_u:f_d:f_s:f_{\rm baryon}$ 
$= 0.375\pm0.023:0.375\pm0.023:0.160\pm0.044:0.090\pm0.029$, 
with the assumption that $f_u=f_d$.  If they release this
assumption they obtain, $f_d/f_u=0.84\pm0.16$.
It is generally presumed that, except for threshold effects,
the fragmentation process is independent of the production 
process and is roughly independent of energy.  For comparison
the same production fractions measured at LEP and SLD 
are\cite{lepprodfrac},
$f_u:f_d:f_s:f_{\rm baryon}$ 
$=0.401\pm0.010:0.401\pm0.010:0.100\pm0.012: 0.099\pm0.017$.
In both of these measurements, the production of $B_c$
mesons is too small to be significant.
In the standard event generators
the choice of hadron species for the $b$ quark
is independent of the choice of hadron species for the $\bar{b}$
quark.  This cannot be exactly true since there is presumably some
production via the $\Upsilon(4S)$ resonance, which decays only
to $B^0\overline{B}^0$ or $B^+B^-$. Moreover the 
$B^0\overline{B}^0$ production from the $\Upsilon(4S)$ is coherent.
While it is likely that these effects do occur, they can be safely
ignored for purposes of this workshop.  If there is enough resonant
production to affect the physics results, the amount of such
production can be easily measured with the Run~II data.

The major points of this section are summarized in
Table~\ref{2:tab:bbbarprop}.

\begin{table}
\begin{enumerate}
\hrule\vspace*{2pt} \hrule
\item The mechanisms which produce heavy flavors produce
      $b\bar{b}$ pairs but not single $b$ or $\bar{b}$ quarks.
      Similarly for charm production.
\item The cross-section for $b\bar{b}$ production, integrated over
      all $\eta$ and $p_T$,  is about 100~$\mu$b
      and that for $c\bar{c}$ production is about 1~mb.
\item The $b\bar{b}$ cross-section is approximately flat in $\eta$ over the
      central region, and falls off at large $|\eta|$.
      See figure~\ref{2:fig:vseta}.
\item $b$ hadrons produced in the forward region have a higher momentum
      than those produced in the central region.
      See figure~\ref{2:fig:bgvseta}.
\item The pair of $b$ hadrons from one $b\bar{b}$ pair are approximately
      approximately back to back in $\varphi$ and have an RMS separation
      in $\eta$ of about one unit of $\eta$.
\item The production ratio of $B_u:B_d:B_s:$baryons is,
      $f_u:f_d:f_s:f_{\rm baryon}                           
       = 0.375\pm0.023:0.375\pm0.023:0.160\pm0.044:0.090\pm0.029$
      \cite{cdfprodfrac}.
\vspace*{4pt}
\hrule\vspace*{2pt} \hrule
\end{enumerate}
\caption{Summary of the important properties of $b\bar{b}$ production. }
\label{2:tab:bbbarprop}
\end{table}

\section{Production Rates and Interactions Per Crossing}
\label{2:sec:lumi}
\index{interactions!mean rates of}

The design value for the
peak instantaneous luminosity during Run~II
is $2\times 10^{32}\;{\rm cm}^{-2}\;{\rm s}^{-1}$.
This specifies the luminosity at the start of a fill, when
the beam intensities are greatest.  As a fill progresses
the instantaneous luminosity will drop.  Also there will
be shutdowns, both planned and unplanned, throughout the running
period.
The rule of thumb for converting the peak
instantaneous luminosity to the yearly integrated luminosity
is to assume that a year contains $10^7$ seconds of running
at the peak instantaneous luminosity.   This is about one third 
of the actual number of seconds in a year, which 
accounts both for the drop in luminosity as a fill
progresses and for a normal amount of down-time.
Therefore a peak instantaneous luminosity of 
$2\times 10^{32}\;{\rm cm}^{-2}\;{\rm s}^{-1}$ 
corresponds to 2~fb$^{-1}$/year.

   Given $\sigma_{b\bar{b}}=100\; \mu$b from Table~\ref{2:tab:xsecs},
the above luminosities imply $b\bar{b}$ yield of
20,000/s or $2\times 10^{11}$/year, about 3 to 4 orders of magnitude
larger than the projected yields at the $e^+e^-$ $B$ factories.

   Given $\sigma_{tot}=75$~mb, from Table~\ref{2:tab:xsecs},
a luminosity of $2\times 10^{32}\;{\rm cm}^{-2}\;{\rm s}^{-1}$
implies a total interaction rate of
$1.5\times 10^7$/s.  During the first few years of Run~II the
bunch structure of the Tevatron will be 396~ns between
bunch crossings.   At the design luminosity this would
correspond to about 6 interactions per crossing but it is 
not expected that the design luminosity will be achieved 
this early in the run.  After the first few years of Run~II the
bunch structure of the Tevatron will be changed to have
132~ns between bunch crossings.  The purpose of this change
is to allow an increase in luminosity while reducing the
number interactions per beam crossing.  At 132~ns between
bunches, the design luminosity corresponds to about 2 interactions per 
bunch crossing.

   The above discussion, along with the corresponding numbers
for $c\bar{c}$ production, is summarized in Table~\ref{2:tab:rates}.

The presence of multiple background interactions has 
many consequences for the design of the detector.  It was
already mentioned that the trigger must be robust against
multiple background interactions in one beam crossing.
The presence of multiple interactions must also be considered
when designing the granularity of detectors to ensure that
the occupancy is acceptably low.

\begin{table}
\begin{center}
\begin{tabular}{lccc} \hline
Rate        & $b\bar{b}$ & $c\bar{c}$ & Total \\ \hline \hline
Interactions/s    & $2\times 10^4$ & $2\times 10^5$       &$1.5\times 10^7$ \\
Interactions/year & $2\times 10^{11}$ & $2\times 10^{12}$ &
                                                         $1.5\times 10^{14}$ \\
Interactions/crossing @ 396~ns & $0.008$ & $0.08$  & 6 \\
Interactions/crossing @ 132~ns & $0.003$ & $0.03$  & 2 \\ \hline
\end{tabular}
\medskip
\caption[Interactions per beam crossing]
{Summary of production rates for $b\bar{b}$ pairs,
$c\bar{c}$ pairs and total interactions for the design
peak luminosity of the Tevatron during Run~II,
$2\times 10^{32}\;{\rm cm}^{-2}\;{\rm s}^{-1}$
(2~fb${-1}$/year). The interactions
per bunch crossing are given twice, once for the bunch structure
planned for early in Run~II, 396~ns between bunch crossings, and once
for the bunch structure planned for later in Run~II, 132~ns
between bunch crossings.
}
\label{2:tab:rates}
\end{center}
\end{table}

\subsection{The Distribution of Interactions Per Crossing}
\label{2:sec:multi}
\index{interactions!Poisson distribution of}

To a good approximation, if there are multiple interactions
in one beam crossing, they are statistically
independent of each other. This is not 
strictly true because, once the first interaction takes place, 
there are fewer beam particles left to participate in future
interactions.  However, in the limit that the number of
particles per bunch is much larger than the number of
interactions per crossing, each interaction can be treated
as independent of all others.

Each of these independent interactions has some probability
to produce a signal interaction and some probability to produce
a background interaction.  There are two, equivalent ways of
looking at the distribution of signal and background interactions
among the multiple interactions in one event.  These equivalent
ways are related to each other by the following identity.
Given two independent
Poisson processes, signal and background for example,
the probability to observe $n_1$
interactions from the first process and $n_2$ interactions from the second
process is,
\begin{eqnarray}
P(n_1,n_2) &=& \frac{(\mu_1)^{n_1}}{n_1!}\; e^{-\mu_1}\;   
                \frac{(\mu_2)^{n_2}}{n_2!}\; e^{-\mu_2} \nonumber \\
           &=& \left[ \frac{\mu^n}{n!}
                \; e^{-\mu} \right] \;
               \left[f^{n_1} (1-f)^{n-n_1}
               \frac{n!}{n_1!(n-n_1)!} \right],
\end{eqnarray}
where $f=\mu_1/\mu$ and $\mu=\mu_1+\mu_2$.
The first factor in $[]$ is the Poisson probability to observe
\hbox{$n=n_1+n_2$} interactions in total, while the second factor in $[]$ is
the binomial probability that the $n$ interactions are split into $n_1$
from the first process and $n-n_1$ from the second process.

One can also show the general case,
that the sum of $M$ independent Poisson processes is itself a Poisson
process with a mean $\mu=\mu_1+\mu_2+\dots+\mu_M$.
In the general case, the factor multiplying the overall Poisson 
distribution will be a multinomial distribution, 
with the $M-1$ independent parameters, $\mu_1/\mu$, $\mu_2/\mu$,
\dots $\mu_{M-1}/\mu$.

The two equivalent descriptions are:
first, one can say that 
the total number of interactions within a beam crossing is
Poisson distributed with a mean of $\mu$
and that within each beam crossing the interactions are distributed
among the possible types according to a multinomial distribution.
Second, one can say that there
are $M$ independent pieces to the cross-section and that 
each piece contributes to each beam crossing
a Poisson distributed number of interactions with mean 
$\mu_M$.

This second description is less well known but it allows one
to more easily answer the following question: describe a
typical beam crossing which produces a $b\bar{b}$ pair.
For definiteness, consider the numbers summarized in 
Table~\ref{2:tab:rates} for the case of 132~ns bunch spacing;
$\mu_{b\bar{b}}=0.003$, $\mu_{c\bar{c}}=0.03$, and $\mu_{BG}=2.0$\ .
Clearly most beam crossings will contain no $b\bar{b}$ pairs.
An event which contains a typical $b\bar{b}$ pair will contain
exactly one such pair and it will be accompanied by
a Poisson distributed number of $c\bar{c}$ interactions
with a mean of 0.03 interactions per crossing and by a
Poisson distributed number of background interactions
with a mean of 2.0 interactions per crossing.

\section{Flavor Tagging}
\label{2:sec:exp_common_tagging}
\index{flavor tagging!motivation for}

One of the main $B$ physics goals of all three experimental
programs is to make precision measurements of mixing mediated CP 
violating effects, some of which are 
discussed in chapter~6 of this report.  Also, $x_s$
has yet to be measured and that
is interesting to measure in its own right.   In order to
perform any mixing related study it is
necessary to know whether a particular meson was produced as 
a $B^0(B_s)$ or as a $\bar{B}^0(\bar{B}_s)$.  Making
such a determination is called flavor tagging the $B$ 
meson.\footnote{There is another, and very different concept
called $b$ tagging.
If a lepton is part of a jet, and if the $p_T$ of the lepton with 
respect to the jet axis is sufficiently large, 
then that lepton is most probably from the 
decay of a $b$ quark within the jet.  A sample of jets containing 
such a lepton will be heavily enriched in $b$ jets.
This technique was used extensively in Run~I
to tag samples of $b$ jets which were used in the $W$ boson
and top quark physics programs.  This technique is mostly
of interest for top physics, not for $B$ physics itself.
}

Every tagging method sometimes produces the wrong answer
and the effectiveness of flavor tagging is characterized by an
effective tagging efficiency $\epsilon\,D^2$,
where
$\epsilon=(N_R+N_W)/N$, 
$D=(N_R-N_W)/(N_R+N_W)$, $N$ is the number of reconstructed 
signal events before tagging, $N_R$ the number of right flavor tags,
and where $N_W$ is the number of wrong flavor tags.  
Another useful expression is $D=(1-2w)$ where $w=N_W/(N_R+N_W)$ is the
fraction of wrong sign tags; from this expression it is clear that 
the tagging power goes to
zero when the wrong sign tag fraction reaches 50\%.
Maximizing
$\epsilon D^2$ is critical to the design of every experiment.

  The quantity $D$ is known as the dilution.  This choice
of nomenclature has the anti-intuitive result
that a large dilution is good while a small dilution is bad.
Never-the-less it is the standard nomenclature.
\index{$D$(tagging dilution)}
\index{$\epsilon D^2$ (tagging power)}
\index{flavor tagging!dilution ($D$)}
\index{flavor tagging!power ($\epsilon D^2$)}

Tagging algorithms can be broken down into two classes, away
side tagging and same side tagging.  In away side tagging,
or opposite side tagging,
one looks at some property of the other $b$ hadron in the event
to determine its $b$ quantum number.   Since $b$ quarks are 
produced as $b\bar{b}$ pairs, one can infer the flavor of
the signal $B$ meson.   In same side tagging one uses
the correlations which exist between the signal $B$ meson
and the charge of nearby tracks produced either in the fragmentation
chain or in the decay of $B^{**}$ resonances.  For tagging $B^0$
mesons the correlation is with a charged pion, while for $B_s$
mesons the correlation is with a charged kaon.

\subsection{Away Side Tagging}
\index{flavor tagging!description of!away side}

The perfect away side tag would be to fully reconstruct the
other $b$ hadron in the event and to discover that it is a
$B^-$ or a $\Lambda_b$, neither of which undergoes flavor mixing.
In this case one knows that the other $b$ hadron contains
a $b$ quark and that the signal $B$ meson must have been born
with a $\bar{b}$ quark.  So the signal $B$ is tagged as being born
as a $B^0$ or as a $B_s$.  
In practice the efficiency for
reconstructing a complete $b$ hadron on the away side is much
too small to be useful.  Instead one looks for inclusive properties
of $b$ hadrons which are different from those of $\bar{b}$ hadrons.
Four such properties have been explored: lepton tagging,
kaon tagging, jet charge tagging and vertex charge tagging.

Lepton tagging exploits the sign of the lepton in the
decays $b\to X \ell^-$ compared to
$\bar{b}\to X \ell^+$, where $\ell$ is either an electron
or a muon.  The branching fractions for these decays is roughly 10\%
into each of the $e$ and $\mu$ channels.  There is some dilution
in this tag from the decay chain $b\to c \to X \ell^+ $
compared to
$\bar{b}\to \bar{c} \to X \ell^- $.   However the two different
sources of leptons have different kinematic properties and
different vertex topology properties.  So good separation between
these two sources of leptons can be achieved.  Another factor
causes further dilution.  In an ensemble of tags, the away 
side $b$ hadron will be some mixture of $B^+$, $B^0$, $B_s$ and 
several $b$ baryons.  The $B^+$ and the $b$ baryons do not mix and so
the observation of the sign of the lepton is a clear tag.  However,
17.4\% of the $B^0$ mesons will oscillate to $\bar{B}^0$ mesons
before decaying\cite{PDG} and will, therefore, give an incorrect tag.
The $B_s$ system, which is fully mixed, 
provides no tagging power at all.

  Kaon tagging exploits the charge of the kaon in the away side
decay chain,
$b\to c \to X K^-$ compared with $\bar{b}\to \bar{c} \to X K^+$.
Because of the large product branching fraction this tag has a much
higher efficiency than lepton tags but historically has had 
worse dilution.  With the improved vertexing power and
particle identification capabilities of the of the
Run~II detectors one expects significantly improved dilutions.
As with lepton tagging, there is tagging dilution from the mixing
of the away side $B^0$ and $B_s$.   It is often noted that a typical 
$B_s$ decay contains two kaons of opposite strangeness and so 
contributes no power to kaon tagging.  While this is true, one
must remember that the $B_s$ system is fully mixed and had no
tagging power to start with.

  A method called ``jet charge tagging''
exploits the fact that the sign of the momentum weighted sum of the
particle charges of the opposite side $b$~jet is the same as the
sign of the charge of the $b$~quark producing this jet.
In a simple version, the jet charge $Q_{\rm jet}$ can be calculated as
\begin{equation}
  Q_{\rm jet} = \frac{ \sum_i q_i\; (\vec{p}_i \cdot\hat{a}) }
                     { \sum_i \vec{p}_i \cdot\hat{a}  },
\end{equation}
where $q_i$ and $\vec{p}_i$ are the charge and momentum
of track $i$ in the jet and $\hat{a}$ is
a unit~vector along the jet axis.
On average, the sign of the jet charge is the same as the sign
of the $b$~quark that produced the jet.

  Vertex charge tagging involves reconstructing the full 
vertex topology of the away side.  This does not necessarily
constitute full reconstruction of the away side since the
away side decay will usually contain $\pi^0$'s, photons,
$K^0_S$ and $K^0_L$.  However these missing particles do
not modify the charges of the remnant vertices.
If the vertices have been correctly reconstructed, and if
the away side secondary vertex has a charge of $\pm 1$, then the 
flavor of the away side $b$ is known.  If the charge of the
away side secondary vertex is zero, then there is no tagging
power.  Also, if the away side tertiary vertex has charge
$\pm 1$, one can infer the flavor of the away side $b$.

\subsection{Same Side Tagging}
\index{flavor tagging!description of!same side}

\begin{figure}[thb]
\centerline{\epsfig{file=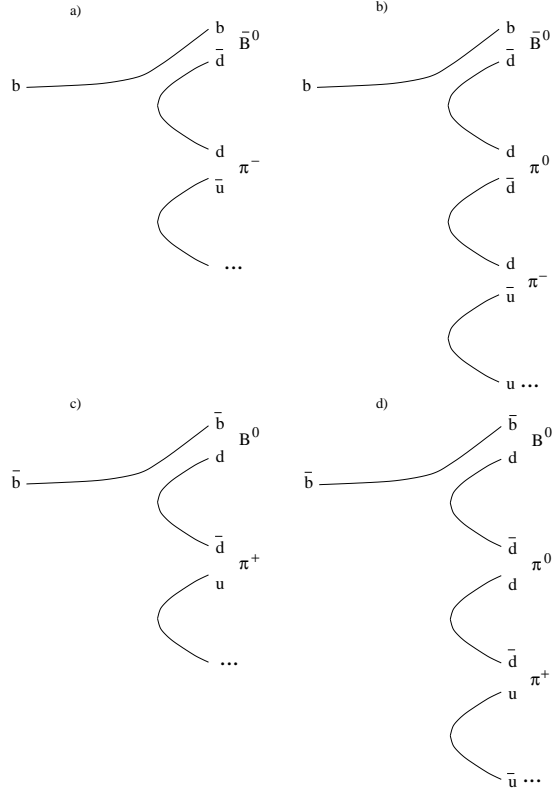, height=10.5cm}}
\medskip
\caption[Quark diagrams for same side tagging]
{Four quark diagrams for the  fragmentation of $b$ and $\bar{b}$
quarks to $B^0$ and $\bar{B}^0$ mesons.  The charged pion which
is nearest in the fragmentation chain to the $B$ meson tags the
birth flavor of the $B$ meson.  
The notation \dots indicates that the fragmentation
chain continues out of the picture.}
\label{2:fig:frag}
\end{figure}

Same side tagging exploits charge correlations between the
a $B^0$, or $\bar{B}^0$, and the nearest pion in the fragmentation
chain.  Fig.~\ref{2:fig:frag} illustrates the idea behind 
the method.
 
One can think of the hadronization, or fragmentation,
processes as pulling light quark pairs from the vacuum and forming hadrons
from nearby quarks.  
In order to form a $B^0$ or a $\bar{B}^0$ meson 
the light quark pair which is nearest in the fragmentation
chain to the initial heavy quark must have been a $d\bar{d}$
pair.  This leaves a $d$ or $\bar{d}$ quark at the dangling
end of the fragmentation chain.
If the second nearest light quark pair is $u\bar{u}$ pair
then the nearest meson in the fragmentation chain
will be a $\pi^-$ or $\pi^+$, which can
be used to tag the flavor of the initial $b$ or $\bar{b}$.  
If the second nearest light quark pair is a $d\bar{d}$ pair then
the nearest meson is a $\pi^0$, which itself has no tagging power.
However the dangling end of the fragmentation chain remains 
a $d$ or $\bar{d}$ and, if the third nearest light quark pair is a $u\bar{u}$
pair, then the second nearest meson will be a $\pi^-$ or $\pi^+$
which can be used as a flavor tag.  The bottom line is that the
charge of the nearest charged pion tags the birth flavor of the
$B^0$ or $\bar{B}^0$ meson.

The question now is to discover an algorithm
that will identify the charged pion that is the
nearest charged pion in the fragmentation chain.  
CDF successfully developed such an algorithm in Run~I.
To select the same side tag pion,
all tracks within a cone 
of radius $0.7$ in $\eta\,\varphi$~space, centered
around the direction of the $B$ meson, were
considered.
Same side tag candidate tracks were required to originate from
the $B$ production point (the primary event vertex),
and were therefore required 
to satisfy $d_0/\sigma_{d_0}<3$,
where $\sigma_{d_0}$ is the uncertainty on the track
$r\,\varphi$~impact parameter $d_0$. 
This selection produced, on average, 2.2 same side tag candidate
tracks per $B$~candidate. 
String fragmentation models indicate 
that particles produced in the $b$~quark hadronization chain 
have a small momenta transverse to the direction of the $b$~quark momentum.
CDF thus selected as the tag the track that had the minimum
component of momentum, $p_T^{\rm rel}$, orthogonal to the momentum sum of
the track and the $B$~meson.

  The same fragmentation chain argument can be used to show
that the nearest charged kaon in the fragmentation chain can be used to 
tag the flavor of $B_s$ and $\bar{B}_s$ mesons. 
If the nearest kaon is a $K^+$, then the meson is a $B_s$ but if
the nearest kaon is a $K^-$, then the meson is a $\bar{B}_s$.
If, however, the nearest kaon is neutral, then there 
is no kaon
tagging power because the nearest charged neighbor will be a pion.
While there does remain a charge correlation with the nearest
charged pion, the author is not aware of any work done to exploit this.

  Compared with away side tagging methods, the same side
tagging methods have a higher efficiency but a worse dilution.
That is, they almost always find a candidate charged track but it 
is not always the correct one.  Because of its high
efficiency, same side tagging makes an important contribution to the total
tagging power of an experiment.

\subsection{Overall Tagging Strategy}
\index{flavor tagging!description of!overall strategy}

  The various methods described above have quite different
properties.  For example lepton tagging has a relatively
low efficiency but a very good dilution.  Same
side tagging and jet charge tagging, on the other hand, 
are more efficient but 
have  poorer dilutions.  At CDF in Run\,I Kaon tagging
was intermediate in both efficiency and dilution; better particle
identification capability in the CDF Run~II detector and in BTeV will 
significantly improve the dilution for this tag.  The optimal 
tagging strategy is some method which involves all of the the
tagging techniques.
Any such strategy must account for the correlations among the away
side tagging methods; same side tagging is statistically
independent of all away side methods.  
One very simple strategy is to poll each method in order 
of decreasing dilution and to accept the first method that gives
an answer.  A more powerful idea is to combine all of the 
methods into an overall likelihood ratio, a linear discriminant or 
a neural net.  The strategy employed by CDF in their Run\,I 
analysis of $\sin2\beta$ is described in reference~\cite{cdftagging}.

  For further details one should consult the chapters for 
the specific experiments and the references therein.

\section{The Measurement Error on Proper Decay Times}
\label{2:sec:properdecaytime}
\index{proper decay time}

 It is instructive to describe how one measures the proper decay time of
a $b$ hadron and to see that, for the decays of interest, the error
on the proper decay time is, to a good approximation, 
independent of momentum.

To measure the proper decay time one reconstructs the primary 
interaction vertex at which the $b$ hadron
was produced and the secondary vertex at which the $b$ hadron decayed.
The proper decay time, $t$,  is then given by $t=Lm/pc=L/\beta\gamma c$,
where $L$ is the decay length measured by the separation of the vertices,
$p$ is the measured momentum of the $b$ hadron, $m$ is the mass of
the $b$ hadron, $c$ is the speed of light, and where
$\beta$ and $\gamma$ are the usual Lorentz parameters for the $b$ hadron.
The uncertainty
on the decay length contains contributions from the error on the
primary vertex position, the error on the secondary vertex position
and the error on the momentum of the $b$.  In all three  experiments the
contributions from the position errors are much larger than
those from the error on the momentum.  And the multiplicity of 
the primary vertex is usually much higher than that of the 
secondary vertex, making the error on the primary vertex position
smaller than that of the secondary vertex;
therefore the error on the proper decay time
is dominated by the error on the position of the secondary vertex.
The relevant part of the error on the secondary vertex is the 
projection of its error ellipse onto the flight direction of
the $b$ hadron.  That is the dominant contribution to the
error is just,
\begin{equation}
\sigma_t({\rm dominant\ contribution}) = \sigma^{(2ndry)}_{L}/(\beta\gamma c),
\label{2:eq:sigt}
\end{equation}
where $\sigma^{(2ndry)}_L$ is the contribution of the secondary vertex
to the decay length.

There is one familiar exception to this rule.  When using semi-leptonic 
decays to reconstruct the $b$, the momentum carried 
by the missing neutrals is poorly known 
and the error in the proper
decay time has important contributions from the error in the momentum.

To understand how $\sigma_t$ depends on momentum, consider
two different instances of the decay $B^0\to\pi^+\pi^-$.  
In the the first case the $B^0$ has some definite momentum and
the decay takes place at a particular point in some detector.
Suppose that the momentum is large enough that both pions are boosted 
forward along the $B$ flight direction in the lab.
In the second case,  the $B^0$ decays at exactly the same space point
and with the same center-of-mass decay angles as in case 1, but
it has a larger momentum.
The decay products of this decay are then measured in the same detector.
Further suppose that the detector is sufficiently uniform that each
track from these two decays is equally well measured.
The main difference between these two cases is that
the lab frame opening angle between the two pions will be smaller in 
case 2 than in case 1.  Because the opening angle is smaller, the point
at which the tracks intersect is more poorly known.  In particular
the component of the vertex position along the $b$ flight
direction in the lab is more poorly known.  This is purely a geometric
effect.  One result of this geometric effect
is that the error on the secondary vertex position grows
like $\gamma$; that is
$\sigma^{(2ndry)}_L \propto \gamma $.  Plugging
this into Eq.~\ref{2:eq:sigt} gives $\sigma_t \propto  1/(\beta c)$.
Therefore, for $\beta\simeq1.0$, 
the error on the proper decay time is independent of momentum.
This property has been exploited by experiments such as E687, E791,
SELEX and FOCUS to make precision measurements of the charmed hadron
lifetimes.

The above analysis holds approximately for multi-body decays of $b$ 
hadrons.  It will fail for very small boosts, in which case 
the decay products travel both forwards and backwards along 
the $b$ flight direction.  It will also fail if the decay products
of the $b$ hadron are slow enough that their errors are dominated
by multiple scattering and not by the measurement errors in the
apparatus.

The above analysis is only valid if the two decays are measured
by the same detector.  It is not useful for
comparing two very different detectors; in that case there are
no short cuts and one must compute the resolution of each detector.

\section{Properties of a Good $B$ Physics Detector}
\index{detector! desired properties}

A detector for doing $b$ and $c$ physics at a hadron  
collider must have the following components, 
a high precision vertex detector,
a tracking system giving excellent momentum resolution,
excellent particle identification (ID) capability,
and a  robust trigger integrated into a high bandwidth DAQ.
And it is very desirable to have electromagnetic calorimetry so
that modes containing photons and $\pi^0$'s can be measured.

It would also be useful to have 
hadronic calorimetry which is precise enough to reconstruct a
$K^0_L$.  Because the $K^0_L$ has opposite $CP$ quantum numbers to
the $K^0_S$, many tests of the weak interactions can be made by
comparing exclusive final states which differ by substituting a $K^0_L$
for a $K^0_S$.  None of the three experiments, however, anticipate
a significant ability to reconstruct $K^0_L$ mesons.

There are many other constraints on the detector design.
For example, all of the detector elements must have sufficiently fine 
granularity to deal with the high multiplicities which occur 
in $p\bar{p}$ collisions.  
The detector must be able to deal with several such interactions 
in one beam crossing.  And it is important to design the
detector with as little mass as possible in the fiducial
volume.  

  The three experiments have approached these challenges
from different directions and with different constraints.

\subsection{Forward vs Central}
\label{2:sec:centfor}
\index{detector!central vs forward}

 At first glance the most striking difference among the three 
detectors is that CDF and D\O\ are central detectors while BTeV is a
forward detector.
But there is a much more important distinction --- BTeV
is a dedicated $B$ physics detector while CDF and D\O\ are multipurpose
detectors whose primary mission is high $p_T$ physics, including
precision top quark physics, the search for the Higgs boson and
the search for supersymmetric particles.  Some of
the constraints imposed on CDF and D\O\ are not
intrinsic limitations of the central geometry; rather they are 
consequences of their optimization for a different spectrum of physics.

But there do remain some issues for which either the forward
or central geometry has an advantage.
First, BTeV has a harder particle ID job than either CDF or
D\O\ because BTeV  must identify tracks over a much wider range of
momentum.  However the forward geometry allows for a RICH
detector which gives BTeV better overall hadronic particle ID.
Second, BTeV has a somewhat higher efficiency for reconstructing 
the decay products of the second $B$ in the event, given that the 
first $B$ has already been reconstructed.  The reasons behind
this involve the interplay of production dynamics with the
myriad constraints of detector design.  Third, the forward
geometry is more open than the central geometry, thereby simplifying
the mechanical design and maintenance.
In a central geometry, on the other hand, one unit
of $\eta$ is much more opened up in space
than in the forward region.  Since multiplicities are approximately
uniform in $\eta$, this allows a device with coarse granularity 
to have the same multi-track separation power as does a fine granularity
device in the forward region; this has advantages in channel count.
Many of the advantages discussed in this paragraph are tied
to available technologies and the situation might well change
with new developments in detector technologies.

The $B$ mesons produced in the forward region have higher momenta,
and consequently longer decay lengths, than do those produced in 
the central region.  Before drawing any conclusions about the merits of
forward produced $B$'s, one must take many other things into
account.  Not all $B$'s produced in the central region have low momentum
and
the ones which pass all analysis cuts have much higher momentum than the
average $B$ meson.  There are more $B$ mesons produced in the 
central region than in the forward region so
central detectors can tolerate a smaller efficiency for their 
topological cuts and still have a comparable event yield.
Higher momentum $B$ mesons have poorer resolution 
on their decay vertex positions (see Section~\ref{2:sec:properdecaytime});
this cuts the advantage of the highest momentum $B$ mesons in forward
detectors.
The decay products of higher momentum $B$ mesons undergo less multiple
scattering than do those of lower momentum $B$ mesons; this
helps to improve resolutions.
And the details of the detector design turn out to be the
critical.  The net result is that, after all analysis cuts, 
the early designs for the Run~II CDF detector had a significantly 
poorer resolution on proper decay time than does BTeV.  But with the 
addition of Layer-00, CDF now has a resolution on proper decay time,
after all analysis cuts, which is comparable to that of BTeV.

At trigger time a different set of priorities is present.
For example, in the lowest level of the BTeV trigger, the track 
fitting algorithms are crude and it is important that
most $B$ meson daughters are of high enough momentum that multiple
scattering is a small enough effect to treat in a crude fashion.

The most important difference which arises from of BTeV being a dedicated 
experiment, while CDF and D0 are not, is in the trigger 
and DAQ systems.  The BTeV trigger and DAQ system reconstructs
tracks and makes a detachment based trigger decision 
at the lowest trigger level.  Every beam crossing
is inspected in this way.  CDF and D\O,
on the other hand, must live within bandwidth budgets
that were established before this sort of trigger was
feasible.  Therefore they have detachment information 
available only at level 2 and higher.  For similar reasons their
triggers have a higher $p_T$ cut than does the BTeV trigger.

\subsection{A Precision Vertex Detector }
\index{vertex detector! desired properties}

   First and foremost it is necessary to have a high precision vertex 
detector.  The importance of the vertex detector to the trigger
has already been emphasized.  Also, excellent resolution on 
proper decay time, which results from excellent vertex resolution, 
is necessary to study the time dependence of mixing mediated $CP$ 
violating effects in the $B_s$ system.   

In a typical offline analysis chain, the vertex resolution 
appears to a high power.  A typical candidate-driven 
$B^0\to \pi^+\pi^-$ analysis might proceed as follows.
First one finds a $B^0$ candidate and demands that the candidate
$B$ have a well defined vertex with a good $\chi^2$.
Next, one must find the primary vertex of the $b\bar{b}$ interaction
and care must be taken to ensure that this vertex is not contaminated
by tracks from the other $b$ hadron.  
One demands that the secondary vertex be well separated from the primary
vertex.  For some analyses
it will be necessary to exclude $B$ candidates if other tracks
from the event are consistent with coming from the $B$ decay vertex.
Finally, one applies the available tagging methods.
Each of these steps exploits the vertexing power of the
experiment in a slightly different way.  With so many steps, poor
resolution has many chances to strike.

The vertex detector must have as low a mass as possible.
Less mass implies less multiple scattering and better vertex resolution. 
But a more important effect is that less mass reduces the number of
interactions of signal tracks in the detector materials.
When a signal track interacts in the detector, it is often unusable
for physics and the event is lost.  Examples include tracks which
undergo inelastic hadronic interactions before reaching the particle
ID device and photons which pair convert in the detector material.

A final consideration is the occupancy of the vertex detector.
The occupancy is defined as the fraction of channels which are 
hit during a typical beam crossing.  As the occupancy rises, the number 
of hit combinations which must be considered grows exponentially
and pattern recognition becomes more difficult.  
If the occupancy is less than a few percent, offline pattern 
recognition is straight forward and standard algorithms compute 
sufficiently quickly.  The driving
factor in behind BTeV's choice of a pixel detector, rather than
a strip detector, was to reduce the occupancy to approximately
$10^{-4}$.  With this very low occupancy, even very simple, 
pattern recognition algorithms are efficient and
produce low background levels; this allows their use at the
lowest level of the trigger.

\subsection{Tracking}
\index{tracking system!required properties}

The vertex detector must be supplemented by a tracking system with excellent
momentum resolution.  For most decay modes of interest, the mass
resolution on the $b$ hadron is dominated by the momentum resolution 
of the apparatus.  If the mass resolution can be decreased by, say,
10\%, one will get a 10\% improvement in signal-to-background
ratio without loss of signal efficiency.  In a decay chain with
several intermediate mass constraints this can add up.

 Again it is important to minimize the mass in the tracking system
and pay careful attention to the expected occupancy.

\subsection{Particle ID}
\index{particle identification!overview}

 It is important to have a excellent particle identification with the 
ability to separate, with high efficiency, all of $e,\mu,\pi,K,p$ over
a broad momentum range.  All of the detectors have triggering
modes which require lepton identification (lepton ID).  Particle ID
is also critical for reducing backgrounds
which arise when one $B$ decay mode is mistaken for another, such
as $B_s\to D_s \pi$ being mistaken for $B_s\to D_s K$.  Finally,
excellent particle ID is crucial for a 
large $\epsilon D^2$ for kaon tagging.

  All of the experiments have excellent lepton ID.  Muon ID is
done by finding tracks which penetrate a hadron 
shield.    Electron ID is done by matching tracks in the
tracking system with clusters of energy in the electromagnetic 
calorimeter.   

  BTeV is a dedicated $B$ physics experiment and one of the factors 
driving the decision to build a forward spectrometer, not a central
one, was that the forward geometry has room for a 
Ring Imaging Cherenkov counter (RICH).  This provides the 
power to separate $\pi,K,p$ from one another.
On the other hand, the CDF and D\O\ detectors were originally
optimized for high $p_T$ physics, which did not require
powerful $\pi,K,p$ separation.  Therefore the early designs for the 
CDF and D\O\ Run~II detectors did not include any device to do 
hadronic particle ID.    Since then CDF has added a time of 
flight (TOF) system to perform $\pi,K,p$ separation.  

   The reader is referred to the chapters~3 to~5 for
further details.

\subsection{Trigger and DAQ}
\index{triggering!overview of}

In order to have a broad based $B$ physics program, it is important
to have an open trigger which is able to trigger on many 
$B$ decay modes.  This must be accompanied by a 
high bandwidth DAQ system which can move the data off the detector,
move it between trigger levels and store it until a trigger
decision is made.

The job of the trigger is to 
sort through the much more copious background
interactions and extract a high purity $b$ sample to write to tape.
Ideally the trigger should be sensitive to some general property 
of $b$ events, and have a high efficiency for a wide variety
of $B$ decay modes; it is not enough that the trigger performs
well on some list of benchmark decay modes.  This allows the greatest
flexibility to explore ideas which are first thought of long after
the trigger design was frozen.  Of course one must verify that 
the trigger works well on the modes which we know now to be important. 

   A detachment based trigger meets all of these requirements; in particular
it can trigger on all hadronic decay modes, a capability which was missing
from the previous generation of experiments.  
A lepton based trigger, while missing the all hadronic modes, 
does meet many of the requirements and it will provide a 
redundant triggering method to calibrate the detachment based triggers.

  The background rejection needed by the trigger is set by several
things.  Each level of the trigger must reduce the background to
a low enough level that the bandwidth to the next level is not
saturated.  One must also consider the total amount of data which
is written to tape; if too much data is written to tape, the main 
data reconstruction pass will take too long and the production 
of physics papers will be delayed.  
The cost of archival media is also an issue.

  A final consideration is projecting results to higher 
instantaneous luminosities.  As the luminosity increases,
several limiting effects arise: one might reach the
bandwidth limit of the DAQ system; one might exceed the amount
of buffering at some level of the DAQ; the dead time might become
too large.  Once one of these limits is reached, the normal
response is to raise some trigger threshold or to prescale some 
trigger.  Typically one tries to sacrifice either the triggers which carry
the least interesting physics or the triggers with poor
signal to background ratios.  

 During an extended Run~II but it is very likely that some of the $B$ 
physics triggers will need to be modified to deal with the
increased luminosity but it is difficult to project in detail
what might need to be done. 
These decisions will depend, in part, on an understanding
of backgrounds which is not yet available.  Depending on the 
characteristics of the backgrounds, the trigger efficiency for 
some $B$ physics channels may be unaffected while the trigger 
efficiency for other $B$ physics channels may drop significantly.

  The collaborations have concentrated their $B$ physics
trigger simulations on the conditions which will be present
early in Run~II, up to a luminosity of around
$2\times 10^{32}\, {\rm cm^2 s^{-1}}$.  So they have decided not
to present projections for integrated luminosities of 10 and 
30~fb$^{-1}$.

\subsection{EM Calorimetry}
\index{EM Calorimeter!overview of}

A good electromagnetic calorimeter (EMcal) is necessary for
the reconstruction of decay modes which contain final state
photons and $\pi^0$'s.  It is also necessary for electron ID,
which can be used in triggering, in flavor tagging, in many
searches for physics beyond the standard model.

  There is one high profile decay mode for which the EMCal is
critical, the analysis of the Dalitz plot for the decay
$B^0\to \pi^+\pi^-\pi^0$, a mode which measures the CKM
angle $\alpha$.   The $\Upsilon(4S)$ machines are rate
limited and are likely not to have sufficient statistics to
make a definitive measurement of this quantity.  The Tevatron
detectors have the rate and, provided the EMCal technology
is good enough, the measurement can be done.

 A final use for an EMCal is to help sort out strong interaction
effects which are entangled with the weak interaction physics
that it the main goal.   It is most straightforward to disentangle
the strong interaction effects when all isospin permutations 
of the final state can be measured.  The classic example
of this are the decays $B^0\to \pi^+\pi^-$, 
$B^0\to \pi^0\pi^0$, and $B^+\to \pi^+\pi^0$.  While this
complete set of decay modes is probably not measurable at the
Tevatron, it illustrates the point.

  All of the detectors have electromagnetic calorimetry.
Both BTeV and CDF have discussed a $B$ physics program which exploits it.

  The EMCal system has a unique sensitivity to the issue of
track density.  As the number of tracks in the detector goes up,
the occupancy of the calorimeter goes up.  In the case of the tracking
detectors one can compensate for high occupancy by making a more
granular detector.  This works because each track usually makes
a small, localized signal in a tracking detector.  In the case of a
calorimeter, showers are extended objects which, by design,
deposit energy in neighboring crystals or cells.  Increasing
the granularity of the detector will not make the showers any smaller.
One can compensate for high occupancy by choosing
calorimeter materials in which showers are contained in a smaller 
volume; if such materials are available, have sufficient energy
resolution and radiation tolerance, and are affordable, then
they can be used to make a calorimeter which can tolerate
higher multiplicities.

\subsection{Muon Detector}
\index{muon detection!overview of}
 
The final major hardware component in a $B$ physics experiment
is a muon detector system.  This provides muon identification (ID)
for such purposes as flavor tagging, 
reconstruction of $J/\psi$ candidates,
reconstruction of semileptonic decays,
and searches for rare or forbidden decays.  These last two classes
include modes such as $B^0\to \mu^+\mu^-$ and $B^0\to e^+\mu^-$.
A second job of these system is to provide an element of the trigger
system; in some cases this is a stand alone trigger element and in
other cases it is used in conjunction with other detector components
to make a trigger decision.  

The basic design of a muon detector is an iron or
steel shield, many hadronic interaction lengths thick, which 
absorbs hadrons.  Muons which penetrate this shield are detected 
by some sort of wire chamber tracking detector, or perhaps a scintillator, 
placed behind the shield.  When possible other detector components, such 
as calorimeters and flux returns are used as part of the shielding.
In another variation, iron shielding can be magnetized to allow
measurement of the muon momentum in the muon system alone.
This gives several benefits: it allows one to design a muon
based trigger with a well defined $p_T$ cut and it allows better
matching between tracks in the muon system and tracks in the
main tracking system.

\section{Software}
\label{2:sec:software}

The software used by CDF, D\O\ and BTeV can be thought of in
four classes, event generators, the $B$ decay code, 
detector simulation tools
and reconstruction code.  The three experiments use common
tools for the first two classes of software but generally
use their own, detector specific software for the last two classes.
The one exception is the MCFast fast simulation package which was
used by both BTeV and CDF.  MCFast is described extensively
in the BTeV TDR\cite{btev_tdr}.  BTeV is a new experiment which
has tools which are quite advanced for such a young experiment
but which are still primitive on an absolute scale.  
After Run~I, CDF and D\O\ embarked on a major retooling of their software 
infrastructure, which was incomplete at the time of the workshop.
So all of the results presented here use preliminary versions of
code.  At the level of precision required for the studies performed
at this workshop, all of these tools are good enough.

  The event generators are programs which generate the physics
of a $p\bar{p}$ interaction; its output is usually just a list
of vertices and particles which come out of those vertices.
These programs are typically the intellectual property of
the theoretical physicists who developed the model which is implemented
in the program.  The generators used by CDF, D\O\ and BTeV
are {\tt PYTHIA}\cite{pythia}, 
{\tt ISAJET}\cite{isajet} and {\tt HERWIG}\cite{herwig}. Generally 
these programs are used only to predict the shapes of 
the differential cross-sections, not for the absolute cross-sections.
CDF and D\O\ have tuned the parameters of their event generators to
match their Run~I data.  BTeV, on the other hand, has no such data
against which to tune the codes.  Therefore BTeV is using the programs
as is.

   The event generators have been carefully developed to simulate
the properties of $p\bar{p}$ collisions but much less care was taken
in their model of how $b$ hadrons decay.  To circumvent this, the
$b$ and $c$ hadrons produced by the event generators are handed
to a separate code to simulate their decay.  Until recently this code
was the QQ code, which was developed and maintained by CLEO,
and which contains their integrated knowledge about the decays
of $B$'s and $D$'s.  The BaBar collaboration also has such
a program, {\tt EVTGEN} which will soon replace QQ.  The results
of the workshop were obtained using QQ.

   The next step in a typical simulation is to compute the
detector response to the simulated events.  All of the experiments
have both a fast simulation program and a detailed simulation
program.  A typical fast simulation program uses a simplified and/or
parameterized description of the detector response and directly
produces smeared 4-vectors for the tracks which were input to it.
It may also declare that a track is outside of the fiducial volume
and is not reconstructible.  The output of the fast simulation can
usually be used as is to perform the simulated analysis.

  A typical full simulation is based on the {\tt GEANT}~3 program
from CERN.  This is a program which knows how to describe a detector
by building it up from a library of known shapes.  It also has
extensive knowledge of the interactions of particles with materials.
It takes tracks from the event generator and propagates them through
the detailed description of the detector, at each step checking to
see how the track interacted with the material.  If a particle
interacts in the detector material to produce new particles, those
new particles are also propagate through the detector.  If a shower
starts in material, {\tt GEANT}~3 will follow the daughters through
each stage of the shower, and deposit the energy of the shower
in the appropriate detector cells.  The output of
this simulation is typically a list of pulse heights or arrival
times for hits in individual detector cells.   This information
is then passed to the reconstruction program and the trigger
simulation codes.

  At the time of the workshop, the experiments were still evaluating 
at the {\tt GEANT}~4 program, the C++ based successor to
{\tt GEANT}~3.  

     CDF and D\O\ have data samples from Run~I
which can be used.  Signal yields can be projected from the
Run~I signal yields by computing the ratio of efficiencies in the
old and new detectors.   This avoids the need to make assumptions
about the total cross-section, as BTeV must do.  CDF and D\O\ use
background samples from Run~I to estimate background levels in
Run~II.  BTeV must rely entirely on simulations for this purpose.

  The reconstruction code starts with raw hits, either from the
detector or from a simulation of the detector, calibrates them,
find tracks, fits them, finds showers, applies the particle
ID algorithms and so on.  The output of this step is the measured
properties of tracks and showers, which can be used directly
for physics analysis.

  The trigger simulation codes start with the same raw hits as
the reconstruction code.  In some cases the codes emulate the 
trigger hardware and produce trigger decisions which should
very closely represent the real trigger behavior.  In other
cases the simulation codes 

  For more details on the software of each experiment,
consult their 
TDRs~\cite{cdf_tdr}\cite{cdf_tof_l00}\cite{d0_tdr}\cite{btev_tdr}
and chapters~3 to~5 in this report.

\section{Comparison with \lowercase{$e^+e^-$} Machines }
\index{$e^+e^-$ machines!comparison with}

  For most of their lifetime the Tevatron experiments will
be in competition with the detectors from the
$\Upsilon(4S)$ $e^+e^-$ factories, BaBar and Belle.
The charm physics program of BTeV will also face competition from
CLEO-c.  While these programs are competition,
they also complement the Tevatron program. 
Some precision measurements will be best done in the cleaner
environment of $e^+e^-$: they have a well determined initial state, 
either pure $B^0\bar{B}^0$ or $B^+B^-$, with no additional tracks
in the event.  And the production of $B$ mesons represents about
20\% of their total cross-section, which greatly simplifies
triggering and removes many trigger biases.  Similar advantages
hold for the open charm program at CLEO-c.

  On the other hand, the Tevatron experiments have a significant
rate advantage which give it the advantage for many rare decay
modes and in those measurements which are statistically limited.  
Only the Tevatron experiments have access to the decays of the
$B_s$ and $b$ baryons, which are necessary to complete the
program of over constraining the CKM matrix.  See, for example,
the discussion of $B_s$ mixing in chapter~8.

\clearpage{\pagestyle{empty}\cleardoublepage}
\chapter{The CDF Detector}

\authors{M.~Paulini, A.B.~Wicklund}

\section{Introduction}

The CDF detector has evolved over a twenty year period. 
CDF was the first experiment at the Tevatron to perform
quantitative measurements of $b$-quark production,
using single-lepton and $J/\psi$ samples in
Run 0 (1988-1989).  CDF was the first hadron collider
experiment to successfully employ a silicon vertex
detector (SVX).  The four layer, axial readout 
SVX (Run~Ia) and SVX' (Run~Ib) detectors were
used to discover the top quark through detection
of the $b$-quark decay chain.  They were also used for
a systematic program of $b$-physics studies, including
$B$-lifetimes, $B\overline{B}$ mixing, discovery
of the $B_c$, and measurement of $CP$ violation 
in the $B^0 \rightarrow J/\psi K_S^0$ mode.  In the course
of this program, CDF has developed the techniques to
identify $B$ hadron final states in $J/\psi$ and 
$\ell\nu D$ semileptonic modes, and to flavor tag
these states using away-side lepton and jet-charge
tags and toward side fragmentation correlations.
As a byproduct, CDF has developed control sample
strategies to calibrate particle identification using
relativistic rise d$E$/d$x$, to optimize flavor tagging
efficiency, and to measure material effects
(energy loss and radiation length corrections) needed
for precision mass measurements.  CDF has published over fifty
papers on $B$ physics with the Run 0 and Run~I data. In addition
to Run~I physics data, CDF recorded data on a variety
of specialized triggers in order to estimate rates and
backgrounds for the Run~II program.  Although the Run~II
$B$-physics program at CDF will be technically more
challenging than Run~I, CDF starts with the advantage
of extensive experience and benchmark data.

The upgraded CDF detector and physics program for Run~II is described in detail in
the CDF Technical Design Report~\cite{cdf_tdr}. The
additional upgrades for time-of-flight and innermost
silicon detector are described in the PAC proposal P909~\cite{cdf_p909}.
Below we summarize the Run~II detector configuration, 
with expected performance including particle identification,
issues for central solenoidal detectors, CDF $B$-physics trigger 
plans, and offline analysis issues.
 
\section{CDF Run~II Detector}
 
  The main upgrades to the CDF detector for Run~II
can be summarized as follows:
\index{CDF!detector improvements}
\begin{itemize}
\item[-]{Fully digital DAQ system designed for 132 ns bunch crossing times}
\item[-]{Vastly upgraded silicon detector}
\subitem{707,000 channels compared with 46,000 in Run~I}
\subitem{Axial, stereo, and 90$^\circ$ strip readout}
\subitem{Full coverage over the luminous region along the beam axis}
\subitem{Radial coverage from 1.35 to 28 cm over $|\eta| < 2$}
\subitem{Innermost silicon layer(``L00'') on beampipe with 6 $\mu$m axial hit resolution}
\item[-]{Outer drift chamber capable of 132 ns maximum drift}
\subitem{30,240 sense wires, 44-132 cm radius, 96 d$E$/d$x$ samples possible per track}
\item[-]{Fast scintillator-based calorimetry out to $|\eta| \simeq 3$}
\item[-]{Expanded muon coverage out to $\eta \simeq 1.5$}
\item[-]{Improved trigger capabilities}
\subitem{Drift chamber tracks with high precision at Level-1}
\subitem{Silicon tracks for detached vertex triggers at Level-2}
\item[-]{Expanded particle identification via time-of-flight and d$E$/d$x$}
\end{itemize}

\begin{figure}[t]
\centerline{\epsfxsize=13cm \epsfbox{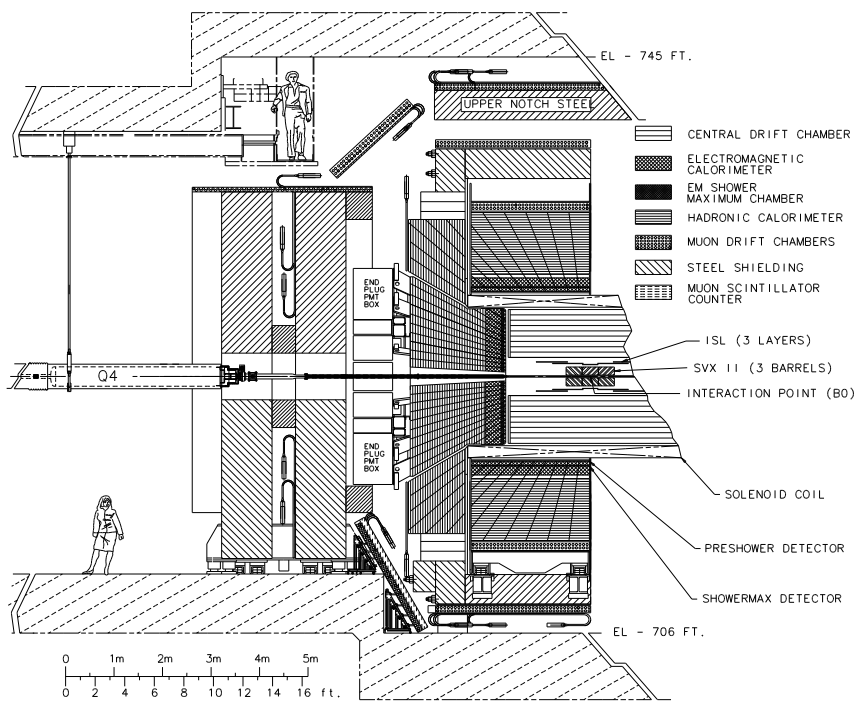}}
\vspace*{.2cm}
\caption[CDF detector elevation view]{Elevation view of the Run~II CDF detector}
\label{fig:cdfelev}
\end{figure}

\begin{table}[htbp]
\begin{center}
\begin {tabular} {l c}
\hline \hline
COT                         &    \\ \hline
Radial coverage             & 44 to 132 cm \\
Number of superlayers       &  8  \\
Measurements per superlayer & 12  \\
Readout coordinates of SLs  & +2$^\circ$ 0$^\circ$ -2$^\circ$ 0$^\circ$ +2$^\circ$ 0$^\circ$ -2$^\circ$
 0$^\circ$ \\
Maximum drift distance      & 0.88 cm \\
Resolution per measurement  & 180 $\mu$m \\
Rapidity coverage           &  $|\eta| \leq 1.0$ \\
Number of channels          & 30,240 \\
Layer 00                     &                        \\ \hline
Radial coverage             &1.35 to 1.65 cm \\
Resolution per measurement  & 6 $\mu$m (axial) \\
Number of channels          & 13,824 \\
SVX II                       &                        \\ \hline
Radial coverage             &   2.4 to 10.7 cm, staggered quadrants      \\
Number of layers            &   5          \\
Readout coordinates         &r-$\phi$ on one side of all layers \\
Stereo side                 &r-z, r-z, r-sas, r-z, r-sas (sas $\equiv
                             \pm 1.2^\circ$ stereo) \\
Readout pitch               & 60-65 $\mu$m r-$\phi$; 60-150 $\mu$m stereo\\
Resolution per measurement  & 12 $\mu$m (axial) \\
Total length                &  96.0 cm  \\
Rapidity coverage           &  $|\eta| \leq 2.0$ \\
Number of channels          & 423,900    \\
ISL                         &       \\ \hline
Radial coverage             &  20 to 28 cm \\
Number of layers            &  one for $|\eta|<1$; two for $1<|\eta|<2$  \\
Readout coordinates         &  r-$\phi$ and r-sas (sas$\equiv \pm 1.2^\circ$ stereo)
                              (all layers)\\
Readout pitch               &  110 $\mu$m (axial); 146 $\mu$m (stereo) \\
Resolution per measurement  & 16 $\mu$m (axial) \\
Total length                &  174 cm  \\
Rapidity coverage           &  $|\eta| \leq 1.9$ \\
Number of channels          & 268,800    \\
\hline \hline
\end{tabular}
\end{center}
\vspace*{.2cm}
\caption[CDF tracking systems]{Design parameters of the CDF tracking systems}
\label{tb_tracking}
\end{table}

The CDF detector features excellent charged particle tracking
and good electron and muon identification in the central region.
The detector is built around a 3 m diameter 5 m long superconducting
solenoid operated at 1.4 T.  The overall CDF Run~II detector 
schematic is shown in elevation view in Fig.~\ref{fig:cdfelev}.
The CDF tracking system includes a central outer drift chamber (COT),
a double-sided five layer inner silicon detector (SVX II), 
a double-sided two layer intermediate silicon tracker (ISL),
and a single layer rad-hard detector mounted on the beampipe (L00).
COT tracks above 1.5 GeV/c are available for triggering at Level-1 (XFT);
SVX layers 0-3 are combined with XFT tracks at Level-2 (SVT).
The main parameters of the CDF tracking system are summarized in
Table~\ref{tb_tracking}.

Outside the solenoid, Pb-scintillator
electromagnetic (EM) and Fe-scintillator hadronic (HAD) calorimeters cover
the range $|\eta| < 3.6$.
\index{CDF!calorimetry} 
Both the central ($|\eta| < 1.1$) and plug ($1.1 < |\eta| < 3.6$)
electromagnetic calorimeters have fine grained shower profile detectors
at electron shower maximum, and preshower pulse height detectors
at approximately $1 X_o$ depth. 
Electron identification is
accomplished using $E/p$ from the EM calorimeter and also in the shower maximum and
preshower detectors; using $HAD/EM\sim 0$; and using shower shape and position
matching in the shower max detectors.  Together with COT d$E$/d$x$, CDF gets
$\sim$10$^{-3}$ $\pi/e$ rejection in the central region.
The calorimeter cell segmentation is summarized in
Table~\ref{calseg}.  A comparison of the central and plug
calorimeters is given in Table~\ref{plugcomp}.

\begin{table}
\begin{center}
\begin {tabular} {l|c|c}
\hline \hline
$|\eta|$ Range & $\Delta\phi$ & $\Delta\eta$ \\
\hline
0. - 1.1 (1.2 h) & $15^{\circ}$ &  $\sim0.1$ \\
1.1 (1.2 h) - 1.8 & $7.5^{\circ}$ & $\sim0.1$ \\
1.8 - 2.1 & $7.5^{\circ}$ & $\sim0.16$ \\
2.1 - 3.64 & $15^{\circ}$ & 0.2 $-$ 0.6 \\
\hline \hline
\end{tabular}
\vspace*{.2cm}
\caption[CDF calorimeter segmentation]{CDF II Calorimeter Segmentation}
\label{calseg}
\end{center}
\end{table}

The calorimeter steel serves
as a filter for muon detection in the central (CMU) and
extension (CMX) muon  proportional chambers, over the
range $|\eta| < 1$, $p_T > 1.4$ GeV/c.
\index{CDF!muon systems}  Additional iron
shielding, including the magnet yoke, provides a muon filter
for the upgrade muon chambers (CMP) in the range $|\eta| < 0.6$, $p_T >2.2$ GeV/c.
The (non-energized) forward toroids from Run~I
provide muon filters for intermediate $1.0 < |\eta| < 1.5$ 
muon chambers (IMU) for $p_T > 2$ GeV/c.  Scintillators for
triggering are included in CMP, CMX, and IMU.
Muon identification is accomplished by matching track segments in the muon chambers
with COT/SVX tracks; matching is available in $R\Phi$ for all detectors and
in the $Z$ views in CMU and CMX. 
The muon systems are summarized in Table~\ref{tab:overcmu}.

\begin{table}
\begin{center}
\begin {tabular} {l|c|c}
\hline \hline
 & Central  & Plug \\
\hline
EM: & &\\
Thickness & $19 X_0, 1 \lambda$ & $21 X_0, 1 \lambda$ \\
Sample (Pb) & $0.6 X_0$ & $0.8 X_0$  \\
Sample (scint.) & 5 mm & 4.5 mm \\
WLS & sheet & fiber \\
Light yield & $160$ pe/GeV & $300$ pe/GeV \\
Sampling res. & $11.6\%/\sqrt{E_T}$ & $14\%/\sqrt{E}$ \\
Stoch. res. & $14\%/\sqrt{E_T}$ & $16\%/\sqrt{E}$ \\
Shower Max. seg. (cm)& 1.4$ \phi \times $(1.6-2.0) Z& $0.5\times0.5$ UV\\
Pre-shower seg. (cm)& $1.4 \phi\times 65$ Z & by tower \\
\hline
Hadron: &&\\
Thickness& $4.5 \lambda$  & $7 \lambda$\\
Sample (Fe)&1 to 2 in.& 2 in.\\
Sample (scint.)& 10 mm & 6 mm\\
WLS & finger & fiber\\
Light yield& $\sim40$ pe/GeV  & $39$ pe/GeV\\
\hline \hline
\end{tabular}
\vspace*{.2cm}
\caption[CDF central/plug calorimeters]{Central and Plug Upgraded Calorimeter
Comparison}
\label{plugcomp}
\end{center}
\end{table}

\begin{table}[t]
\begin{center}
\begin{tabular}{lcccc}
\hline \hline
 & CMU   & CMP   & CMX & IMU \\ \hline
Pseudo-rapidity coverage
  & $|\eta|\leq 0.6$ & $|\eta|\leq 0.6$
  & $0.6 \leq |\eta|\leq 1.0 $
  & $1.0 \leq |\eta|\leq 1.5 $ \\
Drift tube cross-section
 & 2.68 x 6.35~cm & 2.5 x 15~cm & 2.5 x 15~cm & 2.5 x 8.4~cm \\
Drift tube length   & 226~cm   & 640~cm  & 180~cm & 363~cm \\
Max drift time
 & ${800~{\rm ns}}$ & 1.4~$\mu {\rm s}$
 & 1.4~$\mu {\rm s}$ &  800~{\rm ns} \\
Total drift tubes (present)     & 2304 & 864     & 1536  & none \\
Total drift tubes (Run~II)      & 2304 & 1076    & 2208  & 1728 \\
Scintillation counter thickness &      & 2.5~cm  & 1.5~cm & 2.5~cm \\
Scintillation counter width     &      & 30~cm   & 30-40~cm & 17~cm \\
Scintillation counter length    &      &  320~cm & 180~cm & 180~cm \\
Total counters (present)        &      & 128     & 256 & none \\
Total counters (Run~II)         &      & 269     & 324 & 864 \\
Pion interaction lengths        & 5.5  & 7.8     & 6.2 & 6.2-20 \\
Minimum muon $p_T$
  & 1.4~GeV/c & 2.2 ~GeV/c & 1.4~GeV/c & 1.4-2.0~GeV/c\\
Multiple scattering resolution
  & 12~cm/$p$  & 15~cm/$p$ & 13~cm/$p$ &13-25~cm/$p$\\ \hline \hline
\end{tabular}
\vspace*{.2cm}
\caption[CDF muon detectors]{Design Parameters of the CDF II Muon Detectors.
Pion interaction lengths and multiple scattering are computed at a reference
angle of $\theta =90^{\circ}$ in CMU and CMP, at an angle of
$\theta=55^{\circ}$ in CMX, and for a range of angles for the IMU.}
\label{tab:overcmu}
\end{center}
\end{table}

 The Run~II CDF detector configuration
allows electron and muon identification with drift chamber tracking
over the range $|\eta| < 1.0$, with additional coverage out to
$|\eta| \sim 1.5$ using stand-alone silicon tracking.  Typical
thresholds are $p_T > 1$ GeV/c (electrons), $p_T > 1.5$ GeV/c (muons).
Calibration of electron and muon identification is accomplished
{\it in situ} with large samples of $J/\psi$'s and photon conversions.

 Photon identification is done using the EM calorimetry, using the preshower
and shower maximum detectors to separate $\pi^0/\gamma$.  Channels like
$B\rightarrow J/\psi\eta$ can be reconstructed with the calorimeter. Rare decays
such as $B\rightarrow K^*\gamma$ use photon conversions to obtain
precision $\gamma$ reconstruction.

Asymptotic tracking resolutions are $\sigma(p_T) \simeq 0.0007~p_T^2$ 
($|\eta < 1.1$), and $\sigma(p_T) \simeq 0.004~p_T^2$
($|\eta| < 2$ -stand-alone SVX tracking). The SVX detectors provide
typical impact parameter resolution of
15 ($R-\phi$- view) and 30 ($Z$- view) $\mu m$. For example, this results in
mass resolution of around 20 MeV for $B^0\rightarrow\pi^+\pi^-$
and proper time resolution of 45 fs for $B_s^0 \rightarrow D_s^-\pi^+$.

\begin{figure}[t]
\centerline{\epsfxsize=4.5in \epsfbox{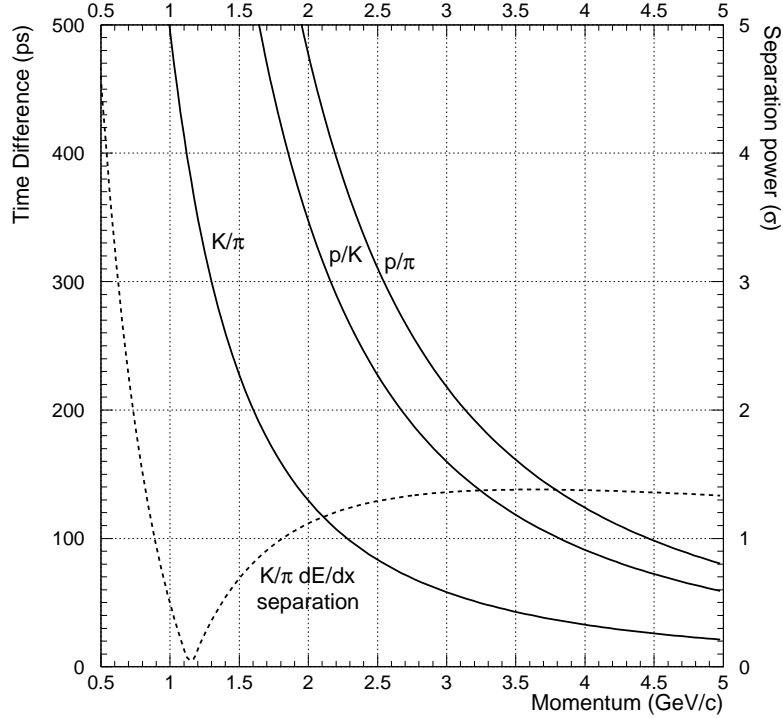}}
\vspace*{.2cm}
\caption[CDF particle identification]{Time difference as a function of momentum
for $\pi, K, p$ at a radius of 140 cm, expressed in ps and separation power,
assuming 100 ps resolution. The dashed line shows the $K/\pi$ separation using
$dE/dx$ in the COT.}
\label{fig:pid}
\end{figure}

Particle identification to separate pions, kaons, protons,
and electrons is provided by d$E$/d$x$ and time-of-flight
(TOF) detectors ($|\eta| < 1$):
\begin{itemize}
\item {$1/\beta^2$ d$E$/d$x$ in both COT and SVX}
\item {Relativistic rise d$E$/d$x$ in COT}
\subitem{$\simeq$1 $\sigma$ $\pi-K$ separation, $p > 2$ GeV/c}
\item {TOF bars outside the COT (1.4 m radius)}
\subitem{$\simeq$100 ps resolution}
\subitem{2$\sigma$ $K-\pi$ separation at 1.5 GeV/c}
\end{itemize}
Figure~\ref{fig:pid} shows the expected separation
as a function of momentum.  
The low momentum particle identification provides flavor
tagging with kaons, both on the away side $\overline{b}$-jet, and
on the same side as the trigger $B$ (e.g., $B^0_s$ correlated with
$K^+$, $\overline{B^0_s}$ with $K^-$); it should also provide
useful kaon separation for charm reconstruction, e.g. $B^0_s\rightarrow D_s^-\pi^+$.
Above 2 GeV/c, the COT $dE/dx$ provides a statistical
separation of pions and kaons.  This will be crucial for
determination of $CP$ asymmetries in $B^0\rightarrow\pi^+\pi^-$
and $B^0_s\rightarrow K^+K^-$.  The d$E$/d$x$ and TOF calibrations can
be determined {\it in situ} using pions and protons
from $K_S$ and $\Lambda$ decays, kaons from $\phi \rightarrow K^+K^-$,
electrons from conversion photons, and muons from $J/\psi$ decay.

\section[Issues for Central Solenoidal Detectors]
{Issues for Central Solenoidal Detectors at the Tevatron}

In the Run~II detector 
full charged particle tracking using the SVX and the COT 
is confined to $\eta < 1.1$.  Since all $B$-physics
triggers are track based, and rely on COT tracks in the Level-1
trigger, the $B$ triggers are basically limited to this
$\eta$ range.  Furthermore, in order to achieve a reasonable
Lorentz boost for the triggered $B$ hadrons, it is necessary
to select events with $p_T(B)$ above some minimum cutoff, typically
of order 4-6 GeV/c.  In terms of production rates, this is not
really a severe limitation.  At 10$^{32}$ luminosity, the total
$b\overline{b}$ production rate is around 10 kHz at the
Tevatron (eg. 10$^{11}$ events per fb$^{-1}$), much more than the
CDF data acquisition rate of $\simeq$75 Hz to tape. Since
$B$ production at high $p_T$ peaks at small $|y|$, the kinematic 
restriction is such that about 6\% of $B$-hadrons are produced in
the region $|y|<1$ and $p_T>6$ GeV/c (e.g., 1.2 kHz of $B$+$\overline{B}$ at
10$^{32}$.  
  
Thus, the basic trigger strategy is to select $B$ decays to 
specific final states in the central region, for example
$B^0\rightarrow J/\psi K_S$ or $B^0_s\rightarrow D^-_s\pi^+$,
with charged particle tracks confined to $|\eta| < 1$.  For mixing
and $CP$ studies, flavor tagging is accomplished using either the away-side
$\overline{B}$ or same-side fragmentation correlations.  The 
fragmentation correlations give a highly efficient flavor tag
($\simeq$ 70\% in Run~I), since the tagging tracks are guaranteed
to be in the central region.  The away side tagging would be limited
if tracking were confined to $|\eta| < 1$, but the stand-alone
SVX tracking extends the coverage in $\eta$, and so is expected to
greatly improve efficiency for jet-charge and lepton
tagging of away-side $b$-jets.  
 
There are some advantages to the central detector
configuration:
\begin{itemize}
\item{The Tevatron beamspot is small ($\sim 25 \mu$m) in
the transverse plane. This feature is exploited at the trigger level in CDF Run~II,
using the SVX detector to identify large-impact parameter tracks
from $B$ decays in the level-2 trigger. This requires the
Tevatron beam to be aligned parallel to the detector axis,
due to the long ($\sim 60$cm) luminous region in $Z$.
The $SVT$- trigger will yield a highly pure sample of inclusive
$B$ decays, approximately 10$^8$ $b\overline{b}$ events per fb$^{-1}$.}
 
\item{With the $p_T(B)$ cut, the proper time resolution
is adequate for demanding analyses such as $B_s$ mixing
(e.g., 45 fs on $B_s\rightarrow D_s^-\pi^+$)}
 
\item{The $p_T(B)$ requirement also improves the ratio
of $B$ production to QCD background.  While the total $b\overline{b}$
cross section is of order 0.2\% of the minimum bias rate, the
high-$p_T$ $b$-jet cross section is measured to be 2\% of the
QCD jet rate, in the $p_T$ range of interest for CDF.}
 
\item{In the central region, tracks are well spread out in $Z$,
and CDF has demonstrated highly efficient track reconstruction
for $b\overline{b}$ events.}
\end{itemize}

\section{Trigger Strategies}
 
The main $B$-physics goals for CDF Run~II make use of the following
final states.  Of course in each case, other final states, both
inclusive and exclusive, are used as calibration samples, and so the triggers
must be designed to be inclusive.
\begin{itemize}
\item{$B^0\rightarrow J/\psi K_S$ for $\sin{2\beta}$}
\item{$B^0_s\rightarrow J/\psi\phi$ for $CP$ and $\Delta\Gamma$}
\item{$B\rightarrow \mu^+\mu^- K^{(*)}$ rare decays}
\item{$B^0_s\rightarrow D_s^-\pi^+$ for $B_s$ mixing}
\item{$B_s\rightarrow D_s^-\ell^+\nu$ for $B_s$ mixing}
\item{$B^0_{(s)}\rightarrow K^{*0}\gamma, \phi\gamma$ using photon conversion $\gamma
\rightarrow e^+e^-$}
\item{$B^0\rightarrow \pi^+\pi^-$, $B^0_s\rightarrow K^+K^-$ for CKM
angle $\gamma$}
\end{itemize}
This is of course only a partial list.  CDF has a three-level trigger
design.  Level-1 is pipelined so as to be deadtimeless. Digitized 
data are stored in 42$\times$132 ns buffers to create and count trigger 
primitives (electrons, muons, COT tracks, and combinations of these).
Level-2 is a global trigger that allows finer grained electron and
muon track matching, and more sophisticated
combinations of trigger primitives (for example, opposite charged
lepton pairs with invariant mass requirements).  
Level-2 finds SVT-tracks using four SVX II layers; these are
matched to the XFT/COT tracks found in level-1 to define high
resolution, large impact parameter $b$-decay candidates. Level-2
also matches electrons to the central shower-maximum detector. Finally, Level-3 
provides offline quality tracking and calorimeter reconstruction.
 
Using Run~I data, CDF has carefully optimized the use of bandwidth
for Run~II triggers.  The basic trigger requirements can be
summarized as follows:
\begin{itemize}
\item{ Dimuons and dielectrons}
\subitem{Level-1: Two leptons with COT matched to muon chambers or central calorimeter}
\subitem{Level-2,3: Additional mass, charge, and $\Delta\Phi$ cuts}
\item{Single Leptons}
\subitem{Level-1: Single leptons matched to muon chambers or central calorimeter}
\subitem{Level-2,3: Additional requirement of accompanying SVT track}
\item{$B$ hadronic triggers}
\subitem{Level-1: two tracks, opposite charge, $\Delta\Phi$ cuts}
\subitem{Level-2,3: two SVX tracks, impact parameter and $ct$ cuts}
\end{itemize}
Thresholds for dilepton and two-track triggers are typically 1.5-2.0 GeV/c,
and for single leptons 3-4 GeV/c.  Level-3 can make additional cuts in order
to divide the data into manageable data sets; for example, a high mass
cut will be used to define a $B\rightarrow\pi\pi$ stream based on
the two-track triggers, while other SVX track requirements will be used
to make a $b\rightarrow c$ generic hadronic sample.
 
The total trigger bandwidth, including $B$ physics, is
designed to be 40 kHz (level-1), 300 Hz (Level-2) and 75 Hz Level-3).
Thus, the maximum rate level-3 would be 750 nb at 10$^{32}$ luminosity.
$B$ triggers are expected to require about half of that bandwidth.
Table~\ref{tab:rate} gives the expected bandwidth requirements for
the trigger streams.

\begin{table}
\begin{center}
\begin{tabular}{l|r|r|r}
\hline\hline
Trigger & L1 $\sigma$ [nb] & L2 $\sigma$ [nb] & L3 $\sigma$ [nb]\\
\hline
$B\rightarrow h^+h^-$  & 252,000 & 560 & 100 \\
$B\rightarrow \mu \mu (X)$ & 1100 & 90 & 40\\
$J/\psi \rightarrow ee$ & 18000    & 100 & 6\\
Lepton plus displaced track     & 9000     &130     & 40\\
\hline \hline
\end{tabular}
\vspace*{.2cm}
\caption[CDF $B$-physics trigger rates]{Trigger rates for the main CDF II $B$-physics triggers}
\label{tab:rate}
\end{center}
\end{table}

\section{Offline Analysis and Simulation}
CDF has estimated rates for literally hundreds of 
reconstructable decay modes that come in on the lepton or
SVT two-track triggers.  Table~\ref{tab:cdfyield} lists
some examples.

\begin{table}
\begin{center}
\begin{tabular}{l|r}
\hline\hline
Physics Channel & Event yield (2 fb$^{-1}$)\\
\hline
$b\rightarrow J/\psi X$ & 28,000,000\\
$B^0\rightarrow J/\psi K^0_S$ & 28,000\\
$B_s\rightarrow D_s\pi$ & 10,000\\
$B_s\rightarrow D_s\ell\nu$ & 30,000\\
$B\rightarrow \mu^+\mu^- K^{(*)}$ & 50\\
$B\rightarrow K^*\gamma$ & 200\\
$B\rightarrow \pi\pi,K\pi, KK$  & 30,000\\
\hline \hline
\end{tabular}
\vspace*{.2cm}
\caption[CDF $B$-physics event yields]{Event yields expected from CDF Run~II
$B$-physics triggers}
\label{tab:cdfyield}
\end{center}
\end{table}

Detailed studies of the physics reach expected for CDF may be found elsewhere
in this document. We discuss briefly two issues that enter into calculations of
sensitivity, namely flavor tagging and generic simulation.


\subsection{Flavor Tagging}
CDF measured the flavor tagging efficiencies and dilutions for
three methods in the Run~I mixing and $\sin{2\beta}$ analyses:
``same side'' (fragmentation correlation) tagging, jet-charge
tagging, and soft electron and muon tagging.  To understand
the ``same side'' tagging at least qualitatively, CDF 
analyzed $B^{**}$ production using large samples
of semileptonic $B\rightarrow D^{(*)}\ell\nu$ decays
and applied these to tune the PYTHIA event generator.
This gave qualitative agreement with observed same-side
tagging efficiencies, including the observed differences between
$B^0$ and $B^+$ tagging caused by kaons and protons.  For jet-charge
algorithms, the tagging was optimized initially using
PYTHIA and HERWIG event generators, but actual efficiencies and
dilutions were measured with data, for example, by fitting the
$B^0\overline{B^0}$ mixing amplitude or by direct comparison
of tagging in $B^{\pm}\rightarrow J/\psi K^{\pm}$.  Soft lepton
tags were calibrated similarly.  For the $\sin{2\beta}$ analysis the
combined tagging efficiency, including correlations between tags,
was $\epsilon D^2$ = 6.3$\pm$1.7\%.
 
CDF extrapolates the Run~I efficiencies and dilutions to
Run~II conditions, taking into account the standalone silicon
tracking to $|\eta|\sim 2$, and including kaon tagging based on
TOF.  This gives estimates of $\epsilon D^2 \sim$ 9.1\% for $B^0\rightarrow J/\psi K_S$
and $\epsilon D^2 \sim$ 11.3\% for $B_s\rightarrow D_s^-\pi^+$.
The basic strategies for calibrating and optimizing the tagging
efficiencies for each method would be similar to Run~I.  For example,
the same side tags can be optimized using semileptonic decays
$B^{0,+}\rightarrow D^{(*)}\ell\nu$ and $B^0_s\rightarrow D_s^{(*)}\ell\nu$,
including the effects of fragmentation kaons.
Opposite side tags can be optimized on the same semileptonic channels,
or on inclusive samples such as $b\rightarrow J/\psi X$ (e.g., by
comparing tagging rates for prompt and long lived $J/\psi$'s).
Same-side tags for $\sin{2\beta}$ can be calibrated using
$B^{\pm}\rightarrow J/\psi K^{\pm}$ and $B^0\rightarrow J/\psi K^{*0}$.
Opposite side tags would be calibrated using $B^{\pm}\rightarrow J/\psi K^{\pm}$.
Thus, there are a variety of data channels and cross checks
that are available to understand each tagging method.  Event generator
Monte Carlo's play a role in helping to model same-side tags, including
the effects of kaons, and in understanding correlations and variations
of dilution with kinematics.

\subsection{Monte Carlo Issues}
\index{CDF!detector simulation}
CDF has adopted a GEANT based detector simulation for Run~II,
which is used for optimization of track reconstruction,
muon matching, jet energy analysis, SVT response, etc.  The calorimeter response
has to be tuned to match data from testbeam, Run~I, and eventually Run~II.
 It is obviously straightforward to understand the efficiency
for signals such as $B^0\rightarrow \pi^+\pi^-$, where the generation is
trivial, and the detector response depends only on tracking and SVT efficiency.
The main concern for $B$ triggers is backgrounds.  For lepton-based
triggers, these can be estimated from Run~I data.  For hadronic triggers
such as $B^0\rightarrow \pi^+\pi^-$, CDF can set limits on backgrounds
using Run~I data, and can use the PYTHIA generator (plus SVT simulation)
to estimate the backgrounds from real $b\overline{b}$ events.  The backgrounds
that dominate $B^0\rightarrow \pi^+\pi^-$ at the trigger level do not
appear to come from real $b\overline{b}$ events but from QCD backgrounds that
fake the impact parameter trigger; this conclusion comes from comparison
of trigger rates with $b\overline{b}$ Monte Carlo,
and also by examining Run~I data, comparing 
high mass two-track rates rates from QCD and real $b\overline{b}$ events.
Thus, while CDF can certainly estimate backgrounds from real $b\overline{b}$ events
in the SVT two-track trigger sample (these come in due to real
$b$-hadron lifetimes), it cannot reliably simulate ``QCD'' backgrounds
in this sample (these depend critically on offline SVX reconstruction).

\clearpage{\pagestyle{empty}\cleardoublepage}
\chapter{The D\O\ detector}

\authors{R.~Jesik, F.~Stichelbaut, K.~Yip, A.~Zieminski}

\def\Do{\mbox{D\O}}
\def\pp{$p\bar{p}$}
\def\ptmu{$p_T^{\mu}$}
\def\pt{$p_T$}
\def\ptb{$p_T^b$}
\def\ptc{$p_T^c$}
\def\etamu{$\eta^{\mu}$}
\def\dsdptmu{\mbox{d$\sigma_b^{\mu}/\mathrm{d}p_T^{\mu}$}}
\def\dsdptb{\mbox{d$\sigma^b/\mathrm{d}p_T^b$}}
\def\bb{\mbox{$b\bar{b}$}}
\def\cc{\mbox{$c\bar{c}$}}
\def\QQ{\mbox{$Q\bar{Q}$}}
\def\ptuu{\mbox{$p_T^{\mu\mu}$}}
\def\bsdsphi{\mbox{$B_s^0 \rightarrow D_s^- + \mu^+ + \nu_{\mu}$}}
\def\bsds3pi{\mbox{$B_s^0 \rightarrow D_s^- + 3\pi^{\pm}$}}
\def\bsdslX{\mbox{$B_s^0 \rightarrow D_s^- + \ell^+ + X$}}
\def\bsdsnpi{\mbox{$B_s^0 \rightarrow D_s^- + n\pi^{\pm}$}}
\def\bks{\mbox{$B_d^0 \rightarrow K_s + J/\psi$}}
\def\jpsi{\mbox{$J/\psi$}}
\def\bkspsi{\mbox{$B_d^0 \rightarrow J/\psi K_S^0$}}
\def\bkst{$B_d^0 \rightarrow K^{*\circ} \mu^+\mu^-$}
\def\bkstee{\mbox{$B_d^0 \rightarrow K^{*\circ} e^+e^-$}}
\def\buu{\mbox{$B_s^0 \rightarrow \mu^+\mu^-$}}
\def\bsuu{ \mbox{$b \rightarrow s \mu^+\mu^-$}}
\def\dskkp{\mbox{$D_s^- \rightarrow K^+ K^- \pi^-$}}
\def\dsphi{\mbox{$D_s^- + \mu^+$}}
\def\dsmunu{\mbox{$D_s^-\, \mu^+\, \nu_{\mu}$}}
\def\ds3pi{\mbox{$D_s^- + 3\pi^{\pm}$}}
\def\kspsi{\mbox{$K_s + J/\psi$}}
\def\QQmumu{\mbox{$Q\bar{Q}\rightarrow \mu \mu$}}
\def\QQmu{\mbox{$Q\bar{Q}\rightarrow \mu$}}
\def\QQee{\mbox{$Q\bar{Q}\rightarrow e e$}}

\begin{figure}[t]
\centerline{\psfig{figure=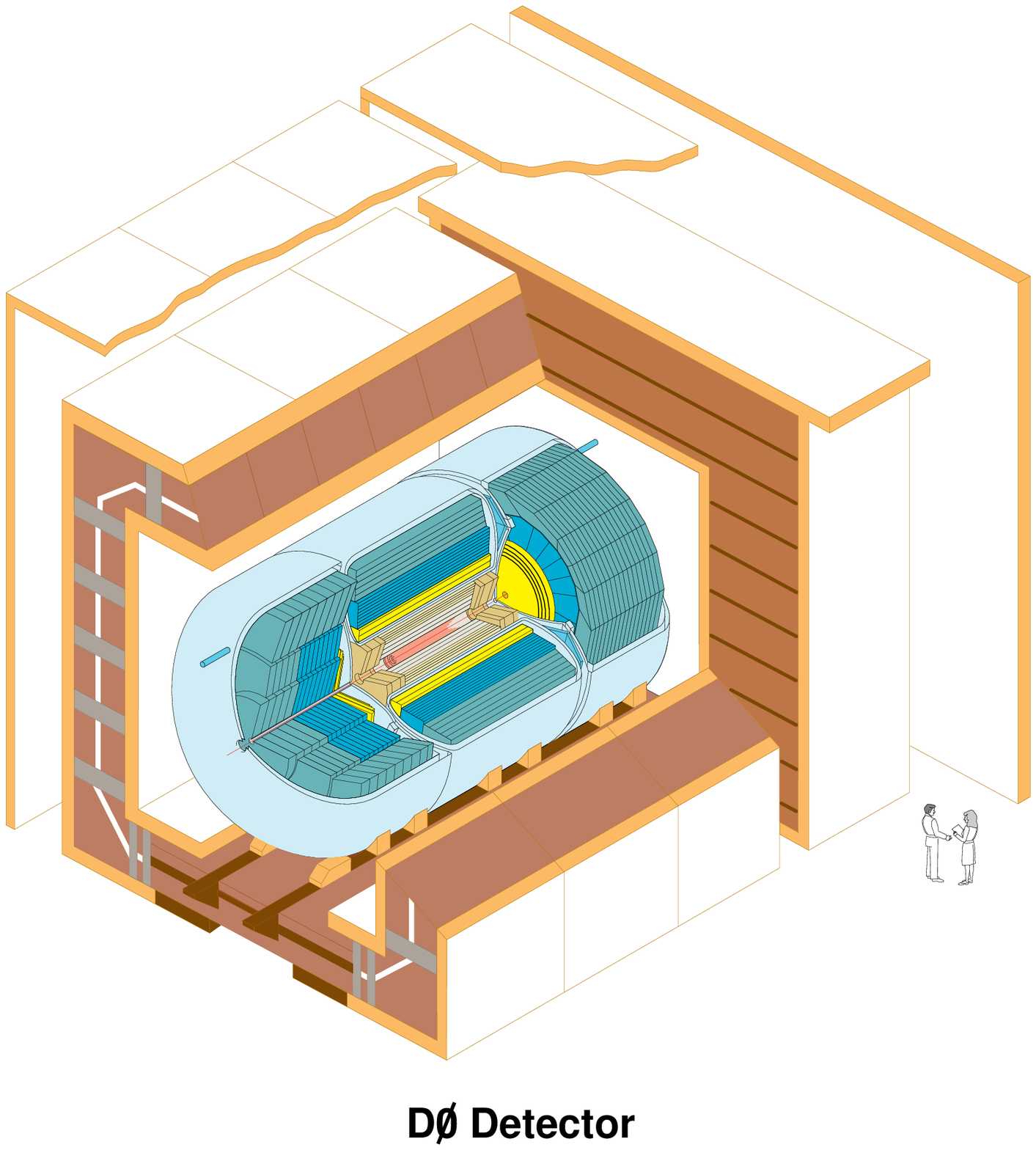,height=9.5cm}}
\caption{The D\O\ detector}
\label{d0det}
\end{figure}

The D\O\ Run II detector \cite{upgrade} shown in Fig.~\ref{d0det}
builds on the detector's previous strengths of excellent calorimetry
and muon detection in an extended rapidity range. The major addition to
the apparatus is a precision tracking system,
consisting of a silicon vertex detector surrounded by an eight layer central
fiber tracker. These detectors are located inside a 2 T solenoid magnet.

\section{The silicon vertex detector}

\begin{figure}[bh]
\centerline{\psfig{figure=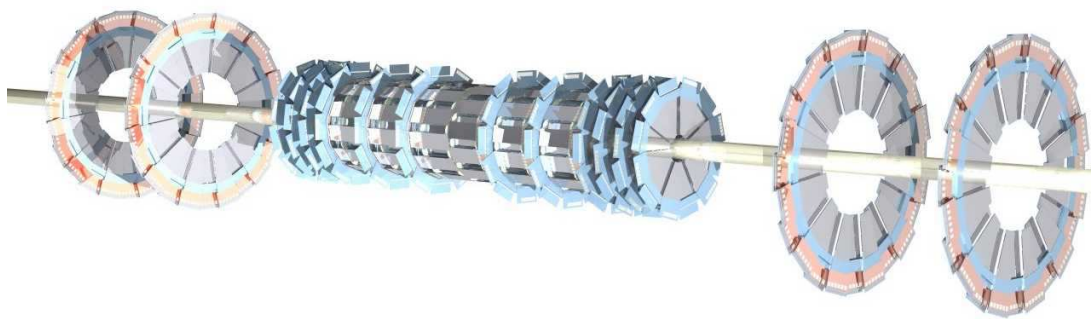,width=0.95\textwidth}}
\caption{The D\O\ silicon vertex detector}
\label{SMT}
\end{figure}

The silicon vertex detector\index{\Do\ silicon vertex detector}(SMT), shown in
Fig.~\ref{SMT}, is a hybrid barrel and disk design. The central detector,
covering $|z| < 32$ cm, consists of six  barrels with disks interspersed
between them.
Each barrel module consists of four radial layers of detector
ladders located at a radius of 2.7, 4.5, 6.6, and 9.5 cm,
respectively. Layers one and three have 50 $\mu$m pitch double sided
silicon microstrip detectors (with axial and $90^\circ$ strips) in
the four inner barrels, and have single sided (axial strip) detectors
in the two end barrels. Layers two and four have double sided detectors
with axial and 2$^\circ$ stereo strips. Each disk module has twelve
wedge-shaped double sided
detectors (extending radially from 2.7 to 10 cm) with 30$^\circ$ stereo angle.
The forward detector
consists of six disks of similar design (three located near each end of 
the central detector) plus four larger 24-wedge disks made of single sided
detectors, located at z = $\pm$ 110 cm and $\pm$ 120 cm. The detector
is read out using SVX-II chips and has 800,000 total channels.   
The silicon vertex detector provides tracking information out
to $|\eta|$ = 3 and gives a reconstructed vertex position resolution
of 15-40 $\mu$m in $r-\phi$ and 75-100 $\mu$m in $z$, depending on
the track multiplicity of the vertex.  

\section{The central fiber tracker}

The central fiber tracker\index{\Do\ central fiber tracker} consists of 74,000 scintillating fibers
mounted on eight concentric carbon fiber cylinders at radii from
19.5 to 51.5 cm. Each cylinder supports four layers of fibers, one
doublet in the axial direction and one doublet oriented at a 
$3^\circ$ helical stereo angle for odd numbered cylinders and at
$-3^\circ$ for even ones. The fibers are multi-clad and
have a diameter of
830 $\mu$m. Clear fiber waveguides carry the light 
for about ten meters from the scintillating fibers to the 
visible light photon counters situated in cryostats under the
calorimeter. These are silicon devices that have a
quantum efficiency of over $80\%$ and a gain of about 50,000.
They operate at a temperature of about 10 K. Digitization is
also performed with SVX-II chips and the axial fibers are used
to form fast level 1 trigger track objects.

The combined silicon vertex detector and central fiber tracker
have excellent tracking performance. The full coverage of the
entire combined detector extends out to $|\eta|$ =  1.6.
Tracks will be reconstructed with high efficiency ($\sim 95\%$ 
in the central region) with a resolution of 
$\sigma_{p_T}/p_{T}^2 \sim 0.002$. The silicon detector disks
allow efficient tracking out to an $|\eta|$ of 3.  
     
\section{Electromagnetic pre-shower detectors}
Another important piece of the upgrade is the new pre-shower 
detectors attached to the inner surfaces of the calorimeter
cryostats. These scintillating strip detectors greatly improve
electron identification by providing a finer grain spatial match
to the inner tracker than is obtained by the calorimeter alone.
This reduces electron trigger backgrounds by a factor of 5-10.
We will be able to trigger on electrons with $p_T$ down to 1.5 
GeV/$c$, and out to an $|\eta|$ of 2.5.

The calorimeter\index{\Do\ calorimetry} itself remains unchanged from its Run I
configuration. It is a uranium-liquid argon sampling calorimeter with fine
longitudinal and transverse segmentation, $\Delta\eta \times \Delta\phi = 0.1
\times 0.1$, that allows electromagnetic showers to be distinguished from
hadronic jets. It has full coverage out to $|\eta| = 4$ with energy resolutions
of $15\%/\sqrt{E}$ for electromagnetic showers and $80\%/\sqrt{E}$ for hadronic
jets.

\section{Muon Spectrometers}

The central part of the muon system\index{\Do\ muon detector} (covering the $|\eta| < 1$ range) includes 94 proportional drift tubes (PDT), barrel scintillator counters (A-PHI) and cosmic ray veto scintillator counters. There are three PDT layers (A, B, and C) with three or four 5-cm-thick, $\pm 5$-cm-wide drift cells per layer. A $r - \phi$ magnetic field, with an average magnitude of 16.7 kGauss, is contained in  an iron toroid sandwiched between layers A and B of the PDT system. The A-PHI counters have nine segments in the $z$ direction and 80 segments in the $\phi$ direction. There are 630 A-PHI counters in total.

	The forward muon system (located at both endcaps) covers the $\eta$ [$\pm1, \pm2$] region of the \Do\ detector. It includes mini-drift tubes (MDT's with A, B, and C layers) with three or four decks of 1-cm square cells per layer. Iron toroids are located between the A and B layers of the MDT systems. The average magnitude of the field is 16.0 kGauss. Three layers of PIXEL scintillating counters, located next to corresponding MDT stations, have a $4.5^\circ$ $\phi$-segmentation and a 0.1 $\eta$-segmentation.

 The upgraded muon system offers excellent efficiency, purity, and coverage. We are 
able to trigger on muons with $p_T$ down to 1.5 GeV/$c$, and
out to an $|\eta|$ of 2.

\section{Trigger system}

Significant upgrades to the trigger\index{\Do\ trigger system} and data
 acquisition systems are also required to operate in the high luminosity ($L = 2 \times 10^{32}$cm$^{-2}$
S$^{-1}$), high rate (7 MHz) environment of the upgraded TeVatron. The D\O\
Run II trigger consists of three levels.

\subsection{Trigger levels}

The first level (L1) trigger consists of
signals from the axial layers of the CFT, the pre-shower detectors, the 
calorimeter, and the muon scintillators and tracking chambers. The level 1
 trigger hardware for each of these systems examine the event and report
information to an array of front-end digitizing crates which have sufficient
memory to retain data from 32 crossings. Trigger decisions are made in less
than 4.2 $\mu$s at this level, providing deadtimeless operation with a maximum
accept rate of 10 kHz. Upon a level 1 trigger accept, the entire event is 
digitized and moved into a series of 16 event buffers to await a level 2
decision.

The level 2 (L2) trigger will reduce the L1 accept rate of 10 kHz by 
roughly a factor of 10 within 100 $\mu$s by correlating multi-detector
objects in an event. In the first stage, or preprocessor stage, each
detector system builds a list of trigger information. This information
is then transformed into physics objects such as energy clusters or 
tracks. The time required for the formation of these objects is about
50 $\mu$s. These objects are then sent to the level 2 global processor
where they are combined and trigger decisions are made. For example, 
spacial coorelations between tracking segments, pre-shower 
depositions, and calorimeter energy depostions may all be used to 
select electron candidates. The deadtime of the level 2 trigger is
expected to be less than $5\%$.

The third and final stage (L3) of the trigger will reconstruct events
in a farm of PC processors with a final accept rate of 50 Hz.

\subsection{Level 1 muon triggers}
\label{sec_trigger}

D\O\ Level 1 muon trigger
\index{\Do\ muon trigger! Level 1! description} 
(L1MU)\cite{muon} identifies muon candidates by using combinatorial logic that makes use of tracks from L1 Central Fiber Tracker (L1CFT) trigger and hits from all muon detector elements: drift chambers and scintillation counters.	The central, north and south regions of the detector are divided into octants. The L1MU triggers are formed locally in each octant. For the purpose of this report the L1MU trigger terms are labelled by two digits and two letters
\index{\Do\ muon trigger! Level 1! notation}
: L1MU(i, j, A, B).
\begin{itemize}
   \item The first number refers to the number of muons requested:
   \item The second number corresponds to an approximate value of 
                    the muon \pt{} threshold in GeV/$c$. 
   The nominal values are 2,4,7 and 11 GeV/c.
   \item The first letter refers to the covered $\eta^{\mu}$ region:
   \begin{itemize}
      \item C = Central, with $\mid \eta^{\mu} \mid \leq 1$;
      \item A = All, with $\mid \eta^{\mu} \mid \leq 1.6$;
      \item X = eXtended, with $\mid \eta^{\mu} \mid \leq 2$.
   \end{itemize}
   \item Finally, the second letter describes the muon tag criterium:
   \begin{itemize}
      \item M stands for Medium tag, requiring a L1CFT track matched with the PDT or MDT centroids, and with at least one layer-scintillator confirmation.
      \item T stands for Tight tag, requiring a L1CFT track combined with the PDT or MDT centroids, and with a two layer-scintillator confirmation.
   \end{itemize}
\end{itemize}

\subsection{Level 1 muon trigger rates and efficiencies}

    Performance of various muon triggers was evaluated in Ref.~\cite{FS_note}.
For this analysis we concentrated on:
   \begin{itemize}
      \item the single muon trigger L1MU(1,4,A,T) 
      \item and the dimuon trigger L1MU(2,2,A,M). 
    \end{itemize}

The $\eta^{\mu}$
coverage of both triggers extends to $\mid \eta^{\mu} \mid \leq 1.6$ and 
effective
muon \pt{} thresholds are 2 GeV/c for the dimuon trigger and 4 GeV/c for the
single muon trigger. Muons with \pt{} $>$ 1.5 GeV/c have a chance to penetrate
the calorimeter, muons with \pt{} $>$ 3.5 GeV/c have a chance to be detected in
the entire muon detector, including B and C layers outside of the toroid
magnet.
 
    To estimate the QCD  trigger rates, we used the 
dijet 
ISAJET events generated in 6 \pt{} intervals (2--5, 5--10, 10--20, 40--80
and 80--500) with 1, 3, 5 or 7 additional minimum bias interactions
(\pt{} between 1 and 100 GeV/$c$) per event. 
The background trigger rates were computed
with the assumptions of an instantaneous luminosity of 
$2\times10^{32}$ cm$^{-2}$s$^{-1}$, a beam crossing time interval of
132 ns and a dijet total cross section of 57 mb.
    Studies of
Ref.\cite{FS_note} indicate that the expected Level 1 trigger rates at this 
instantaneous
luminosity are approximately
400 Hz and 80 Hz for the L1MU(2,2,A,T) and L1MU(1,4,A,T) triggers, 
respectively.
\index{\Do\ muon trigger! Level 1! trigger rates}%
The event samples used in these studies  contained about 1000 events each. 
Therefore some rate calculation could
suffer from large fluctuation due to the small number of selected events.
The absolute rates obtained from these samples 
could be underestimated by a factor up to two.

    For the trigger efficiency studies we have used a sample of \bks\ events processed through the latest version
of D0GEANT and the trigger simulator. Trigger efficiencies are normalized to the numbers of events with $\mid \eta^{\mu}\mid < 1.6$ and 
    kinematic cuts imposed on $p_T^{\mu}$, and $p_T^{\mu\mu}$ as
    specified in the Table~\ref{ta:trigger}.
\index{\Do\ muon trigger! Level 1! efficiency for dimuon events}%
 A large difference between efficiencies for the L1MU(1,4,A,M) and  
L1MU(1,4,A,T) triggers reflects the fact that the "tight" tag condition
requires two scintillator signals for each muon. This requirement reduces the
geometrical acceptance of the trigger due to a limited coverage of the 
scintillating
counters at the bottom of the detector.

\begin{table}
\centering
\begin{tabular}{l||c||c|c} \hline \hline
   $p_T^{\mu\mu}$ & $ > 2.0$ GeV/c & $> 5.0$ GeV/c& $> 5.0$ GeV/c \\
   $p_T^{\mu}$ & $    > 1.5$ GeV/c & $> 1.5$ GeV/c& $> 3.0$ GeV/c \\ \hline
    L1MU(2,2,A,M)  &27\% & 46\% & 57\%  \\
    L1MU(1,4,A,M)  &33\% & 69\% & 78\%  \\
    L1MU(1,4,A,T)  &15\% & 32\% & 49\%  \\
  dimuon or single (M) &41\% & 71\% & 80\%  \\
  dimuon or single (T) &32\% & 55\% & 67\%  \\
\hline \hline
\end{tabular}\vspace*{4pt}
\caption[Trigger efficiencies for dimuon events preselected with the kinematic
cuts listed~]{Trigger efficiencies for dimuon events preselected with the
kinematic cuts listed}
\label{ta:trigger}
\end{table}

 The trigger rates at the instantenous luminosity of 2 $10^{32} cm^{-2}s^{-1}$
due to dimuons from the
    genuine $Q{\bar Q}$ signal
\index{\Do\ muon trigger! Level 1! $Q{\bar Q}$ dimuon rates} 
    are $\approx$ 13 Hz ($\approx$ 4 Hz for $p_T^{\mu\mu}> $ 5 GeV/c). The 13
    Hz combines contributions from: \cc\ pair production (2.5 Hz), \bb\ pair
    production (9.5 Hz) and $b \rightarrow J/\psi +X$ decays (1.0 Hz). A
    requirement of dimuon $p_T^{\mu\mu} > 2$ GeV/$c$ reduces the rate to 9~Hz.

To estimate the contribution from $J/\psi$ production
\index{\Do\ muon trigger! Level 1! $J/\psi$ event rates} 
to the single muon 
trigger rate, we normalized the relevant Monte Carlo samples 
($b\rightarrow J/\psi + X$ and prompt $J/\psi$'s)
to the CDF measurement of the 
$J/\psi$ cross sections in the kinematic range $p_T^{J/\psi} > 5$ GeV/$c$ 
and $\mid \eta(J/\psi) \mid <0.6$~\cite{CDFpsi}.  
With this normalization we find that the irreducible L1 trigger rate 
for signal events is at least 4.0 Hz for prompt $J/\psi$'s 
and 0.8 Hz for $J/\psi$'s from $b$-quark decays
($J/\psi$ with $p_T^{J/\psi} > 2$ GeV/$c$ and $\mid \eta^{J/\psi} \mid <1.5$).   


\subsection{Level 2 and Level 3 muon triggers}

	The second level of the muon trigger
\index{\Do\ muon trigger!Level 2 and 3! description} 
(L2MU) uses calibration and more precision timing information to improve the quality of muon candidates. Fast processors and a highly parallelized data pathway are the basis of the L3 muon trigger (L3MU). L3MU improves the resolution and the rejection efficiency of L2MU candidates. This is accomplished by performing local muon tracking, by adding inputs form the calorimeter and the sillicon micro-strip tracker (SMT) and by performing more analytical calculations on CFT tracks and PDT and A-PHI hits with the calibrated data.
The performance of the  L2MU and L3MU triggers has not been fully evaluated at the time of this writing.

    In addition, we expect to have the STT (Silicon Tracker Trigger)
\index{\Do\ Silicon Tracker Trigger} 
  preprocessor installed
in the middle of 2002. The STT will be part of the Level 2 trigger and will
provide an option of triggering on displaced vertices (impact parameter
significance). It will also further improve momentum resolution of muon tracks.

   The STT will allow to tag $B$ decays using displaced secondary
         vertices or tracks with large impact 
         parameters. The expected impact parameter resolution in 
the transverse plane can be parametrized as~\cite{FS_note}:
         \begin{equation}
          \sigma^2(d_0) = (12.6 \ \mu\mathrm{m})^2 +
                       \left(\frac{36.6\ \mu\mathrm{m}\, GeV}
          {p\times\sin^{3/2}\theta}\right)^2
         \end{equation}

\begin{figure}[t]
\centerline{\psfig{figure=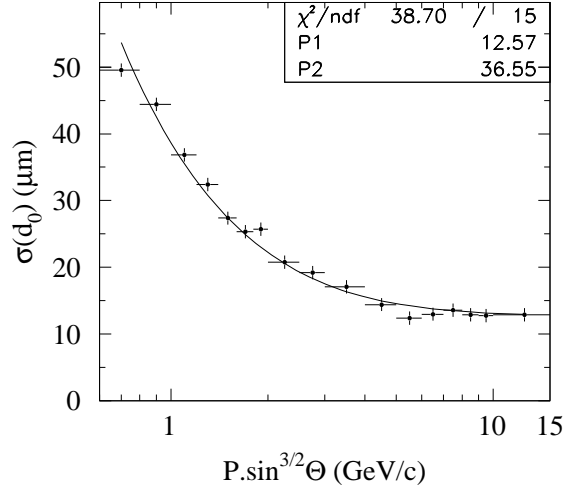,height=7cm}}
\caption{Expected impact parameter uncertainty dependence on track momentum and
polar angle.}
\label{IMPACT} 
\end{figure}

This dependence on particle momentum and polar angle is illustrated
in Fig.~\ref{IMPACT}.
\index{\Do\ Silicon Tracker Trigger!
 impact parameter resolution}

\begin{figure}[bh]
\centerline{\psfig{figure=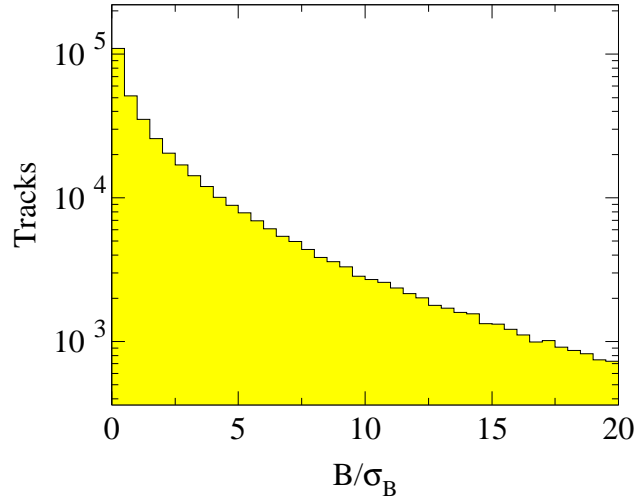,height=7cm}}
\caption{Significance distribution for STT tracks reconstructed in \bkst{}
events with all four charge particles produced with $p_T \geq 0.5$ GeV/$c$.}
\label{IMPACT_KST}
\end{figure}

In Fig.~\ref{IMPACT_KST} we show the impact parameter  significance, $S =
d_0/\sigma(d_0)$,  for the \bkst{} decay products, under the condition that all
four charged particles are produced with $p_T \geq 0.5$ GeV/$c$ (muons with
$p_T \geq 1.5$ GeV/$c$). On the average 1.8   particles from the \bkst{} decay
will have an impact parameter significance greater than 2. This number
increases to 3.2 for events preselected by a request of a 400 $\mu$m
transversal separation between primary and secondary vertices.

\begin{figure}[t]
\centerline{\psfig{figure=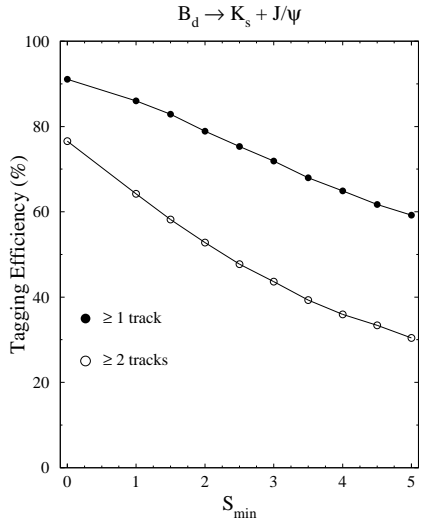,height=7cm}}
\caption{\bb{} event tagging efficiency as a function of $S_{min}$ with either
one or two tracks in the event satisfying the condition $S > S_{min}$.}
\label{BTAG_EFF}
\end{figure}

    Therefore, once STT becomes operational, we intend to use a simple $B$ tagging algorithm based on 
the number of tracks in the event 
         with a significance greater than $S_{min}$ to improve our Level 2
         trigger rates for selecting \bb\ events. The algorithm was tested on 
charged tracks with
        $p_T > 0.5$ GeV/$c$;
       $\mid \eta \mid < 1.6$;
	and with hits in at least 3 layers of the SVX~\cite{FS_note}. As shown
	in Fig.~\ref{BTAG_EFF}  an
        efficiency of 50\% can be achieved by requesting at least two tracks 
    per event
        with an impact parameter significance greater than 2. Similarily, requiring at least one track with an  impact parameter significance
greater than 5 permits the reduction of the background rate by a factor
10 while keeping 50\% of the signal (assuming a primary vertex smearing
of 30 $\mu$m). 
\index{\Do\ Silicon Tracker Trigger!
 efficiency} 

Finally, we have studied the effect of the STT trigger on the separation of the
$J/\psi$ signal due to $b$-quark decays from the prompt $J/\psi$ production.
In case we need to suppress the prompt $J/\psi$ signal, requiring at least 
one track with $\mid S_B \mid > 2.5$ provides a factor 7
reduction at the 30\% cost to the non-prompt $J/\psi$ signal.
\index{\Do\ Silicon Tracker Trigger! 
 prompt $J/\psi$ separation}

\subsection {Low \pt\ dielectron trigger}

 \Do\ has also introduced a dielectron trigger\cite{electron} aimed at
detection of soft electron pairs, primarily from $J/\psi$ dielectron decays.
\index{\Do\ dielectron trigger}%
Level 1 electron candidates are selected with a transverse energy deposit $E_T>2.0~GeV$ in the 
EM calorimeter trigger towers, and with a low $\rm p_T>1.5~GeV/c$ track coincident 
with a pre-shower cluster. The two ``electrons'' are required to have opposite signs 
and to match  within a quadrant in $\phi$ with the EM deposits. 
In the forward rapidity region ($1.5<|\eta|<2.5$) the trigger is based on EM deposits 
and preshower clusters only, since no tracking 
coverage is available.
 
 The dielectron Level 2 trigger is based on a refined spatial matching between tracks and
$\rm 0.2\times 0.2$ EM trigger towers as well as on the information on the  EM fraction of the clusters and their isolation. Finally, invariant mass and angular criteria are applied 
to select $J/\psi$ dielectron  decays. 

	Efficiencies and rates for this trigger have been estimated using {\it ISAJET} generated events 
overlaid with 2 extra interactions. Trigger rates at the nominal Run II luminosity 
are expected to be  below 1~kHz at Level 1 and about 150~Hz at Level 2.
 The expected yield of
 triggered $B_d^0~\to~J/\psi~K^0_s$ events is  
$15\times 10^3$ for an integrated luminosity of 2\,fb$^{-1}$.



\clearpage{\pagestyle{empty}\cleardoublepage}
\chapter{The BTeV Detector}
\authors{R. Kutschke, for the BTeV Collaboration\cite{btev_auth}}

\section{Introduction}
\index{BTeV!detector}

On May 15, 2000 the BTeV Collaboration submitted their
proposal\cite{btev_prop} to the Fermilab management and on 
June 27, 2000 the Fermilab director announced that 
BTeV experiment had been given Stage I approval.
The information given in this chapter is abstracted from
that proposal.   Much additional information
is available in the proposal, including
the physics case for the experiment, a detailed
description of the proposed detector, a description
of the simulation tools used to evaluate the detector design,
a summary of the physics reach, a cost
estimate and extensive appendices.

This chapter will discuss the reasons for choosing a forward
detector, followed by overviews of the detector, the
simulation tools and the analysis software.  
Some illustrative results will also be included.
As discussed in section~\ref{2:ssec:xsec} of this report, all BTeV 
event yields were computed under the presumption that the 
cross-section for $b\bar{b}$ production at the Tevatron is~100~$\mu$b.
All effieciencies and background levels
were computed for
an average 2 interactions per beam crossing, which
corresponds to a luminosity of \hbox{2$\times 10^{32}$ cm$^{-2}$s$^{-1}$}
and an interval of 132~ns between beam crossings.
\index{BTeV!presumed $b\bar{b}$ cross-section}
\index{BTeV!presumed luminosity}
\index{luminosity}

\section{Rationale for a Forward Detector at the Tevatron}
\index{BTeV!detector!forward vs central}

The BTeV detector is a double arm forward spectrometer which 
covers from 10 to 300~mrad, with respect to the colliding
beam axis.\footnote{BTeV refers to both the proton direction and the
antiproton directions as forward.}  It resembles a pair of
fixed target detectors arranged back-to-back.
In Section 2.1 of the BTeV proposal\cite{btev_prop} the reasons
for this choice are explained in detail; a summary is presented here.

According to QCD calculations of $b$ quark production, 
there is a strong correlation between the $B$ momentum and pseudorapidity,
$\eta$.  Near $\eta$ of zero, $\beta\gamma\approx 1$, while
at larger values of $|\eta |$, $\beta\gamma$ can easily reach values 
well beyond~6.  This is important because the mean decay length increases
with $\beta\gamma$ and, furthermore, the momenta of the 
decay products are larger, suppressing multiple scattering errors.
As discussed in section~\ref{2:sec:centfor} this is most important
in the trigger.

A crucially important attribute of
$b\bar{b}$ production at hadron colliders is that the $\eta$ of
the $b$ hadron and that of its companion $\bar{b}$ hadron are strongly 
correlated: 
when  the decay products of a $b$-flavored hadron are within
the acceptance of one arm 
of the spectrometer, the decay products of the accompanying 
$\bar{b}$ are usually within the acceptance of the {\em same} arm
of the detector.  This allows for reasonable levels of flavor tagging.

The long $B$ decay length, the correlated
acceptance for both $b$ hadrons and the suppression of multiple scattering
errors make the forward direction an ideal choice.

\section{Detector Description}
\label{5:sec:detector}

\begin{figure}[tbp]
\centerline{\epsfig{figure=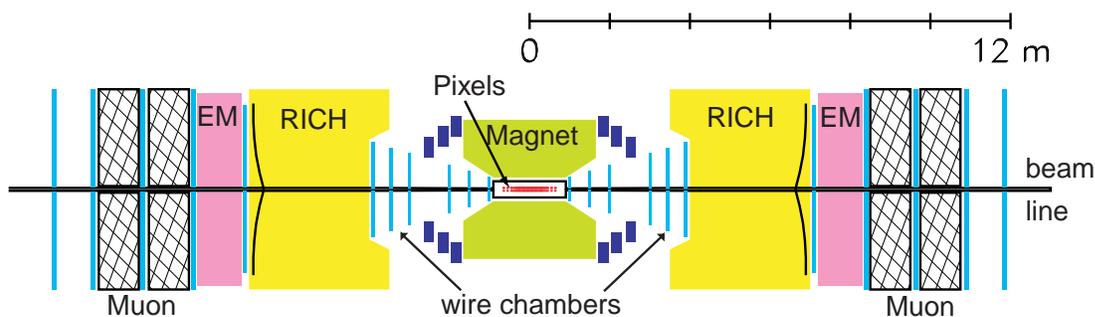,width=.95\hsize}}
\caption[A Sketch of the BTeV Detector]
{\label{5:fig:btev_det_simp}A sketch of the BTeV detector. The two 
arms are identical.
\index{BTeV!detector!sketch}
}
\end{figure}

A sketch of the BTeV detector is shown in Fig.~\ref{5:fig:btev_det_simp}.
The geometry is complementary
to  that used in current collider experiments. 
The detector looks similar to a fixed target experiment, but
has two arms, one along the proton direction and the other along the
antiproton direction.

The key design
features of BTeV include:
\index{BTeV!detector!key features}
\begin{itemize}
\item A dipole located on the IR, which  gives BTeV an effective ``two arm''
           acceptance;
\item A precision vertex detector based on planar pixel arrays;
\item A detached vertex trigger at Level 1 that makes BTeV efficient for
   most final states, including purely hadronic modes;
\item Excellent particle identification using a Ring Imaging Cherenkov Detector
(RICH);
\item A high quality PbWO$_4$ electromagnetic calorimeter capable of
reconstructing 
final states with single photons, $\pi^o$'s, $\eta$'s or $\eta'$'s;
it can also identify electrons;
\item Precision downstream 
      tracking using straw tubes and silicon microstrip detectors,
      which provide excellent momentum and mass resolution;
\item Excellent identification of muons using a dedicated detector with the
ability to supply a dimuon trigger; and
\item A very high speed and high throughput data acquisition system which
eliminates the need to tune the experiment to specific final states.
\end{itemize}                  
Each of these key elements of the detector is discussed in detail
in Part~II of the proposal\cite{btev_prop} and is discussed briefly
below.

\subsection{Dipole Centered on the Interaction Region}

A large dipole magnet, bending vertically and with a 1.6 T central field, 
is centered on the
interaction region.  This is the most compact way to provide momentum 
measurements in both arms of the spectrometer.  Moreover the
pixel detector is inside the magnetic field, which
gives the Detached Vertex Trigger the capability of rejecting low
momentum tracks.  Such tracks undergo large multiple Coulomb 
scattering and might sometimes be misinterpreted as detached tracks.

\subsection{The Pixel Vertex Detector}
\index{BTeV!detector!pixel vertex detector}
\index{pixel detector}
\index{vertex detector! BTeV pixel}

In the center of the magnet there is a silicon pixel vertex detector.
This detector serves two functions: 
it is 
an integral part of the charged particle tracking system, providing
accurate vertex information for the offline analysis; and 
it delivers very clean, precision space points to the BTeV vertex trigger.

\begin{figure}[tbp]
\centerline{\epsfig{file=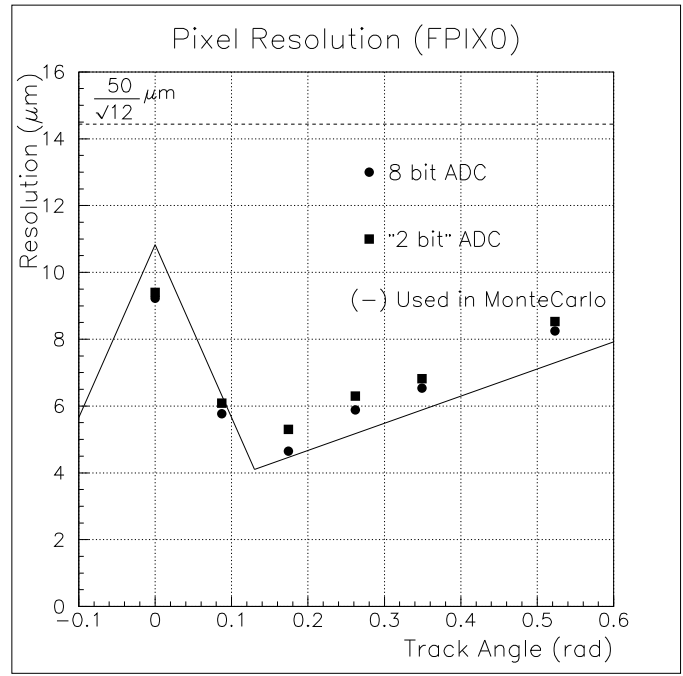, height=3.5in}}
\vspace{0.5cm}
\caption[The Resolution of the Pixel Devices]
{\label{5:fig:pixel_res}The resolution achieved in the test beam run
using 50 $\mu$m wide pixels and an 8-bit ADC (circles) or a 2-bit ADC 
(squares), compared with the simulation (line).}
\end{figure}

\begin{sloppypar}
BTeV has tested prototype pixel devices in a test beam at Fermilab. 
These devices consist
of \hbox{50 $\mu$m $\times$ 400 $\mu$m} pixel sensors
bump-bonded to custom made 
electronics chips, developed at Fermilab.  The 
position resolution achieved in the test beam 
is shown in Fig.~\ref{5:fig:pixel_res}; overlayed on that figure
is resolution function used in the Monte Carlo simulation.
The measured resolution is excellent and exceeds the requirement of 9~$\mu$m.
\end{sloppypar}

\begin{figure}[tbp]
\vspace{-1.5cm}
\centerline{\epsfig{file=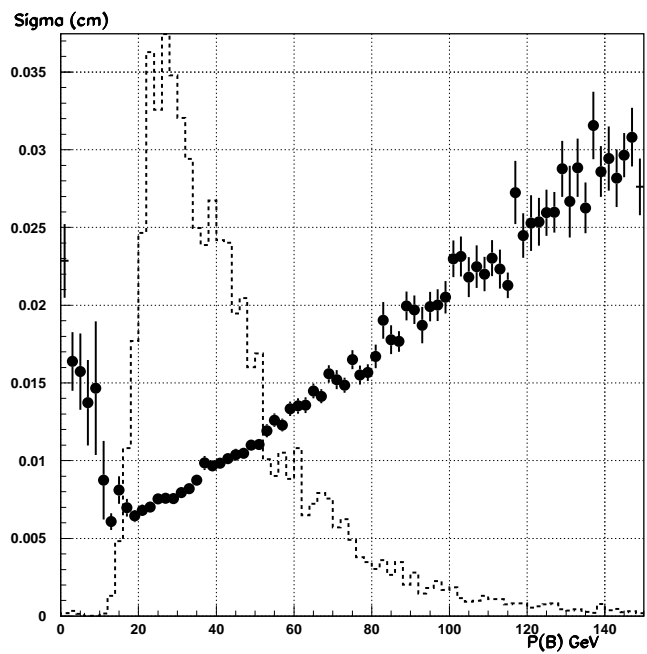, height=3.5in}}
\caption[The $B$ Momentum Distribution]
{\label{5:fig:das_pipi}The $B$ momentum distribution for
$B^o\to\pi^+\pi^-$ events (dashed) and the error in decay length
$\sigma_L$ as a function of momentum.}
\end{figure}

The critical quantity for a $b$ experiment is $L/\sigma_L$, where $L$ is
the distance between the primary (interaction) vertex and the secondary (decay)
vertex, and $\sigma_{L}$ is its error.
The efficacy of this geometry is illustrated by considering the distribution
of the resolution on the $B$ decay length, $L$, for the decay 
$B^o\to\pi^+\pi^-$.
Fig.~\ref{5:fig:das_pipi} shows
the r.m.s. error in the decay length as a function of
momentum;  it also shows the momentum distribution of the $B$'s accepted
by BTeV.
The following features are noteworthy:
\begin{itemize}
\item The $B$'s used by BTeV peak at $p$ = 30 GeV/c and average about 40 GeV/c.
\item The mean decay length is equal to $450~\mu$m $\times\, p/M_B$.
\item The error on the decay length is smallest near the peak of the
accepted momentum distribution. It increases at lower values of $p$, due to
multiple scattering, and increases at larger values of $p$ due to the 
smaller angles of the Lorentz-boosted decay products. 
\end{itemize}

\subsection{The Detached Vertex Trigger}
\index{BTeV!detector!detached vertex trigger}
\index{triggering!BTeV}
\index{triggering!detached vertex}

It is impossible to record data from each of the 7.5 million
beam crossings per second. A prompt decision, colloquially called a 
``trigger,'' must be made to record or discard the data from each crossing.
The main BTeV trigger is provided by the silicon pixel detector.
The Level 1 Vertex Trigger inspects every beam crossing and, using only
data from the pixel detector, reconstructs the primary vertices and
determines whether there are detached tracks which could signify a $B$
decay. Since  the $b$'s are at high momentum, 
the multiple scattering of the decay products is minimized, allowing for
triggering on detached heavy quark decay vertices. 

With outstanding pixel resolution, it is possible to trigger efficiently
at Level 1 on a variety of $b$ decays.  The trigger has been fully
simulated, including the pattern 
recognition code.   Table~\ref{5:tab:trigeff} summarizes the
results of the trigger simulations.  The trigger efficiencies
are generally above 50\% for the $b$ decay states of interest and at
the 1\% level for minimum bias background.
The efficiencies for signal channels are quoted as a percentage of 
events which pass the offline analysis cuts.  This is an
appropriate statistic because most of the $b$ decays are not
useful for doing physics: their decay products lie
outside of the fiducial volume of the detector, their decay
lengths are not long enough, their decay products can be ambiguously
assigned to several candidate vertices and so on.   
Separate Monte Carlo runs were performed to measure
overall event rates and bandwidth requirements and to
ensure that the trigger is efficient for generic $b$ and $c$ decays.

\begin{table}
\begin{center}
\begin{tabular}{lc}
\hline\hline
Process & Eff. (\%) \\
\hline\hline
Minimum bias & 1 \\
\hline
$B_s\to D_s^+ K^-$         & 74  \\
$B^0\to D^{*+} \rho^-$     & 64  \\
$B^0\to \rho^0 \pi^0$      & 56  \\
$B^0\to J/\psi K_s$        & 50  \\
$B_s\to J/\psi K^{*o}$     & 68  \\
$B^-\to D^0 K^- $          & 70  \\
$B^-\to K_s \pi^- $        & 27  \\
$B^0\to$ 2-body modes      & 63  \\
$(\pi^+\pi^-, K^+\pi^-,K^+K^-)$  & \\
\hline\hline
\end{tabular}
\end{center}
\caption[Level 1 Trigger Efficiencies in BTeV]
{Level 1 trigger efficiencies for minimum-bias events and for various
         processes of interest.
 For the minimum bias events the efficiency is quoted as a percentage
 of all events but for the signal channels the
 efficiencies are quoted as a percentage of those events which
 pass the offline analysis cuts.  
 All trigger efficiencies are determined
 for an average of two interactions per crossing.
        }
\label{5:tab:trigeff}
\end{table}

\subsection{Charged Particle Identification}
\index{BTeV!detector!RICH}
\index{BTeV!detector!charged particle identification}
\index{particle identification!BTeV RICH}

Charged particle identification is an absolute requirement for an
experiment designed to study the decays of $b$ and $c$ quarks. 
The relatively open
forward geometry has sufficient space to install a Ring Imaging
Cherenkov detector (RICH), which provides powerful particle ID 
capabilities over a broad range of momentum.
The BTeV RICH detector must separate pions from kaons and protons in 
a momentum range from $3$ to $70$~GeV/$c$. The lower momentum limit
is determined by soft kaons useful
for flavor tagging, while the higher momentum limit
is given by two-body $B$ decays.
Separation is accomplished using a gaseous freon radiator
to generate Cherenkov light in the optical frequency range. The light is then
focused by mirrors onto Hybrid Photo-Diode (HPD) tubes. 
To separate kaons from
protons below 10 GeV/c an aerogel radiator will be used. 

\begin{figure}[tbp]
\vspace{-0.8cm}
\centerline{\epsfig{file=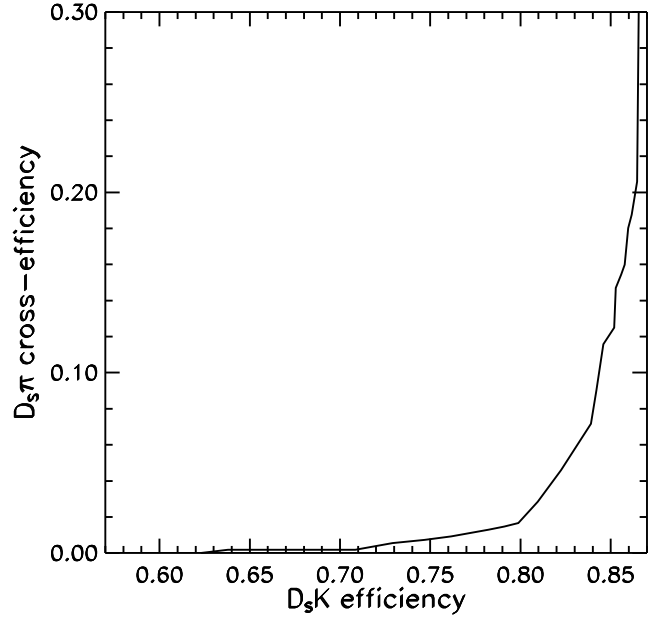, height=3.6in}}
\caption[Separation of $B_s\to D_s^+K^-$ and $B_s\to D_s^+\pi^-$ with the RICH]
{The efficiency to detect the fast $K^-$ in the
reaction $B_s\to D_s^+K^-$ versus the rate to misidentify the $\pi^-$ from
$B_s\to D_s^+\pi^-$ as a $K^-$.\label{5:fig:dskvsdspi}}
\end{figure}

As an example of the usefulness of this device,
Fig.~\ref{5:fig:dskvsdspi} shows the efficiency for
detecting the $K^-$ in the decay $B_s\to D_s^+K^-$ versus the rejection for
the $\pi^-$ in the decay $B_s\to D_s^+\pi^-$.  One sees that high 
efficiencies can be obtained with excellent rejections. 

\subsection{Electromagnetic Calorimeter}
\index{BTeV!detector!electromagnetic calorimeter}
\index{EM Calorimeter!BTeV}

In BTeV, photons and electrons are detected when they create an electromagnetic
shower in crystals of PbWO$_4$, a dense and transparent medium that 
produces scintillation light.
The amount of light is proportional to the incident
energy. The light is sensed by photomultiplier tubes (or possibly hybrid
photodiodes). The crystals are 22 cm long and have a small transverse 
cross-section, 
26~mm~$\times$~26~mm, providing excellent segmentation.
The energy and position resolutions are exquisite,
\begin{eqnarray}
\frac{\sigma_{E}}{E} &= & \sqrt{{{(1.6\%)^2}\over{E}}+(0.55\%)^2}~, \\
\sigma_x &=& \sqrt{{{(3500~\mu m)^2}\over{E}}+(200~\mu m)^2}~,
\end{eqnarray}
where $E$ is in units of GeV. This leads to an r.m.s. $\pi^o$ mass
resolution between 2 and 5 MeV/$c^2$ over the $\pi^o$ momentum range 1 to 40 GeV/c.

The crystals are designed to point at the center of the interaction region.
They start at a radial distance of 10 cm with respect to the beam-line and
extend out to 160 cm. They cover $\sim$210 mrad. This is smaller
than the 300 mrad acceptance of the tracking detector; 
the choice was made to reduce costs. For
most final states of interest, this leads to a loss of approximately 20\%
in signal. 

\begin{figure}[tbp]
\centerline{\epsfig{file=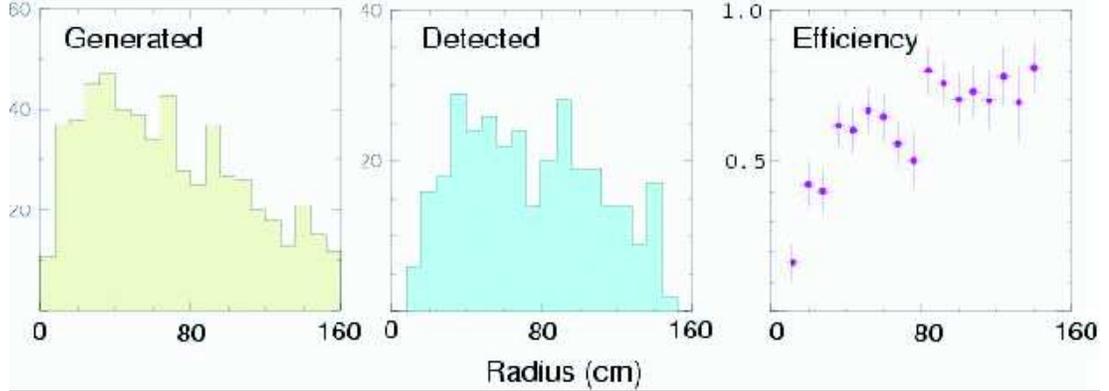, width=.95\textwidth}}
\caption[Radial Distributions of Generated and Detected Photons]
{The radial distribution of generated and detected
photons from $B^o\to K^*\gamma$ and the resulting efficiency. The detector
response was simulated by GEANT and clusters of hit crystals were
formed by the BTeV clustering software.  This software is derived
from software used for the Crystal Ball and CLEO experiments.
The charged tracks from the $K^*$ were required to 
hit the RICH. The simulation was run at 2 interactions/crossing.}
\label{5:fig:eff_btev_3sigma}
\end{figure}

At 2 interactions per crossing the calorimeter has a high rate close to the
beam pipe, where the reconstruction efficiency and resolution
is degraded by overlaps with
other tracks and photons. As one goes
out to larger radius, the acceptance becomes quite good. This can be seen
by examining the efficiency 
to reconstruct the $\gamma$ in the decay $B^o\to K^*\gamma$,
$K^*\to K^- \pi^+$.  For this study the decay products of 
the $K^*$ are required 
to reach the RICH detector. Fig.~\ref{5:fig:eff_btev_3sigma} shows the radial 
distributions of the generated
$\gamma$'s, the reconstructed $\gamma$'s and the $\gamma$ efficiency.
The shower reconstruction code, described in Chapter 
12 of the proposal, was developed from that used for the CLEO
CsI calorimeter; for reference, the efficiency of the CLEO barrel 
electromagnetic calorimeter is 89\%.

\subsection{Forward Tracking System}
\index{BTeV!detector!forward tracking system}
\index{tracking system!BTeV}

The other components of the charged-particle tracking system are straw-tube
wire proportional chambers and, near the beam where
occupancies are high, silicon microstrip detectors. These devices are used 
primarily for track momentum
measurement, $K_s$ detection and the Level 2 trigger. These detectors
measure the deflection of charged particles by the BTeV analyzing magnet and
give BTeV excellent mass and momentum resolution for charged particle
decay modes.

\subsection{Muon Detection}
\index{BTeV!detector!muon detection}
\index{muon detection!BTeV}
Muon detection is accomplished by insisting that the candidate charged track
penetrate several interaction lengths of magnetized iron and insuring that the
momentum determined from the bend in the toroid matches that 
given by the main spectrometer tracking system. The
muon system is also used to trigger on the dimuon decays of the $J/\psi$. This
is important not only to gather more signal but as a cross check on the
efficiency of the main trigger, the  Detached Vertex Trigger. 

\subsection{Data Acquisition System}
\index{BTeV!detector!data acquisition system}

BTeV has a data acquisition system (DAQ) which is capable of recording
a very large number of events. The full rate of $B$'s whose decay products
are in the detector is very high, over 1 kHz. The rate from
direct charm is 
similar.  Some other experiments are forced by the limitations of their 
data acquisition system to make very harsh decisions on which $B$ events to 
accept. BTeV can record nearly all the potentially interesting $B$ and charm 
candidates in its acceptance. Therefore it can address many topics that might
be discarded by an experiment whose DAQ is more restrictive.
Since nature has a way of surprising us, the openness of the
BTeV trigger and the capability of the DAQ are genuine strengths which
permit the opportunity to learn something new and unanticipated.

\section{Simulation and Analysis Tools}
\index{BTeV!software!simulation tools}
\index{BTeV!software!analysis tools}

The physics reach of BTeV has been established by an extensive and 
sophisticated program of simulations, which is described in detail
in Part~III of the BTeV proposal~\cite{btev_prop}.
For this study $p\bar{p}\to b\bar{b}X$ events were generated using
{\tt PYTHIA}\cite{pythia}
and the $b$ hadrons were decayed using {\tt QQ}. 
These packages are discussed in chapter~2 of this report.
To model the detector response to these events,
two detector simulation packages have been used, BTeVGeant and MCFast.
BTeVGeant is a GEANT\cite{geant} based simulation of the BTeV detector 
which contains a complete description of the BTeV geometry including
the materials needed for cooling, support and readout.  GEANT
models all physical interactions of particles with material and allows 
us to see the effects of hard to calculate backgrounds.
Most of the results presented in this report were obtained using BTeVGeant.
Some of the results presented here, and all earlier BTeV results,
were obtained using MCFast\cite{MCFast}, a fast parameterized simulation
environment which allows the user to quickly change the detector
design without the need to do any coding.
BTeVGeant writes an MCFast geometry file which describes the
same detector in simplified fashion; in this way the number, size
and resolution 
of detector elements is synchronized between the two simulation tools.
MCFast models the most of the processes that GEANT does, 
energy loss, multiple scattering, pair creation, bremstrahlung,
and hadronic interactions, but in
a simplified fashion; for example some of the detector components
are described by simpler shapes, the model of multiple scattering
is purely gaussian and the model of energy deposition in the
calorimeter is parametric.  

\begin{sloppypar}
Chapter~13 of the proposal shows quantitative comparisons between
MCFast and BTeVGeant.  From these studies one sees that
MCFast is a reliable tool for computing
resolutions, efficiencies and the level of backgrounds which  
arise from real tracks; the slower, more complete, BTeVGeant is 
necessary when occupancies and event confusion are the critical issues.
\end{sloppypar}

In most circumstances the simulations are done at the hit level,
not at the digitization level.  That is, the simulation
packages produce smeared measurements, not a stream of device addresses
and
digitized pulse heights thta simulate the raw, experimental data stream.
Digitization level simulations have been done to address the
issue of required bandwidth at various levels of the trigger and
DAQ systems.

For some of the physics studies presented in this report
very high statistics Monte Carlo runs were required to reliably 
estimate the background level.
These studies were performed on a farm of
500~MHz dual Pentium~III Linux machines; over a period of 3 months
an average of about 30 machines (~60~CPUs~) per day were available.
To give one example, over a period of about 1~month, 4.5 million
generic $b\bar{b}$ events were produced to investigate backgrounds
in the channel $B^0\to\rho\pi$.

Brief descriptions of simulation and analysis software for some
specific subsystems were given in section~\ref{5:sec:detector}
of this report.  References to the relevant TDR sections were
also given.  In order to make this report a little more stand
alone, a few more details are given below.

\subsection{Tracking and Track Fitting Software}
\index{BTeV!software!tracking}

In both packages the simulation code keeps track of which tracking
hits were created by which particles.  This information is not used 
by the trigger code, which does full pattern recognition, but it is
used
to check the results of the trigger package and to debug it.
Both simulation packages smear pixel hits using resolution functions
with non-gaussian tails that were measured in the test beam.  Hits in 
the other tracking detectors are modeled with gaussian resolution 
functions.  

  In the offline analysis package no pattern recognition is 
done; instead the Monte Carlo truth table is used to collect all of 
the hits which belong on a track.  Each hit list is then Kalman 
filtered to give the track parameters and their covariance matrix
in the neighborhood of the interaction region.  Extrapolating from the 
excellent performance of the pattern recognition in the trigger,
the final pattern recognition codes are expected to be
be highly efficient and to find few false tracks; therefore the
approximation of perfect pattern recognition gives a reliable
estimate of what BTeV will achieve.

\subsection{Electromagnetic Calorimeter Software}
\index{BTeV!software!calorimeter clustering}

BTeVGeant does a complete development of electromagnetic and 
hadronic showers in all materials and follows the products of these 
showers into neighbouring detector volumes.  When a track or photon
within a shower traverses one of the PbWO$_4$ crystals, 
BTeVGeant deposits energy from that track or photon in the crystal.
The total energy deposited in crystal is summed over each
beam crossing and a parametric function is then used to convert
the total deposited energy into a measured energy plus the error on
the measured energy.  For each crystal, a record is kept of how much 
energy was deposited for which track.  This information is only
used to characterize and debug the reconstruction code --- it is not
used by the reconstruction code.

When modelling shower development, it is necessary
to stop tracing new particles when their energy drops below some
cut-off.  If these cut-off values are set too high, then the showers
are too narrow, resulting in artificially clean events and 
artificially good energy resolution.  If, on the other hand,
these cut-off values are set too low, then simulations
are prohibitively slow.  The studies done to resolve of these 
tradeoffs are discussed in detail in Section~12.1.5 of the BTeV 
proposal.  The final result is that, because adequate CPU power was
available, no significant compromises in the quality of the
simulation were necessary.

MCFast does the same calorimeter bookkeeping as BTeVGeant but 
the model of shower development is parametric, rather than a detailed
following of each generation of particles.  Both packages create identical
data structures so that the same shower reconstruction
and user analysis codes will work on 
events from both simulation packages.

\subsection{Trigger Simulation Software}
\index{BTeV!software!trigger simulation}

Chapter~9 of the proposal describes the overall plan for the trigger
and it describes in detail the algorithm for the Level~1 Detached Vertex
Trigger.   This algorithm has been coded in C and is callable from
by user code within either BTeVGeant or MCFast.  A few pieces of
the code have been ported to the target DSP's and carefully timed.
The rest is written in a high level language for ease of algorithm
development.  

Similarly a prototype for the Level~2 trigger has also been coded.
Since Level~2 will run on standard processes, the code represents
a true prototype, and is not just a clone of the algorithm.

The physics analyses presented in the BTeV Proposal~\cite{btev_prop}
and in this report have
been used as models for possible Level~3 algorithms.

The trigger results presented in Table~\ref{5:tab:trigeff}, and the
results presented in chapters~6 through~8 were obtained using this
trigger simulation software.

\subsection{RICH Software}
\index{BTeV!software!RICH simulation}

For most of the results presented in this workshop, nominal
RICH efficiencies and misidentification probabilities 
were used.  It was assumed that if a reconstructed track contained
a hit in the chambers between the RICH and the EMCal, then that
track could be identified by the RICH.  

In a few selected analyses, notably, $B_s\to D_s K$, a more detailed
simulation was used.  For this analysis tracks passing through the
RICH generated photons which were propagated through the aerogel and
chamber gas and reflected from the mirror.  They were then propagated
to the detector plan where a model of the HPD efficiency was applied.
A pattern recognition algorithm was then run to find Cherenkov
rings from the list of HPD hits.  Particle ID decisions
were made using these reconstructed rings.  The physics reach
predicted by these simulations is presented in chapter~6.



\section{Flavor Tagging}
\index{BTeV!flavor tagging}
\index{flavor tagging!BTeV}

Section~\ref{2:sec:exp_common_tagging}
of this report describes the flavor tagging strategies which
are available to $B$ physics experiments at the Tevatron.
In Chapter~15 of the BTeV proposal presents a preliminary study of the
tagging power which can be achieved with the BTeV detector.
This section will summarize the results of that study.
Mixing of the opposite side $B$ meson has not yet been included in the 
results shown here.  The upcoming sections describe the algorithms
used for tagging, and tagging powers which they achieve are summarized in 
Table~\ref{5:tab:tag_effic}.

\subsection{Away Side Tagging}

Three different away side tagging methods have been studied,
lepton tagging, kaon tagging and vertex charge tagging.
The first step in all three methods is to select
tracks which are detached from all primary vertices
in the event.  In events with multiple primary vertices, detached
tracks are only considered if they are associated with the
same primary vertex as is the signal candidate.

\subsubsection{Lepton Tagging}
\index{BTeV!flavor tagging!away side lepton}

 The lepton tagging algorithm must
deal with possible wrong-sign tags which result 
from the cascade $b\to c\to\ell^+$.  Because
leptons from $b\to\ell^-$ and $b\to c\to\ell^+$ 
have quite different transverse momentum ($p_T$) distributions,
good separation can be achieved.
If there was more than one lepton tag candidate in an event,
the highest $p_T$ lepton was chosen to be the tag.

Candidates for muon tags were selected from the detached track list
if they had a momentum greater than 4.0~GeV/c.
A tagging muon with $p_T>1.0$\ GeV/c was 
considered to be from the process $b\to \ell^-$,
while one with $p_T<0.5$\ GeV/c was considered to be from 
the process $b\to c\to\ell^+$, thereby flipping the sign of the tag.

 Candidates for electron tags were selected using a parametrized
electron efficiency and hadron misidentification
probability.  The tagging lepton was required to have
$p_T>1.0$\ GeV/c and was always assumed to come from the 
process $b\to \ell^-$.  There were not enough
MC events to study electrons from $b\to c\to\ell^+$.
  
\subsubsection{Kaon Tagging}
\index{BTeV!flavor tagging!away side kaon}

Because of the large branching ratio for $b\to c\to K^- X$,
kaon tagging is the most potent tagging method at $e^+e^-$ $B$~factories.
At BTeV, in which the multiplicity of the underlying event is much
greater, excellence in both particle identification and 
vertex resolution is required to exploit kaon tagging.
Both are strong points of the forward detector geometry.  

Candidates for kaon tags were selected from the secondary track list if
they were identified as kaons in the RICH detector.  If there was more 
than one kaon tag candidate in an event, the kaon with the
largest normalized impact parameter with respect to the
primary vertex was selected.

\subsubsection{Vertex Charge Tagging}
\index{BTeV!flavor tagging! away side vertex charge}
 In this method a search was made for a detached vertex which is
consistent with being from the charged decay products of
the other $b$.  The charge of that vertex determines the charge of
the $b$.  When the opposite side $b$ hadronizes into a $\bar{B}^0$
or a $\bar{B}_s^0$, the
tagging vertex has a neutral charge and there is
no useful vertex tagging information in the event.  However this method
has the advantage that it is not affected by mixing of
the away side $b$.

\begin{sloppypar}
Tracks from the secondary list were accepted provided
they had \hbox{$p_T>100$~MeV/c} and provided they
had  $\Delta\eta<4$ with respect to the direction
of the signal $B^0$ candidate.  The tracks from the secondary
list were sorted into candidate vertices and only vertices
with a detachment of at least $1.0\, \sigma$ from the primary vertex 
were accepted.
If more than one vertex was found in an event, 
the one with the highest transverse momentum was selected;
if no secondary vertices passed the selection cuts and if
there was at least one track with $p_T>1.0$\ GeV/c, 
then the highest $p_T$ track was selected.  If the charge of the selected
vertex is non-zero, then it determines the flavor of the 
away side $b$.
\end{sloppypar}

  This tagging method is similar to jet charge tagging used by 
other experiments but BTeV has not yet investigated the possibility
of weighting the tracks by their momenta.
   
\subsubsection{Combining Away Side Tagging Methods}
\index{BTeV!flavor tagging! away side combined}
In many events, several of the same side tagging
methods may give results; moreover it can happen that two methods
will give contradictory answers.
BTeV has not yet optimized the method of combining all tagging
information but have used the following simple algorithm.  The
methods were polled in decreasing order of dilution and the first
method to give an answer was accepted.  That is, if  lepton
tagging gave a result, the result was accepted; if not, and if
kaon tagging gave a result, the kaon tag was accepted; if not,
and if the vertex charge tagging gave a result, the vertex
charge tag was accepted.

\subsection{Same Side Tagging}
\index{BTeV!flavor tagging!same side}
BTeV has studied the power of
same side kaon tagging for $B_s$ mesons.
For this study tracks were selected provided
they had a momentum greater than 3.0~GeV/c,
were identified as kaons in the RICH and
had an impact parameter with respect to the primary vertex
less than 2~$\sigma$.  It was further required that 
the system comprising the $B_s$ candidate plus the candidate
tagging track have an
invariant mass less than 7.0~GeV/c$^2$.
If more than one track passed these cuts, then the track
closest in $\phi$ to the $B_s$ direction was selected.

For same side tagging of $B_d$ events BTeV expects to use $B^{**}$ 
decays.  This study is at an early, conceptual stage.
The flag in Pythia to turn on $B^{**}$ production has not been used.
Instead BTeV has used a sample of simulated $B\to \psi K_s$ decays
and has selected events in which the $B$ and the next pion
in the generator track list have an invariant mass in the
range 5.6 - 5.8~GeV/c$^2$.   It was assumed that 30\% of these
events will come from $B^{**}$ decays and therefore be
right sign tags.  The remainder 70\% of the events were assumed to have
an equal number of right sign and wrong sign tags.
 
\subsection{Summary of Tagging}
\label{5:ssec:tagging_summary}
\index{BTeV!flavor tagging! same and away combined}

The results from the tagging study are
summarized in Table~\ref{5:tab:tag_effic}.
These results are preliminary and one should be aware that
all algorithms  have yet to exploit the full power
available to them.  In particular, the vertexing information
has yet to be fully exploited.  For example, the
$b\to\ell^-$ and $b\to c\to\ell^+$ samples differ not only in 
their $p_T$ spectra; they have distinctly different topological 
properties.  Similarly, kaon tagging can be improved if there
is evidence that the kaon comes from a tertiary vertex, indicative
of the $b\to c\to K^-$ cascade.  Finally, the vertex charge
algorithm should expect to find two vertices on the away side,
the $b$ decay vertex and the $c$ decay vertex; the charge of
both vertices provides tagging power.
Other tagging methods have yet to be studied such as using 
a $D^*$ from the decay of the opposite side $B$.
Finally, we have not yet explored the optimal use
of the correlations among all of the methods.
Therefore, the results quoted in 
Table~\ref{5:tab:tag_effic} probably underestimate the 
tagging power of BTeV, even though they do not yet incorporate
mixing on the away side. Therefore the BTeV results quoted in
this report are presented using nominal 
values of $\epsilon=0.7$ and $D=0.37$, giving $\epsilon D^2=0.1$.
The studies presented in the present chapter should be regarded as 
evidence that these nominal values lie well within the ultimate 
reach of the experiment.

\begin{table}[tb]
\begin{center}
\begin{tabular}[t]{|l|c c c|}
\hline
 Tag Type         & $\epsilon$  & $D$      &  $\epsilon{D^2}$ \\  \hline
 Muon             &  4.5\%      & 0.66        & 2.0\%              \\
 Electron         &  2.3\%      & 0.68        & 1.0\%              \\
 Kaon             &  18\%       & 0.52        & 4.9\%            \\
 Vertex Charge    &  32\%       & 0.36        & 4.1\%            \\
 Same Side Kaon   &  40\%       & 0.26        & 2.6\%            \\ 
 Same Side Pion   &  88\%       & 0.16        & 2.2\%            \\ \hline
 Total for $B_s$  &             &             & 14.6 \%    \\
 Total for $B_d$  &             &             & 14.2 \%    \\
 Total for $B_s$ with overlaps  
                  & 65\%        & 0.37        & 8.9\% \\ \hline\hline
\end{tabular}
\end{center}
\caption[Summary of Tagging Power in BTeV]
{Results of first generation studies of tagging power in BTeV.
         In the text it is discussed that these studies are incomplete
         and that they likely underestimate the tagging power which
         can be realized.
          }
\label{5:tab:tag_effic}
\end{table}

\section{Schedule}
\index{BTeV!schedule}

The BTeV program is an ambitious one.  Its goal is to begin data taking in 
2005/6. This timing is well-matched to the world $B$ physics program. 
The $e^{+}e^{-}$  $B$ factories and the Fermilab collider experiments will 
have had several years of running, the first results
in, and their significance thoroughly digested. It should be clear what the 
next set of goals is and BTeV will be guaranteed to be well-positioned to 
attack them.  This schedule also  gives BTeV a good opportunity
to have a head start in its inevitable competition with LHC-b, especially 
since BTeV can be installing and operating components of its detector in the 
collision hall well in advance of 2005. Finally, this schedule is sensibly 
related to BTeV's plan to conduct a rigorous R\&D program which includes a 
sequence of engineering runs to test the technically  challenging
systems in the BTeV design.  We believe that the scale of the BTeV construction
effort is comparable to the scale of one of the current detector upgrades.
The time scale is comparable as well.

\section{Conclusions}

BTeV is a powerful and precise scientific instrument capable of exquisite
tests of the Standard Model. It has great potential to discover new physics 
via rare or CP violating decays of heavy quarks.  Details of
the physics reach of BTeV and can be found in chapters~6 through~9
of this report and a summary table can be found in the
executive summary~\cite{execsum} of the BTeV Proposal.

\clearpage{\pagestyle{empty}\cleardoublepage}
\newcommand{\ra}{\rightarrow}
\newcommand{\stb} {\ifmmode \sin 2\beta \else 
                           $\sin 2\beta$\fi}
\newcommand{\jpks} {\ifmmode J/\psi K^0_S \else 
                            $J/\psi K^0_S$\fi}
\newcommand{\bjpks} {\ifmmode B^0 \ra J/\psi K^0_S \else 
                            $B^0 \ra J/\psi K^0_S$\fi}
\newcommand{\bpipi} {\ifmmode B^0 \ra \pi^+\pi^- \else 
                            $B^0 \ra \pi^+\pi^-$\fi}
\newcommand{\bkk} {\ifmmode B^0 \ra K^+ K^- \else 
                            $B^0 \ra K^+ K^-$\fi}
\newcommand{\tfb} {2~fb$^{-1}$}
\def\gevc {\ifmmode {\rm GeV}/c \else GeV$/c$\fi}
\newcommand{\eD} {\ifmmode \varepsilon{\cal D}^2 \else 
                          $\varepsilon{\cal D}^2$\fi}
\newcommand{\bskk} {\ifmmode B_s^0 \ra K^+ K^-
                        \else $B_s^0 \ra K^+ K^-$\fi}
\newcommand{\bsdsk} {\ifmmode B_s^0 \ra D_s K
                        \else $B_s^0 \ra D_s K$\fi}
\newcommand{\Pt} {\ifmmode p_T \else $p_T$\fi}
\def\bstopik {B_s^0\rightarrow \pi^+ K^-}
\def\bstokk  {B_s^0\rightarrow K^+ K^-}

\newcommand{\Bs} {\ifmmode B_{s}^{0}
                       \else $B_{s}^{0}$\fi}
\def\btopipi {B^0\rightarrow \pi^+ \pi^-}
\def\bstopik {B_s^0\rightarrow \pi^+ K^-}
\def\btokpi  {B^0\rightarrow K^+ \pi^-}
\def\mevcc {\ifmmode {\rm MeV}/c^2 \else MeV$/c^2$\fi}
\def\gevcc {\ifmmode {\rm GeV}/c^2 \else GeV$/c^2$\fi}
\def\pipi    {\pi^+ \pi^-}
\def\pik     {\pi^+ K^-}
\def\kpi     {K^+ \pi^-}
\def\kk      {K^+ K^-}
\newcommand{\roarrow}[1]{\stackrel{\rightharpoonup}{#1}}
\def\psietp {B_s^0 \rightarrow J\!/\psi\,\eta^\prime}
\def\psiea  {B_s^0 \rightarrow J\!/\psi\,\eta}
\def\mumu   {\mu^+\,\mu^-}


\chapter{\protect\boldmath $CP$ Violation}

\authors{D.~Atwood,
S.~Bailey, 
W.~Bell, 
J.~Butler,
A.~Cerri,
S.~Donati,
A.~Falk,
S.~Gardner,
Y.~Grossman,
R.~Jesik,
P.A.~Kasper,
R.~Kutschke,
G.~Majumder,
P.~Maksimovi\'c,
Y.~Nir,
L.~Nogach,
M.~Paulini,
M.~Petteni,
M.~Procario,
G.~Punzi,
H.~Quinn,
K.~Shestermanov,
G.~Signorelli,
J.~Silva,
T.~Skwarnicki,
S.~Stone,
A.~Vasiliev,
B.~Wicklund,
F.~W\"urthwein,
J.~Yarba,
K.~Yip}

\section[Introduction]
{Introduction
$\!$\authorfootnote{Authors:  A.~Falk, Y.~Grossman, Y.~Nir, H.~Quinn.}
}

\label{ch6:intro}

$CP$~violation is still one of the least tested aspects of the
Standard Model. Many extensions of the Standard Model predict
that there are new sources of $CP$~violation, beyond the single
Kobayashi-Maskawa phase in the CKM mixing matrix for quarks.
Considerations related to the observed baryon asymmetry of the
Universe imply that such new sources must exist. The experimental
study of $CP$~violation is then highly motivated.

For 37 years, $CP$~violation has only been observed in the neutral $K$-meson
system. Very recently, the first observations of 
$CP$~violation in the $B$-meson system 
have been reported by the $e^+e^-$~$B$~factories~\cite{cpbbstbaug}
providing the
first tests of the Standard Model picture of $CP$~violation.
In the near future, more experimental tests will be performed including the
Tevatron experiments. 
The greater the variety of $CP$~violating observables measured,
the more stringently will the Standard Model be tested.
If deviations
from the Standard Model predictions are observed, the information from
different meson decays will provide crucial clues for the type
of new physics that can account for such deviations. 

This situation makes the search for $CP$~violation in the $B_s^0$ decays
highly interesting. $B_s^0$ mesons cannot be studied at the
$B$-factories operating at the $\Upsilon(4S)$ resonance. Hadron
colliders, on the other hand, with their high statistics, provide
an opportunity to measure $CP$~violation in the $B_s^0$ system
with high accuracy 
in addition to allowing studies of certain $B^0$ modes.

In the context of the
\index{CP violation@$CP$ violation!Standard Model}%
Standard Model, the main goal is to measure
the phases of CKM elements accurately. These are conveniently described
as angles of unitarity triangles. In particular, all relevant phases
can be expressed in terms of two large angles,
\beq\label{6:defgam}
\beta\equiv\arg\left(-{V_{cd}V_{cb}^*\over V_{td}V_{tb}^*}\right), \qquad 
\gamma\equiv\arg\left(-{V_{ud}V_{ub}^*\over V_{cd}V_{cb}^*}\right),
\eeq
and two small angles,
\beq\label{6:betas}
\beta_s\equiv\arg\left(-{V_{ts}V_{tb}^*\over V_{cs}V_{cb}^*}\right)
={\cal O}(\lambda^2), \qquad
\beta_K\equiv\arg\left(-{V_{cs}V_{cd}^*\over
V_{us}V_{ud}^*}\right)={\cal O}(\lambda^4),
\eeq
where $\lambda=0.22$ is the Wolfenstein parameter. $CP$~violation in $B_s^0$
decays allows, in particular, a determination of $\gamma$ and $\beta_s$. 

Much of the following discussion is based on 
Refs.~\cite{BLS,HaQu,nirssi}\ where more details can be found.

\boldmath
\subsection{$B_s^0$--$\bar B_s^0$ Mixing}
\unboldmath
\index{Bs mixing@$B_s^0$ mixing}%

Here we introduce only what is needed to define notations that are
important for $CP$~violation. $B_s^0$ mixing and measurements to determine
it are discussed in Chapter~8. A $B_s^0$ meson
is made from a $b$-type antiquark and an $s$-type quark, while
the $\bar B_s^0$ meson is made of a $b$-type quark and an $s$-type
antiquark.  The heavy, $B_s^H$, and light, $B_s^L$, 
\index{mass eigenstates} 
mass eigenstates 
can be written as linear combinations of $B_s^0$ and $\bar B_s^0$:
\begin{eqnarray}
\label{6:Bsmaei}
|B_s^L\rangle\ &=&\ p|B_s^0\rangle+q|\bar B_s^0\rangle,\nonumber \\
|B_s^H\rangle\ &=&\ p|B_s^0\rangle-q|\bar B_s^0\rangle,
\end{eqnarray}
with 
\beq\label{6:qpnorm}
|q|^2+|p|^2=1.
\eeq
In writing (\ref{6:Bsmaei}), we assume $CPT$ conservation and use of part of
the freedom to re-phase the meson states:
\beqa
\label{6:chopha}
|B_s\rangle\ &\ra&\ e^{i\zeta}|B_s\rangle,\nonumber \\
|\bar B_s\rangle\ &\ra&\ e^{i\bar\zeta}|\bar B_s\rangle.
\end{eqnarray}
The \index{mass difference \dm!$\dm_s$}%
mass difference $\Delta m_s$ and 
\index{width difference!$B_s$}%
width difference 
$\Delta\Gamma_s$ are defined as follows:
\index{mixing parameter!$x$}%
\index{mixing parameter!$y$}%
\beq\label{6:delmg}
\Delta m_s\equiv M_H-M_L, \qquad 
\Delta\Gamma_s\,\equiv \Gamma_L-\Gamma_H,
\eeq
so that $\Delta m_s>0$ by definition and the Standard Model prediction
is that $\Delta\Gamma_s\,>0$. The average mass and width are given by
\beq\label{6:avemg}
M_{\Bs}\equiv{M_H+M_L\over2}, \qquad 
\Gamma_s\,\equiv {\Gamma_H+\Gamma_L\over2}.
\eeq
It is useful to define dimensionless ratios $x_s$ and $y_s$:
\beq\label{6:defxy}
x_s\equiv{\Delta m_s\over\Gamma_s\,}, \qquad 
y_s\equiv{\Delta\Gamma_s\,\over{  2\Gamma_s} \,}.
\eeq

The time evolution of the mass eigenstates is simple, following from
the fact that they have well defined masses and decay widths:
\beqa\label{6:maeite}
|B_s^H(t)\rangle\ &=&\ e^{-iM_Ht}\,e^{-\Gamma_Ht/2}|B_s^H\rangle,\nonumber\\
|B_s^L(t)\rangle\ &=&\ e^{-iM_Lt}\,e^{-\Gamma_Lt/2}|B_s^L\rangle.
\eeqa
The time evolution of the strong interaction eigenstates is complicated
and obeys a Schr\"o\-dinger-like equation:
\beq\label{6:teinei}
i{d\over dt}\pmatrix{B_s\cr \bar B_s\cr}=\left(M-{i\over2}
\Gamma\right)\pmatrix{B_s\cr \bar B_s\cr},
\eeq
where $M$ and $\Gamma$ are Hermitian $2\times2$ matrices.
The off-diagonal elements in these matrices are not
invariant under the re-phasing (\ref{6:chopha}),
\beq\label{6:chphmg}
M_{12}\ra e^{i(\bar\zeta-\zeta)}M_{12}, \qquad 
\Gamma_{12}\ra e^{i(\bar\zeta-\zeta)}\Gamma_{12}.
\eeq
Therefore, physical parameters can only depend on $|M_{12}|$,
$|\Gamma_{12}|$ and arg($M_{12}\Gamma_{12}^*$). Indeed,
the relations between the parameters in the mass eigenbasis
and in the interaction eigenbasis can be written as follows:
\beqa\label{6:masint}
(\Delta m_s)^2-{1\over4}\, (\Delta\Gamma_s\,)^2\ &=&\ 
4|M_{12}|^2-|\Gamma_{12}|^2, \nonumber\\
\Delta m_s\Delta\Gamma_s\ &=&\ -4\Re(M_{12}\Gamma_{12}^*),
\eeqa
and
\beq\label{6:pqint}
{q\over p}=-{\Delta m_s+{i\over2}\Delta\Gamma_s\,\over
2M_{12}-i\Gamma_{12}}.
\eeq

\boldmath
\subsection{$B_s^0$ Decays}
\unboldmath

We define the \index{decay amplitude}
decay amplitudes for $B_s^0$ and $\bar B_s^0$ into a final 
state $f$:
\beq\label{6:defAf}
A_f\ \equiv\ \langle f|B_s^0\rangle, \qquad 
\bar A_f\ \equiv\ \langle f|\bar B^0_s\rangle.
\eeq
In addition to their dependence on the re-phasing (\ref{6:chopha}), these
amplitudes are affected by re-phasing of $|f\rangle$,
\beq\label{6:rephf}
|f\rangle\ra e^{i\zeta_f}|f\rangle.
\eeq
Under (\ref{6:chopha}) and (\ref{6:rephf}), we have
\beq\label{6:AAqprp}
A_f\ \ra\ e^{i(\zeta-\zeta_f)}A_f, \qquad 
\bar A_f\ \ra\ e^{i(\bar\zeta-\zeta_f)}\bar A_f, \qquad 
q/p\ \ra\ e^{i(\zeta-\bar\zeta)}q/p.
\eeq
We learn that of the three complex parameters, $A_f$, $\bar A_f$
and $q/p$, one can construct three real quantities,
\beq\label{6:phyabs}
|A_f|, \quad |\bar A_f|, \quad |q/p|,
\eeq
and one phase, that is the phase of
\beq\label{6:deflam}
\lambda_f\equiv{q\over p}{\bar A_f\over A_f},
\eeq
that are phase-convention independent and, consequently, could
be observable. Note that $|\lambda_f|=|q/p|\times|\bar A_f/A_f|$
is not independent of the parameters of (\ref{6:phyabs}), but $\arg(\lambda_f)$ is.

\boldmath
\subsection{$CP$~Violation}
\label{sec:cp_type}
\unboldmath

The $CP$~transformation interchanges $B_s^0$ and 
$\bar B_s^0$:\footnote{Unless specified otherwise we use the phase
  convention $\xi=\pi$ throughout this report, see Sect.~\ref{subs:disc}.}
\beq\label{6:CPonBs}
CP|B_s^0\rangle\ =\ e^{i\xi}|\bar B_s^0\rangle, \qquad 
CP|\bar B_s^0\rangle\ =\ e^{-i\xi}|B_s^0\rangle.
\eeq
The phase $\xi$ is not invariant under the re-phasing (\ref{6:chopha}),
\beq\label{chphxi}
\xi\ra\xi+\zeta-\bar\zeta.
\eeq
We also define $\bar f$ to be the $CP$~conjugate state of $f$:
\beq\label{6:CPonf}
CP|f\rangle\ =\ e^{i\xi_f}|\bar f\rangle, \qquad 
CP|\bar f\rangle\ =\ e^{-i\xi_f}|f\rangle.
\eeq

$CP$~is a good symmetry if there exist some phases $\xi$ and $\xi_f$ 
such that the Lagrangian is left invariant under (\ref{6:CPonBs}) and 
(\ref{6:CPonf}). For $CP$~to be a good symmetry of the mixing process, 
it is required then that 
\beq\label{6:goCPmiin}
M_{12}^*=e^{2i\xi}M_{12}, \qquad \Gamma_{12}^*=e^{2i\xi}\Gamma_{12}.
\eeq
In terms of the observable parameters in Eq.~(\ref{6:phyabs}), 
this gives the condition 
\beq\label{6:goCPmima}
|q/p|=1.
\eeq
For $CP$~to be a good symmetry of the decay processes, 
it is required that 
\beq\label{6:goCPdein}
\bar A_{\bar f}=e^{i(\xi_f-\xi)}A_f, \qquad 
A_{\bar f}=e^{i(\xi_f+\xi)}\bar A_f.
\eeq
In terms of the observable parameters in Eq.~(\ref{6:phyabs}),
this results in the condition
\beq\label{6:goCPdema}
|\bar A_{\bar f}/A_f|=|\bar A_f/A_{\bar f}|=1.
\eeq
Finally, for $CP$~to be a good symmetry of processes that involve both
mixing and decay, it is required that
\beq\label{6:CPlam}
\lambda_f\lambda_{\bar f}=1.
\eeq
For final $CP$~eigenstates $f_{CP}$, such that $|\bar f_{CP}\rangle=\pm 
|f_{CP}\rangle$, the condition (\ref{6:CPlam}) translates into 
$|\lambda_{f_{CP}}|=1$, which just combines (\ref{6:goCPmima}) and 
(\ref{6:goCPdema}), and
\beq\label{6:CPcpes}
\Im\lambda_{f_{CP}}=0.
\eeq

Violation of each of the three conditions for $CP$~symmetry, 
(\ref{6:goCPmima}), (\ref{6:goCPdema}) and (\ref{6:CPcpes}), 
corresponds to a different type of $CP$~violation:
\par {1.} \index{CP violation@$CP$ violation!in mixing}%
$CP$~violation in mixing, which occurs when the $B_s^H$
and $B_s^L$ mass eigenstates cannot be chosen to be $CP$~eigenstates:
\beq\label{6:CPmix}
|q/p|\neq1.
\eeq
\par {2.} \index{CP violation@$CP$ violation!in decay}%
$CP$~violation in decay, when the $B_s^0\ra f$ and $\bar B_s^0\ra\bar f$
decay amplitudes have different magnitudes:
\beq\label{6:CPdec}
|\bar A_{\bar f}/A_f|\neq1.
\eeq
\par {3.} \index{CP violation@$CP$ violation!interference type}%
$CP$~violation in interference between decays with and without
mixing, which occurs in decays into final states that are common to
$B_s^0$ and $\bar B_s^0$:
\beq\label{6:CPint}
\Im(\lambda_f\lambda_{\bar f})\neq0.
\eeq
In particular, for final $CP$~eigenstates,
\beq\label{6:CPfCP}
\Im\lambda_{f_{CP}}\neq0.
\eeq

The effects of $CP$~violation in mixing in the $B_s^0$ system are small.
The lower bound on $\Delta m_s$~\cite{leposc} as of August 2001,
\beq\label{6:expdelm}
\Delta m_s\geq14.6\ {\rm ps}^{-1},
\eeq
and the measured $B_s^0$ lifetime \cite{blay},
\beq\label{6:expG}
\tau(B_s^0)=(1.46\pm0.06)\ {\rm ps},
\eeq
imply that $|\Gamma_{12}/M_{12}|\leq{\cal O}(0.05)$. For
$|\Gamma_{12}/M_{12}|\ll1$, we have (see (\ref{6:pqint}))
\beq\label{6:qpmG}
\left|{q\over p}\right|-1=-{1\over2}\, \Im\left(
{\Gamma_{12}\over M_{12}}\right).
\eeq
Therefore, experimental data give $|q/p|-1\leq{\cal O}(0.1)$.
Moreover, $\Gamma_{12}$ comes from long distance contributions,
where effects of new physics are expected to be negligible.
Consequently, the Standard Model calculation of 
$\Gamma_{12}$~\cite{BBGLN},
which yields values of
$\Delta\Gamma_s\,/\Gamma_s$ between 
${\cal O}(0.15)$~\cite{HIOY} and
${\cal O}(0.05)$~\cite{beci}, is expected to hold model 
independently. Within the Standard Model, $\Im(\Gamma_{12}/M_{12})$ is further
suppressed by the smallness of $\beta_s$, the relative phase between 
$\Gamma_{12}$ and $M_{12}$ defined in Eq.~(\ref{6:betas}). We conclude that the
deviation of $|q/p|$ from unity is very small:
\beq\label{6:qpmodind}
\Im(\Gamma_{12}/M_{12})\cases{\leq{\cal O}(10^{-2})& 
model independent,\cr ={\cal O}(10^{-4})&standard model.\cr}
\eeq
We can therefore safely 
neglect $CP$~violation in mixing, and we do so from here on.

\subsection{Tagged Decays}
\index{CP violation@$CP$ violation!tagged decay}%

We consider the time evolution of a state $|B_s(t)\rangle$
($|\bar B_s(t)\rangle$) which was tagged as $|B_s\rangle$
($|\bar B_s\rangle$) at time $t=0$. The time evolution can be read
from Eqs. (\ref{6:Bsmaei}) and (\ref{6:maeite}):
\beqa\label{6:tagte}
|B_s(t)\rangle\ &=&\ g_+(t)|B_s\rangle+(q/p)g_-(t)|\bar B_s\rangle,\nonumber\\
|\bar B_s(t)\rangle\ &=&\ (p/q)g_-(t)|B_s\rangle+g_+(t)|\bar B_s\rangle,
\eeqa
where
\beq\label{6:defgpm}
g_\pm(t)\equiv{1\over2}\left(
e^{-iM_Lt}e^{-\Gamma_Lt/2}\pm e^{-iM_Ht}e^{-\Gamma_Ht/2}\right).
\eeq
Then, we get the following decay rates:
\beqa\label{6:tagdr}
\Gamma[B_s(t)\ra f]\ &=&\ |A_f|^2\left\{|g_+(t)|^2+|\lambda_f|^2|g_-(t)|^2
+2\Re [\lambda_f g_+^*(t)g_-(t)]\right\},\nonumber\\
\Gamma[B_s(t)\ra\bar f]\ &=&\ |\bar A_{\bar f}|^2\left|{q\over p}\right|^2
\left\{|g_-(t)|^2+|\lambda^{-1}_{\bar f}|^2|g_+(t)|^2
+2\Re \Big[\lambda^{-1}_{\bar f} g_+(t)g_-^*(t)\Big]\right\},\nonumber\\
\Gamma[\bar B_s(t)\ra f]\ &=&\ |A_f|^2\left|{p\over q}\right|^2
\left\{|g_-(t)|^2+|\lambda_f|^2|g_+(t)|^2
+2\Re [\lambda_f g_+(t)g_-^*(t)]\right\},\nonumber\\
\Gamma[\bar B_s(t)\ra\bar f]\ &=&\ |\bar A_{\bar f}|^2
\left\{|g_+(t)|^2+|\lambda^{-1}_{\bar f}|^2|g_-(t)|^2
+2\Re\Big[\lambda^{-1}_{\bar f} g_+^*(t)g_-(t)\Big]\right\}.
\eeqa
Assuming $|q/p|=1$, we find
\beqa\label{6:diovsu}
{\cal A}_f(t)\ &=&\ {\Gamma[\bar B_s(t)\ra f]-\Gamma[B_s(t)\ra f]\over
\Gamma[\bar B_s(t)\ra f]+\Gamma[B_s(t)\ra f]}\nonumber\\ 
&=&\ -{\left(1-|\lambda_f|^2\right)\cos(\Delta m_s\,t)
-2\,\Im\lambda_f\sin(\Delta m_s\,t)\over
\left(1+|\lambda_f|^2\right)\cosh(\Delta\Gamma_s\, t/2)
-2\,\Re\lambda_f\sinh(\Delta\Gamma_s\, t/2)}.
\eeqa

Consider cases where the decay amplitudes are each dominated
by a single weak phase. Then 
\beq\label{6:singleamp}
|A_f|=|\bar A_{\bar f}|, \qquad |A_{\bar f}|=|\bar A_f|,
\eeq
and
\beq\label{6:singleCKM}
\lambda_f=|\lambda_f|e^{i(\phi_f+\delta_f)}, \qquad 
\lambda^{-1}_{\bar f}=|\lambda_f|e^{i(-\phi_f+\delta_f)},
\eeq
where $\phi_f$ ($\delta_f$) is the relevant weak (strong) phase.
Eqs. (\ref{6:tagdr}) can be rewritten for this case as follows:
\beqa\label{6:tagsing}
\Gamma[B_s(t)\ra f]\ &=&\ {|A_f|^2e^{-\Gamma_s\,t}\over2}\left\{
(1+|\lambda_f|^2)\cosh(\Delta\Gamma_s\,t/2)+
(1-|\lambda_f|^2)\cos(\Delta m_s\,t)\right.\nonumber\\
& &\quad \left. -2|\lambda_f|\cos(\delta_f+\phi_f)\sinh(\Delta\Gamma_s\,t/2)
-2|\lambda_f|\sin(\delta_f+\phi_f)\sin(\Delta m_s\,t) \right\},\nonumber\\
\Gamma[B_s(t)\ra\bar f]\ &=&\ {|A_f|^2e^{-\Gamma_s\,t}\over2}\left\{
(1+|\lambda_f|^2)\cosh(\Delta\Gamma_s\,t/2)-
(1-|\lambda_f|^2)\cos(\Delta m_s\,t)\right.\nonumber\\
& &\quad \left. -2|\lambda_f|\cos(\delta_f-\phi_f)\sinh(\Delta\Gamma_s\,t/2)
+2|\lambda_f|\sin(\delta_f-\phi_f)\sin(\Delta m_s\,t) \right\},\nonumber\\
\Gamma[\bar B_s(t)\ra f]\ &=&\ {|A_f|^2e^{-\Gamma_s\,t}\over2}\left\{
(1+|\lambda_f|^2)\cosh(\Delta\Gamma_s\,t/2)-
(1-|\lambda_f|^2)\cos(\Delta m_s\,t)\right.\nonumber\\
& &\quad \left. -2|\lambda_f|\cos(\delta_f+\phi_f)\sinh(\Delta\Gamma_s\,t/2)
+2|\lambda_f|\sin(\delta_f+\phi_f)\sin(\Delta m_s\,t) \right\},\nonumber\\
\Gamma[\bar B_s(t)\ra\bar f]\ &=&\ {|A_f|^2e^{-\Gamma_s\,t}\over2}\left\{
(1+|\lambda_f|^2)\cosh(\Delta\Gamma_s\,t/2)+
(1-|\lambda_f|^2)\cos(\Delta m_s\,t)\right. \nonumber\\
& &\quad \left. -2|\lambda_f|\cos(\delta_f-\phi_f)\sinh(\Delta\Gamma_s\,t/2)
-2|\lambda_f|\sin(\delta_f-\phi_f)\sin(\Delta m_s\,t) \right\}. \nonumber\\
\eeqa

When the final state is a $CP$~eigenstate, $CP$~symmetry requires
$\lambda_{f_{CP}}=\eta_f=\pm1$, where $\eta_f$ is the $CP$~parity
of the final state. Since the ratio (\ref{6:diovsu}) vanishes for 
$\lambda_f=\pm1$, we conclude that ${\cal A}_{f_{CP}}$ is an appropriate
definition of the $CP$~asymmetry in the $B_s^0\ra f_{CP}$ decay.
When the decay process into a final $CP$~eigenstate is dominated by a 
single $CP$~violating phase or by a single strong phase, we have
no $CP$~violation in decay, $|\bar A_{f_{CP}}/A_{f_{CP}}|=1$.
Consequently, for such modes, $CP$~violation is purely a result of
interference between decays with and without mixing and the expression 
for the $CP$~asymmetry simplifies considerably:
\beqa\label{6:AfCP}
{\cal A}_{f_{CP}}(t)\ &=&\ {\Gamma[\bar B_s(t)\ra f_{CP}]-\Gamma[B_s(t)\ra f_{CP}]\over
\Gamma[\bar B_s(t)\ra f_{CP}]+\Gamma[B_s(t)\ra f_{CP}]}\nonumber\\ &=&\ 
{\Im\lambda_{f_{CP}}\sin(\Delta m_s\,t)\over\cosh(\Delta\Gamma_s\, t/2)
-\Re\lambda_{f_{CP}}\sinh(\Delta\Gamma_s\, t/2)}.
\eeqa

Experimentally, the value of $y_s\equiv\Delta\Gamma_s\,/(2 \Gamma_s)$ is 
not yet known. As long as experimental errors are large compared to 
$\Delta\Gamma_s\,/\Gamma_s$, it is valid to use the simpler formulae that apply for 
the case $y_s=0$. In this approximation using, for consistency, $|q/p|=1$, 
Eqs. (\ref{6:tagdr}) simplify as follows:
\beqa\label{6:tagdrysz}
\Gamma[B_s(t)\ra f]\ &=&\ |A_f|^2 e^{-\Gamma_s t}
\left\{\cos^2\!\bigg({\Delta m_s\over2}t\bigg)
+|\lambda_f|^2\sin^2\!\bigg({\Delta m_s\over2}t\bigg)
-\Im(\lambda_f)\sin(\Delta m_s t)\right\},\nonumber\\
\Gamma[B_s(t)\ra\bar f]\ &=&\ |\bar A_{\bar f}|^2e^{-\Gamma_s t}
\left\{\sin^2\!\bigg({\Delta m_s\over2}t\bigg)
+|\lambda^{-1}_{\bar f}|^2\cos^2\!\bigg({\Delta m_s\over2}t\bigg)
+\Im(\lambda^{-1}_{\bar f})\sin(\Delta m_s t)\right\},\nonumber\\
\Gamma[\bar B_s(t)\ra f]\ &=&\ |A_f|^2e^{-\Gamma_s t}
\left\{\sin^2\!\bigg({\Delta m_s\over2}t\bigg)
+|\lambda_f|^2\cos^2\!\bigg({\Delta m_s\over2}t\bigg)
+\Im(\lambda_f)\sin(\Delta m_s t)\right\},\nonumber\\
\Gamma[\bar B_s(t)\ra\bar f]\ &=&\ |\bar A_{\bar f}|^2e^{-\Gamma_s t}
\left\{\cos^2\!\bigg({\Delta m_s\over2}t\bigg)
+|\lambda^{-1}_{\bar f}|^2\sin^2\!\bigg({\Delta m_s\over2}t\bigg)
-\Im(\lambda^{-1}_{\bar f})\sin(\Delta m_s t)\right\}. \nonumber\\
\eeqa

If, in addition, Eqs. (\ref{6:singleamp}) and (\ref{6:singleCKM}) hold, 
that is for decay channels that are dominated by a single weak phase,
the expressions (\ref{6:tagdrysz}) for the decay rates are further simplified:
\beqa\label{6:tagsingle}
\Gamma[B_s(t)\ra f]\ &=&\ |B_f|^2e^{-\Gamma_s t}
\left\{1+a_{\rm dir}\cos\left(\Delta m_s t\right)
-\sqrt{1-a_{\rm dir}^2}\, \sin(\phi_f+\delta_f)\sin(\Delta m_s t)\right\},
\nonumber\\
\Gamma[B_s(t)\ra\bar f]\ &=&\ |B_f|^2e^{-\Gamma_s t}
\left\{1-a_{\rm dir}\cos\left(\Delta m_s t\right)
-\sqrt{1-a_{\rm dir}^2}\, \sin(\phi_f-\delta_f)\sin(\Delta m_s t)\right\},
\nonumber\\
\Gamma[\bar B_s(t)\ra f]\ &=&\ |B_f|^2e^{-\Gamma_s t}
\left\{1-a_{\rm dir}\cos\left(\Delta m_s t\right)
+\sqrt{1-a_{\rm dir}^2}\, \sin(\phi_f+\delta_f)\sin(\Delta m_s t)\right\},
\nonumber\\
\Gamma[\bar B_s(t)\ra\bar f]\ &=&\ |B_f|^2e^{-\Gamma_s t}
\left\{1+a_{\rm dir}\cos\left(\Delta m_s t\right)
+\sqrt{1-a_{\rm dir}^2}\, \sin(\phi_f-\delta_f)\sin(\Delta m_s t)\right\}, 
\nonumber\\
\eeqa
where
\beq\label{6:defBR}
B_f={1\over2}(1+|\lambda_f|^2)A_f, \qquad 
a_{\rm dir}={1-|\lambda_f|^2\over1+|\lambda_f|^2}.
\eeq

Finally, if $f$ in (\ref{6:tagsingle}) is a $CP$~eigenstate, then 
\beq\label{6:singleCP}
|A_{f_{CP}}|=|\bar A_{f_{CP}}|, \qquad |\lambda_{f_{CP}}|=1, \qquad 
\delta_{f_{CP}}=0.
\eeq
Consequently, we get:
\beqa\label{6:tagsingcp}
\Gamma[B_s(t)\ra {f_{CP}}]\ &=&\ |A_{f_{CP}}|^2e^{-\Gamma t}
\left\{1-\sin(\phi_{f_{CP}})\sin(\Delta m_s t)\right\},\nonumber\\
\Gamma[\bar B_s(t)\ra {f_{CP}}]\ &=&\ |A_{f_{CP}}|^2e^{-\Gamma t}
\left\{1+\sin(\phi_{f_{CP}})\sin(\Delta m_s t)\right\}.
\eeqa
The \index{CP asymmetry@$CP$ asymmetry!in Bs decay@in $B_s$ decay}%
$CP$~asymmetry defined in Eq.~(\ref{6:AfCP}) is then given by
\beqa\label{6:AfCPsim}
{\cal A}_{f_{CP}}(t)\ &=&\ -\Im\lambda_{f_{CP}}\sin(\Delta m_s\,t),\nonumber\\
\Im\lambda_{f_{CP}}\ &=&\ \sin\phi_{f_{CP}}.
\eeqa

\subsection{Untagged Decays}
\index{CP violation@$CP$ violation!untagged decay}%

The expectation that $y_s\equiv\Delta\Gamma_s\,/(2\Gamma_s)$ is
not negligible, opens up the interesting possibility to learn about
$CP$~violation from untagged $B_s^0$ decays \cite{DunBs}.
The untagged decay rates are given by
\beqa\label{6:untagged}
\Gamma_f(t)\ &\equiv&\ 
\Gamma[B_s(t)\ra f]+\Gamma[\bar B_s(t)\ra f]\nonumber\\
&=&\ {1\over2}|A_f|^2 e^{-\Gamma_s\, t}\left\{
\left(1+\left|{p\over q}\right|^2
\right)\left[\left(1+|\lambda_f|^2\right)\cosh{\Delta\Gamma_s\,t\over2}
-2\Re\lambda_f\sinh{\Delta\Gamma_s\,t\over2}\right]\right.\nonumber\\ 
& &\left.
+\left(1-\left|{p\over q}\right|^2\right)\left[\left(1-|\lambda_f|^2\right)
\cos(\Delta m_s\,t)-2\Im\lambda_f\sin(\Delta m_s\,t)\right]\right\}\nonumber\\
&=&\ |A_f|^2 e^{-\Gamma_s\, t}
\left[\left(1+|\lambda_f|^2\right)\cosh{\Delta\Gamma_s\,t\over2}
-2\Re\lambda_f\sinh{\Delta\Gamma_s\,t\over2}\right],
\eeqa
where for the last equality we used $|q/p|=1$. 

Consider now the case of an untagged decay into a final $CP$~eigenstate.
For channels that are dominated by a single weak phase, we have
$|\lambda_{f_{CP}}|=1$. For time $t\lsim1/\Gamma_s$,
we can rewrite (\ref{6:untagged}) to first order in $y_s$:
\beq\label{6:untag}
\Gamma_f(t)=2|A_f|^2 e^{-\Gamma_s\, t}
\left[1- y_s \Re\lambda_f(\Gamma_s\, t)\right].
\eeq
The sensitivity to $CP$~violation is through the dependence on $\Re\lambda_{f_{CP}}$, 
and therefore requires that $y_s$ is not very small.

\subsection{Some Interesting Decay Modes}

In this section we describe several $B_s^0$ decay channels that will
provide useful information on $CP$~violation. We give examples of  $CP$~violation 
in the interference of decays with and without mixing for both final
$CP$~eigenstates and final non $CP$~eigenstates, and $CP$~violation in decay
for final $CP$~eigenstates. We do not discuss $CP$~violation in mixing
in semileptonic decays, because the effect is expected to be very small.
A recent review of many interesting aspects of $CP$~violation
in $B_s^0$ decays can be found in \cite{flebhm}.

\boldmath
\subsubsection{$B_s^0\ra J/\psi\phi$}
\label{ch6:intro_psiphi}
\unboldmath
\index{decay!$B_s^0 \to J/\psi\phi$}

The $CP$~asymmetry in the $B_s^0\ra J/\psi\phi$ decay is subject to a clean
theoretical interpretation because it is dominated by $CP$~violation
in interference between decays with and without mixing. 
The branching ratio has been measured \cite{CDFpsph}:
\beq\label{6:exppsph}
{\cal B}(B_s^0\ra J/\psi\phi)=(9.3\pm3.3)\times10^{-4}.
\eeq

The quark sub-process $\bar b\ra\bar cc\bar s$ is dominated
by the $W$-mediated tree diagram:
\beq\label{6:Apsiphi}
{\bar A_{ J/\psi\phi}\over A_{ J/\psi\phi}} = - \eta_{ J/\psi\phi}
\left({V_{cb}V_{cs}^*\over V_{cb}^*V_{cs}}\right).
\eeq
The penguin contribution carries a phase that is similar to (\ref{6:Apsiphi})
up to effects of ${\cal O}(\lambda^2)\sim0.04$. Hadronic uncertainties
enter the calculation then only at the level of a few percent.

Note that since $ J/\psi$ and $\phi$ are vector-mesons, the $CP$~parity
of the final state, $\eta_{ J/\psi\phi}$, depends on the relative
angular momentum, and the asymmetry may be diluted by the cancellation
between even- and odd-$CP$~contributions. It is possible to use the angular
distribution of the final state to separate the $CP$~parities.
The decay may be dominated by the $CP$~even final state. If this is 
established, the $CP$~asymmetry is more readily interpreted.

As concerns the mixing parameters, the Standard Model gives
\beq\label{6:qpBssm}
{q\over p}= - \left({V_{ts}V_{tb}^*\over V_{ts}^*V_{tb}}\right).
\eeq
Deviations from a pure phase are of ${\cal O}(10^{-4})$ and were
neglected in (\ref{6:qpBssm}).

Combining (\ref{6:Apsiphi}) and (\ref{6:qpBssm}) into (\ref{6:deflam}), we find
\beq\label{6:Imlpsph}
\Im\lambda_{ J/\psi\phi}=  (1-2f_{\rm odd}) \sin2\beta_s\,,
\eeq
where $\beta_s$ is defined in Eq.~(\ref{6:betas}) and $f_{\rm odd}$ is
the fraction of $CP$~odd final states. We learn the following:
\par {(i)} A measurement of the $CP$~asymmetry in $B_s^0\ra J/\psi\phi$
will determine the value of the very important CKM phase $\beta_s$
(see (\ref{6:AfCP}) or (\ref{6:AfCPsim})) \cite{DDF}.
\par {(ii)} The asymmetry is small, of order of a few percent, and may be even 
further diluted by cancellation between $CP$~odd and $CP$~even contributions.
\par {(iii)} An observation of an asymmetry that is significantly
larger than ${\cal O}(\lambda^2)$ will provide an unambiguous signal
for new physics. Specifically, it is likely to be related to new,
$CP$~violating contributions to $B_s^0$-$\bar B_s^0$ mixing \cite{NiSi}. 

\boldmath
\subsubsection{$B_s^0\ra J/\psi K_S^0$}
\label{sec:cpintrokos}
\unboldmath
\index{decay!$B_s^0 \to J/\psi K_S^0$}%

Interference between tree and penguin contributions to $B_s^0$ decays
is often sensitive to the CKM phase $\gamma$ of Eq.~(\ref{6:defgam}).
Since the angle $\gamma$ is much more difficult to determine than
$\beta$ ($\sin2\beta$ will be determined cleanly from the
$CP$~asymmetry in $B^0\ra J/\psi K_S^0$), the sensitivity of $B_s^0$
decays to this angle is highly interesting. On the other hand, since
this interference effect is a manifestation of $CP$~violation in
decay, its calculation involves hadronic parameters that are poorly
known. It is possible, however, to use various $B^0$ decays that are
related by flavour $SU(3)$ symmetry to the corresponding $B_s^0$ decays
to determine both $\gamma$ and the hadronic parameters. More
precisely, the relevant symmetry is $U$-spin, that is an $SU(2)$
subgroup that interchanges $d$ and $s$ quarks. $U$-spin breaking
effects, like all $SU(3)$ breaking effects, are not particularly small
($\sim m_{K}/\Lambda_{\chi SB}$) or well known, and will limit the
accuracy of this determination. Note, however, that since the $s$ and
$d$ quarks have both charge $-1/3$, electroweak penguins do not break
this symmetry.

Proposals for such a determination of CKM phases and hadronic
parameters have been made for $\bar b\ra \bar cc\bar d(\bar s)$ decays,
such as $B_s^0\ra J/\psi K_S^0$ ($B^0\ra J/\psi K_S^0$) \cite{flepks},
for $\bar b\ra \bar cc\bar s(\bar d)$ decays,
such as $B_s^0\ra D_s^+D_s^-$ ($B^0\ra D^+D^-$) \cite{flepks}, and for
$\bar b\ra \bar uu\bar s(\bar d)$ decays,
such as $B_s^0\ra K^+K^-$ ($B^0\ra\pi^+\pi^-$) \cite{flekk}.
To demonstrate the sensitivity of $B_s^0$ decays to $\gamma$ and the
need to use additional information to overcome the hadronic uncertainties,
we will discuss the $B_s^0\ra J/\psi K_S^0$ mode and mention only very briefly
the other two channels.

Measuring $CP$~violation in the $B_s^0\ra J/\psi K_S^0$ decay will be useful
for the extraction of the CKM phase $\gamma$ and will provide an estimate 
of the size of penguin uncertainties in the extraction of $\beta$ from
$B^0\ra J/\psi K_S^0$ \cite{flepks}. There is no experimental information 
on this mode yet. Theoretical estimates give
\beq\label{6:exppsks}
{\cal B}(B_s^0\ra J/\psi K_S^0)={\cal O}(2\times10^{-5}).
\eeq

The quark sub-process, $\bar b\ra\bar cc\bar d$ has contributions
from a tree diagram with a $CP$~violating phase $\arg(V_{cb}^*V_{cd})$,
and three types of penguin diagrams with $CP$~violating phases
$\arg(V_{qb}^*V_{qd})$, for $q=u,c,t$. Using CKM unitarity, one can write
\beq\label{6:Apsiks}
{\bar A_{ J/\psi K_S^0}\over A_{ J/\psi K_S^0}} = - \eta_{ J/\psi K_S^0}
\left({A_1 V_{cb}V_{cd}^*+A_2e^{i\theta}V_{ub}V_{ud}^*
\over A_1 V_{cb}^*V_{cd}+A_2e^{i\theta}V_{ub}^*V_{ud}}\right)
\left({V_{ud}V_{us}^*\over V_{ud}^*V_{us}}\right).
\eeq
Here, $A_1$ and $A_2$ are real and $\theta$ is the relative strong phase shift.
The last factor on the right hand side of Eq.~(\ref{6:Apsiks}) comes from
$K$-$\bar K$ mixing, since that is essential in producing a $K_S^0$~meson from
the outgoing $K^0$ and $\bar{K^0}$ mesons. The small measured value
of $\epsilon_K$ guarantees that this factor is essentially
model independent \cite{NiSi}.

Since $A_2/A_1$ is not particularly small, and there is no reason
to assume that $\theta$ is small, $|\lambda_{ J/\psi K_S^0}|=
|\bar A_{ J/\psi K_S^0}/A_{ J/\psi K_S^0}|$ (we use $|q/p|=1$)
could significantly differ from unity:
\beq\label{6:devuni}
| \lambda_{ J/\psi K_S^0} |^2 - 1 \approx 4 \, \frac{A_2}{A_1}
\left|{V_{ub}V_{ud}\over V_{cb}V_{cd}}\right|\sin\theta\sin\gamma.
\eeq
The deviation of $|\lambda_{ J/\psi K_S^0}|$ from unity can be measured 
(see Eq.~(\ref{6:diovsu})). We learn from Eq.~(\ref{6:devuni}) that 
any non-zero value of $|\lambda_{ J/\psi K_S^0}|^2-1$ requires non-zero
$\sin\gamma$, but that to extract the value of this fundamental
parameter, we need to know the hadronic parameters, $A_2/A_1$
and $\sin\theta$.  $U$-spin symmetry relates these hadronic parameters
to corresponding ones in the $B^0\ra J/\psi K_S^0$ decay. Consequently,
measurements of various observables in both the $B_s^0\ra J/\psi K_S^0$ and 
$B^0\ra J/\psi K_S^0$ decays will allow us to extract the phase $\gamma$
as well as the hadronic parameters \cite{flepks}. This extraction is model 
independent, with accuracy that depends on the size of $U$-spin breaking.

A similar analysis applies to the $B_s^0\ra D_s^+D_s^-$ and $B^0\ra
D^+D^-$ decays \cite{flepks}, and to the $B_s^0\ra K^+K^-$ and $B^0\ra\pi^+\pi^-$ 
decays \cite{flekk}. For the $B_s^0\ra D_s^+ D_s^-$ decay, the experimental  
upper bound is \cite{aledd}\
\beq\label{6:expdd}
{\cal B}(B_s^0\ra D_s^+ D_s^-)\leq0.218,
\eeq
while theoretical estimates give \cite{flepks}\
\beq\label{6:thedd}
{\cal B}(B_s^0\ra D_s^+ D_s^-)={\cal O}(8\times10^{-3}).
\eeq
For the $B_s^0\ra K^+ K^-$ decay, the experimental upper bound is \cite{alekk}\
\beq\label{6:expkk}
{\cal B}(B_s^0\ra K^+ K^-)\leq5.9\times10^{-5},
\eeq
while theoretical estimates give \cite{flekk}\
\beq\label{6:thekk}
{\cal B}(B_s^0\ra K^+ K^-)={\cal O}(1.4\times10^{-5}).
\eeq

\boldmath
\subsubsection{$B_s^0\ra D_s^\pm K^\mp$}
\label{sec:cp_bsdskintro}
\unboldmath
\index{decay!$B_s^0 \to D_s^- K^+$}%

Final $D_s^\pm K^\mp$ states are different from the states that we discussed
so far in this section because they are not $CP$~eigenstates. Yet, both
$B_s^0$ and $\bar B_s^0$ can decay into either of these states, and therefore
$CP$~violation in the interference of decays with and without mixing affects 
the time dependent decay rates. Consequently, it is possible to use the four 
time dependent decay rates to extract the angle $\gamma$ \cite{ADK}.

The quark sub-processes are $\bar b\ra\bar cu\bar s$, $\bar b\ra\bar uc\bar s$,
and the two $CP$-conjugate processes. These are all purely tree-level processes. 
It is important that the ratio between the magnitudes of the CKM combinations
is of order one:
\beq\label{6:CKMmag}
R_u\equiv\left|{V_{ub}V_{cs}\over V_{cb}V_{us}}\right|=0.41\pm0.05.
\eeq
The interference effects, which are crucial for this measurement, are large. 

For the $CP$~violating parameters, we have:
\beqa\label{6:lamdk}
\lambda_{D_s^+ K^-}\ &=&\ \rho
\left({V_{ts}V_{tb}^*\over V_{ts}^*V_{tb}}\right) 
\left({V_{cb}V_{us}^*\over V_{ub}^*V_{cs}}\right),\nonumber\\
\lambda_{D_s^- K^+}\ &=&\ {1\over\rho}
\left({V_{ts}V_{tb}^*\over V_{ts}^*V_{tb}}\right) 
\left({V_{ub}V_{cs}^*\over V_{cb}^*V_{us}}\right),
\eeqa
where $\rho$ is related to strong interaction physics.
From Eq.~(\ref{6:diovsu}) (or from (\ref{6:tagdr})) it is clear that 
measurements of the four time dependent decay rates would allow a determination 
of both $\lambda_{D_s^+ K^-}$ and $\lambda_{D_s^- K^+}$. Then we can find
\beq\label{6:prolam}
\lambda_{D_s^+ K^-}\lambda_{D_s^- K^+}=
\left({V_{ts}V_{tb}^*\over V_{ts}^*V_{tb}}\right)^2
\left({V_{cb}V_{cs}^*\over V_{cb}^*V_{cs}}\right)
\left({V_{ub}V_{us}^*\over V_{ub}^*V_{us}}\right)=
\exp[-2i(\gamma - 2\beta_s-\beta_K)].
\eeq 
We learn that a measurement of the four decay rates will
determine $\gamma - 2\beta_s $, up to very small corrections of
${\cal O}(\beta_K)$.

There is no experimental information on this mode at present.
The theoretical estimates give \cite{ADK}:
\beqa\label{6:thedk}
{\cal B}(B_s^0\ra D_s^- K^+)\ &=&\ {\cal O}(2.4\times10^{-4}),\nonumber\\
{\cal B}(B_s^0\ra D_s^+ K^-)\ &=&\ {\cal O}(1.4\times10^{-4}).
\eeqa

\boldmath
\subsection{Penguins in $B$ Decays: General Considerations}
\label{sec:cpintropenguin}
\unboldmath
\index{penguins!in $B$ decay}%

As discussed above, $CP$~violating asymmetries are often of particular 
experimental interest because of their simple dependence on the weak 
phase of the quantum mechanical amplitude of a decay.  This is most 
useful for probing fundamental physics if this weak phase 
can be related reliably to the phase of an element of the CKM matrix.  
This is difficult to do if there are two or more distinct 
quark-level transitions with different CKM structure which can mediate 
the decay. For reasons which will be clear momentarily, this problem 
is commonly known as  
\index{penguins!pollution}%
``penguin pollution.''

To illustrate the problem, let us take a simplified version of a 
concrete example. Consider the decay $B^0\to\pi^+\pi^-$, which 
requires the quark-level transition $\bar b\to u\bar u\bar d$.  The 
leading contributions to this transition are from a product of two 
weak currents, $\bar b_L\gamma^\mu u_L\,\bar u_L\gamma_\mu d_L$, and 
from a one-loop operator induced by a virtual $t$ quark, $\bar 
b\gamma^\mu T^ad\,\bar u\gamma_\mu T^au$. These two pieces carry distinct 
weak phases, and the overall amplitude is of the form (notation in 
this section is adapted from Ref.~\cite{charles})
\beq\label{6:ampa}
A(B^0\to\pi^+\pi^-)=V_{ud}V_{ub}^*M^{(u)} 
+V_{td}V_{tb}^*M^{(t)}=e^{i\gamma} T+e^{-i\beta}P\,.
\eeq
Here the notations $T$ and $P$ are inspired by the fact that the 
leading contributions to the two terms have tree and penguin 
topologies, but it is important to understand that ({\ref{6:ampa}) is 
in fact a {\it general\/} decomposition of the amplitude in terms of 
the weak phases $e^{i\gamma}$ and $e^{-i\beta}$.
Note that $M^{(u)}$ and $M^{(t)}$ depend on both short-distance and 
long-distance physics.  The long-distance parts, for which the 
leading contributions are
\beqa\label{6:mprop}
M^{(u)}&\propto&\langle\pi^+\pi^-|\,\bar 
b_L\gamma^\mu u_L\,\bar u_L\gamma_\mu d_L\,|B^0\rangle\,,\nonumber\\
M^{(t)}&\propto&\langle\pi^+\pi^-|\,\bar b\gamma^\mu T^ad\,\bar 
u\gamma_\mu T^au\,|B^0\rangle\,,
\eeqa
depend on nonperturbative strong interactions and are not yet 
amenable to calculation from first principles.  Since the two 
contributions to the amplitude have different weak phases and, in general, 
different strong phases, there is the possibility not only of
$CP$~violation in the 
\index{CP violation@$CP$ violation!interference type}%
interference between decays with and without mixing, but also of
\index{CP violation@$CP$ violation!in decay}%
$CP$~violation in the decay itself.  The 
\index{CP asymmetry@$CP$ asymmetry!time dependent}%
time-dependent $CP$~violating 
asymmetry takes the general form
\beq\label{6:cpasym}
{\cal A}_{CP}(t)=a_{\rm dir}\cos\Delta mt-\sqrt{1-a_{\rm dir}^2}\,
\sin2\alpha_{\rm eff}\,\sin\Delta mt\,,
\eeq
where $a_{\rm dir}$ was defined in Eq.~(\ref{6:defBR}).
In the limit $P=0$, we have $a_{\rm dir}=0$ and $\alpha_{\rm 
eff}=\alpha=\pi-\beta-\gamma$.  As can be seen from Eq.~(\ref{6:cpasym}),
the quantities $a_{\rm dir}$  and $\alpha_{\rm eff}$ may be extracted 
directly from the time-dependent experimental analysis. To determine
$\alpha$ from these measurements one needs to know also the ratio 
$|P/T|$~\cite{charles}:
\beq\label{6:cosdiff}
\cos(2\alpha-2\alpha_{\rm eff})={1\over\sqrt{1-a_{\rm dir}^2}}
\left[1-\left(1-\sqrt{1-a_{\rm dir}^2}\,\cos2\alpha_{\rm eff}\right)
\left|{P\over T}\right|^2\right].
\eeq
In the absence of either an experimental bound on or a theoretical 
calculation of $|P/T|$, it is not possible to extract $\alpha$ 
cleanly from a measurement of ${\cal A}_{CP}(t)$.

Whether or not it is possible to constrain $|P/T|$ in some way 
depends entirely on the process under consideration.  The literature 
on proposals for doing so is extensive.  At this point, we make a 
number of general comments:
\par {(i)} The essential problem is that the $CP$~violating phase of 
the decay amplitude is not known, because it depends on $|P/T|$, 
which depends in turn on hadronic physics.  (Statements about overall 
weak phases should be understood in the context of some definite 
phase convention.)
\par{(ii)} The ratio $|P/T|$ itself depends on CKM matrix elements, 
but this only complicates the form of the constraints on the 
Unitarity Triangle without introducing further uncertainties.
\par {(iii)} The two contributions with different weak phases, denoted 
$T$ and $P$ above, are commonly called ``tree" and ``penguin" contributions. 
This is something of a misnomer.   There are three penguin diagrams, each 
with a different weak phase, but one of these weak phases can be rewritten 
in terms of the other two phases using unitarity of the CKM matrix. Thus 
the charm quark ``penguin'' contribution to $B^0\to\pi^+\pi^-$, 
proportional to $V_{cd}V_{cb}^*=-V_{ud}V_{ub}^*-V_{td}V_{tb}^*$, is 
absorbed into both $T$ and $P$ in (\ref{6:ampa}), while the up ``penguin'' 
provides a contribution solely to $T$.
\par {(iv)} Similarly, it is irrelevant whether a penguin with a 
light quark in the loop is thought of as a ``penguin'' or a 
``rescattering'' contribution.  This terminology is often used in the 
context of modeling hadronic matrix elements, but in fact there is no 
physically meaningful distinction between the two processes.
\par {(v)} There are, in fact, two sorts of penguin diagrams which 
contribute to $B$ decays: ``gluonic'' penguins and ``electroweak'' 
penguins.  Although the electroweak penguins are typically much 
smaller, in general they may not be neglected.  The two types of 
penguins typically induce transitions with distinct flavour ({\it 
e.g.}~isospin) structures, which can complicate or even 
invalidate proposals to bound penguin contributions through flavour 
symmetries.  The relative importance of electroweak penguins depends 
on the decay under consideration.
\par {(vi)} In the case above, the contributions to $P$ are suppressed by 
\beq\label{6:rPT}
r_{PT}\approx{\alpha_s\over12\pi}\ln{m_t^2\over m_b^2}={\cal O}(0.1).
\eeq
Note that $\alpha_s\ln(m_t^2/m_b^2)$ is not a small factor and appears
at leading logarithmic order in RG-improved perturbation theory.
In other cases, penguin contributions might also be suppressed 
by powers of the CKM suppression factor $\lambda$.  If $|P/T|$ can be 
shown to be very small, then it is not necessary to know it 
precisely.  However, typically even $|P/T|$ of the order of 10-20\%
is significant enough to require a constraint or calculation with 
high confidence.
\par {(vii)} Any new physics terms, whatever their weak phases, can always
be written as a sum of two terms with weak phases $\gamma$ and $-\beta$. 
The impact of new physics is thus only to change the ratio of $|P/T|$ from 
that expected in the SM. We learn from this that we are only sensitive to 
new physics in cases where we have some knowledge of the ratio $|P/T|$.
For example, in cases where relationships between channels, such as those
from isospin or $SU(3)$, can be used to determine or constrain the ratio $P/T$
in a given channel from that in another, one is sensitive to any new
physics that does not respect this flavour symmetry~\cite{GKNtr}.  

\boldmath
\subsection{Penguins in $B^0\to J/\psi K_S^0$}
\unboldmath
\label{sec:cpintrokpen}
\index{decay!$B^0 \to J/\psi K_S^0$}%
\index{penguins!in $B^0\to J/\psi K_S^0$}%

The process $B^0\to J/\psi K_S^0$ is one in which the penguin 
contribution turns out to be relatively harmless, and it is 
instructive to begin by seeing why this is so.

The decay is mediated by the quark transition $\bar b\to c\bar c\bar s$.  
The dominant contribution is from tree level $W$ exchange, proportional 
to $V_{cs}V_{cb}^*$.  In the Wolfenstein 
parameterization, $V_{cs}V_{cb}^*$ is real and of order $\lambda^2$. 
In analogy to (\ref{6:ampa}) it is convenient to choose the decomposition
\beq\label{6:psiks}
A(B^0\to J/\psi K_S^0)=T_{ J/\psi K}+e^{i\gamma}P_{ J/\psi K}\,.
\eeq
The leading penguin diagram has a virtual $t$ quark in the loop and 
is proportional to $r_{PT}V_{ts}V_{tb}^*$ (see Eq.~(\ref{6:rPT})), which 
up to the $r_{PT}$-factor is the same size as $V_{cs}V_{cb}^*$.  However, 
if we use unitarity to write 
$V_{ts}V_{tb}^*=-V_{cs}V_{cb}^*-V_{us}V_{ub}^*$, we see that 
$\beta_s=\arg(-V_{ts}V_{tb}^*/V_{cs}V_{cb}^*)$ is small, of order 
$|V_{us}V_{ub}^*/V_{cs}V_{cb}^*|={\cal O}(\lambda^2)$.  Therefore 
this penguin diagram actually contributes mostly to $T_{ J/\psi K}$ in 
the decomposition (\ref{6:psiks}); the contribution to $|P_{ J/\psi
K}/T_{ J/\psi K}|$ is of order $r_{PT}\lambda^2$, below the level of $1\%$. 
The other potentially dangerous contribution is from the $u$ penguin, 
proportional to $V_{us}V_{ub}^*$.  The weak phase of this term is 
$e^{i\gamma}$, but its magnitude is ${\cal O}(\lambda^4)$.  Hence its 
contribution to $|P_{ J/\psi K}/T_{ J/\psi K}|$ is also of order 
$r_{PT}\lambda^2$. Finally, the $c$~penguin diagram is 
proportional to $V_{cs}V_{cb}^*$ and contributes only to $T_{ J/\psi K}$.

The ``penguin pollution'' in $B^0\to J/\psi K_S^0$ is thus below the level 
of $1\%$, even though penguin diagrams themselves contribute at a 
higher level.  Since the weak phase of $A(B^0\to J/\psi K_S^0)$ is 
known to high accuracy, the time-dependent $CP$~asymmetry in this mode 
provides a clean extraction of a parameter in the CKM matrix (in this 
case, $\sin2\beta$).  Only new physics effects could lead to a significant
difference between the asymmetry measured in this decay and $\sin2\beta$.
This example illustrates nicely the fact that 
the real issue is how well we know the weak phase of the decay 
amplitude.  The inclusion of electroweak penguins, which have the 
same phase structure, does not change the argument.

\boldmath
\subsection{Penguins in $B^0\to\pi\pi$}
\unboldmath
\index{decay!$B^0 \to \pi \pi$}%
\index{penguins!in $B^0\to\pi\pi$}%

The penguin contributions in $B^0\to\pi^+\pi^-$ are a much more difficult 
problem, one which has received intense attention in recent years. 
Much of what has been learned is collected in Ref.~\cite{charles}.  We 
parameterize
\beq\label{6:pipi}
A(B^0\to\pi^+\pi^-)=e^{i\gamma} T_{\pi\pi}+e^{-i\beta}P_{\pi\pi}\,.
\eeq
The leading contribution to $T_{\pi\pi}$ comes from $W$ exchange and 
is proportional to $V_{ud}V_{ub}^*$; in addition, $T_{\pi\pi}$ gets a 
contribution from penguin diagrams with a virtual $u$ quark.  The 
leading contribution to $P_{\pi\pi}$ is from a $t$ penguin diagram, 
proportional to $V_{td}V_{tb}^*$.  Since both $|V_{ud}V_{ub}^*|$ and 
$|V_{td}V_{tb}^*|$ are of order $\lambda^3$, 
$|P_{\pi\pi}/T_{\pi\pi}|$ is suppressed only by the the factor 
$r_{PT}$.  If nonperturbative QCD enhances the hadronic matrix 
element in $P_{\pi\pi}$ as compared to that in $T_{\pi\pi}$, then the 
penguin contribution might be significant enough to pollute the 
extraction of $\alpha$.

One may make a rough estimate of $|P_{\pi\pi}/T_{\pi\pi}|$ from the 
\index{decay!$B^0 \to K \pi$}%
decay $B^0\to K^+\pi^-$, which is convenient to parameterize by
\beq\label{6:kpiparam}
A(B^0\to K^+\pi^-)=e^{i\gamma} T_{K\pi}+P_{K\pi}\,.
\eeq
In this case, the leading contribution to $T_{K\pi}$ is of order 
$|V_{us}V_{ub}^*|={\cal O}(\lambda^4)$, while the $t$~penguin piece 
of $P_{K\pi}$ is of order $|V_{ts}V_{tb}^*|={\cal O}(\lambda^2)$, 
times a loop factor.  Hence one might expect that if QCD enhances the 
penguin contribution to $B\to\pi\pi$, then $B\to K\pi$ would be 
dominated by penguin processes.  Let us make the following 
assumptions for the moment: (i) flavour $SU(3)$ symmetry in the QCD 
matrix elements; (ii) electroweak penguins and ``color suppressed'' 
processes are negligible; (iii) penguins dominate $B\to K\pi$, so 
$T_{K\pi}$ may be ignored in ${\cal B}(B^0\to K^+\pi^-)$; 
(iv)~penguins make a small enough contribution to $B\to\pi\pi$ that 
$P_{\pi\pi}$ may be ignored in ${\cal B}(B^0\to \pi^+\pi^-)$.  Then 
it is straightforward to derive the relation
\beq\label{6:pengest}
\left|{P_{\pi\pi}\over T_{\pi\pi}}\right|=\left|{P_{\pi\pi}\over P_{K\pi}}
\right|\left|{P_{K\pi}\over T_{\pi\pi}}\right|\simeq
\left|{V_{td}\over V_{ts}}\right|\sqrt{{{\cal B}(B^0\to 
K^+\pi^-)\over{\cal B}(B^0\to \pi^+\pi^-)}}\,.
\eeq
The current constraints on the Unitarity Triangle yield roughly~\cite{psckm}
\beq\label{6:vtdvts}
0.1\lsim|V_{td}/V_{ts}|\lsim0.25\,.
\eeq
A recent CLEO measurement of the $B$ branching ratios gives~\cite{cleobranch}
\beqa\label{6:cleobranch}
{\cal B}(B^0\to \pi^+\pi^-)&=&(4.3^{+1.6}_{-1.4}\pm0.5)\times 10^{-6},\nonumber\\
{\cal B}(B^0\to K^\pm\pi^\mp)&=&(17.2^{+2.5}_{-2.4}\pm1.2)\times 10^{-6}\,.
\eeqa
Thus we obtain the rough estimate
\beq\label{6:ptest}
0.2\lsim|P_{\pi\pi}/T_{\pi\pi}|\lsim0.5\,.
\eeq
More elaborate analyses can somewhat lower the upper bound, but
it is clear that penguin effects are unlikely to be negligible in 
$B\to\pi\pi$.  In view of the shift (\ref{6:cosdiff}) of the measured 
$\alpha$ to $\alpha_{\rm eff}$, the problem of 
\index{penguins!pollution}%
``penguin pollution'' in the 
extraction of $\alpha$ is a serious one.

A variety of solutions to this problem have been proposed, falling 
roughly into two classes.  Each class requires assumptions, and each
has implications for the $B$ physics goals at Tevatron Run\,II and beyond.

The first type of approach is to convert the estimate given above 
into an actual measurement of $|P_{K\pi}|$ from the process $B\to 
K\pi$. (The list of papers on this subject is long. Early works include 
\cite{kpiNQ,kpiSW,kpiGr}.  For a much more complete list of references, 
see Ref.~\cite{charles}.) Once $|P_{K\pi}|$ is known, flavour $SU(3)$ is used 
to relate $|P_{K\pi}|$ to $|P_{\pi\pi}|$.  One must then include a number of 
additional effects:
\par {(i)} Electroweak penguins.  The effects of these are 
calculable~\cite{neroewp}.
\par {(ii)} Color suppressed and rescattering processes.  These must 
be bounded or estimated using data and some further assumptions.
\par {(iii)} $SU(3)$ corrections.  Some, such as $f_K/f_\pi$, can be 
included, but $SU(3)$ corrections generally remain a source of 
irreducible uncertainty.
\par {(iv)} Better knowledge of  $|V_{td}/V_{ts}|$.  This will be 
forthcoming from $\Delta m_s/\Delta m_d$, a crucial measurement which 
should be made during Run\,II.

The 
\index{SU3 relations@$SU(3)$ relations}%
$SU(3)$ relations typically take as inputs a variety 
of modes related to $B\to\pi\pi$ by $SU(3)$ symmetries, such as 
$B^0\to (K^\pm\pi^\mp,K^0\pi^0)$, $B^\pm\to (K^\pm\pi^0,K^0\pi^\pm)$, 
and $B_s^0\to (K^\pm\pi^\mp,K^+K^-,K^0K^0)$.  Both $CP$-averaged rates 
and $CP$~asymmetries can play a role.  The implication for Run\,II is 
that it is very important to measure accurately as many of these 
branching fractions, both tagged and untagged, as is possible.  Upper 
bounds on branching ratios are also important.  The choice of the 
most useful analysis will depend ultimately on which modes can be 
measured most accurately.

The second type of approach is to exploit the fact that the penguin 
contribution $P_{\pi\pi}$ to $B\to\pi\pi$ is pure $\Delta 
I={1\over2}$, while the tree contribution $T_{\pi\pi}$ contains a 
piece which is $\Delta I={3\over2}$.  This is not true of the electroweak 
penguins~\cite{fleiewp}, but these and other isospin violating corrections 
such as $\pi^0$-$\eta$ mixing are expected to be small and only become the 
dominant corrections in the case that the penguin effects are also 
small~\cite{gardner}.  Isospin 
symmetry allows one to form a relation among the amplitudes for 
\index{decay!$B^0 \to \pi \pi$}%
$B^0\to\pi^+\pi^-$, $B^0\to\pi^0\pi^0$ and $B^+\to\pi^+\pi^0$,
\beq\label{6:amprel}
{1\over\sqrt2}\,A(B^0\to\pi^+\pi^-)+A(B^0\to\pi^0\pi^0)
=A(B^+\to\pi^+\pi^0).
\eeq
There is also a relation for the charge conjugate processes.  A 
simple geometric construction then allows one to disentangle the 
unpolluted $\Delta I={3\over2}$ amplitudes, from which $\sin2\alpha$ 
may be extracted cleanly~\cite{grlon}.

The key experimental difficulty is that one must measure accurately 
the flavour-tagged rate for $B^0\to\pi^0\pi^0$.  Since the final state 
consists only of four photons, and the branching fraction is expected 
to be approximately at the level of $10^{-6}$, this is very hard. 
There is as yet no proposal to accomplish this measurement with any 
current or future detector.  It has been noted that an upper bound on 
this rate, if sufficiently strong, would also allow one to bound 
$P_{\pi\pi}$ usefully~\cite{grquinn}.

An alternative is to perform an isospin analysis of the process 
\index{decay!$B^0 \to \rho \pi$}%
$B^0\to\rho\pi\to\pi^+\pi^-\pi^0$~\cite{lnqs,qusny}.  Here one must study the 
time-dependent asymmetry over the entire Dalitz plot, probing variously the 
intermediate states $\rho^\pm\pi^\mp$ and $\rho^0\pi^0$.  The 
advantage here is that final states with two $\pi^0$'s need not be 
considered.  On the other hand, thousands of cleanly reconstructed 
events would be needed.  A very important question for any future $B$ 
experiment is whether it will be capable of performing this measurement.

Finally, one might attempt to calculate the penguin matrix elements, 
at which point only more precise information on $V_{td}$ is needed in 
order to know the level of contamination.  Model-dependent analyses 
are not really adequate for this purpose, since the goal is the 
extraction of fundamental parameters.  Precise calculations of such 
matrix elements from lattice QCD are far in the future, given the 
large energies of the $\pi$'s and the need for an unquenched 
treatment. Lattice calculations performed in the Euclidean regime 
also have difficulty including final state interactions. 
Recently, a new QCD-based analysis of the $B\to\pi\pi$ matrix elements 
has been proposed~\cite{bbns}.  The idea originates in the suggestion that 
these matrix elements factorize, in a novel sense, for asymptotically
large values of $m_B$, an idea with its roots in the ``color 
transparency'' picture of Bjorken.  This method is based on classifying
the diagrams in terms of a limited number of unknown functions with
calculable short distance corrections. At present, the
phenomenological relevance of this technique for realistic
$m_B=5.28~\gevcc$ is not yet well understood. In particular, it is not
yet clear whether $m_B$ is really in the regime where both soft final
state interactions and Sudakov logarithms may be neglected. Furthermore,
another recent analysis~\cite{KLS} based on similar ideas seems to be in 
substantial disagreement about the details of this factorization.    
One may hope that additional  progress on this front will be forthcoming.

\boldmath
\subsection{Penguins in $B\to K\pi$}
\unboldmath
\index{decay!$B^0 \to K \pi$}%
\index{penguins!in $B^0\to K\pi$}%

Analyses analogous to those which constrain $|P_{\pi\pi}|$ through 
the measurement of $|P_{K\pi}|$ may allow one to extract the CKM 
matrix element $\gamma$ through studies of direct 
$CP$~violation (see the reviews in \cite{kpigamBF,kpiNeu,kpigamGr}\  
and references therein).  For example, the ratio~\cite{neroewp,neurosrstar}
\beq\label{6:rstar}
R_*={{\cal B}(B^+\to K^0\pi^+)+{\cal B}(B^-\to K^0\pi^-)
\over 2\left[{\cal B}(B^+\to K^+\pi^0)+{\cal B}(B^-\to K^-\pi^0)\right]}
\eeq
is directly sensitive to $\cos\gamma$, and~\cite{fleimann}
\beq\label{6:rflma}
R={{\cal B}(B^0\to K^+\pi^-)+{\cal B}(\bar B{}^0\to 
K^-\pi^+)\over {\cal B}(B^+\to K^0\pi^+)+{\cal B}(B^-\to K^0\pi^-)}
\eeq
can be sensitive to $\sin\gamma$ if $R<1$.  The spirit of these 
analyses is to disentangle tree and penguin contributions through the 
use of $SU(3)$ symmetry and additional dynamical assumptions.  The 
theoretical issues are much the same as before:  one must find a way 
to control electroweak penguins, avoid making too many dynamical 
assumptions such as the neglect of rescattering or color suppressed 
processes, and include $SU(3)$ corrections.  The number of such 
proposals is extensive and growing.  What they typically have in 
common is that, as before, they profit from the accurate measurement 
of a wide variety of charmless hadronic two-body $B^0$, $B^+$ and 
$B_s^0$ decays.  In addition to those mentioned above, the modes 
$B\to\eta^{(\prime)}K$ have been proposed for the extraction of 
$\gamma$~\cite{grorosgam}. The experimental challenge is to measure or bound as 
many of these decays as possible, with as much precision as can be obtained.

\boldmath
\subsection{New Physics in $B_s^0$ Mixing}
\unboldmath
\index{new physics!in $B_s$ mixing}%

The SM predicts that the $CP$~asymmetries in the leading $B_s^0$ decays are all
very small. Consequently, these asymmetries will constitute good probes
of new physics. Since the reason for the SM prediction is the smallness
of the relative phase between the mixing amplitude and the leading decay
amplitudes ($\beta_s$), there are two possible sources for deviations 
from this predictions: new contributions to the decays or new contributions
to the mixing. The leading $B_s^0$ decay amplitudes are tree level, CKM favored, 
and therefore relatively large. In most new physics scenarios there
are no competing new contributions to these amplitudes. In contrast, the mixing 
amplitude is an electroweak loop and thus relatively small.
Indeed, many new physics models accommodate, or even predict, large
new $CP$~violating contributions to $B_s^0$ mixing 
\cite{NPBsa,NPBsn,nirssi,NPBsb,NPBsg,CKLN,GNR,RaSu,BBMR,BaRa}. 

Since in the SM the $B_s^0$ mixing amplitude is much larger than the $B^0$ mixing 
amplitude, roughly by a factor of order $|V_{ts}/V_{td}|^2$,
it may seem that a significant new physics contribution to $B_s^0$ mixing is always 
associated with a relatively much larger new contribution to $B^0$ mixing.
This, however, is not always the case. The new contributions to the mixing  
are often flavour dependent and might have a hierarchy that is similar
to (or even stronger than) the SM Yukawa structure. 

The question that we would like to answer in this section is the following:
If there is a contribution from new physics to $B_s^0$ mixing that is of magnitude
similar to the SM and relative phase of order one, how can we find it?
There are, in principle, many ways to demonstrate the presence of new physics in 
$B_s^0$ mixing. Which ones will be useful with realistic experimental analyses and
theoretical uncertainties depends on some (as yet) unknown parameters, both of 
Nature ({\it e.g.} $\Delta m_s$) and of the experiments.
In the following we discuss several observables that are
sensitive to new physics in $B_s^0$ mixing. For each of them we explain what 
are the requirements for the method to be interesting in practice.

New physics effects in $B_s^0$ mixing can also be found indirectly. 
A measurement of $\Delta m_{d}/\Delta m_s$ determines one
side of the unitarity triangle in the SM. With new physics, it may be 
inconsistent with other constraints on the unitarity triangle.
In such a case one does not know which of the observables are modified by 
new physics. In the discussion below we do not elaborate on indirect effects
and focus our attention on direct indications of new physics in $B_s^0$ mixing.
 
The relevant effects of new physics can be described by two new parameters, 
$r_s$ and $\theta_s$ \cite{SoWo,DDO,SiWo,GNW}, defined by
\beq
r_s^2 e^{2i\theta_s}\equiv{
\langle{B^0|H^{\rm full}_{\rm eff}|\bar B^0} \rangle\over
\langle{B^0|H^{\rm SM}_{\rm eff}|\bar B^0}\rangle} \,,
\eeq
where $H^{\rm full}_{\rm eff}$ is the effective Hamiltonian
including both SM and new physics contributions, and
$H^{\rm SM}_{\rm eff}$ includes only the SM box diagrams.
We work in the Wolfenstein parametrization where, to a good approximation,
both $V_{cb}V_{cs}^*$ and $V_{tb}V_{ts}^*$ are real. In other words, we
take $\beta_s=0$. With these convention and approximation, $\theta_s$
is the relative phase between the $B_s^0$ mixing amplitude and any 
real amplitude. In particular, the $CP$~asymmetry for decays into final 
$CP$~eigenstates that are mediated by $b \to c\bar c s$ is given by
\beq
a_{CP} = \pm\sin2\theta_s,
\eeq
and also
\beq
\arg( -\Gamma_{12}^* M_{12}) = 2\theta_s = \phi_s,
\eeq
where $\phi_s$ is defined in \eq{defphi}.

\boldmath
\subsubsection{Time Dependent $CP$~Asymmetries}
\unboldmath
\index{CP asymmetry@$CP$ asymmetry!time dependent}%

The most promising way to discover new physics contributions to $M_{12}$
is through measurements of the mass difference $\Delta m_s$ and
various time dependent $CP$~asymmetries. Note that while 
in the SM $\Delta m_s\lsim30\ {\rm ps}^{-1}$, this may not be the case 
in the presence of new physics. A larger value of $\Delta m_s$ makes 
its measurement more difficult. For example, a measurement of the
time dependent $CP$~asymmetry in the 
\index{decay!$B_s^0 \to J/\psi\phi$}%
$B_s^0 \to J/\psi\phi$ channel 
will directly determine $\sin2\theta_s$. If a value that is above the few percent 
level is found, it would provide a clean signal of new physics.
Note that $J/\psi\phi$ is not a pure $CP$~eigenstate, and therefore an angular 
analysis is required to project out the $CP$~even and $CP$~odd parts and to measure 
$\sin2\theta_s$. However, it may be the case that the presence of new physics
can be demonstrated even without such an analysis. Other time dependent $CP$~asymmetries 
for transitions  mediated by real quark level decay amplitudes, {\it e.g.} 
\index{decay!$B_s^0 \to D_s^{+(*)} D_s^{-(*)}$}%
$B_s^0 \to D_s^{+(*)} D_s^{-(*)}$, can  provide similar tests. 
Again, we emphasize that a non vanishing $CP$~asymmetry in the 
$D_s^{(*)} D_s^{(*)}$ channel, which is not a $CP$~eigenstate, is 
a clean signal for new physics in the $B_s^0$ mixing amplitude.

If $B_s^0$ oscillations turn out to be too fast to be traced,
the above methods cannot be applied. Below we describe various other
methods that are sensitive to $\theta_s$ and do not require that the
fast oscillations are traced.

\boldmath
\subsubsection{Time Integrated $CP$~Asymmetries}
\unboldmath
\index{CP asymmetry@$CP$ asymmetry!time integrated}%

For the $B^0$ system, one can use time integrated asymmetries. The dilution factor 
due to the time integration, $D \sim x_q/(1+x_q^2)$ is not very small for $x_d \sim 
0.7$. For the $B_s^0$ system, however, $x_s \gg1$, leading to a strong dilution of the
time integrated asymmetries, $D \sim 1/x_s$. In principle, however, the time 
integrated asymmetry can be measured. Since expected SM effects are small, any 
non vanishing asymmetry would be an indication for new physics. The goal here is not 
necessarily to make a precise  measurement of the asymmetry, but rather to demonstrate 
that it is not zero. Assuming, for example, $x_s\sim40$, and $\sin \theta_s\sim 0.8$, 
the time integrated asymmetry in $B_s^0 \to  J/\psi\phi$ is of ${\cal O}(0.02)$. 
If the combined statistical and systematic experimental error on such asymmetry 
measurements is below $1\%$,  
the presence of a non vanishing asymmetry can be established.

\subsubsection{The Width Difference}
\index{width difference!$B_s$}%

If the $B_s^0$ width difference ($y_s$) can be measured, there are 
more ways to see the effects of $\theta_s$ \cite{DunBs,yuval-Bs}.
Note that new physics in the mixing amplitude always reduces $y_s$
compared to its SM value. This fact can be readily seen from the
following equation:
\beq
\Delta \Gamma_s\, = 2 |\Gamma_{12}| \cos 2 \theta_s.
\eeq
Since we assume that new physics affects $M_{12}$ but not $\Gamma_{12}$,
the only modification of the right hand side can be a reduction of $\cos2\theta_s$
compared to its SM value of one. The reduction of $y_s$ can be understood intuitively 
as follows. In the absence of $CP$~violation, the two mass eigenstates are also
$CP$~eigenstates. The large $\Delta \Gamma_s$ in the SM is an indication that
most of the $b \to c \bar c s$ decays are into $CP$~even final states.
With $CP$~violation, the mass eigenstates are no longer approximate $CP$~eigenstates. 
Then, both mass eigenstates decay into $CP$~even final states. Consequently, 
$\Delta \Gamma_s$ is reduced.

A large enough $y_s$, say ${\cal O}(0.1)$, would allow various ways of finding
a non vanishing $\theta_s$ \cite{yuval-Bs}. We now discuss one such method
which makes use of both flavour specific decays (semileptonic decays are
flavour specific; $b\to c\bar ud$ decays are also flavour specific to a
very good approximation) and decays into final $CP$~eigenstates.

The time dependent decay rate of a flavour specific mode, $f$, is given by:
\beq 
\Gamma\left[f\left(t\right)\right]=
\Gamma\left[\bar f\left(t\right)\right] =
\frac{\Gamma\left(B_s \rightarrow f\right)}{2} \;
\bigg\{e^{-\Gamma_L t} +e^{-\Gamma_H t} \bigg\} \;.
\eeq
Both $\Gamma_H$ and $\Gamma_L$ and, therefore, also $\Delta\Gamma_s$, 
can be extracted from such a measurement.
The time dependent decay rate into a $CP$~even final state 
from a $b \to c \ov{c} s$ transition is given by: 
\beq \label{method}
\Gamma(B \to CP_{\rm even},t) \propto
\cos^2\theta_s \; e^{-\Gamma_L t} +
\sin^2\theta_s \; e^{-\Gamma_H t}.
\eeq
For a decay into a $CP$~odd state, $\Gamma_L$ and $\Gamma_H$ are interchanged.
In principle, a three parameter fit of a decay into a $CP$~even state can be used 
to measure $\Gamma$, $\Delta\Gamma$ and $\theta_s$ using Eq.~(\ref{method}). 
Even if this cannot be done in practice, $\theta_s$ can be measured by comparing 
the measurements of $\Delta\Gamma$ from flavour specific decays and $CP$~eigenstate 
decays. Experimentally, most of the data are expected to be taken for small
$\Gamma\, t$. Then, using $\Delta\Gamma \, t \ll 1$, Eq.~(\ref{method}) becomes
\beq \label{sltimp}
\Gamma(B \to CP_{\rm even},t) \propto e^{-\Gamma_+ \,t}\,, \qquad
\Gamma_+ \equiv \left(\Gamma + {\Delta\Gamma\,|\cos 2\theta_s| \over 2} \right)\,.
\eeq
Using $\Gamma$ and $\Delta\Gamma$ as measured from the flavour specific data,
a one parameter fit to the decay rate gives $\theta_s$.
Actually, such a fit determines
\beq
2(\Gamma_+ - \Gamma) = 
\Delta\Gamma\, |\cos( 2 \theta_s )|\,.
\eeq
By comparing it to the real width difference, $\Delta\Gamma$, we get
\beq \label{combi}
|\cos2\theta_s| = { 2(\Gamma_+ - \Gamma) \over \Delta\Gamma}\,. 
\eeq
This method would be particularly useful if $\theta_s$ is neither very
small nor very large. For $\theta_s \sim \pi/4$ the width difference becomes 
too small to be measured. For $\theta_s \sim 0$ the required precision of the
measurement is very high.

\boldmath
\subsubsection{The Semileptonic $CP$~Asymmetry}
\unboldmath
\index{CP asymmetry@$CP$ asymmetry!semileptonic}%

The semileptonic asymmetry, $a_{\rm sl}$, which is sensitive to
$\theta_s$ \cite{SLref,SaXi,CaWo,BEN,Wora}, does not require a
measurement of either $x_s$ or $y_s$. In the SM, $a_{\rm sl}$ is very
small: \beq a_{\rm sl} \approx \Im(\Gamma_{12}/M_{12}) =
|\Gamma_{12}/M_{12}| \times \sin2\theta_s={\cal O}(10^{-4}).  \eeq
With new physics, the first factor, $|\Gamma_{12}/M_{12}|={\cal
  O}(10^{-2})$, cannot be significantly enhanced, but the second,
$\sin2\theta_s$, could. Actually, if $\sin2\theta_s \sim 1$ the
semileptonic asymmetry is expected to be of ${\cal O}(10^{-2})$.
Since in the SM $a_{\rm sl}$ is negligibly small, any observation of a
non vanishing asymmetry is a clear signal for new physics. Whether
such a measurement is possible depends, among other things, on the
actual value of the asymmetry: a factor of a few in one or the other
direction can make a significant difference as the purely experimental
systematic uncertainties are expected to be at the percent level.

%
%
%
%


\boldmath
\section{Study of $B^0 \rightarrow J/\psi K^0_S$}
\label{sec:psiphi}
\unboldmath
\index{decay!$B^0 \to J/\psi K_S^0$}%

In the following sections, we report the results of studying the prospects
of the CDF, D\O\ and BTeV experiments for measuring $CP$~violation in
different $B$~decay modes. The outline of the following sections consists
of a brief theoretical introduction to the particular decay modes of
interest, the prospects of the three Tevatron experiments (not all detectors
are capable of measuring all modes and we do not necessarily have always
reports from all three experiments) followed by a brief summary. We start
with the study of $B^0 \rightarrow J/\psi K^0_S$.

\boldmath
\subsection[$B^0 \rightarrow J/\psi K^0_S$: Introduction]
{$B^0 \rightarrow J/\psi K^0_S$: Introduction
$\!$\authorfootnote{Authors: S.~Gardner and R.~Jesik.}
}
\unboldmath

As discussed in the introduction in Sec.~\ref{ch6:intro} 
(see Sec.~\ref{sec:cpintrokos} and \ref{sec:cpintrokpen}), a single weak phase
dominates the decay $B^0\ra J/\psi K_S^0$, so that the $CP$~asymmetry
in this channel is dominated by the interference
between decays with and without $B$-${\bar B}$ mixing. 
Identical considerations
apply to the study of $B_s^0\ra J/\psi \phi$.
Assuming the CKM matrix to be unitary, 
there are two distinct decay topologies, 
characterized by the CKM matrix elements
$V_{cs} V_{cb}^\ast$ 
and $V_{us} V_{ub}^\ast$, indicating
$CP$~violation in direct decay to be suppressed by
${\cal O}(\lambda^2)$. 
Nevertheless, the two decays are sensitive to different
CKM information. 
We find for $B_s^0 \ra  J/\psi \phi$
\beq\lambda_{ J/\psi \phi}= \eta_{ J/\psi \phi} 
\left( \frac{V_{tb}^\ast V_{ts}}{V_{tb} V_{ts}^\ast}\right)
\left( \frac{V_{cb} V_{cs}^\ast}{V_{cb}^\ast V_{cs}}\right)
\Rightarrow \Im\lambda_{ J/\psi\phi}= \sin2 \beta_s \;.
\eeq
The first set of CKM factors reflects $B_s^0$-${\bar B_s^0}$ 
mixing in the
Standard Model, whereas the second set reflects
those of the assumed dominant decay topology in 
${\bar b}\ra{\bar c} c {\bar s}$. 
As discussed in Section~\ref{sec:cpintrokos},
we obtain for $B^0 \ra  J/\psi K_S^0$ 
\beq
\lambda_{ J/\psi K_S^0}= \eta_{ J/\psi K_S^0} 
\left( \frac{V_{tb}^\ast V_{td}}{V_{tb} V_{td}^\ast}\right)
\left( \frac{V_{cb} V_{cs}^\ast}{V_{cb}^\ast V_{cs}}\right)
\left( \frac{V_{cd}^\ast V_{cs}}{V_{cd} V_{cs}^\ast}\right) 
\Rightarrow \Im\lambda_{ J/\psi K_S^0}= \sin2 \beta \;, 
\eeq
where the first set of CKM factors now reflects $B^0$-${\bar B^0}$ 
mixing and the second set reflects those of the dominant decay topology in 
${\bar b}\ra{\bar c} c {\bar s}$. 
Finally, the third set reflects
$K$-${\bar K}$ mixing  necessary 
to realize the $K_S^0$ final state. Indeed, $K$-${\bar K}$ mixing must
occur in order to generate interference
between the $B^0 \ra J/\psi K^0$ and ${\bar B^0} \ra J/\psi {\bar K^0}$
decay channels. 
We have assumed, as in the $B_s^0$ case, that
$B^0$-${\bar B^0}$ mixing is controlled by a pure phase. 
The quality of this assumption is likely to be less impressive than in the
$B_s^0$ case. Nevertheless, it still ought to be good with 
$(|q/p|-1) < {\cal O}(10^{-2})$~\cite{Qpdg1}. 
In the case of $K$-${\bar K}$ mixing, the 
deviation of $|q/p|$ from unity is empirically known; the non-zero 
semileptonic asymmetry 
$[\Gamma(K_L\ra \pi^- \ell^+ \nu_{\ell})-\Gamma(K_L\ra \pi^+ \ell^- 
{\bar \nu_{\ell}})]/[\Gamma(K_L\ra \pi^- \ell^+ \nu_{\ell})+\Gamma(K_L\ra \pi^+ \ell^- 
{\bar \nu_{\ell}})]$ implies that $|q/p|-1 \sim -3\cdot 10^{-3}$~\cite{pdg98}. 
Thus $K$-${\bar K}$ mixing can also be typified by a pure 
phase. The top quark contribution to $K$-${\bar K}$ mixing 
is strongly suppressed by CKM
factors, so that the charm quark determines $(q/p)_K$.
Note that $\beta_s$ is itself ${\cal O}(\lambda^2)$, whereas $\beta$ is ${\cal O}(1)$. 
Thus, an asymmetry ${\cal A}_{CP}$ in $B_s^0\ra\phi K_S^0$ 
considerably larger than ${\cal O}(\lambda^2)$ would signal 
the presence of new physics in $B_s^0$-${\bar B_s^0}$ mixing. 

In the case of $B_s^0 \ra  J/\psi \phi$, the $CP$~of the final state
depends on the partial wave in which the vector mesons sit, so that 
an analysis of the angular distribution is required in order to 
extract weak phase information~\cite{DDF}. The 
information encoded in the time-dependent angular distributions of
$B\ra VV$ decays can be quite rich, and an angular analysis of 
$B^{0/+}(t) \ra  J/\psi (\ra \ell^+\ell^-)  
K^\ast (\ra \pi^0 K_S^0)$~\cite{Kakst,Dukst,Dikst,DDF}
is sensitive to $\cos2\beta$
as well~\cite{DDF,DDF2}. The expected 
determination of $\sin 2\beta$ from ${\cal A}_{CP}$ in 
$B^0 \ra  J/\psi K_S^0$ leaves a four-fold discrete ambiguity in the angle 
$\beta$, so that the determination of 
$\cos2\beta$~\cite{GNW,bto3pi:GQ97,dambig}
plays an important role in resolving the value of $\beta$ itself. 
Unfortunately, $\cos2\beta$ 
appears in conjunction with a signed 
hadronic parameter. However, under the assumption of $U$-spin symmetry, 
the latter can be extracted from the $CP$~asymmetry in $B_s^0 \ra  J/\psi \phi$,
so that $\cos2\beta$ can be determined as well~\cite{DDF2}. 

Since both decay modes $B^0\ra J/\psi K^0_S$ and $B^0_s\ra J/\psi \phi$ 
are very similar from an experimental point of view (trigger and
reconstruction efficiencies), we will focus in the following experimental
sections on describing the strategies to reconstruct $B^0\ra J/\psi K^0_S$
and give estimates for $\sin2\beta$. We will add the estimates for 
\index{decay!$B_s^0 \to J/\psi \phi$}%
$B^0_s\ra J/\psi \phi$ event yields as appropriate.

\boldmath
\subsection[$B^0 \rightarrow J/\psi K^0_S$: CDF Report]
{$B^0 \rightarrow J/\psi K^0_S$: CDF Report
$\!$\authorfootnote{Authors: M.~Paulini and B.~Wicklund.}
}
\label{sec:cdf-sin2beta}
\unboldmath
\index{decay!$B^0 \to J/\psi K_S^0$}%
\index{CDF!$\sin 2\beta$ prospects}%

For the measurement of \stb\ in the $B^0 \ra \jpks$ channel~\cite{tdr}, CDF
expects 
to reconstruct in \tfb\ of data in Run\,II about 20,000 \jpks\ events
with $J/\psi \ra \mu^+\mu^-$ and  $K^0_S\ra \pi^+\pi^-$. 
Starting with $\sim\!400$ \jpks\ events~\cite{sin2b_prl} reconstructed in
110~pb$^{-1}$ in Run\,I, this number is obtained in the following way.
To estimate the increase in $J/\psi$ and $J/\psi K_S^0$ signals,
we first measure the inclusive $J/\psi$ signal yields in
each of the Level\,2 trigger paths used in Run\,Ib. 
We scale these to Run\,II conditions with the following
modifications:
\begin{itemize}
  \item[--] 2~fb$^{-1}$/110~pb$^{-1}$ for the total Run\,II integrated
        luminosity $\Rightarrow \times$ 20 gain in event yield
\item[--]{Assume increase of $\times 1.1$ from $\sqrt{s}=1.8$~TeV
$\rightarrow$ 2.0~TeV} 
\item[--]{Wider muon stub gates $\Rightarrow \times$ 1.36 gain in efficiency}
\item[--]{Increased muon coverage with CMX miniskirt $\Rightarrow
\times$1.396 increase} 
\item[--]{Remove Run\,I wedge cuts $\Rightarrow \times$ 1.1 gain in efficiency}
\item[--]{Add Run\,II trigger cuts on $m_T^{\mu\mu}$ and
$\Delta\phi^{\mu\mu}$ $\Rightarrow 
\times$ 0.85 loss in efficiency} 
\item[--]{Add lower $p_T^{\mu\mu}$ threshold of 2.0~$\rightarrow$ 1.5~\gevc\
for central muons (CMU) 
$\Rightarrow \times$ 2
for CMU-CMU dimuons} 
\end{itemize}
\index{CDF!dimuon trigger}%

The effects of these cuts were modeled for $J/\psi K_S^0$ Monte Carlo
events, to get the relative change in yield for each modification. 
Figure~\ref{psiyields}(a) shows the dependence of lowering the muon
$p_T$~threshold 
for the $J/\psi K_S^0$ yields in CMU-CMU from a generator-level Monte Carlo
study.
The upper histogram is for the proposed Run\,II trigger with a $p_T$ 
threshold of 1.5~\gevc, while the lower histogram is the convolution of the
Run\,I CMU-CMU trigger with the Level\,1 stub gate.
The solid points are the sideband subtracted yields for
the $B^+ \rightarrow J/\psi K^+$ CMU-CMU signal in Run\,I.

\begin{figure}[tbp]
\centerline{
\put(55,210){\large\bf (a)}
\put(265,210){\large\bf (b)}
\epsfxsize=3in
\epsffile{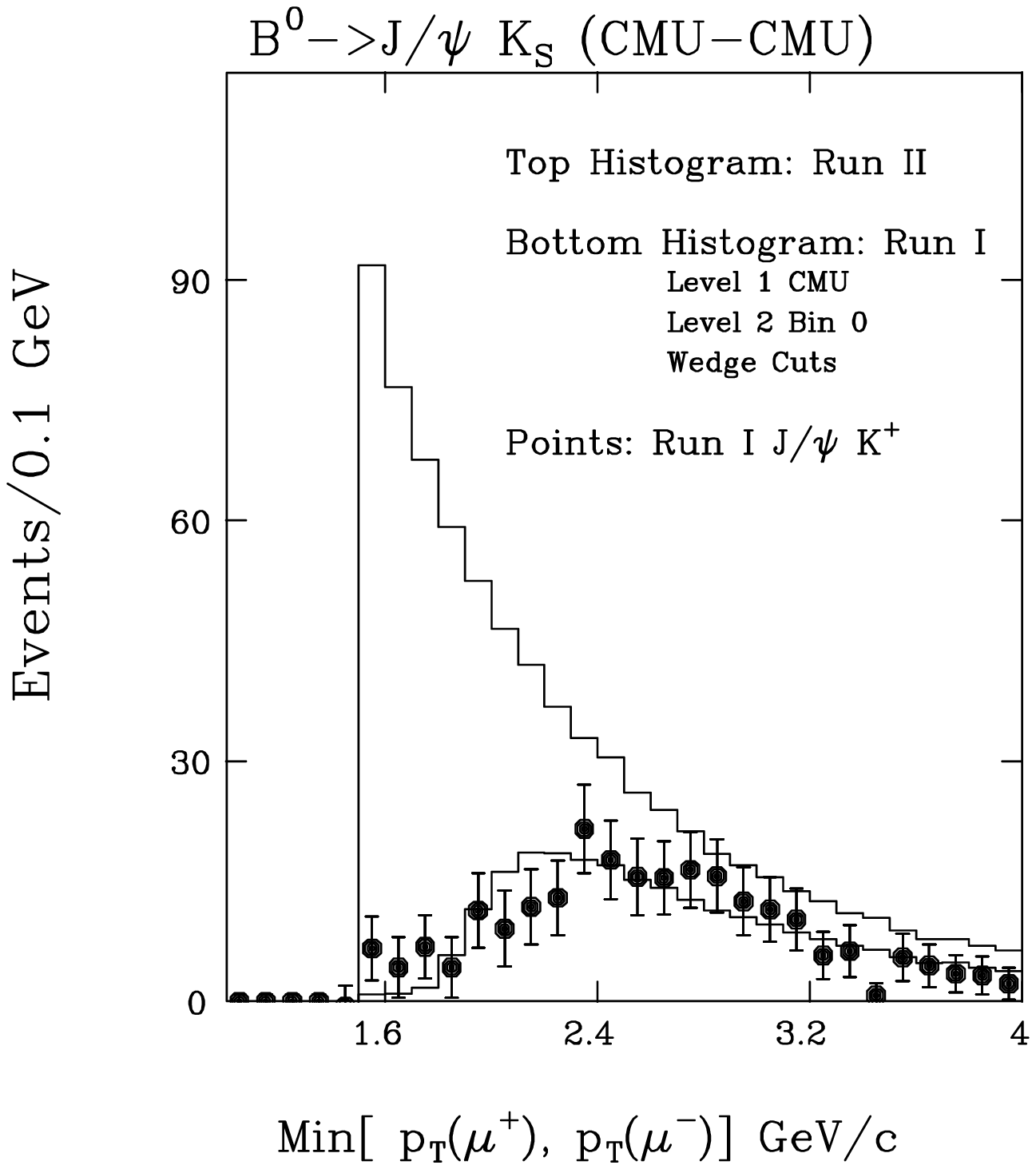}
\epsfxsize=3in
\epsffile{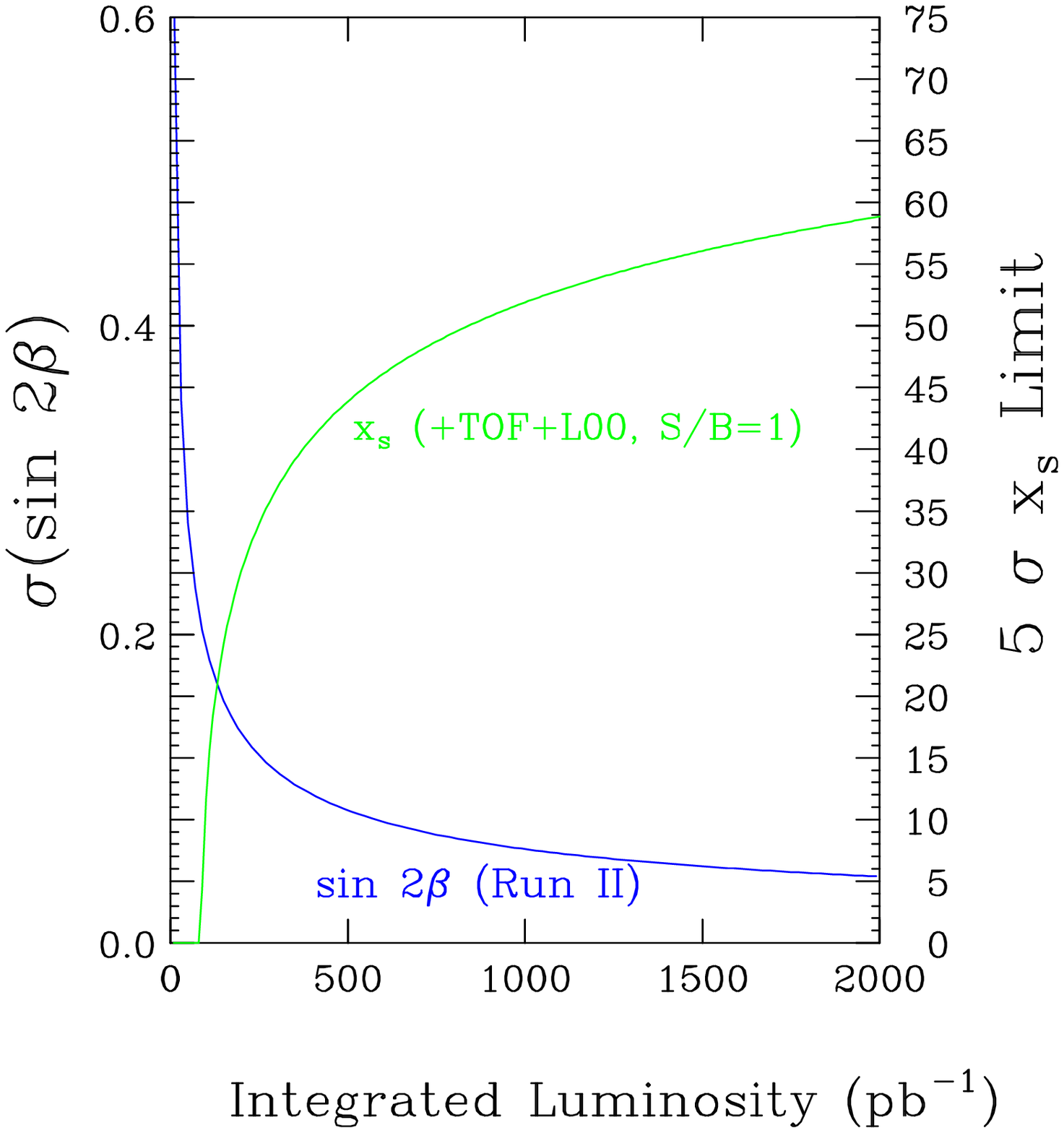}
}
\vspace*{0.3cm}
\caption[Dependence of CMU-CMU $J/\psi K_S^0$ yields on the 
lower $p_T$ muon threshold at CDF.]
{Dependence of CMU-CMU $J/\psi K_S^0$ yields on the 
lower $p_T$ muon threshold at CDF: Run\,II trigger (top histogram), Run\,I
trigger (bottom histogram).  
The points are the $B^+ \rightarrow J/\psi K^+$ CMU-CMU signal in Run\,I.
(b) Uncertainty on $\sin2\beta$ (left scale)
and 5$\sigma$ reach
for $x_s$ (right scale) as functions of integrated luminosity.}
\label{psiyields}
\end{figure}

For \tfb\ luminosity, this gives a net increase of a factor
of $\times 50$ in the $J/\psi K_S^0$ yield over the 400 events found in
Run\,I.
Assuming the same $K^0_S$ finding efficiency as in Run\,I, this yields
20,000 fully reconstructed $B^0\ra\jpks$ events. 
CDF also plans to 
trigger on $J/\psi \ra e^+e^-$, which would increase the number of \jpks\
events by $\sim\!50$\%~\cite{tdr}. The yield of 20,000 \jpks\ events thus
represents a conservative estimate.

In Run\,II, CDF expects to improve the effective tagging efficiencies \eD\ of
the $B$~flavour tagging methods, as summarized in 
\index{CDF!flavor tagging}%
Table~\ref{ed_run2}. 
The extended lepton coverage with the completed muon extension systems and
the plug 
calorimeter results in a total \eD\ of 1.7\% for lepton tagging.   
A significant improvement in $\eD \sim 3\%$ is possible for jet charge
tagging. The extended coverage of the SVX\,II detector together with ISL as
well as their 
improved pattern recognition capabilities will substantially enhance the
purity of the jet charge algorithm. Together with a value of $\eD\sim2\%$  
assumed for same side tagging, this yields a total $\eD\sim\!9.1\%$ in
Run\,II including opposite side kaon tagging made possible with a
Time-of-Flight detector~\cite{pac_prop}. This results in an
error of $\sigma(\stb)\sim\!0.05$ on a measurement of the $CP$~violation
parameter \stb.

Starting with nominal assumptions on flavour tagging
efficiencies and signal-to-back\-ground ratios ($S/B$), 
the reach on \stb\ can be calculated as a
function of integrated luminosity.  This is shown in
Figure~\ref{psiyields}(b) together with the 5\,$\sigma$ reach
for the $B_s^0\bar B_s^0$ oscillation parameter $x_s$ (right scale). 

\begin{table}
\begin{center}
\begin{tabular}[t]{l|cccc}
\hline
 Flavour tag & \eD\ Run\,I &\eD\ Run\,II & Calib. sample & Sample size \\
\hline
Same side tag & $(1.8\pm0.4\pm0.3)\%$~\cite{sin2b_prl} 
        & 2.0\%~\cite{tdr} & $J/\psi K^{*0}$ & $\sim\!30,000$ \\
Jet charge tag & $(0.78 \pm 0.12 \pm 0.08)\%$~\cite{qjet_mix} 
        & 3.0\%~\cite{tdr} & $J/\psi K^+$ & $\sim\!50,000$ \\
Lepton tag & $(0.91 \pm 0.10 \pm 0.11)\%$~\cite{qjet_mix}  
        & 1.7\%~\cite{tdr} & $J/\psi K^+$ & $\sim\!50,000$ \\
Kaon tag & --
        & 2.4\%~\cite{pac_prop} & $J/\psi K^+$ & $\sim\!50,000$ \\
\hline
\end{tabular}
\vspace*{0.3cm}
\caption[Summary of flavour tagging methods used in the measurement
of \stb\ at CDF.]
{Summary of flavour tagging methods used in the measurement
of \stb, the measured \eD\ values from Run\,I and the data samples
used to calibrate the tagging algorithms in Run\,II.} 
\label{ed_run2}
\end{center}
\end{table}

With respect to estimating the yield of $B_s^0 \ra J/\psi \phi$ events in
\tfb\ in Run\,II, we compare the number of observed events in 
$B_s^0 \ra J/\psi \phi$ to the number of $B^0 \ra J/\psi K^0_S$ events
with comparable signal-to-noise in Run\,I data. Here, we restrict our
estimate to $J/\psi$~events fully reconstructed in the Run\,I silicon
vertex detector. We observe a signal of about 80~$B_s^0 \ra J/\psi \phi$
events in Run\,I as shown in Figure~\ref{psiphiyield}. With about 200
$B^0 \ra J/\psi K^0_S$ events~\cite{sin2b_prl} reconstructed in CDF's
Run\,I silicon detector, we find the number of $B_s^0 \ra J/\psi \phi$ is
approximately 40\% the number of $B^0 \ra J/\psi K^0_S$. With 20,000 
$J/\psi K^0_S$ events estimated above, we expect about 8000 $B_s^0 \ra
J/\psi \phi$ events in \tfb\ in Run\,II.

\begin{figure}[tb]
\centerline{
\epsfxsize=4.0in
\epsffile{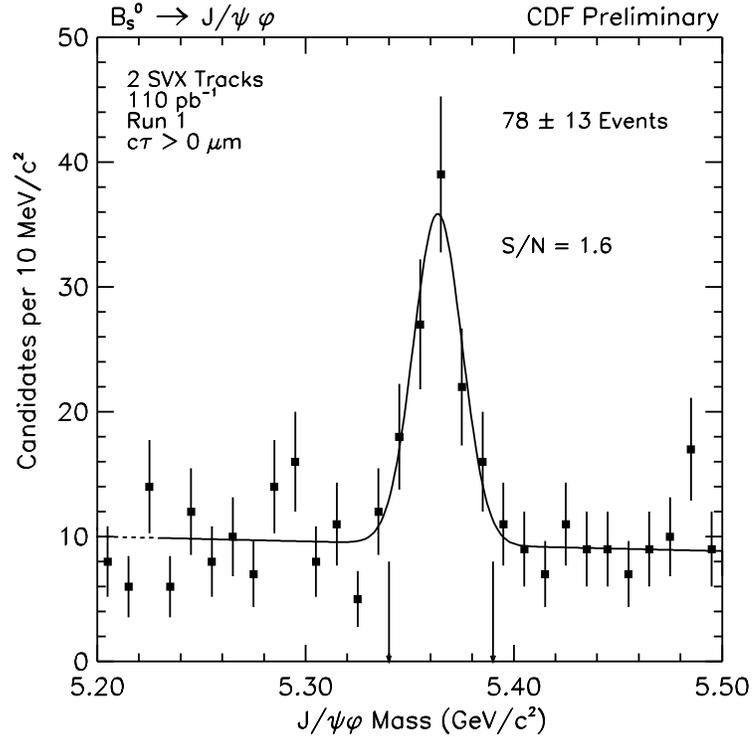}
}
\vspace*{0.3cm}
\caption[
Reconstructed $B_s^0 \ra J/\psi \phi$ events from CDF Run\,I data.]
{Reconstructed $B_s^0 \ra J/\psi \phi$ events from CDF Run\,I data with
positive $B_s^0$ lifetime. 
}
\label{psiphiyield}
\end{figure}

\boldmath
\subsection[$B^0 \rightarrow J/\psi K^0_S$: D\O\ Report]
{$B^0 \rightarrow J/\psi K^0_S$: D\O\ Report
$\!$\authorfootnote{Authors:  R.~Jesik and K.~Yip.}
}
\label{sec:d0-sin2beta}
\unboldmath
\index{decay!$B^0 \to J/\psi K_S^0$}%
\index{D\O\!$\sin 2\beta$ prospects}%

One of D{\O}'s primary physics goals is a measurement of $CP$~violation
in the golden mode $B^0 \rightarrow J/\psi K^0_S$, with $J/\psi\rightarrow
\mu^{+} \mu^{-}$ and $K^0_S\rightarrow\pi^{+} \pi^{-}$.
The measured asymmetry is defined by
\begin{equation}
{\cal A}_{CP} = \frac{ \Gamma ( \overline{B}^0 \rightarrow J/\psi  K^0_S )  -  \Gamma ( B^0 \rightarrow  J/\psi  K^0_S ) }
     { \Gamma ( \overline{B}^0  \rightarrow \ J/\psi  K^0_S )  +  \Gamma ( B^0  \rightarrow  J/\psi  K^0_S ) }.
\end{equation}
Measured as a function of time, the asymmetry is directly related to
the CKM angle~$\beta$:
\begin{equation}
{\cal A}_{CP}(t) =  \sin 2 \beta \cdot \sin\Delta m_d\, t.
\end{equation}
This measurement involves the full reconstruction of the final state,
the reconstruction of the primary and $B$ decay vertices, and a determination
of the $B$ flavour at production. The $J/\psi$~decay into dimuons
provides a 
relatively clean trigger signature. With D\O's upgraded muon scintillation
counter 
arrays, these events can be triggered on at the $30\%$ level (see Chapter 4).

This study is based on a sample of 10,000 Monte Carlo events generated by
Pythia plus QQ. The D\O\ detector response was obtained with a full GEANT simulation. An average of 1.1 additional minimum bias interactions were
added to the generated events. 
This sample was also analyzed using MCFAST 
for comparison. 

All of the four tracks comprising the candidate $B$~meson are required to
have a 
hit in each of the 16 layers of the Central Fiber Tracker (CFT). This 
effectively forces the tracks to be confined in the central rapidity range
$| \eta | < 1.6$. 
The tracks are also required to have at least 8 hits in the silicon
detector out of a maximum number of 10 hits possible on average.
The CFT hit requirement is dropped for the other tracks in the events.
These tracks, which are used for primary vertex finding and
flavour tagging, are reconstructed out to $|\eta | < 3.0$. 

The trigger for these events requires at least two oppositely charged
tracks in the muon system with matching tracks in the CFT 
with $p_T > 1.5$ GeV/$c$. The muon tracks must pass the track
quality cuts mentioned above during offline reconstruction, and the pair
must form a common vertex. 
The $J/\psi$ vertex defines the $B$~decay vertex in these events.
The reconstructed invariant mass of the muon pairs is shown in
Fig.~\ref{d0_MJpsi_MKs}(a). The momentum of combinations with a
reconstructed invariant 
mass within $3\,\sigma$ of the nominal $J/\psi$ mass is
re-determined in a kinematic fit with the $J/\psi$ mass constraint imposed.

\begin{figure}[tbp]
\centerline{
\put(30,200){\large\bf (a)}
\put(250,200){\large\bf (b)}
\epsfxsize=3in
\epsffile{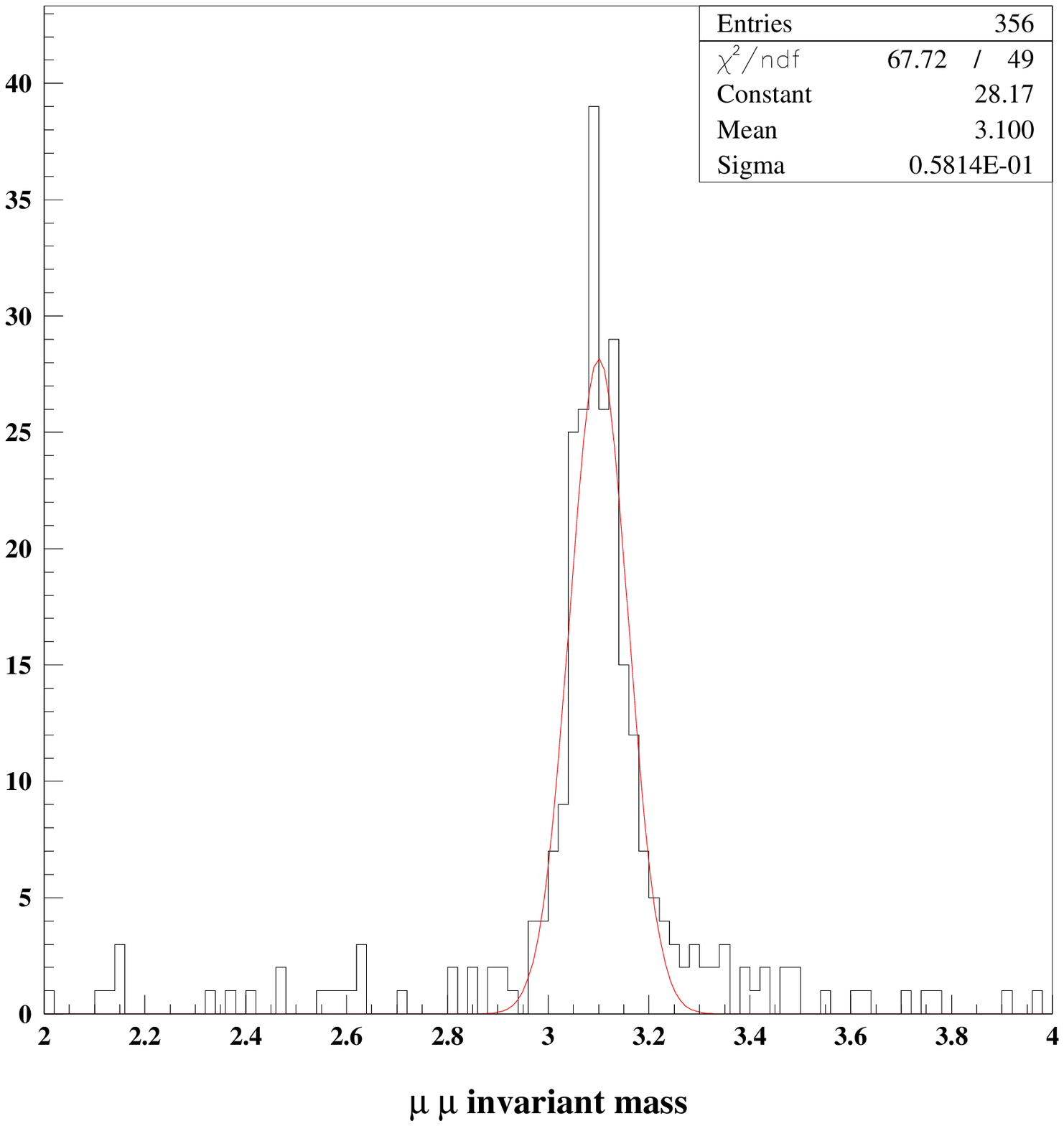}
\epsfxsize=3in
\epsffile{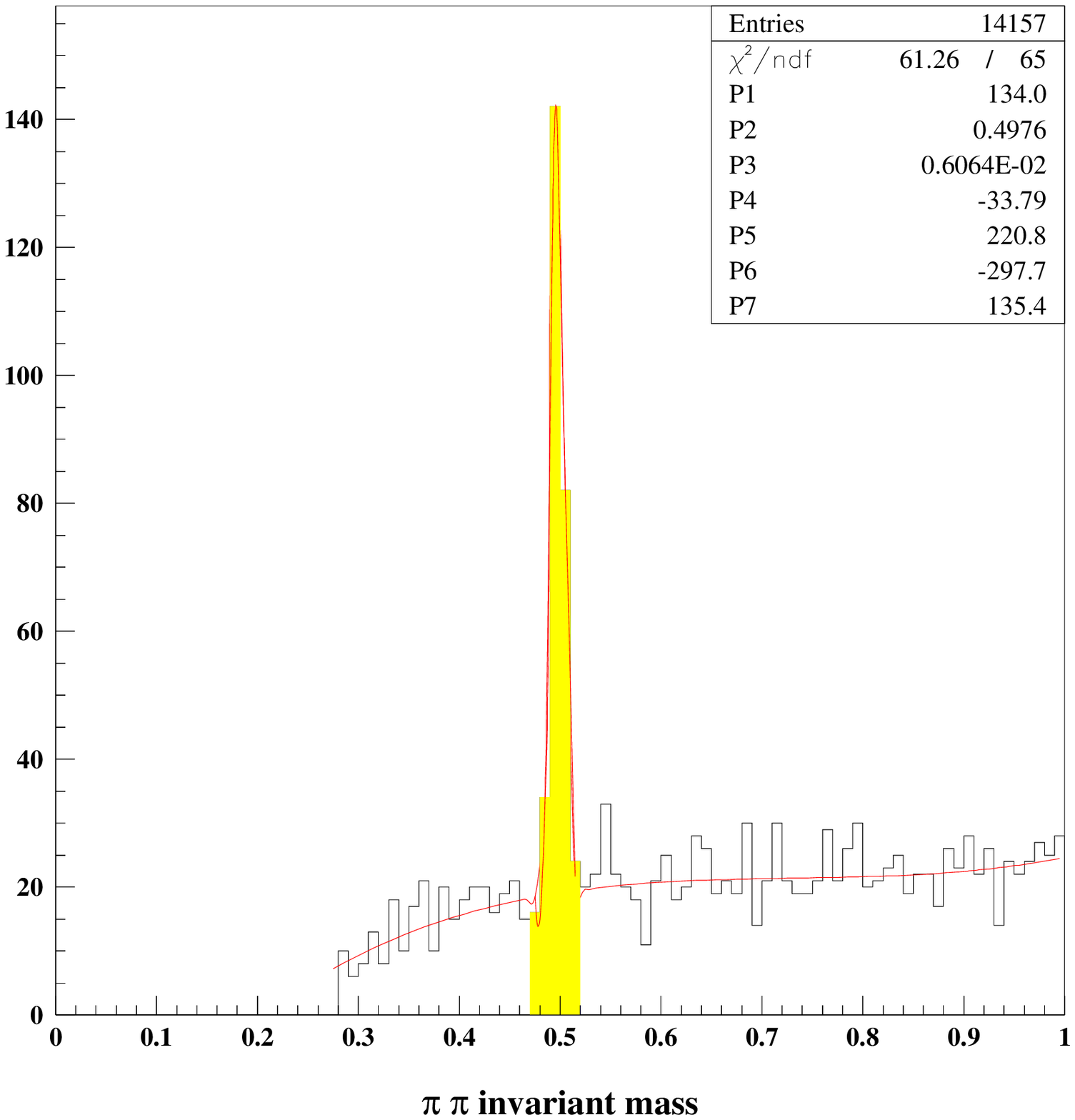}
}
\vspace*{0.3cm}
\caption[
Reconstructed $J/\psi\ra\mu\mu$ and 
$K^0_S \ra \pi^{+} \pi^{-}$ invariant mass in
          $B^0\rightarrow J/\psi K^0_S$ events.]
{Reconstructed (a) $J/\psi\ra\mu\mu$ and 
(b) $K^0_S \ra \pi^{+} \pi^{-}$ invariant mass in
          $B^0\rightarrow J/\psi K^0_S$ events.} 
\label{d0_MJpsi_MKs}
\end{figure}

The most difficult part of the analysis is the reconstruction of the two 
soft pions from the $K^0_S$ decay in the hadronic $p\bar p$~environment with a 
detector designed to do high $p_T$~physics. At present, D\O's
track finding software only reconstructs tracks with $p_T$ greater than 
0.5~\gevc. This is a stringent cutoff for $K^0_S$ detection. Lowering
this threshold has been shown to be viable for $B$ physics events.
It remains to be seen if it will be possible to lower the momentum
threshold for  
more complicated events, such as $t\bar{t}$.  Thus, we will use the default
cutoff of 0.5~GeV/$c$ for this study. 
$K^0_S$ candidates are formed by
combining pairs of oppositely charged tracks which do not point back to
the primary vertex -- an impact parameter significance of at least three is
required for each track. The track pairs are also required to form a
common vertex downstream of that of the $J/\psi$. The invariant mass
of these pairs (assuming they are pions) is shown in 
Fig.~\ref{d0_MJpsi_MKs}(b).
A clear $K^0_S$ peak is observed, and track pairs with a reconstructed mass
within $3\,\sigma$ of the actual $K^0_S$ mass undergo a kinematic fit
determining 
new momentum vectors after imposing the $K^0_S$ mass constraint. The
$K^0_S$ candidate's momentum vector is then required to point back to
the $J/\psi$ vertex to within $3\, \sigma$, and is combined with that of
the mass constraint $J/\psi$ to form the candidate $B$~momentum, which
is then required to point back to the primary vertex.  

The invariant mass spectrum of $B$ candidates which pass
these criteria is shown in Fig.~\ref{d0_MJpsiKs}. A clear signal is 
obtained with a width of about 10~MeV/$c^2$. The corresponding proper
decay time resolution is 90~fs.
We obtain a reconstruction efficiency for the entire decay chain
of $8.5\%$, resulting in 34,000 fully reconstructed 
$B^0\rightarrow J/\psi K^0_S$
($J/\psi \rightarrow \mu^+ \mu^-$, $K^0_S\rightarrow\pi^{+} \pi^{-}$) 
decays in 2 fb$^{-1}$ 
\index{D\O\!$\sin 2\beta$ prospects}%
(see Table \ref{expBJpsiKs}). For comparison, the
MCFAST study gives an efficiency of $10\%$.

\begin{figure}[tbp]
\centerline{
\epsfxsize=4.5in
\epsffile{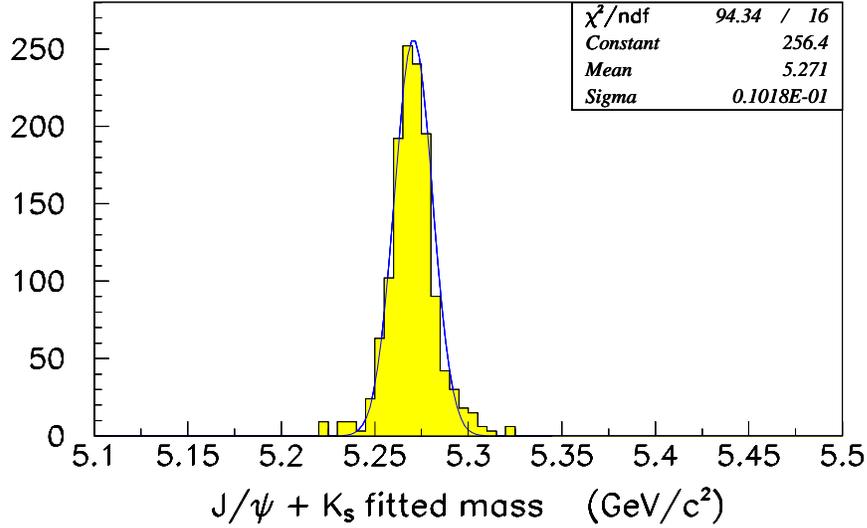}
}
\vspace*{0.3cm}
\caption[Reconstructed $B$~mass in $B\rightarrow J/\psi K^0_S$ events.]
{Reconstructed $B$~mass in $B\rightarrow J/\psi K^0_S$ events
          after mass and vertex constraints.}
\label{d0_MJpsiKs}
\end{figure}

\begin{table}[tb]
\begin{center}
\begin{tabular}{l|l} 
\hline
Integrated luminosity                        & 2 fb$^{-1}$ \\
$\sigma_{b\bar{b}}$                          & $158\,\mu$b  \\
$f(b\bar{b} \rightarrow B^0, \bar{B^0})$     & 0.8 \\
Kinematic acceptance                         & 0.31 \\
${\cal B}(B^0 \rightarrow \mu^+ \mu^- \pi^+\pi^-)$   & $2.0 \times 10^{-5}$ \\
Trigger efficiency                           & $0.30$  \\   
Reconstruction efficiency                    & 0.085  \\     
\hline
Number of reconstructed $B^0 \rightarrow J/\psi K^0_S$ &  34,000 \\
Effective tagging efficiency ($\eD$)  & 0.10 \\
\hline
 \end{tabular} 
\vspace*{0.2cm}
 \caption
[The expected number of $B^0 \rightarrow J/\psi K^0_S$ events at D\O.]
{The expected number of $B^0 \rightarrow J/\psi K^0_S$ events at D\O.}
 \label{expBJpsiKs}
 \end{center}
 \end{table}   
 
The other crucial element in this analysis is tagging the initial
flavour of the decaying $B^0$~meson. One method for doing this makes use of the
correlation between the charge of a nearby pion and the $B$ flavour due
to fragmentation or $B^{**}$ production.  This requires the
reconstruction of soft pions from the primary vertex.  Two other methods 
use information from the other $B$~hadron in the event. If the $B$ decays 
semileptonically, its flavour is determined by the charge of the lepton. 
If not, its flavour can be determined by the $p_T$ weighted net charge of
its jet. The effectiveness of a tagging method is quantified by
the effective tagging efficiency $\eD$, 
where $\varepsilon$ is the tagging efficiency and 
$\cal D$ is the dilution factor. $\cal D$ is equal 
to $2 P - 1$, where $P$ is the probability that the method tags the $B$~flavour
correctly. Extrapolating from the effective tagging efficiencies measured
by CDF in Run\,I (see Section~\ref{sec:cdf-sin2beta}), 
D\O\ expects to achieve an effective tagging
efficiency of $\eD \sim 10\%$. The breakdown of the effective
tagging efficiency for each of the flavour tagging methods is shown in
\index{D\O\!flavor tagging}%
Table \ref{flavtag}. The increase over CDF Run\,I efficiencies is primarily
due to D{\O}'s extended rapidity range for tracking and lepton
identification.     

\begin{table}[tb]
\begin{center}
\begin{tabular}{l|l|c} 
\hline
Flavour tag     & $\eD$ CDF Run\,I      & $\eD$ D\O\ Run\,II \\
\hline
Same side tag  & $(1.8 \pm 0.4 \pm 0.3)\%$    & $2.0\%$ \\ 
Jet charge tag & $(0.78 \pm 0.12 \pm 0.08)\%$ & $3.1\%$ \\
Lepton tag     & $(0.91 \pm 0.10 \pm 0.11)\%$ & $4.7\%$ \\
\hline
 \end{tabular} 
\vspace*{0.3cm}
 \caption
[Summary of flavour tagging methods at D\O.]
{Summary of flavour tagging methods at D\O.}
 \label{flavtag}
 \end{center}
 \end{table}   

The accuracy of a time dependent $\sin 2 \beta$ measurement is given by:
\begin{equation}
 \sigma ( \sin 2 \beta ) \approx 
{\rm e}^{x^2_d {\Gamma}^2 {\sigma}^2_t}\,
	\sqrt{ \frac{1+4 x^2_d}{2 x_d^2} }\,
	\frac{1}{\sqrt{\eD\,N}} \sqrt{ 1 + \frac{B}{S} }\,,
\end{equation}
where $x_d$ and $\Gamma$ are the mixing parameter and decay width of the 
$B^0$, ${\sigma}_t$ is the proper time resolution 
(which is about 90~fs),
$N$ is the number of reconstructed signal events, 
and ${S}/{B}$ is the signal to background ratio
(extracted from Run\,I data to be about 0.75). With these considerations,
D\O\ will be able to measure $\sin 2 \beta$ in the dimuon mode with an
uncertainty of 0.04 in 2 fb$^{-1}$ of data. Similar accuracy will be
achieved in the dielectron mode. This precision is quite competitive with
CDF's projections and both experiments will reach $B$~factory 
sensitivities with further data taking.

Similarly, D\O\ will look for $CP$~violation in $B_s^0 \rightarrow J/\psi \phi$
decays. D\O\ expects a sample of 1400 fully reconstructed events in \tfb\
in Run\,II.
Although the expected Standard Model asymmetry in this channel is not
within our experimental reach, an observation would be a clear 
signal of new physics.

\boldmath
\subsection[$B^0 \rightarrow J/\psi K^0_S$: BTeV Report]
{$B^0 \rightarrow J/\psi K^0_S$: BTeV Report
$\!$\authorfootnote{Authors: P.A.~Kasper and R.~Kutschke.}
}
\label{sec:btev-sin2beta}
\unboldmath
\index{decay!$B^0 \to J/\psi K_S^0$}%
\index{BTeV!$\sin 2\beta$ prospects}%

As discussed in Section~\ref{ch6:intro},
the decay $B^{0} \rightarrow J/\psi K^0_S$ is the golden mode
for measuring the angle $\beta$ of the unitarity triangle.
While $\sin\,2\beta\,$ has been measured before the
BTeV experiment begins operation, 
the collaboration aims to significantly improve that measurement.
This section will present the reconstruction efficiency, trigger 
efficiency and signal to background ratio for the decay chain
$B^{0} \rightarrow J/\psi K^0_{S}$,
$J/\psi\rightarrow \mu^{+} \mu^{-}$ and $K^0_S\rightarrow\pi^{+} \pi^{-}$.

For this study, Monte Carlo events were
generated using Pythia and QQ 
and the detector
response was simulated using BTeVGeant.
The output of BTeVGeant was analyzed as would be real data. 
When designing analysis cuts, it is important to understand both the
efficiency of the cuts on signal events and the
power of the cuts to reject background.  Because of the narrow
widths of the $J/\psi$ and the $K^0_S$, the dominant source of 
background entries is combinations of real $J/\psi\to \mu^+\mu^-$
decays with real $K^0_S\to\pi^+\pi^-$ decays.
CDF found that prompt $J/\psi$'s constitute a large fraction of the 
total $J/\psi$ 
production \cite{cdf_prompt} and, extrapolating from their results,
one expects that $J/\psi$'s from $B$~decays comprise only about 5\% of 
the total $J/\psi$ production including the regions of high pseudorapidity.
However, the background from 
prompt $J/\psi$ production is strongly suppressed by the topological
cuts, leaving decays of the type $b\rightarrow J/\psi X$ as the
dominant source of background.

The analysis was performed as follows.
Each event was required to have an identified primary vertex that
was successfully fitted.
A track was identified as a muon candidate provided the Monte Carlo 
truth table indicated that it was a muon, it had a momentum of
more than 5.0~GeV$/c$ and it had a  hit in the most
downstream muon detector.
$J/\psi$ candidates were formed by combining pairs of oppositely 
charged muon candidates and requiring that the invariant mass of 
the $\mu^+\mu^-$ pair be within $3\,\sigma$ of the known mass of the $J/\psi$.
It was also required that the $\mu^+\mu^-$
pair pass a fit to a common vertex and 
the vertex be detached from the primary
vertex by at least $L/\sigma_L>4$, where $L$ is the distance
between the two vertices and $\sigma_L$ is the error on $L$.
As illustrated in Figure~\ref{fig:psiks_los}, this
cut rejects 99.95\% of the background from 
prompt $J/\psi$'s while keeping 80\% of the signal. 
A fit was performed to constrain the $\mu\mu$ mass to that of the $J/\psi$.
\begin{figure}[t]
\centerline{\epsfxsize=6.in\epsffile{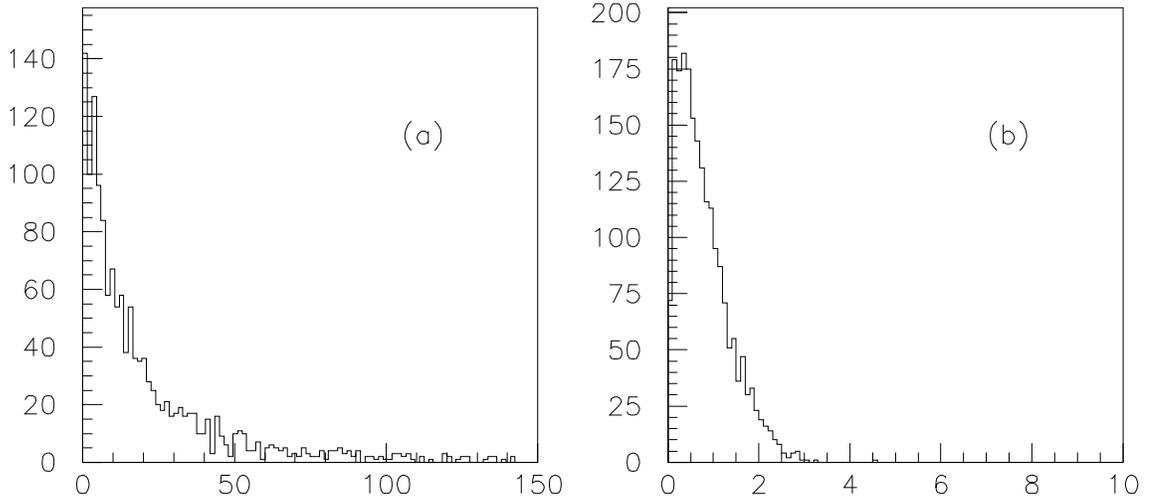}}
\vspace*{0.2cm}
\caption
[Distributions of $L/\sigma_L$ for $J/\psi$ candidates
         from the decays of $b$~hadrons
         and prompt $J/\psi$ candidates.]
{Distributions of $L/\sigma_L$ for (a) $J/\psi$ candidates
         from the decays of $b$~hadrons
         and (b) prompt $J/\psi$ candidates.  The prompt candidates
         are suppressed by requiring $L/\sigma_L>4$.}
\label{fig:psiks_los}
\end{figure}

All other tracks with a momentum of at least 0.5~GeV/$c$ were accepted
as pion candidates, provided they missed the primary vertex by
$d>3\,\sigma_d$, where $d$ is the impact parameter between the track 
and the primary vertex, while $\sigma_d$ is the error on $d$.
$K^0_S$ candidates were selected by combining oppositely charged
pairs of pion candidates and requiring 
that the $\pi^+\pi^-$ invariant mass be within $3\,\sigma$ of the known
$K^0_S$ mass.
It was also required that $K^0_S$ candidates pass a fit to a common 
vertex.
Finally, the mass of the $K^0_S$ 
candidate was constrained to that of the known $K^0_S$ table mass. 

A $B^0$ candidate was defined as the combination of a 
$J/\psi$ candidate and a $K^0_S$ candidate which pointed back to
the primary vertex.
To reduce combinatorial background, it was required
that the $K^0_S$ candidate points back to the $J/\psi$ vertex 
within $3\,\sigma$ and that 
the $K^0_S$ impact parameter with respect to the $J/\psi$ vertex divided 
by its impact parameter with respect to the primary vertex be less than 2.0. 

The invariant mass spectrum of $B$ candidates which pass
the above criteria is shown in Figure~\ref{fig:B_psiks_mass}.
A clear signal with a width of  $\sigma=9.3~\mevcc$
is seen at the mass of the $B^0$.  The efficiency for
a $B^0\to J/\psi K^0_S$ decay to fall into the mass peak is
$0.040\pm0.002$ and the mean resolution on the proper decay time
is 40-50~fs.
\begin{figure}[t]
\centerline{\epsfxsize=3.5in\epsffile{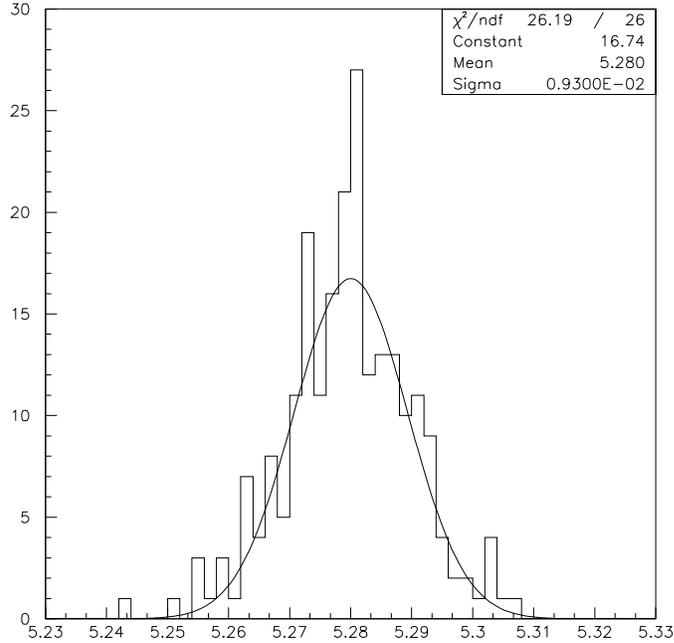}}
\vspace*{0.2cm}
\caption[The $J/\psi K^0_S$ invariant mass distribution at BTeV.]
{The $J/\psi K^0_S$ invariant mass distribution for
         candidates which survive the selection criteria described in
         the text.}
\label{fig:B_psiks_mass}
\end{figure}

As mentioned above, the dominant source of background arises from
decays of the type $b\rightarrow J/\psi X$.
This background was studied by generating large samples of such decays, 
using Pythia and QQ. These samples were passed through the MCFast based 
detector simulation and analyzed as real data.
This study predicted that the signal 
to background ratio in this channel is approximately $S/B=10$.

The BTeV trigger simulation (see Sec.~5.4.3)
was run on events which passed 
the analysis cuts, and the Level\,1 trigger 
was found to have an efficiency of 
$(52\pm3)$\%.
This decay mode can also be triggered by muon and dimuon triggers
with an estimated trigger efficiency of 50\%. 
%
Furthermore, it is estimated that the combined Level\,2 
trigger efficiency is 90\%.  

In Section~5.5, it is estimated that the effective tagging efficiency~\eD\
for $B^0$ decays is 0.10.
There are two methods which can be used to extract
$\sin2\beta$ from the reconstructed, tagged $J/\psi K^0_S$
candidates, a time integrated method and a time dependent method.
The sensitivity of the time integrated method is given by,
\begin{equation}
\sigma(\sin2\beta) = \frac{1+x_d^2}{x_d}
                         \sqrt{\frac{1}{\eD N}}
                         \sqrt{\frac{S+B}{S}}\,,
\end{equation}
while the sensitivity of the time dependent method is given by,
\begin{equation}
\sigma ( \sin 2 \beta ) \approx 
{\rm e}^{x^2_d {\Gamma_d}^2 {\sigma}^2_t}
                         \sqrt{\frac{1+4x_d^2}{2x_d^2}}
                         \sqrt{\frac{1}{\eD N}}
                         \sqrt{\frac{S+B}{S}}\,,
\end{equation}
where $N$ is the number of tagged decays,
$x_d=0.723\pm0.032$~\cite{pdg98} is the $B^0$ mixing parameter, 
$\sigma_t$ is the
resolution on the proper decay time and where
$\Gamma_d=(0.641\pm0.016)\times 10^{12}\, \rm{s}^{-1}$~\cite{pdg98}
is the natural width of the $B^0$.  For the $B^0$, the time dependent 
method yields a sensitivity which is about 20\% better than
that given by the time integrated method.
In previous documents the BTeV collaboration has reported the 
sensitivity on $\sin2\beta$ using the time integrated method
but in this document the time dependent method will be quoted.
The above discussion is summarized in
\index{BTeV!$\sin 2\beta$ prospects}%
Table~\ref{tab:psiks_sin2b}
which reports a sensitivity of $\sigma(\sin2\beta)= 0.025$.

\begin{table}[tb]
\begin{center}
\begin{tabular}[t]{c|c}
\hline\hline
Luminosity                     & $2\times 10^{32}\,{\rm cm}^{-2}\,s^{-1}$  \\
Running time                                 & $ 10^7$ s         \\
$\sigma_{b\bar{b}}$                     & $100\,\mu$b  \\
Number of $B\bar{B}$ events             & $2\times 10^{11}$ \\
${\cal B}(\bar{b}\to B^0)$                   & 0.4 \\
Number of $B^0$ or $\bar B^0$                 & $1.6\times 10^{11}$  \\
${\cal B}$($B^0 \rightarrow J/\psi K^0_S$) & $4.45\times 10^{-4}$ \\
${\cal B}(J/\psi\rightarrow \mu^+\mu^-)$     & 0.061 \\
${\cal B}(K^0_S\rightarrow \pi^+\pi^-)$      & 0.6861 \\
$\epsilon$(Geometric + Cuts)                 & 0.04 \\
Level\,1 Trigger efficiency                   & 0.75 \\
Level\,2 Trigger efficiency                   & 0.90 \\
Number of reconstructed $B^0 \rightarrow J/\psi K^0_S$ &  80,500  \\
Tagging efficiency $\eD$                       &  10\%  \\
$S/B$                                                    &  10 \\ 
Resolution on proper decay time                          & 0.043~ps \\ 
\hline
$\sigma({\sin2\,\beta})$, time integrated                & 0.030 \\ 
$\sigma({\sin2\,\beta})$, time dependent                 & 0.025 \\  
\hline\hline
\end{tabular}
\vspace*{0.2cm}
\caption
[Summary of the
         sensitivity to $\sin2\,\beta$ using
         $B\rightarrow J/\psi K^0_{s}$ at BTeV.]
{Summary of the
         sensitivity to $\sin2\,\beta$ using
         $B\rightarrow J/\psi K^0_{s}$ at BTeV.}
\label{tab:psiks_sin2b}
\end{center}
\end{table}

\boldmath
\subsection[$B^0 \rightarrow J/\psi K^0_S$: Summary]
{$B^0 \rightarrow J/\psi K^0_S$: Summary
$\!$\authorfootnote{Author: M.~Paulini.}
}
\unboldmath
\index{decay!$B^0 \to J/\psi K_S^0$}%

The main goal of measuring the $CP$~violating asymmetry in the so-called
golden-plated decay mode $B^0\ra J/\psi K^0_S$ is to determine the phase
$\beta$ within the
Standard Model. It is given in terms of CKM matrix elements as 
$\beta\equiv\arg (-{V_{cd}V_{cb}^* / V_{td}V_{tb}^*})$.
Evaluating the sensitivity of the Tevatron experiments towards
measuring \stb\ was motivated by using $B^0\ra J/\psi K^0_S$ as a benchmark
process for all three experiments and as a comparison with the expectations
of the $B$~factories.

With \tfb\ of integrated luminosity, CDF expects to reconstruct 
20,000 $B^0 \ra \jpks$ events with $J/\psi \ra \mu^+\mu^-$ and 
$K^0_S \ra \pi^+\pi^-$, a net increase of a factor
of $\sim 50$ compared to the $J/\psi K_S^0$ yield in Run\,I.
Assuming a total effective tagging efficiency of
$\eD\sim\!9.1\%$, 
as discussed in Sec.~\ref{sec:cdf-sin2beta}, 
this results in an
error on a measurement of \stb\ of 
$\sigma(\stb)\sim\!0.05$ at CDF.
The D\O\ experiment expects to measure \stb\ with similar precision.
D\O\ will reconstruct about
34,000 $B^0 \ra \jpks$ events with $J/\psi \ra \mu^+\mu^-$ and 
$K^0_S \ra \pi^+\pi^-$ in \tfb.
D\O\ uses a total effective tagging efficiency of
$\eD\sim\!10\%$ derived from CDF's Run\,I experience of $B$~flavour tagging
(see Sec.~\ref{sec:cdf-sin2beta}). 
This gives D\O\ an uncertainty of 
$\sigma(\stb)\sim\!0.04$.

While $\sin\,2\beta\,$ will have been measured before the
BTeV experiment will turn on, the goal of the BTeV 
collaboration is to significantly improve the precision of that
measurement. Within one year of running at design luminosity, BTeV 
expects to reconstruct about 
80,000 $B^0 \ra \jpks$ events with $J/\psi \ra \mu^+\mu^-$ and 
$K^0_S \ra \pi^+\pi^-$. Together with an 
effective tagging efficiency of
$\eD\sim\!10\%$, 
as discussed in Sec.~\ref{sec:btev-sin2beta}, 
this will allow BTeV to measure \stb\ with an error of 
$\sigma(\stb)\sim\!0.025$. At that point in time, the $B$~physics community
will clearly have entered the area of precision CKM measurements.

\boldmath
\section{Study of $B \rightarrow \pi\pi/KK$}
\unboldmath

\boldmath
\subsection[$B \rightarrow \pi\pi/KK$: Introduction]
{$B \rightarrow \pi\pi/KK$: Introduction
$\!$\authorfootnote{Author: F.~W\"urthwein.}
}
\label{ch6:pipi_intro}
\unboldmath
\index{decay!$B^0 \to \pi \pi$}%

One of the key physics goals of Run\,II is the study of $CP$~violation in 
$B$~meson decays. At the time the CDF Technical Design 
Report~\cite{tdr} was written,
the most important decay modes were
believed to be \bjpks\ and \bpipi.
Time dependent $CP$~violation in the former mode measures \stb~\cite{bisa},
while
the decay $B^0\to\pi^+\pi^-$ usually appears in the literature as 
a tool to determine $\alpha=180^\circ-\beta-\gamma$. 
Using standard phase conventions, $\beta$ and $\gamma$ are the phases of the
CKM matrix elements $V_{td}^*$ and $V_{ub}^*$, respectively. 

As discussed in Section~\ref{sec:cpintropenguin} from a theoretical aspect, 
penguin contributions are expected to affect the
determination of $\alpha$ severely~\cite{alph}.
Experimentally,
the CLEO collaboration~\cite{cleobranch} has shown that 
\index{penguins!pollution}%
"penguin pollution" in \bpipi\ is sufficiently large to make the
extraction of fundamental physics parameters from the measured
$CP$~asymmetry rather difficult.  
Any evaluation of the physics reach in measuring $CP$~violation in \bpipi\
does therefore 
require a strategy to dis-entangle ``penguin'' contributions  from 
``tree'' diagrams in order to lead
to a meaningful measurement of short distance physics.

Figure~\ref{fig:pengtree} shows the two dominant Feynman diagrams in
charmless hadronic $B$ decays contributing to $B^0\to\pi^+\pi^-$ and
$B_s^0\to K^+K^-$.  
Simple counting of vertex factors indicates
that $b\to s \bar{u}u$ ``penguin'' and  $b\to u \bar{u}d$ ``tree'' transitions
are roughly of the same magnitude, while 
$b\to d \bar{u}u$ ``penguin'' and  $b\to u \bar{u}s$ ``tree'' transitions
are suppressed by ${\cal O}(\lambda )$ with respect to these dominant amplitudes. 
Defining $\Delta S$ as the change in strangeness quantum number,
it is thus expected that transitions with $\Delta S = 0$ are dominated
by external $W$-emission (``tree'') decays. In contrast, $\Delta S =1$
transitions generally receive their dominant contributions from gluonic
penguin decays.  


\begin{figure}[tb]
\centerline{
\epsfxsize=4in
\epsffile{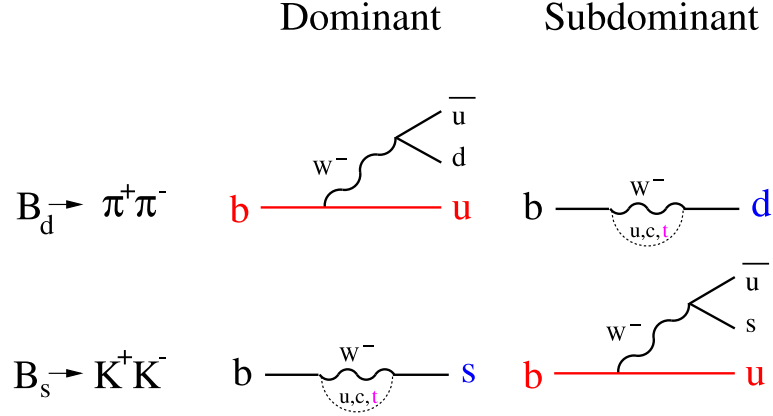} 
}
\vspace*{0.3cm}
\caption
[Feynman diagrams in charmless hadronic $B$~meson decays
contributing to $B^0\to\pi^+\pi^-$ and $B_s^0\to K^+K^-$.]
{Feynman diagrams in charmless hadronic $B$~meson decays
contributing to $B^0\to\pi^+\pi^-$ and $B_s^0\to K^+K^-$. 
}
\label{fig:pengtree}
\end{figure}

A large number of
strategies to disentangle penguin and tree contributions can be found in
the literature~\cite{revs,alph}.
However, they generally require either very large data sets
or involve hard to quantify theoretical uncertainties.
In the following, we evaluate a strategy of measuring the 
CKM angle~$\gamma$~\cite{flekk} which is
particularly well matched to the capabilities of the Tevatron as 
it relates $CP$~violating 
observables in \bskk\ and \bpipi.
Combining the $CP$~violating observables in these two decays
with the $CP$~violation measured in $B^0\to J/\psi K^0_s$ allows
for a measurement of $\gamma$ up to a fourfold ambiguity.
The
utility of $B_s^0\to K^+K^-$ to probe $\gamma$ was already pointed out
in several previous publications \cite{BsKK}, and the use
of $CP$~violating asymmetries in $B$ to $K^\pm\pi^\mp$ decays is
discussed in Ref.~\cite{gammaFromDirectCP}.

The decays $B^0\to\pi^+\pi^-$ and 
$B_s^0\to K^+K^-$ are related to each other by interchanging all
down and strange quarks, i.e.\ through the so-called ``U-spin''
subgroup of the SU(3) flavour symmetry of strong interactions. 
The strategy proposed in Ref.~\cite{flekk} uses this
symmetry to relate the ratio of hadronic matrix elements
for the penguin and tree contributions, and thus uses \bskk\ to correct for the
penguin pollution in \bpipi.

This strategy
does not rely on certain ``plausible'' dynamical or 
model-dependent assumptions, nor are final-state interaction effects 
\cite{FSI} of any concern. These led to considerable attention in the
recent literature  
on measuring $\gamma$ from $B\to\pi K$ decays 
\cite{BpiK}.
The theoretical accuracy is
only limited by U-spin-breaking effects.
We evaluate the likely size of these effects and find them to be small
compared to the expected experimental error on $\gamma$ in Run\,II.

\boldmath
\subsection[$B \rightarrow \pi\pi/KK$: CDF Report]
{$B \rightarrow \pi\pi/KK$: CDF Report
$\!$\authorfootnote{Author: F.~W\"urthwein.}
}
\unboldmath
\index{decay!$B^0 \to \pi \pi$}%
\index{CDF!$B^0 \to \pi \pi$ prospects}%
\index{CDF!$\sin\gamma$ prospects}%

\subsubsection{Trigger Issues}
\label{sec:cdf_pipi_trigger}
\index{CDF!hadronic trigger}%

The key to measuring the $CP$~asymmetry in $B^0 \ra \pi^+\pi^-$ is to
trigger on this decay mode in hadronic collisions. 
CDF will do this with its three level trigger system where the
through-put of each level will be increased by
more than an order of magnitude from the Run\,I trigger scheme
to accommodate the shorter $p\bar{p}$ crossing interval
(initially 396~ns and later in Run\,II 132~ns), and the increase in
instantaneous luminosity by one order of magnitude. 
The maximum output of Level\,1 and Level\,2 will be 50~kHz and 300~Hz,
respectively. The trigger rates presented in the following have been
studied using minimum bias data 
for Level\,1 and data sets collected with specialized test triggers taken
during Run\,Ib for Level\,2.

At Level\,1, two oppositely charged tracks found by the XFT track
processor~\cite{tdr} are used.
The XFT can find tracks of $\Pt >1.5$~\gevc\ that traverse the full radius
of the COT with a momentum resolution $\Delta \Pt/\Pt^2<0.015~(\gevc)^{-1}$
and an azimuthal resolution at superlayer~6 ($r=106$~cm)
of $\Delta\phi_6<0.0015$~rad.
The two-track module compares the values of \Pt\ and $\phi_6$ from all pairs
of tracks to valid trigger patterns in a lookup table.
Three sets of two-track trigger criteria~\cite{pac_jan} are listed in
Table~\ref{trig_rates} 
corresponding to three possible operating conditions
of the Tevatron. Scenarios A, B and C
cover the possible bunch separations ($T_{\rm bunch}$),
instantaneous luminosity ($\cal{L}$) and mean number of interactions
per crossing ($\langle N_{p\bar{p}}\rangle$). The Level\,1 trigger
cross sections are listed in Table~\ref{trig_rates}. CDF expects to
allocate a maximum of 30~kHz to the two-track trigger at Level\,1. 

\begin{table}[tb]
\begin{center}
\begin{tabular}{c|ccccccc}
\hline
Scenario & $T_{\rm bunch}$  & $\cal{L}$  & $\langle N_{p\bar{p}}\rangle$ &
 L1 cross & L1 rate & L2 cross & L2 rate \\
          & [ns] & [cm$^{-2}$s$^{-1}$]  & & 
 section [$\mu$b]     & [kHz] &  section [nb] &  [Hz] \\
\hline
 A & 396 & $0.7\times10^{32}$ & 2 & $252\pm18$ & 18  & $360\pm100$ & 25 \\
 B & 132 & $2.0\times10^{32}$ & 2 & $152\pm14$ & 30  & $196\pm74$  & 39 \\
 C & 396 & $1.7\times10^{32}$ & 5 & $163\pm16$ & 28  & $ 84\pm48$  & 14 \\
\hline
\end{tabular}
\vspace*{0.3cm}
\caption
[Level\,1 trigger criteria and event rates 
for three operating scenarios of the Tevatron.]
{Level\,1 trigger criteria and event rates as well as Level\,2 trigger cross
sections and event rates for three operating
scenarios of the Tevatron during Run\,II~\cite{pac_jan}.} 
\label{trig_rates}
\end{center}
\end{table}

At Level\,2, CDF uses the SVT~\cite{tdr}, which associates clusters formed from
axial strips 
in the SVX\,II with tracks of $\Pt>2$~\gevc\ found by the XFT.
This provides a measurement of the impact parameter
of the track in the plane transverse to the beam axis.
This measurement is sufficiently precise to
resolve the true large impact parameters of tracks coming
from the decays of heavy flavour from the impact parameters of
tracks originating from QCD jets, which have non-zero impact parameter
only due to measurement resolution. The assumed impact parameter
resolutions for the SVT~\cite{pac_jan} are $\sigma_d =
(19+40~\gevc/\Pt)~\mu$m   
for tracks that miss the hybrid in Layer\,0 of SVX\,II and
$\sigma_d = (19+80~\gevc/\Pt)~\mu$m 
for tracks that pass through the hybrid in Layer\,0.
The expected Level\,2 trigger rates are
given in Table~\ref{trig_rates} and are well below the total Level\,2
bandwidth of 300~Hz. At Level\,3, the full event information 
is available further reducing the trigger rate.

The data collection of \Bs~decay modes for the measurement of \Bs\
flavour oscillations at CDF in Run\,II is also based on the two-track
hadronic trigger. The Level\,1 two-track
trigger scheme is the same as for $B^0 \ra \pi\pi$ as summarized in
Table~\ref{trig_rates}. The Level\,2 trigger selection requirements have
been slightly adjusted~\cite{pac_jan} to achieve a better efficiency for
triggering on any two tracks from the hadronic \Bs~decay (see also
Section~8.6 and in particular Sec.~8.6.2).

\subsubsection{Expected Rates}

As discussed in Section~\ref{ch6:pipi_intro}, the decays $B^0\to K^+\pi^-$
and \bskk\ are $\Delta S = 1$ transitions, 
and are expected to be dominated
by gluonic penguin decays. In contrast, $\bpipi$ and 
$\bstopik$ are expected to receive their dominant contributions from 
external $W$-emission (``tree''). 
For the decays 
\index{decay!$B^0 \to KK$}%
$B^0\to K^+K^-$ and 
\index{decay!$B_s^0 \to \pi \pi$}%
$\Bs\to\pi^+\pi^-$ 
neither of the initial quarks is present in the final state.
These transitions are thus expected to be highly suppressed
as they require either $W$-exchange
or in-elastic final state re-scattering.

Experimental information on these decays comes from the CLEO
experiment~\cite{cleobranch}. They measured 
\index{decay!$B^0 \to K \pi$}%
${\cal B}(B^0\to K^+\pi^-) = (17.2^{+2.5}_{-2.4}\pm 1.2) \times 10^{-6}$,
\index{decay!$B^0 \to \pi \pi$}%
${\cal B}(B^0\to \pi^+\pi^-) = (4.3^{+1.6}_{-1.4}\pm 0.5)\times 10^{-6}$,
and ${\cal B}(B^0\to K^+K^-) < 1.9\times 10^{-6}\ \mathrm{at}\ 90\%$ Confidence
Level.
Average over charge conjugate decays is implied in all three of these
measurements. In addition, CLEO measured
$[{\cal B}(B^0\to K^+\pi^-) - {\cal B}\bar B^0\to K^-\pi^+)]
/ 
[{\cal B}(B^0\to K^+\pi^-) + {\cal B}(\bar B^0\to K^-\pi^+)]
= -0.04 \pm 0.16 $~\cite{cleoacp}.
More recent results from BaBar and Belle~\cite{BaBarBelle} might point
towards a more favorable ratio of $\btopipi/\btokpi$.
To be conservative, we base our projections on the published CLEO
numbers~\cite{cleobranch}. 

The corresponding \Bs\ decays have not been observed. However, we can make
an educated guess regarding their branching fractions by
assuming $SU(3)$ flavour symmetry as follows:
\begin{eqnarray}
{\cal B}(\Bs\to K^+K^-) &=& (F_K/F_\pi )^2\times {\cal B}(B^0\to K^+\pi^-)\,,
  \nonumber\\
{\cal B}(\Bs\to \pi^+K^-) &=& (F_K/F_\pi )^2\times{\cal B}(B^0\to \pi^+\pi^-)\,.
\end{eqnarray}
The factor  $(F_K/F_\pi )^2$ accounts for $SU(3)$ breaking.
Assuming factorization
$F_K (F_\pi)$ is given by the $B\to K (B\to\pi)$ form factor, and thus
 $(F_K/F_\pi )^2 \sim 1.3$.
Taking into account the production ratio of $f_s/f_d \sim 0.4$~\cite{cdf_frag},
we expect the following relative yields: 
\beq
(B^0\to K\pi) : (B^0\to\pi\pi) : (\Bs\to KK) : (\Bs\to \pi K)\ \sim\ 
4\ :\ 1\ :\ 2\ :\ 0.5.
\label{ch6:pipiratio}
\eeq
The $B^0 \ra \pi^+\pi^-$ signal yield is obtained from Monte Carlo
simulation taken from Ref.~\cite{pac_jan}. We rescale the yield cited there
by the CLEO branching fractions quoted above and the updated measurement of
the $B$ cross section $\sigma_B = (3.35\pm 0.46\pm 0.50)~\mu$b~\cite{todcross}
using fully reconstructed $B^+\to J\psi K^+$ decays.
From this estimate, CDF expects 5060 to 9160 fully reconstructed \bpipi\
events in 
\tfb. To be conservative, we choose
5000 $B^0\to\pi^+\pi^-$ and 
20,000 $B^0\to K^+\pi^-$ events for this study.
With the event ratio given in Eq.~(\ref{ch6:pipiratio}), we
arrive at an 
expected $B_s^0\to K^+K^-$ and $\pi^+K^-$ yield of
10,000 and 2500 events, respectively. Yields in the two $K\pi$ final states
refer to the sum of $K^+\pi^-$ and $K^-\pi^+$.

To answer the question whether
CDF will be able to extract these large signals from potentially enormous
backgrounds, we discuss physics backgrounds such as $B\ra K\pi$ and
combinatorial background separately. A study using specialized test trigger
data, described in Ref.~\cite{pipibg}, addresses the issue of
combinatorial background. This study finds two events in
a region of $\pm 500$~\mevcc\ around the nominal $B$~mass.
Based on trigger simulations and the branching fractions listed above, CDF
expects 0.08 signal events in the
sum of all two track decays of the $B^0$ and $B_s^0$ within a signal
window of $\pm 50$~\mevcc\ around the nominal $B$~mass. From this we
conclude a signal-to-background ratio ($S/B$) not worse than 0.4.

Based on the measured cross sections and Monte Carlo simulation of
the trigger efficiency for generic $B$ decays, CDF expects that roughly 1/4
of the two-track hadronic trigger rate is from $b\bar b$ and $c\bar c$ each. 
Backgrounds from these two sources result in a two-track invariant mass
spectrum far away from the $B$~signal region. We thus expect the dominant
backgrounds to come from mis-measured tracks without true lifetime.
Detailed studies of this type of background can only be done
once data with the new Run\,II silicon detector is available. However, 
it is not unreasonable 
to expect the 3-dimensional vertexing capabilities of SVX\,II to improve upon
the $S/B$ of 0.4 obtained from the Run\,I estimates.

\boldmath
\subsubsection{Disentangling $\pi\pi$, $K\pi$, $KK$ and $\pi K$ Final States}
\label{sec:cdf-pipi-untagged}
\unboldmath

Figure~\ref{fig:mball}(a) shows the expected invariant mass peaks for
20,000 $B^0\to K^\pm\pi^\mp$, 5000 $B^0\to\pi^+\pi^-$, 10,000 $\Bs\to
K^+K^-$ and 
2500 $\Bs\to K^\mp\pi^\pm$, on top of 56250 events of combinatorial
background.  
In each case the pion mass is used to calculate the track energy.
The four mass peaks are not particularly distinct and are shown separately
in Figure~\ref{fig:mball}(b). 
This initial simulation indicates a $\pi\pi$ invariant mass resolution of
about 25~\mevcc. 
The flat background generated is equivalent to a signal/background ratio of
3/1 over 
the region $5.2 < m_{\pi\pi} < 5.3$~\gevcc, rather than the
$S/B\sim 0.4$ from the
previous section. 

A \bpipi\ signal can be extracted from the
physics backgrounds from $B \ra K \pi$ and $\Bs \ra K K$ decays by making
use of the invariant $\pi\pi$ mass 
distribution as well as the d$E$/d$x$ information provided by the COT. 
Using the specific energy loss d$E$/d$x$,
we expect a $K$-$\pi$ separation of 1.3~$\sigma$ for track momentum
$\Pt>2$~\gevc. 
Note, the $\Bs\ra K^+ K^-$ 
peak lies directly under the $B^0\ra\pi^+\pi^-$ signal requiring particle
identification 
through d$E$/d$x$.

\begin{figure}[tb]
\centerline{
\epsfysize=2.4in
\epsffile{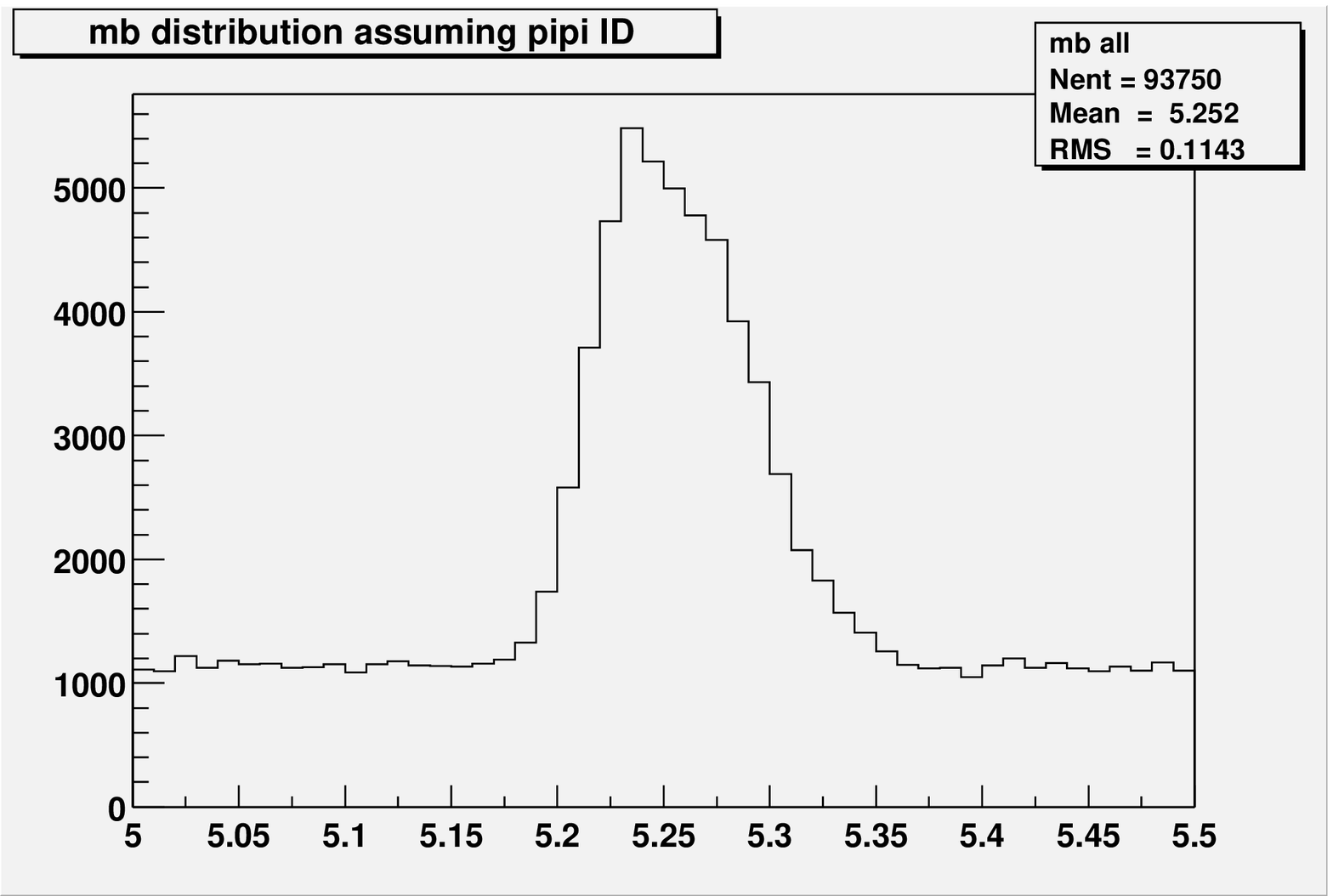} 
\epsfysize=2.4in
\epsffile{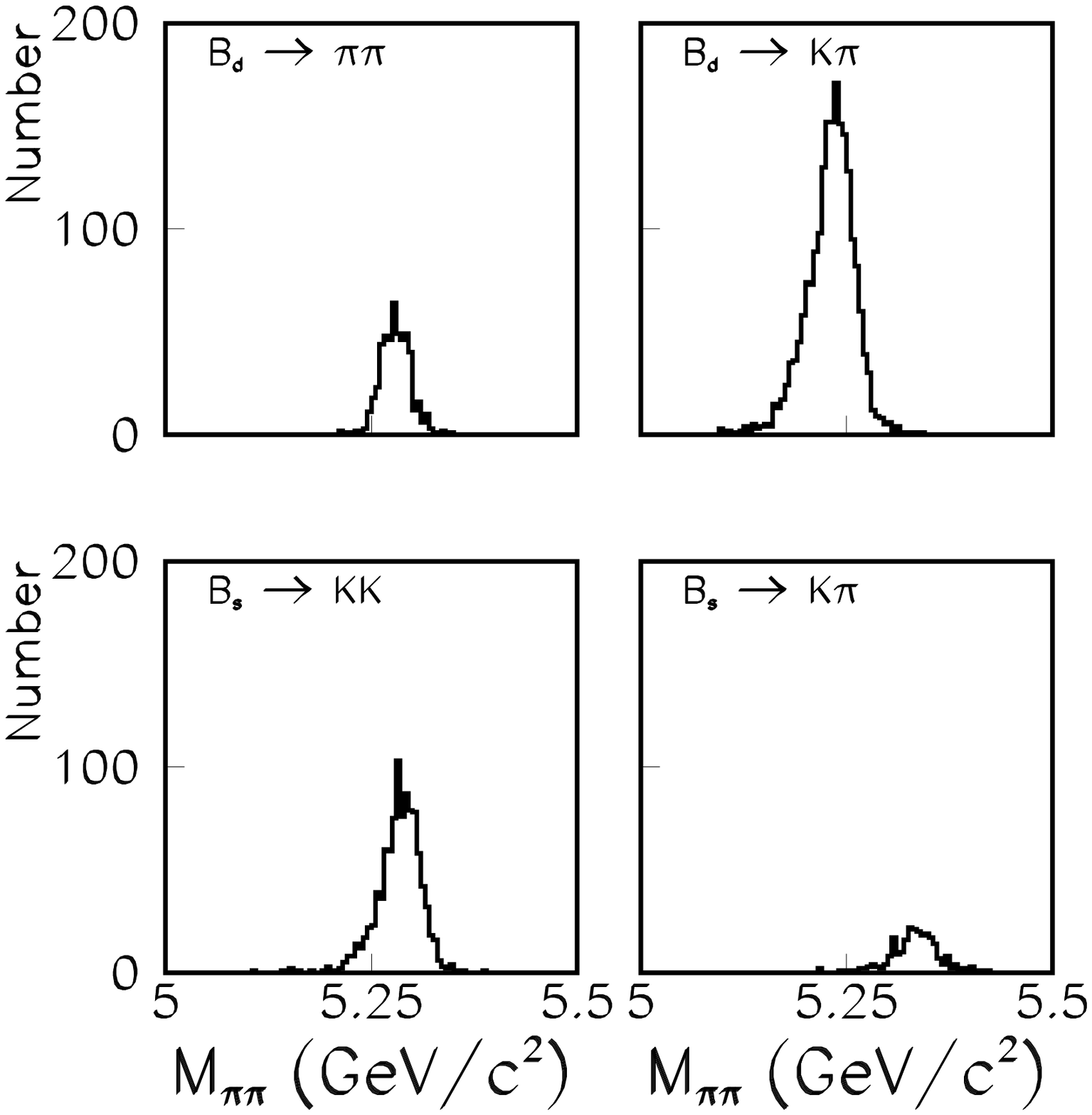} 
\put(-135,145){\large\bf (b)}
\put(-395,135){\large\bf (a)}
} 
\vspace*{0.3cm}
\caption[Two-track invariant mass for 
$B \ra \pi\pi$, $K\pi$, $KK$ and $\pi K$ at CDF.]
{Two-track invariant mass assuming pion hypothesis for 
$B \ra \pi\pi$, $K\pi$, $KK$ and $\pi K$
final states (a) added together and
(b) shown separately.}
\label{fig:mball}
\end{figure}

Given the limited particle identification capabilities provided by invariant
mass resolution and d$E$/d$x$, it is
important to demonstrate how well CDF can separate the four final states using
mass and d$E$/d$x$ alone. To assess this issue, we generate a sample of
93,750 events 
drawn from the four signal hypotheses as shown in Table~\ref{tab:untagged}. 
We also include combinatoric background, where the ratio of $K\pi : \pi\pi
: KK = 1 : 2 : 1$ in the background sample  
is a completely arbitrary choice. 
We then perform a Maximum Likelihood fit to determine the yields for the
four signal and three background hypotheses. 
Comparing the errors on the yields as returned from the fit
with $1/\sqrt{N}$ of the number of generated events,  we can calculate an
``effective'' 
signal/background ratio $=:$ $S/B$ for the four signal samples as follows:
\begin{equation}
\sigma_{\mathrm{yield}}/\mathrm{yield} = \sqrt{ 1 + {B/S} \over \mathrm{yield}}
\end{equation}
The relative errors on the yields and  the effective
signal/background are listed in Table~\ref{tab:untagged}.

\begin{table}
\begin{center}
\begin{tabular}{c|ccc}
\hline
       & $K\pi$ & $\pi\pi$ & $KK$ \\
\hline
$B^0$  & 20,000    & 5000       & 0    \\
$\sigma$ & 0.95\% & 2.8\% & - \\
$B_s^0$  & 2500   & 0        & 10,000  \\
$\sigma$ & 4.8\% & - & 1.6\%  \\
bkg    & 14,000    & 28,000      & 14,000  \\
\hline
 &  \multicolumn{3}{c}{``Effective'' $S/B$}\\ 
\hline
$B_s^0$:  & 0.21 & - & 0.64 \\ 
$B^0$:  & 1.24 & 0.34 & - \\ 
\hline
\end{tabular}
\vspace*{0.3cm}
\caption
[Parameters used in the Toy Monte Carlo study to
determine the errors on the 
$B \ra \pi\pi$, $K\pi$, $KK$ and $\pi K$
yields.]
{Parameters used and results obtained in the Toy Monte Carlo study to
determine the errors on the 
$B \ra \pi\pi$, $K\pi$, $KK$ and $\pi K$
yields in the untagged sample.}
\label{tab:untagged}
\end{center}
\end{table}

In summary, we expect the $B\ra \pi\pi$, $K\pi$, $KK$ and $\pi K$
yields in
the untagged sample to be measured with an uncertainty of only 
a few percent.
In the absence of exact knowledge of relative production cross
sections for $B^0$ and $B_s^0$, as well as branching fractions
this fit to the untagged sample is crucial in determining the 
denominator for the measured $CP$~asymmetry. 
Separating $\pi\pi$, $K\pi$ and $KK$ is less of a problem for the
numerator as we are helped here by the vast difference in 
oscillation frequency.

From the Monte Carlo exercise described above, we conclude that
separating the various $B$ decays into two track hadronic final states 
is not a limiting factor in the 
measurement of the time dependent $CP$~asymmetries.

\boldmath
\subsubsection{$CP$ Violating Observables}
\label{sec:cdf-pipi-cpvars}
\unboldmath

Two of the four signal modes of interest 
($B^0\to K^\pm\pi^\mp$ and $\Bs\to K^\mp\pi^\pm$)
are self-tagging, two ($B^0\to \pi^+\pi^-$ and $\Bs\to K^+K^-$) are
$CP$ eigenstates for which we expect sizable yields, and two 
($\Bs\to \pi^+\pi^-$ and $B^0\to K^+K^-$) are unlikely
to be observed at CDF during Run\,II, unless final state re-scattering
and/or new physics effects in these decays are sizable.
For the self-tagging decay modes, one can distinguish in principle two
$CP$~violating observables, depending on whether or not the $B$ has mixed
before it decayed:
\begin{eqnarray}
\mbox{unmixed:}\quad 
\frac{(B^0\to K^+\pi^-) - (\bar B^0\to \pi^+ K^-)}{(B^0\to K^+\pi^-) +
(\bar B^0\to \pi^+ K^-)}
&\ =\ &  { |A|^2 - |\bar A|^2 \over |A|^2 + |\bar A|^2} \nonumber\\*[4pt]
\mbox{mixed:}\quad 
{(B^0\to K^-\pi^+) - (\bar B^0\to \pi^- K^+) \over (B^0\to K^-\pi^+) +
(\bar B^0\to \pi^- K^+)}
&\ =\ & - { |A|^2 - |q/p|^4 |\bar A|^2 \over |A|^2 + |q/p|^4 |\bar A|^2 }
\end{eqnarray}

In practice, i.e. within the Standard Model where $|q/p| - 1 \ll 1 $, and
even for many
reasonable extensions of the Standard Model, we expect at most $|q/p|-1
\sim {\cal O}(10^{-2})$.  
Furthermore, $|p/q| \neq 1$ is probably better searched for
with doubly tagged inclusive $b\bar b$ samples. The classic example analysis is
to search for a charge asymmetry $(\ell^+\ell^+ - \ell^-\ell^-)/(\ell^+\ell^+ + \ell^-\ell^-)$
in events where both $b$ and $\bar b$ decay semileptonically.
In the following, we therefore will not consider
a time dependent analysis nor tagging for the two self-tagging decay modes.

For the decays into $CP$~eigenstates there are three $CP$~violating observables
\index{CP violation@$CP$ violation!in decay}%
\index{CP violation@$CP$ violation!in mixing}%
${\cal A}^{dir}_{CP}$, ${\cal A}^{mix}_{CP}$, and ${\cal A}_{\Delta \Gamma}$. Either ${\cal A}^{dir}_{CP}\neq 0$, or
${\cal A}^{mix}_{CP}\neq 0$, or $|{\cal A}_{\Delta \Gamma}| \neq 1 $ would indicate $CP$~violation.
In fact, the three observables are related for each decay mode separately by:
\begin{equation}
({\cal A}^{dir}_{CP})^2 +  ({\cal A}^{mix}_{CP})^2 +  ({\cal A}^{\Delta\Gamma}_{CP})^2 = 1.
\end{equation}
The time dependent rate asymmetry is given by:
\begin{eqnarray}
{(B_s^0\to K^+K^-) - (\bar B_s^0\to K^+K^-) \over
 (B_s^0\to K^+K^-) + (\bar B_s^0\to K^+K^-) } 
&\ =\ &    
{2 e^{-\langle \Gamma \rangle t}\over 
e^{-\Gamma_H t} + e^{-\Gamma_L t} 
+ {\cal A}^{\Delta\Gamma}_{CP} (e^{-\Gamma_H t} - e^{-\Gamma_L t})} 
  \nonumber\\[4pt]
&&{} \times ({\cal A}^{mix}_{CP} \sin \Delta m t  
  + {\cal A}^{dir}_{CP}\cos\Delta m t)
\end{eqnarray}

In other words, the oscillation amplitude $ {\cal A}_{CP} = \sqrt{({\cal A}^{dir}_{CP})^2 +  ({\cal A}^{mix}_{CP})^2} $
is modulated by an exponentially rising (or falling) 
``pre-factor'' as shown in 
Figure~\ref{fig:deltaGamma}. The size of this effect depends on the size
of the lifetime difference, $\Delta \Gamma = \Gamma_H - \Gamma_L \neq 0$ and on
$|{\cal A}^{\Delta\Gamma}_{CP}| \neq 1$. 
For $B^0$ we can safely assume $\Delta\Gamma/\Gamma = 0$, and ignore this modulation.
For $B_s^0$ we expect $\Delta\Gamma/\Gamma\sim20\%$. 
Figure~\ref{fig:deltaGamma}
shows that this may lead to an $\sim 7\%$ change of the oscillation 
amplitude per unit of lifetime. 
Given the experimental sensitivity discussed in 
Section~\ref{sec:cdf-pipi-tagged}, we do not expect to observe this effect
in the first \tfb\ of data in Run\,II. 
We therefore ignore it in the present discussion.
The analysis in the two decay modes into $CP$~eigenstates thus reduces to a
fit of the time dependence of the $CP$~violating rate asymmetries to the
sum of a sine and a cosine term. 

\begin{figure}
\centerline{
\epsfxsize=3.5in
\epsffile{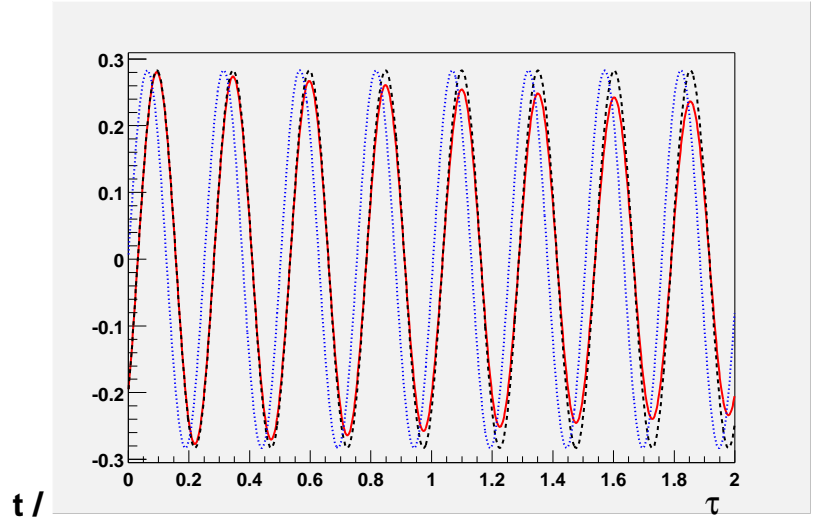}}
\vspace*{0.3cm}
\caption
[Expected time-dependent $CP$~violation in $B_s^0\to K^+K^-$ for
different values of 
${\cal A}^{dir}_{CP}$, ${\cal A}^{mix}_{CP}$
and $\Delta\Gamma$.]
{Red (solid), black (dashed), and blue (dotted) curves show
the expected time-dependent $CP$~violation in $B_s^0\to K^+K^-$ for
different values of 
${\cal A}^{dir}_{CP}$, ${\cal A}^{mix}_{CP}$
and $\Delta\Gamma$.
The red (black) curve assumes 0.2 (0.2) and $-0.2$ (0.0) for
${\cal A}^{dir}_{CP}$ (${\cal A}^{mix}_{CP}$), whereas
the blue curve assumes that both 
${\cal A}^{dir}_{CP}$ and
$\Delta\Gamma$ are zero.
}
\label{fig:deltaGamma}
\end{figure}

\subsubsection{Measurements on the Tagged Sample}
\label{sec:cdf-pipi-tagged}

As discussed in Section~\ref{sec:cdf-pipi-cpvars} above, the time dependent
$CP$~asymmetry 
in $B^0\to \pi^+\pi^-$ and $B_s^0\to K^+K^-$ is given by:
\begin{equation}
{\cal A}_{CP} = {\cal A}^{dir}_{CP} \cos \Delta m t + 
{\cal A}_{CP}^{mix} \sin \Delta m t  
\end{equation}
It is straightforward to derive the expected errors on the coefficients 
${\cal A}_{CP}^{mix}$ and ${\cal A}_{CP}^{dir}$
analytically~\cite{blocker-fkw}. For simplification, we use the
abbreviations $A = {\cal A}_{CP}^{dir}$ and $B = {\cal A}_{CP}^{mix}$ in the
following:
\begin{eqnarray}\label{eq:cdf-pipi-errmatrix}
G_{AA} &\ =\ & N\times e^{-t_0}( {  1 + f(t_0)}) \nonumber\\
G_{BB} &\ =\ & N\times e^{-t_0}( {  1 - f(t_0)}) \nonumber\\
G_{AB} &\ =\ & N\times e^{-t_0}( 2 x \cos (2x t_0) + 
   \sin ( 2x t_0))/(1 + 4x^2) \nonumber\\ 
N &\ =\ & 0.5 \times N_{t_0 = 0} \times \eD \times {S/B \over S/B + 1 }\, 
        e^{-(x \sigma_t/\tau)^2}\nonumber\\
f(t_0) &\ =\ & (\cos 2x\, t_0 - 2x\sin2x\, t_0)/(1+4x^2).
\end{eqnarray}

Here $G_{AA}, G_{BB}$, and $G_{AB}$ are the three elements of the
inverse of the covariance 
matrix, $t_0$ is the minimum lifetime cut implied by the trigger, ``$S/B$''
is the 
signal/background ratio, and $x = \Delta m/\Gamma$, while $\sigma_t$
is the expected proper 
time resolution. 
While deriving Equation~(\ref{eq:cdf-pipi-errmatrix}), 
we made the approximation 
${\cal A}_{CP}\times {\cal D} \ll 1$. 

Figure~\ref{fig:trigTau} shows the proper time in units
of $B_s^0$ lifetime for a Geant based 
Monte Carlo simulation
of $\Bs\to K^+K^-$, followed by track reconstruction.
The depletion for small lifetimes is due to the impact parameter
requirements in the trigger
(scenario A, see Table~\ref{ed_run2})~\cite{pac_jan}.
This shows 
that $t_0 = 0.5$ is a reasonable value to pick for our estimates.

\begin{figure}
\centerline{
\epsfxsize=3.5in
\epsffile{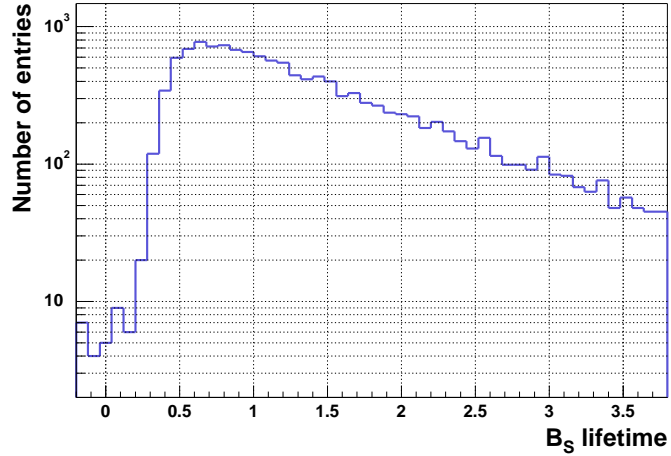}}
\vspace*{0.1cm}
\caption
[Effect of two-track trigger on $B_s^0\to K^+K^-$ lifetime distribution.]
{Effect of two-track trigger on $B_s^0\to K^+K^-$ lifetime distribution.} 
\label{fig:trigTau}
\end{figure}

Table~\ref{tab:cdf-pipi-tagged} shows the values that we consider 
for the various 
parameters entering the equations above. 
It is probably worthwhile mentioning that the effective signal/background from
the untagged study is not relevant here. The oscillation frequencies are
sufficiently different between $B_s^0$ and $B^0$ that $KK\leftrightarrow
\pi\pi$ misidentification does not 
enter the numerator of ${\cal A}_{CP}$ in any significant way. 
We verified this with a 
fit to a Toy Monte Carlo that uses only $m_{\pi\pi}$ and proper time as
input, and assumes 
the relative yields to be known, e.g. from a fit to the untagged sample.
The correlation coefficient between $CP$~violating asymmetries in
$B^0\to\pi^+\pi^-$ 
and $B_s^0\to K^+K^-$ is negligible, despite the fact that the two signal peaks overlap
almost exactly in the invariant two-track mass $m_{\pi\pi}$.

\begin{table}
\begin{center}
\begin{tabular}{|ccc|ccc|}
\hline
 $N^{t_0=0.5}_{\pi\pi} $&$=$&$ 5000$   &  $N^{t_0=0.5}_{KK}$ &$=$&$ 10,000$ \\ 
$t_0$ & = & 0.5 & $\sigma_t/\tau$ &=& 0.03 \\
$x_d $& = & 0.7 & $x_s$ &=& 25 \\
$(\eD)_{\pi\pi}$ &=& 0.091 &  $(\eD)_{KK}$ &=& 0.113 \\  
$(S/B)_{\pi\pi}$ &=& 1/2 & $(S/B)_{KK}$ &=& 1 \\
\hline
\end{tabular}
\begin{tabular}{|c|cc|}
\hline
\multicolumn{3}{|c|}{Resulting values for $G_{AA},G_{BB},G_{AB}$}\\
\hline
      & $B^0\to\pi^+\pi^-$ & $\Bs\to K^+K^-$ \\
$G_{AA}$ & 72 & 161 \\
$G_{BB}$ & 79 & 161 \\
$G_{AB}$ & 44 & 3.2 \\
\hline
\end{tabular}
\end{center}
\vspace*{0.1cm}
\caption
[Inverse of covariance matrix based on analytical calculations.]
{Inverse of covariance matrix based on analytical calculations.}
\label{tab:cdf-pipi-tagged}
\end{table}

\boldmath
\subsubsection{Extracting $CP$ Violating Phases from ${\cal A}^{dir}_{CP}$ and 
${\cal A}^{mix}_{CP}$}
\label{sec:cdf-pipi-gammafit}
\unboldmath

Let us define $\vartheta = arg(\bar{A}/A)/2 $, the $CP$~violating phase in
the decay, and 
$\phi = arg(q/p)/2 $, the $CP$~violating phase in mixing for some phase convention.
$CP$~violation in the interference of mixing and decay is then given by:
\begin{equation}
{\cal A}^{mix}_{CP} (t) = 
   {\Gamma(\bar{B}^0\to f_{CP}) - \Gamma(B^0\to f_{CP})\over
    \Gamma(\bar{B}^0\to f_{CP}) + \Gamma(B^0\to f_{CP})}
    = - \sin 2 (\phi +\vartheta )\times \sin \Delta m t\,.
\end{equation}
In the limit where we ignore anything but the dominant contribution to the
decay amplitude 
${\cal A}^{mix}_{CP}(J/\psi K^0)$ and
${\cal A}^{mix}_{CP}(\pi^+\pi^-)$
measure 
$\sin 2\beta$ and
$\sin 2(\beta +\gamma )$,
respectively, while ${\cal A}^{dir}_{CP} = 0$ in both cases.
If nature was that simple then a non-zero ${\cal A}^{mix}_{CP}(K^+K^-)$ or
${\cal A}^{mix}_{CP}(J/\psi\phi)$
would be a clear sign
of new physics, and any difference between e.g.
${\cal A}^{mix}_{CP}(K^+K^-)$ and ${\cal A}^{mix}_{CP}(J/\psi\phi)$ would signal new physics
in penguin loops. 
Allowing for gluonic penguins in $B^0\to\pi^+\pi^-$ and $b\to u\bar{u}d$ contributions to
$B_s^0\to K^+K^-$ leads to non-zero ${\cal A}^{dir}_{CP}$ if and only if there is also a $CP$~conserving 
phase difference between dominant and sub-dominant decay processes, i.e. ``penguins'' and 
``trees''. 

In the following, we
discuss one particular suggestion by Fleischer~\cite{flekk} that
relates 
$B_s^0\to K^+K^-$ to $B^0\to \pi^+\pi^-$ using $U$-spin symmetry, a subgroup
of flavour $SU(3)$. 
This is neither the only nor necessarily the most promising use of
experimental information but is
meant to give a flavour of what can be achieved with Run\,II data at CDF.
The basic idea is as follows. 
We decompose the two decay amplitudes into the sum
of a part that has the $CP$~violating phase of $\bar{b}\to\bar{c}c\bar{d}$ , 
and a part 
that has the same $CP$~violating phase as $\bar{b}\to\bar{u}u\bar{d}$.
For the standard phase conventions these are $0$ and $\gamma$, respectively.
We then rewrite the four $CP$~violating asymmetries in terms of the 
modulus $d$, the $CP$~conserving
phase $\theta$ describing the ratio of hadronic matrix elements for these
two parts, 
the $CP$~violating phase $\gamma$ and the two $CP$~violating phases
for $B^0$ and $B_s^0$, $\phi_d$ and $\phi_s$, respectively.

In the limit of $U$-spin symmetry
the two sets of $d$ and $\theta$ 
in $B^0\to\pi^+\pi^-$
and $\Bs\to K^+K^-$ (denoted by $^\prime$) are related via:
\begin{eqnarray}
\theta^\prime &=& \theta \,, \nonumber\\
d^\prime &=& d\times \bigg({1-\lambda^2\over \lambda^2}\bigg) \,.
\label{eq:cdf-pipi-d-theta}
\end{eqnarray}
To be specific:
\begin{eqnarray}
{\cal A}^{dir}_{CP} &=& \pm {2{d}\, {  \sin\theta \sin\gamma}\over 
   1-2d \cos\theta \cos\gamma + d^2} \,, \nonumber\\
{\cal A}^{mix}_{CP} &= & {{  \sin 2(\phi +\gamma)} 
             - 2{d}\, {  \cos\theta \sin (2\phi +\gamma)} + d^2 \sin 2\phi 
           \over 1 - 2d \cos\theta \cos\gamma + d^2}\,.
\label{eq:four}
\end{eqnarray}

Here, $2\phi = arg(q/p)$ is the $CP$~violating phase of mixing. 
The equations for $B^0$ and $B_s^0$ are thus identical except for the replacement of
$d,\theta ,\phi_d\ \leftrightarrow\ d^\prime,\theta^\prime ,\phi_s$, and
${\cal A}^{dir}_{CP}(\pi^+\pi^-) = - {\cal A}^{dir}_{CP}(K^+K^-)$.
The latter sign change being due to $V_{us}/V_{cd} = -1$.

In principle, this leads to a system of four equations with the five unknowns $d,\theta ,\phi_s ,
\phi_d$, and $\gamma$. Furthermore, if $\theta \sim 0$ then two of the four equations
are degenerate within our experimental sensitivity 
(${\cal A}^{dir}_{CP}(\pi^+\pi^-) \sim {\cal A}^{dir}_{CP}(K^+K^-) \sim 0$), leading to only three independent
equations and five unknowns. To arrive at a system of equations that is
solvable, 
we add ${\cal A}^{mix}_{CP}(J/\psi K^0) = \sin 2\phi_d$ as an additional
constraint, and 
fix $\phi_s = 0$, which is correct for the Standard Model to within 
${\cal }O(\lambda^2)$.

We then perform a $\chi^2$ fit of hypothetical measurements
of the two asymmetries ${\cal A}^{dir}_{CP}$ and the three asymmetries 
${\cal A}^{mix}_{CP}$ and their errors to the 
corresponding theoretical expressions that relate them to the fit parameters
$\beta ,\gamma ,\theta$ and $d$. 
We choose the following nominal values:
\begin{eqnarray}
\beta  &=& 22.2^\circ \pm 2.0^\circ\,, \nonumber\\
\gamma &=& 60^\circ\,, \nonumber\\
\theta &=& 0\,,\nonumber\\
d &=& 0.3\,.
\label{eq:cdf-gamma-fit-nominal-truth}
\end{eqnarray}

This results in the expected ``measurements''
${\cal A}^{dir}_{CP}(\pi^+\pi^-) = 0$, 
${\cal A}^{mix}_{CP}(\pi^+\pi^-) = -0.316$,
${\cal A}^{dir}_{CP}(K^+K^-) = 0$ and
${\cal A}^{mix}_{CP}(K^+K^-) = 0.266$.
The error on $\beta$ is slightly larger than the CDF projections as discussed
in Section~\ref{sec:cdf-sin2beta}. For the errors on ${\cal A}^{dir}_{CP}$ and ${\cal A}^{mix}_{CP}$
in $\pi^+\pi^-$ and $K^+K^-$, we choose the inverse error matrices as quoted in
Table~\ref{tab:cdf-pipi-tagged}. This nominal fit returns:
\begin{eqnarray}
\gamma & = & (60.0^{+ 5.4}_{-6.8})^\circ\,, \nonumber\\
\beta &=& (22.2\pm 2.0)^\circ\,, \nonumber\\
\theta &=& (0.0^{+10.8}_{-10.5})^\circ\,,  \nonumber\\
d &=& 0.3^{+0.11}_{-0.07}\,.
\end{eqnarray}

An exhaustive scan of the parameter space showed that the error on 
$\gamma$ changes by a factor of $\sim 3$ over the range $d = 0.1$ to
$0.5$. Variations in the other parameters are less important.
Further details may be found in reference~\cite{fkwGamma}. 

\subsubsection{Theoretical Error due to $SU(3)$ Breaking}
\index{SU3 relations@$SU(3)$ relations}%

In this section, we study the dependence of the fit for $\gamma$
on the assumption of $SU(3)$ symmetry. This is done by calculating the
``measured'' values for the four $CP$ violating asymmetries in 
$\btopipi$ and $\bstokk$ with 
$d e^{i\theta}\neq d^\prime e^{i\theta^\prime}$, while strict
$SU(3)$ symmetry is used in the fit.

$SU(3)$ breaking for form factors or decay constants is known to be a 
10-15\% effect.
Both of these are ``long distance'' effects in the sense that they describe
meson formation, rather than 
physics at the weak scale. This type of $SU(3)$ breaking affects 
amplitudes but tends to
cancel in appropriately chosen ratios of amplitudes. For the rate
asymmetries that we care about here
such ``long distance'' $SU(3)$ breaking corrections do indeed cancel,
e.g.~within factorization
models Eq.~(\ref{eq:cdf-pipi-d-theta}) is exact.
An $SU(3)$ breaking effect that matters would have to alter the ratio of
hadronic
matrix elements for penguin and tree diagrams. This means, it would have
to invalidate Equation~(\ref{eq:cdf-pipi-d-theta}). 
To what extend such effects should be expected, remains an open question.
Future data for these and other processes will tell us the range of such
effects. 

We can model a potential effect of this type by using different sets
of $d,\theta$ for
$B_s^0$ and $B^0$ when calculating the four hypothetical $CP$~violating
asymmetries, but
using the same $d,\theta$ for $B_s^0,B^0$ when minimizing the $\chi^2$. 
In principle, one might expect an increase in $\chi^2$ at the minimum, 
i.e. a poorer fit,
as well as a systematic shift in $\gamma$ returned by the fit.
To be conservative, we chose 20\% $SU(3)$ breaking and implement it as follows:
\begin{equation}
\overrightarrow{\Delta d} = (d\times e^{i\theta})_{B_s^0} 
  - (d\times e^{i\theta})_{B^0} 
= |\overrightarrow{\Delta d} |\times e^{i\phi}
= 0.2\times d \times e^{i\phi}\,.
\end{equation}  

In other words, the set of possible $SU(3)$ breaking effects that we consider is given by
a circle with radius $0.2\times d$. We can then plot $\gamma_{\rm measured}$ as a function of $\phi$
for fixed $\gamma_{\rm true}$. This is shown in Figure~\ref{fig:cdf-pipi-su3breaking} for our
nominal fit parameters. We conclude that 20\% $SU(3)$ breaking leads to a systematic error
on $\gamma$ of at most $\pm 3$ degrees for our nominal set of parameters.

\begin{figure}
\centerline{
\epsfxsize=3.5in
\epsfbox{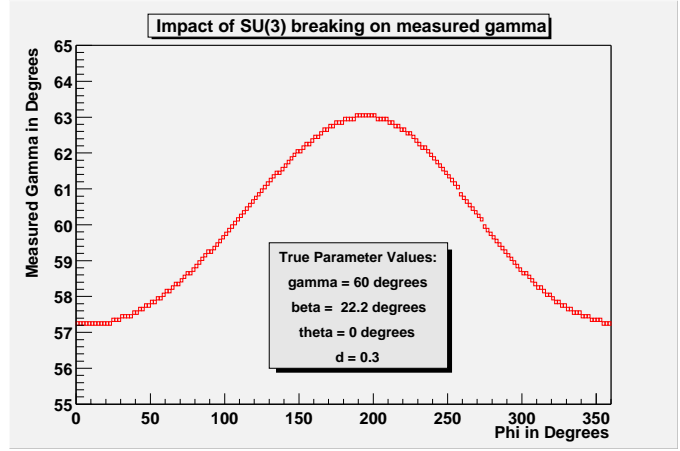}}
\vspace*{0.3cm}
\caption[Impact of $SU(3)$ breaking.]
{Impact of $SU(3)$ breaking.} 
\label{fig:cdf-pipi-su3breaking}
\end{figure}

\boldmath
\subsection[$B \rightarrow \pi\pi/KK$: D\O\ Report]
{$B \rightarrow \pi\pi/KK$: D\O\ Report
$\!$\authorfootnote{Authors: R.~Jesik and M.~Petteni.}
}
\unboldmath
\index{decay!$B^0 \to \pi \pi$}%
\index{D\O\!$B^0 \to \pi \pi$ prospects}%

As discussed in Section~\ref{sec:cpintropenguin} and \ref{ch6:pipi_intro},
a measurement of the $CP$~asymmetry in the decay 
$B^0 \rightarrow \pi^+ \pi^-$ was once thought to be the ``golden''
mode to determine the CKM angle $\alpha$. But an unexpectedly small
branching ratio and large penguin contributions have made this analysis,
however, much more difficult than expected. The situation is more complicated
without significant $\pi/K$ separation, as the
decay $\Bs \rightarrow K^+ K^-$ lies in the same 
reconstructed mass range as the $\pi^+\pi^-$ signal.
In addition, the fully hadronic final state poses
another problem for D\O\, as it is not possible to trigger on the $\pi\pi$ 
final state directly. 
The background rate for two tracks with $p_T$ thresholds set low
enough to collect these events is well above the maximum Level\,1 trigger rate 
of 10 kHz. However, it will be possible to
trigger on these decays for the case of the other $B$~hadron in the
event decaying semileptonically. 
Due to the semileptonic branching ratio, this requirement has an efficiency
of 10\% at best. But since
the initial flavour of the $B\ra\pi\pi$ decay has to be tagged in order to
measure the $CP$~asymmetry,  
an opposite side lepton tag is one of the most effective ways to do this.

The trigger requires one lepton with $p_T$ greater then 3.0 \gevc\
plus two other tracks with $p_T$ greater than 1.5 \gevc. In order to minimize
the number of fake tracks, all three tracks must have a hit in each of the Central
Fiber Tracker's (CFT) eight axial layers. To limit background rates, an 
isolation cut is made in which two of the tracks are required to have no
other tracks with $p_T$ above 1.5 \gevc\ within the same, or adjacent, CFT
sectors (the CFT is  divided into 80 equal sectors at the trigger level). 
To further lower background rates, multiple interactions are removed by 
rejecting events which have more than 68 sectors exceeding a threshold of
$12\%$ occupancy.

This study is based on a Monte Carlo sample of $B^0$ and $\Bs$ decays generated
by Pythia plus QQ. The $B^0$ mesons were forced to decay into $\pi^+ \pi^-$
and  
$K^+ \pi^-$ final states with proportion according to branching
ratios as measured by CLEO~\cite{cleobranch}. The $\Bs$ mesons were forced
into $K^+ K^-$ and $K^- \pi^+$ final states. The branching ratios for the
$\Bs$~decays were extrapolated from the measured values for $B^0$ using
spectator 
quark flavour invariance. The relative fraction of $B^0$ to $\Bs$ meson events
in this sample was as generated by Pythia, which agrees with
Run\,I measurements from CDF~\cite{cdf_frag}. Kinematic cuts of 
$p_T > 4$~\gevc\ and $|\eta| < 3$ were made on the $B$ mesons at 
generator level, 
leaving a final sample of about 300,000 events.

The D\O\ detector acceptance was simulated using MCFAST. Imposing the trigger
$p_T$, isolation, and hit requirements on this sample leaves a trigger
acceptance 
of $0.76\%$ for these events. Since the D\O\ muon system is not represented in
MCFAST, the trigger acceptance is corrected by a factor of $78\%$
to account for the holes in the bottom of the detector. This efficiency 
is determined using a full GEANT simulation. An additional efficiency of 
$98\%$ per track is imposed in order to take into account hit 
in-efficiencies not present in the MCFAST analysis. The efficiency of the
high occupancy rejection of this trigger was found to be $80\%$ using a full
GEANT simulation. These factors bring the trigger
efficiency to a $0.45\%$ level.  

The offline reconstruction of these events is simply a refinement of the 
trigger requirements using information from the full detector. All tracks 
are required to have a hit in each of the 8 stereo layers of the CFT, in
addition to the 8 axial hits required by the trigger. The tracks are also 
required to have at least 8 hits in the silicon detector (the 
maximum number of hits is 10 on average). The efficiency of these requirements
is $90\%$. With these considerations, summarized in Table \ref{expected}, we
expect to reconstruct 1400 $B^0 \rightarrow \pi^+ \pi^-$ events in 2 fb$^{-1}$
of data. Similarly, D\O\ expect 5600 $B^0 \rightarrow K^+ \pi^-$, 2500 $\Bs
\rightarrow K^+ K^-$, and 600 $\Bs \rightarrow K^+ \pi^-$ events in this
sample. 
 
\begin{table}[tbp]
\begin{center}
\begin{tabular}{l|l} 
\hline
Integrated luminosity                         & 2 fb$^{-1}$ \\
$\sigma_{b\bar{b}}$                          & $158\,\mu$b  \\
$f(b\bar{b} \rightarrow B^0, \bar{B^0})$     & 0.8 \\
Kinematic acceptance                         & 0.31 \\
${\cal B}(B^0 \rightarrow \pi^+\pi^-)$       & $4.3 \times 10^{-6}$ \\
Trigger efficiency                           & $4.5 \times 10^{-3}$  \\   
Reconstruction efficiency                    & 0.9  \\     
Number of reconstructed $B^0 \rightarrow \pi^+\pi^- $ &  1400 \\
\hline
Effective tagging efficiency ($\eD$)  & 0.40 \\
\hline
 \end{tabular} 
\vspace*{0.3cm}
\caption
[Expected number of $B^0 \rightarrow \pi^+ \pi^-$ events at D\O.]
{Expected number of $B^0 \rightarrow \pi^+ \pi^-$ events at D\O.}
\label{expected}
\end{center}
\end{table}

The mass resolution of the $B^0$ meson in this channel is 44 MeV/$c^2$
as can be seen in Fig.\ref{fig:d0_mpipi}(a). 
Figure~\ref{fig:d0_mpipi}(b) shows the
mass distributions for all 
four channels assuming that the final state particles are pions. From this plot
it can be seen that it is not possible to separate the
$B^0 \rightarrow \pi^+ \pi^-$ decays from 
 $\Bs \rightarrow K^+ K^-$ based on the reconstructed mass. The situation is
further complicated by the fact that the $B^0 \rightarrow K^+ \pi^-$ decay
lies directly over the two channels of interest. Fortunately, $\Bs$~mesons 
oscillate at a much faster frequency than $B^0$~mesons.
With the use of a multi-variant fit it could be possible to separate all
the contributions. It should be noted that the reconstructed samples are
already flavour tagged by the requirement of the lepton in the trigger.
The soft lepton tag has a dilution of $63\%$
and $\varepsilon$ will be very near unity, leading to an effective tagging 
efficiency of $\eD =  0.40$. Work is progressing on how well the
$CP$~asymmetries can be measured and on how well they can be translated into
extracting CKM parameters.

\begin{figure}[tbp]
\centerline{
\put(45,175){\large\bf (a)}
\put(265,175){\large\bf (b)}
\epsfxsize=2.9in
\epsffile{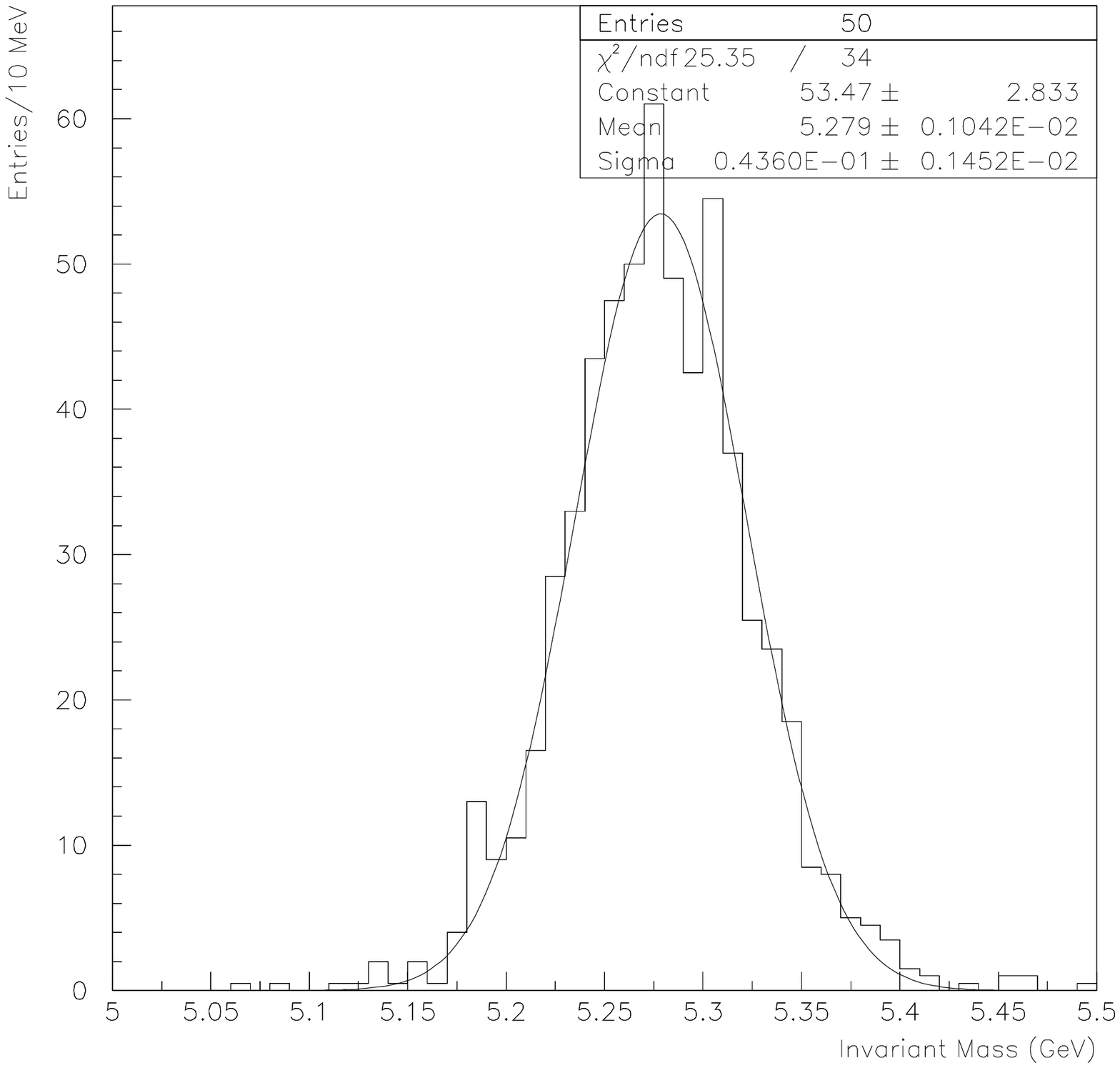}
\epsfxsize=3.0in
\epsffile{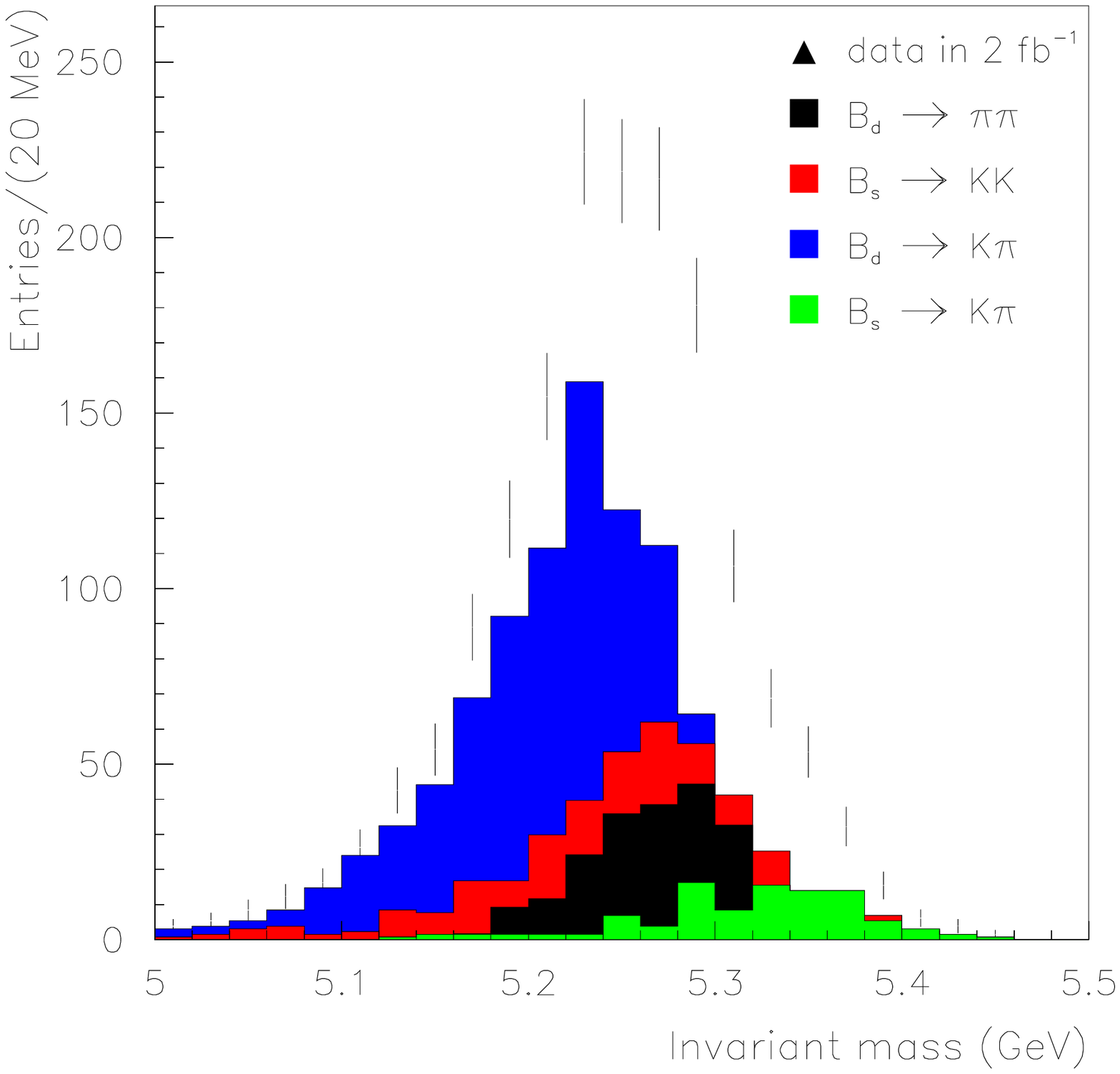}
}
\vspace*{0.3cm}
\caption
[Reconstructed invariant mass for $B^0 \rightarrow \pi^+ \pi ^-$ at D\O.]
{(a) Reconstructed invariant mass for $B^0 \rightarrow \pi^+ \pi ^-$ at D\O.
(b) Expected two track invariant mass signal assuming both tracks
are pions.}
\label{fig:d0_mpipi}
\end{figure}

\boldmath
\subsection[$B \rightarrow \pi\pi/KK$: BTeV Report]
{$B \rightarrow \pi\pi/KK$: BTeV Report
$\!$\authorfootnote{Authors: G.~Majumder, M.~Procario, S.~Stone.}
}
\label{sec:b0_pipi}
\unboldmath
\index{decay!$B^0 \to \pi \pi$}%
\index{BTeV!$B^0 \to \pi \pi$ prospects}%

The decay of $\btopipi$ is the traditional choice for measuring 
$\sin2\alpha$, but the evidence of large penguin amplitudes in
the observation of $\btokpi$ by the CLEO 
collaboration~\cite{cleobranch}
implies that a simple extraction of $\sin2\alpha$
from this mode is no longer likely. However, since this mode has been used 
to benchmark so many experiments, it is still worthwhile to understand.
In addition, it may be useful for the extraction of $\gamma$ when 
combined with a measurement of $\bstokk$ as
explained in Sections~\ref{ch6:intro} and \ref{ch6:pipi_intro}.

  The data for this study are generated using Pythia while QQ is used to decay the
heavy particles. The detector simulation is performed using the
BTeVGeant simulation package. We also compare our result with the 
result obtained using MCFast.  Each signal event which is simulated by
BTeVGeant (or MCFast) contains one signal interaction ($b\bar{b}$) and 
$n$ background  interactions (minimum bias), where $n$ has a Poisson 
distribution of mean 2. This corresponds to the BTeV design luminosity of 
$\rm 2\,\times\,10^{32}\:cm^{-2}\:s^{-1}$.

To find this decay, BTeV selects two oppositely charged 
tracks with a displaced vertex and an invariant mass close to the $B^0$ 
mass.  Most of the background rejection against random combinations
comes from the displaced $B$ vertex and the momentum balance of 
the $\pipi$ combination
with respect to the direction of the $B$. While particle 
identification is vital to reject backgrounds from decays like 
\index{decay!$B^0 \to K \pi$}%
\index{decay!$B_s^0 \to K K$}%
$\btokpi$, $\bstopik$ and $\bstokk$, it has a small effect on random 
combinations since most particles are pions.

To start this analysis, BTeV first fits the primary vertices using all tracks
which have at least 4 silicon pixel hits. 
%
For the two tracks to be considered as $B$~daughter candidates, they must satisfy
the following criteria.
Each track must have $p_T>0.5$~\gevc\ and at least one track must
       have $p_T>1.5$~\gevc. 
Each track must project into the
       RICH detector acceptance, because particle identification
       is required later. 
The distance of closest approach (DCA) of the 
       track with respect to the primary vertex must be less than 1\,cm, 
       which reduces backgrounds from long lived particles, e.g. $K_{S}^0$,
	$\Lambda$, \dots\ 
       It is also required that the DCA divided by
       its error of each track be $>$ 3 which 
       removes tracks from the primary vertex.

BTeV attempts to fit a secondary vertex with pairs of tracks that satisfy the
above criteria.
For each secondary
vertex found, the following selection criteria are applied:
%
The absolute distance between the primary and secondary vertices ($L$) must
       be greater than 0.5~mm and $L/\sigma_L > 4$. 
Considering all other tracks that do not come from this
       primary vertex and forming a $\chi^2$ with each of these tracks and the
       selected two tracks for a 
       secondary vertex, combinations with $\chi^2\:<\:10$ are rejected,
       since this might indicate a many-body $B$~decay.
The $B^0$ direction is calculated from the primary and 
       secondary $B$ vertex positions and
the invariant mass of the two tracks (assumed to be $\pi^\pm$) must be
       within 
2~$\sigma$ of $m_{B^0}$. 
Using the selection criteria defined above, gives an acceptance and
reconstruction 
efficiency of 8\% for $\btopipi$, not
including trigger efficiency or particle identification.

Figure~\ref{fig:sigbkg} shows a comparison of signal and 
background for several of the variables used above. The background
distributions are generated considering all oppositely charged
two-track combinations except for the signal $\pi^+\pi^-$.
  
\begin{figure}[htbp]
\centerline{
\epsfxsize=6.0in
\epsffile{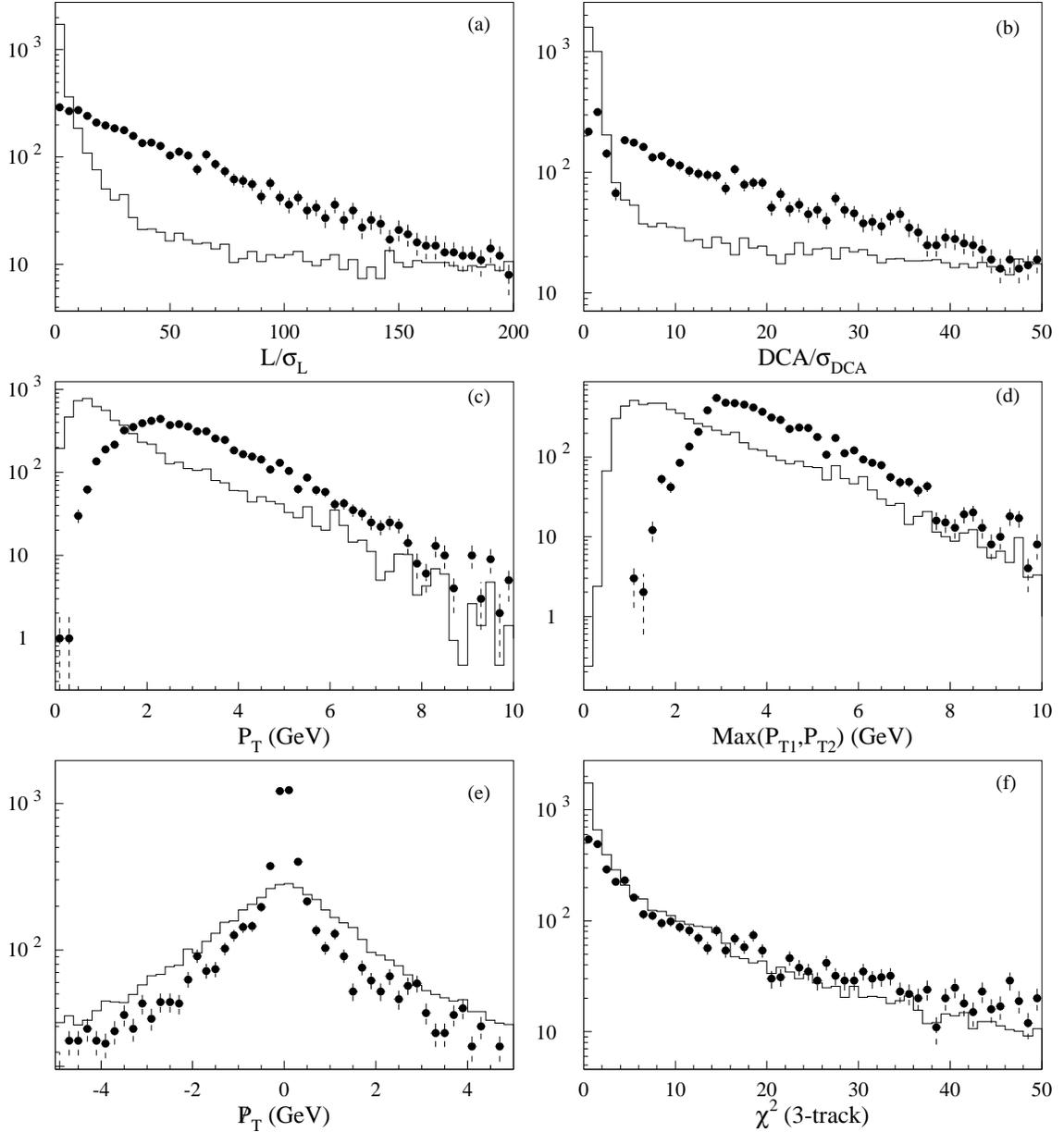}}
\vspace*{0.3cm}
\caption[Distribution of signal and background for the
         most important vertex and kinematic variables at BTeV.]
{Distribution of signal (circles) and background (line) for the
         most important vertex and kinematic variables.
         (a) Normalized distance between primary and secondary vertex,
	 $L/\sigma_L$,
         (b) normalized DCA of track with respect to the primary vertex, 
         $DCA/\sigma_{DCA}$, 
         (c) transverse momentum of a track,
         (d) maximum value of transverse momentum of two tracks,
         (e) $p_T$ imbalance of $\pipi$ with respect to the $B^0$
         direction and	 
         (f) $\chi^2$ of secondary vertex using the $\pi^+\pi^-$ with
	     an additional track candidate.}
\label{fig:sigbkg}
\end{figure}


   It has been shown by the BCD group~\cite{bcdgroup} that the dominant
background to $\btopipi$ comes from random combinations of tracks in 
events coming from $B$'s. Tracks from real $B$'s 
are already displaced from the primary vertex and have a higher 
probability of faking a secondary vertex compared to $c\bar{c}$ and 
minimum bias events. 

  In addition to background from generic $b\bar{b}$ events, there are 
several exclusive decay modes of $B$ mesons that can mimic a 
$\btopipi$ decay. The decay 
$\bstokk$, which is due to a hadronic penguin 
decay mechanism, is the most important, along with other contributions from 
$\btokpi$ and $\bstopik$. 
Recent CLEO measurements of some of the $B^0$ decay modes give
$\cal{B}$ $(\btopipi)~=~0.43\,\times\,10^{-5}$ and 
$\cal{B}$ $(\btokpi)~=~1.7\,\times\,10^{-5}$~\cite{cleobranch}.
In order to normalize the $B_s^0$ contribution, we use a $B_s^0$ production rate 
which is 35\% of the $B^0$ rate~\cite{CDFpsph} and assume
that the penguin and $b \rightarrow u$ decays of the $B_s^0$
have the same branching ratios as the $B^0$.
Using these results as input, and without $\pi /K$ discrimination,
the two-pion mass plots for the four different two-body decay modes
are shown in Fig.~\ref{fig:masspp}(a).
These plots indicate that kinematic separation is inadequate
to discriminate among these decays. 


\begin{figure}[tbp]
\centerline{
\put(40,130){\large\bf (a)}
\epsfig{figure=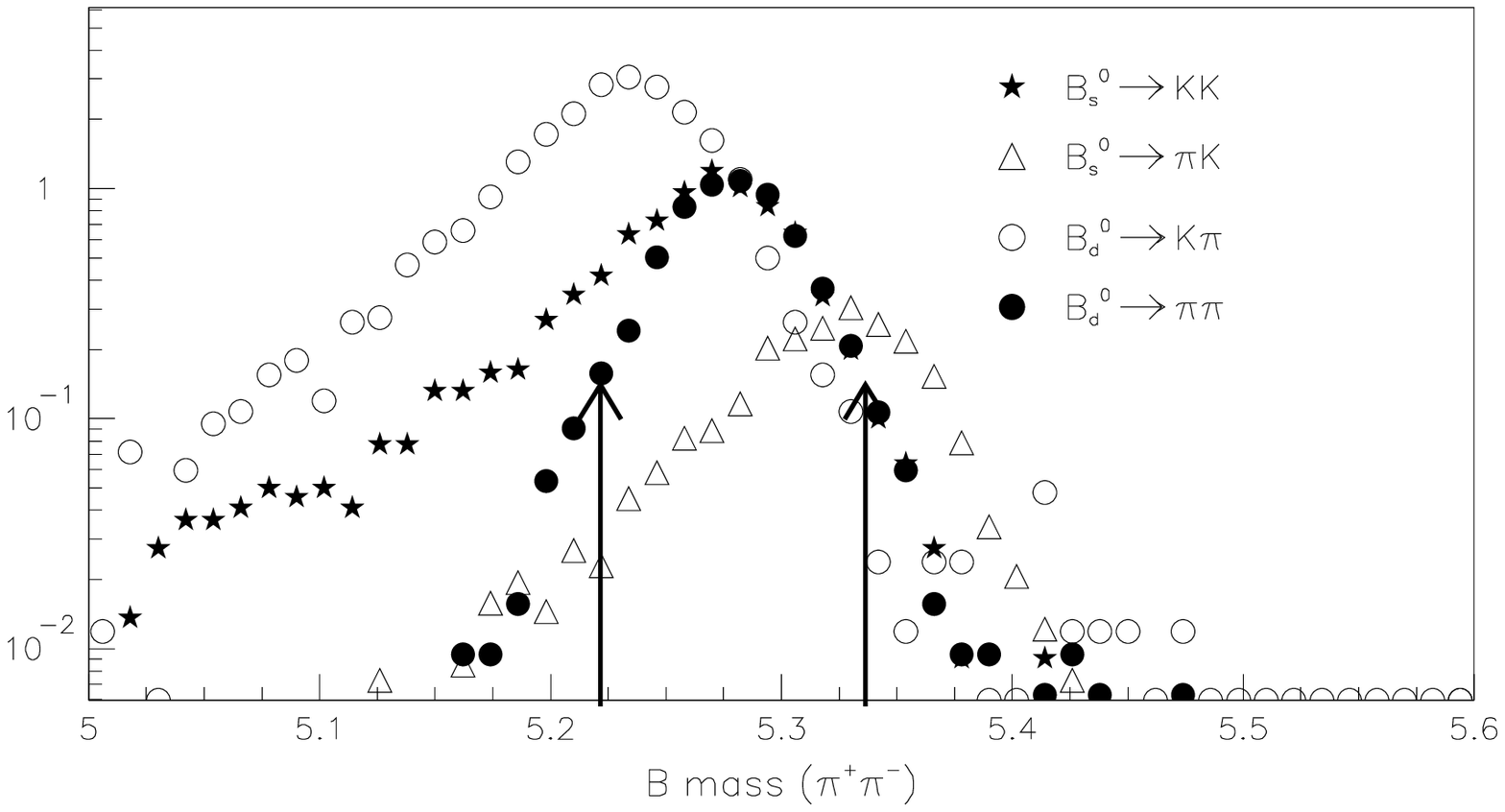,width=4in}
}
\centerline{
\put(40,130){\large\bf (b)}
\epsfig{figure=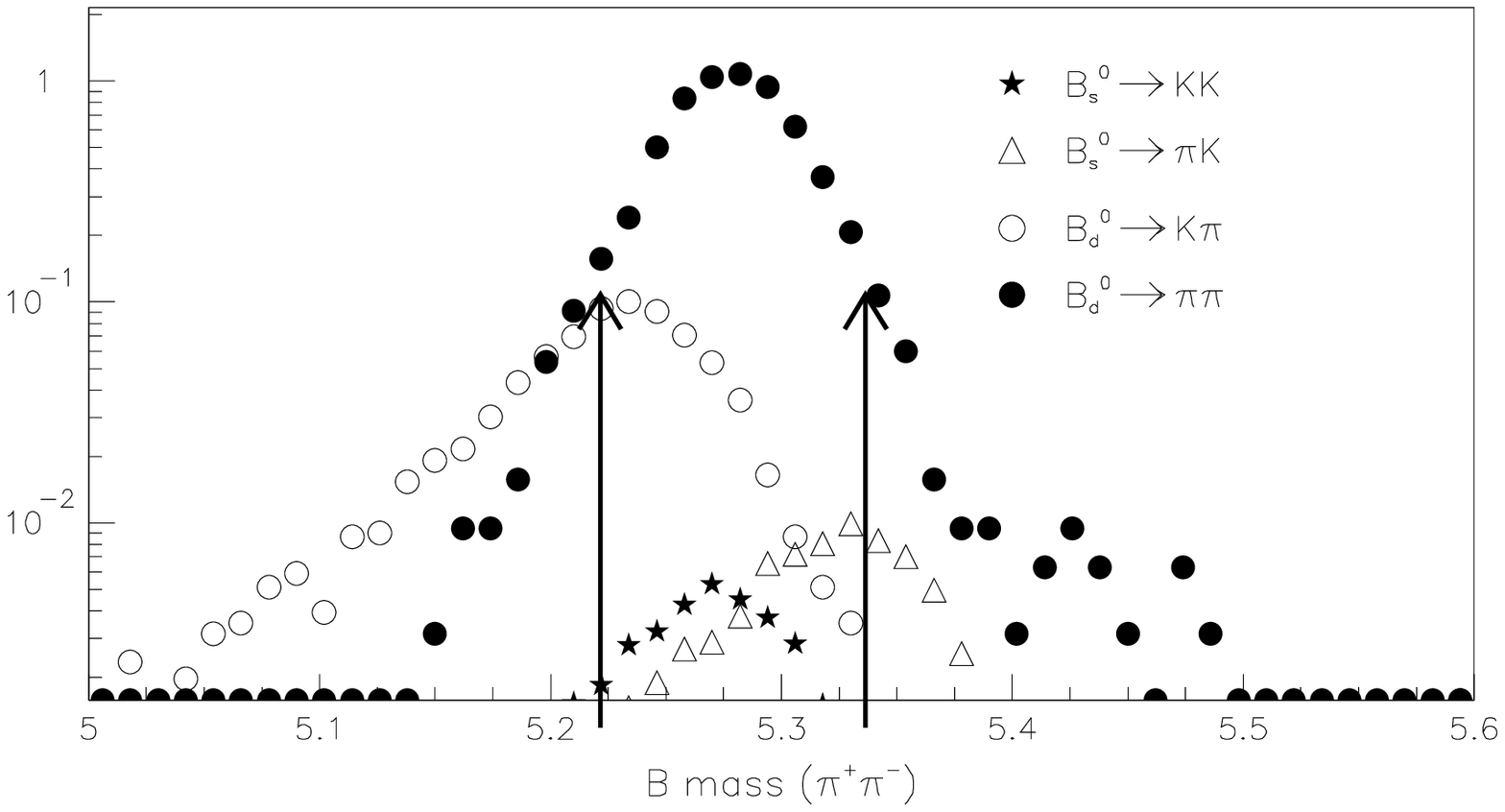,width=4in}
}
\caption
[Two body $\pipi$ mass without and with particle 
         identification at BTeV.]
{Two body ($\pipi$) mass plot (a) without and (b) with particle 
         identification. Different decay channels are normalized by 
         their production cross sections. The arrows indicate the range
         of the signal mass window. (Note the log scale.)}
\label{fig:masspp}
\end{figure}

  The BTeV detector will have an excellent RICH detector for particle
identification. BTeV can virtually eliminate two-body backgrounds using 
the RICH. The simulated background tracks (all
tracks including all other interactions in that event) were passed through
the RICH  
simulation code. The efficiency versus background 
contamination is shown in Fig.~\ref{fig:eff_con}.
For an 80\% $\pipi$ signal efficiency, the contamination 
from $\pi^\pm\,K^\mp~(K^+\,K^-)$ is 4.0\% (0.5)\%.

\begin{figure}[tbp]
\begin{center}
\epsfig{figure=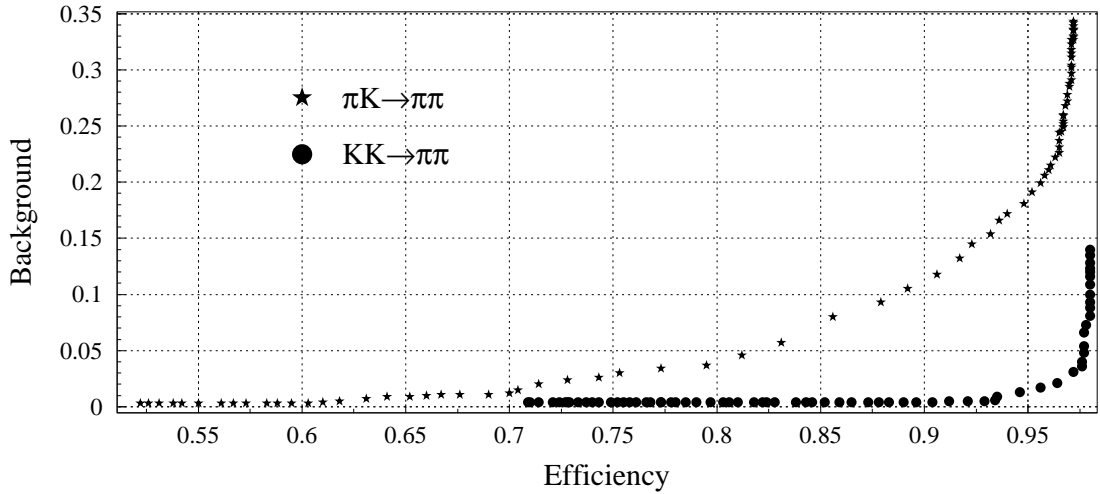,width=6in}
\end{center}
\vspace{-6mm}
\caption[RICH event selection at BTeV.]
        {RICH event selection: $\pipi$ signal efficiency versus 
         contamination from other two-body decay modes.}
\label{fig:eff_con}
\end{figure}

Since the primary purpose of the Level\,1 trigger is to reject light quark
backgrounds, there is a strong correlation between triggered events and 
reconstructed events. The BTeVGeant simulation shows that 64\% of the 
selected events pass the Level\,1 trigger condition.  Given a Level\,2
efficiency of 90\%, this leaves 
23,700 events per year of running after applying the acceptance, reconstruction
efficiency, particle ID efficiency, and trigger efficiency but
before flavour tagging.

  Besides the two-body $B$ decay background samples, a full BTeVGeant 
simulation of $b\bar{b}$ backgrounds was performed. 
In order to reduce the CPU time required
to simulate a sufficiently large data sample of $b\bar{b}$ decays, 
a method to preselect events at the generator level which are likely to cause 
difficulties, was investigated. BTeV found that the difference 
between the reconstructed and generated $p_T$ of the tracks is fairly
small and Gaussian.
On the basis of the small 
observed differences, BTeV preselected the generator events before the
BTeVGeant    
simulation. The preselection criteria are based on the $p_T~(>$0.4\,GeV/c) of 
each track, the sum of the $p_T~(>$1.8\,GeV/c) of two tracks, the opening angle of
the tracks, the extrapolation of tracks to the RICH chamber, etc. 
In order to reject background at the generator level, 
a small fraction of event selection efficiency had to be sacrificed.


These preselection requirements reduce the generic $b\bar{b}$ event sample by a factor of 100.
From this sample, only 4 events (two $\pipi$, one $\kpi$ and one $\pik$)
have a  
$\pipi$ mass that lies within 200\,MeV/c$^2$ of $m_{B^0}$. Applying the RICH 
identification leads to an 80\% efficiency for the two $\pipi$ events and
a 4\% efficiency for each of the $\kpi$ and $\pik$ events. Thus, there are
1.68 background events.  If we scale to the
$B$ signal region which is 115 MeV/c$^2$ and multiply by the combined
Level\,1 and Level\,2 trigger efficiency (64\% $\times$ 90\%), we expect 
$\approx$ 4,600 $b\bar{b}$ background events from one year 
of running BTeV at the design luminosity of $2 \times
10^{32}$~cm$^{-2}$s$^{-1}$. 
  The remaining contributions (from the two-body decay channels) are listed in
Table~\ref{tab:pipitable} and add up to 3,200 events per year. Therefore,
the total  
background is 7,600 events per year leading to a signal-to-background ratio
of 3:1  with a 25\% error. 
 
\begin{table}[tbp]
\begin{center}
\begin{tabular}{|l|c|} \hline
Luminosity            & $2\,\times\,10^{32}\,{\rm cm}^{-2}\,{\rm s}^{-1}$ \\
Running time          & 10$^7$ sec \\
Integrated Luminosity & 2000 pb$^{-1}$ \\
$\sigma_{b\bar{b}}$   & 100 $\mu$b \\
Number of $B\bar{B}$ events       & $2\,\times\,10^{11}$ \\
Number of $B^0$ events               & $1.5\,\times\,10^{11}$ \\
${\cal{B}}(B^0\,\rightarrow\,\pi^+\,\pi^-)$ & $0.43\,\times\,10^{-5}$ \\ \hline
Reconstruction efficiency            & 8.0\% \\
Trigger efficiency (Level\,1)                   & 64\% \\ 
Trigger efficiency (Level\,2)                   & 90\% \\ 
RICH I. D. efficiency                & 80\%\\
Number of reconstructed 
$\btopipi$                             & 2.37$\,\times\,10^4$ \\ \hline
Background after RICH rejection        &                       \\ 
$\btokpi$                              & 0.27$\,\times\,10^4$ \\
$\bstopik$                             & 0.03$\,\times\,10^4$ \\
$\bstokk$                              & 0.02$\,\times\,10^4$ \\
$B$-generic                            & 0.46$\,\times\,10^4$ \\ \hline
$S/B$                                     & 3  \\
Tagging efficiency $\eD$ & 10.0\% \\
$\sigma({\cal A}_{CP})$                        & 2.36$\times\,10^{-2}$ \\ \hline
\end{tabular}
\vspace*{0.3cm}
\caption[Projected yield of $\btopipi$ and the 
         uncertainty on ${\cal A}_{CP}$ at BTeV.]
{Projected yield of $\btopipi$ and the 
         uncertainty on ${\cal A}_{CP}$ from a BTeV\-Geant simulation.}
\label{tab:pipitable}
\end{center}
\end{table}

The effective tagging efficiency ($\eD$), discussed in 
Section~5.5, is 
estimated to be 10\%. Using the tagging efficiency and the $\btopipi$ yield,
we can obtain an 
uncertainty on the $CP$~asymmetry. Based on one year of running at design
luminosity, BTeV expects an uncertainty on ${\cal A}_{CP}$ of 0.024, as
summarized in Table~\ref{tab:pipitable}.


As mentioned in Section~\ref{ch6:pipi_intro}, measuring both $\btopipi$
and $\bstokk$ may allow 
an extraction of $\gamma$.  To this end, BTeV has also looked for
$\bstokk$ signal events.  This analysis is nearly identical to the
$\btopipi$ analysis after interchanging
$\btopipi$ and $\bstokk$ 
samples from signal to background (and vice versa). As in the $\btopipi$
analysis, other two-body decay modes can
mimic the signal as shown in Fig.~\ref{fig:masskk}(a). 

\begin{figure}[tb]
\centerline{
\put(40,130){\large\bf (a)}
\epsfig{figure=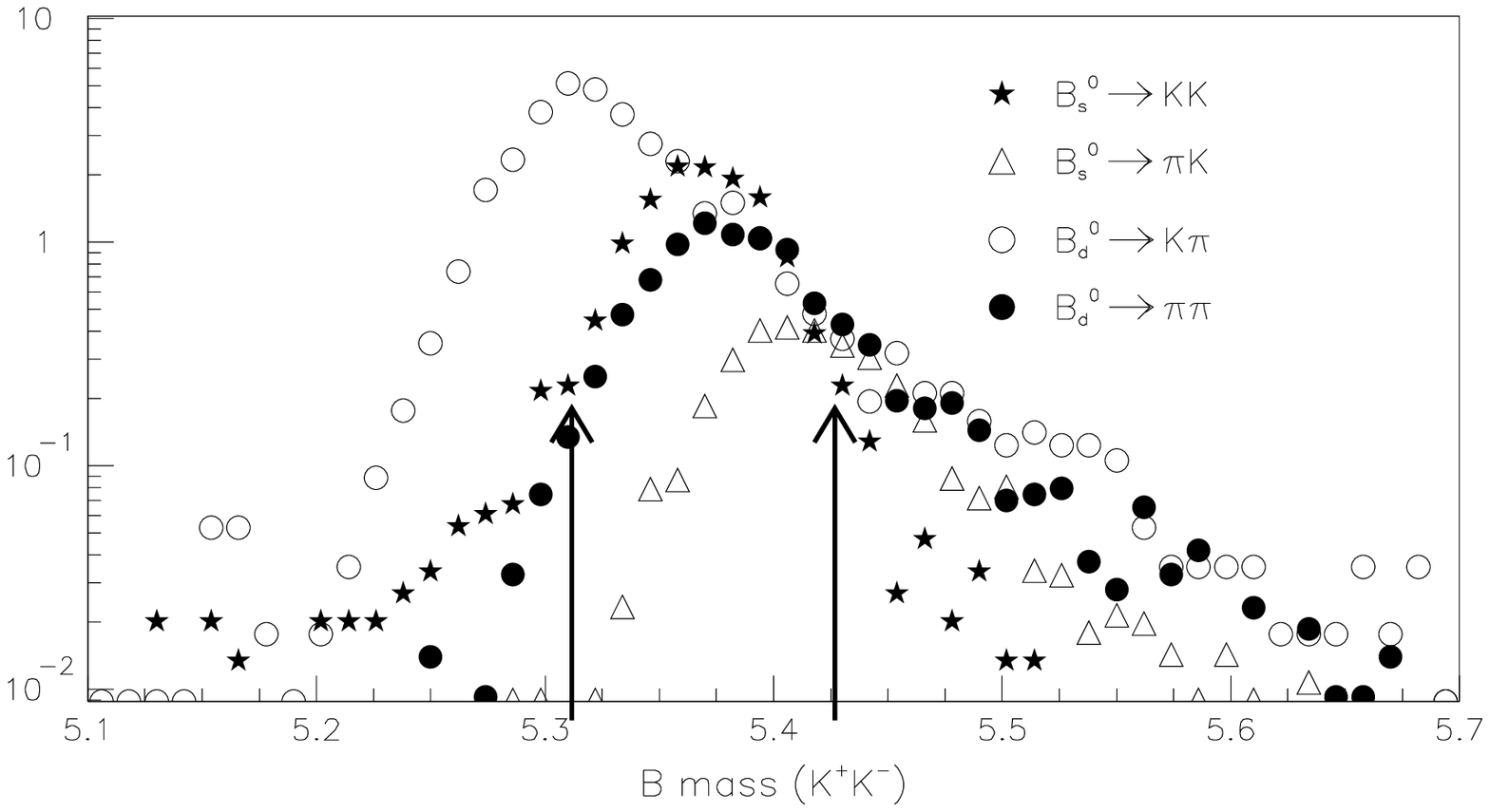,width=4in}
}
\centerline{
\put(40,130){\large\bf (b)}
\epsfig{figure=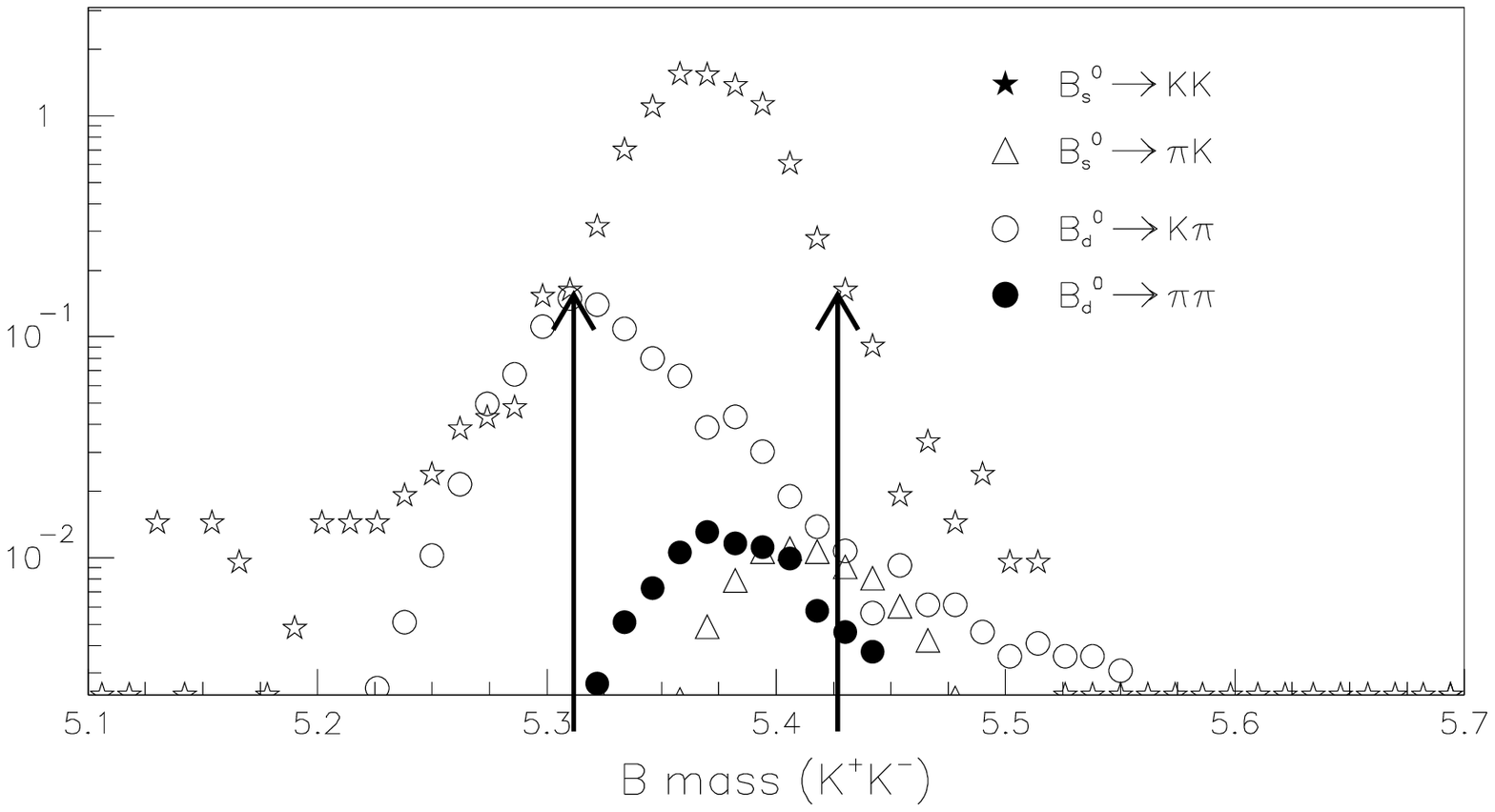,width=4in}
}
\caption[Two body $\kk$ mass without and with particle 
         identification at BTeV.]
{Two body ($\kk$) mass plot (a) without and (b) with particle 
         identification. Different decay channels are normalized by 
         their production cross sections. The arrows indicate the range
         of the signal mass window. (Note the log scale.)}
\label{fig:masskk}
\end{figure}

From the RICH simulation, BTeV finds that at an 80\% signal efficiency
for $\bstokk$, they accept 5\% (1.5)\% $\pipi (\kpi, \pik)$ background events as $\kk$. 
It is clear from Fig.~\ref{fig:masskk}(b) that by using the 
RICH information, BTeV can reject most of the backgrounds which are coming 
from other two-body decay modes. 

The expected $\bstokk$ yield, including the acceptance, reconstruction efficiency,
trigger efficiency, and particle ID efficiency is 33,000 events per year at
the design luminosity.  This is summarized in Table~\ref{tab:kktable}.

\begin{table}[tb]
\begin{center}
\begin{tabular}{|l|c|} \hline
Luminosity            & $2\,\times\,10^{32}\,{\rm cm}^{-2}\,{\rm s}^{-1}$ \\
Running time          & 10$^7$ sec \\
Integrated Luminosity & 2000 pb$^{-1}$ \\
$\sigma_{b\bar{b}}$   & 100 $\mu$b \\
Number of $B\bar{B}$ events       & $2\,\times\,10^{11}$ \\
Number of $B_s^0$ events               & $0.52\,\times\,10^{11}$ \\
${\cal{B}}(B_s^0\,\rightarrow\,K^+\,K^-)^{\dagger}$ & $1.7\,\times\,10^{-5}$ \\ \hline
Reconstruction efficiency            & 8.1\% \\
Trigger efficiency (Level\,1)                   & 64\% \\ 
Trigger efficiency (Level\,2)                   & 90\% \\ 
RICH I. D. efficiency                & 80.0\% \\

Number of reconstructed 
$B_s^0\,\rightarrow\,K^+\,K^-$           & 3.29$\,\times\,10^4$ \\ \hline
Background after RICH rejection        &                       \\ 
$B^0\,\rightarrow\,K^+\,\pi^-$         & 0.39$\,\times\,10^4$ \\
$B_s^0\,\rightarrow\,\pi^+\,K^-$         & 0.04$\,\times\,10^4$ \\
$B^0\,\rightarrow\,\pi^+\,\pi^-$       & 0.04$\,\times\,10^4$ \\
$B$-generic                               & 0.04$\,\times\,10^4$ \\ \hline
$S/B$                                     & 6.6  \\ \hline
\end{tabular}
\vspace*{0.3cm}
\caption
[Projected yield of $B_s^0\,\rightarrow\,K^+\,K^-$ and fake rates at BTeV.]
{Projected yield of $B_s^0\,\rightarrow\,K^+\,K^-$ and fake rates
($^{\dagger}$ indicates estimated branching fractions.)}
\label{tab:kktable}
\end{center}
\end{table}

\boldmath
\subsection[$B \rightarrow \pi\pi/KK$: Summary]
{$B \rightarrow \pi\pi/KK$: Summary
$\!$\authorfootnote{Author: M.~Paulini.}
}
\unboldmath
\index{decay!$B^0 \to \pi \pi$}%

Several years ago, 
the most important decay modes for the study of $CP$~violation in the
$B$~system were
believed to be \bjpks\ and \bpipi.
As discussed in Sec.~\ref{sec:psiphi}, 
time dependent $CP$~violation in the former mode measures \stb~\cite{bisa},
while
the decay $B^0\to\pi^+\pi^-$ usually appears in the literature as 
a tool to determine $\alpha=180^\circ-\beta-\gamma$. 
However, the CLEO collaboration~\cite{cleobranch} has shown that the
so-called penguin pollution  in \bpipi\ is sufficiently large to make the
extraction of fundamental physics parameters from the measured
$CP$~asymmetry rather difficult. 
An evaluation of measuring $CP$~violation in \bpipi\
does therefore 
require a strategy to distinguish penguin contributions  from 
tree diagrams.
A large number of
strategies to disentangle both contributions is discussed in
the literature~\cite{revs,alph}.
However, they generally require either very large data sets
or involve hard to quantify theoretical uncertainties.

For this workshop, CDF
evaluated a strategy of measuring the 
CKM angle~$\gamma$ as suggested by Fleischer in Ref.~\cite{flekk}.
This method is
particularly well matched to the capabilities of the Tevatron as 
it relates $CP$~violating 
observables in 
\index{decay!$B^0 \to \pi \pi$}%
\index{decay!$B_s^0 \to K K$}%
\bskk\ and \bpipi.
Both decays are related to each other by interchanging all
down and strange quarks, i.e.\ through the so-called ``U-spin''
subgroup of the SU(3) flavour symmetry of strong interactions. 
The strategy proposed in Ref.~\cite{flekk} uses this
symmetry to relate the ratio of hadronic matrix elements
for penguins and trees, and thus uses \bskk\ to correct for the
penguin pollution in \bpipi .

With the two-track hadronic trigger, CDF expects to reconstruct at least 
5000 $B^0\to\pi^+\pi^-$ and 20,000 $B^0\to K^\pm\pi^\mp$ events in \tfb\ of
data assuming the branching ratios measured by CLEO~\cite{cleobranch}, in
particular 
${\cal B}(B^0\to \pi^+\pi^-) = (4.3^{+1.6}_{-1.4}\pm 0.5)\times 10^{-6}$.
The question whether
CDF will be able to extract these signals from potentially enormous
backgrounds, has been studied throughout this workshop.
With respect to combinatorial background, a signal-to-background ratio not
worse than $S/B \sim 0.4$ can be expected.
Regarding physics backgrounds from $B \ra K \pi$ and $\Bs \ra K K$ decays,
a \bpipi\ signal can be extracted by exploiting
the invariant $\pi\pi$ mass 
distribution as well as the d$E$/d$x$ information provided by CDF's
Central Outer Tracker. From this, CDF expects the $B\ra \pi\pi$, $K\pi$,
$KK$ and $\pi K$ yields in the untagged sample to be measured with an
uncertainty of only a few percent.

Measurements on the tagged sample determines the time dependent
$CP$~asymmetry for $B^0\to \pi^+\pi^-$ and $B_s^0\to K^+K^-$ which is given by:
$
{\cal A}_{CP} = {\cal A}^{dir}_{CP} \cos \Delta m t + 
{\cal A}_{CP}^{mix} \sin \Delta m t  
$.
With the strategy suggested in Ref.~\cite{flekk}, the studies performed
during this workshop indicate that a measurement of the CKM angle
$\gamma$ to better than $10^{\circ}$ could be feasible at CDF with \tfb\ of
data. 
The utility of these modes depends on how well the uncertainty
from flavour $SU(3)$ breaking can be controlled.
Data for these and other processes should tell us the range of such
effects. The resulting Standard Model constraints could be quite
stringent. 
CDF estimates of possible SU(3) breaking effects show that 20\% SU(3)
breaking leads to a systematic error of less than half the statistical
precision given above. This encouraging result might allow CDF to make a
significant contribution to our understanding of the CKM unitarity triangle
within the first \tfb\ of Tevatron data in Run\,II.

Since the BTeV experiment will operate a RICH detector for particle
identification, excellent $\pi$-$K$ separation can be achieved and two-body
physics backgrounds can virtually be eliminated at BTeV.
Based on one year of running at design luminosity,
BTeV expects to reconstruct about 20,000~$\btopipi$ events with small
background contamination at the $10^{-4}$~level from 
$\btokpi$, $\bstopik$ and $\bstokk$.
With this event yield, BTeV expects an
uncertainty on the $CP$~asymmetry 
${\cal A}_{CP}$ of 0.024, as
summarized in Table~\ref{tab:pipitable}.
BTeV did not study a possible extraction of $\gamma$ using the method
proposed in Ref.~\cite{flekk} as discussed above, but has estimated
the yield for a $\bstokk$ signal to be 
33,000 events per year at
design luminosity (see Sec.~\ref{sec:b0_pipi}).








\boldmath
\section{Study of $B \ra D K$}

\subsection[$B \ra D K$: Introduction]
{$B \ra D K$: Introduction
$\!$\authorfootnote{Author: D.~Atwood.}
}
\unboldmath

The 
\index{CKM angle!$\gamma$}%
CKM angle $\gamma$ can be extracted via two related sets of four
decay processes, $B^-\to K^- D^0(\bar{D^0})$ and the $CP$~conjugate 
decays, or $\Bs(\bar\Bs)\to K^\pm D_s^\mp$. In both of these cases,
the sensitivity to $CP$~violation is achieved through the interference of
the two quark level processes $b\to c \bar u s$ and $b\to u \bar c s$. 

The final state particles for the most interesting decay
channels in this category contain  combinations of $K$'s and $\pi$'s.
Hence, an important feature of any detector is its ability to identify
these particles, resolve their momenta and perform $K$-$\pi$ separation.  
In addition, backgrounds, often from decay modes with branching fractions
that are orders of magnitude larger, 
must be well controlled. Otherwise, the $CP$~asymmetry will be diluted 
and the precision of measuring $\gamma$ will suffer. 

We briefly review first the extraction of $\gamma$ from $\Bs$ decays and
then summarize how the angle $\gamma$ can be obtained from $B \ra D^0 K$.

\boldmath
\subsubsection{$B_{s}^{0} \ra D_s^- K^+$: Introduction}
\label{sec:cp_bsdsk_intro}
\unboldmath
\index{decay!$B_s^0 \to D_s^- K^+$}%

The necessary interference effect is achieved through mixing of the initial
state via $\Bs\bar\Bs$ oscillation. For example, we could have either
a direct decay amplitude for $\Bs\to D_s^- K^+$ ($\bar b\to c \bar u s$ channel) 
or first a $\Bs\to\bar\Bs$ transition and then the $\bar\Bs\to D_s^- K^+$ 
($b\to c \bar u s$ channel) decay. Note that the two decay amplitudes
are not $CP$~conjugates (in contrast to the case of final $CP$~eigenstates)
and therefore carry different strong phases. These phases cannot be
reliably calculated with currently available theoretical methods. 
Therefore enough data must be gathered to fit simultaneously for $\gamma$
and the strong phase difference $\delta$.  

The time dependent decay rates for the four relevant processes
are given in Eq.~(\ref{6:tagsing}) and reproduced here using 
$\phi_{D_s^+ K^-}=-\gamma$.
\beqa\label{6:tagsing_rep}
\Gamma(B_s^0 \to D_s^- K^+)\ &=&\ {|A_f|^2e^{-\Gamma_s\,t}\over2}\Big\{
(1+|\lambda_f|^2)\cosh(\Delta\Gamma_s\,t/2)+
(1-|\lambda_f|^2)\cos(\Delta m_s\,t)\nonumber\\
& &\ \ \  -2|\lambda_f|\cos(\delta+\gamma)\sinh(\Delta\Gamma_s\,t/2)
-2|\lambda_f|\sin(\delta+\gamma)\sin(\Delta m_s\,t) \Big\},\nonumber\\
\Gamma(B_s^0 \to D_s^+ K^-)\ &=&\ {|A_f|^2e^{-\Gamma_s\,t}\over2}\Big\{
(1+|\lambda_f|^2)\cosh(\Delta\Gamma_s\,t/2)-
(1-|\lambda_f|^2)\cos(\Delta m_s\,t)\nonumber\\
& &\ \ \  -2|\lambda_f|\cos(\delta-\gamma)\sinh(\Delta\Gamma_s\,t/2)
+2|\lambda_f|\sin(\delta-\gamma)\sin(\Delta m_s\,t) \Big\},\nonumber\\
\Gamma(\bar B_s^0 \to D_s^- K^+)\ &=&\ {|A_f|^2e^{-\Gamma_s\,t}\over2}\Big\{
(1+|\lambda_f|^2)\cosh(\Delta\Gamma_s\,t/2)-
(1-|\lambda_f|^2)\cos(\Delta m_s\,t)\nonumber\\
& &\ \ \  -2|\lambda_f|\cos(\delta+\gamma)\sinh(\Delta\Gamma_s\,t/2)
+2|\lambda_f|\sin(\delta+\gamma)\sin(\Delta m_s\,t) \Big\},\nonumber\\
\Gamma(\bar B_s^0 \to D_s^+ K^-)\ &=&\ {|A_f|^2e^{-\Gamma_s\,t}\over2}\Big\{
(1+|\lambda_f|^2)\cosh(\Delta\Gamma_s\,t/2)+
(1-|\lambda_f|^2)\cos(\Delta m_s\,t)\nonumber\\
& &\ \ \  -2|\lambda_f|\cos(\delta-\gamma)\sinh(\Delta\Gamma_s\,t/2)
-2|\lambda_f|\sin(\delta-\gamma)\sin(\Delta m_s\,t) \Big\}.\nonumber\\
\label{eq:bdk_gamma14}
\eeqa
Here, we abbreviated $A_f$ for 
$A_{D_s^-K^+}$ and $\lambda_f$ for $\lambda_{D_s^-K^+}$.
The primary concern is to extract  $\gamma$ from these rates. 
In the following, we will assume that $\Delta m_s$ and $\Delta\Gamma_s$ 
are already known since they can be determined more accurately with 
other modes. All four parameters, $|A_{D_s^-K^+}|$, $|\lambda_{D_s^-K^+}|$ 
and $\delta\pm\gamma$, can, in principle, be extracted from the time dependent 
data for the four decay processes. For example, the overall normalization 
$|A_{D_s^-K^+}|^2$ can be extracted from $\Gamma[\Bs(t=0)\to D_s^-K^+]$
and $\Gamma[\bar\Bs(t=0)\to D_s^-K^+]$, and the value of
$|\lambda_{D_s^-K^+}|$ can then be obtained from $\Gamma[\Bs(t=0)\to D_s^-K^+]$
and $\Gamma[\bar\Bs(t=0)\to D_s^-K^+]$. In actuality, one performs a simultaneous 
fit for $|\lambda|$, $|A|$, $\delta$ and $\gamma$ from the experimental data on 
the four channels.  Note, the measurements determine only 
$\sin(\delta\pm\gamma)$ and $\cos(\delta\pm\gamma)$. This 
determines both $\delta$ and $\gamma$ (which we are most interested 
in) up to the two fold ambiguity,
\begin{equation}
(\delta,\gamma);~~~~(\delta+\pi,\gamma+\pi).
\end{equation}

Aside from the issue of gathering enough statistics to obtain accurate time 
dependent rates, there are two situations for which data may not be able to
 unambiguously fit all the coefficients as suggested above:
\begin{enumerate}
\item[(1)]
$\Delta m_s$ is so large that the time resolution is insufficient to
extract the  ``$\sin$'' and ``$\cos$'' terms.
\item[(2)]
$\Delta\Gamma_s/\Gamma_s$ is so small that the ``$\sinh$'' term does not
become large enough to be distinguished.
\end{enumerate}

In case (1) the crucial problem is the finite time resolution of the 
detector. To get a feeling for how this affects the data, let us assume 
 the time resolution of the detector has a Gaussian spread with a width 
$\sigma/\Gamma_s$. If $x_s\sigma\gg1$, the oscillating terms will be 
damped due to the smearing by $\sim\exp(-x_s^2\sigma^2/2)$ and only the 
``cosh" and ``sinh" terms survive. 
In this regime we are, in effect, seeing $\Bs$ states as incoherent 
mixtures of $B^{L}_s$ and $B^{H}_s$, without the knowledge of the coherence 
between the states encoded in the oscillatory terms. If the data allows us to 
isolate the ``sinh" and ``cosh" terms, we will be able to extract 
$\cos(\delta+\gamma)$ and $\cos(\delta-\gamma)$. This then allows us to
 determine $(\delta,\gamma)$ up to the following ambiguity:
\begin{eqnarray}
(\pm\delta,\pm\gamma);~~~(\pm\gamma,\pm\delta);~~~(\pi\pm\delta,\pi\pm\gamma);
~~~(\pi\pm\gamma,\pi\pm\delta).
\end{eqnarray}
In particular, $\gamma$ has an 8-fold ambiguity between
$\{\pm\gamma,\pi\pm\gamma,\pm\delta,\pi\pm\delta\}$.
This could be reduced to a 4-fold ambiguity if a second final state, such as 
$D_s^- K^*$, is also analyzed in a similar fashion, provided the two values of 
$\delta$ are significantly different. 

In case (2), that is, if $\Delta\Gamma_s/\Gamma_s$ is so small that the ``sinh" and 
``cosh" terms cannot be measured, we are in a similar situation except 
that we now can only determine $\sin(\delta+\gamma)$ and $\sin(\delta-\gamma)$.
In this case, a given solution $(\delta,\gamma)$ produces the same results as:
\begin{eqnarray}
(\delta,\gamma);~~~(\pi+\delta,\pi+\gamma);~~~(\pi-\delta,-\gamma);~~~
(-\delta,\pi-\gamma);~~~({\pi\over 2}-\gamma,{\pi\over 2}-\delta);\nonumber\\
(-{\pi\over 2}-\gamma,-{\pi\over 2}-\delta);~~~
({\pi\over 2}+\gamma,-{\pi\over 2}+\delta);~~~
(-{\pi\over 2}+\gamma,{\pi\over 2}+\delta).
\end{eqnarray}
Consequently, $\gamma$ has an 8-fold ambiguity between
$\{\pm\gamma,\pi\pm\gamma,{\pi\over 2}\pm\delta,-{\pi\over 2}\pm\delta\}$
and again an additional mode such as $D_s^- K^*$ will  reduce this to a 4-fold 
ambiguity if the two modes have significantly different values of $\delta$.


\boldmath
\subsubsection{$B^- \ra D^0 K^-$: Introduction}
\label{sec:cp_bdk_intro}
\unboldmath
\index{decay!$B^0 \to D^0 K^-$}%

In the Standard Model $b \ra c\bar{u}s$ and $b \ra \bar{c}us$ transitions
have a relative 
\index{CKM angle!$\gamma$}%
CKM phase $\gamma$. 
In the case of the $B^-\to K^- D^0(\bar{D^0})$ decay mode,
the sensitivity to $\gamma$ is achieved through the 
interference of common decay modes of the $D^0$ and $\bar{D^0}$ channels.  
The Gronau-London-Wyler (GLW) method \cite{glw} extracts $\gamma$ by
measuring the  
$B^{\pm}$ decay rates to
$D^0/\bar{D}^0$ mesons. If the  $D^0$ and $\bar{D}^0$ decay to a 
$CP$~eigenstate, then 
the two decays $B^- \ra K^- D^0$ and $B^- \ra K^- \bar{D}^0$ lead to a
common final 
   state and can give rise to $CP$~violating effects. However, the two interfering amplitudes
   are very different in magnitude and thus the interference effects are limited to
   ${\cal O}(10\%)$. Another problem is that it is necessary to measure separately the branching
   ratios ${\cal B}(B^- \ra K^- D^0)$ and ${\cal B}(B^- \ra K^- \bar{D}^0)$. While the former can
   be measured in a straightforward way, the latter is very difficult to measure.
   
   Recently Atwood, Dunietz and Soni \cite{ads} have pointed out that $CP$~violation
   can be greatly enhanced for decays to final states that are common to both $D^0$
   and $\bar{D}^0$ but are not $CP$~eigenstates. In particular, large asymmetries are
   possible for final states $f$ such that $D^0 \ra f$ is doubly Cabibbo suppressed 
   and $\bar{D}^0 \ra f$ is Cabibbo allowed.
 
   The Atwood, Dunietz and Soni method requires the determination of branching ratios 
   for at least two distinct final states $f_1$ and $f_2$.
   
We define the following quantities :
   
   \begin{equation} a = {\cal B}(B^- \ra K^- D^0) \end{equation}
\begin{equation} 
b = {\cal B}(B^- \ra K^- \bar{D}^0)  \end{equation}
   \begin{equation} c(f_1) = {\cal B}(D^0 \ra f_1) ,\qquad  c(f_2) = {\cal B}(D^0 \ra f_2)    \end{equation}
   \begin{equation} c(\bar{f_1}) = {\cal B}(D^0 \ra \bar{f_1}) , \qquad
    c(\bar{f_2}) = {\cal B}(D^0 \ra \bar{f_2})    \end{equation}
   \begin{equation} d(f_1) = {\cal B}(B^- \ra K^- f_1) , \qquad
    d(f_2) = {\cal B}(B^-  \ra K^- f_2)    \end{equation}
   \begin{equation} \bar{d}({f_1}) = {\cal B}(B^+ \ra K^+ f_1) , \qquad
    \bar{d}({f_2}) = {\cal B}(B^+ \ra K^+ f_2)    \end{equation}
   Assume that we can measure the quantities $a$, $c(f_1)$, $c(f_2)$, $c(\bar{f_1})$, $c(\bar{f_2})$,
   $d(f_1)$, $d(f_2)$, $\bar{d}({f_1})$ and  $\bar{d}({f_2})$ 
   but not $b$.
   
   We can express $d(f_1)$ in terms of $a$, $b$, $c(f_1)$, $c(\bar{f_1})$,  the strong phase $\xi_1$ and
   the weak phase $\gamma$. 
   \begin{equation} 
d(f_1) = a \times c(f_1) + b \times c(\bar{f_1}) + 
           2\sqrt{a \times b \times c(f_1) \times c(\bar{f_1})}\cos(\xi_1 + \gamma)  
\label{eq:bdk_intro_d12_1}
\end{equation}
   \begin{equation} \bar{d}(f_1) = a \times c(f_1) + b \times c(\bar{f_1}) +
              2\sqrt{a \times b \times c(f_1) \times c(\bar{f_1})}\cos(\xi_1 - \gamma)  \end{equation}
   \begin{equation} d(f_2) = a \times c(f_2) + b \times c(\bar{f_2}) + 
              2\sqrt{a \times b \times c(f_2) \times c(\bar{f_2})}\cos(\xi_2 + \gamma)  \end{equation}
   \begin{equation} \bar{d}(f_2) = a \times c(f_2) + b \times c(\bar{f_2}) + 
              2\sqrt{a \times b \times c(f_2) \times c(\bar{f_2})}\cos(\xi_2 - \gamma)  
\label{eq:bdk_intro_d12_4}
\end{equation}

   These four equations contain the four unknowns $\xi_1, \xi_2, b$
   and $\gamma $ 
   which can be determined up to discrete ambiguities. 
    Adding additional decay modes will reduce the ambiguities.
    The strong phases $\xi_i$ are related to the $D$ decay phase shifts $\delta_i$
    by the relation :
     \begin{equation} \xi_1 - \xi_2 = \delta_1 -\delta_2. \end{equation}
    If the  $D$ decay phase shifts can be determined elsewhere then we have an extra constraint
    on the equations.
    This method measures direct $CP$~violation and does not require tagging nor time-dependent 
    measurements.
   If we add a third decay mode we have six equations with five unknowns
   which will help 
   to resolve ambiguities.






\boldmath
\subsection{$B \ra D K$: CDF Report}
\unboldmath
\index{CDF!$\sin\gamma$ prospects}%

We summarize the study of measuring the unitarity triangle angle $\gamma$
at CDF in Run\,II, first using the decay mode $\Bs\ra D_s^- K^+$
and second exploiting the decay $B^-\to D^0 K^-$. 

\boldmath
\subsubsection[$B_s^0 \ra D_s^- K^+$: CDF Report]
{$B_s^0 \ra D_s^- K^+$: CDF Report
$\!$\authorfootnote{Authors: S.~Bailey and P.~Maksimovi\'c.}
}
\label{sec:bdkcdf}
\unboldmath
\index{decay!$B_s^0 \to D_s^- K^+$}%
\index{CDF!$B_s^0 \to D_s^- K^+$ prospects}%



As outlined in Sec.~\ref{sec:cp_bsdsk_intro} above,
the decay mode $B_s^0 \ra D_s^- K^+$
probes the
unitarity triangle angle $\gamma$ by $CP$ violation due to
interference of decays with and without mixing~\cite{ADK, pursuit}
(see also Sec.~\ref{sec:cp_type}).
The weak amplitude of $B_s^0 \leftrightarrow \bar B_s^0$ mixing is
approximately 
real, as is the weak amplitude of the decay $B_s^0 \to D_s^- K^+$. But
the decay $\bar B_s^0 \to D_s^- K^+$ has a non-zero phase which
is approximately the angle $\gamma$ of the unitarity triangle.
Thus, the overall $CP$ violating weak phase of this decay is
$\gamma$ to the accuracy of the Wolfenstein
parameterization of the CKM matrix $({\cal O}(10^{-4})).$

The decay rates given in Eq.~(\ref{6:tagsing_rep}) 
allow the extraction of $\sin(\gamma\pm\delta)$.
If $\Delta\Gamma_s/\Gamma_s$ is large
enough, $\cos(\gamma\pm\delta)$ may additionally be
extracted~\cite{DunBs}.
Since the $\cos(\gamma\pm\delta)$ terms are identical for the
same final states, tagging is unnecessary to measure
$\cos(\gamma\pm\delta)$ and a much larger untagged
sample may be used.  Extracting $\cos(\gamma\pm\delta)$ with
the untagged sample has the additional benefit of not needing to
resolve the rapid $B_s^0 \leftrightarrow \bar B_s^0$ oscillations.
Unfortunately, the two measurements cannot extract $\gamma$
separately, but they can be used to
constrain the tagged fit and resolve discrete ambiguities in
extracting $\gamma$ from $\sin(\gamma\pm\delta)$.

If $\Delta\Gamma_s/\Gamma_s$ is too small to allow an extraction
of $\cos(\gamma\pm\delta)$, theoretical input on
$\delta$ will likely be necessary.  Although a measurement
of $\sin(\gamma\pm\delta)$ may
exclude much of the $(\gamma, \delta)$ plane, the discrete
ambiguities are such that projecting onto the $\gamma$ axis
usually does not exclude much of $\gamma$, even with fairly
small errors on $\sin(\gamma\pm\delta)$.  The current theoretical prediction of
$|\delta|<5^\circ$~\cite{delta}, however, is sufficient to exclude most
discrete ambiguities.

An additional subtlety which must be considered is the possibility of
measuring an unphysical value of $\sin(\gamma\pm\delta)>1$.  If either
$\sin(\gamma\pm\delta)$ is very near or at 1, even measurements with
small errors would frequently produce unphysical results of 
$\sin(\gamma\pm\delta) > 1$.  Thus a
technique such as the unified approach of Feldman and
Cousins~\cite{Feldman:1998qc}  must be
used to convert the measured amplitude of $\sin(\gamma\pm\delta)$
to the quantities of interest, $\gamma$ and $\delta$, rather
than relying upon a straightforward trigonometric transformation.

\subsubsection*{Background Studies}

The reduction of backgrounds will be one of the primary
challenges for using the $\Bs \ra D_s^- K^+$ mode at CDF.  
The physics backgrounds which
closely mimic the signal are given below where the branching ratios used in
this study are
estimated branching fractions.

\begin{center}
\begin{tabular}{l|l||l|l}
Background Mode & ${\cal B}\ \times 10^{-3}$ & Signal Mode & ${\cal B}\ \times 10^{-3}$ \\ 
\hline
$B_s^0 \to D_s^- \pi^+$ & \hskip 0.15in 3.0 &
	$B_s^0 \to D_s^- K^+$   & \hskip 0.15in 0.2 \\
$B_s^0 \to D_s^{*-} \pi^+$ & \hskip 0.15in 3.0 &
	$B_s^0 \to D_s^+ K^-  $ & \hskip 0.15in 0.1 \\
$B_s^0 \to D_s^{*-} K^+$   & \hskip 0.15in 0.2 & & \\
$B_s^0 \to D_s^{*+} K^-  $ & \hskip 0.15in 0.1 & & \\
$B^0 \to D_s^- \pi^+$ & \hskip 0.15in 0.1 & & \\
$B^0 \to D_s^{*-} \pi^+$ & \hskip 0.15in 0.1 & & \\
\end{tabular}
\end{center}

As shown in Figure~\ref{fig:sbmass}, reconstructing the physics backgrounds
as $B_s^0 \to D_s^- K^+$
produces a mass shift away from the $B_s^0$ mass such that the
$S/B$ in the $B_s^0$ mass region is $1/3$ even though the ratio of
branching fractions is much worse.

\begin{figure}
\begin{center} \includegraphics{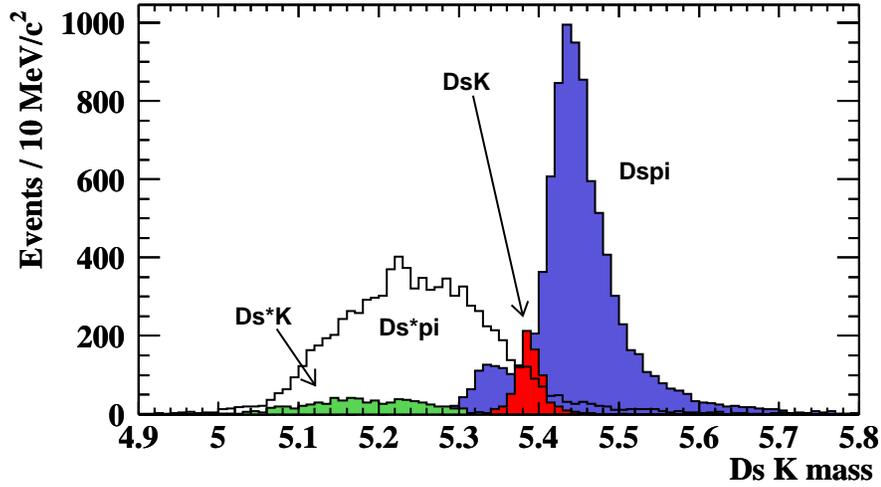} \end{center}
\caption
[Mass plot for the $B_s^0 \to D_s^- K^+$ signal and various
physics backgrounds at CDF.]
{Mass plot for the $B_s^0 \to D_s^- K^+$ signal (marked DsK) and various
physics backgrounds.  The $S/B$ in the signal region is 1/3 before any
particle identification.
}
\label{fig:sbmass}
\end{figure}

Combinatoric backgrounds are expected to be the primary concern.
A $S/B$ study for 
\index{decay!$B_s^0 \to D_s^- \pi^+$}%
$B_s^0 \to D_s^- \pi^+$ using CDF Run I data concluded
that a $S/B$ in the range $1/2$ to $2/1$ was reasonable for that mode.
That study was statistics limited and did not consider the $S/B$
improvements that will be achieved using the 3-dimensional vertexing
capabilities of the SVX\,II detector and d$E$/d$x$ cuts.
Without including those improvements,
scaling for branching fractions produces
a nominal combinatoric $S/B$ for $B_s^0 \to D_s^- K^+$ of $1/15$.

Figure \ref{fig:sbvssep} shows the resulting $S/B$ (physics and combinatoric)
after applying d$E$/d$x$ cuts as a function of the d$E$/d$x$ separation power.
The cuts used here have a constant signal efficiency corresponding
to 850 signal events.
For this study we use a nominal $S/B$ of 1/6 which corresponds to a
d$E$/d$x$ separation power of 1.1\,$\sigma$. 

\begin{figure}[tb]
\begin{center} \includegraphics{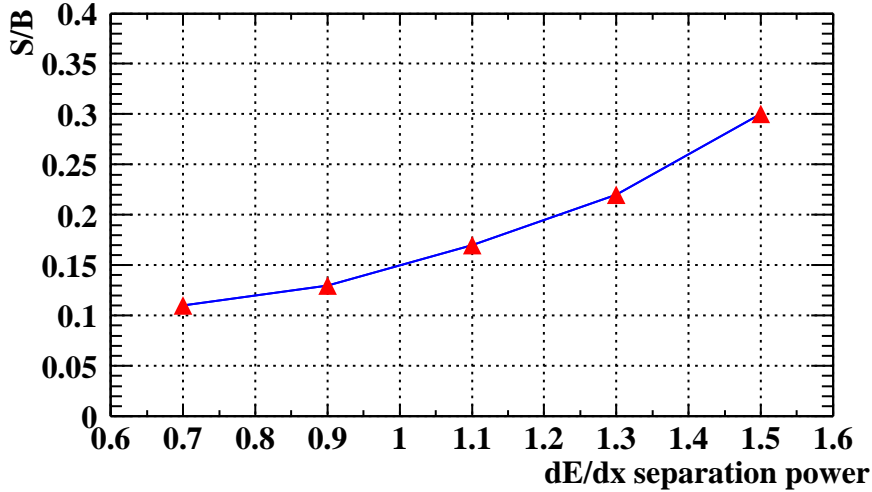} \end{center}
\caption
[Example signal to background ratio as a function of the d$E$/d$x$ separation
power at CDF.]
{Example signal to background ratio $S/B$ as a function of the d$E$/d$x$
separation power between kaons and pions.}
\label{fig:sbvssep}
\end{figure}

\subsubsection*{Results of Toy Monte Carlo Study}
\label{sec:tmc}

To study CDF's sensitivity to measuring $\gamma$ using this mode,
we wrote a Toy Monte Carlo plus fitter.  We generated signal events
according to the decay rate Equations~(\ref{eq:bdk_gamma14})
and added background events
with appropriate proper time dependencies.  The events
were smeared by a Gaussian resolution function and (mis)assigned
an observed flavour according to a mistag probability.  We fit these data
using an unbinned likelihood method and compared the results and
their errors with the input values.


The central values used as input parameters for this study are given in
Table~\ref{tab:cdf_bdk_input}. 
The left table lists physical parameters to be measured over which
we have no control. The chosen values are based upon Standard Model
predictions \cite{ali}.
The right table lists parameters which are
CDF dependent and may be improved with effort. Their values
are chosen based upon other CDF\,II studies.  $N$ is the number
of reconstructed events before flavour tagging is applied. 
Our study shows that CDF expects to reconstruct about 850 
$B_s^0 \ra D_s^- K^+$ signal events in \tfb\ of Run\,II data.
While studying
the dependence of the error upon a given parameter, we kept the
rest of the parameters fixed at these values.

\begin{table}[tb]
\begin{center}
\begin{tabular}{l|l||l|l}
\hline
Parameter & Standard Model Estimate & Parameter & CDF\,II Estimate \\
\hline
$\gamma$ & $90^\circ$ & $\sigma_t$ & 0.03 \\
$\delta$ & $10^\circ$ & $\eD$ & 0.113 \\
$|A_f|/|\bar A_f|$	& $\sqrt{1.4/2.4}$ & $N(B_s^0 \to D_s^- K^+)$	& 850 \\
$x_s$ & 20 & $S/B$ & 1/6 \\
$x_d$		& 0.723 & & \\
$\Delta\Gamma_s/\Gamma_s$	& 0.16 & & \\
\hline
\end{tabular}
\vspace*{0.3cm}
\caption
[Central values of parameter used in the study of $B_s^0 \ra D_s^- K^+$ at
CDF.] 
{Central values of parameter used in the study of $B_s^0 \ra D_s^- K^+$ at
CDF.} 
\label{tab:cdf_bdk_input}
\end{center}
\end{table}

\begin{figure}[tbp]
\begin{center} \includegraphics{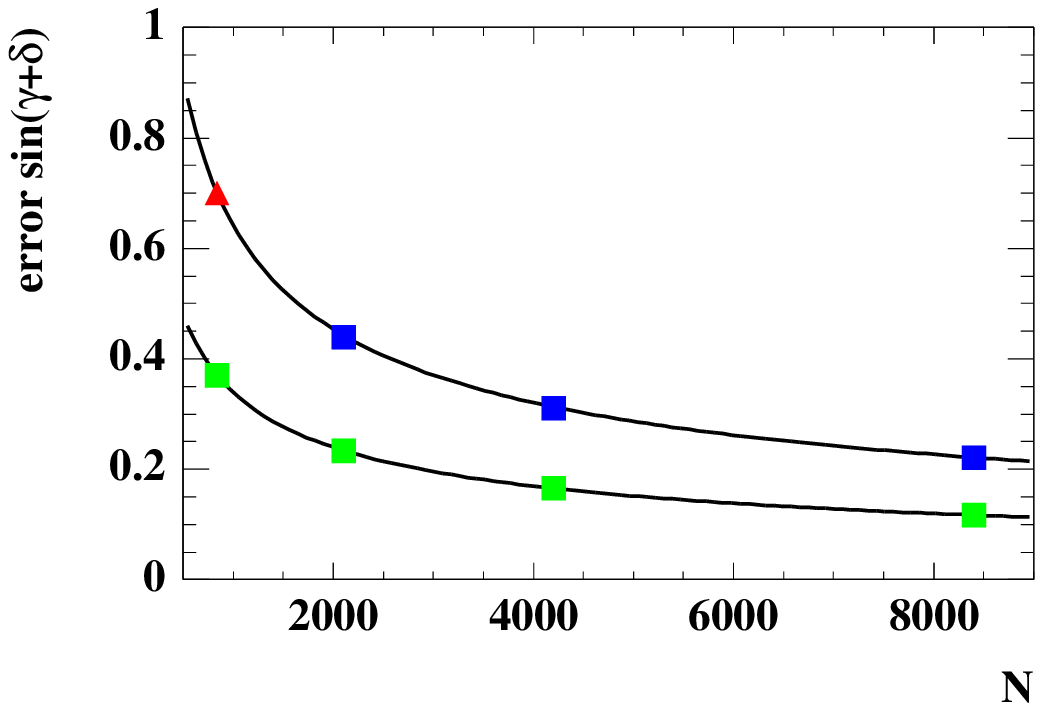} \end{center}
\caption
[The error on $\sin(\gamma\pm\delta)$ 
as a function of the number of $B_s^0 \to D_s^- K^+$ events.]
{The error on $\sin(\gamma\pm\delta)$ from Toy Monte Carlo experiments
as a function of the number of observed $B_s^0 \to D_s^- K^+$ events $N$.
The points correspond to approximately 2, 5, 10, and 20 fb$^{-1}$ of data.
}
\label{fig:errVsN}
\end{figure}

Figure \ref{fig:errVsN} shows the dependence of the error on
the number of pre-tagged signal events for both $S/B = 1/6$ (upper points)
and $S/B = 1/1$ (lower points).  The points correspond to approximately
2, 5, 10, and 20 fb$^{-1}$ of data.

Figure \ref{fig:err} shows how the
errors scale with the proper time resolution $\sigma_t$, 
the effective tagging efficiency $\eD$, 
the \Bs~mixing parameter $x_s$
and the ratio of decay amplitudes $\rho$.
The triangles represent the error using
the central values of the input parameters, while the squares are the errors
from varying one parameter while leaving the others fixed.  The
curves are the theoretical errors discussed below.
The lower points and curves are for $S/B = 1/1$ for comparison.

\begin{figure}[tbp]
\centerline{
\epsfysize=3.6in
\epsffile{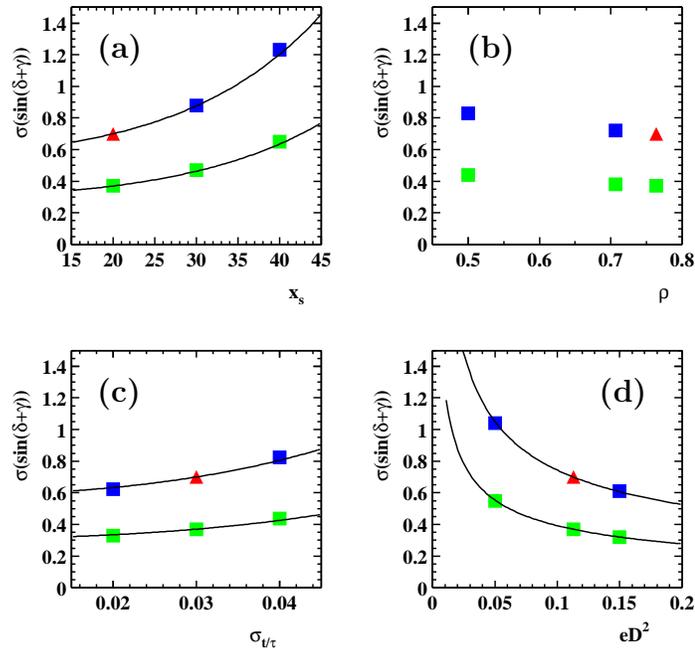}
\put(-235,228){\large\bf (a)}
\put(-95,228){\large\bf (b)}
\put(-235,98){\large\bf (c)}
\put(-45,98){\large\bf (d)}
}
\caption
[The error on $\sin(\gamma\pm\delta)$ 
as a function of different parameters at CDF.]
{The error on $\sin(\gamma\pm\delta)$ from Toy Monte Carlo experiments
as a function of (a) $x_s$, (b) the ratio of decay amplitudes $\rho$,
(c) the proper time resolution $\sigma_t$ and (d) the effective flavour
tagging efficiency $\eD$.  The
triangle is the error using the central values of all parameters with
a $S/B = 1/6$.  The curve is the theoretically expected error.
The lower points and curves are for $S/B = 1/1$.
}
\label{fig:err}
\end{figure}


The expected error on $\sin(\gamma\pm\delta)$ is closely modeled by the
following expression:
\begin{equation}
\label{eq:sigma}
\sigma(\sin(\gamma\pm\delta)) = {1 \over {\cal D}_{res}} {1 \over {\cal D}_{bkg}}
  {1 \over {\cal D}_{fit}}
  {1 \over \sqrt{ \eD N}}
\end{equation}
where ${\cal D}_{res} = e^{-\sigma_t^2 x_s^2 / 2}$, ${\cal D}_{bkg} = \sqrt{S \over S+B}$,
$\eD$ is the effective flavour tagging efficiency and ${\cal D}_{fit}$ is
normalized to the error obtained using the central values of the input
parameters. A discussion of the terms of this equation may be found in
Ref.~\cite{mcdonald}.
There was very little dependence of the errors upon the values of
$\gamma$, $\delta$ and $\Delta\Gamma_s/\Gamma_s$.


In conclusion, an initial measurement
of $\gamma$ using $B_s^0 \to D_s^- K^+$ should be possible with
CDF in Run\,II.  Within the first 2 fb$^{-1}$, the expected error on
$\sin(\gamma\pm\delta)$ is around 
0.4 to 0.7 depending upon what the background levels turn out to be.
By the end of Run\,II an uncertainty near $0.1$ may be achievable.
The most limiting factors for CDF\,II are the background levels and
the overall signal size.  There are significant uncertainties on
these parameters, but our Toy Monte Carlo studies indicate that
Eq.~(\ref{eq:sigma}) is an accurate predictor of the error over
a wide range of input parameters.

\boldmath
\subsubsection[$B^- \ra D^0 K^-$: CDF Report]
{$B^- \ra D^0 K^-$: CDF Report
$\!$\authorfootnote{Authors: A.~Cerri, G.~Punzi and G.~Signorelli.}
}
\unboldmath
\index{decay!$B^0 \to D^0 K^-$}%
\index{CDF!$B^0 \to D^0 K^-$ prospects}%

In this section, we evaluate the prospects of measuring the CKM angle
$\gamma$ using the decay channel $B^- \ra D^0 K^- \ra [K\pi] K^-$ at
CDF in Run\,II.
This requires the knowledge of all branching fractions involved, 
where we list the estimated branching ratios used as input for this
study in Table~\ref{tab:cdf_bdkm_in}.

\begin{table}[tb]
\begin{center}
\begin{tabular}{llc}
\hline
${\cal B}(B^+ \to K^+ \bar D^0  $) & $=2.6\pm 0.08 \times 10^{-4}$ & CLEO  \\
${\cal B}(B^+ \to K^+      D^0  $) & $\approx 2 \times 10^{-6}$   & 
Estim. \cite{ads} \\
${\cal B}(\bar D^0 \to K^- \pi^+$) & $=1.3\pm 0.3 \times 10^{-4}$ & CLEO   \\
${\cal B}(\bar D^0 \to K^+ \pi^-$) & $=3.8\pm 0.1 \times 10^{-2}$ & PDG  \\
\hline
\end{tabular}
\vspace*{0.3cm}
\caption
[Estimated branching ratios of decays involved in the analysis
of $B^- \ra D^0 K^-$ at
CDF.]
{Estimated branching ratios of decays involved in the analysis
of $B^- \ra D^0 K^- \ra [K\pi] K^-$ at
CDF.}
\label{tab:cdf_bdkm_in}
\end{center}
\end{table}

Beginning with Equations~(\ref{eq:bdk_intro_d12_1}) - 
(\ref{eq:bdk_intro_d12_4}) as shown in Sec.~\ref{sec:cp_bdk_intro}, 
the number of events in each channel, which we will shortly 
refer to as $X_{1/2}$ and $Y_{1/2}$, is given by
\beq
\begin{array}{c}
X_{1/2} = d_{1/2}\cdot [\sigma_B\,\eta_{1/2}\,\epsilon_{1/2}\,{\cal  L} ] 
\hspace*{0.5cm} {\rm and} \hspace*{0.5cm}
Y_{1/2} = \bar d_{1/2}\cdot [\sigma_B\,\eta_{1/2}\,\epsilon_{1/2}\,{\cal L} ] 
\end{array}
\eeq
where $\sigma_B$ is the  $B^+$ production cross section,
$\epsilon_{1/2}$ is the detector acceptance times the trigger efficiency 
for the corresponding channel, 
$\eta_{1/2}$ is the efficiency on the signal from offline
requirements and
$\cal L$ is the integrated luminosity.
From the measurement of $X_1$, $X_2$, $Y_1$ and $Y_2$, as well as knowing 
$[\sigma_B\, \eta_{1/2}\, \epsilon_{1/2}\, {\cal L} ]$,
it is formally possible to 
invert the relations given in Eqs.~(\ref{eq:bdk_intro_d12_1}) - 
(\ref{eq:bdk_intro_d12_4}) to
obtain a value for $\cos( \xi_{1/2} + \gamma)$ and $\cos (\xi_{1/2} - \gamma)$.

As a first step, we evaluate the resolution on the angle $\gamma$ when
$\gamma$  lies in the range  $60^\circ < \gamma
< 100^\circ$ and $\xi$ in the range $-10^\circ < \xi
< 30^\circ$, as suggested by Standard Model fits~\cite{PRS}.
We use a Toy Monte Carlo to estimate the resolution on the studied parameters
in the following way. We extract $\gamma$ and $\xi$ within their range
and the values of all branching fractions from Gaussian distributions 
around their nominal values. With these parameters, and a given signal to
noise ratio, we calculate the expectation values of the number of 
events in each channel, $\bar x$ and $\bar y$. $X$ and $Y$ are then
obtained from a Poisson distribution around those values.
From such ``pseudo-measurements'' we obtain the values of $\widehat \gamma$
and  $\widehat \xi$ that maximize the likelihood. 
We then plot the distribution of the experimental error  $\widehat \gamma -
\gamma$, averaged over the whole range of $\gamma$ and $\xi$ considered, and
extract its sigma by a Gaussian fit.
In Figure~\ref{fig:cdf_bdk_pull} we show an example distribution using 
140 observed events, zero background, and a $10\%$ uncertainty on all
branching ratios
involved. The sigma of this distribution is about $9^\circ$.

\begin{figure}[tb]
\centerline{
\epsfxsize=3.2in
\epsffile{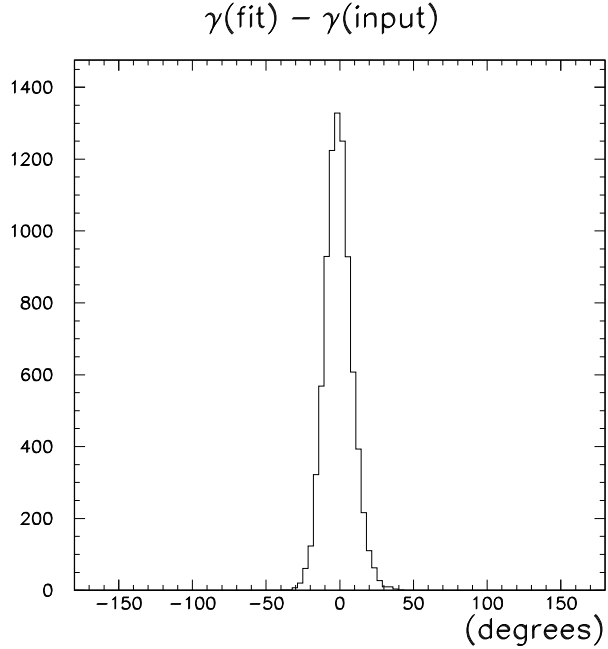}
}
\vspace*{0.3cm}
\caption
[Deviation of the value of  $\gamma$ obtained from
the fit versus the value used as input in the Monte Carlo study.]  
{Deviation of the value of  $\gamma$ obtained from
the fit and the value of $\gamma$ used as input in the Monte Carlo study.  
We neglected backgrounds and assumed a $10\%$ uncertainty on 
the four branching ratios. Note, one of the four branching fractions,
namely $D^0 \ra K^- \pi^+$, is   
at present known to better than $3\%$.}
\label{fig:cdf_bdk_pull}
\end{figure}

Given the good behavior of the resolution function even with this small
sample, 
we decided it was more convenient to replace the Monte Carlo
method by a semi-analytical
calculation of the resolution using the standard approximation based on the
Hessian matrix of the Likelihood function. This makes it easier to plot the
dependence on various parameters. We explicitly checked that this method
gives the same results as the Toy Monte Carlo.

\subsubsection*{Collection of Data Sample}

\begin{table}[tbp]
\begin{center}
\begin{tabular}{c}
\hline 
2 SVT tracks with $p_T > 2$ \gevc \\
100 $\mu$m $< d <$ 1 mm for the two tracks \\
$\vec p_T \cdot \vec X_V > $ 0.2 \gevc\ $\cdot$ cm \\
\hline 
\end{tabular}
\vspace*{0.3cm}
\caption
[L2 trigger cuts proposed for multibody $B$ decay selection.]
{L2 trigger cuts proposed for multibody $B$ decay selection.}
\label{tab:newtrigger}
\end{center}
\end{table}

\begin{table}[tbp]
\begin{center}
\begin{tabular}{c|ccc} 
\hline 
Int. Luminosity & Scenario A & \hspace*{0.5cm} Scenario B  \hspace*{0.5cm} 
& Scenario C \\
\hline
2 fb$^-1$ & 135 & 120 & 90 \\
 & (1:14) & (1:12) & (1:9) \\
10 fb$^-1$ & 675 & 585 & 450 \\
 & (1:70) & (1:60) & (1:45) \\
30 fb$^-1$ & 2025 & 1755 & 1350 \\
 & (1:200) & (1:175) & (1:135) \\
\hline
\end{tabular}
\vspace*{0.3cm}
\caption
[Expected event yields for 
$B^- \to D^0 K^-$ for different Tevatron operation scenarios. ]
{Expected event yields for 
$B^- \to [K \pi] K^-$ for different Tevatron operation scenarios. 
The worst
 S\,:\,B ratio that can be tolerated when requiring a
  resolution on $\gamma$ better than $\approx 30^\circ$ is given in
parenthesis.} 
\label{tab:cdf_bdk_yields}
\end{center}
\end{table}

The data sample considered here, $B^- \ra D^0 K^- \ra [K\pi] K^-$, will be
accumulated with the two-track hadronic trigger used for the collection of 
$B\ra\pi\pi/KK$ events (see Sec.~\ref{sec:cdf_pipi_trigger}). 
To optimize the event selection efficiency, we performed a study of varying 
Level\,2 trigger
requirements and ended up with a slightly 
modified version of the hadronic two-track trigger.
In Table~\ref{tab:newtrigger} we show the L2 selection requirements 
as proposed for the multi-body $B$ decay selection.
For the determination of the corresponding number of expected signal events,
we use a $B^+$ production cross
section of $(3.35\pm 0.68)~\mu$b and integrated luminosities of 2, 10 and
30~fb$^{-1}$ (see Table~\ref{tab:cdf_bdk_yields}). 
The Level\,2 trigger efficiencies for the $[K\pi] K^-$ final state
are 0.59\%, 0.52\% and 0.40\% for
the three different Tevatron operating scenarios 
A, B and C, respectively.

\subsubsection*{Background}

The reduction of backgrounds is the most important issue to address at CDF.
Note, the signal we are considering here
 is two orders of magnitude smaller than the number of $B^0 \to \pi^+ \pi^-$
events. 
A detailed study of the contribution of the combinatoric background 
has not been performed. 
To obtain a reliable background estimate,
we will need real Run\,II data.
We therefore concentrate on the  ``physics background'' consisting of 
$B$~decay channels which are difficult to separate from the signal. 
Most of them differ from the signal only in the identity of the final sate
particles. Some of them are given in 
Table~\ref{tab:fondofisico}.
The channel $B^\pm \to D^0 \pi^\pm$ is kinematically almost identical
to the signal $B^\pm \to D^0 K^\pm$ and its branching ratio is an order of
magnitude larger.
The decay $B^0 \to D^{*-}\pi^+ \to \bar{D}^0(\pi^-)\pi^+$ is similar to the
previous one, with the difference that the reconstructed fake $B^+$ meson
has a reduced mass.  
$\bar D^0 \to \pi^+ \pi^- $ decay modes are potential backgrounds.
The decay $B^+ \to [K^+ \pi^-] K^+$ results
from combining the two Cabibbo-allowed decays, 
and is potentially the most dangerous channel, being two orders of
magnitude larger than our signal. 

\begin{table}[tbp]
\begin{center}
\begin{tabular}{lcr}
\begin{tabular}{lc}
\hline
Channel & ${\cal B}$\\
\hline
$B^+ \to \bar D^0 K^+ $ & $2.6 \times 10^{-4}$\\
$B^+ \to D^0 K^+ $ & $2 \times 10^{-6}$ \\
$B^+ \to \bar D^0 \pi^+ $ & $5 \times 10^{-3}$\\
$B^0 \to \bar D^0 \pi^+ (\pi^-) $ & $2.1 \times 10^{-3}$\\
\hline
$\bar D^0 \to K^- \pi^+ $ & $1.3 \times 10^{-4}$ \\
$\bar D^0 \to \pi^- K^+ $ & $3.8 \times 10^{-2}$ \\
$\bar D^0 \to \pi^- \pi^+ $ & $1.5 \times 10^{-3}$\\
\hline
\end{tabular}
& \hspace{1.5cm} &
\begin{tabular}{lc}
\hline
Channel & yield/S \\
\hline
$B^+ \to [K^- \pi^+]  K^+ $ & $ 1$ \\
\hline
$B^+ \to [\pi^- K^+]  K^+ $ & $ 47$ \\
$B^+ \to [\pi^- \pi^+]  K^+ $ & $ 2$ \\
$B^+ \to [K^- \pi^+]  \pi^+ $ & $ 3$ \\
$B^+ \to [\pi^- K^+]  \pi^+ $ & $ 910$ \\
$B^+ \to [\pi^- \pi^+]  \pi^+ $ & $ 36$ \\
\hline
\end{tabular} 
\end{tabular}
\vspace*{0.3cm}
\caption
[Branching ratios of 
potential physics backgrounds at CDF.]
{Branching ratios of 
potential physics backgrounds.
The right table lists the relative abundance of each final state configuration 
with respect to the signal.
Note, the channel $B^+ \to  [\pi^- K^+] \pi^+ $ is about 1000
times larger than the signal.
}
\label{tab:fondofisico}
\end{center}
\end{table}

A detailed description of CDF's capability to separate signal from 
background is beyond the scope of this report, but we want to give the reader
an idea of possible methods for signal to background reduction. 
Figure~\ref{fig:cdf_bdk_md} shows the invariant mass distribution of pairs
of $D$~daughter particles, obtained by assigning the pion mass 
to the particle with the same charge as the $B^-$~meson 
and the kaon mass to the other particles. 
In Figure~\ref{fig:cdf_bdk_md}(a) the scale is arbitrary, while in (b)
the correct  normalization 
between physics backgrounds and signal is used.

\begin{figure}[tbp]
\centerline{
\put(45,180){\large\bf (a)}
\put(255,180){\large\bf (b)}
\epsfxsize=3.0in
\epsffile{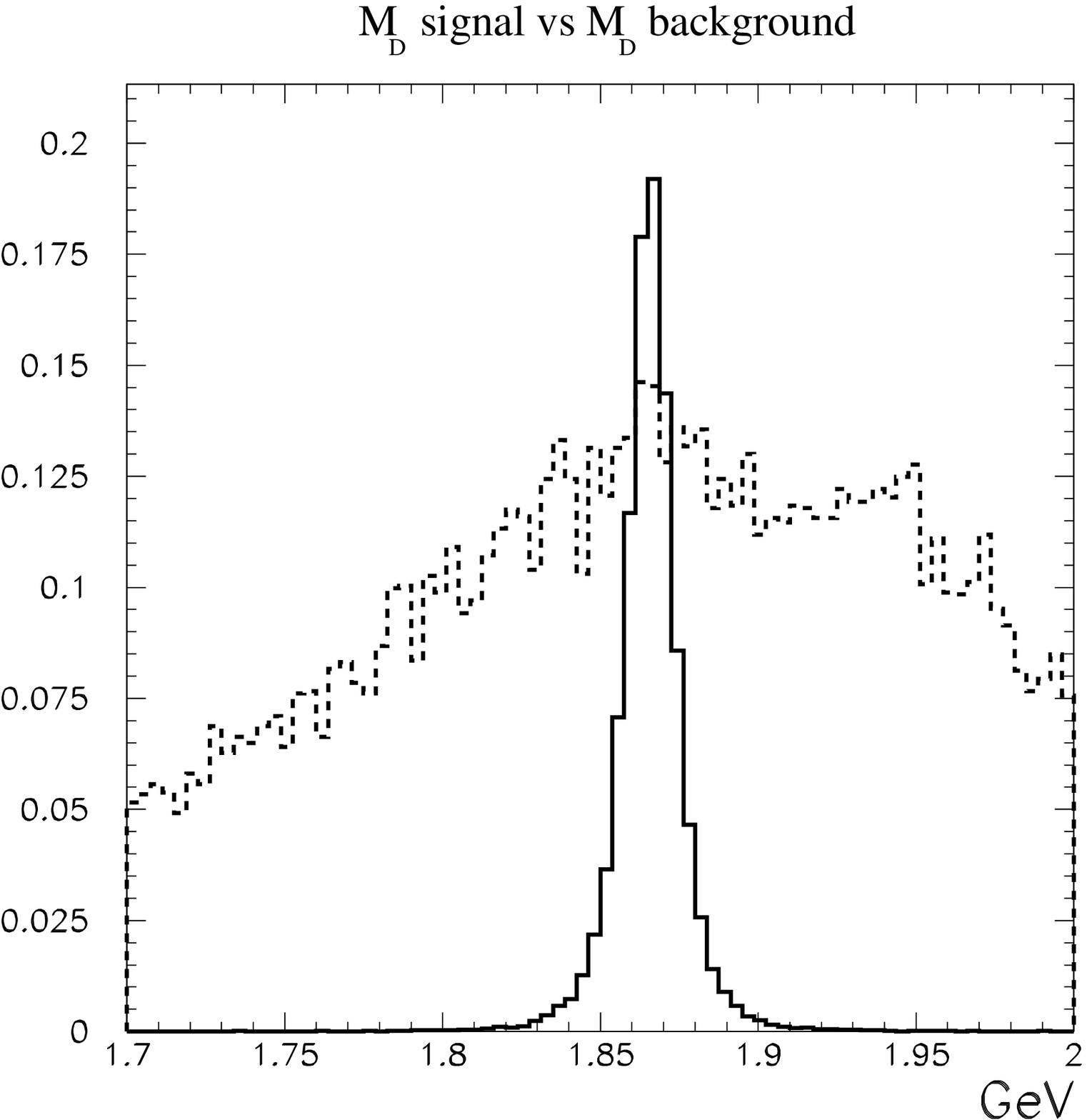}
\epsfxsize=2.85in
\epsffile{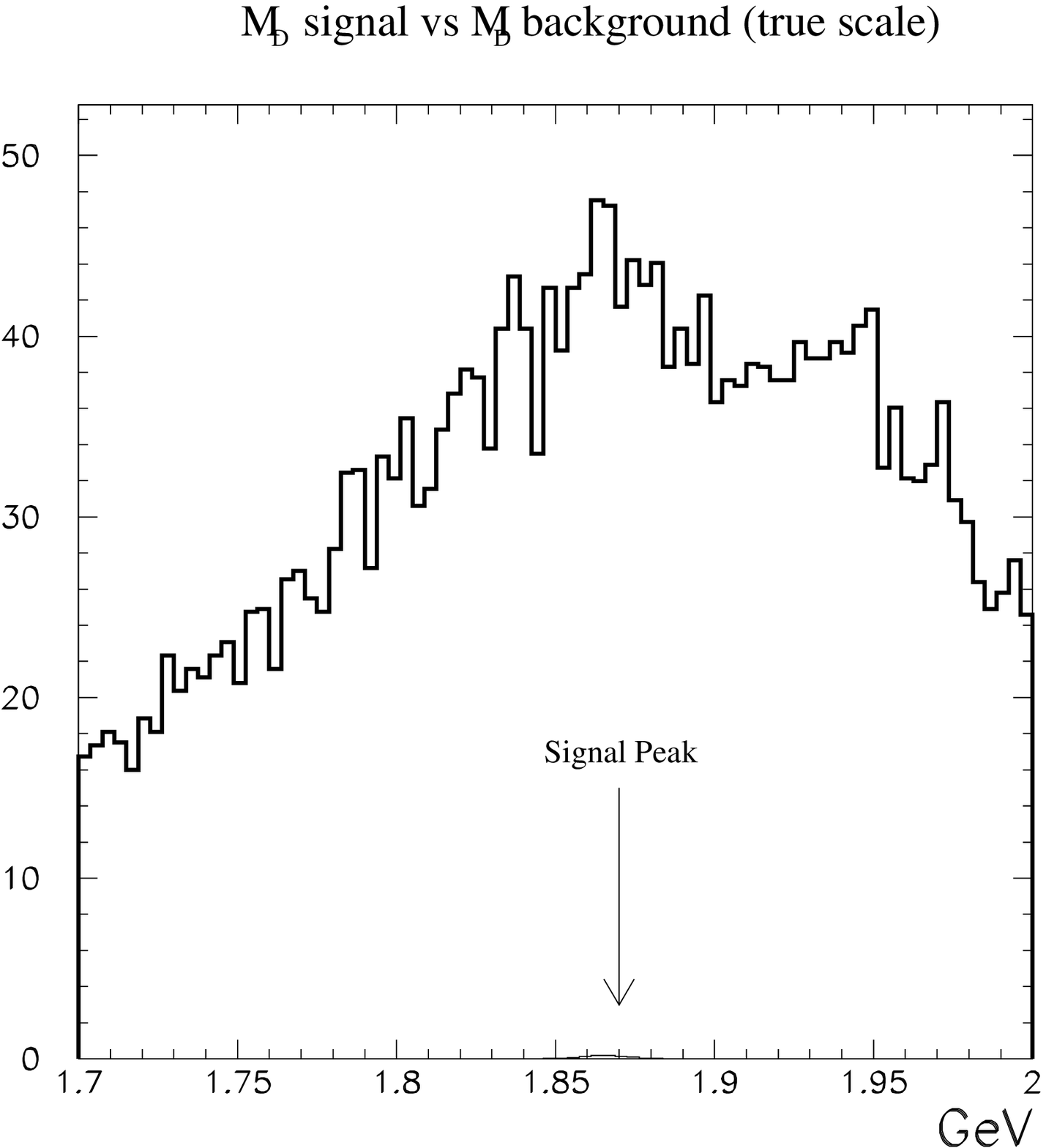}
}
\vspace*{0.3cm}
\caption
[Invariant mass distribution of pairs
of $D$ daughter particles at CDF.]
{Invariant mass distribution of pairs
of $D$ daughter particles, obtained by assigning the pion mass 
to the particle with the same charge as the $B^-$ meson 
and the kaon mass to the other particles. 
In (a) the scale is arbitrary, while in (b)
the correct  normalization 
between physics backgrounds and signal is used.}
\label{fig:cdf_bdk_md}
\end{figure}

We plan to perform the signal to physics background separation both with
particle identification  and  kinematics.
If we assign incorrect rest masses to the final state particles, both
the $D$ and $B$ mass distributions will appear wider and/or shifted.
A special case is the contribution of $B^0$ mesons, where a charged pion is
lost and 
the reconstructed ``$B^-$" has a significantly lower mass. We see from 
Table~\ref{tab:fondofisico} that the size of this background is 40\% of the
corresponding contribution from real $B^-$.
However, in a window of $\pm 50~\mevcc$ around the nominal $B^-$~mass, only
a fraction of 3.9\% of $B^0$~decays remain. 
We therefore neglect the contribution of $B^0$ with respect 
to real $B^-$.

A more refined analysis is needed to reject real $B^-$ background by exploiting
 the mass differences due to
missassigned particle identities. Many different methods of
various degree of refinement
can be used. Here we only want to give an example illustrating that 
a powerful background rejection is achievable.
Let's assume we consider final states with three particles,
$[a^+ b^-] c^+$, and want to identify $a$, $b$ and $c$.
We can formulate several hypotheses, e.g.~$I = \{ a=K; \:\: b=\pi 
;\:\: c=K \}$. Given a set of hypotheses  ${\cal I} = \{ I_1, \ldots, 
I_N \}$, we can compute the distances $d$ from the true PDG
masses~\cite{pdg98}  
\begin{equation}
d_D(I_k) = \Big| M(D | I_k ) - M(D)_{\rm true} \Big| 
~~~{\rm and}~~~
d_B(I_k) = \Big| M(B | I_k ) - M(B)_{\rm true} \Big| \,. 
\end{equation}
We call
\begin{equation}
d_T(I_k) = \sqrt{d_D(I_k)^2 + d_B(I_k)^2}
\end{equation}
and consider the right hypothesis $I_k$ for which $d_T(I_k)$ is 
smallest.

With this algorithm we obtain signal efficiencies of $(90\pm1)$\% and
$(0.8\pm0.2)$\% for background events. This method provides more than a
factor of 100 in background rejection, reducing
the physics background to a level of $B/S=9:1$.

The $B/S$ ratio can be further improved by using CDF's particle
identification capabilities from the energy loss measurement d$E$/d$x$ in
the COT.
From this study we expect the combined application of kinematical selections
and particle identification to have a sufficient rejection power against
physics backgrounds. However, we 
expect the combinatoric background to be an important issue.
From the numbers in Tab.~\ref{tab:cdf_bdk_yields} we see that 
if the combinatoric background were negligible, 
a resolution of $15^\circ$ on $\gamma$ can be achieved assuming
${\cal B}(B^+ \to K^+ D^0)$ is determined with sufficient precision ($\approx 20\%$).

In conclusion,
we discussed a method for measuring $\gamma$ in Run\,II using charged 
$B$~decays. We expect to collect a small but significant sample of both
 candidate channels for this analysis by using the two-track hadronic
trigger at CDF. The physics
background can be brought down to the same level as the signal, but there
could be considerable combinatoric background.
If we are able to reduce the combinatoric background to a level
comparable to the signal, we expect a significant measurement of $\gamma$
with this method in Run\,II. 

\boldmath
\subsubsection[Fully Hadronic $B$~Decays Accessible at CDF in Run\,II]
{Fully Hadronic $B$~Decays Accessible at CDF in Run\,II
$\!$\authorfootnote{Author: A.~Cerri.}
}
\unboldmath
\index{CDF!hadronic $B$ decays}%

The selection of the decay modes $\Bs\ra D_s^- K^+$ and $B^-\to D^0 K^-$ is
based on 
collecting these events with the 
\index{CDF!hadronic trigger}%
two-track hadronic trigger which was
originally designed to select a large sample of $B^0 \ra \pi^+\pi^-$ decays
but is
also used to obtain $B_s^0 \ra D^-_s \pi^+$ events for the measurement of
\Bs~flavour 
oscillations (see Sec.~\ref{sec:cdf_pipi_trigger}). In the context of
evaluating the yield of fully reconstructed $B^-\to D^0 K^-$ events, a more
systematic study has been performed to explore the event yields of other
potential $CP$~modes that could be collected with the two-track hadronic
trigger at CDF. The list of decay modes compiled was assembled
under the aspect of some interest being expressed in the literature for a
particular decay mode. Because of CDF's poor efficiency to reconstruct
decays involving photons, decay modes with neutral particles in the final
state were not considered in this study. 
The list of decays has been completely specified up to 
the final state daughters and a rough estimate of the involved branching
fractions was determined.
We briefly want to summarize the results
of this study to give the reader an idea about event yields for potential
$CP$~modes that could be collected at CDF with the two-track hadronic
trigger. 

The study of the different decay modes used  
a Monte Carlo generator that simulates only a single 
$B$~hadron and its decay products which was completely appropriate for this
study. The final event yield is the
result of an event selection based on a parametric 
simulation of the two-track trigger path and a rough geometric acceptance
calculation for the whole event, including parametrized detector and
trigger efficiencies. 
The estimate of 
the total number of expected events assumes a $B^+$ production cross 
section of $(3.35\pm0.68)~\mu b$ for $\left|y\right|\le 1$ and
$p_T(B)\ge6~\gevc$.

In Table~\ref{tab:cdf_survey_bn} we list the estimated total branching
ratio and the expected 
number of events per 1~fb$^{-1}$ for several neutral 
$B$~decay modes. The corresponding numbers of events for
$B^+$ and $B_s^0$ decay modes are listed in 
Table~\ref{tab:cdf_survey_bc} and
Table~\ref{tab:cdf_survey_bs}, respectively. It is clear from that study
that the two-track hadronic 
trigger will allow CDF to collect significant datasets of fully hadronic
$B$~decays. This will be the source of a rich $B$~physics program at CDF
involving many different $B$~decay modes.

\begin{table}
\begin{center}
\begin{tabular}[t]{|ll||c|c|}
\hline
Decay & Subsequent Decay & Total ${\cal B}$ & N per 1~fb$^{-1}$ \\
\hline
$B^0 \rightarrow \pi^+\pi^- $ & 
 &
$ 4.3 \cdot 10^{-6} $ & 
$4900\pm2100$ 
\vspace*{-3pt} 
\\ 
\hline 
$B^0 \rightarrow D^{\pm} \pi^\mp $ & 
$D^\pm \rightarrow K^\mp \pi^\pm \pi^\pm$ &
 $ 2.7 \cdot 10^{-4} $ & 
$81000\pm18000$ 
\vspace*{-2pt} 
\\ 
\hline 
$B^0 \rightarrow D^{\ast \pm} \pi^\mp $ & 
$D^{\ast-}\rightarrow \bar{D}^0 \pi^-,D^0 \rightarrow K^-\pi^+$ &
$ 7.9 \cdot 10^{-5} $ & 
$20000\pm4600$ 
\vspace*{-2pt} 
\\ 
$B^0 \rightarrow D^{\ast \pm} \pi^\mp $ & 
$D^{\ast-}\rightarrow \bar{D}^0 \pi^-,D^0 \rightarrow K_S^0\pi^+\pi^-$ &
$ 4 \cdot 10^{-5} $ & 
$7100\pm1600$ 
\vspace*{-2pt} 
\\ 
$B^0 \rightarrow D^{\ast \pm} \pi^\mp $ & 
$D^{\ast-}\rightarrow \bar{D}^0 \pi^-,D^0 \rightarrow K^-\pi^+\pi^+\pi^-$ &
$ 1.5 \cdot 10^{-4} $ & 
$17000\pm4200$ 
\vspace*{-2pt} 
\\ 
\hline 
$B^0 \rightarrow D^0 K_S^0 $ & 
$D^0 \rightarrow K^-\pi^+,K_S^0 \rightarrow \pi^+\pi^-$ &
$ 5 \cdot 10^{-7} $ & 
$92\pm21$ 
\vspace*{-2pt} 
\\ 
$B^0 \rightarrow D^0 K_S^0 $ & 
$D^0 \rightarrow K_S^0\pi^+\pi^-,K_S^0 \rightarrow \pi^+\pi^-$ &
$ 2.5 \cdot 10^{-7} $ & 
$21\pm5.3$
\vspace*{-2pt} 
\\ 
$B^0 \rightarrow D^0 K_S^0 $ & 
$D^0 \rightarrow K^-\pi^+\pi^+\pi^-,K_S^0 \rightarrow \pi^+\pi^-$ &
$ 9.7 \cdot 10^{-7} $ & 
$74\pm19$ 
\\ 
\hline 
$B^0 \rightarrow D^0 K^{\ast 0} $ & 
$D^0 \rightarrow K^-\pi^+,K^{\ast 0} \rightarrow K^+\pi^-$ &
$ 2.5 \cdot 10^{-7} $ & 
$71\pm16$ 
\vspace*{-2pt} 
\\ 
$B^0 \rightarrow D^0 K^{\ast 0} $ & 
$D^0 \rightarrow K_S^0\pi^+\pi^-,K^{\ast 0} \rightarrow K^+\pi^-$ &
$ 1.3 \cdot 10^{-7} $ & 
$17\pm4.1$ 
\vspace*{-2pt} 
\\ 
$B^0 \rightarrow D^0 K^{\ast 0} $ & 
$D^0 \rightarrow K^-\pi^+\pi^+\pi^-,K^{\ast 0} \rightarrow K^+\pi^-$ &
$ 4.9 \cdot 10^{-7} $ & 
$60\pm14$ 
\vspace*{-2pt} 
\\ 
\hline 
$B^0 \rightarrow D_{1,2} K^{\ast 0} $ & 
$D_{1,2} \rightarrow \left(\pi^+\pi^-,K^+K^-\right)
K^{\ast 0} \rightarrow K^+\pi^-$ &
$ 1 \cdot 10^{-8} $ & 
$2.3\pm.5$ 
\vspace*{-2pt} 
\\ 
\hline 
$B^0 \rightarrow D^{\ast 0}K^{\ast 0} $ & 
$D^{0\ast}\rightarrow D^0 \pi^0,
D^0 \rightarrow K^-\pi^+$ &
$ 1 \cdot 10^{-7} $ & 
$22\pm5$ 
\vspace*{-2pt} 
\\ 
$B^0 \rightarrow D^{\ast 0}K^{\ast 0} $ & 
$D^{0\ast}\rightarrow D^0 \pi^0,
D^0 \rightarrow K_S^0\pi^+\pi^-$ &
$ 5 \cdot 10^{-8} $ & 
$7.4\pm1.8$ 
\vspace*{-2pt} 
\\ 
$B^0 \rightarrow D^{\ast 0}K^{\ast 0} $ & 
$D^{0\ast}\rightarrow D^0 \pi^0,
D^0 \rightarrow K^-\pi^+\pi^+\pi^-$ &
$ 2 \cdot 10^{-7} $ & 
$21\pm5.2$ 
\vspace*{-2pt} 
\\ 
\hline 
$B^0 \rightarrow D_1 K_S^0 $ & 
$D_1 \rightarrow \left(\pi^+\pi^-,K^+K^-\right),
K_S^0 \rightarrow \pi^+\pi^-$ &
$ 4 \cdot 10^{-8} $ & 
$6\pm1.4$ 
\\ 
\hline 
$B^0 \rightarrow \phi K_S^0 $ & 
$\phi \rightarrow K^+K^-,
K_S^0 \rightarrow \pi^+\pi^-$ &
$ 3 \cdot 10^{-6} $ & 
$350\pm85$ 
\\ 
\hline 
$B^0 \rightarrow D^+ D^-  $ & 
$D^\pm \rightarrow K^\mp \pi^\pm \pi^\pm$ &
$ 3 \cdot 10^{-6} $ & 
$560\pm130$ 
\vspace*{-2pt} 
\\ 
\hline 
$B^0 \rightarrow D^{\ast+}D^{\ast-} $ & 
$D^{\ast-}\rightarrow \bar{D}^0 \pi^-,D^0 \rightarrow K^-\pi^+$ &
$ 4 \cdot 10^{-7} $ & 
$69\pm16$ 
\vspace*{-2pt} 
\\ 
$B^0 \rightarrow D^{\ast+}D^{\ast-} $ & 
$D^{\ast-}\rightarrow \bar{D}^0 \pi^-,D^0 \rightarrow K_S^0\pi^+\pi^-$ &
$ 2 \cdot 10^{-7} $ & 
$13\pm3.4$ 
\vspace*{-2pt} 
\\ 
$B^0 \rightarrow D^{\ast+}D^{\ast-} $ & 
$D^{\ast-}\rightarrow \bar{D}^0 \pi^-,D^0 \rightarrow K^-\pi^+\pi^+\pi^-$ &
$ 7.8 \cdot 10^{-7} $ & 
$49\pm13$ 
\vspace*{-2pt} 
\\ 
\hline 
$B^0 \rightarrow D^{\ast+}D^{\ast-}K_S^0 $ & 
$D^{\ast-}\rightarrow \bar{D}^0 \pi^-,D^0 \rightarrow K^-\pi^+$ &
$ 4.5 \cdot 10^{-6} $ & 
$450\pm110$ 
\vspace*{-2pt} 
\\ 
$B^0 \rightarrow D^{\ast+}D^{\ast-}K_S^0 $ & 
$D^{\ast-}\rightarrow \bar{D}^0 \pi^-,D^0 \rightarrow K_S^0\pi^+\pi^-$ &
$ 2.3 \cdot 10^{-6} $ & 
$86\pm27$ 
\vspace*{-2pt} 
\\ 
$B^0 \rightarrow D^{\ast+}D^{\ast-}K_S^0 $ & 
$D^{\ast-}\rightarrow \bar{D}^0 \pi^-,D^0 \rightarrow
K^-\pi^+\pi^+\pi^-$ &
$ 8.8 \cdot 10^{-6} $ & 
$260\pm86$ 
\\ 
\hline 
$B^0 \rightarrow \rho^0 \rho^0 $ & 
$\rho^0 \rightarrow \pi^+\pi^-$ & 
$ 1 \cdot 10^{-6} $ & 
$330\pm72$ 
\vspace*{-2pt} 
\\ 
\hline 
$B^0 \rightarrow D^+ D^- K_S^0  $ & 
$D^\pm \rightarrow K^\mp \pi^\pm \pi^\pm$, $K_S^0 \rightarrow \pi^+\pi^-$ &
$ 7 \cdot 10^{-6} $ & 
$630\pm160$ 
\vspace*{-2pt} 
\\ 
\hline 
$B^0 \rightarrow D^\pm \pi^\mp K_S^0  $ & 
$D^\pm \rightarrow K^\mp \pi^\pm \pi^\pm$, $K_S^0 \rightarrow \pi^+\pi^-$ &
$ 1 \cdot 10^{-5} $ & 
$1000\pm260$ 
\vspace*{-2pt} 
\\ 
\hline 
$B^0 \rightarrow D_{CP}^0 \pi^+\pi^- $ &
$D^0_{CP} \rightarrow \pi^+\pi^-,K^+K^-$ & 
$ 1 \cdot 10^{-5} $ & 
$2900\pm640$ 
\vspace*{-1pt} 
\\ 
\hline 
$B^0 \rightarrow K^{\ast +} \pi^- $ & 
$K^{\ast+} \rightarrow K_S^0 \pi^+ \rightarrow \pi^+\pi^-\pi^+$ & 
$ 2 \cdot 10^{-6} $ & 
$400\pm91$ 
\\ 
\hline 
$B^0 \rightarrow D_s^\pm K^\mp $ & 
$D_s^\pm \rightarrow \phi \pi^\pm$, $\phi \rightarrow K^+K^-$ &
$ 4.1 \cdot 10^{-6} $ & 
$1000\pm220$ 
\vspace*{-1pt} 
\\ 
\hline 
$B^0 \rightarrow D^0 \rho^0 $ & 
$D^0 \rightarrow K^-\pi^+$ &
$ 1.5 \cdot 10^{-5} $ & 
$3900\pm870$ 
\vspace*{-2pt} 
\\ 
$B^0 \rightarrow D^0 \rho^0 $ & 
$D^0 \rightarrow K^0_S\pi^+\pi^-$ &
$ 7 \cdot 10^{-6} $ & 
$1100\pm250$ 
\vspace*{-2pt} 
\\ 
$B^0 \rightarrow D^0 \rho^0 $ & 
$D^0 \rightarrow K^-\pi^+\pi^+\pi^-$ &
$ 3 \cdot 10^{-5} $ & 
$4300\pm1000$ 
\vspace*{-2pt} 
\\ 
\hline 
$B^0 \rightarrow \rho^0 K_S^0 $ & 
$\rho^0 \ra \pi^+\pi^-$, $K_S^0 \ra \pi^+\pi^-$ &
$ 2.6 \cdot 10^{-5} $ & 
$2400\pm620$ 
\vspace*{-2pt} 
\\ 
\hline 
$B^0 \rightarrow D_s^- K^+ $ & 
$D_s^\pm \rightarrow \phi \pi^\pm$, $\phi \rightarrow K^+K^-$ &
$ 7 \cdot 10^{-6} $ & 
$1700\pm380$ 
\\ 
\hline 
 \end{tabular} 
\vspace*{0.3cm}
\caption[Estimated total branching ratio and expected
number of events per 1~fb$^{-1}$ for several hadronic $B^0$ decay modes.]
{Estimated total branching ratio and expected
number of events per 1~fb$^{-1}$ for several hadronic $B^0$ decay modes.} 
\label{tab:cdf_survey_bn}
\end{center}
\end{table}

\begin{table}
\begin{center}
\begin{tabular}[t]{|ll||c|c|}
\hline
Decay & Subsequent Decay & Total ${\cal B}$ & N per 1~fb$^{-1}$ \\
\hline
$B^\pm \rightarrow D^0 K^\pm $ & 
$D^0 \rightarrow K^-\pi^+$ &
$ 7.5 \cdot 10^{-8} $ & 
$28\pm6.1$ \\ 
$B^\pm \rightarrow D^0 K^\pm $ & 
$D^0 \rightarrow K_S^0\pi^+\pi^-$ &
$ 3.8 \cdot 10^{-8} $ & 
$5.4\pm1.3$ \\ 
$B^\pm \rightarrow D^0 K^\pm $ & 
$D^0 \rightarrow K^-\pi^+\pi^+\pi^-$ &
$ 5.2 \cdot 10^{-8} $ & 
$8.2\pm1.9$ \\ \hline 
$B^\pm \rightarrow K^{\ast\pm} \rho^0 $ & 
$\rho^0 \rightarrow \pi^+\pi^-, K^{\ast+}\rightarrow K_S^0\pi^+,K_S^0\rightarrow\pi^+\pi^-$ &
$ 1.7 \cdot 10^{-6} $ & 
$180\pm45$ \\ \hline 
$B^\pm \rightarrow \pi^{\pm} \rho^0 $ & 
$\rho^0 \rightarrow \pi^+\pi^-$ &
$ 9 \cdot 10^{-6} $ & 
$3400\pm730$ \\ \hline 
$B^\pm \rightarrow \pi^+\pi^-\pi^\pm $ & 
 &
$ 3.5 \cdot 10^{-5} $ & 
$16000\pm3500$ \\ \hline 
$B^\pm \rightarrow K^+K^-\pi^\pm $ & 
 &
$ 1.5 \cdot 10^{-5} $ & 
$5300\pm1200$ \\ \hline 
$B^\pm \rightarrow K^\pm K_S^0 $ & 
$K_S^0 \rightarrow \pi^+\pi^-$ &
$ 6.8 \cdot 10^{-6} $ & 
$200\pm67$ \\ \hline 
$B^\pm \rightarrow \pi^\pm K_S^0 $ & 
$K_S^0 \rightarrow \pi^+\pi^-$ &
$ 1.6 \cdot 10^{-5} $ & 
$340\pm130$ \\ \hline 
$B^\pm \rightarrow \phi K^\pm $ & 
$\phi \rightarrow K^+K^-$ &
$ 1.2 \cdot 10^{-5} $ & 
$3800\pm830$ \\ \hline 
 \end{tabular} 
\vspace*{0.3cm}
\caption[Estimated total branching ratio and expected
number of events per 1~fb$^{-1}$ for several hadronic $B^+$ decay modes.]
{Estimated total branching ratio and expected
number of events per 1~fb$^{-1}$ for several hadronic $B^+$ decay modes.} 
\label{tab:cdf_survey_bc}
\end{center}
\end{table}

\begin{table}
\begin{center}
\begin{tabular}[t]{|ll||c|c|}
\hline
Decay & Subsequent Decay & Total ${\cal B}$ & N per 1~fb$^{-1}$ \\
\hline
$B_s^0 \rightarrow K^* \bar{K}^* $ & 
$K^{\ast 0} \rightarrow K^{\pm}\pi^\mp$ &
$ 1 \cdot 10^{-6} $ & 
$110\pm24$ \\ \hline 
$B_s^0 \rightarrow K^{*+} K^{*-} $ & 
$K^{\ast \mp} \rightarrow K^0 \pi$, $K^0 \rightarrow \pi\pi$ &
$ 1 \cdot 10^{-6} $ & 
$78\pm18$ \\ \hline 
$B_s^0 \rightarrow \bar{D}^0 \phi $ & 
$D^0 \rightarrow K^-\pi^+$, $\phi \rightarrow K^+K^-$ &
$ 1.1 \cdot 10^{-7} $ & 
$12\pm2.7$ \\
$B_s^0 \rightarrow \bar{D}^0 \phi $ & 
$D^0 \rightarrow K_S^0\pi^+\pi^-$, $K_S \rightarrow \pi^+ \pi^-$ &
$ 5.3 \cdot 10^{-8} $ & 
$2.2\pm.5$ \\
$B_s^0 \rightarrow \bar{D}^0 \phi $ & 
$D^0 \rightarrow K^-\pi^+\pi^-\pi^+$, $\phi \rightarrow K^+K^-$ &
$ 2.2 \cdot 10^{-7} $ & 
$14\pm3.3$ \\ \hline 
$B_s^0 \rightarrow \bar{D}^0 \bar{K}^{*0} $ & 
$D^0 \rightarrow K^-\pi^+$, $K^{\ast 0} \rightarrow K^{\pm}\pi^\mp$ &
$ 4.6 \cdot 10^{-6} $ & 
$430\pm96$ \\
$B_s^0 \rightarrow \bar{D}^0 \bar{K}^{*0} $ & 
$K^{\ast 0} \rightarrow K^{\pm}\pi^\mp$ &
$ 2.2 \cdot 10^{-6} $ & 
$140\pm33$ \\
$B_s^0 \rightarrow \bar{D}^0 \bar{K}^{*0} $ & 
$K^{\ast 0} \rightarrow K^{\pm}\pi^\mp$ &
$ 9 \cdot 10^{-6} $ & 
$600\pm140$ \\ \hline 
$B_s^0 \rightarrow D_s^\pm \pi^\mp $ & 
$D_s^\pm \rightarrow \phi \pi^\pm$, $\phi \rightarrow K^+K^-$ &
$ 5.3 \cdot 10^{-5} $ & 
$6200\pm1400$ \\ \hline 
$B_s^0 \rightarrow D_s^\pm \pi^\mp \pi^+\pi^- $ & 
$D_s^\pm \rightarrow \phi \pi^\pm$, $\phi \rightarrow K^+K^-$ &
$ 1.4 \cdot 10^{-4} $ & 
$7700\pm1800$ \\ \hline 
 \end{tabular} 
\vspace*{0.3cm}
\caption[Estimated total branching ratio and expected
number of events per 1~fb$^{-1}$ for several hadronic $B^0_s$ decay modes.]
{Estimated total branching ratio and expected
number of events per 1~fb$^{-1}$ for several hadronic $B^0_s$ decay modes.} 
\label{tab:cdf_survey_bs}
\end{center}
\end{table}

\boldmath
\subsection[$B \ra D K$: BTeV Report]
{$B \ra D K$: BTeV Report
$\!$\authorfootnote{Author: P.A.~Kasper.}
}
\unboldmath
\index{BTeV!$\sin\gamma$ prospects}%

  Several suggestions on how to measure the CKM angle $\gamma$
  have been discussed in Section~\ref{ch6:intro}.
  While discrete ambiguities are inherent in each of these methods, using
  several methods will help remove some of these ambiguities as well as help
  control systematic errors.
  We report first the BTeV studies for $CP$~Violation in 
  $B_s^0 \ra D_{s}^- K^+$ followed by $B^- \ra D^0 K^-$ in
  Section~\ref{sec:cp_bdk_btev}.

\boldmath
\subsubsection{$B_s^0 \ra D_s^- K^+$: BTeV Report}
\unboldmath
\label{sec:btev_bdsk}
\index{decay!$B_s^0 \to D_s^- K^+$}%
\index{BTeV!$B_s^0 \to D_s^- K^+$ prospects}%


 
 A study of the reconstruction efficiency has been performed for the decay
 modes  
\beq
\begin{array}{c}
B_s^0\ra D_s^- K^+, \; D_s^- \ra \phi \pi^-, \; \phi \ra K^+ K^-\ {\rm and} \\
B_s^0\ra D_s^- K^+, \; D_s^- \ra K^{*0} K^- , \; K^{*0} \ra K^+ \pi^-. 
\end{array}
\eeq
  The events were generated with Pythia and the detector modeled using
  BTeVGeant. 
  Each event consists of a $b\bar{b}$ interaction and a mean of two minimum bias
  interactions, to simulate 
  a luminosity of $2\times10^{32}$~cm$^{-2}$s$^{-1}$.
  Loose cuts were applied initially and the tighter cuts were chosen after
  the background was studied.

  For the $D_s^- \ra \phi \pi^-$ decay mode the following requirements were used.
At least one of the kaons from the $\phi$ decay and also the $K^+$
  from the $B_s^0$ decay were required to be identified in the RICH.
The impact parameter with respect to the primary vertex had to be
  $> 3\,\sigma$
  for all four charged tracks.
To reduce the background due to ``detached'' tracks that come from other
  interactions, we require that the impact parameter with respect to the
  primary  
  vertex be less than 0.2 cm for all tracks.
The $\phi$ and $D_s^-$ were required to be within $\pm{2.5\,\sigma}$ of their
  nominal mass.  
The distance between the primary vertex and $D_s^-$ decay vertex has to be
  $L<8.0$ cm and  
  $L/\sigma_{L}(D_s^-)>10.0$.
We also require $L/\sigma_{L}(B_s^0)>4.0$.
The transverse momentum of the $B_s^0$ with respect to its line of
  flight from the primary vertex  was required to be less than 1.0~\gevc. 
The impact parameter with respect to the primary vertex was required
   to be less than $3\,\sigma$ for the reconstructed $B$.
 
\begin{figure}[tb]
\centerline{\epsfysize=4.5in \epsffile{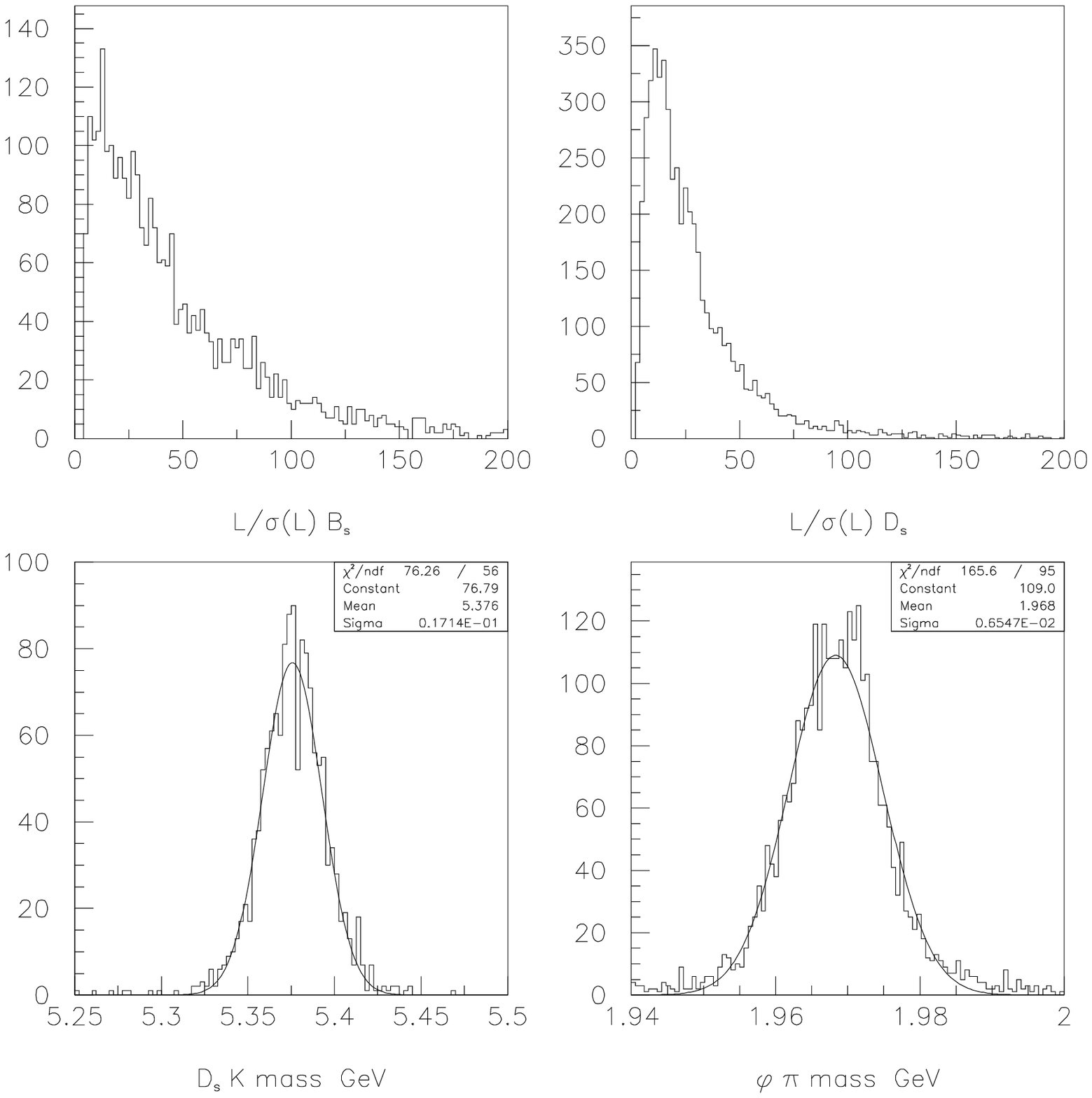}}
\caption
[$L/\sigma_L$ and mass peaks for $B_s^0$ and $D_s^-$ at BTeV.]
{$L/\sigma_L$ and mass peaks for $B_s^0$ and $D_s^-$ at BTeV.}
\label{fig:bsmass}
\end{figure}
        
  The distributions of $L/\sigma_L$  and
  the mass peaks for the  $D_s^-$ and $B_s^0$ are shown in
  Fig~\ref{fig:bsmass}.
  The combined geometric acceptance and reconstruction efficiency was found to be
   4.5\%. If we require both kaons from the $\phi$ decay to be identified
   in the RICH, the efficiency drops to  2.5\%.
    Of the events that passed these analysis cuts, 74\% passed the
  secondary vertex trigger. 
  For the $D_s^- \ra K^{*0} K^-$ mode, we used the same cuts except that both
  kaons from the $D_s^-$ decay were required to be identified in the RICH.
     The combined reconstruction
  efficiency and geometric acceptance for the $D_s^- \ra K^{*0} K^-$ mode
  was found to be 2.3\%, and the trigger efficiency for
  the events passing the analysis cuts was 74\%.

    The results of the tagging study described in Sec.~5.5
    indicate that we can expect a  tagging efficiency  $\varepsilon=0.70$
     and a  dilution ${\cal D}=0.37$
    giving an effective tagging efficiency $\eD = 0.10$.
   The expected number of events in $10^7$ seconds is shown in 
Table~\ref{tab:bs_dsk_nevents}.
   
\begin{table}
\begin{center}
\begin{tabular}[t]{|c|c c|}
\hline
Luminosity   &  \multicolumn{2}{|c|}{  $2 \times 10^{32}\,$cm$^{-2}$s$^{-1}$}  \\

Running time &  \multicolumn{2}{|c|}{  $10^7$ s }        \\

Integrated Luminosity & \multicolumn{2}{|c|}{ 2 fb$^{-1}$}  \\

$\sigma_{b\bar{b}}$ & \multicolumn{2}{|c|}{ $100~\mu$b } \\

Number of $b\bar{b}$ events & \multicolumn{2}{|c|}{$2 \times 10^{11}$ }\\

Number of $B_s^0$ + $\bar{B}_s^0$   &  \multicolumn{2}{|c|}{$5 \times 10^{10}$ } \\

${\cal{B}}$($B_s^0 \rightarrow D_s^- K^+$)$^{\dagger}$ & \multicolumn{2}{|c|}{$2 \times 10^{-4}$} \\

${\cal{B}}$($B_s^0 \rightarrow D_s^+ K^-$)$^{\dagger}$ & \multicolumn{2}{|c|}{$1 \times 10^{-4}$} \\

${\cal{B}}(D_s^- \rightarrow \phi \pi^-)\times{\cal{B}}(\phi \rightarrow K^+K^-)$ & $1.8
\times 10^{-2}$  &\\

${\cal{B}}(D_s^- \rightarrow \bar{K}^{*0} K^-)\times {\cal{B}}( \bar{K}^{*0}\rightarrow K^-\pi^+)$ &   &$2.2
\times 10^{-2}$  \\

Reconstruction efficiency & 0.045 & 0.023 \\

Trigger efficiency L1 & 0.74  & 0.74 \\

Trigger efficiency L2 & 0.90  & 0.90 \\

Number of reconstructed $B_s^0(\bar{B}_s^0) \rightarrow D_s^- K^+$ & 8000  & 5100 \\

Tagging efficiency $\varepsilon $ & \multicolumn{2}{|c|}{  0.70}  \\

Number of tagged events & 5600 &  3570 \\
\hline
\end{tabular}
\vspace*{0.3cm}
\caption
[Projected number of reconstructed $B_s^0 \ra D_s^- K^+$ decays at BTeV.]
{Projected number of reconstructed $B_s^0 \ra D_s^- K^+$ decays 
($^{\dagger}$ indicates estimated branching fractions).}
\label{tab:bs_dsk_nevents}
\end{center}
\end{table}

  As the $CP$~asymmetry is diluted by a factor of e$^{-\sigma_t^2{x_s^2}/2}$,
  good time resolution is important.
   Fig~\ref{fig:tres} is a plot of the generated proper time 
($t_{\rm gen}$) minus the reconstructed proper time 
($t_{\rm rec}$) for 
   events passing the selection criteria described above.
    A Gaussian
   fit to the residual $t_{\rm gen}-t_{\rm rec}$ distribution gives 
a proper time resolution $\sigma_t=0.043$~ps.
    Given $\tau_{B_s^0}= 1.54$~ps, we obtain 
   $\sigma_t/\tau = 0.03$.
       
\begin{figure}
\centerline{\epsfysize=3.0in \epsffile{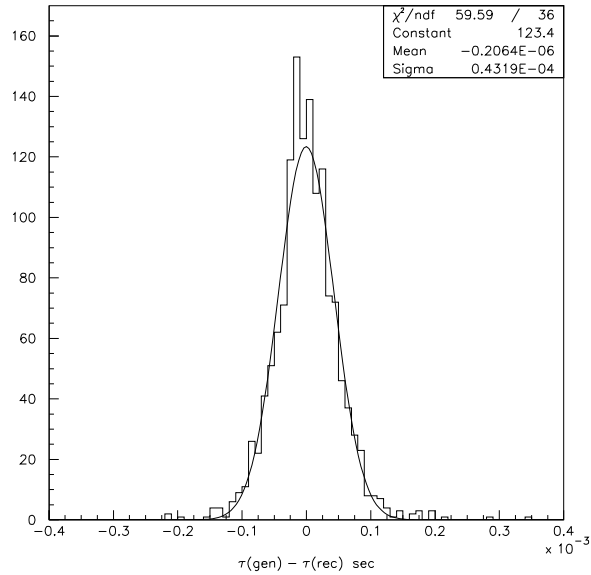}}
\vspace{0.1cm}
\caption
[Proper time resolution for $B_s^0$ : $t_{\rm gen}-t_{\rm rec}$ (ns).]
{ Proper time resolution for $B_s^0$ : $t_{\rm gen}-t_{\rm rec}$ (ns).}
\label{fig:tres}
\end{figure}

 \subsubsection*{Background Studies}
 Background can arise from real physics channels such as $B_s^0 \ra D_s^- \pi^+$ and
 $B_s^0 \ra D_s^{*-} \pi^+$ where the pion is misidentified as a kaon or
 comes from
 random combinations of a real $D_s^-$ with a $K$ from the other $B$~hadron
 in the event or the primary interaction vertex.
 
  The combinatoric  background was studied in two steps. 
  First, generic $b\bar{b}$ events were 
  generated in order to study the  signal to background
  of $D_s^- \ra \phi \pi^-$.
  Preliminary results indicate we can achieve $S/B \sim 1$
 and  estimate that most
  of the combinatoric background will come from real $D_s^-$.

   Second,
  $``B" \ra D_s^- X, D_s^-\ra \phi \pi^-$ events 
  were generated to determine the
  background from real $D_s^-$  combinations with other tracks in the event.
   The $D_s^-$ can be from directly produced charm
 or from $B$~decays. Although the charm production cross-section is expected to be about a factor
 of 10 higher than the $b\bar{b}$ production cross-section, the trigger efficiency for charm
 events is  much lower.
     
    The background events were reconstructed as described above for the signal 
   except that all pion 
   tracks were used as kaon candidates to simulate misidentification in the
   RICH.
   A pion misidentification rate was imposed later.
   
   For 900,000 $``\bar{B}" \ra D_s^- X, D_s^- \ra \phi \pi^-$ events, 
   10 events remained in the mass window 5.0 - 6.0~\gevcc\
   after all the cuts above were applied. In all these events the kaon candidate
   was really a pion. We then use a pion misidentification rate of 2\% and estimate that the
   combinatoric  background is about 1\% of the signal. 

  Background can also come from decays such as $B_s^0\ra D_s^- \pi^+$, $B_s^0\ra D_s^{*-} \pi^{+}$ 
  where the pion is
  misidentified as a kaon. Most of the background comes from $B_s^0\ra D_s^- \pi^+$. For decays
  where there is a missing particle there is very little overlap of the reconstructed
  mass with the signal region. The signal and scaled background are shown
   in Fig.~\ref{fig:misid}.
  We expect that this will be the largest source of background and estimate
  $S/B \sim 7$. 
  These results assume that pions are misidentified as kaons at a rate of 2\%.
  We have used the stand-alone simulation of the RICH detector 
  described in Sec.~5.4 to study
  the efficiency of the signal versus efficiency of  the background from misidentified
  pions. The results are shown in Table~\ref{tab:RICH_eff}.

\begin{figure}[tbp]
\centerline{\epsfxsize=3.4in \epsffile{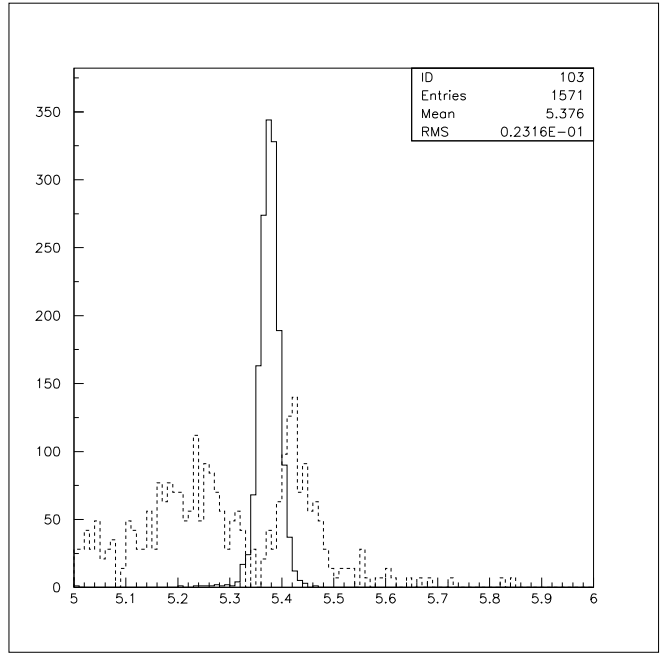}}
\vspace{0.3cm}
\caption
[Comparison of $B_s^0\ra D_s^- K^+$ signal and background at BTeV.]
{Comparison of $B_s^0\ra D_s^- K^+$ signal and background 
 from $B_s^0\ra D_s^- X$, where $X$ contains at
 least one pion misidentified as a $K$.}
\label{fig:misid}
\end{figure}

\begin{table}[tbp]
\centering
\begin{tabular}[t]{|c c|}
\hline
 $B_s^0\ra D_s^- K^+$  &  $B_s^0\ra D_s^- \pi^+$  \\
\hline
    0.62     &0.00000   \\
    0.66     &0.00184    \\
    0.73     &0.00551   \\
    0.75     &0.00735   \\
    0.76     &0.00919   \\
    0.78     &0.01287   \\
    0.79     &0.01471   \\
    0.80     &0.01654   \\
    0.81     &0.01838   \\
    0.82     &0.04596   \\
    0.84     &0.07700   \\
    0.85     &0.12132   \\
    0.86     &0.17647   \\
\hline       
\end{tabular}
\vspace*{0.3cm}
\caption
[Comparison of RICH efficiency at BTeV.] 
{Comparison of RICH efficiency for  $B_s^0\ra D_s^- K^+$  versus  
$D_s^- \pi^+$  at BTeV.} 
\label{tab:RICH_eff}
\end{table}
  
\boldmath
\subsubsection*{Extracting $\rho$ and $\sin\gamma$ 
from a Toy Monte Carlo Study}
\unboldmath
\index{BTeV!$\sin\gamma$ prospects}%

 A Toy Monte Carlo study was performed to determine the expected error on
 $\gamma$. 
  For the first study, the input values of the  parameters were chosen to be
    $x_s=30.0$, $\rho = |A_f|/|\bar A_f| = 0.7$,
  $\sin\gamma = 0.75$,
   $\delta=10^{\circ}$ and $\Delta\Gamma/\Gamma=0.16$.
With the Toy Monte Carlo,  
    a set of ``events'' (i.e.~proper times) was generated and split
    into  
    the four decay modes 
   with correct time 
  distributions. 
  The proper times were  then smeared with a Gaussian of width $\sigma_t =
    0.03\,\tau$,  
  and a cutoff at low $t$ which simulated a $L/\sigma_L$ cut: $t_{\rm min}=
  0.25\,\tau$.
 A fraction of the events
   were assigned to come from the ``wrong flavour''
  parent. A mistag fraction of 32\% is used. 
    Background events with a pure exponential time distribution are added 
    to the ``signal'' events.
     The background is assumed to have the same  lifetime
   as the signal.
   
   A maximum likelihood fit was used to find the values of    $\rho$,
 $\gamma$, $\delta$  and $\Delta\Gamma$. 
   One thousand trials were done, each of 6,800 events. The fitted values of 
  the parameters are shown in Figure~\ref{fig:dg12} .
 The values of the input parameters were varied to study the impact on the
 error.  
   The results of the fits are shown in Table~\ref{tab:fit1}. 

\begin{figure}[tb]
\centerline{\epsfxsize=4.0in \epsffile{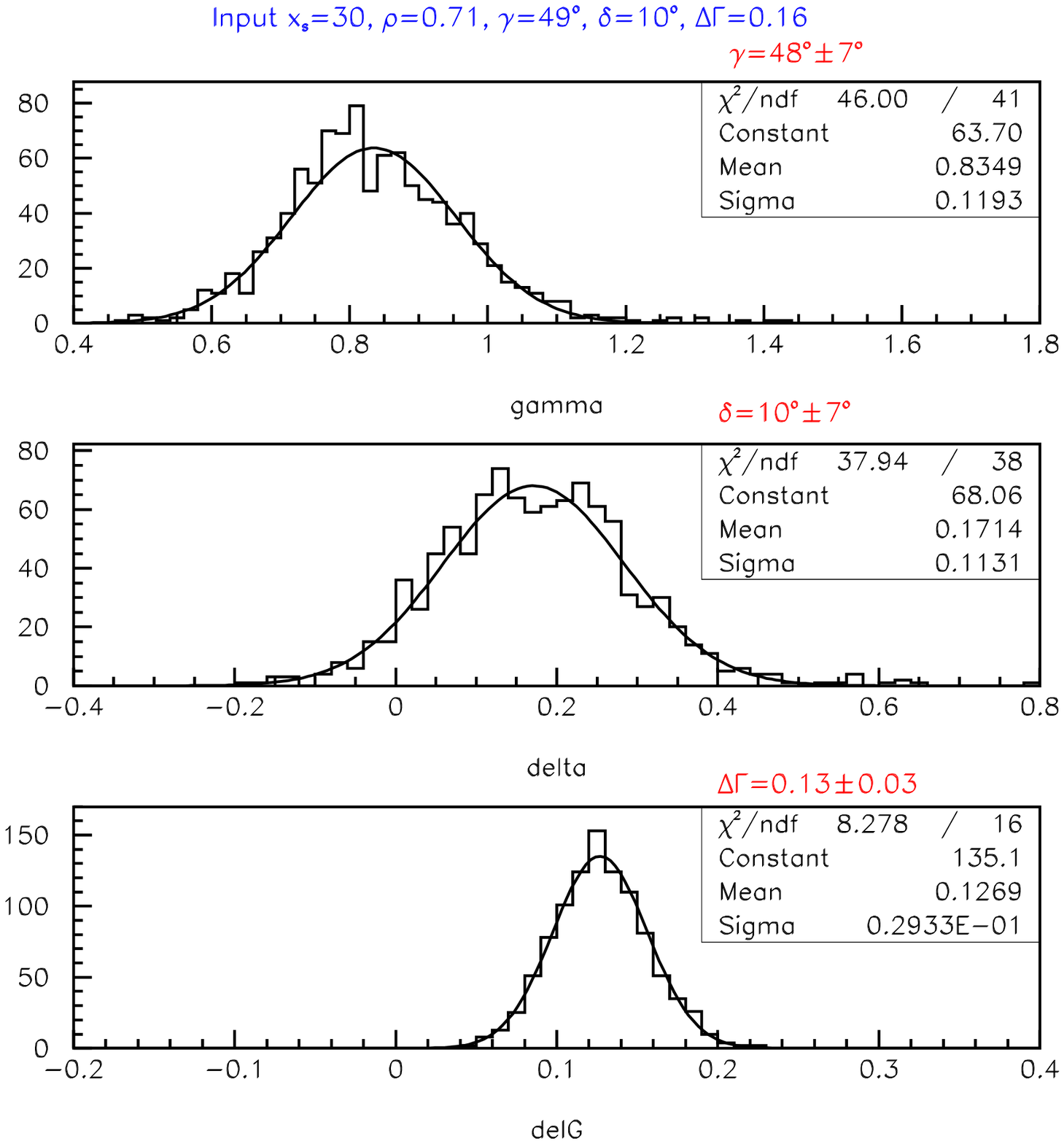}}
\caption
[Fitted values of $\gamma$, $\delta$,
 and $\Delta\Gamma$.]
{Fitted values of $\gamma$, $\delta$,
 and $\Delta\Gamma$. }
\label{fig:dg12}
\end{figure}

\begin{table}[htb]
\centering
\begin{tabular}[t]{|c c c c c|c c|}
\hline
 $x_s$ & $\rho$ & $\delta$ & $\gamma$ & $\Delta\Gamma$ & $\sigma(\gamma)$ & $\sigma(\Delta\Gamma)$ \\
\hline
  20   & 0.71   & $10^{\circ}$ & $49^{\circ}$ & 0.16            & 6$^{\circ}$  	  & 0.03 \\
  30   & 0.71   & $10^{\circ}$ & $49^{\circ}$ & 0.16            & 7$^{\circ}$  	  & 0.03 \\
  40   & 0.71   & $10^{\circ}$ & $49^{\circ}$ & 0.16            & 8$^{\circ}$  	  & 0.03 \\
  30   & 0.50   & $10^{\circ}$ & $49^{\circ}$ & 0.16            & 8$^{\circ}$  	  & 0.03 \\
  30   & 0.71   & $10^{\circ}$ & $30^{\circ}$ & 0.16            &  6$^{\circ}$  	  & 0.03 \\
  30   & 0.71   & $10^{\circ}$ & $90^{\circ}$ & 0.16            &  15$^{\circ}$ 	  & 0.04  \\
  30   & 0.71   & $0^{\circ}$  & $49^{\circ}$ & 0.16            &  6$^{\circ}$ 	  & 0.03 \\
  30   & 0.71   & $20^{\circ}$ & $49^{\circ}$ & 0.16            &  6$^{\circ}$ 	  & 0.03 \\
  30   & 0.71   & $10^{\circ}$ & $49^{\circ}$ & 0.06            &  8$^{\circ}$ 	  & 0.04  \\
  30   & 0.71   & $10^{\circ}$ & $49^{\circ}$ & 0.26            &  6$^{\circ}$ 	  & 0.03 \\
\hline
\end{tabular}
\vspace*{0.3cm}
\caption
[Results of fits with variation of input parameters at BTeV.]
{Results of fits with variation of input parameters at BTeV.} 
\label{tab:fit1}
\end{table}
     
In conclusion, 
   the ability of BTeV to measure the angle $\gamma$ of the unitarity 
   triangle  depends
   on several factors which are not well known at the moment, in particular 
   the branching fractions
   for $B_s^0 \ra D_s^- K^+$ and the $B_s^0$ mixing parameter $x_s$.
     
     Using the estimates of branching fractions given in 
     Ref.~\cite{ADK}, 
   we expect to have about 
    9200 reconstructed and
    tagged events per year at a luminosity 
    of $2\times 10^{32}$~cm$^{-2}$s$^{-1}$. The study of the sensitivity
    to $\gamma$ presented above was done
    assuming 6800 tagged events and 
    gave error on $\gamma$
    of about $7^{\circ}$. We expect that this will improve with the increased number
    of events.

\boldmath
\subsubsection{$B^- \ra D^0 K^-$: BTeV Report}
\label{sec:cp_bdk_btev}
\unboldmath
\index{decay!$B^0 \to D^0 K^-$}%
\index{BTeV!$B^0 \to D^0 K^-$ prospects}%


  The  reconstruction efficiency of the proposed BTeV detector 
  for $B^- \ra K^- D^0$ has been studied
 for two  $D^0$  decay modes: $ D^0 \ra K^+ \pi^-$ and 
 $ D^0 \ra K^- K^+ $.
   Note that the $K^- K^+ $ decay mode represents a  $CP$~eigenstate.
   In this case, even though the branching fraction for 
    $B^- \ra K^- \bar{D^0}, \bar{D^0} \ra K^+ K^-$
   is expected to be only
   1\% of $B^- \ra K^- D^0$, $D^0 \ra K^+ K^-$, 
    we could still obtain  a $CP$~asymmetry up to 20\%.
    The events are generated with PYTHIA and the detector is modeled with
    MCFAST.

  The  reconstruction efficiency is determined requiring that all 
  tracks be reconstructed and  can be identified in the RICH with
  momentum between
  3 and 70~\gevc\ hitting the forward tracking 
  plane downstream of the RICH. We assume that 98\% of tracks in this
  momentum range are correctly identified. 
     The final analysis cuts are selected to give a clean $D^0$ signal
   and reduce background from random combinations with kaons.  
   The selection requirements are shown in Table~\ref{tab:rec1}. 
   The reconstructed signal is shown in Fig~\ref{fig:btev_sig}. The fitted 
   Gaussian has a width of 17~\mevcc.
 
    The combined geometric acceptance and reconstruction efficiency 
    is 2.6\% for the $D^0\ra K^+\pi^-$ mode 
and 2.3\% for the $D^0\ra K^+ K^- $ mode.
    The trigger efficiency for events that pass the
    final analysis cuts is about 60\% for both modes.
     The expected number of events  is shown in Table~\ref{tab:nevents}.
    
\begin{figure}[tbp]
\centerline{
\put(40,120){\large\bf (a)}
\epsfxsize=4.0in
\epsffile{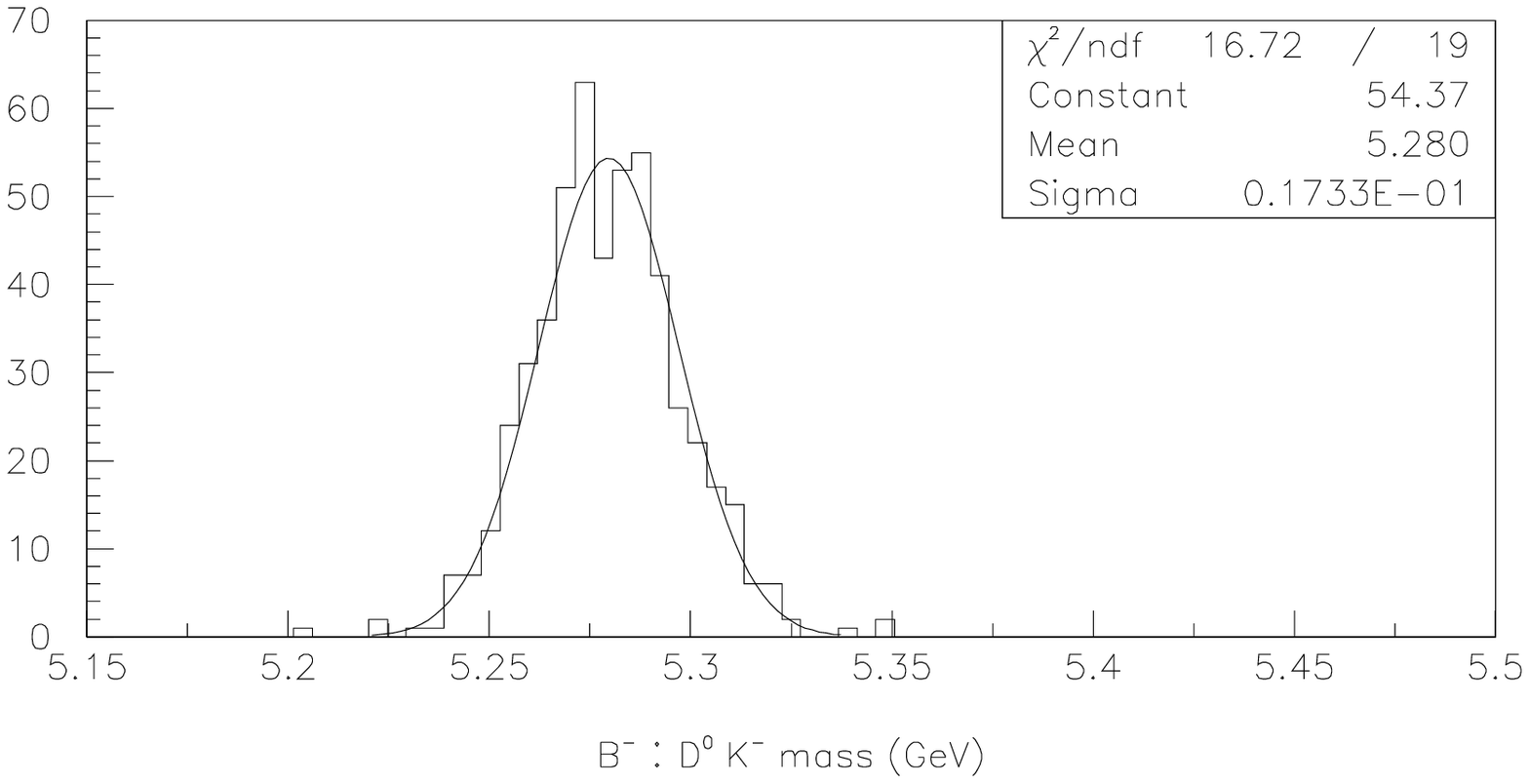}
}
\centerline{
\put(40,105){\large\bf (b)}
\epsfxsize=4.0in
\epsffile{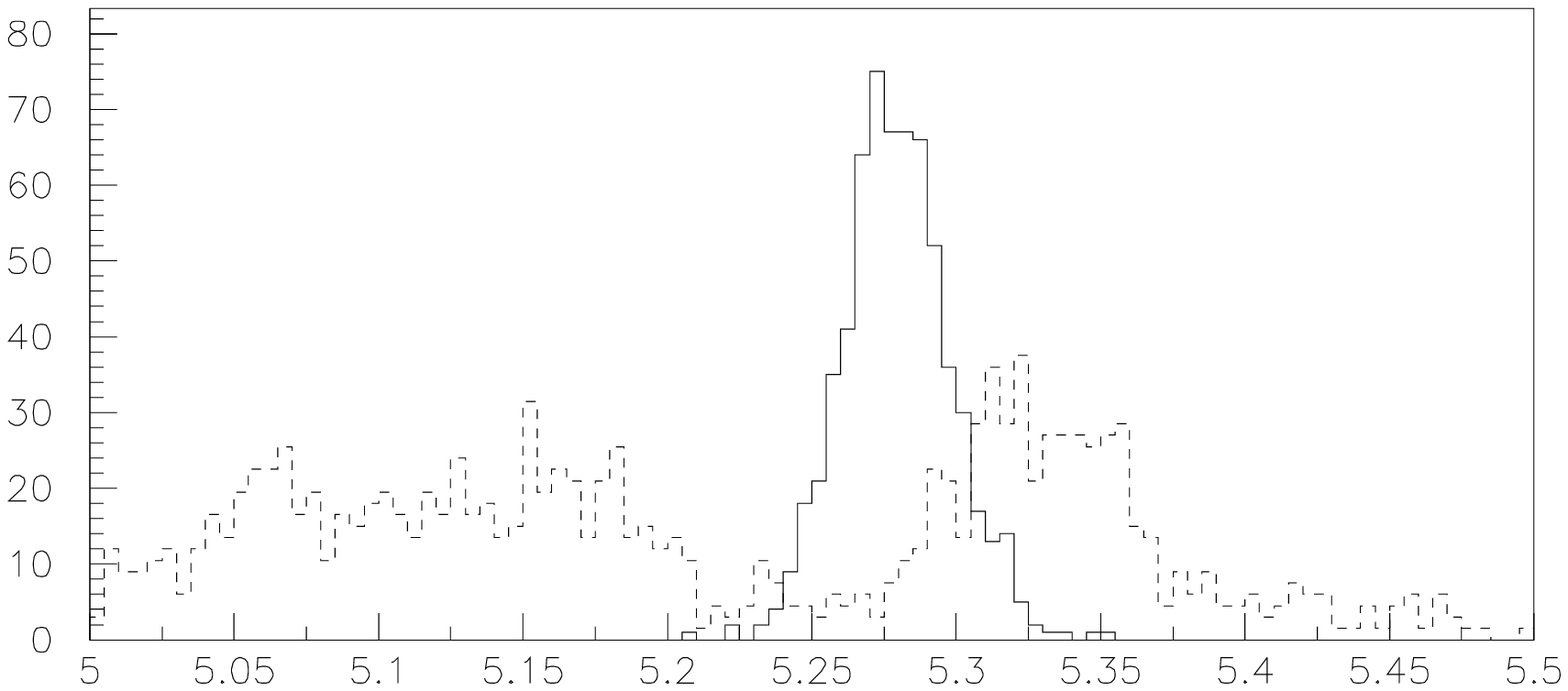}
}
\vspace*{0.3cm}
\caption
[$B^- \ra D^0 K^-$ signal mass and background at BTeV.]
{(a) $B^- \ra D^0 K^-$ mass [\gevcc].
(b)
Signal (solid line) and background (dashed line) from  $B^- \ra \pi^- D^0$ and
 $B^- \ra \pi^- D^0 X$ where the $\pi^-$ is misidentified as a $K^-$. } 
\label{fig:btev_sig}
\end{figure}

\begin{table}[btp]
\centering
\begin{tabular}[t]{|c|c|}
\hline
 $L/\sigma (B^-)$  &  $>10.0$ \\
 $L/\sigma (D^0)$  & $>4.0$ \\
%
%
 $\chi^2$ ($B$ vertex) & $<5.0$ \\
 $\chi^2$ ($D$ vertex) & $<10$ \\
 $B$ point back to prim.~vertex &   \\
%
  $D^0$ mass window & 1.85 - 1.88~\gevcc\ \\
\hline
\end{tabular}
\vspace*{0.3cm}
\caption
[Selection requirements for $D^0 \ra K^- \pi^+$ and $D^0 \ra K^+ K^-$.]
{Selection requirements for $D^0 \ra K^- \pi^+$ and $D^0 \ra K^+ K^-$.}
\label{tab:rec1}
\end{table}

\begin{table}[tbp]
\centering
\begin{tabular}[t]{|c|c c |}
\hline
Decay Mode & $K^- (K^+ \pi^-)$& $K^- (K^+ K^-)$  \\

Luminosity   &  \multicolumn{2}{|c|}{  $2\times 10^{32}$~cm$^{-2}$s$^{-1}$}  \\

Running time &  \multicolumn{2}{|c|}{  $10^7$ sec }        \\

Integrated Luminosity & \multicolumn{2}{|c|}{ 2 \tfb}  \\

$\sigma_{b\overline{b}}$ & \multicolumn{2}{|c|}{ $100~\mu$b } \\

Number of $B^{\pm}$   &  \multicolumn{2}{|c|}{$1.5 \times 10^{11}$ } \\

Branching ratio & $1.7\times10^{-7}$ &  $1.1\times10^{-6}$   \\

Reconstruction efficiency & 0.026  & 0.022\\

Trigger efficiency & 0.6  & 0.6 \\

Number of reconstructed $B^{\pm}$ & 410 & 2500  \\
\hline
\end{tabular}
\vspace*{0.3cm}
\caption
[Projected number of reconstructed $B^- \ra K^- D^0$ events at BTeV.]
{Projected number of reconstructed $B^- \ra K^- D^0$ events at BTeV.}
\label{tab:nevents}
\end{table}


 \subsubsection*{Background Studies}

	Generic $b\bar{b}$ and $c\bar{c}$ events were studied and it was found that
	for both types of events
    the $D^0 \ra K^- \pi^+$ and $D^0 \ra K^- K^+$ signals  had $S/B > 5$ using the 
    same cuts as for the 
    $D^0$ in the $B^- \ra K^- D^0$ decays.
     Therefore only background arising
    from real $D^0$~mesons need to be considered.
    
    Charm events with a $D^0 \ra K^- \pi^+$ have a probability of 3.3\% of passing
    the $D^0$ analysis cuts.
    The events which pass the cuts have a trigger efficiency
    of 10\% and 0.6\% of these events have another detached $K$.
    Generic $b\bar{b}$ events 
    with a $D^0$ have a 7.0\% probability of 
     passing the $D^0$ analysis cuts. These events have a trigger efficiency of 35\% 
     and 4.0\% of these
    have another detached $K$. 
    Therefore we estimate that a generic  $b\bar{b}$ event is 50 times more likely
    to contribute to background than a $c\bar{c}$ event.
    Thus even though the charm production cross-section
    is much larger than the  $b\bar{b}$ cross-section, more background will
    come from $b\bar{b}$~events.

   Background in both  modes $B^- \ra K^- [ K^+ \pi^-]$ and  $B^- \ra K^- [ K^+ K^-]$ 
   could arise from:
\begin{itemize}   
   \item $B^- \ra \pi^- D^0$ where the $\pi^-$ 
   is misidentified as a $K^-$, and similar  decays such as $B^- \ra \pi^- D^{*0}$ and
   $B^- \ra \rho^- D^0$ where there is a missing $\pi^0$ and the $\pi^-$ is misidentified.
   These decays all have a significantly higher branching fraction than the signal.
   If we assume that the probability of misidentifying a $\pi^-$ as a $K^-$
   is  2\%, the relative signal and background from these modes is shown in Fig~\ref{fig:btev_sig}(b). 
   This is the most significant source of background for the $D^0 \ra K^+ K^-$ mode.   
   \item $"B" \ra \bar{D^0} X$ events where the $\bar{D^0}$ forms a good vertex
    with a $K^-$ from the other $B$~hadron or from the underlying event.
    This was studied by generating $"B" \ra \bar{D^0} X$ with 
$\bar{D}^0 \ra K^+ \pi^-$ events
using the same reconstruction as for the signal.
We generated 1.6 million $"B" \ra \bar{D^0} X, \bar{D}^0 \ra K^+ \pi^-$
   events. After applying the selection requirements,
no events remained in the
     mass window 5.0 - 5.5~\gevcc, while one event was found in
     the 5.5 - 6.0~\gevcc\ mass window. 
     
     We assume this type of background has the same trigger efficiency as the signal.
     We estimate, we can
     achieve $S/B\sim1$ in the $D^0 \ra K^+ \pi^-$ mode, and we expect this to be  
      the dominant source of background for this mode.
     This type of background will
     be insignificant  
     in the $D^0 \ra K^+ K^-$ mode because both the signal and background
     come from singly Cabibbo suppressed decays.
\end{itemize}

\boldmath
\subsubsection*{ Extracting $\gamma$ from Toy Monte Carlo Studies }
\unboldmath

To estimate our ability to measure $\gamma$, several sets of input parameters
  ($b$, $\gamma$, $\xi_1$, $\xi_2$) were chosen 
(see Equations~(\ref{eq:bdk_intro_d12_1}) to (\ref{eq:bdk_intro_d12_4}))
and for each set the expected number of
  events in each channel was calculated.
  Then 1000 trials were done for each set, 
  smearing the number of events by $\sqrt{N+B}$.
 For each trial  values for $b$ and $\gamma$ are calculated.
  The  fitted values of $b$ and $\gamma$ 
  are shown in Table~\ref{tab:gam_param}
  and Fig.\ref{fig:gam5}.

\begin{figure}[tb]
\centerline{
\epsfxsize=4.0in
\epsffile{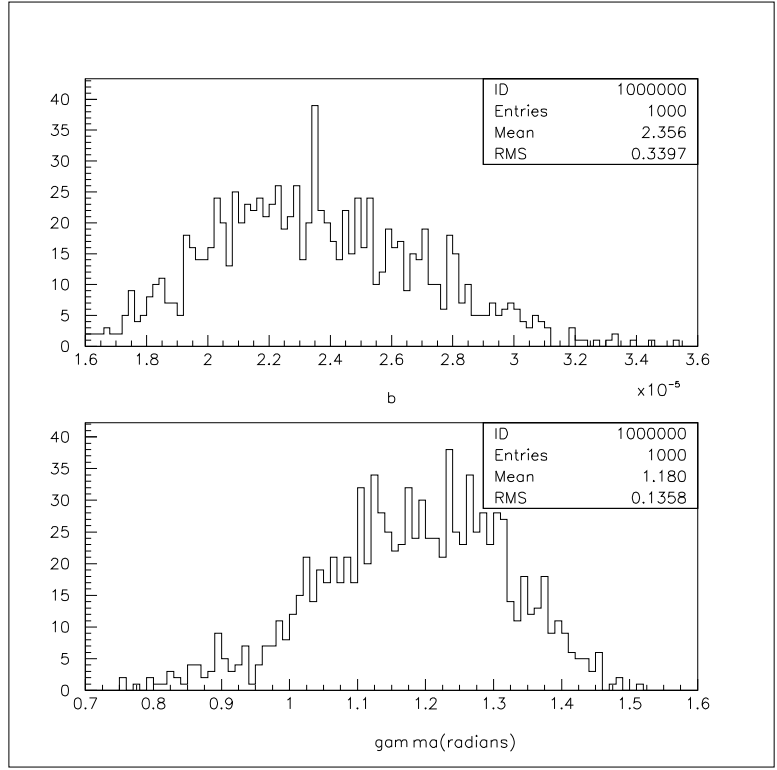}
}
\vspace*{0.3cm}
\caption
[Fitted values of $\gamma$ and $b$ at BTeV.]
{Fitted values of $\gamma$ and $b$ for input values
 $\gamma=65^{\circ}$ (1.13 rad) and $b=2.2\times10^{-5}$.}
\label{fig:gam5}
\end{figure}

   \begin{table}[btp]
   \centering
   \begin{tabular}{|c|c c c c|}
   \hline
           & test 1 &  test 2  & test 3  & test 4 \\
   \hline
     $b$ ($\times 10^{-5})$    & 2.2   & 2.2   & 2.2   & 2.2 \\
    $\xi_1$                & $45^{\circ}$  & $0^{\circ} $  & $90^{\circ}$  & $70^{\circ}$  \\
    $\xi_2$                & $30^{\circ}$  & $45^{\circ}$  & $10^{\circ}$  & $30^{\circ}$    \\
    $\gamma$               & $65^{\circ}$  & $75^{\circ}$  & $85^{\circ}$  & $50^{\circ}$     \\
    $\gamma$  fit          & $(67\pm 10)^{\circ}$  & $(75\pm 7)^{\circ}$  & $(85.0\pm 2.4)^{\circ}$ &
                              $(50.0\pm 3.2)^{\circ}$   \\
   \hline
    \end{tabular}
\vspace*{0.3cm}
  \caption
[Input Values of parameters and results of fit for $\gamma$ at BTeV.]
{Input Values of parameters and results of fit for $\gamma$ at BTeV.}
   \label{tab:gam_param}
   \end{table}

In conclusion,
  we expect to reconstruct about 400 $B^{\pm} \ra (K\pi) K^{\pm}$
  and 2500 $B^{\pm} \ra (K K) K^{\pm}$ events per year at design luminosity. 
   With this number of events, $\gamma$
 can  be measured to $\pm 10^{\circ}$ for most values of $\gamma$, $\xi_1$ and $\xi_2$.
 The  error on $\gamma$ depends on the value of $\gamma$ and the strong phases,
 in particular the error decreases with increasing difference in the strong phases.
 If we assume that the ratio of Cabibbo favored to doubly Cabibbo suppressed branching
 fractions is the same for the two decay modes, then the equations have no
 solution for $|\xi_1| = |\xi_2|$.

\boldmath
\subsection[$B \rightarrow D K$: Summary]
{$B \rightarrow D K$: Summary
$\!$\authorfootnote{Author: M.~Paulini.}
}
\unboldmath
\index{decay!$B_s^0 \to D_s^- K^+$}%
\index{decay!$B^0 \to D^0 K^-$}%

The CKM angle $\gamma$ can be extracted via related sets of $B\ra DK$
decay processes. The two decay modes 
$\Bs \ra D_s^- K^+$ and
$B^-\ra D^0 K^-$ have been studied in this section as
an alternative method of measuring $\gamma$.
The ability to measure the 
\index{CKM angle!$\gamma$}%
angle $\gamma$ in the decay mode $B_s^0 \to
D_s^- K^+$ depends
on several factors which are not well known at the moment, in particular 
the branching fractions for $B_s^0 \ra D_s^- K^+$ and the $B_s^0$ mixing
parameter $\Delta m_s$. The lack of knowledge of certain branching
fractions creates similar uncertainties to evaluate the prospects of
determining the angle $\gamma$ from $B^- \ra D^0 K^-$ decays. 

The reduction of backgrounds, in
particular physics backgrounds from  the Cabibbo allowed process 
$B_s^0 \to D_s^- \pi^+$, is the primary
challenge for CDF in extracting the $B_s^0 \to D_s^- K^+$ signal.
Exploiting the $D_s^- K^+$ invariant mass as well as d$E$/d$x$ information of
the final state particles, the performed studies show that a
signal-to-background ratio of 1/6 can be achieved. 
Assuming branching fractions as outlined in Sec.~\ref{sec:bdkcdf},
a nominal signal of
850 $B_s^0 \to D_s^- K^+$ events can be expected at CDF in \tfb. 
Thus, an initial measurement
of $\gamma$ should be possible at CDF in the beginning of Run\,II.
Within the first \tfb\ of data, the expected error on
$\sin(\gamma\pm\delta)$ is
0.4 to 0.7 depending on the assumed background levels.
By the end of Run\,II an uncertainty near $0.1$ for $\gamma$ may be achievable.
The most limiting factors for CDF\,II are the background levels and
the overall signal size. 

Since the BTeV detector will have a 
RICH detector providing excellent $\pi$-$K$ separation, 
physics backgrounds and a clean extraction of the 
$B_s^0 \to D_s^- K^+$ signal
will play a minor role for BTeV.
With the caveats mentioned in Sec.~\ref{sec:btev_bdsk}, BTeV
expects to collect about 9200 reconstructed events per year at design
luminosity  
of $2\times 10^{32}$~cm$^{-2}$s$^{-1}$. The study of the sensitivity
to $\gamma$ presented above was done assuming 6800 tagged events and 
gave error on $\gamma$ of about $7^{\circ}$. 

A similar conclusion can be drawn for the CDF and BTeV prospects of 
measuring the angle $\gamma$ with charged $B$~decays using 
$B^- \ra D^0 K^-$. 
CDF expects to collect a small sample of about 100 signal 
candidates with the two-track hadronic trigger in \tfb\ in Run\,II. 
There is optimism that the physics
background can be brought down to the same level as the signal, but there
could be considerable combinatoric background.
If the combinatoric background can also be reduced to a level
comparable to the signal, CDF would be in the position to measure 
$\gamma$ with an uncertainty in the order of 10-20$^{\circ}$
in Run\,II. 
A study to explore the event yields of other potential $CP$~modes that can
be collected with the two-track hadronic trigger showed that this device
will allow CDF to accumulate significant datasets of fully hadronic
$B$~decays. The two-track hadronic trigger will be the source of a
rich $B$~physics program involving many different $B$~decay modes at CDF in
Run\,II. 

BTeV expects to reconstruct about 400 $B^{-} \ra [K\pi] K^{-}$ events
per year at design luminosity. 
With this number of events, $\gamma$
can  be measured to $\pm 10^{\circ}$ 
for most values of $\gamma$, $\xi_1$ and $\xi_2$.
In summary, comparing both decay channels 
$\Bs \ra D_s^- K^+$ and
$B^-\ra D^0 K^-$ considered for extracting the angle $\gamma$, it appears
that the $\Bs$ decay mode offers better prospects of determining $\gamma$
from the four time-dependent asymmetries.

%
%
%
%
%
%
%
%
%
%
%
%
%
 

\boldmath
\section{Study of $B \rightarrow \rho\pi$}

\subsection[$B \rightarrow \rho\pi$: Introduction]
{$B \rightarrow \rho\pi$: Introduction
$\!$\authorfootnote{Authors: H.R.~Quinn and J.P.~Silva.}
}
\unboldmath
\index{decay!$B^0 \to \rho \pi$}%
 
%
%

Snyder and Quinn \cite{qusny} have proposed a method to measure the CKM
phase $\alpha = \pi - \beta - \gamma$ using the decays
$B^0 \rightarrow 
\{ \rho^+ \pi^-, \rho^0 \pi^0, \rho^- \pi^+ \}
\rightarrow
\pi^+ \pi^- \pi^0$
and $CP$~conjugate.
The method consists in constructing the Dalitz plot for the three 
pions in the final state \cite{bto3pi:Dalitz53,bto3pi:Fabri54}.
This is then fitted for the expression of the rate as a function
of all amplitudes, 
relative weak phases and relative strong phases for this system.
The $\rho$-resonances are described by a Breit-Wigner function.
The presence of non-zero decay widths is a source of $CP$-even
phases which interfere with the $CP$-odd and $CP$-even phases already present
in the $B \rightarrow \rho \pi$ decay amplitudes and $B^0$-$\bar B^0$
mixing.
The rich interference patterns that arise are the hallmark of this
method.

The decay amplitudes may be written as
\begin{equation}
a(B^0 \rightarrow \pi^+ \pi^- \pi^0)
=
f_+ a_{+-}
+
f_- a_{-+}
+
f_0 a_{00},
\end{equation}
where $a_{ij} = a(B^0 \rightarrow \rho^i \pi^j)$,
with $(i,j) = (+,-), (-,+)$ \mbox{or} $(0,0)$,
and similarly for the $CP$-conjugate mode.
From the Dalitz plot,
the coefficient of $|f_i|^2$ fixes $|a_i|^2$,
the coefficient of $f_+\, f_0^\ast$ fixes $\arg{(a_+\, a_0^\ast)}$,
the coefficient of $f_-\, f_0^\ast$ fixes $\arg{(a_-\, a_0^\ast)}$,
and
the coefficient of $f_+\, f_-^\ast$ fixes
$\arg{(a_+\, a_-^\ast)}$.
Each individual $B \rightarrow \rho \pi$ ($|a_i|^2$)
band lies close to the edges of the Dalitz plot,
because the mass of the $\rho$ meson is much smaller than the mass of
the $B$ meson.
Moreover,
since the $B$ and $\pi$ are spinless,
the $\rho$ must have helicity zero.
As a result,
the functions $f_k$ contain the Breit-Wigner resonance multiplied
by the cosine of the helicity angle~$\theta_k$:
\begin{equation}
f_k (s) = \frac{1}{s - m_{\rho}^2 + i\, \Pi(s)} \cos{\theta_k}.
\label{f-function}
\end{equation}
This throws the events into the corners of the Dalitz plot,
which contain the overlap ($a_i a_j^\ast$)
regions between the different channels.
In the Breit-Wigner form in Eq.(\ref{f-function}),
$s$ is the square of the invariant mass of the $\rho$,
$\theta_k$ is the angle between the line of flight of the $\rho$
and the direction of a daughter pion (in the $\rho$ rest frame),
and the choice of the exact form for the function $\Pi(s)$
is the source of systematic uncertainties.
The form advocated by the BaBar Physics Book \cite{HaQu}
is
\begin{equation}
\Pi(s) = \frac{m_\rho^2}{\sqrt{s}} \left( \frac{p(s)}{p(m_\rho^2)} \right)^3
\Gamma_\rho (m_\rho^2),
\end{equation}
where $p(s) = \sqrt{s/4 - m_\pi^2}$ is the momentum of the daughter
pion in the $\rho$ rest frame.

Using the unitarity of the CKM matrix, we may write all decay amplitudes as
a sum of two terms.
The first term is proportional to $|V_{ub}^\ast V_{ud}|$
and receives contributions from tree level and penguin diagrams.
The second term is proportional to $|V_{tb}^\ast V_{td}|$
and receives contributions from penguin diagrams alone.
Combining this with the isospin decomposition of the decay amplitudes
\cite{lnqs},
one may write \cite{qusny,lnqs}
\begin{eqnarray}
a_{+-}
&=&
e^{i \gamma}\, T_{+-} + e^{- i \beta} (P_1 + P_0),
\nonumber\\
a_{-+}
&=&
e^{i \gamma}\, T_{-+} + e^{- i \beta} (- P_1 + P_0),
\nonumber\\
a_{00}
&=&
e^{i \gamma}\, T_{00} + e^{- i \beta} (- P_0).
\end{eqnarray}
There are also electroweak penguin diagrams,
but these are expected to be very small in these channels
\cite{HaQu,bto3pi:electroweak}.
$P_0$ and $P_1$ describe the penguin contributions to the final state
with isospin $0$ and $1$, respectively.
The $T$ and $P$ amplitude parameters contain magnitudes and $CP$-even phases,
and the relative weak phase between their terms is
$\alpha = \pi - \beta - \gamma$.
The amplitudes for the $CP$~conjugate decays are obtained simply
by changing the signs of the weak phase.

There are ten observables in these decay amplitudes:
nine parameters are the magnitudes and $CP$-even phases in the $T$ and $P$ terms,
except for an irrelevant overall phase;
the last parameter is $\alpha$.
Eight of the amplitude parameters may be fixed using untagged data alone,
with the ninth one fixed by the tagged time-integrated data \cite{bto3pi:QS00}.
Nevertheless,
time-dependent data are needed to fix the $CP$~violating phase $\alpha$.
For example,
one may construct \cite{qusny,bto3pi:GQ97}
\begin{equation}
a_{\rm sum} = a_{+-} + a_{-+} + 2 a_{00}
= e^{i \gamma} (T_{+-} + T_{-+} + 2 T_{00}).
\end{equation}
Therefore,
using $q/p = e^{- 2 i \beta}$,
one obtains for the interference $CP$~violating quantity
present in the time-dependent decay rate,
\begin{equation}
{\cal I}m \frac{q}{p} \frac{\bar a_{\rm sum}}{a_{\rm sum}}
= \sin{2 \alpha}.
\end{equation}
Since $q/p$ was used,
any new phase due to new physics contributions
to $B^0$-$\bar B^0$ mixing will affect this determination of $\alpha$.
In contrast,
the relative weak phase between the $T$ and $P$ terms
($\alpha$)
appears in direct $CP$~violating observables,
which are not affected by any new physics contributions
to $B^0$-$\bar B^0$ mixing.
Unfortunately,
these direct $CP$~violating observables are
always affected by the unknown hadronic matrix elements in the $T$
and~$P$~terms.




\boldmath
\subsection[$B \rightarrow \rho\pi$: BTeV Report]
{$B \rightarrow \rho\pi$: BTeV Report
$\!$\authorfootnote{Authors: J.~Butler, G.~Majumder, L.~Nogach, 
K.~Shestermanov, S.~Stone, A.~Vasiliev and J.~Yarba.}
}
\unboldmath
\index{decay!$B^0 \to \rho \pi$}%
\index{BTeV!$B^0 \to \rho \pi$ prospects}%

%

There are three final states in $B^0 \rightarrow \pi^+ \pi^- \pi^0$ decays:
$B^0 \rightarrow \rho^0 \pi^0$, $B^0 \rightarrow \rho^+ \pi^-$ and 
$B^0 \rightarrow \rho^- \pi^+$.  CLEO has measured the average branching ratio of the latter two modes to
be $(2.8^{+0.8}_{-0.7}\pm 0.4)\times 10^{-5}$ and limits the $\rho^0\pi^0$
branching fraction 
to $< 5.1\times 10^{-6}$ at 90\% confidence level \cite{bto3pi:cleo_rhopi}.  
The energy and angular resolution 
of the CDF and D\O\ electromagnetic calorimeters 
is not good enough to detect
$\pi^0$'s produced in these decays with 
good efficiency and low background.
Even though detection of converted photons may provide sufficient resolution,
the reconstruction efficiency of this method is too low to accumulate large
statistics samples in this rare decay mode. 
Large statistics is necessary for the analysis of the interfering amplitudes.
Furthermore, one of the charged pions is soft in the 
kinematic regions where the $\rho^0 \pi^0$ interferes 
with the $\rho^\pm\pi^\mp$,
which makes it more difficult to trigger on these events.
BTeV, with its crystal calorimeter and generic vertex trigger, 
should be able to collect and reconstruct 
a substantial sample of $B \rightarrow \rho \pi $ events.

     The reconstruction efficiencies for $B \rightarrow \rho \pi$ and backgrounds were 
studied by BTeV using a full GEANT simulation for $\rho^{\pm}\pi^{\mp}$ and
$\rho^0\pi^0$ separately. All signal and background samples were generated
with a mean of two interactions per crossing.
      While signal events are relatively easy to generate, 
backgrounds are more difficult to estimate.
For channels with branching ratios on the order of $10^{-5}$ and efficiencies on the
order of 1\%, it is necessary to generate at least $10^7$ $b\bar{b}$ background
events. This is a difficult task that requires large amounts of CPU time and
data storage. Since almost 90\% of the time spent in generating the events is
in the electromagnetic calorimeter, BTeV passes all the
generated events through the tracking system and performs a preliminary
analysis on the charged tracks \textit{before} generating the calorimeter information.
The output of this procedure is as realistic as running all the events through the entire
GEANT process but saves a factor of three in computing time. 

BTeV looks for events containing a secondary vertex formed by two oppositely charged
tracks.   One of the most important
selection requirements for discriminating the signal from the 
background is that the events have well measured, and separated primary and 
secondary vertices. Both the primary and the secondary 
vertex fits are required to have a small chisquare  ($\chi^2/{\rm dof} < 2$).
The distance between the primary and the secondary 
vertices, divided by the error, must be large ($L/\sigma_L > 4$).
The two vertices must also be separated from each other
in the plane transverse to the beam.
BTeV defines $r_T$ in terms of the primary interaction vertex position
$(x_P, y_P, z_P)$ and the secondary decay vertex position $(x_S, y_S, z_S)$
as $r_{T}=\sqrt{(x_P-x_S)^2+(y_P-y_S)^2}$ and removes events
where the secondary vertex is close to the reconstructed primary vertex.
Furthermore, to insure that the charged tracks do not originate from the
primary vertex, 
both the $\pi^+$ and the $\pi^-$ candidates are required to
have a large impact parameter with respect to 
the primary vertex (DCA $>100~\mu$m).

Events passing these selection criteria are passed through the
electromagnetic calorimeter simulation which uses
GEANT.
To find photons from the $\pi^0$ decay energies detected in
the calorimeter are clustered. 
Local energy maxima are taken for photon candidates.
The photon candidates are
required to have a minimum energy of 1 GeV and pass the
shower shape cut which requires $E9/E25 > 0.85$.  The shower
shape cut is used to select electromagnetic showers. 
We reduce the background rate by ensuring
that the photon candidates are not too close to the projection 
of any charged tracks on the calorimeter.
For $\rho^{\pm}\pi^{\mp}$, the minimum distance requirement is $>$ 2 cm,
while for $\rho^0\pi^0$, we require the minimum distance $>$ 5.4 cm. 
Candidate $\pi^0$'s are two-photon combinations with invariant
mass between 125 and 145~MeV/$c^2$.

Kinematic cuts can greatly reduce the background to $B \rightarrow \rho \pi $
while maintaining the
signal efficiency.   Minimum energy and transverse momentum ($p_T$)
requirements are placed on each of the three pions.  
Here $p_T$ is
defined with respect to the $B$ direction which is defined by the
position of the primary and secondary vertices.
We demand that the momentum vector of the reconstructed $B$~candidate
points back to the primary vertex.  The cut is 
implemented by requiring $p_T$ balance among the
$\pi^+, \pi^-$ and $\pi^0$ candidates relative to the $B$~meson direction and
then divided by 
the sum of the $p_T$ values for all three particles 
($\Delta p_T/\Sigma {p_T}$).
BTeV also applies a cut on the $B$ decay time
requiring the $B$~candidate to live less than
5.5 proper lifetimes ($t/\tau_B < 5.5$).
The selection criteria for the two modes are summarized 
in Table~\ref{tab:rhopicuts}.

\begin{table}
\begin{center}
\begin{tabular}{|l|c|c|}
\hline
Criteria         & $\rho^{\pm}\pi^{\mp}$ & $\rho^0\pi^0$    \\ 
\hline   
 Primary vertex criteria   & $\chi^2<2$   & $\chi^2<2$ \\
 Secondary vertex criteria & $\chi^2<2$   & $\chi^2<2$ \\
 $r_T$ [cm]   &   0.0146  & 0.0132  \\
 Normalized distance  $L/\sigma$ & $>4$   & $>4$ \\
 Distance $L$ [cm] & $<5$         & $<5$          \\
 DCA of track [$\mu$m] & $>100$  & $>100$        \\
 $t/\tau_B$ &    $<5.5$       & $<5.5$          \\
 $E_{\pi^+}$ [GeV] & $>4$         & $>4$          \\
 $E_{\pi^-}$ [GeV] & $>4$         & $>4$          \\
 $p_T(\pi^+)$~[GeV/$c$] & $>0.4$  & $>0.4$        \\ 
 $p_T(\pi^-)$~[GeV/$c$] & $>0.4$  & $>0.4$        \\
 Isolation for $\gamma$~[cm] & $>2.0$& $>5.4$        \\  
 $E_{\pi^0}$ [GeV] & $>5$         & $>9$          \\  
 $p_T(\pi^0)$~[GeV/$c$] & $>0.75$  & $>0.9$        \\ 
 $\Delta p_T/\Sigma p_T$ & $<0.06$& $<0.066$      \\  
 $m_{\pi^0}$ [MeV/$c^2$]  & $125-145$ & $125-145$ \\ 
 $m_{\rho}$ [GeV/$c^2$]   & $0.55-1.1$ & $0.55-1.1$\\ 
\hline
\end{tabular}
\vspace*{0.3cm}
\caption
[Selection Criteria for $B\ra\rho\pi$ at BTeV.]
{Selection Criteria for $B\ra\rho\pi$ at BTeV.}
\label{tab:rhopicuts}
\end{center}
\end{table}

For this study, we generated three large samples of events using
BTeVGeant:
      125,000 $B^0 \rightarrow \rho^0 \pi^0$ events, 125,000 
$B^0 \rightarrow \rho^+ \pi^-$ events, and 4,450,000 generic $b \bar b$
background events.  
The results of the analysis after applying 
the cuts in Table~\ref{tab:rhopicuts} are presented in
Figure~\ref{bto3pi:btev1}(a) and (b) for 
$\rho^0\pi^0$ and Fig~\ref{bto3pi:btev1}(c) and (d) for $\rho^+\pi^-$. 
The background mass spectra are Fig.~\ref{bto3pi:btev1}(a) and (c), while 
the signal events are 
Fig.~\ref{bto3pi:btev1}(b) and (d).

\begin{figure}[tbp]
\centerline{
\put(75,175){\large\bf (a)}
\put(185,175){\large\bf (b)}
\put(295,175){\large\bf (c)}
\put(405,175){\large\bf (d)}
\epsfig{file=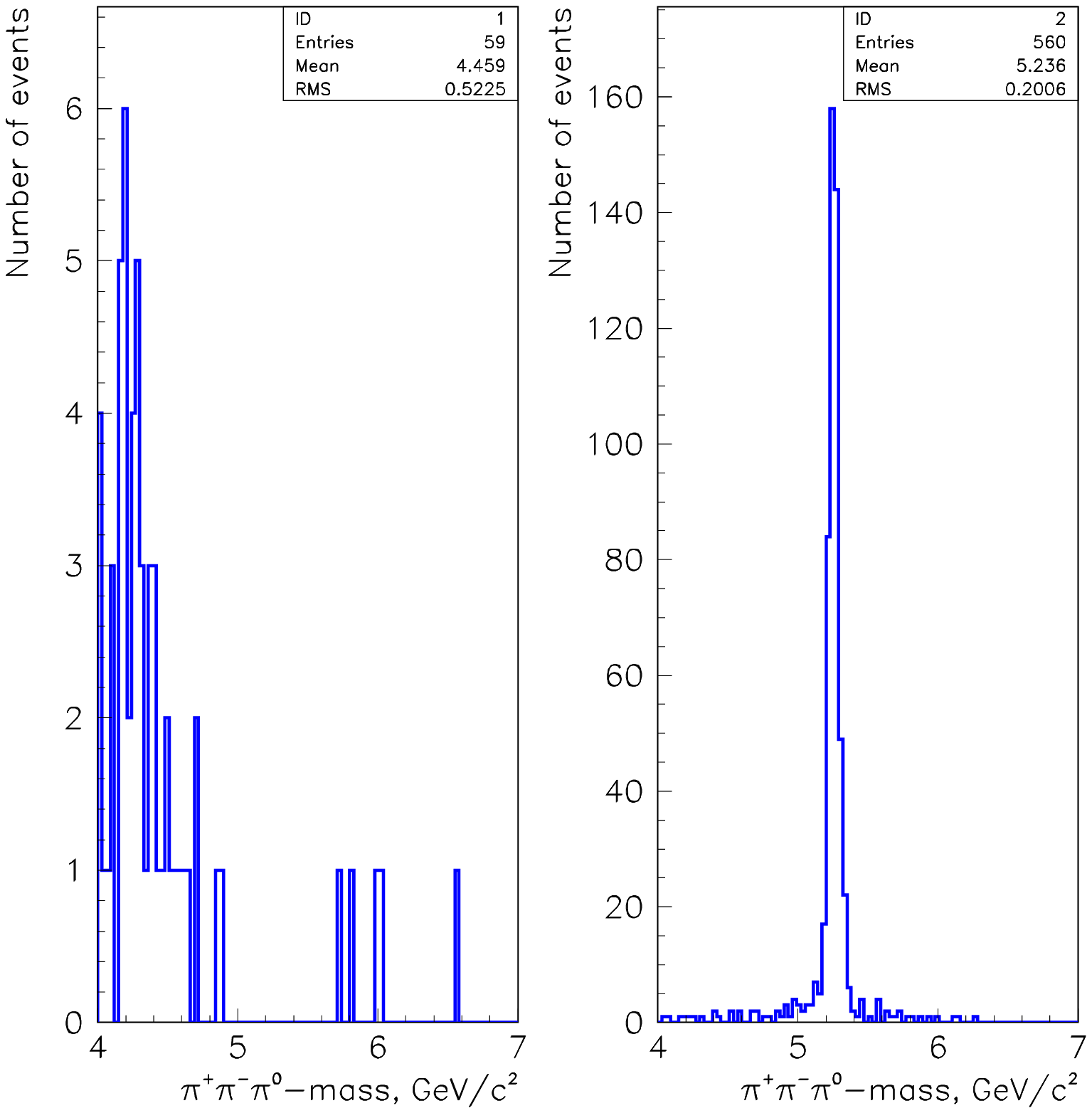,width=3in} 
\epsfig{file=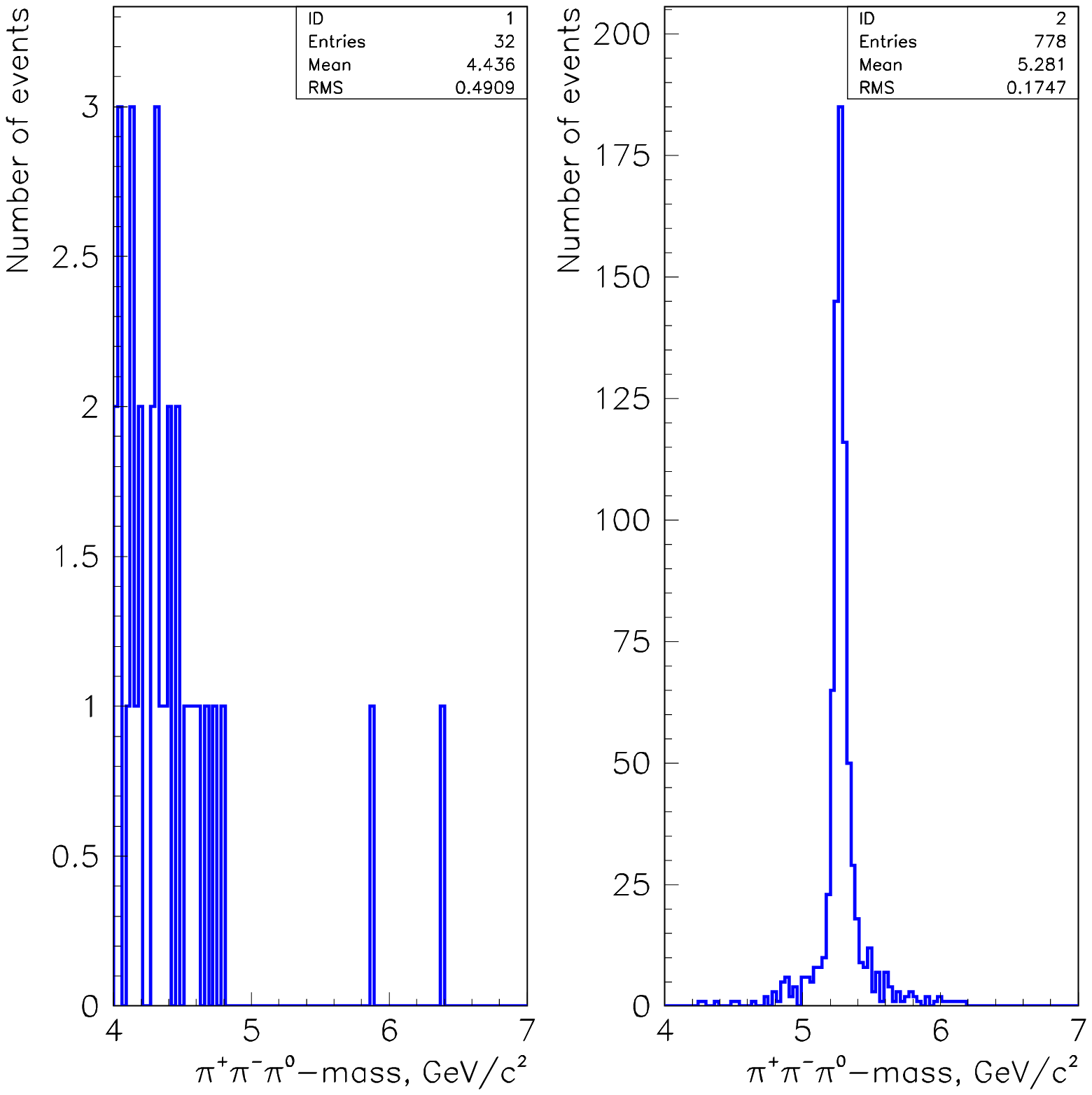,width=3in} }
\vspace*{0.3cm}
\caption
[Invariant $\pi^+\pi^-\pi^0$ mass distributions for background
          and signal events for 
          $B \rightarrow \rho^0 \pi^0$ at BTeV.] 
{ Invariant $\pi^+\pi^-\pi^0$ mass distributions for (a) background
          and (b) signal events for 
          $B \rightarrow \rho^0 \pi^0$. Invariant $\pi^+\pi^-\pi^0$ mass
          for (c) background 
          and (d) signal events for 
          $B \rightarrow \rho^+ \pi^-$. 
}
\label{bto3pi:btev1}
\end{figure}


The mass resolution for the $B$~meson is approximately 28~\mevcc.
The mean $\pi^0$ mass value in the $B \rightarrow \rho\pi$ 
events is 135~MeV/$c^2$ with a resolution of about $3$~MeV/$c^2.$ 
The relevant yields for $\rho\pi$ are shown in Table~\ref{tab:rhopi2}.  
The reconstruction efficiency is $(0.36 \pm 0.02)\%$ for $\rho^0\pi^0$
and $(0.44 \pm 0.02)\%$ for $\rho^+\pi^-.$  The background was 
obtained by considering the mass interval between 5 and
7 GeV/$c^2$. The signal interval is taken as $\pm 2\sigma$ around the
$B$ mass or $\pm 56~\mevcc$.

\begin{table}
\begin{center}
\begin{tabular}{|l|c|c|}
\hline
Quantity &{$\rho^{\pm}\pi^{\mp}$}& {$\rho^0\pi^0$}\\\hline
Branching ratio & 2.8$\times 10^{-5}$ & 0.5$\times 10^{-5~(\dagger)}$\\
Efficiency & 0.0044  &   0.0036 \\
Trigger efficiency (Level\,1) & 0.6 &   0.6 \\
Trigger efficiency (Level\,2) & 0.9 &   0.9 \\
$S/B$   &   4.1 & 0.3 \\
Signal/$10^7$ s & 9,400  & 1,350\\
$\eD$   & 0.10 & 0.10 \\
Flavour tagged yield& 940 & 135 \\\hline
\end{tabular}
\vspace{0.3cm}
\caption
[Summary of BTeV $B\to\rho\pi$ event yields.]
{Summary of BTeV $B\to\rho\pi$ event yields
($^{\dagger}$ indicates estimated branching fractions).}
\label{tab:rhopi2}
\end{center}
\end{table}
\index{BTeV!$B^0 \to \rho \pi$ prospects}%

The final numbers of both signal and background events are reduced by 
including the Level\,1 and
Level\,2 trigger efficiency, but the $S/B$ ratio
is not significantly changed.   From this study BTeV 
expects to reconstruct 
about 9,400 $\rho^{\pm}\pi^{\mp}$ events
and 1,350 $\rho^0\pi^0$ events per year
(940 and 135 fully tagged events),
with signal-to-background levels of approximately
4:1 and 1:3, respectively.

BTeV has not yet done a full simulation of the sensitivity to $\alpha$.
Final results will depend on several unknown quantities including the branching
ratio for $\rho^0\pi^0$ and the ratio of tree to penguin amplitudes. The
analysis by Snyder and Quinn \cite{qusny} showed that with 2,000
background free events they could always
find a solution for $\alpha$ and the accuracy was in the range of 
5-6$^{\circ}$.
BTeV can collect these 2,000 events in $2\times 10^7$ seconds, but 
some backgrounds will be present. 
The effect of backgrounds,
including contributions from other $B$ decays into three pions,
and the influence of experimental cuts need to be addressed.
One example of the former could arise from the decay chains 
$B \rightarrow B^\ast \pi \rightarrow \pi \pi \pi$ \cite{bto3pi:DeAndrea00}.
One example of the latter is the experimental inability to access the corner
of the Dalitz plot containing the $f_+ f_-^\ast$ interference term. 
This corner is lost because soft $\pi^0$ mesons have large
backgrounds which must be eliminated. 
Fortunately, this region probes
$\arg{(a_+\, a_-^\ast)} = \arg{(a_+\, a_0^\ast)} - \arg{(a_-\, a_0^\ast)}$,
and the right-hand side can be obtained from the 
$f_+ f_0^\ast$ and $f_- f_0^\ast$ interference regions \cite{bto3pi:Helen00}
in which the $\pi^0$ is energetic.
Assuming that the background presence will dilute experimental
sensitivity by a factor 2, BTeV should be able to measure $\alpha$ 
with an accuracy of about $10^{\circ}$.
As described in the previous section, 
Quinn and Silva~\cite{bto3pi:QS00} have proposed using 
non-flavour tagged
rates as additional input,
which should improve the accuracy of the $\alpha$
determination.



\boldmath
\section[Study of $B_{\lowercase{s}}^0\ra J/\psi\,\eta^{(\prime)}$]
{Study of $B_s^0\ra J/\psi\,\eta^{(\prime)}$}

\subsection{{$B_s^0\ra J/\psi\,\eta^{(\prime)}$}: Introduction}
\unboldmath
\index{decay!$B_s^0 \ra J/\psi \eta^{(\prime)}$}%

The $CP$~asymmetry in the decay $B_s^0\ra J/\psi\eta^{(\prime)}$ is subject 
to a clean theoretical interpretation because it is dominated by 
$CP$~violation from interference between decays with and without mixing. 
The branching ratio has not yet been measured:
\beq\label{6:exppset}
{\cal B}(B_s^0\ra J/\psi\eta)<3.8\times10^{-3}.
\eeq

The calculation of the $CP$~asymmetry is very similar to that
of the $B_s^0\ra J/\psi\phi$ mode which is discussed in
Section~\ref{ch6:intro_psiphi}. 
The quark subprocess $\bar b\ra\bar cc\bar s$ is dominated
by the $W$-mediated tree diagram:
\beq\label{6:Apsieta}
{\bar A_{ J/\psi\eta^{(\prime)}}\over A_{ J/\psi\eta^{(\prime)}}}
=\eta_{ J/\psi\eta^{(\prime)}}
\left({V_{cb}V_{cs}^*\over V_{cb}^*V_{cs}}\right).
\eeq
The penguin contribution carries a phase that is similar to 
Eq.~(\ref{6:Apsieta}) up to effects of ${\cal O}(\lambda^2)\sim0.04$. 
Hadronic uncertainties
enter the calculation then only at the level of a few percent.

Unlike the $ J/\psi\phi$ mode, here the final state consists of
a vector meson and a pseudoscalar. Consequently, the final state
is a $CP$~eigenstate, $\eta_{ J/\psi\eta}=-1$, and there is no dilution
from cancellation between $CP$-even and odd contributions. 

The $CP$~asymmetry is then given by
\beq\label{6:Imlpset}
\Im\lambda_{ J/\psi\eta^{(\prime)}}=-\sin2\beta_s.
\eeq
From a study of $B_s^0\ra J/\psi\eta^{(\prime)}$ we will learn the
following:
\begin{itemize}
\item[(i)] A measurement of the $CP$~asymmetry in 
$B_s^0\ra J/\psi\eta^{(\prime)}$
will determine the value of the very important CKM phase $\beta_s$.
\item[(ii)] The asymmetry is small, of the order of a few percent. 
\item[(iii)] An observation of an asymmetry that is significantly
larger than ${\cal O}(\lambda^2)$ will provide an unambiguous signal
for new physics. Specifically, it is likely to be related to new,
$CP$~violating contributions to $B_s^0\bar B_s^0$ mixing.
\end{itemize}


\boldmath
\subsection[$B_s^0\ra J/\psi\eta^{(\prime)}$: CDF Report]
{$B_s^0\ra J/\psi\eta^{(\prime)}$: CDF Report
$\!$\authorfootnote{Authors: W.~Bell, M.~Paulini, B.~Wicklund.}
}
\label{sec:cdfpsieta}
\unboldmath
\index{decay!$B_s^0 \ra J/\psi \eta^{(\prime)}$}%
\index{CDF!$B_s^0 \ra J/\psi \eta^{(\prime)}$ prospects}%
\index{CDF!$\sin 2\beta_s$ prospects}%

Although the CDF detector is equipped with a well-segmented
calorimeter for the detection of electrons, it is less suited for the
detection of low energy photons. However, at CDF it is not impossible to
reconstruct neutral mesons such as $\pi^0$ or $\eta$ decaying into 
two photons from energy depositions in CDF's electromagnetic
calorimeter. Although a measurement of the $CP$~violating angle $\beta_s$
will probably be best approached using the $B_s^0$~decay mode into
$J/\psi\phi$, where CDF will accumulate a large statistics sample in
Run\,II, we present here a preliminary study 
for the event yield of $B_s^0\ra J/\psi\eta^{(\prime)}$. We will
concentrate only on the decay mode $B_s^0\ra J/\psi\eta$ followed by
$\eta\ra\gamma\gamma$. In this section, we will estimate the event yield in
\tfb\ of data using Run\,I data as well as Monte Carlo extrapolations, show
the feasibility of reconstructing $\eta\ra\gamma\gamma$ with the CDF
calorimeter using Run\,I data and estimate the expected background for
reconstructing a $B_s^0$ signal. 

\subsubsection{Expected Signal}

To estimate the expected signal of $B_s^0\ra J/\psi\eta$ in \tfb\ in Run\,II,
we normalize this $B_s^0$~decay mode to the $B^+ \rightarrow J/\psi K^+$
channel as
many uncertainties such as production cross sections or trigger
efficiencies cancel in the ratio
and relative acceptances are more reliably calculated using Monte Carlo
studies.  
We can then use the ratio of the two expected data signals to obtain the
number of $B_s^0\ra J/\psi\eta$ events from the expected number of 
$B^+ \rightarrow J/\psi K^+$ 
in Run\,II.

\begin{table}[b]
\begin{center}
\begin{tabular}{c|c}\hline 
$p_T$ of both muons & $\geq 2.0~\gevc$ \\ 
$\eta$ of both muons & $\leq 0.6$  \\ 
$\eta$ of both photons & $\leq 1.0$ \\ 
$E_T$ of both photons & $\geq 1.0$~GeV \\ 
\hline 
$p_T$ of both muons & $\geq 2.0~\gevc$ \\ 
$p_T$ of $K^+$ & $\geq 1.25~\gevc$ \\ 
$\eta$ of both muons & $\leq 0.6$  \\ 
$\eta$ of $K^+$ & $\leq 1.0$ \\ \hline
\end{tabular}
\vspace*{0.3cm}
\caption
[Constraints used for the generation of Monte Carlo data for $J/\psi \eta$
at CDF.]
{Constraints used for the generation of Monte Carlo data.  
At the top the constraints for $J/\psi \eta$ are described, while $J/\psi
K^+$ is listed at the bottom.}   
\label{ch6:cdf_psietacuts}
\end{center}
\end{table}

The starting point for this analysis is the generation of 
$B_s^0\ra J/\psi\eta$ where $\eta \rightarrow \gamma \gamma$ is 
chosen as the most favourable $\eta$ decay mode accounting for 
$(39.3 \pm 0.3)\%$~\cite{pdg98} of the decay width. In addition, the 
decay channel $B^+ \rightarrow J/\psi K^+$ was also produced. 
Table~\ref{ch6:cdf_psietacuts} gives a summary of the kinematic constraints
applied to the generated Monte Carlo data.
The photon resolution in the CDF calorimeter was assumed to be
$\sigma(E_T) = 0.136 \sqrt{E_T}$~\cite{emcalor} for this study. 
The four-momenta of the daughter particles were then combined to obtain the
invariant mass of the $B_s^0$ candidates. In order to improve the mass
resolution effected by the energy resolution, the $B_s^0$ four momentum can
be corrected using the following relation: 
\begin{equation}
\roarrow{B_s^0}\ =\ \roarrow{J/\psi}\ +\
\frac{m_{\eta}^{\rm PDG}}{m_{\eta}^{\rm rec}}
\times(\roarrow{\gamma_1}+\roarrow{\gamma_2}) 
\label{ch6:eq_fudge}
\end{equation}
Here, $\roarrow{B_s^0}$, $\roarrow{J/\psi}$ and $\roarrow{\gamma}$ refer to
the  
four vector quantities of the respective particles while 
$m_{\eta}^{\rm PDG}$ and ${m_{\eta}^{\rm rec}}$ are the $\eta$ table mass
from the PDG~\cite{pdg98} and the reconstructed diphoton mass, respectively.
After mass
constraining the $J/\psi\ra\mu^+\mu^-$ dimuon combination to the nominal
$J/\psi$~mass and applying the correction given
in Eq.~(\ref{ch6:eq_fudge}), a $B_s^0$ mass resolution of better than
40~\mevcc\ can be achieved at CDF.
The improvement from the uncorrected to the corrected $B_s^0$ invariant mass 
minus the nominal $B_s^0$ PDG mass value is illustrated in
Figure~\ref{ch6:cdf_psieta_1}. 

\begin{figure}[tbp]
\centerline{
\epsfxsize=3.5in
\epsffile{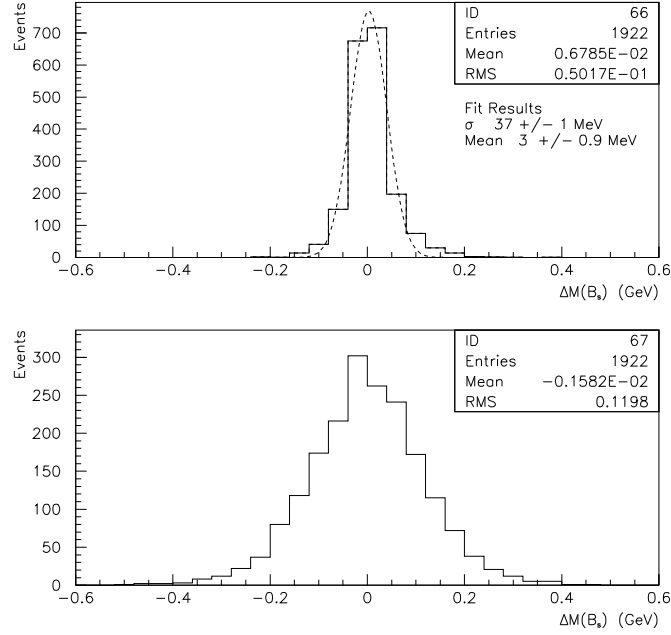}
}
\vspace*{0.3cm}
\caption
[Distribution of $B_s^0$ invariant mass 
minus the nominal PDG mass.]
{Distribution of $B_s^0$ invariant mass 
minus the nominal PDG mass value after (top) and before (bottom) the
correction described in the text.
}
\label{ch6:cdf_psieta_1}
\end{figure}

To determine the expected signal for $B_s^0\ra J/\psi\eta$ in Run\,II, we used
the ratio 
\begin{equation}
\frac{f_s}{f_u}\frac{{\cal B}(B_s^0 \rightarrow J/\psi \eta){\cal B}(\eta \rightarrow \gamma \gamma)}{{\cal B}(B^+ \rightarrow J/\psi K^+)}\frac{N(J/\psi \eta)}{N(J/\psi K^+)} 
\label{ch6:eq_cdf_psieta_ratio}
\end{equation}
relating the $J/\psi \eta$ signal rate to the number of 
$B^+ \rightarrow J/\psi K^+$ events.
The ratio of the fragmentation fractions 
${f_s}/{f_u} = 0.427$  
is taken from~Ref.~\cite{cdf_frag} and
the $B_s^0 \rightarrow J/\psi \eta$ branching fraction is estimated to
${\cal{B}}(B_s^0\to J/\psi \eta)= 4.8\times 10^{-4}$ from corresponding
$B^0$ decays. The number of reconstructed $J/\psi \eta$ and $J/\psi K^+$
events starting with $1 \times 10^6$ $B^+/B_s^0$ mesons were approximately
1800 versus 6700. 
The ratio in Eq.~(\ref{ch6:eq_cdf_psieta_ratio}) finally yields
approximately 0.022.

The expected number of fully reconstructed $B^+\ra J/\psi K^+$ events in
\tfb\ of data has been estimated in Section~\ref{sec:cdf-sin2beta} to be
approximately 50,000 (see also Table~\ref{ed_run2}). With this number and 
the ratio from Eq.~(\ref{ch6:eq_cdf_psieta_ratio}) we estimate to observe
about 1100 $B_s^0\ra J/\psi\eta$ decays in \tfb\ in Run\,II.
The $B^+ \rightarrow J/\psi K^+$ Monte Carlo generation was also checked
against the  observed number of Run\,Ib signal events including
acceptance factors.   

\subsubsection{Reconstruction of Neutrals at CDF}

To demonstrate the feasibility of observing neutral particles such as
$\pi^0\ra\gamma\gamma$ or $\eta\ra\gamma\gamma$ with the CDF calorimeter,
we investigated the reconstruction of low energy photons using Run\,I
data. Using the Run\,I exclusive electron trigger data, which represent a
data sample 
enhanced in $b\bar b$ events, we combined photon candidates in separate
calorimeter towers with $E_T^{\gamma} > 1$~GeV. Using requirements on
$E_{\rm had}/E_{\rm em}$, isolation and the pulse height in the strip
chambers, we find almost 18,000 $\pi^0\ra\gamma\gamma$ candidates on a low
background as shown in Figure~\ref{ch6:cdf_psieta_2}(a). A similar search
for $\eta\ra\gamma\gamma$ candidates yields a signal of about 1600 events
as can be seen in Figure~\ref{ch6:cdf_psieta_2}(b).

\begin{figure}[tbp]
\centerline{
\put(175,220){\large\bf (a)}
\put(395,220){\large\bf (b)}
\epsfxsize=3in
\epsffile{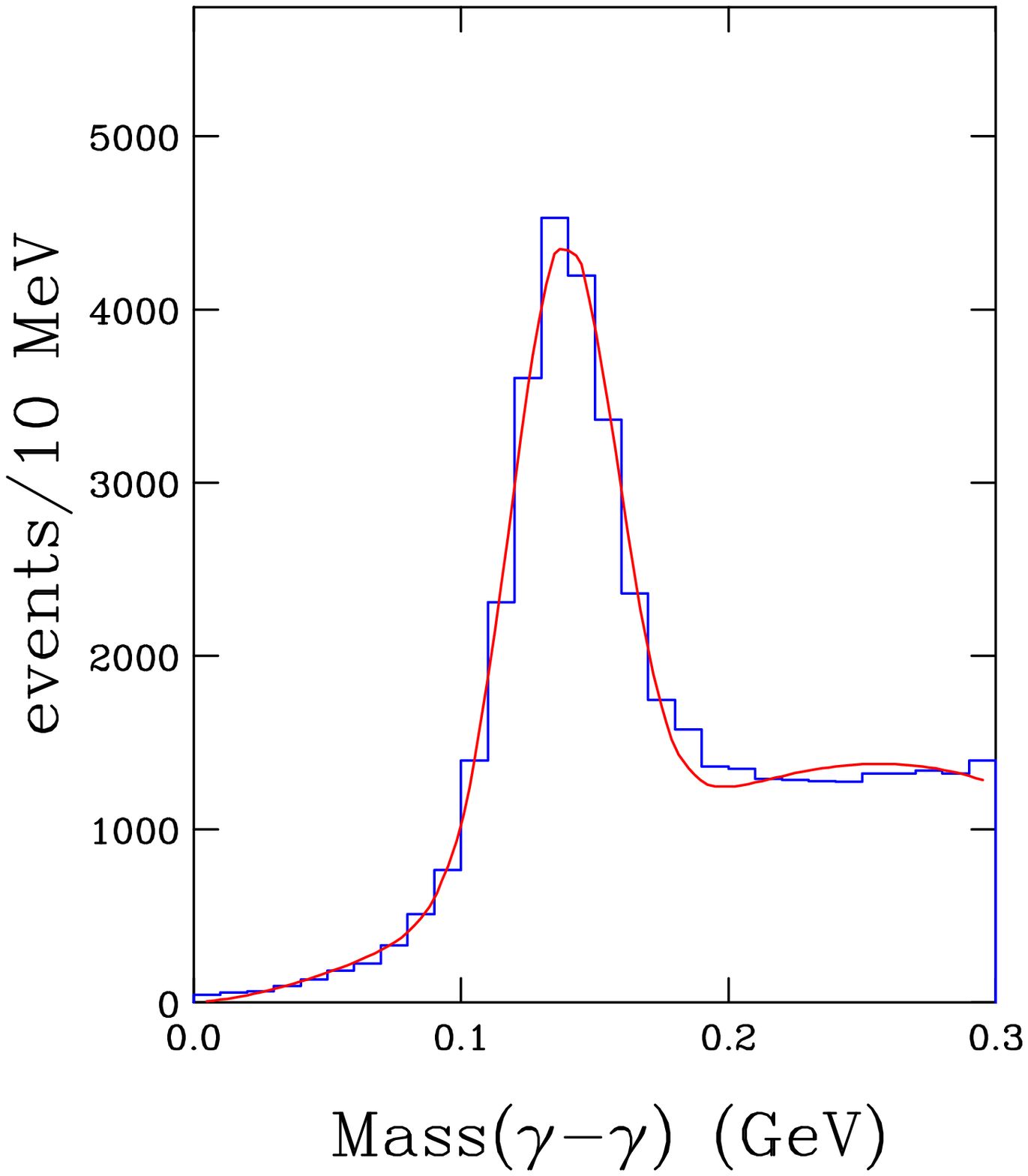}
\epsfxsize=3in
\epsffile{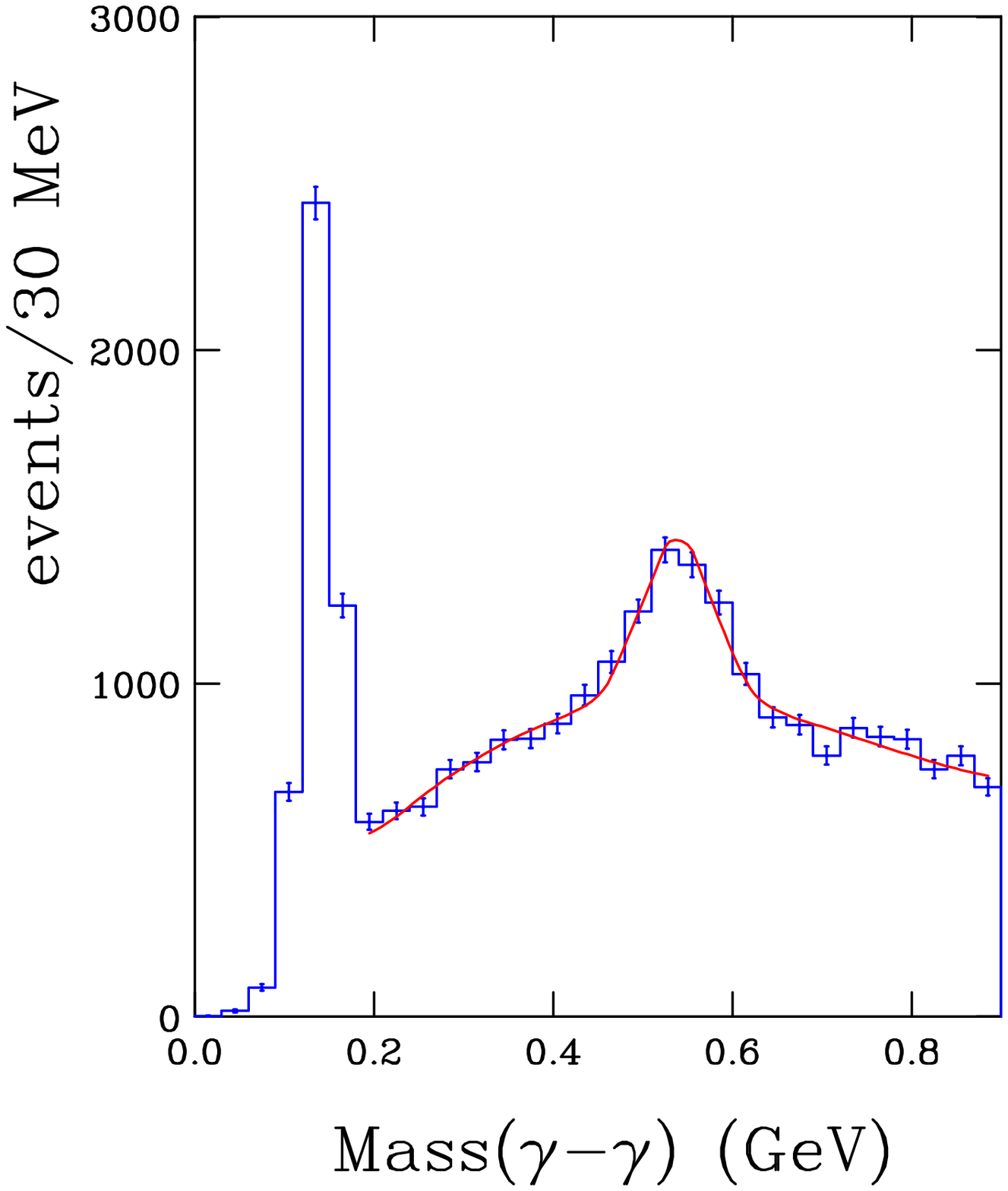}
}
\vspace*{0.3cm}
\caption
[Invariant diphoton mass distribution showing a
$\pi^0\ra\gamma\gamma$ and $\eta\ra\gamma\gamma$ 
signal in CDF Run\,I data.]
{Invariant diphoton mass distribution showing a
(a) $\pi^0\ra\gamma\gamma$ and (b) $\eta\ra\gamma\gamma$ 
signal in CDF Run\,I data.
}
\label{ch6:cdf_psieta_2}
\end{figure}

\subsubsection{Expected Background}

To estimate the expected background rate for $B_s^0\ra J/\psi\eta$ in Run\,II,
$J/\psi\ra\mu\mu$ data from Run\,I were used. These data were also exploited to
improve the $\eta\ra\gamma\gamma$ selection as suggested by the Monte Carlo.
We again use $B^+\ra J/\psi K^+$ as the reference mode and estimate from
the observed $J/\psi K^+$ signal together with
Eq.~(\ref{ch6:eq_cdf_psieta_ratio}) to detect six $B_s^0\ra J/\psi\eta$
events in the Run\,I $J/\psi$ data. 
To obtain an idea about the shape of the background underneath a potential 
$B_s^0\ra J/\psi\eta$ signal, 
the two-dimensional distribution of $m(\gamma\gamma)$
versus $m(B_s^0)$ is plotted in Figure~\ref{ch6:cdf_psieta_3}(a).
It appears from that figure that a large proportion of the  
background can be excluded by a cut around the $\eta$
invariant mass. Using a $\pm 120$~\mevcc\ window around the nominal $\eta$
mass,  we observe the distribution of $J/\psi\eta$
background events from Run\,I $J/\psi$ data shown in 
Figure~\ref{ch6:cdf_psieta_3}(b).
Overlaid onto the data is the Monte Carlo expectation scaled to the six
signal events estimated. The Monte Carlo
expectation is plotted as points and as a
Gaussian fit to the MC data.  

\begin{figure}[tbp]
\centerline{
\epsfxsize=3.1in
\epsffile{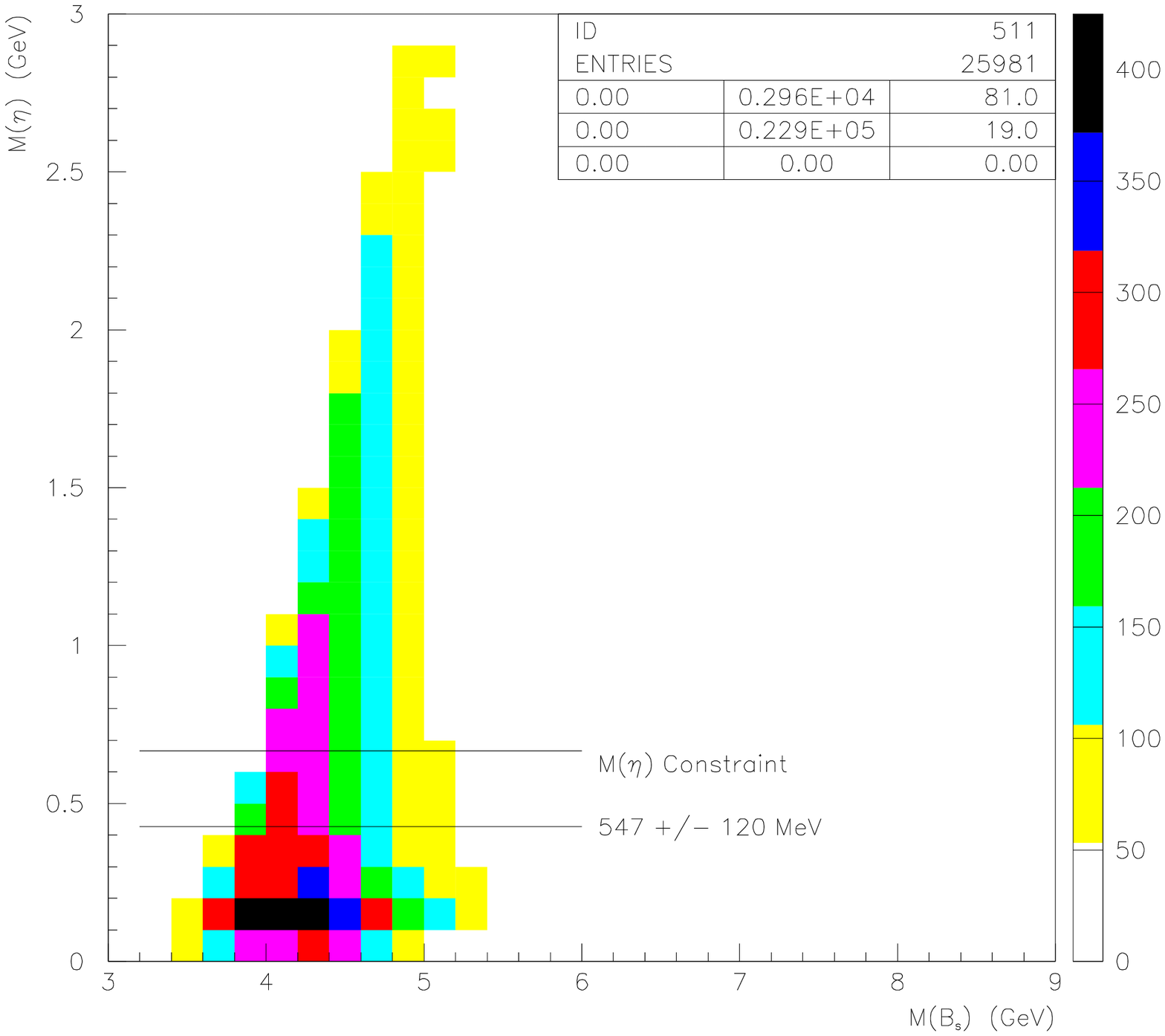}
\epsfxsize=2.9in
\epsffile{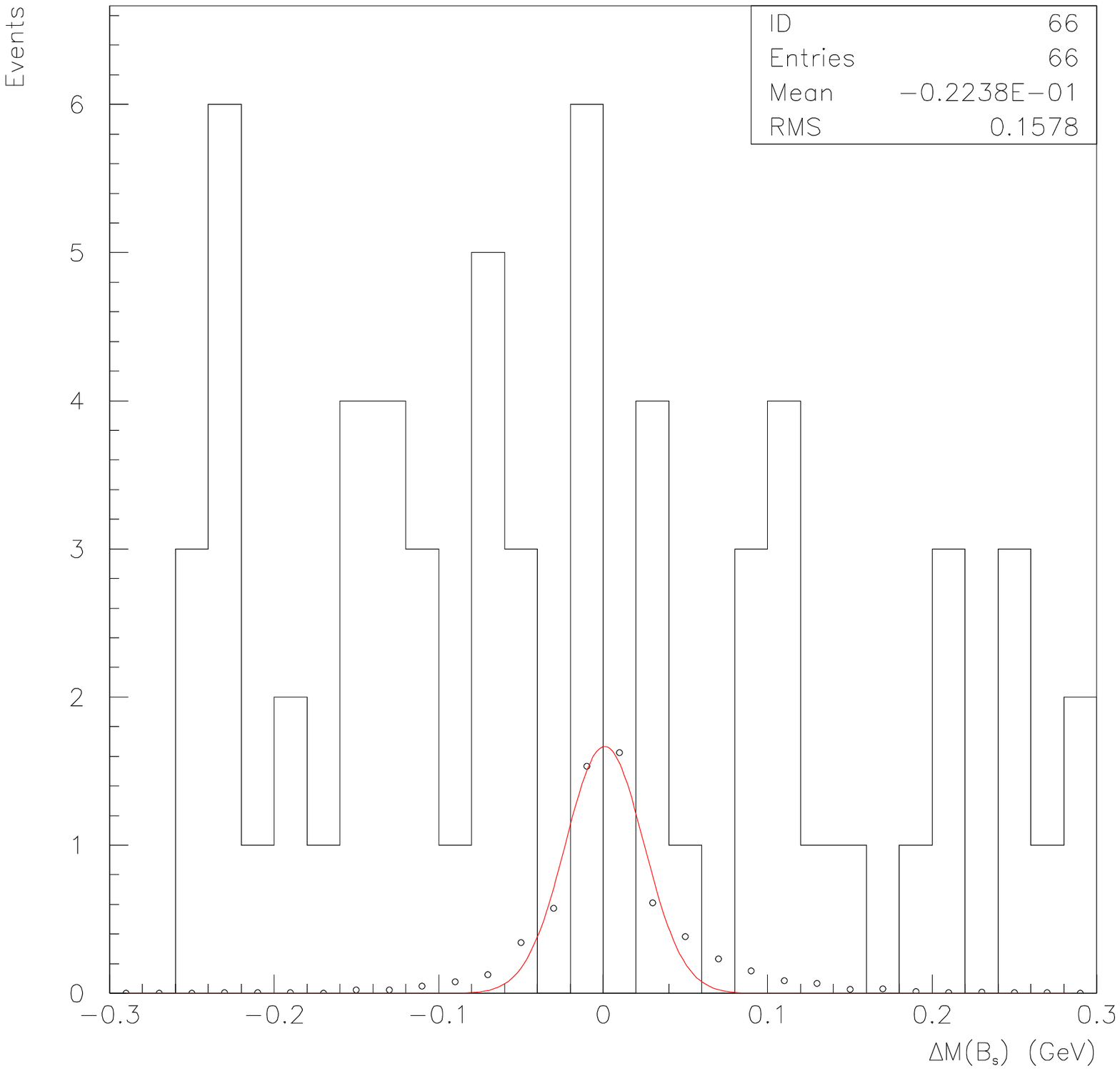}
\put(-145,175){\large\bf (b)}
\put(-405,175){\large\bf (a)}
}
\vspace*{0.3cm}
\caption
[$B_s^0 \rightarrow J/\psi\eta$ background study using CDF Run\,I $J/\psi$
data.]  
{$B_s^0 \rightarrow J/\psi\eta$ background study using CDF Run\,I $J/\psi$
data.  
(a) Two-dimensional
distribution of $m(\gamma\gamma)$ versus $m(B_s^0)$ before a cut around the  
$\eta$ invariant mass. 
(b) Background events from Run\,I $J/\psi$ data passing the $J\psi \eta$
selection. Overlaid is the Monte Carlo expectation scaled to the six events
expected. 
}
\label{ch6:cdf_psieta_3}
\end{figure}

To summarize this preliminary study, CDF expects to reconstruct a
signal of 
approximately 1000 $B_s^0\ra J/\psi\eta$ events in \tfb\ under Run\,II running 
conditions. A resolution of the $B_s^0$ signal of better than 40~\mevcc\
can be expected. Using Run\,I $J/\psi$ data to study the background, we
observed a combinatoric background at the level of six events per 40~\mevcc\
bin. Further background reduction using CES and CPR should be possible.

\boldmath
\subsection[$B_s^0 \ra J/\psi\eta^{(\prime)}$: BTeV Report]
{$B_s^0 \ra J/\psi\eta^{(\prime)}$: BTeV Report
$\!$\authorfootnote{Authors: G.~Majumder, S.~Stone.}
}


\unboldmath
\index{decay!$B_s^0 \ra J/\psi \eta^{(\prime)}$}%
\index{BTeV!$B_s^0 \ra J/\psi \eta^{(\prime)}$ prospects}%
\index{BTeV!$\sin 2\beta_s$ prospects}%

The $CP$~violating angle, $\beta_s$, defined in Section~\ref{ch6:intro}, can be
measured by using $B_s^0$ decay modes.  The all-charged mode 
$B_s^0\,\rightarrow\,J/\psi \phi$ 
is one way to measure this, but due
to the fact that this is a vector-vector final state of mixed-$CP$,
a complicated angular analysis is required and therefore 
a very large data sample must be obtained.
 The channels $\psietp$ and $\psiea$, can be used to 
determine the angle $\beta_s$ from a simple asymmetry measurement.

We estimate the relevant branching ratios using the quark model. 
The $\eta$ and $\eta'$ wave functions are given in terms of the quark
wave functions as:
\begin{eqnarray}
\Psi(\eta) & = & (u\bar{u} +d\bar{d} -s\bar{s})/\sqrt{3},\\
\Psi(\eta') & = & (u\bar{u} +d\bar{d} +2s\bar{s})/\sqrt{6}.
\end{eqnarray}
Thus the branching ratios are related to the measured decay $B^0\to J/\psi
K^0$, taking equal $B$~lifetimes as
\begin{eqnarray}
{\cal{B}}(B_s^0\to J/\psi \eta) & = & {1\over 3} {\cal{B}}(B^0\to J/\psi K^0),\\
{\cal{B}}(B_s^0\to J/\psi \eta') & = & {2\over 3} {\cal{B}}(B^0\to J/\psi K^0).
\end{eqnarray}
It should be noted that a large enhancement in one of these rates is possible,
as implied by the large branching fraction for $B\to \eta' K$. 
 
 We consider only the decays $\eta\to\gamma\gamma$, $\eta'\to\rho^0\gamma$
 and $\eta'\to\pi^+\pi^-\eta$. The
 $J/\psi$ will be reconstructed in the $\mumu$ decay mode.  All input
 branching ratios used for this study are listed 
 in~Table~\ref{tab:epbrs}.

\begin{table}[tb]
\begin{center}
\begin{tabular}{|l|c|} \hline
Decay & Branching Fraction \\\hline
${\cal{B}}(B_s^0\to J/\psi \eta)^{\dagger}$ & 3.3$\times 10^{-4}$\\
${\cal{B}}(B_s^0\to J/\psi \eta')^{\dagger}$ & 6.7$\times 10^{-4}$ \\
$J/\psi\to \mu^+\mu^-$  & 0.059\\
$\eta\to\gamma\gamma$ & 0.392 \\
$\eta'\to\rho\gamma$ & 0.308 \\
$\eta'\to\pi^+\pi^-\eta$ & 0.438\\\hline
\end{tabular}
\vspace*{0.3cm}
\caption
[Input branching fractions for $B_s^0\ra J/\psi\eta^{(\prime)}$ used for BTeV
study.] 
{Input branching fractions for $B_s^0\ra J/\psi\eta^{(\prime)}$ used for 
the BTeV study. Note,
$^{\dagger}$ indicates estimated branching fractions.}
\label{tab:epbrs}
\end{center}
\end{table}

\subsubsection{Signal Selection}  

We now discuss selection requirements for $B_s^0\ra J/\psi\eta^{(\prime)}$
 signal events.  
 First of all, the signal channels contain photons. They are
selected as isolated energy depositions in the PbWO$_4$ calorimeter that are at least 
7 cm away from any track intersection and satisfy the following criteria:
 $E_{\gamma}$  $>$  0.5 GeV,
 E9/E25        $>$  0.95,
 and the second moment mass is required to be less than  100~\mevcc.

%
We now list the criteria for the individual particles.
\begin{description}
\item [$J/\psi\to\mumu$]~
 \begin{itemize} 
  \item Both muons should have hits in the rear end of the RICH and at 
        least one must be identified in the muon system.
  \item $p_T$ of each muon $>$ 0.2~\gevc\ and at least one with
        $p_T>$ 1.0~\gevc.
  \item $\chi^2$ of common vertex of both muons $<$ 4.
  \item Invariant mass within 100~\mevcc\ of the $J/\psi$ mass.
 \end{itemize}

\item [$\eta\to\gamma\gamma$]~ 

\begin{itemize}
 \item Each photon has $E_\gamma~>$ 4~GeV and $p_T>$ 0.4~\gevc.
 \item Invariant mass of two-photon combinations must be within 15~\mevcc\ of
       the $\eta$ mass.
\end{itemize}

\item [$\eta^\prime\to\rho^0\gamma$]~
 
 \begin{itemize}
  \item Two oppositely charged tracks, each with momenta
        greater than 1~\gevc\ are taken as $\pi^+\pi^-$ candidates.
  \item The $\pi^+\pi^-$ invariant mass must be within 0.55~\gevcc\ of the $\rho$ mass.
  \item The $\pi^+\pi^-$ must form a common secondary vertex with the
   $\mumu$ from the $J/\psi$ with a fit $\chi^2<$10.
  \item Addition of a single photon ($p_T>$ 0.3~\gevc) to these tracks
       produces an invariant mass within 15~\mevcc\ of the ${\eta^\prime}$ mass.
\end{itemize}

\item [$\eta^\prime\to\pi^+\pi^-\eta$]~ 

 \begin{itemize}
  \item The same selection criteria as for $\eta$ defined above,
         except that for each photon $p_T>$ 0.2~\gevc\ is required.
  \item Two oppositely charged tracks, each with momenta
        greater than 1~\gevc\ are taken as $\pi^+\pi^-$ candidates.
  \item The $\pi^+\pi^-$ must form a common secondary vertex with the
   $\mumu$ from the $J/\psi$ with a fit $\chi^2<$10.
  \item The $\eta$ and the $\pi^+\pi^-$ have
       an invariant mass within 15~\mevcc\ of the ${\eta^\prime}$ mass.
\end{itemize}

\end{description}

  Signal events are also 
required to satisfy the following  general criteria.
%
A good primary vertex must exist.
The distance between the primary and secondary vertices must be 
$L >$ 50~$\mu$m  
       for $\eta^\prime$ and $>$~100~$\mu$m for $\eta$.
We require $L/\sigma_L>$3.
The normalized distance of closest approach with respect to the primary vertex
   (DCA/$\sigma_{DCA}$) of each charged track must be greater than 3.
No additional track is consistent with the $B_s^0$ vertex.
The opening angle between the `$B$'-direction and the particle direction
is required to be $<10$\,mrad and $<15$\,mrad for $J/\psi\eta^\prime$
       and $J/\psi\eta$, respectively. Here the `$B$'-direction is defined
by the vector joining the primary and secondary vertices and the particle
direction is  
       defined as the vector sum of the momenta of all measured particles.
The invariant mass of $J/\psi\,\eta$ or $J/\psi\,\eta^\prime$ have to be within
       $\pm 40~\mevcc$ of the $B_s^0$ mass ($\sigma_{M_B} = 19~\mevcc$).

\begin{figure}[tbp]
\begin{center}
\epsfig{figure=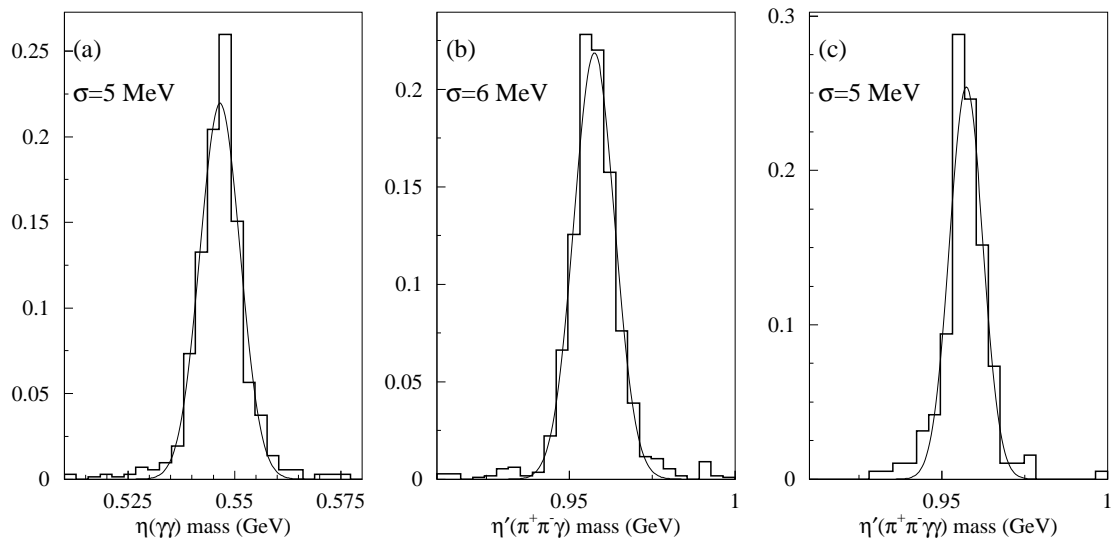, width=6in}
\caption
[The invariant mass distributions for $\eta\to\gamma\gamma$,
(b) $\eta'\to\pi^+\pi^-\gamma$, and $\eta'\to\pi^+\pi^-\eta$ with 
$\eta\to\gamma\gamma$ at BTeV.]
{The invariant mass distributions for (a) $\eta\to\gamma\gamma$,
(b) $\eta'\to\pi^+\pi^-\gamma$, and $\eta'\to\pi^+\pi^-\eta$ with
$\eta\to\gamma\gamma$ at BTeV. The Gaussian mass resolutions are indicated.}
\label{etamass}
\end{center}
\end{figure}

We show in Fig.~\ref{etamass} the invariant mass distributions of signal candidates
for $\gamma\gamma$, $\rho^0\gamma$ and $\pi^+\pi^-\eta$.
The $\mu^+\mu^-$  mass distribution from $J/\psi$ decays
is shown in Fig.~\ref{psibmass}(a).  We can 
improve the $B$~mass distributions by constraining the dimuons 
 to be at the nominal $J/\psi$ mass. This greatly improves the
four-vector of the reconstructed $J/\psi$. After applying this constrained
fit we find the $B_s^0$ mass distributions shown in Fig.~\ref{psibmass}(b). Note, that we could
also constrain the $\eta$ and $\eta'$ masses to their nominal values using the same
fitting technique. This will be done for future analyses.

\begin{figure}[tbp]
\begin{center}
\epsfig{figure=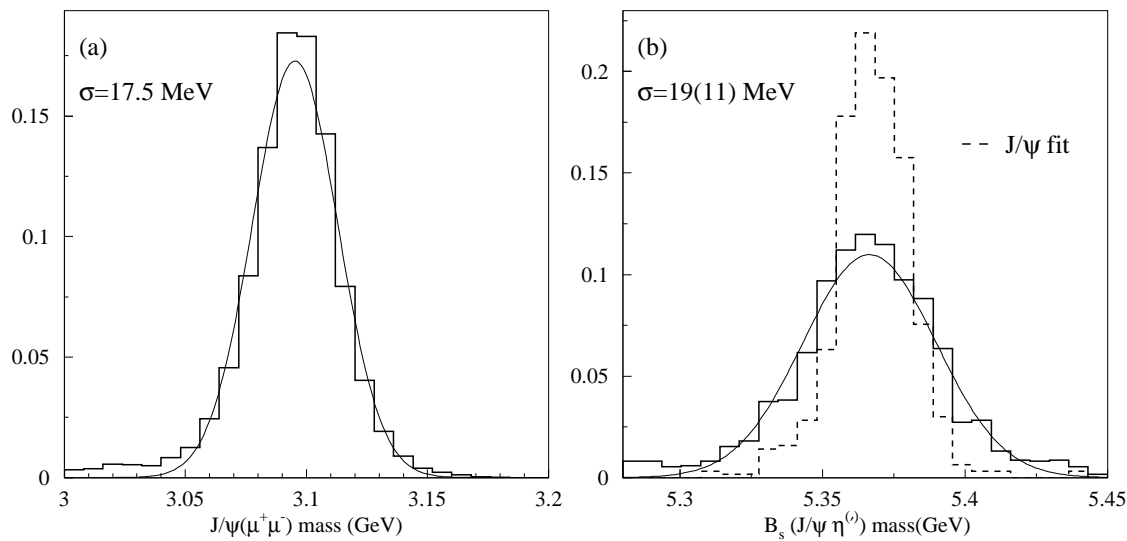, width=6in}
\caption
[The dimuon invariant mass and reconstructed $B_s^0$ mass
for all three final states of $\eta$ and $\eta'$ summed together.]
{(a) The dimuon invariant mass. (b) The reconstructed $B_s^0$ mass
for all three final states of $\eta$ and $\eta'$ summed together.
The solid curve is without constraining the $\mu^+\mu^-$ to the
$J/\psi$ mass, while the dashed curve is with this constraint.
The $B_s^0$ mass resolution improves from 19 to 11~\mevcc.}
\label{psibmass}
\end{center}
\end{figure}

\subsubsection{Background Estimation}

 The dominant background to these decay modes is from
$b(\bar{b})\rightarrow J/\psi\,X$. 
  To calculate reconstruction efficiencies of signals and 
of background, Monte Carlo events were generated using Pythia and
QQ to decay the heavy particles. Only events with real 
$J/\psi\to \mu^+\mu^-$ decays were kept for further analysis.
The events were traced through the BTeV detector simulation
using the GEANT simulation package. We add to the $b\bar{b}$
background events another set of light quark background distributed
with a mean Poisson multiplicity of two.
   Distributions of several variables for both signal and 
background are compared in Fig.~\ref{psimatch}.

\begin{figure}[tbp]
\begin{center}
\epsfig{figure=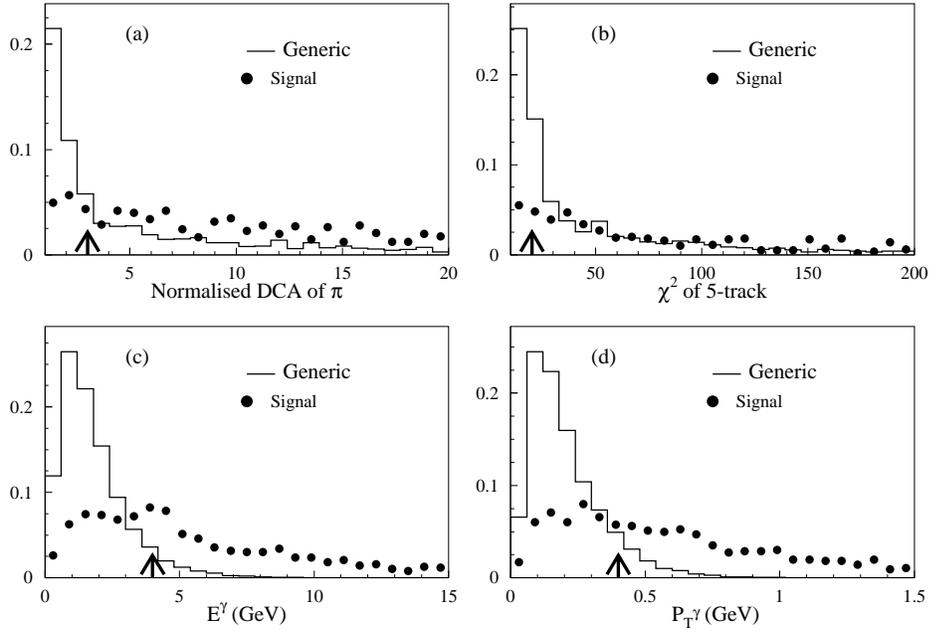, width=5.0in}
\end{center}
\caption
[Distributions of backgrounds in several variables compared
with the signal for $\eta'\to\rho\gamma$.]
{Distributions of backgrounds in several variables compared
with the signal. For $\eta'\to\rho\gamma$ (a) DCA$/\sigma_{DCA}$, (b)
$\chi^2$ of adding an additional track to the $J/\psi \pi^+ \pi^-$ vertex.
For $\eta\to\gamma\gamma$ 
(c) energy of the photons and (d) the transverse momentum of the photons
with respect to the beam direction. The arrows show the position of the
selection requirements.}
\label{psimatch}
\end{figure}

The results discussed below are based on $\sim$ 4,500 detector simulated
signal events (each 
channel), which  were preselected in generator level using the criteria
that all particles of these signals are within the geometrical acceptance 
region of the detector. Similarly, 40,000 background events are also 
preselected from 5.8 million generic $b\bar{b}$ events. To determine backgrounds we only
looked at the dimuon channels, and the $\gamma\gamma$ decay of the $\eta$
and the $\rho^0\gamma$ decay of the $\eta'$.

  After all selection criteria, one event survived in each of the
$J/\psi\,\eta$ and $J/\psi\,\eta^\prime$ channels within a wide $B_s^0$ 
mass window of 400~\mevcc\ (signal mass window is 44~\mevcc). This leads to a 
signal-to-background expectation for 
$J/\psi\,\eta$ and $J/\psi\,\eta^\prime$, of 15:1 and 30:1.
It is not surprising  that the backgrounds are so low. We therefore feel
confident that we can add the $\eta'\to \pi^+\pi^-\eta$
modes in without adding significant background.

\boldmath
\subsubsection{Sensitivity to $\sin2\beta_s$}
\unboldmath
\index{BTeV!$\sin 2\beta_s$ prospects}%

 The expected yield of signal events and the resulting asymmetry measurement are given
in Table~\ref{tab:psietpexp}. The trigger
efficiency consists of Level\,1 efficiencies from the detached
vertex trigger, the dimuon trigger and the Level\,2 trigger.
 
\begin{table}[btp]
\begin{center}
\begin{tabular}{|l|c|c|c|} \hline
Luminosity            & 
\multicolumn{3}{c|}{$2\,\times\,10^{32}\,{\rm cm}^{-2}\,{\rm s}^{-1}$} \\
Running time          & \multicolumn{3}{c|}{10$^7$ sec} \\
Integrated Luminosity & \multicolumn{3}{c|}{2000 pb$^{-1}$} \\
$\sigma_{b\bar{b}}$   & \multicolumn{3}{c|}{100 $\mu$b} \\
Number of $b\bar{b}$ events       & \multicolumn{3}{c|}{$2\,\times\,10^{11}$} \\
Number of $B_s^0$ events               & 
\multicolumn{3}{c|}{$0.5\,\times\,10^{11}$} \\ \hline
    &  \multicolumn{2}{c|}{$\psietp$}  & $\psiea $ \\\cline{2-4}
    & $\eta'\to\rho^0\gamma$ & $\eta'\to\pi^+\pi^-\eta$ &
    $\eta\to\gamma\gamma$\\\cline{2-4}

Reconstruction efficiency [\%]       & 1.2 &0.60 & 0.71  \\
$S/B$                                  &  30:1 & -  & 15:1 \\
Level\,1 Trigger efficiency     [\%]          & 85 & 85 & 75  \\ 
Level\,2 Trigger efficiency     [\%]          & 90 & 90 & 90  \\ 
Number of reconstructed signal events  & 5670 & 1610 & 1920  \\\hline
Tagging efficiency $\eD$ & \multicolumn{3}{c|}{0.1} \\
Total Number tagged   &\multicolumn{3}{c|}{994}\\
$\sigma(\sin2\beta_s)$                        
& \multicolumn{3}{c|}{0.033} \\ \hline

\end{tabular}
\vspace*{0.3cm}
\caption
[Projected yield of $B_s^0\ra J/\psi\eta^{(\prime)}$ and 
         uncertainty on $\sin2\beta_s$ at BTeV.]
{Projected yield of $B_s^0\ra J/\psi\eta^{(\prime)}$ and 
         uncertainty on $\sin2\beta_s$ at BTeV.}
\label{tab:psietpexp}
\end{center}
\end{table}

The accuracy on $\sin2\beta_s$ is not precise enough to measure the Standard
Model predicted value, which is comparable to the error, in $10^7$ seconds of
running. 
The low background level makes it
possible to loosen the cuts and gain acceptance. We could also add in
the $J/\psi\to e^+e^-$ decay mode. This will not be as efficient as $\mu^+\mu^-$
due to radiation of the electrons, but will be useful. We also believe that ways can be found to
improve flavour tagging efficiency, especially for $B_s^0$. Furthermore, we will
have many years of running, and we can expect some improvement from
the use of $B_s^0\to J/\psi \phi$.

\boldmath
\subsection[$B_s^0\ra J/\psi\eta^{(\prime)}$: Summary]
{$B_s^0\ra J/\psi\eta^{(\prime)}$: Summary
$\!$\authorfootnote{Author: M.~Paulini.}
}
\unboldmath
\index{decay!$B_s^0 \ra J/\psi \eta^{(\prime)}$}%

A measurement of the $CP$~asymmetry in 
$B_s^0\ra J/\psi\eta^{(\prime)}$
will determine the value of the CKM phase $\beta_s$.
The asymmetry is expected to be small within the Standard Model, of the
order of a few percent. This means that an observation of an asymmetry that
is significantly 
larger than ${\cal O}(\lambda^2)$, will provide an unambiguous signal
for new physics which  is likely to be related to new
$CP$~violating contributions to $B_s^0$-$\bar B_s^0$ mixing.

Although the CDF detector is not ideally suited for the
detection of low energy photons, it is not impossible 
to reconstruct neutral mesons such as $\pi^0$ or $\eta$ decaying into
two photons from energy depositions in CDF's electromagnetic
calorimeter. Although a measurement of the $CP$~violating angle $\beta_s$
will probably be best approached using the $B_s^0$ decay mode into
$J/\psi\phi$, a preliminary study 
for the event yield of $B_s^0\ra J/\psi\eta^{(\prime)}$ has been performed.
To estimate the expected signal of $B_s^0\ra J/\psi\eta$, CDF
normalized this decay mode to the $B^+ \rightarrow J/\psi K^+$ channel. 
As discussed in Sec.~\ref{sec:cdfpsieta}, CDF
estimates to observe about 1000 $B_s^0\ra J/\psi\eta$ decays in \tfb\ in
Run\,II. To demonstrate the feasibility of observing neutral particles such as
$\pi^0\ra\gamma\gamma$ or $\eta\ra\gamma\gamma$ with the CDF calorimeter,
the reconstruction of low energy photons using Run\,I data has been
investigated (see Sec.~\ref{sec:cdfpsieta}). 
From this preliminary study, a resolution on the $B_s^0$ signal of better
than 40~\mevcc\ can be expected.
To estimate the expected background rate for $B_s^0\ra J/\psi\eta$ in Run\,II,
$J/\psi\ra\mu\mu$ data from Run\,I were used.
A combinatoric background at the level of 6 events per 40~\mevcc\
bin were observed, while further background reduction using CES and CPR
should be possible. 

Photons are reconstructed as isolated energy depositions in BTeV's
fine segmented PbWO$_4$ calorimeter. 
For the signal selection of $B_s^0\ra\psi\eta^{(\prime)}$, BTeV considered
the decays $\eta\to\gamma\gamma$, $\eta'\to\rho^0\gamma$ 
and $\eta'\to\pi^+\pi^-\eta$.
From these decays modes, BTeV expects to reconstruct almost 10,000
$B_s^0$~signal events with a mass resolution of about 20~\mevcc.
For the resulting asymmetry measurement an uncertainty 
$\sigma(\sin2\beta_s)$ of about 0.03 is expected from this signal yield. 
Although this accuracy is not precise enough to measure
the value of $\sin2\beta_s$ predicted by 
the Standard Model, which is comparable to the error, this is an
encouraging result. It 
gives optimism to probe physics beyond the Standard Model with 
$B_s^0\ra J/\psi\eta^{(\prime)}$ within a few years 
of running at design luminosity at BTeV.

\begin{boldmath}
\section[$CP$ Violation: Summary]
{$CP$ Violation: Summary
$\!$\authorfootnote{Author: M.~Paulini.}
}
\end{boldmath}

Since the time the Workshop on $B$~Physics at the Tevatron was held in
September 1999 and February 2000 and the time this write-up is coming to a
completion, a significant amount of time has elapsed. It therefore
constitutes a non-trivial task to report the findings of the workshop but
to also include actual updates that the heavy flavour physics community has
witnessed. An incredibly successful turn-on of both $B$~factories together
with an exceptional performance of both their detectors, BaBar and Belle,
has already produced a wealth of new measurements including 
the first observation of $CP$~violation in the $B^0$~meson
system~\cite{cpbbstbaug}. 
A compilation of our current
knowledge on the value of \stb\ is shown in Figure~\ref{fig:cpstb}. 
The individual measurements are listed in Refs.~\cite{cpbbstbaug,cpstb}
while the quoted 
average is taken from Ref.~\cite{olsen}. 
Clearly, the recent measurements of \stb\ from BaBar and Belle establish
$CP$~violation in $B^0$~decays while the results from OPAL, CDF
and ALEPH~\cite{cpstb}  were still compatible with \stb\ being zero.

\begin{figure}[btp]
\centerline{
\epsfxsize=3in
\epsffile{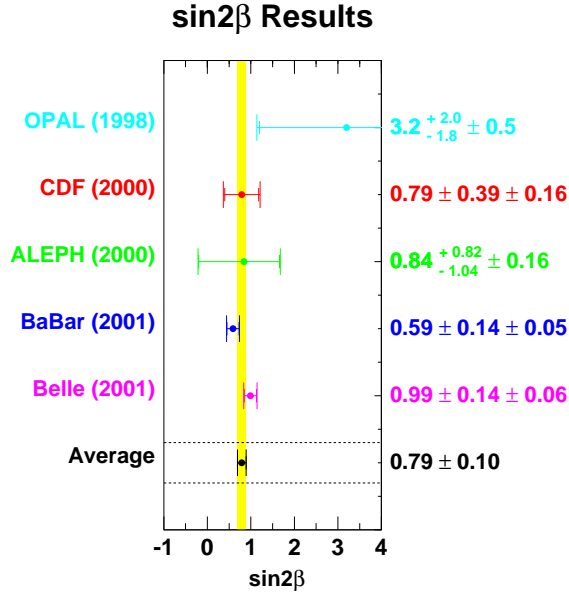}
}
\vspace*{0.3cm}
\caption
[Compilation of measurements of \stb\ as of August 2001.] 
{Compilation of measurements of \stb\ as of August
2001~\cite{cpbbstbaug,cpstb}.  
The displayed average on \stb\ is taken from Ref.~\cite{olsen}.
}
\label{fig:cpstb}
\end{figure}

With the official start of Run\,II in March 2001, the Tevatron aims to
turn the findings of 
the $B$~Physics at the Tevatron
Workshop into real measurements of $CP$~violation in the
$B$~meson system confirming the exciting results on $\sin 2\beta$ from the
$B$~factories. 

Evaluating the sensitivity of 
measuring \stb\ was motivated by using 
\index{decay!$B^0 \to J/\psi K_S^0$}%
$B^0\ra J/\psi K^0_S$ as a benchmark
process and as a comparison to the expectations (and presented measurements)
of the $B$~factories.
With \tfb\ of integrated luminosity, CDF expects to measure \stb\ with
a precision of
$\sigma(\stb)\sim\!0.05$.
The D\O\ experiment estimates to obtain a similar precision on \stb\
quoting $\sigma(\stb)\sim\!0.04$.
While $\sin\,2\beta\,$ will have been measured to a fair accuracy before the
BTeV experiment will turn on, the goal of the BTeV 
collaboration is to significantly improve the precision of that
measurement. 
Within one year of running at design luminosity, BTeV 
expects to measure \stb\ with an error of 
$\sigma(\stb)\sim\!0.025$. 

Considering the status of the CKM unitarity triangle in a couple of years
from now, the angle $\beta$ is measured from $B^0 \rightarrow J/\psi
K^0_S$ decays by the $B$~factories now,
assisted by complimentary measurements at CDF and D\O\ in the near future.  
In addition, we will have more information about the leg of the unitarity
triangle opposite the angle $\beta$: $V_{ub}/V_{cb}$ will be measured more
precisely by the observation of higher statistics $b \rightarrow u$
transitions at CLEO and the $B$~factories. However, the ultimate precision
on determining $V_{ub}$ from data will probably be limited by theoretical
uncertainties. 
The information that will finally allow to over-constrain the CKM triangle, is
the observation of $B_s^0 \bar B_s^0$ oscillations anticipated at CDF if the
oscillation parameter $\Delta m_s$ is less than 40~ps$^{-1}$.
The question might then be, what will be the next ``precision CKM
measurement'' after $\stb$ and $\Delta m_s$? 

Several years ago, 
the decay 
\index{decay!$B^0 \to \pi \pi$}%
$B^0\to\pi^+\pi^-$ appeared in the literature as 
a tool to determine $\alpha=180^\circ-\beta-\gamma$ as the second CKM angle
to be measured after $\beta$ had been determined.
As we now know, the so-called "penguin pollution" in \bpipi\ is sufficiently
large and introduces a significant theoretical uncertainty in the
extraction of fundamental physics parameters  
from the measured $CP$~asymmetry in this channel.
BTeV studied a method to measure the CKM
phase $\alpha = \pi - \beta - \gamma$ using the decays
\index{decay!$B^0 \to \rho \pi$}%
$B^0 \rightarrow 
\{ \rho^+ \pi^-, \rho^0 \pi^0, \rho^- \pi^+ \}
\rightarrow
\pi^+ \pi^- \pi^0$ as proposed by Snyder and Quinn \cite{qusny}. 
From this study BTeV 
expects to reconstruct 
about 9,400 $\rho^{\pm}\pi^{\mp}$ events
and 1,350 $\rho^0\pi^0$ events per year with reasonable
signal-to-background levels. 
CDF evaluated for this workshop a strategy of measuring the 
CKM angle~$\gamma$ using $\btopipi$ and $\bstokk$
as suggested by Fleischer in Ref.~\cite{flekk}.
This method is
particularly well matched to the capabilities of the Tevatron as 
it relates $CP$~violating 
observables in $B^0$ and $B_s^0$ decays.
The studies performed
during this workshop indicate that a measurement of the CKM angle
$\gamma$ to better than $10^{\circ}$ could be feasible at CDF with \tfb\ of
data. 
The utility of these modes depends on how well the uncertainty
from flavour $SU(3)$ breaking can be controlled.
Data for these and other processes should be able to tell us the range of such
effects.
A~study by CDF shows that 20\% effects from $SU(3)$ breaking
lead to an uncertainty of only $\sim 3^\circ$ on $\gamma$.
Of course, BTeV will also be able to exploit this method. 
Based on one year of running, 
BTeV expects to reconstruct about 20,000~$\btopipi$ events with small
background contamination  from 
\index{decay!$B^0 \to K \pi$}%
\index{decay!$B_s^0 \to K \pi$}%
$\btokpi$, $\bstopik$ and $\bstokk$ and estimates 
an uncertainty on the $CP$~asymmetry 
${\cal A}_{CP}^{\pi\pi}$ of 0.024.

Another well suited method of determining the unitarity triangle angle $\gamma$
has been studied by measuring $CP$~violation 
in the decay mode 
\index{decay!$B_s^0 \to D_s^- K^+$}%
$B_s^0 \to D_s^- K^+$. This will allow a clean
measurement of $\gamma + 2\beta_s$ in a tree-level process. 
Four time-dependent asymmetries need to be measured in the presence of
large physics backgrounds, in particular from the Cabibbo allowed process
$B_s^0 \to D_s^- \pi^+$.
An initial measurement
of $\gamma$ should be possible at CDF in Run\,II.
Within the first \tfb\ of data, the expected error on
$\sin(\gamma\pm\delta)$ is
0.4 to 0.7 depending on the assumed background levels.
By the end of Run\,II an uncertainty on $\gamma$ near $0.1$ may be achievable.
Since the BTeV detector will have excellent $\pi$-$K$ separation provided
by a RICH detector, physics backgrounds will play a minor role and a
$B_s^0 \to D_s^- K^+$ signal of
about 9200 reconstructed events can be collected per year. This will allow
a determination of the angle $\gamma$ to better than $10^{\circ}$.

A similar conclusion can be drawn for the CDF and BTeV prospects of 
measuring the angle $\gamma$ with charged $B$~decays using 
\index{decay!$B^0 \to D^0 K^-$}%
$B^- \ra D^0 K^-$. 
CDF expects to collect a small sample of $D^0 K^-$ 
candidates with the two-track hadronic trigger in \tfb\ in Run\,II
while BTeV will reconstruct about 400 $B^{-} \ra [K\pi] K^{-}$ events
per year at design luminosity.
With this number of events, BTeV can 
measure $\gamma$ with an uncertainty of about $\pm 10^{\circ}$ for most of
the assumed parameter space.  
There is optimism at CDF that the physics
background can be brought down to the same level as the signal, but there
could be considerable combinatoric background which is difficult to
evaluate without Run\,II collision data.
Comparing both decay channels, 
$\Bs \ra D_s^- K^+$ and
$B^-\ra D^0 K^-$, considered for extracting the angle $\gamma$, 
the $\Bs$ decay mode offers better prospects of determining $\gamma$
from the four time-dependent asymmetries.

Looking for physics beyond the Standard Model, 
measuring the $CP$~asymmetry in 
\index{decay!$B_s^0 \ra J/\psi \eta^{(\prime)}$}%
$B_s^0\ra J/\psi\eta^{(\prime)}$
has been evaluated. This decay mode 
will determine the value of the CKM phase $\beta_s$ but
the asymmetry is expected to be small within the Standard Model.
Although the CDF detector is not ideally suited for the
detection of low energy photons, CDF estimates to observe about 1000
$B_s^0\ra J/\psi\eta$ decays with a resolution on the $B_s^0$ signal of better
than 40~\mevcc\ in \tfb\ in Run\,II.
However, BTeV will probably be the experiment to probe 
the $CP$~asymmetry in $B_s^0\ra J/\psi\eta^{(\prime)}$,
achieving an uncertainty 
$\sigma(\sin2\beta_s)$ of about 0.03 in one year of data taking. 
This precision approaches the level of
the value of $\sin2\beta_s$ predicted by the Standard Model.

Even after the discovery of $CP$~violation in the $B$~system by BaBar and Belle~\cite{cpbbstbaug},
$CP$~violation is still one of the least tested aspects of the
Standard Model.
It is clear that Run\,II at the Tevatron will offer many important
$CP$~violation measurements which will be complementary to the results that
we expect from the $e^+ e^-\ B$~factories. 
After $CP$~violation had been observed only in the neutral $K$~meson
system for 37 years, 
the discovery of $CP$~violation in the neutral $B$~meson system
has been made at the $B$~factories awaiting confirmation at 
the Tevatron. The next few years will 
provide further tests of the Standard Model picture of $CP$~violation and
will hopefully unveil the holy grail of heavy flavour physics in its entire
beauty.

\clearpage{\pagestyle{empty}\cleardoublepage}
\def\chsevenfigs{ch7/wg2figs/}
\chapter{\protect\boldmath Rare and Semileptonic Decays}

\def\be{\begin{equation}}
\def\ee{\end{equation}}
\newcommand{\bm}[1]{\mbox{\boldmath ${#1}$}}
\def\GeV{{\rm GeV}}
\def\MeV{{\rm MeV}}
\def\ltap{\mathrel{\mathpalette\vereq<}}
\def\vereq#1#2{\lower3pt\vbox{\baselineskip1pt\lineskip1pt
     \ialign{\\$#1\hfill##\hfil\\$\crcr#2\crcr\sim\crcr}}}
\def\Do{\mbox{D0}}
\def\bentarrow{\relax\ifmmode{
     \raise 1.13ex\hbox{$\lfloor$}\kern-.4em\rightarrow\kern-.2em}
  \else{
     $\raise 1.13ex\hbox{$\lflambdaloor$}\kern-.4em\rightarrow\kern-.2em$}\fi}
\def\fm{{\rm fm}}

\authors{Christian~W.~Bauer, Gustavo~Burdman, Aida~X.~El-Khadra, 
JoAnne~Hewett, Gudrun~Hiller, Michael~Kirk, Jonathan~Lewis, 
Heather~E.~Logan, Michael~Luke, Ron~Poling, Alex~Smith, 
Ben~Speakman, Kevin~Stenson, Masa~Tanaka, Andrzej~Zieminski}

\section{Rare Decays: Theory}
\subsection{Preliminaries}
\label{sec:motivation}

\index{FCNC transitions}
The Flavor-changing-neutral-current (FCNC) transitions, such as $b\to s$ 
and $b\to d$, arise only at the loop level in the Standard Model (SM).  
These decays provide tests of the detailed structure of the theory at 
the level of radiative corrections where Glashow-Iliopoulos-Maiani (GIM) 
cancellations are important, and are sensitive to CKM matrix elements:  
the flavor structure of a generic $b \to s$ amplitude 
$T$ is $T=\sum_i \lambda_i T_i$, where $\lambda_i=V_{ib} V_{is}^\ast$
and the sum runs over all up-quark flavors $i=u,c,t$. Using CKM unitarity
$\sum_i \lambda_i=0$ and $\lambda_u \ll \lambda_t$ we obtain $T=\lambda_t
(T_t-T_c)$.

Furthermore, in many extensions of the Standard Model, loop graphs with 
new particles (such as charged Higgses or supersymmetric partners) 
contribute at the same order as the SM contribution.  Precision 
measurements of these rare processes therefore  provides a complementary 
probe of new physics to that of direct collider searches.  
Finally, these rare decays are subject to both perturbative and 
non-perturbative QCD effects, which can be studied here.

The most interesting FCNC $B$ decays at the Tevatron are $B\to 
X_s\gamma$, $B\to X_s\ell^+\ell^-$, $B_{s,d}\to \ell^+\ell^-$, 
and the corresponding exclusive modes for the first two.  
(A fourth decay, $B\to X_{s,d}\nu\bar\nu$, is theoretically cleaner, 
but because of the neutrinos in the final state is not likely to be 
accessible at a hadron collider.)  
Of these decays, the exclusive modes $B\to K^{(*)}\ell^+\ell^-$ are likely 
to be the most important at the Tevatron in the near future:  
inclusive $B\to X_s\gamma$ is difficult to measure at a hadron collider, 
while the SM branching fraction for $B_s\to\mu^+\mu^-$ is at the 
$10^{-9}$ level.  
Furthermore, as we shall discuss, the theoretical prediction for inclusive 
$B\to X_s\ell^+\ell^-$ is poorly behaved in the large $q^2$ region, 
where it is easiest to measure. In this section, we focus on tests of 
the SM via these decays.
%
%
%

\subsubsection{The effective Hamiltonian} \label{sec:effham}

\index{effective Hamiltonian}
Radiative corrections to the FCNC decay amplitudes contain 
terms of
order $\alpha_s\ln{m_W^2/m_b^2}$, which are enhanced by the large 
logarithm of
$m_W/m_b$ and make perturbation theory poorly behaved.  To make precision
calculations, these terms must be summed to all orders.  This is most
conveniently performed using an effective field theory and the 
renormalization
group, as discussed in Chapter 1.\index{effective Hamiltonian!renormalization group}

The effective field theory for $b\to s$ transitions is thoroughly 
summarized in
a review article by Buchalla \etal \cite{burasrev}.  Here we briefly 
outline the
general features which are universal for the channels discussed in this 
chapter.
The effective Hamiltonian is obtained by integrating out heavy degrees of
freedom (the top quark and $W^\pm$ bosons in the SM) from the full
theory \cite{heff}:
\begin{equation}
 {\cal{H}}_{\rm eff} = -4 \frac{G_F}{\sqrt{2}}  V_{t s}^\ast V_{tb}
                       \sum_{i=1}^{10} C_i (\mu) O_i(\mu)
         \label{eq:heffequation}
\end{equation}
where $\mu$ is the renormalization scale, and the operators $O_i$ are
\begin{equation}\label{eq:operatorbasis}\index{effective Hamiltonian!
Wilson coefficients defined}
\begin{array}{*{2}{l}}
   O_1 = (\bar{s}_{L \alpha} \gamma_\mu b_{L \alpha})
   (\bar{c}_{L \beta} \gamma^\mu c_{L \beta}),
   &O_6 = (\bar{s}_{L \alpha}
   \gamma_\mu b_{L \beta})
   \sum_{q=u,d,s,c,b}
   (\bar{q}_{R \beta} \gamma^\mu q_{R \alpha}),   \\
   O_2 = (\bar{s}_{L \alpha} \gamma_\mu b_{L \beta})
   (\bar{c}_{L \beta} \gamma^\mu c_{L \alpha}),
   &O_7 = \frac{e}{16 \pi^2}m_b
   \bar{s}_{L \alpha} \sigma_{\mu \nu}  b_{R \alpha}
   F^{\mu \nu},   \\
   O_3 = (\bar{s}_{L \alpha} \gamma_\mu b_{L \alpha})
   \sum_{q=u,d,s,c,b}
   (\bar{q}_{L \beta} \gamma^\mu q_{L \beta}),
   &O_8 = \frac{g}{16 \pi^2}m_b
   \bar{s}_{L \alpha} T_{\alpha \beta}^a \sigma_{\mu \nu}
   b_{R \beta} G^{a \mu \nu},   \\
   O_4 = (\bar{s}_{L \alpha} \gamma_\mu b_{L \beta})
   \sum_{q=u,d,s,c,b}
   (\bar{q}_{L \beta} \gamma^\mu q_{L \alpha}),
   &O_9 = \frac{e^2}{16 \pi^2}
   \bar{s}_{L \alpha} \gamma^{\mu}  b_{L \alpha}
   \bar{\ell} \gamma_{\mu} \ell,   \\
   O_5 = (\bar{s}_{L \alpha} \gamma_\mu b_{L \alpha})
   \sum_{q=u,d,s,c,b}
   (\bar{q}_{R \beta} \gamma^\mu q_{R \beta}),\qquad
   &O_{10} = \frac{e^2}{16 \pi^2} \bar{s}_{L \alpha} \gamma^{\mu}
   b_{L \alpha} \bar{\ell} \gamma_{\mu}\gamma_5 \ell
\end{array}
\end{equation}
(note that these operators are not the same as the $O_i$'s for the 
$|\Delta B|=1$ Hamiltonian discussed in Chapter 1).
The subscripts $L$ and $R$ denote left and right-handed components, and 
we have
neglected the strange quark mass $m_s \ll m_b$. The coefficient 
$C_i(m_W)$ are
systematically calculable in perturbation theory, and the 
renormalization group
equations are used to lower to renormalization scale to $\mu=m_b$. The
renormalization group scaling is a significant effect, enhancing (for 
example)
the \bsg\ rate by a factor of $\sim 2$. Details on the renormalization 
scale
dependence,  the renormalization group equations and analytical formulae 
can be
found in \cite{heff}.  The SM values at $\mu=4.8$ GeV of the $C_i$ at 
NLO are
given in Table~\ref{tab:wilson}.

\index{effective Hamiltonian!Wilson coefficients in SM}
\begin{table}
         \begin{center}
         \begin{tabular}{|c|c|c|c|c|c|c|c|c|c|}
         \hline
         \multicolumn{1}{|c|}{ $C_1$}       &
         \multicolumn{1}{|c|}{ $C_2$}       &
         \multicolumn{1}{|c|}{ $C_3$}       &
         \multicolumn{1}{|c|}{ $C_4$}       &
         \multicolumn{1}{|c|}{ $C_5$}       &
         \multicolumn{1}{|c|}{ $C_6$}       &
         \multicolumn{1}{|c|}{ $C_7^{\rm eff}$}       &
         \multicolumn{1}{|c|}{ $C_9$}       &
                 \multicolumn{1}{|c|}{$C_{10}$}     \\
         \hline
         $-0.25$ & $+1.11$ & $+0.01$ & $-0.03$ & $+0.01$ & $-0.03$ &
    $-0.31$ &   $+4.34$ &    $-4.67$       \\
         \hline
         \end{tabular}
         \end{center}
\caption{SM values of the Wilson coefficients at NLO
($C_7^{\rm eff} \equiv C_7 -C_5/3 -C_6$).}
\label{tab:wilson}
\end{table}

While $C_7^{\rm eff}$ measures the $bs\gamma$ coupling strength, an 
analogous
correspondence can be made for $C_{10}$: comparing the charge 
assignments of
lepton-Z-couplings $|(\bar{\ell} \ell Z|_V)/(\bar{\ell} \ell Z|_A)|= |1-4
\sin^2\Theta_W| \simeq 0.08 $ shows that the $Z$-penguin contribution to 
$C_9$
(V) is suppressed with respect to $C_{10}$ (A) and can be neglected as a 
first
approximation: $C_{10}$ probes the effective $\bar{s}_L  Z b_L $ vertex 
modulo
the box contribution \cite{GGG}.

Different FCNC decays are sensitive to different linear combinations of 
the $C_i$'s, and so each of the decays of interest provides independent 
information.
  The quark-level transition $b\to s\gamma$ is largely governed by $O_7$, 
while $b\to s\ell^+\ell^-$ receives dominant contributions from $O_7$, $O_9$ 
and $O_{10}$, and $B_s\to\ell^+\ell^-$ is primarily due to $O_{10}$.  As 
discussed in the next section, the current measurement of $\bsg$ is in excellent 
agreement with theory, but this is only sensitive to the magnitude of the photon 
penguin $C_7^{\rm eff}$.  In contrast, $b\to s\ell^+\ell^-$ is sensitive to the 
sign of this coefficient, as well as to $O_9$ and $O_{10}$.

\subsubsection{Inclusive vs.\ Exclusive Decays}

The Wilson coefficients in Eq.~(\ref{eq:heffequation}) can be measured 
in either exclusive or inclusive decays of $b$ flavored hadrons.
The theoretical tools used to study exclusive and inclusive decays are 
very different. Experimental measurements of exclusive and inclusive 
decays are also faced with different challenges. Hence, it is convenient
to consider them separately.

In inclusive decays one can avoid the theoretical difficulties 
associated with the physics of hadronization by using 
\index{quark-hadron duality}
quark-hadron
duality together with the operator product expansion (OPE)
\cite{ope}. Quark-hadron 
duality allows us to relate inclusive decays of $B$ hadrons into 
hadronic final states to decays into partons (see Section~5.3 of Chapter~1). 
Using an OPE it can be shown that the $B$ decay is given by the 
corresponding parton-level decay. There are perturbative and 
nonperturbative corrections which must be taken into account. The 
leading nonperturbative corrections to this expression scale like 
$(\lqcd/m_b)^2$, which is of order a few percent. 
There are some caveats, both in the application of the OPE and in the 
assumption of quark-hadron duality.

The size of the corrections in the OPE typically grow as the final state 
phase space is restricted. If the phase space is restricted to too small 
a region the OPE breaks down entirely. This is an important consideration 
when experimental cuts are taken into account. A familiar example is the 
endpoint region above the $b\to c$ kinematic limit of the charged lepton 
spectrum in semileptonic $b\to u$ decay, which is important for measuring 
$|V_{ub}|$.  In this region the standard OPE breaks down, and a class of
leading twist \index{twist expansion}
operators in the OPE must be resummed to all orders. As we 
will discuss in Section~(\ref{sec:bsllinclusive}), the OPE also breaks down 
in the high lepton $q^2$ region of $B\to X_s\ell^+\ell^-$, but in this region 
the twist expansion also break down.

The range of validity of quark-hadron duality and the size of the
corrections which violate it are unknown, at present. There 
are theoretical reasons to believe that these corrections are small
(and it has been suggested that duality violation is reflected in the 
asymptotic nature of the OPE) \cite{opeduality}. However, it has also been 
argued \cite{isg98} that duality violations are much larger than 
commonly expected. As the data improve and more inclusive quantities 
are measured, the comparison between theory and experiment will provide 
an indication of the size of duality violations.

While theoretically appealing, inclusive rare decays are very difficult 
to measure, particularly in a hadronic environment. It is likely 
that they will be constructed by measuring a series of exclusive decays.
Hence, it will be much easier to measure exclusive rare decays
at the Tevatron.

In theoretical studies of exclusive decays, we must deal with 
nonperturbative QCD corrections to the quark-level process, 
as manifest in hadronization effects, for example. 
Lattice QCD is the only first principles tool for calculations
of nonperturbative QCD effects. Unfortunately, results from
lattice QCD calculations are incomplete, at present. Furthermore, 
numerical simulations based on lattice QCD are time consuming and 
expensive. The prospects for lattice QCD calculations of rare
exclusive decays with small and controlled errors are
excellent, as discussed in detail in Section~\ref{sec:lattice}.
At present, however, we have to deal with hadronic uncertainties 
which result in a loss of sensitivity to the interesting short 
distance physics. 
It is therefore important to use a variety of theoretical strategies 
for calculations of these decays. We include model-independent
approaches based on approximate symmetries as well as calculations
which use a variety of different models in our discussion
of exclusive decays in Section~\ref{sec:exclusivedecays}.

\OMIT{
The only operator which has non vanishing tree-level matrix
elements for $b\to s\gamma$ decays is
the $\bar{s} \gamma b$ vertex $O_7$, while $b\to s\ell^+\ell^-$ also
receives contributions from the 4-Fermi operators $O_9$
and $O_{10}$.  The one-loop matrix elements of $O_1-O_8$ are the same
order as the tree-level matrix element of $O_7$ and so must also be
included at NLO.  A subtlety which arises is that these matrix elements
are renormalization scheme dependent, depending on the scheme chosen to
define $\gamma_5$ in $d$ dimensions.
It is convenient to deal with this by
defining the ``effective coefficient" \cite{buraasetal} $C_7^{\rm eff}$,
\be
C_7^{\rm eff}(\mu)\equiv C_7(\mu)+\sum_{i=1}^6 y_i C_i
\ee}

\subsection{Inclusive Decays}

\subsubsection{$B\rightarrow X_s\gamma$} \label{sec:btosg}\index{$B\to X_s\gamma$}

As discussed in Chapter 1, the theoretical description of the inclusive 
decay
\bsg\ is particularly clean as it is essentially given by the partonic 
weak
decay $b\to s\gamma$ with small corrections of order $1/m_b^2$ \cite{ope} 
in the HQET expansion (although as the photon energy cut is raised above 
$\sim 2\,\gev$ the nonperturbative Fermi motion of the $b$ quark becomes a 
significant effect \cite{fermimotion}).  Although it is a difficult 
task for hadron colliders to measure the photon energy spectrum governing 
the inclusive channel, it is discussed here for completeness.

The radiative decay is a magnetic dipole transition and is thus mediated 
by the
operator $O_7$.  The corresponding Wilson coefficient $C_7(\mu)$ 
is
evolved to the $b$-quark scale via the effective Hamiltonian of
Eq.~(\ref{eq:heffequation}), with the basis for this decay consisting of 
the
first eight operators in the expansion. The perturbative QCD corrections 
to the
coefficients introduce large logarithms of the form 
$\alpha_s^n(\mu)\log^m
(\mu/M_W)$, which are resummed order by order via the RGE.\index{effective 
Hamiltonian!renormalization group}  The 
next-to-leading
order logarithmic QCD corrections have been computed and result in a much
reduced dependence on the renormalization scale in the branching fraction
compared to the leading-order result.  The inclusion of the QCD 
corrections
enhance the rate by a factor of $\sim 2$, yielding agreement with the 
present
experimental observation.

The higher-order QCD calculation to NLO precision
involves several steps, requiring
corrections to both the Wilson coefficients and the matrix element of
$O_7$ in order to ensure a scheme independent result.
For $C_7$, the NLO computation entails the calculation of the ${\cal O}
(\alpha_s)$ terms in the matching conditions \cite{yao}, and the
renormalization group evolution of $C_7(\mu)$ must be computed
using the ${\cal O}(\alpha_s^2)$ anomalous dimension
matrix \cite{misiaketal}.  For the
matrix element, this includes the QCD bremsstrahlung
corrections \cite{aligreub} $b\to s\gamma +g$, and the NLO virtual
corrections \cite{greubhurth}.  Summing these contributions to the
matrix element and expanding them around $\mu=m_b$, one arrives
at the decay amplitude\index{$B\to X_s\gamma$!decay amplitude}
\be
{\cal M}(b\to s\gamma)=-{4G_FV_{tb}V^*_{ts}\over\sqrt 2}D
\langle s\gamma|O_7(m_b)|b\rangle_{tree}\,,
\ee
with
\be
D=C_7(\mu)+{\alpha_s(m_b)\over 4\pi}\left( C_i^{(0)eff}(\mu)
\gamma_{i7}^{(0)}\log{m_b\over\mu}+C_i^{(0)eff}r_i\right)\,.
\label{7:D}
\ee
Here, the quantities $\gamma^{(0)}_{i7}$ are the entries of the
effective leading order anomalous dimension matrix, the $r_i$
are computed in \cite{greubhurth}, and the index $i$ sums
over the operator basis.  The first term in
Eq. (\ref{7:D}), $C_7(\mu)$, must be computed at NLO precision, whereas
it is consistent to use the leading order values of the other 
coefficients.
The NLO expression for $C_7(\mu)$ is too complicated to present here,
however, for completeness, we give the leading order result,
\be\label{c7eff}
C_7^{(0)eff}=\eta^{16/23}C_7(M_W)+{8\over 3}\left( \eta^{14/23}
-\eta^{16/23}\right) C_8(M_W)+C_2(M_W)\sum^8_{i=1}h_i\eta^{a_i}\,,
\label{7:LOC7}
\ee
where $\eta\equiv\alpha_s(M_W)/\alpha_s(\mu)$ and $h_i\,, a_i$ are
known numerical coefficients \cite{burasrev}.  The form of this result
will be relevant for our discussion of new physics contributions to \bsg,
and clearly demonstrates the mixing between $O_7$ and the
chromomagnetic dipole operator as well as the four quark operator.

There are also long-distance effects arising from emission of a gluon 
from
a charm loop which are only suppressed by powers $\lqcd m_b/m_c^2$.  The
effects of these operators has been estimated to be small, contributing 
to the rate at the few percent level \cite{charmlongdistance}.

After employing an explicit lower cut on the photon energy in the
gluon bremsstrahlung correction, the partial width is given by
\be
\Gamma(\bsg)=\Gamma(b\to s\gamma)+\Gamma(b\to s\gamma
+g)^{E_\gamma>(1-\delta)E_\gamma^{max}}\,,
\ee
where $E_\gamma^{max}=m_b/2$, and $\delta$ is a parameter defined by
the condition that $E_\gamma$ be above the experimental threshold.
In addition, the 2-loop electroweak corrections have been
computed \cite{czar} and are found to reduce the rate
by  $\sim 3.6\%$.  The resulting branching fraction is then obtained
by scaling the partial width for \bsg\ to that for $B$ semileptonic
decay as the uncertainties due to the values of the CKM matrix
elements and the $m_b^5$ dependence of the widths cancel in
the ratio.  The Standard Model prediction for the branching faction 
is then found to be\index{$B\to X_s\gamma$!SM prediction}
\footnote{Ref. \cite{bsgcharmmass} 
argues that the running charm quark mass rather than the pole mass should be used in the
two loop matrix element;  this results in a slightly higher central value 
$(3.73\pm 0.30)\times 10^{-4}$.}
\begin{equation}
B(\bsg)=(3.28\pm 0.30)\times 10^{-4} \,.
\end{equation}
This is in good agreement with the observations by CLEO and ALEPH
\index{$B\to X_s\gamma$!measured branching fraction}
\cite{cleo_bsginc}
which yield $B=(3.15\pm 0.35\pm 0.41)\times 10^{-4}$ and $B=(3.38\pm
0.74\pm 0.85)\times 10^{-4}$, respectively, with the $95\%$ C.L.
bound of $2\times 10^{-4}< B(\bsg)<4.5\times 10^{-4}$.
The inclusive decays are measured by analyzing
the high energy region of the photon energy spectrum.  A good theoretical
description of the spectral shape is thus essential in order to perform a
fit to the spectrum and extrapolate to the total decay rate.  Higher 
order
analyses of the spectrum within HQET have been performed in 
Ref.~\cite{alex,bfl},
where it is found that the shape of the spectrum is
dominated by QCD dynamics and is insensitive to the
presence of new physics. Measurement of the spectral moments of
the photon energy distribution can also be used to determine the HQET
parameters $\bar\Lambda$ and $\lambda_1$ with small theoretical
uncertainty \cite{spect}.

The CKM suppressed mode, $B\to X_d\gamma$,
\index{$B\to X_d\gamma$} is computed in
similar fashion with the substitution $s\to d$ in the above formulae
and in the complete set of operators.  There is also a slight 
modification
of the 4-quark operators $O_1$ and $O_2$ to include the
contributions from $b\to u$ \cite{bdg} transitions.
The NLO predicted branching fraction spans the range
$6.0\times 10^{-6}\leq B(B\to X_d\gamma)\leq 2.6\times 10^{-5}$ with
the main uncertainty arising from the imprecisely determined values of
the CKM elements.
This CKM suppressed channel populates the high
energy region of the photon energy spectrum and hence \bsg\
constitutes the main background source.   Observation thus requires
a veto of strange hadrons in the hadronic $X_d$ system.

\subsubsection{$B \to X_s\ell^+\ell^-$}
\label{sec:bsllinclusive}\index{$B \to X_s\ell^+\ell^-$}

The decay $B \to X_s \ell^+ \ell^-$ is suppressed relative to
$B \to X_s \gamma$ by an additional factor of the electromagnetic 
coupling constant $\alpha \simeq 1/137$, and has not yet been observed.
The SM prediction for the branching fraction is
\index{$B \to X_s\ell^+\ell^-$!SM prediction}
\begin{equation}
B(B\to X_s e^+e^-)=(8.4\pm 2.3)\cdot 10^{-6},\qquad
B(B\to X_s\mu^+\mu^-)=(5.7\pm 1.2)\cdot 10^{-6}
\end{equation}
which may be compared with the current experimental 90\% C.L. upper 
bounds of
$5.7\cdot 10^{-5}$ and $5.8\cdot 10^{-5}$ \cite{CLEO98} 
respectively.
Unlike \bsg, which is only sensitive to the magnitude of $C_7^{\rm 
eff}$, this
decay has the appeal of
being sensitive to the signs and magnitudes of the Wilson coefficients
$C_7^{\rm eff}$, $C_9$ and $C_{10}$, which can all be affected by
physics beyond the standard model. To extract the
magnitudes and phases of all three Wilson coefficients,
several different measurements must be performed. It has been shown
in \cite{newphysics1,newphysics2} that information from the dilepton invariant mass
spectrum and the differential forward-backward asymmetry is sufficient
to extract these parameters.

\subsubsection*{The decay amplitude}

Since over most of phase space the differential rate is well 
approximated by the
parton model, we 
first
consider the parton level results.
 From the effective Hamiltonian (\ref{eq:heffequation}) one easily 
obtains the
parton level decay amplitude\index{$B \to X_s\ell^+\ell^-$!decay amplitude}
\begin{eqnarray}
\label{btosamplitude}
{\cal A} (b \to s \ell^+ \ell^-) = \frac{G_F \alpha}{\sqrt{2}\pi}
V_{ts}^* V_{tb} && \bigg[ \left(C_9^{\rm eff} - C_{10} \right) (\bar{s}
   \gamma_\mu L b) (\bar{\ell} \gamma^\mu L \ell)
  \nonumber\\
&&{} + \left(C_9^{\rm eff} + C_{10} \right) (\bar{s}
   \gamma_\mu L b) (\bar{\ell} \gamma^\mu R \ell) \nonumber\\
&&{}
   - 2 C_7^{\rm eff}
   \left( \bar{s} i \sigma_{\mu\nu} \frac{q^\nu}{q^2} (m_s L + m_b R) b
     \right) (\bar{\ell} \gamma^\mu \ell) \bigg]\,.
\end{eqnarray}
where $C_7^{\rm eff}$ is defined in Eq.~(\ref{c7eff}).
\OMIT{Similarly, $C_9^{\rm
   eff}$ includes contributions from all four-fermion operators
$O_{1-6}$.
As explained in {\bf ????}, the large difference between
the electroweak scale $\mu = m_W$, where the matching onto the
operators $O_i$ is performed, and the typical energy scale of the decay
$\mu = m_b$ requires to use the RGE's to run the operators from $\mu =
m_W$ to $\mu = m_b$, summing logarithms of the form $\alpha_s^n
\log(m_b/m_W)^m$. This has been performed at leading logarithmic order
$(n = m)$ and the results for the coefficients of $C_{1-8}$ have been
given in (\ref{???}). }
The additional operators $O_9$ and $O_{10}$
receive contributions only from penguin and box diagrams in the
matching and are therefore of order $\alpha$. The coefficient $C_9$
contains a term proportional to $\alpha \log(\mu/m_W)$ at one loop,
and so logarithms of the form $\alpha_s^{n+1} \log^n(m_b/m_W)$ must
be summed to obtain leading logarithmic accuracy. Thus, the one loop 
matrix
element of
$O_9$ is required  as well as the two loop running of $C_9$. This amount 
to the
identification $C_9^{\rm eff} \equiv C_9^{\rm eff}(\hat s)$, where
\begin{eqnarray}
C_9^{\rm{eff}}(\hat s)=C_9 \eta(\hat s) + Y(\hat s) \, .
\end{eqnarray}
The one-loop matrix elements of the
four-Fermi operators are represented by the function
$Y(\hat{s})$, which in the NDR scheme is given by \cite{heff,burasmuenz}
\begin{eqnarray}
\label{Ypert}
         Y(\hat s) & = & g(\hat m_c,\hat s)
                 \left(3 \, C_1 + C_2 + 3 \, C_3
                 + C_4 + 3 \, C_5 + C_6 \right)
\nonumber \\
         & & - \frac{1}{2} g(1,\hat s)
                 \left( 4 \, C_3 + 4 \, C_4 + 3 \,
                 C_5 + C_6 \right)
          - \frac{1}{2} g(0,\hat s) \left( C_3 +
                 3 \, C_4 \right) \nonumber \\
         & &     + \frac{2}{9} \left( 3 \, C_3 + C_4 +
                 3 \, C_5 + C_6 \right)
                 \label{eqn:y} \; ,
\end{eqnarray}
where
\begin{eqnarray}
\label{gpert}
g(z,\hat{s}) &=& -\frac{8}{9}\ln (\frac{m_b}{\mu})
  -\frac{8}{9} \ln z + \frac{8}{27} +\frac{4}{9}y
-\frac{2}{9}(2 + y) \sqrt{\vert 1-y \vert}\nonumber\\
&\times & \Bigg[\Theta(1-y)\bigg(\!\ln\frac{1+\sqrt{1-y}}{1-\sqrt{1-y}} -i\pi
\bigg)
+\Theta(y-1)\, 2 \arctan \frac{1}{\sqrt{y-1}} \Bigg] \,, \\
g(0,\hat{s})& =& \frac{8}{27}-\frac{8}{9}\ln \bigg(\frac{m_b}{\mu}\bigg)
               -\frac{4}{9}\ln \hat{s} + \frac{4}{9}i\pi \,,
\end{eqnarray}
with  $y=4z^2/\hat{s}$.
The one loop matrix element of $ O_9$ as a function of the dilepton
invariant mass is written as
\begin{eqnarray}
         \eta(\hat s)  =  1 + \frac{\alpha_s(\mu)}{\pi}
                 \omega(\hat s) \; ,
\end{eqnarray}
where
\begin{eqnarray}
\omega(\hat{s}) &=& -\frac{2}{9}\pi^2 -\frac{4}{3}{\rm Li}_2(\hat{s})
-\frac{2}{3}
\ln \hat{s} \ln(1-\hat{s}) -
\frac{5+4\hat{s}}{3(1+2\hat{s})}\ln(1-\hat{s})\nonumber\\
&-& \frac{2\hat{s}(1+\hat{s})(1-2\hat{s})}{3(1-\hat{s})^2(1+2\hat{s})}
\ln \hat{s} + \frac{5 + 9\hat{s} -6\hat{s}^2}{6(1-\hat{s})(1+2 
\hat{s})} \;
\label{omegahats}
\end{eqnarray}
and we have neglected the strange quark mass.


It is convenient to normalize the rate of $b \to s \ell^+ \ell^-$ to that for semileptonic $b \to c
\ell \bar \nu$ decay
\begin{equation}
d {\cal B}(B \to X_s \ell^+ \ell^-) = {\cal B}_{sl} \frac{d
   \Gamma(B \to X_s \ell^+ \ell^-)}{\Gamma(B \to X_c \ell
   \nu_{\ell})}\,.
\end{equation}
This introduces the normalization constant
\begin{equation}
{\cal B}_0 = {\cal B}_{sl} \frac{3 \alpha^2}{16 \pi^2} \frac{|V_{ts}^*
   V_{tb}|^2}{|V_{cb}|^2} \frac{1}{f(\hat{m}_c)  
   + [\alpha_s(m_b)/\pi] A_0(\hat{m}_c)}\, .
\end{equation}
In this expression $f(\hat{m}_c)$ is the well known phase space factor
for the parton decay rate  $b \to c  \ell \bar{\nu}$
\begin{equation}
f(\hat{m}_c) = 1-8 \hat{m}_c^2 + 8 \hat{m}_c^6 - \hat{m}_c^8 - 24
\hat{m}_c^4 \log \hat{m}_c,
\end{equation}
and $A_0(\hat{m}_c)$ is the ${\cal O}(\alpha_s)$ QCD
correction to the semileptonic $b\to c$ decay rate \cite{niralphas}. 

\subsubsection*{Parton model differential decay rate and forward-backward asymmetry}
\index{$B \to X_s\ell^+\ell^-$!forward-backward asymmetry}

The forward-backward asymmetry in inclusive $b\to s\ell^+\ell^-$
has been studied in detail~\cite{newphysics1}. 
{}From the amplitude of the decay $b \to s \ell^+ \ell^-$ (\ref{btosamplitude}) the dilepton
invariant mass distribution in the parton model can easily be calculated
\begin{eqnarray}
\label{diffbr}
\frac{d{\cal B}}{d\hat s} = \frac{4}{3}\, {\cal B}_0 &&\bigg[ (1-\hat
   s)^2(1+\hat s) \left(|C_9^{\rm eff}|^2 + C_{10}^2 \right)
   \nonumber\\
&&{} + 2(1-\hat s)^2(2+\hat s) \frac{|C_7^{\rm eff}|^2}{\hat
     s} + 6(1-\hat s)^2\, {\rm Re}\left( C_9^{\rm eff} \right) C_7^{\rm
     eff} \bigg]\,.
\end{eqnarray}
A plot of this distribution is shown by the solid
line in Figure~\ref{diff_spec}. The divergence at $\hat{s}=0$ is due to the intermediate photon going on shell an is a well known feature of this decay. In this limit the differential decay rate reduces to the $B \to s \gamma$ rate with an on-shell photon in the final state, convoluted with the fragmentation function which describes the probability for the photon to fragment into a lepton pair.

\begin{figure}[htbp]
\centerline{\epsfxsize=10 cm \epsfbox{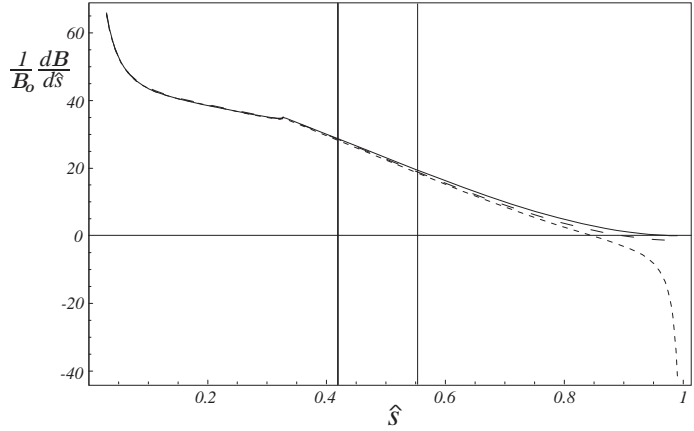}}
\caption[The differential decay spectrum for the decay $B \to X_s \ell^+
     \ell^-$. ]{The differential decay spectrum $\frac{1}{{\cal B}_0}
   \frac{d{\cal B}}{d\hat{s}}$ for the decay $B \to X_s \ell^+
     \ell^-$. The solid line shows the
   free quark prediction, the long-dashed line includes the
   ${\cal O}(\Lambda/m_b^2)$
   corrections and the short-dashed line contains all corrections up to $
   {\cal O}(\Lambda/m_b^3)$.}
\label{diff_spec}
\end{figure}

The differential forward-backward asymmetry is defined by
\begin{equation}
\label{fbaDefinition}
\frac{d {\cal A}}{d\hat{s}} = \int_0^1 \; dz \;
\frac{d\cal{B}}{dx\,d\hat{s}}
   - \int_{-1}^0 dx \; \frac{d\cal{B}}{dx\,d\hat{s}}
\end{equation}
where
\begin{equation}
\label{xsub}
x = cos\theta = \frac{\hat{u}}{\hat{u}(\hat{s},\hat{m}_s)}
\end{equation}
parameterizes the angle between the $b$ quark and the
$\ell^+$ in the dilepton CM frame. An experimentally more useful
quantity is the normalized FB asymmetry defined by
\begin{equation}
\frac{d \bar{{\cal A}}}{d\hat{s}} = \frac{d{\cal
     A}}{d\hat{s}}\bigg/\frac{d{\cal B}}{d\hat{s}}\,.
\end{equation}

In the parton model the differential forward-backward asymmetry is
given by
\begin{equation}
\frac{d {\cal A}}{d\hat{s}} = -4 \, {\cal B}_0 (1-\hat s)^2 \left[\hat 
s\,
   {\rm Re}\left(C_9^{\rm eff}(\hat s) \right) C_{10} + 2
   C_{10}  C_7^{\rm eff} \right]\,,
\end{equation}
which is shown by the solid line in Figure~\ref{asymm}.
\begin{figure}[htbp]
\centerline{\epsfxsize=10 cm \epsfbox{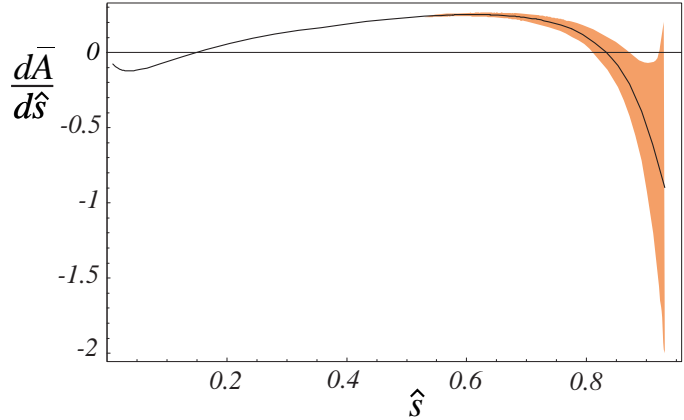}}
\caption[The normalized forward backward asymmetry.]
{The normalized forward backward asymmetry.  The three curves 
show
the mean value and the $1 \sigma$ uncertainty of the forward
backward asymmetry. }
\label{asymm}
\end{figure}

\subsubsection*{Charmonium resonances}

\index{$B \to X_s\ell^+\ell^-$!charmonium resonances}
Both the dilepton invariant mass spectrum and the differential
forward-backward asymmetry contain a cusp at the threshold of $c
\bar c$ pair production. For such values of $\hat s$ long distance
contributions from tree level processes $B \to B^{(*)}
\psi^{(')}$, followed by $\psi^{(')} \to \ell^+ \ell^-$, are important,
which can not be calculated perturbatively. The location of the first two 
$c \bar{c}$ resonances  are indicated in Figs.~\ref{diff_spec} and \ref{asymm} by the two vertical lines.

Since the $c \bar c$ resonance contributions cannot be
calculated model independently, suitable cuts on the dilepton
invariant mass are conventionally applied to eliminate these resonance
contributions.
Such cuts naturally divide the available phase space into two separate
regions: a
low $\hat s$ region for $s \le M_\psi^2 - \delta$ and a high $\hat s$
region for $s \ge M_{\psi'}^2 + \delta'$, where $\delta^{(')}$
depends on the exact values of the experimental cuts. The region of
phase space below the $\psi$ resonance is contaminated by background
from sequential $B$ decays. This background can only be suppressed if
the inclusive process is measured by summing over a large number of
individually reconstructed final
states. In the region above the $\psi'$ resonance, there is almost no
background from other $B$ decays, making the measurement much
easier. It is the latter region of phase space that is accessible to
experiments at the Tevatron.

\subsubsection*{Power corrections to the dilepton invariant mass
spectrum and the forward-backward asymmetry}

{}Nonperturbative physics can be parameterized by matrix
elements of higher dimensional operators. This is done by performing
an OPE as described in Chapter 1. The leading corrections
arise at order ${\cal
   O}(\Lambda_{QCD}/m_b)^2$ and can be parameterized by the matrix
elements of two dimension five operators. Both matrix elements,
$\lambda_1$ and $\lambda_2$ have
been measured, albeit with large uncertainties for $\lambda_1$. At
order ${\cal O}(\Lambda_{QCD}/m_b)^3$ there are seven operators
contributing, with none of the matrix elements known. Calculating the
contributions from these dimension six operators therefore do not
improve the theoretical accuracy, but can be used to investigate the
convergence of the OPE and estimate theoretical
uncertainties.
The contributions of the dimension five operators to the differential 
decay rate and the forward backward asymmetry were calculated in 
Ref. \cite{ahhmNP} and the calculation including all power corrections 
up to order $\Lambda_{\rm QCD}^3/m_b^3$ is presented in Ref.~\cite{bauerburrell}.

\index{$B \to X_s\ell^+\ell^-$!convergence of OPE}
The effects of these corrections on the
differential decay rate are shown
in Figure~\ref{diff_spec} by the long and short dashed lines,
respectively. It is obvious from Figure~\ref{diff_spec} that the
effect of higher dimensional operators is negligible below the $\psi$
resonance, whereas it is large in the large $\hat s$
region. This can also be seen by calculating the branching ratio with an 
upper cut on the dilepton invariant mass 100 MeV below $m_\psi$, 
$\hat s = 0.35$. Including a cut $\hat s > 0.01$ to eliminate the 
fragmentation divergence at low $q^2$ the expansion in $1/m_b$ yields
\begin{equation}
\int_{0.01}^{0.35}\,d\hat{s} \frac{d{\cal B}}{d\hat{s}} = 22.0 \bigg[
   1 + 0.5\frac{\lambda_1}{m_b^2}  
   + 1.2 \bigg( \frac{\lambda_2}{m_b^2}  - \frac{\rho_2}{m_b^3} \bigg) 
   - 3.7 \frac{\rho_1}{m_b^3} \bigg]
\end{equation}
where the $\rho_i$'s and $f_1$ are unknown matrix elements of order $\lqcd^3$, and we have
neglected contributions of a comparable size coming from T-products, which may be 
absorbed into a redefinition of $\lambda_1$ and $\lambda_2$.
All numerical coefficients are of order unity and the OPE is therefore
converging well. In the region above the resonances the situation looks 
quite different. 
Imposing a lower cut 100 MeV above the $\psi'$ resonance, 
$\hat s = 0.59$,  the partially integrated branching ratio is
\begin{equation}
\label{CLEObranchingratio}
   \int_{0.59}^{1}\,d\hat{s} \frac{d{\cal B}}{d\hat{s}} = 3.8 \bigg[ 1+
   0.5 \frac{\lambda_1}{m_b^2}  
   - 35.4 \bigg( \frac{\lambda_2}{m_b^2} -\frac{\rho_2}{m_b^3}\bigg) 
 + 161.8 \frac{\rho_1}{m_b^3}+ 147.4 \frac{f_1}{m_b^3} \bigg]\,.
\end{equation}
 From this expression it is clear that the convergence of the OPE is
very poor and the branching ratio above the $\psi'$ resonance can not
be calculated well.

Higher dimensional operators can also be included for the
forward-backward asymmetry. This leads to
\begin{eqnarray}
\frac{d{\cal A}}{d\hat{s}} &=& C_7^{\mbox{eff}} C_{10} \bigg[ -8
         \left( 1 - \hat{s} \right)^2-
      {\frac{4\left( 3 + 2\hat{s} + 3{{\hat{s}}^2} \right)
          \lambda_1}{3{m_b^2}}} +
      {\frac{4 \left( 7 + 10\hat{s} - 9{{\hat{s}}^2} \right)
          \lambda_2}{{m_b^2}}} \nonumber\\
&&\qquad\qquad
   +
      {\frac{4\left( 5 + 2\hat{s} + {{\hat{s}}^2} \right)
          \rho_1}{3{m_b^3}}} -
      {\frac{4\left( 7 + 10\hat{s} - 9{{\hat{s}}^2} \right)
          \rho_2}{{m_b^3}}} \bigg] \nonumber \\
&+&
    C_{9}^{\mbox{eff}}(\hat{s}) C_{10}\bigg[ -4\hat{s}
          {{\left( 1 - \hat{s} \right) }^2}
             - {\frac{2\hat{s}\left( 3 + 2\hat{s} +
              3{{\hat{s}}^2} \right) \lambda_1}
        {3{m_b^2}}} + {\frac{2\hat{s}\left( 9 + 14\hat{s} -
            15{{\hat{s}}^2} \right) \lambda_2}
        {{m_b^2}}} \nonumber\\
&&\qquad\qquad\quad
   - {\frac{2\hat{s}\left( 1 + 2\hat{s} + 5{{\hat{s}}^2} \right)
       \rho_1}
        {3{m_b^3}}} - {\frac{2\hat{s}\left( 1 + 6\hat{s} -
            15{{\hat{s}}^2} \right) \rho_2} {{m_b^3}}} \bigg]
\end{eqnarray}
It is clear from this expression that the
third order terms do not have abnormally large coefficients, and
therefore introduce only small variations relative to the second
order expressions. The normalized forward-backward asymmetry, however,
inherits the poor behavior of the differential branching ratio in
the endpoint region $\hat s \to 1$. The normalized asymmetry is shown
in Figure~\ref{asymm}, with the shaded region representing the
uncertainties due to $(\Lambda_{QCD}/m_b)^3$ terms. From this Figure
it is clear that the nonperturbative uncertainties on the differential
asymmetry are small below the $\psi$ resonance, whereas they are large
above the $\psi'$.

\OMIT{\subsubsection*{The relationship between $B \to X_s \ell^+ \ell^-$ 
and
   $B \to X_u \ell \bar \nu$ and the determination of  $|V_{ub}|$}
{}Even though the OPE converges very poorly in the region accessible
to the Tevatron, for values of $\hat s$  above the $\psi'$ resonance,
a measurement of the resulting branching ratio is
still very useful. This is due to the fact that the numerical values
of the Wilson coefficients $C_9^{\rm eff}$, $C_7^{\rm eff}$ and
$C_{10}$ are such that the decay amplitude given in 
(\ref{btosamplitude}) is
very close to the $(V-A)\times(V-A)$ current mediating semileptonic
$B$ decays. This implies that in the large $\hat s$ region, where the
contribution from $O_7$ is negligible, the differential decay rates
for $B \to X_u \ell \bar \nu$ and $B \to X_s \ell^+ \ell^-$
have almost the same functional dependence on $\hat s$. Since this
argument holds for the nonperturbative corrections suppressed by
powers of $\Lambda_{QCD}/m_b$, the ratio of these two spectra can be
predicted well even in a region of phase space where the OPE does not
converge. The similarity of the two spectra in the large $\hat s$
region can be seen from Figure~\ref{two_spect}, where the differential
branching ratio for $B \to X_s \ell^+ \ell^-$ and $B \to X_u \ell \bar
\nu$ is shown for $\lambda_1 = 0.19 {\rm GeV}^2$, $\lambda_2 = 0.12
{\rm GeV}^2$, $\rho_i = \tau_i = f_i = (0.5 {\rm GeV})^3$.
The dilepton invariant mass spectrum of $B \to X_u \ell \bar\nu$ was
shown to be very useful for a model independent extraction of $|
V_{ub}|$, since a lower cut on the dilepton invariant mass can be used
to eliminate the charm background in inclusive semileptonic $B$ decays
without destroying the convergence of the OPE. If experimental
considerations force this cut to be above $\sim 15 \, {\rm GeV}^2$,
however, theoretical uncertainties become large. It is in this
situation where the information from $B \to X_s \ell^+ \ell^-$ can be
used to reduce these uncertainties considerably. The relation
\begin{equation}
{ {\cal B}(B \to X_u\ell\bar\nu)|_{q^2 > q_0^2} \over
   {\cal B}(B \to X_s\ell^+\ell^-)|_{q^2 > q_0^2} }
= {|V_{ub}|^2 \over |V_{ts} V_{tb}|^2}\, {8\pi^2 \over \alpha^2}\,
   { 1 \over |\widetilde C_9|^2 + |C_{10}|^2
   + 12\, {\rm Re}\, (C_7 \widetilde C_9)\, B(q_0^2) } \,,
\end{equation}
for $q_0^2 > m_{\psi'}^2 + \delta$, is expected to hold at the
$\sim15\%$ level.  }

\subsection{Exclusive Decays \label{sec:exclusivedecays}}

The Wilson coefficients defined in Section~\ref{sec:effham} 
contain the short distance information that allows us to test the one 
loop structure of the standard model. The exclusive decays we will 
consider in this section can be used to determine these coefficients. 
However, these decays also depend on the hadronic matrix elements of 
the operators in Eqs.~(\ref{eq:operatorbasis}), which describe the 
transition from the initial state $b$ flavored hadron to the final 
state hadron. These hadronic matrix elements are dominated by 
nonperturbative QCD effects.
They are calculable in principle from lattice QCD, the only
{\em ab initio} framework available for quantitative calculations of
nonperturbative QCD.

At present, lattice QCD calculations of these processes are
incomplete. This results in important uncertainties
in theoretical predictions of exclusive rare decays, with a
corresponding loss of sensitivity to the interesting short distance
physics. We therefore need a variety of other theoretical tools
at our disposal. These include model independent approaches based
on approximate symmetries, such as heavy quark and chiral symmetry,
and model-dependent approaches based on phenomenologically motivated 
models. While not rigorous, model calculations can serve to guide 
lattice calculations and provide a simple framework for studying 
these processes.

The rest of this section is organized as follows. After introducing 
the matrix elements and form factor parameterizations in 
Section~\ref{sec:meff}, we discuss results and prospects from 
lattice QCD in Section~\ref{sec:lattice}.
Rare semileptonic decays are discussed in Section~\ref{sec:btokll}
which describes results and constraints from model independent approaches 
first, followed by a summary of model-dependent results. 
Section~\ref{sec:radiative} gives a discussion of the status of 
theoretical predictions for exclusive radiative decays. Finally,
Section~\ref{sec:btoll} discusses results for $B_{s,d} \to l^+l^-$
decays.

\subsubsection{Hadronic Matrix Elements and Form Factors \label{sec:meff}}

The hadronic matrix elements can be parametrized in terms of
form factors which are functions of the momentum transfer between
the initial and final state hadrons.

\index{$B\to K\ell^+\ell^-$!form factors}
For the $B\to K\ell^+\ell^-$ decay
the hadronic matrix elements of the operators $O_7$, $O_9$ and $O_{10}$ 
(which were defined in Eqs.~(\ref{eq:operatorbasis})) 
are parametrized as
\begin{eqnarray}
\langle K(k)|\bar{s}\sigma_{\mu\nu}q^\nu b|B(p)\rangle &=& 
i\,\frac{f_T}{m_B+m_K}
\left\{(p+k)_\mu q^2-  q_\mu(m_B^2-m_K^2)\right\}\,, \label{bp_sig} \\*[4pt]
\langle K(k)|\bar{s}\gamma_\mu b|B(p)\rangle &=& f_+ (p+k)_\mu +f_-\,
q_\mu \,,
\label{bp_sem}
\end{eqnarray}
with $f_T(q^2)$ and $f_{\pm}(q^2)$ unknown functions of $q^2=(p-k)^2=
m_{\ell^+\ell^-}^2$.
In the SU(3) limit $f_{\pm}$ in (\ref{bp_sem}) are the same as the
form factors entering in the semileptonic decay $B\to\pi\ell\nu$.

\index{$B\to K^*\ell^+\ell^-$!form factors, defined}
For the vector meson mode, $B\to K^*\ell^+\ell^-$, we have the
``semileptonic'' matrix element
\begin{eqnarray}
\langle K^*(k,\epsilon)|\bar{s}_L\gamma_\mu b_L|B(p)\rangle
&=&\frac{1}{2}\Big\{ ig\,\epsilon_{\mu\nu\alpha\beta}\epsilon^{*\nu}
(p+k)^\alpha (p-k)^\beta
-f\,\epsilon^*_\mu \nonumber\\
& &{} - a_+\, (\epsilon^* \cdot p)\,(p+k)_\mu
-a_-\, (\epsilon^* \cdot p)\, (p-k)_\mu \Big\} \,,
\label{vacur}
\end{eqnarray}
where $\epsilon_\mu$ is the $K^*$ polarization four-vector.
The form factors defined in (\ref{vacur}) can be identified, in the
$SU(3)$ limit, with those appearing in the semileptonic transition
$B\to\rho\ell\nu$.
The matrix element of the penguin operator takes the form
\begin{eqnarray}
\langle K^*(k,\epsilon)|\bar{s}_L\sigma_{\mu\nu} q^\nu 
b_R|B(p)\rangle &=&
i\epsilon_{\mu\nu\alpha\beta}\epsilon^{*\nu}p^\alpha k^\beta\, 
2T_1\nonumber\\
& &{} + T_2 \left\{ \epsilon_\mu^*(m_B^2-m_{K^*}^2) - (\epsilon\cdot
p)\,(p+k)_\mu
\right\}\nonumber \\
& &{} +T_3 (\epsilon\cdot p)\,\bigg\{q_\mu -\frac{q^2}{m_B^2-m_{K^*}^2}\, 
(p+k)_\mu
\bigg\}\,, \label{sig2kst}
\end{eqnarray}

Radiative decays receive a contribution from the local operator
$O_7$ -- also called the magnetic dipole operator -- and its associated 
Wilson coefficient. This is also sometimes called the short distance
contribution, since another contribution comes from non-local operators.
These non-local operators are due to (i) the process $B\to V V^*$ 
with the subsequent conversion of the neutral vector meson $V^*$ to a 
real photon, and (ii) weak annihilation and $W$ exchange diagrams with 
subsequent $\gamma$ radiation. 
Contributions from non-local operators are sometimes also called 
long distance contributions.

\index{$B\to K^*\gamma$!form factors, defined}
The hadronic matrix element of the magnetic dipole operator for the
$B\to V \gamma$ decay, were $V$ represents a vector meson, is 
generally written in terms of three form factors, with
\begin{eqnarray}
\langle V(p, \epsilon)|\bar s\sigma_{\mu\nu}q^\nu(1+\gamma_5)b|B(p_B) \rangle & = &
2i\epsilon_{\mu\nu\rho\sigma}\epsilon^{*\nu}p_B^\rho p^\sigma T_1(q^2) \\
& &{} +\left[\epsilon^*_\mu(m_B^2-m_V^2)-(\epsilon^*\cdot 
q)(p_B+p)_\mu\right]
T_2(q^2)\nonumber\\
& &{} +(\epsilon^*\cdot q)\bigg[ q_\mu-{q^2\over m^2_B-
m^2_V}\,(p_B+p)_\mu\bigg]
T_3(q^2)\,,\nonumber
\end{eqnarray}
where $\epsilon^*$ represents the polarization vector of the vector 
meson, and $q$ corresponds to the momentum of the outgoing photon.
This simplifies for the case of an on-shell photon, where the 
coefficient of $T_3$ vanishes and $T_2(0)=-iT_1(0)$.
Hence in the physical cases of interest here, the
decay width can be expressed in terms of a single form factor,
\begin{equation}\label{7:excbr}
\Gamma(B\to V+\gamma)={\alpha G_F^2\over 32\pi^4}|V_{tb}V^*_{ts}|^2
(m_b^2+m_s^2)m_B^3\left( 1-{m_V^2\over m_B^2}\right)^3|C_7^{(0)eff}(m_b)
|^2\, |T_1^{B\to V}(q^2=0)|^2 \,,
\end{equation}
and the branching fraction is computed by scaling to the semileptonic 
rate as usual.

\subsubsection{Lattice QCD} \label{sec:lattice}

\index{lattice QCD}
Lattice QCD methods are well suited for theoretical calculations
of the hadronic matrix elements (and form factors) which describe
rare decays. However, because of the lack of experimental information
on rare semileptonic decays (such as $B \to K \ell^+ \ell^-$),
they have not been studied on the lattice to date.
Results from lattice calculations of rare radiative decays do exist, 
and are discussed in more detail in Section~\ref{sec:radiative}.

The hadronic matrix elements which describe rare semileptonic
decays (as shown in Eqs.~(\ref{bp_sem}--\ref{sig2kst})) are similar
to the matrix elements for semileptonic decays, such as
$B \to \pi \ell \bar\nu$.
These have been studied extensively using lattice methods.
We can use the existing results from lattice calculations of
$B \to \pi \ell \bar\nu$ to discuss the prospects for lattice
calculations of rare decays, like $B \to K^{(*)} \ell^+ \ell^-$.

\index{lattice QCD!sources of uncertainty}
Current lattice calculations of the $B \to \pi \ell \bar\nu$
form factors are accurate to about $15-20 \%$ \cite{btopi_lattice}.
It would be relatively straightforward to perform a lattice QCD
calculation of the form factors for $B \to K \ell^+ \ell^-$
with a similar accuracy using current technology.
The quoted uncertainty includes all systematic errors except for 
the quenched approximation -- unquenched results do not yet exist
for these decays.
From unquenched calculations of other quantities, we can estimate 
the expected size of the effect to be in the range of $10-15 \%$.
Apart from the quenched approximation, the most important 
errors in lattice QCD calculations are due to statistics (from 
the Monte-Carlo integration), the chiral extrapolation, the lattice 
spacing, and perturbation theory (see Section~5.3 of Chapter~1
for a detailed discussion of how these errors arise).

Lattice QCD calculations are in principle improvable to
arbitrary precision. In practice, the accuracy of lattice
calculations depends on the computational effort and available
technology. Numerical simulations based on lattice QCD are time
consuming and computationally expensive. In the following, we shall 
discuss the prospects for reducing the total uncertainty in lattice 
QCD calculations to a few percent. We assume that there will be 
reasonable growth in the computational resources available for 
these lattice QCD calculations.

Improving the statistical and chiral extrapolation
errors is straightforward; it just requires more computer time.
This is within reach of the computational resources which should
become available to lattice QCD calculations within the next
few years.

\index{lattice QCD!lattice spacing errors}
Lattice spacing errors can be reduced by explicitly reducing the
lattice spacing ($a \to 0$) used in the calculations.
However, the computational cost of a lattice calculation scales
like $1/a^{6-10}$. 
In general, lattice spacing errors are proportional to terms
which grow like $(a\Lambda)^n$ where $\Lambda$ is the typical
momentum scale of the process in question. Typical lattice
spacings used in numerical simulations are in the range 
$0.05\, \fm \, \ltap \, a \, \ltap \, 0.2 \, \fm$, 
so that $a\Lambda \ll 1$ for momenta of order $\Lambda_{\rm QCD}$.
The power $n$ (and hence the size of lattice spacing errors) 
depends on the discretization used in the calculation. With the 
(improved) lattice actions currently in use, $n=2$.
With highly improved lattice actions we can increase the power
to $n=4$. 
The situation is a bit more complicated in the presence of heavy
quarks. However, as discussed in detail in Section~1.5.3, the 
lattice spacing errors associated with heavy quarks can be as 
easily controlled as the errors associated with the light degrees 
of freedom. In summary, we can keep lattice spacing errors
under control at the few percent (or less) level, by using
highly improved actions in simulations at relatively coarse
lattice spacings. The big advantage of this strategy is its
low computational cost. It is therefore also the best
strategy for realistic unquenched calculations.

\index{lattice QCD!momentum dependent errors}
There is a further restriction for semileptonic decays which
arises from the need to control lattice spacing errors.
The hadronic matrix elements (and form factors) for semileptonic
decays are functions of the daughter recoil momentum.
Since lattice spacing errors increase with increasing recoil
momentum, the momentum range accessible to lattice QCD calculations
is limited.
At present, in order to keep lattice errors under control,
most calculations impose an upper momentum cut of
\begin{equation}
  \bm{p}_{\rm recoil} < 1 \, \GeV \,.
\end{equation}
For decays like $B \to D \ell \bar\nu$ this is not a problem, as the
allowed recoil range is also small. However, for decays of $B$ mesons
into light hadrons we can obtain the matrix elements and form factors
only over part of the allowed range. In particular, the high recoil
region, $\bm{p}_{\rm recoil} \propto  m_B/2$, which corresponds
to $q^2$ small or near zero, is not directly accessible to lattice QCD
calculations.

A remedy used in early calculations is to extrapolate the form factors
from the high $q^2$ region to low $q^2$ assuming a functional form for
the shape of the form factors. This procedure introduces a model
dependent systematic error into the calculation which can't be 
quantified.
This was an acceptable compromise for early lattice calculations
intended to establish the method. However, it is certainly undesirable
for first principles calculations designed to test the standard model.

If there is significant overlap between the recoil momentum
ranges accessible in lattice QCD calculations and experimental
measurements, then we can avoid model dependent extrapolations and
limit the comparison between theory and experiment to the common
recoil momentum range. Indeed, this appears to be the case for
$B \to \pi \ell \bar\nu$, and will most likely also be true for
rare semileptonic decays such as $B \to K \ell^+ \ell^-$.
However, as discussed in the following section, the high recoil 
region is of particular phenomenological interest in
$B \to K^{(\ast)} \ell^+ \ell^-$ decays.

It is possible to increase the recoil momentum range accessible 
to lattice calculations by using highly improved actions, especially 
in combination with asymmetric lattices \cite{lepage}.
However, lattice calculations of radiative decays such as 
$B \to K^\ast \gamma$ remain problematic. With a real
photon, the two body decay takes place at maximum recoil or $q^2 = 0$.
If we want to avoid model-dependent extrapolations, we will need
to develop better techniques for dealing with this high recoil 
physics.

\subsubsection{$B \to K^{\ast}\ell^+\ell^-$ and 
               $B \to K \ell^+\ell^-$} \label{sec:btokll}

\index{$B\to K^*\ell^+\ell^-$}
\index{$B\to K\ell^+\ell^-$}
In the following, we review results for calculations of the form
factors in Eqs.~(\ref{bp_sem}--\ref{sig2kst}) from different
theoretical approaches. The prospects for lattice QCD results
were already discussed in the previous section.
We first present the constraints derived from heavy quark symmetry (HQS),
followed by results from SU(3) symmetry and finally from the large energy
limit (LEL).

We then review results obtained from calculations using
phenomenological models.
A vast variety of models are available for these calculations.
Here, we present the relevant features of current models.
Although predictions from different models still disagree with each
other, the situation has greatly improved since the
experimental observation of exclusive $B$ decays to light hadrons.

\subsubsection*{Predictions from HQET}

\label{HQET}
As discussed in Chapter 1, the Dirac structure of $b$ quarks simplifies in the Heavy Quark Limit, $m_b\gg \lqcd$, allowing relations between different form factors to be derived.
For example, in the rest frame of the heavy quark, $v^\mu=(1,\vec 0)$,
the heavy quark field obeys $\gamma_0 h_0=h_0$, and so 
\begin{equation}\label{dirac}
\bar h_b i\sigma_{0i} h_b=\bar h_b\gamma_i h_b, \qquad 
\bar h_b i\sigma_{0i}\gamma_5 h_b=-\bar h_b \gamma_i\gamma_5 h_b.
\end{equation}
By making use of (\ref{dirac}) we can now obtain relations among the form
factors in (\ref{bp_sig}) and (\ref{bp_sem}). They are \cite{iw_1}
\index{HQET!spin symmetry relations}
\begin{eqnarray}
f_T(q^2)&=&-\frac{m_B+m_K}{2m_B} \left(f_+(q^2) - f_-(q^2)\right)\,,
\label{rel_1}\\
T_1(q^2)&=&\frac{f(q^2)-2(q\cdot p)\,g(q^2)}{2m_B}\,, \label{rel_2}\\
T_2(q^2)&=&T_1(q^2)-\frac{f(q^2)-2(m_B^2+k\cdot p)\, g(q^2)}{2m_B}\,
\left(\frac{q^2}{m_B^2-m_{K^*}^2}\right)\,,
\label{rel_3}\\
T_3(q^2)&=&\frac{m_B^2-m_{K^*}^2}{2m_B}\left\{a_+(q^2)-a_-(q^2) +2g(q^2)
\right\}\,. \label{rel_4}
\end{eqnarray}
In terms of the symmetries of the HQET, Eqs.~(\ref{rel_1}--\ref{rel_4})
are a result of the Heavy Quark {\em Spin} Symmetry (HQSS) that arises 
in the
heavy quark limit due to the decoupling of the spin of the
heavy quark~\cite{hqs}.
This is a very good symmetry when considering $B$ decays, a measure of
which is
for instance the quantity
\begin{equation}
\frac{m_{B^*}-m_B}{m_B}\simeq 0.009\sim \left({0.45\over 4.8}\right)^2
\end{equation}
which is in agreement with the HQET prediction of $O(\lqcd^2/m_b^2)$.
Thus the relations (\ref{rel_1}-\ref{rel_4}), which  are
valid over the entire physical
region~\cite{bd_1}, will receive only small corrections.
They allow us to express all the hadronic matrix elements entering in
$B\to K^{(*)}\ell^+\ell^-$ processes, in terms of the ``semileptonic''
form factors $f$, $g$ and $a_{\pm}$.

\index{HQET!flavor symmetry relations}
Furthermore, there is an additional $SU(2)_F$ flavor symmetry
in the heavy quark limit, leading to relations among form factors
occurring in the decays of charm and bottom hadrons~\cite{hqs,iw_1}.
For instance in $H\to\pi\ell\nu$ one obtains
\begin{eqnarray}
(f_+ - f_-)^{B\to\pi}&=&C_{BD}\,\sqrt{\frac{m_B}{m_D}}\,
(f_+ - f_-)^{D\to\pi}\,,\label{fmf}\\
(f_+ + f_-)^{B\to\pi}&=&C_{BD}\,\sqrt{\frac{m_D}{m_B}}\,
(f_+ + f_-)^{D\to\pi}\,,\label{fpf}
\end{eqnarray}
where $C_{BD}=(\alpha_s(m_B)/\alpha_s(m_D))^{-6/25}$ is a leading 
logarithmic
QCD correction to the heavy quark currents~\cite{pw}.
Similar scaling relations are obtained for $H\to\rho\ell\nu$,
\begin{eqnarray}
f^{B\to\rho} &=& C_{BD}\,\sqrt{\frac{m_B}{m_D}}\,
f^{D\to\rho}\,,\\
g^{B\to\rho} &=& C_{BD}\,\sqrt{\frac{m_D}{m_B}}\,
g^{D\to\rho}\,,\\
(a_+ - a_-)^{B\to\rho} &=& C_{BD}\,\sqrt{\frac{m_D}{m_B}}\,
(a_+ - a_-)^{D\to\rho}\,.\label{apmam}
\end{eqnarray}
In the above relations the form factors must be evaluated at the same
value of the hadronic energy recoil $v\cdot k$, not the same value of 
$q^2$.
The semileptonic $D$ decays have a maximum recoil energy of about 
$1~$GeV,
whereas $B$ decays go up to $\simeq m_B/2$.
Thus the use of data from $D$ decays requires an extrapolation from the
low to the high recoil regions of phase space, for which the
$v\cdot k$ dependence of the form factors must be assumed.
In addition, the relations (\ref{fmf}-\ref{apmam}) are valid in the
leading order in the HQET and will receive corrections of the order of
$\bar\Lambda/2m_c\simeq0.15$, with $\bar\Lambda$
the effective mass of the light degrees of freedom.
Then, the corrections to the flavor symmetry are likely
to be larger than those to the spin symmetry.
For instance, in the $H\to\pi\ell\nu$, corrections as large as
$20\%-30\%$ are possible~\cite{blnn}.
On the other hand, it was shown in Ref.~\cite{zlmw} that the 
$B\to K^*\gamma$ rate \index{$B\to K^*\gamma$!HQET prediction}
can be well reproduced by using both the spin and the flavor symmetries
in HQET to relate the $D\to K^*\ell\nu$ form factors to  $T_1(0)$ 
determining
the radiative branching fraction, with the additional assumption of
a monopole $q^2$ dependence for the form factors $f$ and $g$ all the way
from $q^2=16.5~{\rm GeV}^2$ to $q^2=0$.

\subsubsection*{Other Theoretical Approaches}

{\em\bf SU(3):}

\index{$SU(3)$}
A necessary ingredient in the application  of the HQSS relations
(\ref{rel_1}-\ref{rel_4}) to predictions for $B\to K^{(*)}\ell^+\ell^-$
making use of the semileptonic form factors in $B\to(\pi\rho)\ell\nu$
is the assumption of well-behaved $SU(3)$ symmetry relations.
Intuitively, and since the form factors are determined by the strong
interactions,
we expect that at very high recoil energies $SU(3)$ is a very good
approximation.
For rare $B$ decays, where most of the events occur in this region of
phase space,
we should be confident that $SU(3)$ corrections are small.
However, it is difficult to make a quantitative statement about the
size of the
$SU(3)$ breaking in a completely model independent way.
For instance, in the constituent quark model picture, a relevant quantity
parameterizing $SU(3)$ breaking could be
\index{$SU(3)$!breaking}
\begin{equation}
\delta_3\equiv\frac{\tilde m_s - \tilde m_d}{E_h}\,,
\end{equation}
where $\tilde m_q$ are constituent quark masses and $E_h$ is the
recoil energy of the hadron. Thus, for standard values of the
strange and down constituents masses this suggests and $SU(3)$
breaking below $10\%$ in most of phase space.
On the other hand, the deviations from $1$ of the double ratio
$F^{(B\to\rho)}/F^{(B\to K^*)}/F^{(D\to\rho)}/F^{(D\to K^*)}$, with
$F$ some arbitrary form factor, were estimated in Ref.~\cite{lsw98}
by calculating the effects of chiral loops.
The effect was found to be smaller than $3\%$ and, although there
could be contributions from higher orders, adds credibility to
the use of $SU(3)$ relations.

The short distance structure of the $B_s$ meson decays
$B_s \to (\eta^{(')},\phi) \ell^+ \ell^-$ is the same 
as that of $B\to K^{(*)}\ell^+\ell^-$. In the SU(3) limit the 
branching ratios should be the same. Thus, although departures 
from the SU(3) predictions could be as large as $20-30\%$, our 
understanding of the $B$ modes gives us a very good starting point 
for the $B_s$ decays. 

\noindent{\em\bf Large Energy Limit (LEL):}

\index{large energy limit}
In addition the symmetries of the heavy quark limit, additional 
simplifications occur for exclusive decays in which the recoil energy 
of the light meson is large, the so-called Large Energy
Limit (LEL) 
\cite{bfl,leet1,leet2,afb0,leleft,benfel00,bfps01,ABHH00,burdmanhiller00,
benfelseid01,boschbuchalla01}.   
In this limit, interactions of the light quark with soft or hard collinear 
gluons do not change its helicity, giving rise to additional symmetries, 
and corresponding additional relations 
between form factors.  These were first noted in \cite{leet2}, based on 
symmetries of the "large energy effective theory" (LEET) \cite{leet1}.  
(Although LEET is not a well-defined effective theory, these relations 
remain true in the LEL \cite{bfl,leleft,bfps01}.)
In addition to the heavy quark symmetry relations in 
Eqs.~(\ref{rel_1}--\ref{rel_4}), the additional symmetries of the LEL 
gives new relations among the form factors  defined in 
Eqs.~(\ref{bp_sig}--\ref{vacur}). 

The main result of the LEL which is important for our discussion here,
is the fact that all of the form factors in
$H\to (P,V)\ell^+\ell^-$ can be expressed by a total of {\em three} 
functions of the heavy mass $M$ and the recoil energy $E$.
For example, the $H\to P\ell^+\ell^-$ form factors can be written
as~\cite{leet2}
\index{$B\to K^*\ell^+\ell^-$!in large energy limit}
\index{$B\to K\ell^+\ell^-$!in large energy limit}
\begin{eqnarray}
f_+(q^2) &=& \xi(M,E)\,,\label{le1}\nonumber \\
f_-(q^2) &=& -\xi(M,E)\,,\label{le2}\nonumber \\
f_T(q^2) &=& \left(1+\frac{m_P}{M}\right) \xi(M,E)\,, \label{le3}
\end{eqnarray}
where $\xi(M,E)$ is an unknown function of $M$ and $E$. Simple 
inspection shows that the previously derived HQSS relation 
Eq.(\ref{rel_1}) is satisfied. 
For the vector meson final state, the form factors obey 
\begin{eqnarray}
g(q^2) &=& \frac{1}{M}\,\xi_{\perp}\,,\label{le4}\nonumber \\
f(q^2) &=& -2E \,\xi_{\perp}\,,\label{le5}\nonumber \\
a_+(q^2) &=& \frac{1}{M}\left\{\xi_\perp -
           \frac{m_V}{E}\xi_\parallel\right\}\,,
\label{le6}\nonumber \\
a_-(q^2) &=& \frac{1}{M}\left\{-\xi_\perp +
              \frac{m_V}{E}\xi_\parallel\right\}\,,
\label{le7}
\end{eqnarray}
where $\xi_\perp(M,E)$ and $\xi_\parallel(M,E)$ 
refer to the transverse and longitudinal polarizations, respectively.

Additionally, there will be expressions for the ``penguin'' form factors
$T_i(q^2)$, $i=1,2,3$, in terms of $\xi_\perp$ and $\xi_\parallel$, which
satisfy the HQSS relations in Eqs.~(\ref{rel_2}--\ref{rel_4}).
The power of the predictions in Eqs.~(\ref{le1}--\ref{le7}) will become
apparent later when computing observables in 
$B\to K^{(*)}\ell^+\ell^-$ decays.
Let us now only note in passing one example: the ratio of the vector form
factor $g(q^2)$ to the axial-vector form factor $f(q^2)$
\begin{equation}
R_V\equiv \frac{g(q^2)}{f(q^2)} \simeq -\frac{1}{2\,E_{K^*}\,m_B}\,,
\label{rvlet}
\end{equation}
only depends on kinematical variables and is unaffected by hadronic 
uncertainties.
This ratio determines, for instance the ratio of the two transverse
polarizations
in $B\to (K^*,\rho)$ decays. 

\index{large energy limit!corrections}
Corrections to the LEL relations arise from (i) radiative 
corrections to the heavy light vertex, (ii) hard gluon exchange 
with the spectator quark and (iii) nonperturbative corrections which
scale like $\lqcd/E_h$.  $E_h\sim O(m_b)$ is the recoil energy of the 
light hadronic state.  The leading contribution to (i) was calculated 
in Refs.~\cite{benfel00,bfps01}, while (ii) was calculated in 
Ref.~\cite{benfel00}.  An effective field theory formulation of the LEL
appears to be much more complicated than HQET, but there has been much recent
work in this direction \cite{bfl,leleft,bfps01}.  Such a formulation should
allow the nonperturbative corrections (iii) to be parametrized, but thus far
this has not been done.  The theory of exclusive decays in the 
LEL is currently a very active field, and much additional theoretical 
work on this subject is to be expected in the future, in particular 
clarifying the size of the corrections to the limit.

The various LEL relations may be experimentally tested:  
for example, an experimental measurement of the ratio  of the transverse 
polarizations $\Gamma_+/\Gamma_-$ in the semileptonic decay 
$B\to\rho\ell\nu$ will provide a test of the relation (\ref{rvlet}).
In addition,
the relation (\ref{rvlet}), together with the experimental data on 
$b\to s\gamma$ decays, has been used to put constraints 
on the form-factors entering the $B\to K^*$ matrix element at 
$q^2=0$. This can  be seen in Figure~\ref{vvsa1}, from which we 
can fit the vector and axial-vector form-factors giving
$V(0)=0.39\pm0.06$ and $A_1(0)=0.29\pm0.02$. Here, 
\begin{equation}
V\equiv -(m_B+m_{K^*}) g\,, \qquad A_1\equiv \frac{f}{m_B+m_{K^*}}\,,
\label{conversion}
\end{equation}
\begin{figure}[ht]
\leavevmode
\centering
\epsfig{file=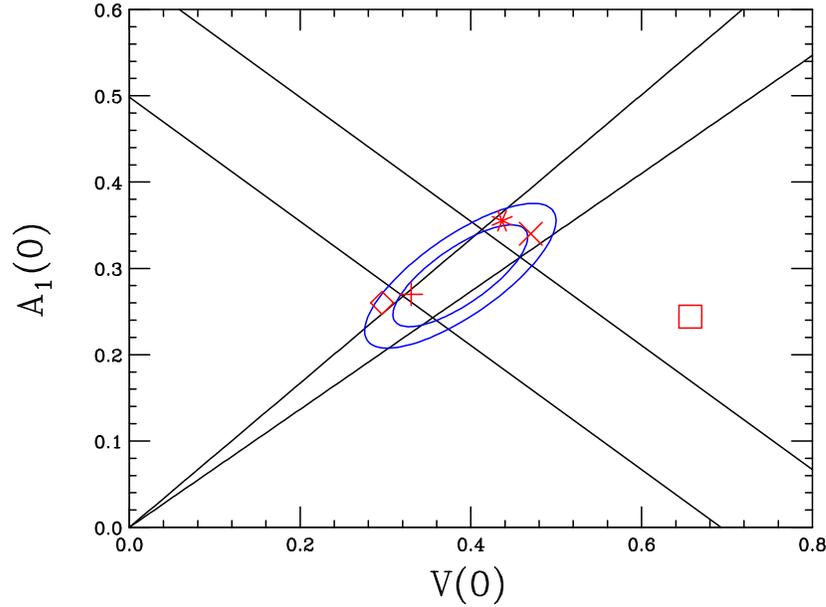,width=8cm,angle=90}
\caption[Constraints on the semileptonic form-factors $V(0)$ and $A_1(0)$ from 
$B\to K^*\gamma$ data plus HQSS together with the relation from the
LEL.]{Constraints on the semileptonic form-factors $V(0)$ and $A_1(0)$ from 
$B\to K^*\gamma$ data plus HQSS (thicker band) together with the relation from
the LEL (cone). The ellipses correspond to $68\%$ and $90\%$ confidence level
intervals.  Central values of model predictions are also shown and correspond
to  BSW~\cite{bsw} (vertical cross), ISGW2~\cite{isgw2} (diamond), MS~\cite{ms}
(star), LCSR~\cite{ABHH00} (diagonal cross) and LW~\cite{zlmw} (square),
respectively.}
\label{vvsa1}
\end{figure}

\subsubsection*{Model Calculations}

\index{quark models}

The model independent statements described above are not enough
to specify all the hadronic matrix elements needed in the decays
of interest. Furthermore, lattice QCD calculations of the rare decay 
form factors are incomplete, at present. 

Model calculations are much easier. On the one hand, they tend to be
based on uncontrolled approximations. This leads to uncertainties which
are difficult (if not impossible) to estimate.
On the other hand, models can provide very useful parameterizations
of the physics and may help us understand the region of validity
of some of those assumptions. 

In heavy-to-light transitions, such as $B\to K^{(*)}\ell^+\ell^-$ 
and $B\to (\pi\rho)\,\ell\nu$, the rate receives most of its 
contributions from the large recoil region where $E_h>1~$GeV. 
We therefore expect those models 
that incorporate -- in one way or another -- our understanding
of the hadronization of a light quark with relatively large energies
to be best suited for these processes.
For example, calculations in the light-cone, performed at $q^2<0$, 
and matched at $q^2=0$ with the physical region~\cite{bsw,fqd}, as well as
light-cone sum rule (LCSR) calculations~\cite{ball} will give the
correct asymptotic behavior of the form factors in QCD as one takes
$q^2\to -\infty$.
Relativistic quark models such as Refs.~\cite{stech,joao,meli}, 
include all relativistic effects from the start, instead of
treating them as corrections.
An important aspect of the transition form factors in these decays, is 
that their $q^2$-dependence may not be trivial. The widely used 
assumption of monopole behavior
\begin{equation}
F(q^2)=\frac{F(0)}{1-q^2/M_*^2}\,,
\label{monop}
\end{equation}
where $F(q^2)$ is a given form factor and $M_*$ is the mass of a nearby
resonance, may receive large corrections in heavy-to-light transitions.
This is not the case in $D$ and $B\to D^{(*)}\ell\nu$ decays, where
the energy release is small compared to the mass gap to heavier 
resonances, and the nearest (or single) pole approximation is good.
In $B\to {\rm light}$ transitions the form factors are sensitive
to the influence of additional resonances at high recoil. In fact, 
a sum rule for the resonance contributions can be derived~\cite{bkdr} 
once the asymptotic behavior in the $q^2\to -\infty$ limit, which
is known in  QCD, is imposed.
This leads to a suppression of the monopole behavior in favor
of a mixed $q^2$ dependence in agreement with QCD 
predictions in the appropriate limit. This also agrees with 
results from LCSR calculations.\index{light-cone sum rules}

In summary, the phenomenological models we consider, capture at least
some of the important physics (especially at high recoil). Until
reliable lattice QCD results come on line, we can combine the model 
results with model independent results from HQET and LEL, as well as 
bounds on form factors from dispersion relations~\cite{drbounds}. 
All of this taken together results in rather constrained form 
factors.
Another strategy for reducing theoretical uncertainties, is the
identification of observables which are insensitive to differences
in the model predictions.

\subsubsection*{The issue with $c \bar{c}$ resonances, cuts 
\label{sec:cc}}

%

\index{$B\to K^*\ell^+\ell^-$!charmonium resonances}
Rare decays receive a contribution from diagrams which contain
$q\bar{q}$ loops.
The $q \bar{q}$ loops can hadronize into vector mesons before decaying 
electromagnetically. The contribution of $c \bar{c}$ loops 
at $q^2$ values near the resonance masses, $q^2 \sim m_V^2$,
where $V=J/\psi, \psi^\prime, \psi^{\prime \prime} \ldots $
is an important background to rare decays. It contributes via
$B \to K,K^* V \to K,K^*  \ell^+ \ell^-$.
Of the six charmonium resonances \cite{pdg98}, the dominant ones, 
$J/\psi(3097)$ and $\psi^\prime(3686)$, divide the spectra
naturally into three regions: a low $q^2$-region below the $J/\psi$, 
a mid $q^2$-region between the $J/\psi$ and the $\psi^\prime$, and a 
high $q^2$-region above the $\psi^\prime$., This is shown in 
Figure~\ref{fig:BKsusy}. The resonance regions can be included into 
the calculation by the parameterization given in \cite{ks96} which is
based on dispersion relations and experimental data on 
$e^+ e^- \to \mbox{hadrons}$. For a discussion of other approaches 
\cite{amm91,lsw98} see Refs.~\cite{ahccbar98,GGG}. 
All methods result in a modification of the function $Y$ in
$C_9^{\rm eff}$ and rely on factorization.

Kinematic cuts in $q^2$ are required to allow a reliable
extraction of the short distance coefficients from experimental
measurements.
Figures~\ref{fig:BKsusy} and \ref{fig:BKstsusy} show the difference
between the differential decay rate for $B \to K,K^\ast \mu^+ \mu^-$
with and without inclusion of the resonant $c\bar{c}$ states.
The lower curves only include non-resonant (or pure short distance)
contributions, while the upper curves also include the contribution
from resonant $c\bar{c}$ states (according to Ref.~\cite{ks96}).
It is clear from these figures that the low $q^2$-region is
the preferred region for comparing theory and experiment, because
this region i) receives the largest contribution to the rate
and ii) is not affected by higher $c \bar{c}$ resonances.


\begin{figure}[p]
\centerline{\epsfysize=3.3in
{\epsffile{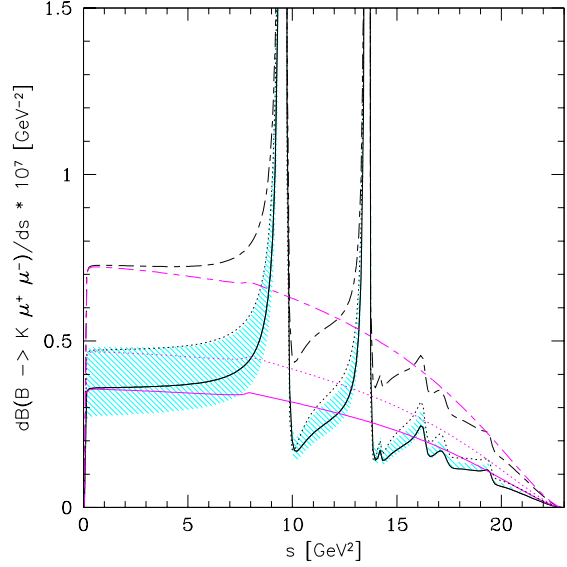}}
}
\caption[The dilepton invariant mass distribution in
$B \to K \mu^+ \mu^-$ decays, using the form factors from LCSR.]
{The dilepton invariant mass distribution in
$B \to K \mu^+ \mu^-$ decays, using the form factors from LCSR.
Resonant $c \bar{c}$ states are parametrized as
in Ref.~\protect\cite{ks96}.
The solid line represents the SM and the shaded area
depicts the form-factor related uncertainties.
The dotted line corresponds to the SUGRA model with
$R_7=-1.2,~R_9=1.03$ and $R_{10}=1$. The long-short dashed
lines correspond to an allowed point in the parameter space of the
MIA-SUSY model, given by $R_7=-0.83$, $R_9=0.92$ and $R_{10}=1.61$.
The corresponding pure short-distance spectra are shown in the 
lower part of the plot. Figure taken from Ref.~\cite{ABHH00}.} 
\label{fig:BKsusy}
\end{figure}
\begin{figure}
$$
\centerline{
\epsfysize=3.3in
{\epsffile{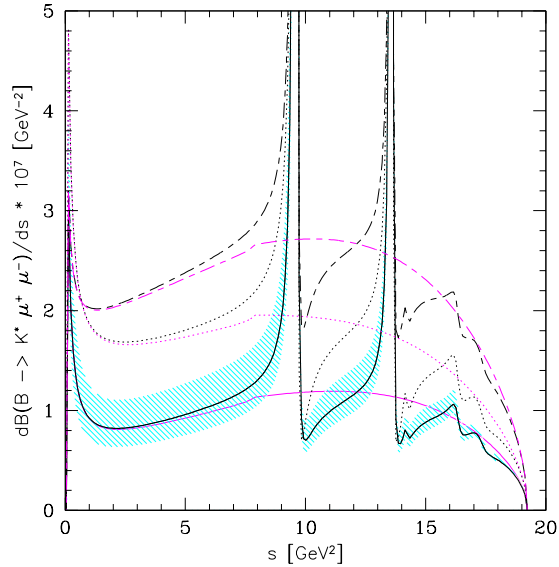}}
}
$$
\vspace*{-30pt}
\caption[The dilepton invariant mass distribution in
$B \to K^* \mu^+ \mu^-$ decays.]
{The dilepton invariant mass distribution in
$B \to K^* \mu^+ \mu^-$ decays.
Legends are the same as in Figure~\protect\ref{fig:BKsusy}.
Figure taken from Ref.~\cite{ABHH00}.} \label{fig:BKstsusy}
\end{figure}


\subsubsection*{Branching ratios and invariant dilepton mass 
distributions}

\index{$B\to K^*\ell^+\ell^-$!SM prediction}
Table~\ref{tab:bounds} lists the non-resonant branching ratios 
for the various $B \to K,K^\ast \ell^+ \ell^-$ channels in the
standard model\footnote{More stringent experimental bounds have recently
been published by CLEO \cite{CLEOBK01}.}. The kinematic range of the dilepton mass is
$4 m_\ell^2 \leq q^2 < (m_B-m_{(K,K^*)})^2$. 
The $B \to K^\ast \ell^+ \ell^-$ decays receive a contribution
from the photon pole, $|C_7^{\rm eff}|^2/q^2$. 
The rate for $B \to K^\ast e^+ e^-$ is enhanced compared to the 
rate for the corresponding decay into muons, because of the
greater sensitivity to the photon pole in the decay into electrons.
(The photon pole is absent for decays into pseudoscalar $K$
mesons, see Figure~\ref{fig:BKsusy}).
\begin{table}[ht]
         \begin{center}
         \begin{tabular}{|l|c|c|r|}
  \hline
     \multicolumn{1}{|c|}{mode}
       & \multicolumn{1}{|c|}{SM branching ratio}
       & \multicolumn{1}{|c|}{data}
       & \multicolumn{1}{|c|}{Exp. reference} \\
         \hline
   $b \to s e^+ e^-$
&$8.4 \pm 2.3 \times 10^{-6}$
& $<10.1 \times 10^{-6}$& BELLE \cite{Abe:2001qh}\\
   $b \to s \mu^+ \mu^-$
&$5.7 \pm 1.2 \times 10^{-6}$
&$<19.1 \times 10^{-6} $ & BELLE \cite{Abe:2001qh} \\
\hline
   $B \to K e^+ e^-$
&$5.7 \pm 2.0 \times 10^{-7}$ &$(0.48^{+0.32+0.09}_{-0.24-0.11}) \times 10^{-6}$
& BELLE \cite{Abe:2001dh} \\
   $B \to K \mu^+ \mu^-$
&$5.7 \pm 2.0 \times 10^{-7}$ &$(0.99^{+0.40+0.13}_{-0.32-0.14})\times 10^{-6}$
&BELLE \cite{Abe:2001dh}  \\
\hline
   $B \to K \ell^+ \ell^-$
&$5.7 \pm 2.0 \times 10^{-7}$ &$(0.75^{+0.25}_{-0.21} \pm 0.09 )\times 10^{-6}$
&BELLE \cite{Abe:2001dh}  \\
   $B \to K \ell^+ \ell^-$
&$5.7 \pm 2.0 \times 10^{-7}$ &$ < 0.6\times 10^{-6}$
&BABAR  \cite{Aubert:2001xt} \\
\hline
   $B \to K^{ \ast} e^+ e^-$ &
$2.3 \pm 0.8 \times 10^{-6}$ &$<5.1 \times 10^{-6} $ &
BELLE \cite{Abe:2001qh}\\
   $B \to K^{ \ast} \mu^+ \mu^-$
&$1.9 \pm 0.7 \times 10^{-6}$
&$<3.0 \times  10^{-6} $& BELLE \cite{Abe:2001qh}\\
         \hline
         \end{tabular}
         \end{center}
\caption[Current status of rare semileptonic $B$ decays.]
{Current status of rare semileptonic $B$ decays. 
SM branching ratios are taken from \cite{ABHH00,ahhmNP}, and 
         upper bounds are given at $ 90 \%$ C.L. 
}
\label{tab:bounds}
\end{table}

The dilepton invariant mass distributions for the 
$B \to K,K^\ast \mu^+ \mu^-$ decays are shown in
Figures~\ref{fig:BKsusy} and \ref{fig:BKstsusy}, respectively.
Imposing the cuts $0.25\, {\mbox{GeV}}^2 \leq q^2 <8.0\, {\mbox{GeV}}^2$
(low $q^2$ region), and including the charmonium resonances according to
Ref.~\cite{ks96} we obtain the following partially integrated standard
model branching ratios: $\Delta {\cal{B}}_H$ for $B \to H \mu^+ \mu^-$:
$\Delta {\cal{B}}_K=2.90 \times 10^{-7}$ and
$\Delta {\cal{B}}_{K^\ast}=7.67 \times 10^{-7}$.
The theoretical uncertainty in these branching fractions has been estimated
to be $\pm 30 \%$ \cite{ABHH00}.

\index{supersymmetry}
For comparison, in a generic non-standard model scenario,
choosing $C_7^{\rm eff}=-C_7^{\rm eff}|_{SM}$ and 
$C_9,C_{10}$ equal to their standard model values, we obtain
$\Delta {\cal{B}}_K=3.63 \times 10^{-7}$ and
$\Delta {\cal{B}}_{K^\ast}=13.09 \times 10^{-7}$. The enhancement results
from constructive interference of $C_7^{\rm eff}$ with $C_9$.

\subsubsection*{The Forward-Backward Asymmetry $A_{FB}(q^2)$ \label{sec:AFB}}

\def\OMIT#1{{}}

As discussed in Sec.~\ref{sec:bsllinclusive}, the forward-backward asymmetry of the leptons in inclusive $b\to s\ell^+\ell^-$ provides a means of measuring the Wilson coefficients
$C_7$, $C_9$ and $C_{10}$.  The latter two
may be sensitive to different aspects of the physics at short distances
and disentangling their contributions, as well as the sign
of $C_7$, will result in additional constraints on New Physics.

In this section we discuss the forward-backward asymmetry for exclusive decays.  
In addition to the branching ratios and the decay distributions, 
exclusive decays to vector mesons carry angular information sensitive 
to the short distance physics.
Here we are concerned with the
potential for cleanly extracting short distance physics from the
asymmetry in exclusive modes, such as $B\to K^*\ell^+\ell^-$,
$B_s\to\phi\ell^+\ell^-$, etc.
In principle, one might expect that theoretical predictions for
exclusive modes are much more uncertain than predictions for inclusive 
decays due to the presence of hadronic form factors.  
However, as we will discuss, the LEL relations~(\ref{le1}-\ref{le7}) 
allow for a clean determination of the Wilson coefficient 
$C_9^{\rm eff}$ in terms of $C_7^{\rm eff}$, through a measurement
of the position of the zero of $A_{FB}(q^2)$.

\index{$B\to K^*\ell^+\ell^-$!angular distribution}
The angular distribution in $B\to K^*\ell^+\ell^-$ is 
given by
\begin{eqnarray}
\frac{d^2\Gamma}{dq^2 
d\cos\theta}&=&\frac{G_{F}^{2}\alpha^2\left|V^{*}_{tb}V_{ts}\right|^2}
{768\pi^5m_{B}^{2}}\,\bm{k}q^2\bigg\{(1+\cos\theta)^2
\left[|H_{+}^{L}|^2+|H_{-}^{R}|^2\right] \nonumber \\*
&&{} +(1-\cos\theta)^2\left[|H_{-}^{L}|^2+|H_{+}^{R}|^2\right]
+2\sin^2\theta\, |H_{0}|^2\bigg\}\,,\label{dg2}
\end{eqnarray}
where $\bm{k}$ is the $K^*$ spatial momentum, and $\theta$ is the angle
between the $\ell^+$ and the $B$ meson in the dilepton center-of-mass 
frame.
The transverse helicity amplitudes in terms of the form factors
take the form~\cite{test}
\begin{mathletters}
\label{helamps}
\begin{eqnarray}
H_{\alpha}^{L}&=& \left[C_7\frac{m_b\left(m_B-E_{K^*}
+\eta_\alpha \bm{k} \right)}{q^2}
+\frac{C_9-C_{10}}{2}\right]\left(f+\eta_\alpha 2m_B \bm{k}g\right)\,,
\label{heleft}\\
H_{\alpha}^{R}&=& \left[C_7\frac{m_b\left(m_B-E_{K^*}+
\eta_\alpha \bm{k}\right)}{q^2}
+\frac{C_9+C_{10}}{2}\right]\left(f+\eta_\alpha 2m_B \bm{k}g\right)\,,
\label{heright}
\end{eqnarray}
\end{mathletters}%
where $\alpha=+,-$, $\eta_\alpha=(1,-1)$, and $E_{K^*}$ is the $K^*$ 
energy in the $B$ rest frame. The index $\alpha$ in Eqs.~(\ref{helamps})
refers to the $+,-$ polarizations of the $K^*$, and the $L,R$ subscripts
refer to left and right-handed leptons.
The longitudinal helicity amplitude is described by
\begin{eqnarray}
H_0^L= \frac{m_{B}^2}{m_{K^*}\sqrt{q^2}}&&\left\{C_7\frac{m_b}{q^2m_B}
\left\{f[E_{K^*}(m_B-E_{K^*})- \bm{k}^2]
+2g\,m_B{\bf k}^2(m_B-2E_{K^*})\right\} \right. \nonumber \\
& &{}+\left. \frac{(C_9-C_{10})}{2} 
\left[2\bm{k}^2a_+-\frac{E_{K^*}}{m_B}f\right]
\right\}\,, \label{hel_long}
\end{eqnarray}
and $H_0^R$ is given by replacing $(C_9-C_{10})/2$ with
$(C_9+C_{10})/2$ in Eq.~(\ref{hel_long}).

The forward-backward asymmetry for leptons as a function of the
dilepton mass squared $m_{\ell\ell}^2=q^2$ is now defined as
\index{$B\to K^*\ell^+\ell^-$!forward-backward asymmetry}
\begin{equation}
A_{FB}(q^2)=\frac{\displaystyle
\int_{0}^{1}\frac{d^2\Gamma}{dxdq^2}\, dx -
\int_{-1}^{0}\frac{d^2\Gamma}{dxdq^2}\, dx
}
{\displaystyle \frac{d\Gamma}{dq^2}}  \,,
\label{afbdef}
\end{equation}
where $x\equiv\cos\theta$.
We can write $A_{FB}$ in terms of the helicity amplitudes defined
in Eqs.~(\ref{helamps}) and (\ref{hel_long}):\footnote{The sign of 
$A_{FB}(q^2)$ defined in this way will change when considering 
$\bar B^0$ or $B^-$ decays.}
\begin{equation}
A_{FB}(q^2)=\frac{3}{4}\,\frac{|H_{-}^{L}|^2+|H_{+}^{R}|^2-|H_{+}^{L}|^2
-|H_{-}^{R}|^2}
{|H_{-}^{L}|^2+|H_{+}^{R}|^2+|H_{+}^{L}|^2+|H_{+}^{R}|^2+|H_{0}^L|^2
+|H_{0}^R|^2}\,.
\label{afb_hel}
\end{equation}
As it can be seen from Eqs.~(\ref{helamps}) and (\ref{afb_hel}),
the asymmetry is proportional to the Wilson coefficient $C_{10}$
and vanishes with it.
Furthermore, it is proportional to a combination of $C_9^{\rm eff}$ and
$C_7^{\rm eff}$ such that it has a zero in the physical region
if the following condition is satisfied~\cite{afb0}
\begin{equation}
\Re[C_9^{\rm eff}]=-\frac{m_b}{q^2_0}\,C_7^{\rm eff} \left\{
\frac{T_1}{g} + (m_B^2-m_{K^*}^2)\,\frac{T_2}{f}\right\}\,,\label{afbz}
\end{equation}
where $q^2_0$ is the position of the zero of $A_{FB}$ and all 
$q^2$-dependent quantities are evaluated at $q^2_0$. 
This relation depends on the form factors $T_1$ and $T_2$; however, 
it was noted in Ref.~\cite{afb0} that the location of the zero of 
the asymmetry was approximately constant in a variety of form-factor models,
as shown in Figure~\ref{afbps}.   This is a consequence of helicity conservation of the $K^*$
in the large energy limit in these models, arising from the relativistic treatment of quark spin.
\begin{figure}[ht]
\leavevmode
\centering
\epsfig{file=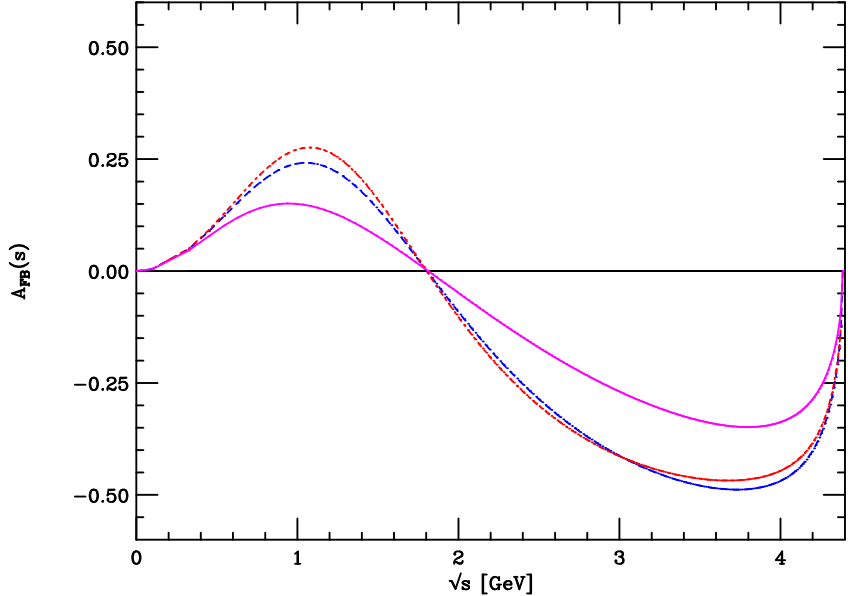,width=8cm,angle=90}
\caption[The non-resonant forward-backward asymmetry of leptons $A_{FB}$ for
$B\to K^* e^+ e^-$ as a function of the dilepton mass $s$.]{The non-resonant
forward-backward asymmetry of leptons $A_{FB}$ defined in (\ref{afbdef}), for
$B\to K^* e^+ e^-$ as a function of the dilepton mass $s$, from
Ref.~\cite{afb0}. The asymmetry is computed by making use of the semileptonic
form factors from the BSW* model of Ref.~\cite{bsw}  (solid line), the
light-cone QCD sum rule calculation of Ref.~\cite{ball} (dashed line) and the
relativistic quark model  of Ref.~\cite{meli} (dot-dashed line). 
}
\label{afbps}
\end{figure}

\index{large energy limit!zero in forward-backward asymmetry}
The model independence of the location of the zero in the asymmetry was 
shown to be a consequence of the large energy limit of QCD in 
Ref.~\cite {ABHH00}. After we apply the 
HQSS and LEL relations (\ref{le7}) to (\ref{afbz}) we find
\OMIT{Making use of the HQSS relations (\ref{rel_2}) and 
(\ref{rel_3})
for $T_1(q^2)$ and $T_2(q^2)$, the form of Eq.~(\ref{afbz}) simplifies
to
\begin{equation}
Re[C_9^{\rm eff}]=-\frac{m_b}{2\,q_0^2}\,C_7^{\rm eff}
\left\{4m_Bk^2\,R_V+\frac{1}{m_B\,R_V}+4(m_B-E_V)\right\}\,,
\label{con2}
\end{equation}
where $R_V$ is the ratio of vector to axial-vector form factors
defined in (\ref{rvlet}) and evaluated at $q_0^2$. Thus, the 
determination
of the zero of $A_{FB}(q^2)$ in $B\to K^*\ell^+\ell^-$
gives a relation between the short distance
Wilson coefficients $C_9^{\rm eff}$ and $C_7^{\rm eff}$ where the only
uncertainty
from form factors is in the ratio $R_V$.
As mentioned earlier, since (\ref{con2}) was derived using the heavy 
quark
{\em spin} symmetry, it is expected receive small corrections.
In principle, information on $R_V(q^2)$ could be extracted from
the $B\to\rho\ell\nu$ decay.
On the other hand, the LEL predicts $R_V(q^2)=-1/(2m_BE_{K^*})$ as 
extracted
from Eqs.~(\ref{le4}) and (\ref{le5}). Then, the condition for the
vanishing of $A_{FB}(q^2)$ now reads}
\begin{equation}
Re[C_9^{\rm eff}]=-{2 m_B m_b\over q_0^2} C_7^{\rm eff}+\dots
\label{condlet}
\end{equation}
where we have neglected the mass of the $K^*$, which is formally subleading.

Radiative corrections to the location of the asymmetry zero (\ref{condlet}) were calculated in
\cite{benfel00,bfps01,benfelseid01} and are at the few percent level.  
Ref.~\cite{benfelseid01} finds
the location of the zero in the SM to be
\begin{equation}
q_0^2=4.2\pm 0.6\,\gev^2
\end{equation}
where the largest uncertainty comes from the scale dependence of the Wilson coefficients
and the estimate of formally subleading $O(m_{K^*}^2/m_B^2)$ terms.

We conclude that the measurement of the zero of the forward-backward 
asymmetry for leptons in $B\to K^*\ell^+\ell^-$ constitutes a test of 
the short distance
structure of the Wilson coefficients $C_9$ and $C_7$ through
Eq.~(\ref{condlet}).  It should be stressed that there are unknown 
nonperturbative corrections
to the relation (\ref{condlet}) which formally scale as $\lqcd/E_H$, but 
whose size is unknown.
Thus, experimental tests of LEL relations will be important in establishing
the reliability and accuracy of this approach.

\subsubsection*{The Forward-Backward CP Asymmetry \label{subsec:AFBCP}}

\index{$B\to K^*\ell^+\ell^-$!$CP$ asymmetries}
The forward-backward CP asymmetry ($A^{CP}_{FB}$) has been proposed in
Ref.~\cite{GGG} as an observable to probe non-standard CP violation in 
FCNC Z-penguins. It is defined from the forward-backward asymmetry of
the previous subsection as
\begin{eqnarray}
  A^{CP}_{FB}(s) = \frac{ A^{(\bar B)}_{FB}(s) +
  A^{(B)}_{FB}(s)}{A^{(\bar B)}_{FB}(s) - A^{(B)}_{FB}(s)}\,.
  \label{eq:FBCPdef}
\end{eqnarray}
This definition isolates the phase of $C_{10}$ and the effect scales
in units of the phase of $C_9^{\rm eff}$, which has a CP conserving
phase encoded in the function $Y$ from the 4-Fermi operators
$ A^{CP}_{FB}(s) \sim (\Im C_{10} /\Re C_{10}) (\Im C_9^{\rm eff}/\Re
C_9^{\rm eff})$, see Ref.~\cite{GGG} for details.
Using the high $q^2$ integration region above the $\psi^\prime$
(only here $\Im C_9^{\rm eff}$ is sizeable)
$14.5 \mbox{GeV}^2 \leq q^2 < (m_B-m_{K^*})^2$
yields
$\Delta A^{CP}_{FB}= (0.03 \pm 0.01) \times  \Im C_{10} /\Re C_{10}$.
Despite the substantial uncertainties related
with higher $\psi^{\prime \prime ..}$ resonances, the forward-backward CP
asymmetry is a unique probe of the
flavor sector, since the SM background due to
CKM phases is very small $<10^{-3}$ and $\Delta A^{CP}_{FB}$ can be
sizeable in case of large CP violating phases of $C_{10}$.

\subsubsection*{CP asymmetries in the rate}

\index{$B\to K^*\ell^+\ell^-$!$CP$ asymmetries}
\index{$B\to K\ell^+\ell^-$!$CP$ asymmetries}
We define a direct CP violating asymmetry distribution between
the dilepton mass spectra in $\bar{B} \to \bar{H} \ell^+\ell^-$ decays
and the CP conjugate process $B \to H \ell^+\ell^-$ as \cite{KR00}%
\footnote{An alternative definition is
$A^{CP\, \prime}_H(s) =
\Big(\frac{d \Gamma_{H}^{\bar{B}}}{d s }
-\frac{d \Gamma_{H}^B}{ds } \Big)/(\Gamma_{H}^{\bar{B}}+\Gamma_{H}^{B})$
\cite{alihillerckm,LS99}. \label{foot:acp}}
\begin{eqnarray}
A^{CP}_{H} (s)=
\frac{\displaystyle\frac{d \Gamma_{H}^{\bar{B}}}{d s }
-\frac{d \Gamma_{H}^B}{ds } }
{\displaystyle \frac{d \Gamma_{H}^{\bar{B}}}{ds }
+\frac{d \Gamma_{H}^B}{ds }}
\end{eqnarray}
Here $H$ can be a pseudoscalar or vector final state meson, for
example, $K$ or $K^*$.
For a non-zero $A^{CP}_H$ in the SM we have to reintroduce the CKM
suppressed piece $\lambda_u (T_u-T_c)$ (see Section~\ref{sec:motivation})
into the amplitude, e.g. \cite{alihillerckm}.
In addition to the charmonium background discussed in Section~\ref{sec:cc} 
now
intermediate $u \bar{u}$
resonances $\rho,\omega$ have to be taken into account \cite{KSCP}.
To reduce the related uncertainties one uses kinematical cuts
$q^2 \gsim m_{\rho}^2, m_{\omega}^2$
analogous to the $c\bar{c}$ sector.

Unlike the radiative modes induced by $b \to s \gamma$ where
$A^{CP}_{\gamma} \sim \alpha_s$ \cite{bsgCP}, the
CP asymmetry in $b \to s \ell^+ \ell^-$ transitions starts at the lowest 
order:
the SM contribution to $A^{CP}_H$ stems from interference between the
weak phase and the CP conserving imaginary part of $C_9^{\rm eff}$.
Both lead to very small values of $A^{CP}_H$ in the SM:
as in any $b \to s$ transition CKM structure dictates
$A^{CP} \sim \Im(\lambda_u/\lambda_t)=\lambda^2
\eta< 2 \%$,
where $\lambda,\eta$ are Wolfenstein parameters.
The second suppression comes from the strong phase $\Im Y(q^2) \ll C_9$,
which holds everywhere except at $q^2 \sim m_V^2$.
Integrating $A^{CP}_H(s)$ over the low $q^2$-region
$1.4 \mbox{GeV}^2 \leq q^2 < 8.4 \mbox{GeV}^2$
yields
$A^{CP}_{K,K^\ast}\simeq 0.1 \%$ in the SM \cite{KR00}, comparable with 
the
findings for the inclusive $B \to X_s  \ell^+ \ell^-$ decays
$(a_{CP})_s=-0.19^{+0.17}_{-0.19} \%$
\cite{alihillerckm}, which uses
different cuts ($1 \mbox{GeV}^2 \leq q^2 < 6 \mbox{GeV}^2$) and a
slightly different CP asymmetry (sign and
normalization), see footnote \ref{foot:acp} and Ref.~\cite{alihillerckm}
for details.

Supersymmetric effects in the CP asymmetry in exclusive
$B \to K,K^* \ell^+\ell^-$ decays have been
studied in Refs.\cite{KR00,LS99}. The presence of non-SM CP phases
can change the sign and magnitude of $A^{CP}_H$: In the low $q^2$-region,
$1.4 \mbox{GeV}^2 \leq q^2 < 8.4 \mbox{GeV}^2$, the integrated asymmetry
is still not
large $|A^{CP}_{K,K^\ast}| \lsim 1 \%$ \cite{KR00}, but can exceed the 
SM asymmetry.

\subsubsection{$B\rightarrow K^*\gamma$ and Related Decays \label{sec:radiative}}

\index{$B\to K^*\gamma$}
Exclusive radiative decays are experimentally relatively easily 
accessible, since their final states can be completely reconstructed. 
The study of these decays is well motivated as they can provide 
information on the ratio of CKM elements $V_{td}/V_{ts}$, and assist 
in the reduction of the theoretical error on the determination of 
$V_{ub}$ from $B\to\rho\ell\nu$. In addition they are sensitive to 
loop effects of new interactions which may result in CP violating 
effects in the charge asymmetry of $B\to K^*\gamma$. 
Unfortunately, these transitions are also sensitive to theoretical 
uncertainties of two different origins.
First, there is the uncertainty due to the poorly known hadronic matrix 
elements of the short distance operators which contribute to the rate. 
The second uncertainty is due to long distance contributions (see the discussion in Section~\ref{sec:meff}). 
More theoretical effort is needed in this area. 
At present, CLEO has observed two channels \cite{cleobsg}, 
with the branching fractions, \index{$B\to K^*\gamma$!measured branching fractions}
$B(B^0\to K^{*0}\gamma)=(4.55~^{+0.72}_{-0.68}\pm 0.34)\times 10^{-5}$, 
$B(B^+\to K^{*+}\gamma)=(3.76~^{+0.89}_{-0.83}\pm 0.28)\times 10^{-5}$, 
and 
$B(B\to K^*_2(1430)\gamma)=(1.66~^{+0.59}_{-0.53}\pm 0.13)\times 10^{-5}$.

As shown in Section~\ref{sec:meff}, the short distance contribution to 
radiative decays depends on only one form factor, $T_1(q^2=0)$. This 
form factor has been calculated from a wide variety of theoretical 
approaches. A sampling of some more recent results \cite{excff} is 
given in Table~\ref{7:excradb} for the case of $B\to K^*\gamma$. 
The LCSR 
\index{light-cone sum rules}
results listed there are in good agreement with the CLEO data. 
However, Table~\ref{7:excradb} also shows that there are significant 
differences among the theoretical predictions of the form factor 
$T_1^{B\to K^\ast}(q^2=0)$, and the related ratio of rates for exclusive 
to inclusive decays, $R_{K^*}$.

\index{lattice QCD!predictions for $B\to K^*\gamma$}
We note that the lattice results shown in Table~\ref{7:excradb}
do not contain a complete analysis of all systematic errors.
The calculations date back to 1994 and 1995, a time when improved
actions and heavy quark methods were just being established.
The results were obtained in the quenched approximation. 
Both calculations use pole dominance to extrapolate the form
factors from the high $q^2$ region (where it was calculated)
to the physical $q^2 = 0$ point. Both calculations are performed
at heavy quark masses below the $b$ quark mass, and they both
rely on heavy quark extrapolations to obtain results for 
the $B$ meson decay. 

Estimates for the rates of decays into higher $K^*$ resonances 
are cataloged in Ref.~\cite{highk}.
In the case of the $K_2^*(1430)$ mode, the CLEO data appears to
favor the model of Veseli and Olsson, which predicts 
$B(B\to K^*_2(1430)\gamma)=(1.73\pm 0.80) \times 10^{-5}$.
\index{$B\to K_2^*(1430)\gamma$}  
We note that theoretical predictions do not yet exist for the decay
$B_s\to\phi\gamma$, which is not accessible to the $B$ factories.

\begin{table}
\centering
\begin{tabular}{c|c|c|c|c} \hline\hline
Ref. & $T_1^{B\to K^*}(0)$ & ${T_1^{B\to K^*}(0)\over T_1^{B\to\rho}(0)}$
& $B(B\to K^*\gamma) (\times 10^{-5})$ & $R_{K^*}$ \\ \hline
LCSR & $0.32\pm 0.05$ & $1.32\pm 0.1$ & $4.8\pm 1.5$ &
$0.16\pm 0.05$\\
LCSR & $0.31\pm 0.04$ & $1.14\pm 0.02$ & $4.45\pm 1.13$ &
$0.16\pm 0.05$ \\
LCSR & $0.38\pm 0.06$ & $1.33\pm 0.13$ & $--$ &
$0.20\pm 0.06$ \\
LQCD & $0.10\pm 0.01\pm 0.3$ &$--$ &$--$ &
$0.060\pm 0.012\pm 0.034$ \\
LQCD & $0.16^{+0.02}_{-0.01}$ & $--$ & $--$ &
$0.16^{+0.04}_{-0.03}$\\
\hline\hline
\end{tabular}\vspace*{4pt}
\caption[Form-factor predictions from Ref.~\cite{excff}.]
{Form-factor predictions from Ref.~\cite{excff}. LCSR denotes
calculations based on light-cone sum rules, and LQCD denotes
calculations based on lattice QCD.} 
\label{7:excradb}
\end{table}

We can determine the ratio of CKM elements, $V_{td}/V_{ts}$ from
the ratio of exclusive decay rates,
\index{$B\to\rho\gamma$}
\begin{equation}
\label{7:vtd}
{\Gamma(B\to\rho\gamma)\over\Gamma(B\to K^*\gamma)}=\Phi 
{|T_1^{B\to\rho}(0)|^2
\over |T_1^{B\to K^*}(0)|^2} {|V_{td}|^2\over |V_{ts}|^2}\,.
\end{equation}
$\Phi$ is a phase space factor.
The ratio of form factors in Eq.~(\ref{7:vtd}), 
$T_1^{B\to\rho}(0)/T_1^{B\to K^*}(0)$, is mostly sensitive to
SU(3) breaking effects. Since other theoretical uncertainties
are likely to cancel, the ratio may be more accurately calculated 
than the form factors themselves.

Eq.~(\ref{7:vtd}) assumes that the decay rates are dominated by 
contributions from the short distance operator. This is
the case for $B\to K^*\gamma$, where long distance effects have been
estimated to be no more than $\simeq 5\%$ \cite{ld_bkg,grinpir}, and 
where the theoretical estimates of $R_{K^*}$ are tend to be
consistent with experiment.
\index{$B\to\rho\gamma$!long distance contribution}
However, the long distance contributions to $B\to \rho\gamma$ can be
large \cite{grinpir,ld_rhog} and can potentially destroy the 
validity of Eq.~(\ref{7:vtd}), since they have a different CKM dependence.
These contributions arise from (i) the decay $B\to\rho V^*$ 
(which is due to the contributions of internal $c$- and $u$-quark loops) 
with the subsequent conversion of the neutral vector meson $V^*$ to a 
photon, (ii) weak annihilation and $W$ exchange diagrams with subsequent 
$\gamma$ radiation, and (iii) final state interactions. 
If these these effects are included, the resulting theoretical error 
in determinations of $V_{td}$ from Eq.~(\ref{7:vtd}) has been
estimated to be $\sim 35\%$ \cite{grinpir}.

Finally, we consider the radiative baryon decay, $\Lambda_b\to\Lambda\gamma$.
This decay 
is well suited for the hadron collider environment, and has an estimated Standard 
Model branching ratio of 
$B(\Lambda_b\to\Lambda \gamma)\sim 5 \times 10^{-5}$
\cite{lambda_b, hklambda_b}.
Like the corresponding decays of $B$ mesons, the underlying quark 
transition is $b \to s \gamma$ and described
by the short distance effective Hamiltonian in 
Eq.~(\ref{eq:heffequation}).
However, the spin $1/2$ baryons makes more degrees of freedom 
accessible to experiments.
In particular, one can probe the $V-A$ structure of the Standard Model
and search for contributions of non-standard helicity in the
FCNC dipole operator.

Measurement of final state polarization in 
$\Lambda_b\to (\Lambda\to \pi p) \gamma$ decays has been recently discussed
in \cite{hklambda_b} along with asymmetries related to 
initial $\Lambda_b$ polarization. This work corrected the expression
for the $\Lambda$ polarization asymmetry of the original work
\cite{lambda_b}. (The older calculation was not in agreement with
existing calculations of general spin correlations 
for baryon $\to$ baryon-vector decays).
Note that the $\Lambda$ asymmetry observable is theoretically simple,
since the amplitude into an on-shell photon involves only one form
factor, which drops out in the asymmetry \cite{hklambda_b}.
The relevant single form
factor can be extracted from $\Lambda_c\to\Lambda\ell\bar\nu_\ell$  decays
using heavy quark spin and flavor symmetry \cite{lambda_b} and can
be used for an estimate of the new physics reach 
\cite{hklambda_b}.
In addition, the long distance effects due to vector-meson dominance
and weak annihilation diagrams are estimated to be small
\cite{lambda_b}. 
Hence, $\Lambda_b\to\Lambda\gamma$  
decays is dominated by short distance physics and is particularly 
clean, theoretically. It offers unique opportunities to test the
helicity structure of the underlying theory, but also to study CP violation
\cite{hklambda_b}
by comparing decays of the $\Lambda_b$ and its conjugate
 $\bar{\Lambda}_b$.

\subsubsection{$B_{s,d} \to \ell^+ \ell^-$ \label{sec:btoll}}

\index{$B_{s,d} \to \ell^+ \ell^-$}
The decay $B_q \to \ell^+ \ell^-$, where $q = d$ or 
$s$ and $\ell = e$, $\mu$ or $\tau$,
proceeds through loop diagrams and is of fourth order in the weak
coupling.  In the SM, the dominant contributions to this decay come
from the $W$ box and $Z$ penguin diagrams shown in Figure~\ref{Bllfigure}.
\begin{figure}[bthp]
\centerline{\epsfysize=3.3in
{\epsffile{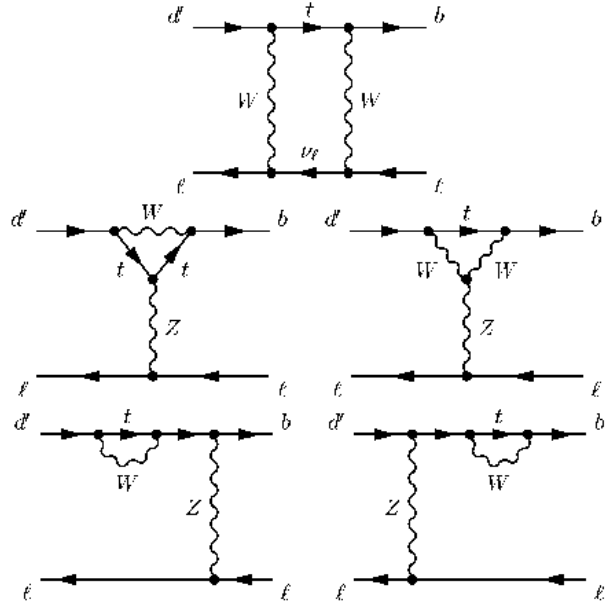}}
}
 \caption{Dominant SM diagrams for $B_{s,d} \to \ell^+ \ell^-$.}
 \label{Bllfigure}
\end{figure}
Because the contributions with a top quark in the loop are dominant,
at low energies of order $m_b$ the decay can be described by a local
$\bar b q \bar \ell \ell$ coupling via the effective 
Hamiltonian,
\begin{equation}
	\mathcal{H}_{\rm eff} = -4 \frac{G_F}{\sqrt{2}} V^*_{tq}
	V_{tb} \left[C_{10} O_{10} + C_S O_S + C_P O_P \right],
\end{equation}
where $O_{10}$ is given in Eq.~\ref{eq:operatorbasis} and the other
two operators are
\index{effective Hamiltonian!for $B_{s,d} \to \ell^+ \ell^-$}
\begin{equation}
	O_S \, = \, -\frac{e^2}{16 \pi^2} \bar q_{L\alpha}
	b_{R\alpha} \bar \ell \ell, \qquad
	O_P \, = \, -\frac{e^2}{16 \pi^2} \bar q_{L\alpha}
	b_{R\alpha} \bar \ell \gamma_5 \ell,
\end{equation}
where we have neglected contributions proportional to the $q$ mass.
The vector leptonic operator $\bar \ell \gamma^{\mu} \ell$
does not contribute for on--shell leptons because it gives zero when
contracted with the $B_q$ momentum.

The diagrams in Figure~\ref{Bllfigure} were calculated
in \cite{InamiLim} and contribute only to the
Wilson coefficient $C_{10}$.  There is no contribution from
a photonic penguin because of the photon's purely vector coupling to 
leptons.
There are also contributions to the Wilson coefficient $C_S$ from a
SM Higgs penguin \cite{GK}
and to the Wilson coefficient $C_P$ from the would--be neutral Goldstone
boson penguin \cite{Krawczyk},
but these contributions to the amplitude are suppressed by a factor
of $m_b^2/M_W^2$ relative to the dominant contributions and can be 
ignored.  We keep $C_S$ and $C_P$ here for completeness because
they can be significant in some extensions of the SM.
A recent review may be found in \cite{bbl}.

The Wilson coefficients are evaluated at the high scale 
$\sim \mathcal{O}(M_W)$ and then run down to the low scale 
$\sim \mathcal{O}(m_B)$, where the hadronic matrix elements
of the operators are evaluated.  This running in general leads to 
QCD corrections enhanced by large logarithms of the ratio of scales, which
must be resummed. \index{effective Hamiltonian!renormalization group}
The operator $O_{10}$ 
has zero anomalous dimension because it is a
$(V-A)$ quark current, which is conserved in the limit of vanishing
quark masses.  
Thus the renormalization group evolution of $C_{10}$ is trivial.
The operators $O_S$ and $O_P$ have the same form as a quark mass term
and thus have the anomalous dimension of a quark mass.  In the SM and many
extensions,
$C_S$ and $C_P$ are proportional to $m_b$.  Thus the running 
of these Wilson coefficients is properly taken into account by replacing
$m_b(M_W)$ with $m_b(m_B)$ in $C_S$ and $C_P$.

\index{$B_{s,d} \to \ell^+ \ell^-$!branching ratio}
Evaluating the hadronic matrix elements, the resulting branching ratio is
\begin{eqnarray}
	\mathcal{B}(B_q \to \ell^+\ell^-)
	&=& 
	\frac{G_F^2 \alpha^2 m_{B_q}^3 
        \tau_{B_q} f^2_{B_q}}
        {64 \pi^3}
        |V^*_{tb} V_{tq}|^2 
        \sqrt{1 - \frac{4 m_{\ell}^2}{m_{B_q}^2}} \nonumber \\
        &&\hspace{-3cm} \times 
        \left[ \left(1 - \frac{4 m_{\ell}^2}{m_{B_q}^2} \right)
        \left| \frac{m_{B_q}}{m_b + m_q} C_S 
        \right|^2
        + \left| \frac{2 m_{\ell}}{m_{B_q}} C_{10}
        - \frac{m_{B_q}}{m_b + m_q} C_P \right|^2
        \right],
\end{eqnarray}
where $\tau_{B_q}$ is the $B_q$ lifetime,
$f_{B_q}$ is the $B_q$ decay constant
normalized according to $f_{\pi} = 132$ MeV, and we retain the 
Wilson coefficients $C_S$ and $C_P$ for completeness.

The SM decay amplitude is given by the Wilson coefficient \cite{InamiLim}
\index{effective Hamiltonian!Wilson coefficients in SM}
\begin{equation}
	C_{10} = - Y(x_t) / \sin^2 \theta_W,
\end{equation}
where $x_t = m_t^2(m_t)/M_W^2 = 4.27 \pm 0.26$ with $m_t$ evaluated in 
the
$\overline{\rm MS}$ scheme at $\mu = m_t$ (giving $m_t(m_t) = 166$ GeV).
The function $Y(x_t)$ is given by
$Y(x_t) = Y_0(x_t) + \frac{\alpha_s}{4\pi}Y_1(x_t)$ at NLO.
At LO \cite{InamiLim},
\begin{equation}
	Y_0(x_t) = \frac{x_t}{8} \left[ \frac{x_t-4}{x_t-1}
	+ \frac{3x_t}{(x_t-1)^2} \log x_t \right]
	= 0.972 \left[ \frac{m_t(m_t)}{166 \, {\rm GeV}} \right]^{1.55},
\end{equation}
where we have taken the central value of $x_t$ and parameterized
the remaining $m_t$ dependence.

As explained above, the operator $O_A$ has zero anomalous dimension and 
so the
QCD running of the Wilson coefficient from the electroweak scale
to the $B_q$ mass scale is trivial.
Nontrivial QCD corrections first arise at
NLO and require the calculation of two-loop diagrams \cite{bb,mu}.
The result of the two-loop calculation in the $\overline{\rm MS}$
scheme is \cite{mu}
\begin{eqnarray}
	Y_1(x) &=& \frac{x^3 + 2x}{(x-1)^2}{\rm Li}_2(1-x)
	+ \frac{x^4 - x^3 + 14x^2 - 2x}{2(x-1)^3} \log^2 x \nonumber \\
	& & + \frac{-x^4 - x^3 - 10x^2 + 4x}{(x-1)^3} \log x
	+ \frac{4x^3 + 16x^2 + 4x}{3(x-1)^2} \nonumber \\
	& & + \bigg[ \frac{2x^2 - 4x}{(x-1)} + \frac{-x^3 + 7x^2}{(x-1)^2}
	+ \frac{-6x^2}{(x-1)^3} \log x \bigg]
	\log \bigg(\frac{\mu^2}{M_W^2}\bigg).
\end{eqnarray}
Here $\mu$ is the renormalization scale at which the top quark mass is
renormalized.
Numerically, $Y_1(x_t) = 2.65$, and the NLO contributions give (taking
the central value of $x_t$)
$Y(x_t) = 1.026 \times Y_0(x_t) = 0.997$.

\index{$B_{s,d} \to \ell^+ \ell^-$!SM prediction}
The SM predictions for the branching fractions are given in 
Table~\ref{table:BllSMBRs}, where parameter and hadronic uncertainties
have been taken into account.\footnote{We use the following parameters: 
$\alpha = 1/128$ (at $M_Z$), 
$s^2_W = 1 - M_W^2/M_Z^2 = 0.2222$, 
$\bar m_t(m_t) = 166 \pm 5$ GeV,
$m_{B_d} = 5279.4$ MeV,
$m_{B_s} = 5369.6$ MeV,
$\tau_{B_d} = 1.548 \pm 0.032$ ps,
$\tau_{B_s} = 1.493 \pm 0.062$ ps,
$|V_{tb}| = 0.999$,
$|V_{td}| = 0.009 \pm 0.003$,
$|V_{ts}| = 0.039 \pm 0.002$,
$f_{B_d} = 208 \pm 10 \pm 11$ MeV, and
$f_{B_s} = 250 \pm 10 \pm 13 ^{+8}_{-0}$ MeV.
All numbers are taken from the PDG~\cite{pdg} except for $f_{B_d}$ 
and $f_{B_s}$ which are taken from Ref.~\cite{CPPACS}.
For $f_{B_d}$ and $f_{B_s}$ the statistical and systematic errors
are listed separately, and the third error for $f_{B_s}$ comes from 
the uncertainty in the strange quark mass.}

\begin{table}
\begin{center} 
\begin{tabular}{l|cc} \hline\hline
& $B_d$ & $B_s$ \\ \hline
$\tau^+\tau^-$ & $3.4^{+2.7}_{-2.0} \times 10^{-8}$
	& $9.2^{+1.9}_{-1.8} \times 10^{-7}$ \\
$\mu^+ \mu^-$ & $1.6^{+1.3}_{-0.9} \times 10^{-10}$ 
	& $4.3^{+0.9}_{-0.8} \times 10^{-9}$ \\
$e^+ e^-$ & $3.8^{+3.0}_{-2.2} \times 10^{-15}$ 
	& $1.0 \pm 0.2 \times 10^{-13}$ \\ \hline\hline
\end{tabular} 
\end{center}
\caption[SM branching ratios for $B_d$ and $B_s$ into  $\tau^+\tau^-$, $\mu^+
\mu^-$ and $e^+ e^-$.]{SM branching ratios for $B_d$ and $B_s$ into 
$\tau^+\tau^-$, $\mu^+ \mu^-$ and $e^+ e^-$. The difference in the relative
size of the errors in the $B_d$ and $B_s$ branching ratios  is due primarily to
the difference in  the relative size of the errors in $V_{td}$ and $V_{ts}$.}
\label{table:BllSMBRs} 
\end{table}

The uncertainties in the branching ratios are due primarily to the 
uncertainties in $|V_{td}|$, $|V_{ts}|$, and  $f_{B_q}$. An
additional uncertainty in the branching ratios  due to scheme dependence in the
definition of $\sin^2 \theta_W$ is not taken into account;  we estimate it to
be about 8\%.

\subsection{Expectations for Physics Beyond the Standard Model} %
\label{sec:beyondSM}

\subsubsection*{Supersymmetry}

\label{supersymmetry!effects on Wilson coefficients}
It is customary to define ratios of Wilson coefficients renormalized
at a scale $\mu=m_b$
\begin{eqnarray}
R_i \equiv  \frac{C_i}{C_{i \, SM}}
\end{eqnarray}
parameterizing possible enhancement/decrease w.r.t. the SM
Wilson coefficients.
Analytical expressions of the MSSM $C_i$ are given in \cite{bertolini},
\cite{cmw}.

Supersymmetric effects on $R_7,R_9,R_{10}$ are studied in three
scenarios \cite{ABHH00}, respecting bounds on
$b \to s \gamma $ and direct searches:
an effective SUSY model based on minimal flavor violation (MFV)
\cite{hw97},\cite{MFVbsg}, where there are no extra sources of flavor
violation besides the ones present already in the Yukawa couplings of the SM,
a (minimal and/or relaxed) supergravity (SUGRA)
scenario with universal initial conditions at the GUT scale \cite{goto96}
which is effectively MFV like, and
a model with generic squark off-diagonal entries parametrized by
the mass insertion approximation (MIA) \cite{LMSS}.

To summarize: In MFV and SUGRA only very small deviations from the
SM in $C_{9,10}$ are possible:
$R_9, R_{10} \sim 1$, while $C_7^{\rm eff}$ can vary a lot. However,
imposing the $b \to s \gamma$ constraint on the modulus we get
$0.8 < |R_7| < 1.2$ allowing roughly for two solutions:
$R_7 \sim 1$ (SM like sign) and $R_7 \sim -1$.
Note that the opposite-of-the-SM-sign for $C_7^{\rm eff}$ is only
possible for large $\tan \beta$ \cite{ABHH00}, \cite{goto96}.
Effects of non SM valued $R_i$ in SUGRA and MIA-SUSY on the invariant
mass spectra in $B \to K,K^* \mu^+ \mu^-$ decays are shown in
Figure~\ref{fig:BKsusy} and Figure~\ref{fig:BKstsusy}, respectively.
Figure~\ref{Fig:AFBSusy} shows a comparison of the forward-backward
asymmetry in the standard model with SUGRA and MIA-SUSY models. 

\begin{figure}
\epsfxsize=3in
\centerline{\epsffile{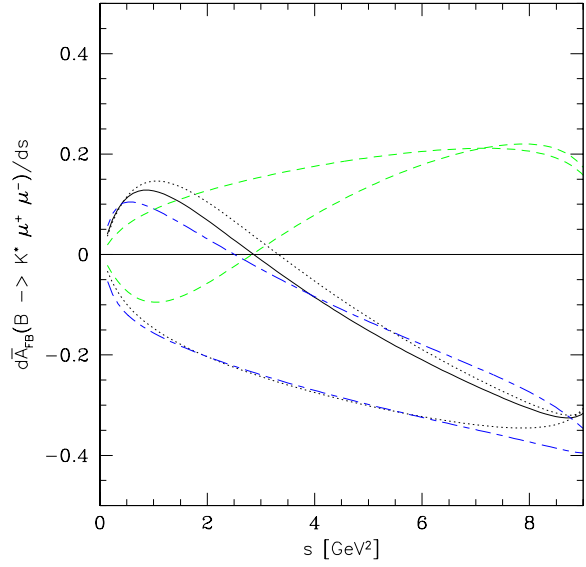}}
\vspace{0.5cm}
\caption[The forward-backward asymmetry in $B_d \to K^{*0}\mu\mu$ decay
as a function of $s = M_{\mu\mu}^2$ predicted with the Standard Model, SUGRA
and MIA-SUSY.]{The forward-backward asymmetry in $B_d \to K^{*0}\mu\mu$ decay
as a function of $s = M_{\mu\mu}^2$ predicted with the Standard Model 
(solid line), the SUGRA (dotted), and MIA-SUSY (long-short dashed 
line)~\cite{ABHH00}. }
\label{Fig:AFBSusy}
\end{figure}

The MIA-SUSY scenario is one example of a model with non-standard FCNC
Z-cou\-plings \cite{GGG}. Here drastic effects are possible in $C_{10}$
\cite{LMSS}, which at present is best constraint by $|C_{10}| < 10$
\cite{ABHH00,GGG} or equivalently $|R_{10}| < 2$-3, namely
\begin{itemize}
\item An enhanced $|C_{10}|$, which results in enhanced branching ratios
${\cal{B}}(b \to s \ell^+ \ell^-)$
\item $sign(C_{10})=-sign(C_{10}^{\rm SM})$ causing a sign flip in the
forward-backward asymmetry (see Section~\ref{sec:AFB}).  This is not
measurable in the rate, which is proportional to $\supset |C_{10}|^2$.
\item A non zero forward-backward-CP asymmetry \cite{GGG}, if $C_{10}$
has a ${\cal{O}}(1)$ phase (see Section~\ref{subsec:AFBCP}).
\end{itemize}
All of these effects are currently not excluded, but none of them can be
saturated in a MFV  scenario with family universal initial 
conditions.

Let us put this quite strong statement into a broader context.
SUSY as a realistic extension of the SM has to be broken, which is
supposed to happen at energies  much higher than the weak scale.
Experiments in $b$-physics now have the power to probe the flavor
structure of SUSY breaking, i.e.~to discriminate between scenarios which
are MFV like, and those who are not.
The popular models of SUGRA,
gauge mediation GMSB,
anomaly mediation AMSB and the non-supersymmetric 2HDM are all MFV, but
in a general MSSM this does not have to be the case. One example which is
non-MFV is given in e.g.\cite{Masiero:1999ub}.
We compile some powerful observables and experimental signatures,
which could decide this.
MFV is ruled out, if
\begin{itemize}
\item $\sin 2 \beta$ is small \cite{Buras:2001xq}
\item $a^{Dir}_{CP}(B \to X_s \ell^+ \ell^-) > {\cal{O}}(1) \%$ for low
dilepton mass
\item $A_{FB}(B \to K^* \ell^+ \ell^-)$ flips sign
\item $A_{FB}^{CP}(B \to K^* \ell^+ \ell^-)$ is significant
\item there a large 'wrong' (opposite to the SM ones)
helicity contributions found e.g.~in $b \to s \gamma$
\end{itemize}
Finally, at large $\tan\beta$ there can be large supersymmetric
contributions to the scalar operators $C_S$ and $C_P$ (see
Sec.~\ref{sec:btoll}) leading to large enhancements of
$\mathcal{B}(B_{s,d} \to \ell^+\ell^-)$ by orders of magnitude
\cite{susybll}.

\subsubsection*{Anomalous Triple Gauge Boson Couplings}
\label{sec:atgc}

\index{anomalous triple gauge boson couplings}
The Triple Gauge Boson Couplings (TGC) are an important feature of
the gauge sector of the SM. In principle, they may be affected by
new physics coming from a scale $\Lambda$ where this may be, for 
instance,
the scale at which the dynamics responsible for electroweak symmetry 
breaking
becomes apparent.
Imposing $CP$ conservation,
the most general form of the
$WWN$ ($N=\gamma, Z$)
couplings can be written as~\cite{dieter}
\begin{eqnarray}
{\cal L}_{WWN}&=&g_{WWN}\bigg\{i\kappa_N W_{\mu}^{\dagger}W_\nu
N^{\mu\nu}
+ig_1^N \left(W_{\mu\nu}^{\dagger}W^\mu N^\nu -
W_{\mu\nu}W^{\dagger\mu} N^\nu\right) \nonumber\\
& & +g_5^V\epsilon^{\mu\nu\rho\sigma}(W^\dagger_\mu\partial_\rho
W_\nu
-W_\mu\partial_\rho W^\dagger_\nu)N_\sigma
+i\frac{\lambda_N}{M_W^2}W^\dagger_{\mu\nu}W^\nu_{~\lambda}
N^{\nu\lambda}
\bigg\}\,,
\label{wwnc}
\end{eqnarray}
with the conventional choices being $g_{WW\gamma}=-e$ and
$g_{WWZ}=-g\cos\theta$.
Deviations from the SM values for the TGC are constrained
directly from LEPII~\cite{tgclep2} and Tevatron~\cite{tgctev}
measurements of gauge boson production.
On the other hand,  FCNC decay processes
at low energies, such as loop-induced $B$ and $K$ decays, probe these
vertices indirectly.
The effects of anomalous TGC in rare $B$ decays have been extensively
studied in the literature.
For instance, the effects of the
dimension four anomalous $WW\gamma$ coupling $\Delta\kappa_\gamma$
in $b\to s\gamma$
transitions
were first considered in~\cite{chia}, whereas this plus the dimension six
coupling $\lambda_\gamma$ where studied in~\cite{numata,rizzo}.
These plus the corresponding CP violating  couplings and their effects 
in the
$b\to s\gamma$ branching fractions
were also considered in~\cite{he}. Finally, the anomalous $WWZ$ couplings
and their effects in $b\to s\mu^+\mu^-$ were  studied in
Ref.~\cite{baillie}.
In Ref.~\cite{gbtgc} the effects in $b\to s\ell^+\ell^-$ are correlated
with those in $K\to\pi\nu\bar\nu$ decays.
There it is shown that
$50\%$ deviations in these branching fractions are possible. This remains
the case even after we consider the latest results from 
LEPII~\cite{tgclep2}.
For instance, from the two-parameter fits in Ref.~\cite{tgclep2} with
$\Delta\kappa_\gamma$ and $\Delta g_1^Z$, the $95\%~$C.L. bound for
$\Delta g_1^Z$ is $[-0.08,0.025]$.
The largest contributions come from $\Delta g_1^Z$.
This sensitivity stems from the fact that the effects induced by
$\Delta g_1^Z$ in rare $B$
and $K$ decays are logarithmically sensitive to the high energy scale
$\Lambda$~\cite{gbtgc}.
In addition, this kind of values for $\Delta g_1^Z$ would induce
an enhancement of $\epsilon'/\epsilon$~\cite{he3}.

Since $\Delta g_1^Z$ affects almost exclusively the Wilson
coefficient $C_{10}$, it will not change the position of the
zero of the forward-backward asymmetry
in $B\to K^*\ell^+\ell^-$. However, the overall value of $A_{FB}$
will be affected.

Finally, we comment on the CP violating anomalous TGC.
The main effect there comes from the dimension-four
$\gamma W^+W^-$ coefficient $\tilde\kappa$. This is bound
from $b\to s\gamma$ to be in the interval $(-0.60,0.60)$.
For instance, this bound translates into~\cite{gbtgc}
$A_{CP}(B^{\pm}\to K^{\pm}\ell^+\ell^-)<1\%$.

\subsubsection*{Anomalous Couplings of Fermions to SM Gauge Bosons}
\label{sec:acqw}

\index{anomalous fermion couplings}
The new physics above the energy scale $\Lambda$ may also
modify the effective interactions of the SM fermions to the
electroweak gauge bosons. In principle, this also has a parallel
in low energy QCD, as it is pointed out in Ref.~\cite{peccei},
where symmetry alone is not enough to determine the axial
coupling of nucleons to pions. In fact, the departure of this
coupling from unity is a non-universal effect, only determined by
the full theory of QCD.  Thus, in Ref.~\cite{peccei} it is
suggested that in addition to the effects in the EWSB sector of
the theory, it is possible that the interactions of fermions with
the NGBs are affected by the new dynamics above $\Lambda$,
resulting in anomalous interactions with the electroweak gauge
bosons. This is particularly interesting if fermion masses are
dynamically generated, as is the case with the nucleon mass.
Interestingly, the proximity of the top quark mass to the
electroweak scale $v=246~$GeV, hints the possibility the top mass
might be a dynamically generated ``constituent'' mass.  Thus, it
is of particular interest to study the couplings of third
generation quarks to electroweak gauge bosons.

The anomalous couplings
of third generation quarks to the $W$ and the $Z$ can come from
dimension-four and dimension-five operators.  The indirect effects of
the dimension-four operators have been considered in relation to
electroweak observables in Ref.~\cite{peccei,yuan}, as well as
the $b\to s\gamma$ transitions~\cite{fuji}.  The constraints on
dimension-five operators from electroweak physics have been studied
in Ref.~\cite{ewd5}.
In Ref.~\cite{actg} the effects of all dimension-four
and dimension-five operators in $B$ FCNC transitions such
as $b\to s\gamma$ and $b\to s\ell^+\ell^-$ were considered.

{\em Dimension-four Operators}:\\
In a very general parameterization, the dimension-four anomalous
couplings of third generation quarks can be written in terms of
the usual physical fields as,
\begin{eqnarray}
{\cal L}_{{4}} &=&-\frac{g}{\sqrt{2}}\, \Big[ C_L\,(\bar t_L\gamma_\mu b_L)
+C_R\,(\bar t_R\gamma_\mu b_R) \Big]\, W^{+\mu} \nonumber\\
&& -\frac{g}{2\,c_W}\left[ N_L^t \,(\bar t_L\gamma_\mu t_L)
+ N_R^t \,(\bar t_R\gamma_\mu t_R)\right]Z^\mu  + \mbox{h.c.} \,,
\label{lfer}
\end{eqnarray}
where $s_W$ ($c_W$) is the sine (cosine) of the
weak mixing angle, $\theta_W$.
The dimension-four operators defined in Eq.~(\ref{lfer}) induce new
contributions  to the $b\to s\gamma$
and $b\to s Z$ loops
as well as the box diagram. They appear in the effective Hamiltonian
formulation as shifts of the Wilson coefficients $C_7(M_W)$,
$C_9(M_W)$ and $C_{10}(M_W)$.

The measured $b\to s\gamma$ branching ratio imposes a stringent
bound on $C_R$ as its contribution to $C_7$ is enhanced by
the factor $m_t/m_b$. This has been discussed in the
literature~\cite{fuji}, where the obtained bounds on $C_R$  :
$-0.05<C_R<0.01$.
In principle, this appears to make $C_R$ unnaturally small if it
were to be generated by some strong dynamics at the scale $\Lambda$.
However, it is
possible to generate such value for $C_R$ in a large variety of
generic strongly coupled theories.
For instance, the pseudo-Nambu-Goldstone Bosons (pNGBs)
of Extended Technicolor (ETC) that result from the breaking of the
various fermion chiral symmetries, generate at one loop a small $C_R$
proportional to $m_b$~\cite{actg}:
\begin{equation}
C_R \simeq \frac{1}{4\pi} \frac{m_bm_t}{f_\pi^2}
\log\bigg(\frac{m_\pi^2}{m_t^2}\bigg)\,.
\label{cretc}
\end{equation}
This is well within experimental bounds in all ETC incarnations, and it 
is
even smaller in modern ETC theories such as
Topcolor-assisted Technicolor~\cite{tcolor},
where the top quark mass entering in~(\ref{cretc})
is only a few GeV.
Thus, here the fact that $C_R$ is small reflects its origin in
the explicit ETC-breaking of
chiral symmetry responsible for $m_b$.
Another hint of this, is the fact that in general
$C_R$ contributes to the renormalization of the
$b$-quark line with a term which does not vanish with $m_b$:
\begin{equation}
\Sigma(m_b)=\frac{g^2}{32 \pi^2} \, C_R \, m_t\, (x-4)\,
\log\bigg(\frac{\Lambda^2}{M_W^2}\bigg)\,.
\end{equation}
Thus if we take into account the potential role of chiral symmetry
in suppressing $C_R$
and we rescale  this coefficient by
defining $\hat C_R$ as
\begin{equation}
C_R = \frac{m_b}{\sqrt{2}v}\hat C_R ,
\label{crst}
\end{equation}
(where $v=246$~GeV),
the rescaled bounds on  $\hat C_R$ are ${\cal O}(1)$,
leaving the possibility that
natural values of  this coefficient may still lead to deviations in 
these decay
modes.

On the other hand, the effect in $b\to s\ell^+\ell^-$ is
dominated by the coefficients $C_L$, $N^t_L$ and $N^t_R$.
In principle, these coefficients are constrained  by electroweak
precision measurements, most notably $\epsilon_1=\Delta\rho=\alpha T$ and
$R_b$ \cite{yuan}.
Once these constraints are taken into account, the effects in FCNC $B$
decays~\cite{actg} are below $15\%$.

{\em Dimension-five Operators}:\\
Although in principle dimension-five operators -which involve two gauge
bosons or one gauge boson and one derivative- are suppressed by the new 
physics
scale $\Lambda$, it is possible that they may have important effects
in $b\to s\ell^+\ell^-$ decays. In Ref.~\cite{actg} all 17 independent
operators are considered. Even after the constraints from electroweak
precision measurements and $b\to s\gamma$ are included $50\%$ to $75\%$
deviations in the branching ratios are possible.

\subsubsection*{New Physics in the Higgs Sector}

\index{Higgs sector}
The sector responsible for Electroweak Symmetry Breaking (EWSB)
is the least understood aspect of the SM. The simplest picture,
where one Higgs doublet gives rise to $M_W$ and $M_Z$, and its
Yukawa couplings to fermions give them their masses, is likely
to be an effective picture only valid at low energies.
Besides the extension of the Higgs sector necessary in supersymmetric
theories, it is possible to imagine various more exotic scalar sectors.
The simplest extension to a  two-Higgs doublet sector results in
three possible realizations. In the first one only one doublet gives
masses to the fermions (Model~I).
Another possibility is that each doublet is responsible
for giving masses for either the up or the down type fermions (Model~II).
Both these models avoid tree-level FCNCs in the scalar 
sector~\cite{gw77}.
Model~II is also the Higgs sector of the MSSM.
Finally, the more general possibility (Model~III) allows for such FCNC
interactions
to take place~\cite{model3}.
The presence of the additional scalar states will in general contribute
to FCNC processes. In the case of Models~I and~II, this happens through 
the
one-loop  contributions of charged scalars. These have been studied 
extensively
in the literature~\cite{2hdms}. For instance, $b\to s\gamma$ constraints 
the
mass of the charged Higgs in Model~II to be roughly $m_{H^\pm}>300$~GeV,
almost independently of the values $\tan\beta$~\cite{hw97}.
For $m_{H^{\pm}} > 300$ GeV and large $\tan\beta$, $\mathcal{B}(B_{s,d} \to
\ell^+\ell^-)$ can vary by a factor of two from its SM value in Model~II
\cite{2HDMbll}.  The phenomenology of Model~III has been studied in
Ref.~\cite{ars}. Experimental measurements in $b\to s\ell^+\ell^-$ modes
such as $B\to K^{(*)}\ell^+\ell^-$ and $B_s \to \ell^+\ell^-$ are going to
have an important impact on the parameter space of these models.

\subsubsection*{Strong Dynamics}

\index{Higgs sector!strong dynamics}
If strong dynamics were responsible for the breaking of the electroweak
symmetry at the ~TeV scale, there could be remnant effects
at the weak scale. These could manifest as small deviations in the SM
model couplings. In this case, the EWSB sector of the SM can be described
by an effective Lagrangian~\cite{leff} where the leading order 
corresponds
to the SM and higher order corrections come in through
higher dimensional operators, and are therefore suppressed by the scale
$\Lambda\simeq {\cal O(1)}~$TeV.
Among the possible effects relevant for FCNC $B$ decays
are the anomalous triple gauge boson couplings
discussed in Section~\ref{sec:atgc} and  the anomalous couplings of fermions
to SM gauge bosons of Section~\ref{sec:acqw}.
Additionally, corrections to the Nambu-Goldstone boson (NGB) propagators
lead, at next to leading order in ${\cal L}_{\rm eff.}$, to
non-standard four fermion operators~\cite{pich}.
These are constrained by measurements of $Z\to b\bar b$ and
$B^0-\bar{B^0}$ mixing.
They also contribute at one-loop to $b\to s$ as well as $s\to d$
transitions which were studied in Ref.~\cite{bews}.
Their contribution to $b\to s\gamma$ is negligible since it only starts
at two loops. However, the  $b\to s\ell^+\ell^-$ processes receive
potentially large deviations, which are correlated with similar
deviations in $K^{(+,0)}\to\pi^{(+,0)}\nu\bar\nu$.

Finally, many specific scenarios of strong dynamics in the EWSB sector
have relatively light scalar states some of which may contribute
to FCNC through loops, or even in some cases at tree level.
To a large extent, the phenomenology relevant to $b\to s\ell^+\ell^-$
decay modes is similar to that of multi-Higgs models.
Model-dependent specifics can be see in Ref.~\cite{balaji}
for extended technicolor and in Ref.~\cite{btcolor} the
topcolor flavor signals were extensively studied.
In most cases the power of $b\to s\gamma$ to constrain the masses and
couplings of these scalar states is limited due to the possibility of
cancellations. Modes such as $B\to K^*\ell^+\ell^-$ will be much more
constraining.

\section{Rare Decays: Experiment}

\subsection{Rare Decays at \Do\ }

\def\pp{$p\bar{p}$}
\def\ptmu{$p_T^{\mu}$}
\def\pt{$p_T$}
\def\ptb{$p_T^b$}
\def\ptc{$p_T^c$}
\def\etamu{$\eta^{\mu}$}
\def\dsdptmu{\mbox{d$\sigma_b^{\mu}/\mathrm{d}p_T^{\mu}$}}
\def\dsdptb{\mbox{d$\sigma^b/\mathrm{d}p_T^b$}}
\def\bb{\mbox{$b\bar{b}$}}
\def\cc{\mbox{$c\bar{c}$}}
\def\QQ{\mbox{$Q\bar{Q}$}}
\def\ptuu{\mbox{$p_T^{\mu\mu}$}}
\def\bsdsphi{\mbox{$B_s^0 \rightarrow D_s^- + \mu^+ + \nu_{\mu}$}}
\def\bsds3pi{\mbox{$B_s^0 \rightarrow D_s^- + 3\pi^{\pm}$}}
\def\bsdslX{\mbox{$B_s^0 \rightarrow D_s^- + \ell^+ + X$}}
\def\bsdsnpi{\mbox{$B_s^0 \rightarrow D_s^- + n\pi^{\pm}$}}
\def\bks{\mbox{$B_d^0 \rightarrow K_s + J/\psi$}}
\def\jpsi{\mbox{$J/\psi$}}
\def\bkspsi{\mbox{$B_d^0 \rightarrow J/\psi K_S^0$}}
\def\bkst{\mbox{$B_d^0 \rightarrow K^{*0} \mu^+\mu^-$}}
\def\bbkst{\mbox{\boldmath $B_d^0 \rightarrow K^{*0} \mu^+\mu^-$}}
\def\bkstee{\mbox{$B_d^0 \rightarrow K^{*0} e^+e^-$}}
\def\bbuu{\mbox{\boldmath $B_s^0 \rightarrow \mu^+\mu^-$}}
\def\buu{\mbox{$B_s^0 \rightarrow \mu^+\mu^-$}}
\def\bbsuu{\mbox{\boldmath $b \rightarrow s \mu^+\mu^-$}}
\def\bsuu{\mbox{$B \rightarrow X_s \mu^+\mu^-$}}
\def\dskkp{\mbox{$D_s^- \rightarrow K^+ K^- \pi^-$}}
\def\dsphi{\mbox{$D_s^- + \mu^+$}}
\def\dsmunu{\mbox{$D_s^-\, \mu^+\, \nu_{\mu}$}}
\def\ds3pi{\mbox{$D_s^- + 3\pi^{\pm}$}}
\def\kspsi{\mbox{$K_s + J/\psi$}}
\def\QQmumu{\mbox{$Q\bar{Q}\rightarrow \mu \mu$}}
\def\QQmu{\mbox{$Q\bar{Q}\rightarrow \mu$}}
\def\QQee{\mbox{$Q\bar{Q}\rightarrow e e$}}

    We have investigated \Do\ options to study several rare B-decay
processes in Run II:

\begin{itemize}
        \item \bkst\ decay followed by $K^{*o}$ $\rightarrow K^{\pm}
     \pi^{\mp}$, with the expected combined branching ratio
     of 0.67 $\times$1.5$\times 10^{-6}. $
	\item inclusive $b\rightarrow s\mu^+\mu^-$ decay with the expected SM
    branching ratio of 6$\times 10^{-6}$.
    \item exclusive \buu\ decay, with the expected SM
    branching ratio of 4$\times 10^{-9}$.

\end{itemize}

    With its extended muon coverage and excellent muon identification, 
\Do{} can easily trigger on the semileptonic decay of $B$ mesons
into muons. In particular, we expect that the dimuon trigger, with an effective transverse
momentum ($p_T$) threshold for
individual muons of 1.5-2 GeV/c in the pseudorapidity range $|\eta^{\mu}|< 1.6$, 
will run
unprescaled even at the highest luminosities. 
Thanks to the installation of the central and forward
preshowers in Run II, \Do{} will also be able to trigger on low $p_T$ dielectrons.
However, because of the limited bandwidth available 
at the level one (10 kHz) and level two (1 kHz) of the current trigger system, 
the rate of low threshold lepton triggers could become unacceptable.
We are protecting ourselves against this possibility 
by adding a level 2 trigger preprocessor using 
the data from 
the Silicon Vertex Detector (SMT). The processor will allow to trigger on events 
containing tracks with large impact parameters in the transversal plane, 
coming from the decay of $B$/$D$ mesons. 

    Various trigger combinations and kinematic cuts have been considered to optimize
selection of the rare decay processes. The
expected numbers of events are quoted for an 
integrated luminosity of 2 $fb^{-1}$ and the $B$ meson production 
cross section normalized to $\sigma (B_d^0) = 3.2 \mu b$
for $p_T^B > 6$ GeV,  $|$y(B)$|< 1$. The 
combined trigger efficiency for the proposed dimuon 
and the single muon trigger is 55\% for events with two muons with 
$p_T^{\mu}> 1.5$
    GeV/c, $|\eta^{\mu})|< 1.6$, and $p_T^{\mu\mu} > 5$ GeV/$c$. 
We have verified
    that, with these kinematic cuts, trigger efficiency is independent of the
    dimuon mass.

    We found in our Run I analysis\cite{Daria} that for the inclusive $B\rightarrow X_s\mu^+\mu^-$
decay it would be necessary to restrict the search to a limited dimuon mass range of
(3.9-4.4) GeV, representing $\approx$ 7\% of the decays\cite{Baer}, in order to avoid
the sequential decays $B\rightarrow D +
\mu + X_1 ; D\rightarrow \mu + X_2$ and $J/\psi$ and $\psi(2s)$
resonances.  However, even in this limited dimuon
mass window, we expect only 1000 signal events compared to 100,000 \QQ\
$\rightarrow \mu^-\mu^+ +X$ physics background events, with muons
originating from two different $b$ quarks. Some additional kinematic cuts on the
event topology and multivertex searches could improve the signal to 
background ratio, however, it does not seem worth the effort.
There is only a limited interest in measuring the small and least
theoretically known part of the dimuon mass spectrum.

    The process \buu\ is also rather hopeless to measure, unless the
branching ratio is boosted by some additional, non SM contributions, like
Higgs doublet exchanges. We expected fewer than 5 recorded \buu events in 2
$fb^{-1}$ of data.

        On the other hand, \Do\ has a fair chance to make a competitive
measurement of the \bkst\ decay, including the rate, and the decay
asymmetry dependence on the dimuon mass. We have
generated relevant Monte Carlo events combining the ISAJET production
information with the predicted decay distributions, taken from Ref.
\cite{Greub:1995pi}. A simple analysis of the 
Monte Carlo events was based on the
CDF experiences from their attempt to isolate this channel in the Run I
data~\cite{CDF_rare1,Affolder:1999eb}. Details of the investigations 
are described in the next sections.

\subsubsection{Monte-Carlo Samples}
\label{sec_samples}


    This study is based on various Monte-Carlo samples generated with
the ISAJET program at $\sqrt{s}=2$ TeV, with events selected by the presence of two muons
in the final state.  Only a small sample of events has been processed through
the current \Do\ event simulator, D0RECO and the current Level 1 trigger
simulator. For the remaining events, the detector response was simulated
using an older version of the muon trigger simulator.

	The physics background is primarily due to $Q\bar{Q} \rightarrow\, \mu \mu X$ events, where $Q$ stands for a $c$ or $b$
quark. A large sample of such events was generated with the NLO-QCD ISAJET version 7.22
         in FOUR bins of \ptb{}: (2--3) GeV/c, (3--5) GeV/c, (5--10) GeV/c,
 and (5--80) GeV/c. Only 80K events with
         2 muons satisfying the acceptance cuts $p_T^{\mu} > 1.5$ GeV/$c$
         and $\mid \eta^{\mu} \mid < 2$ were kept for Geanting. 
We compared the \ptb{} differential spectrum for all events,  
         $d\sigma^b/dp_T^b$, to the MNR~\cite{MNR} 
         prediction and renormalized the  
         ISAJET weigths to match the MNR \ptb{} and \ptc{} spectra 
	In addition, the absolute normalization
         of this sample was done using the CDF measurement of $b\rightarrow 
        J/\psi$
         production cross section at 1.8 TeV~\cite{CDF_bjpsi}, extrapolated to the c.m. energy
         of 2 TeV (cross section increased by 25\%). 
\index{dilepton mass spectrum!$Q\bar{Q}$ production}
         A smaller sample of $Q\bar{Q} \rightarrow\, \mu \mu X$, again 
in the \ptb{}
         ranges from 2 GeV/c to 80 GeV/c, was generated with ISAJET 7.37 to confirm previous
         results.

    The expected dimuon mass distribution due to $Q\bar{Q}$ production is shown
in Fig.~\ref{QQ_MASS} for events with both muons satisfying the conditions 
$p_T^{\mu} > 1.5$ GeV/$c$ and and $\mid
\eta^{\mu} \mid < 1.6$. The dimuon mass spectrum for muons originating from
different $b$ quarks is relatively flat between 2 GeV and 7 GeV,
 where the dominant
process is the gluon splitting into \bb\ pairs (Fig.~\ref{QQ_MASS} (a)). 
\index{dilepton mass spectrum! sequential decays}
The mass spectrum resulting from sequential $b\rightarrow c \rightarrow s$ 
quark 
decays has a maximum around $m_{\mu\mu}$ = 2 GeV and does not extend beyond 
the $m_{\mu\mu}$ = 4 GeV (Fig.~\ref{QQ_MASS} (b)).

\begin{figure}[t]
\centering \leavevmode 
\epsfxsize=3.0in
\epsffile{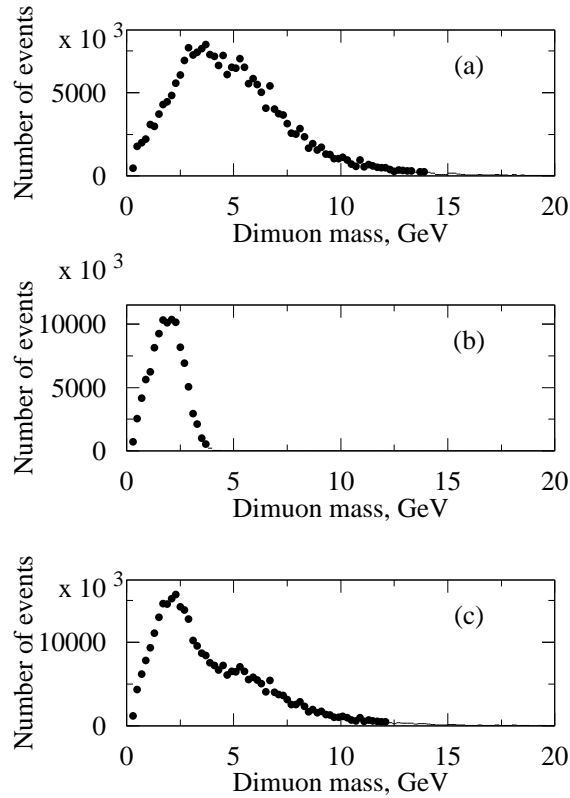}
\caption[Expected number of dimuon events in 2\,fb$^{-1}$ data, due to
$Q\bar{Q}$ production]{Expected number of dimuon events in 2 $fb^{-1}$ of data,
due to $Q\bar{Q}$ production, as a function of the dimuon mass for muons with
$p_T^{\mu\mu} > 2.0$ GeV/$c$,  $p_T^{\mu} > 1.5$ GeV/$c$  and $\mid \eta^{\mu}
\mid < 1.6$: (a) muons from different $b$ quarks, (b)sequential $b$ quark
decays, and  (c) total. Trigger and reconstruction efficiencies are not
included. Events were generated with ISAJET V7.22 and normalized to the MNR
differential $p_T^Q$ distributions with the absolute normalization based on the
measured $b\rightarrow J/\psi + X$ cross section.} 
\label{QQ_MASS}
\end{figure}

    The signal samples of events were generated with ISAJET 7.37, using the
    leading order
         only and a single \ptb{} bin between 2 and 80 GeV/$c$.  ISAJET
         decays $B_d^0$ mesons into $K^* \mu^+\mu^- $ system according to the
         three-body phase space. Therefore ISAJET events had to be
 weighted to match
         expected decay spectra, as calculated in Ref.~\cite
         {Greub:1995pi}. The event weight depends on two observables: the dimuon mass
         and the energy of the negative muon in the $B_d^0$ rest frame. We have verified
         that the distributions of weighted Monte Carlo events are consistent
         with predictions of Ref.\cite{Greub:1995pi}. 
\index{$B\to K^*\ell^+\ell^-$!phase space}
The expected non-resonant
dimuon mass
         distributions for the \bkst\ process from phase space and those predicted by
         the theory are shown in Fig.~\ref{PHASE_SPACE} (a). The predicted asymmetry
         plot is shown in Fig.~\ref{PHASE_SPACE} (b).

\begin{figure}[thb]
\begin{center}
         \centerline{\psfig{figure=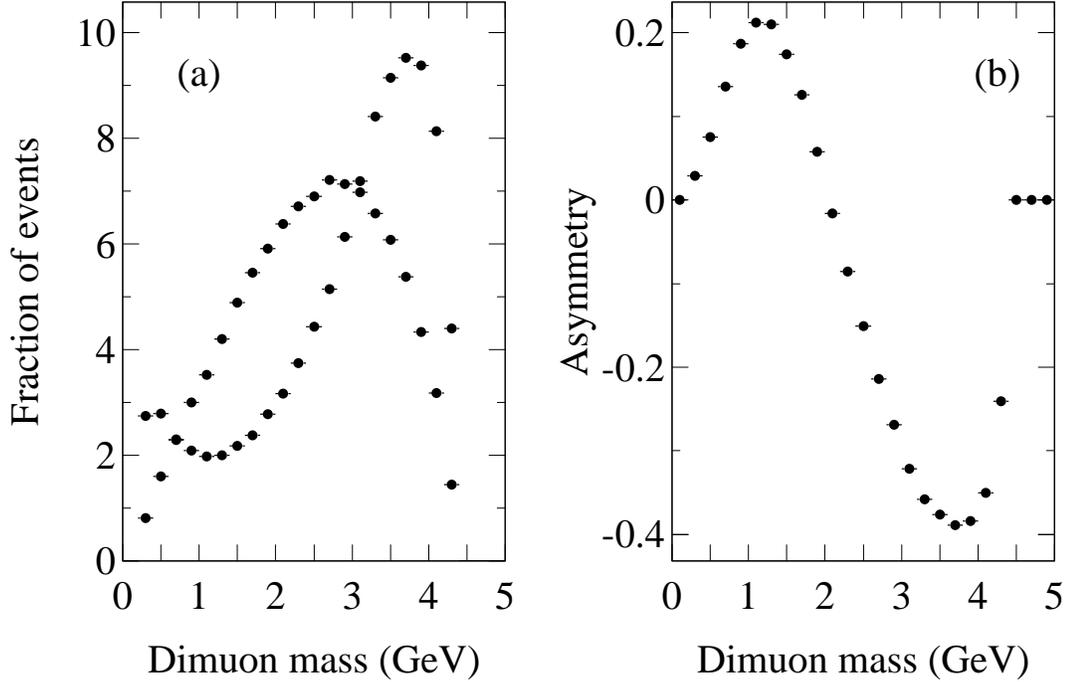,height=10cm}}
\caption[Comparison of the expected dimuon mass distributions for
\bkst\ from phase space and predicted by theory.]{Comparison of the expected
non-resonant dimuon mass distributions for the \bkst\ process from phase space
and predicted by theory (a). Predicted asymmetry as a function of dimuon mass
(b).}
\label{PHASE_SPACE}
\end{center}
\end{figure}

The primary vertex 
position was generated at
(0,0,$z_0$), with $z_0$ following a Gaussian distribution with
a width of 25 cm.

\index{dilepton mass spectrum!trigger rates, \Do}
 The combined single muon/dimuon  trigger rates at the instantaneous luminosity of 2 $10^{32} cm^{-2}s^{-1}$
due to dimuons from the
    genuine $Q{\bar Q}$ signal
    are $\approx$ 13 Hz ($\approx$ 4 Hz for $p_T^{\mu\mu}> $ 5 GeV/c). The 13
    Hz combines contributions from: \cc\ pair production (2.5 Hz), \bb\ pair
    production (9.5 Hz) and $b \rightarrow J/\psi +X$ decays (1.0 Hz). A
    requirement of $p_T^{\mu\mu}> 2$ GeV/$c$ reduces the rate to 9 Hz (see Chapter 4.5.2).

    It turns out that there is little trigger efficiency dependence on the
dimuon mass for events selected with  kinematic cuts used in this analysis. This
is illustrated in Fig.~\ref{TRIG_EFF}, where we plot results of our
investigations from early 1998.
Therefore, muon/dimuon trigger 
efficiencies for the inclusive $b \rightarrow J/\psi$ production, discussed in Chapter 4.5.3 apply to the entire 
\bkst\ sample and the trigger
does not significantly distort the dimuon spectrum once the kinematic cuts are
introduced.

\begin{figure}[thb]
\begin{center}
\psfig{figure=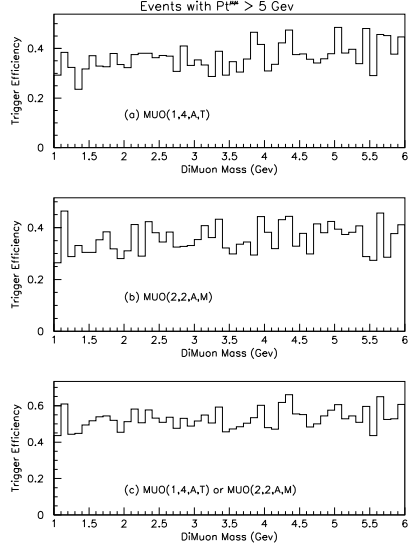,height=8cm}
\caption[Trigger rate dependence on the dimuon mass for events preselected with
kinematic cuts $\mid \eta^{\mu} \mid < 1.6$,  $p_T^{\mu} >$ 1.5 GeV/c and
$p_T^{\mu\mu} >$ 5 GeV/c.]{Trigger rate dependence on the dimuon mass for
events preselected with kinematic cuts $\mid \eta^{\mu} \mid < 1.6$, 
$p_T^{\mu} >$ 1.5 GeV/c and $p_T^{\mu\mu} >$ 5 GeV/c. These results were
obtained with the Sep. 97 version of the muon trigger simulator.}
\label{TRIG_EFF}
\end{center}
\end{figure}

\subsubsection{The Exclusive Channel \bkst }

    In this section we summarize results for the process \bkst with 
$K^{*o}$ $\rightarrow \pi^{\pm} K^{\mp}$.
  Expected numbers
of events were obtained under the following assumptions:

\begin{itemize}
    \item  integrated luminosity of 2 $fb^{-1}$

    \item  production cross section normalized to $\sigma (B_d^0) = 3.2 \mu b$
 for $p_T^B > 6$ GeV,
     $|$y(B)$|< 1$. This assumption results in the predicted number of
 produced $B^0_d$ or $\bar{B^0_d}$ equal to 1.4$\times10^{11}$,  
the same number as obtained 
 assuming $\sigma_{b\bar{b}} = 100 \mu $b and 
{\it B}($\bar{b}\rightarrow B^0$) = 0.35 (see also 
Tables 6.4 and 7.8).

    \item  $B_d$ decay branching ratio {\it B}(\bkst ) = 1.5 $\times 10^{-6}$ and {\it B}($K^{*o}\rightarrow \pi^{\pm}K^{\mp}$) = 0.67.

      \item two muons with $p_T^{\mu}> 1.5~$ GeV/c and  $\mid 
\eta^{\mu} \mid< 1.6$
      \item dimuon pair transverse momentum $p_T^{\mu\mu} > 5.0 $ GeV/$c$.
 
    \item  Level 1 trigger efficiencies for a combined trigger L1MU(2,2,A,M)
 and 
    L1MU(1,4,A,T) as discussed in Chapter 4.5.3 (Table 4.1). Level 2 and Level 3 trigger efficiencies are unknown at the time of this writing and are set to 1.0.
 
    \item  track reconstruction efficiency of 95\% per track (81\% per event).
\end{itemize}
    
\index{$B\to K^*\ell^+\ell^-$!analysis cuts, \Do}
Analysis cuts included: 

   \begin{itemize}

      \item  primary - secondary vertex
separation in the transverse plane of 400 $\mu$m

      \item charged particles from the $K^{*o}\rightarrow \pi^{\pm}K^{\mp}$ 
            decay with transverse momenta: $p_T^{\pi (K)}> 0.5$ GeV/c and 
 $\mid 
\eta^{\pi (K)} \mid< 1.6$

      \item	$K^{*o}$ transverse momentum $>$ 2 GeV/c

      \item dimuon invariant mass outside the \jpsi\ ($(3.05 - 3.15)$ GeV)
 or $\psi(2s)$
            ($(3.62-3.76)$ GeV) mass windows.

      \item isolation $I > 0.6$, where $I$ is the transverse momentum of the
  $B$ candidate divided by the scalar sum of transverse momenta of the $B$ and
  all other tracks. CDF has established efficiency for this cut as $0.92\pm0.
  06$\cite{Affolder:1999eb}.

	\item  the transverse plane impact parameter significance $>$ 2 requirement for either three out of four tracks or 
all four tracks from  the \bkst\ decay. 
   \end{itemize}


\index{$B\to K^*\ell^+\ell^-$!event rates, \Do }
    Table \ref{ta:bkst} lists expected event rates for various  kinematic
cuts. The inclusion of the $B^0_d \rightarrow e^+e^-K^{*}$ decay mode could result in a 50\% increase in the number of observed events.  

\begin{table}
\centering
\begin{tabular}{|l||c|c|} \hline \hline
  $p_T^{\mu\mu} >$ & $ 5.0$ GeV/c & $ 5.0 $ GeV/c \\
  $p_T^{\mu} >$    & $ 1.5$ GeV/c & $ 3.0 $ GeV/c \\ \hline

Muon (dimuon) kinematic acceptance,$\epsilon_{\mu}$ & 0.052 & 0.014  \\    
Level 1 trigger efficiency, $\epsilon_{Lev1}$       & 0.55  & 0.67 \\
Level 2 \& 3 trigger efficiency,$\epsilon_{Lev23}$  & 1.00  & 1.00 \\
Number of recorded events           & 4000 & 1350  \\
  $\epsilon_{analysis}$         &  0.17 &  0.22  \\
  $\epsilon_{reco}$                     &  0.81 & 0.81  \\
Number of reco. events prior to the IP cuts & 550 & 250\\ 
IP signif. $>2$ for at least 3 tracks  &  490 &  220  \\
IP signif. $>2$ for all 4 tracks       &  310 &  130  \\ \hline \hline
\end{tabular}\vspace*{4pt}
\caption{Expected numbers of events for the \bkst\ process with different
analysis cuts.}
\label{ta:bkst}
\end{table}

\begin{figure}[t]
\begin{center}
         \centerline{\psfig{figure=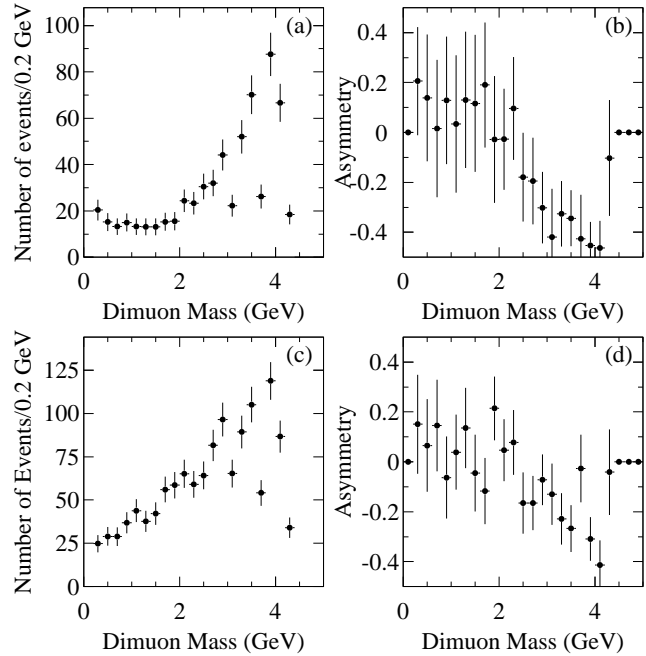,height=9cm}}
\caption[Dimuon mass distribution in the decay \bkst]{(a) Dimuon mass
distribution in decay \bkst , and (c) assuming 1:1 signal to background ratio.
(b) Predicted asymmetry signal as function of dimuon mass, and (d) assuming the
background level as in (c).}
\label{KSTUU_DATA}
\end{center}
\end{figure}

    As an illustration, the dimuon mass distribution for the  reconstructed
sample of 630 events is shown in Fig.~\ref{KSTUU_DATA}(a). The minima in the
distribution are due to the removal of the \jpsi\ and $\psi(2s)$ mass bands. A
corresponding plot, assuming a 1:1 signal to background ratio, is shown in
Fig.~\ref{KSTUU_DATA} (c). The background was distributed according to the
three body phase space and its rate estimate is based on the CDF extrapolations
from their run I experience. An independent MCFAST Monte Carlo background
evaluation has not yet been completed.

    We conclude that the number of expected events, combined with relatively
modest background level, will enable \Do\ to establish the signal and to measure
its $q^2 = m^2_{\mu\mu}$ dependence. However, the numbers quoted in 
Table \ref{ta:bkst} represent an optimistic scenario, based on the \Do\ 
nominal trigger and track reconstruction efficiencies. A reduction in the per 
track reconstruction efficiency from 95\% to 88\% (a value used by CDF) 
and an inclusion of a Level 2\&3 trigger efficiency of 50\% (the STT preprocessor - see Chapter 4.5.4) would drop the number of expected events  
listed in Table \ref{ta:bkst} by a factor of three. The clue to a successful measurement of the \bkst\ process with the \Do\ detector is our claimed ability to trigger on low
mass dimuons over large rapidity range. 


\index{$B\to K^*\ell^+\ell^-$!forward-backward asymmetry, \Do }
    The \bkst\ process is expected to exhibit an asymmetry in the
$cos\theta$ distribution, where $\theta$ is the $\mu^+$ decay angle between the direction of the $B_d$ and the direction of $\mu^+$ in the rest frame of the 
$\mu^+\mu^-$
 rest frame. This
asymmetry manifests itself as a difference in the energy distributions of
$\mu^+$ and $\mu^-$ in the $B_d$ rest frame, at a given dimuon mass. The
asymmetry is expected to vary with the dimuon mass from approximately 0.2
at small masses to -0.4 around 3.5 GeV. The sign reversal of the asymmetry 
occurs at the dimuon mass of $\approx$ 2 GeV and turns out to be
relatively model independent.

   The predicted asymmetry signal as a function of the dimuon mass for the
same sample of events  is shown in Fig.
\ref{KSTUU_DATA}(b).
A related plot, assuming a 1:1 signal to background ratio, is shown in
Fig.~\ref{KSTUU_DATA} (d). 
    The observed asymmetries, corrected for the  assumed background
contribution are 0.13$\pm$0.13 and -0.31$\pm$0.06 for $m_{\mu\mu}< 2$ GeV and
$m_{\mu\mu}> 2$ GeV, respectively.

\subsubsection{The Inclusive Decay $B \to X_s\ell^+\ell^-$ }

    This process, although theoretically the most interesting to measure, 
is very difficult to separate experimentally in hadronic collisions. 
\index{$B \to X_s\ell^+\ell^-$!$Q\bar{Q}$ background, \Do }
The expected $\mu^+\mu^-$
spectrum due to the heavy quark production is shown in Fig.~\ref{QQ_CUT}
  (onia states are
removed). The muons are selected with transverse momenta greater than 
1.5 GeV/c  and $\mid \eta^{\mu} \mid<$ 1.6. The mass region below
3.9 GeV is dominated by the sequential b-quark decays, for which one of the muon tracks
originates from the B hadron vertex, whereas the other from the charm decay
vertex. At larger masses muon
pairs are produced predominantly by semi leptonic decays of $b$ and ${\bar b}$
quarks with the \bb\ pair resulting from a gluon splitting. Therefore the muon
tracks
are expected to point to two different vertices. 

\index{$B \to X_s\ell^+\ell^-$!dimuon mass spectrum, \Do }
    The expected dimuon mass spectrum for the \bsuu\ process 
 smeared by the expected experimental resolution 
is shown in Fig. 
\ref{BSUU}. The dimuon mass range (3.9 -
4.4) GeV represents only 7\% of the spectrum (expected number of  events is 30
\% larger for (3.8 $< m_(\mu\mu)<
4.4$) GeV). The dominant \QQ\ physics background could be slightly reduced by
increasing the required transverse momentum of the muon pair, as shown in Fig.
\ref{QQ_CUT}.
\index{$B \to X_s\ell^+\ell^-$!event rates, \Do }
 Imposing kinematic cuts
similar to those described for the \bkst\ analysis leads to the expected 
number of events listed in the Table \ref{ta:bsuu}. The numbers are quoted for
the assumed branching ratio for the \bsuu{} decay of 6$\times 10^{-6}$.

\begin{figure}
\begin{center}
         \centerline{\psfig{figure=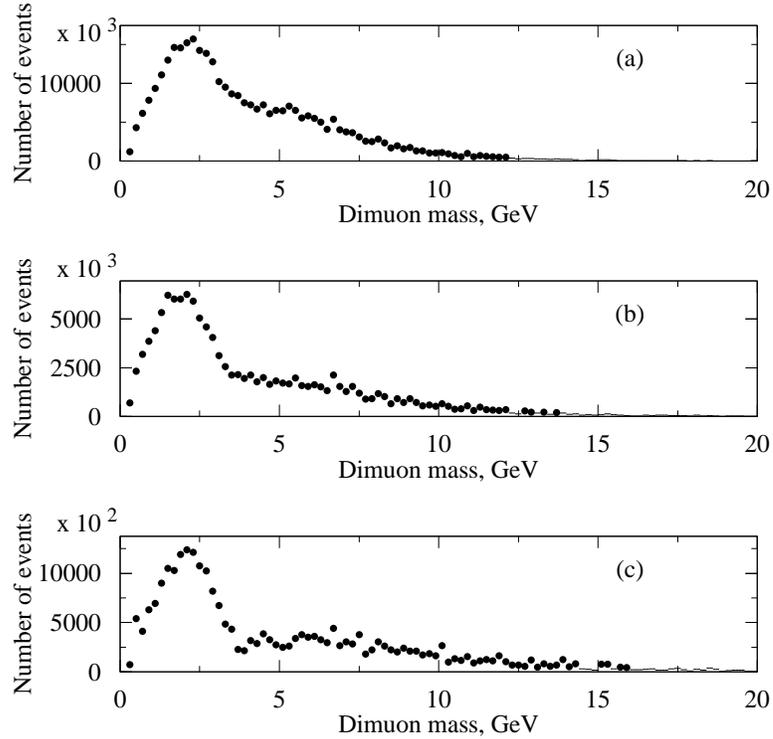,
height=10cm}}
\caption[Expected dimuon mass distributions due to non-resonant $Q{\bar Q}$
production.]{Expected dimuon mass distributions due to the non-resonant $Q{\bar
Q}$ production. Both muons are required to have $p_T^{\mu} > 1.5$ GeV/$c$ (3.0
GeV/$c$ in (c)) and $\mid \eta^{\mu} \mid < 1.6$ and the dimuon $p_T$ must be
greater than: (a) 2 GeV/c, (b,c) 5 GeV/c. Trigger and reconstruction
efficiencies are not included. Events were generated with ISAJET V7.22 and
normalized to the MNR differential $p_T^Q$ distributions with an absolute
normalization based on the measured $b\rightarrow J/\psi + X$ cross section.}
\label{QQ_CUT}
\end{center}
\end{figure}

\begin{figure}
\begin{center}
         \centerline{\psfig{figure=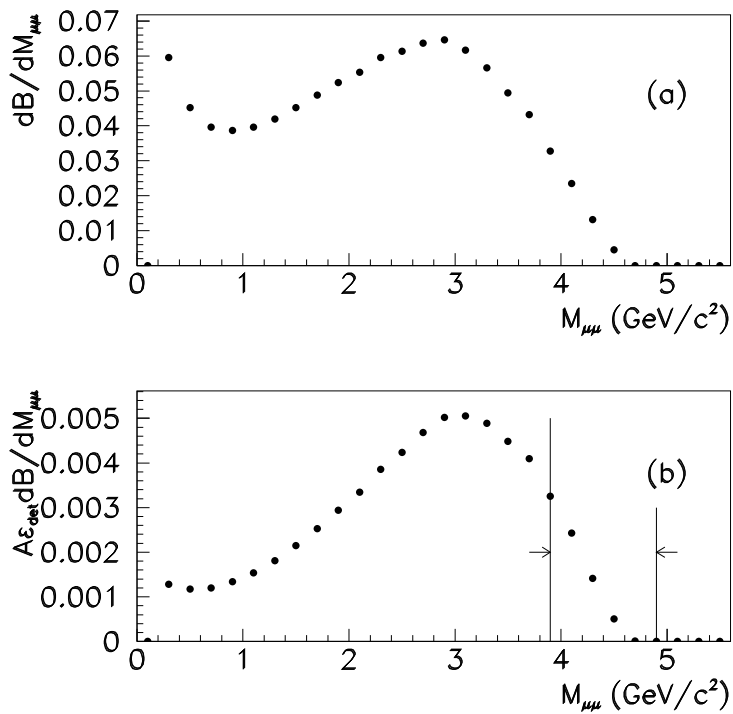,height=10cm}}
\caption[The calculated differential branching fraction for the decay \bsuu ,
as a function of $m_{\mu\mu}$.]{The calculated differential branching fraction
for the decay \bsuu , as a function of $m_{\mu\mu}$. (b) the same differential
branching fraction modified by the response of the \Do\ detector. The arrows
indicate the search window used in this analysis.}
\label{BSUU}
\end{center}
\end{figure}

\begin{table}
\centering
\begin{tabular}{|l||c||c|c|} \hline \hline
   $p_T^{\mu\mu}$ & $ > 2.0$ GeV/c & $> 5.0$ GeV/c& $> 5.0$ GeV/c \\
   $p_T^{\mu}$ & $    > 1.5$ GeV/c & $> 1.5$ GeV/c& $> 3.0$ GeV/c \\ \hline
Trigg. effic. (\%)   &32 & 55 & 67  \\ \hline
Recorded     &  2300 & 1750 & 1000  \\ \hline
Vtx separation cut of 400 $\mu$m      &  1200 & 1050 &  650  \\ \hline\hline
$b {\bar b}$ Bkgd (no analysis cuts) & 1,900,000 & 400,000 & 100,000\\
 \hline \hline
\end{tabular}\vspace*{4pt}
\caption{Expected number of recorded \bsuu\ events in the mass window 3.9$<
m_{\mu\mu} < 4.4 $ GeV with different analysis cuts.}
\label{ta:bsuu}
\end{table}

    The signal is overwhelmed by the physics background, with a 1:1000 ratio.
The background estimates are based on the ISAJET 7.37 version. The earlier
ISAJET versions, like V7.22, predict the background level twice as large.
Additional cuts on the event topology and a requirement of  a common muon
vertex will reduce the background by factors 3-10, not sufficient to establish the
\bsuu\ signal.

\subsubsection{The Exclusive Channel \buu }

\index{$B_{s,d} \to \ell^+ \ell^-$!event rates, \Do }
    The expected number of events for the \buu{} mode is summarized in
Table~\ref{ta:buu}.
We assume that $B_s$ mesons are produced with a rate equal to 40\% of that for
$B_d$ mesons\cite{Affolder:2000iq}. The quoted numbers are for the \buu{} branching
ratio of 4$\times 10^{-9}$. The analysis cuts include: 
(i) the $B_s$ isolation cut, I$>$ 
0.6, (ii) the requirement that the transverse decay length in the $B_s$ rest
frame exceeds 100 $\mu$m, and an
impact parameter significance for each muon track greater than 2. 
A 95\% reconstruction efficiency per track is also assumed.

\begin{table}
\centering
\begin{tabular}{|c||c|c|c|} \hline \hline
   $p_T^{\mu\mu}$ & $ > 2.0$ GeV/c & $> 5.0$ GeV/c& $> 5.0$ GeV/c \\
   $p_T^{\mu}$ & $    > 1.5$ GeV/c & $> 1.5$ GeV/c& $> 3.0$ GeV/c \\ \hline
Signal events after cuts      &  6 & 3 &  1.5  \\ \hline\hline
\end{tabular}\vspace*{4pt}
\caption{Expected number of \buu\ events.}
\label{ta:buu}
\end{table}

    With the expected number of events there are limited chances to measure
this branching ratio, unless its actual value is significantly boosted up
by some non-SM processes. The background rates have not yet been estimated.

\subsection{Rare Decays at CDF}

In Run~1I CDF expects to collect a large enough $b$ sample to observe
rare $b$ decays with branching ratios of order $10^{-6}$, including $b
\rightarrow s \gamma$ (radiative) decays and $b \rightarrow s \mu\mu$
decays.  In this section, we describe the prospects for CDF in Run~1I
for measurements in several channels: $B_{d,s} \rightarrow
K^{*0}\gamma$, $\Lambda_b \rightarrow \Lambda\gamma$, $B_d \rightarrow
K^{*0}\mu\mu$, and $B_{d,s} \rightarrow \mu\mu$.  We will discuss
trigger selections for these channels and estimates signal yields.  We
also study the potential to measure the forward-backward asymmetry
$A_{FB}$ in the $B_d \rightarrow K^{*0}\mu\mu$ decays and show some
ideas to extract the zero position of $A_{FB}$ as a function of
$M_{\mu\mu}$.

\subsubsection{Radiative $B$ Meson Decays}


BaBar and Belle are expected to observe approximately 20 
$B \rightarrow K^{*}\gamma$ decays per 1 fb$^{-1}$ of $\Upsilon (4S)$
data.  Each experiment plans to obtain or order 100 fb$^{-1}$ within 
3 years (1000 $B \rightarrow K^{*}\gamma$). However, these projections
are recently getting much better, and each experiment may obtain 
several 100 fb$^{-1}$ of data by 2004. Our goal is to implement a 
trigger to collect of order 1000 $B \rightarrow K^{*}\gamma$ events 
during Run~1I.  Our studies of radiative decays of $B_s$ and 
$\Lambda_b$ are unique to the Tevatron.

In Run~1, CDF included a dedicated trigger for radiative $b$ decays,
searching for a photon associated with a nearby pair of
tracks\cite{Ref:PengTrg}.  In this trigger, we required two energetic
oppositely-charged tracks, each with $p_{T}>$ 2 GeV/$c$, in the
vicinity of the photon.  We collected 22.3 pb$^{-1}$ of data in Run~1B
with $E_T(\gamma)>10$\,GeV, and 6.6 pb$^{-1}$ in Run~1C with
$E_T(\gamma)>6$\,GeV and obtained upper limits on the branching
fractions for $B_d$ and $B_s$ radiative decays to be $1.6\times
10^{-4}$ and $2.5\times 10^{-4}$, respectively.  Another search for
radiative $b$ decays used photon conversions. One of the conversion
electrons was triggered with an 8 GeV
threshold~\cite{Ref:ConvTrg,Ref:MethodII}.  The two methods had
similar acceptance after all cuts, but because it relied on tracking
information to reconstruct the photon, the conversion method had
superior $B$-mass resolution and a more straightforward analysis
procedure.  Also, the conversion method has a ready normalization in
the kinematically similar $B^0\rightarrow J/\psi\,K^{*0},
J/\psi\rightarrow e^+e^-$ mode.

\index{$B\to K^*\gamma$!trigger, CDF}
In order to trigger on radiative $B$ meson decays, we can take
advantage of the long lifetime of $b$ hadrons and use the SVT track
processor to find charged-particle tracks significantly displaced from
the beamline.  We will use the same trigger selection as for
semileptonic decays discussed in Section~\ref{cdfsemi}, requiring a
4\,GeV electron associated with a track of $p_T>$ 2~GeV/$c$ found by
the SVT to have an impact parameter $d_0$ greater than 120\,$\mu$m.
Furthermore, we require the angle between the electron and track to be
less than $90^\circ$ and the transverse mass to be less than
5\,GeV/$c^2$.  Because of the lower electron threshold, after
kinematic cuts made in the Run~I analysis to purify the sample, this
trigger selection has about a factor of 3 greater acceptance for
radiative decays than the inclusive 8\,GeV electron requirement that
was used in Run~I but with a substantially lower trigger rate.  The
rate for the same-side electron plus displaced track selection is
expected to be 9\,Hz at a luminosity of $10^{32}\,\rm{cm}^{-2}\,\rm{s}^{-1}$.
\index{$B\to K^*\gamma$!trigger, CDF}

An estimate of the signal yield for $B_d \rightarrow K^{*0}\gamma$ is obtained by scaling the Run~1 analysis results with 
the ratio of the acceptances between Run~1 and Run~1I.
The Run~1 analysis yield can be described as~\cite{Ref:MethodII},
\begin{equation}
N({\rm Run~I}) = \frac{Br(B_d \rightarrow K^{*0}\gamma)}{(4.36 \pm 1.13) \times 10^{-5}}.
\end{equation}
The expected signal yield of the Run~1 analysis was $1.03 \pm 0.17$
events with $Br(B_d \rightarrow K^{*0}\gamma)= 4.5 \times 10^{-5}$.
To calculate the acceptance ratio between Run~1 and Run~1I, we require
all the tracks ($e^+$, $e^-$, $K^+$, and $\pi^-$) to have $p_T > 400$
MeV/$c$ and to pass through the full tracking volume to ensure high
resolution and reconstruction efficiency.  We also require the same
offline selection cuts as the Run~1 analysis ($ct(B) > 100$ $\mu$m and
$|d_0(K,\pi)|>$ 4.5$\sigma$ = 100 $\mu$m).  We correct for the
improved SVX acceptance in Run~II and the relative efficiencies of the
Run~I and Run~II track processors, and we assume the SVT tracking
efficiency to be 0.88 per track.  We also assume the photon conversion
probability before the central drift-chamber tracking volume to be 6\%
in Run~1 and 8\% in Run~1I.  Other efficiencies are assumed to cancel
in the ratio.  For the $B_s \rightarrow K^{*0}\gamma$ channel, we
expect the branching fraction to be scaled by $|V_{td}|^2/|V_{ts}|^2
\sim 0.16$ relative to $Br(B_d \rightarrow K^{*0}\gamma)$, and the
ratio of the production rates for $B_s$ and $B_d$ mesons is $f_s/f_s =
0.426 \pm 0.07$ \cite{Affolder:2000iq}. Thus the expected yield is
\begin{eqnarray}
N(B_s \rightarrow K^{*0}\gamma) \sim \frac{f_s}{f_d}\frac{|V_{td}|^2}{|V_{ts}|^2}N(B_d \rightarrow K^{*0}\gamma) \sim
0.07 \times N(B_d \rightarrow K^{*0}\gamma).
\end{eqnarray}
\index{$B\to K^*\gamma$!event rates, CDF}
For the same-side 4\,GeV electron plus displaced track selection, we
expect the following signal yields after all cuts:
\begin{eqnarray}
N(B_d \rightarrow K^{*0}\gamma) &=& (170 \pm 40) 
\times \frac{\int{\cal L}~({\rm fb}^{-1})}{2 {\rm ~fb}^{-1}}
\times \frac{Br(B_d \rightarrow K^{*0}\gamma)}{4.5 \times 10^{-5}}
\\
N(B_s \rightarrow K^{*0}\gamma) &=& (12 \pm 4) 
\times \frac{\int{\cal L}~({\rm fb}^{-1})}{2 {\rm ~fb}^{-1}}
\times \frac{Br(B_d \rightarrow K^{*0}\gamma)}{4.5 \times 10^{-5}}.
\end{eqnarray}
Note that lowering the electron threshold to 3\,GeV would
increase the acceptance by about 50\% but would lead to significantly
higher trigger rates.

\subsubsection{Radiative $b$ Baryon Decays}

Since the $\Lambda$ baryon has a long lifetime ($c\tau=8$\,cm), most of the
 $\Lambda$ decays from $\Lambda_b \rightarrow \Lambda
\gamma \rightarrow p\pi ee$ events are expected to be outside of the
SVX fiducial volume, so there would be a low probability for the
proton from the $\Lambda$ to be reconstructed by the SVT.  A way to
trigger on this channel is to find an electron from the conversion and
find a displaced track that originates from the opposite $b$ quark.
This electron plus opposite-side displaced track selection is also
described in detail in Section~\ref{cdfsemi}.  We would require an
electron with a 4\,GeV threshold and a displaced track found by the
SVT with $p_T>$ 2~GeV/$c$ and $d_0>120\,\mu$m with a large opening
angle between the two ($\Delta\phi>90^\circ$) and transverse mass
$M_T>5$\,GeV/$c^2$ such that the electron and track not come from the
decay of a single $b$ hadron.
\index{$\Lambda_b\to\Lambda\gamma$!trigger strategy, CDF}

The expected yield for the Run~1 $\Lambda_b \rightarrow \Lambda\gamma$
search can be summarized in terms of the acceptance as~\cite{Ref:Lamb},
\begin{equation}
N({\rm Run~I}) = \frac{Br(\Lambda_b \rightarrow \Lambda\gamma)}{(2.80 \pm 0.95) \times 10^{-4}}.
\end{equation}
Thus the expected signal events of the Run~1 analysis is $0.16 \pm 0.06$ events
with $Br(\Lambda_b \rightarrow \Lambda\gamma)= 4.5 \times 10^{-5}$.
To calculate the acceptance ratio between Run~1 and Run~1I, 
we require all the tracks ($e^+$, $e^-$, $p$, and $\pi^-$) to have
$p_T > 400$ MeV/$c$ and pass through the full tracking volume.

In the Run~1 analysis, we required $p_T(\Lambda)>4$ GeV$/c$
for the $\Lambda$ decays reconstructed without SVX tracks and
 $p_T(\Lambda)>$ 2 GeV/$c$ with $|d_0(\Lambda)|>$ 70 $\mu$m 
for those reconstructed with SVX tracks.
For the Run~1I estimate, the $\Lambda$ is required to decay before the ISL 
(Radius $<$ 20 cm) to improve signal purity.  This allows us to
lower the $p_T(\Lambda)$ threshold to 2\,GeV/$c$.
The opposite-side SVT track required to be in the tracking fiducial with
$p_T>$ 2 GeV$/c$ and 120 $\mu$m $<|d_0|<$ 2 mm.
The signal yield with the opposite-side 4 GeV electron plus
displaced-track trigger is found to be
\index{$\Lambda_b\to\Lambda\gamma$!event rate, CDF}
\begin{equation}
N({\rm Run~II}) = (5.0 \pm 2.1) 
\times \frac{\int{\cal L}~({\rm fb}^{-1})}{2 {\rm ~fb}^{-1}}
\times \frac{Br(\Lambda_b \rightarrow \Lambda\gamma)}{4.5 \times 10^{-5}}.
\end{equation}

\subsubsection{$B_d \rightarrow K^{*0}\mu\mu$ Decays}

Because the trigger rate for dimuon events peaks at low $\mu\mu$ mass,
to trigger on $J/\psi\rightarrow\mu^+\mu^-$ decays at high
luminosity, CDF expects to impose a cut on transverse mass
$2<M_T<4$\,GeV$/c^2$ for $J/\psi$ selections.  However, the low-mass
region is needed in the study of $B_d \rightarrow K^{*0}\mu\mu$
decays.  Since the looser dimuon transverse mass window cut ($M_T<5$
GeV/$c^2$) for rare decays increases the Level 2 trigger rate by about
factor of 4, we need further background reductions.  We plan two
complementary trigger options:
\begin{itemize}
\item  Improving muon purity by requiring one of the trigger muons to
be found in the outer (CMP) muon chambers
\item  Requiring there to be a track of
2.0 GeV/$c$ found to be displaced by the SVT with $|d_0|>120\,\mu$m.
\end{itemize}
In the second case, the SVT-selected track can be one of the two muons
or a hadron track.  We expect the combined trigger rate for the two
selections to be about 10\,Hz at a luminosity of
$10^{32}\,\rm{cm}^{-2}\,\rm{s}^{-1}$. 
\index{$B\to K^*\ell^+\ell^-$!trigger strategy, CDF}

The signal yield is obtained by using the same procedure as the
radiative decays.  The expected yield of the Run~1 analysis can be described as~\cite{Affolder:1999eb}
\begin{equation}
N({\rm Run~I}) = \frac{Br(B_d \rightarrow K^{*0}\mu\mu)}{(1.65 \pm 0.33) \times 10^{-6}},
\end{equation}
and the expected signal events of the Run~1 analysis is $0.91 \pm
0.18$ events with $Br(B_d \rightarrow K^{*0}\mu\mu)= 1.5 \times
10^{-6}$.  To calculate the acceptance ratio between Run~1 and Run~1I,
we make the kinematic and geometric fiducial cuts as with the
radiative decays.  We also require the same offline selection cut as
Run~1 analysis ($L_{XY}(B) > 400$ $\mu$m and $|d_0(\mu,K,\pi)|>$
2$\sigma\simeq 50\,\mu$m).  We correct for the increased acceptance of
the muon triggers in Run~II.  For the dimuon + SVT trigger, we assume
the SVT tracking efficiency to be 0.88 per track. Any other
efficiencies are assumed to be canceled in the ratio.  Thus for an
assumed branching ratio of $1.5\times10^{-6}$, in 2\,fb$^{-1}$ CDF
expects to observe $44\pm9$ events with the tight muon selection and
$36\pm7$ events with the dimuon plus SVT selection for a combined
yield of $61\pm12$ events.
\index{$B\to K^*\ell^+\ell^-$!event rates, CDF}

\subsubsection*{Forward-Backward Asymmetry}

The Forward-Backward asymmetry in the $B_d \rightarrow \mu\mu K^{*0}$ decay
is defined as
\begin{equation}
A_{FB} = 
\frac{N(\cos\Theta>0) - N(\cos\Theta<0)}
{N(\cos\Theta>0) + N(\cos\Theta<0)}
=\frac{N_F-N_B}{N_F+N_B}
\end{equation}
where $\Theta$ is the angle between the direction of the $B_d$ and the
direction of the $\mu^+$ in the rest frame of the $\mu^+\mu^-$ system.
Note that the definition for the $B_d$ meson is the same as that for
the $\overline{B}_d$ meson so flavor tagging is not necessary to
measure $A_{FB}$.  In general $A_{FB}$ depends on the decay
kinematics.  Standard Model calculations predict the distribution of
$A_{FB}$ as a function of the dimuon mass to cross the zero around
$\sqrt{s}=M_{\mu\mu}=2$ GeV/$c$.  As discussed in Section~\ref{sec:btokll},
the $A_{FB}$ distribution strongly depends on the 
$B \rightarrow K^{*}$ form factor; however, the zero
position ($M_0$) is stable under various form-factor
parameterizations. Figure~\ref{Fig:AFBSusy} compares the $A_{FB}$ 
distributions predicted by the Standard Model with several SUSY 
models.  Some new physics models
predict there to be no zero in the $A_{FB}$ distribution.

Figure~\ref{Fig:Afbfinal} shows the expected $A_{FB}$ distributions
with 50 and 400 $B_d \rightarrow K^{*0}\mu\mu$ events after all the
trigger and offline requirements. The solid line in the figure
corresponds to the Monte Carlo generated distribution.
\index{$B\to K^*\ell^+\ell^-$!forward-backward asymmetry, CDF}

\begin{figure}[thbp]
\epsfxsize=0.95\textwidth
\centerline{\epsffile{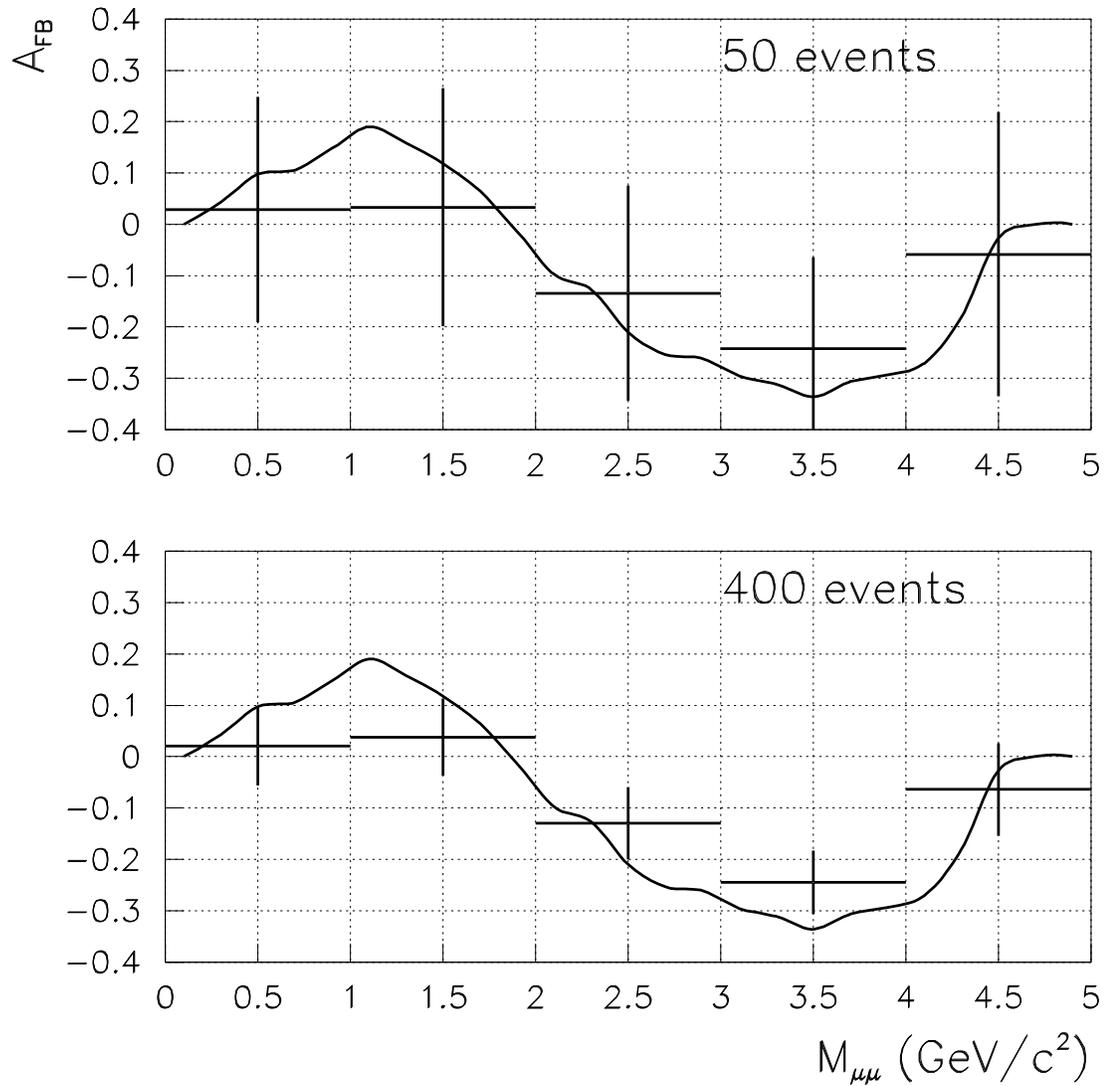}}
\vspace{0.2cm}
\caption{$A_{FB}$ with 50 and 400 events of the $B_d \rightarrow K^{*}\mu\mu$ 
signal and $S/B=1$.}
\label{Fig:Afbfinal}
\end{figure}

\subsubsection*{Asymmetry in Background Events}

Figure~\ref{Fig:AfbBg} shows the $A_{FB}$ distribution as a function of
$M_{\mu\mu}$ for the background, estimated from the same dataset as 
the Run~1 $B_d \rightarrow K^*\mu\mu$ search~\cite{Affolder:1999eb}.
We define four background regions,
\begin{itemize}
\item non-$b$ SR : (non-$b$-like $B$ mass signal region event),
\item non-$b$ SB : (non-$b$-like $B$ mass side-band event),
\item $B$ mass $b$ SR : ($b$-like $B$ mass signal region event),
\item $B$ mass $b$ SB : ($b$-like $B$ mass side-band event),
\end{itemize}
where the cuts are defined as,
\begin{itemize}
\item non-$b$-like : prompt; specifically $L_{XY}$, $d_0(\mu)$, $d_0(K)$, and $d_0(\pi) < 1\sigma$,
\item $b$-like : displaced; specifically $L_{XY} > 2\sigma$, and $d_0(\mu)$, $d_0(K)$, and $d_0(\pi) > 1\sigma$,
\item signal region : $|M_{B_d}-M_{\mu\mu K\pi}|<100$ MeV/$c^2$,
\item side-band : 100 MeV/$c^2$ $<|M_{B_d}-M_{\mu\mu K\pi}|<600$ MeV/$c^2$.
\end{itemize}
In the above $\sigma$ indicates the r.m.s. uncertainty of each quantity.
There is no significant forward-backward asymmetry in any of the background samples.\\
\index{$B\to K^*\ell^+\ell^-$!forward-backward asymmetry, CDF}.

\begin{figure}[thbp]
\epsfxsize=0.95\textwidth
\centerline{\epsffile{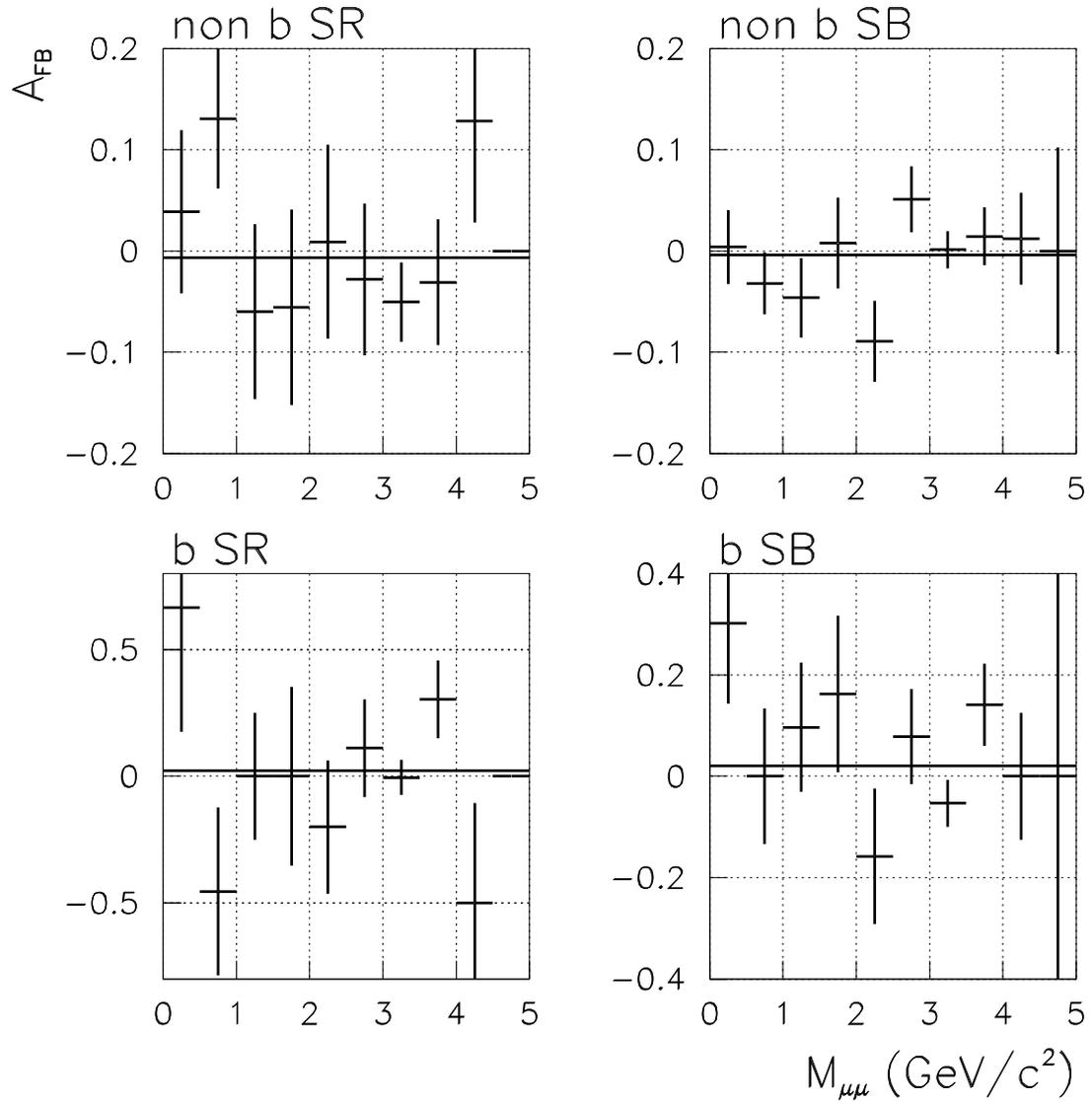}}
\vspace{0.2cm}
\caption{The forward-backward asymmetry for the background events
obtained from the Run~1 data.}
\label{Fig:AfbBg}
\end{figure}

\subsubsection*{Extraction of $A_{FB}$ Zero Point}

To extract the zero-point of the asymmetry with respect to $M_{\mu\mu}$, we define
the significance of $A_{FB}$ as
\begin{equation}
{\cal S} = \frac{N_{F}-N_{B}}{\sqrt{N_{F}+N_{B}+N_{BG}}}.
\end{equation}
We define a likelihood function to extract the zero position:
\begin{equation}
{\cal L} = {\cal S}(M_{\mu\mu}<M) - {\cal S}(M_{\mu\mu}>M)
= {\cal S}^- - {\cal S}^+.
\end{equation}
The likelihood is expected to be maximal at a mass $M_0$ where
$A_{FB}(M_0)=0$.  Figure~\ref{Fig:AfbZero} shows the $A_{FB}$ and
likelihood distributions in a Monte Carlo sample of 10000 signal
events and no background events.
\index{$B\to K^*\ell^+\ell^-$!forward-backward asymmetry, zero-point extraction, CDF}
We repeated the same analysis for the case of 50 (400) signal events
and a 1:1 signal-to-background ratio under the assumption there is no
background asymmetry.  The results are shown in
Figure~\ref{Fig:AfbZerofinal}.  The histograms show the distribution
of $M_0$ values for 1000 trials with signal sizes of 50 and 400
events.  The points are results for a generated samples with no
forward-backward asymmetry.  Therefore, it appears that it will be
difficult to extract the asymmetry zero point after only 2\,fb$^{-1}$
in Run~IIa, but the prospects are much more promising for
15\,fb$^{-1}$ in Run~IIb.  However, more work needs to be done on
defining an asymmetry significance.

\begin{figure}[thbp]
\epsfxsize=0.95\textwidth
\centerline{\epsffile{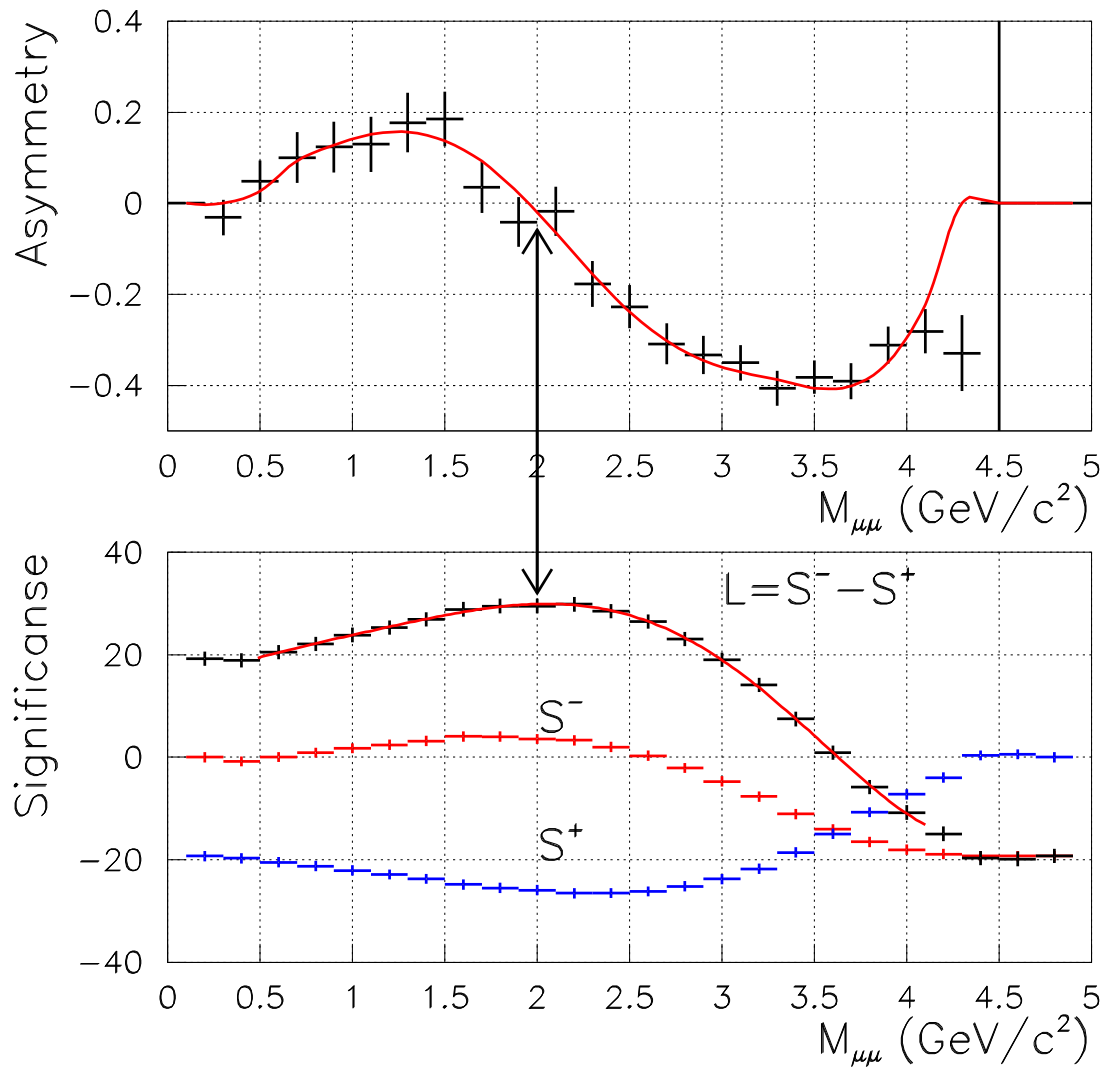}}
\vspace{0.2cm}
\caption{The forward-backward asymmetry in Ref~\cite{Greub:1995pi},
and a likelihood function for extracting the asymmetry zero-point $M_0$.}
\label{Fig:AfbZero}
\end{figure}

\begin{figure}[thbp]
\epsfxsize=0.95\textwidth
\centerline{\epsffile{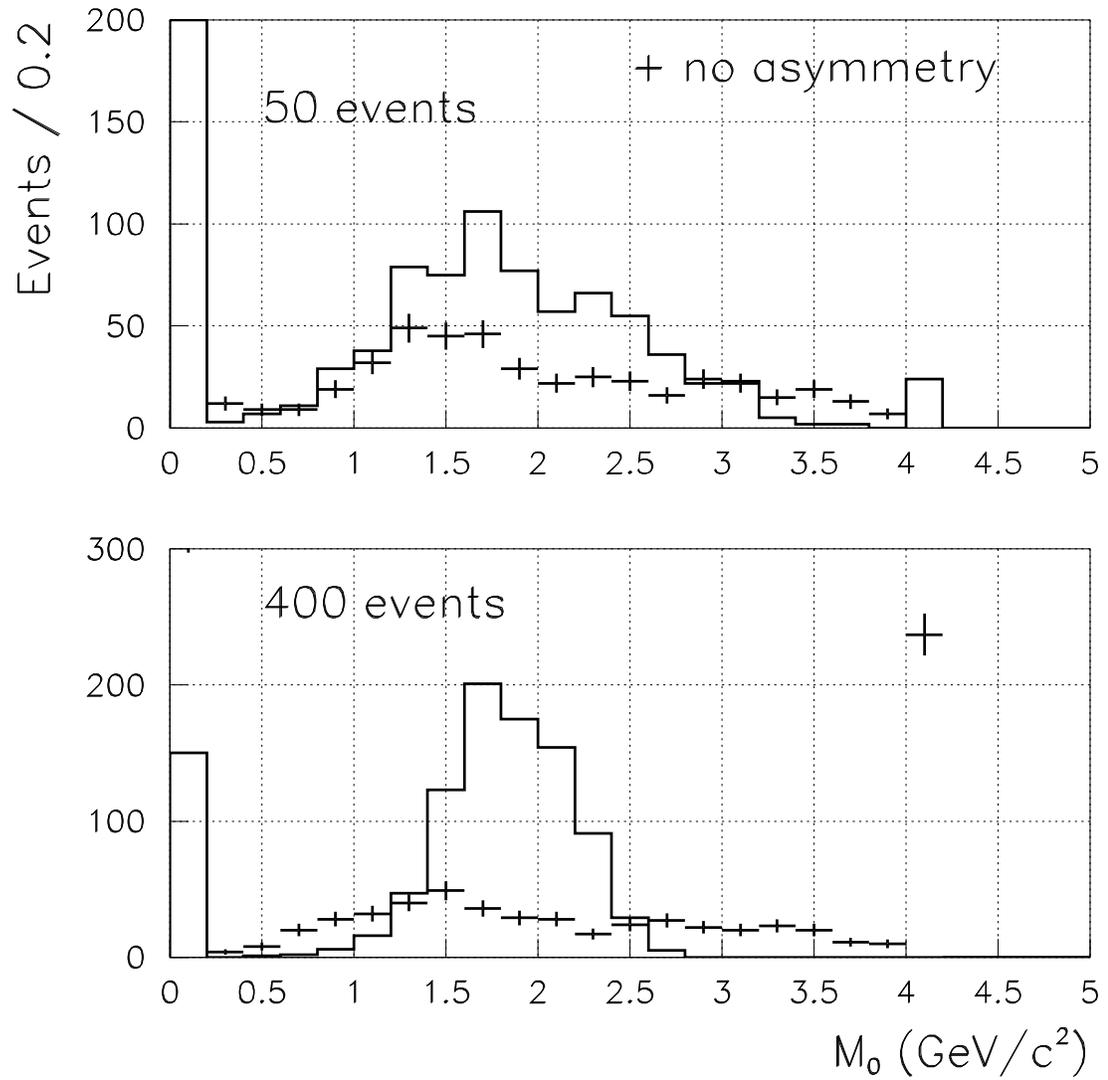}}
\vspace{0.2cm}
\caption{$A_{FB}$ and $M_{0}$ with 1000 $B_d \rightarrow K^{*}\mu\mu$ 
events and $S/B=1$.}
\label{Fig:AfbZerofinal}
\end{figure}

\subsubsection{$B \rightarrow \mu^+\mu^-$}

The dimuon triggers are also useful to study the two-body decay
$B_{d(s)} \rightarrow \mu^+\mu^-$.  Since the Standard Model predicts
the branching fraction of $B_d \rightarrow \mu^+\mu^-$ to be much
lower than the reach of CDF in Run~II, we give an expected
``single-event sensitivity'' instead of the signal yield.
Single-event sensitivity is defined as branching ratio for which we
would expect to observe one event in 2\,fb$^{-1}$.

The sensitivity is obtained by using the same procedure as the $B_d \rightarrow \mu\mu K^{*0}$ decays. 
The result of the Run~1 analysis is~\cite{Affolder:1999eb},
b\begin{eqnarray}
{\cal S}(B_d \rightarrow \mu\mu) &=& (2.0 \pm 0.5) \times 10^{-7}\\
{\cal S}(B_s \rightarrow \mu\mu) &=& (6.0 \pm 1.6) \times 10^{-7}.
\end{eqnarray}
The Run~1I expectation is obtained by scaling the Run~1 sensitivity for
the same trigger selections as CDF plans to use for $B^0\rightarrow\mu^+\mu^-
K^{*0}$.  Combining the results for the two trigger paths, we find the
sensitivities to be
\begin{eqnarray}
{\cal S}(B_d \rightarrow \mu\mu) &=& 3.5 \times 10^{-9}
\times \frac{2 {\rm ~fb}^{-1}}{\int{\cal L}~({\rm fb}^{-1})}\\
{\cal S}(B_s \rightarrow \mu\mu) &=& 1.0 \times 10^{-8}
\times \frac{2 {\rm ~fb}^{-1}}{\int{\cal L}~({\rm fb}^{-1})} .
\end{eqnarray}
Given the Standard module prediction $B_d$ and $B_s$ branching
fractions of $1.5\times10^{-10}$ and $3.5\times10^{-9}$ respectively,
we would expect a few $B_s \rightarrow \mu\mu$ signal in 15 fb$^{-1}$
of Run~IIb.
\index{$B_{s,d} \to \ell^+ \ell^-$!event rates, CDF}

\subsubsection{Summary}

We have examined the sensitivity of Run~1I CDF for the four rare-decay
modes $B_d(s) \rightarrow K^{*0}\gamma$, $\Lambda_b \rightarrow
\Lambda\gamma$, $B_d \rightarrow K^{*0}\mu\mu$, and $B_d(s)
\rightarrow \mu\mu$. The expected signal yields are obtained by
scaling the results of the Run~1 analyses:

\begin{eqnarray}
N(B_d \rightarrow K^{*0}\gamma) &=& (170 \pm 50) 
\times \frac{\int{\cal L}}{2 {\rm ~fb}^{-1}}
\times \frac{Br(B_d \rightarrow K^{*0}\gamma)}{4.5 \times 10^{-5}},\\
N(B_s \rightarrow K^{*0}\gamma) &=& (12 \pm 4) 
\times \frac{\int{\cal L}}{2 {\rm ~fb}^{-1}}
\times \frac{Br(B_d \rightarrow K^{*0}\gamma)}{4.5 \times 10^{-5}},\\
N(\Lambda_b \rightarrow \Lambda\gamma) &=& (4.0 \pm 1.7) 
\times \frac{\int{\cal L}}{2 {\rm ~fb}^{-1}}
\times \frac{Br(\Lambda_b \rightarrow \Lambda\gamma)}{4.5 \times 10^{-5}},\\
N(B_d \rightarrow K^{*0}\mu\mu) &=& (59 \pm 12) 
\times \frac{\int{\cal L}}{2 {\rm ~fb}^{-1}}
\times \frac{Br(B_d \rightarrow K^{*0}\mu\mu)}{1.5 \times 10^{-6}},\\
{\cal S}(B_d \rightarrow \mu\mu) &=& 3.5 \times 10^{-9}
\times \frac{2 {\rm ~fb}^{-1}}{\int{\cal L}},\\
{\cal S}(B_s \rightarrow \mu\mu) &=& 1.0 \times 10^{-8}
\times \frac{2 {\rm ~fb}^{-1}}{\int{\cal L}} .
\end{eqnarray}

 We also studied the forward-backward asymmetry in the $B_d
 \rightarrow K^{*0}\mu\mu$ decay and showed some ideas to extract the
 zero position of the $A_{FB}$ distribution.

\subsection{Rare Decays at BTeV}

        Because the Tevatron produces more than 10$^{11}$ $b$ hadrons per 
year, we should be able to observe some of these decays and to set 
stringent limits on others.  The precise vertexing of the BTeV silicon
pixel detector will allow us to easily differentiate $b$ decays from
non-$b$ backgrounds in the Tevatron environment.  We present the
expected sensitivities from studies of some of these decay channels.


\subsubsection{The Exclusive Channel $B^{0} \to K^{*0}\mu^{+}\mu^{-}$}
\label{sec:bkpmm}

\index{$B\to K^*\ell^+\ell^-$!backgrounds, BTeV}
Since we expect large backgrounds under the signal, an understanding of
these backgrounds is critical to understanding our
sensitivity. The various sources of background are:
\begin{itemize}
\item $b$-events where portions of the two $b$ hadrons in the event
      appear to form a vertex downstream of the production point. In
      approximately 1\% of all $b\bar{b}$ events
      both $B$ hadrons decay semileptonically producing two real muons.
      In addition, there is a charged kaon in at least one of the  
      $b$'s over  90\%  of the time.
\item Minimum bias events where three particles conspire to fake a 
      secondary vertex and two of the particles either decay 
      downstream of the magnet or make hadronic showers which leave a
      signal in the muon detector (hadron punch-through).
\item Charm events with one or more real muons and kaons.
\item More generally, any admixture of $b$, charm, minimum bias 
      events, primary interactions and secondary decays, combined with 
      hadronic punch-through.
\item Decays from single $B$ mesons where two charged pions fake muons.
\end{itemize}

The basic weapons to combat these backgrounds are:
\begin{itemize}
\item Excellent discrimination between the primary and secondary
  vertex, which eliminates backgrounds from minimum bias
  events and from the underlying event within a true $b$ event.
  Tracks which are not part of the $b$ vertex are easily rejected by
  requiring a non-zero vertex fit probability, as shown in
  Fig.~\ref{fig:kpmm_sg_bg}(a).  Also, the normalized decay length
  ($L/\sigma_{L}$), shown in Fig.~\ref{fig:kpmm_sg_bg}(c), provides
  additional discrimination against background. 
\item Excellent mass resolution (of order 17~MeV) on the final state, as
  shown in Fig.~\ref{fig:kpmm_fit}.
\item Excellent ``point-back'' resolution of the reconstructed $b$
  candidate with respect to the primary vertex.  This will help to
  reject vertices that have been artificially pieced together from
  particles from the two separate $b$'s in the event.  The normalized
  $B$ impact parameter ($b_{B}/\sigma_{b_{B}}$) with respect to the primary
  vertex is quite different for signal and background events, as shown in
  Fig.~\ref{fig:kpmm_sg_bg}(b).
\item The ability to reject combinations which include tracks
  that are from the primary vertex or other vertices in the event, by
  cutting on the impact parameter of the track with respect to that vertex.
  Figures~\ref{fig:kpmm_sg_bg}(e) and (f) show the normalized impact
  parameter of the kaon and pion with respect to the primary vertex 
  ($b_{K}/\sigma_{b_{K}}$ and $b_{\pi}/\sigma_{b_{\pi}}$).
\end{itemize}
In addition, the signal-to-background depends on the quality of both the
muon detector and the particle identification.

\begin{figure}[t]
\centering \leavevmode 
\epsfxsize=0.92\textwidth
\centerline{\epsffile{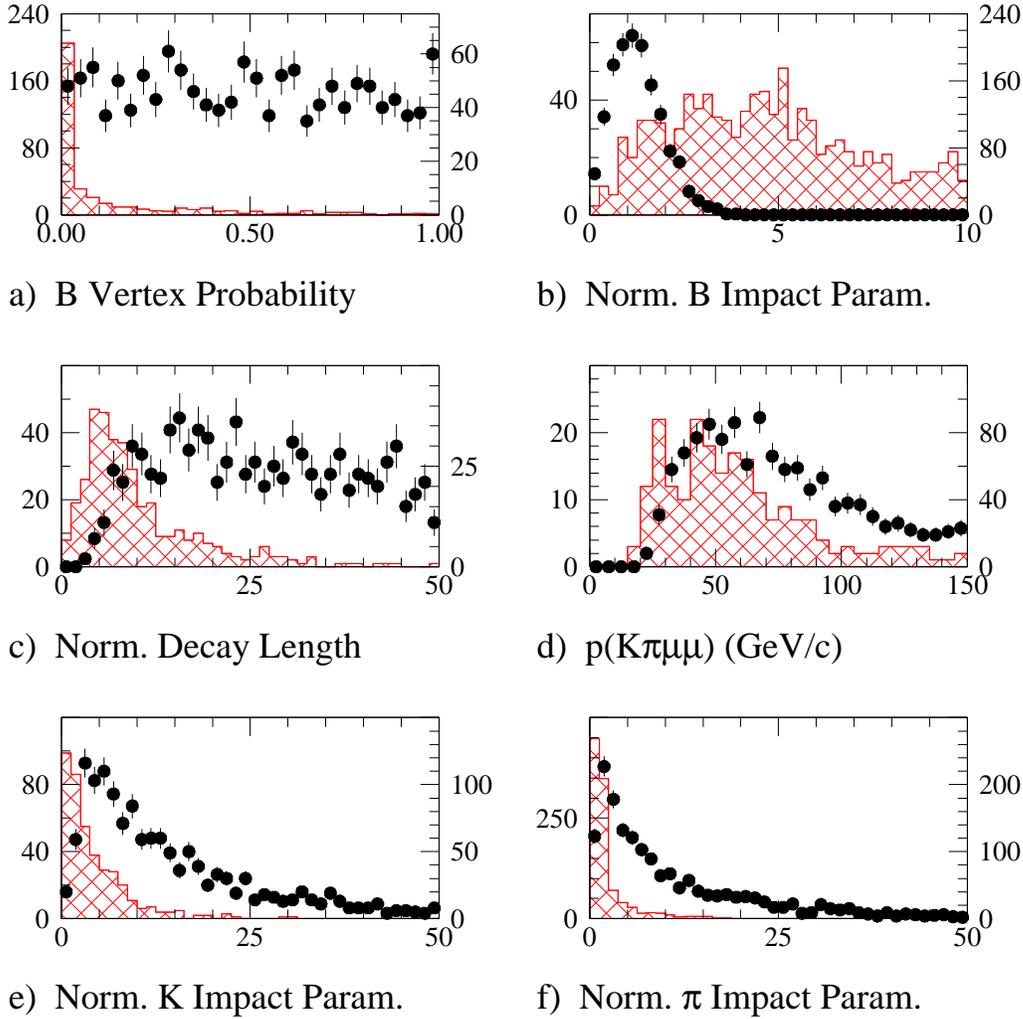}}
\caption{Distributions of cut variables for signal (points) and
  $b\overline{b}$ background (hatched) MCFAST events.}
\label{fig:kpmm_sg_bg}
\end{figure}

$B^{0} \rightarrow K^{*0}\mu^{+}\mu^{-}$ events were selected using
the following criteria: 
\index{$B\to K^*\ell^+\ell^-$!analysis cuts, BTeV}

\begin{itemize}
  \item Two muons of opposite charge, each with momentum greater than
    $5$ G$e$V$/c$.  Both muon tracks were required to have at least
    one hit in the muon chambers. 
  \item $K$ track momentum greater than $3$ G$e$V$/c$.  The kaon track
    was also required to have at least one hit in the forward drift
    chamber between the RICH and calorimeter.  Perfect $\pi/K$
    separation and 100\% efficiency for reconstructing the
    Cherenkov photons of tracks which traverse the RICH is assumed. 
  \item Good primary vertex with probability greater than 0.01.
  \item Good $b$ vertex with probability greater than 0.01.
  \item Decay length greater than $7\sigma$.
  \item $B$ impact parameter with respect to the primary vertex
    less than $2.5\sigma $.
  \item $K$ impact parameter with respect to the primary vertex
    greater than $2.5\sigma $. 
  \item $\pi$ impact parameter with respect to the primary vertex
    greater than $2.5\sigma $. 
  \item $B$ momentum greater than $20$ G$e$V$/c$.
  \item $|m(K\pi)-m_{K^{*0}}| < 50$ M$e$V$/c^{2}$.
  \item Cut 100 M$e$V$/c^{2}$ about the $J/\psi$ and $\psi^{\prime}$
    nominal masses to remove regions dominated by 
    $B \rightarrow \psi K^{*}$ and $B \rightarrow \psi^{\prime}
    K^{*}$, which interfere with the signal.
\end{itemize}

Of 4.4~pb$^{-1}$ of MCFAST $b\overline{b}$ background events generated
(about one million events), nine pass the selection criteria.  
For 2~fb$^{-1}$ of data (one year of running at a luminosity of
$2 \times 10^{32}$~cm$^{-2}$~s$^{-1}$), this would correspond to
$4090$ events in the range  
$4.7\ {\rm G}e{\rm V}/c^{2} <  m(K\pi\mu\mu) < 5.7\ {\rm G}e{\rm
  V}/c^{2}$, shown in Fig.~\ref{fig:kpmm_fit}.  The width of the
$B^{0}$ mass peak obtained from the 
MCFAST signal Monte Carlo sample is $17\ {\rm M}e{\rm V}/c^{2}$.  Thus,
we can expect about 280 background events from semileptonic
$b\overline{b}$ decays under the $B^{0}$ mass peak, as shown in 
Table~\ref{tab:sens_bkpmm}.  Considering that
we expect about 2240 signal events, this corresponds to a signal to
background ratio of about 8.
\index{$B\to K^*\ell^+\ell^-$!backgrounds, BTeV} 
\index{$B\to K^*\ell^+\ell^-$!event rates, BTeV}

We did not include the decay $B^{-}\rightarrow \psi K^{-}$ as a background.
That decay is large compared to the rare decay being considered here
and will interfere with the rare decay and distort the dimuon
mass distribution in the vicinity of 3 GeV/c$^{2}$.
This, however, is a physics contribution and will
certainly be observed and studied based on a mass cut on the dimuon.
In fact, this state can be used to calibrate the efficiency of the
analysis and can be used as a normalization for a measurement
of the relative branching fraction.

A sample of 2~fb$^{-1}$ of signal MCFAST Monte Carlo events were generated
according to the Standard Model prediction for $A_{fb}$ and $Q^{2}$
\cite{Greub:1995pi}.  Figure~\ref{fig:asym} shows the distributions
of $A_{fb}$ and number of events as a function of $m(\mu^{+}\mu^{-})$
for this sample, after all cuts have been
applied.  With our estimated signal to background, we should be able
to easily observe and measure the position of a zero in the asymmetry 
if it exists, or make a strong case for non-Standard Model physics, 
if it does not.
\index{$B\to K^*\ell^+\ell^-$!forward-backward asymmetry, BTeV}

\begin{figure}[htp]
\centering \leavevmode 
\epsfxsize=2.5in
\epsffile{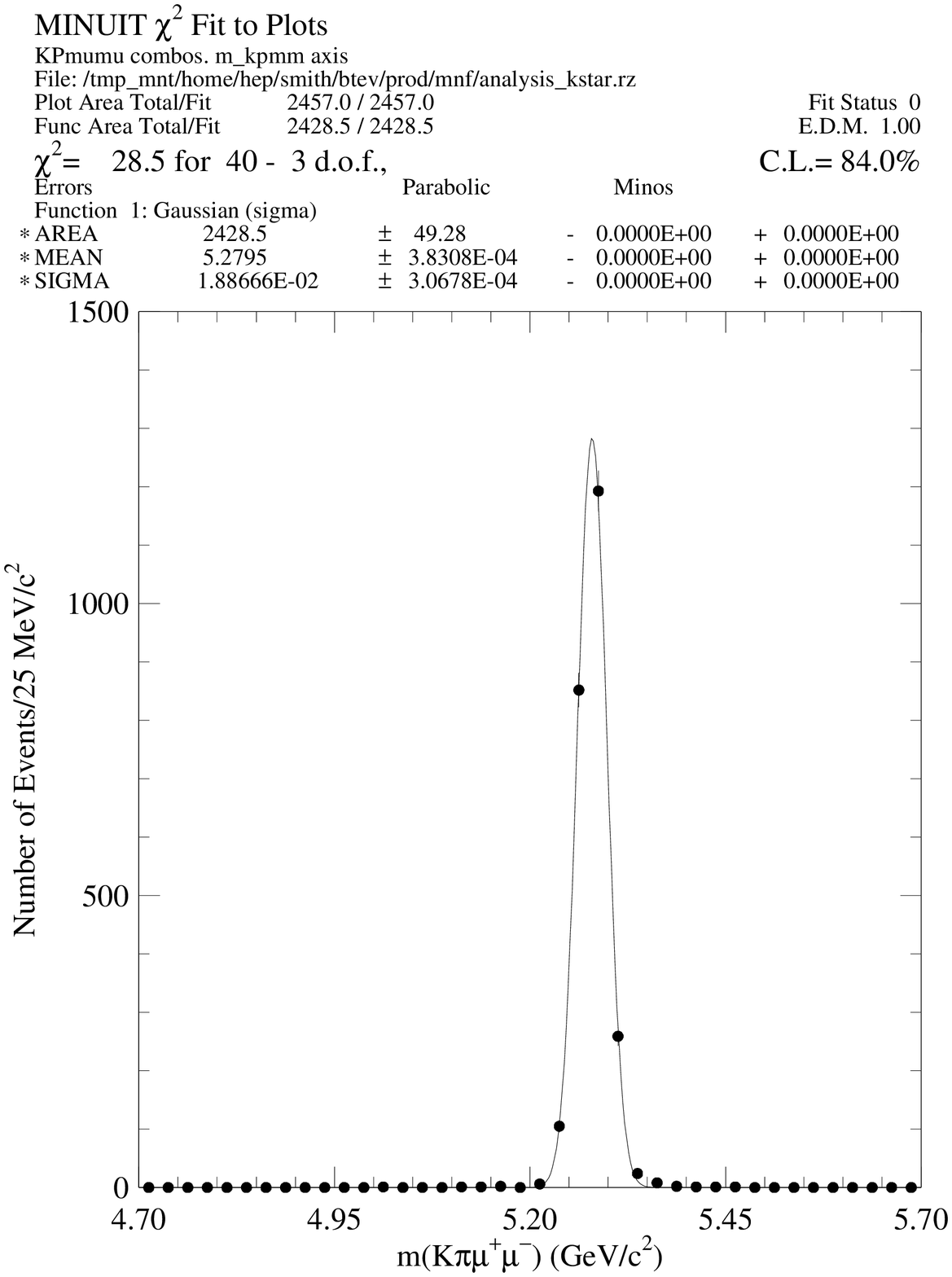}
\epsfxsize=2.5in
\epsffile{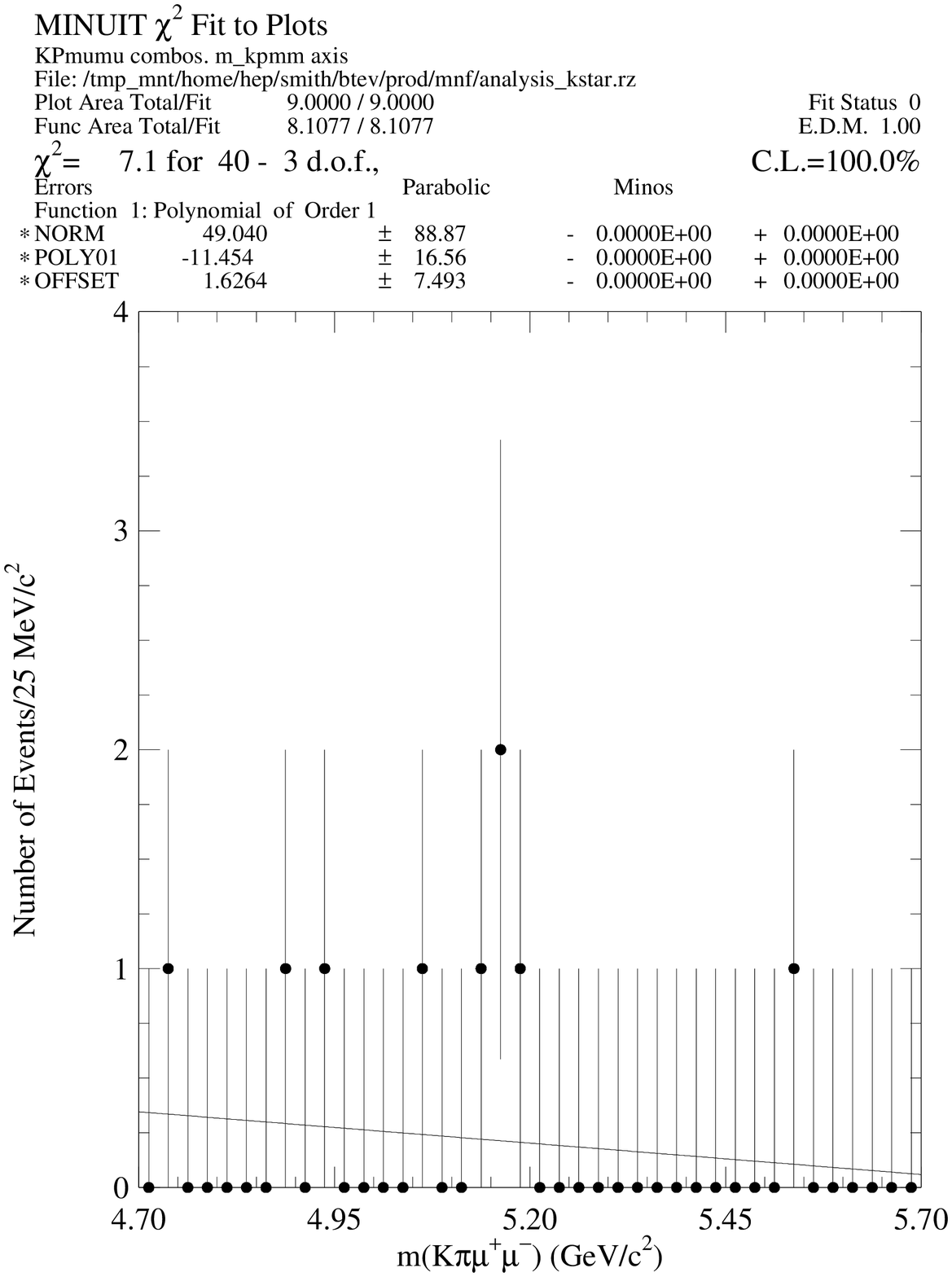}
\caption{Distributions of tagged $B^{0} \rightarrow
  K^{*0}\mu^{+}\mu^{-}$ signal (left) and 4.4~pb$^{-1}$ of 
  $b\overline{b}$ background (right) MCFAST events.}
\label{fig:kpmm_fit}
\end{figure}

\renewcommand{\arraystretch}{1.1}
\begin{table}[htp]
\begin{center}
\begin{tabular}{|c|c|} \hline \hline
Integrated Luminosity & 2~fb$^{-1}$ \\ \hline 
$b\overline{b}$ Cross Section & $100\mu$b \\ \hline 
Number of $b\overline{b}$ Pairs Produced & $2\times10^{11}$ \\ \hline 
$N_{B^{0}} + N_{\B0bar}$ Produced & $1.4\times10^{11}$ \\ \hline 
Est. ${\cal B}(B^{0}\rightarrow K^{*0}\mu^{+}\mu^{-})$ & $(1.5\pm
0.6)\times 10^{-6}$ \\ \hline 
${\cal B}(K^{0*}\rightarrow K^{+}\pi^{-})$ & $0.67$ \\ \hline 
Number of Signal Events Produced & $1.4 \times 10^{5}$ \\ \hline 
$\epsilon_{\rm trig}$ & $80 \%$ \\ \hline
$\epsilon_{\rm cuts}$ & $2.0 \%$ \\ \hline
Number of Signal Events & $2240$ \\ \hline 
Number of Background in Signal Box & $280$ \\ \hline
Signal$/$Background & $8$ \\ \hline \hline
\end{tabular}
\end{center}
\caption[Estimate of sensitivity to $B^{0}\rightarrow K^{*0}\mu^{+}\mu^{-}$ for
an integrated luminosity of 2\,fb$^{-1}$.  Only backgrounds from 
$b\overline{b}$ semileptonic decays were included.]{Estimate of sensitivity to
$B^{0}\rightarrow K^{*0}\mu^{+}\mu^{-}$ for an integrated luminosity of
2\,fb$^{-1}$.  Only backgrounds from $b\overline{b}$ semileptonic decays were
included in this study.} 
\label{tab:sens_bkpmm}
\end{table}
\renewcommand{\arraystretch}{1.0}

\begin{figure}[htp]
\centering \leavevmode 
\epsfxsize=7.0cm
\epsffile{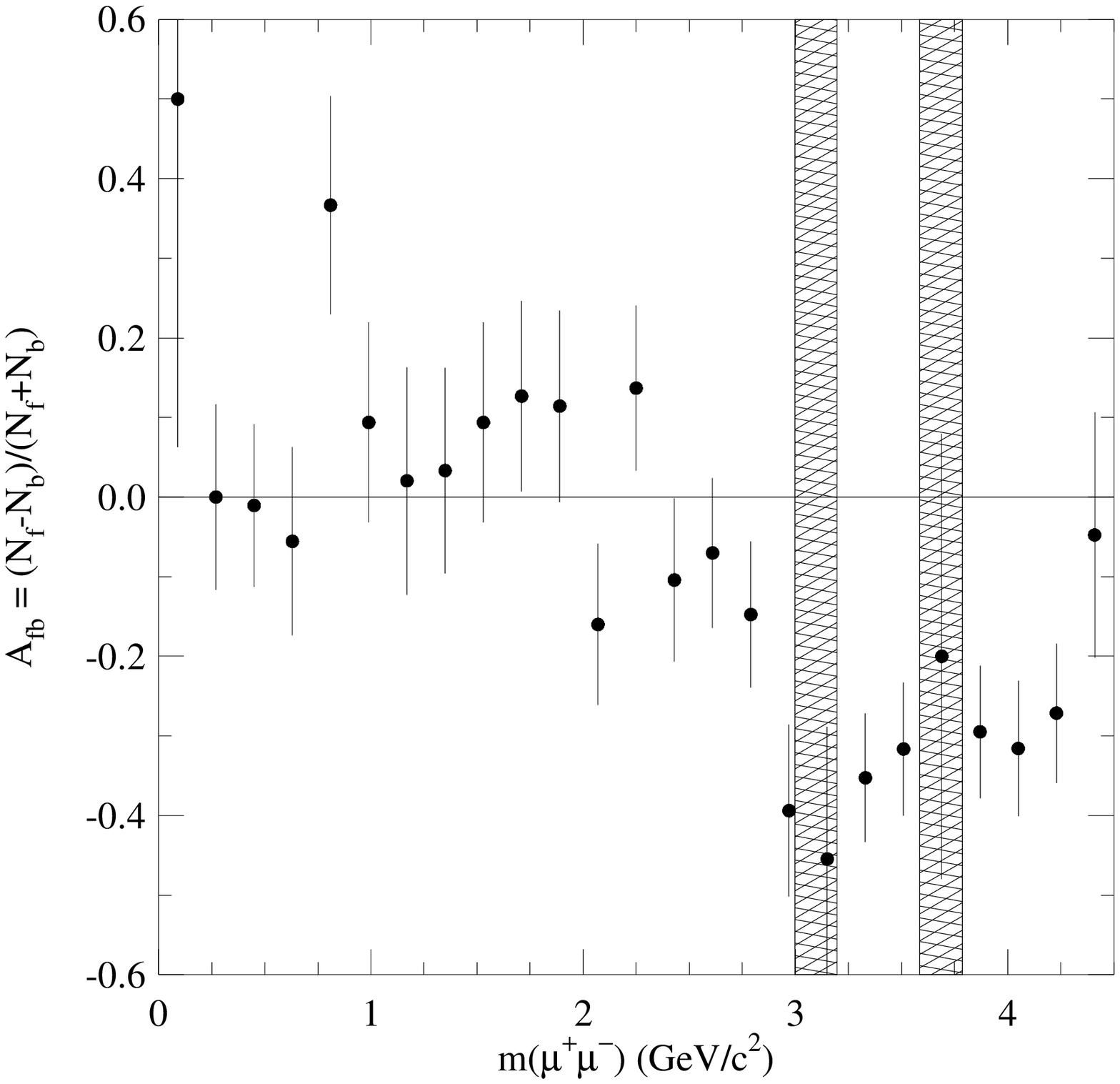}
\epsfxsize=7.0cm
\epsffile{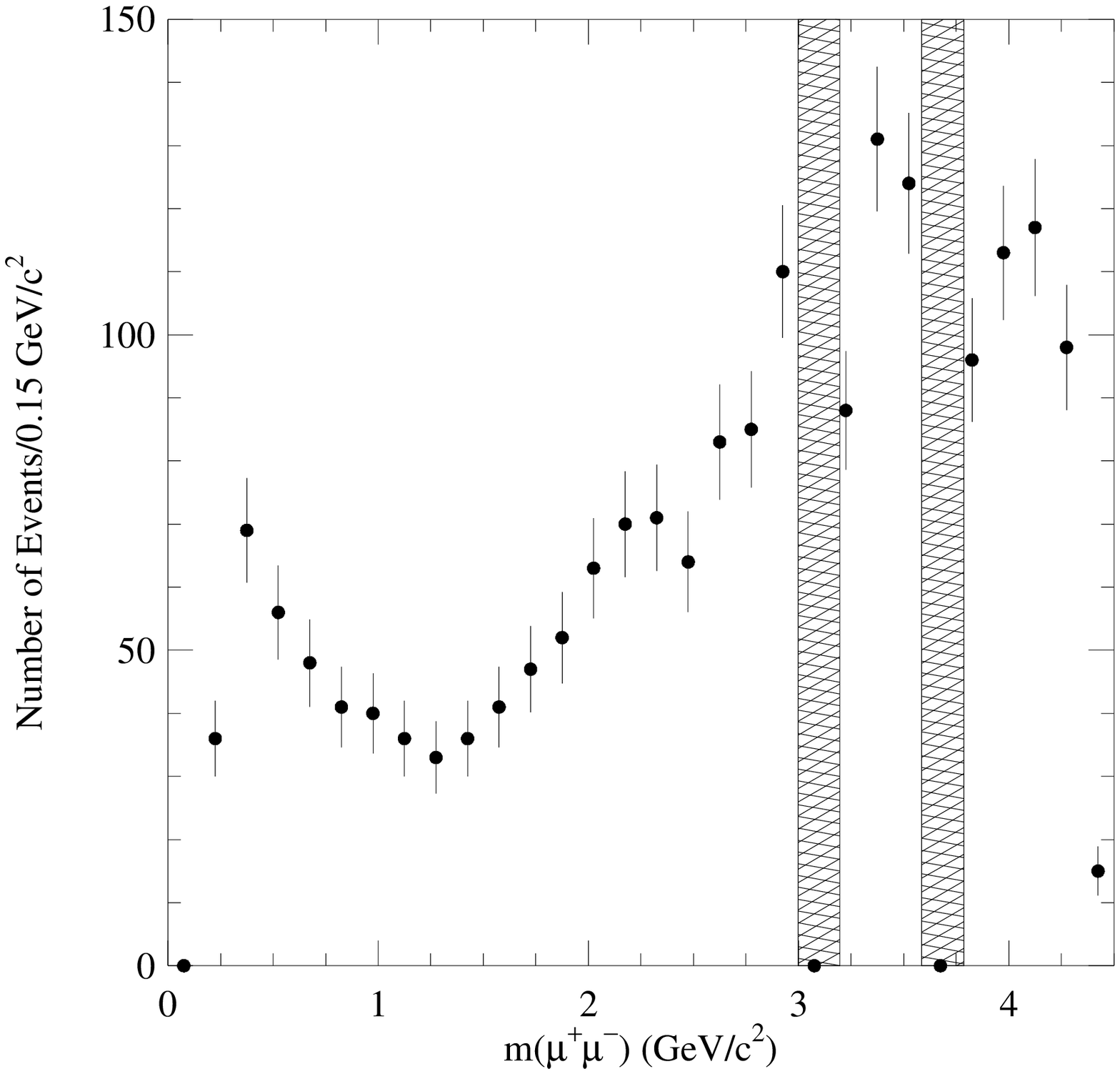}
\caption[Expected forward-backward asymmetry and number of events as a function
of $m(\mu^{+}\mu^{-})$ for signal events after one year of running.]{Expected
forward-backward asymmetry (left) and number of events (right) as a function of
$m(\mu^{+}\mu^{-})$ for signal events after one year of running. No background
is included in these plots.}
\label{fig:asym}
\end{figure}


\subsubsection{The Exclusive Channel 
               $B^{+}\rightarrow K^{+}\mu^{+}\mu^{-}$}

While the channel $B^{+} \rightarrow K^{+}\mu^{+}\mu^{-}$ is not as rich as 
$B^{0} \rightarrow K^{*0}\mu^{+}\mu^{-}$, in that the asymmetry
$A_{fb}$ is expected to be small within the Standard Model and beyond the
Standard Model, a measurement of the decay rate
is still a sensitive probe of new physics.
In particular, measurement of the differential decay rate will provide
input to determine the magnitude and sign of the Wilson coefficients $C_{7}$,
$C_{9}$, $C_{10}$.

\index{$B\to K\ell^+\ell^-$!analysis cuts, BTeV}
Most of the backgrounds to this channel are the same as those listed
for the $B^{0} \rightarrow K^{*0}\mu^{+}\mu^{-}$ analysis in
Section~\ref{sec:bkpmm}.  
Events for this study were selected using nearly the same criteria as
the $B^{0} \rightarrow K^{*0}\mu^{+}\mu^{-}$ analysis:
\begin{itemize}
  \item $K$ track momentum greater than $4$ G$e$V$/c$.  The kaon track
    was also required to have at least one hit in the forward drift
    chamber between the RICH and calorimeter.  Perfect $\pi/K$
    separation and 100\% efficiency for reconstructing the
    Cherenkov photons is assumed. 
  \item Two muons with momentum greater than $5$ G$e$V$/c$.  Both muon
    tracks were required to have at least one hit in the muon chambers.
  \item Good primary vertex with probability greater than 0.01.
  \item Good $b$ vertex with probability greater than  0.01.
  \item Decay length greater than $7\sigma$.
  \item $B$ impact parameter with respect to the primary vertex
    less than $2.5\sigma $.
  \item $K$ impact parameter with respect to the primary vertex
    greater than $2.5\sigma $. 
  \item $B$ momentum greater than $20$ G$e$V$/c$.
  \item Cut 100 M$e$V$/c^{2}$ about the $J/\psi$ and $\psi^{\prime}$
    nominal masses to remove regions dominated by 
    $B \rightarrow \psi K^{*}$ and $B \rightarrow \psi^{\prime}
    K^{*}$, which interfere with the signal.
\end{itemize}

We have not simulated all sources of background. Our estimates
indicate that the most serious background is from events with pairs of $b$'s, 
each of which undergoes semileptonic decay. The background contribution 
was estimated by applying the selection criteria
to a sample of 2.5 million MCFAST semileptonic $b\overline{b}$ events, 
corresponding to a luminosity of 10~pb$^{-1}$.  Of these events, 41
passed the selection cuts and fall within a 1 G$e$V$/c^{2}$ window
centered on the $B^{+}$ nominal mass.  Extrapolating to an integrated 
luminosity of  2~fb$^{-1}$,
we expect about 8200 events in this window. 
Assuming a uniform distribution across the 
$B$ mass window (this is conservative, since it is actually falling,
as shown in Fig.~\ref{fig:kmm_fit}), one can expect about 560 events
within the $2\sigma$ of the $B^{+}$ mass.  
\index{$B\to K\ell^+\ell^-$!backgrounds, BTeV}

The overall efficiency for this state, with cuts designed to achieve
good background rejection, is about 3.0\%. 
Table~\ref{tab:sens_bkmm} gives a calculation of the yield obtained
for an integrated luminosity of 2~fb$^{-1}$.
We include in this 
calculation a triggering efficiency of 80\% for those events which
satisfy all the analysis cuts. This is consistent with what we expect to
get from the dimuon trigger (70\%) `or-ed' with the vertex trigger
which recovers almost half of what the muon trigger failed to accept.
The number of signal events passing the trigger and all selection
criteria is approximately 1680.  This gives an impressive
signal-to-background ratio of 3. 
\index{$B\to K\ell^+\ell^-$!event rates, BTeV}

The reason that BTeV can achieve excellent signal-to-background is due in
a large part to a powerful particle identification system. For example,
the version of the CDF detector described in the
CDF~II Technical Design Report \cite{bib:CDFTDR}, 
lacks particle
identification for tracks above 1 GeV/c. 
So although CDF expects
a signal of 100-300 $B^{+}\rightarrow K^{+} \mu^{+}\mu^{-}$ events in 
Run~II for that version of the detector,
they would be exposed to background
from all pions in the event conspiring with the muons to create
background. It is unlikely that CDF's signal-to-background
in Run~II (0.1 in Run~I) will approach that expected at BTeV.
In BTeV, because of the RICH, only the kaons can contribute to the
background and there are fewer of them.

\begin{figure}[t]
\centering \leavevmode 
\epsfxsize=2.5in
\epsffile{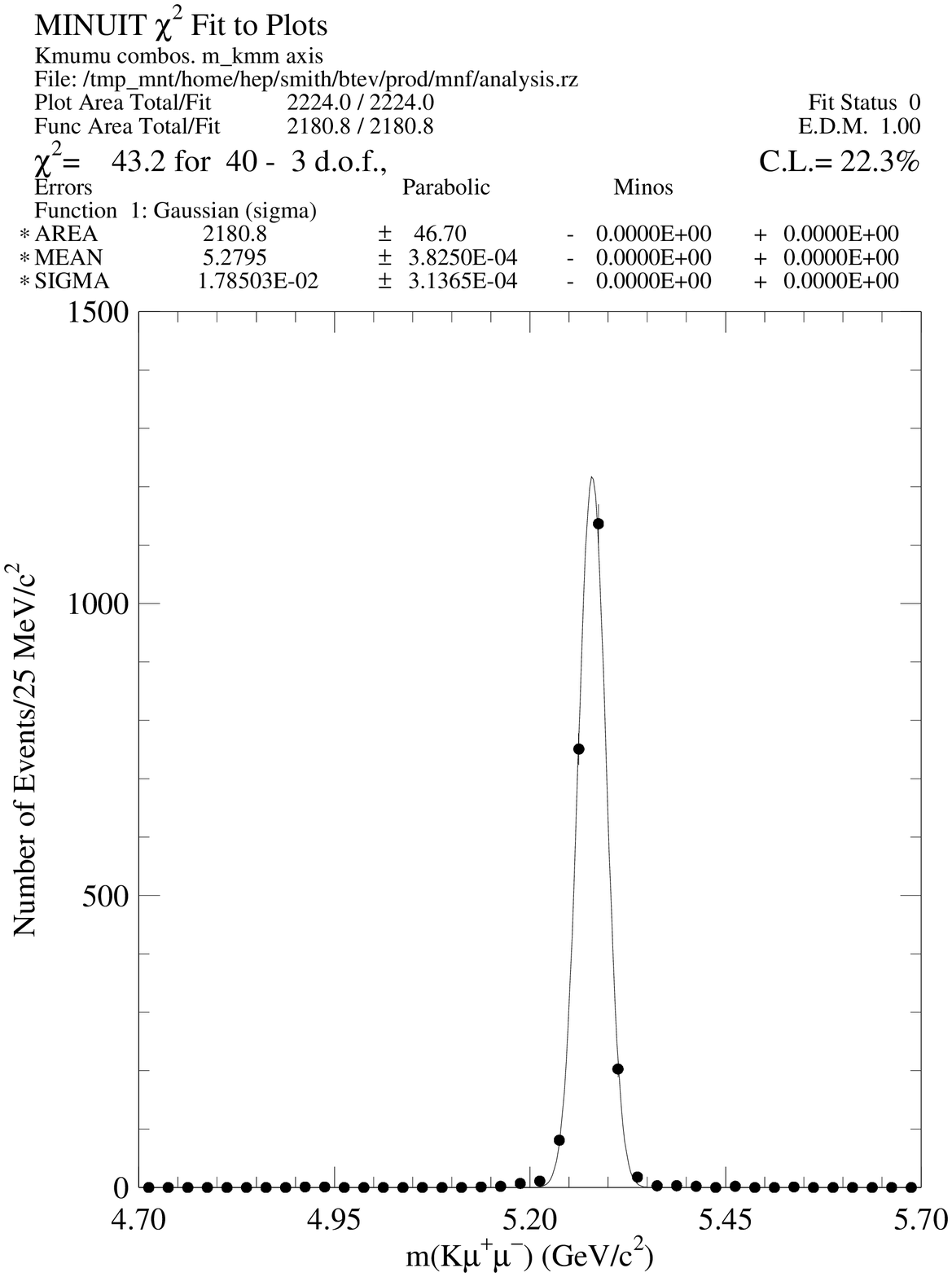}
\epsfxsize=2.5in
\epsffile{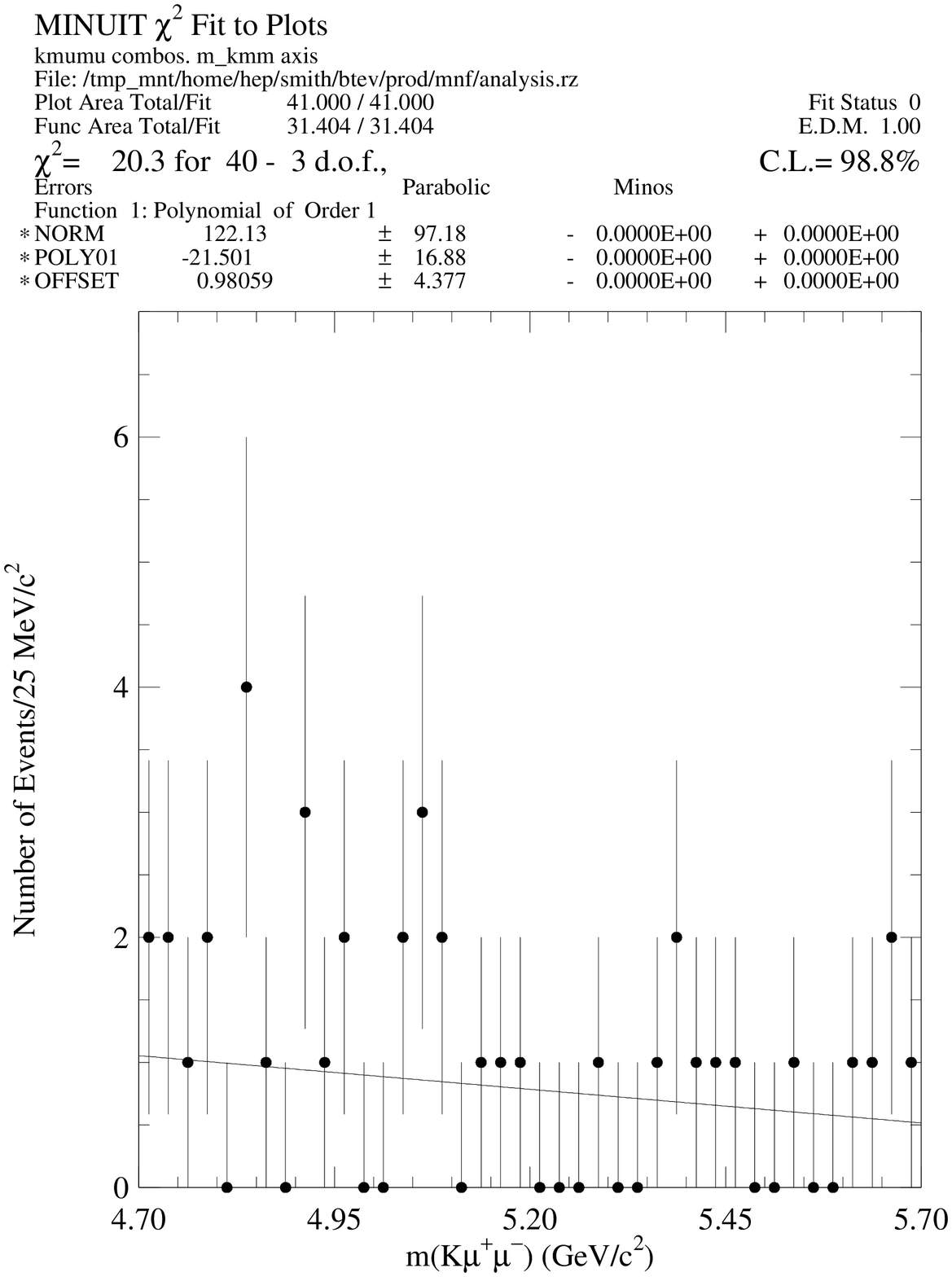}
\caption{Distributions of $B^{+}\rightarrow K^{+} \mu^{+}\mu^{-}$
  signal (left) and 10~$pb^{-1}$ of $b\overline{b}$ background (right)
  events.}
\label{fig:kmm_fit}
\end{figure}

\renewcommand{\arraystretch}{1.1}
\begin{table}[htp]
\begin{center}
\begin{tabular}{|c|c|} \hline \hline
Integrated Luminosity & 2~fb$^{-1}$ \\ \hline 
$b\overline{b}$ Cross Section & $100\mu$b \\ \hline 
Number of $b\overline{b}$ Pairs Produced & $2\times10^{11}$ \\ \hline 
Number of $B^{+}/B^{-}$ Produced & $1.4\times10^{11}$ \\ \hline 
Est. ${\cal B}(B^{+}\rightarrow K^{+}\mu^{+}\mu^{-})$ & $(4.0\pm
1.5)\times 10^{-7}$ \\ \hline 
Number of Signal Events Produced & $5.6 \times 10^{4}$ \\ \hline 
Trigger Efficiency & $80 \%$ \\ \hline
Selection Cut Efficiency & $3.0 \%$ \\ \hline
Number of Background Events in Signal Box & $560$ \\ \hline
Number of Signal Events & $1680$ \\ \hline 
Signal$/$Background & 3 \\ \hline \hline
\end{tabular}
\end{center}
\caption{Estimate of sensitivity to $B^{+}\rightarrow K^{+}\mu^{+}\mu^{-}$ for
an integrated luminosity of 2\,fb$^{-1}$.} 
\label{tab:sens_bkmm}
\end{table}
\renewcommand{\arraystretch}{1.0}


\subsubsection{The Inclusive Decay $b \rightarrow s\ell^{+}\ell^{-}$}

\index{$B \to X_s\ell^+\ell^-$!intended approach, BTeV}
Studies are underway to determine if the BTeV detector will provide
enough background rejection to make a competitive measurement of
inclusive $b \rightarrow s\ell^{+}\ell^{-}$.  The method under study is 
similar to that used by CLEO
\cite{bib:CLEObsll}\cite{bib:CLEObsgamma}, in which a kaon and 0-4
pions are combined with the dilepton pair.  For the purposes of this
study, no neutral pion candidates are allowed in the combination.  

The ability to precisely reconstruct $b$ vertices in BTeV will be
instrumental in removing combinations involving non-$b$ daughters.

Unlike the exclusive
modes, an inclusive measurement would provide a {\em model independent}
determination of the Wilson coefficients.  It is important to avoid
restricting this measurement to only the high $m(\ell^{+}\ell^{-})$
region above the $\psi^{\prime}$, as this introduces model dependence.

\section{Summary of Rare Decays}

Rare $b$ decays provide detailed tests of the flavor structure of the 
SM at the loop level, and as
such provide a complementary probe of new physics to that of direct 
collider searches.
While radiative $b\to s\gamma$ decays are sensitive only to the 
magnitude of the Wilson coefficient $C_7$, the semileptonic rare decays 
$b\to s\ell^+\ell^-$ and the purely leptonic decays 
$B_{d,s}\to\ell^+\ell^-$ are sensitive to additional operators, and so 
further constrain new physics.

Inclusive decays are in general cleaner theoretically than exclusive, 
while experimentally the difficulty is in the opposite order.  Because 
of the difficulty of inclusive measurements, theoretical techniques to 
handle exclusive modes in a model-independent fashion are extremely 
important.  There has been much recent theoretical interest in the large 
energy limit (LEL) of QCD, which simplifies exclusive heavy-light decays 
in the limit that the decay products are produced at large recoil.   
This has particular application to $B\to K^*\ell^+\ell^-$ decays.  In 
particular, the position of the zero in the forward-backward asymmetry 
in this decay has been shown to be model-independent.  We look forward 
to additional advances in the theoretical understanding of the LEL.

In Run IIa, the radiative $b$ decays $B_d\to K^{*0}\gamma$, $B_s\to 
K^{*0}\gamma$
and $\Lambda_b\to\Lambda\gamma$ are expected to be measured, while the 
purely leptonic decays $B_d\to\mu^+\mu^-$ and
$B_s\to\mu^+\mu^-$ are not expected to be visible at the SM level.  
Probably the most important decay studied in this section for Run II is 
$B\to K^*\ell^+\ell^-$.  While this decay should be seen at 
2\,fb$^{-1}$, precision study (particularly of the zero in the 
forward-backward asymmetry) will require larger integrated luminosity.   
The inclusive measurement $b\to X_s\ell^+\ell^-$ is most easily seen for 
large dimuon mass (above the $\psi(2S)$), but in this region the theory 
breaks down, and so the measurement is of limited interest.  BTeV is 
currently studying the feasibility of building up an inclusive 
measurement for lower dimuon invariant mass from exclusive measurements 
involving a kaon and 0-4 pions.

\section{Semileptonic Decays: Theory}
\subsection{Introduction}

Semileptonic decays have long been used to determine elements of the
CKM matrix. Examples are the determination of $|V_{ud}|$ from nuclear
$\beta-$decay, $|V_{us}|$ from $K_{l3}$ decays, and $|V_{cb}|$ from
$B \to D^{(*)} \ell \bar{\nu}$ \cite{pdg}. In every one of these three cases
a flavor symmetry (isospin, SU(3) flavor, and HQS, respectively)
greatly simplifies the theoretical understanding of the hadronic 
matrix element in question. In the symmetry limit, and at zero recoil,
current conservation ensures that the matrix element is exactly
normalized. While the deviations from the symmetry limit may be
difficult to calculate, they tend to be small. Hence, the overall
theoretical uncertainty on the decay process is under control.
Given good experimental measurements, the associated CKM element 
can be reliably determined.

For semileptonic decays of $b$ flavored hadrons to light mesons 
flavor symmetries are not sufficient to constrain the hadronic matrix 
elements. Ultimately, theoretical predictions based on lattice QCD 
will allow for an accurate determination of $|V_{ub}|$ from
measurements of exclusive decays. Currently, the best determination 
of $|V_{ub}|$ comes from measurements of the inclusive decay spectrum of 
$b \to u \ell \bar{\nu}$ \cite{pdg}. However, the kinematic cuts which are 
necessary to eliminate the huge charm background introduce additional 
theoretical difficulties, similar to those discussed in 
Section~\ref{sec:bsllinclusive}. As a result, theoretical
uncertainties, which are not well known, currently dominate 
the determination of $|V_{ub}|$ \cite{pdg}.

It is important that the Tevatron experiments fully
explore their accessible range of exclusive semileptonic 
$B$ (and $B_s$) decays to light hadrons. While semileptonic $B$ 
decays will also be measured at the $B$ factories, the hadronic 
environment has the advantage that not only $B$ meson decays but 
also $B_s$ and $\Lambda_b$ decays may be studied.
In particular, measurements of semileptonic $B_s$ and $\Lambda_b$ 
decays can provide additional information on the parameters of the 
heavy quark and chiral expansions. For example, a measurement
of the decay $\Lambda_b \to \Lambda_c \ell \bar{\nu}$, can test HQET
predictions at ${\cal O}(1/m_{b,c})$. Measurements of $B_s$
decays provide tests of SU(3) symmetry violations.

Since inclusive semileptonic decays are notoriously difficult to
study at hadron colliders, we focus our discussion in this section
on exclusive decays. In Section~\ref{sec:btod} we first review the 
determination of $|V_{cb}|$ from $B \to D^{(*)} \ell \bar{\nu}$ decays,
and then discuss the decay $\Lambda_b \to \Lambda_c \ell \bar{\nu}$. 
Section~\ref{sec:btopi} contains our discussion of semileptonic 
$B$ decays to light hadrons.

\subsection{Decays to Charm Flavored Final States} \label{sec:btod}

\subsubsection{$B\to D^{(*)}\ell\bar\nu$} 

As discussed in Chapter 1, heavy quark symmetry allows all the form 
factors, which appear in semileptonic $B\to D^{(*)}$ decay, to be related, 
at leading order in $1/m_{b,c}$, to a single universal function, 
the Isgur-Wise function $\xi_(w)$. Corrections to these relations 
have been calculated to ${\cal O}(1/m_{b,c}^2)$ and $O(\alpha_s^2)$.
See, for example, Ref.~\cite{corrensiw}.

\index{HQET!constraints on semileptonic form factors}
We can write the differential decay rate as
\begin{eqnarray}
{d\Gamma(B\to D^*\ell\bar\nu)\over dw}&=&{G_F^2 |V_{cb}|^2\over
48\pi^3}\left(m_B-m_{D^*}\right)^2 
m_{D^*}^3\sqrt{w^2-1}(w+1)^2\nonumber \\
&\times&\left(1+{4w\over w+1}{m_B^2-2w m_B m_{D^*}+m_{D^*}^2\over
\left(m_B-m_{D^*}\right)^2}\right){\cal F}(w)^2
\end{eqnarray}
where the corrections to the symmetry limit are included in the
form factor ${\cal F}(w)$. At zero recoil ${\cal F}(1)$ coincides 
with the Isgur-Wise function up to perturbative and $O(1/m_{b,c}^2)$ 
corrections, which can be parametrized as follows \cite{babarbook}:
\begin{equation}
{\cal F}(1)=\eta_A \left( 1 + \delta_{1/m^2} \right)\;.
\end{equation}
\index{HQET!perturbative corrections}
$\eta_A$ contains the perturbative QCD (and QED) corrections which have 
been calculated to ${\cal O}(\alpha_s^2)$\cite{czar96}. $\delta_{1/m^2}$ contains
the power corrections, which start at ${\cal O}(1/m^2_{b,c})$ for this
case. The power corrections must be calculated from nonperturbative
methods. They have have been estimated from a number of different
approaches, which include non-relativistic quark models and QCD sum
rules.  Once the perturbative and nonperturbative corrections are
included, Ref.~\cite{babarbook} gives the value
\begin{equation}
{\cal F}(1)=0.91\pm 0.04 \;,
\label{eq:f1}
\end{equation}
where the error is dominated by the uncertainty in the nonperturbative 
corrections. The uncertainty in Eq.~(\ref{eq:f1}) leads to a
theoretical error on $V_{cb}$ which is similar in size to the
current experimental error. 
Hence, a significant reduction of the uncertainty on $V_{cb}$ will 
require a more accurate theoretical calculation of ${\cal F}(1)$.

\index{$B\to D^{(*)}\ell\bar\nu$}
\index{lattice QCD!semileptonic decays}
\index{lattice QCD!$B\to D^{(*)}\ell\bar\nu$}
The $B \to D^{(*)} \ell \bar{\nu}$ transition has also been studied in
lattice QCD calculations. The first calculations concentrated
on the slope of the Isgur Wise function \cite{earlybtod}.
At that point, the errors on the form factors were too large
to be competitive with the results shown above.
Since the experimental results have to be extrapolated to zero
recoil, theoretical predictions of the slope can help reduce
the error associated with the extrapolation.

\index{$V_{cb}$}
Ref.~\cite{btod_fnal} introduces a new method based on ratios of 
matrix elements, which exploits heavy quark flavor symmetry 
to calculate the form factors at zero recoil with high precision. 
The ratios from which the form factors are obtained become exactly 
equal to unity in the flavor symmetry limit, where all errors
cancel. Away from the symmetry limit, the errors are proportional
to ${\cal F}(1) - 1$ (instead of ${\cal F}(1)$). 
As a result, as shown in Ref.~\cite{btod_fnal}, the statistical
and systematic errors on ${\cal F}(1)$ are small, $2-3 \%$.
The results are obtained in the quenched approximation. 
Given a sufficient computational effort, the prospects
for improved theoretical predictions of ${\cal F}(1)$ are
excellent.

% ...

The form factors in semileptonic $B_s$ decay are related to those in $B$ 
decay via SU(3).  The leading SU(3)-breaking chiral corrections to the 
Isgur-Wise function were calculated in Ref.~\cite{jenkinssavage}.

\subsubsection{$\Lambda_b\rightarrow \Lambda_c\ell\bar\nu$}

Semileptonic $\Lambda_b\to\Lambda_c$ decays, which cannot be studied at 
the $\Upsilon$(4S), not only provide an alternate means to obtain 
$|V_{cb}|$, but more importantly provide a test of the heavy quark 
expansion at subleading order.

\index{$\Lambda_b\to\Lambda_c\ell\bar\nu_{\ell}$}
The most general expressions for the matrix element of the vector and
axial vector currents between $\Lambda_b$ and $\Lambda_c$ states are
\begin{eqnarray}\label{lambdaform}
\langle\Lambda_c(v^\prime,s^\prime)|\bar c\gamma^\mu b|\Lambda_b(v,s)&=&
\bar u(v^\prime, s^\prime)\left[g_1\gamma^\mu+g_2 v^\mu+g_3 v^{\prime
\mu}\right]
u(v,s)\\
\langle\Lambda_c(v^\prime,s^\prime)|\bar c\gamma^\mu \gamma_5
b|\Lambda_b(v,s)&=&
\bar u(v^\prime, s^\prime)\left[g_1\gamma^\mu+g_2 v^\mu+g_3 v^{\prime
\mu}\right]u(v,s)
\nonumber
\end{eqnarray}
where the states have been labelled with their four-velocities instead 
of their
momenta, and the form factors $F_i$ and $G_i$ are functions of $w\equiv v\cdot
v^\prime$. At leading order in $1/m_c$ and $\alpha_s$ all six form factors are related 
to a universal form factor,
\begin{eqnarray}
f_1(w)&=&g_1(w)=-f_2(w)=-g_2(w)=\xi_\Lambda(w)\nonumber\\
f_3(w)&=&g_3(w)=0
\end{eqnarray}
where $\xi_\Lambda(1)=1+O(\alpha_s(m_c))$.

Because the light degrees of freedom in a $\Lambda_Q$ baryon are in a 
spin 0 state, the subleading corrections to the heavy quark limit take a 
simpler form than for mesons \cite{ggw90}.  In contrast with $B\to D^{(*)}$ decay, 
in which three new form functions and one constant (in addition to the 
Isgur-Wise function) are required to specify the form factors at 
$O(1/m_{b,c})$, the form factors for $\Lambda_b\to\Lambda_c$ transitions 
are determined at $O(1/m_{b,c})$ in terms of
the Isgur-Wise function and one additional parameter,
\begin{equation}
\bar\Lambda_\Lambda= m_{\Lambda_b}-m_b+O(1/m_b^2)=m_{\Lambda_c}-m_c+O(1/m_c^2)
\simeq 700\,\MeV.
\end{equation}
Since $m_b$ may be determined in a number of ways (such as Upsilon sum
rules\cite{upsrules}, moments of spectra in inclusive $B$ decays
\cite{mbbdecays} and lattice calculations of the $\bar b b$ spectrum
\cite{mblattice}), precision measurements of the
$\Lambda_b\to\Lambda_c\ell\bar\nu$ form factors provide a stringent test 
of HQET at subleading order.

Including corrections up to $O(1/m_{b,c})$ in the heavy quark
expansion, the form factors (\ref{lambdaform}) satisfy the relations
\cite{ggw90}
\begin{eqnarray}
f_1(w)&=&\left[1+\left({\bar\Lambda_\Lambda\over 2 m_c}+{\bar\Lambda_\Lambda\over 2 m_b}
\right)\right] \xi_\Lambda(w),\nonumber\\
f_2(w)&=&g_2(w)=-{\bar\Lambda_\Lambda\over m_c}\left({1\over 1+w}\right) \xi_\Lambda(w),\nonumber\\
f_3(w)&=&-g_3(w)=-{\bar\Lambda_\Lambda\over m_b}\left({1\over 1+w}\right) \xi_\Lambda(w),\nonumber\\
g_1(w)&=&\left[1-\left({\bar\Lambda_\Lambda\over 2 m_c}+{\bar\Lambda_\Lambda\over 2 m_b}
\right)\left({1-w\over 1+w}\right)\right] \xi_\Lambda(w).
\end{eqnarray}
Thus, measuring the form factors in $\Lambda_b\to\Lambda_c$ decay 
provides a stringent test of the subleading corrections to HQET.
Complete differential distributions for these decays are given in 
Ref.~\cite{kornerkramer}, including the effects of $\Lambda_b$ 
polarization.

\index{$\Lambda_b\to\Lambda_c\ell\bar\nu_{\ell}$!background from excited hadrons}
An important background to $\Lambda_b\to\Lambda_c$ semileptonic decay 
comes from $\Lambda_b$ decays to excited charmed hadrons, which then 
decay via emission of a soft photon or pion to a $\Lambda_c$.  At leading 
order in the heavy quark
expansion this branching fraction would be predicted to be small, since 
the light degrees of freedom in an excited baryon are orthogonal to those 
in a $\Lambda_b$ in the heavy quark limit, but, as discussed in 
Ref.~\cite{leibstewart}, there are large $O(\lqcd/m_c)$ corrections to this 
statement (note that because $\bar\Lambda$ for baryons is roughly twice 
that in mesons, $1/m_c$ effects are expected to be correspondingly larger 
in baryons).  These authors considered the HQET expansion for semileptonic 
$\Lambda_b$ decays to the spin 1/2 $\Lambda_c(2593)$ and its spin symmetry 
partner the spin 3/2 $\Lambda_c(2625)$.  Using large $N_c$ arguments to 
determine the corresponding matrix elements, they estimated  the branching 
fraction to these two states to be
\begin{equation}
{\Gamma(\Lambda_b\to(\Lambda_c^*(2593)+\Lambda_c^*(2625))\ell\bar\nu_\ell)
\over
\Gamma(\Lambda_b\to X\ell\bar\nu_\ell)}\sim 25-33\% .
\end{equation}
Decays from excited baryons are therefore expected to provide a 
significant background to semileptonic $\Lambda_b\to\Lambda_c$ decay.

To date, most of the lattice QCD calculations of beauty systems
have concentrated on the meson sector. Lattice QCD calculations
of $\Lambda_b \rightarrow \Lambda_c l \nu$ do not yet exist.
However, it should be straightforward to extend the lattice analysis 
of $B \to D^{(*)} l \nu$ decays described in the previous section
to the baryon decay $\Lambda_b \rightarrow \Lambda_c l \nu$.

\subsection{$B\to \pi(\rho)\ell\bar\nu$} \label{sec:btopi}

\index{lattice QCD!$B\to \pi(\rho)\ell\bar\nu$}
\index{$B\to \pi(\rho)\ell\bar\nu$}
The best determination of $|V_{ub}|$ comes at present from the 
measurements of the inclusive decay spectrum of 
$b \to u \ell \bar\nu$ \cite{pdg}. However, in order to reduce 
the huge charm background, one has to impose kinematic cuts
on the charged lepton energy, for example.
Because such cuts restrict the available final state phase
space, they can introduce large nonperturbative corrections
in the OPE, or cause the OPE to break down entirely.
Kinematic cuts in different variables, such as the hadronic 
invariant mass \cite{flw} or the $q^2$ spectrum \cite{bzl}, 
have been proposed in order to reduce the theoretical uncertainties, 
which currently dominate the errors on $|V_{ub}|$.
This work together with improved experimental measurements
of the inclusive $b \to u \ell \bar\nu$ decay at the $B$ factories 
will lead to a better determination of $|V_{ub}|$.

Here, we explore the potential of accurate 
determinations of $|V_{ub}|$ via exclusive decays.
In contrast to the cases discussed in the previous section, 
in the case of exclusive heavy hadron decays to light hadrons
flavor symmetries alone do not provide sufficient constraints on
the hadronic matrix elements (and form factors).
Heavy quark spin and flavor symmetries and SU(3) symmetry yield 
relations among the form factors for $B \to \pi(\rho) \ell \bar\nu$, 
$D \to \pi(\rho) \ell \bar\nu$, $D \to K^{(\ast)} \ell \bar\nu$, 
$B\to K^{(\ast)} \ell^+ \ell^-$, $B\to K^{(\ast)} \gamma$, and 
related $B_s$ and $D_s$ decays. The expected corrections to these 
relations vary from a few to 20 $\%$.
This is discussed in more detail in Section~\ref{sec:btokll}.
If we want to get absolute predictions for the form factors,
we must rely on nonperturbative methods such as lattice 
QCD.

A number of improved lattice QCD calculations of the exclusive 
semileptonic decay $B \to \pi \ell \bar\nu$ have recently become 
available \cite{btopi_lattice}. At present, the uncertainties
in the lattice QCD calculations are still large; the errors
on the form factors are roughly $15-20 \%$ (see 
Section~\ref{sec:lattice}). 
Reducing these theoretical errors will require a significant 
effort and the commitment of sufficient computational resources
to such calculations. 
Ultimately, lattice QCD calculations will provide accurate 
predictions of the hadronic form factors in the high recoil
momentum region. In order to use these predictions for
determinations of $|V_{ub}|$, we need experimental measurements 
of partial differential decay rates, with matching precision.

\section{Semileptonic Decays: Experiment}
\subsection{Semileptonic Decays at CDF}\label{cdfsemi}

\subsubsection{Introduction}

In this report, we describe CDF's prospects for study of semileptonic
decays.  Specifically, we focus on the decay of the $\Lambda_b$ baryon
which is not produced at the $e^+e^-$ $B$ factories.  The primary
interest is total and differential decay rates.  These measurements
are limited by statistical uncertainties.  Therefore, our studies have
largely focussed on trigger strategies to optimize event yields
versus trigger bandwidth.  We have considered the possibility of
measuring the differential decay rate $(1/\Gamma) d\Gamma/dQ^2$.
Semileptonic $B$ decay events are also useful as a control sample for
study of tagging methods or as a backup sample for measuring
$B_s$-$\overline{B}_s$ flavor oscillations.  However, except for a
discussion of possible trigger selections, we leave discussion of
these topics to other sections of the Workshop report.

\index{semileptonic decays!trigger strategy, CDF}
The strategy for extracting semileptonic decay events is to
take advantage of the high purity of lepton triggers as well as the
significant impact parameters of $B$ decay daughters.  
CDF's three-level trigger system in Run~II will provide the tools necessary to
maintain the trigger rate at a manageable level while maintaining a
high efficiency for $B$-decay events.  Specifically, the eXtremely Fast Tracker
(XFT) will offer a significant improvement for the Run~II trigger over
the Run~I trigger by providing tracking information in Level 1.  This
capability enables a track to be matched to an electromagnetic
calorimeter cluster for improved electron identification or to be
matched to a track segment in the muon system for better muon identification as
well as a track-only trigger.  In Level 2 the Silicon Vertex Tracker
(SVT) will add SVX information to the XFT tracks and provide impact
parameter information and thus provide the possibility of a
displaced-track trigger.

\subsubsection{Physics Goals}

The semileptonic decay of heavy baryons can be described by five form
factors.  However, in Heavy Quark Effective Theory (HQET) these reduce
to a single universal form factor in next-to-leading order.  This is
to be contrasted to meson decays in which the form factors reduce to a
single form factor only at leading order.  Measuring the differential
decay rate as a function of the momentum transfer $Q^2$ in $\Lambda_b$
decays can provide stringent tests of HQET.  Because $Q^2$ is the mass
of the lepton-neutrino pair, we must know the neutrino momentum.  With
the possibility of using
3D vertex reconstruction in Run~2, we can find the $\Lambda_b$ direction and 
derive the neutrino momentum up to a quadratic ambiguity.  This is
described in more detail in Section ~\ref{sec:qsqr} where we describe the
potential for measuring the differential decay rate
$(1/\Gamma)d\Gamma/dQ^2$ in $\Lambda_{b} \rightarrow\Lambda_{c}
\ell \nu$ decays.  In the Run~I $\Lambda_b$ lifetime analysis, 197
$\pm$ 25 semileptonic $\Lambda_{b} \rightarrow
\Lambda_{c} l \nu$, $\Lambda_{c} \rightarrow pK\pi$, were
partially reconstructed \cite{run1lambda}.  We use the cuts and
yield from that analysis to provide a basis for Run~II yields.


Semileptonic decays may also provide a good sample for measuring $B_s$ mixing
for lower values of $x_{s}$.  Using semileptonic decays provides a
fall-back position if the yield is high and the all-hadronic trigger
can not collect enough data.  Two possibilities exist for studying
$B_{s}$ mixing through semileptonic decay channels.  The lepton may be
used to tag the event and then one must fully reconstruct the
away-side $B_{s}$, or Same-Side-Tagging, Jet-Charge or Soft-Lepton
tagging may be used to measure $x_{s}$ from the ``first wiggle'' in
$B_{s} \rightarrow D_{s} \ell \nu X$ decays.  360 events were
reconstructed in Run~I in the $8$ GeV inclusive lepton trigger data
through the decay $B_{s} \rightarrow D_{s} \ell \nu X$ where $D_{s}
\rightarrow \phi \pi$ or $K K^{*0}$. Details of the expected time
resolution and $x_s$ reach are discussed in the report of Working
Group 3.  Because
the neutrino momentum is unknown, mixing measurement using
semileptonic decays suffer from poor resolution of the decay time.
The all-hadronic decay $B_s\rightarrow D_{s} \pi $, where $D_{s}
\rightarrow \phi \pi$ or $K K^{*0}$ and the $B_s$ is fully
reconstructed, offers the prospect for the greatest reach in $\Delta
m_s$.  Therefore, we have studied the prospect of a trigger on a
tagging lepton and an opposite-side displaced track which could come
from a hadronic $B_s$ decay.

\subsubsection{Simulations}

\index{semileptonic decays!simulation method, CDF}
To study the efficiency of the possible trigger selections for the
signals of interest, we use a simple parametric Monte Carlo
simulation to compare the Run~II geometric and kinematic acceptance to
that of published Run~I physics analyses. 
We use the measured yields for normalization.
In our Monte Carlo studies, we generate single $b$ quarks according to
next-to-leading-order QCD.  The $B$ hadrons that result after Peterson
($\epsilon=0.006$)\cite{Peterson} fragmentation smearing are forced to
decay to modes of interest using the CLEO Monte Carlo QQ \cite{cleoQQ}.
To model detector performance, we apply Gaussian smearing to the
generated quantities in these Monte Carlo events.

We assume that the offline track-reconstruction and
analysis-cut efficiencies will be the same as in Run~I.  Since we
determine our yields relative to Run~I, we do not correct for these
effects.  The part of the detector and the trigger that is
substantially different from Run~I is the silicon detector (SVXII).  We model
the SVXII as 5 concentric cylinders at the mean radii of the 5 layers.
We account for the gaps between silicon sensors and assume that there is
an additional 2$\%$ hit inefficiency per layer.  

In this study we want to compare the acceptance for decays in the
reconstruction fiducial.  Therefore, we require all charged particles
in a final state to have transverse momentum $P_T$ exceeding
0.5\,GeV/$c$ and to leave the COT drift chamber at its outside
radius.  Furthermore, after accounting for geometry and expected hit
efficiency, we require all tracks to have hits in 4 of 5 SVX layers,
and if a track is to be considered fiducial for the SVT, it must be of
$P_T>2$\,GeV$/c$ and have hits in the 4 inner layers.  We also require
electrons and muons to project to the fiducial regions of the central
calorimeter and muon detectors, respectively.  We also model the
trigger efficiencies with parameterizations.

Since the output of the trigger is dominated by backgrounds, it is
not possible to determine trigger rates from pure Monte Carlo
samples.  Instead, we simulate the performance of the Run~II trigger
system using data taking Run~I using trigger thresholds significantly
lower than were used in normal operation and below the cuts we intend
to apply in Run~II.  We model the performance of the Run~II trigger
electronics using a version of the Run~II simulations modified for the
Run~I detector configurations.  The instantaneous 
luminosities of the test runs correspond to 
 0.4 to 1.4 $\times 10^{32}\,\rm{cm}^{-2}\,\rm{s}^{-1}$ with 36-bunch
operation, allowing us to model the change in trigger performance as a
function of instantaneous luminosity.  We correct the results for the
increased muon and silicon detector acceptances.

\subsubsection{Selection Criteria}

There are various event properties that can be examined in the Level 1
and Level 2 trigger systems.  Because the Level~1 bandwidth is large
and there is substantial overlap with other proposed trigger
selections, we propose using only a single-lepton selection in Level~1
which limits systematic effects.  For Level~2, we want to take
advantage of the decay properties of $b$ hadrons, especially the long
lifetime.  The SVT allows us to select displaced tracks.  We can also
take advantage of $b$ production and decay kinematics to select tracks
associated with the lepton in the Level~2 trigger.  Our proposed
trigger signature is a lepton with a displaced track.   Additional
handles include the angle between the lepton and the displaced track
found by SVT $\Delta \phi(\ell, SVT)$ and the transverse mass $M_T$
of the lepton track pair.  ($M_T^2\simeq
p_{T,1}p_{T,2}(1-\cos\Delta\phi)/c^2$)  For tracks coming from the decay of
a single $B$, we expect the angle to be small and the two-particle
mass to be less than the $B$ mass.  
\index{$\Lambda_b\to\Lambda_c\ell\bar\nu_{\ell}$!trigger, CDF}

\begin{figure}[t]
\begin{center}
\epsfxsize=9.0cm
\epsfxsize=9.0cm
\mbox{\epsffile{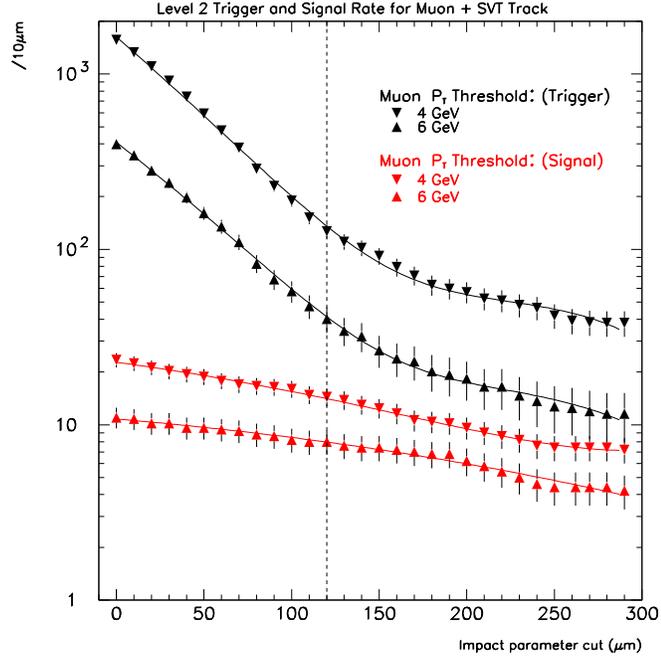}}
\caption[Dependence of the trigger rate on the impact parameter cut for the
muon-trigger test data and the $b \bar{b} \rightarrow \Lambda_b X \rightarrow
\Lambda_c \mu \nu$ Monte Carlo.]{Dependence of the trigger rate on the impact
parameter cut on tracks found by the SVT for the muon-trigger test data (top
two curves) and the $b \bar{b} \rightarrow \Lambda_b X \rightarrow \Lambda_c
\mu \nu$ Monte Carlo (bottom two curves).  The vertical scale is arbitrary. The
dashed line shows an impact parameter cut of 120 $\mu$m.}
\label{implambdab}
\end{center}
\end{figure}

Figure~\ref{implambdab} shows the dependence of Level 2 trigger rate
on the impact parameter cut applied to tracks found by SVT in events
with trigger muons for the
test-run data and for the Monte Carlo simulations of the benchmark
channel the $b \bar{b} \rightarrow \Lambda_b X \rightarrow \Lambda_c
\mu\nu, \Lambda_c\rightarrow pK\pi$ mode.  The trigger rate falls
sharply with impact parameter up to about 120\,$\mu$m.  Above
150\,$\mu$m, the decrease in trigger rate is approximately equal to
that for the signal, indicating a background of real semileptonic
$B$-decay events.  Therefore cutting on impact parameter beyond
150\,$\mu$m will not increase purity.  For consistency with other CDF
selections, we expect to cut at 120\,$\mu$m.
Figure~\ref{dphi_lambdab_2d} shows the distributions of $\Delta\phi$
versus transverse mass for events in test run data in which a track
has been found by the SVT simulation with $|d_0|>120\,\mu$m.
Requiring $\Delta\phi<90^\circ$ and $M_T<5$\,GeV$/c^2$ gives
substantial background reduction without loss of the
semileptonic-decay signal.  For the proposed trigger selection with a
4\,GeV$/c$ cut on lepton momentum, we expect a trigger cross section
of $53\pm8$\,nb for muons and $90\pm36$\,nb for electrons.  At a
luminosity of $10^{32}\,\rm{cm}^{-2}\,\rm{s}^{-1}$ ({\it i.e.}
100\,$\mu{\rm b}^{-1}{\rm s}^{-1}$) this corresponds to a rate of
$14\pm4$\,Hz out of a total Level~2 trigger budget of 300\,Hz.  
Note that this trigger rate is about a factor of 3 lower than would be
achieved with an inclusive 8\,GeV lepton sample as was used in Run~I.
\index{$\Lambda_b\to\Lambda_c\ell\bar\nu_{\ell}$!trigger, CDF}

\begin{figure}[thbp]
\begin{center}
\epsfxsize=9.0cm
\mbox{\epsffile{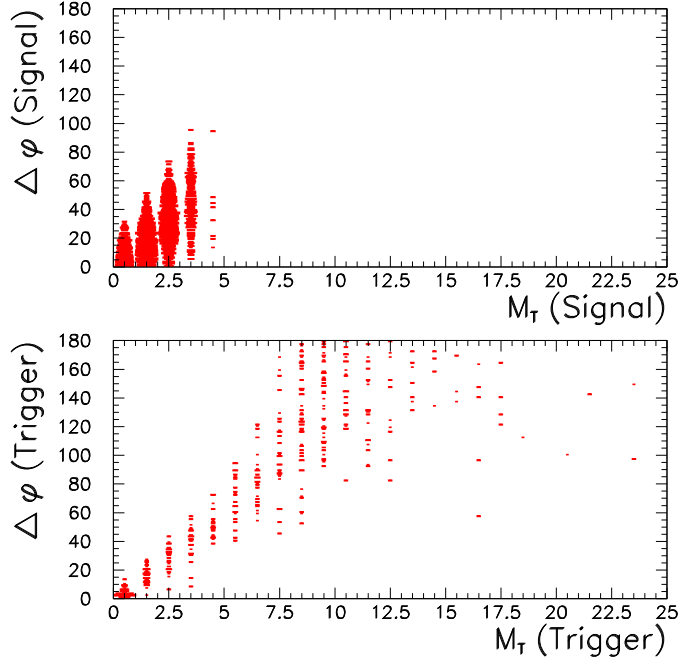}}
\caption[The distributions of $\Delta \phi(\mu, SVT)$, and $M_{T}(\mu, SVT)$
for the test-run data and the $b \bar{b} \rightarrow \Lambda_b \rightarrow
\Lambda_c\mu\nu$ signal after impact parameter cut.]{The distributions of
$\Delta \phi(\mu, SVT)$, and $M_{T}(\mu, SVT)$ for the test-run data (bottom
plot) and the $b \bar{b} \rightarrow \Lambda_b \rightarrow \Lambda_c\mu\nu$
signal (top plot) after impact parameter cut.}
\normalsize
\label{dphi_lambdab_2d}
\end{center}
\end{figure}

We have also considered the effectiveness of a trigger using a lepton
and a displaced track in order to extract a sample of $B$ decays that
can be fully reconstructed and have an away-side lepton tag.  Our
benchmark process for this type of event is $B_s \rightarrow D_s \pi,
D_s \rightarrow \phi\pi$ or $K K^{*0}$.  As with the lepton and
displaced track from the decay of a single $B$, the effectiveness of a
cut on the impact parameter does not improve beyond
$|d_0|\simeq150\,\mu$m and is independent of the muon $P_T$.  The
$\Delta \phi(\mu, SVT)$ and $M_T$ distributions for simulation of the
benchmark process are compared to the trigger background from the
simulation on the test data in Figure $\ref{dphi_bs_2d}$ after
requiring $|d_0| > 120 \mu$m.  The requirements $\Delta \phi(\mu,
SVT)>90^\circ$ and $M_T>5$\,GeV/c$^2$ reduce backgrounds with only a
small reduction in the singal.  For the proposed trigger selection
with a 4\,GeV$/c$ cut on lepton momentum, we expect a trigger cross
section of $48\pm8$\,nb for muons and $120\pm40$\,nb for electrons.
At a luminosity of $10^{32}\,\rm{cm}^{-2}\,\rm{s}^{-1}$ this
corresponds to a rate of $16\pm4$\,Hz.
\index{semileptonic decays!lepton-displaced track trigger, CDF}

\begin{figure}[htb]
\begin{center}
\epsfxsize=9.0cm
\mbox{\epsffile{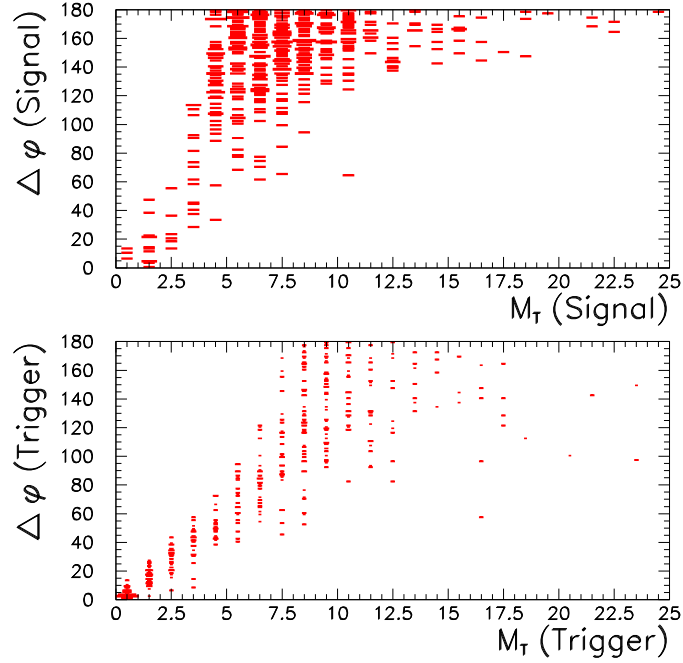}}
\caption{Distributions of $\Delta \phi(\mu, SVT)$, and $M_{T}(\mu,
SVT)$ for the test-run data (bottom plot) and the $b \bar{b} \rightarrow B_s 
X \rightarrow D_s \pi$ signal (top plot).}
\label{dphi_bs_2d}
\end{center}
\end{figure}

\subsubsection{Signal Rate Expectations}\label{sigrate}

To complete the study, we evaluate the various trigger criteria in
terms of the expected Run~II event yield.  Monte Carlo samples are
generated for the semileptonic decays of interest and passed through
the trigger scenarios. The number of Run~II events expected is based
on the ratio of the acceptance for the proposed Run~II semileptonic
triggers and the Run~I semileptonic trigger scaled by acceptance and
efficiency ratios.

The Run~I acceptance $A_I$ is the acceptance for the Run~I 8 GeV
inclusive lepton trigger.  The Run~II acceptance $A_{II}$ corresponds
to the Run~II  trigger of interest.  
To account for the Run~I trigger efficiency, the Monte Carlo events were
required to pass a trigger model based on the Run~I Level-2 inclusive
lepton trigger efficiency \cite{eleeff}\cite{muoneff}.
  There are additional factors which affect the
number of events expected in Run~II. The SVX acceptance will be greater in 
Run~II: 
$A_{(SVXII)} = 1.4 \times A_{(SVXI)}$.  In addition, we assume the total
acceptance gained by the central muon-system upgrade to be a factor of 1.4. The
increase in the instantaneous luminosity in Run~II also increases the
number of expected events, we assume the ratio between Run~I and Run~II to be:
\begin{equation}
\mathcal{L}_{II} / \mathcal{L}_I = \rm{2 \;fb^{-1} / 100 \; pb ^{-1} = 20\,.}
\end{equation}

To normalize our sample with the Run~I events sample, the same offline 
selection cuts are applied. For $\Lambda_{b} \rightarrow
\Lambda_c \ell \nu$ decay the cuts are:
	\begin{center}
	\begin{itemize}
	\item $P_{T}(K,\pi,p) \geq (0.7,0.6,1.5)$ GeV$/c$; 
	\item 3.5 $\leq$ M($\ell \Lambda_{c}$) $\leq 5.6$ GeV/$c^{2}$;
	\item $P_{T}(\Lambda_c) \geq 5.0$ GeV$/c$.
	\end{itemize}
	\end{center}
We also require the kaon, pion and proton to be within a cone of 0.8
in $\eta$-$\phi$ space.
The decay channel $B_s \rightarrow D_s \pi$
has not been reconstructed so we only estimate the efficiency
of the lepton + track trigger selection relative to the all-hadronic trigger
\cite{twotrack}.  This estimate does not include the reduction in
yield from ``analysis'' cuts.
\index{$\Lambda_b\to\Lambda_c\ell\bar\nu_{\ell}$!analysis cuts, CDF}

The number of events expected in Run~II $N_{II}$ with respect to the 
number of events reconstructed in Run~I $N_I$ for each decay channel
is calculated using the following relation:
\begin{equation}
\frac{N_{II}}{N_I} = \frac{A_{II}}{A_I} \cdot
\frac{\mathcal{L}_{II}}{\mathcal{L}_I}  \cdot A_{(SVXII)}\,.
\end{equation}
The expected yield  for $\Lambda_{b} \rightarrow
\Lambda_c \ell \nu$ decays
with the same-side 4\,GeV$/c$ lepton plus displaced-track selection described
above is 25000 in 2\,fb$^{-1}$ of Run~II.  For  $B_s \rightarrow D_s
\ell \nu$ decays, we expect 33000 events.     
\index{$\Lambda_b\to\Lambda_c\ell\bar\nu_{\ell}$!event rates, CDF}

The yield of $B_s\rightarrow D_s \pi$ decay events from the opposite-side
4\,GeV lepton trigger selection described above is 106 events in
2\,fb$^{-1}$ without correction for reconstruction and analysis cut
efficiencies.  This number can be directly compared to the yield of
10600 expected under the same assumptions from required two displaced
tracks with $P_T>2$\,GeV$/c$ as described in Proposal
P-909\cite{twotrack}.  Although tagging dilutions and efficiencies are
outside the scope of this section, it is clear that including the
tagging lepton as the primary trigger element is an inferior procedure
to a trigger on particles of the signal decay.

\subsubsection{$Q^2$ Spectrum}\label{sec:qsqr}

\begin{figure}[bhtp]
\begin{center}
\epsfxsize=9.0cm
\mbox{\epsffile{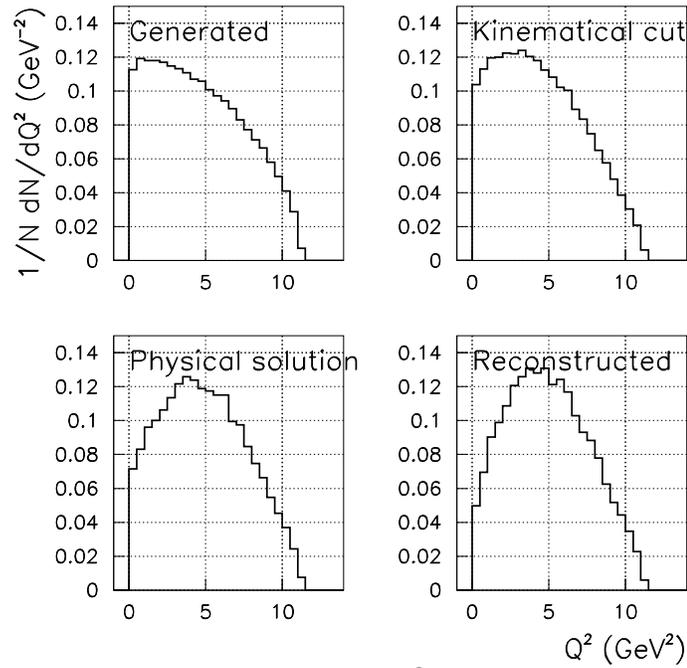}}
\caption[Monte Carlo simulation of the $Q^2$ distribution in $\Lambda_b \to
\Lambda_c\ell\bar\nu$ decay.]{Monte Carlo simulation of the $Q^2$ distribution. 
The generated distribution is shown in the top left.  The distribution after
the kinematic cuts is in the top right.  The distribution after the detector
resolution has been applied is shown in the lower right. The distribution using
the kinematic constraints and 3D-vertexing is shown in the lower right.} 
\label{qdist}
\end{center}
\end{figure}

As discussed in Section 2, semileptonic decays of B baryons present the
possibility of measuring the momentum transfer $Q^2$ in $\Lambda_b$
decays. To study the feasibility of such a measurement, we
generate $\Lambda_{b} \rightarrow \Lambda_{c} \ell \nu$ decays.
For $\Lambda_{b} \rightarrow \Lambda_{c} \ell X$, where $X$ is not observed,
we can describe the kinematics using the following energy and 
momentum conservation rules:
\begin{eqnarray}
& E_{\Lambda_b} = E_{\ell \Lambda_{c}} + E_{X};& \nonumber\\
& p_{X} = |p_{X}|^2 = |p_{\Lambda_{b}} - p_{\ell \Lambda_c}|^2
	= p_{\Lambda_b}^2 + p_{\ell \Lambda_{c}}^2 -
2p_{\Lambda_{b}}p_{\ell \Lambda_{c}}\cos\theta\,.& 
\end{eqnarray}
\index{$\Lambda_b\to\Lambda_c\ell\bar\nu_{\ell}$!$Q^2$ reconstruction, CDF}%
This method is described in more detail in \cite{masabs}.
In our toy Monte Carlo sample, we use $P_{\Lambda_b}$, $P_{\Lambda_c}$, and
$P_{\ell}$ and  3D-vertex and kinematic constraints to reconstruct 
$P_{\nu}$.  The $Q^2$ distributions from the Monte Carlo event generator, after
kinematic cuts, and after detector smearing as well as the reconstructed $Q^2$
distribution are shown in Figure $\ref{qdist}$.
The resolution of the $Q^2$ reconstruction is shown in Figure $\ref{qreso}$
(top plot). The ratio of the generated $Q^2$ distribution
to the physical solution gives the reconstruction efficiency
and is shown in Figure $\ref{qreso}$.  
\index{$\Lambda_b\to\Lambda_c\ell\bar\nu_{\ell}$!$Q^2$ resolution, CDF}
\index{$\Lambda_b\to\Lambda_c\ell\bar\nu_{\ell}$!$Q^2$ efficiency, CDF}%

\begin{figure}[htbp]
\begin{center}
\epsfxsize=9.0cm
\mbox{\epsffile{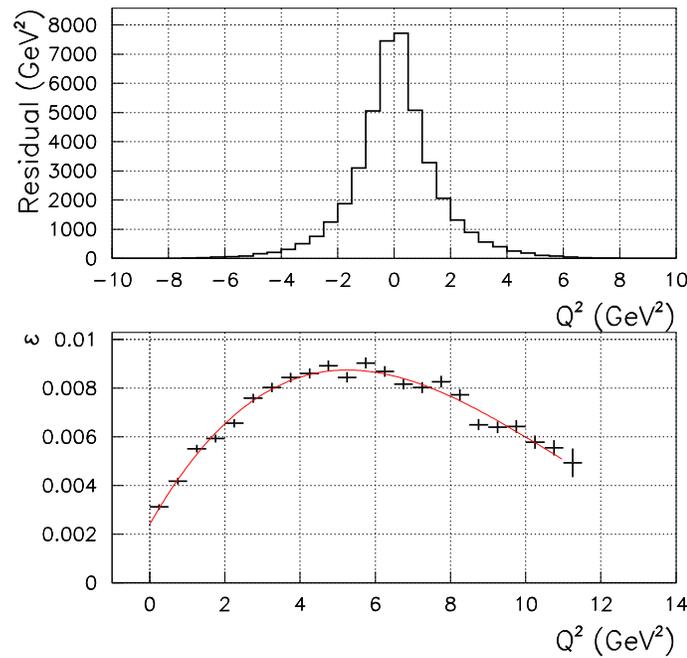}}
\caption{The resolution of the reconstructed $Q^2$ distribution is
shown in the top plot.  The reconstruction efficiency is shown in the
bottom plot.} 
\label{qreso}
\end{center}
\end{figure}

  As a check of our the ability to measure the $Q^2$ distribution in Run~II, we
perform a simple Monte Carlo experiment.  We generate two
independent samples of $\Lambda_{b} \rightarrow \Lambda_{c} \ell \nu$.
One sample is used to make a generator level $Q^2$ distribution.  The
other sample is normalized to the expected Run~II yield.
Kinematic and resolution smearing were applied to the normalized sample.
The $Q^2$ distribution is calculated after correcting for the
reconstruction efficiency.  The generated and reconstructed
distribution is shown in Figure~\ref{q2corrected}.  This shows
that given the number of expected events in Run~II we can reasonably
expect to reconstruct the $Q^2$ distribution.  However, there
are higher-order $\Lambda_b$ decays which complicate the reconstruction. 

\begin{figure}[ht]
\begin{center}
\epsfxsize=9.0cm
\mbox{\epsffile{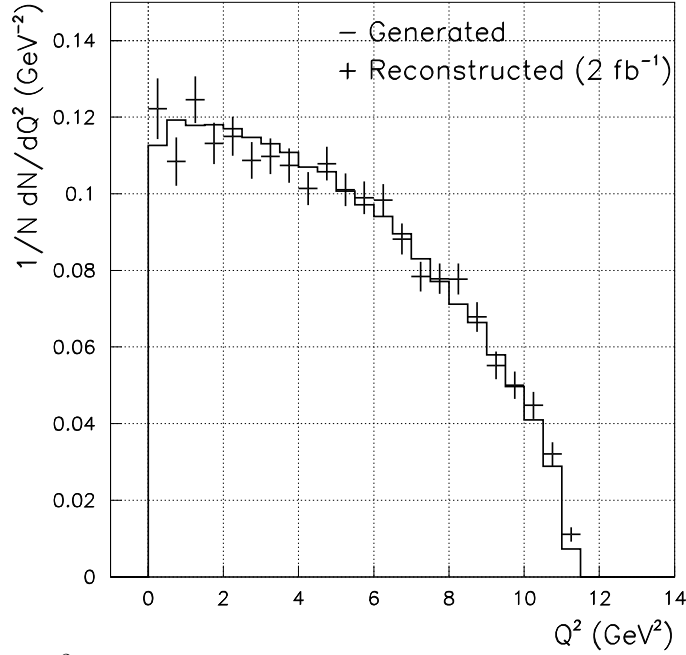}}
\caption{$Q^2$ distribution for generator level and reconstructed
after correcting for efficiency. } 
\label{q2corrected}
\end{center}
\end{figure}

\subsubsection*{Higher-order Contamination}

The branching ratio for semileptonic $\Lambda_b$ decays to
higher-order baryons may not be not negligible and can
contaminate the decay channel of interest. We investigate the
kinematic properties of these higher-order decays to determine a method
to reject these events efficiently.  Presumably the dominant source of
contamination is decays with a $\Lambda_c$ and two pions.  The
possible states are:
\begin{equation}\label{eqone}
\begin{array}{rl}
\Lambda_b \rightarrow & [\Sigma_{c}
\pi^+]_{I=0} \ell \bar{\nu_{\ell}}  \\
	& \hspace{.1in} \bentarrow \hspace{.1in} \Lambda_c^+  \pi^-  \\
	& \hspace{.45in} \bentarrow \hspace{.1in}  p  K^- \pi^+ \\ 
\end{array}
\end{equation}
\begin{equation}\label{eqtwo}
\begin{array}{rll}
\Lambda_b \rightarrow & [\Sigma_{c}^{++} \pi^{-}]_{I=0} \ell \bar{\nu_{\ell}} \\ 
	& \hspace{.1in} \bentarrow \hspace{.1in}\Lambda_c^+ \pi^+ \\
	&  \hspace{.45in}\bentarrow \hspace{.1in} p K^- \pi^+   \\
\end{array}
\end{equation}
\begin{equation}\label{eqthree}
\begin{array}{rll}
\Lambda_b \rightarrow & \Lambda_{c}^{+}
\pi^{+} \pi^{-} \ell \bar{\nu_{\ell}} \hspace{.55in} \\
	& \hspace{.1in}\bentarrow \hspace{.1in}    p K^- \pi^+ \\
\end{array}
\end{equation}
\begin{equation}\label{eqfour}
\begin{array}{rl}
\Lambda_b \rightarrow & [\Sigma_{c}^+
\pi^0]_{I=0} \ell \bar{\nu_{\ell}}  \\
	& \hspace{.1in} \bentarrow \hspace{.1in} \Lambda_c^+  \pi^0  \\
	& \hspace{.45in} \bentarrow \hspace{.1in}  p  K^- \pi^+ \\ 
\end{array}
\end{equation}
The $\Lambda_b$ channels (\ref{eqone})--(\ref{eqthree}) contain
charged pions  and can potentially be identified while channel (\ref{eqfour})
will be impossible to see.  However, if we can identify events from mode
(\ref{eqthree}) we can use them with proper normalization subtract the effect
of mode (\ref{eqfour}).
\index{$\Lambda_b\to\Lambda_c\ell\bar\nu_{\ell}$!backgrounds, CDF}

To study rejection methods, we generate a sample of the decays which
have additional charged pions and apply trigger requirements and the
offline cuts detailed in Section~\ref{sigrate}.  In addition, we
require that the pions from higher order decays be within a cone in
$\eta$-$\phi$ of 0.8 centered on the $\Lambda_c$ direction.  The $P_T$
distribution of these charged pions from higher-order decays is shown
in Figure~\ref{higherpt}.  Estimating that the minimum $P_T$ which
will be reconstructed is 0.3 GeV$/c$, then approximately 75$\%$ of the
charged higher order pions will be reconstructed.

\begin{figure}[t]
\begin{center}
\epsfxsize=9.0cm
\mbox{\epsffile{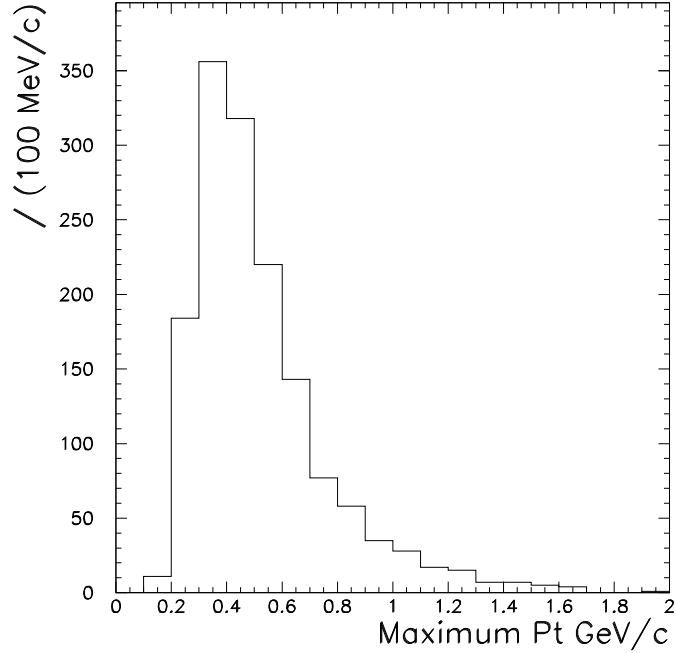}}
\caption{Distribution of the maximum $P_T$ of $\pi$'s from higher order
decays.} 
\label{higherpt}
\end{center}
\end{figure}

\index{$\Lambda_b\to\Lambda_c\ell\bar\nu_{\ell}$!backgrounds, CDF}
In addition to the extra decay pions, we also study tracks coming from
primary interaction which might be confused with $\Lambda_b$-daughter
tracks and cause too many events to be rejected.  To model these
tracks we generate a Monte Carlo sample using PYTHIA tuned to reflect
prompt particle distributions in CDF $B$ events \cite{tunedpyth}. For
events that pass the same trigger and offline requirements as in the
trigger study, we compare the impact parameter of the three categories
of tracks: $\Lambda_c$-daughter tracks, charged pions from
higher-order decays and tracks from the primary vertex.  The
impact-parameter significance ($d/\sigma_d$) with respect to the
primary vertex and to the $\Lambda_b$-decay vertex for the three
categories of tracks is shown in Figure $\ref{d0_combo}$.  If we
reject events with an additional track associated with the
$\Lambda_c$-lepton vertex The requirement that the impact parameter
significance $(d^{\Lambda_b}_0/\sigma^{\Lambda_b}_{d_0}<2)$, random
associations with primary tracks would cause about 10\% of the good
($\Lambda_b \rightarrow \Lambda_c,
\ell, \nu)$ events to be eliminated, but would tag almost 100$\%$ of the
higher order decays.  Thus a small fraction of the time real
$\Lambda_b \rightarrow \Lambda_c,
\ell, \nu$ events will be thrown out but the remaining events will be
a relatively pure sample free of higher order $\Lambda_b$ decays.
This coupled with demonstrated ability to calculate the missing
neutrino momentum leaves us optimistic about measuring $Q^2$ in Run~II.

\begin{figure}[t]
\begin{center}
\epsfxsize=9.0cm
\mbox{\epsffile{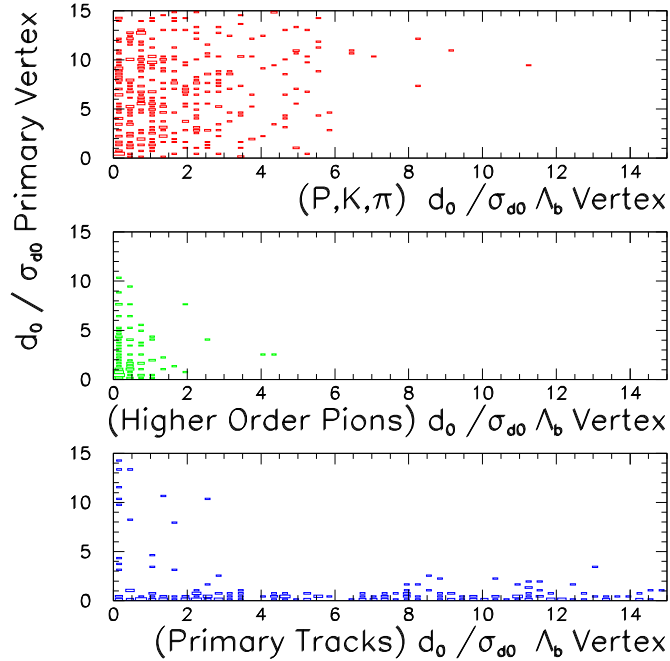}}
\caption[Comparison of impact parameter significance of all 3 categories of
tracks (protons, kaons and pions).]{Comparison of impact parameter
significance of all 3 categories of tracks (protons, kaons and pions) from
$\Lambda_c$ (top plot), higher order pions (middle plot), and tracks from
primary vertex (bottom plot)) with respect to $\Lambda_b$ vertex and the
Primary Vertex.  All tracks must be within a cone of $\Delta R < 0.8$ with $P_T
> 0.3$ GeV/c.}
\label{d0_combo}
\end{center}
\end{figure}

\subsubsection{Summary}

We have shown that a lepton + displaced track trigger can produce
substantial samples of semileptonic $b$ hadron decays for study and
fits well within CDF's overall trigger budget for Run~II.
We expect $\sim$ 25000 $\Lambda_{b}$ decays in 2 fb$^{-1}$.
We have shown that a measurement $Q^2$ decay  in
$\Lambda_b$ decay  using the
impact parameter information to reject tracks from higher order decays 
and the primary vertex is feasible. We recognize that fact that
further studies are needed to solidify this claim.

\subsection{Estimating the BTeV Potential for Semileptonic Decays}

Using techniques developed for fixed-target charm experiments (including 
E691\,\cite{e691_semi}, E687\,\cite{e687_semi}, and E791\,\cite{e791_semie,e791_semimu}) 
we demonstrate that BTeV has the necessary 
capability to extract information from semileptonic decays.  Given the large
number of $b$-hadrons reconstructed by BTeV, the semileptonic reach will be
extraordinary.

\subsubsection{Signal and Background}

The signal and background were generated using the MCFast Monte Carlo program.
A full description of this program can be found elsewhere\,\cite{MCFast,btevprop}.
MCFast is designed to be a fast and accurate detector simulation with speed
and flexibility achieved through parameterization.  The MCFast tracing includes
the effect of magnetic fields, multiple Coulomb scattering, bremsstrahlung, 
dE/dx, decays in flight, pair conversions and secondary hadronic interactions.
The simulation assumed a luminosity of $2\!\times\!10^{32}$\,cm$^{-2}$s$^{-1}$
and included multiple interactions per event.
The muon identification code used in this analysis starts by making an acceptance
cut.  Potential muon tracks must have momentum greater than 5 GeV/$c$.  
All tracks are projected through the three muon stations using the track
parameters determined from the Kalman filter.  If the projection misses any of
the stations the track is thrown out.  If the track is associated with a muon
particle it is identified as a muon.  If the track is not a muon then a 
misidentification probability is determined and a random number is generated
to determine if the particle is misidentified as a muon.  
\index{semileptonic decays!muon identification, BTeV}%

The misidentification probability decreases as the momentum increases and
decreases as the radius increases.  The misidentification rate varies from 7\%
for 5\,GeV/$c$ tracks near the beam to 0.2\% for 50\,GeV/$c$ tracks at the
outer edge of the muon system.
The misidentification rate (away from the central region) is loosely based on
the measured misidentification  rate from the FOCUS experiment. FOCUS is a
fixed-target charm experiment which used a $\sim$180 GeV photon beam at a rate
of approximately 10\,MHz.  BTeV and FOCUS have similar muon rates and
momenta.  The BTeV detector has two advantages over the FOCUS muon system. 
The BTeV detector  has much finer granularity and the shielding is magnetized
which, by allowing a momentum measurement, provides another handle to
distinguish real muons from fakes.


The signal modes analyzed were $B^0 \rightarrow D^{*-}
(\overline{D}^0(K^+\pi^-,K^+\pi^-\pi^-\pi^+)\pi^-)\mu^+ \nu$  and $\Lambda_b^0
\rightarrow \Lambda_c^-(pK^-\pi^+)\mu^+\nu$.  In each case, $\sim$120,000 
events were simulated.  Three sources of background were simulated: minimum
bias events, charm events, and generic $b$ events (without the signal mode).
The cross sections for minimum bias, charm, and $b$ events obtained from 
Pythia\,\cite{Pythia} are shown in Table\,\ref{tab:cross} along with the
predicted numbers of events from one year ($10^{7}$\,s) of running at a
luminosity of $2\!\times\!10^{32}$\,cm$^{-2}$s$^{-1}$.

\begin{table}[t]
\begin{center}
\begin{tabular}{lcccc} \hline \hline
Species  & Quark cross  & Hadron   & Branching & Total produced  \\
         & section (mb) & fraction & Ratio     & for 2~fb$^{-1}$\\ [2pt] \hline
Min bias & 75 & 100\% & 100\% & $1.5\times 10^{14}$ \\
Charm    & 0.75 & 100\% & 100\% & $1.5\times 10^{12}$ \\
generic $b$ & 0.10 & 100\% & 100\% & $2.0\times 10^{11}$ \\
$B^0\!\rightarrow\!D^*\mu\nu$ & 0.10 & 75\% & 
$0.35\%$ & $5.3\times 10^8$ \\
$\Lambda_b\!\rightarrow\!\Lambda_c\mu\nu$ & 0.10 & 10\% & 
$0.20\%$ & $4.0\times 10^7$ \\ \hline\hline
\end{tabular}
\end{center}
\caption[Production cross sections and expected generation rates for signal and
background for semileptonic $B$ and $\Lambda_b$ decay.]{Production cross
sections and expected generation rates for signal and background.  Cross
sections for $b\bar{b}$ production are estimated from D0 data. The minimum bias
cross section is taken to be the  $p\bar{p}$ total cross section at 2~TeV. The
charm cross section is assumed to be 1\% of the minimum bias cross section. 
Branching ratios are from Ref.~\cite{PDG} except $\Lambda_b \rightarrow
\Lambda_c\mu\nu$ which is estimated at 4\%.}
\label{tab:cross}
\end{table}
\index{$\Lambda_b\to\Lambda_c\ell\bar\nu_{\ell}$!backgrounds, BTeV}%
\index{$B\to D^{(*)}\ell\bar\nu$!background rates, BTeV}%

Clearly it is impossible to simulate $10^{14}$ events given the current state of 
computing; simulating more than $10^8$ events is prohibitive.  Therefore we try to 
estimate the background based on a simulation of 4.2 million
minimum bias events, 4.8 million $c\bar{c}$ events, and 1.5 million $b\bar{b}$ events.
Given the large number of produced $b$-hadrons we can certainly make stringent cuts and
still retain a large sample of events.  Unfortunately, using these stringent cuts 
eliminates nearly all of our (limited) background which makes it difficult to determine
the significance or signal-to-noise ratio of the signal.  Since the signals analyzed 
require detached vertices, reconstructed charm particles, and muons, we assume that the
background will be dominated by $c$ and $b$ events, not minimum bias events.  Therefore,
we can safely tighten our cuts enough to eliminate all of the minimum bias events which 
were simulated.  

Using these cuts keeps 4 (1) $c\bar{c}$ and 26 (13) $b\bar{b}$ events for the
$B^0\!\rightarrow\!D^*(D^0(K\pi,K3\pi)\pi)\mu\nu$ 
($\Lambda_b\!\rightarrow\!\Lambda_c(pK\pi)\mu\nu$) decay mode.  
Figure\,\ref{fig:masspl_scal} shows the result of scaling the signal
and background events to an integrated luminosity of 2~fb$^{-1}$
and distributing the background events evenly
through the mass plot.  The yield, significance, signal-to-background, and efficiency
is tabulated in Table\,\ref{tab:yields}.  These results include a conservative 
trigger efficiency (50\%) which is what is expected from the detached vertex trigger.  
A detached muon trigger is also planned which will increase the trigger 
efficiency.
\index{$\Lambda_b\to\Lambda_c\ell\bar\nu_{\ell}$!event rates, BTeV}%
\index{$B\to D^{(*)}\ell\bar\nu$!event rates, BTeV}%

\begin{figure}[t]
\centerline{\epsfig{file=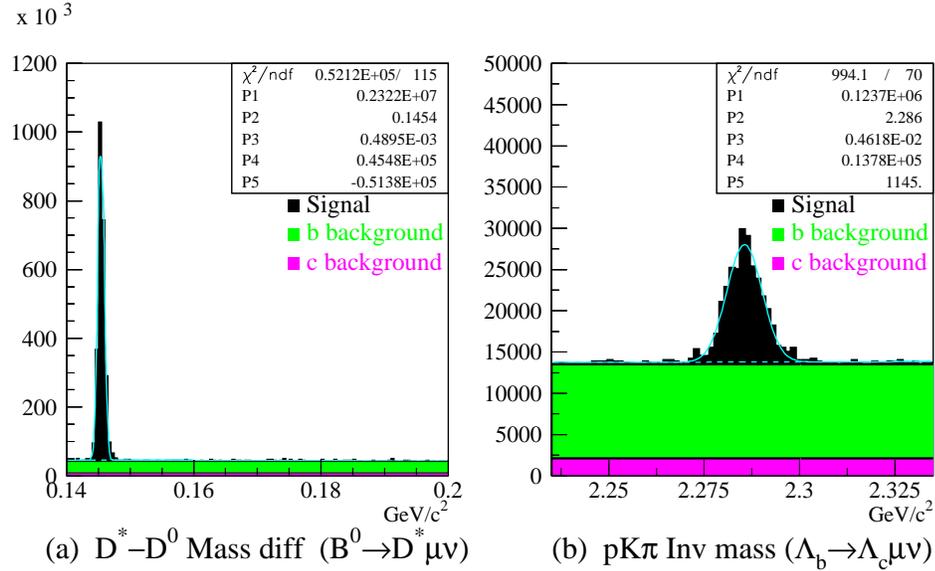,height=3.0in}}
\vspace*{4pt}
\caption[$D^*-D^0$ mass difference distribution for $B^0 \rightarrow
D^*\mu\nu$ signal and backgrounds; pK$\pi$ invariant mass distribution for
$\Lambda_b \rightarrow \Lambda_c\mu\nu$ signal and backgrounds.]{(a) $D^*-D^0$
Mass difference distribution for  $B^0 \rightarrow
D^*(D^0(K\pi,K3\pi)\pi)\mu\nu$ signal events and $b\bar{b}$ and $c\bar{c}$
background events.  (b) pK$\pi$ invariant mass distribution for $\Lambda_b
\rightarrow \Lambda_c(pK\pi)\mu\nu$ signal events and $b\bar{b}$ and $c\bar{c}$
background events. In both plots, the background events have been spread evenly
through the mass range.}
\label{fig:masspl_scal}
\end{figure}

\begin{table}[bht]
\begin{center}
\begin{tabular}{lcccc} \hline\hline \\ \vspace*{-24pt} \\
Decay mode  & Efficiency & Yield & $S/\sqrt{S+B}$   & $S/B$   \\[2pt] \hline
$B^0 \rightarrow D^*(D^0(K\pi,K3\pi)\pi)\mu\nu$ & 0.44\% & 2,300,000 & 1,430 & 21 \\
$\Lambda_b \rightarrow \Lambda_c(pK\pi)\mu\nu$ & 0.31\% & 120,000 & 210 & 1.0 \\
\hline\hline
\end{tabular}
\end{center}
\caption[Efficiency, expected yields, signal-to-background, and significance
for an integrated luminosity of 2\,fb$^{-1}$.]{Efficiency, expected yields,
signal-to-background, and significance for an integrated luminosity of
2\,fb$^{-1}$.  Efficiency includes acceptance, trigger efficiency,
reconstruction efficiency and cut efficiency.  Significance and
signal-to-background are calculated by integrating over a $\pm$2$\sigma$ region
around the mass peak.}
\label{tab:yields}
\end{table}

\subsubsection{Semileptonic Reach}

To determine the form factors associated with a particular semileptonic decay we would
like to have all the kinematic information associated with the decay chain.  The most
important quantity is $q^2$ which is the square of the virtual $W$ mass; i.e. the 
invariant mass of the lepton and neutrino.  Reconstructing the momentum vector is not 
a trivial exercise, however.  The technique used to reconstruct the neutrino momentum, 
pioneered by E691 and used by E687 and E791 among 
others\,\cite{e691_semi,e687_semi,e791_semie,e791_semimu}, is particularly suited to
BTeV as it requires good vertex resolution compared to the vertex separation.  The 
production and decay vertex of the $b$-hadron gives the $b$-hadron momentum vector 
direction.  The neutrino momentum perpendicular to the $b$-hadron momentum vector is
easily measured because it must balance all of the other decay products.  The neutrino
momentum parallel to the $b$-hadron momentum can be determined (up to a quadratic ambiguity)
by assuming the invariant mass of the $b$-hadron.  We pick the low momentum solution for
the parallel component of the neutrino momentum as Monte Carlo studies indicate this is
correct more often.
\index{$\Lambda_b\to\Lambda_c\ell\bar\nu_{\ell}$!$Q^2$ reconstruction, BTeV}%
\index{$B\to D^{(*)}\ell\bar\nu$!$Q^2$ reconstruction, BTeV}%

The most recent published results using this method come from E791\,\cite{e791_semimu}.  Using
a 500\,GeV/$c$ $\pi^-$ beam, they reconstruct over
6,000 $D^+\!\rightarrow\!\overline{K}^{*0}\ell\nu$ decays.  From this sample
they obtain form factor measurements of $r_V = V(0)/A_1(0) = 1.87\pm 0.08\pm 0.07$ and
$r_2 = A_2(0)/A_1(0) = 0.73\pm 0.06\pm 0.08$.  From the 3,000 muon decays, they also measure
$r_3 = A_3(0)/A_1(0) = 0.04\pm 0.33\pm 0.29$.  Defining the $q^2$ resolution as the RMS of 
the generated $q^2$ minus the reconstructed $q^2$ divided by $q^2_{max}$, E791 had a $q^2$
resolution of 0.17.  From the MCFast simulation with the standard selection criteria
and reconstructing the neutrino momentum as described above, BTeV has a $q^2$ resolution of 
approximately 0.14 as shown in Fig.\,\ref{fig:qres}.
With 6,000 events, the E791 results give smaller errors than most lattice QCD calculations.
With a similar $q^2$ resolution and 100 times more data, BTeV will also be able to challenge
theoretical predictions or provide values which can be input into other calculations.
\index{$\Lambda_b\to\Lambda_c\ell\bar\nu_{\ell}$!$Q^2$ resolution, BTeV}%
\index{$B\to D^{(*)}\ell\bar\nu$!$Q^2$ resolution, BTeV}%

\begin{figure}[htb]
\centerline{\epsfig{file=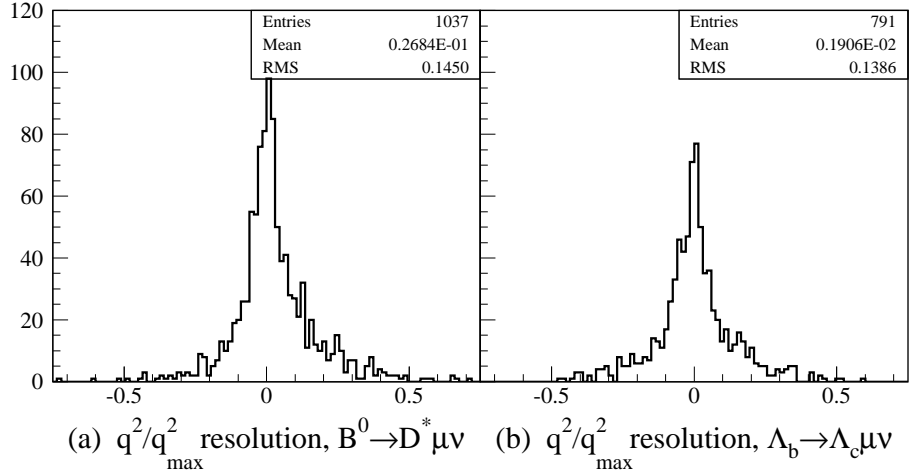,height=2.5in}}
\caption[$q^2 / q^2_{max}$ resolution for (a) $B^0 \rightarrow
D^*\mu\nu$ and (b) $\Lambda_b \rightarrow
\Lambda_c\mu\nu$.]{$q^2 / q^2_{max}$ resolution for (a) $B^0 \rightarrow
D^*(D^0(Kn\pi)\pi)\mu\nu$ and (b) $\Lambda_b \rightarrow
\Lambda_c(pK\pi)\mu\nu$.}
\label{fig:qres}
\end{figure}

One additional difficulty in extracting information from these semileptonic decays comes
from $b$ semileptonic decays into charm excited states which decay into the state being
investigated.  For example, in the decay 
$\Lambda_b^0\!\rightarrow\!\Sigma_c^+\mu\nu$, the $\Sigma_c^+$ can decay to 
$\Lambda_c^+\pi^0$.  Assuming the $\pi^0$ is lost, this event will be reconstructed as
a signal $\Lambda_b^0\!\rightarrow\!\Lambda_c^+\mu\nu$ event and the neutrino reconstruction
(which assumes the invariant mass of the $\Lambda_c^+\mu\nu$ is equal to the $\Lambda_b^0$)
will be inaccurate.  The $q^2$ resolution for these events is shown in Fig.\,\ref{fig:qresbg}a.
Assuming an equal mixture of $\Lambda_b\!\rightarrow\!\Lambda_c$ and 
$\Lambda_b\!\rightarrow\!\Sigma_c$ decays gives the $q^2$ resolution shown in 
Fig.\,\ref{fig:qresbg}b.  This shows a resolution only slightly degraded (0.14 to 0.15)
but with a bias equal to 1/3 of the RMS.  BTeV has excellent $\pi^0$ reconstruction 
capabilities\,\cite{btevprop} which should make it possible to measure the relative 
branching ratios and correct for this bias.
\index{$\Lambda_b\to\Lambda_c\ell\bar\nu_{\ell}$!backgrounds, BTeV}%

\begin{figure}[htb]
\centerline{\epsfig{file=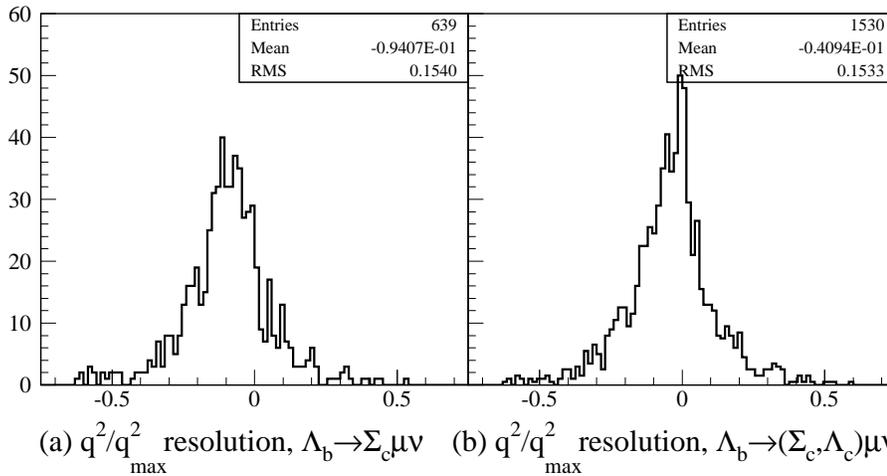,height=2.5in}}
\caption[$q^2 / q^2_{max}$ resolution for $\Lambda_b \rightarrow
\Lambda_c\mu\nu$ where the $\Lambda_c$ comes from $\Sigma_c$, and 
$\Lambda_b \rightarrow \Lambda_c\mu\nu$ where half of the $\Lambda_c$'s come
from $\Sigma_c$'s and half come from $\Lambda_b$'s.]{$q^2 / q^2_{max}$
resolution for (a) $\Lambda_b \rightarrow \Lambda_c(pK\pi)\mu\nu$ where the
$\Lambda_c$ comes from a $\Sigma_c$ and (b)
$\Lambda_b \rightarrow \Lambda_c(pK\pi)\mu\nu$ where half of the
$\Lambda_c$'s come from $\Sigma_c$'s and half come directly from
$\Lambda_b$'s.}
\label{fig:qresbg}
\end{figure}

Since BTeV has a very efficient vertex trigger at Level 1, the semielectronic
decays can also be studied. Even though the acceptance of the 
ECAL is much smaller than that of the muon detectors we expect significant
numbers of reconstructed decays in the electronic modes which can be used
for systematic studies as well as just increasing the statistics.

\subsubsection{Summary}

This study only provides a cursory look at some of the semileptonic physics available with BTeV.
There are many other semileptonic decay modes of $b$-hadrons which are well within the 
grasp of BTeV.  These decay modes include $B\!\rightarrow\!\rho\ell\nu$ to determine 
$V_{ub}$ and $B_s$ semileptonic decay modes to check SU(3).  In addition, BTeV will have
many semileptonic charm decays available for study.

\section{Summary of Semileptonic Decays}

The study of exclusive semileptonic decays complements the
studies on CP violation and mixing discussed in Chapters~6 and 8,
since semileptonic decays can, in principle, provide determinations
of CKM elements such as $|V_{ub}|$ and $|V_{cb}|$.

For any useful comparison between theory and experiment, we
need experimental measurements of $q^2$ (and other) distributions. 
This is a challenging task especially in a hadron collider 
environment, because it requires the reconstruction of the neutrino 
momentum. Studies at both, CDF and BTeV, have established the 
feasibility of neutrino momentum reconstruction and subsequent
measurement of $q^2$ distributions with good resolution.

The decay 
$\Lambda_b \to \Lambda_c \ell \bar\nu$ is of particular interest.
It can only be be studied at hadron colliders and provides 
information on the parameters of the heavy quark expansion at 
sub-leading order. 

Tevatron experiments should explore the full range of semileptonic
decays, including $B$ and $B_s$ meson decays to light hadrons. It 
is important to have measurements of many different semileptonic 
heavy-to-light decays, with as high an accuracy as possible,
for two reasons. First, these measurements can provide tests of 
heavy flavor and SU(3) symmetry relations. Second, we should expect 
significant improvements in theoretical predictions of exclusive
heavy-to-light form factors, based for example, on lattice QCD,
in the time frame for Run II, and certainly by the time BTeV comes
on line. These measurements will help to establish the reliability 
of lattice QCD calculations, and of the corresponding determinations 
of CKM elements, such as $|V_{ub}|$, from them.

\clearpage
\def\PRL #1 #2 #3 {{\em Phys. Rev. Lett.} {\bf #1},\ #2 (#3)}
\def\PRD #1 #2 #3 {{\em Phys. Rev. } {\bf #1},\ #2 (#3)}
\def\PLB #1 #2 #3 {{\em Phys. Lett.} {\bf #1},\ #2 (#3)}
\def\ZPC #1 #2 #3 {{\em Z. Phys.} {\bf #1},\ #2 (#3)}
\def\EPJ #1 #2 #3 {{\em Eur. Phys. J.} {\bf #1},\ #2 (#3)}
\def\NPB #1 #2 #3 {{\em Nucl. Phys.} {\bf #1},\ #2 (#3)}
\def\NPproc #1 #2 #3 {{\em Nucl. Phys. B} (Proc. Suppl.){\bf #1},\ #2 (#3)}
\def\RMP #1 #2 #3 {{\em Rev. Mod. Phys.} {\bf #1},\ #2 (#3)}

\clearpage{\pagestyle{empty}\cleardoublepage}
\newcommand{\bep}{\epsilon}
\newcommand{\ibid}{{\it ibid}.}
\newcommand{\atversim}[2]{\lower0.7ex\vbox{\baselineskp\zatskip\lineskip
\zatskip\lineskiplimit 0pt\ialign{$\matth#1\hfil##hfil$\crcr#2\crcr\sim\crcr}}}

\newcommand{\brunts}[1]{\ensuremath{\BR \, \big(\!\Bsun \rightarrow #1 \,\big)}}
\newcommand{\gqtfb}{\ensuremath{\Gamma (B_q^0(t) \rightarrow \ov{f} )}}
\newcommand{\gqbtf}{\ensuremath{\Gamma (\ov{B}{}_q^0(t) \rightarrow f )}}
\newcommand{\gqtf}{\ensuremath{\Gamma (B_q^0(t) \rightarrow f )}}
\newcommand{\gqbtfb}{\ensuremath{\Gamma (\ov{B}{}_q^0(t) \rightarrow \ov{f} )}}
\newcommand{\bbs}{$B_s$--$\ov{B}{}_s\,$}

\chapter{Mixing and Lifetimes}

%
\authors{K.~Anikeev, F.~Azfar, N.~Cason, H.W.K.~Cheung, A.~Dighe, I.~Furic,
  G.~Gutierrez, J.~Hewett, R.~Jesik, M.~Jones, P.~Kasper, J.~Kroll,
  V.E.~Kuznetsov, R.~Kutschke, G.~Majumder, M.~Martin, U.~Nierste, Ch.~Paus,
  S.~Rakitin, S.~Stone, M.~Tanaka, W.~Taylor, J.~Tseng, M.~Voloshin, J.~Wang,
  N.~Xuan}
%
%
\newcommand{\ipb}{\ensuremath{\mathrm{pb^{-1}}}}
\newcommand{\ifb}{\ensuremath{\mathrm{fb^{-1}}}}
\newcommand{\um}{\ensuremath{\mathrm{\mu m}}}
\newcommand{\mm}{\ensuremath{\mathrm{mm}}}
\newcommand{\ten}[1]{\ensuremath{\times 10^{#1}}}
%
%
\newcommand{\dzero}{D\O\,\,}
\newcommand{\emis}{/\!\!\!\!E}

\section{Overview} 

In hadron colliders $b$-flavored hadrons are produced with a large boost.
Therefore they are a fertile ground for measurements of decay time
distributions. The neutral $B_d^0$ and $B_s^0$ mesons mix with their
antiparticles, which leads to oscillations between the flavor eigenstates. A
measurement of the oscillation frequency allows to
determine the mass difference%
\index{B meson@$B$ meson!mass difference \dm}\index{mass difference \dm}
$\dm_q$, $q=d,s$, between the two physical mass eigenstates. The rapid
oscillations in \bbms\ %
\index{B mixing@$B$ mixing!$B_s$ mixing}%
have not been resolved yet and their discovery has a high priority for the $B$
physics program at Run~II.  Once this has been achieved, the mass difference
$\dm_s$ will be known very precisely. By combining this information with the
already measured $\dm_d$ one will precisely
determine the length of one side of the unitarity triangle.%
\index{unitarity triangle!and \bbmd}%
\index{unitarity triangle!and \bbms}%
$\dm_d/\dm_s$ will have a larger impact on our knowledge of the unitarity
triangle than any previously measured quantity and even than a precisely
measured $\sin (2 \beta)$.  Accurately measured decay distributions will further
reveal the pattern of $b$-hadron lifetimes.  
\index{lifetime!$b$-flavored hadron}%
The large mass of the $b$ quark compared to the QCD scale parameter
$\Lambda_{QCD}$ allows to expand the widths in terms of $\Lambda_{QCD}/m_b$.
Differences among the total widths are dominated by terms of order $16 \pi^2
(\Lambda_{QCD}/m_b)^3$, the measurements of lifetime differences therefore probe
the heavy quark expansion at the third order in the expansion parameter. From
Run~II we expect valuable new information on the lifetimes of the $B^+$, $B_d^0$
and $B_s^0$ mesons, the width difference $\dg_s$ between the two physical $B_s$
meson eigenstates the lifetimes of the $\Lambda_b$ and eventually also of other
b-flavored baryons.  \index{lifetime!$B^+$} \index{lifetime!$B_d^0$}
\index{lifetime!$B_s^0$} \index{lifetime!$\Lambda_b$} From the current
experimental situation it is not clear whether the heavy quark expansion can be
applied to baryon lifetimes, and Run~II data will help to settle this question.

This chapter first discusses the theory predictions for the various quantities
in great detail. Where possible, we derive simple `pocket-calculator' formulae
to facilitate the analysis of the measurements.  It is described which
fundamental information can be gained from the various measurements.  Some
quantities are sensitive to physics beyond the Standard Model and we show how
they are affected by new physics. Then we summarize the experimental techniques
and present the results of the Monte Carlo simulations.

\section[Theory of heavy hadron lifetimes]{Theory of heavy hadron
lifetimes$\!$ \authorfootnote{Author: Mikhail Voloshin}}
\label{chmix:thlife}

The dominant weak decays of hadrons containing a heavy quark, $c$ or $b$, are
caused by the decay of the heavy quark. In the limit of a very large mass $m_Q$
of a heavy quark $Q$ the parton picture of the hadron decay should set in, where
the inclusive decay rates of hadrons, containing $Q$, mesons ($Q\bar q$) and
baryons ($Qqq$), are all the same and equal to the inclusive decay rate
$\Gamma_{parton}(Q)$ of the heavy quark. Yet, the known inclusive decay rates
\cite{pdg} are conspicuously different for different hadrons, especially for
charmed hadrons, whose lifetimes span a range of more than one order of
magnitude from the shortest $\tau(\Omega_c)=0.064 \pm 0.020$ ps to the longest
$\tau(D^+)=1.057 \pm 0.015$ ps, while the differences of lifetime among $b$
hadrons are substantially smaller. The relation between the relative lifetime
differences for charmed and $b$ hadrons reflects the fact that the dependence of
the inclusive decay rates on the light quark-gluon `environment' in a particular
hadron is a pre-asymptotic effect in the parameter $m_Q$, which effect vanishes
as an inverse power of $m_Q$ at large mass.

A theoretical framework for systematic description of the leading at $m_Q \to
\infty$ term in the inclusive decay rate $\Gamma_{parton}(Q) \propto m_Q^5$ as
well as of the terms relatively suppressed by inverse powers of $m_Q$ is
provided \cite{sv:81,sv:85,sv:86} by the operator product expansion (OPE) in
$m_Q^{-1}$. Existing theoretical predictions for inclusive weak decay rates are
in a reasonable agreement, within the expected range of uncertainty, with the
data on lifetimes of charmed particles and with the so far available data on
decays of $B$ mesons.  The only outstanding piece of present experimental data
is on the lifetime of the $\Lambda_b$ baryon: $\tau(\Lambda_b)/\tau(B_d) \approx
0.8$, for which ratio a theoretical prediction, given all the uncertainty
involved, is unlikely to produce a number lower than 0.9.  The number of
available predictions for inclusive decay rates of charmed and $b$ hadrons is
sufficiently large for future experimental studies to firmly establish the
validity status of the OPE based theory of heavy hadron decays, and, in
particular, to find out whether the present contradiction between the theory and
the data on $\tau(\Lambda_b)/\tau(B_d)$ is a temporary difficulty, or an
evidence of fundamental flaws in theoretical understanding.

It is a matter of common knowledge that application of OPE to decays of charmed
and $b$ hadrons has potentially two caveats. One is that the OPE is used in the
Minkowski kinematical domain, and therefore relies on the assumption of
quark-hadron duality at the energies involved in the corresponding decays. In
other words, it is assumed that sufficiently many exclusive hadronic channels
contribute to the inclusive rate, so that the accidentals of the low-energy
resonance structure do not affect the total rates of the inclusive processes.
Theoretical attempts at understanding the onset of the quark-hadron duality are
so far limited to model estimates \cite{cdsu:97,bsuv:99}, not yet suitable for
direct quantitative evaluation of possible deviation from duality in charm and
$b$ decays. This point presents the most fundamental uncertainty of the OPE
based approach, and presently can only be clarified by confronting theoretical
predictions with experimental data. The second possible caveat in applying the
OPE technique to inclusive charm decays is that the mass of the charm quark,
$m_c$, may be insufficiently large for significant suppression of higher terms
of the expansion in $m_c^{-1}$.  The relative lightness of the charm quark,
however, accounts for a qualitative, and even semi-quantitative, agreement of
the OPE based predictions with the observed large spread of the lifetimes of
charmed hadrons: the nonperturbative effects, formally suppressed by $m_c^{-2}$
and $m_c^{-3}$ are comparable with the `leading' parton term and describe the
hierarchy of the lifetimes.

Another uncertainty of a technical nature arises from poor knowledge of matrix
elements of certain quark operators over hadron, arising as terms in OPE. These
can be estimated within theoretical models, with inevitable ensuing model
dependence, or, where possible, extracted from the experimental data.  With
these reservations spelled out, we discuss here the OPE based description of
inclusive weak decays of charm and $b$ hadrons, with emphasis on specific
experimentally testable predictions, and on the measurements, which would less
rely on model dependence of the estimates of the matrix elements, thus allowing
to probe the OPE predictions at a fundamental level.

\subsection{OPE for inclusive weak decay rates}

The optical theorem of the scattering theory relates the total decay rate
$\Gamma_H$ of a hadron $H_Q$ containing a heavy quark $Q$ to the imaginary part
of the `forward scattering amplitude'.
\index{inclusive weak decay rates!operator product expansion}%
For the case of weak decays the latter amplitude is
described by the following effective operator
\begin{equation}
L_{eff}=2 \,{\rm Im} \, \left [ i \int d^4x \, e^{iqx} \, T \left \{
L_W(x),
L_W(0) \right \} \right ]\,,
\label{leff}
\end{equation}
in terms of which the total decay rate is given by\footnote{We use here the
  non-relativistic normalization for the {\it heavy} quark states: $\langle Q |
  Q^\dagger Q | Q \rangle =1$.}
\begin{equation}
  \Gamma_H=\langle H_Q | \, L_{eff} \, | H_Q \rangle\,.
  \label{lgam}
\end{equation}
The correlator in equation (\ref{leff}) in general is a non-local operator.
However at $q^2=m_Q^2$ the dominating space-time intervals in the integral are
of order $m_Q^{-1}$ and one can expand the correlator in $x$, thus producing an
expansion in inverse powers of $m_Q$. The leading term in this expansion
describes the parton decay rate of the quark. For instance, the term in the
nonleptonic weak Lagrangian $\sqrt{2} \, G_F \, V ({\overline q}_{1 L}
\gamma_\mu \, Q_L)({\overline q}_{2 L} \gamma_\mu \, q_{3 L})$ with $V$ being
the appropriate combination of the CKM mixing factors, generates through
Eq.~(\ref{leff}) the leading term in the effective Lagrangian
\begin{equation}
L^{(0)}_{eff, \, nl} = |V|^2 \, {G_F^2 \, m_Q^5 \over 64 \, \pi^3} \,
\eta_{nl} \, \left ( {\overline Q} Q \right )\,,
\label{lef0}
\end{equation}
where $\eta_{nl}$ is the perturbative QCD radiative correction factor.  This
expression reproduces the well known formula for the inclusive nonleptonic
decay rate of a heavy quark, associated with the underlying process $Q \to q_1
\, q_2 \, {\overline q}_3$, due to the relation $\langle H_Q | {\overline Q} Q |
H_Q \rangle \approx \langle H_Q | Q^\dagger Q | H_Q \rangle =1$, which is valid
up to corrections of order $m_Q^{-2}$. One also sees form this example, that in
order to separate individual semi-inclusive decay channels, e.g., nonleptonic
with specific flavor quantum numbers, or semileptonic, one should simply pick
up the corresponding relevant part of the weak Lagrangian $L_W$, describing the
underlying process, to include in the correlator (\ref{leff}).

The general expression for first three terms in the OPE for $L_{eff}$
has the form\index{inclusive weak decay rates!effective Lagrangian}
\begin{eqnarray}
L_{eff} &=&
L_{eff}^{(0)} + L_{eff}^{(2)} + L_{eff}^{(3)} \\
&=& c^{(0)} \, {G_F^2 \, m_Q^5 \over 64 \, \pi^3} \left ( {\overline
Q} Q \right ) +  c^{(2)} \,{G_F^2 \, m_Q^3 \over 64 \, \pi^3} \left
({\overline Q}\, \sigma^{\mu \nu} G_{\mu \nu} \, Q \right ) 
+ {G_F^2 \, m_Q^2 \over 4\, \pi} \, \sum_i c_i^{(3)}\, 
({\overline q}_i \Gamma_i q_i) ({\overline Q} \Gamma^\prime_i Q)\,,\nonumber 
\label{first3}
\end{eqnarray}
where the superscripts denote the power of $m_Q^{-1}$ in the relative
suppression of the corresponding term in the expansion with respect to the
leading one, $G_{\mu \nu}$ is the gluon field tensor, $q_i$ stand for light
quarks, $u,\, d, \, s$, and, finally, $\Gamma_i$, $\Gamma^\prime_i$ denote spin
and color structures of the four-quark operators. The coefficients $c^{(a)}$
depend on the specific part of the weak interaction Lagrangian $L_W$, describing
the relevant underlying quark process.

One can notice the absence in the expansion (\ref{first3}) of a term suppressed
by just one power of $m_Q^{-1}$, due to non-existence of operators of suitable
dimension. Thus the decay rates receive no correction of relative order
$m_Q^{-1}$ in the limit of large $m_Q$, and the first pre-asymptotic corrections
appear only in the order $m_Q^{-2}$.

\begin{figure}[bt]
\vspace*{-.25cm}
\begin{center}
\unitlength 0.9mm
\thicklines
\begin{picture}(140.00,100.00)(10,-15)
\put(10.00,70.00){\line(1,0){20.00}}
\put(30.00,70.00){\line(1,0){30.00}}
\put(60.00,70.00){\line(1,0){20.00}}
\bezier{164}(30.00,70.00)(45.00,84.00)(60.00,70.00)
\bezier{164}(30.00,70.00)(45.00,56.00)(60.00,70.00)
\put(30.00,70.00){\circle*{2.00}}
\put(60.00,70.00){\circle*{2.00}}
\put(20.00,72.00){\makebox(0,0)[cb]{$Q$}}
\put(70.00,71.00){\makebox(0,0)[cb]{$Q$}}
\put(49.00,78.00){\makebox(0,0)[cb]{$q_1$}}
\put(49.00,71.00){\makebox(0,0)[cb]{$q_2$}}
\put(49.00,62.00){\makebox(0,0)[ct]{${\overline q_3}$}}
\put(10.00,43.00){\line(1,0){20.00}}
\put(30.00,43.00){\line(1,0){30.00}}
\put(60.00,43.00){\line(1,0){20.00}}
\bezier{164}(30.00,43.00)(45.00,57.00)(60.00,43.00)
\bezier{164}(30.00,43.00)(45.00,29.00)(60.00,43.00)
\put(30.00,43.00){\circle*{2.00}}
\put(60.00,43.00){\circle*{2.00}}
\put(20.00,45.00){\makebox(0,0)[cb]{$Q$}}
\put(70.00,44.00){\makebox(0,0)[cb]{$Q$}}
\put(49.00,51.00){\makebox(0,0)[cb]{$q_1$}}
\put(49.00,44.00){\makebox(0,0)[cb]{$q_2$}}
\put(49.00,35.00){\makebox(0,0)[ct]{${\overline q_3}$}}
\put(45.00,36.00){\line(0,-1){2.00}}
\put(45.00,33.00){\line(0,-1){2.00}}
\put(45.00,30.00){\line(0,-1){2.00}}
\put(47.00,28.00){\makebox(0,0)[cc]{$g$}}
\put(10.00,12.00){\line(1,0){20.00}}
\put(30.00,12.00){\line(1,0){30.00}}
\put(60.00,12.00){\line(1,0){20.00}}
\bezier{164}(30.00,12.00)(45.00,26.00)(60.00,12.00)
\put(30.00,12.00){\circle*{2.00}}
\put(60.00,12.00){\circle*{2.00}}
\put(20.00,14.00){\makebox(0,0)[cb]{$Q$}}
\put(70.00,13.00){\makebox(0,0)[cb]{$Q$}}
\put(49.00,20.00){\makebox(0,0)[cb]{$q_1$}}
\put(49.00,13.00){\makebox(0,0)[cb]{$q_2$}}
\put(10.00,5.50){\line(3,1){20.00}}
\put(60.00,12.00){\line(4,-1){20.00}}
\put(20.00,7.00){\makebox(0,0)[ct]{$q_3$}}
\put(70.00,8.00){\makebox(0,0)[ct]{$q_3$}}
\put(87.00,70.00){\vector(1,0){11.00}}
\put(87.00,43.00){\vector(1,0){11.00}}
\put(87.00,12.00){\vector(1,0){11.00}}
\put(110.00,70.00){\line(1,0){30.00}}
\put(116.00,71.00){\makebox(0,0)[cb]{$Q$}}
\put(134.00,71.00){\makebox(0,0)[cb]{$Q$}}
\put(110.00,43.00){\line(1,0){30.00}}
\put(116.00,44.00){\makebox(0,0)[cb]{$Q$}}
\put(134.00,44.00){\makebox(0,0)[cb]{$Q$}}
\put(125.00,43.00){\line(0,-1){2.00}}
\put(125.00,40.00){\line(0,-1){2.00}}
\put(125.00,37.00){\line(0,-1){2.00}}
\put(127.00,35.00){\makebox(0,0)[cc]{$g$}}
\put(110.00,12.00){\line(1,0){30.00}}
\put(116.00,13.00){\makebox(0,0)[cb]{$Q$}}
\put(134.00,13.00){\makebox(0,0)[cb]{$Q$}}
\put(110.00,7.00){\line(3,1){15.00}}
\put(125.00,12.00){\line(3,-1){15.00}}
\put(117.00,9.00){\makebox(0,0)[lt]{$q$}}
\put(132.00,9.00){\makebox(0,0)[rt]{$q$}}
\put(125.00,62.00){\makebox(0,0)[cc]{ $m_Q^5 \, ({\overline Q} Q)$}}
\put(125.00,27.00){\makebox(0,0)[cc]{ $ m_Q^3 \, \left ({\overline Q}
(\vec \sigma \cdot \vec B) Q \right )$}}
\put(125.00,0.00){\makebox(0,0)[cc]{ $m_Q^2 \, ({\overline q} \Gamma
q)({\overline Q} \Gamma^\prime Q)$}}
\put(123.50,70.00){ \circle*{2.83}}
\put(123.50,43.00){ \circle*{2.83}}
\put(123.50,12.00){ \circle*{2.83}}
\end{picture}
\vspace*{-.75cm}
\caption{Graphs for three first terms in OPE for inclusive decay rates:
  the parton term, the chromomagnetic interaction, and the four-quark term.}
\end{center}
\label{fig:ope}
\end{figure}
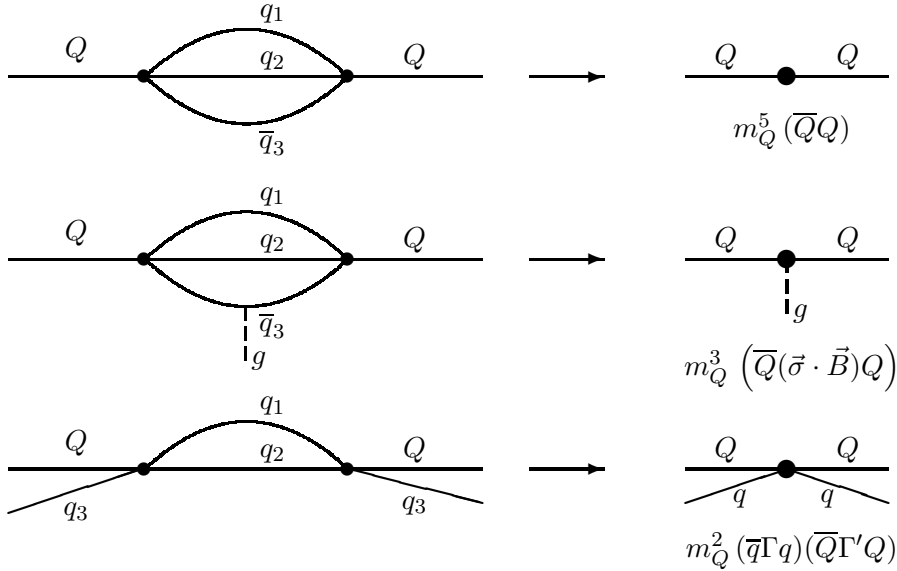

The mechanisms giving rise to the three discussed terms in OPE are shown in
Figure~8.1. The first, leading term corresponds to the parton decay, and does
not depend on the light quark and gluon `environment' of the heavy quark in a
hadron. The second term describes the effect on the decay rate of the gluon
field that a heavy quark `sees' in a hadron. This term in fact is sensitive only
to the chromomagnetic part of the gluon field, and contains the operator of the
interaction of heavy quark chromomagnetic moment with the chromomagnetic field.
Thus this term depends on the spin of the heavy quark, but does not depend on
the flavors of the light quarks or antiquarks. Therefore this effect does not
split the inclusive decay rates within flavor $SU(3)$ multiplets of heavy hadrons,
but generally gives difference of the rates, say, between mesons and baryons.
The dependence on the light quark flavor arises from the third term in the
expansion (\ref{first3}) which explicitly contains light quark fields.
Historically, this part is interpreted in terms of two mechanisms
\cite{sv:81,gnpr:79,bgt:84}: the weak scattering (WS) and the Pauli interference
(PI).\index{inclusive weak decay rates!Pauli interference}
\index{inclusive weak decay rates!weak scattering}
The WS corresponds to a cross-channel of the
underlying decay, generically $Q \to q_1 \, q_2 \, {\overline q}_3$, where
either the quark $q_3$ is a spectator in a baryon and can undergo a weak
scattering off the heavy quark: $q_3 \, Q \to q_1 \, q_2$, or an antiquark in
meson, say ${\overline q}_1$, weak-scatters (annihilates) in the process
${\overline q}_1 \, Q \to q_2 \, {\overline q}_3$. The Pauli interference effect
arises when one of the final (anti)quarks in the decay of $Q$ is identical to
the spectator (anti)quark in the hadron, so that an interference of identical
particles should be taken into account. The latter interference can be either
constructive or destructive, depending on the relative spin-color arrangement of
the (anti)quark produced in the decay and of the spectator one, thus the sign of
the PI effect is found only as a result of specific dynamical calculation. In
specific calculations, however, WS and PI arise from the same terms in OPE,
depending on the hadron discussed, and technically there is no need to resort to
the traditional terminology of WS and PI.

In what follows we discuss separately the effects of the three terms in the
expansion (\ref{first3}) and their interpretation within the existing and future
data.

\subsection{The parton decay rate}

The leading term in the OPE amounts to the perturbative expression for the decay
rate of a heavy quark. In $b$ hadrons the contribution of the subsequent terms
in OPE is at the level of few percent, so that the perturbative part can be
confronted with the data in its own right. In particular, for the $B_d$ meson
the higher terms in OPE contribute only about 1\% of the total nonleptonic as
well as of the semileptonic decay rate.  Thus the data on these rates can be
directly compared with the leading perturbative term in OPE.

\index{B decay@$B$ decay!semileptonic} The principal theoretical topic,
associated with this term is the calculation of QCD radiative corrections, i.e.
of the factor $\eta_{nl}$ in Eq.~(\ref{lef0}) and of a similar factor,
$\eta_{nl}$, for semileptonic decays. It should be noted, that even at this,
perturbative, level there is a known long-standing problem between the existing
data and the theory in that the current world average for the semileptonic
branching ratio for the $B$ mesons, $B_{sl}(B)= 10.45 \pm 0.21 \%$, is somewhat
lower than the value $B_{sl}(B) \ge 11.5$ preferred from the present knowledge
of theoretical QCD radiative corrections to the ratio of nonleptonic to
semileptonic decay rates (see, e.g., \cite{bbsv:94}). However, this apparent
discrepancy may in fact be due to insufficient `depth' of perturbative QCD
calculation of the ratio $\eta_{nl}/\eta_{sl}$. In order to briefly elaborate on
this point, we notice that the standard way of analyzing the perturbative
radiative corrections in the nonleptonic decays is through the renormalization
group (RG) summation of the leading log terms and the first next-to-leading
terms \cite{ap:91,bbbg:95} in the parameter $L \equiv \ln(m_W/m_b)$. For the
semileptonic decays the logarithmic dependence on $m_W/m_b$ is absent in all
orders due to the weak current conservation at momenta larger than $m_b$, thus
the correction is calculated by the standard perturbative technique, and a
complete expression in the first order in $\alpha_s$ is available both for the
total rate \cite{hp:83,nir:89} and for the lepton spectrum \cite{cj:94}.  In
reality however the parameter $L \approx 2.8$ is not large, and non-logarithmic
terms may well compete with the logarithmic ones. This behavior is already seen
from the known expression for the logarithmic terms: when expanded up to the
order $\alpha_s^2$ the result of Ref.\cite{bbbg:94} for the rate of decays with
single final charmed quark takes the form
\begin{equation}
{{\Gamma(b \to c \bar u d) +
\Gamma(b \to c \bar u s)} \over {3 \, \Gamma(b \to c e
\bar \nu )}}=
 1+ {\alpha_s \over \pi} + {\alpha_s^2 \over \pi^2} \left [ 4 \, L^2
+ \left
( {7 \over 6} + { 2 \over 3} \, c(m_c^2/m_b^2) \right ) L \right ]\,,
\label{as2l}
\end{equation}
where, in terms of notation of Ref.\cite{bbbg:94}, $c(a)=c_{22}(a)-c_{12}(a)$.
The behavior of the function $c(a)$ is known explicitly \cite{bbbg:94} and is
quite weak: $c(0)=19/2$, $c(1) = 6$, and $c(m_c^2/m_b^2) \approx 9.0$ for the
realistic mass ratio $m_c/m_b \approx 0.3$. One can see that the term with the
single logarithm $L$ contributes about two thirds of that with $L^2$ in the term
quadratic in $\alpha_s$. Under such circumstances the RG summation of the terms
with powers of $L$ does not look satisfactory for numerical estimates of the QCD
effects, at least at the so far considered level of the first next-to-leading
order terms, and the next-to-next-to-leading terms can be equally important as
the two known ones, which would eliminate the existing impasse between the
theory and the data on $B_{sl}(B)$. One can present some arguments \cite{mv:96}
that this is indeed the case for the $b$ quark decay, although a complete
calculation of these corrections is still unavailable.

\subsection{Chromomagnetic and time dilation effects in decay rates}

The corrections suppressed by two powers of $m_Q^{-1}$ to inclusive decay rates
arise from two sources \cite{buv:92}: the $O(m_Q^{-2})$ corrections to the
matrix element of the leading operator, $({\overline Q} Q$, and the second term
in OPE (\ref{first3}) containing the chromomagnetic interaction. The expression
for the matrix element of the leading operator with the correction included is
written in the form
\begin{equation}
\langle H_Q| {\overline Q} Q | H_Q \rangle = 1-
{\mu_\pi^2(H_Q)-\mu_g^2(H_Q) \over 2 \, m_Q^2} + \ldots ,
\label{qbarq}
\end{equation}
where $\mu_\pi^2$ and $\mu_g^2$ are defined as
\begin{eqnarray}
&&\mu_\pi^2=\langle H_Q| {\overline Q}\, (i {\vec D})^2 \, Q | H_Q
\rangle \,, \nonumber \\
&&\mu_g^2=\langle H_Q| {\overline Q}\, {1 \over 2} \sigma^{\mu \nu}
G_{\mu \nu} Q | H_Q \rangle\,,
\label{mus}
\end{eqnarray}
with $D$ being the QCD covariant derivative. The correction in equation
(\ref{qbarq}) in fact corresponds to the time dilation factor $m_Q/E_Q$, for the
heavy quark decaying inside a hadron, where it has energy $E_Q$, which energy is
contributed by the kinetic part ($\propto \mu_\pi^2$) and the chromomagnetic
part ($\propto \mu_g^2$). The second term in OPE describes the effect of the
chromomagnetic interaction in the decay process, and is also expressed through
$\mu_g^2$.\index{kinetic energy of a heavy quark}\index{chromomagnetic energy of
  a heavy quark}

The explicit formulas for the decay rates, including the effects up to the order
$m_Q^{-2}$ are found in \cite{buv:92} and for decays of the $b$ hadrons read as
follows. For the semileptonic decay rate
\begin{equation}
\Gamma_{sl}(H_b)={|V_{cb}|^2 \, G_F^2 \, m_b^5 \over 192 \, \pi^3}
\,\langle H_b | {\overline b} b | H_b \rangle \,\left [1+
 {{\mu_g^2} \over m_b^2} \left ( {x \over 2}
\,{ d \over
{dx}} -2\right)  \right ] \,\eta_{sl} \, I_0 (x,\,0,\,0)\,,
\label{gamsl1}
\end{equation}
and for the nonleptonic decay rate
\begin{equation}
\Gamma_{nl}(H_b)={|V_{cb}|^2 \, G_F^2 \, m_b^5 \over 64 \, \pi^3} \,
\langle H_b | {\overline b} b | H_b \rangle \, \left \{\left [1+
 {{\mu_g^2} \over m_b^2} \left ( {x \over 2}
\,{ d \over
{dx}} -2\right)  \right ] \,\eta_{nl} \, I (x)-8 \eta_2 \, {{\mu_g^2}
\over m_b^2} \, I_2(x) \right \}\,.
\label{gamnl1}
\end{equation}
These formulas take into account only the dominant CKM mixing $V_{cb}$ and
neglect the small one, $V_{ub}$. The following notation is also used:
$x=m_c/m_b$, $I_0(x,y,z)$ stands for the kinematical suppression factor in a
three-body weak decay due to masses of the final fermions.  In
particular,\index{inclusive weak decay rates!kinematical integrals}
\begin{eqnarray}
&&I_0(x,0,0)=(1-x^4)(1-8\, x^2+ x^4)-24 \, x^4 \, \ln x\,, \\ \nonumber
&&I_0(x,x,0)=(1-14\, x^2-2 \, x^4 -12 x^6)\sqrt{1-4\, x^2} + 24 \,
(1-x^4) \, \ln {1+ \sqrt{1-4\, x^2} \over 1- \sqrt{1-4\, x^2}}\,.
\label{kints}
\end{eqnarray}
Furthermore, $I(x)=I_0(x,0,0)+I_0(x,x,0)$, and
$$
I_2(x)=(1-x^2)^3+\left ( 1+{1 \over
2} x^2+ 3 x^4 \right ) \, \sqrt{1-4\, x^2}-3x^2 \, (1-2x^4)
\ln{{1+\sqrt{1-4\, x^2}} \over {1-\sqrt{1-4\, x^2}}}\,.
$$
Finally, the QCD radiative correction factor $\eta_2$ in Eq.~(\ref{gamnl1}) is
known in the leading logarithmic approximation and is expressed in terms of the
well known coefficients $C_+$ and $C_-$ in the renormalization of the
nonleptonic weak interaction: $\eta_2=(C_+^2(m_b)-C_-^2(m_b))/6$
with\index{nonleptonic weak interaction!QCD renormalization}
\begin{equation}
C_-(\mu)=C_+^{-2}(\mu)=\left [ {\alpha_s(\mu) \over \alpha_s(m_W)}
\right ]^{4/b}\,,
\label{cpm}
\end{equation}
and $b$ is the coefficient in the QCD beta function. The value of $b$ relevant
to $b$ decays is $b=23/3$.

Numerically, for $x \approx 0.3$, the expressions for the decay rates can be
written as
\begin{eqnarray}
&&\Gamma_{sl}(H_b)=\Gamma_{sl}^{parton} \, \left ( 1-
{\mu_\pi^2(H_b)-\mu_g^2(H_b) \over 2 \, m_b^2}-2.6 {\mu_g^2(H_b) \over
m_b^2} \right )\,, \nonumber \\
&&\Gamma_{nl}(H_b)=\Gamma_{nl}^{parton} \, \left ( 1-
{\mu_\pi^2(H_b)-\mu_g^2(H_b) \over 2 \, m_b^2}-1.0 {\mu_g^2(H_b) \over
m_b^2} \right )\,,
\label{gamnum}
\end{eqnarray}
where $\Gamma^{parton}$ is the perturbation theory value of the corresponding
decay rate of $b$ quark.

The matrix elements $\mu_\pi^2$ and $\mu_g^2$ are related to the spectroscopic
formula for a heavy hadron mass $M$,
\begin{equation}
M(H_Q)=m_Q + {\overline \Lambda}(H_Q)+{\mu_\pi^2(H_Q)-\mu_g^2(H_Q) \over
2 \, m_Q} + \ldots
\label{massf}
\end{equation}
Being combined with the spin counting for pseudoscalar and vector mesons, this
formula allows to find the value of $\mu_g^2$ in pseudoscalar mesons from the
mass splitting:
\begin{equation}
\mu_g^2(B)={3 \over 4} \left ( M_{B^*}^2-M_B^2 \right ) \approx 0.36 \,
GeV^2\,.
\label{mugn}
\end{equation}
The value of $\mu_\pi^2$ for $B$ mesons is less certain. It is constrained by
the inequality \cite{mv:95}, $\mu_\pi^2(H_Q) \ge \mu_g^2(H_Q)$, and there are
theoretical estimates from the QCD sum rules \cite{bb:94}: $\mu_\pi^2(B) = 0.54
\pm 0.12 \, GeV^2$ and from an analysis of spectroscopy of heavy hadrons
\cite{neubert:00}: $\mu_\pi^2(B) = 0.3 \pm 0.2 \, GeV^2$. In any event, the
discussed corrections are rather small for $b$ hadrons, given that
$\mu_g^2/m_b^2 \approx 0.015$. The largest, in relative terms, effect of these
corrections in $B$ meson decays is on the semileptonic decay rate, where it
amounts to 4 -- 5 \% suppression of the rate, which rate however is only a
moderate fraction of the total width. In the dominant nonleptonic decay rate
the effect is smaller, and, according to the formula (\ref{gamnum}) amounts to
about 1.5 -- 2 \%.

The effect of the $m_Q^{-2}$ corrections can be evaluated with a somewhat better
certainty for the ratio of the decay rates of $\Lambda_b$ and $B$ mesons. This
is due to the fact that $\mu_g^2(\Lambda_b)=0$, since there is no correlation of
the spin of the heavy quark in $\Lambda_b$ with the light component, having
overall quantum numbers $J^P=0^+$. Then, applying the formula (\ref{gamnum}) to
$B$ and $\Lambda_b$, we find for the ratio of the (dominant) nonleptonic decay
rates:
\begin{equation}
{\Gamma_{nl}(\Lambda_b) \over \Gamma_{nl}(B)} = 1
-{\mu_\pi^2(\Lambda_b)-\mu_\pi^2(B) \over 2 \, m_b^2}+0.5 {\mu_g^2(B)
\over m_b^2}\,.
\label{lambdab}
\end{equation}
The difference of the kinetic terms, $\mu_\pi^2(\Lambda_b)-\mu_\pi^2(B)$, can be
estimated from the mass formula:
\begin{equation}
\mu_\pi^2(\Lambda_b)-\mu_\pi^2(B)={2 \, m_b \, m_c \over m_b-m_c} \left
[ {\overline M}(B)-{\overline M}(D)-M(\Lambda_b)+ M(\Lambda_c) \right] =
0 \pm 0.04 \, GeV^2\,,
\label{difmu}
\end{equation}
where ${\overline M}$ is the spin-averaged mass of the mesons, e.g., ${\overline
  M}(B)=(M(B)+3 M(B^*))/4$. The estimated difference of the kinetic terms is
remarkably small. Thus the effect in the ratio of the decay rates essentially
reduces to the chromomagnetic term, which is also rather small and accounts for
less than 1\% difference of the rates. For the ratio of the semileptonic decay
rates the chromomagnetic term is approximately four times larger, but then the
contribution of the semileptonic rates to the total width is rather small. Thus
one concludes that the terms of order $m_b^{-2}$ in the OPE expansion for the
decay rates can account only for about 1\% difference of the lifetimes of
$\Lambda_b$ and the $B$ mesons.

The significance of the $m_Q^{-2}$ terms is substantially different for the
decay rates of charmed hadrons, where these effects suppress the inclusive
decays of the $D$ mesons by about 40\% with respect to those of the charmed
hyperons in a reasonable agreement with the observed pattern of the lifetimes.

It should be emphasized once again that the $m_Q^{-2}$ effects do not depend on
the flavors of the spectator quarks or antiquarks. Thus the explanation of the
variety of the inclusive decay rates within the flavor $SU(3)$ multiplets,
observed for charmed hadrons and expected for the $b$ ones, has to be sought
among the $m_Q^{-3}$ terms.

\subsection{$L_{eff}^{(3)}$ Coefficients and operators}

Although the third term in the expansion (\ref{first3}) is formally suppressed
by an extra power of $m_Q^{-1}$, its effects are comparable to, or even larger
than the effects of the second term. This is due to the fact that the diagrams
determining the third term (see Fig.~8.1) contain a two-body phase space, while
the first two terms involve a three-body phase space. This brings in a numerical
enhancement factor, typically $4 \pi^2$. The enhanced numerical significance of
the third term in OPE, generally, does not signal a poor convergence of the
expansion in inverse heavy quark mass for decays of $b$, and even charmed,
hadrons the numerical enhancement factor is a one time occurrence in the series,
and there is no reason for similar `anomalous' enhancement among the higher
terms in the expansion.

Here we first present the expressions for the relevant parts of $L_{eff}^{(3)}$
for decays of $b$ and $c$ hadrons in the form of four-quark operators and then
proceed to a discussion of hadronic matrix elements and the effects in specific
inclusive decay rates. The consideration of the effects in decays of charmed
hadrons is interesting in its own right, and leads to new predictions to be
tested experimentally, and is also important for understanding the magnitude of
the involved matrix elements using the existing data on charm
decays.\index{inclusive weak decay rates!spectator quark effects}

We start with considering the term $L_{eff}^{(3)}$ in $b$ hadron nonleptonic
decays, $L_{eff, nl}^{(3,b)}$, induced by the underlying processes $b \to c \,
{\overline u} \, d$, $b \to c \, {\overline c} \, s$, $b \to c \, {\overline u}
\, s$, and $b \to c \, {\overline c} \, d$. Unlike the case of three-body decay,
the kinematical difference between the two-body states $c \overline c$ and $c
\overline u$, involved in calculation of $L_{eff, nl}^{(3,b)}$ is of the order
of $m_c^2/m_b^2 \approx 0.1$ and is rather small. At present level of accuracy
in discussing this term in OPE, one can safely neglect the effect of finite
charmed quark mass\footnote{The full expression for a finite charmed quark mass
  can be found in \cite{ns:97}}. In this approximation the expression for
$L_{eff, nl}^{(3,b)}$ reads as \cite{sv:86}
\begin{eqnarray}
\label{l3nlb}
L_{eff, \, nl}^{(3, b)} &=&  |V_{cb}|^2 \,{G_F^2 \, m_b^2 \over 4
\pi} \,
\bigg\{
{\tilde C}_1 \, (\overline b \Gamma_\mu b)(\overline u \Gamma_\mu u) +
{\tilde C}_2  \,
(\overline b \Gamma_\mu u) (\overline u \Gamma_\mu b) \nonumber\\
&&{} + {\tilde C}_5 \, (\overline  b \Gamma_\mu b +
{2 \over 3}\overline b \gamma_\mu \gamma_5 b) (\overline q \Gamma_\mu
q)+ {\tilde C}_6 \, (\overline  b_i \Gamma_\mu b_k +
{2 \over 3}\overline b_i \gamma_\mu \gamma_5 b_k)
(\overline q_k \Gamma_\mu q_i) \nonumber\\ 
&&{} + {1 \over 3} \,
{\tilde \kappa}^{1/2} \, ({\tilde \kappa}^{-2/9}-1) \, \bigg[ 2 \,
({\tilde C}_+^2 - {\tilde C}_-^2) \,
 (\overline b \Gamma_\mu t^a b) \,
j_\mu^a  \nonumber\\ 
&&{} -
(5{\tilde C}_+^2+{\tilde C}_-^2 - 6 \, {\tilde C}_+ \, {\tilde C}_-)
(\overline  b \Gamma_\mu t^a b +
{2 \over 3}\overline b \gamma_\mu \gamma_5 t^a b) j_\mu^a \bigg]
\bigg\}\,,
\end{eqnarray}
where the notation $(\overline q \, \Gamma \, q)= (\overline d \, \Gamma \, d) +
(\overline s \, \Gamma \, s)$ is used, the indices $i, \, k$ are the color
triplet ones, $\Gamma_\mu=\gamma_\mu \, (1-\gamma_5)$, and $j_\mu^a=\overline u
\gamma_\mu t^a u + \overline d \gamma_\mu t^a d + \overline s \gamma_\mu t^a s$
is the color current of the light quarks with $t^a = \lambda^a /2$ being the
generators of the color $SU(3)$. The notation ${\tilde C}_\pm$, is used as
shorthand for the short-distance renormalization coefficients $C_\pm(\mu)$ at
$\mu=m_b$: ${\tilde C}_\pm \equiv C_\pm(m_b)$. The expression (\ref{l3nlb}) is
written in the leading logarithmic approximation for the QCD radiative effects
in a low normalization point $\mu$ such that $\mu \ll m_b$ (but still, at least
formally, $\mu \gg \Lambda_{QCD}$). For such $\mu$ there arises so called
`hybrid' renormalization \cite{sv:87}, depending on the factor ${\tilde
  \kappa}=\alpha_s(\mu)/\alpha_s(m_b)$. The coefficients ${\tilde C}_A$ with
$A=1, \ldots , 6$ in Eq.~(\ref{l3nlb}) have the following explicit expressions in
terms of ${\tilde C}_\pm$ and ${\tilde \kappa}$:
\begin{eqnarray}
&&{\tilde C}_1= {\tilde C}_+^2+{\tilde C}_-^2 + {1 \over 3} (1 -
\kappa^{1/2}) ({\tilde C}_+^2-{\tilde C}_-^2)\,,
\nonumber \\
&&{\tilde C}_2= \kappa^{1/2} \, ({\tilde C}_+^2-{\tilde C}_-^2)\,,
\nonumber \\
&&{\tilde C}_3=- {1 \over 4} \, \left [ ({\tilde C}_+-{\tilde C}_-)^2 +
{1 \over 3}
(1-\kappa^{1/2})
(5{\tilde C}_+^2+{\tilde C}_-^2+6{\tilde C}_+{\tilde C}_-) \right] \,,
\nonumber \\
&&{\tilde C}_4=-{1 \over 4} \, \kappa^{1/2} \, (5{\tilde C}_+^2+{\tilde
C}_-^2+6{\tilde C}_+{\tilde C}_-)\,, \nonumber
\\
&&{\tilde C}_5=-{1 \over 4} \, \left [ ({\tilde C}_++{\tilde C}_-)^2 +
{1 \over 3} (1-\kappa^{1/2})
(5{\tilde C}_+^2+{\tilde C}_-^2-6{\tilde C}_+{\tilde C}_-) \right]\,,
\nonumber \\
&&{\tilde C}_6=-{1 \over 4} \, \kappa^{1/2} \, (5{\tilde C}_+^2+{\tilde
C}_-^2-6{\tilde C}_+{\tilde C}_-)\,.
\label{coefs}
\end{eqnarray}

The expression for the CKM dominant semileptonic decays of $b$ hadrons,
associated with the elementary process $b \to c \, \ell \, \nu$ does not look to
be of an immediate interest. The reason is that this process is intrinsically
symmetric under the flavor $SU(3)$, and one expects no significant splitting of
the semileptonic decay rates within $SU(3)$ multiplets of the $b$ hadrons. The
only possible effect of this term, arising through a penguin-like mechanism can
be in a small overall shift of semileptonic decay rates between $B$ mesons and
baryons. However, these effects are quite suppressed and are believed to be even
smaller than the ones arising form the discussed $m_b^{-2}$ terms.

For charm decays there is a larger, than for $b$ hadrons, variety of effects
associated with $L_{eff}^{(3)}$ that can be studied experimentally, and we
present here the relevant parts of the effective Lagrangian. For the CKM
dominant nonleptonic decays of charm, originating from the quark process $c \to
s \, u \, {\overline d}$, the discussed term in OPE has the form
\begin{eqnarray}
\label{l3nl}
&&L_{eff, nl }^{(3, \, \Delta C = \Delta S)}= \cos^4\theta_c \,{G_F^2 \,
m_c^2 \over 4 \pi} \,
\bigg\{
C_1 \, (\overline c \Gamma_\mu c)(\overline d \Gamma_\mu d) + C_2  \,
(\overline c \Gamma_\mu d) (\overline d \Gamma_\mu c)
\nonumber \\
&&{} + C_3 \, (\overline  c \Gamma_\mu c +
{2 \over 3}\overline c \gamma_\mu \gamma_5 c) (\overline s \Gamma_\mu
s)+ C_4 \, (\overline  c_i \Gamma_\mu c_k +
{2 \over 3}\overline c_i \gamma_\mu \gamma_5 c_k)
(\overline s_k \Gamma_\mu s_i)
\\ \nonumber
&&{} + C_5 \, (\overline  c \Gamma_\mu c +
{2 \over 3}\overline c \gamma_\mu \gamma_5 c) (\overline u \Gamma_\mu
u)+ C_6 \, (\overline  c_i \Gamma_\mu c_k +
{2 \over 3}\overline c_i \gamma_\mu \gamma_5 c_k)
(\overline u_k \Gamma_\mu u_i) \\ \nonumber
&&{} + {1 \over 3} \,
\kappa^{1/2} \, (\kappa^{-2/9}-1) \, \bigg[ 2 \, (C_+^2 - C_-^2) \,
 (\overline c \Gamma_\mu t^a c) \,
j_\mu^a - (5C_+^2+C_-^2)
(\overline  c \Gamma_\mu t^a c +
{2 \over 3}\overline c \gamma_\mu \gamma_5 t^a c) j_\mu^a \bigg]
\bigg\}\,,
\end{eqnarray}
where, $\theta_c$ is the Cabibbo angle, and the coefficients without the tilde
are given by the same expressions as above for the $b$ decays (i.e., those with
tilde) with the replacement $m_b \to m_c$. The part of the notation in the
superscript $\Delta C = \Delta S$ points to the selection rule for the dominant
CKM unsuppressed nonleptonic decays.  One can rather realistically envisage
however a future study of inclusive rates for the once CKM suppressed decays of
charmed hadrons\footnote{Even if the inclusive rate of these decays is not to be
  separated experimentally, they contribute about 10\% of the total decay rate,
  and it is worthwhile to include their contribution in the balance of the total
  width.}, satisfying the selection rule $\Delta S=0$ and associated with the
quark processes $c \to d \, u \, {\overline s}$ and $c \to d \, u \, {\overline
  d}\,$. The corresponding part of the effective Lagrangian for these processes
reads as
\begin{eqnarray}
&&L_{eff, \, nl}^{(3, \Delta S=0)}= \cos^2 \theta_c \, \sin^2 \theta_c
\,{G_F^2 \, m_c^2 \over 4 \pi} \,
\bigg\{
C_1 \, (\overline c \Gamma_\mu c)(\overline q \Gamma_\mu q) + C_2  \,
(\overline c_i \Gamma_\mu c_k) (\overline q_k \Gamma_\mu q_i)
\nonumber \\*
&&{} + C_3 \, (\overline  c \Gamma_\mu c +
{2 \over 3}\overline c \gamma_\mu \gamma_5 c) (\overline q \Gamma_\mu
q)+ C_4 \, (\overline  c_i \Gamma_\mu c_k +
{2 \over 3}\overline c_i \gamma_\mu \gamma_5 c_k)
(\overline q_k \Gamma_\mu q_i) \\*
&&{} + 2 \, C_5 \, (\overline  c \Gamma_\mu c +
{2 \over 3}\overline c \gamma_\mu \gamma_5 c) (\overline u \Gamma_\mu
u)+ 2 \, C_6 \, (\overline  c_i \Gamma_\mu c_k +
{2 \over 3}\overline c_i \gamma_\mu \gamma_5 c_k)
(\overline u_k \Gamma_\mu u_i)  \nonumber\\
&&{} + {2 \over 3} \,
\kappa^{1/2} \, (\kappa^{-2/9}-1) \, \bigg[ 2 \, (C_+^2 - C_-^2) \,
 (\overline c \Gamma_\mu t^a c) \,
j_\mu^a - (5C_+^2+C_-^2)
(\overline  c \Gamma_\mu t^a c +
{2 \over 3}\overline c \gamma_\mu \gamma_5 t^a c) j_\mu^a \bigg]
\bigg\}\,, \nonumber
\label{l3nl1}
\end{eqnarray}
where again the notation $(\overline q \, \Gamma \, q)= (\overline d \, \Gamma
\, d) + (\overline s \, \Gamma \, s)$ is used.

The semileptonic decays of charm, the CKM dominant, associated with $c \to s \,
\ell \, \nu$, and the CKM suppressed, originating from $c \to s \, \ell \, \nu$,
contribute to the semileptonic decay rate, which certainly can be measured
experimentally. The expression for the part of the effective Lagrangian,
describing the $m_Q^{-3}$ terms in these decays is
\cite{mv:96,cheng:97,gm:98}
\index{inclusive weak decay rates!semileptonic decay of charm}%
\begin{eqnarray}
&&L_{eff, \, sl}^{(3)}= 
{G_F^2 \, m_c^2 \over 12 \pi} \, \bigg\{ \cos^2 \theta_c \left[
L_1 \, (\overline  c \Gamma_\mu c +
{2 \over 3}\overline c \gamma_\mu \gamma_5 c) (\overline s \Gamma_\mu
s)+ L_2 \, (\overline  c_i \Gamma_\mu c_k +  {2 \over 3}\overline c_i
\gamma_\mu \gamma_5 c_k)
(\overline s_k \Gamma_\mu s_i) \right ] \nonumber \\
&&{} + \sin^2 \theta_c \left [
L_1\, (\overline  c \Gamma_\mu c +
{2 \over 3}\overline c \gamma_\mu \gamma_5 c) (\overline d \Gamma_\mu
d)+L_2 \, (\overline  c_i \Gamma_\mu c_k +
{2 \over 3}\overline c_i \gamma_\mu \gamma_5 c_k)
(\overline d_k \Gamma_\mu d_i) \right ]  \nonumber \\
&&{} - 2 \, \kappa^{1/2} \, (\kappa^{-2/9}-1) \,
(\overline  c \Gamma_\mu t^a c +
{2 \over 3}\overline c \gamma_\mu \gamma_5 t^a c) j_\mu^a
\bigg\} \,,
\label{l3sl}
\end{eqnarray}
with the coefficients $L_1$ and $L_2$ found as
\begin{equation}
L_1=(\kappa^{1/2}-1), ~~~~~
L_2 = -   3\, \kappa^{1/2}\,.
\label{coefl}
\end{equation}

\subsection{Effects of $L_{eff}^{(3)}$ in mesons}

The expressions for the terms in $L_{eff}^{(3)}$ still leave us with the problem
of evaluating the matrix elements of the four-quark operators over heavy hadrons
in order to calculate the effects in the decay rates according to the formula
(\ref{lgam}). In doing so only few conclusions can be drawn in a reasonably
model independent way, i.e., without resorting to evaluation of the matrix
elements using specific ideas about the dynamics of quarks inside hadrons. The
most straightforward prediction can in fact be found for $b$ hadrons. Namely,
one can notice that the operator (\ref{l3nlb}) is symmetric under the flavor U
spin (an SU(2) subgroup of the flavor $SU(3)$, which mixes $s$ and $d$ quarks).
This is a direct consequence of neglecting the small kinematical effect of the
charmed quark mass. However the usual (in)accuracy of the flavor $SU(3)$ symmetry
is likely to be a more limiting factor for the accuracy of this symmetry, than
the corrections of order $m_c^2/m_b^2$. Modulo this reservation the immediate
prediction from this symmetry is the degeneracy of inclusive decay rates within
U spin doublets:
\begin{equation}
\Gamma(B_d)=\Gamma(B_s)\,,~~~~\Gamma(\Lambda_b)=\Gamma(\Xi_b^0)\,,
\label{upred}
\end{equation}
where $\Gamma(B_s)$ stands for the average rate over the two eigenstates of the
$B_s - {\overline B}_s$ oscillations. The data on decay rates of the cascade
hyperon $\Xi_b^0$ are not yet available, while the currently measured lifetimes
of $B_d$ and $B_s$ are within less than 2\% from one another. Theoretically, the
difference of the lifetimes, associated with possible violation of the $SU(3)$
symmetry and with breaking of the U symmetry of the effective Lagrangian
(\ref{l3nlb}), is expected to not exceed about 1\%.

For the non-vanishing matrix elements of four-quark operators over pseudoscalar
mesons one traditionally starts with the factorization formula and parameterizes
possible deviation from factorization in terms of `bag constants'. Within the
normalization convention adopted here the relations used in this parameterization
read as\index{Bag constants}
\begin{eqnarray}
&&\langle P_{Q \overline q} | (\overline Q \Gamma_\mu q) \, (\overline q
\Gamma_\mu Q) | P_{Q \overline q} \rangle = {1 \over 2} \, f_P^2 \, M_P
\,B \,, \nonumber \\
&&\langle P_{Q \overline q} | (\overline Q \Gamma_\mu Q) \, (\overline q
\Gamma_\mu q) | P_{Q \overline q} \rangle = {1 \over 6} \, f_P^2 \, M_P
\,{\tilde B}\,,
\label{bagc}
\end{eqnarray}
where $P_{Q \overline q}$ stands for pseudoscalar meson made of $Q$ and
${\overline q}$, $f_P$ is the annihilation constant for the meson, and $B$ and
${\tilde B}$ are bag constants. The parameters $B$ and ${\tilde B}$ generally
depend on the normalization point $\mu$ for the operators, and this dependence
is compensated by the $\mu$ dependence of the coefficients in $L_{eff}^{(3)}$,
so that the results for the physical decay rate difference do not depend on
$\mu$. If the normalization point $\mu$ is chosen at the heavy quark mass (i.e.
$\mu=m_b$ for $B$ mesons, and $\mu=m_c$ for $D$ mesons) the predictions for the
difference of total decay rates take a simple form in terms of the corresponding
bag constants (generally different between $B$ and $D$):
\begin{eqnarray}
\Gamma(B^\pm) - \Gamma(B^0) &=& |V_{cb}|^2 \,{G_F^2 \, m_b^3 \, f_B^2
\over 8
\pi} \left [ ({\tilde C}_+^2-{\tilde C}_-^2) \, B(m_b) + {1 \over 3}\,
({\tilde C}_+^2+{\tilde C}_-^2) \,  {\tilde B}(m_b) \right] \nonumber \\
&\approx& - 0.025 \, \left ( {f_B \over 200 \, MeV } \right )^2 \,
ps^{-1}\,, \label{bpred} \\[6pt]
\Gamma(D^\pm) - \Gamma(D^0) &=&  \cos^4 \theta_c \,{G_F^2 \, m_c^3 \,
f_D^2 \over 8
\pi} \left [ (C_+^2-C_-^2) \,  B(m_c) + {1 \over 3}\, (C_+^2+C_-^2) \,
{\tilde B}(m_c) \right] \nonumber \\
&\sim& - 0.8 \, \left ( {f_D \over 200 \, MeV } \right )^2 \, ps^{-1}\,,
\label{dpred}
\end{eqnarray}
where the numerical values are written in the approximation of exact
factorization: $B=1$, and ${\tilde B}=1$. It is seen from the numerical
estimates that, even given all the theoretical uncertainties, the presented
approach is in reasonable agreement with the data on the lifetimes of $D$ and
$B$ mesons. In particular, this approach describes, at least qualitatively, the
strong suppression of the decay rate of $D^\pm$ mesons relative to $D^0$, the
experimental observation of which has in fact triggered in early 80-s the
theoretical study of preasymptotic in heavy quark mass effects in inclusive
decay rates.  For the $B$ mesons the estimate (\ref{bpred}) is also in a
reasonable agreement with the current data for the discussed difference ($-0.043
\pm 0.017 \, ps^{-1}$).

\subsection{Effects of $L_{eff}^{(3)}$ in baryons}

The weakly decaying heavy hyperons, containing either $c$ or $b$ quark are:
$\Lambda_Q \sim Qud,~ \Xi_Q^{(u)} \sim Qus,~ \Xi_Q^{(d)} \sim Qds$, and
$\Omega_Q \sim Qss$. The first three baryons form an $SU(3)$ (anti)triplet. The
light diquark in all three is in the state with quantum numbers $J^P = 0^+$, so
that there is no correlation of the spin of the heavy quark with the light
component of the baryon.\index{heavy baryons!$SU(3)$ antitriplet} On the contrary,
in $\Omega_Q$ the two strange quarks form a $J^P = 1^+$ state, and a correlation
between the spins of heavy and light quarks is present. The absence of spin
correlation for the heavy quark in the triplet of hyperons somewhat reduces the
number of independent four-quark operators, having nonvanishing diagonal matrix
elements over these baryons. Indeed, the operators entering $L_{eff}^{(3)}$
contain both vector and axial bilinear forms for the heavy quarks. However the
axial part requires a correlation of the heavy quark spin with that of a light
quark, and is thus vanishing for the hyperons in the triplet.  Therefore only
the structures with vector currents are relevant for these hyperons. These
structures are of the type $(\overline c \, \gamma_\mu \, c) (\overline q \,
\gamma_\mu \, q)$ and $(\overline c_i \, \gamma_\mu \, c_k) (\overline q_k \,
\gamma_\mu \, q_i)$ with $q$ being $d$, $s$ or $u$.  The flavor $SU(3)$ symmetry
then allows to express, for each of the two color combinations, the matrix
elements of three different operators, corresponding to three flavors of $q$,
over the baryons in the triplet in terms of only two combinations: flavor octet
and flavor singlet. Thus all effects of $L_{eff}^{(3)}$ in the triplet of the
baryons can be expressed in terms of four independent combinations of matrix
elements.  These can be chosen in the following way:
\begin{eqnarray}
\label{msh:defxy}
x&=&\left \langle  {1 \over 2} \, (\overline Q \, \gamma_\mu \, Q) \left
[
(\overline u \, \gamma_\mu u) - (\overline s \, \gamma_\mu s) \right]
\right \rangle_{\Xi_Q^{(d)}-\Lambda_Q}  \nonumber\\
&=& \left \langle  {1 \over 2} \,
(\overline Q \, \gamma_\mu \, Q) \left [ (\overline s \, \gamma_\mu s) -
(\overline d \, \gamma_\mu d) \right] \right \rangle_{\Lambda_Q -
\Xi_Q^{(u)}}\,,   \nonumber\\
y&=&\left \langle  {1 \over 2} \, (\overline Q_i \, \gamma_\mu \, Q_k)
\left [ (\overline u_k \, \gamma_\mu u_i) - (\overline s_k \, \gamma_\mu
s_i) \right ] \right \rangle_{\Xi_Q^{(d)}-\Lambda_Q}  \nonumber\\
&=& \left \langle  {1
\over 2} \, (\overline Q_i \, \gamma_\mu \, Q_k) \left [ (\overline s_k
\, \gamma_\mu s_i) - (\overline d_k \, \gamma_\mu d_i) \right] \right
\rangle_{\Lambda_Q - \Xi_Q^{(u)}}\,,
\end{eqnarray}
with the notation for the differences of the matrix elements: $\langle {\cal O}
\rangle_{A-B}= \langle A | {\cal O} | A \rangle - \langle B | {\cal O} | B
\rangle$, for the flavor octet part and the matrix elements:
\begin{eqnarray}
&&x_s = {1 \over 3} \, \langle H_Q | ({\overline Q} \, \gamma_\mu \, Q)
\left ( ({\overline u} \, \gamma_\mu \, u)+({\overline d} \, \gamma_\mu
\, d)+ ({\overline s} \, \gamma_\mu \, s) \right ) | H_Q \rangle
\nonumber \\
&&y_s = {1 \over 3} \, \langle H_Q | ({\overline Q}_i \, \gamma_\mu \,
Q_k) \left ( ({\overline u}_k \, \gamma_\mu \, u_i)+({\overline d}_k \,
\gamma_\mu \, d_i)+ ({\overline s}_k \, \gamma_\mu \, s_i) \right ) |
H_Q \rangle
\label{xys}
\end{eqnarray}
for the flavor singlet part, where $H_Q$ stands for any heavy hyperon in the
(anti)triplet.

The initial, very approximate, theoretical estimates of the matrix elements
\cite{sv:86} were essentially based on a non-relativistic constituent quark
model, where these matrix elements are proportional to the density of a light
quark at the location of the heavy one, i.e., in terms of the wave function,
proportional to $|\psi(0)|^2$. Using then the same picture for the matrix
elements over pseudoscalar mesons, relating the quantity $|\psi(0)|^2$ to the
annihilation constant $f_P$, and assuming that $|\psi(0)|^2$ is approximately
the same in baryons as in mesons, one arrived at the estimate
\index{heavy baryons!matrix elements}%
\begin{equation}
  y=-x=x_s=-y_s \approx {f_D^2 \, M_D \over 12} \approx 0.006 \, GeV^2\,,
\label{xyold}
\end{equation}
where the sign relation between $x$ and $y$ is inferred from the color
antisymmetry of the constituent quark wave function for baryons. Since the
constituent picture was believed to be valid at distances of the order of the
hadron size, the estimate (\ref{xyold}) was applied to the matrix elements in a
low normalization point where $\alpha_s(\mu) \approx 1$. For the matrix elements
of the operators, containing $s$ quarks over the $\Omega_Q$ hyperon, this
picture predicts an enhancement factor due to the spin correlation:
\begin{equation}
\langle \Omega_Q | ({\overline Q} \, \Gamma_\mu \, Q)
 ({\overline s} \, \Gamma_\mu \, s)  |\Omega_Q \rangle = - \langle
\Omega_Q | ({\overline Q}_i \, \Gamma_\mu \, Q_k)
 ({\overline s}_k \, \Gamma_\mu \, s_i)  |\Omega_Q \rangle = {10 \over
3}\, y
\label{omegaold}
\end{equation}
Although these simple estimates allowed to correctly predict \cite{sv:86} the
hierarchy of lifetimes of charmed hadrons prior to establishing this hierarchy
experimentally, they fail to quantitatively predict the differences of lifetimes
of charmed baryons. We shall see that the available data indicate that the color
antisymmetry relation is badly broken, and the absolute value of the matrix
elements is larger, than the naive estimate (\ref{xyold}), especially for the
quantity $x$.

It should be emphasized that in the heavy quark limit the matrix elements
(\ref{msh:defxy}) and (\ref{xys}) do not depend on the flavor of the heavy
quark, provided that the same normalization point $\mu$ is used. Therefore,
applying the OPE formulas to both charmed and $b$ baryons, one can extract the
values for the matrix elements from available data on charmed hadrons, and then
make predictions for $b$ baryons, as well as for other inclusive decay rates,
e.g., semileptonic, for charmed hyperons.

The only data available so far, which would allow to extract the matrix
elements, are on the lifetimes of charmed hyperons. Therefore, one has to take
into account several essential types of inclusive decay, at least those that
contribute to the total decay rate at the level of about 10\%. Here we first
concentrate on the differences of the decay rates within the $SU(3)$ triplet of
the hyperons, which will allow us to extract the non-singlet quantities $x$ and
$y$, and then discuss the $SU(3)$ singlet shifts of the rates.

The differences of the dominant Cabibbo unsuppressed nonleptonic decay rates
are given by
\begin{eqnarray}
\delta_1^{nl, \,0} \equiv \Gamma^{nl}_{\Delta S =
\Delta C}(\Xi_c^0)-\Gamma^{nl}_{\Delta S = \Delta C}(\Lambda_c) = \cos^4
\theta_c \, {G_F^2 \,
m_c^2 \over 4 \pi} \left [ (C_5 - C_3) \, x + (C_6 - C_4) \, y \right
]\,, \nonumber \\
\delta_2^{nl, \,0} \equiv \Gamma^{nl}_{\Delta S =
\Delta C}(\Lambda_c)-\Gamma^{nl}_{\Delta S =\Delta C}(\Xi_c^+) = \cos^4
\theta_c \, {G_F^2 \,
m_c^2 \over 4 \pi} \left [ (C_3 - C_1) \, x + (C_4 - C_2) \, y \right
]\,.
\label{dnl0}
\end{eqnarray}
The once Cabibbo suppressed decay rates of $\Lambda_c$ and $\Xi_c^+$ are equal,
due to the $\Delta U =0$ property of the corresponding effective Lagrangian
$L_{eff, nl}^{(3,1)}$ (Eq.~(\ref{l3nl1})). Thus the only difference for these
decays in the baryon triplet is
\begin{eqnarray}
\delta^{nl,1} & \equiv & \Gamma^{nl}_{\Delta S =0}
(\Xi_c^0)-\Gamma^{nl}_{\Delta S = 0 }(\Lambda_c) \nonumber \\
& = & \cos^2 \theta_c \,
\sin^2 \theta_c \, {G_F^2
\, m_c^2 \over 4 \pi} \left [ (2\, C_5 - C_1 - C_3) \, x + (2 \, C_6 -
C_2 - C_4) \, y \right ]\,.
\label{dnl1}
\end{eqnarray}
The dominant semileptonic decay rates are equal among the two $\Xi_c$ baryons
due to the isotopic spin property $\Delta I =0$ of the corresponding interaction
Lagrangian, thus there is only one non-trivial splitting for these decays:
\begin{equation}
\delta^{sl,0} \equiv \Gamma^{sl}_{\Delta S = -1}(\Xi_c) -
\Gamma^{sl}_{\Delta S = -1} (\Lambda_c) = - \cos^2 \theta_c {G_F^2 \,
m_c^2 \over 12
\pi} \left [ L_1 \, x + L_2 \, y \right ]\,.
\label{dsl0}
\end{equation}
Finally, the Cabibbo suppressed semileptonic decay rates are equal for
$\Lambda_c$ and $\Xi_c^0$, due to the $\Delta V =0$ property of the
corresponding interaction. Thus the only difference for these is
\begin{equation}
\delta^{sl,1} \equiv \Gamma^{sl}_{\Delta S = 0}(\Lambda_c) -
\Gamma^{sl}_{\Delta S = 0}(\Xi_c^+) = - \sin^2 \theta_c {G_F^2 \, m_c^2
\over 12
\pi} \left [ L_1 \, x + L_2 \, y \right ]\,.
\label{dsl1}
\end{equation}

Using the relations (\ref{dnl0})--(\ref{dsl1}) on can find expressions for two
differences of the measured total decay rates, $\Delta_1 = \Gamma (\Xi_c^0) -
\Gamma (\Lambda_c)$ and $\Delta_2= \Gamma (\Lambda_c) - \Gamma (\Xi_c^+)$, in
terms of the quantities $x$ and $y$:
\begin{eqnarray}
\Delta_1 &=& \delta_1^{nl, \,0} + \delta^{nl,1} + 2 \, \delta^{sl,0}
\nonumber \\
&=& {G_F^2 \, m_c^2 \over 4 \pi} \, \cos^2 \theta \, \bigg\{ x
 \left [ \cos^2 \theta \, (C_5-C_3) + \sin^2 \theta \, (2 \, C_5 - C_1
-C_3)- {2 \over 3} \, L_1 \right ] \nonumber \\
&&{} + y \left[ \cos^2 \theta \,
(C_6-C_4) + \sin^2 \theta \, (2 \, C_6 - C_2 -C_4)- {2 \over 3} \,
L_2\right ] \bigg\}\,,
\label{d1}
\end{eqnarray}
and
\begin{eqnarray}
\Delta_2 &=& \delta_2^{nl, \,0} - 2 \, \delta^{sl,0} + 2 \, \delta^{sl,1}
\nonumber \\
&=& {G_F^2 \, m_c^2 \over 4 \pi}\, \bigg\{ x \left[ \cos^4
\theta \, (C_3-C_1) + {2 \over 3} \, (\cos^2 \theta - \sin^2 \theta) \,
L_1 \right] \nonumber \\
&&{} + y \left[ \cos^4 \theta \, (C_4-C_2) + {2 \over 3} \,
(\cos^2 \theta - \sin^2 \theta) \, L_2 \right] \bigg\}\,.
\label{d2}
\end{eqnarray}

By comparing these relations with the data, one can extract the values of $x$
and $y$. Using the current data for the total decay rates:
$\Gamma(\Lambda_c)=4.85 \pm 0.28 \, ps^{-1}$, $\Gamma(\Xi_c^0)= 10.2 \pm 2 \,
ps^{-1}$, and the updated value \cite{pdg1} $\Gamma(\Xi_c^+)=3.0 \pm 0.45 \,
ps^{-1}$, we find for the $\mu$ independent matrix element $x$
\begin{equation}
x = -(0.04 \pm 0.01) \, GeV^3 \, \left ( 1.4 \, GeV \over m_c \right
)^2\,,
\label{resx}
\end{equation}
while the dependence of the thus extracted matrix element $y$ on the
normalization point $\mu$ is shown in Fig.~8.2.\footnote{It should be noted
  that the curves at large values of $\kappa$, $\kappa > \, \sim 3$, are shown
  only for illustrative purpose. The coefficients in the OPE, leading to the
  equations (\ref{d1},\ref{d2}), are purely perturbative.  Thus, formally, they
  correspond to $\alpha_s(\mu) \ll 1$, i.e., to $\kappa \ll 1/\alpha_s(m_c) \sim
  (3 - 4)$.}

\begin{figure}[ht]
  \begin{center}
    \leavevmode
    \epsfbox{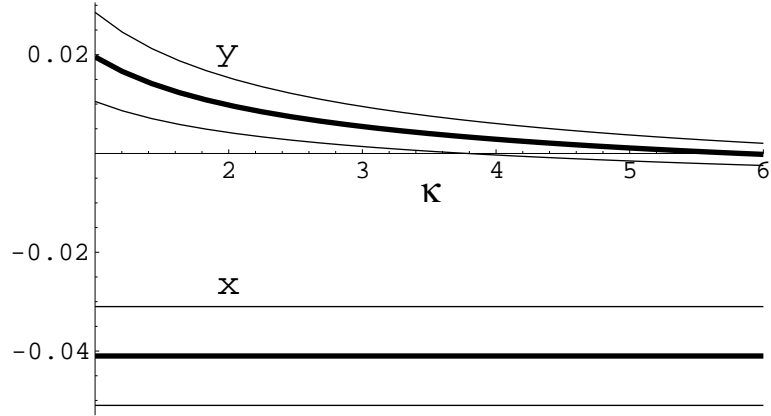}
    \vspace*{6pt}
\caption[The values of the extracted matrix elements $x$ and $y$ in $GeV^3$
vs.\ the normalization point parameter
$\kappa=\alpha_s(\mu)/\alpha_s(m_c)$.]{The values of the extracted matrix
elements $x$ and $y$ in $GeV^3$ vs.\ the normalization point parameter
$\kappa=\alpha_s(\mu)/\alpha_s(m_c)$. The thick lines correspond to the central
value of the data on lifetimes of charmed baryons, and the thin lines show the
error corridors. The extracted values of $x$ and $y$ scale as $m_c^{-2}$ with
the assumed mass of the charmed quark, and the plots are shown for $m_c=1.4 \,
GeV$.}
\label{fig:xy}
  \end{center}
\end{figure}

Notably, the extracted values of $x$ and $y$ are in a drastic variance with the
simplistic constituent model: the color antisymmetry relation, $x=-y$, does not
hold at any reasonable $\mu$, and the absolute value of $x$ is substantially
enhanced\footnote{A similar, although with a smaller enhancement, behavior of
  the matrix elements was observed in a recent preliminary lattice study
  \cite{psm:99}.}

Once the non-singlet matrix elements are determined, they can be used for
predicting differences of other inclusive decay rates within the triplet of
charmed hyperons as well as for the $b$ baryons.
\index{inclusive weak decay rates!splitting for baryons}%
Due to correlation of errors in $x$ and $y$ it
makes more sense to express the predictions directly in terms of the total decay
rates of the charmed hyperons. The thus arising relations between the rates do
not depend on the normalization parameter $\mu$. In this way one finds
\cite{mv:99-1} for the difference of the Cabibbo dominant semileptonic decay
rates between either of the $\Xi_c$ hyperons and $\Lambda_c$:
\begin{equation}
\Gamma_{sl}(\Xi_c)-\Gamma_{sl}(\Lambda_c) \approx \delta^{sl, 0}= 0.13
\, \Delta_1 - 0.065 \, \Delta_2 \approx 0.59 \pm 0.32 \, ps^{-1}\,.
\label{dslr}
\end{equation}
When compared with the data on the total semileptonic decay rate of
$\Lambda_c$, $\Gamma_{sl}(\Lambda_c)=0.22 \pm 0.08 \, ps^{-1}$, this prediction
implies that the semileptonic decay rate of the charmed cascade hyperons can be
2--3 times larger than that of $\Lambda_c$.

The predictions found in a similar way for the inclusive Cabibbo suppressed
decay rates are \cite{mv:99-1}: for nonleptonic decays
\begin{equation}
\delta^{nl,1}=0.082 \, \Delta_1 + 0.054 \, \Delta_2 \approx 0.55 \pm
0.22 \, ps^{-1}
\label{dnl1r}
\end{equation}
and for the semileptonic ones
\begin{equation}
\delta^{sl,1}=\tan^2 \theta_c \, \delta^{sl, 0} \approx 0.030 \pm 0.016
\, ps^{-1}\,.
\label{dsl1r}
\end{equation}

For the only difference of the inclusive rates in the triplet of $b$ baryons,
$\Gamma(\Lambda_b)-\Gamma(\Xi_b^-)$, one finds an expression in terms of $x$ and
$y$, or alternatively, in terms of the differences $\Delta_1$ and $\Delta_2$
between the charmed hyperons,
\begin{eqnarray}
\Gamma(\Lambda_b)-\Gamma(\Xi_b^-) &=& \cos^2 \theta_c \, |V_{cb}|^2
\,{G_F^2 \, m_b^2 \over 4 \pi} \, \left[ (\tilde C_5 - \tilde C_1) \, x
+ (\tilde C_6 - \tilde C_2) \, y \right] \\
&\approx& |V_{cb}|^2  \, {m_b^2 \over m_c^2} \, (0.85 \,
\Delta_1 + 0.91 \, \Delta_2) \approx 0.015 \, \Delta_1 + 0.016 \,
\Delta_2 \approx 0.11 \pm 0.03 \, ps^{-1}\,.\nonumber 
\label{dbres}
\end{eqnarray}
When compared with the data on the total decay rate of $\Lambda_b$ this result
predicts about 14\% longer lifetime of $\Xi_b^-$ than that of $\Lambda_b$.

The singlet matrix elements $x_s$ and $y_s$ (cf. Eq.~(\ref{xys})) are related to
the shift of the average decay rate of the hyperons in the triplet:
\index{heavy baryons!average decay rate}%
\begin{equation}
{\overline \Gamma}_Q= {1 \over 3} \, \left ( \Gamma(\Lambda_Q) +
\Gamma(\Xi_Q^1)+\Gamma(\Xi_Q^2) \right )\,.
\label{avgam}
\end{equation}
For the charmed baryons the shift of the dominant nonleptonic decay rate is
given by \cite{mv:99-2}
\begin{equation}
\delta_{nl}^{(3,0)} {\overline \Gamma}_c= \cos^4 \theta \, {G_F^2 \,
m_c^2 \over 8 \pi} (C_+^2 + C_-^2)\, \kappa^{5/18} \, (x_s-3 \, y_s)\,,
\label{d3gc}
\end{equation}
while for the $b$ baryons the corresponding expression reads as
\begin{equation}
\delta^{(3)} {\overline \Gamma}_b= |V_{cb}|^2 \, {G_F^2 \, m_b^2 \over 8
\pi} ({\tilde C}_+ - {\tilde C}_-)^2\, {\tilde \kappa}^{5/18} \, (x_s-3
\, y_s)\,.
\label{d3gb}
\end{equation}

The combination $x_s-3 \, y_s$ of the $SU(3)$ singlet matrix elements cancels in
the ratio of the shifts for $b$ hyperons and the charmed ones:
\begin{equation}
\delta^{(3)} {\overline \Gamma}_b= {|V_{cb}|^2 \over \cos^4 \theta} \,
{m_b^2 \over m_c^2} \,
{({\tilde C}_+ - {\tilde C}_-)^2 \over C_+^2 + C_-^2 } \,  \left [
{\alpha_s (m_c) \over \alpha_s(m_b)} \right ]^{5/18} \,
\delta_{nl}^{(3,0)} {\overline \Gamma}_c \approx 0.0025 \,
\delta_{nl}^{(3,0)} {\overline \Gamma}_c\,.
\label{rbc}
\end{equation}
(One can observe, with satisfaction, that the dependence on the unphysical
parameter $\mu$ cancels out, as it should.)  This equation shows that relatively
to the charmed baryons the shift of the decay rates in the $b$ baryon triplet is
greatly suppressed by the ratio $({\tilde C}_+ - {\tilde C}_-)^2 / (C_+^2 +
C_-^2)$, which parametrically is of the second order in $\alpha_s$, and
numerically is only about 0.12.

An estimate of $\delta^{(3)} {\overline \Gamma}_b$ from Eq.~(\ref{rbc}) in
absolute terms depends on evaluating the average shift $\delta_{nl}^{(3,0)}
{\overline \Gamma}_c$ for charmed baryons. The latter shift can be
conservatively bounded from above by the average total decay rate of those
baryons: $\delta_{nl}^{(3,0)} {\overline \Gamma}_c < {\overline \Gamma_c} = 6.0
\pm 0.7 \, ps^{-1}$, which then yields, using Eq.~(\ref{rbc}), an upper bound
$\delta^{(3)} {\overline \Gamma}_b < 0.015 \pm 0.002 \, ps^{-1}$. More
realistically, one should subtract from the total average width ${\overline
  \Gamma_c}$ the contribution of the `parton' term, which can be estimated from
the decay rate of $D_0$ with account of the $O(m_c^{-2})$ effects, as amounting
to about $3 \, ps^{-1}$. (One should also take into account the semileptonic
contribution to the total decay rates, which however is quite small at this
level of accuracy). Thus a realistic evaluation of $\delta^{(3)} {\overline
  \Gamma}_b$ does not exceed $0.01 \, ps^{-1}$, which constitutes only about 1\%
of the total decay rate of $\Lambda_b$.  Thus the shift of the total decay rate
of $\Lambda_b$ due to the effects of $L_{eff}^{(3)}$ is dominantly associated
with the $SU(3)$ non-singlet difference (\ref{dbres}). The shift of the
$\Lambda_b$ decay rate with respect to the average width ${\overline \Gamma}_b$
due to the non-singlet operators is one third of the splitting (\ref{dbres}),
i.e., about 5\%. Adding to this the 1\% shift of the average width and another
1\% difference from the meson decays due to the suppression of the latter by the
$m_b^{-2}$ chromomagnetic effects, one concludes that at the present level of
theoretical understanding it looks impossible to explain a more than 10\%
enhancement of the total decay rate of $\Lambda_b$ relative to $B_d$, where an
ample 3\% margin is added for the uncertainties of higher order terms in OPE as
well as for higher order QCD radiative effects in the discussed corrections. In
other words, the expected pattern of the lifetimes of the $b$ hyperons in the
triplet, relative to $B_d$, is
\begin{equation}
\tau(\Xi_b^0) \approx \tau(\Lambda_b) < \tau(B_d) < \tau(\Xi_b^-)\,,
\label{pattern}
\end{equation}
with the ``best" theoretical estimate of the differences to be about 7\% for
each step of the inequality.

For the double strange hyperons $\Omega_c$ and $\Omega_b$ there is presently no
better approach to evaluating the four-quark matrix elements, than the use of
simplistic relations, like (\ref{omegaold}) based on constituent quark model.
Such relations imply that the effects of the strange quark, WS and PI, in the
$\Omega_Q$ baryons are significantly enhanced over the same effects in the
cascade hyperons. In charmed baryons a presence of strange spectator quark
enhances the decay through positive interference with the quark emerging from
the $c \to s$ transition in the decay. For $\Omega_c$ this implies a significant
enhancement of the total decay rate \cite{sv:86}, which is in perfect agreement
with the data on the $\Omega_c$ lifetime. Also a similar enhancement is expected
for the semileptonic decay rate of $\Omega_c$.  In $b$ baryons, on the
contrary, the interference effect for a spectator strange quark is negative.
Thus the nonleptonic decay rate of $\Omega_b$ is expected to be suppressed,
leaving $\Omega_b$ most probably the longest-living particle among the $b$
baryons.

\subsection{Relation between spectator effects in baryons and \\ the decays
  $\Xi_Q \to \Lambda_Q \, \pi$}

Rather unexpectedly, the problem of four-quark matrix elements over heavy
hyperons is related to decays of the type $\Xi_Q \to \Lambda_Q \, \pi$.  The
mass difference between the charmed cascade hyperons $\Xi_c$ and $\Lambda_c$ is
about 180 MeV. The expected analogous mass splitting for the $b$ hyperons should
be very close to this number, since in the heavy quark limit
\begin{equation}
M(\Xi_b)-M(\Lambda_b)=M(\Xi_c)-M(\Lambda_c) + O(m_c^{-2}-m_b^{-2}).
\label{massdiff}
\end{equation}
Therefore in both cases are kinematically possible decays of the type $\Xi_Q \to
\Lambda_Q \, \pi$, in which the heavy quark is not destroyed, and which are
quite similar to decays of ordinary `light' hyperons.  Surprisingly, the rate of
these decays for both $\Xi_c$ and $\Xi_b$ is not insignificantly small, but
rather their branching fraction can reach a level of few per mill for $\Xi_c$
and of one percent or more for $\Xi_b$ \cite{mv:00}.

The transitions $\Xi_Q \to \Lambda_Q \, \pi$ are induced by two underlying weak
processes: the `spectator' decay of a strange quark, $s \to u \, \overline u \,
d$, which does not involve the heavy quark, and the `non-spectator' weak
scattering (WS)
\begin{equation}
s \, c \to c \, d
\label{ws}
\end{equation}
trough the weak interaction of the $c \to d$ and $s \to c$ currents.  One can
also readily see that the WS mechanism contributes only to the decays $\Xi_c \to
\Lambda_c \, \pi$ and is not present in the decays of the $b$ cascade hyperons.
An important starting point in considering these transitions is that in the
heavy quark limit the spin of the heavy quark completely decouples from the spin
of the light component of the baryon, and that the latter light component in
both the initial an the final baryons forms a $J^P=0^+$ state with quantum
numbers of a diquark.  Since the momentum transfer in the considered decays is
small in comparison with the mass of the heavy quark the spin of the amplitudes
with spin flip of the heavy quark, and thus of the baryon, are suppressed by
$m_Q^{-1}$. In terms of the two possible partial waves in the decay $\Xi_Q \to
\Lambda_Q \, \pi$, the $S$ and $P$, this implies that the $P$ wave is strongly
suppressed and the decays are dominated by the $S$ wave.

According to the well known current algebra technique, the $S$ wave amplitudes
of pion emission can be considered in the chiral limit at zero four-momentum of
the pion, where they are described by the PCAC reduction formula (pole terms are
absent in these processes):
\begin{equation}
\langle \Lambda_Q \, \pi_i (p=0) \,| H_W |\, \Xi_Q \rangle = {\sqrt{2}
\over f_\pi} \, \langle \Lambda_Q \, |\left[Q^5_i, \, H_W \right ] |\,
\Xi_Q \rangle\,,
\label{pcac}
\end{equation}
where $\pi_i$ is the pion triplet in the Cartesian notation, and $Q^5_i$ is the
corresponding isotopic triplet of axial charges. The constant $f_\pi \approx 130
\, MeV$, normalized by the charged pion decay, is used here, hence the
coefficient $\sqrt{2}$ in Eq.~(\ref{pcac}). The Hamiltonian $H_W$ in
Eq.~(\ref{pcac}) is the nonleptonic strangeness-changing Hamiltonian:
\begin{eqnarray}
H_W = &&\sqrt{2} \, G_F \cos \theta_c \sin \theta_c \left \{ \left
( C_+ + C_-
\right ) \left [ (\overline u_L \, \gamma_\mu \, s_L)\, (\overline
d_L \, \gamma_\mu \, u_L) - (\overline c_L \, \gamma_\mu \, s_L)\,
(\overline d_L \, \gamma_\mu \, c_L) \right ] \right . \nonumber \\
&&{} + \left. \left ( C_+ - C_- \right ) \left [ (\overline d_L \,
\gamma_\mu \, s_L)\, (\overline u_L \, \gamma_\mu \, u_L) - (\overline
d_L \, \gamma_\mu \, s_L)\, (\overline c_L \, \gamma_\mu \, c_L) \right
] \right \}\,.
\label{hw}
\end{eqnarray}
In this formula the weak Hamiltonian is assumed to be normalized (in LLO) at
$\mu = m_c$.  The terms in the Hamiltonian (\ref{hw}) without the charmed quark
fields describe the `spectator' nonleptonic decay of the strange quark, while
those with the $c$ quark correspond to the WS process (\ref{ws}).

It is straightforward to see from Eq.~(\ref{pcac}) that in the PCAC limit the
discussed decays should obey the $\Delta I =1/2$ rule. Indeed, the commutator of
the weak Hamiltonian with the axial charges transforms under the isotopic SU(2)
in the same way as the Hamiltonian itself. In other words, the $\Delta I=1/2$
part of $H_W$ after the commutation gives an $\Delta I=1/2$ operator, while the
$\Delta I = 3/2$ part after the commutation gives an $\Delta I = 3/2$ operator.
The latter operator however cannot have a non vanishing matrix element between
an isotopic singlet, $\Lambda_Q$, and an isotopic doublet, $\Xi_Q$. Thus the
$\Delta I=3/2$ part of $H_W$ gives no contribution to the $S$ wave amplitudes in
the PCAC limit.

Once the isotopic properties of the decay amplitudes are fixed, one can
concentrate on specific charge decay channels, e.g., $\Xi_b^- \to \Lambda_b \,
\pi^-$ and $\Xi_c^0 \to \Lambda_c \, \pi^-$. An application of the PCAC relation
(\ref{pcac}) with the Hamiltonian from Eq.~(\ref{hw}) to these decays, gives the
expressions for the amplitudes at $p=0$ in terms of baryonic matrix elements of
four-quark operators:
\begin{eqnarray}
&&\hspace*{-.5cm}{} \langle \Lambda_b \, \pi^- (p=0) \,| H_W |\, \Xi_b^- \rangle \\
&=& {\sqrt{2} \over f_\pi} \, G_F \cos \theta_c \sin \theta_c \,
\langle \Lambda_b
\,| \left ( C_+ + C_- \right ) \left [ (\overline u_L \, \gamma_\mu
\, s_L)\, (\overline d_L \, \gamma_\mu \, d_L) - (\overline u_L \,
\gamma_\mu \, s_L)\, (\overline u_L \, \gamma_\mu \, u_L) \right ]
\nonumber \\
&&{} + \left ( C_+ - C_- \right ) \left [ (\overline d_L \, \gamma_\mu \,
s_L)\, (\overline u_L \, \gamma_\mu \, d_L) - (\overline u_L \,
\gamma_\mu \, s_L)\, (\overline u_L \, \gamma_\mu \, u_L) \right] | \,
\Xi_b^- \rangle \nonumber \\
&=& {\sqrt{2} \over f_\pi} \, G_F \cos \theta_c \sin \theta_c \,
\langle \Lambda_b
\,|\, C_- \left [ (\overline u_L \, \gamma_\mu \, s_L)\, (\overline d_L
\, \gamma_\mu \, d_L) - (\overline d_L \, \gamma_\mu \, s_L)\,
(\overline u_L \, \gamma_\mu \, d_L) \right ] \nonumber \\
&&{} - {C_+ \over 3} \left [ (\overline u_L \, \gamma_\mu \, s_L)\,
(\overline d_L \, \gamma_\mu \, d_L) + (\overline d_L \, \gamma_\mu \,
s_L)\, (\overline u_L \, \gamma_\mu \, d_L) +2 \, (\overline u_L \,
\gamma_\mu \, s_L)\, (\overline u_L \, \gamma_\mu \, u_L) \right ] | \,
\Xi_b^- \rangle\,, \nonumber
\label{xib}
\end{eqnarray}
where in the last transition the operator structure with $\Delta I =3/2$ giving
a vanishing contribution is removed and only the structures with explicitly
$\Delta I =1/2$ are retained, and
\begin{eqnarray}
&& \langle \Lambda_c \, \pi^- (p=0) \,| H_W |\, \Xi_c^0 \rangle = \langle
\Lambda_b \, \pi^- (p=0) \,| H_W |\, \Xi_b^- \rangle + 
{\sqrt{2} \over f_\pi} \, G_F \cos \theta_c \sin \theta_c \\
&&{}\qquad \times\langle \Lambda_c \,
| \left ( C_+ + C_- \right ) (\overline c_L \, \gamma_\mu \, s_L)\,
(\overline u_L \, \gamma_\mu \, c_L) + 
\left ( C_+ - C_- \right ) (\overline u_L \, \gamma_\mu \, s_L)\,
(\overline c_L \, \gamma_\mu \, c_L) |\, \Xi_c^0 \rangle\,.\nonumber 
\label{xic}
\end{eqnarray}
In the latter formula the first term on the r.h.s. expresses the fact that in
the heavy quark limit the `spectator' amplitudes do not depend on the flavor or
the mass of the heavy quark. The rest of the expression (\ref{xic}) describes
the `non-spectator' contribution to the amplitude of the charmed hyperon decay.
Using the flavor $SU(3)$ symmetry the latter contribution can be related to the
non-singlet matrix elements (\ref{msh:defxy}) (normalized at $\mu=m_c$) as
\begin{eqnarray}
\Delta A &\equiv& \langle \Lambda_c \, \pi^- (p=0) \,| H_W |\, \Xi_c^0
\rangle - \langle \Lambda_b \, \pi^- (p=0) \,| H_W |\, \Xi_b^- \rangle
\nonumber \\
&=& {G_F \cos \theta_c \sin \theta_c \over 2 \, \sqrt{2} \, f_\pi} \,
\left[ \left ( C_- -
C_+ \right ) \, x - \left ( C_+ + C_- \right ) \, y \right ]\,.
\label{das}
\end{eqnarray}
Furthermore, with the help of the equations (\ref{d1}) and (\ref{d2}) relating
the matrix elements $x$ and $y$ to the differences of the total decay widths
within the triplet of charmed hyperons, one can eliminate $x$ and $y$ in favor
of the measured width differences. The resulting expression has the form
\begin{eqnarray}
\Delta A &\approx& - {\sqrt{2} \, \pi \cos \theta_c \sin \theta_c  \over 
G_F \, m_c^2 \, f_\pi} \left [ 0.45 \left (
\Gamma(\Xi_c^0)-\Gamma(\Lambda_c) \right ) + 0.04 \left (
\Gamma(\Lambda_c)-\Gamma(\Xi_c^+) \right ) \right ] \nonumber \\
&=& -10^{-7} \left [
0.97 \left ( \Gamma(\Xi_c^0)-\Gamma(\Lambda_c) \right ) + 0.09 \left (
\Gamma(\Lambda_c)-\Gamma(\Xi_c^+) \right ) \right ] \left ( {1.4 \,
GeV \over m_c} \right )^2 \, ps \,,
\label{dasm}
\end{eqnarray}
where, clearly, in the latter form the widths are assumed to be expressed in
$ps^{-1}$, and $m_c=1.4 \, GeV$ is used as a `reference' value for the charmed
quark mass. It is seen from Eq.~(\ref{dasm}) that the evaluation of the
difference of the amplitudes within the discussed approach is mostly sensitive
to the difference of the decay rates of $\Xi_c^0$ and $\Lambda_c$, with only
very little sensitivity to the total decay width of $\Xi_c^+$. Using the current
data the difference $\Delta A$ is estimated as
\begin{equation}
  \Delta A= -(5.4 \pm 2) \times 10^{-7}\,,
\label{dasn}
\end{equation}
with the uncertainty being dominated by the experimental error in the
lifetime of $\Xi_c^0$. An  amplitude $A$ of the magnitude,
given by the central value in Eq.~(\ref{dasn}) would produce a decay rate
$\Gamma (\Xi_Q \to \Lambda_Q \, \pi) = |A|^2 \, p_\pi/(2 \pi) \approx
0.9 \times 10^{10} \, s^{-1}$, which result can also be written in a
form of triangle inequality
\begin{equation}
\sqrt{\Gamma(\Xi_b^- \to \Lambda_b \, \pi^-)} + \sqrt{\Gamma(\Xi_c^0 \to
\Lambda_c \, \pi^-)} \ge \sqrt{0.9 \times 10^{10} \, s^{-1}}\,.
\label{tri}
\end{equation}

Although at present it is not possible to evaluate in a reasonably model
independent way the matrix element in Eq.~(\ref{xib}) for the `spectator' decay
amplitude, the inequality (\ref{tri}) shows that at least some of the discussed
pion transitions should go at the level of $0.01 \, ps^{-1}$, similar to the
rates of analogous decays of `light' hyperons.

\subsection{Summary on predictions for lifetimes}

We summarize here specific predictions for the inclusive decay rates, which can
be argued with a certain degree of theoretical reliability, and which can be
possibly experimentally tested in the nearest future.
\index{inclusive weak decay rates!theoretical predictions summary}%

$B$ mesons:
\begin{equation}
\tau(B_d)/\tau(B_s)=1 \pm 0.01\,.
\end{equation}

Charmed hyperons:
\begin{eqnarray}
& \Gamma_{sl}(\Xi_c) = (2 - 3) \, \Gamma_{sl}(\Lambda_c)\, \qquad
\Gamma_{sl}(\Omega_c) > \Gamma_{sl}(\Xi_c)\,,& \nonumber\\[4pt]
& \Gamma^{nl}_{\Delta S =
-1}(\Xi_c^+) \approx \Gamma^{nl}_{\Delta S = -1}(\Lambda_c) \,,& \\[4pt]
&\Gamma^{nl}_{\Delta S =
-1}(\Xi_c^0)-\Gamma^{nl}_{\Delta S = -1}(\Lambda_c) \approx 0.55 \pm
0.22 \, ps^{-1}\,.& \nonumber
\end{eqnarray}

$b$ hyperons:
\begin{eqnarray}
& \tau(\Xi_b^0) \approx \tau(\Lambda_b) < \tau(B_d) <
\tau(\Xi_b^-)<\tau(\Omega_b)\,, &  \nonumber\\[4pt]
& \Gamma(\Lambda_b) - \Gamma(\Xi_b^-) \approx 0.11 \pm 0.03 \,
ps^{-1}\,, & \\[4pt]
& 0.9 < \displaystyle {\tau(\Lambda_b) \over \tau(B_d)} < 1\,.&  \nonumber
\end{eqnarray}

Strangeness decays $\Xi_Q \to \Lambda_Q \, \pi$:
The $\Delta I =1/2$ rule should hold in these decays, so that
$\Gamma(\Xi_Q^{(d)} \to \Lambda_Q \pi^-)  = 2 \, \Gamma(\Xi_Q^{(u)} \to
\Lambda_Q \pi^0)$. The rates are constrained by the triangle inequality
(\ref{tri}).

\boldmath
\section[Theory of \bbm]{Theory of \bbm$\!$ 
\authorfootnote{Author: Ulrich~Nierste}}
\unboldmath
\label{chmix:sec:thbmix}

In Sect.~\ref{subs:mix} the time evolution of the $B^0\!-\!\ov{B}{}^0\,$ system
has been discussed.  \bbm\ involves three physical, rephasing-invariant
quantities: $|M_{12}|$, $|\Gamma_{12}|$ and the phase $\phi$ defined in
\eq{defphi}.  In the following we will discuss how they are related to physical
observables. The discussed quantities are summarized in
Table~\ref{chmix:tab:obs}.  \index{B mixing@$B$ mixing}

\subsection{Mass difference}
\label{chmix:sub:mass}

\index{B meson@$B$ meson!mass difference \dm}\index{mass difference \dm} The
mass difference \dm\ can be measured from the tagged time evolutions in
(\ref{gtfres}-\ref{gbtfbres}) from any decay $B^0\to f$, unless $\lambda_f=\pm
1$, in which case the oscillatory terms vanish.  The time evolution is
especially simple for flavor-specific decays, which are characterized by
$\lambda_f=0$. The corresponding formulae can be found in \eq{gtfs} and
\eq{defa0}. Interesting flavor-specific \index{flavor-specific!decay modes}
modes are tabulated in Table~\ref{chmix:tab:fsd}. Time integrated measurements
determine $x_q=\dm_q/\Gamma_q=\dm_q \tau_{B_q}$, $q=d,s$, defined in \eq{defxy}.
While unfortunately it is common practice to quote measurements of $\dm_q$ in
terms of $x_q$, it should be clear that the measured oscillation frequencies
determine $\dm_d$ and $\dm_s$ and not $x_d$ and $x_s$. Fundamental physics
quantities like CKM elements are related to $\dm_d$ and $\dm_s$, so that the
errors of the lifetimes entering $x_q$ are irrelevant.

\index{B mixing@$B$ mixing!$B_d$ mixing} In order to predict the mass
difference $\dm_q$ within the Standard Model or one of its extensions, one must
first calculate the \dbt\ transition amplitude, which triggers \bbm. The lowest
order contribution to this amplitude in the Standard Model is the box diagram in
Fig.~\ref{ch1:fig:box}. Then one must match the result to an effective field
theory, in which the interactions mediated by heavy particles are described by
local operators represented by pointlike vertices. In the Standard Model only
one operator, $Q$ in \eq{ch1:uli:defq}, emerges.\index{operator!$Q$} This
procedure separates short- and long-distance physics and is described in
Sect.~\ref{1:Heff}. It results in the effective Hamiltonian in \eq{ch1:uli:h2}.
The interesting short-distance physics is contained in the Wilson coefficient
$C$ in \eq{ch1:uli:c2}. New physics can modify $C$ and can introduce new
operators in addition to $Q$ in \eq{ch1:uli:defq}.  \index{Wilson coefficient}
The Standard Model prediction is readily obtained from \eq{ch1:uli:h2} to
\eq{defbb}:
\begin{equation}
\dm_q = 2 |M_{12}^q| \, = \, 
   \frac{|\bra{B_q^0} H^{|\Delta B|=2}  \ket{\ov{B}{}^0_q}| }{m_{B_q}}  
=  \frac{G_F^2}{6 \pi^2}\, \eta_B\, m_{B_q} \, \widehat{B}_{B_q} f_{B_q}^2 
    M_W^2\, S \bigg( \frac{m_t^2}{M_W^2} \bigg) 
    \left| V_{tb} V_{tq}^* \right|^2 . 
    \label{chmix:dmb}
\end{equation}
where $q=s$ or $d$.
\index{B meson@$B$ meson!mass difference \dm}\index{mass difference \dm}%

\begin{table}
\begin{center}
\begin{tabular}{l@{~~~~}l@{~~~~}l@{~~~~}l} \hline
observable & defined in & 
                          \multicolumn{2}{l}{SM prediction for } \\
& & $B_d$ in & $B_s$ in 
                          \\\hline &&&\\[-3mm]
$\dm \simeq 2 |M_{12}| $ & \eq{mgsol:a} & 
 \eq{chmix:dmb},\eq{chmix:dmbrat} & 
 \eq{chmix:dmb},\eq{chmix:dmsnum},\eq{chmix:dmbrat} \\
$\phi$ & \eq{defphi} & \eq{chmix:phinum} & \eq{chmix:phinum} \\
$\dg \simeq 2 |\Gamma_{12}| \cos\phi $ & 
  \eq{mgsol:b} &
  \eq{chmix:dgdm},\eq{chmix:dgdsm}   & 
   \eq{chmix:dgnum2},\eq{chmix:dgdm}  \\
$\dg_{\rm CP} \simeq 2 |\Gamma_{12}|   $ & \eq{chmix:dgcp}
        &   &  \\
$\dg_{\rm CP}^\prime \simeq 2 |\Gamma_{12}| \cos^2\phi$ 
  & \eq{chmix:dgp}  & &  \\
$\dg_{\rm CP}^{cc} =  2 |\xi_c^{d\,2}\Gamma_{12}^{cc}|$ & 
    \eq{chmix:dgcp:d} & \eq{chmix:dgcc}   &  \\
$\ega =  \lt| \frac{\Gamma_{12}}{M_{12}} \rt| \sin \phi  $ & 
   \eq{defepsg} & \eq{chmix:aqnum} & \eq{chmix:aqnum}  \\[4pt] \hline
\end{tabular}
\end{center}
\caption{Observables related to $|M_{12}|$, $|\Gamma_{12}|$ and $\phi$
          discussed in Sect.~\ref{chmix:sec:thbmix}.}
\label{chmix:tab:obs}
\end{table}

Next we discuss the phenomenology of $\dm_d$ in the Standard Model.  We first
insert the numerical values of those quantities which are well-known into
\eq{chmix:dmb}. The QCD factor $\eta_B=0.55$ \cite{bjw} corresponds to the
$\ov{\rm MS}$ scheme for $m_t$. The $\ov{\rm MS}$ value of $m_t=167\,$GeV is
numerically smaller than the pole mass measured at the Tevatron by roughly 7
GeV. Solving
\eq{chmix:dmb} for $|V_{td}|$ one finds:%
\index{B meson@$B$ meson!mass difference \dm!$\dm_d$}%
\index{mass difference \dm!$\dm_d$}
\index{unitarity triangle!and \bbmd}%
\index{Cabibbo-Kobayashi-Maskawa matrix!$V_{td}$}%
\begin{equation}
|V_{td}| = 0.0078 \, \sqrt{\frac{\dm_d}{0.49\, \mbox{ps}^{-1}}} \, 
             \frac{200 \, \mbox{MeV}}{f_{B_d}}\,
             \sqrt{\frac{1.3}{\widehat{B}_{B_d}}} \,.
\end{equation}
The relation of $|V_{td}|$ to the improved Wolfenstein parameters is%
\index{Wolfenstein parameters!for $V_{td}$}%
\begin{equation}
|V_{td}| = A \lambda^3 R_t \lt( 1+ {\cal O} (\lambda^4) \rt)
\, = \, |V_{cb}| \, \lambda \,R_t \lt( 1+ {\cal O} (\lambda^4) \rt)
\end{equation}
and 
\begin{equation}
R_t = \sqrt{ (1- \ov{\rho})^2 + \ov{\eta}^2} \label{chmix:defrt}
\end{equation}
is the length of one side of the unitarity triangle. Hence the measurement of
$\dm_d$ defines a circle in the $(\ov{\rho},\ov{\eta})$ plane centered around
$(1,0)$. Yet the hadronic uncertainties associated with
$f_{B_d}\sqrt{\widehat{B}_{B_d}}$ obscure a clean extraction of $|V_{td}|$ and
$R_t$ from the well-measured $\dm_d$.  The summer 2000 world averages from
lattice calculations are $f_{B_d} = (200 \pm 30)\,$MeV and
$\widehat{B}_{B_d}=1.30 \pm 0.17$ \cite{hash}. This gives $|V_{td}|=0.0078 \pm
0.0013$ and, with $|V_{cb}|= (40.4 \pm 1.8)\times 10^{-3}$, $R_t= 0.88 \pm 0.15$.
\index{hadronic parameter!$f_{B_d}$} \index{hadronic parameter!$\widehat{B}_B$}

\index{B mixing@$B$ mixing!$B_s$ mixing} For the discussion of $\dm_s$ we
first note that the corresponding CKM
element in \eq{chmix:dmb} is fixed from CKM unitarity:%
\index{Cabibbo-Kobayashi-Maskawa matrix!$V_{ts}$}%
\index{Wolfenstein parameters!for $V_{ts}$}%
\index{B meson@$B$ meson!mass difference \dm!$\dm_s$}%
\index{mass difference \dm!$\dm_s$}
\begin{equation}
|V_{ts}| = |V_{cb}| \lt[ 1 + \lambda^2 \lt( \ov{\rho} - \frac{1}{2} \rt) 
               + {\cal O} (\lambda^4) \rt] .
\end{equation}
$|V_{ts}|$ is smaller than $|V_{cb}|$ by roughly 1\%. Hence within the Standard
Model a measurement of $\dm_s$ directly probes the calculation of the hadronic
matrix element. \eq{chmix:dmb} specifies to
\begin{equation}
\dm_s = 17.2 \, \mbox{ps}^{-1} \lt( \frac{|V_{ts}|}{0.04} \, 
          \frac{f_{B_s}}{230 \, \mbox{MeV}} \rt)^2 
             \frac{\widehat{B}_{B_s}}{1.3}
             \label{chmix:dmsnum} .
\end{equation}
This can be rewritten as 
\begin{equation}
f_{B_s}\sqrt{\widehat{B}_{B_s}} =
     \sqrt{\frac{\dm_s}{14.9\, \mbox{ps}^{-1}}} \frac{0.04}{|V_{cb}|} 
     \, 247\, \mbox{MeV} . \label{chmix:dmsbd}
\end{equation}
\index{hadronic parameter!$f_{B_s}$}%
The present 95\% C.L.\ limit of
$\dm_s\geq 14.9\,$ps$^{-1}$ \cite{os} implies a lower bound on
$f_{B_s}\sqrt{\widehat{B}_{B_s}}$ which is only marginally consistent
with some of the quenched lattice calculations. Hence global fits of
the unitarity triangle (using $\dm_d/\dm_s $ to constrain $R_t$) which
use too small values of $f_{B_s}\sqrt{\widehat{B}_{B_s}}$ confine the
apex of the triangle to a too small region of the
$(\ov{\rho},\ov{\eta})$ plane or may even be in conflict with the
measured lower bound on $\dm_s$.

The determination of $R_t$ profits enormously from a measurement of $\dm_s$,
because the ratio of the hadronic matrix elements entering $\dm_d/\dm_s$ can be
calculated with a much higher accuracy than the individual matrix elements:
\begin{equation}  
\xi = \frac{f_{B_s}\sqrt{\widehat{B}_{B_s}}}{f_{B_d}\sqrt{\widehat{B}_{B_d}}}
\label{chmix:defxi}
\end{equation} 
\index{hadronic parameter!and $SU(3)_F$ symmetry}%
\index{hadronic parameter!$\xi$}%
is equal to 1 in the limit of exact $SU(3)_F$ symmetry. Hence the
theorists' task is reduced to the calculation of the deviation from 1. The
current world average from lattice calculations is \cite{hash}
\begin{equation}
\xi = 1.16 \pm 0.05 \label{chmix:xilatt} .
\end{equation}
Further $|V_{cb}|$ drops out from the ratio 
\begin{equation} 
 \frac{\dm_d}{\dm_s} = \lambda^2 R_t^2 \lt( 1 +  
 \lambda^2 (1 - 2 \ov{\rho} ) + {\cal O} (\lambda^4)  \rt) \, 
 \frac{m_{B_d}}{m_{B_s}}\, \frac{1}{\xi^2}  . \label{chmix:dmbrat}
\end{equation} 

With the expected experimental accuracy of $\dm_{d,s}$ and the anticipation of
progress in the determination of $\xi$ in \eq{chmix:xilatt} a determination of
$R_t$ at the level of 1-3\% is possible.  Then, eventually, even the uncertainty
in $\lambda$ cannot be neglected anymore. Keeping the overall factor of
$\lambda^2$ while inserting $\lambda=0.22$ in the subleading terms one finds
from \eq{chmix:dmbrat}:
\begin{equation} 
R_t = 0.880 \, \sqrt{\frac{\dm_d}{0.49 \,\mbox{ps}^{-1}}} \, 
        \sqrt{\frac{17 \,\mbox{ps}^{-1} }{\dm_s}} \,
        \frac{0.22}{\lambda} \,
        \frac{\xi}{1.16} \, 
        \lt( 1 \, + \,  0.05 \, \ov{\rho} \rt). \label{chmix:rtapp}  
\end{equation} 
Here the omission of ${\cal O}(\lambda^4)$ terms induces a negligible error of
less than 0.1\%. At present $R_t$ is obtained from a global fit of the unitarity
triangle and \eq{chmix:dmbrat} is used to predict
$\dm_s = 17.3\epm{1.5}{0.7}$ \cite{cckm}.%
\index{B meson@$B$ meson!mass difference \dm!SM prediction for $\dm_s$}%
\index{mass difference \dm!$\dm_s$!SM prediction} One should be aware that some
of the quantities entering the global fit, especially $\epsilon_K$, are
sensitive to new physics. Hence the measurement of a $\dm_s$ well above the
quoted range would be very exciting. In a large class of extensions of the
Standard model $\dm_d$ and $\dm_s$ change, while their ratio does not. In these
models $\dm_s$ could be in conflict with \eq{chmix:dmsbd} without affecting
$R_t$ in \eq{chmix:rtapp}.  Therefore it is desirable to gain additional
experimental information on $f_{B_s}$, so that the dependence on
non-perturbative methods is reduced. This information can be obtained from the
$B_s$ width difference discussed in Sect.~\ref{chmix:sub:width}.

\subsection{Width difference}
\label{chmix:sub:width}

\subsubsection{Calculation}\label{chmix:subsub:width:calc}

\index{B meson@$B$ meson!width difference $\dg$}%
\index{width difference $\dg$} The two mass eigenstates $B_L$ and $B_H$ in
\eq{defpq} differ not only in their masses but also in their widths. The
prediction of the width difference $\dg = \Gamma_L - \Gamma_H \simeq 2\,
|\Gamma_{12}| \cos \phi $ in \eq{mgsol:b} requires the calculation of
$|\Gamma_{12}|$ and $\phi$. $\Gamma_{12}$ is determined from the absorptive part
of the \dbt\ transition amplitude. It receives contributions from all final
states which are common to $B^0$ and $\ov{B}{}^0$ as shown in the first line of
\eq{1:delgam}. The leading order (LO) diagrams contributing to $\Gamma_{12}$ in
the $B_s$ system are shown in Fig.~\ref{chmix:fig:dga}.
\begin{figure}[tb]
  \centerline{\epsfxsize=0.85\textwidth \epsffile{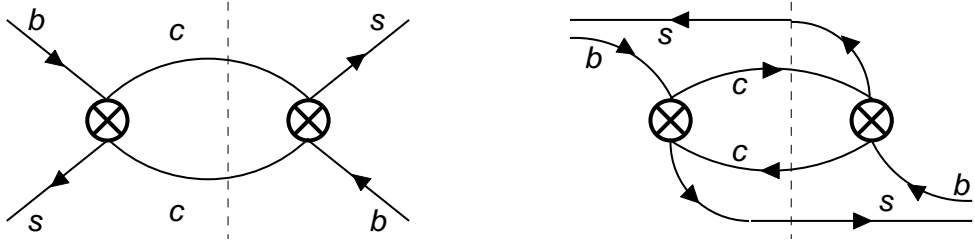}}
  \vspace*{6pt}
\caption[Leading order diagrams determining $\Gamma_{12}^s$. Only the 
CKM-favored contribution $\xi_c^{s*2} \Gamma_{12}^{cc}$ is shown.]{Leading
order diagrams determining $\Gamma_{12}^s$. Only the  CKM-favored contribution
$\xi_c^{s*2} \Gamma_{12}^{cc}$ is shown.  The left diagram is the \emph{weak
annihilation diagram} and the right one is the \emph{spectator interference
diagram}. The dashed line indicates the cut through the final state.}
\label{chmix:fig:dga}
\end{figure}
The dominant contribution comes from the spectator interference diagram, the
weak annihilation diagram is color-suppressed.  We write
\begin{equation}
\Gamma_{12}^q \, = \, -  \lt[  
\xi_c^{q*2} \Gamma_{12,q}^{cc} \, + \, 
               2 \xi_u^{q*} \xi_c^{q*} 
               \Gamma_{12,q}^{uc} \, +  \, \xi_u^{q*2} \Gamma_{12,q}^{uu}
               \rt] , 
               \qquad q=d,s, 
\label{chmix:ga12dec} 
\end{equation}
where the three terms denote the contributions from the diagram with
(anti-)quarks $i$ and $j$ in the final state.
The $\xi_i^{q}$'s denote the corresponding CKM factors:
\begin{equation}  
\xi_u^{q} \,=\, V_{uq} V_{ub}^*, \qquad  
\xi_c^{q} \,=\, V_{cq} V_{cb}^*, \qquad  
\xi_t^{q} \,=\, V_{tq} V_{tb}^*. \label{chmix:xiq}   
\end{equation}  
They satisfy $\xi_u^{q}+\xi_c^{q}+\xi_t^{q}=0$ from CKM unitarity and
read in terms of the improved Wolfenstein parameters \cite{wblo}:
\index{Wolfenstein parameters!for $V_{ud}V_{ub}^*$}
\index{Wolfenstein parameters!for $V_{us}V_{ub}^*$}
\index{Wolfenstein parameters!for $V_{cd}V_{cb}^*$}
\index{Wolfenstein parameters!for $V_{cs}V_{cb}^*$}
\index{Wolfenstein parameters!for $V_{td}V_{tb}^*$}
\index{Wolfenstein parameters!for $V_{ts}V_{tb}^*$}
\begin{eqnarray}
\xi_u^{d} \, = \, A \lambda^3 \lt( \ov{\rho}+ i \ov{\eta} \rt) + 
              {\cal O} (\lambda^5), && \qquad
\xi_c^{d} \, = \,  - A \lambda^3 +{\cal O} (\lambda^5), \nn 
&& \qquad
   \xi_t^{d} \, = \,  A \lambda^3 \lt( 1 - \ov{\rho}- i \ov{\eta} \rt)
           +{\cal O} (\lambda^5), \no \\[4pt] 
\xi_u^{s} \, = \, A \lambda^4 ( 1+\frac{\lambda^2}{2} ) 
              \lt( \ov{\rho}+ i \ov{\eta} \rt) + 
              {\cal O} (\lambda^8), 
&& \qquad
\xi_c^{s} \, = \, A \lambda^2 ( 1 - \frac{\lambda^2}{2} ) 
               +{\cal O} (\lambda^6), \nn 
&& \qquad
   \xi_t^{s} =   -A \lambda^2 \lt[ 1- \lambda^2 ( \frac{1}{2}
              - \ov{\rho} - i \ov{\eta}   )   \rt] 
              +{\cal O} (\lambda^6),  \label{chmix:xis}
\end{eqnarray}
$-\xi_u^{d}/\xi_c^{d}$ and $-\xi_t^{d}/\xi_c^{d}$ define two sides
of the standard unitarity triangle depicted in Fig.~\ref{1:fig:UT}. One has 
\begin{equation}
-\frac{\xi_u^{d}}{\xi_c^{d}} \,=\, R_b e^{i \gamma},\qquad\quad
-\frac{\xi_t^{d}}{\xi_c^{d}} \, = \, R_t e^{-i\beta},
\end{equation}
\index{unitarity triangle!angle $\gamma$}%
\index{unitarity triangle!angle $\beta$}%
\index{unitarity triangle!angle $\beta_s$}%
where $R_b=\sqrt{\ov{\rho}^2+\ov{\eta}^2}$ and $R_t$ is defined in
\eq{chmix:defrt}. The ratio $\xi_t^{s}/\xi_c^{s}$ involves the phase $\beta_s$
measured from the mixing-induced $CP$ asymmetry in $B_s \to D_s^+ D_s^-$:
\index{decay!$B_s \rightarrow D_s^+ D_s^-$}
\begin{equation}
-\frac{\xi_t^{s}}{\xi_c^{s}} \, = \, [1+ \ov{\rho} \lambda^2 
                     +{\cal O}(\lambda^4)] \, 
 e^{i\beta_s}\qquad \mbox{with }\ 
 \beta_s =
  \ov{\eta}\lambda^2 [1+(1-\ov{\rho}) \lambda^2] +{\cal O} (\lambda^6) 
 . \label{chmix:betas}
\end{equation}

The coefficients $\Gamma_{12}^{ij}$ in \eq{chmix:ga12dec} are positive.  They
are inclusive quantities and can be calculated using the heavy quark expansion
described for the heavy hadron lifetimes in Sect.~\ref{chmix:thlife}.  The
leading term of the power expansion in $\Lambda_{QCD}/m_b$ contains two Dirac
structures, each of which can be factorized into a short-distance Wilson
coefficient and the matrix element of a local operator:
\begin{equation}
\Gamma_{12,q}^{ij} = \frac{G_F^2 m_b^2}{6 \pi } f_{B_q}^2 M_{B_q}   
    \lt[ F^{ij}(z) \frac{2}{3} B_{B_q} - 
          F_S^{ij} (z) \frac{5}{12} B_{B_{q}}^{S\prime} \rt] 
          \lt[ 1 + {\cal O} \lt( \frac{\Lambda_{QCD}}{m_b} \rt)\rt].
 \label{chmix:ga12F}
\end{equation}
Here $z= m_c^2/m_b^2$. The operator $Q$, which we
already encountered in the discussion of \dm, was defined in 
\eq{ch1:uli:defq}. When $Q$ occurs in the context of \bbms, 
we understand that the $d$ quarks in \eq{ch1:uli:defq} are
appropriately replaced by $s$ quarks. In $\Gamma_{12,q}$ a second 
operator occurs:
\begin{equation}
 Q_S = \ov{q}_L b_R \, \ov{q}_L b_R, \qquad\quad q=d \mbox{ or } s.
\end{equation}
$B_{B_q}$ and $B_{B_{q}}^{S\prime}$ (or, equivalently, $B_{B_{q}}^{S}$)
are `bag' parameters quantifying the hadronic matrix elements of 
$Q$ and $Q_S$:%
\index{Bag constants!$B_{B_q}$}%
\index{hadronic parameter!$B_{B_q}$}%
\index{Bag constants!$B_{B_q}^S$}%
\index{hadronic parameter!$B_{B_q}^S$}
\begin{eqnarray}
\bra{B_q^0} Q (\mu) \ket{\ov{B}{}_q^0} &=& 
        \frac{2}{3} f_{B_q}^2 m_{B_q}^2 B_{B_q}(\mu)\,,\\
\bra{B_q^0} Q_S (\mu) \ket{\ov{B}{}_q^0} &=& 
        - \frac{5}{12} f_{B_q}^2 m_{B_q}^2 B_{B_q}^{S\prime} (\mu) 
        \, = \, - \frac{5}{12} f_{B_q}^2 m_{B_q}^2 
        \frac{m_{B_q}^2}{\lt[ m_b(\mu)+m_q(\mu) \rt]^2} B_{B_q}^S (\mu) . \no
\label{chmix:Bdef}
\end{eqnarray}
$B_{B_{q}}^S$ and $B_{B_{q}}^{S\prime}$ simply differ by the factor $m_{B_q}^2/[
m_b(\mu)+m_q(\mu)]^2$. While lattice results are usually quoted for
$B_{B_{q}}^S$, the forthcoming formulae are shorter when expressed in terms of
$B_{B_{q}}^{S\prime}$.  These `bag' factors depend on the renormalization scale
$\mu$ and the renormalization scheme. In the literature on \dg\ it is customary
to use $B_{B_q}$ and $B_{B_q}^S$ in the $\ov{\rm MS}$ scheme as defined in
\cite{bbgln}.  Numerical values obtained in lattice calculations are usually
quoted for $\mu=m_b$. The invariant bag factor $\widehat{B}_{B_q}$ defined in
\eq{defbb} is related to $B_{B_q}(\mu)$ by
\begin{eqnarray}
\widehat{B}_{B_q} &=& B_{B_q}(\mu) \, b_B (\mu) \\
 b_B (\mu) &=& \lt[ \alpha_s (\mu ) \rt]^{-6/23} \, 
     \lt[ 1+ \frac{\alpha_s(\mu)}{4 \pi} \frac{5165}{3174}   \rt], 
     \qquad \quad  b_B(m_b) \, = \, 1.52 \pm 0.03 . 
    \no
\end{eqnarray}
\index{hadronic parameter!$\widehat{B}_B$!relation to $B_{B_q}$}%
A recent preliminary lattice calculation with two dynamical flavors found
$B_{B_s} (m_b) = 0.83 \pm 0.08$ and $B_{B_s}^S(m_b) = 0.84\pm 0.08$ \cite{hy}. 
No deviation of $B_{B_s}/B_{B_d}$ and $B_{B_s}^S/B_{B_d}^S$ from 1 is seen. 
The quoted value for $B_{B_q} (m_b)$ corresponds to $\widehat{B}_{B_q}=1.26\pm
0.12$.  The short distance physics entering $\Gamma_{12}^q$ in \eq{chmix:ga12F}
is contained in $F(z)$ and $F_S (z)$. To leading order in $\alpha_s$ they read:
\begin{eqnarray}
F^{cc} (z)   &=& \sqrt{1-4 z} \lt[ (1-z) K_1 + \frac{1- 4z}{2} K_2 \rt] \nn
F_S^{cc} (z) &=& \sqrt{1-4 z}\, (1+2 z) \lt( K_1 - K_2 \rt)\nn
F^{uc} (z)   &=& (1-z)^2 \lt[ \frac{2+z}{2} K_1 + \frac{1-z}{2} K_2
                 \rt] \nn  
F_S^{uc} (z)   &=& (1-z)^2\, (1+2 z) \lt( K_1 - K_2 \rt)\nn
F^{uu} (z)   &=& F^{cc} (0)   \, =\, F^{uc} (0) \, = \, 
                 K_1 + \frac{1}{2} K_2 \nn
F_S^{uu} (z)   &=& F_S^{cc} (0)   \, =\, F_S^{uc} (0) \, = \, 
                 K_1 - K_2 . \label{chmix:theFs}
\end{eqnarray}
$K_1$ and $K_2$ are combinations of the Wilson coefficients $C_1$ and $C_2$,
which are tabulated in Table~\ref{tab}:
\begin{equation}
K_1 \,=\, 3 C_1^2 + 2 C_1 C_2, \qquad \quad 
K_2 \, = \, C_2^2.
\end{equation}
The scale at which the Wilson coefficients are evaluated must equal the scale
used in $B_{B_q} (\mu) $ and $B_{B_q}^S(\mu)$. In \eq{chmix:ga12F} and
\eq{chmix:theFs} we have neglected the small contributions from penguin
coefficients, which can be found in \cite{bbgln}.

It is instructive to eliminate $\xi_c^q=-\xi_t^q-\xi_u^q$ in favor of $\xi_u^q$
and $\xi_t^q$ in \eq{chmix:ga12dec}:
\begin{equation}
\Gamma_{12}^q \, = \, - 
               \xi_t^{q*2} \, \lt[ \Gamma_{12,q}^{cc} \, + \, 
               2\, \frac{\xi_u^{q*}}{\xi_t^{q*}} 
               \lt( \Gamma_{12,q}^{cc} - 
               \Gamma_{12,q}^{uc} \rt) \, +  \, 
               \frac{\xi_u^{q*2}}{\xi_t^{q*2}} 
               \lt( \Gamma_{12,q}^{cc} - 2 \Gamma_{12,q}^{uc} + 
               \Gamma_{12,q}^{uu} \rt) \rt].
\label{chmix:ga12dec2} 
\end{equation}
In the limit $z=0$ all the $\Gamma_{12}^{ij}$'s become equal and $\arg
M_{12}=\arg(-\Gamma_{12}) = \arg (\xi_t^{q*2})$, so that $\phi$ vanishes in this
limit.  With \eq{chmix:xis} one verifies that $\phi_d={\cal O} (\ov{\eta} z)$
and $\phi_s={\cal O} (\lambda^2 \ov{\eta} z)$. This GIM suppression is lifted,
if new physics contributes to $\arg M_{12}^q$ spoiling the cancellation between
$\arg M_{12}$ and $\arg(-\Gamma_{12}^q)$. Therefore $\phi$ is extremely
sensitive to new physics.  \index{phase!of $CP$ violation in mixing, $\phi$}
\index{phase!of $CP$ violation in mixing, $\phi$!SM prediction for $\phi_d$ 
and $\phi_s$}

Combining \eq{chmix:ga12dec2} and \eq{chmix:theFs} we find the Standard Model
predictions for $\phi_d=\phi(B_d)$ and $\phi_s=\phi(B_s)$:
\begin{eqnarray}
\phi_d & = & 
   - \frac{2\ov{\eta}}{R_t^2} \, 
     \frac{\Gamma_{12,d}^{uc} -
   \Gamma_{12,d}^{cc}}{\Gamma_{12,d}^{cc}} 
   \lt[ 1+ {\cal O} \lt( \ov{\eta} z^2, \ov{\eta} \lambda^4\rt) \rt]
      \approx - \frac{24}{5} \, \frac{\ov{\eta}}{R_t^2} 
         \, \frac{B_{B_d}}{B_{B_d}^{S\prime}} \, 
         \frac{K_1+K_2}{K_2-K_1} \, z \, \approx -0.1 \, = \, 
          - 5^\circ \,, \no\\[3mm]
\phi_s & = & 
   2 \lambda^2 \ov{\eta} \, 
     \frac{\Gamma_{12,s}^{uc} -
   \Gamma_{12,s}^{cc}}{\Gamma_{12,s}^{cc}} 
   \lt[ 1+ {\cal O} \lt( \ov{\eta} z^2, \ov{\eta} \lambda^4\rt)
   \rt]
     \approx \frac{24}{5} \, \ov{\eta} \, \lambda^2 
         \, \frac{B_{B_s}}{B_{B_s}^{S\prime}} \, 
         \frac{K_1+K_2}{K_2-K_1} \, z  \, \approx \, 3 \times 10^{-3} 
         \, = \, 0.2^\circ \,. \nonumber\\
         \label{chmix:phinum}
\end{eqnarray}
That is, in the Standard Model, we can safely neglect the factor of $\cos
\phi_q$ in the relation $\dg_q = \ 2 |\Gamma_{12}^q| \cos \phi_q$ (see
\eq{mgsol:b}). For $\Gamma_{12}^s$ the CKM-suppressed contributions
$\Gamma_{12,s}^{uc}$ and $\Gamma_{12,s}^{uu}$ are completely irrelevant.
$\Gamma_{12}^d$ is also dominated by the double-charm contribution
$\Gamma_{12,d}^{cc}$, with up to ${\cal O}(5\% )$ corrections from the second
term in \eq{chmix:ga12dec2}:
\begin{eqnarray} 
\dg_s^{\rm SM} &=& 2 |\Gamma_{12}^s| \, = \, 
        |\xi_t^s|^2 \, 2 |\Gamma_{12,s}^{cc}| \nn 
\dg_d^{\rm SM} &=& 2 |\Gamma_{12}^d| \, = \, 
        |\xi_t^d|^2 \, 2 | \Gamma_{12,d}^{cc} | 
        \, \bigg| 1 + 2 \frac{R_t^2 + \ov{\rho} - 1}{R_t^2} 
        \frac{\Gamma_{12,d}^{uc}-\Gamma_{12,d}^{cc}}{\Gamma_{12,d}^{cc}} 
            + {\cal O} \bigg( \frac{R_B^4}{R_t^4}\, z^2 \bigg) \bigg| \,. 
  \label{chmix:dgsm}
\end{eqnarray} 
\index{B meson@$B$ meson!width difference $\dg$!SM prediction for $\dg_s$}%
\index{width difference $\dg$!SM prediction for $\dg_s$}%
$\Gamma_{12,s}^{cc}$ has been calculated in the next-to-leading order of
$\alpha_s$ \cite{bbgln} and $\Lambda_{QCD}/m_b$ \cite{bbd}. The result gives
the following prediction for $\dg_s^{\rm SM}$:
\begin{equation} 
\frac{\dg_s^{\rm SM}}{\Gamma} \, =\, \frac{2 |\Gamma_{12}^s|}{\Gamma} 
        \, = \, 
        \left( \frac{f_{B_s}}{245~{\rm MeV}} \right)^2 \,
        \left[ \, (0.234\pm 0.035)\, {B_{B_s}^S} - 
        0.080 \pm 0.020 \, \right] .
\label{chmix:dgnum} 
\end{equation} 
Here the coefficient of $B_{B_s}^S$ has been updated to
$m_b(m_b)+m_s(m_b)=4.3\,$GeV (in the $\ov{\rm MS}$ scheme) compared to
\cite{bbgln}. Since the coefficient $F^{cc}(z)$ of $B_{B_s}$ in \eq{chmix:ga12F}
is very small, $B_{B_s}$ in \eq{chmix:dgnum} has been fixed to $B_{B_s} (m_b) =
0.85\pm 0.05$ obtained in quenched lattice QCD \cite{hy}. Recently the
KEK--Hiroshima group succeeded in calculating $f_{B_s}$ in an unquenched lattice
QCD calculation with two dynamical fermions \cite{fbs}. The result is
$f_{B_s}=(245\pm 30)\,$MeV.  With this number and $B_{B_s}^S (m_b) = 0.87\pm
0.09$ from quenched lattice QCD \cite{hy,hioy} one finds from \eq{chmix:dgnum}:
\begin{equation} 
\frac{\dg_s^{\rm SM}}{\Gamma} \,=\, 0.12\pm 0.06 
        . \label{chmix:dgnum2}
\end{equation}
Here we have conservatively added the errors from the two lattice quantities 
linearly. $f_{B_s}$ drops out from the ratio
\begin{equation}  
\frac{\dg_s^{\rm SM}}{\dm_s^{\rm SM}} \,\simeq\, \frac{5 \pi}{6} 
        \frac{m_b^2}{M_W^2\, \eta_B b_B  S (m_t^2/M_W^2) } \, |F_S (z)| 
       \, \frac{B_{B_s}^{S\prime}}{B_{B_s}} 
       \lt[ 1 + {\cal O} \lt( \frac{\Lambda_{QCD}}{m_b} \rt) \rt] .   
\label{chmix:dgdm}
\end{equation}
The full result with next-to-leading order corrections in $\alpha_s$ and
$\Lambda_{QCD}/m_b$ can be found in \cite{bbgln}.  Including these corrections
one finds \cite{bbgln}:
\begin{equation} 
\frac{\dg_s^{\rm SM}}{\dm_s^{\rm SM}} \,=\, \lt( 3.7 \epm{0.8}{1.5} \rt) \times 10^{-3}.
\label{chmix:dgdmnum}
\end{equation}
The uncertainty in \eq{chmix:dgdmnum} is dominated by the renormalization scale
dependence. Its reduction requires a painful three-loop calculation. The
numerical value in \eq{chmix:dgdmnum} is obtained with $B_{B_s}^S/B_{B_s}=1.0\pm
0.1$ \cite{hy}, which is larger than the one used in \cite{bbgln}.

Next we consider the width difference in the $B_d$ meson system: since the
second term in \eq{chmix:dgsm} is negligible in view of the other uncertainties,
\eq{chmix:dgdm} also holds for $\dg_d^{\rm SM}/\dm_d^{\rm SM}$ with the replacement
$B_{B_s}^{S\prime}/B_{B_s} \to B_{B_d}^{S\prime}/B_{B_d}$. The $SU(3)_F$
breaking in these `bag' factors and in the $\Lambda_{QCD}/m_b$ corrections can
safely be neglected, so that the numerical range \eq{chmix:dgdmnum} also holds
for $\dg_d^{\rm SM}/\dm_d^{\rm SM}$. With $\dm_d =0.49\,$ps$^{-1}$ and
$\tau_{B_d}=1.5\,$ps one finds $\dg_d^{\rm SM}\approx 3\times 10^{-3}
\Gamma_d$.%
\index{B meson@$B$ meson!width difference $\dg$!SM prediction for $\dg_d$}%
\index{width difference $\dg$!SM prediction for $\dg_d$} Since $\dg_d$ and $\dm_d$
are affected by new physics in different ways, it is instructive to consider the
ratio of the two width differences: from \eq{chmix:dgsm} and \eq{chmix:dgnum}
one finds
\begin{equation}
\frac{\dg_d^{\rm SM}}{\dg_s^{\rm SM}} \,=\, \frac{|\Gamma_{12}^d|}{|\Gamma_{12}^s|}
      \, \simeq \, \frac{f_{B_d}^2 B_{B_d}^S}{f_{B_s}^2 B_{B_s}^S} 
              \, \lt| \frac{\xi_t^d}{\xi_t^s} \rt|^2 
\, \simeq \,  0.04 \, R_t^2  . \label{chmix:dgdsm}     
\end{equation}
In the last line we have used $f_{B_s}/f_{B_d}=1.16\pm 0.05$ and
$B_{B_s}^S=B_{B_d}^S$. The numerical predictions for $\dg_d$ from
\eq{chmix:dgdmnum} and \eq{chmix:dgdsm} are consistent with each other. Since
$\dg_d$ stems from the CKM-suppressed decay modes, it can be substantially
enhanced in models of new physics.

\boldmath
\subsubsection{Phenomenology of $\dg_s$}\label{chmix:subsub:width:phen}
\unboldmath
\index{B meson@$B$ meson!width difference $\dg$!measurement of $\dg_s$}
\index{width difference $\dg$!measurement of $\dg_s$}

\paragraph{Time Evolution:}
The width difference $\dg_s$ can be measured from the time evolution of an
untagged $B_s$ sample, as shown in sect.~\ref{subs:unt}. In general the decay
$\Bsun \to f$ is governed by the two-exponential formula in \eq{guntfeig}. With
\eq{defpq} and \eq{deflaf} we can calculate $\langle f \ket{B_{L,H}}$ and find
from \eq{guntfeig}:
\begin{equation}
\guntf \,=\,  {\cal N}_f \, \frac{|A_f|^2}{2} \lt( 1+ |\lambda_f|^2 \rt) 
            \lt[ \lt( 1- \adg{f} \rt) e^{- \Gamma_L t} + 
                    \lt( 1+ \adg{f} \rt) e^{- \Gamma_H t} \rt]
\label{chmix:utadg} 
\end{equation}
\index{untagged $B$!two-exponential decay}%
\index{$A_{\Delta \Gamma}$!and untagged decay}%
with $\adg{f}$ defined in \eq{adeg}.  Throughout this
Sect.~\ref{chmix:subsub:width:phen} we neglect small terms of order $\ega$.  The
time-independent prefactor of the square bracket can be eliminated in favor of
the branching ratio $Br \, \big( \!\Bun \rightarrow f \, \big)$ using
\eq{untnorm}.  In principle one could measure $\dg_s=\Gamma_L-\Gamma_H$ by
fitting the decay distribution of any decay with $|\adg{f}|\neq1$ to $\guntf$ in
\eq{chmix:utadg}. In practice, however, one will at best be able to measure the
deviation from a single exponential up to terms linear in $\dg_s t$. In
\eq{guntf} $\guntf$ is expressed in terms of $\Gamma_s=(\Gamma_L+\Gamma_H)/2$
and $\dg_s$. With \eq{untnorm} one finds
\begin{equation}
\guntf \,=\,
2\, \brunts{f} \,
        \Gamma_s 
        \, e^{- \Gamma_s t} \lt[ 
        1 +  \frac{\dg_s}{2} \, \adg{f} \, 
        \lt( t - \frac{1}{ \Gamma_s } \rt) 
        \rt] 
        + {\cal O} \lt(\lt( \dg_s\, t \rt)^2 \rt) . \label{chmix:guntf3} 
\end{equation} 
That is, unless one is able to resolve quadratic ${\cal O} \lt( ( \dg_s)^2 \rt)
$ terms, one can only determine the product $\adg{f} \dg_s $ from the time
evolution. A flavor-specific decay mode like $B_s \to
D_s^- \pi^+$%
\index{decay!$B_s \rightarrow D_s^- \pi^+$} is characterized by $\lambda_f=0$
and therefore has $\adg{f}=0$. In these decay modes the term involving $\dg_s $
in \eq{chmix:guntf3} vanishes. Flavor-specific decays therefore determine
$\Gamma_s$ up to corrections of order $( \dg_s)^2 $. For those decays \guntf\ is
insensitive to new physics in $M_{12}$, because $\lambda_f=0$. In order to gain
information on $\dg_s$ from \eq{chmix:guntf3}, one must consider decays with
$\lambda_f\neq 0$.  But $\lambda_f$ and \adg{f}\ depend on the mixing phase
$\phi_M$ (see \eq{defphm}) and therefore change in the presence of new physics
in $M_{12}$. In the Standard Model we can calculate $\phi_M$ and then extract
$\dg_s$ from the measured $\adg{f} \dg_s $. In the presence of new physics,
however, one needs additional information.  We therefore discuss these two cases
independently below.

Lifetimes are conventionally measured by fitting the decay distribution to a
single exponential.  We now write the two-exponential formula of
\eq{chmix:utadg} as \index{untagged $B$!two-exponential decay}
\begin{eqnarray}
\guntf &=& A \, e^{-\Gamma_L t} \, + \, B \, e^{-\Gamma_H t} \nn
       & = & e^{-\Gamma_s t} \lt[ 
        \lt( A+B \rt) \cosh \frac{\dg_s t}{2} + 
        \lt( B-A \rt) \sinh \frac{\dg_s t}{2}   \rt] \! , 
   \label{chmix:twoex2}
\end{eqnarray}
where $A=A(f)$ and $B=B(f)$ can be read off from \eq{chmix:utadg}.  If one uses
a maximum likelihood fit of \eq{chmix:twoex2} to a single exponential,
\begin{equation}
 F\lt[ f, t \rt] \,=\,
        { \Gamma_f} \,   e^{-\Gamma_f \, t}, \label{chmix:singex} 
\end{equation}
it will yield the following result \cite{hm}:
\begin{equation}
\Gamma_f \,=\,
        \frac{A/\Gamma_L + B/\Gamma_H}{A/\Gamma_L^2 + B/\Gamma_H^2}     
 . \label{fitex}
\end{equation}
We expand this to second order in $\dg_s$:
\begin{equation}
     \Gamma_f \,=\, \Gamma_s \, + \,  
\frac{A-B}{A+B} \, \frac{\dg_s}{2} \, 
                - \frac{2 \, A B}{(A+B)^2}  \frac{(\dg_s)^2}{\Gamma_s} \,  
        + \, {\cal O} \lt( \frac{(\dg_s)^3}{\Gamma_s^2} \rt)
 . \label{chmix:fit}
\end{equation}
In flavor-specific decays we have $A=B$ (see \eq{guntf}). We see from
\eq{chmix:fit} that here a single-exponential fit determines
\begin{equation}
 \Gamma_{\rm fs} \,=\, \Gamma_s \, - \, \frac{(\dg_s)^2}{2\, \Gamma_s} 
             \, + \,  {\cal O} \lt( \frac{(\dg_s)^3}{\Gamma_s^2} \rt)
             . \label{chmix:gafs}
\end{equation}
Heavy quark symmetry predicts that the average \emph{widths}\ $\Gamma_s$ and
$\Gamma_d$ are equal up to corrections of less than one percent
\cite{bbsuv,bbd}. From \eq{chmix:gafs} we then realize that the average $B_s$
\emph{lifetime}\ (defined as $1/\Gamma_f$) can exceed the $B_d$ lifetime by more
than one percent, if $\dg_s$ is sizable.

\paragraph{\boldmath $CP$ Properties and Branching Ratios:}
In the $B_s$ system \dg\ is dominated by $\Gamma_{12}^{s,cc}$. In the following
we will neglect the Cabibbo-suppressed contributions from $\Gamma_{12}^{s,uc}$
and $\Gamma_{12}^{s,uu}$. We also specify to the PDG phase convention for the
CKM matrix, in which $\arg (V_{cb} V_{cs}^*) = {\cal O} (\lambda^6)$, see
\eq{chmix:betas}. For the discussion in the forthcoming paragraphs it will be
useful to define the $CP$ eigenstates
\begin{equation}
\ket{B_s^{\textrm{\scriptsize even}}} \,=\,
 \frac{1}{\sqrt{2}} \lt( \ket{B_s} - \ket{\ov{B}{}_s} \rt), \qquad 
 \mbox{and} \qquad
\ket{B_s^{\textrm{\scriptsize odd}}} \,=\, 
 \frac{1}{\sqrt{2}} \lt( \ket{B_s} + \ket{\ov{B}{}_s} \rt)
 .\label{chmix:cpe}  
\end{equation}
Here we have used the standard convention for the $CP$ transformation:%
\index{CP transformation@\CP\ transformation!B meson@$B$ meson} $CP \ket{B_s} =
- \ket{\ov{B}{}_s}$.

\index{B meson@$B$ meson!width difference $\dg$!$\dg_s$ and branching ratios}%
\index{width difference $\dg$!measurement of $\dg_s$!branching ratios}
Interestingly, one can measure $\dg_s$ from branching ratios, without
information from lifetime fits.  We define
\begin{equation}
\dg_{\rm CP}^s \,\equiv\, 2 |\Gamma_{12}^s| \, = \, 
2 \sum_{f  \in X_{c\ov{c}}} \,  \lt[
\Gamma ( B_s \to f_{\rm CP+} ) \, - \,
        \Gamma ( B_s \to f_{\rm CP-}) \rt]
        . 
        \label{chmix:dgcp}
\end{equation}
Here $X_{c\ov {c}}$ represents the final states containing a
$(c,\ov{c})$ pair, which constitute the dominant contribution to
$\dg_{\rm CP}^s$ stemming from the decay $b \to c\ov{c} s$.  In
\eq{chmix:dgcp} we have decomposed any final state $f$ into its
$CP$-even and $CP$-odd component, $\ket{f}=\ket{f_{\rm CP+}} 
  + \ket{f_{\rm CP-}}$\footnote{The factor of 2 in \eq{chmix:dgcp} is an
  artifact of our normalization of $\ket{f_{\rm CP\pm}}$.} and defined
\begin{equation}
\Gamma ( B_s \to f_{\rm CP\pm} ) \,=\, 
        {\cal N}_f \, | \bra{f_{\rm CP\pm}} B_s \rangle |^2 
  \, = \, 
     \frac{ | \bra{f_{\rm CP\pm}} B_s \rangle |^2}{ | \bra{f} B_s \rangle
     |^2} \, 
        \Gamma ( B_s \to f ) . 
\end{equation}
${\cal N}_f$ is the usual normalization factor originating from the phase-space
integration.  $\dg_{\rm CP}^s$ equals $\dg_s$ in the Standard Model, but these
quantities differ by a factor of $\cos \phi_s $ in models of new physics, see
\eq{mgsol:b}. We will later exploit this feature to probe the Standard Model and
to determine $|\cos \phi_s|$.

We now prove the second equality in \eq{chmix:dgcp} and subsequently discuss how
$\Gamma ( B_s \to f_{\rm CP\pm})$ can be measured.  Start from the definition of
$\Gamma_{12}^s$:
\begin{equation}
\Gamma_{12}^s \,=\, \sum_f {\cal N}_f \, \bra{B_s} f \rangle 
                \bra{f} \ov{B}_s \rangle 
        \, = \, \frac{1}{2} \sum_f {\cal N}_f  
        \lt[ \, \bra{B_s} f \rangle \bra{f} \ov{B}_s \rangle + 
        \bra{B_s} \ov{f} \rangle \bra{{ \ov{f}}} \ov{B}_s \rangle \, \rt]
 .\label{defg12} 
\end{equation}
In the second equation we have paired the final state $\ket{f}$ with its $CP$
conjugate $\ket{\ov{f}}=-CP\ket{f}$.
In the next step we 
trade $f$ for $f_{\rm CP+}$ and $f_{\rm CP-}$ and use the $CP$ transformation%
\index{CP transformation@\CP\ transformation!B meson@$B$ meson}
\begin{equation}
 \bra{f_{\rm CP\pm}} \ov{B}_s \rangle \,=\,
        \mp \, \bra{f_{\rm CP\pm}} B_s \rangle 
\end{equation}
in our phase convention with $\arg (V_{cb} V_{cs}^*) =0$. Then \eq{defg12}
becomes
\begin{eqnarray}
- \, \Gamma_{12}^s &=& 
        \sum_{f\in X_{c\ov{c}}} {\cal N}_f \,
        \lt[ | \bra{f_{\rm CP+}} B_s \rangle |^2 - 
             | \bra{f_{\rm CP-}} B_s \rangle |^2 \rt] \nn
& = & \sum_{f\in X_{c\ov{c}}}
\lt[ \Gamma ( B_s \to f_{\rm CP+} ) \, - \,
        \Gamma ( B_s \to f_{\rm CP-}) \rt] .
\label{chmix:g12oe}
\end{eqnarray}
Interference terms involving both $ \bra{f_{\rm CP+}} B_s \rangle$ and
$\bra{f_{\rm CP-}} B_s \rangle $ drop out when summing the two terms $\bra{B_s}
f \rangle \bra{f} \ov{B}_s \rangle$ and $\bra{B_s} \ov{f} \rangle \bra{{
    \ov{f}}} \ov{B}_s \rangle $.  An explicit calculation of $\Gamma_{12}^s$
reveals that the overall sign of the LHS of \eq{chmix:g12oe} is positive, which
completes the proof of \eq{chmix:dgcp}.

Loosely speaking, $\dg_{\rm CP}^s$ is measured by counting the $CP$-even and
$CP$-odd double-charm final states in $B_s$ decays.  Our formulae become more
transparent if we use the $CP$-eigenstates defined in \eq{chmix:cpe}.  With
$\ket{B_s}= (\ket{B_s^{\textrm{\scriptsize even}}}+\ket{B_s^{\textrm{\scriptsize
      odd}}})/\sqrt{2}$ one easily finds from \eq{chmix:g12oe}:
\begin{equation}
\dg_{\rm CP}^s \,=\, 2 |\Gamma_{12}^s| \, = \, 
\Gamma \lt( B_s^{\textrm{\scriptsize even}} \rt) - 
        \Gamma ( B_s^{\textrm{\scriptsize odd}} ) .
        \label{chmix:dgcp2}
\end{equation}
Here the RHS refers to the total widths of the $CP$-even and $CP$-odd $B_s$
eigenstates.  We stress that the possibility to relate $|\Gamma_{12}^s|$ to a
measurable quantity in \eq{chmix:dgcp} crucially depends on the fact that
$\Gamma_{12}^s$ is dominated by a single weak phase. For instance, the final
state $K^+ K^-$ is triggered by $b \to u \ov{u} s$ and involves a weak phase
different from $b \to c\ov{c} s$.  Although $K^+ K^-$ is $CP$-even, the decay
$B_s^{\textrm{\scriptsize odd}}\to K^+ K^- $ is possible.%
\index{decay!$B_s \rightarrow K^+ K^-$}
An inclusion of such CKM-suppressed modes into \eq{chmix:g12oe}
would add interference terms that spoil the relation to measured
quantities.  The omission of these contributions to $\Gamma_{12}^s$
induces a theoretical uncertainty of order 3--5\% on \eq{chmix:dgcp2}.

A measurement of $\dg_{\rm CP}^s$ has been performed by the ALEPH
collaboration \cite{aleph}.  ALEPH has measured
\index{decay!$B_s \rightarrow D_s^+ D_s^-$}
\index{decay!$B_s \rightarrow D_s^{*+} D_s^-$}
\index{decay!$B_s \rightarrow D_s^+ D_s^{*-}$}
\index{decay!$B_s \rightarrow D_s^{*+} D_s^{*-}$}
\begin{equation}
 2\, \brunts{D_s^{(*)}{}^+ D_s^{(*)}{}^-} \,=\,
        0.26 \epm{0.30}{0.15} \label{chmix:alexp}
\end{equation}
and related it to $\dg_{\rm CP}^s$. For this the following theoretical input has
been used \cite{ayopr}:
\begin{itemize} 
\item[i)] In the heavy quark limit $m_c \to \infty $ and neglecting certain
  terms of order $1/N_c$ (where $N_c=3$ is the number of colors) the decay
  $B_s^{\textrm{\scriptsize odd}} \to D_s^\pm D_s^{*}{}^\mp$ is forbidden. Hence
  in this limit the final state in $\Bsun \to D_s^\pm D_s^{*}{}^\mp$ is $CP$-even.
  Further in $\Bsun \to D_s^{*}{}^+ D_s^{*}{}^-$ the final state is in an
  S-wave.
\item[ii)] In the small velocity limit when $m_c \to \infty $ with $m_b - 2
  m_c$ fixed \cite{sv}, $\dg_{\rm CP}^s$ is saturated by $\Gamma ( \Bsun \to
  D_s^{(*)}{}^+ D_s^{(*)}{}^- )$.  With i) this implies that in the considered
  limit the width of $B_s^{\textrm{\scriptsize odd}}$ vanishes.  For $N_c \to
  \infty$ and in the SV limit, ${ \Gamma ( \Bsun \to D_s^{(*)}{}^+ D_s^{(*)}{}^-
    )}$ further equals the parton model result for $\dg_{\rm CP}^s$
  (quark-hadron duality).
\end{itemize}
Identifying $\Gamma ( B_s^{\textrm{\scriptsize even}} \to D_s^{(*)}{}^+
D_s^{(*)}{}^- ) \simeq \dg_{\rm CP}^s$ and $\Gamma ( B_s^{\textrm{\scriptsize
    odd}} \to D_s^{(*)}{}^+ D_s^{(*)}{}^- ) \simeq 0$ we can integrate
$\gunt{D_s^{(*)}{}^+ D_s^{(*)}} = \Gamma ( B_s^{\textrm{\scriptsize even}}\to
D_s^{(*)}{}^+ D_s^{(*)}{}^- )\exp (-\Gamma_L t) $ over $t$ to find:
\begin{equation}
2\, \brunts{ D_s^{(*)}{}^+ D_s^{(*)}{}^-} \,\simeq\, 
         \, \frac{\dg_{\rm CP}^s}{\Gamma_L}   
. \label{chmix:bdg}
\end{equation}
Thus the measurement in \eq{chmix:alexp} is compatible with the theoretical
prediction in \eq{chmix:dgnum2}.

When using \eq{chmix:bdg} one should be aware that the corrections to the limits
i) and ii) adopted in \cite{ayopr} can be numerically sizeable. For instance, in
the SV limit there are no multibody final states like $D_s^{(*)} \ov{D} X_s$,
which can modify \eq{chmix:bdg}.  As serious would be the presence of a sizeable
$CP$-odd component of the $D_s^{(*)}{}^+ D_s^{(*)}{}^- $ final state, since it
would be added with the wrong sign to $\dg_{\rm CP}^s$ in \eq{chmix:bdg}. A
method to control the corrections to the SV limit experimentally is proposed
below in the paragraph on new physics. One feature of the SV limit is the
absence of $CP$-odd double-charm final states. (Indeed there are only very few
$CP$-odd final states in Table~\ref{chmix:tab:cpe}.) This has the consequence that
$\dg_{\rm CP}^s$ cannot be too small, because for $\Gamma
(B_s^{\textrm{\scriptsize odd}} \to X_{\ov{c} c})$ the spectator contributions
and non-spectator diagrams like those in Fig.~\ref{chmix:fig:dga} must sum to
zero. This favors values of $\dg_{\rm CP}^s = \dg_s^{\rm SM}$ in the upper range of
\eq{chmix:dgnum2}.

\paragraph{Standard Model:}
In the Standard Model the \bbms\ phase $\phi_M^s=-2\beta_s$ can be safely
neglected for the discussion of $\dg_s$. Then the mass eigenstates coincide with
the $CP$ eigenstates defined in \eq{chmix:cpe} with
$\ket{B_L}=\ket{B_s^{\textrm{\scriptsize even}}}$ and
$\ket{B_H}=\ket{B_s^{\textrm{\scriptsize odd}}}$. Any $b\to c \ov{c} s$ decay
into a $CP$-even final state like $D_s^+D_s^-$ stems solely from the $\ket{B_L}$
component in the untagged $B_s$ sample. A lifetime fit to this decay therefore
determines $\Gamma_L$.  Conversely, the $b\to c \ov{c} s$ decay into a $CP$-odd
eigenstate determines $\Gamma_H$. We can easily verify this from
\eq{chmix:utadg} by calculating $\adg{f}$: $q/p$ in \eq{qpsol} equals $-1$ and
$\ov{A}_f/A_f=-\eta_f$, where $\eta_f$ is the $CP$ parity of the final state. Then
\eq{deflaf} yields $\lambda_f=\eta_f$, so that $\adg{f} =-\eta_f$. Hence for any
$b\to c \ov{c} s$ decay the coefficient of $\exp(-\Gamma_H t)$ in
\eq{chmix:utadg} vanishes for a $CP$-even final state, while the $\exp(-\Gamma_L
t)$ term vanishes for a $CP$-odd final state. In practice one will encounter much
more statistics in $CP$-even final states, so that the best determination of
$\dg_s$ will combine
$\Gamma_L$ with $\Gamma_{\rm fs}$ measured in a flavor-specific decay.%
\index{B meson@$B$ meson!width difference $\dg$!$\dg_s$ and lifetimes}%
\index{B meson@$B$ meson!width difference $\dg$!$\dg_s$ and branching ratios}%
\index{width difference $\dg$!measurement of $\dg_s$!lifetimes} From
\eq{chmix:gafs} and $\Gamma_L=\Gamma_s + \dg_s/2$ one finds
\begin{equation}
\dg_s \,=\, 2 \lt( \Gamma_L - \Gamma_{\rm fs} \rt)
             \lt( 1 - 
        2 \frac{\Gamma_L - \Gamma_{\rm fs}}{\Gamma_{\rm fs}} \rt) 
             \, + \,  {\cal O} \lt( \frac{(\dg_s)^3}{\Gamma_s^2}\rt)
\label{chmix:dgressm} .
\end{equation}
Here we have expanded to second order in $\dg_s$, which should be sufficient for
realistic values of $\dg_s$.

It should be stressed that every $b\to c \ov{c} s$ decay encodes the same
information on $\dg_s$, once its $CP$ parity is known. This is also true for $b\to
c \ov{u} d$ decays into $CP$ eigenstates, because the decay amplitude carries the
same phase as the one in $b\to c \ov{c} s$.  Therefore the extracted values for
$\dg_s$ in these decays can be combined to gain statistics. Interesting decay
modes are summarized in Table~\ref{chmix:tab:cpe}. Many of the listed modes, like
$\Bsun \to \psi \phi$, require an angular analysis to separate the $CP$-even from
the $CP$-odd component. This procedure is described in detail in
Sect.~\ref{chmix:twocp}.%

\index{decay!$B_s \rightarrow K^+ K^-$} It is tempting to use $\Bsun \to K^+
K^-$ to measure $\dg_s$ because of its nice experimental signature. But such
CKM-suppressed decay modes cannot be used, because the weak phase of the decay
amplitude is not known. If $B_s \to K^+ K^-$ were dominated by penguin loops and
new physics were absent from these loops, $\lambda_{K^+K^-}$ would indeed be
equal to $+1$ and the coefficient of $\exp(-\Gamma_H t)$ in \eq{chmix:utadg}
would vanish. In practice, however, the tree-level amplitude $b\to u \ov{u} s$
is expected to give a non-negligible contribution. Since this amplitude carries
a different phase, $2 \arg(V_{ub})=-2\gamma$, $\lambda_{K^+K^-}$ deviates from
$\pm 1$ and both exponentials in \eq{chmix:utadg} contribute.

\paragraph{New Physics:}
\index{B meson@$B$ meson!width difference $\dg$!new physics}%
\index{width difference $\dg$!new physics}%
In the presence of new physics the
CP-violating phase $\phi$ in \eq{defphi} and \eq{mgsol:b} can be large. Since
various observables in untagged $B_s$ decays depend on $\cos \phi_s$ in
different ways, one can reveal new physics and determine $|\cos \phi_s|$ by
combining different measurements. We have already seen above that $\dg_{\rm
  CP}^s$ in \eq{chmix:dgcp} does not depend on $\phi_s$ at all, while $\dg_s$ is
diminished in the presence of new physics:
\begin{equation} 
\dg_s \,=\, \dg_{\rm CP}^s \cos \phi_s .
\end{equation}
On the other hand $\sin \phi_s$ can be obtained from $CP$ asymmetries in $B_s$
decays like $B_s \to \psi \phi$.  Therefore measurements of \dg\ are
complementary to the study of $CP$ asymmetries, which require tagging and the
resolution of the rapid \bbs\ oscillation and come with a loss in statistics,
efficiency and purity.  Both avenues should be pursued and their results
combined, because they measure the same fundamental quantities. A detailed
analysis of both tagged and untagged decays can be found in \cite{dfn}.

In our phase convention $\arg (V_{cb} V_{cs}^*) =0$ we simply have 
\begin{equation}
\arg(M_{12}) \,=\, \phi_s .  \label{chmix:dgcph}  
\end{equation}
The mass eigenstates can be expressed as
\begin{eqnarray}
\ket{B_L} & = & \phantom{- \,}
                \frac{1+e^{i\phi}}{2} \, 
                  \ket{B_s^{\textrm{\scriptsize even}}} \, { -} \, 
                  \frac{1-e^{i\phi}}{2} \, 
                  \ket{B_s^{\textrm{\scriptsize odd}}} 
                  , \nn
\ket{B_H} & = & { -} \, \frac{1-e^{i\phi}}{2} \, 
                  \ket{B_s^{\textrm{\scriptsize even}}} \, + \, 
                  \frac{1+e^{i\phi}}{2} \, 
                  \ket{B_s^{\textrm{\scriptsize odd}}} 
         \, . 
        \label{chmix:rot}
\end{eqnarray} 
Whenever we use $B_s^{\textrm{\scriptsize even}}$ and $B_s^{\textrm{\scriptsize
    odd}}$ we implicitly refer to our phase conventions for the CKM matrix and
the $CP$ transformation.  If formulae involving $B_s^{\textrm{\scriptsize even}}$
and $B_s^{\textrm{\scriptsize odd}}$ are used to constrain models with an
extended quark sector, the phase convention used for the enlarged CKM matrix
must likewise be chosen such that $\arg (V_{cb} V_{cs}^*) \simeq 0$.

We next consider the time evolution of a $b\to c\ov{c}s$ decay into a CP
eigenstate with $CP$ parity $\eta_f$. \adg{f}\ reads
\begin{equation}
\adg{f} \,=\, -\eta_f \cos \phi_s . \label{acp2}
\end{equation}
In the Standard Model, where $\phi_s\simeq 0$, \guntf\ simplifies to a
single-exponential law, which can be verified from \eq{chmix:guntf3} or by
inserting \eq{chmix:rot} into \eq{guntfeig}.

Since $\dg_{\rm CP}^s$ is unaffected by new physics and $\dg_{\rm CP}^s > 0$,
several facts hold beyond the Standard Model:
\begin{itemize}
\item[i)] There are more $CP$-even than $CP$-odd final states in $B_s$ decays. 
\item[ii)] The shorter-lived mass eigenstate is always the one with
  the larger $CP$-even component in \eq{chmix:rot}. Its branching ratio into a
  $CP$-even final state $f_{\rm CP+}$ exceeds the branching ratio of the
  longer-lived mass eigenstate into $f_{\rm CP+}$, if the weak phase
  of the decay amplitude is close to $ \arg V_{cb} V_{cs}^*$.
\item[iii)] For $ \cos \phi_s > 0$ $B_L$ has a shorter lifetime than
  $B_H$, while for $ \cos \phi_s < 0$ the situation is the opposite
  \cite{g}.
\end{itemize}
Allowing for a new physics phase $\phi_s$ the result in \eq{chmix:bdg}
is changed. In the SV limit one now predicts:
\begin{eqnarray}
2\, \brunts{ D_s^{(*)}{}^+ D_s^{(*)}{}^-} & \simeq & 
        \dg_{\rm CP}^s \, \lt[ \, \frac{1+\cos \phi_s}{2\, \Gamma_L}  + 
                            \frac{1-\cos \phi_s}{2\, \Gamma_H} \, \rt] 
         \nn
& =&  \,  \frac{\dg_{\rm CP}^s}{\Gamma_s} \lt[ 
  1 + {\cal O} \lt( \frac{\dg_s}{\Gamma_s} \rt) \rt] . \label{chmix:bdg2}
\end{eqnarray}
\index{decay!$B_s \rightarrow D_s^+ D_s^-$}
\index{decay!$B_s \rightarrow D_s^{*+} D_s^-$}
\index{decay!$B_s \rightarrow D_s^+ D_s^{*-}$}
\index{decay!$B_s \rightarrow D_s^{*+} D_s^{*-}$}
The term in square brackets accounts for the fact that in general the $CP$-even
eigenstate $\ket{B_s^{\textrm{\scriptsize even}}}$ is a superposition of
$\ket{B_L}$ and $\ket{B_H}$. It is straightforward to obtain \eq{chmix:bdg}:
inserting \eq{chmix:rot} into \eq{guntfeig} expresses \guntf\ in terms of
$\Gamma ( B_s^{\textrm{\scriptsize even}} \to f )$ and $\Gamma (
B_s^{\textrm{\scriptsize odd}} \to f )$. After integrating over time the
coefficient of $\Gamma ( B_s^{\textrm{\scriptsize even}} \to f )$ is just the
term in square brackets in \eq{chmix:bdg2}.  We verify from \eq{chmix:bdg2} that
the measurement of $\brunts{D_s^{(*)}{}^+ D_s^{(*)}{}^-}$ determines $\dg_{\rm
  CP}^s$.  Its sensitivity to the new physics phase $\phi_s$ is suppressed by
another factor of $\dg_s/\Gamma_s$ and is irrelevant in view of the theoretical
uncertainties.

Next we discuss the determination of $\dg_s$ and $|\cos \phi_s|$.  There are two
generic ways to obtain information on $\dg_s$ and $\phi_s$ :
\begin{itemize}
\item[i)] The measurement of the $B_s$ lifetime in two decay modes
  $\Bsun \to f_1$ and  $\Bsun \to f_2$ with\\
  $\adg{f_1} \neq \adg{f_2}$.
\item[ii)] The fit of the decay distribution of $\Bsun \to f$ to the
  two-exponential formula in \eq{guntf}.
\end{itemize}
As first observed in \cite{g}, the two methods are differently affected by a new
physics phase $\phi_s \neq 0$. Thus by combining the results of methods i) and
ii) one can gain information on $\phi_s$.  In this paragraph we consider two
classes of decays:
\begin{itemize}
\item flavor-specific decays, which are characterized by $\ov{A}_f=0$ {
    implying} $\adg{f}=0$.  Examples are $B_s \to D_s^- \pi^+$ and $B_s \to X
  \ell^+ \nu_{\ell} $,
\index{decay!$B_s \rightarrow D_s^- \pi^+$}%
\index{decay!$B_s \rightarrow X \ell^+ \nu$}
\item the CP-specific decays of Table~\ref{chmix:tab:cpe}, with $\adg{f} = -
  \eta_f \cos \phi_s$.
\end{itemize}
In both cases the time evolution of the untagged sample in \eq{guntf} is not
sensitive to the sign of $\dg_s$ {(or, equivalently, of $\cos \phi_s$)}. For the
CP-specific decays of Table~\ref{chmix:tab:cpe} this can be seen by noticing that
\begin{equation}
\adg{f}\, \sinh \frac{\dg_s \, t}{2} =
        - \, \eta_f \, |\cos \phi_s| \, \sinh \frac{|\dg_s| \, t}{2} .\no
\end{equation}
Here we have used the fact that $\dg_s$ and $\cos \phi_s $ always have the same
sign, because $\dg_{\rm CP}^s>0$.  Hence our untagged studies can only determine
$|\cos \phi_s|$ and therefore lead to a four-fold ambiguity in $\phi_s$.  The
sign ambiguity in $\cos \phi_s$ reflects the fact that from the untagged time
evolution in \eq{guntf} one cannot distinguish, whether the heavier or the
lighter eigenstate has the shorter lifetime.  (Methods to resolve the discrete
ambiguity can be found in \cite{dfn}.)

In order to experimentally establish a non-zero $\dg_s$ from the time evolution
in \eq{guntf} one needs sufficient statistics to resolve the deviation from a
single-exponential decay law, see \eq{chmix:guntf3}.  As long as we are only
sensitive to terms linear in $\dg_s\, t$ and $\dg_s/\Gamma_s$, we can only
determine $\adg{f}\, \dg_s$ from \eq{chmix:guntf3}. $\adg{f}\, \dg_s$ vanishes
for flavor-specific decays and equals $-\eta_f \dg_s\, \cos \phi_s$ for
CP-specific final states.  Hence from the time evolution alone one can only
determine the product $\dg_s\, \cos \phi_s $ in the first experimental stage.

\boldmath
\subparagraph{Determination of $\Gamma_s$ and $\dg_s \cos \phi_s$:}
\unboldmath

In Eqs.~(\ref{chmix:twoex2}) -- (\ref{chmix:fit}) we have related the width
found in a single-exponential fit to the parameters $A(f)$, $B(f)$, $\Gamma_s$
and $\dg_s$ of the two-exponential formula.  \index{untagged $B$!two-exponential
  decay}

In \eq{chmix:gafs} we found that a single-exponential fit in flavor-specific
decays (which have $A=B$) determines $\Gamma_s$ up to corrections of order
$(\dg_s)^2/\Gamma_s^2$.

With \eq{guntf} and \eq{chmix:twoex2} we can read off $A$ and $B$ for the
CP-specific decays of Table~\ref{chmix:tab:cpe} and find $A(f_{\rm CP+})/B(f_{\rm
  CP+}) = (1+\cos \phi)/(1-\cos \phi)$ and $A(f_{\rm CP-})/B(f_{\rm CP-}) =
(1-\cos \phi)/(1+\cos \phi)$ for $CP$-even and $CP$-odd final states, respectively.
Our key quantity for the discussion of $CP$-specific decays 
$\Bsun \to f_{\rm CP}$ is
\begin{equation}
\dg^{s\,\prime}_{\rm CP} \,\equiv\, - \eta_f \adg{f} \, \dg_s \, 
                \, = \, 
              \dg_s \, \cos \phi_s 
          \, = \, 
          \dg_{\rm CP}^s \, \cos^2 \phi_s  
        .\label{chmix:dgp}
\end{equation}   
With this definition \eq{chmix:fit} reads for the decay rate $\Gamma_{\rm CP,
  \eta_f}$ measured in $\Bsun \to f_{\rm CP}$ \cite{dfn}:
\begin{equation} 
     \Gamma_{\rm CP,\eta_f} \,=\, \Gamma_s \, + \,  
        \eta_f \, \frac{\dg^{s\,\prime}_{\rm CP}}{2} 
        { \, - \, \sin^2 \phi_s \, \frac{(\dg_s)^2}{2 \Gamma_s} }
        \, + \, {\cal O} \lt( \frac{(\dg_s)^3}{\Gamma_s^2} \rt) . 
   \label{chmix:gacp} 
\end{equation}
That is, to first order in $\dg_s$, comparing the $\Bsun$ lifetimes measured in
a flavor-specific and a CP-specific final state determines $\dg^{s\,\prime}_{\rm
  CP}$. The first term in \eq{chmix:gacp} agrees with the result in \cite{g},
which has been found by expanding the time evolution in \eq{chmix:twoex2} and
\eq{chmix:singex} for small $\dg_s \, t$.

From \eq{chmix:gafs} and \eq{chmix:gacp} one finds
\begin{equation}
\Gamma_{\rm CP,\eta_f}  \, - \, \Gamma_{\rm fs}  \,=\,
        { \frac{\dg^{s\,\prime}_{\rm CP}}{2}} \lt( \eta_f \, + \, 
                \frac{\dg^{s\,\prime}_{\rm CP}}{\Gamma} \rt) \, + \, 
        {\cal O} \lt( \frac{(\dg)^3}{\Gamma^2}  \rt)  
        . \label{dgp2}
\end{equation}
Hence for a $CP$-even ($CP$-odd) final state the quadratic corrections enlarge
(diminish) the difference between the two measured widths.  A measurement of
$\dg^{s\,\prime}_{\rm CP}$ has a high priority at Run~II of the Tevatron.  The
LHC experiments ATLAS, CMS and LHCb expect to measure $\dg^{s\,\prime}_{\rm
  CP}/\Gamma_s$ with absolute errors between 0.012 and 0.018 for
$\dg^{s\,\prime}_{\rm CP}/\Gamma_s =0.15$ \cite{cern}.  An upper bound on
$\dg^{s\,\prime}_{\rm CP}$ would be especially interesting. If the lattice
calculations entering \eq{chmix:dgnum2} mature and the theoretical uncertainty
decreases, an upper bound on $|\dg^{s\,\prime}_{\rm CP}|$ may { show that}
$\phi_s \neq 0,\pi$ through
\begin{equation}
\frac{\dg^{s\,\prime}_{\rm CP}}{\dg_{\rm CP}^s} \,=\, \cos^2 \phi_s  .
 \label{chmix:cp2}
\end{equation}  
Note that conversely the experimental establishment of a non-zero
$\dg^{s\,\prime}_{\rm CP}$ immediately helps to constrain models of new physics,
because it excludes values of $\phi_s$ around $\pi/2$.

The described method to obtain $\dg_{\rm CP}^{s\,\prime}$ can also be used, if
the sample contains a known ratio of $CP$-even and $CP$-odd components.  This
situation occurs e.g.\ in decays to $J/\psi \phi$, if no angular analysis is
performed or in final states, which are neither flavor-specific nor CP
eigenstates. We discuss this case below for $\Bsun \to D_s^\pm D_s^{(*)}{}^\mp
$.  Further note that the comparison of the lifetimes measured in $CP$-even and
$CP$-odd final states determines $\dg^{s\,\prime}_{\rm CP}$ up to corrections of
order $(\dg_s/\Gamma_s)^3$.

The theoretical uncertainty in \eq{chmix:dgnum2} dilutes the extraction of
$|\cos \phi_s|$ from a measurement of $\dg_{\rm CP}^{s\,\prime}$ alone.  One can
bypass the theory prediction in \eq{chmix:dgnum2} altogether by measuring both
$\dg_{\rm CP}^{s\,\prime}$ and $|\dg_s|$ and determine $|\cos \phi_s|$ through
\begin{equation}
\frac{\dg^{s\,\prime}_{\rm CP}}{\lt| \dg_s \rt|} \,=\, |\cos \phi_s | . 
 \label{chmix:cp3}
\end{equation}
To obtain additional information on $\dg_s$ and $\phi_s$ from the time evolution
in \eq{guntf} requires more statistics: the coefficient of $t$ in
\eq{chmix:guntf3}, $\dg_s\, \adg{f}/2$, vanishes in flavor-specific decays and
is equal to $-\eta_f \dg_{\rm CP}^{s\,\prime}/2 $ in the CP-specific decays of
Table~\ref{chmix:tab:cpe}. Therefore the data sample must be large enough to be
sensitive to the terms of order $(\dg_s \, t)^2$ in order to get new information
on $\dg_s$ and $\phi_s$.  We now list three methods to determine $|\dg_s|$ and
$|\cos \phi_s|$ separately \cite{dfn}. The theoretical uncertainty decreases and
the required experimental statistics increases from method 1 to method 3.  Hence
as the collected data sample grows, one can work off our list downwards. The
first method exploits information from branching ratios and needs no information
from the quadratic $(\dg_s \,t)^2$ terms.

\subparagraph{Method 1:} We assume that $\dg_{\rm CP}^{s\,\prime}$ has been
measured as described on page~\pageref{chmix:dgp}. The method presented now is a
measurement of $\dg_{\rm CP}^s$ using the information from branching ratios.
With \eq{chmix:cp2} one can then find $|\cos \phi_s|$ and subsequently $|\dg_s|$
from \eq{chmix:cp3}.  In the SV limit the branching ratio $\brunts{D_s^{(*)}{}^+
  D_s^{(*)}{}^-}$ equals $\dg_{\rm CP}^s/(2 \Gamma_s)$ up to corrections of
order $\dg_s/\Gamma_s$, as discussed above \cite{ayopr}.  Corrections to the SV
limit, however, can be sizeable. Yet we stress that one can control the
corrections to this limit experimentally, successively arriving at a result
which does not rely on the validity of the SV limit.  For this it is of prime
importance to determine the $CP$-odd component of the final states $D_s^{\pm}
D_s^{*\mp}$ and $D_s^{*+} D_s^{*-}$. We now explain how the $CP$-odd and $CP$-even
component of any decay $\Bsun \to f$ corresponding to the quark level transition
$b \to c\ov{c} s$ can be obtained. This simply requires a fit of the time
evolution of the decay to a single exponential, as in \eq{chmix:singex}.  Define
the contributions of the $CP$-odd and $CP$-even eigenstate to $B_s \to f$:
\begin{equation}
\Gamma (B_s^{\textrm{\scriptsize odd}} \to f) \, \equiv \, {\cal N}_f 
 \, |  \langle f
         \ket{B_s^{\textrm{\scriptsize odd}}} |^2,  \qquad 
\Gamma (B_s^{\textrm{\scriptsize even}} \to f) \, \equiv \, {\cal N}_f 
        \, | \langle f
         \ket{B_s^{\textrm{\scriptsize even}}} |^2 . \label{chmix:defoe}
\end{equation}
It is useful to define the $CP$-odd fraction $x_f$ by
\begin{equation}
\frac{  \Gamma (B_s^{\textrm{\scriptsize odd}}  
                \to f) }{
        \Gamma ( B_s^{\textrm{\scriptsize even}} 
                \to f) }
\,=\, \frac{  |  \langle f
         \ket{B_s^{\textrm{\scriptsize odd}}} |^2 }{
        | \langle f
         \ket{B_s^{\textrm{\scriptsize even}}} |^2 } \, =\, 
\frac{  |  \langle \ov{f}
         \ket{B_s^{\textrm{\scriptsize odd}}} |^2 }{
        | \langle \ov{f}
         \ket{B_s^{\textrm{\scriptsize even}}} |^2 } \, =\,
  \frac{x_f}{1-x_f}  . \label{chmix:x1x}
\end{equation}
The time evolution $(\guntf +\guntfb)/2$ of the CP-averaged untagged decay
$\Bsun \to f,\ov{f}$ is governed by a two-exponential formula:
\begin{equation}
\frac{\guntf +\guntfb }{2} \,=\, 
        A(f) \, e^{-\Gamma_L t} + B(f) \, e^{-\Gamma_H t} .
\label{chmix:twoex3} 
\end{equation}
With \eq{chmix:rot} and \eq{guntfeig} one finds  
\begin{eqnarray}
A(f) & = & 
 \frac{{\cal N}_f}{2} \,  | \langle  f \ket{B_L} |^2 \, +  
 \frac{{\cal N}_f}{2} \,        | \langle  \ov{f} \ket{B_L} |^2 \nn
  & = & \frac{1+ \cos \phi}{2} \, 
        \Gamma ( B_s^{\textrm{\scriptsize even}} \to f) \, + \,
        \frac{1- \cos \phi}{2} \, 
        \Gamma ( B_s^{\textrm{\scriptsize odd}} \to f) , \nn
B(f) & = & \frac{{\cal N}_f}{2} \,
         | \langle f  \ket{B_H } |^2 \, + \, 
        \frac{{\cal N}_f}{2} \, | \langle \ov{f}  \ket{B_H } |^2 \nn
  & = & \frac{ 1- \cos \phi }{2} \, 
        \Gamma ( B_s^{\textrm{\scriptsize even}} \to f) \, + \, 
          \frac{ 1+ \cos \phi }{2} \, 
        \Gamma ( B_s^{\textrm{\scriptsize odd}} \to f) 
        .\label{chmix:abg}
\end{eqnarray}
With \eq{chmix:x1x} we arrive at
\begin{equation}
\frac{A(f)}{B(f)} =  
 \frac{ (1+ \cos \phi) 
        \Gamma ( B_s^{\textrm{\scriptsize even}} \to f) + 
        (1- \cos \phi) 
        \Gamma ( B_s^{\textrm{\scriptsize odd}} \to f)}{
        (1- \cos \phi) 
        \Gamma ( B_s^{\textrm{\scriptsize even}} \to f) + 
        (1+ \cos \phi) 
        \Gamma ( B_s^{\textrm{\scriptsize odd}} \to f) 
        } 
        \, = \, 
        \frac{1+(1- 2 x_f) \cos \phi}{1-(1- 2 x_f) \cos \phi}
\,.\label{chmix:ab}
\end{equation}
In \eq{chmix:abg} and \eq{chmix:ab} it is crucial that we average the decay
rates for $\Bsun \to f$ and the CP-conjugate process $\Bsun \to \ov{f}$. This
eliminates the interference term $\bra{B_s^{\textrm{\scriptsize odd}}} f\rangle
\bra{f} B_s^{\textrm{\scriptsize even}} \rangle $, so that $A(f)/B(f)$ only
depends on $x_f$.  The single-exponential fit with \eq{chmix:singex} determines
$\Gamma_f$. Equations \eq{chmix:fit} and \eq{chmix:ab} combine to give
\begin{equation}
2\, ( \Gamma_f - \Gamma_s ) =
        (1- 2 x_f ) \, \dg_s \, \cos \phi_s 
  \, = \, (1- 2 x_f ) \, \dg_{\rm CP}^s \, \cos^2 \phi_s
  \, = \, (1- 2 x_f ) \, \dg_{\rm CP}^{s\,\prime} \,,
 \label{chmix:dgmx} 
\end{equation}
up to corrections of order $(\dg_s)^2/\Gamma_s$. In order to determine $x_f$
from \eq{chmix:dgmx} we need $\dg_{\rm CP}^{s\,\prime}$ from the
lifetime measurement in a CP-specific final state like $D_s^+ D_s^-$%
\index{decay!$B_s \rightarrow D_s^+ D_s^-$} or from the angular separation of
the $CP$ components in $\Bsun \to \psi
\phi$.%
\index{decay!$B_s \rightarrow \psi \phi$} 
The corrections of order $(\dg_s)^2/\Gamma_s$ to \eq{chmix:dgmx} can be read off
from \eq{chmix:fit} with \eq{chmix:ab} as well.  Expressing the result in terms
of $\Gamma_f$ and the rate $\Gamma_{\rm fs}$ measured in flavor-specific decays,
we find
\begin{equation}
1- 2\, x_f \,=\, 2\, 
        \frac{\Gamma_f-\Gamma_{\rm fs}}{\dg_{\rm CP}^{s\,\prime}} 
        \, \lt[ 1 \, -  \,  
        2\, \frac{\Gamma_f-\Gamma_{\rm fs}}{\Gamma_s} \rt] \, 
        + {\cal O} \lt( \frac{(\dg_s)^2}{\Gamma_s^2} \rt)
        . \label{chmix:xfres} 
\end{equation} 
In order to solve for $\Gamma (B_s^{\textrm{\scriptsize even}} \to f) $ and
$\Gamma (B_s^{\textrm{\scriptsize odd}} \to f) $ we also need the branching
ratio $\brunts{f}+\brunts{\ov{f}}$. Recalling \eq{untnorm} one finds from
\eq{chmix:twoex3} and \eq{chmix:abg}:
\begin{eqnarray}
\brunts{f}+\brunts{\ov{f}} 
&=& \phantom{ + }
\Gamma (B_s^{\textrm{\scriptsize even}}  \to f)
  \lt[    \frac{1+ \cos \phi_s }{2 \Gamma_L}  + 
         \frac{1- \cos \phi_s }{2 \Gamma_H} \rt]  \nn
&&  + \, 
\Gamma (B_s^{\textrm{\scriptsize odd}}  \to f)
  \lt[    \frac{1- \cos \phi_s }{2 \Gamma_L}  + 
         \frac{1+ \cos \phi_s }{2 \Gamma_H} \rt] 
. \label{chmix:boe}
\end{eqnarray}
By combining \eq{chmix:x1x} and \eq{chmix:boe} we can solve for the two CP
components:
\begin{eqnarray}
\Gamma (B_s^{\textrm{\scriptsize even}} \to f)
&=& 
        \lt[ \Gamma_s^2- \lt( \dg_s/2 \rt)^2 \rt]
        \lt( \brunts{ f} + \brunts{\ov{f}} \rt)    
   \frac{1- x_f}{ 2 \Gamma_s \, - \, \Gamma_f } \nn
& = &   (1- x_f) \lt( \brunts{f} + \brunts{\ov{f}} \rt)
        \Gamma_s \,+\, {\cal O} \lt( \dg_s  \rt) \nn  
\Gamma (B_s^{\textrm{\scriptsize odd}}  \to f)
&=&  
        \lt[ \Gamma_s^2- \lt( \dg_s/2 \rt)^2 \rt]
        \lt( \brunts{f} + \brunts{\ov{f}} \rt)
   \frac{x_f}{ 2 \Gamma_s \, - \, \Gamma_f } \nn
& = &   x_f  \lt( \brunts{ f} + \brunts{\ov{f}} \rt)   \,
         \Gamma_s \,  
        +\, {\cal O} \lt( \dg_s \rt) .
\end{eqnarray}
From \eq{chmix:dgcp2} we now find the desired quantity by summing over all final
states $f$:
\begin{eqnarray} 
\dg_{\rm CP}^s & = & 
\Gamma \lt( B_s^{\textrm{\scriptsize even}} \rt) - 
        \Gamma \lt( B_s^{\textrm{\scriptsize odd}} \rt) 
\, = \,
        2 \lt[ \Gamma_s^2- \lt( \dg_s/2 \rt)^2 \rt]
\sum_{f \in X_{c\ov{c}}}  \brunts{ f}    \,
   \frac{1- 2\, x_f}{ 2 \Gamma_s \, - \, \Gamma_f} 
  \nn 
&=& 2\,  \Gamma_s \sum_{f \in X_{c\ov{c}}}  \brunts{ f} 
        \, 
        (1- 2\, x_f) \lt[ 1 \, + \, {\cal O} \lt( \frac{\dg_s}{\Gamma_s}
        \rt) \rt].
\label{chmix:dgcpres} 
\end{eqnarray}
\index{decay!$B_s \rightarrow D_s^+ D_s^-$}%
\index{decay!$B_s \rightarrow D_s^{*+} D_s^-$!$CP$-odd fraction}%
\index{decay!$B_s \rightarrow D_s^+ D_s^{*-}$!$CP$-odd fraction}%
\index{decay!$B_s \rightarrow D_s^{*+} D_s^{*-}$!$CP$-odd fraction}%
It is easy to find $\dg_{\rm CP}^s$: first determine $1-2x_f$ from
\eq{chmix:xfres} for each studied decay mode, then insert the result into
\eq{chmix:dgcpres}. The small quadratic term $( \dg_s/2 )^2=\dg_{\rm CP}^s
\dg_{\rm CP}^{s\,\prime}/4$ is negligible. 
This procedure can be performed
for $\brunts{D_s^{\pm} D_s^{*}{}^{\mp}}$ and $\brunts{D_s^{*}{}^+ D_s^{*}{}^-}$
to determine the corrections to the SV limit. In principle the $CP$-odd P-wave
component of $\brunts{D_s^{*}{}^+ D_s^{*}{}^-}$ (which vanishes in the SV limit)
could also be obtained by an angular analysis, but this is difficult in
first-generation experiments at hadron colliders, because the photon from
$D_s^* \to D_s \gamma$ cannot be detected.  We emphasize
that it is not necessary to separate the $D_s^{(*)}{}^+D_s^{(*)}{}^-$ final
states; our method can also be applied to the semi-inclusive $D_s^{(*)}{}^\pm
D_s^{(*)}{}^\mp $ sample, using $\dg_{\rm CP}^{s\,\prime}$ obtained from an
angular separation of the CP
components in $\Bsun \to \psi \phi$.%
\index{decay!$B_s \rightarrow \psi \phi$} Further one can successively include
those double-charm final states which vanish in the SV limit into
\eq{chmix:dgcpres}.  If we were able to reconstruct all $b \to c \ov{c} s$ final
states, we could determine $\dg_{\rm CP}^s$ without invoking the SV limit.  In
practice a portion of these final states will be missed, but the induced error
can be estimated from the corrections to the SV limit in the measured decay
modes. By comparing $\dg_{\rm CP}^s$ and $\dg_{\rm CP}^{s\,\prime}$ one finds
$|\cos \phi_s|$ from \eq{chmix:cp2}.  The irreducible theoretical error of
method 1 stems from the omission of CKM-suppressed decays and is of order $2
|V_{ub} V_{us}/(V_{cb} V_{cs})| \sim 3-5\%$.

Method 1 is experimentally simple: at the first stage (relying on the SV limit)
it amounts to counting the $\Bsun$ decays into $ D_s^{(*)}{}^+ D_s^{(*)}{}^-$.
The corrections to the SV limit are obtained by one-parameter fits to the time
evolution of the collected double-charm data samples.  This sample may include
final states from decay modes which vanish in the SV limit, such as
multiparticle final states.  No sensitivity to $(\dg_s \, t)^2$ is needed.  A
further advantage is that $\dg_{\rm CP}^s$ is not diminished by the presence of
new physics.
\index{B meson@$B$ meson!width difference $\dg$!$\dg_s$ and branching ratios}%
\index{width difference $\dg$!measurement of $\dg_s$!branching ratios}

\subparagraph{Method 2:} \index{untagged $B$!two-exponential decay} In the
Standard Model the decay into a $CP$ eigenstate $f_{\rm CP}$ is governed by a
single exponential. If a second exponential is found in the time evolution of a
CKM-favored decay $\Bsun \to f_{\rm CP}$, this will be clear evidence of new
physics \cite{dun}. To this end we must resolve the time evolution in \eq{guntf}
up to order $(\dg_s \, t)^2$. At first glance this seems to require a
three-parameter fit to the data, because \guntf\ in \eq{guntf} depends on
$\Gamma_s$, $\dg_s$ and (through $\adg{f}$, see \eq{acp2}) on $\phi_s$. It is
possible, however, to choose these parameters in such a way that one of them
enters \gunt{f_{\rm CP}}\ at order $(\dg_s )^3$, with negligible impact.  The
fit parameters are $\Gamma^\prime$ and $Y$.  They are chosen such that
\begin{equation} 
\guntfcpp \,=\,  2\, \brunts{f_{\rm CP+} } \,
        \Gamma^\prime e^{-\Gamma^\prime t}
        \lt[ 1 + Y \, \Gamma^\prime \, t \, 
        \lt( -1 
                +\frac{\Gamma^\prime t}{2} \rt)
                + {\cal O} \lt( (\dg_s )^3 \rt)
         \rt] 
        . \label{chmix:t2}
\end{equation}
Here we have considered a $CP$-even final state, for which a lot more data are
expected than for $CP$-odd states.  With \eq{chmix:t2} we have generalized the
lifetime fit method described in \eq{chmix:guntf3} -- \eq{chmix:gafs} to the
order $(\dg\, t)^2$.  A non-zero $Y$ signals the presence of new physics.  The
fitted rate $\Gamma^\prime$ and $Y$ are related to $\Gamma_s$, $\dg_s$ and
$\phi_s$ by
\begin{equation} 
Y \, = \, \frac{(\dg_s)^2}{4 \Gamma^{\prime 2}} \sin^2 \phi_s 
        ,  \qquad \qquad
\Gamma^\prime \,=\, \Gamma_s (1-Y) + \frac{\cos \phi_s}{2} \dg_s
      \label{chmix:defgpy} .
\end{equation} 
Note that for $|\cos \phi_s|=1$ the rate $\Gamma^\prime$ equals the rate of the
shorter-lived mass eigenstate and the expansion in \eq{chmix:t2} becomes the
exact single-exponential formula.  After determining $\Gamma^\prime$ and $Y$ we
can solve \eq{chmix:defgpy} for $\Gamma_s$, $\dg_s$ and $\phi_s$. To this end we
need the width $\Gamma_{\rm fs}$ measured in flavor-specific decays. We find
\begin{eqnarray}
|\dg_s| &\,=\,& 2 \sqrt{(\Gamma^\prime-\Gamma_{\rm fs})^2 +\Gamma_{\rm fs}^2} 
        \lt[ 1+ {\cal O} \lt( \frac{\dg_s}{\Gamma_s}  \rt) \rt], \nn
\Gamma_s\, &=& \, \Gamma_{\rm fs} + \frac{(\dg_s)^2}{2 \Gamma_s} + 
                {\cal O} \lt( \lt(\frac{\dg_s}{\Gamma_s}\rt)^3  \rt) \nn 
\dg_{\rm CP}^{s\,\prime} \, &=&\, 
  2 \lt[ \Gamma^\prime - \Gamma_s \lt( 1- Y \rt) \rt] 
        \lt[ 1 + {\cal O} \lt( \lt(\frac{\dg_s}{\Gamma_s}\rt)^2 \rt) \rt], \nn
|\sin \phi_s| \, &=& \, \frac{2\Gamma_s \sqrt{Y}}{|\dg_s|} \, 
        \lt[ 1+ {\cal O} \lt( \frac{\dg_s}{\Gamma_s}  \rt) \rt] . 
\label{chmix:gpyres}
\end{eqnarray} 
The quantity $\dg_{\rm CP}^{s\,\prime}$, which we could already determine from
single-exponential fits, is now found beyond the leading order in
$\dg_s/\Gamma_s$. By contrast, $\dg_s$ and $|\sin \phi_s|$ in \eq{chmix:gpyres}
are only determined to the first non-vanishing order in $\dg_s/\Gamma_s$.

In conclusion, method 2 involves a two-parameter fit and needs
sensitivity to the qua\-dratic term in the time evolution. The presence
of new physics can be invoked from $Y\neq 0$ and does not require to
combine lifetime measurements in different decay modes. 

\subparagraph{Method 3:} \index{untagged $B$!two-exponential decay} Originally
the following method has been proposed to determine $|\dg_s|$ \cite{dun,g}: The
time evolution of a $\Bsun$ decay into a flavor-specific final state is fitted
to two exponentials. This amounts to resolving the deviation of $\cosh (\dg_s\,
t/2)$ from 1 in \eq{guntf} in a two-parameter fit for $\Gamma_s$ and $|\dg_s|$.
If one adopts the same parameterization as in \eq{chmix:t2}, $\Gamma^\prime$ and
$Y$ are obtained from \eq{chmix:defgpy} by replacing $\phi_s$ with $\pi/2$.  The
best suited flavor-specific decay modes at hadron colliders are $\Bsun \to
D_s^{(*)\pm} \pi^\mp $, $\Bsun \to D_s^{(*)\pm} \pi^\mp
\pi^+ \pi^- $ and $\Bsun \to X \ell^{\mp} \nu$.%
\index{decay!$B_s \rightarrow D_s^- \pi^+$}%
\index{decay!$B_s \rightarrow D_s^- \pi^+ \pi^- \pi^+$}%
\index{decay!$B_s \rightarrow D_s^{*-} \pi^+$}%
\index{decay!$B_s \rightarrow D_s^{*-} \pi^+ \pi^- \pi^+$}%
\index{decay!$B_s \rightarrow X \ell^+ \nu$}%
Depending on the event rate in these modes, method 3 could be superior to method
2 in terms of statistics. On the other hand, to find the ``smoking gun'' of new
physics, the $|\dg_s|$ obtained must be compared to $\dg_{\rm CP}^{s\,\prime}$
from CP-specific decays to prove $|\cos \phi_s|\neq 1$ through \eq{chmix:cp3}.
Since the two measurements are differently affected by systematic errors, this
can be a difficult task. First upper bounds on $|\dg_s|$ using method 3 have
been obtained in \cite{semi}.

The L3 collaboration has determined an upper bound $|\dg_s|/\Gamma_s\leq 0.67$
by fitting the time evolution of fully inclusive decays to two exponentials
\cite{l3}. This method is quadratic in $\dg_s$ as well. The corresponding
formula for the time evolution can be simply obtained from \eq{chmix:twoex2}
with $A=\Gamma_L$ and $B=\Gamma_H$.

\boldmath
\subsubsection{Phenomenology of $\dg_d$}
\unboldmath
\label{chmix:subsub:width:phen:d}

\index{B meson@$B$ meson!width difference $\dg$!measurement of $\dg_d$}%
\index{width difference $\dg$!measurement of $\dg_d$} The Standard Model
value $\dg_d^{\rm SM}/\Gamma_d \approx 3\times 10^{-3}$ derived before
\eq{chmix:dgdsm} is presumably too small to be measured from lifetime
fits. In extension of the Standard Model, however, $\dg_d/\Gamma_d$
can be large, up to a few percent. The expected high statistics for
the decay $B_d \to \psi K_S$ can be used to measure the lifetime
$1/\Gamma_{B_d\to \psi K_S}$ in this channel with 
\index{decay!$B_d \to \psi K_S$}
\begin{equation}    
\Gamma_{B_d \to \psi K_S} =
                \Gamma_d - \frac{\dg^d}{2} \cos (2 \beta_{\psi K_S})
\, = \,  \Gamma_d - 
        |\Gamma_{12}^d | \cos (2 \beta_{\psi K_S})   
        \cos  \phi_d 
        .\label{chmix:dgp:d}
\end{equation}
$\sin (2 \beta_{\psi K_S})$ is the quantity characterizing the
mixing-induced $CP$ asymmetry measured from tagged $B_d \to \psi K_S$
decays.  $\Gamma_d$ is obtained from a lifetime measurements in
flavor-specific decay modes.  We stress that this measurement of
$\Gamma_{B_d\to \psi K_S}$ can be done from the \emph{untagged}\ 
$\Bdun \to \psi K_S$ data sample. If $\Gamma_{12}^d$ is dominated by
new physics, its phase and therefore also $\phi_d$ is unknown.  If one
neglects the small SM contribution in \eq{chmix:phinum} to $\phi_d$,
\eq{chmix:dgp:d} reads 
\begin{equation}
\Gamma_{B_d \to \psi K_S} \simeq
   \Gamma_d - |\Gamma_{12}^d | \,\cos (2 \beta_{\psi K_S}) 
              \cos (2 \beta_{\psi K_S} - 2 \beta)  \, = \, 
   \Gamma_d - |\Gamma_{12}^d | \cos (\phi_d + 2 \beta) 
              \cos \phi_d  , 
\end{equation} 
where $\beta$ is the true angle of the unitarity triangle as defined
in \eq{1:eq:beta}. Note that in the presence of new physics $\beta$ is
unknown.  When combined with the $CP$ asymmetry in flavor-specific
decays discussed in Sect.~\ref{chmix:sub:afs} one can determine
$|\Gamma_{12}^d|$ and $\sin \phi_d$.  Then up to discrete ambiguities
also $\beta=\beta_{\psi K_S}-\phi_d/2$ can be determined.  Depending
on whether the enhancement of $\Gamma_{12}^d$ is due to $b\to \ov{c}c
d$, $b\to \ov{u}c d$ or $b\to \ov{u}u d$, transitions, $CP$ asymmetries
in these channels can also help to disentangle $|\Gamma_{12}^d|$ and
$\phi_d$.

Interestingly, one can isolate the contribution to 
$|\Gamma_{12}^d| $ from $b\to
\ov{c} c d$ decays. Define
\begin{equation}
\dg_{\rm CP}^{d,cc} \equiv 2\, | \xi_c^{d\,2} \Gamma_{12}^{d,cc}| \, = \, 
2 \sum_{f  \in X_{c\ov{c}}} \lt[
\Gamma ( B_d \to f_{\rm CP+} ) \, - \,
        \Gamma ( B_d \to f_{\rm CP-}) \rt]
        . 
        \label{chmix:dgcp:d}
\end{equation}
in analogy to \eq{chmix:dgcp}. In the Standard Model $\dg_{\rm CP}^{d,cc} $ is
slightly larger than $2 |\Gamma_{12}^d|$, by a factor of $1/R_t^2$.  From
\eq{chmix:dgdsm} one finds
\begin{equation} 
\frac{\dg_{\rm CP}^{d,cc}}{\dg^s_{\rm CP}} \simeq 
  \frac{f_{B_d}^2 B_{B_d}^S}{f_{B_s}^2 B_{B_s}^S} 
              \, \lt| \frac{\xi_c^d}{\xi_t^s} \rt|^2 \, \simeq \,
0.04, \label{chmix:dgcc}
\end{equation} 
i.e.\ $\dg_{\rm CP}^{d,cc}/\Gamma_d \simeq 5\times 10^{-3}$. Now
$\dg_{\rm CP}^{d,cc} $ can be measured by counting the $CP$-even and
$CP$-odd final states in $b\to c \ov{c} d$ decays, just as described in
Sect.~\ref{chmix:subsub:width:phen} for $b\to c \ov{c} s$ decays of
$B_s$ mesons. Again, in the SV limit the inclusive decay $\Bdun \to
X_{c\ov{c}d}$ is exhausted by $\Bdun \to D^{(*)+} D^{(*)-}$, which is
purely $CP$-even in this limit.  With \eq{chmix:dgcpres} one can find
$\dg_{\rm CP}^{d,cc}$.  That is, in the SV limit one just has to
measure $Br \big(\Bdun \to D^{(*)+} D^{(*)-}\big)$, which equals
$\dg_{\rm CP}^{d,cc}/(2 \Gamma_d)$.  However, the lifetime method
described in `Method 1' above cannot be used to determine the
corrections to the SV limit, because $\dg_d$ is too small. Yet in the
limit of exact U-spin symmetry ($m_d=m_s$) the $CP$-odd components of
$D^{(*)+} D^{(*)-}$ from $B_d$ decay and $D_s^{(*)+} D_s^{(*)-}$ from
$B_s$ decay are the same.  Finally in $b\to c \ov{c} d$ transitions CP
violation in decay could be relevant. It results from penguin loops
involving top- or up-quarks and spoils the relation \eq{chmix:dgcpres}
between branching ratios and $\dg_{\rm CP}^{d,cc} $.  This effect,
however, is calculable for inclusive decays like $\Bdun \to
X_{c\ov{c}d}$. In the Standard Model $CP$ violation in this inclusive
decay is of order $1\%$ and therefore negligible \cite{lno}.
CP-violation from non-standard sources can be revealed by comparing
CP-asymmetries in $b\to c \ov{c} d$ decays with those in $b\to c\ov{c}
s$ decays (namely $\sin 2 \beta$ from $B_d\to \psi K_S$).  Since
\eq{chmix:dgcc} depends on no CKM elements and the hadronic factor is
known exactly in the $SU(3)_F$ limit, a combined measurement of
$\dg_{\rm CP}^{d,cc}$ and $\dg_{\rm CP}^s$ provides an excellent probe
of new physics in $b\to c \ov{c} d$ transitions.

\subsection[$CP$ Asymmetry in Flavor-specific Decays]
{\boldmath $CP$ Asymmetry in Flavor-specific Decays}
\label{chmix:sub:afs}

\index{CP asymmetry@\CP\ asymmetry!in flavor-specific decay} In the preceding
sections we have set the small parameter $a_q\equiv a(B_q)$, $q=d,s$, defined in
\eq{defepsg} to zero. In order to study $CP$ violation in mixing we must keep
terms of order $a_q$. The corresponding ``wrong-sign'' $CP$ asymmetry is measured
in flavor-specific decays $B_q \to f$ and equals
\begin{equation}
a^q_{\rm fs} 
     = \frac{\gqbtf - \gqtfb}{\gqbtf + \gqtfb} 
     = \imag \frac{\Gamma_{12}^q}{M_{12}^q} = a_q \,,
        \qquad  \mbox{for~ }  \ov{A}_f  =  0 
         \mbox{ ~and~ } |A_f| = |\ov{A}_{\ov{f}}|
 . \label{chmix:defafs}
\end{equation}
\index{CP asymmetry@\CP\ asymmetry!semileptonic}%
A special case of $a^q_{\rm fs}$ is the semileptonic asymmetry, where $f=X
\ell^+ \nu$, introduced in Sect.~\ref{ch1:sect:ty}.  A determination of $a_q$
gives additional information on the three rephasing-invariant quantities
$|M_{12}^q|$, $|\Gamma_{12}^q|$ and $\phi_q$ characterizing \bbm.

Observe that $a^q_{\rm fs}$ in \eq{chmix:defafs} is time-independent. While both
numerator and denominator depend on $t$, this dependence drops out from the
ratio.  The ``right-sign'' asymmetry, vanishes:
\begin{equation}
    \gqtf - \gqbtfb \, = \, 0\,,
        \qquad \mbox{for~ } \ov{A}_f  =  0 
        \mbox{ ~and~ }  |A_f| = |\ov{A}_{\ov{f}}|
 . \label{chmix:unm}
\end{equation}
This implies that one can measure $a^q_{\rm fs}$ from \emph{untagged} decays
\cite{y,dfn}. It is easily verified from the sum of \eq{gtfres} and \eq{gbtfres}
that to order $a_q$ the time evolution of untagged decays exhibits oscillations
governed by $\dm_q$.  Since $a$ is small, a small production asymmetry $\epsilon
= N_{\ov{B}}/N_B -1$, which also leads to oscillations in the untagged sample,
could introduce an experimental bias. To first order in the small parameters
$a_q$, and $\epsilon$ one finds
\begin{equation}
a_{\rm fs}^{q,unt} =
   \frac{\guntf - \guntfb}{\guntf + \guntfb}  =
        \frac{a_q}{2} - \frac{a_q+\epsilon}{2} \,  
        \frac{\cos (\dm_q\, t)}{\cosh (\dg_q t/2) }\,,
        \qquad \mbox{for~ } \ov{A}_f  =  0 
        \mbox{ ~and~ } |A_f| = |\ov{A}_{\ov{f}}| 
        . \,  \label{chmix:fsun} 
\end{equation}
Note that the production asymmetry between $B_q^0$ and $\ov{B}{}_q^0$ cannot
completely fake the effect of a non-zero $a_q$ in \eq{chmix:fsun}: while both
$a_q\neq 0$ and $\epsilon\neq 0$ lead to oscillations, the offset from the
constant term indicates $a_q\neq 0$.

The Standard Model predictions for $a_d$ and $a_s$ are 
\index{CP asymmetry@\CP\ asymmetry!in flavor-specific decay!SM prediction}%
\begin{eqnarray}
a_d & \approx & - \frac{\ov{\eta}}{R_t^2} 
   \frac{4 \pi \, \lt( K_1+K_2 \rt) m_c^2}{M_W^2\, \eta_B b_B  
  S (m_t^2/M_W^2) } \, \approx \, - 8 \times 10^{-4} \nn 
a_s & \approx & \ov{\eta} \lambda^2 \, 
   \frac{4 \pi \, \lt( K_1+K_2 \rt) m_c^2}{M_W^2\, \eta_B b_B  
  S (m_t^2/M_W^2) } \, \approx \, 5 \times 10^{-5}. \label{chmix:aqnum}
\end{eqnarray}
The huge GIM suppression factor $m_b^2/M_W^2 \sin \phi_d \propto m_c^2/M_W^2$
leads to these tiny predictions for $CP$ violation in mixing.  $a_d$ plays a
preeminent role in the search for new physics :
\begin{itemize}
\item its sensitivity to new physics is enormous, it can be enhanced
  by two orders of magnitude, 
\item it is affected by a wide range of possible new physics effects: 
      new $CP$ violating effects in $\phi_d$ relax the GIM-suppression
      $ \propto m_b^2/M_W^2 \sin \phi_d$, because $\phi_d$ is no more 
      proportional to $z=m_c^2/m_b^2$. New physics contributions to 
      \emph{any}\ of the CKM-suppressed decay modes $b\to \ov{c}c d$, $b\to
      \ov{u}c d$ or $b\to \ov{u}u d$ can significantly enhance 
      $|\Gamma_{12}^d|$ and thereby $a_d$. 
\end{itemize}
Of course new physics contributions to $\arg M_{12}^d$ will not only affect
$\phi_d$, but also the $CP$ asymmetry in $B_d^0 \to \psi K_S$.\index{decay!$B_d
  \rightarrow \psi K_S$} But from this measurement alone one cannot extract the
new physics contribution, because one will know the true value of $\beta= \arg
(-\xi_{t}^{d*}/\xi_{c}^{d*})$ only poorly, once new physics affects the standard
analysis of the unitarity triangle. For the discussion of new physics it helps
to write
\begin{equation}
a_q  \,=\, \frac{2 |\Gamma_{12}^q|}{\dm_q} \, \sin \phi_q  
\, = \,
\frac{|\dg_q|}{\dm_q} \, \frac{\sin \phi_q}{|\cos \phi_q|}. 
\label{chmix:aqnp}
\end{equation}
If both $|\dg_d|$ and $a_d$ are measured, one can determine both
$|\Gamma_{12}^d|$ and $\sin \phi_d$.

$a_s$ is less interesting than $a_d$, because $\Gamma_{12}^s$ stems from
CKM-favored decays and is not very sensitive to new physics.  The ratio
$\dg_{\rm CP}^s/\Gamma_s \leq 0.2$ from \eq{chmix:dgnum2} and the current
experimental limit $\dm_s \geq 14.9\,$ps${}^{-1}$ \cite{os} imply that $|a_s|
\leq 0.01$. New physics can affect $\phi_s$ only through $\arg M_{12}^s$, but
this new physics can be detected most easily through $CP$ asymmetries in $B_s\to
\psi \phi$ or
\index{decay!$B_s \rightarrow \psi \phi$}
\index{decay!$B_s \rightarrow D_s^+ D_s^-$}
\index{decay!$B_s \rightarrow D_s^{*+} D_s^-$!$CP$-odd fraction}
\index{decay!$B_s \rightarrow D_s^+ D_s^{*-}$!$CP$-odd fraction}
\index{decay!$B_s \rightarrow D_s^{*+} D_s^{*-}$!$CP$-odd fraction}
$B_s \to D_s^{(*)+} D_s^{(*)-}$ decays. Since the Standard Model predictions for
these asymmetries are essentially zero, there is no problem here to disentangle
standard from non-standard physics.  Note that the measurement of $\mbox{sgn}\,
\sin \phi_s$ reduces the four-fold ambiguity in $\phi_s$ from the measurement of
$|\cos \phi_s| $ to a two-fold one.  The unambiguous determination of $\phi_s$ is
discussed in detail in \cite{dfn}.  \index{discrete ambiguity of $CP$ phase}

\subsection[Angular analysis to separate the $CP$ components]{\boldmath Angular analysis to
separate the $CP$ components$\!$ \authorfootnote{Author: Amol Dighe}}

\subsection[$CP$-odd and $CP$-even components in $B_s \to J/\psi\phi$]
{\boldmath $CP$-odd and $CP$-even components in $B_s \to J/\psi \phi$} 
\label{chmix:twocp}

The most general amplitude for $B_s \to J/\psi \phi$ can be written in terms of
the polarization states of the two vector mesons as \cite{rosner,ddlr}
\beq
A(B_s(t) \to J/\psi\, \phi) = \frac{A_0(t)}{x}\, {\bep}^{*L}_{J/\psi}\,
{\bep}^{*L}_{\phi} - A_{\|}(t)\, {\bep}^{*T}_{J/\psi} \cdot {\bep}^{*T}_{\phi} /
\sqrt{2} - i A_{\perp}(t)\, {\bep}^*_{J/\psi} \times {\bep}^*_{\phi} \cdot
\hat{\bf p}_{\phi} / \sqrt{2}\,,
\label{chmix:ampl}
\eeq
where $x\equiv p_{J/\psi}\cdot p_{\phi}/(m_{J/\psi} m_{\phi})$ and $\hat{\bf
  p}_{\phi}$ is the unit vector along the direction of motion of $\phi$ in the
rest frame of $J/\psi$.

Since the ``$CP$ violation in decay'' of $B_s \to J/\psi \phi$ is vanishing,
\beq
\overline{A}_0(0) = A_0(0) \,, \qquad
\overline{A}_{\|}(0) = A_{\|}(0) \,,\qquad
\overline{A}_{\perp}(0) = -A_{\perp}(0)\,.
\label{chmix:cpstates}
\eeq
The final state is thus an admixture of different $CP$ eigenstates: $A_0$ and
$A_\|$ are $CP$-even amplitudes whereas $A_\perp$ is $CP$-odd. The decay rate is
given by \beq \Gamma (t) \propto |A_0(t)|^2 + |A_{\|}(t)|^2 +
|A_{\perp}(t)|^2\,,
\label{chmix:no-angle}
\eeq
where the time evolutions of the individual terms are
\cite{ddf1} 
\beqa
|A_{0, \|}(t)|^2 & =  & |A_{0, \|}(0)|^2 \left[e^{-\Gamma_L t} -
e^{-\overline{\Gamma}t}
\sin(\Delta m_s t)\delta\phi\right], \nonumber \\
|A_{\perp}(t)|^2 & =  & |A_{\perp}(0)|^2 \left[e^{-\Gamma_H t} +
e^{-\overline{\Gamma}t}
\sin(\Delta m_s t)\delta\phi\right] \,.
\label{chmix:time-one}
\eeqa
Here, $\bar{\Gamma} \equiv \Gamma_s = (\Gamma_L + \Gamma_H) /2$. Note that
this is {\it not} the average lifetime of $B_s$ as measured
through its semileptonic decays \cite{paulini}.

The value of 
\beq
\delta \phi \equiv 2 \beta_s \approx 2 \lambda^2 \eta \approx 0.03
\label{chmix:deltaphi}
\eeq is small in the standard model\footnote{Generalizations of the
  formulae to the case of new physics can be found in \cite{dfn}.}, so
that the terms proportional to $\delta \phi$ in (\ref{chmix:time-one})
can be neglected in the first approximation.  The time evolution of
(\ref{chmix:no-angle}) is then a sum of two exponential decays with
lifetimes $1/\Gamma_H$ and $1/\Gamma_L$.

In principle, a fit to the time dependence of
the total decay rate (\ref{chmix:no-angle})
can give the values of $\Gamma_H$ and $\Gamma_L$ separately,
but $\Delta \Gamma_s / \bar{\Gamma}$ is expected to be less than 20\%,
and it is not easy to separate two closely spaced lifetimes.
The inclusion of angular 
information will increase the accuracy in the measurement of 
$\Delta \Gamma_s$ multi-fold, as we'll see in the section
\ref{chmix:transversity} below.

\subsection{The transversity angle distribution}
\label{chmix:transversity}

Since there are four particles in the final state,
the directions of their momenta can define three 
independent physical angles. Our
convention for the definitions of angles \cite{ddlr,ddf1}
is as shown in Fig.~\ref{chmix:ang-conv}.
The $x$-axis is the direction of $\phi$ in  
the $J/\psi$ rest frame, the $z$  axis is perpendicular to 
the decay plane of $\phi \to K^+ K^-$, and $p_y(K^+)
\geq 0$. The coordinates $(\theta, \varphi)$ describe the 
direction of $l^+$ in the $J/ \psi$ rest frame and $\psi$ is 
the angle made by $\vec p(K^+)$ with the $x$ axis in the $\phi$ 
rest frame. With this convention,
\beqa
& {\bf x} = {\bf p}_{\phi} , \qquad
{\bf y} = \displaystyle \frac{ {\bf p}_{K^+} - {\bf p}_{\phi} ( {\bf p}_{\phi}
	\cdot {\bf p}_{K^+} ) }
	{ | {\bf p}_{K^+} - {\bf p}_{\phi} ( {\bf p}_{\phi}
	\cdot {\bf p}_{K^+} ) | } , \qquad
{\bf z} = {\bf x} \times {\bf y}, & \nonumber \\[4pt]
&\sin \theta \cos \varphi =   {\bf p}_{\ell^+} \cdot  {\bf x}, \qquad
\sin \theta \sin \varphi = {\bf p}_{\ell^+} \cdot  {\bf y}, \qquad
\cos \theta =  {\bf p}_{\ell^+} \cdot {\bf z}\,.&
\label{chmix:angdef}
\eeqa
Here, the bold-face characters represent {\it  unit} 3-vectors and
everything is measured in the rest frame of $J/\psi$.
Also
\beq
\cos \psi = - {\bf p}'_{K^+}
                        \cdot {\bf p}'_{J/\psi},
\label{chmix:psidef}
\eeq
where the primed quantities are {\it  unit vectors}
measured in the rest frame of $\phi$.

\begin{figure}[t]
\begin{center}
\epsfig{file=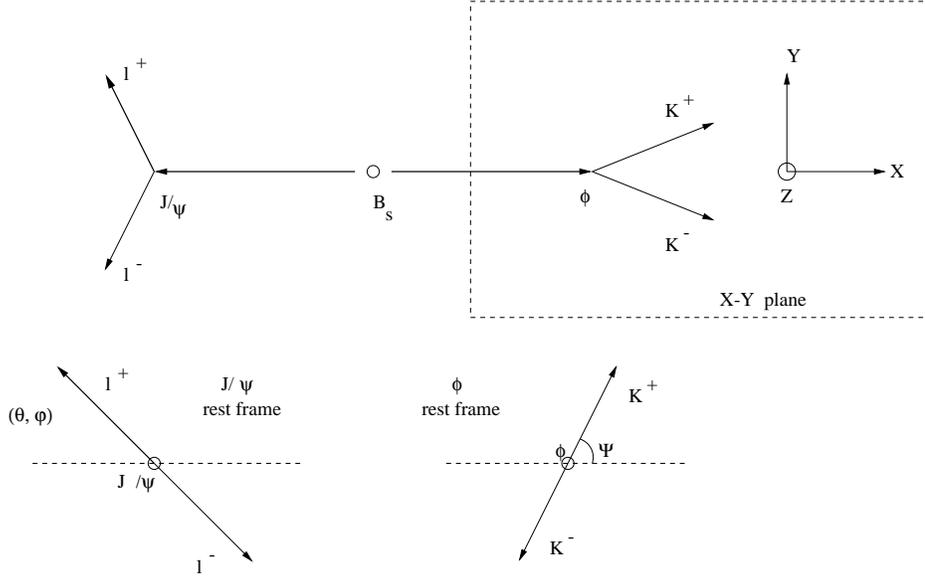,height=3.0in}
\caption{The definitions of angles $\theta, \varphi, \psi$.
Here $\theta$ is the ``transversity'' angle.}
\label{chmix:ang-conv}
\end{center}
\end{figure}

The $\theta$ defined here is the {\it transversity} angle
\cite{dqstl}, which separates out the $CP$-even and
$CP$-odd components. The angular distribution in terms of
$\theta$ is given by \cite{ddlr}:
\beq
\frac{d \Gamma (t)}{d \cos \theta} \propto
   (|A_0(t)|^2 + |A_{\|}(t)|^2)\, \frac{3}{8}\, (1 + \cos ^2 \theta)
                + |A_{\perp}(t)|^2\, \frac{3}{4} \sin^2 \theta \, ,
\label{chmix:one-angle}
\eeq
where the time evolutions of the terms are as given in
(\ref{chmix:time-one}).

The $CP$-even and $CP$-odd components are now separated by not only
their different lifetimes (which are very close) but also by their
decay angular distributions (which are distinctly different). The
study of information content about the value of $\Delta \Gamma_s$ in
the time and angular measurements \cite{sen} suggests that, in order
to get the same degree of accuracy in $\Delta \Gamma_s$ with only time
measurements, one would need about two orders of magnitude more number
of events than if both the time and angular measurements were used
(see Fig.~3 in \cite{sen}).  This indicates that the strategy of
selecting one decay mode ({\it e.g.} $J/\psi \phi$) and studying its
angular distribution will turn out to be more fruitful than trying to
combine all $CP$ eigenstate decay modes and determine $\Gamma_H$ and
$\Gamma_L$ solely from their time evolutions.  Note that, in the
limiting case of $\Gamma_H = \Gamma_L$, the time evolution by itself
{\it cannot} separate the $CP$ even and odd components, whereas the
angular measurements can.

A fit to the transversity angle distribution (\ref{chmix:one-angle}) with its
complete time evolution (\ref{chmix:time-one}) also gives the value of $\delta
\phi$ and $\Delta m_s$, though a better measurement of the latter may be obtained
through other decay channels.

The transversity angle distribution (\ref{chmix:one-angle}) is valid for any
$B_s \to J/\psi(\to \ell^+ \ell^-) C_1 C_2$ decay, where $C_1$ and $C_2$ are (a)
self-conjugate particles, or (b) scalars and $CP$ conjugates of each other
\cite{dqstl}.  The particles $C_1$ and $C_2$ need not be the products of any
resonance, and their total angular momentum is irrelevant. So the time and
transversity angle measurements from all the resonant and non-resonant decays of
this form may be combined to gain statistics. Here the values of $(|A_0(0)|^2 +
|A_\|(0)|^2)$ and $|A_\perp(0)|^2$ are just some {\it effective} average values,
but the decay widths $\Gamma_H$ and $\Gamma_L$ are the same for all such decay
modes, and hence for the whole data sample.

\subsection{Three-angle distribution in
$B_s\to J/\psi(\to l^+l^-)\,\phi(\to K^+K^-)$}

While the one-angle distribution (\ref{chmix:one-angle}) is in principle
sufficient to determine the values of $\Gamma_H$, $\Gamma_L$, $\delta \phi$ and
$\Delta m_s$, using the information present in all the angles $(\theta, \phi,
\psi)$ will improve the measurements. In addition, one also gets access to the
magnitudes of all the three amplitudes $A_0(0), A_\|(0), A_\perp(0)$ and the
strong phases between them, which was not possible with the one-angle
distribution \cite{ddf1}. A method to combine the three-angle distributions of
$B_s \to J/\psi \phi$ and $B_d \to J/\psi K^*$ to resolve a discrete ambiguity
in the CKM angle $\beta$ has also been proposed \cite{ddf2}.

The three angle distribution of an initially present (i.e.\ tagged) $B_s$ meson
is \cite{ddlr,ddf1}
\beqa
&& \frac{d^3 \Gamma [B_s(t) \to J/\psi (\to l^+ l^-) \phi (\to K^+K^-)]}
{d \cos \theta\, d \varphi\, d \cos \psi}
\propto \frac{9}{32 \pi} \bigg\{ 2 |A_0(t)|^2 \cos^2 \psi (1 - \sin^2
\theta
\cos^2 \varphi) \nonumber\\
&&{} + \sin^2 \psi\, \Big[ |A_\parallel(t)|^2  (1 - \sin^2 \theta \sin^2
\varphi)
+ |A_\perp(t)|^2 \sin^2 \theta - \mbox{Im}\,(A_\parallel^*(t)
A_\perp(t))
\sin 2 \theta \sin \varphi \Big] \nonumber\\
&&{} +\frac{1}{\sqrt{2}}\,\sin2\psi\, \Big[\mbox{Re}\,(A_0^*(t) A_\parallel(t))
\sin^2 \theta \sin2\varphi 
+ \mbox{Im}\,(A_0^*(t)A_\perp(t))\sin2\theta\cos\varphi \Big] \bigg\}\,.
\label{chmix:three-angle}
\eeqa 
Note that the same angular distribution (\ref{chmix:three-angle}) is also
valid for $B_d \to J/\psi(\to \ell^+ \ell^-) K^*(\to K^\pm \pi^\mp)$.  The
angular distribution for the $CP$ conjugate decay is obtained simply by
replacing all $A$'s by $\bar{A}$'s \cite{ddf1}.

\begin{table}[ht]
\begin{center}
\begin{tabular}{|c|l|} \hline
Observable & Time evolution \\
\hline
$|A_0(t)|^2$  & $|A_0(0)|^2 \left[e^{-\Gamma_L t} -
e^{-\overline{\Gamma}t}
\sin(\Delta m_s t)\delta\phi\right]$\\
$|A_{\|}(t)|^2$ &$ |A_{\|}(0)|^2 \left[e^{-\Gamma_L t} -
e^{-\overline{\Gamma}t}\sin(\Delta m_s t)\delta\phi\right]$\\
$|A_{\perp}(t)|^2$ & $|A_{\perp}(0)|^2 \left[e^{-\Gamma_H t} +
e^{-\overline{\Gamma}t}\sin(\Delta m_s t)\delta\phi\right]$\\
\hline
Re$(A_0^*(t) A_{\|}(t))$ &  $|A_0(0)||A_{\|}(0)|\cos(\delta_2 -
\delta_1)\left[e^{-\Gamma_L t} - e^{-\overline{\Gamma}t}
\sin(\Delta m_s t)\delta\phi\right]$\\
Im$(A_{\|}^*(t)A_{\perp}(t))$ & $|A_{\|}(0)||A_{\perp}(0)|\left[
e^{-\overline{\Gamma}t}\sin (\delta_1-\Delta m_s t)+\frac{1}{2}\left(
e^{-\Gamma_H t}-e^{-\Gamma_L
t}\right)\cos(\delta_1)\delta\phi\right]$\\
Im$(A_0^*(t)A_{\perp}(t))$ & $|A_0(0)||A_{\perp}(0)|\left[
e^{-\overline{\Gamma}t}\sin (\delta_2-\Delta m_s t)+\frac{1}{2}\left(
e^{-\Gamma_H t}-e^{-\Gamma_L
t}\right)\cos(\delta_2)\delta\phi\right]$\\
\hline
\end{tabular}
\end{center}
\caption{Time evolution of the decay $B_s\to J/\psi(\to l^+l^-)
\phi(\to K^+K^-)$ of an initially (i.e.\ at $t=0$) pure $B_s$ meson.
Here $\bar{\Gamma} \equiv (\Gamma_H + \Gamma_L)/2$.}
\label{chmix:timetable_bs}
\end{table}

\begin{table}[ht]
\begin{center}
\begin{tabular}{|c|l|}
\hline
Observable & Time evolution \\  \hline
$|\overline{A}_0(t)|^2$  & $|A_0(0)|^2 \left[e^{-\Gamma_L t} +
e^{-\overline{\Gamma}t}\sin(\Delta m_s t)\delta\phi\right]$\\
$|\overline{A}_{\|}(t)|^2$ &$ |A_{\|}(0)|^2 \left[e^{-\Gamma_L t} +
e^{-\overline{\Gamma}t}\sin(\Delta m_s t)\delta\phi\right]$\\
$|\overline{A}_{\perp}(t)|^2$ & $|A_{\perp}(0)|^2 \left[e^{-\Gamma_H
t} -
e^{-\overline{\Gamma}t}\sin(\Delta m_s t)\delta\phi\right]$\\
\hline
Re$(\overline{A}_0^*(t)\overline{A}_{\|}(t))$ &
$|A_0(0)||A_{\|}(0)|\cos(\delta_2 -
\delta_1)\left[e^{-\Gamma_L t} + e^{-\overline{\Gamma}t}
\sin(\Delta m_s t)\delta\phi\right]$\\
Im$(\overline{A}_{\|}^*(t)\overline{A}_{\perp}(t))$ &
$-|A_{\|}(0)||A_{\perp}(0)|\left[
e^{-\overline{\Gamma}t}\sin (\delta_1-\Delta m_s t)-\frac{1}{2}\left(
e^{-\Gamma_H t}-e^{-\Gamma_L
t}\right)\cos(\delta_1)\delta\phi\right]$\\
Im$(\overline{A}_0^*(t)\overline{A}_{\perp}(t))$ &
$-|A_0(0)||A_{\perp}(0)|\left[
e^{-\overline{\Gamma}t}\sin (\delta_2-\Delta m_s t)-\frac{1}{2}\left(
e^{-\Gamma_H t}-e^{-\Gamma_L
t}\right)\cos(\delta_2)\delta\phi\right]$\\
\hline
\end{tabular}
\end{center}
\caption{Time evolution of the decay $\overline{B_s}\to J/\psi
(\to l^+ l^-)\phi(\to K^+ K^-)$
of an initially (i.e.\ at $t=0$) pure $\overline{B_s}$ meson.
Here $\bar{\Gamma} \equiv (\Gamma_H + \Gamma_L)/2$.}
\label{chmix:timetable_bsbar}
\end{table}

The time evolution of the observables in the angular distribution
(the coefficients of the angular terms in (\ref{chmix:three-angle})
and its $CP$ conjugate mode)
are given in Table \ref{chmix:timetable_bs} and
\ref{chmix:timetable_bsbar} respectively. 
Here $\delta_1 \equiv {\rm Arg}(A_\perp^* A_\|)$ and
$\delta_2 \equiv {\rm Arg}(A_0^* A_\|)$.
A finite lifetime difference $\Delta \Gamma$ implies that
the $CP$ violating terms proportional to
\beq
\left(e^{-\Gamma_H t} - e^{-\Gamma_L t}\right) \cos(\delta_{1(2)})\,
\delta \phi
\label{chmix:untagged}
\eeq
survive even when the $B_s$ is untagged \cite{df,ddf1}.
An experimental feasibility study for extracting the parameters 
from the time dependent three angle distribution has been performed
for the LHC in \cite{lhcreport}.

\subsection{Angular moments method}

The likelihood fit to the complete angular distribution
(\ref{chmix:three-angle}) -- including the time evolution of
the observables (Tables \ref{chmix:timetable_bs} and
\ref{chmix:timetable_bsbar}) -- is a difficult task due to the
large number of parameters involved. The method of angular
moments proposed in \cite{ddf1} can disentangle
the angular dependences and split up the likelihood fit
into a number of likelihood fits with a smaller number of
parameters.

The angular distributions ((\ref{chmix:one-angle}) or
(\ref{chmix:three-angle})) are of the form
\beq
f( \Theta, {\cal P}; t) = \sum b^{(k)} ( {\cal P}; t)\, g^{(k)}
        (\Theta) \,,
\label{chmix:f=bg}
\eeq
where ${\cal P}$ represents the parameters, 
and $\Theta$ denotes the angles.
If we can find {\it weighting functions} $w^{(i)}$ such that
\beq
\int [{\cal D} \Theta]\, w^{(i)}(\Theta)\, g^{(k)} (\Theta) = 
\delta_{ik}\,,
\label{chmix:wg=delta}
\eeq
then
\beq
b^{(i)} \approx  \sum_{events} w^{(i)}(\Theta)\,, 
\label{chmix:b=sum-w}
\eeq
and the observables are determined directly from the data.
It can be shown \cite{ddf1} that such a set of weighting
functions exists for any angular distribution of the
form (\ref{chmix:f=bg}) and such a set can be determined
without any {\it a priori} knowledge of the values of
the observables $b^{(i)}$. A likelihood fit can then be
performed on each observable $b^{(i)}$ independently in
order to determine the parameters.

The angular moments (AM) method is more transparent and easier to
implement than the complete likelihood fit method. 
Although the AM method in its naive form involves some
loss of information, the extent of this loss of information
in the case of the transversity angle distribution has been
found to be less than 10\% in the parameter range of interest
(see Fig.~5 of \cite{sen}). In its full form, the
AM method can determine the values of parameters almost as well as
the likelihood fit method (see, {\it e.g.} \cite{cleo-drho}).

To conclude, the angular analysis of $B_s \to J/\psi \phi$ decays
can separate the $CP$ even and odd components in the final state, 
and it is perhaps the best way to determine the lifetime difference 
between $B_s^H$ and $B_s^L$. As a byproduct, it also helps  the
measurement of $CP$ odd and even components and their relative 
strong phases, and with enough statistics, the determination of
$\Delta m_s$ and $\delta \phi$. The angular analysis, possibly
employing the angular moments method if the likelihood fit
is inadequate, is highly recommended.

\boldmath
\subsection[\ddm]{\ddm$\!$ \authorfootnote{Author: Ulrich~Nierste}}
\label{chmix:sec:d}
\unboldmath
\index{D mixing@$D$ mixing}%
\index{D meson@$D$ meson!mass difference $\dm_D$}%
\index{D meson@$D$ meson!width difference $\dg_D$}%
We define the mass eigenstates in \ddm\ as 
\begin{eqnarray}
\ket{D_1} &=& 
        p \ket{D^0} + q \ket{\ov{D}{}^0} \,,  \nn
\ket{D_2} &=& 
        p \ket{D^0} - q \ket{\ov{D}{}^0} \,,
\qquad \mbox{with } \lt|p\rt|^2+\lt|q\rt|^2 = 1\,. \label{chmix:d:defpq}
\end{eqnarray}
$q$ and $p$ are obtained from the solution of the eigenvalue problem
for $M - i \Gamma/2$. In \eq{mgqp:c} $q/p$ is determined in terms of 
$M_{12}$ and $\Gamma_{12}$. We define 
\begin{eqnarray}
\dm_D &=& m_2-m_1, \qquad\qquad 
\dg_D \, = \, \Gamma_1-\Gamma_2 \label{chmix:d:dmdg} \\
x_D &=& \frac{\dm_D}{\Gamma} \qquad\qquad
y_D \, = \, - \frac{\dg_D}{2\Gamma} 
\, = \, \frac{\Gamma_2-\Gamma_1}{2\Gamma} \label{chmix:d:xy} .
\end{eqnarray}
The definitions in \eq{chmix:d:defpq} and \eq{chmix:d:dmdg} comply
with the conventions of Sect.~\ref{ch1:sect:gen} for \bbm. In particular
the time evolution formulae of \eq{tgg} - \eq{gpgms} are also valid
for \ddm\ with the replacement $B^0 \to D^0$.  Unlike in the case of
\bbm\ we cannot expand in $\dg_D/\dm_D$. We also refrain from
expanding in $\ega$ defined in \eq{defepsg}. Then \eq{gtfres} - \eq{gbtfbres} 
can be used for $D^0$ mesons, if $1+\ega$ in \eq{gbtfres} is replaced 
by $|p/q|^2$ and $1-\ega$ in \eq{gtfbres} is replaced by $|q/p|^2$.
Note that the definition of $y_D$ in \eq{chmix:d:xy} is opposite in
sign to the one of $y$ in the $B$ meson system in \eq{defxy}. In 
\eq{chmix:d:xy} we have used the sign convention which is usually used
in \ddm. Since \ddm\ is very small, one can expand in $\dm_D t$ and
$\dg_D t$. Using $x_D$ and $y_D$ from \eq{chmix:d:xy} the small $t$
expansion of \eq{gtfres} - \eq{gbtfbres} gives%
\index{time evolution!decay rate!$D$ meson}
\begin{eqnarray} 
\Gamma (D^0(t) \rightarrow f ) &\,=\,& 
 {\cal N}_f \lt| A_f \rt|^2 e^{-\Gamma_D t}\, 
  \Bigg[ 1 + \lt( -\imag \lambda_f \, x_D \, + \, \real \lambda_f \, 
                y_D  \rt) \Gamma_D t 
\phantom{\frac{1}{2}} \nn
&& \phantom{ {\cal N}_f \lt|  A_{f} \rt|^2 e^{-\Gamma_D t}\,} 
  \lt.
        + 
                \lt( \frac{|\lambda_f|^2+1}{4}\, y_D^2 + 
                     \frac{|\lambda_f|^2-1}{4}\, x_D^2  \rt) 
                     \!\lt( \Gamma_D t \rt)^2 \rt] ,
 \label{chmix:d:df} \\[2mm] 
\Gamma (\ov{D}{}^0(t) \rightarrow f ) &\,=\,& 
 {\cal N}_f \lt| A_{f} \rt|^2 e^{-\Gamma_D t}
 \lt| \frac{p}{q} \rt| ^2 \, 
  \Bigg[ 
 |\lambda_f|^2  + \lt( \imag \lambda_f \, x_D \,  + 
                             \real \lambda_f \, y_D \rt) \Gamma_D t \nn
&& \phantom{ {\cal N}_f \lt| \ov{A}_{\ov{f}} \rt|^2 e^{-\Gamma_D t}
 \lt| \frac{p}{q} \rt| ^2 } 
  \lt.
       +
        \lt( \frac{|\lambda_f|^2+1}{4}\, y_D^2 - 
                     \frac{|\lambda_f|^2-1}{4}\, x_D^2  \rt) 
                     \!\lt( \Gamma_D t \rt)^2 \rt] ,
 \label{chmix:d:dbf} \\[2mm] 
\Gamma (D^0(t) \rightarrow \ov{f} ) &\,=\,& 
 {\cal N}_f \lt| \ov{A}_{\ov{f}} \rt|^2 e^{-\Gamma_D t}
  \lt| \frac{q}{p} \rt|^2
   \lt[ |\lambda_{\ov{f}}|^{-2} + 
     \lt( \imag \frac{1}{\lambda_{\ov{f}}} \, x_D + 
          \real \frac{1}{\lambda_{\ov{f}}} \, y_D \rt) \Gamma_D t \rt. \nn
&& \phantom{ {\cal N}_f \lt| \ov{A}_{\ov{f}} \rt|^2 e^{-\Gamma_D t}
 \lt| \frac{p}{q} \rt| ^2} 
  \lt.
       +
    \lt( \frac{|\lambda_{\ov{f}}|^{-2}+1}{4}\, y_D^2 -  
                     \frac{|\lambda_{\ov{f}}|^{-2}-1}{4}\, x_D^2  \rt) 
                     \!\lt( \Gamma_D t \rt)^2 \rt] ,
 \label{chmix:d:dfb} \\[2mm] 
\Gamma (\ov{D}{}^0(t) \rightarrow \ov{f} ) &\,=\,& 
   {\cal N}_f \lt| \ov{A}_{\ov{f}} \rt|^2 e^{-\Gamma_D t}
   \lt[ 1 + 
     \lt( - \imag \frac{1}{\lambda_{\ov{f}}} \, x_D + 
          \real \frac{1}{\lambda_{\ov{f}}} \, y_D \rt) \Gamma_D t \rt. \nn
&& \phantom{ {\cal N}_f \lt| \ov{A}_{\ov{f}} \rt|^2 e^{-\Gamma_D t}} 
  \lt.
       +
    \lt( \frac{|\lambda_{\ov{f}}|^{-2}+1}{4}\, y_D^2 +  
                     \frac{|\lambda_{\ov{f}}|^{-2}-1}{4}\, x_D^2 \rt) 
                     \!\lt( \Gamma_D t \rt)^2 \rt] ,
  \label{chmix:d:dbfb}
\end{eqnarray}
with $\Gamma_D=(\Gamma_1+\Gamma_2)/2$. $\dm_D$ and $\dg_D$ are 
very small, because they are GIM-suppressed by a factor of
$m_s^2/M_W^2$. For this reason they are also difficult to calculate,
because at scales of order $m_s$ non-perturbative effects become
important. A recent calculation, which incorporates non-perturbative
effects with the help of the quark condensate, can be found in \cite{bu}.

\subsection[New Physics Effects in Meson Mixing]{New Physics Effects in Meson
Mixing$\!$ \authorfootnote{Author: JoAnne Hewett}}

The existence of new physics may modify the low-energy effective Hamiltonian
governing $B$ and $D$ physics in several ways: (i) via contributions to the
Wilson coefficients of the Standard Model operators, (ii) by generating new
operators, or (iii) through the presence of new $CP$ violating phases.  These
effects may originate from new interactions in tree-level meson decays or from
the virtual exchange of new physics in loop-mediated processes.  The scale of
new physics is expected to be large compared to $M_W$, and hence it is generally
anticipated that additional tree-level contributions to meson decays are
suppressed\cite{howe}.  However, large new contributions may be present in loop
processes, making meson mixing a fertile ground to reveal the influence of new
interactions.  All three above classes of new physics contributions may play a
role in meson mixing.  Such effects may be discovered in observables which are
suppressed in the Standard Model, such as the asymmetry $a_{\rm fs}^q$ measuring
$CP$ violation in $B_{d,s}$ mixing or in $D^0$ meson mixing, they may modify the
mixing-induced $CP$ asymmetries in $B\to\psi K_S$ and $B\to\pi\pi$ decays, or they
may alter the precisely calculated SM value of the ratio of $B_s$ to $B_d$
mixing.  We will discuss each of these observables in this section.  The effects
of new physics on meson width differences is described in
Sect.~\ref{chmix:subsub:width:phen}.

\subsubsection{$B_d$ Mixing}
\index{B mixing@$B$ mixing!$B_d$ mixing}%
\index{B meson@$B$ meson!mass difference \dm!$\dm_d$}%
\index{B meson@$B$ meson!mass difference \dm!new physics}%
\index{mass difference \dm!$\dm_d$}%
\index{mass difference \dm!$\dm_d$!new physics}%
It is well-known that new physics can play a large role in $B_d$
mixing.  One important consequence of this is that the constraints
in the $\rho-\eta$ plane from $\Delta m_d$
can be altered, resulting in a
significant shift\cite{indirect_rev} of the allowed region in this plane
from its Standard Model range.  This in turn modifies the expected values for
$\sin 2\beta$ and $\sin 2\alpha$, even if new sources of $CP$ violation 
are not present.  In fact, this comprises the most significant effect from
new physics on the $CP$ asymmetries in $B\to\psi K_S$ and $B\to\pi\pi$
decays in a large class of models.%
\index{decay!$B_d \rightarrow \psi K_S$}%
\index{decay!$B_d \to \pi \pi $ }%
\index{unitarity triangle!angle $\beta$}%
\index{unitarity triangle!angle $\alpha$}%

A model independent determination of such effects has been presented
in Ref.~\cite{yossi}.  New physics contributions to $B_d$ mixing can be
parameterized in a model independent fashion by considering the ratio
\begin{equation}
{\langle B_d^0|{\cal H}^{\rm full}|\bar B^0_d\rangle\over
 \langle B_d^0|{\cal H}^{\rm SM}|\bar B^0_d\rangle} 
   = \Big( r_d\, e^{i\theta_d} \Big)^2\,,
\end{equation}
where ${\cal H}^{\rm full(SM)}$ represents the Hamiltonian responsible
for $B_d$ mixing in the case of the Standard Model plus new physics
(just Standard Model), and $r_d(\theta_d)$ represents the new physics
contribution to the magnitude (phase) of $B_d$ mixing.  In the
Standard Model, the unitarity triangle is constrained by measurements
of $\sin 2\beta$, $\sin 2\alpha$, the ratio of semileptonic decays $
{\Gamma(b\to u\ell\nu)/\Gamma(b\to c\ell\nu)}$, and $x_d=C_tR_t^2$,
where $R_t$ is defined in Section~\ref{chmix:sub:mass}.%
\index{B decay@$B$ decay!semileptonic}%
\index{decay!$B_d \rightarrow X \ell^+ \nu$}%
\index{unitarity triangle!and Vub@and $V_{ub}$}%
\index{Cabibbo-Kobayashi-Maskawa matrix!$V_{ub}$}%
These quantities are then modified in the presence of new interactions
via $a_{\psi K_S}=\sin (2\beta+2\theta_d)$,
$a_{\pi\pi}=\sin(2\alpha-2\theta_d)$,\ and $x_d=C_tR_t^2r_d^2$. Note
that the new phase contributions in $a_{\psi K_S}$ and $a_{\pi\pi}$
conspire to cancel in the triangle constraint and the relation
$\alpha+\beta+\gamma=\pi$ is retained.%
\index{unitarity triangle!angle $\alpha$}%
Measurement of these four quantities allows one to disentangle the new
physics effects and fully reconstruct the true unitarity triangle
(\ie, find the true values of $\alpha$, $\beta$, and $R_t$) as well as
determine the values of $r_d$ and $\theta_d$ in a geometrical fashion.
This is depicted in Fig.~\ref{mix_modelindep}.  While this technique
is effective in principle, in practice it is limited by theoretical
uncertainties in $x_d\,, \alpha$, and the ratio of semileptonic
decays, as well as discrete ambiguities.

\begin{figure}[thb]
\centerline{
\epsfig{file=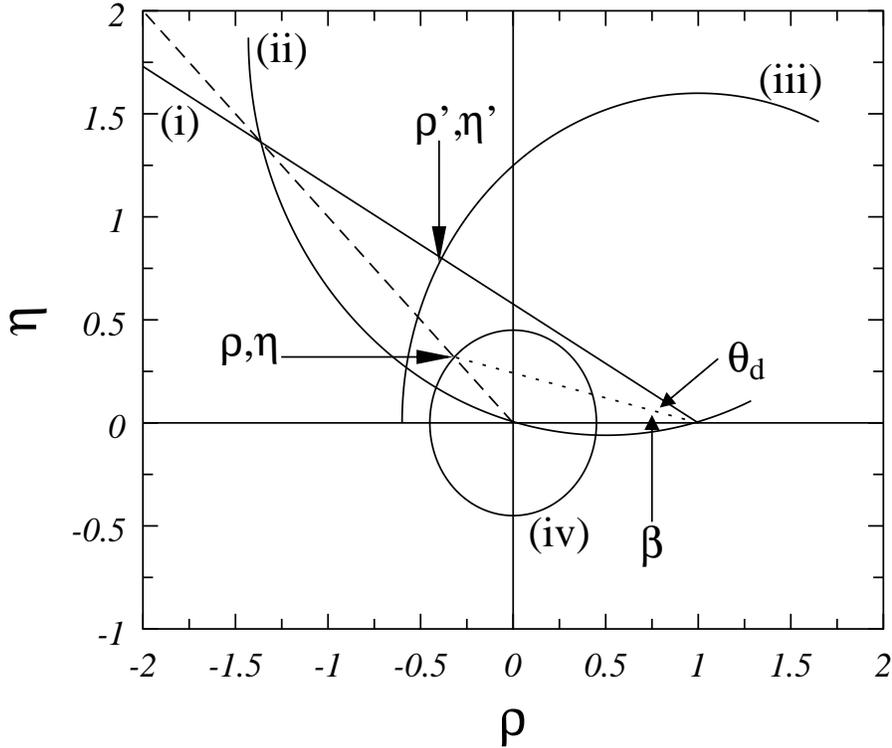,width=12cm}}
\caption[The model independent analysis in the $\rho-\eta$ plane]{The model
independent analysis in the $\rho-\eta$ plane: (i) the $a_{\psi K_S}$ ray, (ii)
the $a_{\pi\pi}$ circle, (iii) the $x_d$ circle, (iv) the semileptonic decay
ratio circle.  The $\gamma$ ray is given by the dashed line and the true
$\beta$ ray corresponds to the dotted line.  The true vertex of the unitarity
triangle is at $(\rho,\eta)$, while the point $(\rho',\eta')$ serves to
determine $r_d$ and $\theta_d$.}
\label{mix_modelindep}
\end{figure}
    
Model independent bounds on new physics contributions to $B_d$ mixing
can also be directly placed from measurements of $\Delta m_d$ and $\sin
2\beta$.  In the class of models
which respect $3\times 3$ CKM unitarity, where tree-level $B$
decays (in particular their phase) are dominated by the SM, and where
$\Gamma_{12}\simeq\Gamma_{12}^{\rm SM}$, the new physics effects in $B_d$
mixing can be isolated. The modification to $M_{12}$ can then be
described as above in terms of $r_d\,, \theta_d$.
The direct determination of $\Delta m_d$ provides a
bound on the magnitude of new physics contributions, $r_d$, while 
measurement of $\sin 2\beta$ constrains the new
phase $\theta_d$.  Taking into account the uncertainties on the values
of the relevant CKM factors and the hadronic matrix elements, present
data constrains $0.5\leq r_d\leq 1.8$ and $\sin 2\theta_d\geq -0.53$
at $95\%$ C.L.  It is clear that large contributions to $B_d$ mixing from new
interactions are still allowed, and may hence admit for an exciting discovery
as future measurements improve.

We note that another useful parameterization of new physics contributions
which is common in the literature is given by
\begin{equation}
M_{12}^{\rm NP}=h\, e^{i\sigma}M_{12}^{\rm SM}\,,
\end{equation}
where these variables are related to the previous ones via
\begin{equation}
r_d^2\, e^{2i\theta_d}=1+h\, e^{i\sigma}\,.
\end{equation}
Constraints on this set of parameters from current data is presented in
\cite{niretal} in various classes of models.

These parameters can be related to new physics contributions in other
observables.  For instance, the $CP$ asymmetry in flavor-specific $B_d$
decays (see Sect.~\ref{chmix:sub:afs}), which is given by%
\index{CP asymmetry@\CP\ asymmetry!in flavor-specific decay}%
\begin{equation}
a^d_{\rm fs}  = \imag {\Gamma^d_{12}\over M^d_{12}}\,,
\end{equation}
and has a value of $-8 \times 10^{-4}$ in the Standard Model (see
\eq{chmix:aqnum}), provides a good opportunity to probe the existence
of new interactions.  Since $\Gamma^d_{12}/M^d_{12}$ is essentially
real in the Standard Model, the contributions of new interactions to
the flavor-specific $CP$ asymmetry can be written as
\begin{equation}
a^d_{\rm fs}=-\left( {\Gamma_{12}^d\over M^d_{12}} \right)_{\rm SM} 
               {\sin 2 \theta_d \over r_d^2}\,.
\end{equation}
The above bounds from present data on the new physics contributions to
$B_d$ mixing restrict
\begin{equation}
-2.1\leq {a_{\rm fs}^d\over (\Gamma^d_{12}/M^d_{12})_{\rm SM}} \leq 4.0\,.
\end{equation}
It is important to note that $a_{\rm fs}^d$ can lie outside this
range, if new new physics enhances $\Gamma_{12}^d$, which is composed
of CKM-suppressed decay modes.

There are a plethora of new physics scenarios which can yield
substantial contributions to $B_d$ mixing; some examples are briefly
cataloged here.  Models which respect the structure of the $3\times 3$
CKM matrix contribute simply to the Wilson coefficient of the Standard
Model operator.  This is best illustrated by the virtual exchange in
the box diagram of charged Higgs bosons which are present in flavor
conserving two-Higgs-doublet models\cite{chHiggs} and by the
contributions of supersymmetric particles\cite{mix_susy} when a
Standard Model-like super-CKM structure is assumed.%
\index{supersymmetry}%
\index{two-Higgs doublet model}%
\index{new physics!supersymmetry}%
If the super-CKM angles ($\tilde V$) are allowed to be arbitrary, the
structure of the Wilson coefficients are altered.  In this case, the
supersymmetric amplitude relative to that of the Standard Model is
roughly given by $\sim (M_W/\tilde m)^n (\tilde V_{td}\tilde
V_{tb}/V_{td}V_{tb})$ and can constitute a flavor problem for
Supersymmetry if the sparticle masses, $\tilde m$, are near the weak
scale.  The existence of a fourth generation would also modify the CKM
structure of the Wilson coefficients.  New $|\Delta B|=2$ operators
are generated in scenarios\cite{newops} such as Left-Right Symmetric
models, theories of strong dynamics, as well as in Supersymmetry.
\index{new physics!left-right symmetry}%
\index{new physics!strong dynamics}%
Tree-level contributions\cite{mix_tree} are manifest in flavor
changing two-Higgs-doublet models, in scenarios with a flavor changing
extra neutral gauge boson, and in Supersymmetry with R-parity
violation.%
\index{two-Higgs doublet model!flavor-changing Higgs}%
\index{new physics!two-Higgs doublet model!flavor-changing Higgs}%
Most of these examples also contain new phases which may
be present in $B_d$ mixing.  It is interesting to note that various
forms of Supersymmetry may affect $B_d$ mixing in all possible
manners!

While it is possible to obtain large effects in $B_d$ mixing in all of
these scenarios, it is difficult to use $\Delta m_d$ at present to
tightly restrict these contributions and constrain the parameter space
in the models, due to the current sizable errors on the Standard Model
theoretical prediction arising from the imprecisely determined values
of the CKM factors and the hadronic matrix elements.  Frequently,
other flavor changing neutral current processes, such as $b\to
s\gamma$, provide more stringent constraints.

\subsubsection{$B_s$ Mixing}
\index{B mixing@$B$ mixing!$B_s$ mixing}
\index{B meson@$B$ meson!mass difference \dm!$\dm_s$}%
\index{B meson@$B$ meson!mass difference \dm!new physics}%
\index{mass difference \dm!$\dm_s$}%
\index{mass difference \dm!$\dm_s$!new physics}%
A similar analysis as employed above may be used to constrain new physics
contributions to $B_s$ mixing.  In the class of models which respect the
$3\times 3$ CKM unitarity and where tree-level decays are dominated by the
Standard Model contributions, we can again parameterize potential new
contributions to $\Delta m_s$ via $M_{12}=(r_se^{i\theta_s})^2M_{12}^{\rm SM}$.
This gives
\begin{equation}
{\Delta m_d\over\Delta m_s} = {\lambda^2R_t^2\over\xi^2}\, 
  {m_{B_d}\over m_{B_s}}\, {r^2_d\over r^2_s}\,,
\end{equation}
which reduces to the Standard Model expression when $r_d=r_s$.  The parameter
$\theta_s$ can be constrained once $CP$ asymmetries in the $B_s$ system are
measured or, if $\theta_s$ is large, from measurements of $\dg_s$ as described
in Sect.~\ref{chmix:subsub:width:phen}.

As discussed above, the ratio $\Delta m_d/\Delta m_s$ yields a good
determination of the CKM ratio $|V_{td}/V_{ts}|$ within the Standard Model,
since the ratio of hadronic matrix elements is accurately calculated in lattice
gauge theory.  Remarkably, this remains true in many scenarios beyond the
Standard Model.  In a large class of models which retain the $3\times 3$ CKM
structure, the virtual exchange of new particles in the box diagram alters the
Inami-Lim function, but not the remaining factors in the expression for $\Delta
m_{d,s}$.  The effects of new physics thus cancel in the ratio.  As an explicit
example, consider charged Higgs exchange in the box diagram within the context
of two-Higgs-doublet models.  The expression for the mass difference in $B_s$
mixing in this case is
\begin{eqnarray}
\Delta m_s & = & {G_F^2M_W^2\over 6\pi^2}\,f_{B_s}^2B_{B_s}\eta_{B_s}
m_{B_s}\, |V_{tb}V^*_{ts}|^2\, 
\Big[S(m_t^2/M_W^2) + F(m_t^2/m^2_{H^\pm},\tan\beta)\Big]
\nonumber\\
& = & \Delta m_d\, \xi^2\, {m_{B_s}\over m_{B_d}}\, 
{|V_{ts}|^2\over |V_{td}|^2} \,,
\end{eqnarray}
where $m_{H^\pm}$ represents the charge Higgs mass and $\tan\beta$ is
the ratio of vevs.  Here,
we see that the charged Higgs contribution is the same for $B_d$
and $B_s$ mixing (neglecting $d$- and $s$-quark masses)
and thus cancels exactly in the ratio.  This cancellation also occurs
in several other classes of models, including minimal 
Supersymmetry with flavor conservation.
Notable exceptions to this
are found in models which (i) change the structure of the 
CKM matrix, such as the addition of a fourth generation, or extra singlet
quarks, or in Left-Right symmetric models, (ii) have sizable Yukawa couplings 
to the light fermions, such as leptoquarks or Higgs models with
flavor changing couplings, 
or (iii) have generational dependent couplings, including supersymmetry with
R-parity violation.

\subsubsection{Mixing in the Charm Sector}
\index{D mixing@$D$ mixing}%
\index{D meson@$D$ meson!mass difference $\dm_D$!new physics}%

The short distance Standard Model contribution to $\Delta m_D$ proceeds through
a $W$ box diagram with internal $d,s,b$-quarks.  In this case the external
momentum, which is of order $m_c$, is communicated to the light quarks in the
loop and can not be neglected.  The effective Hamiltonian is
\begin{equation}
{\cal H}^{\Delta c=2}_{eff} = {G_F\alpha\over 8\sqrt 2x_w}\, \Big[
|V_{cs}V^*_{us}|^2\, (I_1^s\,{\cal O} - m_c^2 I_2^s\,{\cal O}') 
+ |V_{cb}^*V_{ub}|^2\, (I_3^b\, {\cal O} - m_c^2 I^b_4\, {\cal O}') \Big]\,,
\end{equation}
where the $I_j^q$ represent integrals\cite{dmixsm} that are functions of
$m_q^2/M_W^2$ and $m_q^2/m_c^2$, and ${\cal O}=[\bar u\gamma_\mu(1-
\gamma_5)c]^2$ is the usual mixing operator, while ${\cal O}'=[\bar u
\gamma_\mu(1+\gamma_5)c]^2$ arises in the case of non-vanishing external
momentum.  The numerical value of the short distance contribution is $\Delta
m_D\sim 5\times 10^{-18}$ GeV (taking $f_D=200$ MeV).  The long distance
contributions have been computed via an intermediate state dispersive approach
and in heavy quark effective theory, yielding values\cite{dmixld} in the range
$\Delta m_D\sim 10^{-17} - 10^{-16}$ GeV.  The Standard Model predictions are
clearly quite small and allow for a large window for the observation of new
physics effects.

\begin{figure}[t]
\vspace*{-.5cm}
\centerline{\hspace*{20mm}
\epsfig{file=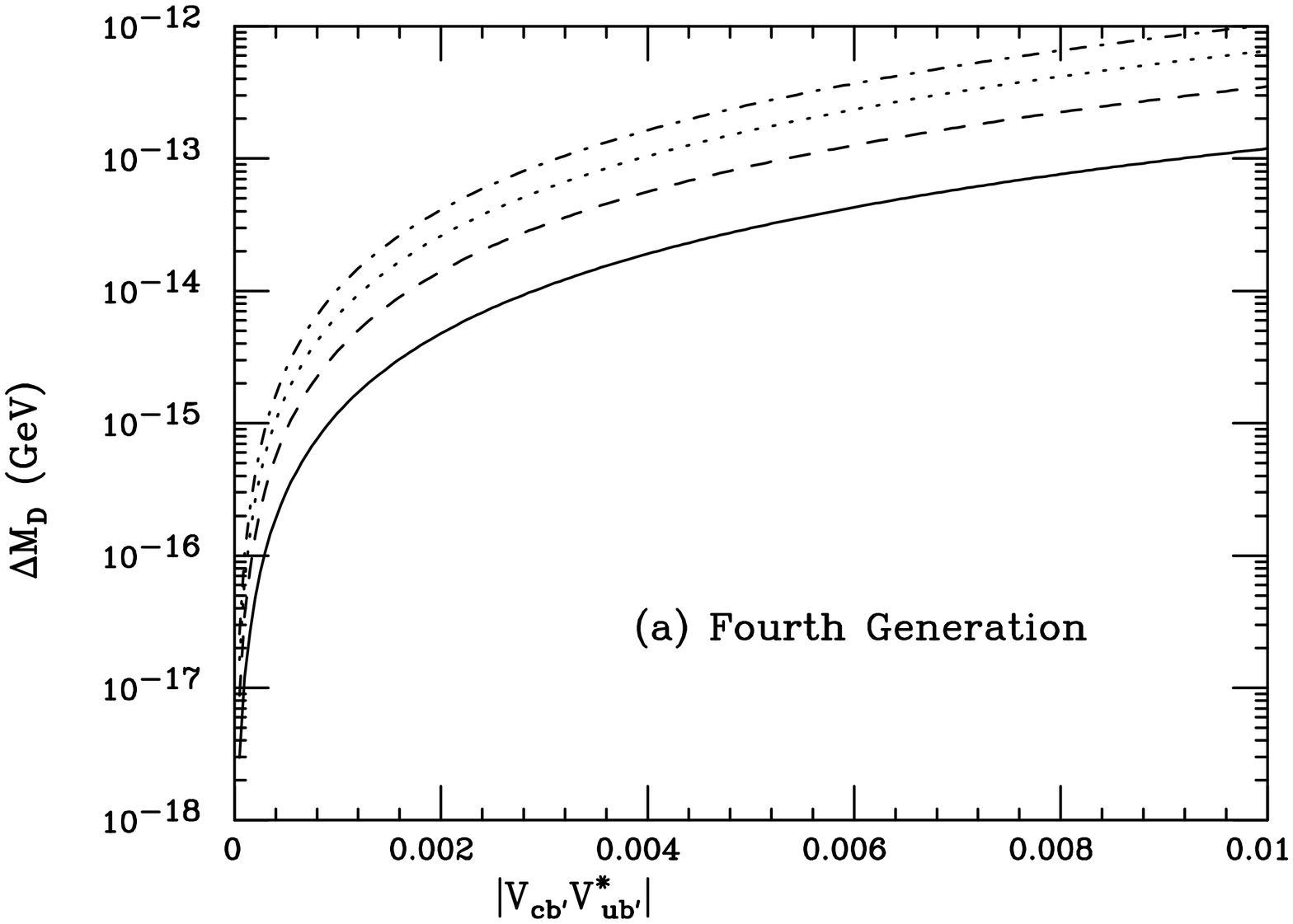,
        height=0.52\textwidth,angle=-90}
\hspace*{1.5cm}
\raisebox{7pt}{\epsfig{file=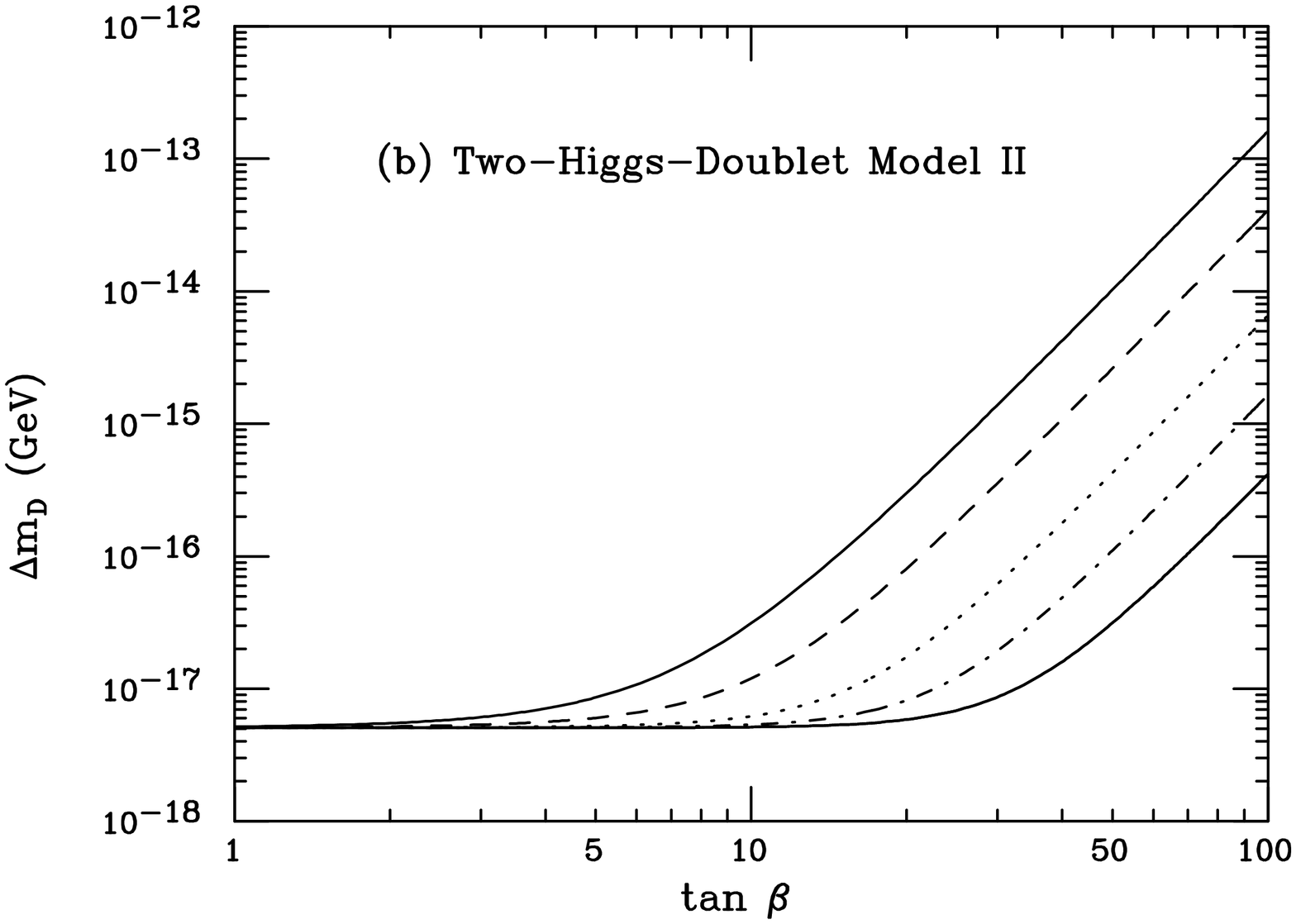,
        height=0.5\textwidth,angle=90}}}
\vspace*{-2cm}
\centerline{
  \psfig{file=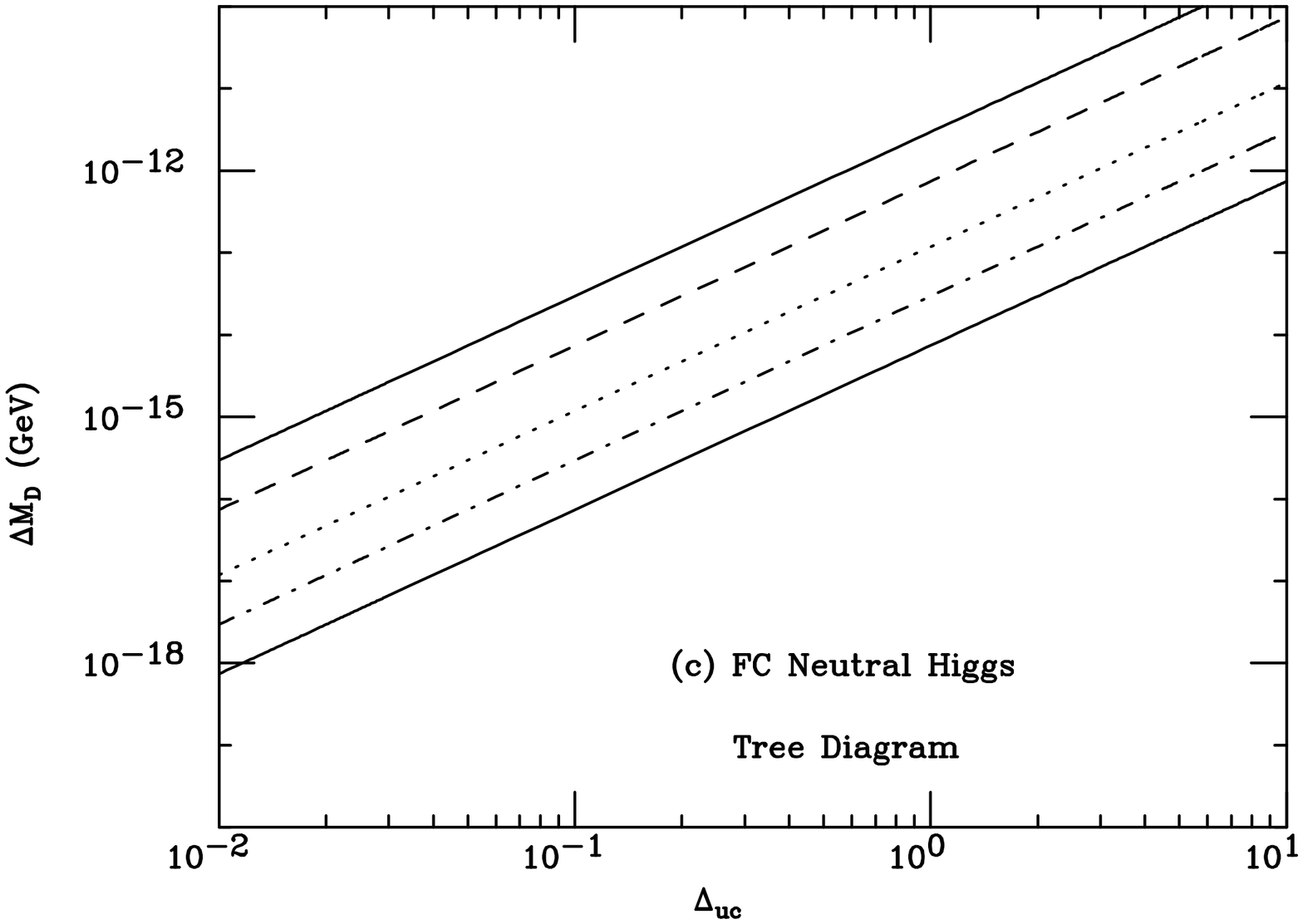,width=6cm,angle=-90}}
\vspace*{-.5cm}
\caption[$\Delta m_D$ in the (a) four generation Standard Model; (b)
two-Higgs-doublet model~II; and (c) flavor-changing Higgs model with
tree-level contributions.]{$\Delta m_D$ in (a) the four generation Standard
Model as a function of the appropriate $4\times 4$ CKM elements, taking the
$b'$-quark mass to be $100,200,300,400$ GeV from top to bottom, (b) the
two-Higgs-doublet model II as a function of $\tan\beta=v_2/v_1$ with
$m_{H^\pm}=50,100,250, 500, 1000$ GeV from top to bottom, and (c) the
flavor-changing Higgs model with tree-level contributions as a function of the
mixing factor, with $m_{h^0}=50,100,250,500,1000$ GeV from top to bottom.}
\label{charm_mix}
\end{figure}

Since the Standard Model expectation is so small, large enhancements in $\Delta
m_D$ are naturally induced by new interactions.  A compilation of such effects
in various models and list of references can be found in \cite{harry}.  This
article shows that the present experimental bound on $D$-mixing already
constrains the parameter space in many scenarios, and an order of magnitude
improvement would exclude (or discover) some models.  Here, for purposes of
illustration, we present the potential enhancements that can occur in three
scenarios\cite{indirect_rev}.  We examine (i) the case of a fourth generation to
demonstrate the effect of heavy fermions participating in the box diagram, (ii)
the contributions from charged Higgs exchange in flavor conserving
two-Higgs-doublet models, which is often used as a benchmark in studying new
physics, and (iii) the tree-level contributions in flavor-changing Higgs models,
where the flavor changing couplings are taken to be $\lambda_{h^0f_if_j}\sim
(\sqrt 2 G_F)^{1/2} \sqrt{m_im_j}\Delta_{ij}$ with $\Delta_{ij}$ being
a combination of mixing angles.%
\index{new physics!two-Higgs doublet model}%
\index{new physics!two-Higgs doublet model!flavor-changing Higgs}%
\index{new physics!fourth generation}%
\index{fourth generation}%
\index{two-Higgs doublet model!flavor-changing Higgs}%
\index{two-Higgs doublet model}%
The mass difference as a function of the model parameters is shown in
Fig.~\ref{charm_mix} for each case.  We see that in each case the
parameter space is already restricted by the current experimental
value, and that an improvement in the bound would provide a sensitive
probe of these models.
  
\section[Interesting decay modes]{Interesting decay modes$\!$ \authorfootnote{Author: Ulrich~Nierste}}

In this section we list the decay modes which are useful for the determination
of $\dm_q$, $\dg_q$ and $B$ meson lifetimes.  Flavor-specific decay modes are
summarized in Table~\ref{chmix:tab:fsd}, CKM-favored decays into $CP$ eigenstates
are listed in Table~\ref{chmix:tab:cpe} and decays which are neither
flavor-specific nor $CP$-specific can be found in Table~\ref{chmix:tab:cpne}.

\begin{table}[thb]
\begin{center}
\begin{tabular}{@{}r|p{0.45\textwidth}|p{0.35\textwidth}} \hline
quark decay & hadronic decay & remarks \\\hline\hline  
$\ov{b} \rightarrow \ov{c} \ell^+  \nu_\ell $ & 
  $B_{d,s}\rightarrow D_{(s)}{}^- \ell^+ \nu_\ell$ & 
\index{decay!$B_s \rightarrow D_s^- \ell^+ \nu$}
\index{decay!$B_d \rightarrow D^- \ell^+ \nu$}
\\
& $B_{d,s}\rightarrow X \ell^+ \nu_\ell $ & 
\index{decay!$B_s \rightarrow X \ell^+ \nu$}
\index{decay!$B_d \rightarrow X \ell^+ \nu$}
\\\hline
$\ov{b} \rightarrow \ov{c} u \ov{d}$ & 
  $B_d \rightarrow D^{(*)}{}^- \pi^+$ & 
\index{decay!$B_d \rightarrow D^- \pi^+$}
\index{decay!$B_d \rightarrow D^{*-} \pi^+$}
\\
& $B_d \rightarrow D^{(*)}{}^- \pi^+ \pi^+ \pi^-$ & 
\index{decay!$B_d \rightarrow D^- \pi^+ \pi^- \pi^+$}
\index{decay!$B_d \rightarrow D^{*-} \pi^+ \pi^- \pi^+$}
\\
& $B_d \rightarrow D^{*}{}^- \pi^+ \pi^+ \pi^- \pi^0$ &
\index{decay!$B_d \rightarrow D^{*-} \pi^+ \pi^+ \pi^- \pi^0$}
\\
& $B_d \rightarrow \ov{D}{}^{(*)0} \rho^0$
\index{decay!$B_d \rightarrow \ov{D}{}^0 \rho^0$}
\index{decay!$B_d \rightarrow \ov{D}{}^{*0} \rho^0$}
  [$\rightarrow  K^+ \pi^-$ or $K^+ \pi^+ \pi^- \pi^-$ 
   \mbox{\hspace{15ex}or} 
   $K^{(*)}{}^+ \ell^- \ov{\nu}_{\ell}$ etc.] &
   $\BR(\rho^0 \rightarrow \pi^+ \pi^-) \approx 100\%$. \hfill\break
   The $\ov{D}{}^{(*)0}$ must be detected in a final state $f$
         such that $D^{(*)0} \rightarrow f$ is forbidden or suppressed. \\
& $B_s \rightarrow D_s^{(*)}{}^- \pi^+$ & 
\index{decay!$B_s \rightarrow D_s^- \pi^+$}
\index{decay!$B_s \rightarrow D_s^{*-} \pi^+$}
\\ 
& $B_s \rightarrow D_s^{(*)}{}^- \pi^+ \pi^+ \pi^-$ & 
\index{decay!$B_s \rightarrow D_s^- \pi^+ \pi^+ \pi^-$}
\index{decay!$B_s \rightarrow D_s^{*-} \pi^+ \pi^+ \pi^-$}
\\
& $B_s \rightarrow D_s^{*}{}^- \pi^+ \pi^+ \pi^- \pi^0$ &
\index{decay!$B_s \rightarrow D_s^- \pi^+ \pi^+ \pi^- \pi^0$}
\index{decay!$B_s \rightarrow D_s^{*-} \pi^+ \pi^+ \pi^- \pi^0$}
\\
& $B_s \rightarrow \ov{D}{}^{(*)0} K_S$
  [$\rightarrow  K^+ \pi^-$ or $K^+ \pi^+ \pi^- \pi^-$ 
   \mbox{\hspace{15ex}or} 
   $K^{(*)}{}^+ \ell^- \ov{\nu}_{\ell}$ etc.] &
\index{decay!$B_s \rightarrow \ov{D}{}^0 K_S $}
\index{decay!$B_s \rightarrow \ov{D}{}^{*0} K_S $}
\\\hline
$\ov{b} \rightarrow \ov{c} c \ov{s}$ & 
   $B_d \rightarrow \psi K^+ \pi^-$ & 
        mainly \mbox{$B_d \rightarrow \psi K^{(*)}$ 
        [$\rightarrow  K^+ \pi^-$]}  
\index{decay!$B_d \rightarrow \psi K^+ \pi^-$}
\\
&  $B_d \rightarrow D^{(*)+} D_s^{(*)-}$ & 
\index{decay!$B_d \rightarrow D^+ D^-$}
\index{decay!$B_d \rightarrow D^{*+} D^-$}
\index{decay!$B_d \rightarrow D^+ D^{*-}$}
\index{decay!$B_d \rightarrow D^{*+} D^{*-}$}
\\\hline
$\ov{b} \rightarrow \ov{c} c \ov{d}$ & 
$B_s \rightarrow \psi K^- \pi^+ $  & 
        mainly \mbox{$B_s \rightarrow \psi \ov{K}{}^{(*)}$ 
        [$\rightarrow  K^- \pi^+$]}  
\index{decay!$B_s \rightarrow \psi K^- \pi^+$}
\\
&  $B_s \rightarrow D_s^{(*)}{}^- D^{(*)}{}^+$
& 
\index{decay!$B_s \rightarrow D_s^+ D_s^-$}
\index{decay!$B_s \rightarrow D_s^{*+} D_s^-$}
\index{decay!$B_s \rightarrow D_s^+ D_s^{*-}$}
\index{decay!$B_s \rightarrow D_s^{*+} D_s^{*-}$}
\\\hline
$\ov{b} \rightarrow \ov{c} X $ 
& $B_d \rightarrow D^{(*)}{}^- X$  & 
        small contamination from \hfill\break 
	$\ov{b} \rightarrow \ov{c} c \ov{d}$ 
\index{decay!$B_d \rightarrow D^- X$}
\index{decay!$B_d \rightarrow D^{*-} X$}
\\\hline  
\end{tabular}
\end{center}
\caption{Interesting flavor-specific decays.}
\label{chmix:tab:fsd}
\end{table}

\begin{table}[thbp]
\begin{center}
\begin{tabular}{@{}r|p{0.49\textwidth}|p{0.31\textwidth}} \hline
quark decay & hadronic decay & remarks \\\hline\hline  
$\ov{b} \rightarrow \ov{c} u \ov{d}$ 
& $B_d \rightarrow \ov{D}{}^{(*)0}  \rho^0 $
  [$\rightarrow  \rho_0 K_S$ or $K \ov{K}$ or $\pi^+ \pi^- $] 
  &   The $\ov{D}{}^{(*)0}$ must be detected in a $CP$-specific final state $f$
      (hence $D^{(*)}{}^0\rightarrow f$ is equally allowed). 
\index{decay!$B_d \rightarrow \ov{D}{}^0  \rho^0 $}
\index{decay!$B_d \rightarrow \ov{D}{}^{*0}  \rho^0 $}
\\
& $B_d \rightarrow \ov{D}{}^{(*)0} \pi^+ \pi^- $
  [$\rightarrow  \rho_0 K_S$ or $K \ov{K}$ or $\pi^+ \pi^- $] 
  & This decay mode has color-unsuppressed contributions.  
\index{decay!$B_d \rightarrow \ov{D}{}^{0} \pi^+ \pi^- $}
\index{decay!$B_d \rightarrow \ov{D}{}^{*0} \pi^+ \pi^- $}
\\
& $B_s \rightarrow \ov{D}{}^{(*)0} K_S$
  [$\rightarrow  \rho_0 K_S$ or $  K \ov{K}$ or $\pi^+ \pi^- $] &  
\index{decay!$B_s \rightarrow \ov{D}{}^0 K_S $}
\index{decay!$B_s \rightarrow \ov{D}{}^{*0} K_S $}
\\\hline
$\ov{b} \rightarrow \ov{c} c \ov{s}$ 
& $B_d \rightarrow \psi K_S$ &
\index{decay!$B_s \rightarrow \psi K_S$}
\\
& $B_d \rightarrow \psi K_S \rho^0$ & 
The $\ov{B}_d$ can decay into the
same final state $K_S \rho^0$. \hfill\break 
Angular analysis separates $CP$-eigenstates.
\index{decay!$B_d \rightarrow \psi K_S \rho^0$}
\\
& $B_d \rightarrow \psi \phi K_S$ or $\psi K_S \rho^0$ & 
                        Angular analysis required.
\index{decay!$B_d \rightarrow \psi \phi K_S$}
\index{decay!$B_d \rightarrow \psi K_S \rho^0$}
\\
& $B_d \rightarrow \psi \phi K^*$ [$\rightarrow K_S \pi^0$] &
                        Angular analysis required, \hfill\break
			$\pi^0$ is problematic. 
\index{decay!$B_d \rightarrow \psi \phi K^*$}
\\
&  $B_d \rightarrow D_s^{(*)}{}^+ D_s^{(*)}{}^- K_S$ &
                        Angular analysis required.
\index{decay!$B_d \rightarrow D_s^+ D_s^- K_S$}
\index{decay!$B_d \rightarrow D_s^{*+} D_s^{*-} K_S$}
\\
& $B_s \rightarrow \psi \phi$ & 
                        Angular analysis required. 
\index{decay!$B_s \rightarrow \psi \phi$}
\\
& $B_s \rightarrow \psi K \ov{K}$ or $\psi K^* \ov{K}{}^*$ &
                        Same remark 
\index{decay!$B_s \rightarrow \psi K \ov{K}$}
\index{decay!$B_s \rightarrow \psi K^* \ov{K}{}^*$}
\\
& $B_s \rightarrow \psi \phi \phi$ &
                        Same remark. 
\index{decay!$B_s \rightarrow \psi \phi \phi$}
\\
& $B_s \rightarrow \psi \eta $& 
\index{decay!$B_s \rightarrow \psi \eta $}
\\
& $B_s \rightarrow \psi \eta^\prime$ & 
\index{decay!$B_s \rightarrow \psi \eta^\prime$}
\\
& $B_s \rightarrow D_s^+ D_s^-$ &
\index{decay!$B_s \rightarrow D_s^+ D_s^-$}
\\
& $B_s \rightarrow D_s^{*}{}^+ D_s^{*}{}^-$ & 
                        Angular analysis required.
\index{decay!$B_s \rightarrow D_s^+ D_s^-$}
\index{decay!$B_s \rightarrow D_s^{*+} D_s^-$}
\index{decay!$B_s \rightarrow D_s^+ D_s^{*-}$}
\index{decay!$B_s \rightarrow D_s^{*+} D_s^{*-}$}
\\
& $B_s \rightarrow D^{(*)+} D^{(*)-}$ or $D^{(*)0}\ov{D}{}^{(*)0}$ &
                        Non-spectator decays.
\index{decay!$B_s \rightarrow D^+ D^-$}
\index{decay!$B_s \rightarrow D^{*+} D^-$}
\index{decay!$B_s \rightarrow D^+ D^{*-}$}
\index{decay!$B_s \rightarrow D^{*+} D^{*-}$}
\\ 
& $B_s \rightarrow \psi f_0 $ &
                        $CP$-odd final state.
\index{decay!$B_s \rightarrow \psi f_0$}
\\
& $B_s \rightarrow \chi_{c0} \phi $ &
                        $CP$-odd final state.
\index{decay!$B_s \rightarrow \chi_{c0} \phi $}
\\\hline 
$\ov{b} \rightarrow \ov{c} c \ov{d}$ 
& $B_d \rightarrow D^+ D^-$ &
                        $CP$-even final state.
\index{decay!$B_d \rightarrow D^+ D^-$}
\index{decay!$B_d \rightarrow D^{*+} D^-$}
\index{decay!$B_d \rightarrow D^+ D^{*-}$}
\index{decay!$B_d \rightarrow D^{*+} D^{*-}$}
\\
& $B_d \rightarrow D^{*}{}^+ D^{*}{}^-$ & 
\\
& $B_d \rightarrow \psi \rho^0$ & 
\index{decay!$B_d \rightarrow \psi \rho^0$}
\\
& $B_s \rightarrow \psi K_S$ &
\index{decay!$B_s \rightarrow \psi K_S$}
\\
& $B_s \rightarrow \psi K_S \pi^0$ &         
    Mainly \hfill\break $B_s \rightarrow \psi \ov{K}{}^{(*)}$ 
                [$\rightarrow  K_S \pi^0$].\hfill
\index{decay!$B_s \rightarrow \psi K_S \pi^0$}%
\\\hline
\end{tabular}
\end{center}
\caption{Interesting CKM-favored decays into $CP$-eigenstates.}
\label{chmix:tab:cpe}
\end{table}

\begin{table}[hbtp]
\begin{center}
\begin{tabular}{@{}r|p{0.49\textwidth}|p{0.31\textwidth}} \hline
quark decay & hadronic decay & remarks \\\hline\hline  
$\ov{b} \rightarrow \ov{c} c \ov{s}$ 
& $B_s \rightarrow \psi K \ov{K}{}^*$ combined with 
        $\psi  K^* \ov{K}$ & angular analysis plus \hfill\break analysis
analogous to \hfill\break $B_s \rightarrow D_s^{\pm} K^{\mp}$ required.
\index{decay!$B_s \rightarrow \psi K \ov{K}{}^*$}
\index{decay!$B_s \rightarrow \psi K^* \ov{K}$}
 \\ \hline
\end{tabular}
\end{center}
\caption{Interesting CKM-favored decays into $CP$ non-eigenstates 
         accessible to $B$ and $\ov{B}$.}
\label{chmix:tab:cpne}
\end{table}

\section{Introduction and Physics Input for CDF}

This section summarizes technical issues and physics inputs that are relevant to
mixing and lifetime measurements at CDF. The CDF detector improvements that are
critical for these measurements are the following
\index{CDF!detector improvements}:
%
\begin{enumerate}
\item The Level-1 and Level-2 trigger systems have been upgraded to allow
  triggers on high momentum tracks at Level-1, using the central drift chamber (``XFT''),
  and on large impact parameter
  tracks at Level-2, using the silicon vertex detector (``SVT''). This in turn allows
  CDF to trigger on all-hadronic decays of $b$ hadrons, such as
  $B_s\to D_s\pi$. Extensive simulations, together with run~I data, have
  been used to estimate trigger rates and event yields for the analyses
  discussed below.
\item The silicon vertex detector has been upgraded for improved silicon tracking and
  extended fiducial coverage.  This upgrade is most relevant to analyses which
  depend critically on vertex position resolution, in particular the measurement of
  $B_s$ mixing.
\item The CDF muon system has been upgraded to allow extended fiducial coverage
 and lower trigger thresholds. The increase of the fiducial volume
  is treated as a simple scale factor for all analyses using muons; the change
  in trigger threshold applies to the analysis of central di-muons only.

\end{enumerate}
The basic information needed to make projections for Tevatron Run~II is
the event yield for the $b$ hadron decay channels in question. Projections
are relatively simple for channels obtained from lepton triggers
using extrapolations from the Run~I data. An example of
this scaling is given in~\ref{sec:cdf-sin2beta}.
The projections for channels which are triggered by the newly implemented
displaced track trigger system, often referred to as the hadronic trigger, are more
difficult and are based primarily on simulations.

The CDF trigger system is organized in three levels of which only the first two
are simulated for the following studies, since the third level should not
reject good signal events.  In addition to the trigger simulation,
physics inputs are needed for the total $B$ cross sections and production
and branching fractions. Both the physics inputs and the description of the
trigger simulation are given below. Some of the issues are already partially
covered in the $CP$ violation chapter in Section~\ref{sec:cdf-sin2beta}. Here
the emphasis is mostly on the hadronic trigger.
\subsection{Physics Input}

The Monte Carlo program Bgenerator~\cite{Ref:BGEN}
\index{CDF!Bgenerator}%
is used to generate $b$ hadrons; it parameterizes the $p_T$ and $y$ distributions for
$b$ quarks according to next-to-leading-order calculations~\cite{Ref:NDE}. The $b$
quarks are fragmented into $b$ hadrons using the
Peterson~\cite{Ref:PETERSON} function with a fragmentation parameter value of
$\epsilon_b=0.006$. The CLEO Monte Carlo Program QQ~\cite{Ref:CLEOMC} is used to
decay the $b$ hadrons.  Events generated with Bgenerator contain
particles only from the decay of the $b$ hadrons, and do not include
particles produced in the $b$ quark fragmentation or the underlying event from the $p\overline{p}$ collision.

The overall production cross section is normalized using the CDF
measurement for $B^0$ production with $\rm p_T > 6 ~\GeV, ~|y| < 1$
~\cite{Ref:CDF-XS-BD}.
The $B_s$ and $\Lambda_b$ production fractions in $p\bar{p}$ collisions
are based on the CDF measurement of $f_s / (f_u + f_d)$~\cite{Ref:CDF-FSFUFD}
and the world average value for $f_{\Lambda_b}$, respectively.
  Assuming the $b$ hadron production spectra follow the
distributions from~\cite{Ref:NDE}, and using the CDF measurement
from Reference~\cite{Ref:CDF-XS-BD}, a total production cross section
for $B^0$ mesons of 50.1 $\rm \mu b$ is obtained.
Table~\ref{tab:br-prod} lists the measured rates and production fractions
assumed in the CDF analyses, together with the relevant hadronic decay
branching fractions that are known. In addition, we have estimated
the branching fractions for $b$-hadron decays that have not been
measured directly, using various symmetry assumptions as
described below.

\begin{table}
\begin{center}
\begin{tabular}{ l  c  l }
\hline 
\hline 
Quantity                  & Value                     & Reference\\
\hline 
{\cal B}($D_s^+ \to \phi \pi^+$)    &$( 3.6  \pm 0.9  )\ten{-2}$& \cite{Ref:PDG}\\
{\cal B}($D_s^+ \to K^+\overline{K^{*0}}$)       &$( 3.3  \pm 0.9  )\ten{-2}$& \cite{Ref:PDG}\\
{\cal B}($D_s^+ \to \pi^+\pi^+\pi^-$)   &$( 1.0  \pm 0.4  )\ten{-2}$& \cite{Ref:PDG}\\
\hline 
{\cal B}($D^0\to K^-\pi^+$)     &$( 3.85 \pm 0.09 )\ten{-2}$& \cite{Ref:PDG}\\
{\cal B}($D^0\to K^-\pi^+\pi^+\pi^-$)     &$( 7.6  \pm 0.4  )\ten{-2}$& \cite{Ref:PDG}\\
\hline 
{\cal B}($\Lambda_c^+ \to p K^-\pi^+$)
                                &$( 5.0  \pm 1.3  )\ten{-2}$& \cite{Ref:PDG}\\
{\cal B}($\Lambda_c^+ \to \Lambda\pi^+\pi^-\pi^+$)
                                &$( 3.3  \pm 1.0  )\ten{-2}$& \cite{Ref:PDG}\\
\hline 
{\cal B}($\Lambda \to p \pi^-$) &$(63.9  \pm 0.5  )\ten{-2}$& \cite{Ref:PDG}\\
\hline 
{\cal B}($\phi(1020) \to K^+K^-$)        &$(49.1  \pm 0.8  )\ten{-2}$& \cite{Ref:PDG}\\
{\cal B}($K^*(892) \to K \pi$)       &$ 1                       $& \cite{Ref:PDG}\\
\hline 
$f_{\Lambda_b}$                 &$(0.116 \pm 0.020)\ten{-2}$& \cite{Ref:PDG}\\
$f_s/(f_u + f_d)$               &$(0.213 \pm 0.038)\ten{-2}$& \cite{Ref:CDF-FSFUFD}\\
$\mathrm{\sigma_B^0}(\mathrm{p_T(B^0)>6~\GeV;~|y|<1})$
                                &$(3.52  \pm 0.61 )\mu b   $& \cite{Ref:CDF-XS-BD}\\
\hline 
\hline 
\end{tabular}
\end{center}
\caption{Physics input used for event yield estimates.}
\label{tab:br-prod}
\end{table}

\paragraph{Branching Fraction Estimates}
Since many of the hadronic decay channels have so far not been measured or even
observed, certain branching ratios have to be estimated.
This is relatively simple for $B_s$ decays, using  related $B^0$
decay modes. These estimates are summarized in
Table~\ref{tab:bs-branchings}.  The related uncertainties should be small, on the
order of roughly 10\% in the form factors or 20\% in the event
yields.

\begin{table}
\begin{center}
\begin{tabular}{ l  c  l }
\hline 
\hline 
Quantity                        & Value               & Reference\\
\hline 
{\cal B}($B_s\to D_s \pi$)      &$(3.0\pm0.4)\ten{-3}$& from $B^0$~\cite{Ref:PDG}\\
{\cal B}($B_s\to D_s \pi\pi\pi$)&$(8.0\pm2.5)\ten{-3}$& from $B^0$~\cite{Ref:PDG}\\
{\cal B}($B_s\to D_s D_s$)      &$(8.0\pm3.0)\ten{-3}$& from $B^0$~\cite{Ref:PDG}\\
{\cal B}($B_s\to D_s^* D_s$)    &$(2.0\pm0.6)\ten{-2}$& from $B^0$~\cite{Ref:PDG}\\
{\cal B}($B_s\to D_s^* D_s^*$)  &$(2.0\pm0.7)\ten{-2}$& from $B^0$~\cite{Ref:PDG}\\
\hline 
\hline 
\end{tabular}
\end{center}
\caption{Branching fraction estimates for $B_s$ decays.}
\label{tab:bs-branchings}
\end{table}
\index{decay!$B_s \rightarrow D_s^- \pi^+$}
\index{decay!$B_s \rightarrow D_s^- \pi^+ \pi^- \pi^+$}
\index{decay!$B_s \rightarrow D_s^+ D_s^-$}
\index{decay!$B_s \rightarrow D_s^{*+} D_s^-$}
\index{decay!$B_s \rightarrow D_s^+ D_s^{*-}$}
\index{decay!$B_s \rightarrow D_s^{*+} D_s^{*-}$}


For $\Lambda_b$ baryons the situation is
more complicated. Several patterns arise when comparing bottom and charm
branching fractions, as well as meson and baryon branching fractions. The most
important is the fact that the branching fractions of $B$ mesons are often quite small
compared to those of the corresponding $D$ decays. For instance
\begin{eqnarray}
  {\cal B}(B^0\to D^-\pi^+) & = & (3.0 \pm 0.4) \ten{-3}\,,\nonumber\\
  {\cal B}(D^0\to K^-\pi^+) & = & (3.83\pm 0.09)\ten{-2}\,.
\end{eqnarray}
Comparing the two gives a $B$ to $D$ ratio of $0.08$.  One might suppose,
neglecting that $c \to s$ involves a light quark, that the widths of these modes
would be similar, but the total widths reflected in the mean lifetimes are
different. Moreover, the $b$ sector involves considerably more decay channels.
The semileptonic decays, however, remain qualitatively different, and do not
scale in the same way.

%

A similar pattern for the hadronic decay modes may reasonably be expected for
baryons.  Indeed, in the one hadronic branching fraction measured for the
$\Lambda_b$, we have:
\begin{eqnarray}
  {\cal B}(\Lambda_b\to J/\psi\Lambda) & = & (4.7\pm 2.8) \ten{-4}\,,\nonumber \\
  {\cal B}(\Lambda_c^+\to p K^{*0})    & = & (1.6\pm 0.5) \ten{-2}\,,\nonumber \\
  {\cal B}(\Lambda_c^+\to p \phi)      & = & (1.2\pm 0.5) \ten{-3}\,.
\end{eqnarray}
\index{decay!$\Lambda_b \rightarrow J\psi\Lambda$}
In comparing the $\Lambda_c^+$ branching fraction to the $\Lambda_b$ branching
fraction, it is assumed that virtual $W^- \to \overline{c}s$ occurs about as
frequently as $W^- \to \overline{u}d$.  The $\Lambda_b$ to $\Lambda_c^+$ ratio is about 0.03 when comparing
to $\Lambda_c^+ \to pK^{*0}$.  A similar comparison can be made with the second
$\Lambda_c^+$ decay mode, which, aside from the $|V_{us}/V_{ud}|^2$ factor, is
most similar to $\Lambda_b \to J/\psi \Lambda$; the ratio is about $0.02$.
However, since applying this factor to a color-suppressed mode is
problematic, the first ratio is preferred
\begin{equation}
  g_{bc} = 0.03
\end{equation}
to multiply $\Lambda_c^+$ hadronic branching fractions to obtain estimates of
corresponding $\Lambda_b$ fractions.

Another difference between charm and bottom decays is that the bottom hadrons
can avail themselves of the virtual $W^- \to \overline{c}s$ transition which is
disallowed for charm decays. The ratio of branching fractions for an external
$W^- \to \overline{c}s$ to $W^- \to \overline{u}d$ is similar to the ratio of
the square of the decay constants
\begin{equation}
  g_{D_s}^2 = \left(\frac{f_{D_s^+}}{f_{\pi^+}}\right)^2
            = \left(\frac{280\;{\rm MeV}}{130.7\;{\rm MeV}}\right)^2
          = 4.58 \,,
\end{equation}
the effect of which is seen in comparing branching fractions of $B^0 \to
D^-D_s^+$ and $D^-\pi^+$, with the usual caveats.

Adding a $\pi^+\pi^-$ to the final state of a decay mode tends to result in a
larger branching fraction. This effect is observed in the mesons, but the ratio
calculated among $\Lambda_c^+$ modes is preferred because of the different
baryon structure:
\begin{equation}
  g_{\pi\pi} = \frac{{\cal B}(\Lambda_c^+\to\Lambda\pi^+\pi^+\pi^-)}
                    {{\cal B}(\Lambda_c^+\to\Lambda\pi^+)}
             = \frac{3.3}{0.9} = 3.7 \,.
\end{equation}

\begin{table}
\begin{center}
\begin{tabular}{ l  l  c  c  c  }
\hline\hline
$\Lambda_b$ decay & $\Lambda_c$ decay &    ${\cal B}(\Lambda_c^+)$ &     estimate & ${\cal B}(\Lambda_b)$ estimate \\
\hline
$\Lambda_c^+\pi^-$&
            $\Lambda \pi^+$           & $b_1=(9.0\pm 2.8)\ten{-3}$ & $b_1 g_{bc}$ & $2.6\ten{-4}$ \\
$\Lambda_c^+\pi^-\pi^+\pi^-$ &
            $\Lambda \pi^+\pi^-\pi^+$ & $b_2=(3.3\pm 1.0)\ten{-2}$ & $b_2 g_{bc}$ & $9.7\ten{-4}$ \\
\hline
$pD^0\pi^-$(nr)   & $p K^- \pi^+$(nr) & $b_3=(2.8\pm 0.8)\ten{-2}$ & $b_3 g_{bc}$ & $8.2\ten{-4}$ \\
\hline
$p K^-$           &                   &                            & $b_1 g_{bc} g_{bu} g_{\pi K}$ & $8.1\ten{-6}$ \\
\hline
$p\pi^-$          &                   &                            & $b_1 g_{bc} g_{bu}$ & $2.0\ten{-6}$ \\
$p\pi^-\pi^+\pi^-$&                   &                            & $b_2 g_{bc} g_{bu}$ & $7.4\ten{-6}$ \\
\hline\hline
\end{tabular}
\end{center}
\caption{
  Branching ratio estimates for $\Lambda_b$ decays using scale factors
  described in the text.}
\label{tab:lb-branchings}
\end{table}
\index{decay!$\Lambda_b \rightarrow \Lambda_c^+\pi^-$}
\index{decay!$\Lambda_b \rightarrow \Lambda_c^+\pi^-\pi^+\pi^-$}
\index{decay!$\Lambda_b \rightarrow pD^0\pi^-$}

Another factor is used to estimate the branching fractions where the external
$W$ yields a $D_s^{*+}$ rather than a $D_s^+$.  Here the $B^0$ branching
fractions are used
\begin{equation}
  g_{*} = \frac{{\cal B}(B^0\to D^{(*)-}D_s^{*+})}
               {{\cal B}(B^0\to D^{(*)-}D_s^+)}
        = \frac{1.0 + 2.0}{0.80 + 0.96}
        = 1.7.
\end{equation}
A similar factor is obtained when comparing analogous decays with $\rho^+$ and
$\pi^+$ final states of $B^0$ decays.  Decay modes such as $\Lambda_c^+ \to
\Lambda\rho^+$ have not been observed.

Once certain $b \to c$ branching fractions have been estimated, they are scaled
by
\begin{equation}
  g_{bu} = |V_{ub}/V_{cb}|^2 \sim 0.0077
\end{equation}
to obtain estimates for corresponding $b \to u$ transitions.  Finally, the
recently measured branching fractions
\begin{eqnarray}
  {\cal B}(B^0 \to \pi^+\pi^-) & = &  (4.3\pm 1.6) \ten{-6},, \nonumber \\
  {\cal B}(B^0 \to K^+\pi^-)   & = & (17.2\pm 2.8) \ten{-6}\,,
\end{eqnarray}
are used to estimate $\Lambda_b \to p K^-$ from $\Lambda_b \to p \pi^-$:
\begin{equation}
  g_{\pi K} = \frac{17.2}{4.3} = 4.
\end{equation}
The resulting branching ratio estimates for the different
$\Lambda_b$ decay modes are summarized in Table~\ref{tab:lb-branchings}.

\subsection{Detector Simulation}
\index{CDF!detector simulation}

\begin{table}
\begin{center}
\begin{tabular}{ l  c  c  c  c  c  c }
\hline 
\hline 
$\rm B_s$ decay & $\rm \epsilon_{L1} $ &
$ \rm \epsilon_{L2Bs}$ & $\rm \epsilon_{L2Bd} $ & $\rm \epsilon_{L2Tot}$
& $ \rm \epsilon_{fid} $ & $\rm N_{RunII} $ \\
\hline 
$B_s \to D_s \pi$       & 0.025 & 0.0045 & 0.0029 & 0.0050 & 0.0027 & 36900 \\
$B_s \to D_s \pi\pi\pi$ & 0.018 & 0.0032 & 0.0018 & 0.0033 & 0.0011 & 38300 \\
\hline 
$B_s \to D_s D_s$       & 0.019 & 0.0035 & 0.0015 & 0.0037 & 0.0014 &  2500 \\
$B_s \to D_s^* D_s$     & 0.016 & 0.0031 & 0.0011 & 0.0032 & 0.0013 &  5700 \\
$B_s \to D_s^* D_s^*$   & 0.014 & 0.0030 & 0.0011 & 0.0031 & 0.0012 &  5200 \\
\hline 
\hline 
\end{tabular}
\end{center}
\caption{
  Event yields for hadronic $B_s$ decays relevant for $CP$ violation and
  $\Delta\Gamma_s$ measurements. Only the feasible $D_s$ decays to $\phi \pi$,
  $K^* K$ and $\pi\pi\pi$ are considered.}
\label{tab:hadrtrig-bs}
\end{table}

\begin{table}
\begin{center}
\begin{tabular}{ l  c  c  c  c  c  r }
\hline 
\hline 
$\Lambda_b$ (sub)decay & $\rm \epsilon_{L1} $   &
$\rm \epsilon_{L2Bs}$  & $\rm \epsilon_{L2Bd} $ &
$\rm \epsilon_{L2Tot}$ & $ \rm \epsilon_{fid} $ &
$\rm N_{RunII} $ \\
\hline 
$\Lambda_b \to \Lambda_c^+ \pi^- (\Lambda_c^+ \to p K^-\pi^+)$
                         & 0.026 & 0.0040 & 0.0029  & 0.0045 & 0.0031 &  2400 \\
$\Lambda_b \to \Lambda_c^+ \pi^-\pi^+\pi^- (\Lambda_c^+ \to p K^-\pi^+)$
                         & 0.017 & 0.0024 & 0.0014  & 0.0026 & 0.0012 &  3400 \\
$\Lambda_b \to p D^0 \pi^- (D^0 \to K^-\pi^+)$
                         & 0.029 & 0.0043 & 0.0036  & 0.0048 & 0.0032 &  6100 \\
$\Lambda_b \to p D^0 \pi^- (D^0 \to K^-\pi^+\pi^-\pi^+)$
                         & 0.020 & 0.0028 & 0.0018  & 0.0030 & 0.0012 &  4300 \\
$\Lambda_b \to p \pi^-$  & 0.056 & 0.0054 & 0.011   & 0.011  & 0.011  &  1300 \\
$\Lambda_b \to p \pi^-\pi^+\pi^-$
                         & 0.030 & 0.0041 & 0.0032  & 0.0046 & 0.0030 &  1300 \\
$\Lambda_b \to p K^-$    & 0.056 & 0.0053 & 0.011   & 0.011  & 0.011  &  5400 \\
\hline 
\hline 
\end{tabular}
\end{center}
\caption{
  Event yields for most sizeable hadronic $\Lambda_b$ decays.}
\label{tab:hadrtrig-lb}
\end{table}

\paragraph{Hadronic Trigger Only}
\index{CDF!hadronic trigger}
The Level-1 track trigger is based on a set of kinematic cuts originally
developed for two-body decays of neutral $B$ mesons. The Level-1 triggering
algorithm is therefore based on pairs of XFT trigger tracks. To reduce the background of
inelastic collisions relative to $B$ hadron production, only track pairs in
which both tracks have transverse momentum $p_T$ greater than a specified value
are considered. Because of the time that would be spent on combinatorics, an
event with more than six such tracks passes Level-1 automatically.
For real $B^0$ and $B^0_s$ decays of interest, the two highest $p_T$ tracks are correlated
in angle and generally have opposite charge; consequently, the Level-1 requirements
are chosen as follows:
%
\begin{list}{$\bullet$}{\setlength{\topsep}{3pt}\setlength{\itemsep}{0pt}
                        \setlength{\parsep}{0pt}}%
\item two tracks having opposite charge
\item individual track $p_T > 2.0~\GeV/c$
\item $p_T,1 + p_T,2 > 5.5~\GeV/c$
\item $\delta \phi   < 135 \deg $
\end{list}

The Level-2 trigger is based on tracking information provided by the
SVT~\cite{Ref:TDR,Ref:PAC}. One application for the hadronic $b$ trigger
is the decay $B^0 \to \pi^+\pi^-$, where the two pions give oppositely-charged
tracks with high transverse momenta and large impact parameters.  The trigger is
also used to select other multibody hadronic $b$ decays, but due to the
different kinematics of these decays, the Level-2 selection criteria are
modified to optimize the efficiencies ~\cite{Ref:PAC}.  An event
passes the Level-2 track trigger if it satisfies either option A) or option
B)\\[10pt]
\phantom{HAL}
\begin{minipage}[t]{0.46\textwidth}
A) Multi hadronic $B$ trigger
\begin{list}{$\bullet$}{\setlength{\topsep}{3pt}\setlength{\itemsep}{0pt}
                        \setlength{\parsep}{0pt}}%
\item$120\um <  | d_0 | <  1 \mm$
\item$2\deg < \delta\phi <  90\deg$
\item$ p_T \cdot X_v > 0$
\end{list}
\end{minipage} \hfill
\begin{minipage}[t]{0.46\textwidth}
B) Hadronic pair trigger
\begin{list}{$\bullet$}{\setlength{\topsep}{3pt}\setlength{\itemsep}{0pt}
                        \setlength{\parsep}{0pt}}%
\item $100\um <  | d_0 | <  1 \mm$
\item $20\deg < \delta\phi < 135\deg$
\item $p_T \cdot X_v > 0$
\item $d_{0,B} < 140 \um$
\end{list}
\end{minipage}
$ $\\[15pt]
For an event to be useful in offline analysis after it passes the trigger, all the
$b$ hadron decay products have to be reconstructible in the detector. The requirement for a $b$
hadron to be considered reconstructible in this simulation is that all its
stable, charged daughter particles are within $|\eta| < 1$ and have transverse
momenta greater than $400~{\MeV}/c$.
These requirements are probably conservative in two ways: first, track
reconstruction in Run~II will be possible over a larger $\eta$ range;
stand-alone silicon tracking will probably be possible up to $|\eta| <
2$. Second, the reconstruction efficiency for the COT will be similar to that for the CTC
during Run~I. The efficiency for track reconstruction
in the CTC extended down to $p_T \simeq 200~\MeV/c$ and rose over the
transverse momentum range $200~\MeV/c < p_T < 400~\MeV/c$, reaching about
93\% for tracks with $p_T > 400~\MeV/c$.

Applying the trigger to various $B_s$ and $\Lambda_b$ decays, we estimate the event yields at
the two trigger levels. A summary of these estimates is given in
Tables~\ref{tab:hadrtrig-bs} and \ref{tab:hadrtrig-lb} for $B_s$ and
$\Lambda_b$, respectively.

\paragraph{Hadronic Trigger Combined with Lepton}
Apart from the purely hadronic trigger there is the possibility of using the
hadronic trigger in conjunction with the lepton triggers. Therefore, a single
lepton requirement is combined with the requirement of a displaced track in the
SVT. This trigger option is studied for the semi-inclusive $B_s \to \nu \ell D_s
X$ sample.

The additional requirement of a displaced track allows a lower threshold for
the lepton momentum, while keeping the trigger rate at a reasonable level. The trigger
cross section for an 8~{\GeV} inclusive electron trigger would need to be prescaled
in Run~II.  However, it is possible to lower the cross section for a 4~{\GeV}
electron trigger below 100~nb by adding the displaced track found by SVT with
$p_T>2~\GeV/c$ and $d_0>120~\mu$m.

Since Run~I data are considered to be most reliable for predictions, the signal
yield for Run~II CDF is obtained by scaling the Run~I analysis results with the
ratio of the acceptances between Run~I and Run~II. The acceptance ratio between
Run~I and Run~II is obtained using a Monte Carlo sample of semileptonic $B_s$
decays containing a $D_s$.
The SVT tracking efficiency is assumed to be $\sim 75\%$ per track, or
$56\%$ for 2-tracks.

The $\ell + D_s$ sample composition is assumed to be,
\begin{list}{$\bullet$}{\setlength{\topsep}{3pt}\setlength{\itemsep}{0pt}
                        \setlength{\parsep}{0pt}}%
\item $e D_s$ : $\mu D_s$ = 50\% : 50\%
\item $B_s \to \ell \nu D_s$ : $\ell \nu D_s^*$ : $\ell \nu D_s^{**}$ =
  2 : 5 : 0; ~~~~ $D_s^{**}$ usually decays to $D^{0,\pm}$
\end{list}
The $E_T$ and SVT $d_0$ resolutions are taken to be of $14\%/\sqrt{E_T}$ and
$35~\mu$m, respectively.
All tracks ($\ell$, $K$, and $\pi$) are required to have $p_T > 400~\MeV/c$, and
to be in the fiducial volume of the tracking detector. The silicon vertex
detector coverage is $|z|<30$~cm in Run~I and $|z|<45$~cm in Run~II. The
standard analysis Run~I cuts are applied to the final state particles, namely $p_T(K) >
1.2~\GeV/c$, $p_T(\pi) > 0.8~\GeV/c$, and $3~\GeV/c^2 < M(\ell D_s) <
5~\GeV/c^2$.

The event yields for different lepton $p_T$ values are summarized in
Table~\ref{tab:hadrlepttrig-bs}, which shows signal yields per 2~{\ifb}.
Choosing a value of $3~\GeV/c$ lepton $p_T$, roughly 40k semileptonic $B_s$
decay are obtained in 2~{\ifb} of integrated luminosity.

\begin{table}
\begin{center}
\begin{tabular}{ l  c  r }
\hline\hline
Trigger             & Run~II/Run~I & $N_\mathrm{Run~II}$ \\
\hline
$8~\GeV \ell$       & 1.0          & 14000 \\
$2~\GeV \ell$ + SVT & 5.9          & 64000 \\
$3~\GeV \ell$ + SVT & 4.0          & 43000 \\
$4~\GeV \ell$ + SVT & 2.7          & 30000 \\
$5~\GeV \ell$ + SVT & 1.9          & 21000 \\
\hline\hline
\end{tabular}
\end{center}
\caption{
  Event yields corresponding to 2~{\ifb} for semileptonic $B_s$ decays ($B_s \to
  \nu \ell D_s X$) for different lepton $p_T$ trigger thresholds.}
\label{tab:hadrlepttrig-bs}
\end{table}

\paragraph{Rate Estimates -- Hadronic Trigger}
Because the trigger rates depend on the way in which the Tevatron is operated in
Run~II, different XFT trigger cuts were considered for three different running
scenarios.  Scenario A corresponds to a luminosity of less than $1\times
10^{32}\;{\rm cm^{-2}s^{-1}}$ with collisions every 396~ns, while scenarios B
and C correspond to luminosities of $1-2\ten{32}~{\rm cm^{-2}s^{-1}}$ with
collisions every 132~ns and 396~ns, respectively.  The cuts considered for each
scenario are listed in Table~\ref{tab:hadrtrig-rate} along with the total cross
section.  These expectations were derived using tracks recorded in Run~I with
additional hit occupancy close to the beam axis generated using the
MBR~\cite{Ref:MBR} Monte Carlo program.
\begin{table}
\begin{center}
\begin{tabular}{ c c c c }
\hline\hline
            & Scenario A & Scenario B & Scenario C \\
\hline
${\cal L}$  &    $<1\times 10^{32}\;{\rm cm^{-1}s^{-1}}$ &
              $1$-$2\times 10^{32}\;{\rm cm^{-1}s^{-1}}$ &
              $1$-$2\times 10^{32}\;{\rm cm^{-1}s^{-1}}$ \\
BX interval           &          396 ns &         132 ns   & 396 ns \\
\hline
$p_T^{(1)},p_T^{(2)}$ &     $>2~\GeV/c$ & $>2.25~\GeV/c$   & $>2.5~\GeV/c$\\
$p_T^{(1)}+p_T^{(2)}$ &   $>5.5~\GeV/c$ & $>6~\GeV/c$      & $>6.5~\GeV/c$ \\
$\delta \phi$         &    $<135^\circ$ & $<135^\circ$     & $<135^\circ$\\
\hline
cross section         & $252\pm18~\mu$b & $152\pm14~\mu $b & $163\pm16~\mu$b \\
\hline\hline
\end{tabular}
\end{center}
\caption{
  Level-1 XFT trigger cuts and cross sections for the three Tevatron operating
  scenarios considered.}
\label{tab:hadrtrig-rate}
\end{table}
The trigger cuts provide trigger rates which are compatible with the total
Level-1 bandwidth of approximately 50~kHz.

At Level-2, the impact parameter information associated with the tracks is
available, and the cuts described above are used to select $b$ hadron decays.
The requirements are that the impact parameters of both tracks satisfy $120~{\rm
  \mu m} < |d| < 1 \;{\rm mm}$, that their point of intersection occurs with a
positive decay length, and that their opening angle is further restricted to
$\delta \phi < 90^\circ$.  The trigger cross sections are reduced to
approximately 489, 386 and 283~nb for scenarios A, B and C, respectively, and
produce Level-2 trigger rates between 38~Hz and 67~Hz. This is well within the
available Level-2 bandwidth of 300~Hz.

For efficiency estimates in the sections below, scenario A has been chosen.
When implementing trigger option B the numbers of expected events do not
vary significantly compared with the uncertainties quoted above. The
trigger scenario C is not likely to be implemented, asssuming that the Tevatron will be
upgraded to 132~ns bunch spacing.

\boldmath
\section{Projections for $\Delta m$}
\unboldmath

\boldmath
\subsection[$B_s$ mixing measurement at CDF]
{$B_s$ mixing measurement at CDF
$\!$\authorfootnote{Authors: M.~Jones, Ch.~Paus, M.~Tanaka.}}
\unboldmath
\index{CDF!$B_s$ mixing}

The probability that a $B_s$ meson decays at proper time $t$ in the same state,
or has mixed to the $\overline{B}_s$ state is given by
\begin{eqnarray}
  P_{\rm unmix}(t) & = & \frac{1}{2}\left(1 + \cos\Delta m_s t\right)\,, \nonumber \\
  P_{\rm mix}(t)   & = & \frac{1}{2}\left(1 - \cos\Delta m_s t\right)\,,
\end{eqnarray}
where the mixing frequency, $\Delta m_s$, is the mass difference between the
heavy and light $CP$ eigenstates.

The canonical $B_s$ mixing analysis, in which oscillations are observed and the
mixing frequency, $\Delta m_s$, is measured, proceeds as follows. The $B_s$
meson flavor at the time of its decay is determined by reconstructing a flavor
specific final state. The proper time, $t=m_{B^0_s}L/p c$, at which the decay
occurred is determined by measuring the decay length, $L$, and the $B_s$
momentum, $p$.  Finally, the production flavor must be tagged in order to
classify the event as being mixed or unmixed at the time of its decay.
Oscillations manifest in a time dependence of, for example, the mixed asymmetry:
\begin{equation}
  A_{\rm mix}(t) = \frac{N_{\rm mixed}(t)-N_{\rm unmixed}(t)}
                        {N_{\rm mixed}(t)+N_{\rm unmixed}(t)}.
\end{equation}

In practice, the production flavor will be correctly tagged with a probability
${\cal P}_{\rm tag}$ which is significantly smaller than unity. The functional
form of the mixed asymmetry follows
\begin{equation}\label{eqn-amix}
  A_{\rm mix}(t) = -D \cos \Delta m_s t
\end{equation}
with the dilution, $D$, related to $P_{\rm tag}$ by $D=2P_{\rm tag}-1$.  The
mixing frequency is determined for example by fitting the measured asymmetry to
a function of this form.

So far, $B_s$ oscillations have not been observed experimentally, and the lower limit
on $x_s$ is above 15. This means that $B_s$ mesons oscillate much more rapidly
than $B^0$ mesons. The rapidity of the $B_s$ oscillation implies a
significant difference in the experimental requirements for the $B^0$ and $B_s$ analyses. The
limiting factor in $B^0$ mixing analyses is solely the effective tagging
efficiency, which is equivalent to the effective statistics. In $B_s$ mixing
measurements the resolution of the proper time becomes another very critical
issue. To determine the proper time, not only the positions of primary and
secondary vertices have to be measured precisely, but also the measurement of the
$B_s$ momentum is crucial. Therefore, it is desirable to have fully exclusive
final states such as $B_s \to D_s \pi (D_s \to \phi \pi, \phi \to KK)$.
Semileptonic $B_s$ decays have the intrinsic disadvantage that the neutrino
momentum is undetected.

In Run~I CDF reconstructed 220 and 125 $B_s$ semileptonic decay events
with fully reconstructed $D_s \to \phi \pi$ and $D_s \to K^* K$ channels,
respectively, in low $p_T$ ($>8~\GeV/c$) inclusive lepton ($e$ and $\mu$)
trigger samples~\cite{Ref:SemiBsLifetime}.  An additional 600 semileptonic $B_s$
decays were used, reconstructed in the $D_s \to \phi X$ + track channel. Those
events were part of the dilepton ($\mu\mu$ and $e\mu$) trigger samples, where
the second lepton was used for the $B$ flavor tag.  The best limit on $x_s$ was
given by the dilepton trigger dataset~\cite{Ref:SemiBsMixing}.

In Run~II much more statistics will be available using the lepton trigger, the
lepton trigger plus one secondary vertex track and the all hadronic trigger.
From the event yield estimates in Tables~\ref{tab:hadrlepttrig-bs} and
\ref{tab:hadrtrig-bs}, we expect 40k events in the lepton plus displaced track and 75k
events in the all hadronic trigger.

In the following four sections the measurements of $B_s$ mixing using
semileptonic or hadronic decays are discussed. Since the flavor tagging and
the sensitivity estimates are very similar for the
semileptonic and hadronic $B_s$ decay samples, they will be discussed first.

\boldmath
\subsubsection{Projections for Sensitivity to $x_s$}
\unboldmath

The mixing frequency
\index{B mixing@$B_s$ mixing!mixing frequency}
can be determined by calculating, for example, a maximum
likelihood function derived from the measured and expected asymmetries and
minimizing this function with respect to $\Delta m_s$.  The significance of an
observation of mixing is quantified in terms of the depth of this minimum
compared with the second deepest minimum or some asymptotic value at large
$\Delta m_s$.  To a good approximation, the {\it average} significance
\index{B mixing@$B_s$ mixing!average significance}
is given
as
\begin{equation}\label{eqn-sig}
  {\rm Sig}(\Delta m_s) = \sqrt{\frac{N\epsilon D^2}{2}}
                          e^{-(\Delta m_s \sigma_t)^2/2}\sqrt{\frac{S}{S+B}}
\end{equation}
where $N=S$ is the number of reconstructed $B_s$ signal events, $S/B$ is the
signal-to-background ratio, $\epsilon$ is the efficiency for applying the flavor
tag with associated dilution $D$, and $\sigma_t$ is the average resolution with
which the proper time is measured. This definition is essentially the same as
what would be used to define $n \sigma$ confidence intervals for a Gaussian
probability density function.

Given estimates for these parameters, the limit of sensitivity is defined as the
maximal value of $\Delta m_s$ for which the significance is above a specified
value. The studies described here use the canonical 5 standard deviations to
define an unambiguous observation of mixing. In the following sections the
estimates for $N$, $\epsilon D^2$, $\sigma_t$ and $S/B$ are described.

\subsubsection{Flavor Tagging Efficiency}
\index{CDF!flavor tagging}
\index{flavor tagging!CDF}

In Run~II, a Time-of-Flight detector will provide CDF with the ability to
distinguish kaons from pions at the $2\sigma$ level below a momentum of about
$1.6~\GeV/c$.  This allows two new flavor tags to be implemented which rely on
the charge of kaons identified in the event to tag the production flavor of the
$B_s$. As a summary, in Table~\ref{tab:cdf-taggers} the tagging efficiency for $B_s \to D_s
\pi$
\index{decay!$B_s \rightarrow D_s^- \pi^+$}
expected for Run~II is compared to equivalent numbers obtained in Run~I.
We compare the standard figure of merit for each tagger, namely $\epsilon D^2$. The two
additional kaon taggers are briefly explained below.

\begin{table}
\begin{center}
\begin{tabular}{ l  r  r  }
\hline\hline
Method     & Run~I -- $\epsilon D^2$ & Run~II -- $\epsilon D^2$ \\
\hline
SLT        &                  1.7\% &                     1.7\% \\
JQT        &                  3.0\% &                     3.0\% \\
SST(kaon)  &                  1.0\% &                     4.2\% \\
OSK        &                    --- &                     2.4\% \\
\hline
Total      &                  5.7\% &                    11.3\% \\
\hline\hline
\end{tabular}
\end{center}
\caption[Comparison of the various flavor taggers in terms of the 
$\epsilon D^2$ parameter between Run~I and expectations for Run~II.]
{Comparison of the various flavor taggers in terms of the $\epsilon D^2$
  parameter between Run~I and expectations for Run~II. The most significant
  differences are the kaon taggers based on the new Time-of-Flight detector.}
\label{tab:cdf-taggers}
\end{table}

\paragraph{Opposite Side Kaon Tag}
\index{CDF!flavor tagging!opposite side kaon}
Due to the $b\to c\to s$ weak decays, $B$-mesons containing a $b$ quark will be
more likely to contain a $K^-$ in the final state than a $K^+$.  As for all the
other opposite side taggers the determination of the quark flavor on the
opposite side determines the flavor on the vertexing side,
since $b\overline b$ quarks are produced in pairs.

For the opposite side kaon tagging, kaon candidates are selected that are well separated from
the reconstructed $B_s$ decay. Kaon candidates coming
from a $b$ hadron decay are separated from prompt kaons by requiring a large
impact parameter. These requirements are implemented by imposing an isolation
cut of
\begin{equation}
  \Delta R_{\eta \phi}=\sqrt{\Delta \eta^2+\Delta \phi^2}>1
\end{equation}
and a cut on the combined transverse and longitudinal impact parameters of
\begin{equation}
  \chi^2_{d_0z_0} = \frac{d_0^2}{\sigma^2_{d_0}}+
                    \frac{z^2_0}{\sigma^2_{z_0}} > 9.
\end{equation}
Tagging the production flavor of the $B_s$ using the charge of the kaon selected
in this way gives a contribution to the tagging efficiency of
\begin{equation}
  \epsilon D^2 = (2.4\pm 0.2)\%.
\end{equation}

\paragraph{Same Side Kaon Tag}
\index{CDF!flavor tagging!same side kaon}
Same side kaon tagging in $B_s$ decays is the equivalent of same side pion
tagging in $B^0$ decays. In the hadronization process, when a $B_s$ meson is
produced, an $s\bar{s}$ pair must be popped from the vacuum during
fragmentation. The remaining $s$ or $\bar{s}$ quark is likely to join with
a $\bar{u}$ or $u$ quark to form a charged kaon. The charge of the kaon thus
depends on the flavor of the $B_s$ meson at production.

To estimate the $\epsilon D^2$ of the same side kaon tagging algorithm the same
side pion tagging algorithm is extended with particle identification
using Time-of-Flight information. It should be noted that $\epsilon D^2$ for this
algorithm is strongly dependent on the momentum spectrum of the $B_s$ meson.
Therefore, a rough simulation of the Level-1 and Level-2 triggers are applied to
the Monte Carlo sample. The $B_s~p_T$ spectrum of this event sample peaks at
around $10~\GeV/c$, and $\epsilon D^2$ is estimated to be
\begin{equation}
  \epsilon D^2 = (4.2\pm 0.3)\%.
\end{equation}

\boldmath
\subsubsection{$B_s$ Mixing with Semileptonic Decays}
\unboldmath
\index{B mixing@$B_s$ mixing!semileptonic}

As discussed in the introduction the key issue for semileptonic $B_s$
decays is the resolution of the proper time measurement. Including the estimates for
flavor-tagged event
yields, the $x_s$ sensitivity determination is straightforward.

\paragraph{Proper Time Resolution}
\index{CDF!proper time resolution!semileptonic}
The proper decay time in semileptonic decays is derived as follows
\begin{equation}
  ct = \frac{L_T(B_s) M(B_s)}{p_T(B_s)} =
           \frac{L_T(B_s) M(B_s)}{p_T(\ell D_s)} \cdot K,, \qquad
  K  = \frac{p_T(\ell D_s)}{p_T(B_s)} \,,
\end{equation}
where the transverse decay length, $L_T(B_s)$, and transverse momentum, $p_T(\ell
D_s)$, are measured from data. The $K$ factor which is used as an average
correction for the the incomplete $B_s$ reconstruction is obtained from Monte
Carlo samples, and $M(B_s)$ is the $B_s$ mass~\cite{Ref:PDG}. The proper time
resolution is given as
\begin{equation}
  \sigma_t = \sigma_{t_0} \oplus t \cdot \frac{\sigma K}{K},
\end{equation}
where the constant term ($\sigma_{t_0} \sim 60$~fs) is due to the beam spot and the vertex
detector resolution, and the $K$ factor resolution ($\sigma K/K \sim 14\%$) is
due to the momentum spectrum of the undetected particles, namely
the neutrino in the $B_s$ decay or the photon/$\pi^0$ in the subsequent $D_s^*$
decay.

\begin{figure}
\begin{center}
  \mbox{\epsfig{file=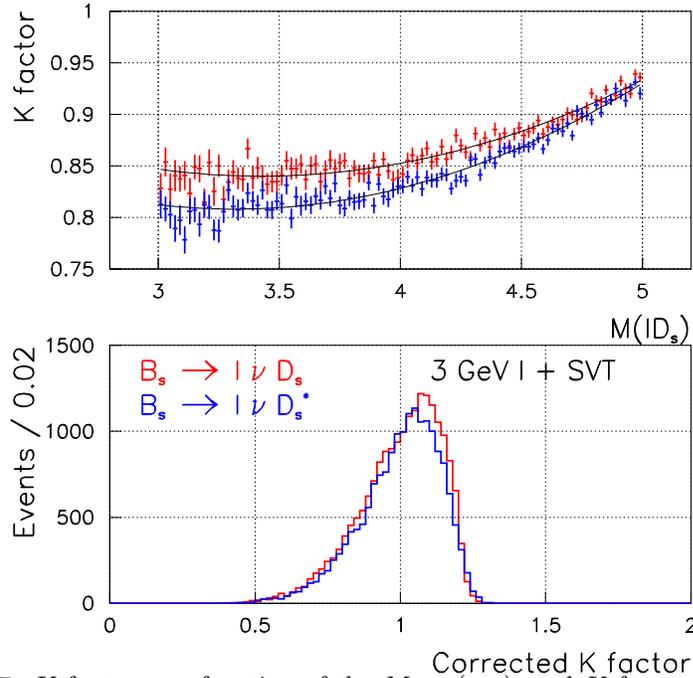,width=0.6\textwidth}}
  \caption{
    $K$ factor as a function of the $M_{\ell D_s}$ (top), and $K$ factor
    distribution after the $M_{\ell D_s}$ correction (bottom).}
  \label{fig:kfac-mld}
\end{center}
\end{figure}

The $K$ factor
\index{B mixing@$B_s$ mixing!K factor}
depends strongly on the lepton
+ $D_s$ invariant mass, $M(\ell D_s)$;
this dependence is shown in Figure~\ref{fig:kfac-mld} for the $D_s$ and $D_s^*$ channels.
Since the invariant lepton +
$D_s$ mass is measured, this dependence is corrected for on an event-by-event
basis to improve the $K$ factor resolution.

For the $B_s \to \ell D_s X$, $X = \nu + x$ channel, the following energy and
\index{decay!$B_s \rightarrow D_s^- \ell^+ \nu$}
momentum conservation rules are given
\begin{eqnarray}
  E_{B_S} & = & E_{\mu D_{S}} + E_{X}, \nonumber\\
  p_X     & = & |\vec{p}_{X}|^2= |\vec{p}_{B_S}-\vec{p}_{\mu D_S}|^2
          = p_{B_S}^2 + p_{\mu D_S}^2 - 2p_{B_S} p_{\mu D_S} \cos{\Theta},
\end{eqnarray}
where $\Theta$ is the angle in the laboratory frame between the $B_s$ and
$\ell+D_s$ directions, which are obtained using the 3D vertex information of the
Run~II SVX. By assuming $M_X=0$ the quadratic equation can be solved exactly.
Notice that the equation gives generally two solutions, $K^- < K^+$. The equation
sometimes has nevertheless no physical solutions due to the finite detector
resolution. Furthermore, the equation does not give the correct answer for the $B_s
\to \ell\nu D_s^*$ channel because of the missing photon,
$M_{X}=M_{\nu\gamma}>0$.

The $K$ factor distributions after all the trigger and offline selection cuts
are shown in Figure~\ref{fig:kfac-mld-3d}.  A vertex resolution of
$\sigma_{t_0}$ = 60~fs and $\sigma_{L_z}$ = 50~$\mu$m has been assumed.
The $D_s$ and the $D_s^*$ channels are displayed in the upper and lower plots,
respectively.  The probability for the quadratic equation to have a real
solution is approximately 50 percent. However, the correction still improves the
$K$ factor resolution if there is a physical solution.

\begin{figure}
\begin{center}
  \mbox{\epsfig{file=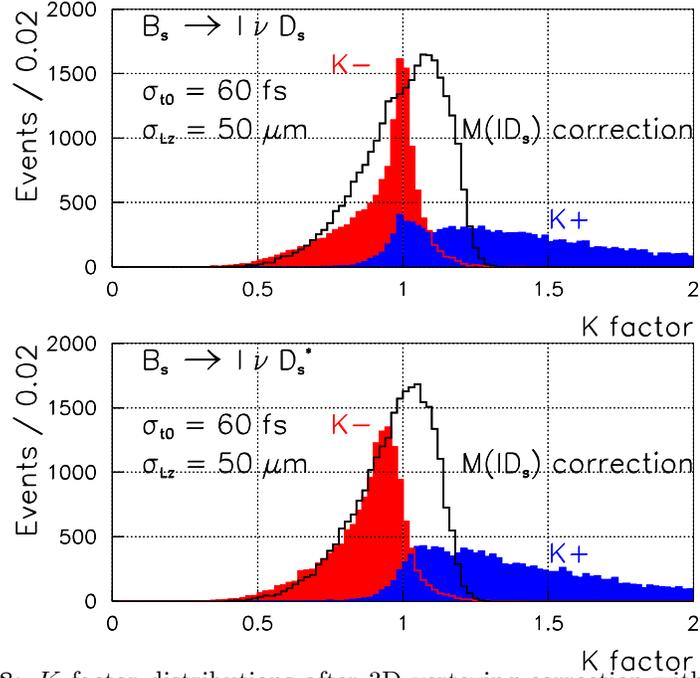,width=0.6\textwidth}}
  \caption{
    $K$ factor distributions after 3D vertexing correction with vertex detector
    resolutions of $\sigma_{t_0}$ = 60~fs and $\sigma_{L_z}$ = 50~$\mu$m.}
  \label{fig:kfac-mld-3d}
\end{center}
\end{figure}

Since the 3D vertexing correction strongly depends on the $\Theta$ resolution,
it improves for longer $B_s$ decay lengths.
In Figure~\ref{fig:time-res} the $K$ factor and time resolutions are shown as
functions of the proper decay time for the $D_s$ channel after the $M_{\ell D_s}$
correction and 3D vertexing correction.

To perform the 3D vertexing correction, it is assumed that the correct solution,
$K^-$ or $K^+$, is known, and that the $M(\ell D_s)$ corrected $K$ factor is
used if there are no physical solutions. The $K$ factor resolution is
significantly improved for the longer decay time events. Both channels $D_s$
and $D_s^*$ show similar results.  Unfortunately, the improved $K$ factor resolution
is not sufficient to greatly improve the sensitivity to $x_s$; practically
all of the sensitivity comes from the very short decay length events.

\begin{figure}
\begin{center}
  \mbox{\epsfig{file=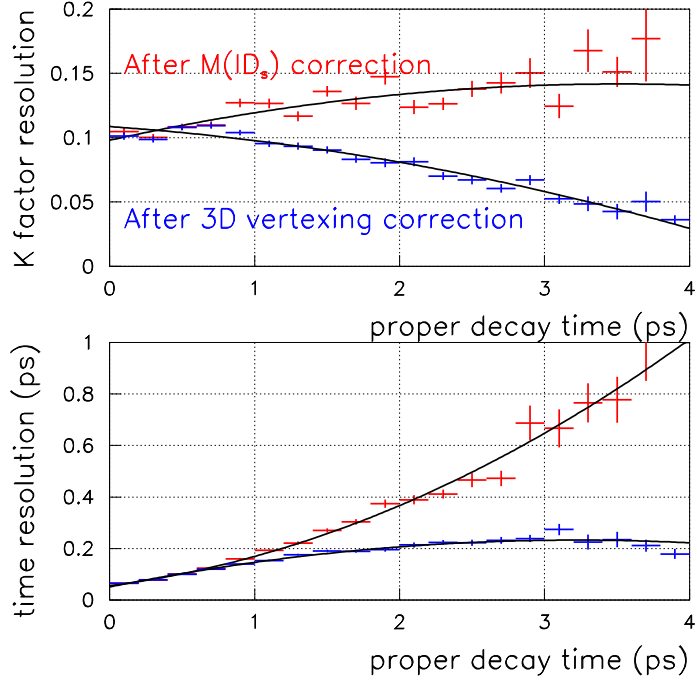,width=0.6\textwidth}}
  \caption{
    $K$ factor and time resolution for $B_s \to \ell\nu D_s$ decays.}
  \label{fig:time-res}
\end{center}
\end{figure}

\paragraph{Backgrounds}
In Run~I the signal to background ratio in $B_s$ reconstruction in the
semileptonic channels was typically 1:1. Since the kinematics of the Run~II
event sample will be somewhat different, a conservative signal to background
ratio of 1:2 is assumed in the following.

\paragraph{Projected Sensitivity}
The parameters which influence the projected $B_s$ mixing sensitivity,
calculated using equation~\ref{eqn-sig}, are summarized as follows:
\begin{eqnarray*}
  N(B_s)       : & &    43k~\mathrm{see~Table~\ref{tab:hadrlepttrig-bs}}\\
  \epsilon D^2 : & & 11.3\%~\mathrm{see~Table~\ref{tab:cdf-taggers}}\\
  \sigma_t     : & &        \mathrm{see~Figure~\ref{fig:time-res}}\\
  S/B          : & &    1:2~\mathrm{as~explained~above}
\end{eqnarray*}

\begin{figure}
\begin{center}
  \mbox{\epsfig{file=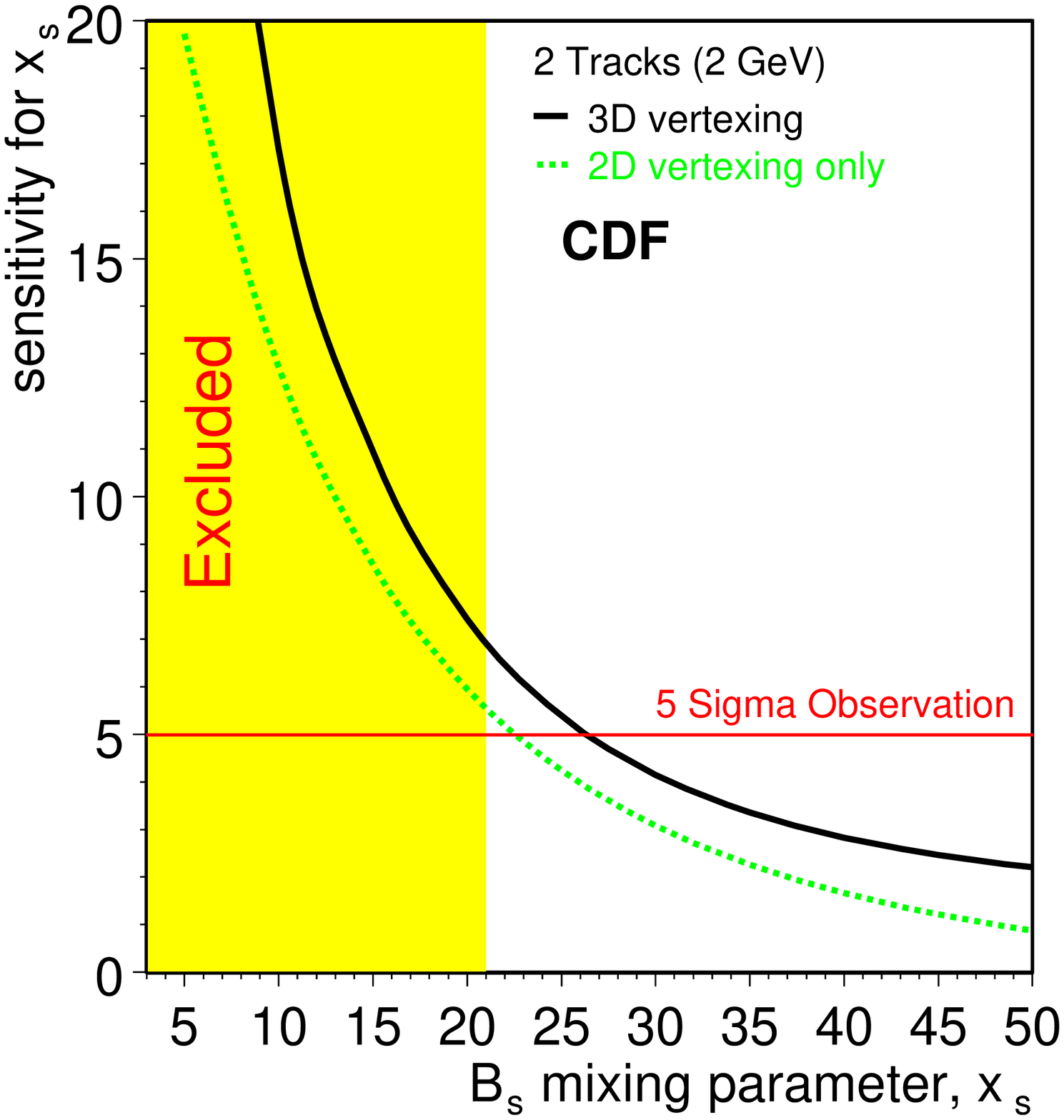,width=0.47\textwidth}}\hfill
  \mbox{\epsfig{file=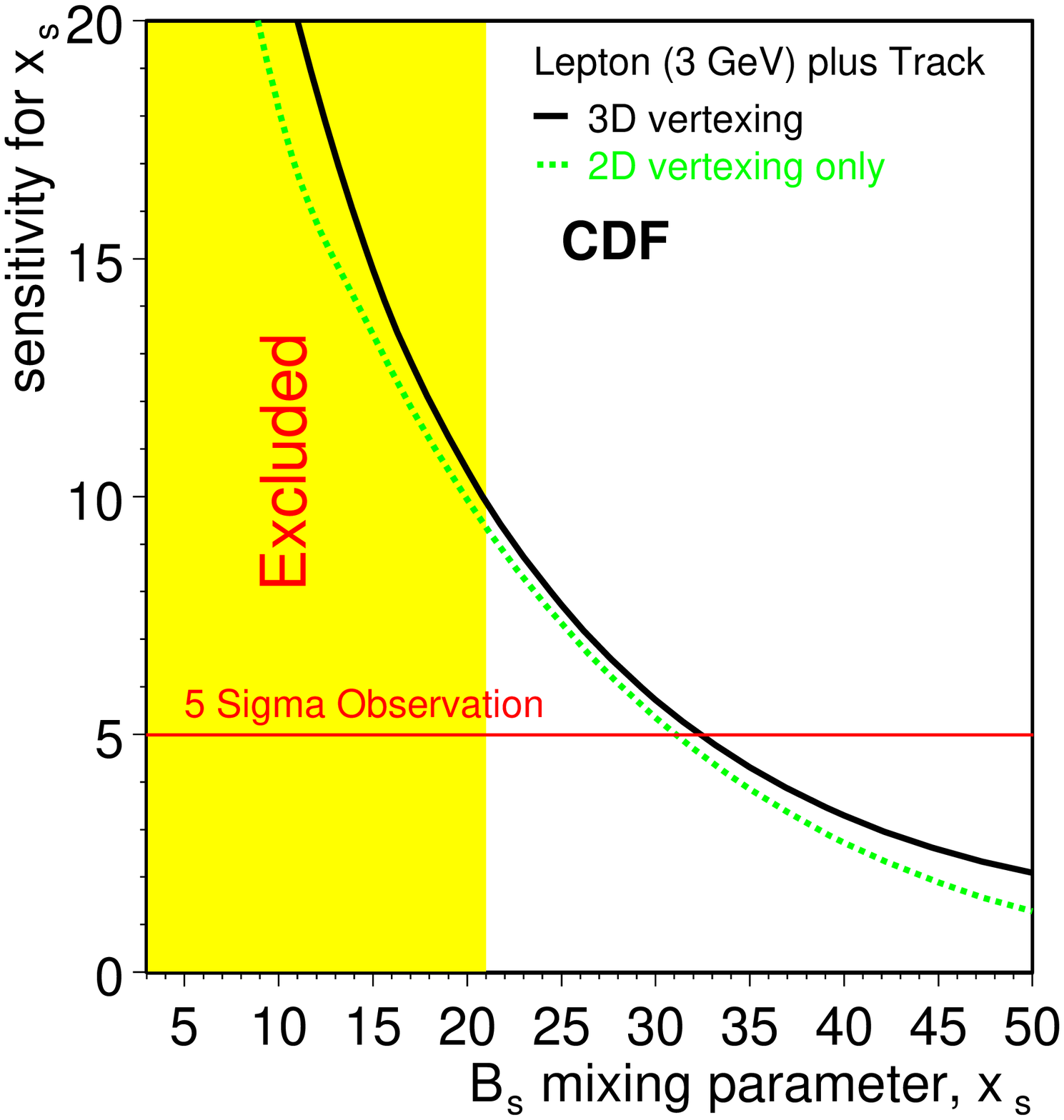,width=0.47\textwidth}}
  \caption[Sensitivity for measuring $x_s$ of the $B_s$ semileptonic decays for
the lepton plus displaced track trigger (left) and the two track trigger
(right).]{
    Sensitivity for measuring $x_s$ using $B_s$ semileptonic decays for the
    hadronic two track trigger (left) and the
    lepton plus displaced track trigger (right).
   The dashed lines show the significance after $M_{\ell D_s}$ correction,
    the solid lines after further applying the 3D vertex correction.}
  \label{fig:bs-semilep-sens}
\end{center}
\end{figure}

The analysis described above for the $3~\GeV/c$ lepton plus displaced track
triggers can easily be extended to the hadronic two-track triggers
defined in Table~\ref{tab:hadrtrig-rate}, where the trigger path
is satisfied by a semileptonic decay; given the large ($\sim 20\%$)
semileptonic combined branching ratio, the hadronic trigger turns out to be
highly efficient for semileptonic decay modes.
The significances for measuring $x_s$ for the two-track trigger (left)
and the lepton plus displaced track (right) are shown in
Figure~\ref{fig:bs-semilep-sens}. The dashed lines show the significance after
the $M_{\ell D_s}$ correction, and the solid lines after
the additional 3D vertexing correction. The $x_s$ reach of the semileptonic decay
sample is estimated to be about 30 for an observation with five standard
deviations. This is significantly less than that for the fully reconstructed hadronic channels,
which are
discussed below, but it does provide an independent trigger path.

\boldmath
\subsubsection{$B_s$ Mixing with Hadronic Decays}
\unboldmath
\index{B mixing@$B_s$ mixing!hadronic}

The fully hadronic event sample is particularly important for $B_s$ mixing
analyses since the fully reconstructed decays $B_s \to D_s^- \pi^+$ and $B_s \to
D_s^- \pi^+ \pi^- \pi^+$ have excellent proper time resolution, much smaller
than the expected period of oscillation.

\paragraph{Proper Time Resolution}
\index{CDF!proper time resolution!hadronic}
The proper time of a $B_s$ decay is calculated from the measured decay length,
and the reconstructed $B_s$ momentum.  The uncertainty on the proper time is
then given by
\begin{equation}
  \sigma_t = t\, \sqrt{\left(\frac{\sigma_L}{L}\right)^2 +
                    \left(\frac{\sigma_P}{P}\right)^2}.
\end{equation}
To resolve the rapid oscillations of the $B_s$ it is generally required that
this resolution not be significantly larger than the period of oscillation.  For
partially reconstructed semileptonic $B_s$ decays, the uncertainty in the
momentum is the limiting factor in the mixing analyses.  However, for the fully
reconstructed $B_s$ decays obtained using the XFT+SVT triggers, the momentum
uncertainty will be less than 0.4\%. This does not contribute significantly to
the overall proper time resolution and has been ignored in the projections
described below.

\paragraph{Backgrounds}
To date, no hadron collider experiment has operated with a displaced track based
trigger. Hence, the level of backgrounds to be expected in the $B_s$ sample is
uncertain. Data recorded by CDF in Run~I based on the single lepton triggers are
used to study the purity of the $B_s$ signal after imposing the XFT and SVT
trigger cuts on the opposite side $b$ hadron decays.

It has been observed that even a modest decay length cut suppresses the light
flavor contribution significantly. Therefore, the main concern is that the
signal is not overwhelmed by background from events containing real $b$- and
$c$-quarks.

Because of the small branching ratios of the $B_s \to D_s^+\pi^-,
D_s^+\pi^+\pi^-\pi^-$ decays, few, if any, such decays are expected to be
present in the Run~I data after imposing the XFT and SVT trigger cuts.  A
similar set of cuts with higher efficiency was used to search for the hadronic
$B^0$ and $B^+$ decays in the $D\pi$ final states. As a result of those studies
it is concluded that a signal to background ratio of 1:1 should be achievable.
To see the dependence of the significance on this parameter, this ratio is
varied between 1:2 and 2:1 in the following projections.

\paragraph{Projected Sensitivity}
The parameters which influence the projected $B_s$ mixing sensitivity,
calculated using equation~\ref{eqn-sig}, are summarized as follows:
\begin{eqnarray*}
  N(B_s)       & = &     75k~\mathrm{see~Table~\ref{tab:hadrlepttrig-bs}}\\
  \epsilon D^2 & = &  11.3\%~\mathrm{see~Table~\ref{tab:cdf-taggers}}    \\
  \sigma_t     & = &   0.045~\mathrm{ps}                                 \\
  S/B          & = & 1:2-2:1~\mathrm{as~explained~above}
\end{eqnarray*}
The results are again presented in terms of the dimensionless mixing parameter
$x_s=\Delta m_s \tau_{B_s}$ where the $B_s$ lifetime of
$1.54~\mathrm{ps}$~\cite{Ref:PDG99} is used.  In addition to the analytic
expression for the sensitivity, Equation~\ref{eqn-sig}, an alternative analysis
has been performed using a series of simulated
Monte Carlo experiments. The lifetime distributions of mixed and
unmixed decays are generated and the mixed asymmetry is fitted to
Equation~\ref{eqn-amix}. An example of the mixed asymmetry distribution and the
negative log-likelihood curve is shown in Figure~\ref{fig:example}. The negative
log-likelihood curve is shown as a function of $x_s$, obtained from one of these
Monte Carlo experiments.
\begin{figure}
  \mbox{\epsfig{file=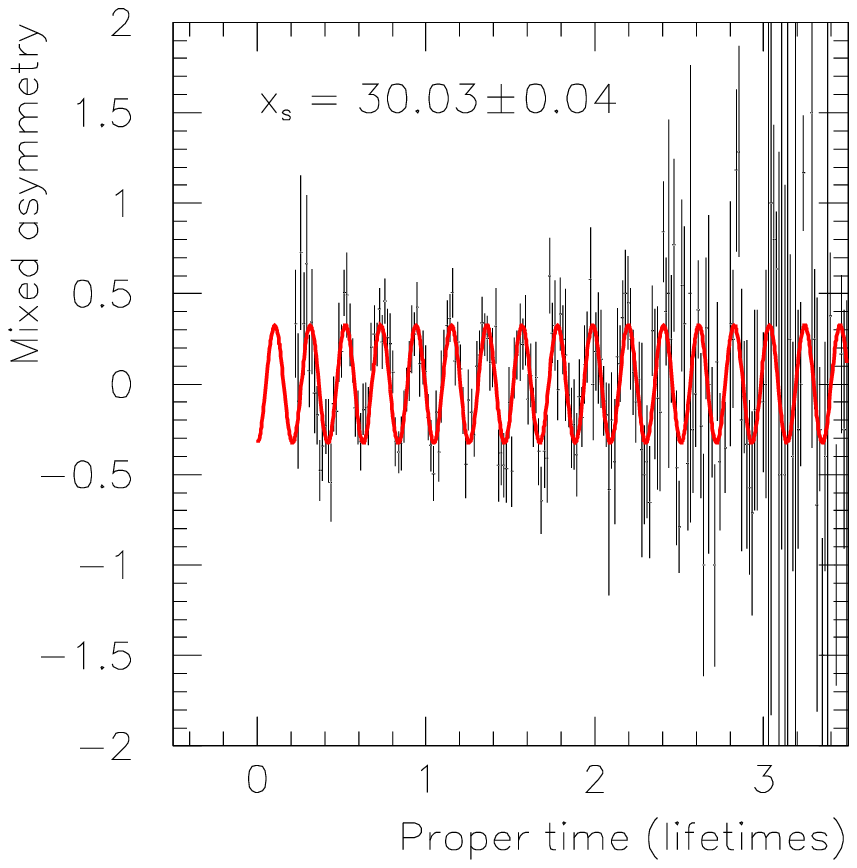,     width=0.47\textwidth}}
  \mbox{\epsfig{file=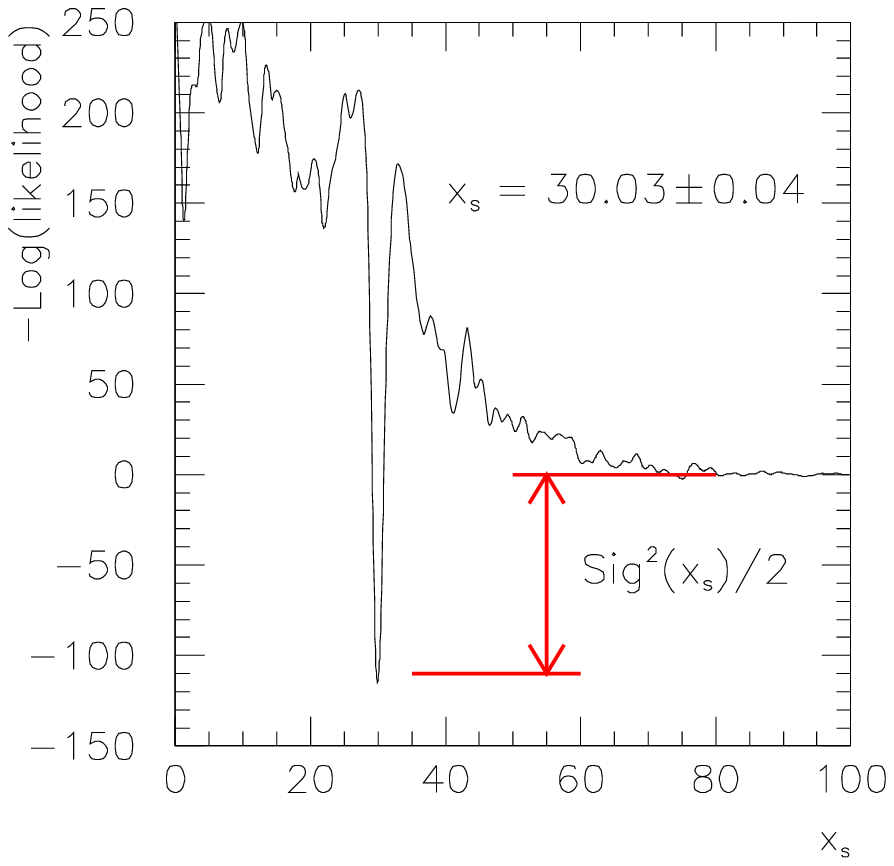,width=0.47\textwidth}}
  \caption[Example of a single toy Monte Carlo experiment:
the mixed asymmetry distribution, and negative log-likelihood
from the fit as a function of $x_s$]{
    Example of a single toy Monte Carlo experiment:
the mixed asymmetry distribution (left) and negative log-likelihood
from the fit as a function of $x_s$ (right)}
\label{fig:example}
\end{figure}
The comparison of the analytic expression with the averages of many Monte Carlo
simulations indicates that the analytic approximation is very good.

The average significance for
\index{B mixing@$B_s$ mixing!average significance in CDF}
$B_s$ oscillation measurements is shown in
Figure~\ref{fig:sig}. Various event yields and signal-to-background scenarios
are considered.  As stated above, the default event yield is 75k
events and the signal-to-background fraction is 1:1.

\begin{figure}
\begin{center}
  \mbox{\epsfig{file=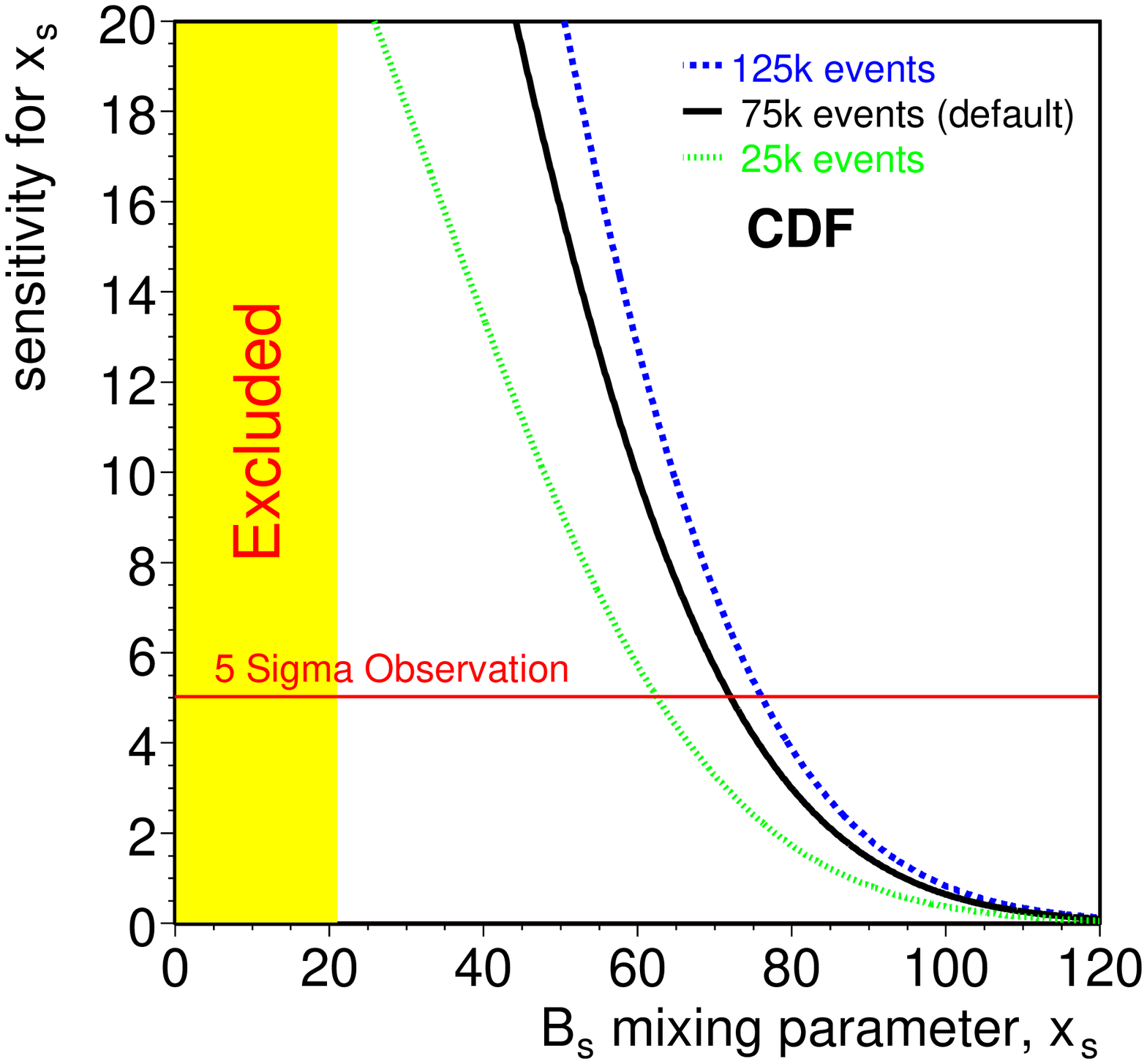,width=0.47\textwidth}}\hfill
  \mbox{\epsfig{file=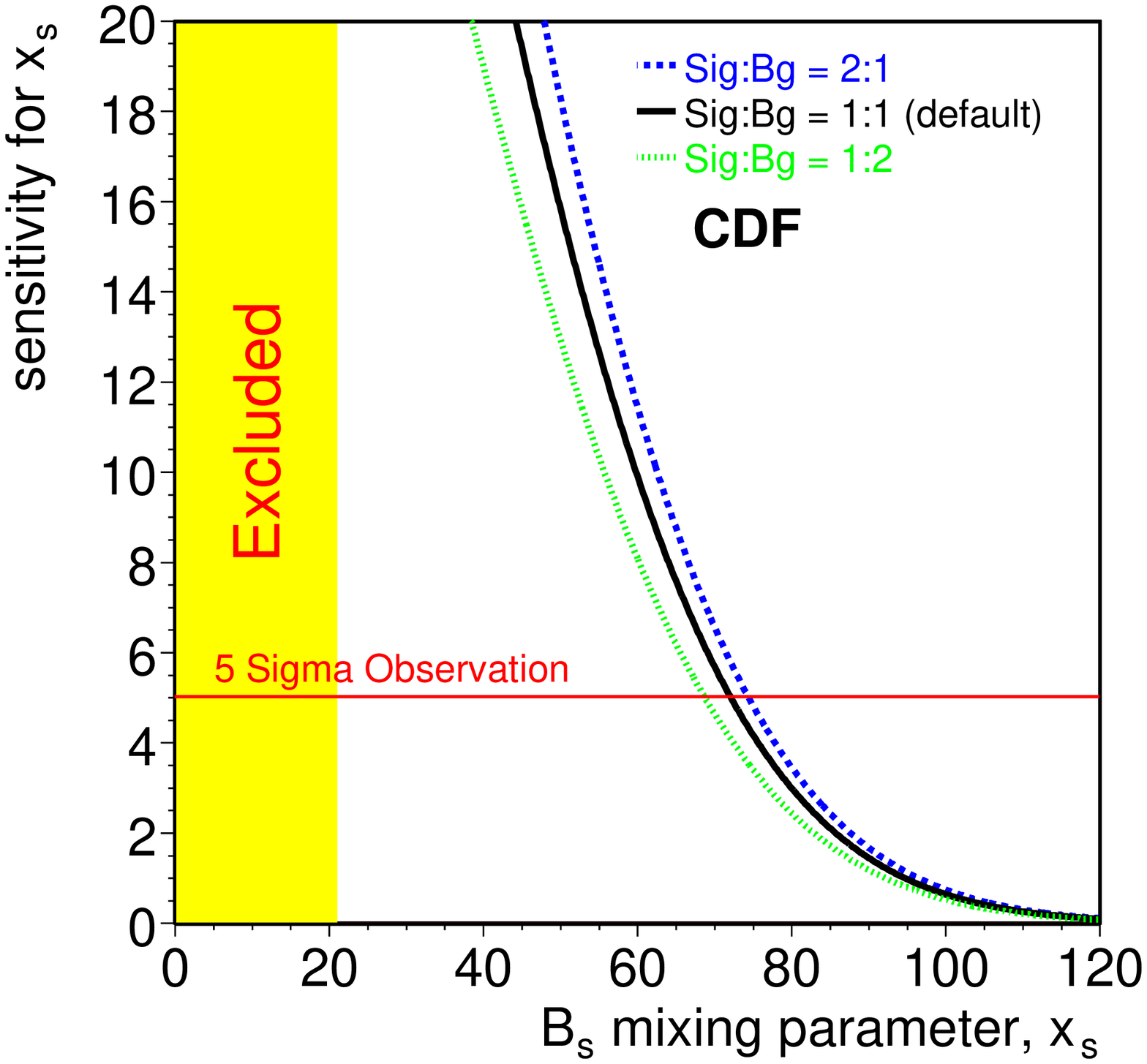,width=0.47\textwidth}}
  \caption[Average significance of mixing measurements expected as a function of
$x_s$ for various event yields and signal-to-background ratios.]{
    Average significance of mixing measurements expected as a function of the
    mixing parameter $x_s$ for various event yields (left) and
    signal-to-background ratios (right). The default is 75k events at a
    signal-to-background ratio of 1:1. The shaded area is excluded by the
    combined world lower limit on $x_s$.}
  \label{fig:sig}
\end{center}
\end{figure}

From the analytic expression, equation~\ref{eqn-sig}, the following 5 standard deviation sensitivity limits
are derived:
\begin{eqnarray*}
  \mbox{Maximum: 75k events}\; x_s &=& \left\{\begin{array}{lcl}
                                                74 & & \mbox{for $S/B=2:1$}\\
                                                69 & & \mbox{for $S/B=1:2$}
                                              \end{array}\right. \\
  \mbox{Maximum: S:B = 1:1}\;  x_s &=& \left\{\begin{array}{lcl}
                                                73 & & \mbox{for 125k events}\\
                                                59 & & \mbox{for  25k events}
                                              \end{array}\right. \\
\end{eqnarray*}
The Monte Carlo samples give very similar results as indicated above. It
is concluded that even in the worst case the reach is $x_s \sim 60$.

Fits to various experimental results which assume the Standard Model indicate
that $22.0<x_s<30.8$ at the 95\% confidence level~\cite{Ref:CKM-MATRIX}.  If
mixing occurs near the frequency expected in the Standard Model, it should be
easily observable by CDF in Run~II.  Figure~\ref{fig:lumi-error} (left) shows the
luminosity required to achieve an observation with an average significance of 5
standard deviations as a function of $x_s$.  The figure indicates that if mixing
occurs within the context of the Standard Model, then it should be observed
with a small fraction of 2 fb$^{-1}$, with CDF in a fully operational state.

\begin{figure}
\begin{center}
  \mbox{\epsfig{file=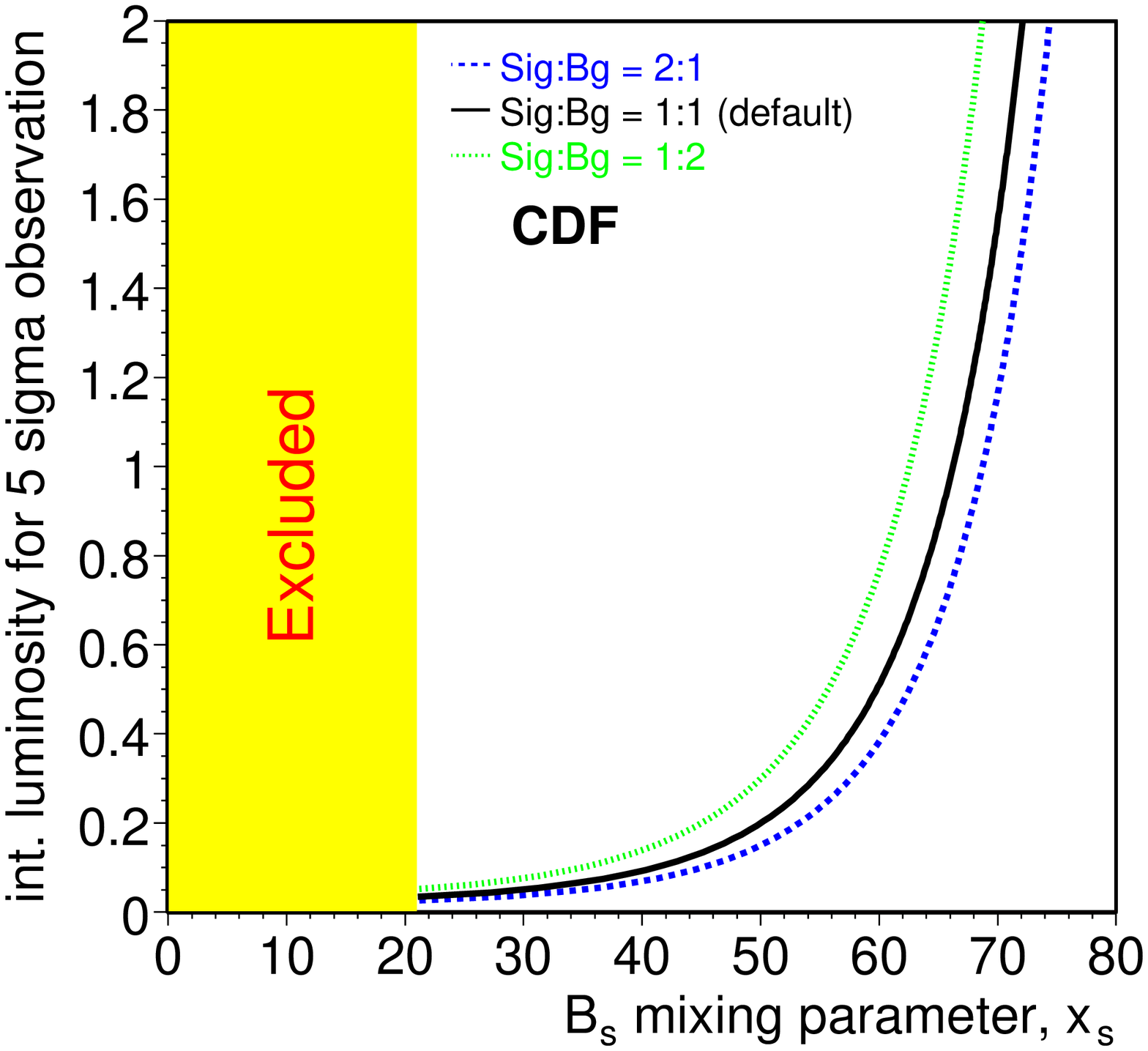,width=0.47\textwidth}}\hfill
  \mbox{\epsfig{file=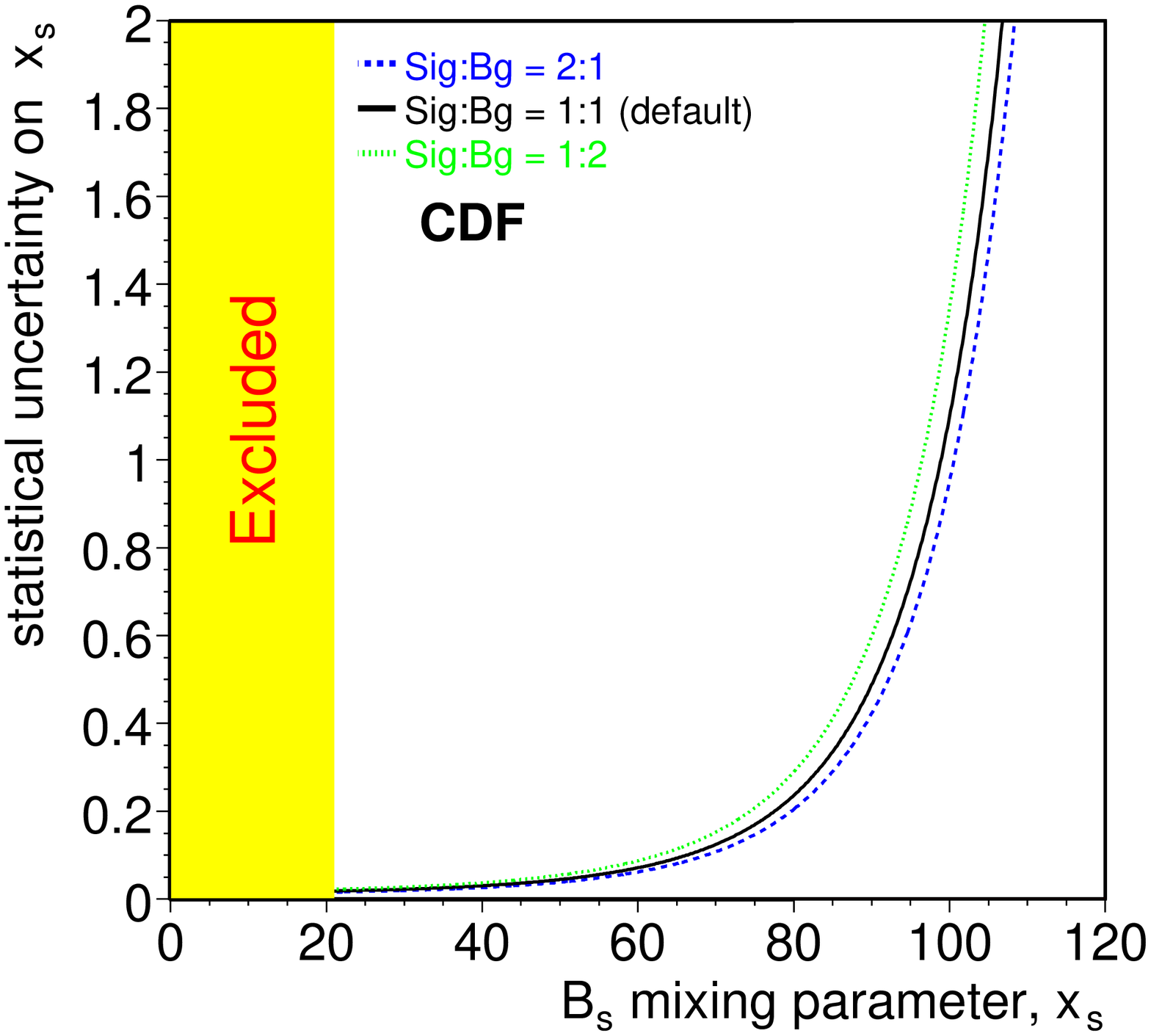,width=0.47\textwidth}}
  \caption[Luminosity required to achieve a 5 standard deviation observation of
mixing (left) and the statistical uncertainty as a function of $x_s$
(right).]{
    Luminosity required to achieve a 5 standard deviation observation of mixing
    (left) and the statistical uncertainty as a function of the mixing
    parameter, $x_s$ (right). The curves on the right are calculated using
    Equation~\ref{eq:error}.}
  \label{fig:lumi-error}
\end{center}
\end{figure}

While the significance of an observation of mixing is determined from the depth
of the minimum in the negative log-likelihood curve, the uncertainty on the
measured value of $x_s$ is determined by how sharp this minimum is.  In the case
of rapid oscillations, many periods will be reconstructed over a few $B_s$
lifetimes and the minimum is expected to be very sharp leading to a small
uncertainty.

The average uncertainty is described approximately by the analytic expression
\begin{equation}\label{eq:error}
  \frac{1}{\sigma_{x_s}} = \sqrt{N \epsilon D^2} e^{-(x_s \sigma_t/\tau)^2/2}
                           \sqrt{\frac{S}{S+B}}.
\end{equation}
The expected uncertainty from the analytic formula versus the mixing parameter,
$x_s$, is shown in Figure~\ref{fig:lumi-error} (right(). This formula is confirmed using
a series of Monte Carlo samples.


\boldmath
\subsection[$B_s$ mixing measurement at \dzero]
{$B_s$ mixing measurement at \dzero
$\!$\authorfootnote{Authors: N.~Cason, R.~Jesik, and N.~Xuan.}}
\unboldmath
\index{\dzero!$B_s$ mixing}

The expected luminosity of the Tevatron, $2\times10^{32}$ cm$^{-2}$s$^{-1}$, in
Run~II will lead to a huge rate for $b\,\bar{b}$ production, $\sim 10^{11}$
events/year. These enormous statistics combined with the upgraded detector will
allow us to search for $B_{s}$ mixing.  The resulting measurement of $\Delta
m_s$, when used to determine the ratio $\Delta m_d$/$\Delta m_s$ using the
well-measured value for $\Delta m_d$, gives a theoretically clean measurement of
$|V_{td}|^2$/$|V_{ts}|^2$. This puts a precise constraint on the CKM parameters
$\rho$ and $\eta$.

For $B^0_s$ mesons, existing data exclude small values of the mixing parameter
$x_s = \Delta m_s / \Gamma_s$, requiring $x_s>19.0$ at the 95\%
CL~\cite{Ref:PDG}.  Consequently the mass difference $\Delta m_s$ is much larger
than $\Delta m_d$, and the $B^0_s-\overline{B^0_s}$ oscillation frequency will
therefore be much higher than that for the $B^0_d$. Excellent decay length and
momentum resolutions are thus essential in order to observe the rapid
oscillations as a function of proper time.

Various decay modes of $B^0_s$ mesons are under investigation by the \dzero
collaboration. Among them are:
\begin{eqnarray}
B^0_s &\rightarrow& D^-_s(K^-K^+\pi^-)\pi^+,\quad\quad\quad\quad\quad\quad 
  (\BR=1.1\times10^{-4}),
  \nonumber\\ 
B^0_s &\rightarrow& D^-_s(K^-K^+\pi^-)3\pi,\,\quad\quad\quad\quad\quad\quad 
  (\BR=2.8\times10^{-4}),
  \nonumber\\
B^0_s &\rightarrow& J/\psi(\mu^+\mu^-,e^+e^-) K^*(K^{\pm}\pi^\mp),\,\quad 
  (\BR=5.1\times10^{-6}),
  \nonumber\\ 
B^0_s &\rightarrow& D^-_s(K^-K^+\pi^-)\ell^+\nu,\,\,\quad\quad\quad\quad\quad 
  (\BR=1.1\times10^{-4}).
\end{eqnarray}
\index{decay!$B_s \rightarrow D_s^- \pi^+$}%
\index{decay!$B_s \rightarrow D_s^- \pi^+ \pi^- \pi^+$}%
\index{decay!$B_s \rightarrow \psi K^* \ov{K}$}%
\index{decay!$B_s \rightarrow D_s^- \ell^+ \nu$}%

Taking advantage of the good SMT resolution, we can select $B_s$ decays by using
displaced secondary vertices or using tracks with large impact parameters. The
$B_s$ final states will be flavor tagged by the charge of the lepton, the charge
of the reconstructed charm meson or the charge of the kaon as appropriate.

The actual measurement strategy for $x_s$ will depend on the frequency of the
oscillation. For smaller oscillation frequencies, semileptonic $B_s$ decays can
be used. The lepton in the final state provides an easy trigger, giving a large
statistics sample.  If nature cooperates, a measurement in this range will come
very early in Run~II.

For higher oscillation frequencies, the measurement becomes more difficult.
Exclusive decays must be used in order to achieve the necessary momentum (and
therefore proper time) resolution. Decays which \dzero has focused on include:
$B_s \rightarrow D_s^- \pi^+(\pi^-\pi^+)$, where the $D_s^-$ decays to $\phi
\pi^-$ or $K^{*-}K^0/K^-K^{*0}$; and $B_s\rightarrow J/\psi \overline{K}{}^{*0}$
followed by $J/\psi\rightarrow e^+e^-$ or $\mu^+\mu^-$ and
$\overline{K}{}^{*0}\rightarrow K^-\pi^+$. We can only trigger on $B_s$ decays
into fully hadronic final states when the other $B$ in the event decays
semileptonically. Using single lepton triggers, we expect to be able to collect
about a thousand reconstructed exclusive $B_s$ decays in each mode in the first
two years, allowing us to measure $x_s$ values up to $\sim 20-30$. 
\index{B mixing@$B_s$ mixing!average significance in \dzero}
We are
presently investigating better trigger scenarios, such as lowering the $p_T$
threshold of the lepton and requiring another moderately high $p_T$ track (or
tracks), which would increase our $\Delta m_s$ reach.

A Monte Carlo study of the $B_s\rightarrow J/\psi \overline{K}{}^{*0}$ decay has
been carried out in order to estimate the number of events which will be in a
data sample based on a 2~fb$^{-1}$ exposure. We have analyzed 20,000 events
using the MCFast program.  The events were generated using Pythia and a
simulation of the upgraded D\O\ detector.  Each event had a $B_s \rightarrow
J/\psi \overline{K}{}^{*0}$ decay as well as a generic $\overline{B}$ decay. The
$J/\psi$ decayed to $\mu^+\mu^-$ (83\% of the time) or $\mu^+\mu^- \gamma$ (17\%
of the time). (The radiative decays are not discussed further here.) The
$\overline{K}{}^{*0}$ decayed to $\pi^+ K^-$.

Event reconstruction efficiency was estimated using the geometric acceptance of
the silicon vertex detector and of the fiber tracker. (Tracking inefficiencies
are not yet included.)  We find that 21\% of the events have all four charged
tracks reconstructed.

\begin{figure}
\centerline{\epsfig{file=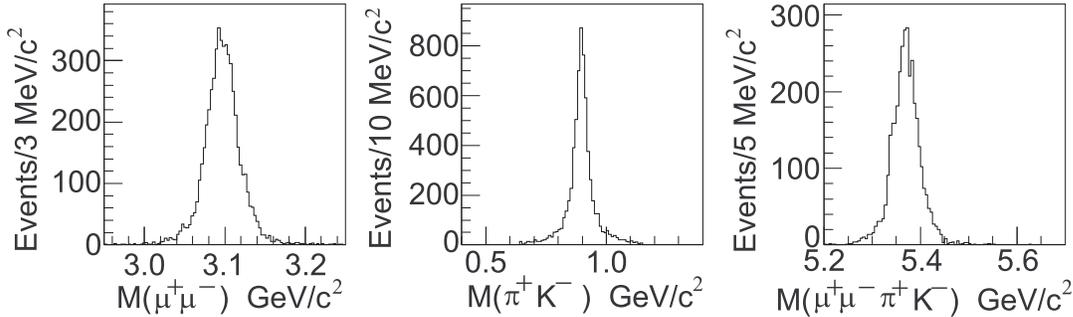,width=0.95\textwidth}}
\caption[Effective mass distributions for the reconstructed (a)~$\mu^+\mu^-$
system; (b)~$\pi^+ K^-$ system; and (c)~$\mu^+\mu^-\pi^+ K^-$
system.]{Effective mass distributions for the reconstructed: (a) $\mu^+\mu^-$
system, (b) $\pi^+ K^-$ system, and (c) $\mu^+\mu^-\pi^+ K^-$ system.  These
distributions are prior to using vertex and mass constraints.}
\label{massdists}
\end{figure}

Shown in Fig.~\ref{massdists} are the reconstructed $\mu^+\mu^-$, $\pi^+ K^-$,
and $\mu^+\mu^-\pi^+ K^-$ effective mass distributions for the reconstructed
tracks.  The mass resolution of the $B_s$ improves by more than a factor of two
when vertex and $J/\psi$ mass constraints are imposed.  Effective mass
resolutions are given in Table \ref{resolution}.

\begin{table}
\begin{center}
\begin{tabular}{lrr}
\hline\hline
Quantity        &             Level & $\sigma~(MeV/\rm{c}^2$) \\
$M(B_s)$        &    Reconstruction &                      37 \\
$M(B_s)$        & $J/\psi$ mass fit &                      15 \\
$M(\mu^+\mu^-)$ & Reconstruction    &                      29 \\
\hline\hline
\end{tabular}
\end{center}
\caption{Mass Resolutions \label{resolution}}
\end{table}

\begin{figure}
\begin{center}
\epsfig{file=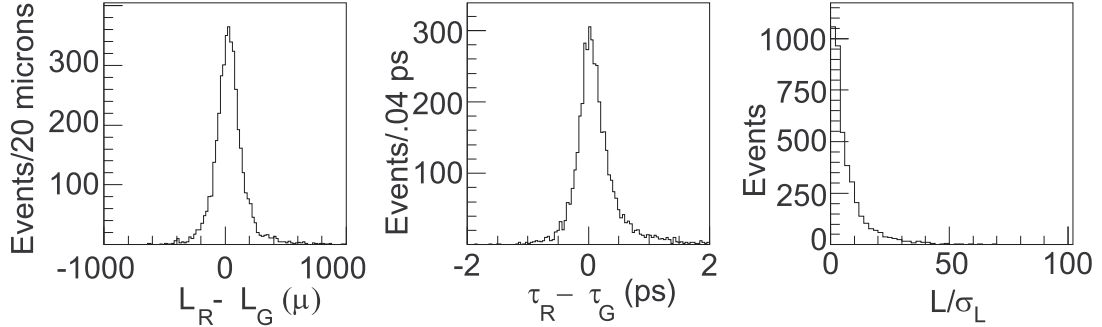,width=0.95\textwidth}
\caption[Reconstructed $B_s$ decay length and proper time
distributions.]{Distributions of: (a) the reconstructed $B_s$ decay length
minus the generated $B_s$ decay length; (b) the reconstructed $B_s$ proper time
minus the generated $B_s$ proper time; and (c) the reconstructed $B_s$ decay
length divided by its error.}
    \label{resolutionfig}
\end{center}
\end{figure}

Resolutions of vertex position, decay length, and proper time were estimated
using the nominal silicon vertex detector and fiber tracker resolutions.  Shown
in Fig.~\ref{resolutionfig} are distributions of the fitted minus measured decay
length of the $B_s$, the fitted minus measured proper decay time of the $B_s$,
and the ratio of the decay length to the error in the decay length ($L/\sigma_L
$). We summarize the resolutions in vertex positions, decay length and proper
time in Table~\ref{resolution2}.

\begin{table}
\begin{center}
\begin{tabular}{lr}
\hline\hline
Quantity &  $\sigma$  \\
\hline
Production vertex (x,y, and z) ($\mu m$) & 34, 34,  80 \\
Decay vertex (x,y, and z) ($\mu m$)      & 50, 50, 140 \\
$B_s$ decay length ($\mu m$)             &         140 \\
Proper time ($ps$)                       &        0.40 \\
\hline\hline
\end{tabular}
\end{center}
\caption{Resolutions \label{resolution2}}
\end{table}

In order to get a realistic estimate of the number of events which would be
available for analysis, additional cuts were placed on the sample.  The muons
were required to have $p_T> 1.5~ GeV/\rm{c}$ and to have $|\eta|< 2$. A total of
24\% of the reconstructed events satisfied these cuts.  Combined with the
reconstruction efficiency of 21\%, we are left with a sample of 5.0\% of the
generated events for further analysis.

For most purposes, additional cuts will be required to obtain a sample of events
with a good signal-to-noise ratio.  To estimate the sample size after such cuts,
we impose a cut on the variable $L/\sigma_L$.  Of the 5.0\% of the events
satisfying all the previous cuts, 83\% have $L/\sigma_L > 2.0$, 73\% have
$L/\sigma_L > 3.0$, and 63\% have $L/\sigma_L > 4.0$. The required cut value
will not be known until the data is in hand, but we use the $L/\sigma_L > 3.0$
cut for further estimates.  Hence the overall combined efficiency which we use
below is (.050)*0.73=0.036.

In order to do mixing studies, it is necessary to tag the flavor of the $B_s$
(or $\overline{B_s}$).  This can be tagged if we identify the sign of the
charged kaon in the $K^*$.  Although D\O\ does not have particle identification
in the traditional sense, it is possible in this event sample to determine
whether the positive or negative particle from the $K^*$ decay is the charged
kaon by calculating the effective mass under the two assumptions ($K^+\pi^-$ and
$K^-\pi^+$) and taking as correct that combination which gives an effective mass
closest to the nominal $K^*$ mass (0.890 $GeV/\rm{c}^2$).  We find that we are
correct using this assignment 67\% of the time. Flavor tagging of the other $b$
quark will have the efficiency and dilution summarized previously in
Table~\ref{flavtag}.

Using estimates of the $b\overline{b}$ production cross section (158~$\mu$b);
the fraction of this cross section producing $B_s$ (0.167); the $B_s\rightarrow
J/\psi \overline{K}{}^{*0}$ branching ratio (5.1 x 10$^{-6}$); the
$J/\psi\rightarrow\mu^+\mu^-$ branching ratio (0.06); the luminosity (2~{\ifb});
and the overall combined efficiency from above (.036), we obtain a signal of
1000 events. We would expect a similar sample will be obtained using the
$J/\psi\rightarrow e^+ e^-$ mode.

\boldmath
\subsection[Measurement of $B_s$ Mixing in BTeV]
{Measurement of $B_s$ Mixing in BTeV
$\!$\authorfootnote{Author: R.~Kutschke.}}
\unboldmath
\index{BTeV!$B_s$ mixing}

In this section, the $x_s$ reach of BTeV will be demonstrated using $B_s\to
D_s^-\pi^+$ and $B_s\to J/\psi \ov{K}{}^{*0}$.  This study was carried out in
several steps, the first step being a simulation of the BTeV detector response
to signal events.  The output of this step was treated as real data and passed
through a physics analysis program to determine the yield, the time resolution
and the signal-to-background ratio in each mode.  This information was then
passed to a separate program which computed the $x_s$ reach; this program is
discussed in section~\ref{sec_btev_xsreach}.  A separate background study was
performed.

\subsubsection{Yields, Resolutions and Signal-to-Background Ratios}

The mode for which BTeV has the most sensitivity to $x_s$ is $B_s\to D_s^-
\pi^+$, 
\index{decay!$B_s \rightarrow D_s^- \pi^+$}
where the $D_s^-$ decays either by $D_s^-\to \phi\pi^-$, $\phi\to
K^+K^-$, or by $D_s^-\to K^{*0}K^-$, $K^{*0}\to K^+ \pi^-$.  Both of these
$D_s^-$ modes have narrow intermediate states and characteristic angular
distributions, both of which can be used to improve the signal-to-background
ratio.

For this study, Monte Carlo events were generated using Pythia and QQ and the
detector response was simulated using BTeVGeant.  The simulated events were
analyzed as real data.  For the $D_s \to \phi \pi$ decay mode the following cuts
were used:
\begin{itemize}
\item All tracks were required to have at least 3 hits in the silicon pixel
  detector.
\item Each of the tracks in the $B_s$ candidate were required to have an impact
  parameter with respect to the primary vertex of $> 3\sigma$.
\item To reduce the background due to tracks that really come from other
  interactions it was required that all 4 tracks have an impact parameter with
  respect to the primary vertex of less than 0.2~cm.
\item At least one of the kaons from the $\phi$ decay was required to be
  strongly identified as a kaon by the RICH detector.  The second kaon was only
  required to be loosely identified.  No particle ID requirements were placed on
  the pion candidates.
\item The $\phi$ and $D_s$ candidates were required to be within
  $\pm{2.5\sigma}$ of their nominal masses.
\item It was required that the distance between the primary vertex and $D_s$
  decay vertex be $L<8.0$ cm and the $D_s$ decay vertex have a decay length
  significance of $L/\sigma_{L}(D_s)>10.0$.
\item It was required that the $B_s$ have decay length significance of
  $L/\sigma_{L}(B_s)>4.0$.
\item The $B_s$ candidate was required to point back to the primary vertex: the
  transverse momentum of the $B_s$ with respect to its line of flight from the
  primary vertex was required to be less than 1.0 GeV/c and the impact parameter
  of the $B_s$ with respect to the primary vertex was required to be less than
  $3\sigma$.
\end{itemize} 
 
The combined geometric acceptance and reconstruction efficiency was found to be
2.7\%.  Of the events that passed these analysis cuts, 74\%\ passed the level 1
trigger.  For the $D_s \to K^* K$ mode we used the same cuts except that both
kaons from the $D_s$ decay were required to be identified in the RICH.  There
was also a broader cut on the intermediate $K^*$ mass.  The combined
reconstruction efficiency and geometric acceptance for the $D_s \to K^* K$ mode
was found to be 2.3\%, and the level 1 trigger efficiency for the events passing
the analysis cuts was 74\%.  For both modes the resolution on the mass of the
$B$ was found to be 18~MeV$/c^2$ and the mean resolution on the proper decay
time was found to be 43~fs.  The nominal acceptance of the BTeV level~2 trigger
for the accepted events is 90\%\ of the events which remain after the level 1
trigger.  The nominal flavor tagging power of BTeV was estimated in chapter~5 to
be $\epsilon D^2=0.1$ which arises from $\epsilon=0.70$ and $D=0.37$.

It is believed that the dominant source of backgrounds will be events of the
form \hbox{$X_b\to D_s^- X$}, where $X_b$ may be any $b$ flavored hadron.  The
background combinations arise when a true $D_s^-$ combination is combined with
some other track in the event.  An MCFast based study of 1~million $B\to D_s^-
X$ events was performed using an older version of the detector geometry, the one
used for the BTeV Preliminary Technical Design Report
(PTDR)~\cite{Ref:BTEV-PTDR}.  Comparisons between BTeVGeant and MCFast, and
comparisons between the old and new detector geometries, show that these
background studies remain valid.  When the 1~million $B\to D_s^- X$ events were
passed through MCFast and analyzed as real data, 8~entries remained in a mass
window 6~times larger than the mass window used to select signal $B_s$
candidates.  From this it is estimated that the signal-to-background ratio in
this channel is 8.4:1.  This study was performed without the proper treatment of
multiple interactions in one beam crossing.  To account for this, the
signal-to-background ratio used in the estimate of the $x_s$ reach is 3:1.

The background from direct charm production has not yet been investigated. While
direct charm production has a cross-section about 10~times higher than that for
production of charm via $B$ decay, it is triggered much less efficiently.
Moreover the the requirement of two, distinct detached vertices greatly reduces
the background from direct charm.  In the end it is expected that the background
from $B\to D_s^- X$ will dominate.

Table~\ref{9:tab:bstodspiyield} gives a summary of the preceding results and
discusses a list of all assumptions which went into the computation of the
yield. The value for ${\cal B}(\bar{b} \to B_s)$ is obtained from
Reference~\cite{Ref:BD-PROD}.  In one year it is expected that 72,000 events
will trigger, survive all analysis cuts and have their birth flavor tagged.

\begin{table}[tb]
\begin{center}
\begin{tabular}{llr}
\hline\hline
 Quantity & Value & 
\multicolumn{1}{c}{Yield} \\
          &       & 
\multicolumn{1}{c}{(Events/year)} \\
\hline
Luminosity:          & $2\times10^{32}\ {\rm cm^{-2} s^{-1}}$ & \\
One Year:            & $10^7\ {\rm s}$ & \\
$\sigma_{b\,\bar{b}}$:                    & $100\ \mu{\rm b}$ &  \\
${\cal B}(B_s \to D_s^-\pi^+)$:             & $3.0\times 10^{-3}$ & \\
${\cal B}(D_s^- \to \phi\pi^-)$:            & $0.030$ & \\
${\cal B}(D_s^- \to K^{*0} K^-)$:     & $0.036$ &\\
${\cal B}(\phi \to K^+ K^-)$:           & $0.49$ & \\
${\cal B}(K^{*0} \to K^+\pi^-)$:  & $0.67$ & \\
${\cal B}(\bar{b}\to B_s)$  & 0.13     &  6,210,000 \\
$\epsilon({\rm Geometry\ +\ cuts}:\ \phi\pi^-)$   & 0.027 &  \\
$\epsilon({\rm Geometry\ +\ cuts}:\ K^{*0}K^- )$  & 0.023 &  \\
$\epsilon({\rm Trigger})$ Level 1 & 0.74      &  \\
$\epsilon({\rm Trigger})$ Level 2 & 0.90      &  \\
$\epsilon({\rm Tag})$     & 0.70      & 72,000  \\
Tagging Dilution          & 0.37      &                \\
$S/B$                     & 3:1       & \\
$\sigma$(Proper Decay time) & 43 fs &  \\
\hline\hline
\end{tabular}
\end{center}
\caption[Projected yield for $B_s\to D_s^-\pi^+$ in one year of BTeV
running.]{Projected yield for $B_s\to D_s^-\pi^+$ in one year of BTeV running.
The numbers in the third column give the expected yield when all of the factors
down to and including that line have been considered.  The branching fraction
${\cal B}(B_s \to D_s^- \pi^+)$ was estimated to be the same as ${\cal
B}(B_d\to D^- \pi^+)$.}
\label{9:tab:bstodspiyield}
\end{table}

Another mode with good $x_s$ sensitivity is $B_s\to J/\psi \ov{K}{}^{*0}$,
\index{decay!$B_s \rightarrow \psi K{}^{*0}$}
$J/\psi\to\mu^+\mu^-$, $\ov{K}{}^{*0}\to K^-\pi^+$.  Although this mode is
Cabibbo suppressed, other factors are in its favor: the final state consists of
a single detached vertex and the state is triggerable with several independent
strategies, including impact parameter triggers, secondary vertex triggers and
dimuon triggers~\cite{Ref:MCBRIDE-STONE}.  While this mode does not have the
$x_s$ reach of $D_s^-\pi^+$ it does cover much of the expected range and it
provides a powerful check with partly independent systematics.

For reasons of time limitations, the simulation of the $J/\psi\ov{K}{}^{*0}$ mode
used MCFast, not BTeVGeant.  The analysis of this mode proceeded as follows.  To
be considered as part of a signal candidate, a track was required to have at
least 20 total hits and at least 4 pixel hits.  The only further requirement
placed on $\pi^{\pm}$ candidates was that they have a momentum greater than
\hbox{0.5~GeV/$c$}.  In order to be considered a muon candidate, a track was
required to have a momentum \hbox{$p>5$~GeV$/c$}, to penetrate the hadron filter
and to leave hits in the most downstream muon chambers.  Kaon candidates were
required to satisfy a simplified model of the RICH system: the track was
required to have a momentum in the range \hbox{$3<p<70\, $~GeV/$c$} and was
required to have hits in the tracking station downstream of the RICH mirror.
True kaons which satisfied this criteria were identified as kaons with an
efficiency of 90\%; other hadrons which satisfied this criteria were
(mis)identified as kaons 3\%\ of the time.

A $\mu^+\mu^-K^-\pi^+$ combination was accepted as a $B_s $ candidate if the
confidence level of fitting all four tracks to a single vertex was greater than
0.005.  It was also required that the resonant substructure requirements be
satisfied.  Combinations were considered for further analysis provided the decay
length of the $B_s$ candidate, $L$, satisfied \hbox{$L/\sigma_L>10$} and the
impact parameter of the $B_s$ candidate with the primary vertex, $d$, satisfied
$d<3\sigma_{d}$.  Each of the four $B_s$ granddaughters were required to have an
impact parameter with the primary vertex, $d$, of $d>2\sigma_d$.  Candidates
with poor time resolution were rejected by demanding $\sigma_{t}\le 0.09$~ps.
Also the mass of the $J/\psi$ was constrained to its PDG value.  The above
procedure found that the efficiency for the 4 tracks to be within the fiducial
volume of the tracking system was $14.2\pm0.3\,$\%\ and the efficiency for the
remaining candidates to pass the analysis cuts was $0.29\pm0.01$.  The
resolution on the mass of the $B_s$ was found to be $8.6\pm 0.3$~MeV$/c^2$ and
the mean resolution on the proper decay time was found to to be 36~fs.

The BTeV Level 1 trigger simulation was run on the $J/\psi\ov{K}{}^{*0}$ sample
and, of the candidates which passed all analysis cuts, $68\pm 2\,$\%\ also
passed the trigger; the error is statistical only.  However, this mode can also
be triggered by the dimuon trigger.  Section~8.3, of the BTeV
proposal~\cite{Ref:BTEV-PROPOSAL}, which describes the algorithms and
performance of the muon trigger, estimates a trigger efficiency of 50\%\ for
this decay mode.  There is, as yet, no calculation of the total Level~1 trigger
efficiency which takes into account the correlations between the two triggers.
For this proposal it will be estimated that the combined Level~1 trigger
efficiency is 85\%.  As for the $D_s\pi$ final state, the Level~2 trigger is
expected to have an efficiency of about 90\%.

By far the dominant background is expected to come from decays of the form
$X_b\to J/\psi X$, $J/\psi\to \mu^+\mu^-$, where $X_b$ is any $b$
flavored hadron.  An MCFast based simulation of 500,000 such decays was
performed and the signal-to-background level was estimated to be about 2:1.
Some sources of background that one might, at first, think to be important turn
out not to be a problem.  First, the more copious $B_s\to J/\psi\,\phi$ final
state is not a significant source of background because of the excellent
particle ID provided by the RICH system.  Second, the mass resolution is
sufficient to separate the decay $B_d\to J/\psi\ov{K}{}^{*0}$.

Finally, the expected yield can be increased by at least 50\%\ by using the
decay mode \hbox{$J/\psi\to e^+ e^-$}.  This mode will have an efficiency for
secondary vertex triggers which is comparable to that for $J/\psi\to \mu^+
\mu^-$ but the acceptance of the ECAL is smaller than that of the muon
detectors. The smaller acceptance of the ECAL is somewhat offset by also using
the RICH for electron identification.

The information reported here is summarized in Table~\ref{9:tab:psiksyield} and
is used in the mini-Monte Carlo described in the next section.  The estimate for
${\cal B}(B_s \to J/\psi \ov{K}{}^{*0})$ is obtained from
Reference~\cite{Ref:MCBRIDE-STONE} and that for ${\cal B}(\bar{b} \to B_s)$ is
obtained from Reference~\cite{Ref:BD-PROD}.

\begin{table}[tb]
\begin{center}
\begin{tabular}{llr}
\hline\hline
 Quantity & Value & 
\multicolumn{1}{c}{Yield} \\
          &       & 
\multicolumn{1}{c}{(Events/year)} \\
\hline
Luminosity:          & $2\times10^{32}\ {\rm cm^{-2} s^{-1}}$ & \\
One Year:            & $10^7\ {\rm s}$ & \\
$\sigma_{b\,\bar{b}}$: & $100\ \mu {\rm b}$ &  \\
${\cal B}(B_s \to  J/\psi \ov{K}{}^{*0})$: & $8.5\times 10^{-5}$ & \\
${\cal B}(J/\psi \to \mu^+\mu^-)$:              & 0.061 & \\
${\cal B}(\ov{K}{}^{*0}\to K^-\pi^+)$:          & 0.667 & \\
${\cal B}(\bar{b}\to B_s)$                      & 0.13      & 180000 \\
$\epsilon({\rm Geometric})$                     & 0.142     & \\
$\epsilon({\rm Analysis\ cuts})$                & 0.26      & 6600 \\
$\epsilon({\rm Trigger})$ Level 1 Tracking only & 0.60      & \\
$\epsilon({\rm Trigger})$ Level 1 Total         & 0.85      &  \\
$\epsilon({\rm Trigger})$ Level 2               & 0.90      & 5100 \\
$\epsilon({\rm Tag})$                           & 0.70      & 3600   \\
Include $J/\psi\to e^+e^-$                      & 1.5       & 5300  \\
Tagging Dilution                                & 0.37      &   \\
$S/B$                                           & 2:1       & \\
$\sigma$(Proper Decay time)                     & 36 fs     &  \\
\hline\hline
\end{tabular}
\end{center}
\caption[Projected yield for $B_s\to J/\psi\ov{K}{}^{*0}$ in one year of BTeV
running.]{Projected yield for $B_s\to J/\psi\ov{K}{}^{*0}$ in one year of BTeV
running. The numbers in the third column give the expected yield when all of
the factors down to and including that line have been considered. The trigger
efficiency is quoted as a fraction of those events which pass the analysis
cuts.}
\label{9:tab:psiksyield}
\end{table}

\boldmath
\subsubsection{Computation of the $x_s$ Reach}
\unboldmath
\label{sec_btev_xsreach}
\index{B mixing@$B_s$ mixing!average significance in BTeV}

The final step in the study was to use a mini-Monte Carlo to study the $x_s$
reach of BTeV.  This mini-Monte Carlo generates two lifetime distributions, one
for mixed events and one for unmixed events, smears the distributions and then
extracts a measured value of $x_s$ from a simultaneous fit of the two
distributions.  The time smearing is a Gaussian of fixed width, using the mean
time resolutions determined above.  The model includes the effects of
mistagging, background under the signal, and the minimum time cut which is
implied by the $L/\sigma_L$ cut.  It is assumed that the lifetime distribution
of the background is an exponential with the same mean lifetime as that of the
$B_s$.

Figures~\ref{9:fig:dspitime}~a) and b) show the proper time distributions which
result from one run of the mini-Monte Carlo for a generated value of $x_s=40$.
The simulation is for the decay mode $B_s\to D_s^-\pi^+$ for one month of BTeV
running.  Part a) shows the proper time distribution for unmixed decays while
part b) shows the distribution for mixed decays.  Part c) of the figure shows,
as a function of $x_s$, the value of the unbinned negative log likelihood
function computed from the simulated events.  A clear minimum near the generated
value of $x_s$ is observed and the likelihood function determines the fitted
value to be $x_s=39.96\pm0.08$.  A step of 0.5 in the negative log likelihood
function determines the 1~$\sigma$ error bounds and a line is drawn across the
figure at the level of the 5~$\sigma$ error bound.

This figure nicely illustrates the distinction between two quantities which are
often confused, the significance of the result and the error on $x_s$.  The
significance of the signal is determined by how far the depth of the global
minimum falls below that of the next most significant minimum.  The error on
$x_s$ is determined by the curvature of the likelihood function at the global
minimum.  While these quantities are clearly related, they are distinct; in
particular, the significance of the signal is not the relative error on $x_s$.

\begin{figure}[tb]
\centerline{\epsfxsize=4.5in\epsfbox{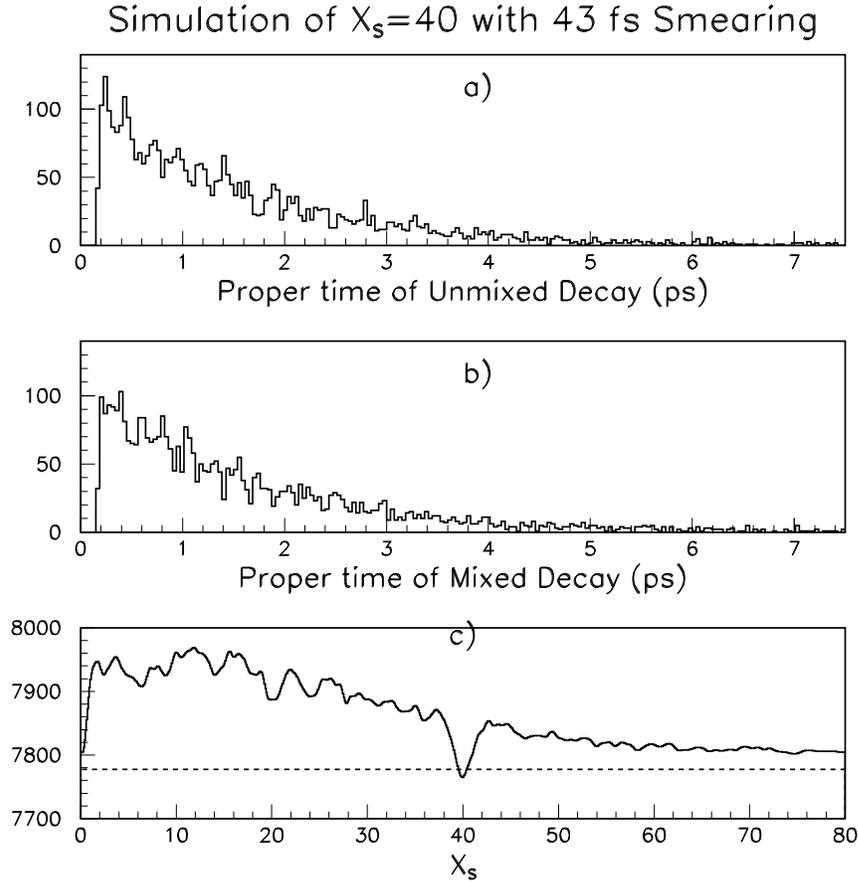}}
\caption[Mini Monte Carlo proper lifetime plots of a) unmixed and b)  mixed
decays for a generated value of $x_s=40$.]{Mini Monte Carlo proper lifetime
plots of a) unmixed and b)  mixed decays for a generated value of $x_s=40$. 
The plots simulate the results of the $B_s\to D_s^- \pi^+$ channel after one
month of running.  The oscillations are prominent.  Part c) shows the negative
log likelihood function which was obtained from the entries in parts a) and
b).  A prominent minimum is seen at the generated value of $x_s$.  The dashed
line marks the level above the minimum which corresponds to 5~$\sigma$
significance.}
\label{9:fig:dspitime}
\end{figure}

The error returned by the fit was checked in two ways. First, an ensemble of
mini-Monte Carlo experiments was performed and the errors were found to
correctly describe the dispersion of the measured values about the generated
ones.  Second, the errors returned by the fit were found to be approximately
equal to the Cramer-Rao minimum variance bound.

The mini-Monte Carlo was also used to study the level of statistics below which
the experiment is unable to measure $x_s$.  As the number of events in a trial
is reduced, the negative log likelihood function becomes more and more ragged
and the secondary minima become more pronounced.  Eventually there are secondary
minima which reach depths within 12.5 units of negative log likelihood
(~5~$\sigma$~) of the global minimum.  When this happens in a sufficiently large
fraction of the trials, one must conclude that only a lower limit on $x_s$ can
be established.  In the region of the parameter space which was explored, the
absolute error on $x_s$ was approximately 0.1 when this limit was reached.  This
was independent of the generated value of $x_s$; that is, the discovery
measurement of $x_s$ will have errors of something like $\pm0.1$, even if $x_s$
is large, say~$40$.

It is awkward to map out the $x_s$ reach of the apparatus by running a large
ensemble of mini-Monte Carlo jobs; instead the following automated procedure was
used.  Following ideas from McDonald~\cite{Ref:KTMCDONALD}, the sum over events
in the likelihood function was replaced with an integral over the parent
distribution.  Because the parent distribution does not have any statistical
fluctuations, the fluctuations in the likelihood function are removed, leaving
only the core information.  An example of such a likelihood function is shown in
Fig.~\ref{9:fig:dspilikelihood}.
\begin{figure}[tb]
\centerline{\epsfxsize=10cm\epsfbox{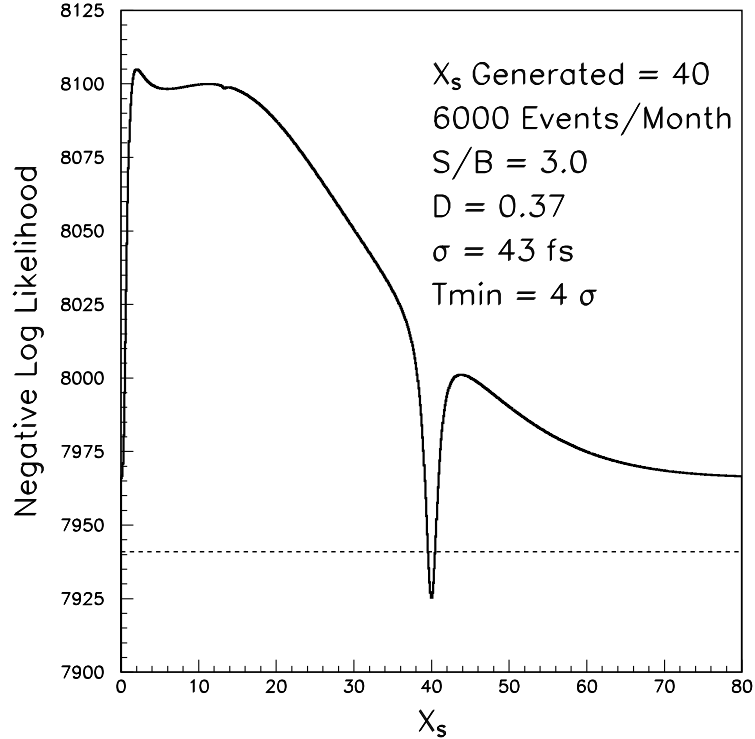}}
\caption[The same likelihood function as in part c) of the previous figure but
obtained using the integral method described in the text.]{The same likelihood
function as in part c) of the previous figure but obtained using the integral
method described in the text.  The overall shape is the same but the
statistical fluctuations have been removed. There is also an overall level
shift which is related to the goodness of fit in the previous figure.}
\label{9:fig:dspilikelihood}
\end{figure}

A likelihood function computed in this way has the property that it scales
linearly with the number of events being simulated.  This can be stated formally
as follows.  Let $x_0$ denote the generated value of $x_s$ and let ${\cal
  L}(x;x_0,N)$ denote the value of the likelihood function, evaluated at $x$,
for a sample which has a true value of $x_0$ and which contains N events.  Then,
\begin{equation}
  {\cal L}(x;x_0,N) = N\, {\cal L}(x;x_0,1)\,.
\end{equation}
Now, one can define the significance of the minimum, $n$, as,
\begin{equation}
  n^2 = 2.0\, N\, \big[{\cal L}(\infty;x_0,1)-{\cal L}(x_0;x_0,1)\big]\,.
  \label{9:eq:runtime}
\end{equation}
For practical purposes $\infty$ was chosen to be 160.  If one did not have to
worry about the missing statistical fluctuations it would be normal to define a
significant signal as $5\,\sigma$, or $n^2=25$.  Instead, sufficient
significance was defined as $n^2=31.25$, by adding a somewhat arbitrary safety
margin; this allows for the usual $5\,\sigma$ plus a downwards fluctuation of up
to $2.5\,\sigma$ anywhere else in the plot.  Equation~\ref{9:eq:runtime} was
solved for $N$, which was then converted into the running time required to
collect $N$ events.  This procedure was repeated for many different values of
$x_0$ to obtain Fig.~\ref{9:fig:xsreach}.  The solid line shows, for the
$D_s^-\pi^+$ mode, the number of years needed to obtain a measurement with a
significance of $5\,\sigma$ plus the safety margin.  The safety margin reduces
the $x_s$ reach at 3 years by only 3 or 4 units of $x_s$.  For small values of
$x_s$, the effect of the safety margin is not visible.  The dashed line shows
the same information but for the $J/\psi \ov{K}{}^{*0}$ mode; for this mode the
effect of the safety margin is similarly small.

\begin{figure}[tb]
\centerline{\epsfxsize=11cm\epsfbox{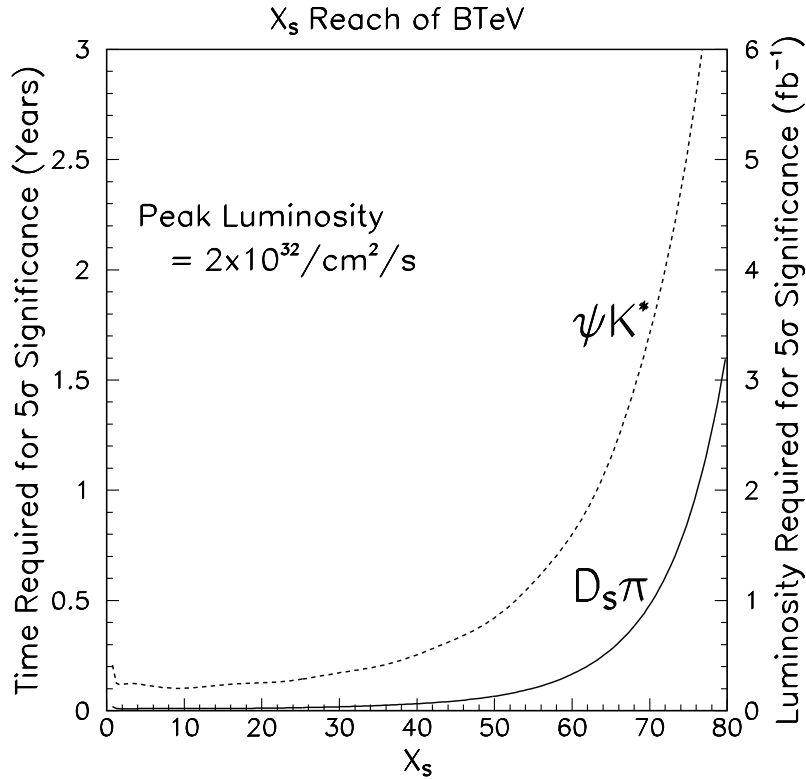}}
\caption[The $x_s$ reach of the BTeV detector.  The curves indicate the number
of years of running required for a measurement with a statistical
significance of $5\,\sigma$]{The $x_s$ reach of the BTeV detector.  The curves
indicate the number of years of running which are required to make a
measurement of $x_s$ with a statistical significance of $5\,\sigma$; a safety
margin, discussed in the text, has been included in the definition of
$5\,\sigma$. The curves are for the two different decay modes indicated on the
figure.}
\label{9:fig:xsreach}
\end{figure}

Inspection of Fig.~\ref{9:fig:xsreach} shows that, using the $D_s^-\pi^+$ mode,
BTeV is capable of observing all $x_s$ less than 75 in one year of running,
which is equivalent to an integrated luminosity of 2~fb$^{-1}$.
\index{B mixing@$B_s$ mixing!average significance in BTeV}

\subsection{Summary of Projections for Mixing}
\index{summary!$B_s$ mixing projections}

The Standard Model expectation for $B_s$ mixing is $22 < x_s <
31$~\cite{Ref:CKM-MATRIX}. All three experiments CDF, D{\O} and BTeV have shown
that they will be able to reconstruct a substantial amount of $B_s$ decays which
will allow for mixing studies.

With 2~fb$^{-1}$ CDF and BTeV safely cover the range for mixing as predicted by
the Standard Model. The CDF sensitivity for a 5 standard deviation observation
reaches from $x_s$ values of 59 to 74 depending on the event yields and the
signal-to-background ratios. BTeV's sensitivity comfortably covers $x_s$ values
of 75 with even some conservative safety margins included. Due to the large
event yields CDF will be able to observe $B_s$ mixing within the first few month
of data taking provided that the displaced track trigger and the silicon
detector work as advertised.

Once the oscillations are observed the statistical uncertainty on $x_s$ will be
small and in conjunction with an accurate $B^0$ mixing measurement it will
constitute a stringent constraint on the unitarity triangle.

\section{Projections for $\Delta\Gamma$}

\boldmath
\subsection[$B_s$ Lifetime Difference in CDF]
{$B_s$ Lifetime Difference in CDF
$\!$\authorfootnote{Authors: Ch.~Paus, J.~Tseng}}
\unboldmath
\index{width difference $\dg$!measurement of $\dg_s$!at CDF}

The promise of large $B_s$ samples in Run~II puts in reach the measurement of
the width difference between the two weak eigenstates of this meson. With its
analysis of semileptonic decays in Run~I, CDF has already published a limit of
$\Delta\Gamma_s/\Gamma_s < 0.85$ at 95\% confidence level~\cite{Ref:tauBs-lDX}.
This limit was established by fitting the lifetime distribution of the $\ell
D_s$ events with two exponentials.

In Run~II, however, it becomes possible to measure the lifetime of samples in
which the weak eigenstates are separated: for instance, 2~fb$^{-1}$ of data
yield approximately 4000 $B_s \to J/\psi\,\phi$ events, which is expected to be
dominated by the shorter-lived eigenstate, $B_s^H$, and for which an angular
analysis is used to separate the two components~\cite{Ref:polBd-Bs}.  Further it
is expected to yield roughly 75,000 of flavor-specific $B_s \to D_s \pi$ and
$B_s \to D_s \pi\pi\pi$%
\index{decay!$B_s \rightarrow D_s^- \pi^+$}%
\index{decay!$B_s \rightarrow D_s^- \pi^+ \pi^+ \pi^-$}%
decays, which are well-defined mixtures of $B_s^H$ and
$B_s^L$.

It is straightforward to show that computing the difference between two
lifetimes has more statistical power than fitting for two exponentials or
fitting for $\Delta\Gamma_s$ in $B_s$ mixing in flavor-specific samples. The
leading term in taking the difference between two measured lifetimes is
$\Delta\Gamma_s$. On the other hand, the lifetime distribution in a
flavor-specific sample is
\begin{equation}
  f(t) = \Gamma_H \left[e^{-\Gamma_Ht} + e^{-\Gamma_Lt}\right] 
       = \Gamma_H\, e^{-\overline{\Gamma}t}\left[e^{-\Delta\Gamma t/2}
                           + e^{+\Delta\Gamma t/2}\right]
       = \Gamma_H\, e^{-\overline{\Gamma}t}\, \bigg[1
                           + \left(\frac{\Delta\Gamma t}{2}\right)^2
                                   + \dots \bigg]\,,
\end{equation}
where $\Delta\Gamma$ enters as a second-order effect.

The $B_s$ decay to $D^+_s D^-_s$ is another valuable source of information to
determine the lifetime difference of the $B_s$ meson. First of all $D^+_s D^-_s$
is a pure $CP$ even eigenstate and thus its lifetime is a clean measurement of
the $CP$ even lifetime. In addition its branching fraction directly measures
$\Delta\Gamma/\Gamma$ under certain theoretical
assumptions~\cite{Ref:Bs-NewPhys}. The complication in this particular decay
mode comes from the associated production of $D^{*\pm}_s$.

\subsubsection{Lifetime Difference Measurements}

\paragraph{\boldmath$B_s \to J/\psi\,\phi$\unboldmath}
The decay mode $B_s \to J/\psi\,\phi$%
\index{decay!$B_s \rightarrow \psi \phi$}%
~has been analyzed at CDF in two Run~I analyses, examining its
lifetime~\cite{Ref:JpsiBsLifetime} and angular distributions~\cite{Ref:polBd-Bs}
separately. The basic strategy for Run~II is to combine these two analyses into
a maximum likelihood fit of the proper decay time, and {\it transversity angle}
of each candidate. In addition to account properly for the background the
invariant mass distribution will be simultaneously fitted.

The transversity angle,%
\index{transversity angle}%
$\theta_T$, is defined by the angle between the $\mu^+$ and the $z$ axis in the
rest frame of the $J/\psi$ decay, where the $z$ axis is orthogonal to the plane
defined by the $\phi$ and the $K^+$ directions.  This angle allows to
distinguish $CP$ even and $CP$ odd components: the probability density function
for the $CP$ even component is $\frac{3}{8}(1+\cos^2\theta_T)$ and for the $CP$
odd component is $\frac{3}{4}\sin^2\theta_T$.  The amplitude of the $CP$ even
component sums the squares of the unpolarized and linearly polarized state
amplitudes, $|A_0|^2+|A_{||}|^2$, of the $\phi$ in the $J/\psi$ rest frame, and
the $CP$ odd component the square of the transversely polarized state amplitude,
$|A_\perp|^2$, of the same.  The analysis depends upon the weak eigenstates
being also $CP$ eigenstates which is a good approximation for the $B_s \to
J/\psi\,\phi$ decays.

A toy Monte Carlo study was performed to estimate the uncertainty in
$\Delta\Gamma_s$ with 4000 $B_s \to J/\psi\,\phi$ decays.  The background shapes
in mass and proper decay time and the relative fraction of the signal to the
background were assumed to be identical to that in the Run~I lifetime analysis.
The mass resolution was assumed to be the same as in Run~I, and all the lifetime
distributions were convoluted with the 18~$\mathrm{\mu m}$ resolution projected
for Run~II. The background was assumed to have a flat transversity angle
distribution.

The error on $\Delta\Gamma_s/\Gamma_s$ depends on the $CP$ admixture of the
final state. The decay $B_s \to J/\psi\,\phi$ is dominated by $CP$ even
eigenstates. Therefore the smaller the admixture of $CP$ odd component the
larger the sensitivity. Assuming the $CP$ composition as measured in
Run~I~\cite{Ref:polBd-Bs} corresponding to a $CP$ even fraction of $0.77\pm0.19$
the expected uncertainty on $\Delta\Gamma_s/\Gamma_s$ is 0.05.  This uncertainty
varies between 0.08 and 0.035 for $CP$ even fractions of 0.5 and 1.0,
respectively.

\paragraph{\boldmath$B_s\to D_s^{(*)+}D_s^{(*)-}$\unboldmath}
The decay modes $B_s\to D_s^{(*)+}D_s^{(*)-}$%
\index{decay!$B_s \rightarrow D_s^+ D_s^-$}%
\index{decay!$B_s \rightarrow D_s^{*+} D_s^-$}%
\index{decay!$B_s \rightarrow D_s^+ D_s^{*-}$}%
\index{decay!$B_s \rightarrow D_s^{*+} D_s^{*-}$}%
are also promising for lifetime difference studies, though with smaller sample
sizes. These decays are expected to be present among the two-track trigger data.
The decay $B_s \to D_s^+D_s^-$, in particular, is purely $CP$ even, and requires
no angular analysis.  Its companion decays, involving $D_s^*$ decays, are
expected in the heavy quark limit, and in the absence of $CP$ violation, to be
sensitive to $CP$-even $B_s$ states as well~\cite{Ref:ALEKSAN-DSSTAR}.  While
these decays are attractive in that they significantly increase the sample size
over that of $B_s \to D_s^+ D_s^-$ alone, their identification is a challenge,
since the missing photons from the decay $D_s^{*+}\to D_s^+\gamma$ and
$D_s^{*+}\to D_s^+\pi^0$ considerably broaden the $D^+_s D^-_s$ invariant mass
distribution. On the other hand, the missing mass introduces only about 3\% to
the proper lifetime resolution.

In a full GEANT based simulation and reconstruction of $B_s \to D^{(*)+}_s
D^{(*)-}_s$ in the CDF detector it is found that the three different cases are
separated quite cleanly by using the invariant mass spectrum of the charged
decay products. Shifts in the invariant $B_s$ mass are due to the neutral
particles which are not reconstructed. PYTHIA has been used to generate
$b\,\bar{b}$ quarks and fragment them to $b$ hadrons. The $B_s$ mesons are decayed
using the CLEOMC program according to the branching ratios given in
Table~\ref{tab:bs-branchings}.

In Figure~\ref{fig:cdf-bsdsds} the invariant mass spectra of the three
different cases are depicted. The spectra are essentially free of combinatoric
background since the reconstructions of the resonances at each step allow
stringent cuts. For this picture the $D_s$ is always decayed into $K^{*0}K$
which has more combinatorics than the $\phi\pi$ decay mode due to the large
width of the~$K^{*0}$.

\begin{figure}[t]
\begin{center}
  \mbox{\epsfig{file=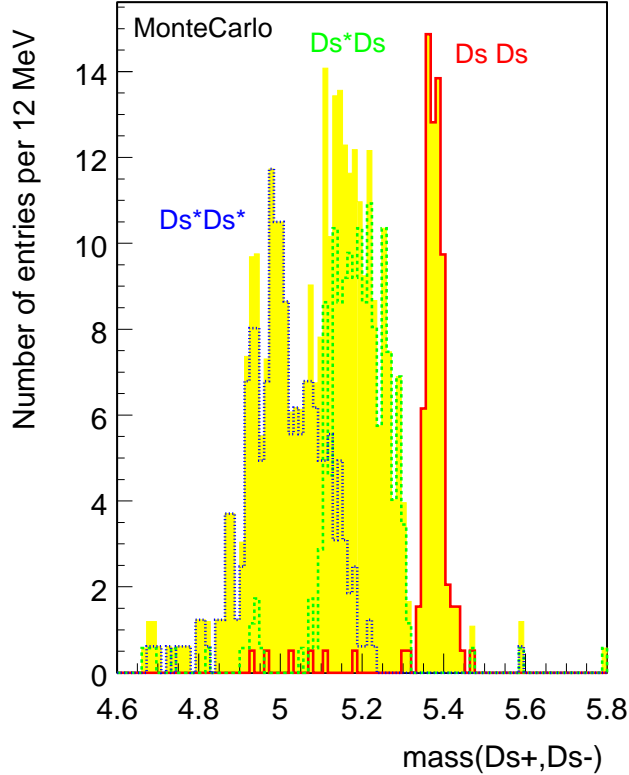,width=0.55\textwidth}}
\caption[$D^+_s,D^-_s$ invariant mass spectra for the three decay modes $B_s
\to D_s D_s$ (solid), $B_s \to D^{(*)}_s D_s$ (large dashes) and $B_s \to
D^{(*)}_s D^{(*)}_s$ (small dashes).]{$D^+_s,D^-_s$ invariant mass spectra for
the three decay modes $B_s \to D_s D_s$ (solid), $B_s \to D^{(*)}_s D_s$ (large
dashes) and $B_s \to D^{(*)}_s D^{(*)}_s$ (small dashes). The $D_s$ is always
decayed into $K^{*0}K$. The shaded background is the sum of the three decay
modes.}
  \label{fig:cdf-bsdsds}
\end{center}
\end{figure}

To estimate the error on the $\Delta\Gamma_s/\Gamma_s$ from this channel
several assumptions have to be made since this mode has not been reconstructed
in Run~I.

In the most conservative estimate only $B_s \to D_s D_s$ is used which is a
clean $CP$ even state. GEANT based Monte Carlo studies indicate that a signal to
background fractions of 1:1 - 1:2 are achievable. This is similar to signal to
background fraction achieved for the decay mode $B_s \to \nu \ell D_s$, measured
in Run~I. The expected error on the lifetime is 0.044~ps is obtained using an
event sample of 2.5k events and a signal to background fraction of 1:1.5. This
converts into an error on $\Delta\Gamma_s/\Gamma_s$ of 0.06.

Assuming that also the other two decay modes involving $D^*_s$ mesons are clean
$CP$ even modes a total of 13k events are available. The estimated error on
$\Delta\Gamma_s/\Gamma_s$ is then reduced to 0.025.

%
%
%
%

\subsubsection{Related Branching Fractions}

The decay modes $B_s \to D_s^{(*)+}D_s^{(*)-}$ are also interesting because it
is expected that they are the largest contribution to the actual difference
between the heavy and light widths. Indeed, the other decay modes are estimated
to contribute less than 0.01 to the projected $\sim 0.15$ value of
$\Delta\Gamma_s/\Gamma_s$~\cite{Ref:ALEKSAN-DGAMMA}. The branching fraction to
this final state is in the small velocity (SV) limit related to $\Delta\Gamma_s$
by
\begin{equation}\label{eq:bs-branching}
  {\cal B}(B_s^H \to D_s^{(*)+} D_s^{(*)-}) =
        \frac{\Delta\Gamma_s}{ \overline{\Gamma_s}
   (1 + \frac{\Delta\Gamma_s}{2\overline{\Gamma_s}})}\,.
\end{equation}
This method has been exploited by ALEPH, using $\phi\phi$ correlations, to
obtain a value of $\Delta\Gamma_s/\Gamma_s =
0.25^{+0.21}_{-0.14}$~\cite{Ref:ALEPH-PHIPHI}. However the small velocity
assumption also referred to a Shifman-Voloshin limit may be very approximative.

The following estimates are made assuming the validity of this limit. Further
when measuring branching fractions many systematic effects have to be
considered. For example the tracking efficiency for the kinematics of the
particular decays has to be determined carefully. Since it is difficult if not
impossible to predict those effects only the statistical uncertainties are
discussed below.

The statistical error on the branching fraction for 13k events (see
Table~\ref{tab:hadrtrig-bs}) with a signal to background ratio of 1:1 is 0.012.
Following Equation~\ref{eq:bs-branching} the turns out to be also the statistic
uncertainty on $\Delta\Gamma_s/\Gamma_s$ since the branching ratio is very
small. This uncertainty deteriorates to 0.015 when making the signal to
background ratio 1:2.

\subsubsection{Combined CDF Projection}

With the analysis possibilities discussed thus far, the lifetime difference
method conservatively yields a statistical uncertainty of 0.04 on
$\Delta\Gamma_s/\Gamma_s$, utilizing both $J/\psi\,\phi$ and $D_s^+ D_s^-$ decays
and just using the lifetime measurements.%
\index{width difference $\dg$!measurement of $\dg_s$!at CDF}%

If one assumes that the decay modes involving $D_s^*$ are also mostly $CP$ even
the sample for the lifetime measurement is extended and the branching ratios can
be used in the SV limit. This decreases the statistical uncertainty on
$\Delta\Gamma_s/\Gamma_s$ to 0.01.

These numbers refer to the projected Run~II luminosity of 2~$\mathrm{fb}^{-1}$
and bear all the caveats mentioned in the text.

\boldmath
\subsection[Estimate of Sensitivity on $\Delta\Gamma$ in BTeV]
{Estimate of Sensitivity on $\Delta\Gamma$ in BTeV
$\!$\authorfootnote{Author: H.W.K.~Cheung.}}
\unboldmath
\index{width difference $\dg$!measurement of $\dg_s$!at BTeV}

Since $\Delta\Gamma_{B_s}$is expected to be much larger than
$\Delta\Gamma_{B^0}$, only projections for measurements of $\Delta\Gamma_{B_s}$
have been studied at BTeV for this report.  The $B_s^0$ decay modes studied
include $CP$-even, $CP$-mixed and
flavor specific decay modes and are listed in Table~\ref{tb_btevdgmodes}.  The
total decay rate for the flavor specific decay $B_s^0\rightarrow D_s^-\pi^+$ 
\index{decay!$B_s \rightarrow D_s^- \pi^+$}
is given by the average of the $CP$-even and $CP$-odd rates.  The decays
$B_s^0\rightarrow J/\psi\eta$, $J/\psi\eta^{\prime}$ 
\index{decay!$B_s \rightarrow \psi \eta $}
\index{decay!$B_s \rightarrow \psi \eta^\prime$}
should be $CP$-even while the
decay to $J/\psi\,\phi$ 
\index{decay!$B_s \rightarrow \psi \phi$}
is predominantly $CP$-even.

\begin{table}[tbh]
\begin{center}
\begin{tabular}{cccc}
\hline\hline
{\boldmath $B_s^0$ \bf{Decay Mode}} & \bf{$CP$ Mode} & 
\bf{Branching Ratio} & \bf{BR Used}\\ \hline
$J/\psi\,\phi$     & Mostly $CP$-even & $(9.3\pm 3.3)\times 10^{-4}$ &
$8.9\times 10^{-4}$\\
$J/\psi\eta$ & $CP$-even & $<0.0038$  &
$3.3\times 10^{-4}$\\
$J/\psi\eta^{\prime}$ & $CP$-even & $<0.0038$  &
$6.7\times 10^{-4}$\\
$D_s^-\pi^+$     & Flavor specific  & $<0.13$  &
$3.0\times 10^{-3}$\\
\hline\hline
\end{tabular}
\end{center}
\caption{The $B_s^0$ decay modes studied for $\Delta\Gamma_{B_s}$
sensitivity studies at BTeV.}
\label{tb_btevdgmodes}
\end{table}

CDF has measured the $CP$-odd fraction of total rate for $J/\psi\,\phi$ to be
$0.229\pm0.188(stat)\pm 0.038(syst)$\cite{Ref:BTEV-CDFJPSIPHI}.  The all charged
mode decay $B_s^0\rightarrow K^+K^-$ has a large enough expected branching
fraction ($\sim 1\times 10^{-5}$) and reconstruction efficiency to get high
statistics. However although the $K^+K^-$ final state is $CP$-even, the decay can
proceed via both a $CP$ conserving Penguin contribution as well as a $CP$ violating
Tree level contribution. Unless the Penguin contribution is completely dominant
or the Penguin and Tree level contributions can be exactly calculated it would
be difficult to use this mode without significant theoretical errors.

\subsubsection{Signal yields and backgrounds}

Estimation of the signal yields and signal/background ratios were determined
using a MCFAST simulation for the $B_s^0$ decay mode to $J/\psi\,\phi$, while all
the other modes were simulated using a Geant simulation of BTeV.  Although it is
easy to simulate the signal to determine the reconstruction efficiency, the
background simulation can be more troublesome. The signal sample always has an
average of two embedded min-bias events. Since it takes too much time to
generate enough background one has to determine which backgrounds are dominant
for a particular decay.

The decay $B_s^0\rightarrow J/\psi\,\phi$ was studied through the decay channels
$J/\psi\rightarrow\mu^+\mu^-$ and $\phi\rightarrow K^+K^-$. Studies show that
since the $J/\psi\rightarrow\mu^+\mu^-$ is expected to be so clean, the dominant
backgrounds come from $b\,\bar{b}\rightarrow J/\psi X$.  For this study only
backgrounds from $b\,\bar{b}\rightarrow J/\psi\,\phi X$ have been studied.  Figure
\ref{btev_psimass1}(a) shows the $\mu^+\mu^-$ invariant mass for a
$b\,\bar{b}\rightarrow J/\psi\,\phi X$ background sample compared to an
appropriately normalized signal sample of $B_s^0\rightarrow J/\psi\,\phi$. The two
muons were required to form a vertex with a confidence level of greater than
1\%.  Figure~\ref{btev_psimass1}(b) shows a similar comparison for the $K^+K^-$
invariant mass, again with a vertex requirement of $CL>1$\%.

\begin{figure}[t]
\centerline{
\epsfxsize 7cm \epsffile{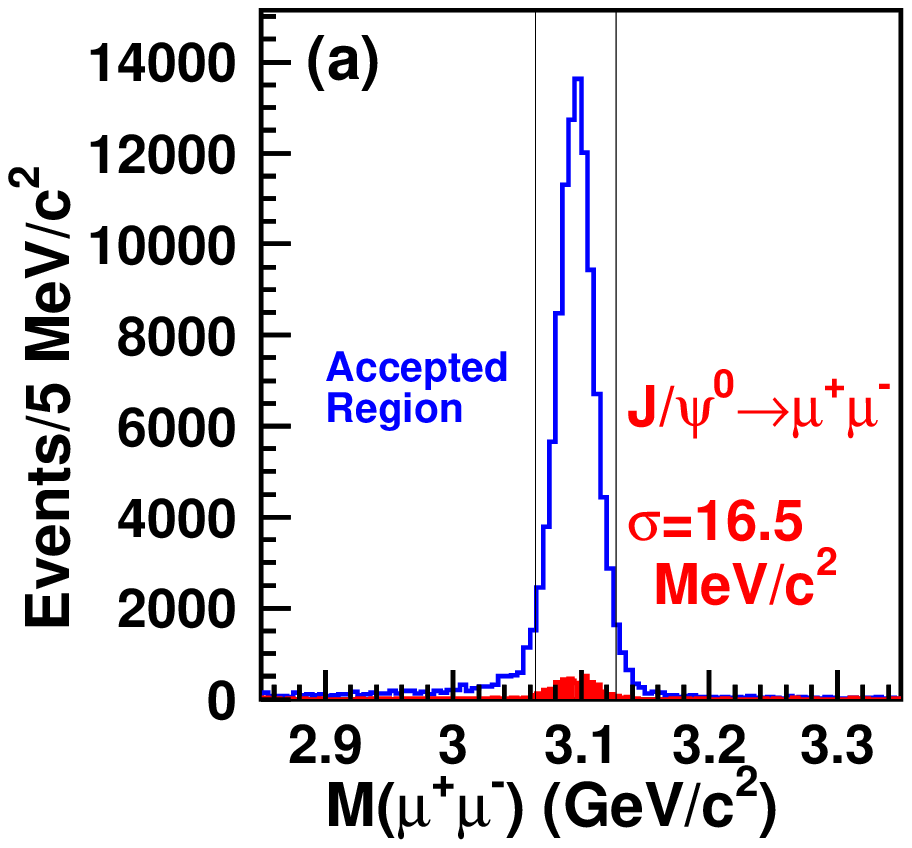}
\epsfxsize 7cm \epsffile{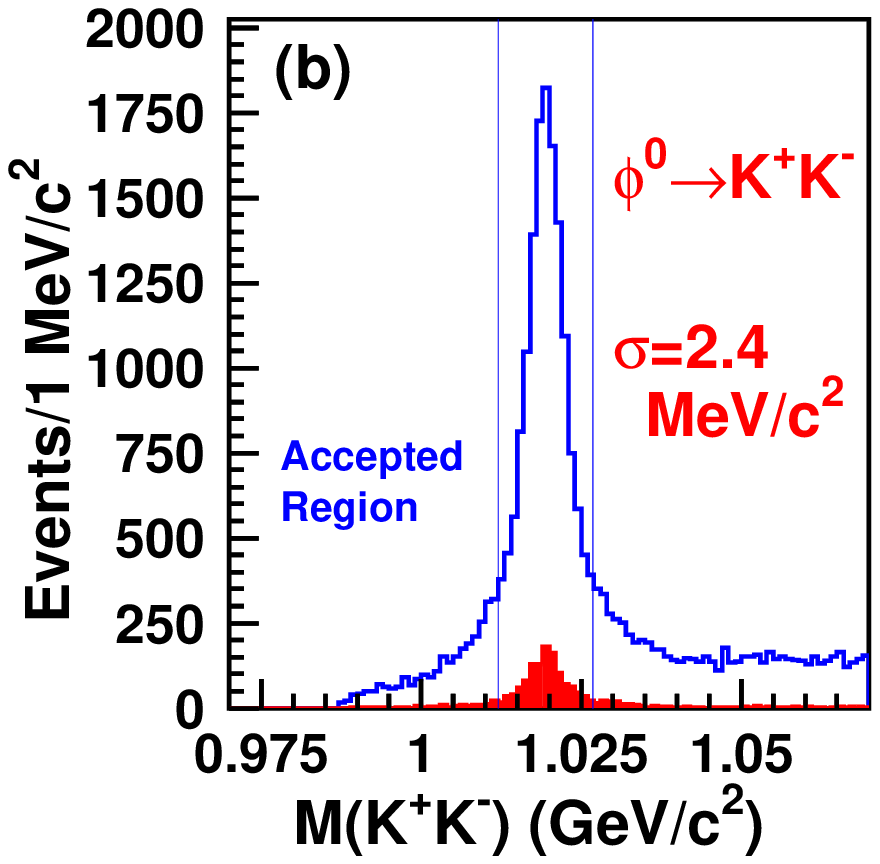}}
\caption[The $\mu^+\mu^-$ invariant mass (a) and the $K^+K^-$ invariant mass
(b) for a $b\,\bar{b}\rightarrow J/\psi\,\phi X$ background samples.]{(a) The
$\mu^+\mu^-$ invariant mass and (b) the $K^+K^-$ invariant mass for a
$b\,\bar{b}\rightarrow J/\psi\,\phi X$ background sample (open histogram) and for
an appropriately normalized $B_s^0\rightarrow J/\psi\,\phi$ signal sample (filled
histogram).}
\label{btev_psimass1}
\end{figure}

Figure~\ref{btev_psiphim1}(a) shows the $\mu^+\mu^-K^+K^-$ invariant mass for
the $b\,\bar{b}\rightarrow J/\psi\,\phi X$ background sample without requiring that
the $\mu^+\mu^-$ and $K^+K^-$ masses are consistent with the $J/\psi$ and $\phi$
masses respectively. The four tracks are required to form a single vertex with a
confidence level greater than 1\%. This is compared in the plot to an
appropriately normalized signal sample.  Figure~\ref{btev_psiphim1}(b) shows the
$J/\psi\,\phi$ invariant mass plot with vertexing and requirements on the
$\mu^+\mu^-$ and $K^+K^-$ masses. The $\mu^+\mu^-$ and $K^+K^-$ masses are
required to be within $\pm 2\sigma$ of the true $J/\psi$ and $\phi$ masses
respectively. Both signal and backgrounds are included in the plot and compared
to the signal only sample.  The signal/background ratio is seen to increase when
one applies a primary-to-secondary vertex detachment requirement of
$L/\sigma_L>15$ in Figure~\ref{btev_psiphim1}(c). Where $L$ is the 3-dimensional
distance between the primary vertex and the $B_s^0$ decay vertex and $\sigma_L$
is the error on $L$ calculated for each candidate $B_s^0$ decay.  For
$L/\sigma_L>15$ the reconstruction efficiency is 6.0\%, and the
signal/background ratio is 47/1 with an error of 32\%\ for the backgrounds
considered.

\begin{figure}[ht]
\centerline{
\epsfxsize 4.67cm \epsffile{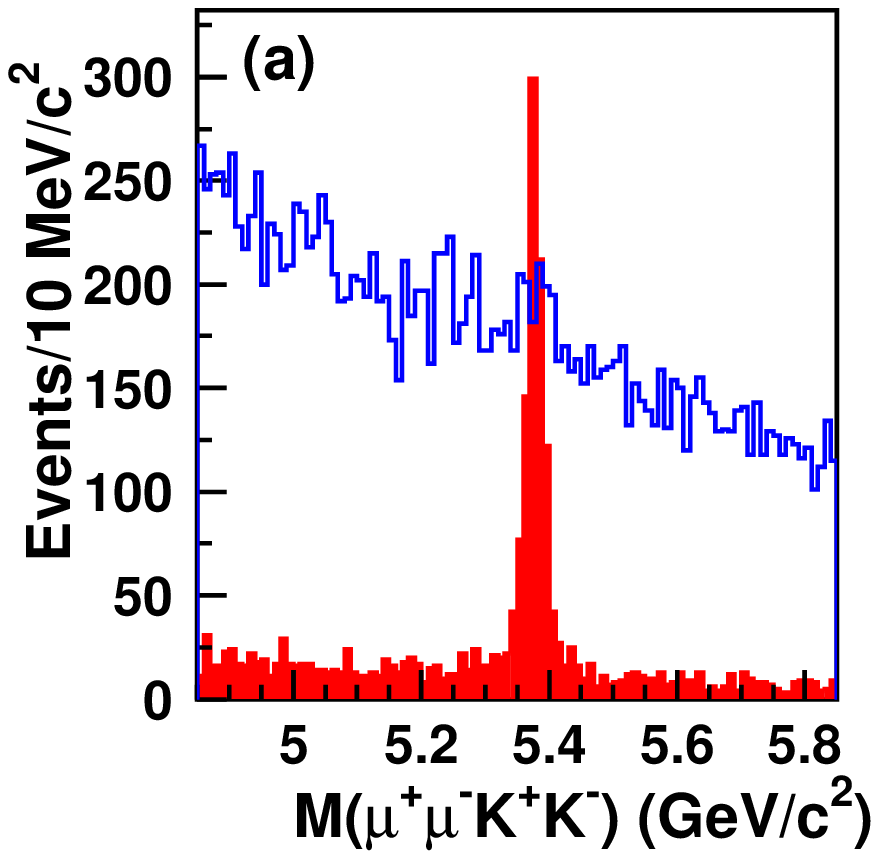}
\epsfxsize 4.67cm \epsffile{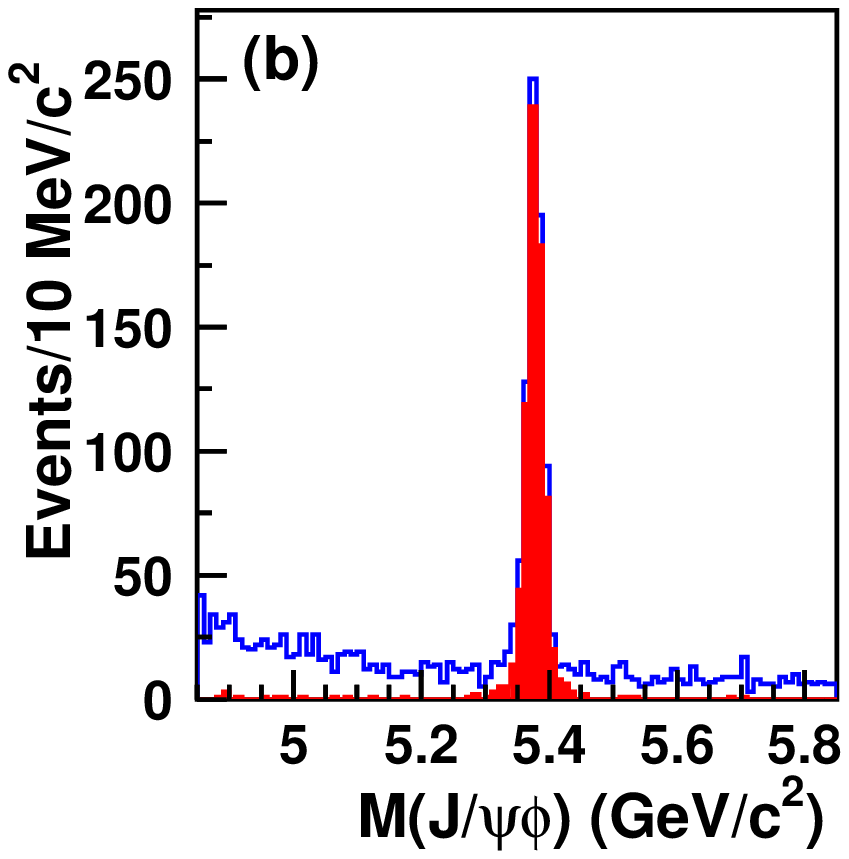} 
\epsfxsize 4.67cm \epsffile{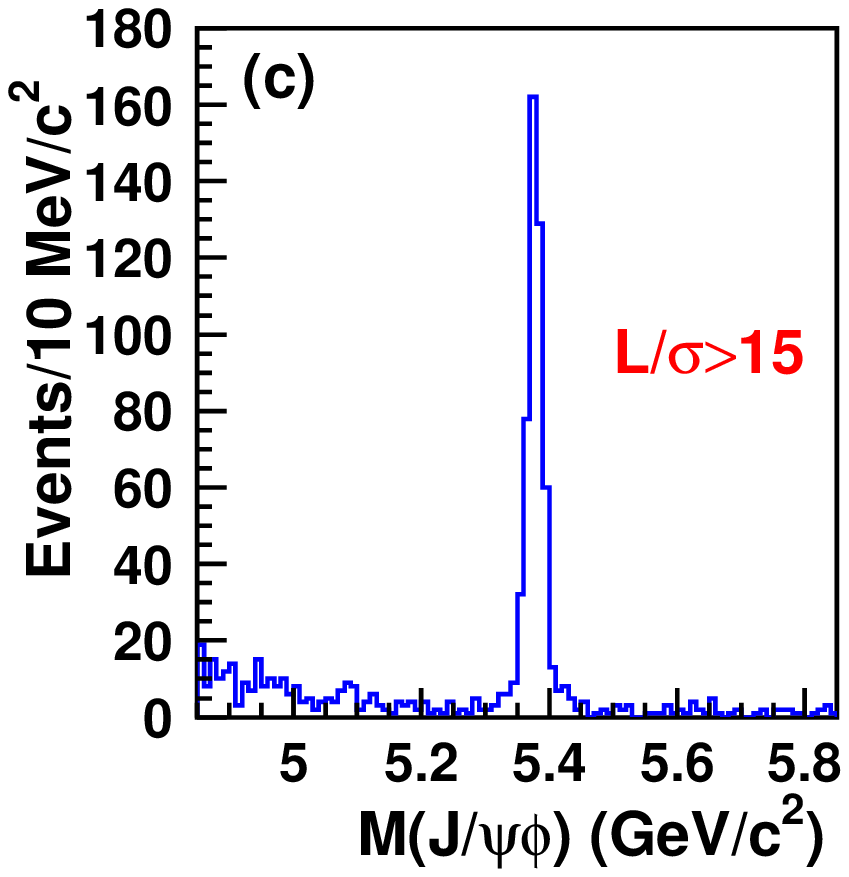}}
\caption[The $\mu^+\mu^-K^+K^-$ invariant mass for $b\,\bar{b}\rightarrow
J/\psi\,\phi X$ background and $B_s^0\rightarrow J/\psi\,\phi$ signal
samples]{(a) The $\mu^+\mu^-K^+K^-$ invariant mass for a 
$b\,\bar{b}\rightarrow J/\psi\,\phi X$ background sample (open histogram) and 
for a correctly normalized $B_s^0\rightarrow J/\psi\,\phi$ signal sample 
(filled histogram). (b) $J/\psi\,\phi$ invariant mass for the background plus
signal sample (open histogram) compared to just the signal sample (filled
histogram). (c) $J/\psi\,\phi$ mass for $L/\sigma_L>15$.}
\label{btev_psiphim1}
\end{figure}

Table~\ref{tb_btevpsiphi1}\ shows projections for the $B_s^0\rightarrow
J/\psi\,\phi$ signal yields for 2~fb$^{-1}$.  A total $b\,\bar{b}$ production
cross-section of 100~$\mu$b is assumed and we take the fraction of
$B_s^0/\overline{B_s^0}$ per $b\,\bar{b}$ from Pythia as 1 in 4.3. The branching
fraction of $B_s^0\rightarrow J/\psi\,\phi$ is taken to be equal to
$BR(B_d^0\rightarrow J/\psi K^0)=8.9\times 10^{-4}$.  A total of $\sim$41400
signal events is expected for 2~fb$^{-1}$.  The expected error on the lifetime
for 2~fb$^{-1}$ was determined with a toy Monte Carlo generating 1000
experiments with 41400 signal events and S/B=47. The background lifetime
distribution was simulated with a short and a long lifetime component as seen in
the background studies and is typical of backgrounds seen in fixed target
experiments. In the toy MC the short and long components were set to be 0.33 and
1.33~ps respectively.  A binned likelihood fit was used to extract the measured
lifetime for each of the 1000 experiments, where the lifetime distribution from
sidebands is used as a measure of the lifetime distribution in the signal
$B_s^0$ mass region. The method is described in
Reference~\cite{Ref:BTEV-E687LIFETIME}.  The expected error is taken to be the
{\it r.m.s.} of the 1000 measured lifetimes and is 0.50\%.

\begin{table}[bht]
\begin{center}
\begin{tabular}{lccc}
\hline\hline
\bf{Quantity}  & \multicolumn{3}{c}{\bf{Value}} \\ \hline
Number of $b\,\bar{b}$ & \multicolumn{3}{c}{$2\times 10^{11}$} \\
Number of $B_s^0/\overline{B_s^0}$ & \multicolumn{3}{c}{$4.7\times 10^{10}$} \\
\hline
           &  $B_s^0\rightarrow J/\psi\,\phi$ & 
              \multicolumn{2}{c}{$B_s^0\rightarrow D_s^-\pi^+$} \\
           & $J/\psi\rightarrow\mu^+\mu^-$ & $D_s^-\rightarrow\phi\pi^-$ &
             $D_s^-\rightarrow K^{\ast 0}K^-$ \\ 
           & $\phi\rightarrow K^+K^-$ & $\phi\rightarrow K^+K^-$ &
             $K^{\ast 0}\rightarrow K^+\pi^-$ \\ \hline
\#\ of Events  & $1.2\times 10^6$  & $2.5\times 10^6$ & $3.1\times 10^6$\\
Reconstruction efficiency (\%) & 6.0 & 2.7  & 2.3 \\
S/B                            & 47:1  & \multicolumn{2}{c}{3:1} \\
L1 Trigger efficiency (\%)     & 70 & \multicolumn{2}{c}{74} \\
L2 Trigger efficiency (\%)     & 90  & \multicolumn{2}{c}{90} \\
\#\ of reconstructed decays    & 41400 & \multicolumn{2}{c}{91700} \\
\hline\hline
\end{tabular}
\end{center}
\caption{Projections for yields of $B_s^0$ decays for 2~fb$^{-1}$ assuming
a total $b\,\bar{b}$ production cross-section of 100~$\mu$b.}
\label{tb_btevpsiphi1}
\end{table}

Although the $J/\psi\,\phi$ signal sample can be obtained through the
$J/\psi\rightarrow \mu^+\mu^-$ trigger, the effect of the Level 1 vertex trigger
\index{width difference $\dg$!measurement of $\dg_s$!effect of vertex trigger}
on this mode was studied to determine the effect of the L1 trigger on the
lifetime analysis. Figure~\ref{btev_ftpsiphi1}(a) shows the proper time
($t=L/\beta\gamma c$) distribution for reconstructed $B_s^0\rightarrow
J/\psi\,\phi$ signal events for a $L/\sigma_L>15$ requirement. The loss of short
lifetime decays is due to the detachment requirement. One can obtain an
exponential distribution if we use the reduced proper time,
$t^{\prime}=t-N\sigma_L/\beta\gamma c$, for a $L/\sigma_L>N$ requirement. This
starts the clock at the minimum required decay time for each decay candidate and
works because the lifetime follows an exponential distribution irrespective of
the when the clock is started. Figure~\ref{btev_ftpsiphi1}(b) shows the reduced
proper time distribution and an exponential fit gives a lifetime of $1.536\pm
0.014$~ps, compared to the generated lifetime of 1.551~ps.

\begin{figure}[t]
\centerline{\epsfxsize 0.95\textwidth \epsffile{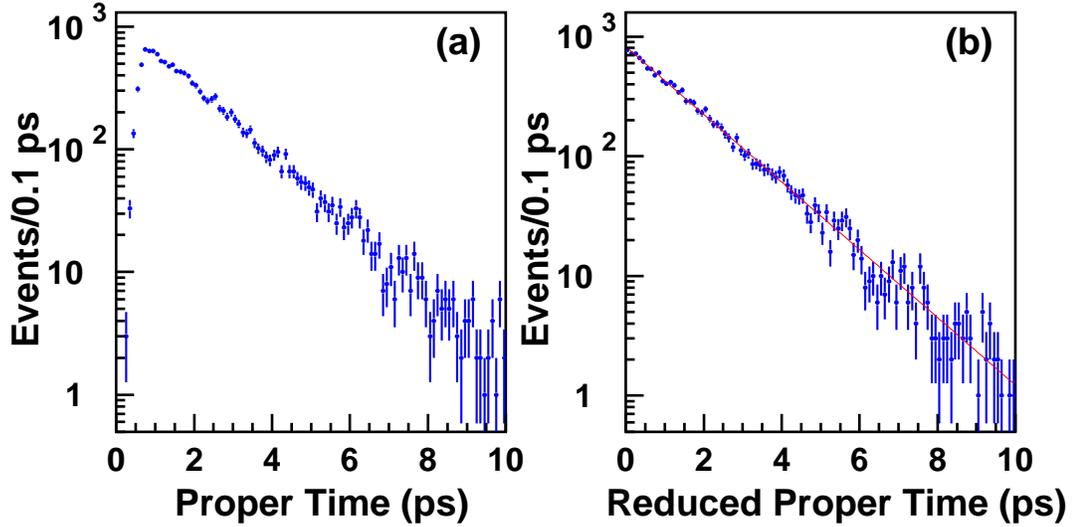}}
\caption{(a) Proper time distribution for reconstructed 
$B_s^0\rightarrow J/\psi\,\phi$; (b) Reduced proper time distribution for
the same decays, the line is an exponential fit.}
\label{btev_ftpsiphi1}
\end{figure}

Figure~\ref{btev_ftpsiphi2}(a) shows the reduced proper time after applying the
Level 1 vertex trigger. Short lifetime decays are again lost because the impact
parameter requirements of the Level 1 trigger effectively gives a larger minimum
required decay time than $N\sigma_L/\beta\gamma c$.  The lifetime acceptance
function is just the observed reduced proper distribution divided by a pure
exponential with the generated lifetime and is given in Figure
\ref{btev_ftpsiphi2}(b).  In order to extract the correct lifetime from the
observed reduced proper time distribution one needs the correct lifetime
acceptance function. This is obtained from Monte Carlo and can be checked by
using decays modes like $J/\psi\,\phi$ that can be obtained with the L1 vertex
trigger and separately through the L1 $J/\psi\rightarrow \mu^+\mu^-$ dimuon
trigger which has no vertexing selection criteria. The L1 trigger lifetime
acceptance correction obtained from MC can also be checked by taking samples of
prescaled triggers that do not have the L1 trigger requirement.

\begin{figure}[bhtp]
\centerline{\epsfxsize 0.95\textwidth \epsffile{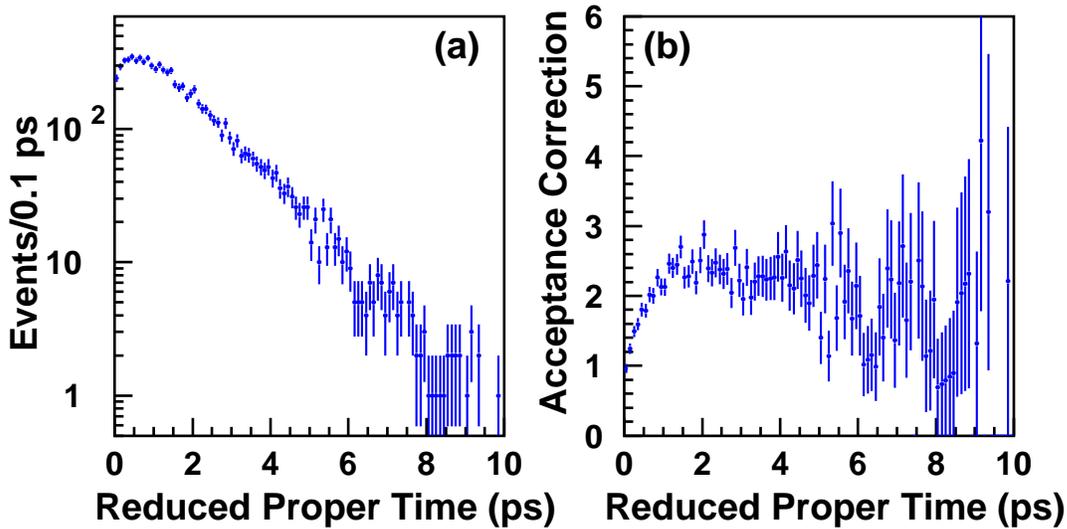}}
\caption[(a) Reduced proper time distribution for reconstructed 
$B_s^0\rightarrow J/\psi\,\phi$ decays that pass the Level 1 vertex trigger;
(b) Lifetime correction function.]{(a) Reduced proper time distribution for
reconstructed  $B_s^0\rightarrow J/\psi\,\phi$ decays that pass the Level 1
vertex trigger; (b) Lifetime correction function, obtained by dividing the
distribution in (a) by a pure exponential distribution with the generated
$B_s^0$ lifetime.}
\label{btev_ftpsiphi2}
\end{figure}

Note that since the L1 trigger can increase the effective minimum required decay
time cut, if one puts on a large impact parameter selection requirement on the
$B$ decay daughters, it may be possible to redefine the reduced proper time to
take this into account and thereby reduce the lifetime acceptance corrections
with some lost of statistics. This still needs to be studied and the details of
the L1 trigger may change since one may be able to redesign it to reduced the
lifetime acceptance correction.

When an acceptance correction like that given in Figure~\ref{btev_ftpsiphi2}(b)
is simulated in the toy Monte Carlo, the expected error on the measured lifetime
increases to 0.58\%, from the previous value of 0.50\%. For the decay mode
$B_s^0\rightarrow D_s^-\pi^+$ where the signal to background is smaller the
effect is similar, the error increases to 0.44\%\ when adding the acceptance
correction effects compared to 0.39\%.

\begin{table}[bht]
\begin{center}
\begin{tabular}{lccc} \hline\hline
\bf{Quantity}  & \multicolumn{3}{c}{\bf{Value}} \\ \hline
Number of $b\,\bar{b}$ & \multicolumn{3}{c}{$2\times 10^{11}$} \\
Number of $B_s^0/\overline{B_s^0}$ & \multicolumn{3}{c}{$4.7\times 10^{10}$} \\
\hline
           &  $B_s^0\rightarrow J/\psi\eta$ & 
              \multicolumn{2}{c}{$B_s^0\rightarrow J/\psi\eta^{\prime}$} \\
           & \multicolumn{3}{c}{$J/\psi\rightarrow\mu^+\mu^-$} \\
           & $\eta\rightarrow \gamma\gamma$ & 
             $\eta^{\prime}\rightarrow \rho^0\gamma$ &
             $\eta^{\prime}\rightarrow\pi^+\pi^-\eta$ \\ \hline
\#\ of Events  & $3.5\times 10^5$  & $5.4\times 10^5$ & $3.2\times 10^5$\\
Reconstruction efficiency (\%) & 0.71 & 1.2  & 0.60\\
S/B                            & 15:1  & \multicolumn{2}{c}{30:1}  \\
L1 Trigger efficiency (\%)     & 75 & \multicolumn{2}{c}{85} \\
L2 Trigger efficiency (\%)     & 90  & \multicolumn{2}{c}{90} \\
\#\ of reconstructed decays    & 1700 & \multicolumn{2}{c}{6400} \\
\hline\hline
\end{tabular}
\end{center}
\caption{Projections for yields of $B_s^0$ decays for 2~fb$^{-1}$ assuming
a total $b\,\bar{b}$ production cross-section of 100~$\mu$b.}
\label{tb_btevpsieta1}
\end{table}

\begin{table}[hbt]
\begin{center}
\begin{tabular}{cccc} \hline\hline
\bf{Decay Mode}      & \multicolumn{3}{c}{\bf{Error on Lifetime (\%)}}\\ \hline
                       & 2~fb$^{-1}$ & 10~fb$^{-1}$ & 20~fb$^{-1}$ \\ \hline
$J/\psi\,\phi$           & 0.50  & 0.23 & 0.16 \\
$J/\psi\eta$           & 2.49  & 1.19 & 0.80 \\
$J/\psi\eta^{\prime}$  & 1.36  & 0.55 & 0.39 \\
$D_s^-\pi^+$           & 0.44  & 0.20 & 0.14 \\
\hline\hline
\end{tabular}
\end{center}
\caption{Projections for statistical errors on lifetimes measured in
different modes for 2, 10 and 20~fb$^{-1}$.}
\label{tb_btevall1}
\end{table}

Tables~\ref{tb_btevpsieta1}\ and \ref{tb_btevpsiphi1}\ show the expected signal
yields and signal/background ratios for the decays modes $B_s^0\rightarrow
J/\psi\eta$, $B_s^0\rightarrow J/\psi\eta^{\prime}$ and $B_s^0\rightarrow
D_s^-\pi^+$.  
\index{decay!$B_s \rightarrow D_s^- \pi^+$}%
\index{decay!$B_s \rightarrow \psi \eta $}%
\index{decay!$B_s \rightarrow \psi \eta^\prime$}%
Backgrounds from $b\,\bar{b}\rightarrow J/\psi X$ were studied for
the decay modes $B_s^0\rightarrow J/\psi\eta^{(\prime)}$, while backgrounds from
$b\,\bar{b}\rightarrow D_s^+X$ were included in the $B_s^0\rightarrow D_s^-\pi^+$
analysis.  Details of the analyses of these modes can be found in
Reference~\cite{Ref:BTEV-PROPOSAL}. Expected errors on the lifetimes were
determined using toy Monte Carlo simulations as before, where acceptance
corrections were also simulated for the $D_s^-\pi^+$ mode.
Table~\ref{tb_btevall1} gives the expected errors on the lifetimes for all
modes.

Note that although only backgrounds from $b\,\bar{b}\rightarrow J/\psi\,\phi X$ were
included in the study of $B_s^0\rightarrow J/\psi\,\phi$ to obtain the value of
S/B=47/1, the effect of lower values of S/B were studied. If the S/B is
decreased to 10/1 which is the expected level for $B_d^0\rightarrow J/\psi
K_s^0$~\cite{Ref:BTEV-PROPOSAL}, the expected error on the measured lifetime
only increases from 0.50\%\ to 0.51\%\ for 2~fb$^{-1}$. (For a much lower
S/B=3/1 the expected error would be 0.58\%.)

\subsubsection{Results for $\Delta\Gamma/\Gamma_{B_s}$ Sensitivity}
\index{width difference $\dg$!measurement of $\dg_s$!sensitivity}
\index{width difference $\dg$!$\Delta\Gamma_{CP}$}

With just two lifetime measurements, like $\tau_{CP+}$ and 
$\tau_{CP-}$, which are defined as $\tau_{CP+}=1/\Gamma(B_s^{even})$
and $\tau_{CP-}=1/\Gamma(B_s^{odd})$, one can determine the error
on $\Delta\Gamma_{CP}/\Gamma$ from the errors on the two lifetimes:
\begin{equation}
\sigma_{\Delta\Gamma_{CP}\over\Gamma} = 4\, {\tau_{CP+}\tau_{CP-}\over
  (\tau_{CP+}+\tau_{CP-})^2}\,
\sqrt{\left({\sigma_{\tau_{CP+}}\over\tau_{CP+}}\right)^2+
  \left({\sigma_{\tau_{CP-}}\over\tau_{CP-}}\right)^2}\,,
\end{equation}
where $\Delta\Gamma_{CP}=\Gamma(B_s^{even})-\Gamma(B_s^{odd})$ and
$\Gamma=(\Gamma(B_s^{even})+\Gamma(B_s^{odd}))/2$.  In the case that one
measures $\tau_{CP+}$ and $\tau_{FS}$ where
$\tau_{FS}=2/(\Gamma(B_s^{even})+\Gamma(B_s^{odd}))$, the error on
$\Delta\Gamma_{CP}/\Gamma$ is
\begin{equation}
\sigma_{\Delta\Gamma_{CP}\over\Gamma}
=2\,{\tau_{FS}\over\tau_{CP+}}\,
\sqrt{\left({\sigma_{\tau_{CP+}}\over\tau_{CP+}}\right)^2+
\left({\sigma_{\tau_{FS}}\over\tau_{FS}}\right)^2}.
\end{equation}

It can be seen that for small $\Delta\Gamma_{CP}/\Gamma$, the error is about 2
times larger when one measures with similar errors $\tau_{CP+}$ and
$\tau_{FS}$ compared to measuring $\tau_{CP+}$ and $\tau_{CP-}$.  Using
$\tau_{CP+}$ from the $J/\psi\eta^{(\prime)}$ decay modes and $\tau_{FS}$
from $D_s^-\pi^+$, we project an error on $\Delta\Gamma_{CP}/\Gamma$ of $0.027$
for $\Delta\Gamma_{CP}/\Gamma=0.15$ for 2~fb$^{-1}$.

Including the lifetime measurement for the decay mode to $J/\psi\pi$ is more
complicated as this is not an equal mixture of $CP$-even and $CP$-odd rates.
Although a combined lifetime and angular analysis should be done to determine
the fraction of $CP$-odd decay in this mode, and the appropriate error on
$\tau_{CP+}$, it was not done for this study.  A simpler determination is made
assuming that the fraction of $CP$-odd has already been determined within some
error in a separate analysis or in some other experiment. If the lifetime for
this decay mode is defined by $\tau_X=1/\Gamma_X$ where
$\Gamma_X=(1-f)\Gamma(B_s^{even})+f\Gamma(B_s^{odd})$, then
\begin{eqnarray}
{\Delta\Gamma_{CP}\over\Gamma} &=& {2(\tau_{FS}-\tau_X)\over (1-2f)\tau_X}\,,
  \\
\sigma_{\Delta\Gamma_{CP}\over\Gamma} &=& {2\tau_{FS}\over (1-2f)\tau_X}\,
\sqrt{\left({\sigma_{\tau_X}\over\tau_X}\right)^2+
  \left({\sigma_{\tau_{FS}}\over\tau_{FS}}\right)^2+
  \left({\tau_{FS}-\tau_X\over\tau_{FS}}\right)^2 {4f^2\over (1-2f)^2}
  \left({\sigma_f\over f}\right)^2}\,. \nonumber
\end{eqnarray}

Setting the value of $f$ to the central valued measured by CDF and assuming the
total error can be improved in Run 2 by a factor of $\sqrt{20}$, then for
$f=0.229\pm 0.043$ and using the $J/\psi\,\phi$ and $D_s^-\pi^+$ modes only, the
projected error on $\Delta\Gamma_{CP}/\Gamma$ is $0.035$ for
$\Delta\Gamma_{CP}/\Gamma=0.15$ and 2~fb$^{-1}$.  Although reducing the error on
$f$ has only a small effect on the projected error on
$\Delta\Gamma_{CP}/\Gamma$, the actual value of $f$ has a huge effect, since as
$f$ approaches 0.5 we lose all sensitivity to $\Delta\Gamma_{CP}$ when only
comparing to the $D_s^-\pi^+$ mode.  It can be seen that even with relatively
low statistics, the $J/\psi\eta^{(\prime)}$ decay modes are just as sensitive to
$\Delta\Gamma_{CP}$.

Table~\ref{tb_btevdgresults}\ shows the projected errors on
$\Delta\Gamma_{CP}/\Gamma$ for different integrated luminosities for the two
different combinations of modes used. The error on $f$ is assumed to reduced by
$\sqrt{R}$ where $R$ is the ratio of integrated luminosities. The total
projected error when all modes are used is also shown. To determine the expected
error when all modes are used, a likelihood fit is used where $f$ is constrained
by a Gaussian probability likelihood term to be within the 0.043 error of the
central value of 0.229. Note that the projected errors have a weak dependence on
the value of $\Delta\Gamma_{CP}/\Gamma$ used but a strong dependence on the
value of $f$ used. It should also be noted that we are assuming that any
systematic errors are insignificant compared to the statistical errors, so that
these projected statistical errors are taken as the total error on
$\Delta\Gamma_{CP}/\Gamma$. For the range of lifetime errors we are considering
this assumption is reasonable since the charm lifetimes can be measured to this
level in fixed target experiments with only small systematics. However the
lifetime acceptance correction in BTeV may be somewhat larger for the hadronic
modes.

\begin{table}[th]
\begin{center}
\begin{tabular}{cccc} \hline\hline
\bf{Decay Modes Used}      & 
       \multicolumn{3}{c}{\bf{Error on $\Delta\Gamma_{CP}/\Gamma$}}\\ \hline
                       & 2~fb$^{-1}$ & 10~fb$^{-1}$ & 20~fb$^{-1}$ \\ \hline
$J/\psi\eta^{(\prime)}$, $D_s^-\pi^+$ & 0.0273 & 0.0135& 0.0081\\
$J/\psi\,\phi$, $D_s^-\pi^+$             & 0.0349 & 0.0158& 0.0082\\
$J/\psi\eta^{(\prime)}$, $J/\psi\,\phi$, $D_s^-\pi^+$ 
                                      & 0.0216 & 0.0095& 0.0067\\ \hline
with $\Delta\Gamma_{CP}/\Gamma=0.03$  & 0.0198 & 0.0088& 0.0062\\
with $f=0.13$                         & 0.0171 & 0.0077& 0.0054\\
with $f=0.33$                         & 0.0258 & 0.0112& 0.0078\\
\hline\hline
\end{tabular}
\end{center}
\caption[Projections for statistical errors on $\Delta\Gamma_{CP}/\Gamma$ for
combining lifetimes from different modes and for using all modes for 2, 10 and
20~fb$^{-1}$.]{Projections for statistical errors on $\Delta\Gamma_{CP}/\Gamma$
for combining lifetimes from different modes and for using all modes for 2, 10
and 20~fb$^{-1}$. The values $\Delta\Gamma_{CP}/\Gamma=0.15$ and $f=0.229$ are
used for the main results and the results for other values of
$\Delta\Gamma_{CP}/\Gamma$ and $f$ are also shown for comparison.}
\label{tb_btevdgresults}
\end{table}

The statistical error can be improved by including also the $J/\psi\rightarrow
e^+e^-$ decay mode for the $J/\psi$ reconstruction. The increase in statistics
is less than by a factor of 2 since the BTeV ECAL acceptance is smaller than the
muon detector and there is no dedicated $J/\psi\rightarrow e^+e^-$ trigger.

It should also be noted that additional measurements of the $CP$-even rate and
especially of the $CP$-odd rate, even with low statistics, can have a very
significant effect on the $\Delta\Gamma_{CP}/\Gamma$ sensitivity. Unfortunately
the $CP$-odd modes look difficult experimentally. For example, two $CP$-odd modes
are $B_s^0\rightarrow J/\psi f_0(980)$ and $B_s^0\rightarrow \chi_{c0}\phi$. 
\index{decay!$B_s \rightarrow \psi f_0$}
\index{decay!$B_s \rightarrow \chi_{c0} \phi $}
The
$f_0(980)$ is relatively broad compared to the $\phi^0$ and decays to $\pi\pi$
or $KK$ and thus will have large backgrounds. The $\chi_{c0}$ has small
branching fractions, and the dominant decays are to non-resonant states with
pions and kaons and thus will also be background challenged. However it will
still be worthwhile looking for these $CP$-odd states.

\boldmath
\subsection{Summary of Projections for $\Delta\Gamma$}
\unboldmath
\index{width difference $\dg$!measurement of $\dg_s$!summary}
\index{summary!width difference $\dg_s$ projections}

The most sensitive direct measurement of $\Delta\Gamma_{B_s}$ will be from
measuring the lifetime differences between decays to $CP$-specific final states,
({\em i.e.} to $CP$-even, $CP$-odd or $CP$-mixed modes) and flavor specific decays.

The decay modes to flavor specific final states like $D_s\pi$ will be the
most precisely measured.  Other decays with larger branching fractions that can
be reconstructed with high efficiency and good signal-to-background will be to
$CP$-mixed final states, like $J/\psi\,\phi$ and $D_s^{(\ast) +}D_s^{(\ast) -}$
involving at least one $D_s^{(\ast)}$. These decays proceed through an unknown
admixture of $CP$-even and $CP$-odd amplitudes that must be determined
experimentally via an angular analysis.  The error on $\Delta\Gamma_s/\Gamma_s$
obtained using these modes will be very sensitive to the actual fractions of
$CP$-even and $CP$-odd, where the sensitivity is poorest for equal mixtures of
$CP$-even and $CP$-odd.

The decays to purely $CP$-even or purely $CP$-odd final states are difficult
experimentally, either because the backgrounds are larger and/or the branching
fractions are small ({\em e.g.} for $D_s^+D_s^-$, $J/\psi K_s$ and $J/\psi
f_0(980)$), or they contain difficult to reconstruct neutrals in the final state
({\em e.g.} like in $J/\psi\eta^{(\prime)}$). However it is important to try to
use these decay modes as even with small samples of events they improve the
error on the $\Delta\Gamma_s/\Gamma_s$ measurement significantly.

With 2~fb$^{-1}$ CDF should be able to determine $\Delta\Gamma_s/\Gamma_s$ with
a statistical error of 0.04 through lifetime measurements, improving to as good
as 0.025 if the decay to $D_s^{\ast +}D_s^{\ast -}$ proceeds through a 100\%\ 
$CP$-even amplitude. CDF can reach a statistical error of 0.01 on a {\em
model dependent} determination of $\Delta\Gamma_s/\Gamma_s$ using just
branching ratio measurements.

With a vertex (hadronic) trigger at Level-1 and excellent particle
identification and neutral reconstruction, the BTeV experiment should be able to
measure the lifetimes of the purely $CP$-even or purely $CP$-odd final states well
enough to give a model independent determination of $\Delta\Gamma_s/\Gamma_s$
with a statistical error of smaller than 0.02 with 2~fb$^{-1}$ of data. This
error is further decreased as low as 0.01 if the decay to $D_s^{\ast +}D_s^{\ast
  -}$ proceeds through a 100\%\ $CP$-even amplitude. With 20~fb$^{-1}$ of data a
statistical error of 0.005 on $\Delta\Gamma_s/\Gamma_s$ should be obtainable.
Systematic uncertainties are expected to be under control to a similar level.

\section{Projections for Lifetimes}

\boldmath
\subsection[$b$ Hadron Lifetimes at CDF]
{$b$ Hadron Lifetimes at CDF
$\!$\authorfootnote{Authors: Ch.~Paus, J.~Tseng.}}
\unboldmath
\index{lifetime!at CDF}

Lifetime measurements have formed an important part of the CDF research program
in $b$ physics since the 1992 introduction of the silicon vertex detector with
its precision tracking capabilities. This part of the research program has been
very successful, producing some of the most precise lifetime measurements with
semi-inclusive data samples, but also the most precise measurements with
exclusive channels. It is expected that the combination of the Tevatron's large
$\overline{b}b$ cross section and precise tracking capabilities will continue to
reap benefits in Run~II, pushing the comparison between experiment and
theoretical calculation to more stringent levels.

A lifetime measurement consists of reconstructing the decay point of a $b$
hadron by tracing back its long-lived charged descendants.  For instance, the
decay $\overline{B}^0\to e^-\overline{\nu}_eD^+X$ 
\index{decay!$B_d \rightarrow D^- \ell^+ \nu$}
is reconstructed by
intersecting the $e^-$ track with the trajectory of the $D^+$ meson; this
trajectory in turn is reconstructed from its own daughters, such as in the decay
$D^+\to K^-\pi^+\pi^-$.  The distance between the primary interaction
vertex and the $b$ decay vertex gives the flight distance of the $b$ hadron. In
Run~I, this flight distance was measured most precisely in the plane transverse
to the beam and the proper time of decay, $ct$, calculated with the formula
\begin{equation}
  ct = \frac{L_{xy}\,m}{p_T} \,,
\end{equation}
where $L_{xy}$ is the transverse flight distance, $m$ is the mass of the $b$
hadron, and $p_T$ is the momentum of the $b$ hadron projected in the with
respect to the beam direction transverse plane. The transverse momentum is
calculated by combining the measured momenta of the charged daughter particles.
Unlike the flight distance, this combination requires either the identification
and reconstruction of all the daughter particles, as in the analyses of an
exclusive hadronic channel, or a correction factor for the particles that are
not reconstructed, as in the analyses of much larger semi-inclusive samples. A
third possibility, applicable in some special circumstances, allows a constraint
to be applied to the momentum of a single particle of known mass; this technique
was not applied in Run~I but may become more important in Run~II with the advent
of three-dimensional silicon tracking.

As opposed to the electron--positron machines at the $\Upsilon(4S)$ where only
$B^+$ and $B^0_d$ mesons are produced, Tevatron produces the full spectrum of
$b$ hadrons. Due to this uniqueness in the following particular emphasize is put
on lifetime measurements of $B^0_s$, $B^0_c$ and $\Lambda_b$.

\subsubsection{Run~I Results at CDF}

The Run~I CDF lifetime results are summarized in
Figure~\ref{fig:cdf-blifetimes}.  They represent a combination of several
analyses of different types but have one thing in common: all data samples have
been obtained by using the trigger on at least one high-momentum lepton
candidate.

\begin{figure}[t]
\begin{center}
  \mbox{\epsfig{file=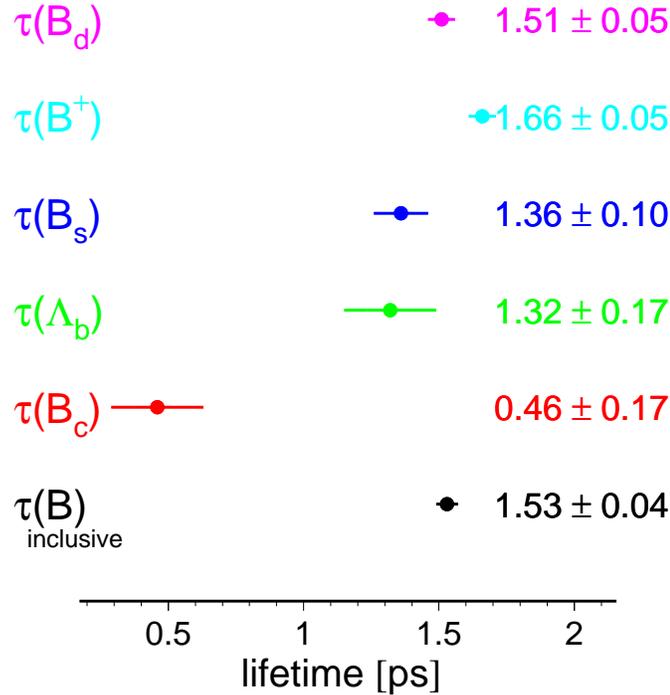,width=0.6\textwidth}}
  \caption{Summary of lifetime measurements from CDF during Run~I.}
  \label{fig:cdf-blifetimes}
\end{center}
\end{figure}

Three different types of analyses are performed. There are the exclusive, the
semi-exclusive and finally the inclusive analyses. While in the exclusive
analyses one or more distinct $b$ hadron decay channels are fully reconstructed,
in the semi-inclusive analyses some neutral particle cannot be reconstructed,
most typically the neutrino from the semileptonic $b$ hadron decay. In the
inclusive analyses the secondary vertex indicates the presence of a $b$ hadron
but no attempt is made to reconstruct the mass explicitly.

The features of the different analysis types are complementary. Exclusive
analyses usually have small data samples and thus a large statistical
uncertainty but the systematic uncertainty is small. Inclusive analyses have
usually very large data samples and thus small statistical uncertainties but the
systematic errors are large.

The exclusive modes measured also used the $J/\psi$ trigger sample: in Run~I,
these analyses yield $436\pm 27$ decays of the type $B^0 \to \psi(1S,2S)
K^{(*)0}$, $824\pm 36$ of the type $B^+ \to \psi(1S,2S) K^{(*)+}$, and $58 \pm
12$ of the mode $B_s \to J/\psi\,\phi$.  \index{decay!$B_s \rightarrow \psi \phi$}
As will be detailed in the following sections, the $B_s$ lifetime thus measured
is expected to mostly reflect the lifetime of the shorter-lived weak eigenstate
of the $B_s$.  A small sample of $\Lambda_b \to J/\psi\Lambda^0$ decays was also
reconstructed but was too small for lifetime analysis in light of the large $B^0
\to J/\psi K^0_S$ backgrounds. The systematic errors are due to background and
resolution modeling and detector alignment.

For the semi-exclusive modes the $B^0$, $B^+$, $B_s$, and $\Lambda_b$ hadrons
are measured using their semileptonic decays. The charm daughters $D^{(*)}$,
$D_s$, and $\Lambda_c^+$ are reconstructed in data. Since these decays contain
neutrinos as well as possible unreconstructed intermediate states such as the
$D^{**}$, they are subject to systematic uncertainties due to production and
decay modeling, as well as uncertainties in background and resolution and
detector alignment.

An inclusive $b \to J/\psi X$ lifetime is also measured, where the
$J/\psi \to \mu^+\mu^-$ is registered on a dilepton trigger, as well as
the lifetime of the $B_c$ meson through its decays to $J/\psi\ell X$.

The lifetime measurements have also yielded measurements of the lifetime ratio
between charged and neutral $B$ mesons. The measurement of lifetime ratios 
\index{lifetime!ratios}
is
particularly interesting since experimental uncertainties and theoretical
uncertainties cancel out and thus allow more sensitive tests of some aspects of
the theory of heavy quarks. Combining the exclusive and the semileptonic decay
modes yields a ratio of $1.09\pm 0.05$ in agreement with the world average
$1.07\pm 0.02$ and both well within the range of the theoretical prediction
$1.0-1.1$.

The lifetime ratio between $B_s$ and $B^0$ has not yet been measured with
sufficient precision to test the prediction which is within $\approx 1$ percent
of unity.  The experimental value of the lifetime ratio of the $\Lambda_b$ to
the $B^0$ is of similar precision, but lies well below the current theoretical
prediction range of $0.9 - 1.0$. More data is needed to clarify whether there is
a discrepancy.  From the theoretical point of view baryons are more difficult to
calculate and thus the experimental data will hopefully shed some light on this
area.

\subsubsection{Run~II Projections for CDF}

To derive the expected lifetime uncertainties for the Run~II data samples it is
assumed that the Run~I measurements are statistically limited. This is certainly
true for the exclusive decay modes. For the other decay channels it is less
clear although experience shows that most systematic errors can be improved when
more statistics is available.  Therefore in the following only the exclusive
measurements are used to estimate the uncertainties on the $b$ hadron lifetime
measurements achievable with the Run~II data samples. This is a conservative
procedure and it is likely that CDF will do better. The increase in statistics
for the exclusive decays involving $J/\psi\to\mu^+\mu^-$ with respect to Run~I
is obtained applying simple scale factors as summarized in
section~\ref{sec:cdf-sin2beta}.

\paragraph{Leptonic Triggers}
\index{lifetime!at CDF!leptonic triggers}
Considering the lifetime measurement capabilities of CDF in Run~II, it is
useful as a first step to make a direct extrapolation from the Run~I lifetime
analyses of exclusive channels. These are shown in Table~\ref{fig:cdf-ltproj},
assuming 2~{\ifb} of $J/\psi$ dimuon triggers with increased muon coverage,
lower muon trigger thresholds, and increased silicon tracking cover. The
projected uncertainties are only statistical, and, given Run~I experience, are
likely to be comparable to the levels of systematic uncertainty.  Other
improvements, not accounted for in the table, include the possibility of a
di-electron trigger, which adds approximately 50\% more data, and the effect of
smaller-radius silicon as well as three-dimensional microstrip tracking, which
improves $S/N$ and hence the lifetime measurements.

\begin{table}
\begin{center}
\begin{tabular}{  l  c c c  } \hline\hline
            &       Run 1 &       Run 2 & projected          \\
Species     & sample size & sample size & $c\tau$ error [ps] \\
\hline
$B^-$       &         824 &       40000 &               0.01 \\
$B^0$       &         436 &       20000 &               0.01 \\
$B_s$       &          58 &        3000 &               0.03 \\
$\Lambda_b$ &          38 &        2000 &               0.04 \\
\hline\hline
\end{tabular}
\end{center}
\caption{Run~II projections for the Run~I exclusive lifetime measurements at
CDF corresponding to 2~{\ifb}.}
\label{fig:cdf-ltproj}
\end{table}

%
%
%
%

It is evident from the table that already the projections of only the exclusive
decay modes as measured by CDF in Run~I improves on the current world-averaged
$B^-/B^0$ lifetime ratio and combined with other experiments, including the $B$
factories, this ratio will be very precisely known.

In the case of the $B_s$, however, since the $J/\psi\,\phi$ final state is mostly
sensitive to one weak eigenstate, it is useful primarily in measuring
$\Delta\Gamma_s$ rather than the average $\Gamma_s$ which is to be compared with
$\Gamma_d$.  The $\Lambda_b \to J/\psi\Lambda^0$ 
\index{decay!$\Lambda_b \rightarrow J\psi\Lambda$}
lifetime is expected to
be known within 0.04~ps --- significantly better than the current world average
--- but depends upon being able to more effectively distinguish the signal from
$B^0 \to J/\psi K^0_S$ decays, which are topologically very similar.

Comparable projections for $B_c$ and $\Xi_b$ exclusive lifetime analyses cannot
be made, since their exclusive decays were not observed in Run~I data.

\paragraph{Hadronic Triggers}
\index{lifetime!at CDF!hadronic triggers}
Beyond the $J/\psi \to \ell^+\ell^-$ trigger samples, there are expected
to be large data samples that will be made available by the hadronic
displaced-track trigger.  Although those samples will also improve on the $B^-$
and $B^0$ lifetimes most interestingly they will improve on the $B_s$ and
$\Lambda_b$ lifetime measurements. Since the hadronic trigger is a new hardware
device all predictions are less certain than the predictions for decay modes
originating from the leptonic triggers.

Most branching fractions for the decay modes used below are not measured and
have to be estimated. This is particularly difficult for the $\Lambda_b$ decay
modes.  Errors of 50 percent for the $B_s$ decays and 100 percent for the
$\Lambda_b$ decays are assumed.

Another principle difficulty in these data samples lies in the understanding of
the effect of the trigger on the lifetime distributions. The displaced vertex
trigger prefers large lifetime events and introduces a bias in the proper time
distribution. In the following it is assumed that it is possible to model the
trigger bias and use all of the statistical power of the projected yields.

%
%
For the $B_s$ decay modes $D_s \pi$ and $D_s \pi\pi\pi$ 
\index{decay!$B_s \rightarrow D_s^- \pi^+$}
\index{decay!$B_s \rightarrow D_s^- \pi^+ \pi^+ \pi^-$}
there are approximately
75k events projected. This includes the $D_s$ decay modes to $\phi \pi$ and
$K^*K$ only, as indicated in Table~\ref{tab:hadrtrig-bs}. Further it is expected
that the signal to background ratios is one, which is rather conservative. This
results in an uncertainty of the $B_s$ lifetime of approximately 0.007~ps.

For the lifetime ratio of $B_s$ and $B^0$ this corresponds to an error of
roughly half a percent and is thus in the same order as the theoretical
prediction for the deviation from unity.

%
%
For the $\Lambda_b$ decay modes $\Lambda_c \pi (\Lambda_c\to pK\pi)$, $pD^0\pi
(D^0 \to K \pi$ and $D^0 \to K \pi\pi\pi)$, 
\index{decay!$\Lambda_b \rightarrow \Lambda_c^+\pi^-$}
\index{decay!$\Lambda_b \rightarrow pD^0\pi^-$}
$p\pi$ and $pK$ there are
approximately 24.4k events projected. Again assuming a signal to background
ratio of one, a statistical uncertainty of 0.01~ps is obtained. This is more
precise than the expectation from the $J/\psi\Lambda^0$ decay mode. A stringent
test will be available for the theoretical predictions of the lifetime ratio of
$\Lambda_b$ to $B^0$.

More inclusive strategies, using triggerable combinations of leptons and
displaced tracks, will also significantly increase the precision of the lifetime
ratios of the rarer $b$ hadrons, such as the $B_c$ and so far by CDF not
observed baryons as $\Xi_b$ might be accessible. These strategies have yet to be
investigated in detail.

\subsection[Lifetime measurements at \dzero]
{Lifetime measurements at \dzero
$\!$\authorfootnote{Author: W.~Taylor.}}

\index{lifetime!at \dzero}

A rich spectroscopy and lifetime measurement program is planned for both beauty
mesons and baryons. We will study any species that has significant decay modes
that result in at least one lepton.  One of the highlights of this program is a
measurement of the $\Lambda_b$ lifetime in the exclusive decay mode $\Lambda_b
\rightarrow J/\psi + \Lambda$. 
This is particularly interesting since the
measured ratio $\tau(\Lambda_b) / \tau(B_d) = 0.78 \pm 0.04$ \cite{Ref:LAMBLIFE}
is significantly different from the naive spectator model prediction of unity.
Current theoretical understanding of non-spectator processes such as final-state
quark interference and $W$ boson exchange cannot account for such a large
deviation.  Another important measurement which we will make is a measurement of
the $B_s$ lifetime using modes such as $B_s\rightarrow J/\psi \phi$.

Measurements of the $\Lambda_b^0$ baryon lifetime have long been hampered by a
dearth of statistics.  For this reason, the lifetime was measured in the
semileptonic decay mode, where the branching fraction is orders of magnitude
higher than for any fully reconstructed
mode~\cite{Ref:ALEPH,Ref:CDFI,Ref:DELPHI,Ref:OPAL}.  The disadvantage of the
semileptonic mode is that the neutrino information is lost.  To obtain the true
decay length in the absence of neutrino information, the ratio $p_T(\Lambda_c
\ell)/p_T(\Lambda_b^0)$ (called the $K$ factor) was obtained from a Monte Carlo
calculation and used in the lifetime fit.  The limited knowledge of the value of
the $K$ factor represented a modest contribution to the uncertainty of this
measurement.

In Run~II at D\O\, we can expect to collect 2~fb$^{-1}$ of $p\overline{p}$
collisions or twenty times the statistics of Run~1. This provides the
opportunity to probe the $\Lambda_b^0$ lifetime in a fully reconstructed mode:
$\Lambda_b^0 \rightarrow J/\psi \Lambda^0$, 
\index{decay!$\Lambda_b \rightarrow J\psi\Lambda$}
where the $J/\psi$ meson decays into
two leptons (muon or electron) and the $\Lambda^0$ baryon decays into a proton
and a pion.  Using a fully reconstructed channel avoids the introduction of the
$K$ factor and any uncertainty associated with it.

To study this mode, we generated 75~000 $\Lambda_b^0$ baryon events using
Pythia, simulated the D\O\ Run~II detector using MCFast, and forced the decay
mode $\Lambda_b^0 \rightarrow J/\psi \Lambda^0$, followed by $J/\psi \rightarrow
\mu^+\mu^-$ and $\Lambda^0\rightarrow p \pi^-$ with QQ, the CLEO Monte Carlo
program.  We use this sample to predict the trigger and reconstruction
acceptances and to estimate the lifetime resolution.  For the purpose of this
study, we assume that electrons from $J/\psi$ decays will be reconstructed with
a similar efficiency to that of the muons, an assumption which is expected to be
approximately true.

The D\O\ $J/\psi$ trigger%
\index{\dzero!$J/\psi$ trigger}%
identifies events resembling $J/\psi$ decays.  The presence of two muon tracks,
each with $|\eta|\leq2.0$ and $p_T \geq 1.5$~GeV/$c$, is sufficient for passing
the $J/\psi$ trigger criteria.  The trigger acceptance is found to be 16.8\% for
the muons in these generated Monte Carlo events.  We introduce by hand an
additional acceptance cut of 55\% for the holes in the muon detection system not
modeled in MCFast.

Track quality cuts are applied to all the tracks in the events passing the
trigger.  The muon tracks from the $J/\psi$ decay are required to have at least
four stereo hits in the tracking system (silicon detector and central fiber
tracker combined) and at least eight hits (stereo and axial combined) in the
silicon detector.  For tracks in the central region ($|\eta|\leq1.7$), at least
fourteen hits are required in the central fiber tracker.  The transverse
momentum is required to be above 400~MeV/$c$.

The $\Lambda^0$ baryon daughters are required to have at least three stereo hits
in the tracking system (silicon detector and central fiber tracker combined) and
a minimum of eight hits in the central fiber tracker.  The transverse momentum
is required to be above 400~MeV/$c$. The long lifetime of the $\Lambda^0$ baryon
prevents the use of strict silicon hit requirements, as the $\Lambda^0$ baryon
often decays outside of the fiducial volume of the silicon detector.

The muons are constrained to come from a common vertex in three dimensions as
are the $\Lambda^0$ daughters.  Finally, the location of the $\Lambda_b^0$ decay
vertex is obtained by extrapolating the $\Lambda^0$ baryon momentum direction in
three dimensions back to the $J/\psi$ decay vertex.  The resulting vertex
defines the decay length of the $\Lambda_b^0$ baryon.

The acceptance for these selection criteria is 9.6\%.  We introduce by hand an
additional acceptance cut of (0.95)$^4$=0.81 for the tracking efficiency not
modeled in MCFast.  The final acceptance times efficiency for the trigger and
reconstruction criteria is 0.72\%.  We therefore predict about 15~000 events to
be reconstructed in 2~fb$^{-1}$. Table~\ref{tab:yield} shows the values used to
obtain this prediction for the yield.

\begin{table}
\centerline{
\begin{tabular}{lr} \hline\hline
  \multicolumn{1}{c}{Multiplier} & \multicolumn{1}{c}{Events} \\
  \hline
  &\\[-1em]
  ${\cal L} = 10^{32}$ cm$^{-2}$s$^{-1}$ =  $10^{36}$ m$^{-2}$s$^{-1}$ & \\
  $t=2\times10^7$ s                                                    & \\
  $\sigma_{b\overline{b}}$ = 158 $\mu$b $\times 10^{-28}$ m$^2$/b      & \\
  $b,\overline{b}$ = 2                               & $6.32\times10^{11}$ \\
  ${\cal B}(b\rightarrow \Lambda_b^0)=0.090$         & $5.7\times10^{10}$ \\
  ${\cal B}(\Lambda_b^0\to J/\psi\Lambda^0)=4.7\times10^{-4}$
                                                     & $2.7\times10^{7}$ \\
 ${\cal B}(J/\psi\rightarrow \mu\mu,ee)=2\times0.06$ & $3.2\times10^6$ \\
  ${\cal B}(\Lambda^0 \rightarrow p \pi)=0.639$      & $2.0\times10^6$ \\
  Detector acceptance = 0.55                         & $1.1\times10^6$ \\
  $\epsilon$(trigger) = 0.168                        & $1.9\times10^5$ \\
  $\epsilon$(track efficiency) = 0.95$^4$            & $1.5\times10^5$ \\
  $\epsilon$(reconstruction) = 0.096                 & 14~850 \\
  \hline\hline
\end{tabular}}
\vspace*{6pt}
\caption{Event yield determination after 1 year or 2~fb$^{-1}$.}
\label{tab:yield}
\end{table}

The mass distributions for the $J/\psi$ and $\Lambda^0$ particles are shown in
Figures~\ref{8fig:jpsimass} and \ref{fig:lambdamass}, respectively.  With the
application of constraints on both the $J/\psi$ mass and the $\Lambda^0$ mass,
we predict the $\Lambda_b^0$ mass resolution to be 16~MeV/$c^2$, as indicated in
Figure~\ref{fig:lambdabmass}.  The lifetime resolution is found to be 0.11~ps
(Figure~\ref{fig:tausig}).

\subsection{Summary of Projections for Lifetimes}
\index{summary!lifetime projections}
\index{lifetime!experimental lifetime}

The exclusive $b$ hadron event samples involving $J/\psi\to\mu^+\mu^-$ alone
will enable CDF and D{\O} during Run~II to push the precision of lifetime
measurements to values close to 1~fs. Further inclusion of other hadronic decay
modes will decrease the statistical error further below 1~fs.

In particular the lifetime measurements of $B_s$ and $\Lambda_b$ which cannot be
measured by the $B$ factories will achieve a statistical precision of about
0.007~ps and 0.01~ps, respectively. Those measurements determine the ratio of
$\tau(B_s)/\tau(B^0)$ to roughly half a percent which allows a first test of the
theoretical predictions.

Lifetimes of other interesting $b$ hadrons like the $B_c$ will be improved
significantly, and there is a good chance to observe $\Xi_b$ for the first time
and measure its lifetime.

\begin{figure}[t]
\centerline{\psfig{figure=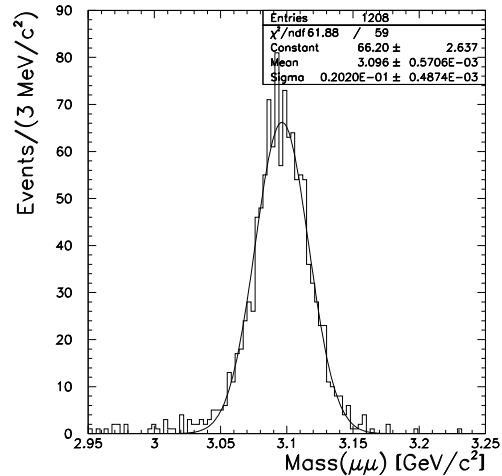,width=6.5cm}}
\caption{\label{8fig:jpsimass} $J/\psi$ mass distribution.}
\end{figure}

\begin{figure}
\centerline{\psfig{figure=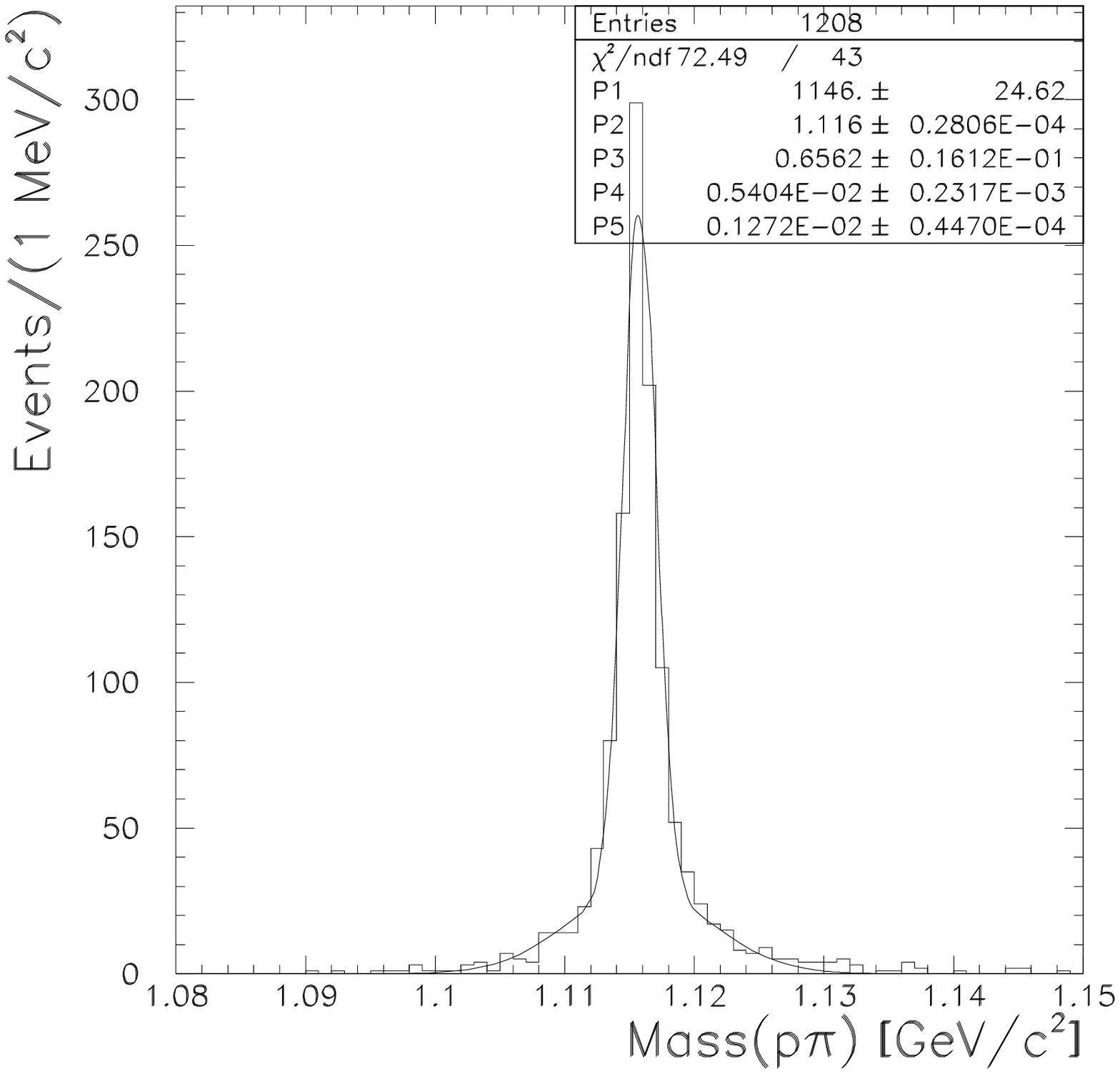,width=6.5cm}}
\caption{\label{fig:lambdamass}$\Lambda^0$ mass distribution.}
\end{figure}

\clearpage

\begin{figure}[thb]
\centerline{\psfig{figure=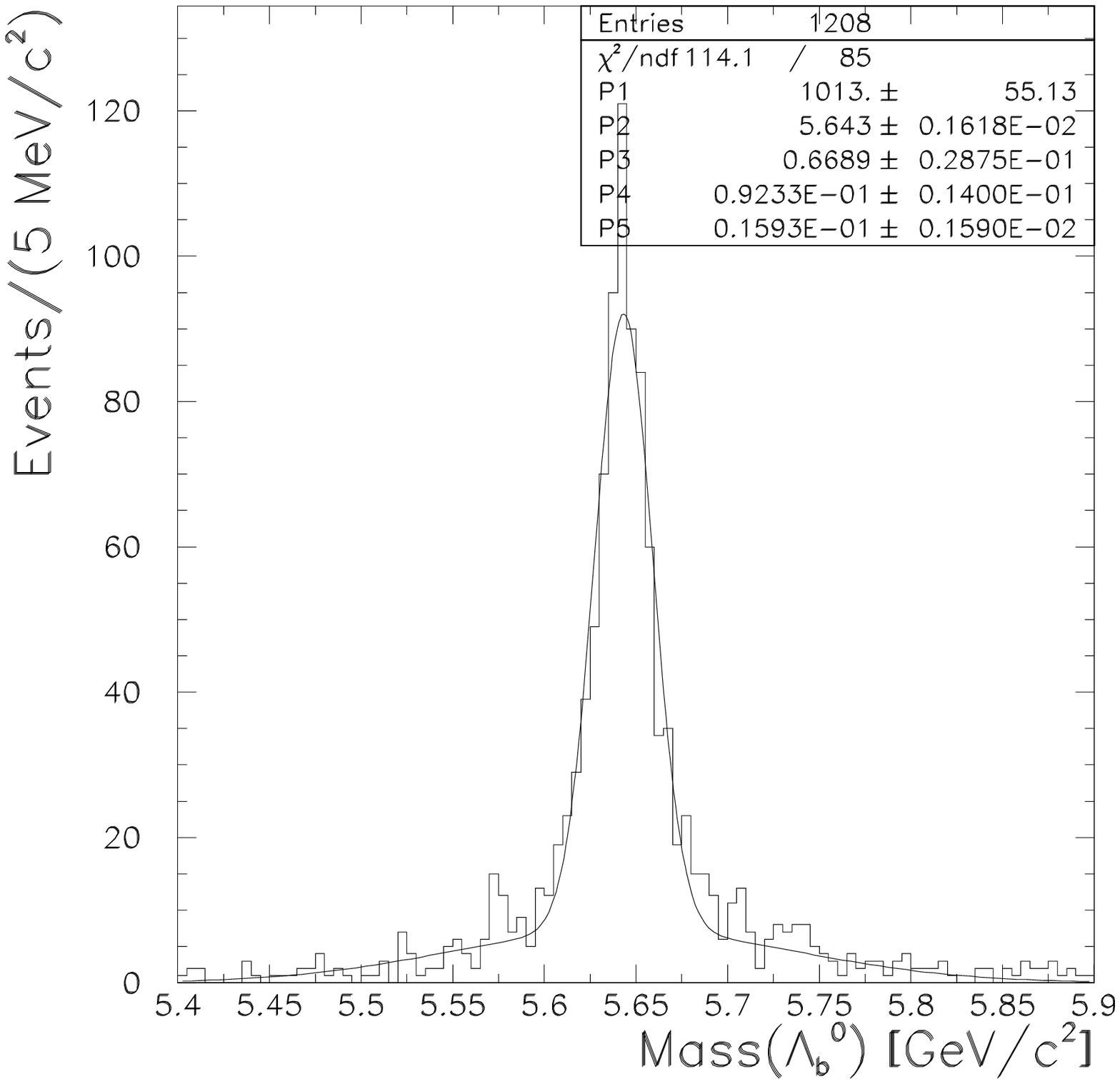,width=6.5cm}}
\caption{\label{fig:lambdabmass}$\Lambda_b^0$ mass distribution.}
\end{figure}

\begin{figure}[thb]
\centerline{\psfig{figure=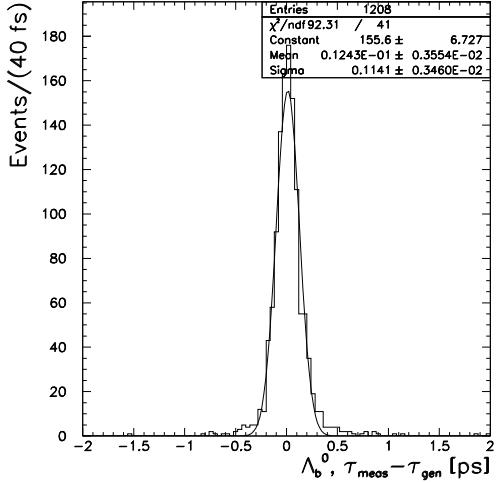,width=6.5cm}}
\caption{\label{fig:tausig}$\tau(\Lambda_b^0)_{meas}-\tau(\Lambda_b^0)_{gen}$.}
\end{figure}


\section{Conclusions}

The Tevatron provides a unique testing ground for mixing and lifetime studies.
For these measurements it is superior to the $B$-factories, because the large
boosts of the produced hadrons and the higher statistics allow to study the
decay distributions more precisely.  Moreover, studies of $B_s$ mesons and
$b$-flavored baryons are not possible at current $B$-factories running at the
$\Upsilon(4S)$ resonance. At Run~II \bbms\ will be discovered and the mass
difference $\dm_s$ will be determined very precisely.  This measurement is of
key importance for the phenomenology of the unitarity triangle.  Our knowledge
of the lifetime pattern of $b$-flavored hadrons will significantly improve, the
yet undetected width difference $\dg_s$ between the $B_s$ mass eigenstates is
within reach of Run~II and the $\Lambda_b$ lifetime puzzle will be addressed.



\clearpage{\pagestyle{empty}\cleardoublepage}
\def\chninehome{ch9/}

\chapter[Production, Fragmentation and Spectroscopy]
{Production, Fragmentation and Spectroscopy}

\def\et{$E_T$} 
\def\ptrel{$P_T^{rel}$}
\def\bbar{$\overline{b}$}
\def\bbbar{$b\overline{b}$}
\def\into{\rightarrow}

\authors{
G.~Bodwin, 
E.~Braaten, 
K.~Byrum, 
R.K.~Ellis, 
G.~Feild, 
S.~Fleming, 
W.~Hao,
B.~Harris, 
V.V.~Kiselev, 
E.~Laenen, 
J.~Lee, 
A.~Leibovich,
A.~Likhoded, 
A.~Maciel, 
S.~Menary,
P.~Nason, 
E.~Norrbin, 
C.~Oleari,
V.~Papadimitriou
G.~Ridolfi, 
K.~Sumorok, 
W.~Trischuk,
R.~Van Kooten
}
\label{ProductionFragmentationandSpectroscopy}

The Tevatron is distinguished from the other current facilities
involved in the study of particles containing heavy quarks by the 
size of the cross sections and the range of hadrons 
containing heavy quarks 
which are produced. The production of heavy quarks at the Tevatron 
and their subsequent fragmentation to a wide range of states is the 
subject of the present chapter. These processes offer a unique chance
to study the dynamics of heavy quark systems in a region where 
perturbative methods can be applied.  

\section{Open beauty production}

We outline the state of the art in theoretical calculations of
$B$ hadron production at a hadron collider. We refer to this as open
beauty production to distinguish it from quarkonium production 
considered in Section~\ref{ch9:quarkonium}. 
We follow up with some considerations of the experimental 
difficulties involved in
measuring open beauty production at the Tevatron. We do not 
draw any firm conclusion beyond the fact that theory and experiment
disagree by a factor of about 3. The discrepancy favors the experiments
by providing three times as many $b$ quarks as theory would expect. 

\subsection[The theory of $b$-quark production]
{The theory of b-quark production\authorfootnote{Author: R.K. Ellis}}
\index{b-quark production@$b$-quark production!theory}
\index{production!$b$-quark}
The cross section for the production of a $b$ quark is calculable in 
perturbative QCD inasmuch as the heavy quark mass, $m$, is larger than
$\Lambda$, the scale of the strong interactions. The results of the 
calculations are described in refs.~\cite{Nason:1988xz,Beenakker:1989bq,
Nason:1989zy,Beenakker:1991ma,Mangano:1992jk}.

The cross section 
\index{QCD improved parton model}
in the QCD improved parton model is
\begin{equation}
\sigma(S)= \sum_{i,j} \int \; \frac{dx_1}{x_1} \frac{dx_2}{x_2} \; 
\hat{\sigma}_{ij}(x_1 x_2 S, m^2, \mu^2) F_i(x_1,\mu^2) F_j(x_2,\mu^2)\; ,
\label{chnine:ppbarcross}
\end{equation}
where the $F_i$ are the 
momentum densities of the partons in the incoming hadrons. The 
quantity $\hat{\sigma}_{ij}$ is the short distance cross section
\index{short distance cross section}
\begin{equation}
\hat{\sigma}_{ij}(\hat{s},m^2,\mu^2)= \sigma_{0} \; 
c_{ij}(\hat{\rho},\mu^2)\; ,
\end{equation}
where $\sigma_0=\as^2(\mu^2)/m^2$, $\hat{\rho}=4 m^2/\hat{s}$
and $\hat{s}=x_1 x_2 S$ is 
the parton total center-of-mass energy squared. 
The function $c_{ij}$ has a perturbative expansion,
\begin{equation}
c_{ij}(\hat{\rho},\mu^2/m^2) = c_{ij}^{(0)}(\hat{\rho}) 
+ 4 \pi \as \Big[c_{ij}^{(1)}(\hat{\rho})  
+ \ln(\frac{\mu^2}{m^2}) \bar{c}_{ij}^{(1)}(\hat{\rho}) \Big] +O(\as^2)\; .
\end{equation}
The lowest order short distance cross section is calculated from the 
diagrams in Fig.~\ref{hvyqrk}.
\begin{figure}[th]
\vspace{9.5cm}
\includegraphics{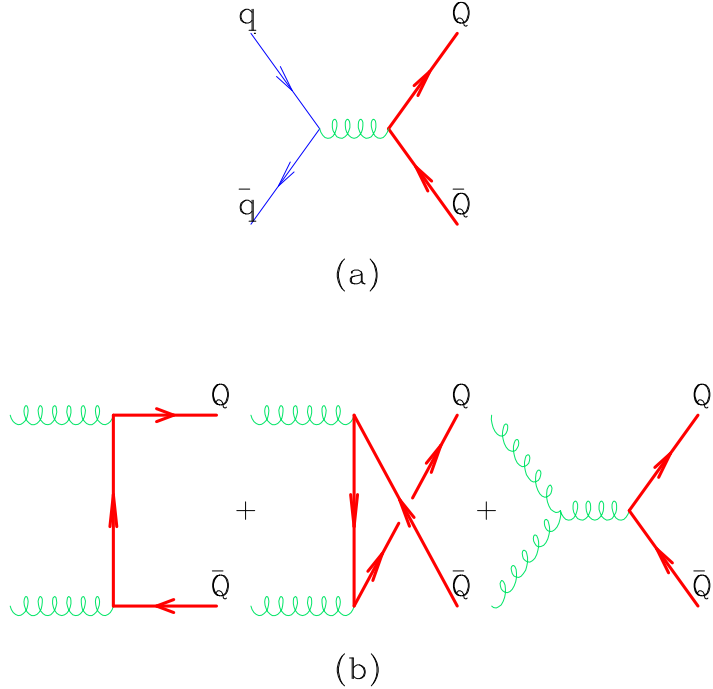}
\caption{Diagrams for heavy quark production at lowest order.}
\label{hvyqrk}
\end{figure}
The results for the lowest order total cross sections derived from these
diagrams are 
\begin{eqnarray}
c_{g g}^{(0)}(\rho) &=&  \frac{\pi \beta \rho}{12}   \Bigg\{
\frac{3}{N} 
\Big[ (1+\rho-\frac{\rho^2}{2} ) L(\beta) -1 -\rho \Big]   
+ \frac{N}{2 V} 
\Big[ 3 \rho^2 L(\beta) -4 -5 \rho \Big] 
\Bigg\}\; ,
\nonumber \\
c_{q \bar{q}}^{(0)}(\rho) &=&  
\frac{V}{N^2} \frac{\pi \beta \rho}{24} [2+\rho] \; ,
\end{eqnarray}
where $V=N^2-1=8,N=3$ and $\beta=\sqrt{1-\rho}, \rho=4 m^2/\hat{s}$ and
\begin{equation}
L(\beta) = \frac{1}{\beta} \log\Big[ \frac{1+\beta}{1-\beta} \Big]\; .
\end{equation}
The quantities $\bar{c}_{ij}^{(1)}$ and $c_{ij}^{(1)}$ are also known
\cite{Nason:1988xz}, 
the former analytically and the latter as a numerical fit. 
The results for these functions as a function of the incoming $\hat{s}$
are shown in Figs.~\ref{chnine:gg},~\ref{chnine:qq} and \ref{chnine:gq}.
\begin{figure}[th]
\vspace{9.5cm}
\includegraphics{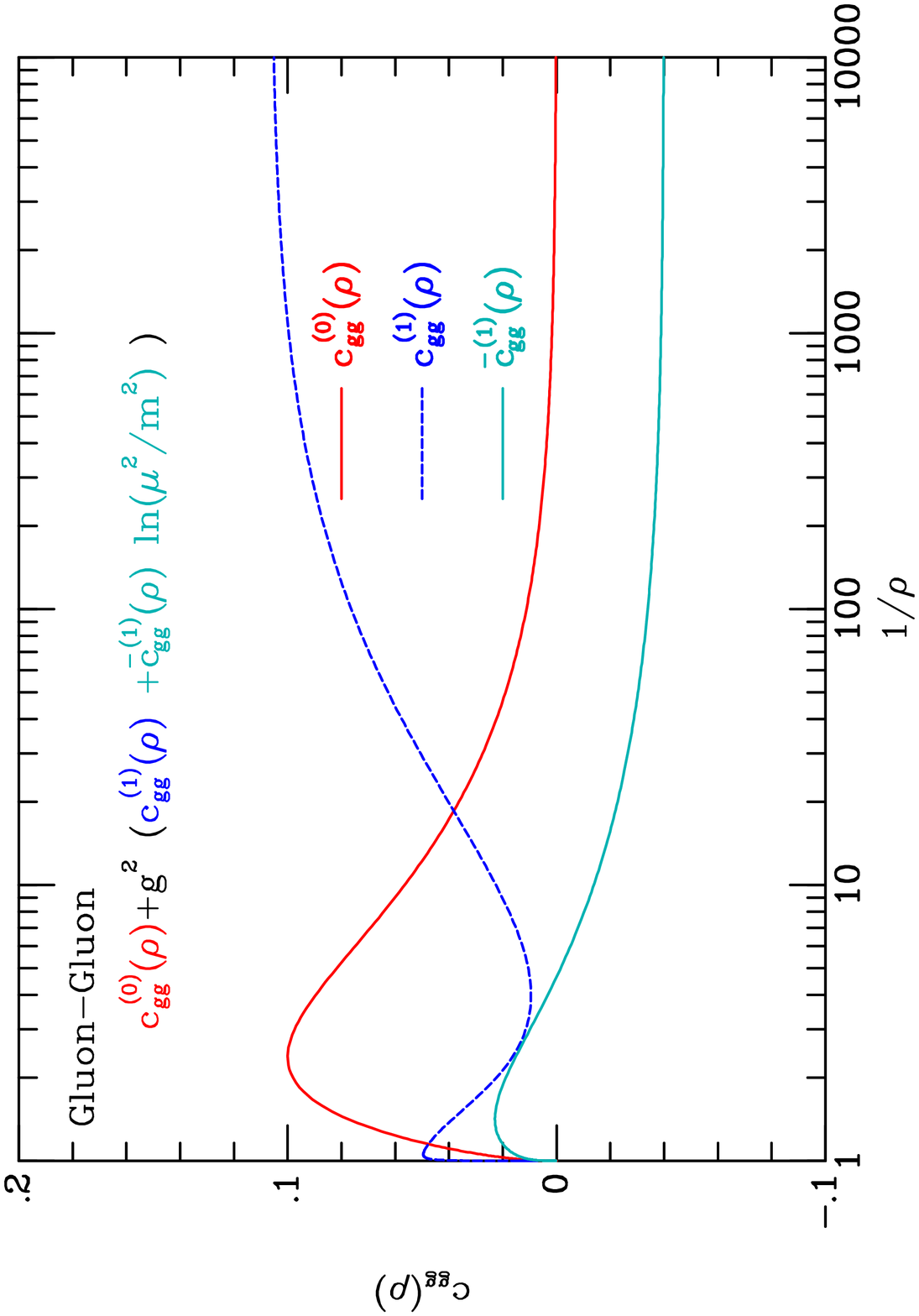}
\caption{The parton level cross section for the $gg$ process.}
\label{chnine:gg}
\end{figure}
\begin{figure}[th]
\vspace{9.5cm}
\includegraphics{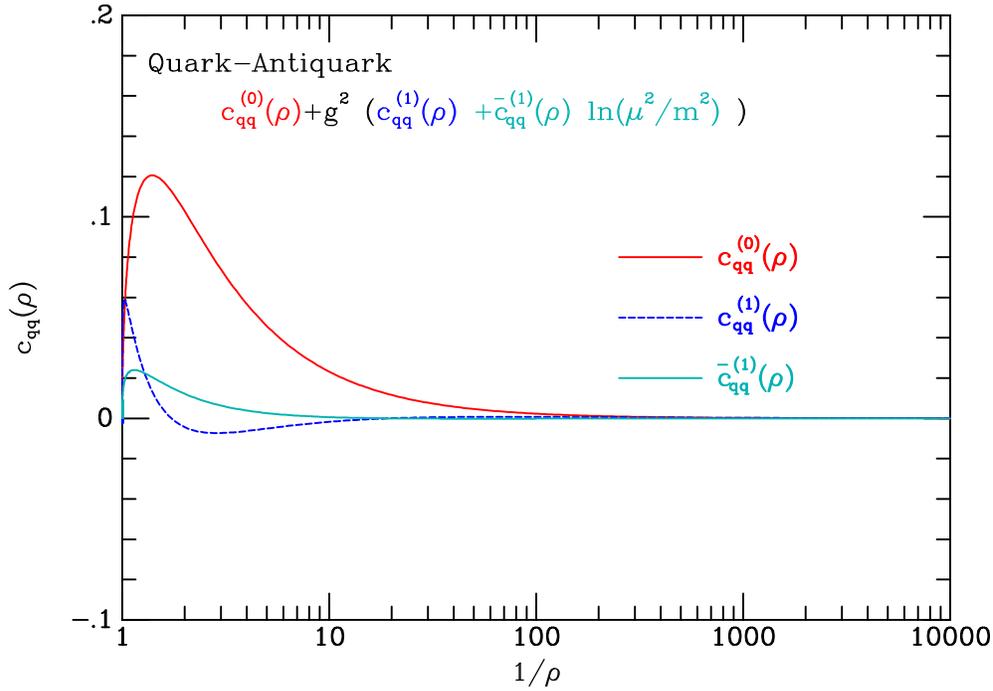}
\caption{The parton level cross section for the $q\bar{q}$ process.}
\label{chnine:qq}
\end{figure}
\begin{figure}[th]
\vspace{9.5cm}
\includegraphics{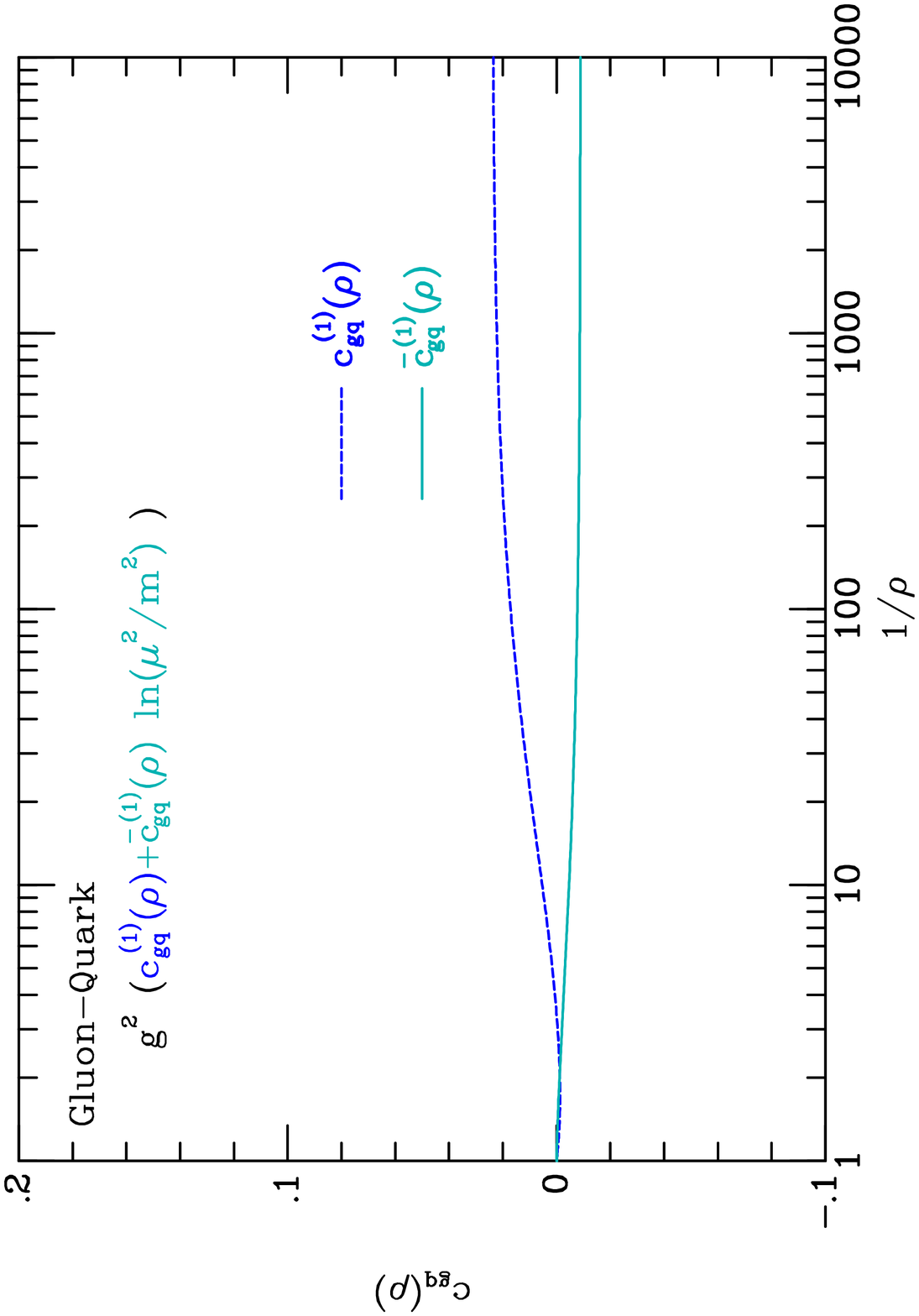}
\caption{The parton level cross section for the $gq$ process.}
\label{chnine:gq}
\end{figure}
\begin{figure}[th]
\vspace{9.5cm}
\includegraphics{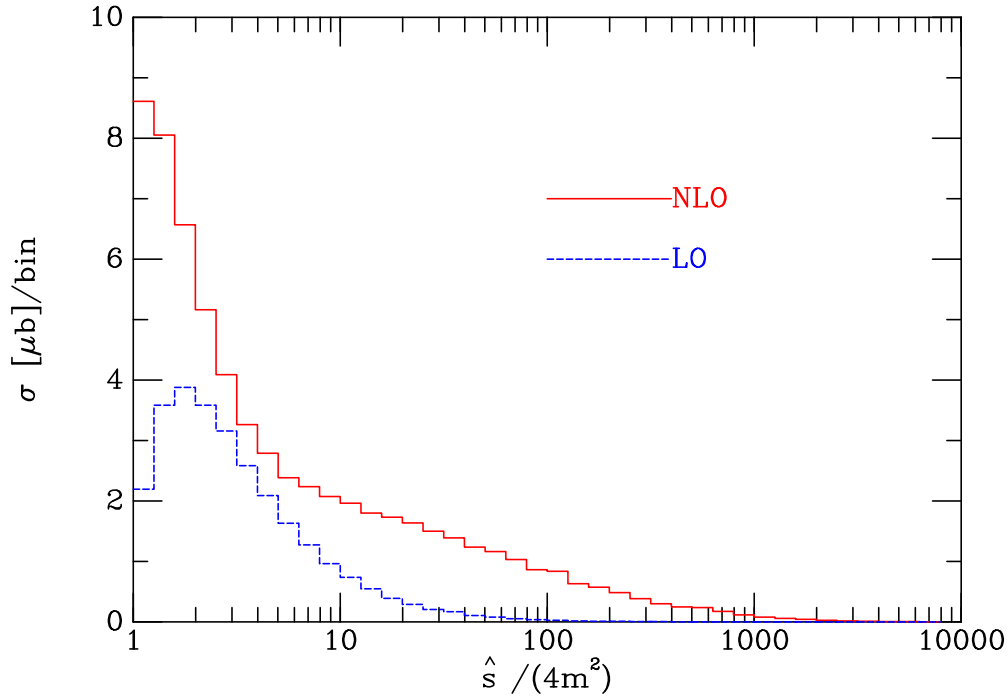}
\caption{Contributions to the total cross section in LO and NLO.}
\label{chnine:comp}
\end{figure}
Two features of $c_{ij}^{(1)}$ as shown in 
Figs.~\ref{chnine:gg},~\ref{chnine:qq} and \ref{chnine:gq} are worthy of note. 
The quark-antiquark and gluon-gluon initiated processes 
have a very rapid rise at threshold:
\begin{eqnarray}
    && c_{q\overline{q}}^{(1)} \rightarrow
       \frac{c^{(0)}_{q\bar{q}}(\rho)}{8\pi^2}
    \Bigg[ - \frac{ \pi^2 }{6 \beta } 
    + \frac{16}{3 } \ln ^2 \big(8 \beta^2 \big) 
    - \frac{82}{3} \ln(8 \beta^2) \Bigg], \nonumber \\
    && c_{gg}^{(1)} \rightarrow \frac{c_{gg}^{(0)}(\rho)}{8\pi^2}
    \Bigg[  \frac{ 11 \pi^2 }{42\beta } + 
    12 \ln ^2 \big(8 \beta^2 \big) 
    - \frac{366}{7} \ln( 8 \beta^2) \Bigg]\; ,
\end{eqnarray}
due\index{b-quark production@$b$-quark production!Sudakov logarithms}
\index{b-quark production@$b$-quark production!Coulomb singularities} to 
Coulomb $1/\beta$ singularities
and to Sudakov double logarithms. Near threshold,
these terms give large contributions and require special treatment, 
such as resummation to all orders.
In addition the $gg$ and $gq$ contributions to $c_{ij}^{(1)}$,
which involve spin-one gluon exchange in the $t$-channel, tend 
to a constant value at large $\hat{s}$.
In agreement with the results in Figs.~\ref{chnine:gg} and \ref{chnine:gq},
the intercept at large $s$ is given by
\index{b-quark production@$b$-quark production!high energy behavior}
\index{b-quark production@$b$-quark production!threshold resummation}
\begin{eqnarray}
c_{gg} &\rightarrow& 6 k \sim 0.1074\; , \\
c_{gq} &\rightarrow& \frac{4}{3} k \sim 0.02388\; ,
\end{eqnarray}
where\cite{Ellis:1990hw} 
\begin{equation}
k= \frac{7291}{16200}\frac{1}{8 \pi} \approx 0.01791 \; .
\end{equation}
The significance of these results for the NLO cross section is 
shown in Fig.~\ref{chnine:comp}, illustrating that, at the 
energy of the Tevatron, both the threshold 
region and the large $\hat{s}$ region are important at NLO.

The resummation of the threshold soft gluon corrections is best carried out
in $\omega$-space, where $\omega$ is the moment variable after
Mellin transform with respect to $\hat{s}$.
In $\omega$-space the threshold region corresponds to the 
limit $\omega \rightarrow \infty$ and the structure
of the threshold corrections is as follows
\begin{equation}
\hat{\sigma} = \sigma^{LO} 
\Big\{ 1 + \sum_{j=1}^{\infty} \sum_{k=1}^{2 j} 
b_{jk} \alpha_s^j \ln^k \omega \Big\} \; .
\end{equation}
Detailed studies\cite{Bonciani:1998vc}
indicate that such threshold resummation is of limited 
importance at the Tevatron where $b$-quarks are normally 
produced far from threshold.
Resummation effects lead to minor changes in the predicted cross section
and only a slight reduction in the scale dependence of the results.

\index{resummation}
At high energy, heavy quark production becomes a two scale 
problem, since $\hat{s} \gg m^2 \gg \Lambda^2$ and the short distance cross
section contains logarithms of $\hat{s}/m^2$.
As we proceed to higher energies such terms can only become 
more important, so an investment
in understanding these terms at the Tevatron will certainly bear fruit
at the LHC. At small $\omega$ the dominant terms 
in the heavy quark production short distance cross section 
are of the form 
\begin{equation}
\hat{\sigma}\sim \frac{\alpha_s^2}{m^2} 
\sum_{j=0}^\infty \as^{j} \sum_{k=0}^{\infty} 
c_{jk}  \Big( \frac{\as}{\omega} \Big)^{k} \; .
\label{logs}
\end{equation}
For $j=0$ we obtain the leading logarithm series.
Resummation of these terms has received considerable theoretical 
attention in the context of the $k_T$-factorization 
formalism\cite{Catani:1990xk,Collins:1991ty,Catani:1991eg,Camici:1996st,
Camici:1997ta,Baranov:2000gv,Ball:2001pq}.

In Ref.~\cite{Berger:2000mp}
an interpretation of the excess bottom-quark
\index{gluino production}
\index{b-quark production@$b$-quark production!via gluino decay}
production rate at the Tevatron is proposed 
via the pair-production of light gluinos, of mass 12 to 16
GeV, with two-body decays into bottom quarks and light bottom squarks.
Among the predictions of this SUSY scenario, the most clearcut is pair
production of like-sign charged $B$ mesons at the Tevatron collider.

\subsection{Run~II reach for $b$-quark production}

We now consider some of the experimental boundary conditions
on the determination of the cross-section for open beauty
production at the Tevatron.
\index{b-quark production@$b$-quark production!Run~II reach}

\subsection[\D0\ Study of $b$ Jet Production]
{ \D0\ Study of $b$ Jet Production and Run~II Projections
\authorfootnote{Author: Arthur Maciel}}

The differential cross section for $b$-jet production as a function of
\index{b-jet@$b$-jet!production}
jet transverse energy is obtained from a sample of muon tagged jets.
We discuss $b$ jets as an added tool for understanding $b$ production, and 
use the present analysis to propose a follow up measurement in Run~II.
Focus on doubly tagged, two $b$-jet events should significantly reduce both
theoretical and experimental uncertainties.

\subsubsection{Introduction}\label{sec1:int}
Previous studies on inclusive $b$ production by the \D0\ collaboration 
have concentrated on the $b$ quarks themselves, either as integrated 
production rates as a function of the $b$ quark $p_T^{min}$~\cite{sgmubib}, 
or the azimuthal correlations between the \bbbar\ pair \cite{dphibib}.

In contrast with such studies, we present 
a complementary measurement with the 
main focus on ``$b$ jets'' rather than $b$ quarks. ``$b$ jets'' are 
defined as hadronic jets 
carrying $b$ flavor. As such, the object of study is directly observable,
and introduces less model dependence when the connection between
observation and theory is made.

This measurement is in direct correspondence with a NLO-QCD calculation
by S. Frixione and M. Mangano\cite{fmnrbib}, who highlight the theoretical
advantages of considering the cross section for producing
jets rather than open quarks. For instance,
large logarithms
that appear at all orders in the open quark calculation (due to hard 
collinear gluons) are naturally avoided when all fragmentation modes are
integrated and it no longer matters what portion of the jet energy is
carried by heavy quarks or by radiation.

Following a brief presentation of our results, we try to assess what
improvements over present measurements can be expected in Run~II,
taking into account the upgraded capabilities of the \D0\ detector.
Moreover, we consider new experimental possibilities that might
contribute to the ongoing process of understanding heavy quark
production in the context of perturbative QCD.

\subsubsection{The $p_t$ Spectrum of $b$ Jets}
\index{b-jet@$b$-jet!$p_t$ spectrum}
\index{b-jet@$b$-jet!production}
\label{sec:bjets}
This measurement uses 5 pb$^{-1}$ of Run~I data from a muon plus jet
trigger, from which an inclusive sample of muon tagged jets is extracted.
Tagged jets have hadronic \et\ above 20 GeV, pseudorapidity $\eta < 0.6$,
and carry within their reconstruction cone 
($\Delta R \equiv \sqrt{\Delta \eta^2+ \Delta \phi^2}= 0.7$) a muon
track with \pt\ above 6 GeV and $\eta < 0.8$. About 36K events are thus 
selected, and less than 0.5\%
of them have either a second tagged jet or a doubly tagged one. This is
because of the relatively high muon threshold, which also
enriches the $b$ flavor content of the sample.

\begin{figure}[htb]
\begin{center}
\mbox{\epsfig{file=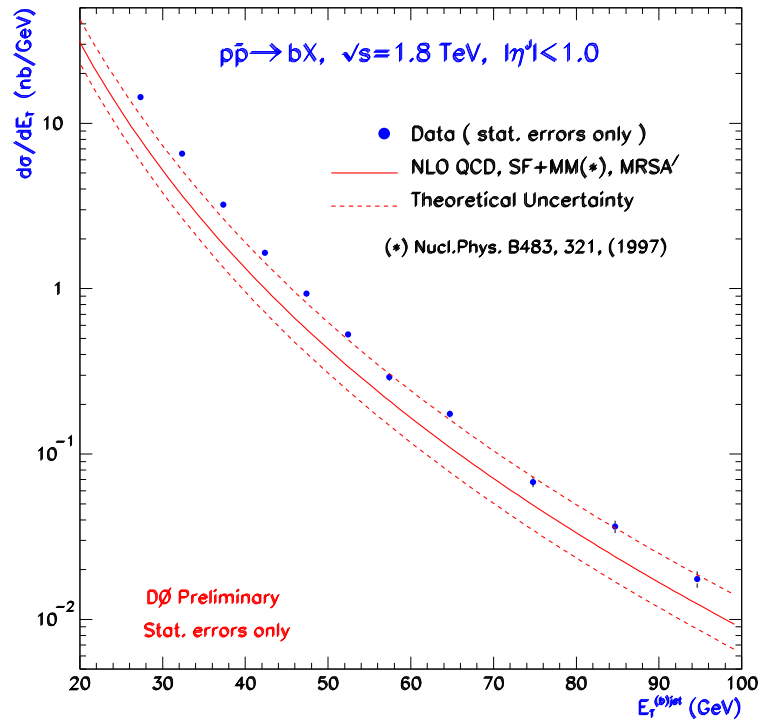,height=11cm,width=11cm}}
\end{center}
\protect\caption[The differential $b$~jet cross section]{The differential 
$b$~jet cross section 
and the next-to-leading order QCD 
prediction \cite{fmnrbib}. Data errors shown here are statistical only.}
\protect\label{fig:bjxs}
\end{figure}

After detection efficiencies and resolution corrections are
applied to the inclusive spectrum, the background of light flavors
decaying to muons needs to be removed. This is done on a statistical
basis, through fits of the distribution of muon transverse momentum
with respect to the associated jet axis (\ptrel). We use
Monte Carlo templates representing light and heavy quark decay patterns.
These templates are floated so as to fit the observed \ptrel\ spectra in
four different transverse energy ranges.
Such fits determine the fraction of muons from $b$ decay as a function of
tagged jet \et.

Tagging acceptance corrections are the last step towards a cross section.
These stem from the tagging muon threshold, pseudorapidity ranges and the
muon-jet association criterion. The unseen lepton energy along the jet 
also needs to be added (statistically) to the hadronic portion registered
in the calorimeter. After such corrections we obtain the $b$ jet differential
cross section shown in Fig.~\ref{fig:bjxs} and defined as the total number
of jets carrying one or more $b$ quarks in a given jet \et\ bin, and
within $|\eta| < 1.$ (\bbar 's are not counted; technically we count tagging
muons of either charge, and divide by twice the inclusive branching ratio of
$b \into \mu + X$).

It is observed that this result repeats the general pattern of past
$b$-production studies where data lies significantly above the
central values of theory prediction, but not incompatibly so, 
considering both the experimental and theoretical uncertainties.

The measurement uncertainties are driven by the \ptrel\ fits for the
$b$-fraction, and to a lesser extent the tagging acceptance 
losses, and calorimeter
energy scale at lower \et.
A summary of the systematic uncertainties (mostly correlated) in this
measurement is shown in Fig.~\ref{fig:berr}, and this leads us naturally
to a discussion of Run~II possibilities.

\begin{figure}
\begin{center}
\mbox{\epsfig{file=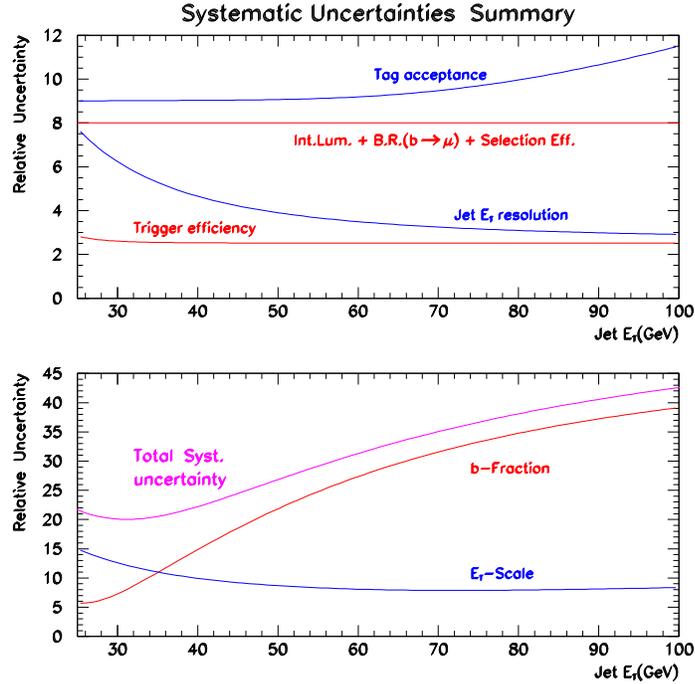,height=10cm,width=10cm}}
\end{center}
\protect\caption{A summary of all systematic uncertainties
expressed as a relative percentage.}
\protect\label{fig:berr}
\end{figure}

\subsubsection{$b$ Tagging for Run~II}
\label{sec:bjr2}

A detector with a precision vertex detector and a magnetic
spectrometer should be able to 
study $b$ jets. The measurement described above
is limited in many directions by its tagging method. First,
the $b$ content of the sample
decreases with jet \et, and the \ptrel\ fit is not sensitive enough to
extract a $b$ signal reliably 
for jets above 100 GeV. Second, part of the
advantage of studying $b$ jets over $b$ quarks is the removal of fragmentation
model dependence. This dependence is partially reintroduced when corrections
for muon tagging acceptances are performed.

The use of displaced vertex distributions for $b$ tagging
not only reduces the $b$ fraction
uncertainty to a secondary level in the analysis, 
but also removes the current
upper limit on jet \et\ imposed by the \ptrel\ fits. Some fragmentation
model dependence still remains associated with the determination of
proper time distributions (decay lengths).

\subsubsection{A Measurement Proposal}

One puzzle that arises in the comparison of heavy flavor production
data to theory is that despite the discrepancy in normalization, 
there is generally fair agreement in angular correlations 
\cite{dphibib,corrcdf,dphicdf}. 
This brings some conflict to the understanding of relative effects 
between the leading- and next-to-leading-order QCD amplitudes.
The goal of resolving this puzzle, coupled with
an interest in $b$ jets, suggests a promising program for Run~II.

In an attempt to minimize the current uncertainties on both 
theory and experiment, we consider a production measurement that
is restricted to a limited
kinematic regime to optimize uncertainties.
We consider the cross section for pairs of $b$ jets which are nearly
back-to-back, have relatively high \et\ and lie in the central 
rapidity region.

From the theoretical point of view, this experiment favors phase space
regions that are far from production thresholds and have high
$x$-values where PDF choices play a much reduced role.
In addition, the back-to-back kinematics suppresses NLO 
amplitudes in the pQCD
calculation~\cite{corrcdf}.

From the experimental point of view there are many advantages.
A muon plus jet trigger, very natural to the \D0\ architecture,
(i) enhances the $b$-content of the sample, (ii) collects an ideal
sample for a second tag by displaced vertex on the non-muon side and
(iii) minimizes the bias towards soft $b$ quarks inside hard jets because 
of the imposed muon \pt\ threshold. 

\subsubsection{Di-$b$-jets with \D0\ in Run~II.}

Inspection of Run~I \D0\ data shows that over 70\% 
of the muons in a muon plus jet trigger are naturally associated
with a jet.
Double tagging (of the $b$ and \bbar\ jets with muon and displaced vertex)
improves $b$ purity, and largely removes the limitations that 
are inherent to the \ptrel\ fits method, in particular the
limitation to low \et\ 
because of increasing systematic uncertainties with $b$ jet \et.

For acceptance projections we take the Run~I measurement as a
starting point: 36K events in 5 pb$^{-1}$ of data. The requirement of a
recoil jet (central, opposite hemisphere) having $E_T \geq 20$ GeV reduces 
the sample to 15K muon tagged (\pt\ above 6 GeV) back-to-back 
dijet events. With a vertex tagging rate near 10\% 
for the opposite jet,
we can  establish a rule of thumb of 300 events per pb$^{-1}$.

To access the transverse momentum of the $b$ quark inside the muon tagged
jet, we can focus on the tagging muon \pt\ spectrum (see next section).
By applying the methods in reference~\cite{sgmubib}
for determining the minimum $b$ quark \pt\ compatible with our muon and
jet thresholds, we find $p_T^{min}(b)$ around 20 GeV.

\subsubsection{$b$ Quark Production}\label{sec:bqxs}
A natural extension to the Run~I $b$-jets study presented above is to use
the same data sample and study $b$ quarks
rather than $b$ jets. This offers a consistency check of 
our previous measurements \cite{sgmubib,dphibib}, 
but also significantly extends the kinematic reach.

Starting from the data sample described in Section \ref{sec:bjets}
we repeat the analysis steps of~\cite{sgmubib} to obtain
the integrated $b$ quark production cross section as a function of
minimum $b$ quark \pt.  Our results are shown in Fig.~\ref{allbxs}
as the higher $p_T^{min}$ points (triangles). In this figure, the present 
(and preliminary) measurement is confronted with previous results from \D0, 
CDF (resonant dimuons), and theory \cite{Mangano:1992jk}.

\begin{figure}[htb] 
\vspace{-1cm}
\begin{center}
\mbox{\epsfig{figure=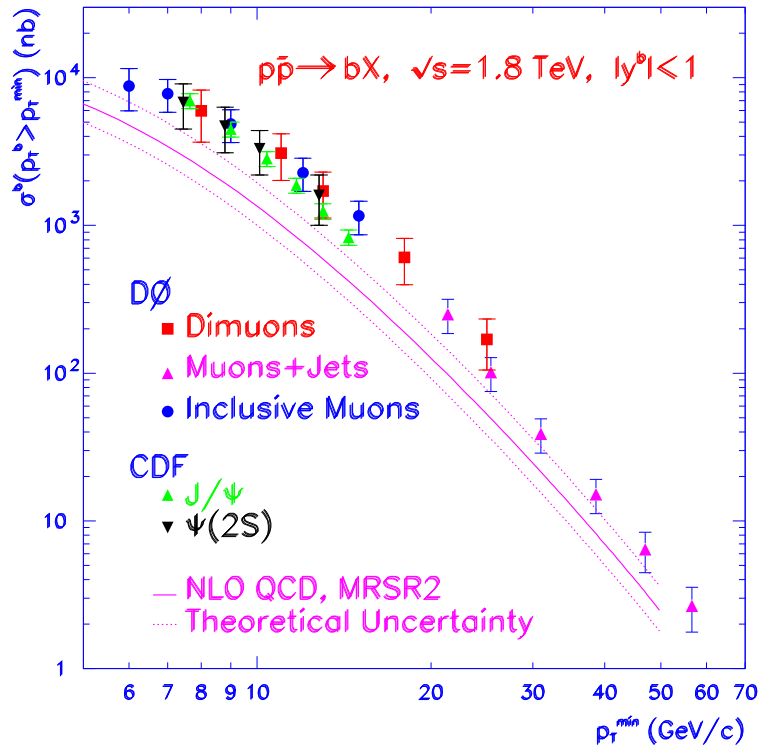,width=11cm,height=11cm}}
\end{center}
\protect\caption[A summary of $b$ production (integrated) cross section
measurements.]
{A summary of $b$ production (integrated) cross section measurements, 
shown as a function of minimum $b$-quark \pt. The preliminary
analysis described here is represented 
by the higher $p_T^{min}(b)$ triangles. The QCD prediction is also shown.}
\protect\label{allbxs}
\end{figure}

The consistency of the measurements is good. Our new
results practically double the $p_T^{min}$ reach to $60$ GeV. 
A general trend that can be extracted from Fig.~\ref{allbxs} is 
that data and theory agree better at higher $p_T^{min}$.

Of course this trend could be confirmed with extended and 
improved measurements in Run~II. But the inclusive muon method described
in this section has some limitations.
Our uncertainties are dominated by systematics
rather than statistics. Because the $b$-quark 
signal increases with muon \pt,
$b$ tagging with \ptrel\ as a function of muon \pt\ is more precise 
than with jet $E_T$. It is the model dependence, introduced in 
relating the $b$-quark thresholds to the muon acceptance, that
completely dominates the uncertainties in this type of measurement.

The full reconstruction of specific $B$-meson final states avoids some
uncertainties. The CDF experiment\cite{mescdf} probes a 
lower $B$-meson \pt\ where the comparison of data to theory
exhibits some of the same characteristics as the earlier measurements 
($p_T^{min}(b) < 30$~GeV) in
Fig.~\ref{allbxs}. It will be interesting to push fully
reconstructed $B$-meson cross-sections to higher meson \pt.
Various triggering and offline selection handles
exist that could significantly extend meson \pt\ range. Such studies
however lie beyond the scope of the present report.

\subsubsection{Conclusion}\label{sec:conc}
We have presented a Run~I cross-section measurement of jets carrying
$b$ flavor, and have illustrated the experimental advantages of 
studying $b$ jets (which are complementary to $b$ quarks). 
The $b$ quark cross section
extracted from the same data sample confirms previous measurements and 
significantly extends the $b$ quark \pt\ reach, while hinting at improved
agreement between data and theory at higher $b$ $p_T$.

For Run~II, a related measurement is envisaged that (i) focuses on 
dijets (doubly tagged) rather than single jets and 
(ii) restricts the kinematic
region of the final states to reduce systematic uncertainties 
both in the theory and the measurements.

A muon plus jet(s) trigger is one of the strengths of the \D0\ experiment,
and a center piece for various other physics projects. In Run~II \D0\ 
will for the first time have an internal magnetic volume for precision tracking
and vertexing. A muon tagged jets sample is an ideal starting point for
the displaced vertex tagging of ``other side'' jets.

Our brief analysis demonstrates the possibility of a very useful and
relatively fast measurement for early Run~II.


\subsection[CDF $P_T$ Reach in the $B$ Cross Section]
{ CDF $P_T$ Reach in the $B$ Cross Section for Run~II
\authorfootnote{Author: Karen Byrum}}
\index{b-quark production@$b$-quark production!Run~II}

The differential cross section for $B \rightarrow D^0+e+X$ 
as a function of $B$-hadron $p_T$ is studied using Monte Carlo events.
We predict the $p_T$ reach for such a measurement in Run~II.

The discrepancy between the next-to-leading order QCD predictions and the 
published Tevatron cross-sections remains unsolved.
Recently, M.~Cacciari, M.~Greco and P.~Nason~\cite{CGN} have presented a 
cross section formalism which is less sensitive to the 
renormalization and factorization scales at larger $b$-quark $p_T$
($p_T^b > 50$~GeV/$c$).   We estimate the $P_T$ reach for the $B$
cross section using the exclusive decay 
$B \rightarrow D^0+e+X$.   

\subsubsection{Run~II $P_T$ Reach}\label{sec:breach}
We use the PYTHIA Monte Carlo program to generate $b\bar{b}$ and use
CLEO QQ~\cite{cleoqq} to decay the $B$ mesons.  
The simulated sample is normalized
to Run~I data to provide a prediction of the 
cross section reach for Run~II.

In Run~II, CDF does not foresee low $E_T$ inclusive 
single lepton triggers as were available in Run~I.  Still,  
we calculate Run~II yields first using Run~I trigger
thresholds and then using the expected Run~II higher 
thresholds. We include the following assumptions:

\begin{itemize}
\item {an electron trigger similar to Run~Ib with kinematic
cuts on the electron of $E_T>8$~GeV and $p_T>$7.5~GeV/$c$ or
the proposed in Run~II threshold of $E_T>12$~GeV;} 
\item {the Run~Ib measured trigger efficiency;}
\item {the electron selection efficiencies predicted by the
CDF-II detector simulation;}
\item {kinematic cuts on the $D^0$ decay products of 
$P_{Tot}^{\pi}>0.5$~GeV/$c$, $P_{Tot}^{K}>$1.5~GeV/$c$, 
$\Delta(R)<~0.6$ for both $\pi,K$; and}
\item {a scale factor equal to the luminosity increase
for Run~I of 20.}
\end{itemize}

Using the above assumptions, we estimate the number of 
$B\rightarrow D^0 e X$ events above some $p_T^{min}$ and use
this number to estimate the size for the statistical 
precision for a given $p_T^{min}$.  The true statistical 
uncertainty will be larger than this and will depend on the 
signal-to-noise ratio of the $D^0$ mass peak.  From the Run~I
data, we estimate that the statistical uncertainty will be a factor of 
two times greater than that determined from a signal-only Monte Carlo.
This scale factor is conservative since we have not
applied lifetime cuts to our Run~I sample and these cuts are
known to greatly increase the signal to noise ratio which 
would reduce the statistical uncertainty, approaching the
predictions from a signal-only Monte Carlo. 

\begin {table}[b]
\begin {center}
\begin{tabular}{|c|c|} 
\hline
\mbox{Source} & \mbox{Uncertainty}  \\ \hline \hline 
Models of $B$ decays & 10$\%$ \\
Underlying event contribution & 8$\%$ \\
Hadron simulation & 10$\%$ \\
$B$ hadron semileptonic decay branching ratio & 10$\%$ \\
Luminosity & 7$\%$ \\ \hline
Total & 20$\%$ \\ \hline
\end{tabular}
\end{center}
\caption[Uncertainties in the published 
$b \rightarrow eX$ measurement]{Uncertainties in the published 
$b \rightarrow eX$ measurement\cite{run0} which are expected to 
contribute at the same level in Run~II.}
\protect\label{tab:cs1}
\end{table}

We estimate the systematic uncertainty using the 
published $b \rightarrow eX$ measurement \cite{run0}.  The
uncertainties from that measurement are shown in Table~\ref{tab:cs1}.
By measuring the $B$ meson cross section as opposed to
the $b$-quark cross section, we should be able to remove the systematics
due to fragmentation of the $b$ quark into a $B$ meson from the
measurement. However there will still be an uncertainty on the
theoretical prediction.
The other uncertainties may be more difficult
to reduce even with large statistics.    For Run~II 
we estimate a total systematic uncertainty of
$20\%$.

\subsubsection {Conclusions}

\begin{figure}[t]
\vspace{1.in}
\begin{center}
\mbox{\epsfig{figure=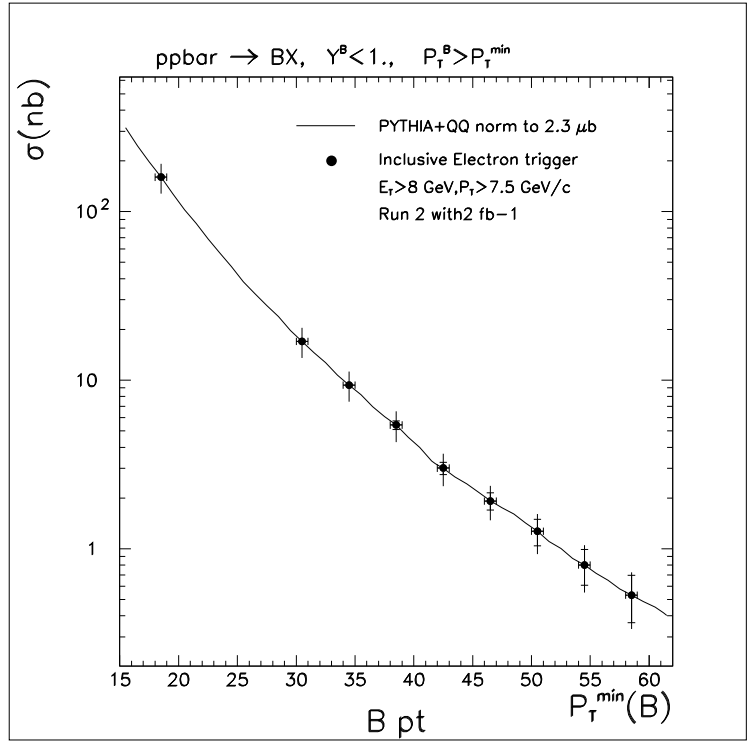,width=10cm}}
\end{center}
\vspace{-1.in}
\protect\caption
[The cross section reach vs. $B$ $P_T$ for electron $E_T>$ 8~GeV ]
{The predicted cross section reach as of function of $B$ $P_T$
for $B \rightarrow D^0 e X$ with
the kinematic cut on the electron of $E_T>$ 8~GeV 
and $P_T>$ 7.5~GeV/$c$ with 2~fb$^{-1}$ of luminosity. 
The statistical errors are 
predicted from Monte Carlo and scaled by a factor of two 
to include the effect of background.  
A constant 20$\%$ systematic error is based on
published results.}
\protect\label{fig:cs1}
\end{figure}

The predicted $p_T$ reach for $B\rightarrow D^0 eX $ with a kinematic 
cut on the electron of $E_T>$ 8 GeV and $p_T > 7.5$ GeV/c for 2 fb$^{-1}$
is shown in Figure~\ref{fig:cs1}.  The statistical uncertainties come
from our signal-only Monte Carlo simulation 
and scaled by a factor of two as an
estimate of the influence of background. 
A flat 20$\%$ systematic error is based on
published results~\cite{run0}.   

We also show the predicted $p_T$ reach for kinematic 
cut on the electron of $E_T>$ 12~GeV for 2 fb$^{-1}$
in Figure~\ref{fig:cs3}.  The $p_T$ requirement reduces
the number of events in the largest $B$ meson $p_T$ bins by
a modest 20$\%$.

\begin{figure}[th]
\centerline{\epsfig{figure=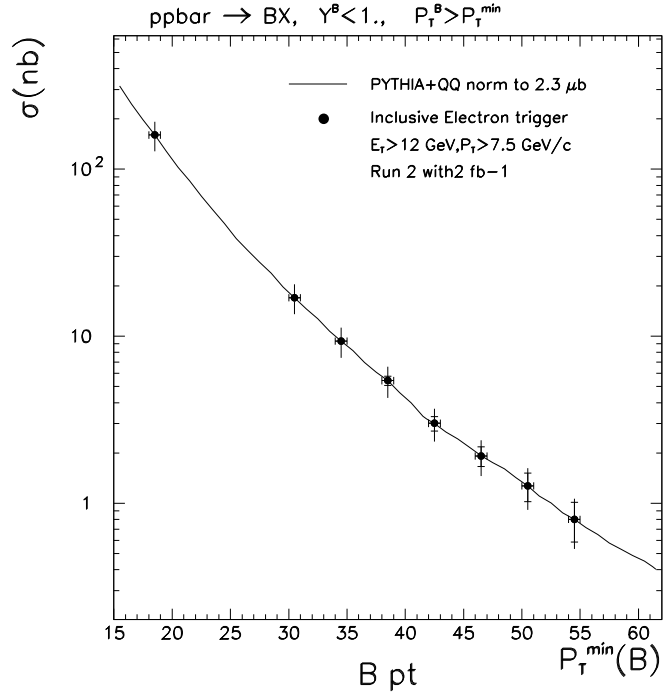,width=9cm}}
\caption[The cross section reach vs.\ $B$ $P_T$ for 
electron $E_T>$ 12~GeV ]
{The predicted cross section reach as of function of $B$ $P_T$ 
for $B \rightarrow D^0 e X$ with
the kinematic cut on the electron of 
$E_T>$ 12~GeV/$c$ with 2~fb$^{-1}$ of luminosity. 
The statistical errors are 
predicted from Monte Carlo and scaled by a factor of two 
to include the effect of background.  
A constant 20$\%$ systematic error is based on
published results.}
\label{fig:cs3}
\end{figure}

\def\GeV{\mbox{GeV}}
\section[Quarkonium production]{Quarkonium production}
\label{ch9:quarkonium}
\index{quarkonium!production}
\index{bottomonium!production}
\index{upsilon@$\Upsilon$!production}
\index{charmonium!production}
The two heavy quarkonium systems provided by nature are 
bottomonium ($b \bar b$ bound states) and 
charmonium  ($c \bar c$ bound states).
The clean signatures provided by the $J^{PC} = 1^{--}$
quarkonium states make them a particularly clear window 
for studying the dynamics of heavy quarks.
Although the major thrust of this report is bottom quark physics,
we will discuss charmonium as well as bottomonium,
since the physics of these two systems is very similar.

\subsection[Spectroscopy]
{Spectroscopy
\protect{\authorfootnote{
Author: Eric Braaten}}}

The charmonium and bottomonium systems have rich spectra 
\index{quarkonium!spectroscopy}
of orbital and angular-momentum excitations.
The masses of these states can be predicted using potential models 
whose parameters are tuned to reproduce the spectrum of observed 
states. Some of the states that have not yet been discovered 
should be produced abundantly at the Tevatron.
The obstacle to their discovery is finding a decay mode with a 
large enough branching fraction that can be used as a trigger.

\begin{figure}[ht] 
\vskip 9.5cm
\begin{center}
\includegraphics{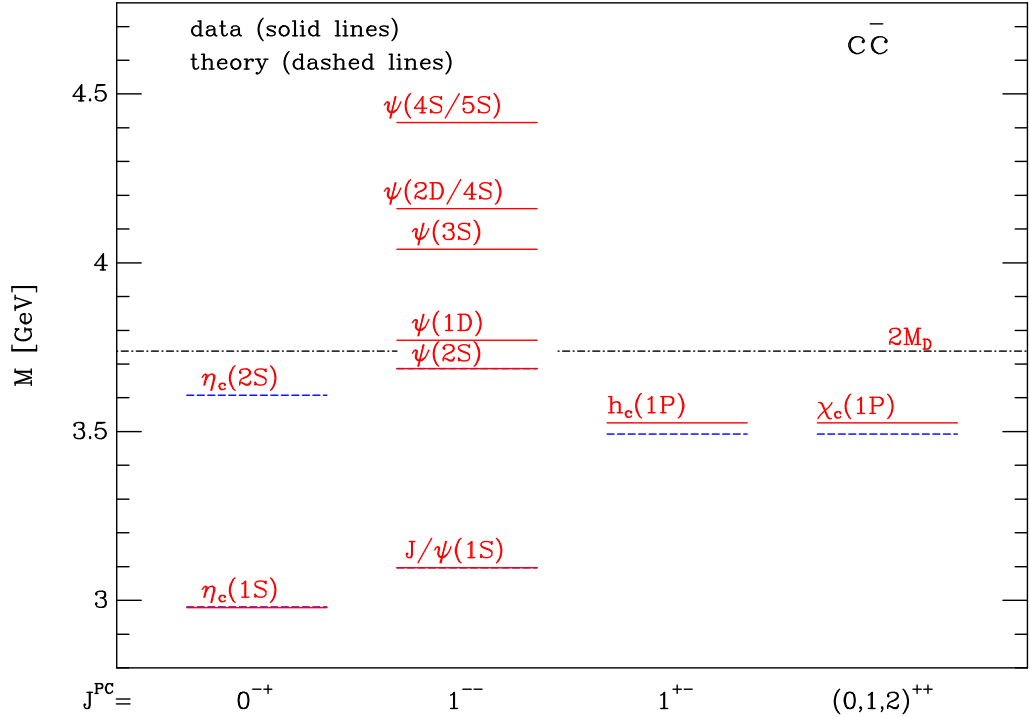}
\end{center}
\caption[Spectrum of $c \bar c$ mesons.]{
Spectrum of $c \bar c$ mesons: observed states (dotted red lines) 
and theoretical predictions (dashed blue lines). 
The $\chi_c(1P)$ lines are averaged over the spin states.}
\label{CCbar}
\end{figure}
The charmonium spectrum is shown in Fig.~\ref{CCbar}.
The observed states are shown as dotted (red) lines 
and the theory predictions are shown as dashed (blue) lines.
\index{eta-c@$\eta_c(2S)$}
\index{eta-c@$\eta_c(1S)$}
\index{charmonium!eta-c@$\eta_c(1S)$}
\index{charmonium!eta-c@$\eta_c(2S)$}
The most prominent of the missing states are the spin-singlet
states $\eta_c(2S)$ and $h_c(1P)$.  A signal for 
$h_c \to J/\psi \pi^0$ 
has been observed in $p \bar p$
annihilation at resonance, but has not been confirmed.
\index{charmonium!decay!$h_c \to J/\psi \pi^0$} 
\index{charmonium!hc@$h_c(1P)$}
\index{hc@$h_c(1P)$!$h_c \to J/\psi \pi^0$}
The states in the spectrum that were observed 
by the CDF collaboration in Run~I of the 
Tevatron are the $J/\psi$ and $\psi(2S)$, 
through their decays into $\mu^+ \mu^-$, and the $\chi_{c1}(1P)$
and $\chi_{c2}(1P)$, through their decays into $J/\psi \gamma$.
The $\chi_{c1}$ and $\chi_{c2}$, which have a mass splitting of 
only 45.7 MeV, were resolved by conversions of the decay photon 
into $e^+ e^-$ in the material of the detector.
The D0 collaboration also observed a $J/\psi$/$\psi(2S)$ signal,
but they were unable to resolve the two states.

The charmonium state with the greatest prospects for discovery 
at the Tevatron is the $\eta_c(2S)$.  
One possibility is through the radiative decay 
$\eta_c(2S) \to J/\psi + \gamma$, whose branching fraction 
is estimated to be about $10^{-3}$ \cite{Eichten:1980ms,D-Silverman}.
\index{charmonium!decay!$\eta_c(2S) \to J/\psi + \gamma$}
This is much smaller than the branching fraction for 
$\chi_{cJ}(1P) \to J/\psi + \gamma$, 
\index{charmonium!decay!$\chi_{cJ}(1P) \to J/\psi + \gamma$} 
which is about 27\% for $J=1$ and 14\% for $J=2$.
In Run~Ia, CDF observed about 1200 $\chi_c(1P)$ or 
$\chi_c(2P)$ candidates in a data sample of about 18 pb$^{-1}$. 
The cross section for $\eta_c(1S)$ is probably a little larger 
than that for $\chi_c(1P)+\chi_c(2P)$.
Thus the rate for $\eta_c(2S) \to J/\psi +\gamma$ should be large enough 
to observe in Run~II.

It is also possible that the $\eta_c(2S)$ could be discovered 
through one of its annihilation decay modes.
However a prerequisite would be the observation of the 
$\eta_c(1S)$, since it has a larger cross section and it 
probably has a larger branching fraction into any 
triggerable decay mode.    
Of the measured decay modes of the $\eta_c(1S)$, the most promising 
for observation at the Tevatron is $\eta_c \to \phi \phi$, 
with one $\phi$ decaying into $\mu^+ \mu^-$ to provide a trigger 
and the other decaying into $K^+ K^-$.
\index{charmonium!decay!$\eta_c \to \phi \phi$}%
This decay path has a 
branching fraction of about $2 \times 10^{-6}$,
which is small compared to the 6\% branching fraction 
for $J/\psi \to \mu^+\mu^-$.
The muons from the decay of the $\phi$ will typically be  
softer than those from the decay of a $J/\psi$ 
with the same $p_T$ as the $\eta_c(1S)$.
The steep dependence of the acceptance of the muons 
on their transverse momentum could give a large reduction factor
compared to the acceptance for $J/\psi \to \mu^+ \mu^-$.
In Run~Ia, CDF observed about $2 \times 10^5$ $J/\psi$ candidates 
in a data sample of about 15 pb$^{-1}$, and the cross section 
for $\eta_c(1S)$ is probably several times larger than that for $J/\psi$.
Thus the rate for $\eta_c \to \phi \phi$ may be large enough 
to observe in Run~II.

The bottomonium spectrum is shown in Fig.~\ref{BBbar}.
\begin{figure}[ht] 
\vskip 9.5cm
\begin{center}
\includegraphics{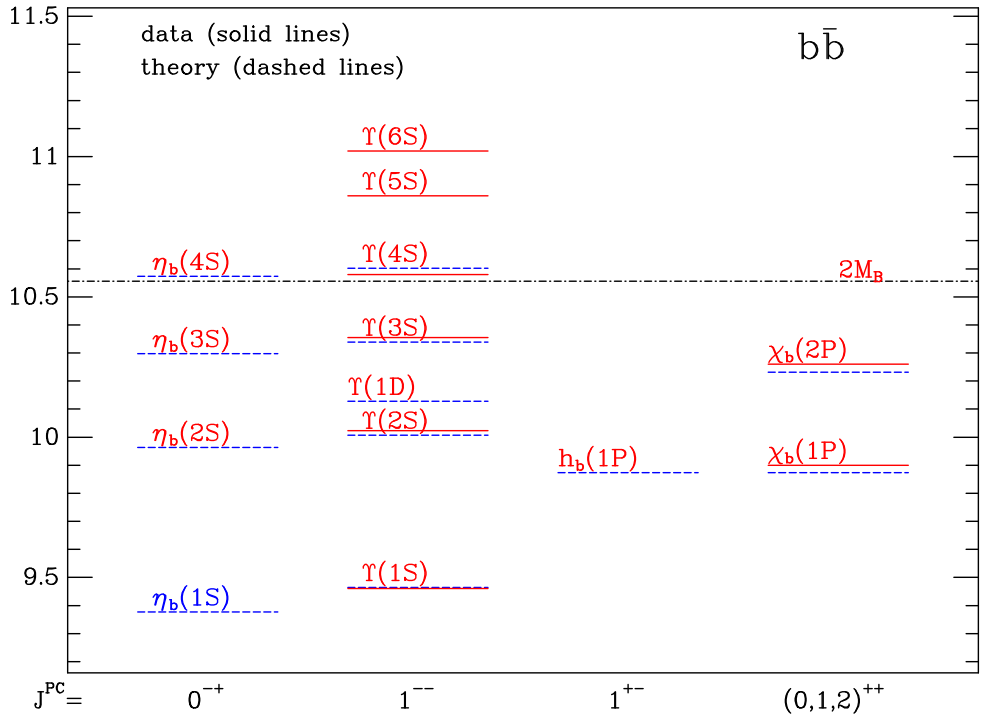}
\end{center}
\caption[Spectrum of $b \bar b$ mesons]
{Spectrum of $b \bar b$ mesons: observed states (solid red lines) 
and theoretical predictions (dashed blue lines). 
The $\chi_b(1P)$ and $\chi_b(2P)$ lines are averaged 
over the spin states.}
\label{BBbar}
\end{figure}
The observed states are shown as dotted (red) lines 
and the theory predictions are shown as dashed (blue) lines.
None of the spin-singlet states in the spectrum have been observed, 
not even the ground state $\eta_b(1S)$.
The states in the spectrum that were observed 
by the CDF collaboration in Run~I of the 
Tevatron are the $\Upsilon(1S)$, $\Upsilon(2S)$ and $\Upsilon(3S)$, 
(through their decays into $\mu^+ \mu^-$), and the $\chi_{bJ}(1P)$
and $\chi_{BJ}(2P)$, (through their decays into $\Upsilon(nS) \gamma$).
The $J=0,1,2$ states of $\chi_{bJ}(nP)$ were not resolved.
The D0 collaboration also observed a $\Upsilon(nS)$ signal,
but they were unable to resolve the $n=1,2,3$ states.

The bottomonium state with the greatest prospects for discovery 
\index{eta-b@$\eta_b(1S)$}
\index{bottomonium!eta-b@$\eta_b(1S)$}
at the Tevatron is the ground state $\eta_b(1S)$.  
The discovery requires finding a triggerable decay mode 
with a large enough branching fraction.
The branching fractions for exclusive decay modes of $\eta_b(1S)$
can not be predicted accurately.
For final states containing light hadrons, a conservative
estimate can be obtained by multiplying the corresponding branching 
fraction for $\eta_c(1S)$ by the appropriate suppression factor 
$(m_c/m_b)^n$ that would hold for asymptotically large quark masses. 
For the vector meson decay mode $\eta_b(1S) \to \rho \rho$,
the appropriate factor is $(m_c/m_b)^4 \approx 10^{-2}$.
A promising discovery mode at the Tevatron is 
$\eta_b(1S) \to J/\psi J/\psi$ \cite{B-F-L}, with one $J/\psi$ 
decaying into $\mu^+ \mu^-$ to provide a trigger 
and the other decaying into either $\mu^+ \mu^-$ or $e^+ e^-$.  
\index{bottomonium!decay!$\eta_b(1S) \to J/\psi J/\psi$}
This decay has almost identical kinematics to 
$\eta_c(1S) \to \phi \phi$, whose branching fraction is 0.7\%,
except that all masses are scaled up by a factor of 3.
The branching fraction for $\eta_b \to J/\psi J/\psi$ is probably 
smaller, but the suppression factor is probably not as severe as 
the asymptotic factor $10^{-2}$.  If the branching fraction
for $\eta_b(1S) \to J/\psi J/\psi$ is greater than $10^{-4}$, 
the prospects for the discovery of $\eta_b(1S)$ 
in Run~II of the Tevatron look bright.

\subsection[Theory of quarkonium production]
{Theory of quarkonium production
\protect{\authorfootnote{
Authors: Eric Braaten, Sean Fleming, Jungil Lee, Adam Leibovich}}}
\label{Quarkonium_production}
\index{quarkonium!production}

Heavy quarkonia (bottomonium and charmonium mesons)
are to a good approximation nonrelativistic bound states 
consisting of a heavy quark and its antiquark.
The bound-state dynamics involves both perturbative 
and nonperturbative aspects of QCD.
Due to the large mass, $m_Q$, of the heavy quark, 
the bound-state dynamics can be described more simply using 
an effective field theory called {\it nonrelativistic QCD} 
(NRQCD)~\cite{LMNMH}.
This effective field theory can be exploited to organize
quarkonium production rates into systematic expansions in $\alpha_s(m_Q)$
and in $v$, the typical relative velocity of the heavy 
quark~\cite{BBL}.\index{NRQCD}

The physical picture of quarkonium production begins with a 
hard scattering that creates a heavy quark-antiquark pair 
in an angular momentum and color state that we will denote 
$b \bar b(n)$.  The color state can be color-singlet ($b \bar b_1$)
or color-octet ($b \bar b_8$).  The angular momentum state is
denoted by the standard spectroscopic notation $^{2S+1}L_J$.
The $b \bar b$ subsequently evolves into the final state quarkonium $H$ 
through nonperturbative dynamics that can be accurately described 
by the effective theory NRQCD.  

The NRQCD factorization formula is the mathematical realization of the
above picture.  To be specific, we will focus on the production of 
spin-triplet bottomonium states.
The inclusive differential cross section for 
producing a bottomonium state, $H$, in a proton-antiproton collision 
can be written as
\begin{equation} \label{chnine:partonmodel}
d\sigma[p + \bar p\to H+X] 
\;=\; \sum_{ij} \int dx_1 dx_2\; f_{i/p}(x_1)f_{j/\bar p}(x_2) 
   \;	d \sigma[ij \to H+X],
\end{equation}
where the functions $f_{k/A}$ are parton distribution functions 
and the sum runs over the partons $i$ and $j$ in the initial state hadrons.
The cross section  $d \sigma[ij\to H+X]$ for the direct production of $H$ by
the collision of partons $i$ and $j$ can be written as the sum of products of
short-distance cross sections and long-distance matrix elements:
\begin{equation}
d \sigma[ij\to H+X] \;=\; \sum_n d\hat{\sigma}[ij\to b \bar b(n)+X]
			\; \langle O^H(n)\rangle.
\label{chnine:partoncross}
\end{equation}
To obtain the completely inclusive cross section, one must also add 
the cross section for indirect production via the decay of higher 
bottomonium states.
The short-distance cross section $d \hat{\sigma}$ can be calculated 
as a perturbative expansion in $\alpha_s$ evaluated at a scale of order 
$m_Q$ or larger.
The {\it NRQCD matrix element} $\langle O^H(n)\rangle$ 
encodes the probability for 
$b \bar b(n)$ to evolve into the quarkonium state $H$.
Each of the NRQCD matrix elements scales as a definite power of $v$, allowing
the sum over quantum numbers $n$ in Eq.~(\ref{chnine:partoncross}) to be
truncated at some order in the $v$ expansion.  
The approximate spin symmetry of NRQCD also implies 
relations between the matrix elements.
Most of these matrix elements can be determined 
only by fitting them to production data.  
The exceptions are the color-singlet matrix elements with the same
angular momentum quantum numbers as the bound state $H$,
which can be related to the wavefunction of the meson.  
For the spin-triplet S-wave and P-wave bottomonium states, 
these color-singlet matrix elements can be expressed as
\begin{eqnarray} \label{chnine:a1}
\langle O_1^{\Upsilon(nS)}(^3S_1)\rangle
   &\simeq& \frac9{2\pi}|R_{nS}(0)|^2,\\
\label{chnine:a2}
\langle O_1^{\chi_{bJ}(nP)}(^3P_J)\rangle
   &\simeq& (2J+1) \frac9{2\pi}|R'_{nP}(0)|^2.
\end{eqnarray}
They can be extracted from data on annihilation decays,
estimated from potential models, or calculated from first principles 
using lattice gauge theory simulations of NRQCD.

The NRQCD factorization formula provides a general framework for analyzing 
inclusive quarkonium production.  Any model of quarkonium production 
that is consistent with QCD at short-distances 
must reduce to assumptions about the NRQCD matrix elements.
For example, the {\it color-singlet model} assumes that only 
the color-singlet matrix elements with the same
angular momentum quantum numbers as the bound state are important,
namely $\langle O_1(^3S_1) \rangle$ for $\Upsilon(nS)$ and 
$\langle O_1(^3P_J) \rangle$ for $\chi_{bJ}(nP)$.
\index{color singlet model}\index{color evaporation model}
To the extent that the {\it color-evaporation model} is consistent with 
QCD at short distances, it reduces to the assumption that the 
S-wave matrix elements $\langle O_1(^1S_0) \rangle$, 
$\langle O_1(^3S_1) \rangle$, $\langle O_8(^1S_0) \rangle$, 
and $\langle O_8(^3S_1) \rangle$ dominate 
and that they are equal up to color and angular momentum factors. 
However the color evaporation model is usually implemented
by starting with the NLO cross section for producing $b \bar b$ pairs 
with invariant mass below the heavy meson threshold, 
imposing an ad hoc cutoff on the invariant mass of exchanged gluons, 
and multiplying by a phenomenological probability factor 
for each quarkonium state.

The effective field theory NRQCD also gives definite predictions for
the relative importance of the NRQCD matrix elements, 
because they scale as definite powers of the typical relative velocity $v$.
Given equal short-distance cross sections,
the most important matrix element for direct
$\Upsilon(nS)$ production should be the parameter 
$\langle O_1(^3S_1) \rangle$ of the color-singlet model.
However the short-distance cross sections are not equal.  From counting 
the color states, we expect the cross section 
for a color-octet $b \bar b$ pair to be about 8 times larger than the 
cross section for a color-singlet pair.
There are also dynamical effects that enhance the color-octet 
cross section relative to that for $b \bar b_1(^3S_1)$
both at small $p_T$ and at large $p_T$.
At small $p_T$, the color-octet cross section is enhanced, 
because it can proceed through order-$\alpha_s^2$ processes
like $i j \to b \bar b$.
At large $p_T$, the color-octet cross sections are enhanced
because the leading order cross section 
for $b \bar b_1(^3S_1)$ is suppressed by $m_b^4/p_T^4$ 
relative to $b \bar b_8(^3S_1)$.
According to NRQCD, the next most important matrix elements for direct
$\Upsilon(nS)$ production are suppressed by $v^4$
and can be reduced by spin-symmetry relations
to three color-octet parameters $\langle O_8(^3S_1) \rangle$, 
$\langle O_8(^1S_0) \rangle$, and $\langle O_8(^3P_0) \rangle$
for each of the radial excitations $\Upsilon(nS)$.
Including only these matrix elements, the NRQCD
factorization formula (\ref{chnine:partoncross}) for direct $\Upsilon(nS)$
production reduces to
\begin{eqnarray}
\sigma[\Upsilon(nS)] &=&
\sigma[b \bar b_1(^3S_1)] \langle O^{\Upsilon(nS)}_1(^3S_1) \rangle
+ \sigma[b \bar b_8(^3S_1)] \langle O^{\Upsilon(nS)}_8(^3S_1) \rangle
\nonumber
\\ 
&& + \sigma[b \bar b_8(^1S_0)] \langle O^{\Upsilon(nS)}_8(^1S_0) \rangle
+ \left( \sum_J (2J+1) \sigma[b \bar b_8(^3P_J)] \right)
        \langle O^{\Upsilon(nS)}_8(^3P_0) \rangle.
\label{chnine:upscross}
\end{eqnarray}
NRQCD predicts that the most important matrix
elements for the direct production of the P-wave states
$\chi_{bJ}(nP)$ are suppressed by $v^2$ relative
to $ \langle O^{\Upsilon(nS)}_1(^3S_1) \rangle$.
They can be reduced by spin symmetry relations to a color-singlet
parameter $\langle O_1(^3P_0) \rangle$ and a single
color-octet parameter $\langle O_8(^3S_1) \rangle$
for each radial excitation.
Including only these matrix elements, 
the NRQCD factorization formula (\ref{chnine:partoncross}) for direct
$\chi_{bJ}(nP)$ production reduces to
\begin{eqnarray}
\sigma[\chi_{bJ}(nP)] &=&
(2J+1) \left( \sigma[b \bar b_1(^3P_J)] 
        \langle O^{\chi_{b0}(nP)}_1(^3P_0) \rangle
+ \sigma[b \bar b_8(^3S_1)] 
         \langle O^{\chi_{b0}(nP)}_8(^3S_1) \rangle \right).
\label{chnine:chicross}
\end{eqnarray}
The factors of $2J+1$ comes from using spin-symmetry relations.

The NRQCD approach to quarkonium production
is a phenomenologically useful framework only if the NRQCD
expansion in (\ref{chnine:partoncross}) can be truncated after
the matrix elements of relative order $v^4$.
For S-waves, the truncation at order $v^2$ essentially reduces to the 
color-singlet model, which has been decisively ruled out by the 
CDF data on charmonium production in Run~I \cite{CDF_TroyD,CDF_chi}.
Truncation at order $v^6$ introduces too many additional adjustable 
parameters to have any predictive power.
In the NRQCD approach truncated at relative order $v^4$,
the direct production of each 
S-wave radial excitation is described by four matrix elements 
$\langle O_1(^3S_1) \rangle$, $\langle O_8(^3S_1) \rangle$, 
$\langle O_8(^1S_0) \rangle$, and $\langle O_8(^3P_0) \rangle$.
The production of each P-wave radial excitation is described 
by two matrix elements 
$\langle O_1(^3P_0) \rangle$ and $\langle O_8(^3S_1) \rangle$.
The NRQCD framework has considerable flexibility, 
because the color-octet matrix elements are all adjustable parameters,
but it is still restrictive enough to have predictive power.
Since the color-octet matrix elements are predicted to dominate
at large $p_T$, the NRQCD approach has sometimes been called the 
{\it color-octet model}.\index{color octet model}
This terminology should be avoided,
because NRQCD predicts that some production processes
are dominated by color-singlet matrix elements.
The phrase {\it NRQCD model} would more
accurately reflect the theoretical assumptions that are involved.

The short-distance cross sections $ d \hat{\sigma}$
in Eq.~(\ref{chnine:partoncross}) can be calculated using perturbative QCD.
Most of the parton cross sections required by the NRQCD approach
have been calculated only to leading order in $\alpha_s$.
There are three regions of $p_T$ that require somewhat different 
treatments: $p_T \ll 2m_b$,  $p_T \sim 2m_b$, and $p_T \gg 2m_b$.
The simplest region is $p_T \sim 2m_b$.
Here the leading order cross sections are of order $\alpha_s^3$
and come from the parton processes $i j \to b \bar b + k$.
For lack of a better term, we refer to these as the 
{\it fusion cross sections}. 
The fusion cross sections are known only to leading order in $\alpha_s$
\cite{ChoLeibovich, Beneke-K-V}.
If they could be calculated to next-to-leading order, 
it would allow for more accurate predictions for quarkonium 
production in the region $p_T \sim 2 m_b$.

That the region $p_T \gg 2m_b$ must be treated differently
can be seen from the behavior of the fusion cross sections 
as $p_T \to \infty$.  The order-$\alpha_s^3$ cross sections 
$d \hat{\sigma}/dp_T^2$ fall like $m_b^4/p_T^8$ for $b \bar b_1(^3S_1)$,
like $m_b^2/p_T^6$ for $b \bar b_1(^3P_J)$, $b \bar b_8(^1S_0)$,
and $b \bar b_8(^3P_J)$, 
and like $1/p_T^4$ for $b \bar b_8(^3S_1)$.
At higher orders in $\alpha_s$, all the $b \bar b$ channels will 
exhibit the scaling behavior $d \hat{\sigma}/dp_T^2 \sim 1/p_T^4$.
The scaling contributions can be expressed in the form
\begin{equation}
d\hat{\sigma}[ij\to b \bar b(n)+X] \;=\; 
\int_0^1 dz \, d\hat{\sigma}[ij\to k+X] \, D_{k \to b \bar b(n)}(z),
\label{chnine:fragcross}
\end{equation}
where $D_{k \to b \bar b(n)}(z)$ is the fragmentation function
that specifies the probability for a parton $k$ produced by a hard 
scattering to ``decay'' into  a $b \bar b$ pair in the state $n$
with a fraction $z$ of the parton's longitudinal momentum.
The fragmentation functions begin at order $\alpha_s$ 
for $b \bar b_8(^3S_1)$, at order $\alpha_s^2$
for $b \bar b_1(^3P_J)$, $b \bar b_8(^1S_0)$,
and $b \bar b_8(^3P_J)$, 
and at order $\alpha_s^3$ for $b \bar b_1(^3S_1)$.
They have all been calculated only to leading order in $\alpha_s$, 
with the exception of the $b \bar b_8(^3S_1)$ fragmentation function, 
which has been calculated to next-to-leading order \cite{Braaten-Lee:NLO}.
If they could all be calculated to order $\alpha_s^3$,
it would allow more accurate predictions for quarkonium 
production at large $p_T$.

That the region $p_T \ll 2m_b$ must be treated differently
can be seen from the behavior of the fusion cross sections 
as $p_T \to 0$.  The order-$\alpha_s^3$ cross sections 
$d \hat{\sigma}/dp_T^2$ for $b \bar b_1(^3S_1)$ 
and for $b \bar b_1(^3P_1)$ are well-behaved in this limit, but those for
$b \bar b_1(^3P_{0,2})$, $b \bar b_8(^3S_1)$, $b \bar b_8(^1S_0)$, 
and $b \bar b_8(^3P_J)$ diverge like $1/p_T^2$.
These are precisely the channels in which a 
$b \bar b$ pair can be created at $p_T = 0$
by the order-$\alpha_s^2$ parton process $i j \to b \bar b$.
If the cross section is integrated over $p_T$,
the singular term proportional to $1/p_T^2$ in the
cross section for $i j \to b \bar b + g$ is canceled 
by a singular term proportional to $\delta(p_T^2)$ in the 
next-to-leading order correction to the cross section for 
$i j \to b \bar b$, so that the cross section integrated 
over $p_T$  is finite.  These next-to-leading order corrections 
have been calculated \cite{PCGMM}, and they allow the cross section 
integrated over $p_T$ to be calculated to next-to-leading order.
However what is really needed to compare with the Tevatron data
is the differential cross section in $p_T$.
To turn the naive perturbative prediction for $d \hat{\sigma}/dp_T^2$,
which includes $1/p_T^2$ terms from $i j \to b \bar b + g$ 
and $\delta(p_T^2)$ terms from $i j \to b \bar b$,
into a smooth $p_T$ distribution requires taking into account 
multiple soft-gluon radiation.  
Reasonable prescriptions for doing this are 
available, but they have not yet been applied to this problem.
This is the biggest obstacle to more quantitative analyses of 
quarkonium production at the Tevatron.

\begin{figure}[th]
\centerline{\epsfxsize 11cm \epsffile{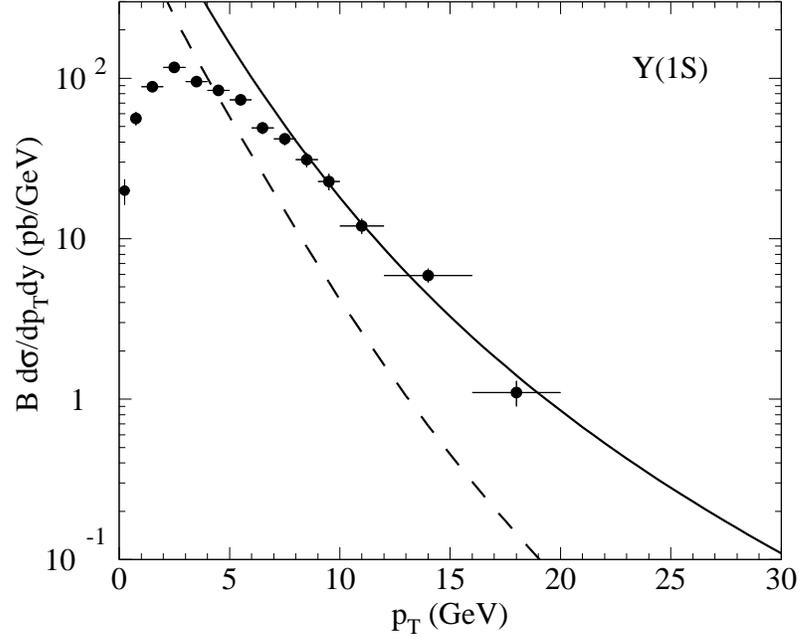}}
\caption[Inclusive cross section for $\Upsilon(1S)$ at Run~I.]
{Inclusive cross section for $\Upsilon(1S)$ at Run~I
	multiplied by its branching fraction into $\mu^+ \mu^-$:  
	CDF data, the NRQCD fit of Ref.~\cite{B-F-L} (solid line), 
	and the color-singlet model prediction (dashed line).}
\label{chnine:ups1plot}
\end{figure}

\begin{figure}[thp]
\centerline{\epsfxsize 11cm \epsffile{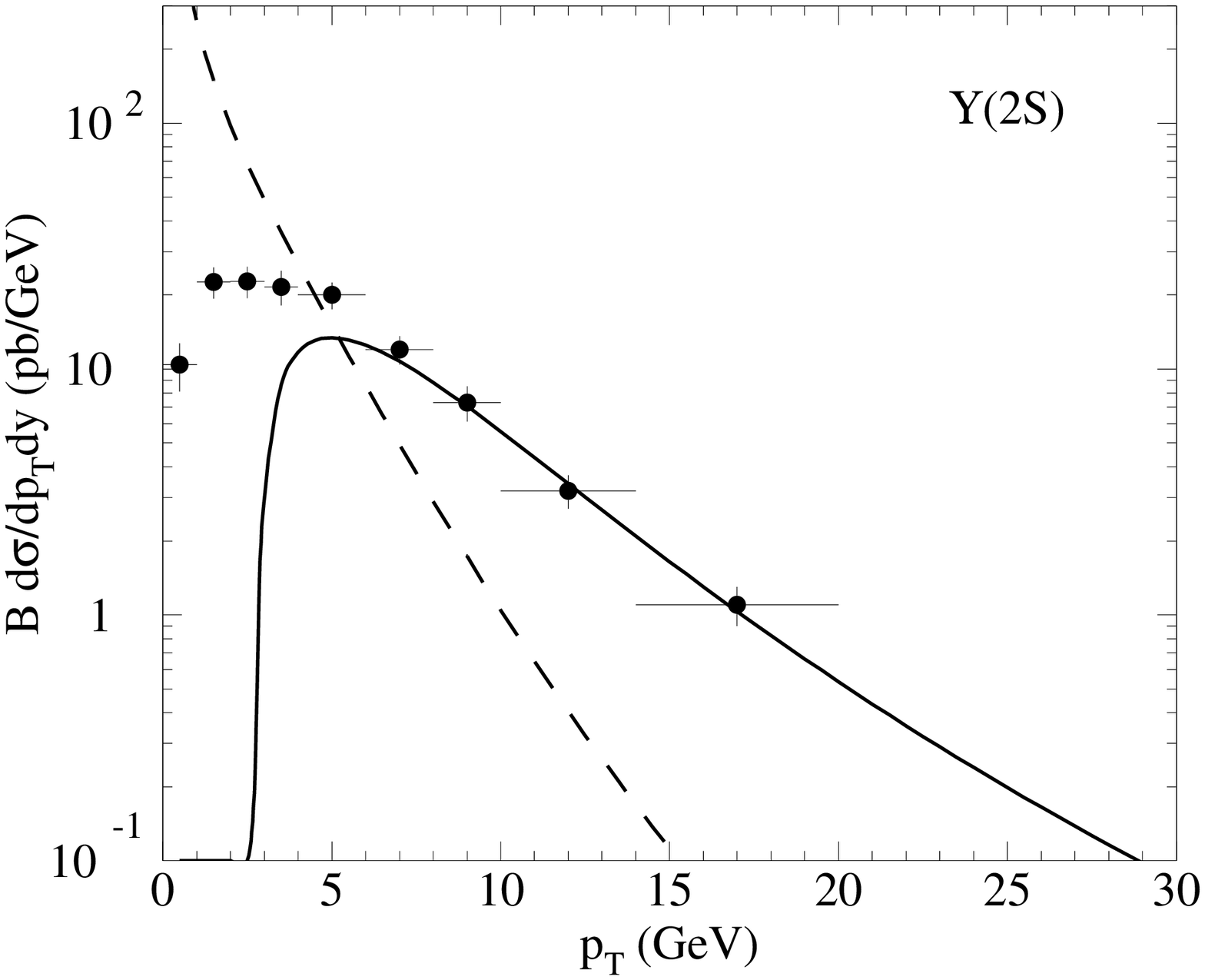}}
\caption[Inclusive cross section for $\Upsilon(2S)$ at Run~I.]
{Inclusive cross section for $\Upsilon(2S)$ at Run~I
	multiplied by its branching fraction into $\mu^+ \mu^-$:  
	CDF data, the NRQCD fit of Ref.~\cite{B-F-L}
	(solid line), and the color-singlet model prediction (dashed line).
\label{chnine:ups2plot}}
\vspace*{.5cm}
\centerline{\epsfxsize 11cm \epsffile{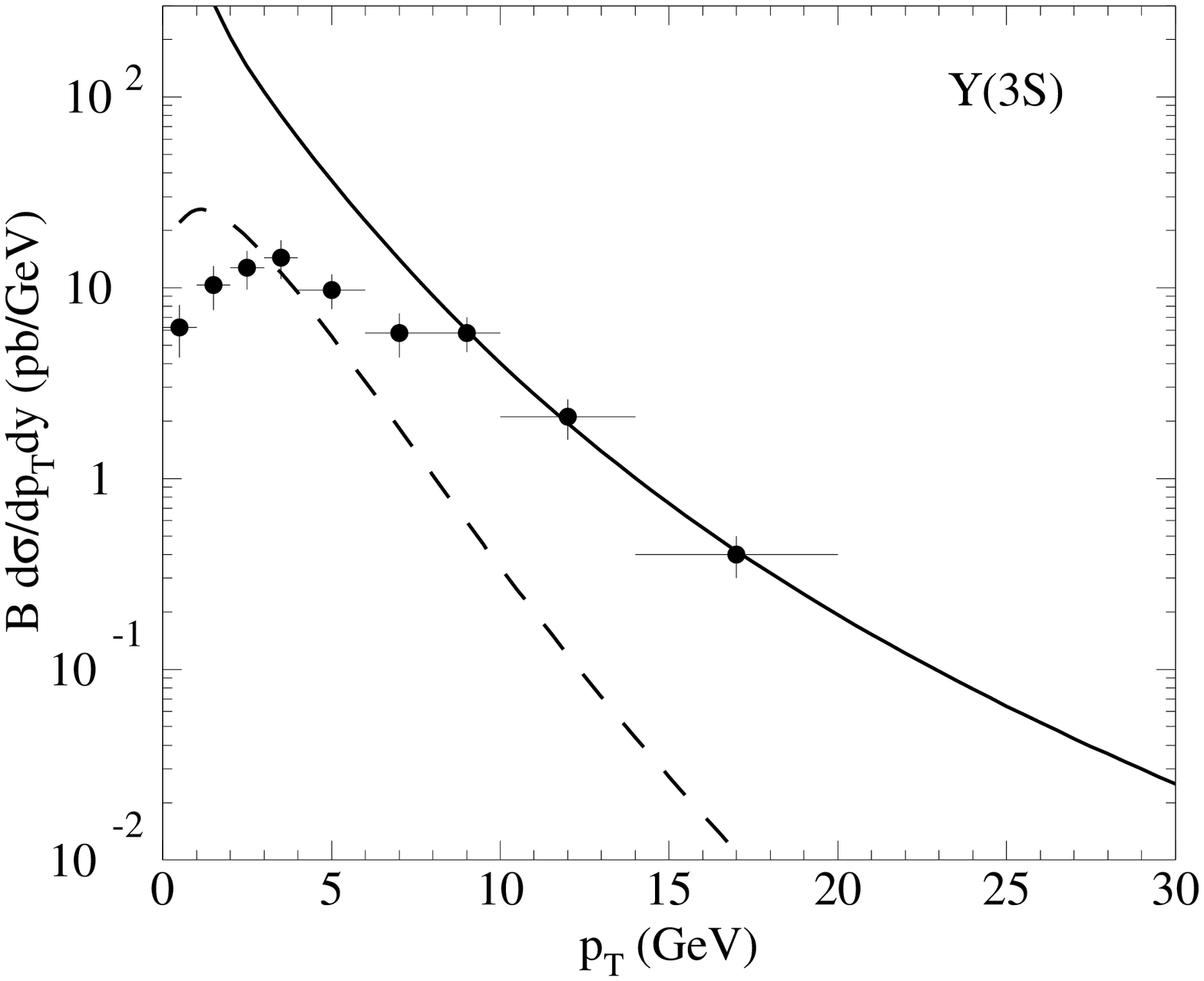}}
\caption[Inclusive cross section for $\Upsilon(3S)$ at Run~I.]
{Inclusive cross section for $\Upsilon(3S)$ at Run~I
	multiplied by its branching fraction into $\mu^+ \mu^-$:  
	CDF data, the NRQCD fit of Ref.~\cite{B-F-L}
	(solid line), and the color-singlet model prediction (dashed line).
\label{chnine:ups3plot}}
\end{figure}

\subsubsection{Bottomonium}
\label{sec:bott_theory}

A pioneering NRQCD analysis of the CDF data on bottomonium 
production in Run~IA of the Tevatron was carried out by 
Cho and Leibovich~\cite{ChoLeibovich}.
An updated NRQCD analysis of the CDF data from Run~Ib was recently 
carried out in Ref.~\cite{B-F-L}.
An alternative analysis that uses PYTHIA to take into account 
some of the effects of soft gluon radiation 
has been presented in Ref.~\cite{DSL}.
In the analysis of Ref.~\cite{B-F-L},
the color-singlet matrix elements were extracted from
$\Upsilon$ decay data or estimated using potential models.  
The color-octet matrix elements were extracted from the CDF 
cross sections for $\Upsilon(1S)$, $\Upsilon(2S)$, and $\Upsilon(3S)$
and the fractions of $\Upsilon(1S)$ coming from $\chi_b(1P)$ and $\chi_b(2P)$.  
To avoid the complications of soft-gluon re-summation at small $p_T$, 
the analysis used only the data for $p_T > 8$ GeV.
Because the cross sections
for $b \bar b_8(^1S_0)$ and $b \bar b_8(^3P_J)$
have similar dependences on $p_T$, only the linear combination
$M_r = \langle O_8(^1S_0)\rangle + r \langle O_8(^3P_0)\rangle/m_b^2$ 
with $r \sim 5$ could be determined. 
Two different sets of parton distribution functions were used for the
analysis: MRST98LO and CTEQ5L.  
The bottom quark mass was set to $m_b = 4.77 \pm 0.11$ GeV.
The resulting color-octet matrix elements have large statistical error bars 
and many of them are consistent with zero.
There are also large uncertainties due to the renormalization 
and factorization scales.
The large errors limit the usefulness of these matrix elements
for predicting bottomonium production in other high energy processes.

In Figs.~\ref{chnine:ups1plot}, \ref{chnine:ups2plot}, and \ref{chnine:ups3plot}, 
we show the CDF data on the differential cross sections for $\Upsilon(1S)$,
$\Upsilon(2S)$, and $\Upsilon(3S)$ at Run~I of the Tevatron.  
The cross sections are integrated over the rapidity range $|y| \leq 0.4$
and then divided by 0.8.
The solid curves are the central curves for the NRQCD fits.
The dashed curves are the leading order predictions of the color-singlet model.
The NRQCD curves agree with the data in the region $p_T > 8$ GeV
used to fit the matrix elements, 
but the agreement deteriorates quickly at lower values of $p_T$.
The data turns over and then approaches 0 at small $p_T$.
This is to be expected, because the differential cross section
$d \sigma/d p_T^2$ should be an analytic function of $p_T^2$,
which implies that there is a kinematic zero in 
$d \sigma/d p_T = 2 p_T d \sigma/d p_T^2$.
In contrast with the data, the NRQCD curves for $\Upsilon(1S)$ 
and $\Upsilon(3S)$ diverge like $1/p_T$ as $p_T \to 0$.
This unphysical behavior is an artifact of fixed-order perturbation 
theory and could be removed by carrying out the appropriate 
resummation of soft gluons. 
The NRQCD curve for $\Upsilon(2S)$ turns over 
and goes negative at small $p_T$.  This unphysical behavior is 
an artifact of the fit, which gives a negative central value 
for $M_5$ that is consistent with zero to within the errors.
The fact that the NRQCD fit fails to describe the data below
$p_T = 8$ GeV emphasizes the need for an
analysis that takes into account soft gluon resummation at low $p_T$.
Such an analysis could use the data from the entire range of $p_T$
that has been measured, and it would therefore give matrix elements 
with much smaller error bars.

\begin {table}[t]
\begin {center}
\begin {tabular}{|l|cc|cc|}
\hline
$H$             & $R^H$(1.8 TeV) && $R^H$(2.0 TeV) & \\
\hline
$\Upsilon(3S)$  & $0.31 \pm 0.14$ && $0.36 \pm 0.16$ & \\
\hline
$\chi_{b2}(2P)$ & $0.44 \pm 0.26$ && $0.52 \pm 0.30$ & \\
$\chi_{b1}(2P)$ & $0.34 \pm 0.16$ && $0.39 \pm 0.19$ & \\
$\chi_{b0}(2P)$ & $0.20 \pm 0.07$ && $0.24 \pm 0.08$ & \\
\hline
$\Upsilon(2S)$  & $0.65 \pm 0.35$ && $0.76 \pm 0.41$ & \\
\hline
$\chi_{b2}(1P)$ & $0.57 \pm 0.26$ && $0.66 \pm 0.31$ & \\
$\chi_{b1}(1P)$ & $0.41 \pm 0.17$ && $0.48 \pm 0.19$ & \\
$\chi_{b0}(1P)$ & $0.23 \pm 0.08$ && $0.26 \pm 0.09$ & \\
\hline
$\Upsilon(1S)$  &        1        && $1.16 \pm 0.02$ & \\
\hline
\end {tabular}
\end {center}
\caption[Inclusive cross sections for spin-triplet bottomonium states]
{ \label {tab:Rincl}
        Inclusive cross sections for spin-triplet bottomonium states $H$
	at 1.8 TeV and 2.0 TeV divided by the 
	inclusive cross section for $\Upsilon(1S)$ at 1.8~TeV. 
	The cross sections are integrated over $p_T>8$~GeV
	and $|y|<0.4$. }
\end {table}

Because of the large uncertainties in the matrix elements,
the only reliable predictions that can be made based on the 
NRQCD analysis of Ref.~\cite{B-F-L} 
are those for which the large errors cancel out.
One such prediction is the increase in the cross section
when the center-of-mass energy is increased from 1.8 TeV 
in Run~I of the Tevatron to  2.0 TeV in Run~II. 
In order to cancel out the large uncertainties in the matrix elements,
it is convenient to normalize the cross section for 
the bottomonium state $H$ at center-of-mass energy $\sqrt{s}$ 
to that for inclusive $\Upsilon(1S)$ at $\sqrt{s}=1.8$ TeV.  
We therefore define the ratio
\begin{eqnarray}
R^H(\sqrt{s})
&=&  {\sigma[H; \; \sqrt{s} \,] \over
	\sigma[{\rm inclusive} \; \Upsilon(1S); \; \sqrt{s}=1.8 \; {\rm TeV}]},
\label{chnine:R-Hb}
\end{eqnarray}
where the cross sections are integrated over $p_T> 8$ GeV 
and $|y|< 0.4$.  In Table~\ref{tab:Rincl}, we give the ratios 
$R^H(\sqrt{s})$ for the inclusive production of spin-triplet 
bottomonium states, both at $\sqrt{s} = 1.8$ TeV and at 
$\sqrt{s} = 2.0$ TeV.
When the center-of-mass energy is increased from 1.8 TeV to 2.0 TeV,
all the cross sections increase by about the same amount.
The increase depends on $p_T$, changing from about 15\% at $p_T = 8$ GeV
to  about 19\% at $p_T = 20$ GeV.
The percentage increase for $\Upsilon(1S)$ is shown as a function of $p_T$
in Fig.~\ref{chnine:ups1inc}.

\begin{figure}[ht]
\centerline{\epsfxsize 11cm \epsffile{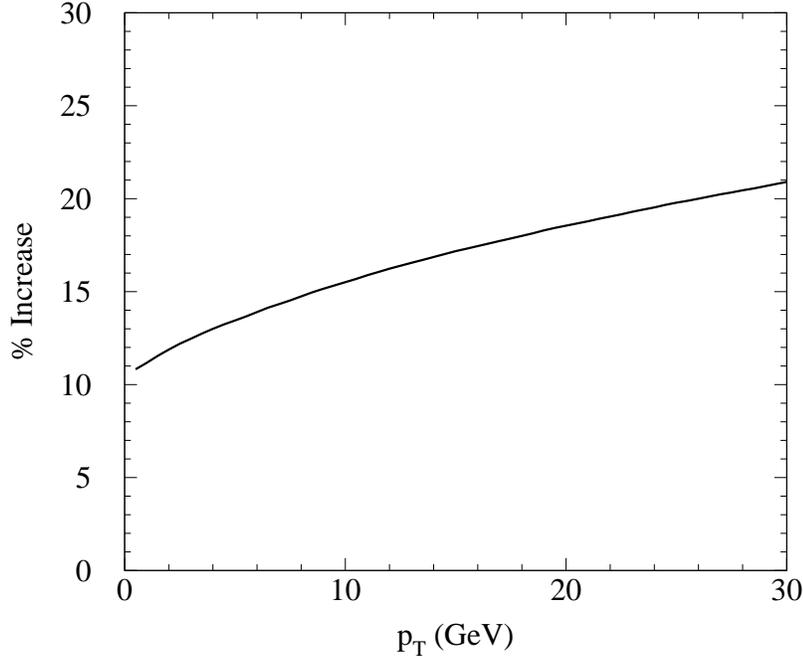}}
\vspace*{4pt}
\caption[Percentage increase in the inclusive $\Upsilon(1S)$ cross section.]
{Percentage increase in the inclusive cross section for $\Upsilon(1S)$
	at 2.0 TeV compared to 1.8 TeV.
	The cross sections are integrated over $p_T>8$ GeV
	and $|y|<0.4$. }
\label{chnine:ups1inc}
\end{figure}

\begin {table}
\begin {center}
\begin {tabular}{|l|cc|cc|}
\hline
$H$             & $R^H$(1.8 TeV) && $R^H$(2.0 TeV) & \\
\hline
$\eta_b(3S)$    & $1.83 \pm 0.54$ && $2.13 \pm 0.62$ & \\
\hline
$h_b(2P)$       & $0.07 \pm 0.07$ && $0.08 \pm 0.09$ & \\
\hline
$\eta_b(2S)$    & $1.60 \pm 0.59$ && $1.87 \pm 0.69$ & \\
\hline
$h_b(1P)$       & $0.11 \pm 0.08$ && $0.13 \pm 0.09$ & \\
\hline
$\eta_b(1S)$    & $4.59 \pm 0.83$ && $5.34 \pm 0.96$ & \\
\hline
\end {tabular}
\end {center}
\caption[Direct cross sections for spin-singlet bottomonium states] 
{\label {tab:Rdirect}
	Direct cross sections for spin-singlet bottomonium states $H$
	at 1.8 TeV and 2.0 TeV divided by the 
	inclusive cross section for $\Upsilon(1S)$ at 1.8 TeV. 
	The cross sections are integrated over $p_T>8$ GeV
	and $|y|<0.4$. }
\end {table}

Another reliable application of the NRQCD analysis is to predict the 
cross sections for the spin-singlet bottomonium states 
$\eta_b(nS)$ and $h_b(nP)$.
The most important matrix elements for $\eta_b(nS)$ and $h_b(nP)$
are related to those for $\Upsilon(nS)$ and $\chi_b(nP)$
by the spin symmetry of NRQCD.
Thus, once the color-octet matrix elements are determined from 
the production of spin-triplet states, 
the NRQCD approach truncated at order $v^4$ predicts the 
cross sections for the spin-singlet states without any new nonperturbative
parameters.  The resulting expressions for the direct cross sections 
of $\eta_b(nS)$ and $h_b(nP)$ are
\begin{eqnarray}
\sigma[\eta_b(nS)] &=&
{1 \over 3} \sigma[b \bar b_1(^1S_0)] 
	\langle O^{\Upsilon(nS)}_1(^3S_1) \rangle
+  {1 \over 3}\sigma[b \bar b_8(^1S_0)] 
	\langle O^{\Upsilon(nS)}_8(^3S_1) \rangle
\nonumber\\* 
&& 
+ \sigma[b \bar b_8(^3S_1)] \langle O^{\Upsilon(nS)}_8(^1S_0) \rangle
+ 3 \sigma[b \bar b_8(^1P_1)] 
   \langle O^{\Upsilon(nS)}_8(^3P_0)\rangle , 
\label{chnine:etacross}
\\[6pt]
\sigma[h_b(nP)] &=&
3 \left( \sigma[b \bar b_1(^1P_1)] \langle O^{\chi_{b0}(nP)}_1(^3P_0) \rangle
+ \sigma[b \bar b_8(^1S_0)] \langle O^{\chi_{b0}(nP)}_8(^3S_1) \rangle \right).
\label{chnine:hcross}
\end{eqnarray}
In Table~\ref{tab:Rdirect}, we give the ratios $R^H(\sqrt{s})$
defined in (\ref{chnine:R-Hb}) for the direct production of the spin-singlet 
states, not including any effects from the feed down 
of higher bottomonium states.
The cross sections for direct $h_b(nP)$ are significantly smaller 
than for inclusive $\Upsilon(1S)$.
Since the $h_b$ does not seem to have any distinctive decay modes 
that can be used as a trigger, its discovery at the Tevatron is unlikely. 
However the cross sections for direct $\eta_b(nS)$, $n=1,2,3$ 
are significantly larger than for inclusive $\Upsilon(1S)$.
Thus it should be possible to discover the $\eta_b$ if one can identify
a suitable decay mode that can be used as a trigger.  
One promising possibility is the decay $\eta_b \to J/\psi + J/\psi$
\cite{B-F-L}.

\subsubsection{Charmonium}
\label{chnine:Ochi}

The NRQCD factorization approach can be applied to 
charmonium production at the Tevatron as well as to bottomonium production.
The differences are only quantitative.
First, the coupling constant $\alpha_s(m_c) \approx 0.36$
is larger than $\alpha_s(m_b) \approx 0.22$, so the radiative corrections 
to the short-distance parton cross sections are larger for charmonium.
Second, the typical relative velocity $v$ of the heavy quark is 
significantly larger for charmonium ($v^2 \sim 1/4$) than for 
bottomonium ($v^2 \sim 1/10$). In the NRQCD approach truncated 
at relative order $v^4$, the terms that are neglected are suppressed 
only by an additional factor of $v^2$ and are therefore larger for 
charmonium than for bottomonium.
A third difference is that the value of $p_T$ above which fragmentation 
contributions become important is smaller by a factor of 3 for charmonium,
simply because $m_c \approx m_b/3$.

An important experimental difference from bottomonium production 
is that charmonium is also produced by decays of $B$ hadrons.
We define the prompt cross section for a charmonium state to be 
the inclusive cross section excluding the contribution from 
$B$ hadron decays.  Thus the prompt cross section is the 
sum of the direct cross section and the indirect contributions 
from decays of higher charmonium states that were produced directly.
The experimental signature of prompt production is the 
absence of a displaced vertex that would signal the 
weak decay of a $B$ hadron.

NRQCD analyses of the CDF data on charmonium production from Run~IA
have been carried out by several 
groups~\cite{ChoLeibovich,C-G-M-P,Kniehl-Kramer,Petrelli,Braaten-K-L}.
We describe briefly the analysis of
Ref.~\cite{Braaten-K-L}. The color-singlet matrix elements
were determined from annihilation decays of the charmonium states.
The color-octet matrix elements were obtained by fitting the CDF 
cross sections for $J/\psi$, $\psi(2S)$, and $\chi_{cJ}$ and by 
imposing the constraint from a preliminary CDF measurement 
of the ratio of $\chi_{c1}$ to $\chi_{c2}$.
Because the cross sections
for $c \bar c_8(^1S_0)$ and $c \bar c_8(^3P_J)$
have similar dependences on $p_T$, only the linear combination
$M_r = \langle O_8(^1S_0)\rangle + r \langle O_8(^3P_0)\rangle/m_b^2$ 
with $r \sim 3.5$ could be determined. 
The sets of parton distribution functions that were used were
MRST98LO and CTEQ5L.  
The charm quark mass was set to $m_c = 1.50 \pm 0.05$ GeV.

\begin{figure}[htp]
\centerline{\epsfxsize 9.75cm \epsffile{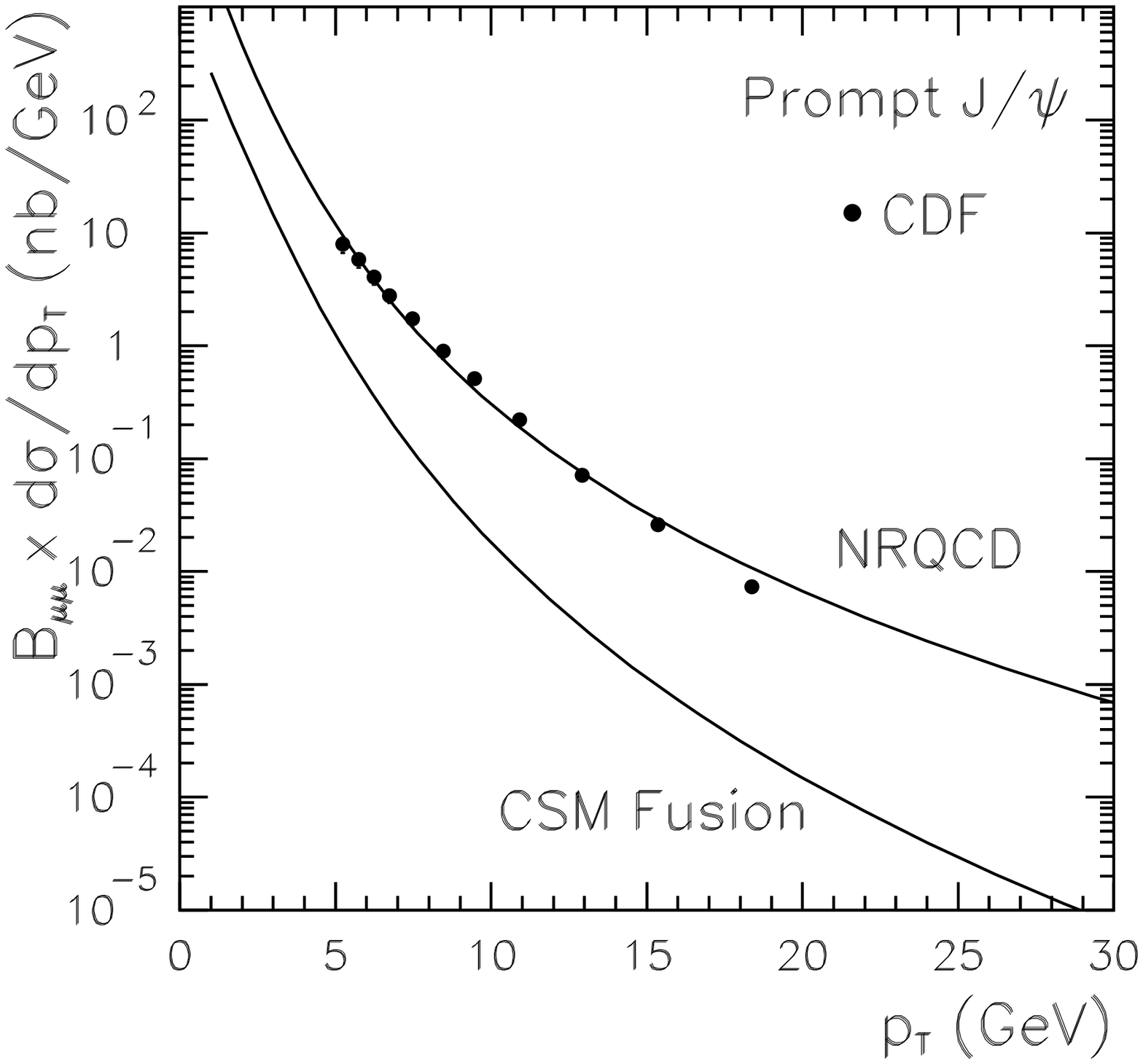}}
\caption[Prompt cross section for $J/\psi$ at Run~I.]
{Prompt cross section for $J/\psi$ at Run~I
	multiplied by its branching fraction into $\mu^+ \mu^-$:  
	CDF data, the NRQCD fit of Ref.~\cite{Braaten-K-L}, 
	and the color-singlet model prediction.}
\label{chnine:psi1}
\vspace*{.5cm}\vfill
\centerline{\epsfxsize 9.75cm \epsffile{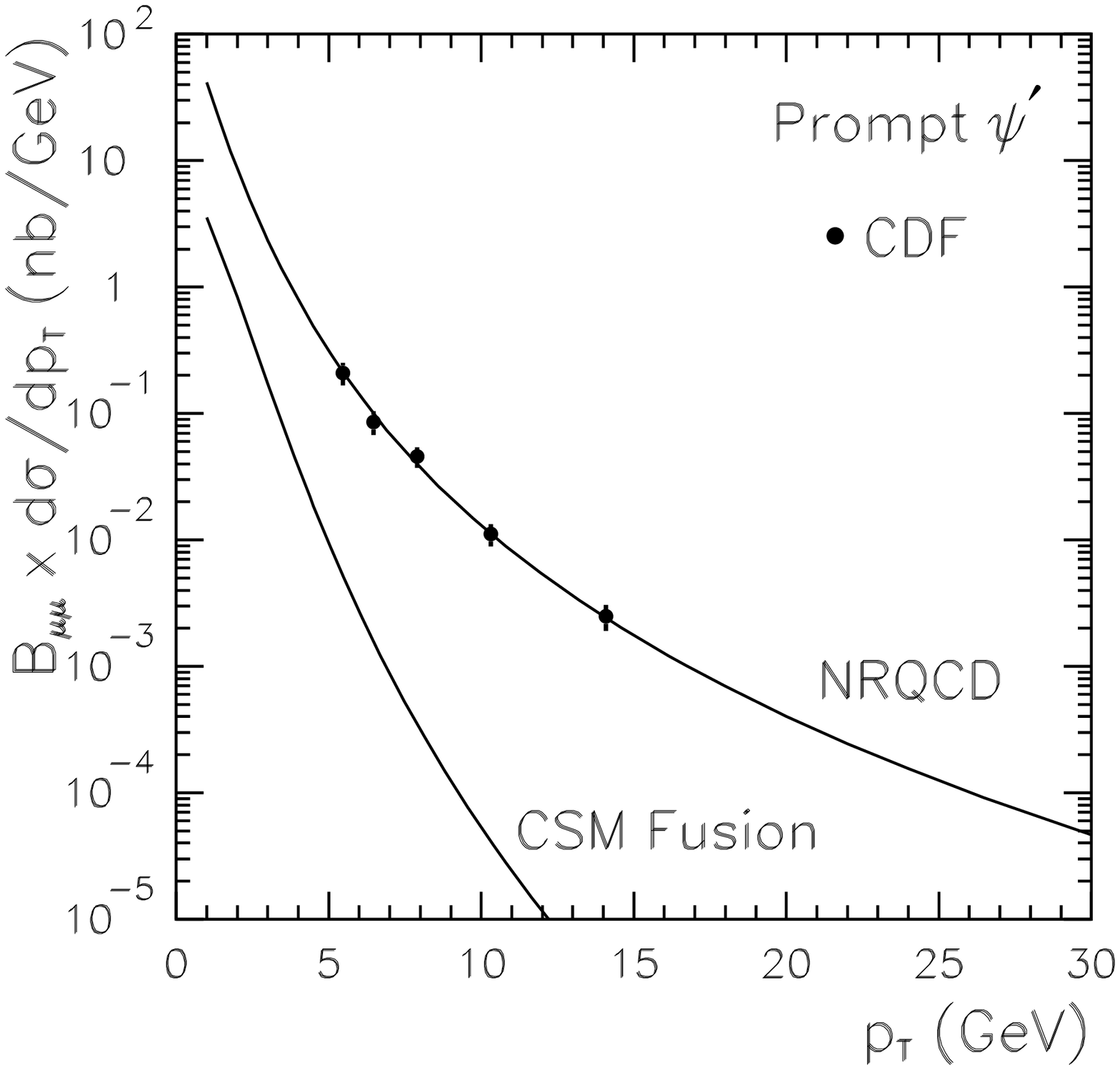}}
\caption[Prompt cross section for $\psi(2S)$ at Run~I.]
{Prompt cross section for $\psi(2S)$ at Run~I
	multiplied by its branching fraction into $\mu^+ \mu^-$:  
	CDF data, the NRQCD fit of Ref.~\protect{\cite{Braaten-K-L}},
        and the color-singlet model prediction.}
\label{chnine:psi2}
\end{figure}

In Figs.~\ref{chnine:psi1} and \ref{chnine:psi2}, we show the CDF data on the 
differential cross sections for $J/\psi$ and $\psi(2S)$ 
from Run~IA \cite{CDF_TroyD,CDF_chi}.
The cross sections are integrated over the pseudorapidity range
$|\eta| < 0.6$.  The curves are the results of the NRQCD fit
and the leading-order predictions of the color-singlet model.
If the cross sections were measured at lower values of $p_T$, 
they would turn over and go to zero like the $\Upsilon(1S)$ cross section
in Fig.~\ref{chnine:ups1plot}.  In contrast, the NRQCD curves continue 
rising and diverge like $1/p_T$ as $p_T \to 0$.  
In order to obtain the correct physical behavior at small $p_T$,
it would be necessary to carry out an analysis that includes 
the effects of soft-gluon radiation.  
Neglecting these effects may introduce large systematic errors 
into the NRQCD matrix elements.  In an analysis using a Monte Carlo 
event generator to take into account initial-state gluon radiation
at the Tevatron, the values of some of the color-octet matrix 
were decreased by an order of magnitude \cite{Sanchis-Lozano}.
One should therefore be wary 
of using the matrix elements from existing analyses of the Tevatron data
to predict production rates in other high energy processes.

\begin {table}[t]
\begin {center}
\begin {tabular}{|l|cc|cc|}
\hline
$H$             & $R^H$(1.8 TeV) && $R^H$(2.0 TeV) & \\
\hline
$\psi(2S)$  & $0.21 \pm 0.07$ && $0.23 \pm 0.08$ & \\
\hline
$\chi_{c2}(1P)$ & $0.44 \pm 0.08$ && $0.50 \pm 0.10$ & \\
$\chi_{c1}(1P)$ & $0.40 \pm 0.08$ && $0.45 \pm 0.09$ & \\
$\chi_{c0}(1P)$ & $0.14 \pm 0.03$ && $0.16 \pm 0.03$ & \\
\hline
$J/\psi$  &        1        && $1.14 \pm 0.08$  & \\
\hline
\end {tabular}
\end {center}
\caption[Prompt cross sections for spin-triplet charmonium states]{ 
\label {tab:Rcpro}
	Prompt cross sections for spin-triplet charmonium states $H$  
	at 1.8 TeV and 2.0 TeV divided by the 
	prompt cross section for $J/\psi$ at 1.8 TeV. 
	The cross sections are integrated over $p_T> 5.5$ GeV
	and $|\eta|<0.6$.}
\end {table}

The safest applications of the NRQCD analysis are to observables
for which the errors associated with the extractions of
the matrix elements tend to cancel out. One such observable is the 
increase in the cross section
when the center-of-mass energy is increased from 1.8 TeV to 2.0 TeV. 
It is convenient to normalize the cross section for 
the charmonium state $H$ at center-of-mass energy $\sqrt{s}$ 
to that for prompt $J/\psi$ at $\sqrt{s}=1.8$ TeV by defining the ratio
(in analogy with Eq.~(\ref{chnine:R-Hb}))
\begin{eqnarray}
R^H(\sqrt{s})
&=&  {\sigma[H; \; \sqrt{s} \,] \over
	\sigma[{\rm prompt} \; J/\psi; \; \sqrt{s}=1.8 \; {\rm TeV}]},
\label{chnine:R-Hc}
\end{eqnarray}
where the cross sections are integrated over $p_T > 5.5$ GeV 
and over $|\eta|< 0.6$.  In Table~\ref{tab:Rcpro}, we give the ratios 
$R^H(\sqrt{s})$ for the inclusive production of spin-triplet 
charmonium states, both at $\sqrt{s} = 1.8$ TeV and at 
$\sqrt{s} = 2.0$ TeV.
When the center-of-mass energy is increased from 1.8 TeV to 2.0 TeV,
all the cross sections increase by about the same amount.
The increase depends on $p_T$, changing from about 13\% at $p_T = 5.5$ GeV
to  about 18\% at $p_T = 20$ GeV.
The percentage increase for $J/\psi$ is shown as a function of $p_T$
in Fig.~\ref{chnine:psi1inc}.

\begin{figure}[ht]
\centerline{\epsfxsize 10cm \epsffile{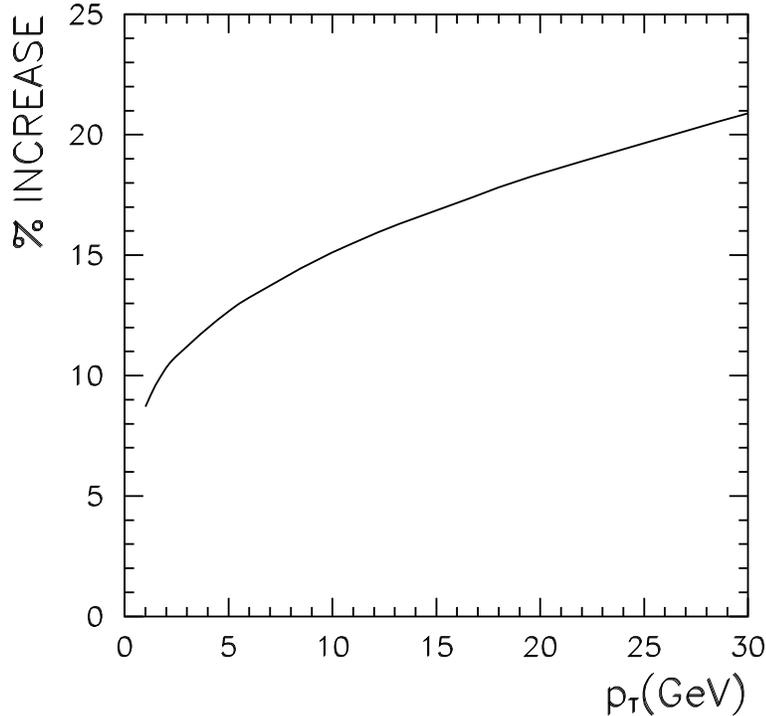}}
\caption[Percentage increase in prompt $J/\psi$ cross section.]
{Percentage increase in prompt cross section for $J/\psi$ 
	at $\sqrt{s} = 2.0$ TeV compared to 1.8 TeV 
	as a function of $p_T$. 
	The cross sections are integrated over $p_T>5.5$ GeV
	and $|\eta|<0.6$.}
\label{chnine:psi1inc}
\end{figure}

\begin {table}
\begin {center}
\begin {tabular}{|l|cc|cc|}
\hline 
$H$             & $R^H$(1.8 TeV) && $R^H$(2.0 TeV) & \\
\hline
$\eta_c(2S)$    & $0.37 \pm 0.11$ && $0.42 \pm 0.12$ & \\
\hline
$h_c(1P)$       & $0.033 \pm 0.006$ && $0.037 \pm 0.007$ & \\
\hline
$\eta_c(1S)$    & $1.73 \pm 0.15$ && $1.97 \pm 0.72$ & \\
\hline 
\end {tabular}
\end {center}
\caption[Direct cross sections for spin-singlet charmonium states]{
 \label {tab:Rcdir}
	Direct cross sections for spin-singlet charmonium states $H$  
	at 1.8 TeV and 2.0 TeV divided by the 
	prompt cross section for $J/\psi$ at 1.8 TeV. 
	The cross sections are integrated over $p_T$ between 5 and 20 GeV
	and $|\eta|<0.6$.}
\end {table}

Another reliable application of the NRQCD analysis is to predict the 
cross sections for the spin-singlet bottomonium states 
$\eta_c(nS)$ and $h_c$.
The most important matrix elements for $\eta_c(nS)$ and $h_c$
are related to those for the spin-triplet states
by the spin symmetry of NRQCD.
Thus, once the color-octet matrix elements are determined from 
the production of spin-triplet states, the NRQCD approach predicts the 
cross sections for the spin-singlet states without any new nonperturbative
parameters. 
In Table~\ref{tab:Rcdir}, we give the ratios $R^H(\sqrt{s})$
defined in (\ref{chnine:R-Hc}) for the direct production of the spin-singlet 
states, not including any effects from the feed-down 
of higher charmonium states.
The cross sections for direct $h_c$ are significantly smaller 
than for prompt $J/\psi$, but the cross sections for direct 
$\eta_c(nS)$, $n=1,2$ are significantly larger.
It may be possible to observe the $\eta_c(1S)$ 
and perhaps even discover the $\eta_c(2S)$ if one can identify
a suitable decay mode that can be used as a trigger.

\subsubsection{Quarkonium polarization}
\index{quarkonium!polarization}

The most dramatic prediction of the NRQCD factorization approach 
for quarkonium production at the Tevatron is that the 
$J ^{PC} = 1^{--}$ states like the $J/\psi$
should be transversely polarized 
at sufficiently large $p_T$.  This prediction follows from 
three simple features of the dynamics of heavy quarks and massless partons.
(1) The inclusive production of quarkonium at large $p_T$ 
is dominated by gluon fragmentation \cite{Braaten-Yuan}.  
(2) At leading order in $\alpha_s$,
the $c \bar c$ pair produced by gluon fragmentation is in
a color-octet $^3S_1$ state \cite{Braaten-Fleming} 
with transverse polarization.
(3) The approximate spin symmetry of NRQCD guarantees that the 
hadronization of a transversely polarized $c \bar c$ pair 
produces a $J/\psi$ that is predominantly transversely polarized 
\cite{Cho-Wise}.

In the NRQCD approach truncated at relative order $v^4$,
the cross sections for polarized quarkonium states can be calculated 
in terms of the matrix elements that describe unpolarized production.  
The only complication 
is that both the parton cross sections and the NRQCD matrix 
elements in the NRQCD factorization formula (\ref{chnine:partoncross})
become density matrices in the angular momentum quantum numbers 
of the $c \bar c$ pairs.  Thus the NRQCD approach gives definite
predictions for polarized quarkonium cross sections, 
without any new nonperturbative parameters.
In contrast, one of the basic assumptions of the color evaporation model 
is that quarkonium is always produced unpolarized.
This model can therefore be ruled out by a measurement of 
nonzero polarization of quarkonium.

\begin{figure}[ht]
\centerline{\epsfxsize 11cm \epsffile{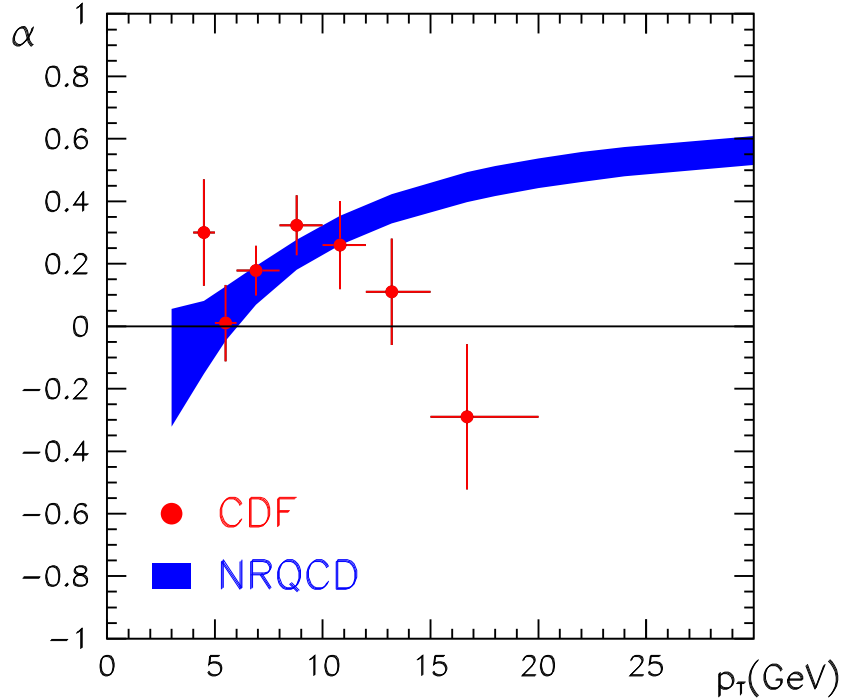}}
\caption[Polarization variable $\alpha$ for prompt $J/\psi$ at Run~I.]
{Polarization variable $\alpha$ for prompt $J/\psi$ at Run~I:  
	CDF data and the NRQCD prediction of Ref.~\cite{Braaten-K-L}.}
\label{chnine:psi1pol}
\end{figure}

\begin{figure}[ht]
\centerline{\epsfxsize 11cm \epsffile{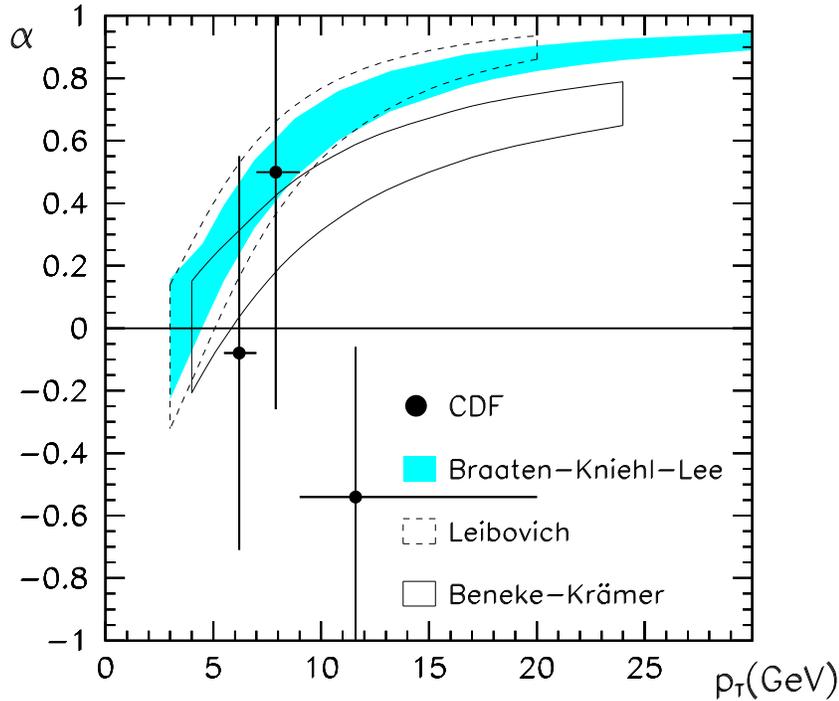}}
\caption[Polarization variable $\alpha$ for prompt $\psi(2S)$ at Run~I.]
{Polarization variable $\alpha$ for prompt $\psi(2S)$ at Run~I:  
	CDF data and the NRQCD predictions of 
	Refs.~\cite{Braaten-K-L} (shaded band), 
	\cite{Leibovich} (dashed box), and \cite{Beneke-Kramer} (solid box).}
\label{chnine:psi2pol}
\end{figure}

Defining the angle $\theta$  by the
direction of the $\mu^{+}$ with respect to the $\psi$
direction in the $\psi$ center of mass frame,
the normalized angular distribution $I(\theta)$ is given by
\begin{eqnarray} \label{Itheta}
  I(\theta) &=& \frac{1}{\sigma_L+\sigma_T}
\Big [\frac{3}{8}(1+\cos^2\theta) \; \sigma_T
+\frac{3}{4}\sin^2\theta \; \sigma_L \Big] \nonumber \\
&\equiv&  \frac{3}{2(\alpha+3)}(1+\alpha\cos^2\theta).
\end{eqnarray}
A convenient measure of the polarization of $J=1$ quarkonium states is 
therefore the  
variable $\alpha = (\sigma_T - 2 \sigma_L)/(\sigma_T + 2 \sigma_L)$,
where $\sigma_T$ and $\sigma_L$ are the cross sections for transversely 
and longitudinally polarized states, respectively.
In Figs.~\ref{chnine:psi1pol} and \ref{chnine:psi2pol},
we show the CDF measurements of $\alpha$ for prompt $J/\psi$ and 
prompt $\psi(2S)$ \cite{CDF-pol}.
In Fig.~\ref{chnine:psi2pol} for $\psi(2S)$,
the bands are the predictions from
the NRQCD analyses of Leibovich~\cite{Leibovich}, 
Beneke and Kramer~\cite{Beneke-Kramer}
and Braaten, Kniehl and Lee~\cite{Braaten-K-L}.  
The widths of the bands and the 
deviations between the predictions are indicators of the size of the 
theoretical errors.  The calculation of $\alpha$ for prompt $J/\psi$ 
is complicated by the need to take into account the feed-down from $\chi_{cJ}$.
The band in Fig.~\ref{chnine:psi1pol} is the prediction by 
Braaten, Kniehl and Lee~\cite{Braaten-K-L}.  For both prompt $J/\psi$ and 
prompt $\psi(2S)$, the theoretical prediction for $\alpha$ is small
around $p_T = 5$ GeV, but then increases steadily with $p_T$.
If the data is taken at its face value, it suggests exactly the opposite trend,
decreasing at the largest values of $p_T$ for which it has been measured.
There are many effects, such as higher order corrections, 
that could dilute the polarization or delay the onset of the rise in $\alpha$,
but the basic conclusion that $\alpha$ should increase to a positive value 
at large $p_T$ seems to follow inescapably from the NRQCD factorization 
approach. If this behavior is not observed in Run~II, 
it will be a serious blow to this  approach to quarkonium production.

\begin{figure}[ht]
\centerline{\epsfxsize 11cm \epsffile{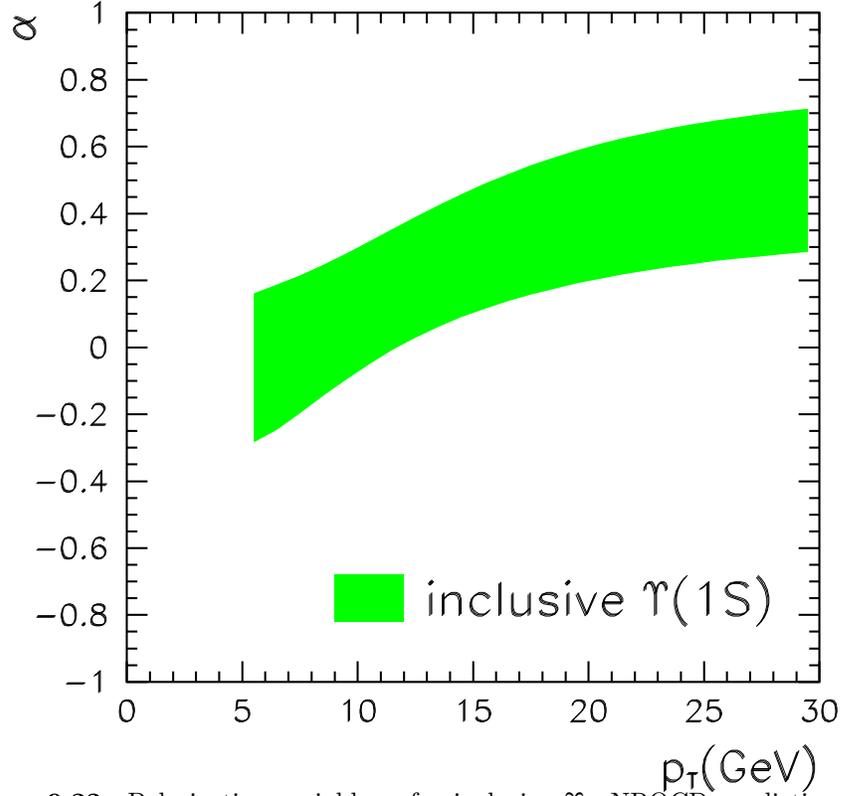}}
\caption[Polarization variable $\alpha$ for inclusive $\Upsilon$ at Run~I.]
{Polarization variable $\alpha$ for inclusive $\Upsilon$:
NRQCD prediction from Ref.~\cite{Braaten-Lee:upspol}.}
\label{chnine:ups1pol}
\end{figure}

In Ref.~\cite{Braaten-Lee:upspol},
the polarization variable $\alpha$ has also been calculated
for inclusive $\Upsilon(nS)$ using the matrix elements from the 
NRQCD analysis of Ref.~\cite{B-F-L}.
The prediction for inclusive $\Upsilon(1S)$ as a function of $p_T$ 
is shown in Fig.~\ref{chnine:ups1pol}.
The variable $\alpha$ is predicted to be small for $p_T$ less than 10 GeV,
but it increases steadily with $p_T$. 
The prediction is compatible with the CDF measurement for $p_T$
in the range from 8 GeV to 20 GeV,
which is $\alpha = 0.03 \pm 0.28$.
The data from Run~II should allow the prediction to be tested.

\subsection[Polarization in quarkonium production]
{Polarization in quarkonium production
\authorfootnote{Authors: G. Feild, K. Sumorok, W. Trischuk}}

\newcommand{\decays}{\rightarrow}
\newcommand{\lxy}{\mbox{$L_{xy} \ $}}
\newcommand{\ctau}{\mbox{$c\tau \ $}}
\newcommand{\ptmin}{\mbox{$P_T^{min} \ $}}
\renewcommand{\Pt}{\mbox{$P_T \ $}}
\newcommand{\Et}{\mbox{$E_T \ $}}
\renewcommand{\etal}{\mbox{\it et al.,}}
Measurements of the $J/\psi$ and $\psi(2S)$ differential cross sections
by CDF~\cite{CDF_TroyD} separate the cross sections into those $\psi$ mesons
coming from $b$-flavored hadron production and those originating from
prompt production mechanisms. This separation is made by 
reconstructing the decay vertex of the $J/\psi$ and $\psi(2S)$ mesons 
using the CDF Silicon Vertex detector (SVX).
The fraction of $J/\psi$ mesons coming from $b$-flavored hadron production
increases from 15$\%$ at 5 GeV/$c$ to 40$\%$ at 18 GeV/$c$
$P_{T}^{J/\psi}$. For $\psi(2S)$ mesons, 
a similar increase is seen in the range
from 5 to 14 GeV/$c$. The prompt production of $J/\psi$ mesons has
three components: a feed-down from $\chi_{c}$ production,
a feed-down from $\psi(2S)$ and a direct component. 
Only the latter occurs in the production of $\psi(2S)$ mesons. 
CDF has measured the fraction of prompt $J/\psi$ mesons 
coming from $\chi_c$ production to be $(29.7\pm1.7\pm5.7)\%$ 
of the total prompt production \cite{CDF_chi}.  From the measured 
$\psi(2S)$ production cross section, the fraction of the
prompt $J/\psi$'s feeding down from $\psi(2S)$ meson decays is calculated to be
$7\pm2\%$ at $P_{T}^{J/\psi}$ =  5 GeV/$c$ and $15\pm5\%$ at $P_{T}^{J/\psi}$
= 18 GeV/$c$. The fraction of directly produced $J/\psi$ mesons 
is $64\pm6\%$, independent of $P_{T}^{J/\psi}$ between 5 and 18 GeV/$c$. 

\par
Having separated out the prompt $J/\psi$ mesons coming from $\chi_c$,
both direct/prompt $J/\psi$ and $\psi(2S)$ production appear
to be $\approx50$ times higher than color-singlet model (CSM)
predictions~\cite{CSM}.

\subsubsection{NRQCD predictions for $\psi$ meson polarization}

\par 
Calculations based on the NRQCD factorization
formalism are able to explain this anomalous 
production~\cite{ChoLeibovich,C-G-M-P}.
The model increases the prompt production cross section by including
color-octet $c\overline{c}$ states $^{1}S_{0}^{(8)}$, $^{3}P_{0}^{(8)}$
and $^{3}S_{1}^{(8)}$ in the hadronization process. 
At leading order in $\alpha_s$, the color-singlet term 
in the parton cross section $d \hat \sigma /d p_T^2$ has
a $1/p_{T}^{8}$ dependence at large $p_T$, 
the $^{1}S_{0}^{(8)}$ and $^{3}P_{0}^{(8)}$
terms have a $1/p_{T}^{6}$ dependence, and the $^{3}S_{1}^{(8)}$ term 
has a $1/p_{T}^{4}$ dependence. This last term,  
which corresponds to the fragmentation of an almost on-shell gluon
into a $c \bar c$ pair, dominates at high $P_{T}$. 
An on-shell gluon is transversely polarized and NRQCD predicts
this polarization is transfered to the $J/\psi$ or $\psi(2S)$ mesons
in the final state.

\par
A consequence of this mechanism  is that the direct prompt $J/\psi$ mesons
and the $\psi(2S)$ mesons 
will approach 100$\%$ transverse  polarization at leading order
in $\alpha_s$ for transverse momenta $p_{T} \gg m_{c}$
where $m_{c}$ is the charm quark mass~\cite{Cho-Wise,Beneke-Rothstein}.
The observation of such a polarization
would be an indication of the NRQCD hadronization mechanism.                                                     

\subsubsection{CDF polarisation measurements from Run~I}
\index{quarkonium!polarization!CDF}
\index{charmonium!polarization!CDF}
\index{CDF!$J/\psi$ polarization}
\label{sec:cdf_pol}

\par
CDF has made a measurement of the production polarization of 
$J/\psi$ and $\psi(2S)$ (collectively $\psi$) mesons by analyzing decays
into $\mu^{+}\mu^{-}$~\cite{CDF-pol}.
Defining the angle $\theta$  by the
direction of the $\mu^{+}$ with respect to the $\psi$
direction in the $\psi$ center of mass frame,
the normalized angular distribution $I(\theta)$ is given by
Eq.~(\ref{Itheta}).
    
Unpolarized $\psi$ mesons would have $\alpha=0$, whereas $\alpha=+1$ and $-1$
correspond to fully transverse and longitudinal polarizations respectively.
The polarization parameter for prompt $\psi$ production can be separated from
the $B$-hadron decay component by fitting the proper decay length distribution
for $\psi$ candidates with both muons reconstructed in the SVX.
Fig.~\ref{J/psi} shows the fitted polarization of $J/\psi$ mesons obtained
by CDF from prompt production and $B$-hadron decay compared with an NRQCD
prediction for the prompt production~\cite{Braaten-K-L}. 
The prediction for $J/\psi$ meson polarization 
includes dilutions due to the  
contributions from $\chi_{c}$ and $\psi(2S)$ decays.  
Fig.~\ref{psi(2S)} shows the fitted polarization for $\psi(2S)$ mesons 
obtained by CDF from prompt production and $B$-hadron decay compared 
with NRQCD predictions for prompt production from Refs.~\cite{Beneke-Kramer}
and~\cite{Braaten-K-L}.

\begin{figure}[th]   
\centerline{\psfig{figure=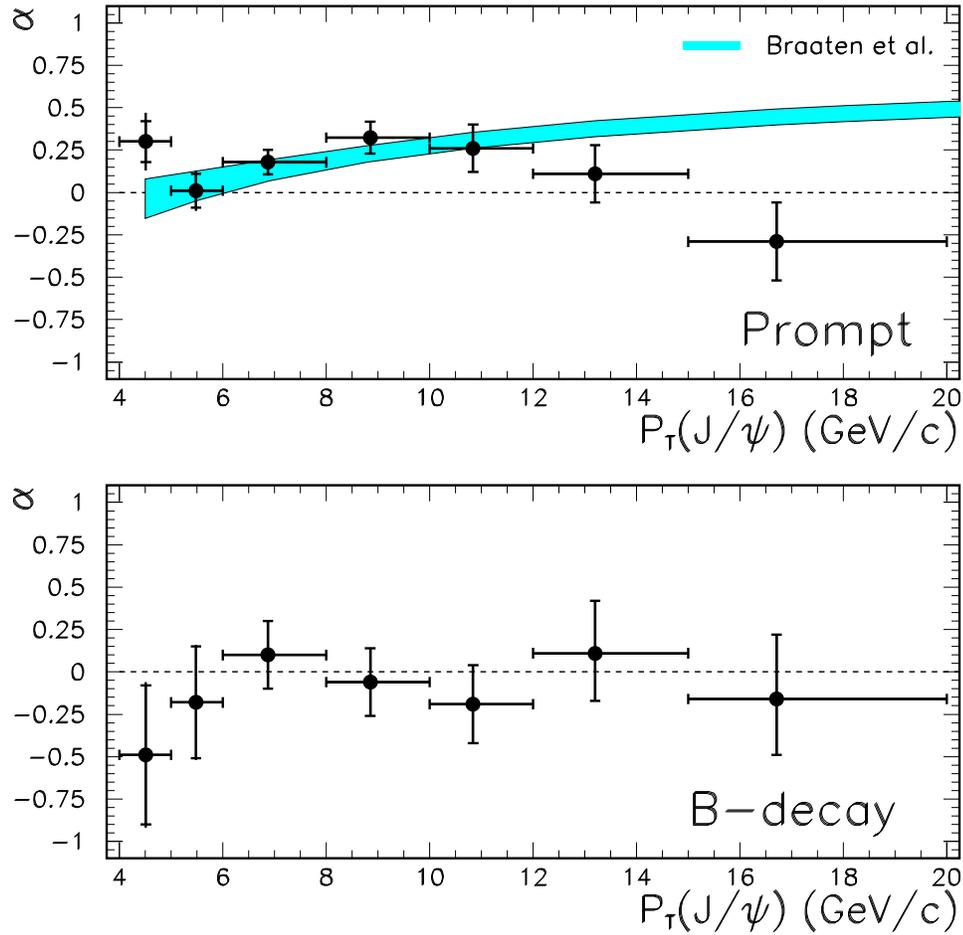,width=13cm}}
\caption[The fitted polarization of $J/\psi$ mesons from prompt production
 and $B$-hadron decay.]
{The fitted polarization of $J/\psi$ mesons from prompt production
 and $B$-hadron decay. 
The shaded curve shows an NRQCD prediction from Ref.~\cite{Braaten-K-L}.}
\label{J/psi}
\end{figure}     

\begin{figure}[htb]   
\centerline{\psfig{figure=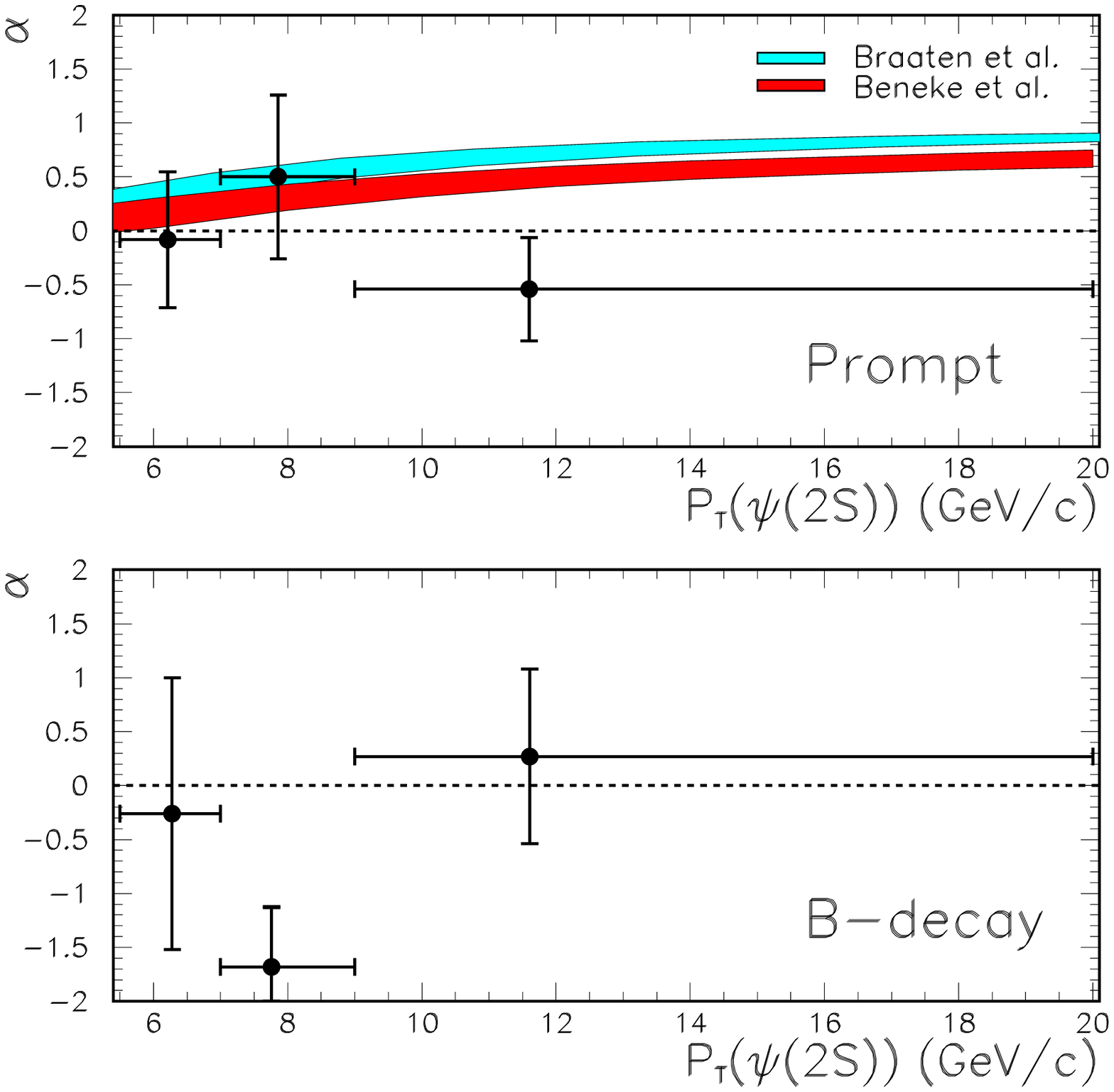,width=13cm}}
\caption[The fitted polarization of $\psi(2S)$ mesons from prompt production
 and $B$-hadron decay.]
{The fitted polarization of $\psi(2S)$ mesons from prompt production 
and $B$-hadron decay. The shaded curves are NRQCD predictions from
 Refs. \cite{Beneke-Kramer} and\cite{Braaten-K-L}.}
\label{psi(2S)}
\end{figure}   

\subsubsection {Polarization predictions for Run~II}

The measurements in Run~I were based on a luminosity of 110~pb$^{-1}$.
Estimates for Run~II are made assuming a factor of 20 increase in
integrated luminosity, or 2~fb$^{-1}$.
CDF is also planning to increase the yield of its dimuon trigger by
50$\%$ by lowering the $p_{T}$ threshold. In addition, the SVX will have
increased coverage along the beam line leading to better acceptance for
dimuons fully reconstructed in the SVX. This is crucial for separation
of prompt and $B$ hadron decay production in this measurement. 
Overall a factor of 50 increase in the effective statistics for Run~II 
is assumed in our projections. We have extrapolated to higher $p_{T}$
by using the fitted shape of the cross section below 20~GeV/$c$$^{2}$
in order to predict the number of events in a given $p_{T}$ bin.
We have estimated the errors by scaling the statistics from 
Run~I measurements.
In Run~I, systematic uncertainties on the polarization measurement were 
small compared to the statistical errors. We anticipate that the systematics
can be improved with the Run~II data samples, but even if this is not
the case systematics should still be negligible at the highest 
transverse momenta.    
Fig.~\ref{Run2 psi} shows the expected precision for a polarization 
measurement in Run~II of promptly produced $J/\psi$ and $\psi(2S)$ mesons. 
Relative to Run~I, the $p_{T}$ range should be extended
and the statistical precision on the measurement improved.  The points are
slightly scattered about zero to ease visibility. 
 
\begin{figure}[htbp]   
\centerline{\psfig{figure=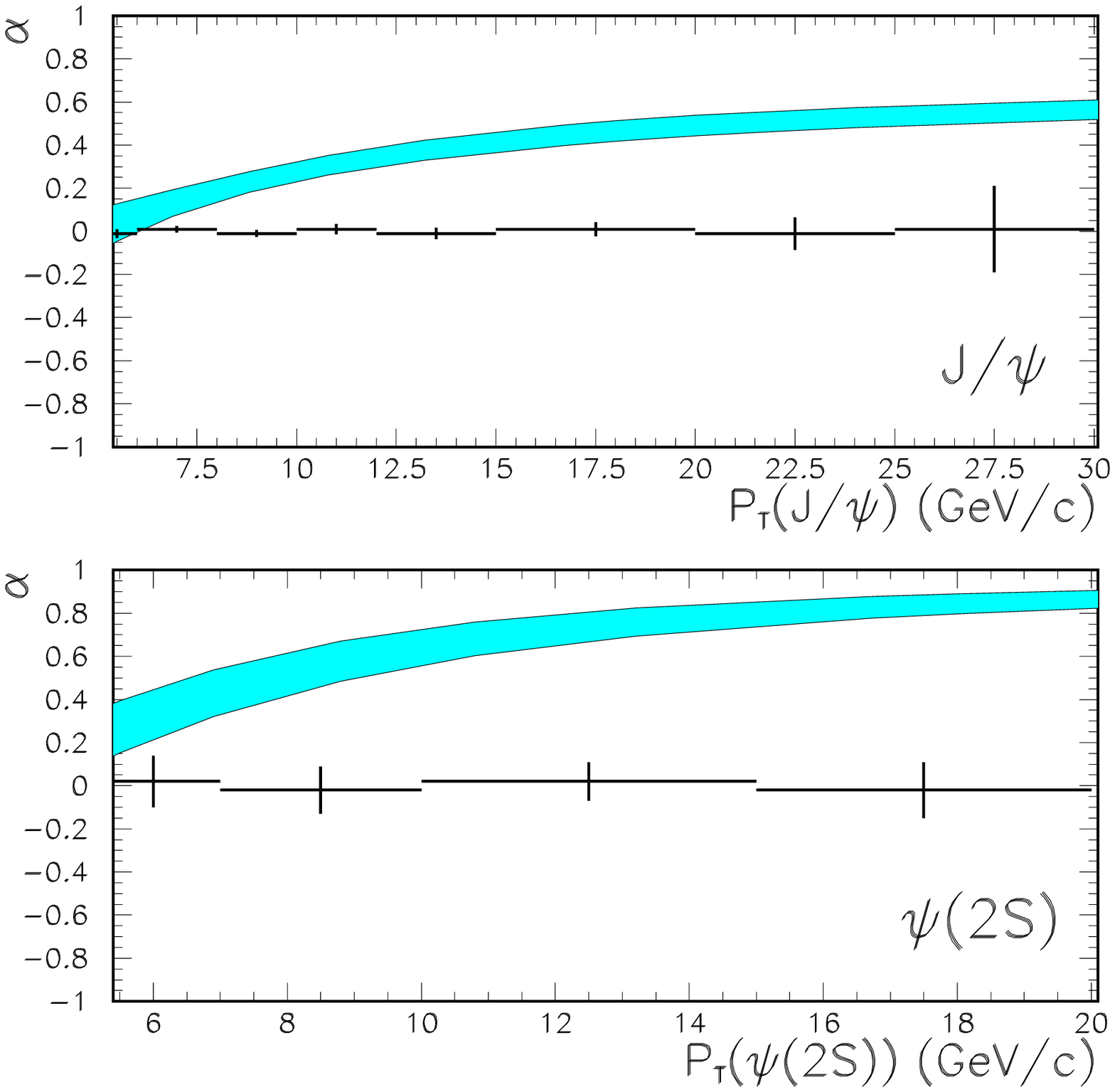,width=13cm}}
\caption[The expected precision for a polarization measurement in Run~II
of promptly produced $J/\psi$ and  $\psi(2S)$ mesons.]
{The expected precision for a polarization measurement in Run~II
of promptly produced $J/\psi$ and  $\psi(2S)$ mesons.
The shaded curves are NRQCD predictions from~\cite{Braaten-K-L}.}
\label{Run2 psi}
\end{figure}

\subsubsection{Upsilon production and polarization in CDF Run~II}

\index{quarkonium!polarization!CDF}
\index{bottomonium!polarization!CDF}
\index{upsilon@$\Upsilon$!polarization!CDF}
\index{CDF!$\Upsilon$ polarization}
The CDF Run~IB $\Upsilon$(1S) differential production cross section, 
measured as a function of transverse energy, is shown in 
Fig.~\ref{fig:ups1s}.  
As discussed in section~\ref{sec:bott_theory}, this cross section
can be described by including NRQCD color-octet matrix elements in
the fit to the 
data. However, due to the free parameters in the fit, 
this compatibility alone is not enough to demonstrate that the 
NRQCD description is correct.  As in the case of the $J/\psi$ and
$\psi$(2S) mesons, observation of transversely polarized $\Upsilon$(1S) 
production due to gluon fragmentation at high $p_T$ would provide 
further evidence
in favor of the NRQCD framework over other proposed models, 
such as the color-singlet model and the color-evaporation model.

The $\Upsilon$(1S) $\rightarrow \mu \mu$ 
production polarization 
for events in the transverse energy range 
8 GeV/$c$ $ < p_T(\Upsilon) < $ 20 GeV/$c$ 
has been determined in Run~IB
by measuring the muon decay angle in the 
$\Upsilon$ rest frame as described in section~\ref{sec:cdf_pol}.
This measurement is done by fitting longitudinally
and transversely polarized Monte Carlo templates to the data shown 
in Fig.~\ref{fig:upol1s}. In this case the longitudinal fraction
$\Gamma_L/\Gamma$ is measured to be $0.32 \pm 0.11$. The longitudinal
fraction can be related to the usual polarisation parameter, $\alpha$,
by: $\Gamma_L/\Gamma = ( 1 - \alpha ) \ ( 3 + \alpha)$. Thus our 
polarisation measurement in $\Upsilon$ decays
yields a value of $\alpha = 0.03 \pm 0.25$. Our data is compatible
with unpolarized production. This result does not contradict the predictions
of NRQCD as the transverse polarisation is only expected to be large
for transverse momenta $p_T \gg m_b$, where $m_b$ is
the $b$ quark mass. We are probably not probing sufficiently high
$p_T$ with our current data.

\begin{figure}[t]
\centerline{\psfig{figure=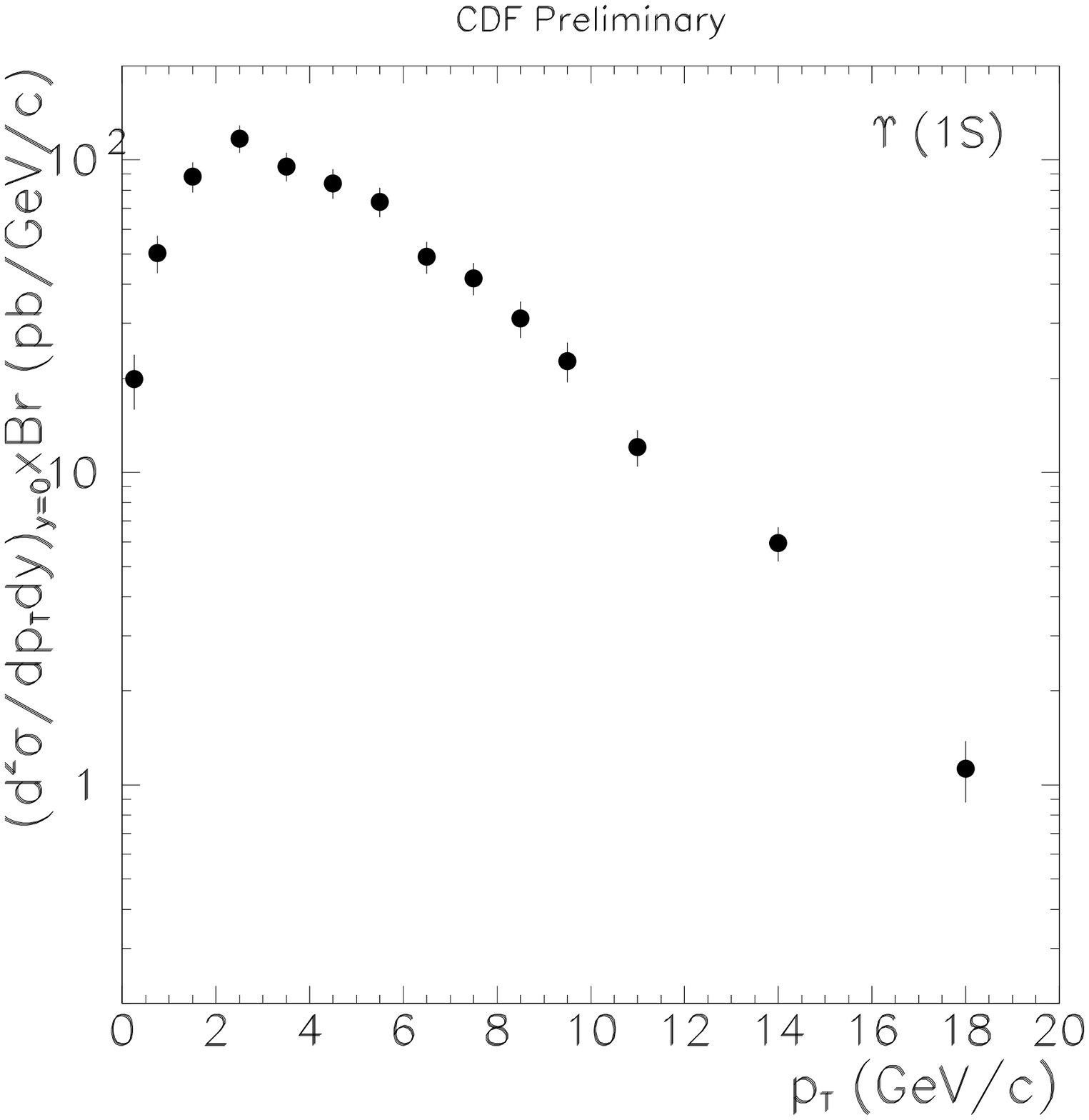,width=11cm}}
\caption{The differential cross-section for $\Upsilon(1S)$ production
measured by CDF in Run~I.}
\label{fig:ups1s}
\end{figure} 

\begin{figure}[thb]
\centerline{\psfig{figure=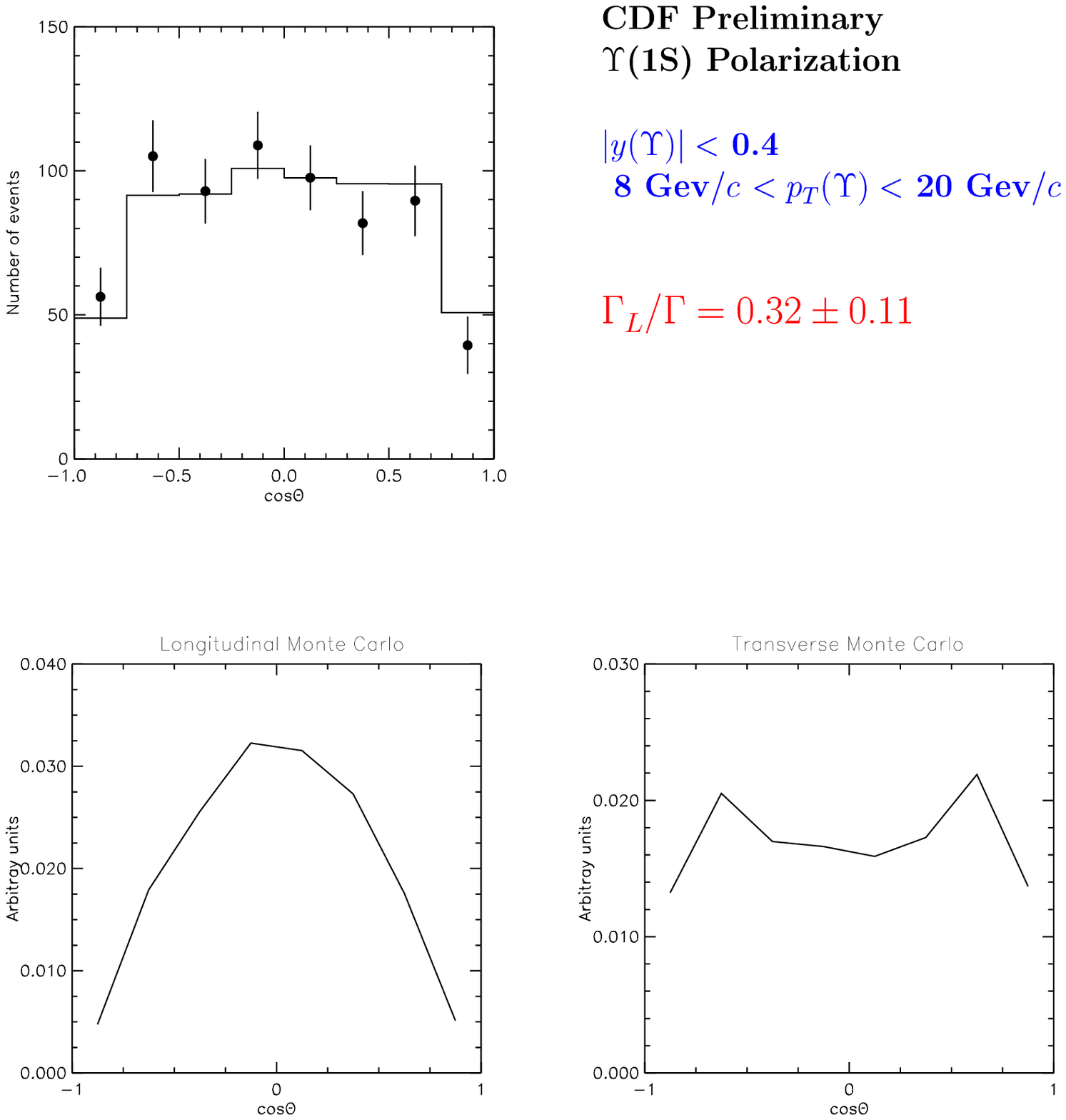,width=13cm}}
\caption{The polarisation measured in $\Upsilon(1S)$ decays by
CDF in Run~I.}
\label{fig:upol1s}
\end{figure} 

An extrapolation of the Run~IB cross section from 
Fig.~\ref{fig:ups1s}
yields no appreciable cross section for $ p_T(\Upsilon) >$ 20 GeV.   
However, assuming a factor 20 increase in data for Run~II, a polarization 
measurement statistically comparable to  that from Run~IB could be 
made in the transverse energy range 14 GeV $ < p_T(\Upsilon) < $ 20 GeV.
Conversely, the prediction from fits to the Run~IB data 
shown in Fig.~\ref{chnine:ups1plot}, 
predict a flattening of the cross section above 20 GeV.  It will be 
quite interesting if such events are seen in Run~II.

\def\D0{D\O}
\section{The $B_{\lowercase{c}}$ Meson}

\index{Bc meson@$B_c$ meson} 
The $B_c$ meson is the ground state of the $\bar b c$ system, 
which in many respects is intermediate between charmonium and bottomonium.
However because $\bar b c$ mesons carry flavor, 
they provide a window for studying heavy-quark dynamics
that is very different from the window provided by quarkonium. 
The observation of approximate 20 $B_c$ events 
in the $B_c\to J/\psi l\nu$ decay mode by the 
CDF-collaboration \cite{cdfbc} 
in Run~I of the Tevatron demonstrates that the experimental study 
of the $\bar b c$ meson system is possible.

\subsection[Spectroscopy]
{Spectroscopy\protect{\authorfootnote{Authors: E. Braaten, R. K. Ellis}}}

\index{bc system@$\bar{b}c$-system!spectroscopy} 
The $\bar b c$ system has a rich spectroscopy 
of orbital and angular-momentum excitations.
The predicted spectrum is shown in Fig.~\ref{grotrian}.
The only state that has been observed is the ground state $B_c$,
\begin{figure}[bht] 
\vskip 11cm
\includegraphics{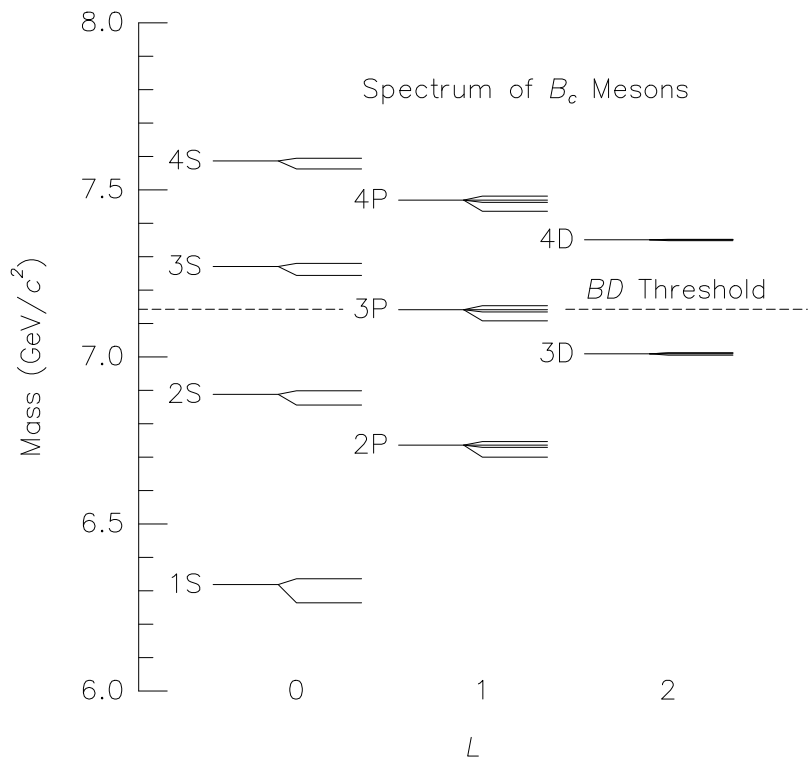}
\caption[Predicted spectrum for the $B_c$ mesons.]
{Predicted spectrum for the $\bar{b} c$ mesons \cite{Eichten:1994gt}.}
\label{grotrian}
\end{figure}
which was discovered by the CDF collaboration in Run~I
of the Tevatron~\cite{cdfbc}.  
They measured the mass to be $6.4 \pm 0.4$ GeV.
\index{Bc meson@$B_c$ meson!mass} 
The masses of the $B_c$ and the other states in the $\bar b c$ spectrum 
can be predicted using potential models whose parameters are tuned to 
reproduce the spectra of the observed charmonium and bottomonium states
\cite{Gershtein:1988jj,Kwong:1991,Chen:1992fq,Eichten:1994gt,Kiselev:1995rc,
Baldicchi:2000cf}. 
The range of the resulting predictions for the $B_c$ mass is 
$6.24 \pm 0.05$ GeV.  The first excited state is the spin-1 state $B_c^*$, 
which is predicted to be heavier by $78 \pm 13$ MeV. 
The mass of the $B_c$ has also been calculated using lattice gauge theory
to be $6.4 \pm 0.1$, where the uncertainty is dominated by the
error from the omission of dynamical quarks \cite{Shanahan:1999mv}.  

The majority of the $B_c$ mesons produced at the Tevatron are produced
indirectly via the decay of excited $\bar b c$ mesons.  
The excited states cascade down through the spectrum via a sequence of
hadronic and electromagnetic transitions, until they reach 
the ground state $B_c(1S)$, which decays via the weak interaction
\cite{Eichten:1994gt}. The $\bar b c$ states with the best
prospects for discovery at the Tevatron are those that
decay into $B_c + \gamma$.  The discovery of the $B_c^*$ is made 
difficult by the very low energy of the photon ($ \sim 80$ MeV).
The first radial excitation $B_c^*(2S)$ decays into $B_c + \gamma$
with a much more energetic photon ($ \sim 600$ MeV),
but a smaller fraction of the decay chains that end in $B_c$
will have the $B_c^*(2S)$ as the next-to-last step.
States in the $\bar b c$ spectrum could also be discovered via 2-pion 
transitions into the $B_c$.  The most promising is the 
first radial excitation $B_c(2S)$, whose mass is higher 
than that of the $B_c^*$ by about 600 MeV.
 

\subsection[Production] 
{Production\footnote{Authors: Eric Braaten, A. Likhoded}} 
\index{Bc meson@$B_c$ meson!production} 
\label{ProB_c}

One can think of the production of a $B_c$ as proceeding in three steps.  
First, a $\bar b$ and $c$
with small relative momentum are created by a parton collision.  Second, the
$\bar b$ and $c$ bind to form the $B_c$ or one of its excited states below the
$BD$ threshold.  Third, the excited states all cascade down to the ground state
$B_c$ via hadronic or electromagnetic transition.  Thus, the total production
cross section for $B_c$ is the sum of the direct production cross sections for
$B_c$ and its excited states.

\index{NRQCD} 
The direct production of the $B_c$ and other $\bar b c$ mesons
can be treated within the NRQCD factorization framework described in
Section~\ref{Quarkonium_production}.
The cross section for the direct production of a $\bar b c$ meson $H$ 
can be expressed in the form of Eqs.~(\ref{chnine:partonmodel}) 
and (\ref{chnine:partoncross}), with $b \bar b(n)$ replaced by $\bar b c(n)$.
The short-distance cross section $d\hat{\sigma}[ij\to  \bar b c(n)+X]$
\index{short distance cross section} 
for creating the $\bar b c$ in the color and angular-momentum state $n$
can be calculated as a perturbative expansion in $\alpha_s$ 
at scales of order $m_c$ or larger.  The nonperturbative matrix element
$\langle O^H(n)\rangle$ encodes the probability for a $\bar b c$ 
in the state $n$ to bind to form the meson $H$. 
The matrix element scales as a definite power 
of the relative velocity $v$ of the charm quark.
For S-wave states, the leading color-octet matrix element 
is suppressed by $v^4$ relative to the leading color-singlet matrix element.
For P-wave states, the leading color-singlet matrix element 
and the leading color-octet matrix element are both
suppressed by $v^2$ relative to the leading color-singlet matrix element
for S-waves.

The production mechanisms for $\bar b c$ differ in an essential way 
from those for $b \bar b$, because two heavy quark-antiquark pairs 
must be created in the collision.  
While a $b \bar b$ pair can be created at order $\alpha_s^2$
by the parton processes $q \bar q, g g \to  b \bar b$,
the lowest order mechanisms for creating $\bar b c$ are the 
order-$\alpha_s^4$ processes $q \bar q, g g \to (\bar b c) + b \bar c$.
At the Tevatron, the $gg$ contribution dominates.  
The parton process $g g \to (\bar b c) + b \bar c$ can create the 
$\bar b c$ in either a color-singlet or color-octet state.
We expect the cross section for color-octet $\bar b c$
to be about a factor of 8 larger than for color-singlet $\bar  b c$,
just from counting the color states.  This factor of 8 can partially
compensate any suppression factors of $v$ from the probability for the
color-octet $\bar b c$ to bind to form a meson.
However, unlike the case of $\bar b b$, there is no dynamical enhancement
of color-octet $\bar b c$ at low $p_T$ or at high $p_T$.
Color-octet production processes should therefore be less important 
for $\bar b c$ mesons than for quarkonium.

All existing calculations for the cross sections of $\bar b c$ mesons 
\index{color singlet model} 
have been carried out within the color-singlet model.  
The $\bar b c$ is assumed to be created in a color-singlet state 
with the same angular-momentum quantum numbers as the meson.
The appropriate long-distance matrix elements can be determined 
from the radial wavefunctions of the mesons as in Eqs.~(\ref{chnine:a1})
and (\ref{chnine:a2}).  For the $B_c$ and the first excited state $B_c^*$,
the matrix element is proportional to $|R_{1S}(0)|^2$.
Since the wavefunctions are known from potential models,
the cross sections for $b \bar c$ mesons in the color-singlet model 
are absolutely normalized.

\begin{figure}[t] 
\vskip 11cm
\includegraphics{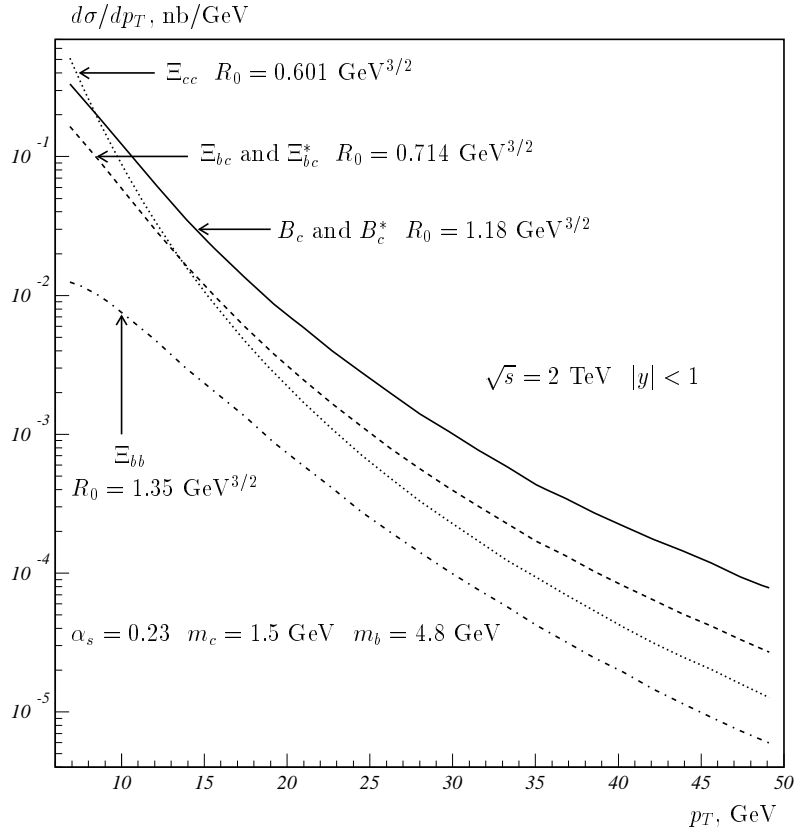}
\caption{Differential cross sections for the $B_c$ meson
	and doubly-heavy baryons.}
\label{tevpic}
\end{figure}

\begin{figure}[ht] 
\vskip 11cm
\includegraphics{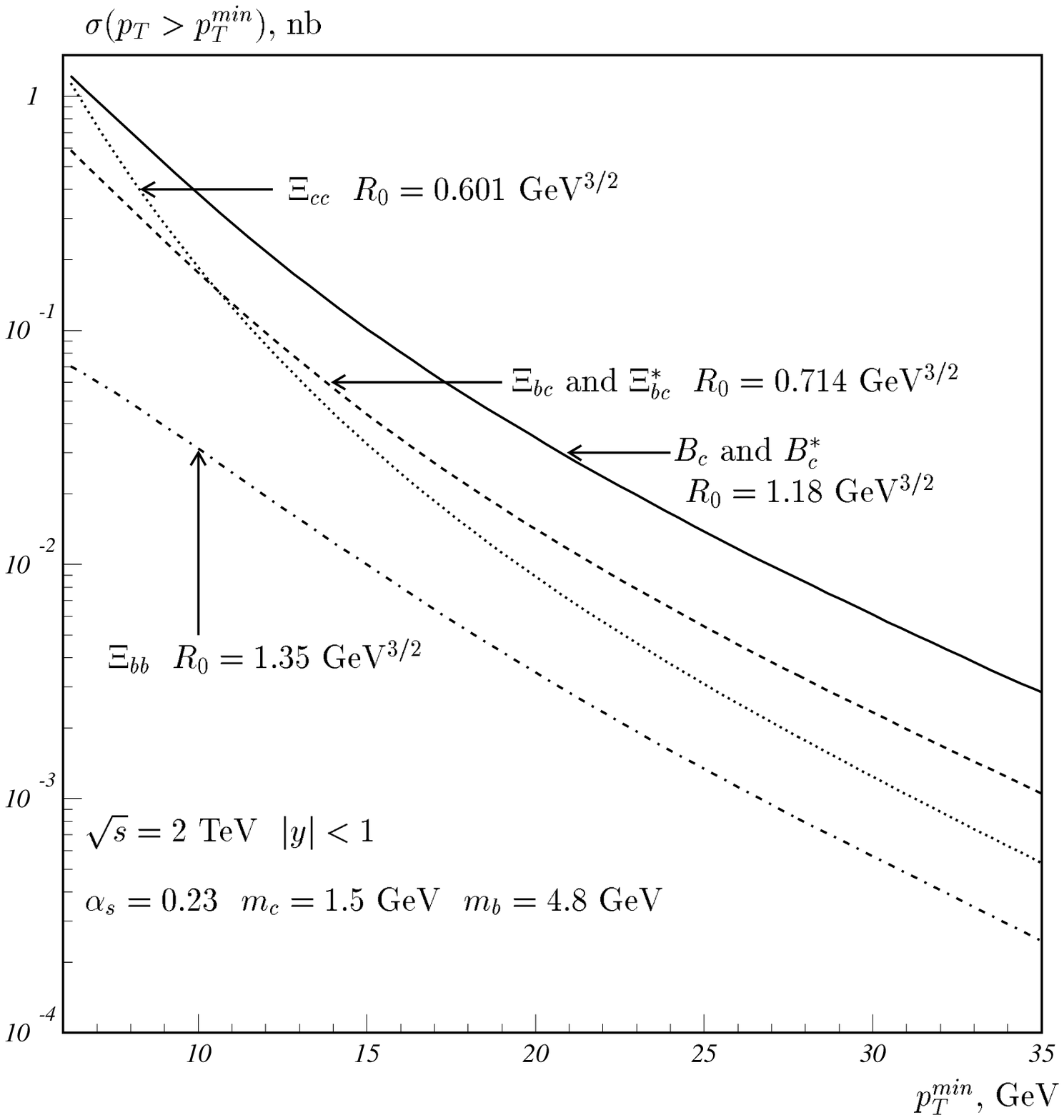}
\caption{Integrated cross sections for the $B_c$ meson
	and doubly-heavy baryons.}
\label{tevpic1}
\end{figure}

The production of $\bar b c$ mesons in the color-singlet model
at leading order in $\alpha_s$ was
studied in detail in the series of papers~\cite{prod}. 
The cross sections are proportional to $\alpha_s^4(\mu)$
and to a wavefunction factor, which is $|R_{nS}(0)|^2$ for S-waves 
and $|R_{nP}'(0)|^2$ for P-waves.
The largest uncertainties in the theoretical predictions arise from the factor
$\alpha_s^4(\mu)$.  There is a large ambiguity in the choice 
of the scale $\mu$, since the short-distance process involves several scales, 
including $m_c$, $m_b$, and $p_T$.
For example, if the scale $\mu$ is varied from $m_c$ up to $2(m_c+m_b)$,
the $\bar b c$ cross-section changes by a factor of 7.
There are additional ambiguities from the 
wave function factors and from the $c$-quark mass, but the resulting 
uncertainties are less than a factor of 2. 

The result of the order-$\alpha_s^4$ color-singlet model calculation for
${d\sigma}/{dp_T}$ of the $B_c$ meson at the Tevatron at 1.8 TeV
are presented in Fig.~\ref{tevpic}. The cross section integrated 
over $p_T$ greater than ${p_T}_{min}$ is shown in Fig.~\ref{tevpic1}.
The prediction includes the feeddown from $B_c^*$, but not from any of the 
higher $\bar b c$ states. 
The cross sections are integrated over $|y|<1$.
In our calculations we used the following values of parameters.
The quark masses were taken to be $m_c=1.5$ GeV and $m_b=4.8$ GeV.
The QCD coupling constant was frozen at the value $\alpha_s=0.23$,
which describes the experimental data of the OPAL Collaboration on the
production of additional $c\bar c$-pairs in 
$e^+e^-$-annihilation~\cite{prod}.
The radial wavefunction at the origin for the $B_c$ and $B_c^*$ was
taken to be $R_{1S}(0)=1.18$ GeV$^{3/2}$.
The cross-section for $B_c$ mesons
is roughly three orders of magnitude smaller than that for $B$ mesons 
due to the presence of two heavy quark-antiquark pairs in the final state. 

To estimate the inclusive $B_c$ cross section, we must take into account
\index{Bc meson@$B_c$ meson!production!cross section} 
the feeddown from all the $\bar b c$ mesons below the $BD$ threshold,
all of which eventually cascade down into the $B_c$ via hadronic 
or electromagnetic transitions.
Including the feeddown from all the higher $\bar b c$ states
and integrating over $p_T>6$ GeV and $|y|<1$,
the inclusive cross section for $B_c$ is predicted to be
\begin{equation}
\sigma_{th}(B^+_c)=2.5\;\;\mbox{nb},
\end{equation}
with an error that is roughly a factor of 3.
This should be compared with the cross section 
obtained from the experimental result\cite{cdfbc}, 
using  the value $(2.5\pm 0.5)\times 10^{-2}$ for 
the branching fraction for $B^+_c\to J/\psi +l\nu$, 
\begin{equation}
\sigma_{exp.}\sim 10\pm 6~\mbox{nb}\; .
\end{equation}

The $B_c$ cross section should be measured much more accurately in Run~II.
If the cross section proves to be significantly higher than 
predicted by the color-singlet model, it may indicate that 
color-octet production mechanisms are also important.

\subsection[Theory of $B_c$ Decays]
{Theory of $B_c$ Decays\footnote{Authors: V.V.Kiselev, A.K.Likhoded}}
\label{KisLik}

Decays of the long-lived heavy meson $B_c$, which contains 
\index{Bc meson@$B_c$ meson!decay} 
two heavy quarks of different flavors, were considered 
in the pioneering paper written by Bjorken
in 1986 \cite{BJ}. Bjorken's report gave a unified view of the decays of
hadrons with heavy quarks: mesons and baryons with a single heavy quark,
the $B_c$ meson, and baryons with two and three heavy quarks. A major 
effort was recently directed to study the long-lived doubly-heavy hadrons 
using modern understanding of QCD dynamics in the weak decays of heavy flavors.
The modern theoretical tools are the Operator Product Expansion, sum rules of
QCD and NRQCD, and potential models adjusted using data from 
hadrons with a single heavy quark. Surprisingly, Bjorken's estimates of
total widths and various branching fractions are close to what is evaluated in
a more strict manner. 

Various hadronic matrix elements enter in the description of weak decays. 
Measuring the lifetimes and branching ratios therefore
provides information about nonperturbative QCD interactions.
This is important in the determination of electroweak parameters,
such as the quark masses and the mixing angles in the CKM matrix, 
which enter into constraints on the physics beyond the Standard Model. 
The accumulation of more data on hadrons with heavy quarks will
provide greater accuracy and confidence in our understanding of the 
QCD dynamics that is necessary to isolate the electroweak parameters. 

A new laboratory for such investigations is 
\index{Bc meson@$B_c$ meson!lifetime} 
the doubly-heavy long-lived quarkonium $B_c$ recently observed 
for the first time by the CDF Collaboration \cite{cdfbc}.
The measured $B_c$ lifetime is equal to
\begin{equation}
\tau[B_c] = 0.46^{+0.18}_{-0.16}\pm 0.03\; {\rm ps}\;.
\end{equation}

In studying the $B_c$ meson, we can take advantage of two features that it 
has in common with the $\bar b b$ and $\bar c c$ quarkonia:
the nonrelativistic motion of the $\bar b$ and $c$ quarks 
and the suppression of the light quark-antiquark sea. 
These two physical conditions imply two small expansion parameters for $B_c$:
the relative velocity $v$ of quarks and
the ratio $\Lambda_{QCD}/m_Q$ of the confinement scale to the heavy quark mass.
The double expansion in $v$ and $1/m_Q$ generalizes the HQET approach
\cite{HQET,Bigi:1997fj} to what is called NRQCD \cite{NRQCD}.  The energy
release in heavy quark decays determines the $1/m_Q$ parameter 
to be the appropriate quantity for the operator product expansion (OPE) 
and also justifies the use of potential models (PM) in the calculations 
of hadronic matrix elements. The same arguments ensure the 
applicability of sum rules (SR) of QCD and NRQCD.

The $B_c$ decays were first calculated using potential models \cite{PM}.
\index{potential models} 
Various models gave similar estimates after adjusting parameters to 
reproduce the semileptonic decay rates of $B$ mesons. 
\index{operator product expansion} 
The OPE evaluation of inclusive decays gave values for 
the lifetime and inclusive widths that agreed with 
the sum of the dominant exclusive modes predicted by the potential models. 
Quite unexpectedly, QCD sum rules gave values for the
semileptonic $B_c$ widths \cite{ColNar} that were an order of
magnitude smaller than predicted by the potential models and by the OPE. 
The reason for this was that Coulomb-like $\alpha_s/v$ corrections 
had been neglected
in the sum rule calculations.  These corrections can be taken into account by 
summing up $\alpha_s/v$ corrections to all orders.  They are
significant both for heavy quarkonia and for the $B_c$ \cite{KT,KL}. 
At present, all three approaches give similar results for the lifetime 
and inclusive decay modes of the $B_c$ for similar sets of parameters. 
Nevertheless, various dynamical questions remain open:
\begin{itemize}
\item
What is the appropriate renormalization scale for
the nonleptonic weak Lagrangian, 
which basically determines the lifetime of the $B_c$?
\item
What are the values of masses for the charmed and beauty quarks?
\item
What are the implications of NRQCD symmetries for the form factors of $B_c$
decays and partial widths?
\item
How consistent is our understanding of the hadronic matrix elements
that characterize $B_c$ decays with the data on the other heavy hadrons?
\end{itemize}
In the present short review of $B_c$ decays, we summarize the theoretical
predictions and discuss how direct experimental measurements can answer the
questions above.

\subsubsection{Lifetime and inclusive decay rates}

The $B_c$-meson decay processes can be subdivided into three classes:
\index{spectator model}
\begin{itemize}
\item 
the $\bar b$-quark decay with a spectator $c$-quark,

\item 
the $c$-quark decay with a spectator $\bar b$-quark and

\item
the annihilation decays
$\bar b c\rightarrow l^+\nu_l,c\bar s, u\bar s$, where $l=e,\; \mu,\; \tau$.
\end{itemize}
In the $\bar b \to \bar c c\bar s$ decays, one separates also the
{\it Pauli interference} with the $c$-quark from the initial state. In
accordance with the given classification, the total width is the sum over the
partial widths from $\bar b c$ annihilation, $\bar b$ decay, $c$ decay,  
and Pauli interference.
\index{Pauli interference}

The annihilation width is the sum of the widths from the annihilation of 
the $\bar b c$ into leptons and quarks.  In the width from annihilation
into quarks, one must take into account the hard-gluon corrections
to the effective four-quark interaction of weak currents, which
give an enhancement factor $a_1=1.22\pm 0.04$. 
The nonperturbative effects of QCD can be absorbed into the
leptonic decay constant $f_{B_c}\approx 400$ MeV. 
This estimate of the contribution from annihilation into quarks 
does not depend on a hadronization model, since a large
energy release of the order of the meson mass takes place. 
Because of helicity suppression, the decay width 
is proportional to the square of the masses of 
leptons or quarks in the final state. 
Thus the only important annihilation channels are
$\bar b c \to \tau^+\nu_\tau$ and $\bar b c \to c \bar s$.

\begin{table}[th]
\begin{center}
\begin{tabular}{|l|c|c|c|}
\hline
$B_c$ decay mode & OPE, \%  & PM, \% & SR, \%\\
\hline
$\bar b\to \bar c l^+\nu_l$ & $3.9\pm 1.0$  & $3.7\pm 0.9$  & $2.9\pm 0.3$\\
$\bar b\to \bar c u\bar d$  & $16.2\pm 4.1$ & $16.7\pm 4.2$ & $13.1\pm 1.3$\\
$\sum \bar b\to \bar c$     & $25.0\pm 6.2$ & $25.0\pm 6.2$ & $19.6\pm 1.9$\\
$c\to s l^+\nu_l$           & $8.5\pm 2.1$  & $10.1\pm 2.5$ & $9.0\pm 0.9$\\
$c\to s u\bar d$            & $47.3\pm 11.8$& $45.4\pm 11.4$& $54.0\pm 5.4$\\
$\sum c\to s$               & $64.3\pm 16.1$& $65.6\pm 16.4$& $72.0\pm 7.2$\\
$B_c^+\to \tau^+\nu_\tau$   & $2.9\pm 0.7$  & $2.0\pm 0.5$  & $1.8\pm 0.2$\\
$B_c^+\to c\bar s$          & $7.2\pm 1.8$  & $7.2\pm 1.8$  & $6.6\pm 0.7$\\
\hline
\end{tabular}
\end{center}
\caption[Branching ratios of $B_c$ decay modes]
{Branching ratios of $B_c$ decay modes calculated using
the operator product expansion (OPE) approach, 
by summing up the exclusive modes from potential models (PM) \cite{PM,KL}, 
and by using sum rules (SR) of QCD and NRQCD \cite{KL}. }
\label{inc}
\end{table}

\begin{figure}[th]
\setlength{\unitlength}{1mm}
\begin{center}
\begin{picture}(130,90)
\put(0,5){\epsfxsize=12cm \epsfbox{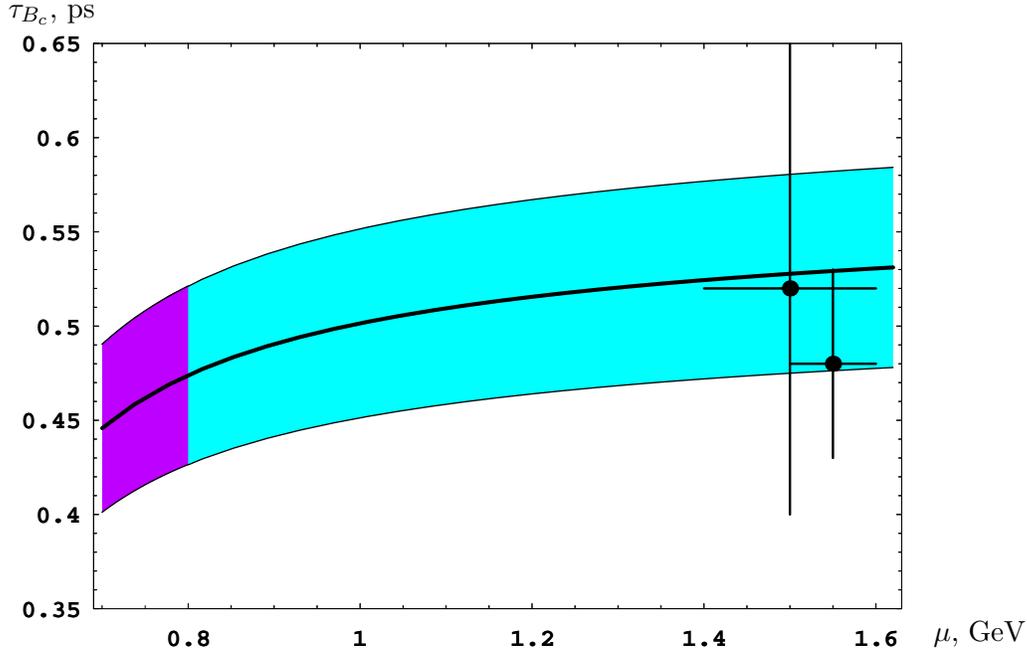}}
\put(0,90){$\tau_{B_c} $, ps}
\put(123,7){$\mu$, GeV}
\end{picture}
\end{center}
\caption[The $B_c$ lifetime calculated in QCD sum rules.]
{The $B_c$ lifetime calculated in QCD sum rules versus the scale of
hadronic weak Lagrangian. The shaded region shows the uncertainty of estimates,
the dark shaded region is the preferable choice as given by the lifetimes of
charmed mesons. The points represent the values in OPE approach taken from
Ref.~\cite{OPE}.}
\label{life}
\end{figure}

For the non-annihilation contributions to the width of the $B_c$, 
\index{operator product expansion} 
one can apply the {\it operator product expansion} (OPE)
for the quark currents of weak decays \cite{OPE}.  One takes
into account the $\alpha_s$-corrections to the free quark decays and uses 
quark-hadron duality for the final states. Then one considers the matrix
element for the transition operator in the bound meson state. The latter
allows one also to take into account the effects caused by the motion and
virtuality of decaying quark inside the meson because of the interaction with
the spectator. In this way the $\bar b\to \bar c c\bar s$ decay mode turns out
to be suppressed almost completely due to the Pauli interference with the charm
quark from the initial state. The $c$-quark decays with a spectator
$\bar b$ are also suppressed compared to the decay of a free $c$-quark,
because of the large binding energy of the initial state. 
Possible effects of interference between the leading-order weak
amplitudes and the penguin corrections in $B_c$ decays were considered 
in the framework of OPE in Ref.~\cite{chang3}, and these corrections were 
estimated to be about 4\%.

To calculate inclusive widths in the potential model approach, 
it is necessary to sum up the widths of exclusive decay modes \cite{PM}. 
For semileptonic decays via the transition $\bar b \to \bar c l^+\nu_l$, 
one finds that the hadronic final state is almost completely saturated 
by the most deeply bound states in the $\bar c c$ system, 
{\it i.e.}~by the $1S$ states $\eta_c$ and $J/\psi$. 
For semileptonic decays via the transition $c\to s l^+\nu_l$, 
one finds that the only $\bar b s$ states that can enter the 
accessible energy gap are the  $B_s$ and $B_s^*$. 
Furthermore, the $\bar b\to \bar c u\bar d$ channel, for example, 
can be calculated by multiplying the decay width for
$\bar b \to \bar c l^+\nu_l$ by a color factor and by taking into account
hard-gluon corrections to the four-quark interaction. 
It can be also obtained as a sum over the widths of decays 
involving specific $u\bar d$ bound states. 

The calculations of the total $B_c$ width in the inclusive OPE approach and
the exclusive potential model approach give consistent values, if one 
takes into account the largest uncertainty, which comes from the quark masses 
(especially the charm quark).  The final result is
\begin{equation}
\tau(B_c)= 0.55\pm 0.15\; \mbox{ps,}
\end{equation}
which agrees with the measured value of the $B_c$ lifetime.

The OPE estimates of inclusive decay rates agree with recent calculations in
the sum rules of QCD and NRQCD \cite{KL}, where one assumed the saturation
of hadronic final states by the ground levels in the $c\bar c$ and $b \bar s$
systems as well as the factorization that allows one to relate the semileptonic
and hadronic decay modes. The Coulomb-like corrections in the heavy quarkonia
states play an essential role in the $B_c$ decays and allow one to remove the
disagreement between the estimates in sum rules and OPE. In contrast to OPE,
where the basic uncertainty is given by the variation of heavy quark masses,
these parameters are fixed by the two-point sum rules for bottomonia and
charmonia, so that the accuracy of sum rule calculations 
for the total width of $B_c$ is determined by the choice of scale $\mu$ 
for the hadronic weak Lagrangian in decays of charmed quark. 
We show this dependence in Figure \ref{life}, where
$m_c/2 < \mu < m_c$ and the dark shaded region corresponds to the
scales preferred by data on the charmed meson lifetimes.
Taking the preferred scale in the $c\to s$ decays of $B_c$
to be equal to $\mu^2_{B_c} \approx (0.85\; {\rm GeV})^2$ and setting
$a_1(\mu_{B_c}) =1.20$ in the charmed quark decays, we predict 
the lifetime to be\cite{KL}
\begin{equation}
\tau(B_c) = 0.48\pm 0.05\;{\rm ps.}
\end{equation}

\subsubsection{Semileptonic and leptonic modes}

\index{Bc meson@$B_c$ meson!decay!leptonic} 
\index{Bc meson@$B_c$ meson!decay!semileptonic} 
The semileptonic decay rates were underestimated in the 
QCD sum rule approach of Ref. \cite{ColNar}, because large Coulomb-like 
corrections were not taken into account. 
The recent sum rule analysis in \cite{KT,KL} 
decreased the uncertainty, so that the estimates agree with the calculations 
in the potential models.
The widths and branching fractions calculated using QCD sum rules are 
presented in Table \ref{semi}.  The expected accuracy is about 10\%.
In practice, the most important semileptonic decay mode is to the $J/\psi$
which is easily detected in experiments via its leptonic decays \cite{cdfbc}. 

The estimates for exclusive semileptonic decay rates
of the $B_c$ into $1S$ charmonium states
obtained from QCD sum rules agree with the values obtained 
from potential models in Ref.~\cite{PM}, which also
considered the contributions of decays 
to the excited $2S$ and $1P$ states.
The direct decay rate into $P$-wave charmonium states 
is about 20\% of the direct decay rate into the $1S$ states. 
The radiative decay of the $\chi_c$ states increases the 
total semileptonic decay rate of $B_c$ to $J/\psi$ by about 5\%.
The exclusive semileptonic decay rates of the $B_c$
into the P-wave states $\chi_c$ and $h_c$
have also been calculated within a framework that involves overlap
integrals of the wavefunctions of the $B_c$ 
and the charmonium states \cite{chang2}.

The dominant leptonic decay of $B_c$ is given by the $\tau \nu_\tau$ mode (see
Table \ref{inc}). However, it has a low experimental efficiency of detection
because of hadronic background in the $\tau$ decays and missing energy.
Recently, in Refs. \cite{radlep} the enhancement of muon and electron channels
in the radiative modes was studied. The additional photon allows one to remove
the helicity suppression for the leptonic decay of a pseudoscalar particle, 
which leads to an increase of the muonic decay rate by a factor of 2.

\begin{table}[th]
\begin{center}
\begin{tabular}{|c|c|c|}
\hline
 mode & $\Gamma$, $ 10^{-14}\ \mbox{GeV}$ & BR, $\%$ \\
\hline
 $B_{s}e^{+}\nu_{e}$ & 5.8 & 4.0 \\
\hline
 $B_{s}^{*}e^{+}\nu_{e}$ & 7.2 & 5.0 \\
\hline 
 $\eta_c e^{+}\nu_{e}$ & 1.1 & 0.75 \\
\hline
 $\eta_c \tau^{+}\nu_{\tau}$ & 0.33 & 0.22  \\
\hline
 $J/\psi e^{+}\nu_{e}$ & 2.8 & 2.1 \\ 
\hline
 $J/\psi \tau^{+}\nu_{\tau}$ & 0.7 & 0.51 \\ 
\hline 
\end{tabular}
\end{center}
\caption[Widths and branching fractions for the semileptonic decay modes 
	of the $B_{c}$]{Widths and branching fractions for the 
semileptonic decay modes of the $B_{c}$ 
meson calculated using QCD sum rules.  (For the 
branching fractions, we set $\tau_{B_{c}}=0.46$ ps.)}
\label{semi}
\end{table}
\index{decay!$B_c \to B_s^{(*)}\ell^{+}\nu$}
\index{decay!$B_c \to \eta_c \ell^+\nu$}
\index{decay!$B_c \to J/\psi \ell^+\nu$}

\subsubsection{Nonleptonic modes}

The inclusive nonleptonic width of the $B_c$ can be
\index{Bc meson@$B_c$ meson!decay!nonleptonic} 
estimated in the framework of quark-hadron duality (see Table \ref{inc}).
However, calculations of exclusive nonleptonic modes usually involve 
the approximation of factorization \cite{fact}.  This approximation
is expected to be quite accurate for the $B_c$, 
since the light quark-antiquark sea is suppressed in $\bar b c$ mesons. 
Thus, the important parameters are the factors $a_1$ and $a_2$ 
in the nonleptonic weak Lagrangian, 
which depend on the renormalization scale for the $B_c$ decays.
The QCD SR estimates for the widths of exclusive modes involving 
the nonleptonic decay of the charmed quark in $B_c$ 
are presented in Table \ref{bs} \cite{KL}.
They agree with the values predicted by the potential models. 

For decays involving large recoils as in $B_c^+\to \psi \pi^+(\rho^+)$, 
the spectator picture of the transition breaks down due to 
hard gluon exchanges.  Taking these nonspectator effects into account 
increases the estimates of potential models by a factor of 4 \cite{Gers}. 
The corresponding estimates in the factorization approach 
are determined by the leptonic decay constants and by the QCD coupling constant 
at the scale of the virtuality of the hard gluon. 
Numerically, one finds the values represented in Table \ref{psi}.
Due to the contribution of a $t$-channel diagram,
the matrix element is {\it enhanced} by a factor of 2 compared to
the potential model value.  
The spin effects in such decays were studied in \cite{Pakh}.
The relative yield of excited charmonium states can be also evaluated.
For example, the branching fraction for $B_c^+\to \psi(2S) \pi^+$
should be smaller than that for $B_c^+\to \psi \pi^+$ 
by about a factor of 0.36.
The contributions to two-particle hadronic decays of $B_c$ 
from $P$-wave states of charmonium were considered in 
Refs.~\cite{saleev2} and \cite{chang2} using methods that involve 
the hard-scattering of constituents with large recoil
and the overlap of wave functions, respectively.
In Ref.~\cite{saleev2},
the form factors have power-law tails that come from one-gluon exchange,
and they therefore obtain larger values for the widths 
than Ref.~\cite{chang2}, in which the form factors fall exponentially. 
The ratios of the widths for $B_c^+\to \psi(2S) \pi^+$ 
in these two approaches agree with each other. 
The estimates of $B_c^+\to \psi(2S) \rho^+$ modes 
are more model dependent. 
The cascade electromagnetic transitions of the $\chi_c$ states increase
the inclusive $B_c\to J/\psi \pi(\rho)$ decay rates by about 8\%.
Finally, suppressed decays caused by flavor-changing neutral currents 
have been studied in Ref.~\cite{rare}.

\begin{table}[th]
\begin{center}
\begin{tabular}{|c|c|c|}
\hline
 mode & $\Gamma$, $ 10^{-14}\ GeV$ & BR, $\%$ \\
\hline
 $B_{s}\pi^{+}$ & 15.8 $a_{1}^{2}$ & 17.5 \\
\hline
$B_{s}\rho^{+}$ & 6.7 $a_{1}^{2}$ & 7.4 \\
\hline
 $B_{s}^{*}\pi^{+}$ & 6.2 $a_{1}^{2}$ & 6.9 \\
\hline
 $B_{s}^{*}\rho^{+}$ & 20.0 $ a_{1}^{2}$ & 22.2 \\
\hline 
\end{tabular}
\end{center}
\caption[Widths and branching fractions of nonleptonic decay modes of 
the $B_c$ meson.]{Widths and branching fractions of nonleptonic 
decay modes of the $B_c$ meson. (For the branching fractions, we set
$\tau_{B_{c}}=0.46$ ps and $a_{1}$=1.26.)}
\label{bs}
\end{table}
\index{decay!$B_c \to B_s^{(*)}\pi^{+}$}
\index{decay!$B_c \to B_s^{(*)}\rho^{+}$}

\begin{table}[th]
\begin{center}
\begin{tabular}{|c|c|}
\hline
 mode & BR, $\%$ \\
\hline
$\psi \pi^+$    & $0.67\pm 0.07$ \\
\hline
$\eta_c \pi^+$  & $0.87\pm 0.09$ \\
\hline
$\psi \rho^+$   & $1.96\pm 0.20$ \\
\hline
$\eta_c \rho^+$ & $2.43\pm 0.24$\\
\hline 
\end{tabular}
\end{center}
\caption
[Widths and branching fractions of charmonium decay modes
of the $B_{c}$ meson.]{Widths and branching fractions of 
charmonium decay modes of the $B_{c}$ meson. (For the branching 
fractions, we set $\tau_{B_{c}}=0.46$ ps and $a_{1}$=1.26.)}
\label{psi}
\end{table}
\index{decay!$B_c \to J/\psi \pi^+$}
\index{decay!$B_c \to \eta_c \pi^+$}
\index{decay!$B_c \to J/\psi \rho^+$}
\index{decay!$B_c \to \eta_c \rho^+$}

{\sf CP}-violation in $B_c$ decays can be investigated in the same
manner as for $B$ decays, although it 
is very difficult in practice because of the
low relative yield of $B_c$ compared to ordinary $B$ mesons:
$\sigma(B_c)/\sigma(B) \sim 10^{-3}$.
The expected {\sf CP}-asymmetry of 
${\cal A}(B_c^\pm \to J/\psi D^\pm)$ is about $4\cdot 10^{-3}$, 
but this decay mode has a very small branching ratio 
of about $10^{-4}$ \cite{CPv}. 
Another possibility is lepton tagging of the $B_s$ 
in $B_c^\pm\to B_s^{(*)}l^\pm \nu$ decays for the study of mixing 
and {\sf CP}-violation in the $B_s$ sector \cite{Quigg}.

\subsubsection{Discussion and conclusions}

We have reviewed the current status of theoretical predictions for the decays
of $B_c$ meson. We have found that the  operator product expansion, 
potential models, and QCD sum rules all
give consistent estimates for inclusive decay rates.
The sum rule approach, which has been
explored for the various heavy quark systems, 
leads to a smaller uncertainty due
to the quite accurate knowledge of the heavy quark masses. 
The dominant contribution to the $B_c$ lifetime is given by the
charmed quark decays ($\sim 70\%$), while the $b$-quark decays and the weak
annihilation add about 20\% and 10\%, respectively. The Coulomb-like
$\alpha_s/v$ corrections play an essential role in the determination of
exclusive form factors in the QCD sum rules. 
The form factors are expected to obey the relations dictated by
the spin symmetry of NRQCD and HQET with quite good accuracy.

The accurate direct measurement of the $B_c$
lifetime can provide us with the information on both the masses of charmed and
beauty quarks and the normalization point of the nonleptonic weak Lagrangian 
that is responsible for the $B_c$ decay (the $a_1$ and $a_2$ factors). 
The experimental study of
semileptonic decays and the extraction of ratios for the form factors can test
the spin symmetry derived in the NRQCD and HQET approaches.
It can also decrease the
theoretical uncertainties in the theoretical evaluation of quark
parameters as well as the hadronic matrix elements that take into account
nonperturbative effects caused by quark confinement. The measurement of
branching fractions for the semileptonic and nonleptonic modes
can give information on the values of factorization parameters, 
which depend again
on the normalization of the nonleptonic weak Lagrangian. 
The counting of charmed quarks in $B_c$ decays is also sensitive
to nonperturbative effects, because it is determined not only by the
contribution from $b$ quark decays, but also by the suppression of 
$\bar b \to c\bar c \bar s$
transitions due to destructive Pauli interference.

Thus, progress in measuring the $B_c$ lifetime and decays should 
enhance the theoretical understanding of what really happens in 
heavy quark decays.


\subsection[\D0\ Study of $B_c$]
{ \D0\ Study of $B_c$: Triggering and Reconstruction
\footnote{Author: R. Van Kooten}}
\index{D0!Bc@$B_c$}

\subsubsection{Introduction}\label{sec:int}

The $B_c$ meson is a particularly interesting system to study since 
\index{Bc meson@$B_c$ meson!triggering} 
\index{Bc meson@$B_c$ meson!reconstruction} 
\index{triggering!$B_c$ meson} 
it contains two different heavy 
quarks that are often in competition regarding subsequent decays. 
As a result, measurements of 
its properties such as mass, lifetime, and decay branching ratios offer a 
unique window into heavy 
quark hadrons. Since it has nonzero flavor, it has no 
strong or electromagnetic annihilation decay channels, and 
it is the heaviest such meson predicted by the Standard Model. 
Its weak decay is expected to 
yield a large branching fraction to final states containing a 
$J/\psi$ which is a useful experimental signature. 
The $B_c$ meson (like the single-$b$ baryons and doubly-heavy 
baryons) is too massive to produce at the $B$ factories running at 
the $\Upsilon(4S)$. LEP has only a few $B_c$
candidates, while CDF isolated a sample of 23 $B_c$ decays in 
approximately 100~pb$^{-1}$  
of data~\cite{cdfbc}
in Run~I resulting in the estimate that
$\sigma(B_c) / \sigma(b) \approx 2 \times 10^{-3}$.

\subsubsection{ \D0\ simulations}\label{sec:simul}

To examine \D0\ prospects for Run~II, simulations were made of 
$B_c$ production using PYTHIA and reweighting the resulting spectrum
to match the differential
$d\sigma/ d p_T$ cross section calculated using code
supplied by the authors of Ref.~\cite{anatoliprod}.
After reweighting, the production spectrum of produced
$B_c$ mesons is shown in Fig.~\ref{fig:bcprod}.
The mass of the $B_c$ meson was set to 6.40~GeV and the
lifetime to 0.50~ps, consistent with CDF and LEP 
measurements~\cite{cdfbc,bcprops}.
The fraction of $b \rightarrow B_c$ was increased to 0.5 while the
fractions of $b \rightarrow B^0_d$, $B^+$, $B^0_s$, and $\Lambda_b$ were
scaled down appropriately. Any events containing two $B_c$ mesons were
discarded in order to get the hadron composition of the ``away-side'' $B$
hadron
correct.  Events with $p_T(B_c) > 3$~GeV and $|\eta (B_c)| < 3$ were 
retained.

\begin{figure}[ht]
\begin{center}
\mbox{\epsfig{file=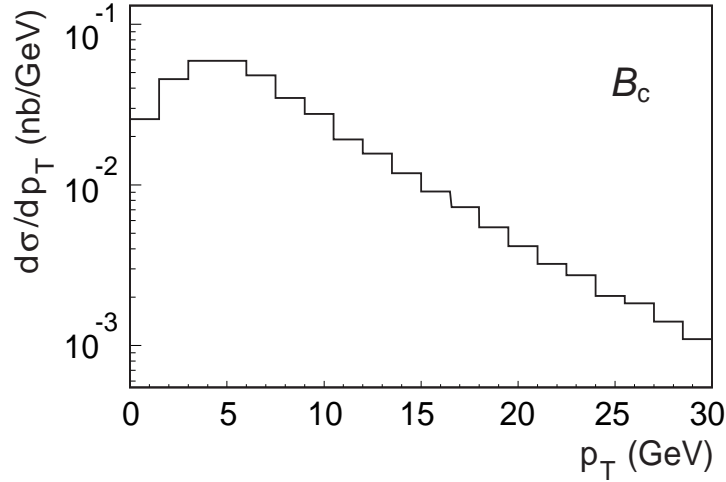,height=7cm,width=10cm}}
\caption{ 
The differential cross section for the $B_c$ meson 
	used in the following studies.}
\protect\label{fig:bcprod}
\end{center}
\end{figure}

\begin{figure}[ht]
\begin{center}
\mbox{\epsfig{file=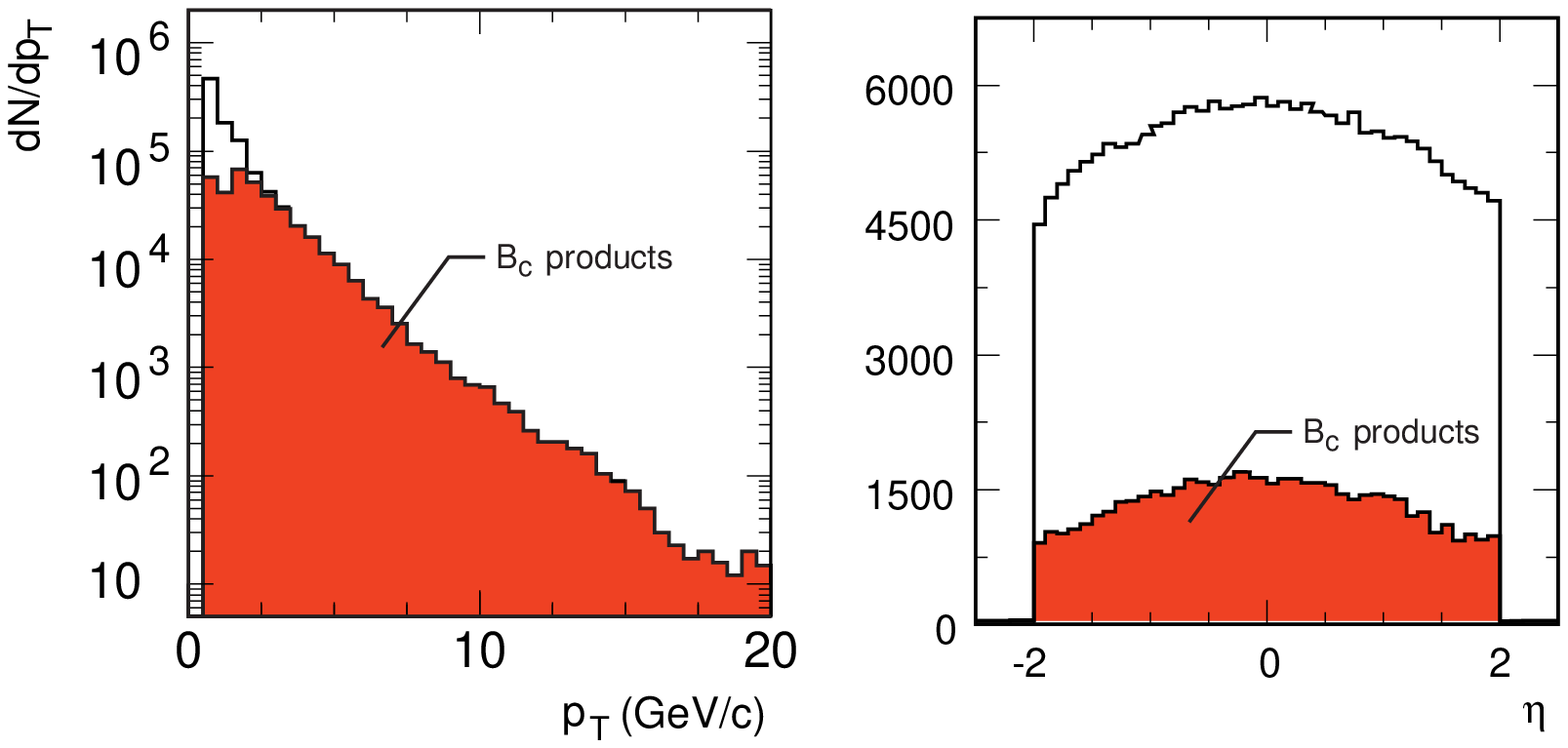,height=6cm,width=14cm}}
\end{center}
\caption{ 
Kinematic distributions of (a) $p_T$ and (b) $\eta$ for all 
particles (open histograms) and for $B_c$ 
semileptonic decay products (dark histograms)}
\protect\label{fig:bcprodd}
\end{figure}

The CLEO QQ package was used to force the semileptonic decay 
$B_c \to J/\psi \ell \nu$ 
(Br $\approx 2.5\%$~\cite{anatolibr}) and 
$B_c \to J/\psi \pi$
(Br $\approx 0.2\%$~\cite{anatolibr})
followed by the subsequent decays of  
$J/\psi \to e^+ e^-$ or 
$J/\psi \to \mu^+ \mu^-$. 
The distinctive trilepton signature of the first decay mode was used 
to allow efficient triggering, 
from the presence of at least two electrons or muons with significant
transverse momentum, with reasonable background rates. 
\index{decay!$B_c \to J/\psi \pi^+$}
\index{decay!$B_c \to J/\psi \ell^+\nu$}

Detector simulations were performed at a number of different
levels of sophistication: MCFAST~\cite{MCFAST}, PMCS (a \D0\ parameterized
fast Monte Carlo), and full GEANT simulated events to allow the use
of a more realistic trigger simulator and reconstruction resolution
determination.  Typical distributions of $B_c$ decay products from
the MCFAST simulation are shown
in Fig.~\ref{fig:bcprodd}.

\subsubsection{Results}\label{sec:results}

\begin{table}[ht]
\begin{center}
\begin{tabular}{|c|c|c|}
\hline
Criteria & Efficiency & \D0\ Designation \\
\hline 
Single Muon         &              &              \\
& & \\
$p_T^{\mu} > 4$~GeV &              & \\ 
$| \eta^{\mu} | < 1.5$ &      0.23    & MTM5, MUO(1,4,A,M)   \\
 ``Medium''        & & \\
& & \\
$p_T^{\mu} > 4$~GeV &              & \\ 
$| \eta^{\mu} | < 1.5$ &      0.08   & MTM6, MUO(1,4,A,T)   \\
 ``Tight''        & & \\ 
\hline
Di-Muon         &              &              \\
& & \\
Both $p_T^{\mu} > 2$~GeV &              & \\ 
$| \eta^{\mu} | < 1.5$ &      0.16    & MTM10, MUO(2,2,A,M)   \\
 ``Medium''        & & \\
& & \\
Both $p_T^{\mu} > 4$~GeV &              & \\ 
$| \eta^{\mu} | < 2.0$ &      0.15    & MTM12, MUO(2,4,A,M)   \\
 ``Medium''        & & \\ 
\hline
``Or'' of above     &       0.28    &                      \\ \hline
\end{tabular}
\end{center}
\caption{ \D0\ Level 1 muon trigger efficiencies for trilepton final
states from semileptonic decay of $B_c$ mesons with
$p_T(B_c) > 3$~GeV and $|\eta (B_c)| < 3$.}
\label{tab:mutrig}
\end{table}

\begin{figure}[ht]
\begin{center}
\mbox{\epsfig{file=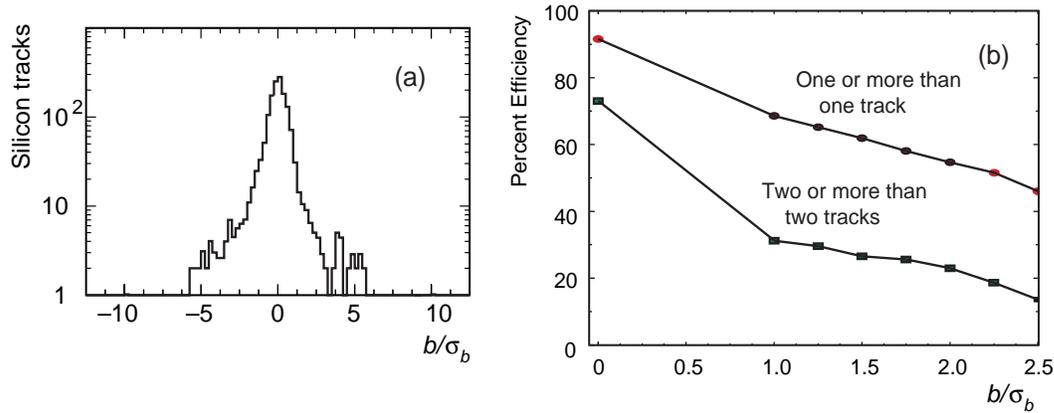,height=6cm,width=14cm}}
\end{center}
\caption{(a) Impact parameter divided by its error for track from
$B_c \to J/\psi \ell \nu$; (b) efficiency of future silicon
track trigger as a function of cut on $b/\sigma_b$.}
\label{fig:stt}
\end{figure}

\begin{figure}[ht]
\begin{center}
\mbox{\epsfig{file=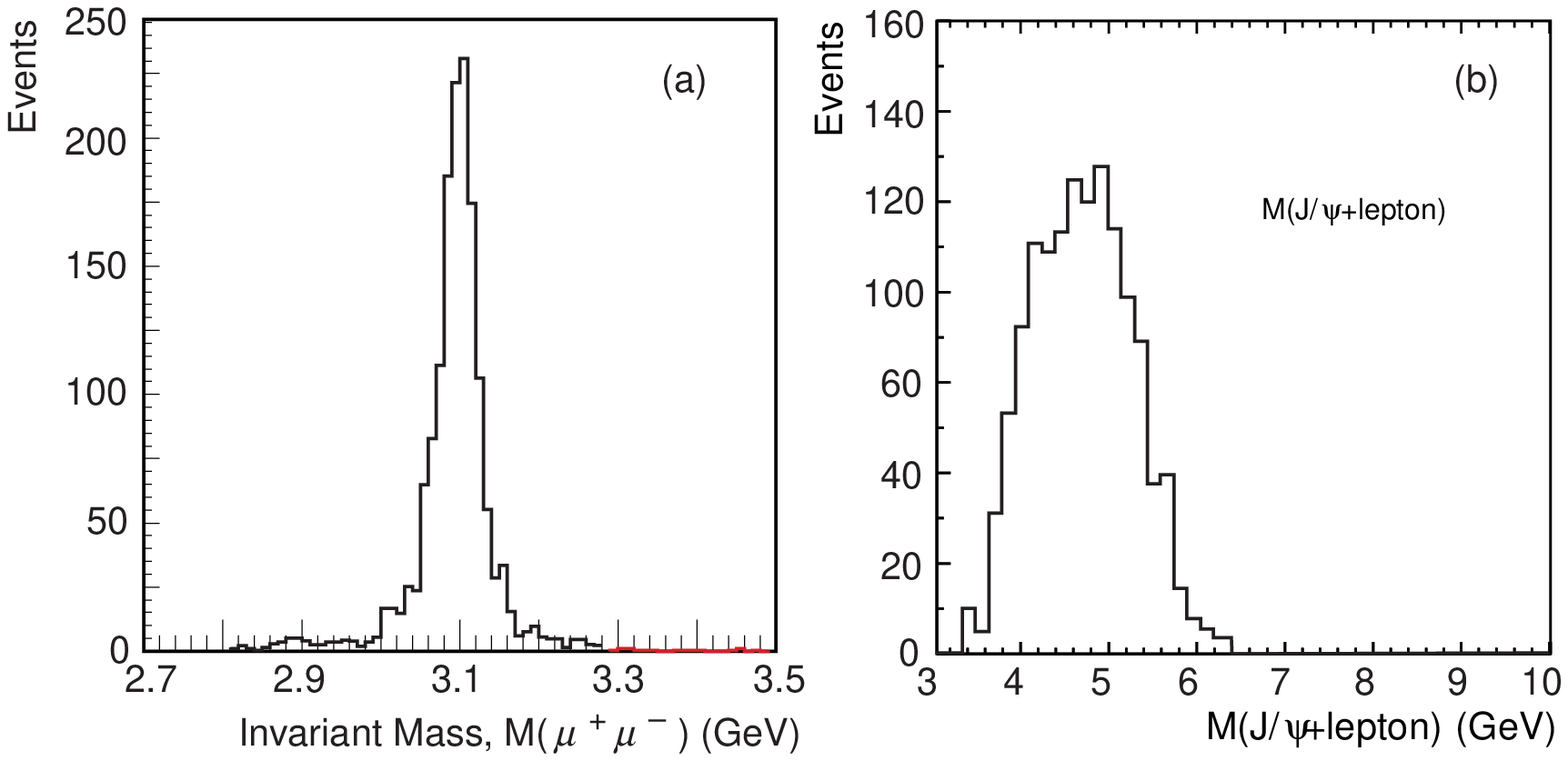,height=6cm,width=14cm}}
\end{center}
\caption{ 
(a) $\mu^+\mu^-$ invariant mass distribution in 
$B_c \to J/\psi \ell \nu$ events; (b) tri-lepton invariant
mass distribution that can be used to extract the $B_c$ mass.}
\label{fig:jpsimass}
\end{figure}

MCFAST simulations could be used to determine kinematic and trigger 
efficiencies; however, the \D0\ trigger simulator was run on GEANT
fully simulated events for a more sophisticated and reliable treatment.
Starting with the case of a semileptonically decaying $B_c$ where all
three leptons are muons, Table~\ref{tab:mutrig} gives the \D0\ muon
trigger efficiency for the indicated criteria. 
``Medium'' and ``Tight'' muon identification refers to the 
correspondingly tighter requirements of coincidence of greater number 
of muon detector layers.

The status of the simulation of triggering on electrons is less complete,
and trigger efficiencies in this case were extrapolated from prior 
studies~\cite{etrig} for the decay $B^0_d \to J/\psi K^0_S$.
\index{decay!$B_d \to J/\psi K^0_S$}
At Level 1, information from the electromagnetic calorimeter, the fiber
tracker, and matches between these and hits in the central preshower
detector are expected to lead to a trigger efficiency of approximately
30\%, but with a substantial background rate. This background rate
is expected to be lowered to a reasonably low value with invariant
mass cuts on the electrons from the $J/\psi \to e^+ e^-$ decay
applied at Level 2 of the trigger.  An overall efficiency for triggering
on states with di-electrons from the $J/\psi$ decay is estimated to
be 12.8\%.

Although \D0\ will initially be running without a vertex silicon track
trigger, studies were made of the potential for such a trigger to help
isolate a large sample of $B_c$ mesons. Note that the events were 
generated with a $B_c$ lifetime set to 0.50 ps rather than the usual
lifetimes of 
1.5--1.6~ps of the other $B$ mesons.  A decay length resolution of 116~$\mu$m
has been measured in MCFAST simulated events. A vertex silicon track 
trigger would operate by examining the impact parameter significance, 
i.e., $b / \sigma_b$, where $b$ is the impact parameter of a track
and $\sigma_b$ its error. The distribution of this quantity for tracks
from $B_c$ decay is shown in Fig.~\ref{fig:stt}(a). This distribution
is substantially narrower than for tracks from other $B$ meson decays
due to the shorter lifetime of the $B_c$.
The trigger efficiency obtained by cutting on the presence of one or
more tracks with large impact parameter significance is shown in
Fig.~\ref{fig:stt}(b). For the same background rate, the 
efficiency is about a factor of 2.5 times smaller than the comparable
rate for $B^0_d \to J/\psi K^0_S$ decays.

In the $B_c \to J/\psi \ell \nu$ channel, a
typical mass reconstruction of the $J/\psi$ through its decay
into $\mu^+ \mu^-$ is shown in Fig.~\ref{fig:jpsimass}(a) indicating
a mass resolution $\sigma_m$ of 29~MeV.  Events are required to have 
the invariant mass of the two
leptons assigned to the $J/\psi$ within
$2 \sigma_m$ of the $J/\psi$ mass.  To reduce backgrounds, the $p_T$ of the
combined ($J/\psi$ plus lepton) system was required to be greater than
8~GeV, and the decay length of the $J/\psi$-lepton vertex was required
to be greater than 50~$\mu$m.  Kinematic quantities such as the 
trilepton invariant mass yield information on the mass 
of the decaying state (Fig.~\ref{fig:jpsimass}). 

Similar studies were extended to the $B_c \to J/\psi \pi$
decay channel that would allow an 
exclusive reconstruction of the $B_c$ meson but
with smaller statistics due to the small (~0.2\%) branching ratio for this
final state.\index{decay!$B_c \to J/\psi \pi^+$}
With only two leptons to trigger on, the overall trigger
efficiency for the di-muon final state analogous to that shown in 
Table~\ref{tab:mutrig} is found to be slightly lower with a value
of 0.24. The mass resolution of the reconstructed $B_c$ without
constraints on the $J/\psi$ mass is found to be 52~MeV.

We proceed to estimate the expected number of $B_c$ events 
reconstructed by \D0\ in Run~II with an integrated luminosity of 2~fb$^{-1}$.
One could use the estimate of 
$\sigma(B_c^+) / \sigma(\bar{b}) = 1.3 \times 10^{-3}$ compared to
the measurement of
$\sigma(B^+) / \sigma(\bar{b}) = 0.378 \pm 0.022$.
It is more straightforward to use the CDF measurement~\cite{cdfbc} of
\begin{equation} \frac{\sigma(B_c^+) \cdot {\mathrm Br}(B_c^+ \to J/\psi \ell \nu)}
{\sigma(B^+) \cdot {\mathrm Br}(B^+ \to J/\psi K^+)}
\end{equation}
and comparisons
of \D0's trigger and reconstruction efficiency with the corresponding CDF 
efficiencies.
This leads to an estimate that approximately 600 identified
$B_c^+ \to J/\psi \ell \nu$ would be produced.  
This sample is large enough to make improvements in the lifetime and
mass measurements that would significantly increase our understanding 
of the $B_c$ system.
In addition, samples of 30--40 fully exclusive decays such as 
$B_c^+ \to J/\psi \pi$ can also be expected that would clearly
supplement the semileptonic decay measurements which suffer from the 
escaping neutrino.  Of course, proportionally much larger samples would
be expected in subsequent runs of the Tevatron beyond Run~IIa.

\subsection[CDF Projections for $B_c^+$ yield]
{CDF Projections for $B_c^+$ yield 
in Run~II\protect{\authorfootnote{Authors: Wei Hao, Vaia Papadimitriou}}}
\index{CDF!Bc@$B_c$}

\subsubsection{Introduction}

We present here the CDF projections for the $B_c^+$ yield 
in Run~II by performing studies with Run~I data and with Monte Carlo 
simulation and by using 
theoretical expectations for the branching ratios of various $B_c^+$ decay
channels. We are making some simple projections for the decays 
$B_c^+ \rightarrow J/\psi \pi^+$, $B_c^+ \rightarrow 
B_s^0 \pi^+$ and $B_c^+ \rightarrow J/\psi l^+ \nu$.
Please keep in mind for the rest of this section that references to
a specific state imply the charge-conjugate state as well. 

\subsubsection{Monte Carlo generation and simulation\label{MC}}

$B_c^+$ Monte Carlo events were generated with a ``toy" Monte Carlo using the
$P_t$ spectrum from a full $\alpha_s^4$ perturbative QCD calculation of
hadronic production of the $B_c^+$ meson \cite{Oakes} and assuming a flat
rapidity $y$ distribution (see Fig.~\ref{pt_bc}).
The $B_c^+$ mesons were generated with the mass set to 6.2 GeV/c$^2$,
lifetime set to 0.3335 $ps$ and in the
region $P_t(B_c^+) >$ 3.0 GeV/c and $|y(B_c^+)|<$ 1.5.
The $B_c^+$ mesons were decayed to $J/\psi \pi^+$ or other decay channels 
using the
CLEO decay table (QQ)  and then they were 
simulated by the CDF Run~I simulation package QFL$^{\prime}$.

$B^+$ events, for comparison, were generated with a Monte Carlo 
which is generating $b$ quarks according to the next-to-leading order 
QCD predictions \cite{nason}, using a scale $\mu = \mu_0/2 \equiv 
\sqrt{(P_t)^2 + (m_b)^2}/2$ where $P_t$ is the transverse momentum
of the $b$ quark and $m_b$ its mass. $m_b$ is set to 4.75 GeV/c$^2$.
The $b$ quark is fragmented into $b$ hadrons using Peterson 
fragmentation \cite{peter} 
with the 
fragmentation parameter, $\epsilon_b$, set to 0.006 . In Fig.~\ref{pt_b}
we show the $P_t$ spectrum of the $B^+$, at the generation level, for 
$P_t(min)$ of the 
$b$-quark equal to 5.0 GeV and in the rapidity region $-1.3 < Y_{RANGE} < 1.3$.
The $B^+$ events were decayed to $J/\psi K^+$ using QQ and then they were
simulated by QFL$^{\prime}$.

\begin{figure}[bhtp]
\begin{center}
\mbox{\epsfxsize=0.65\textwidth\epsffile {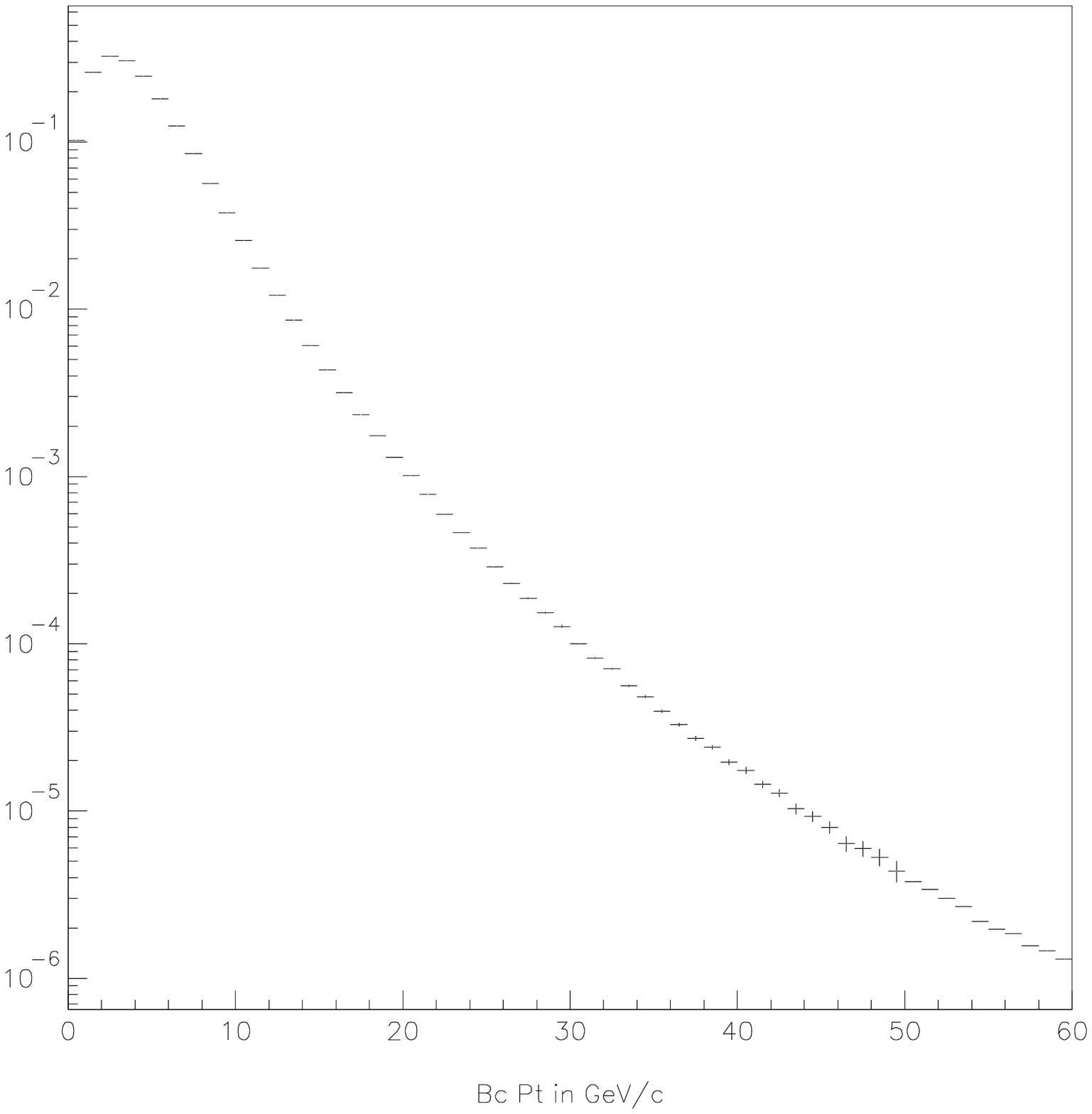}}
\vspace{-0.1cm}
\caption[$P_t$ distribution of $B_c^+$ mesons]
{$P_t$ distribution of $B_c^+$ mesons in GeV/c, at the generation level.}
\label{pt_bc}
\end{center}
\end{figure}

\begin{figure}[bhtp]
\begin{center}
\mbox{\epsfxsize=0.61\textwidth\epsffile{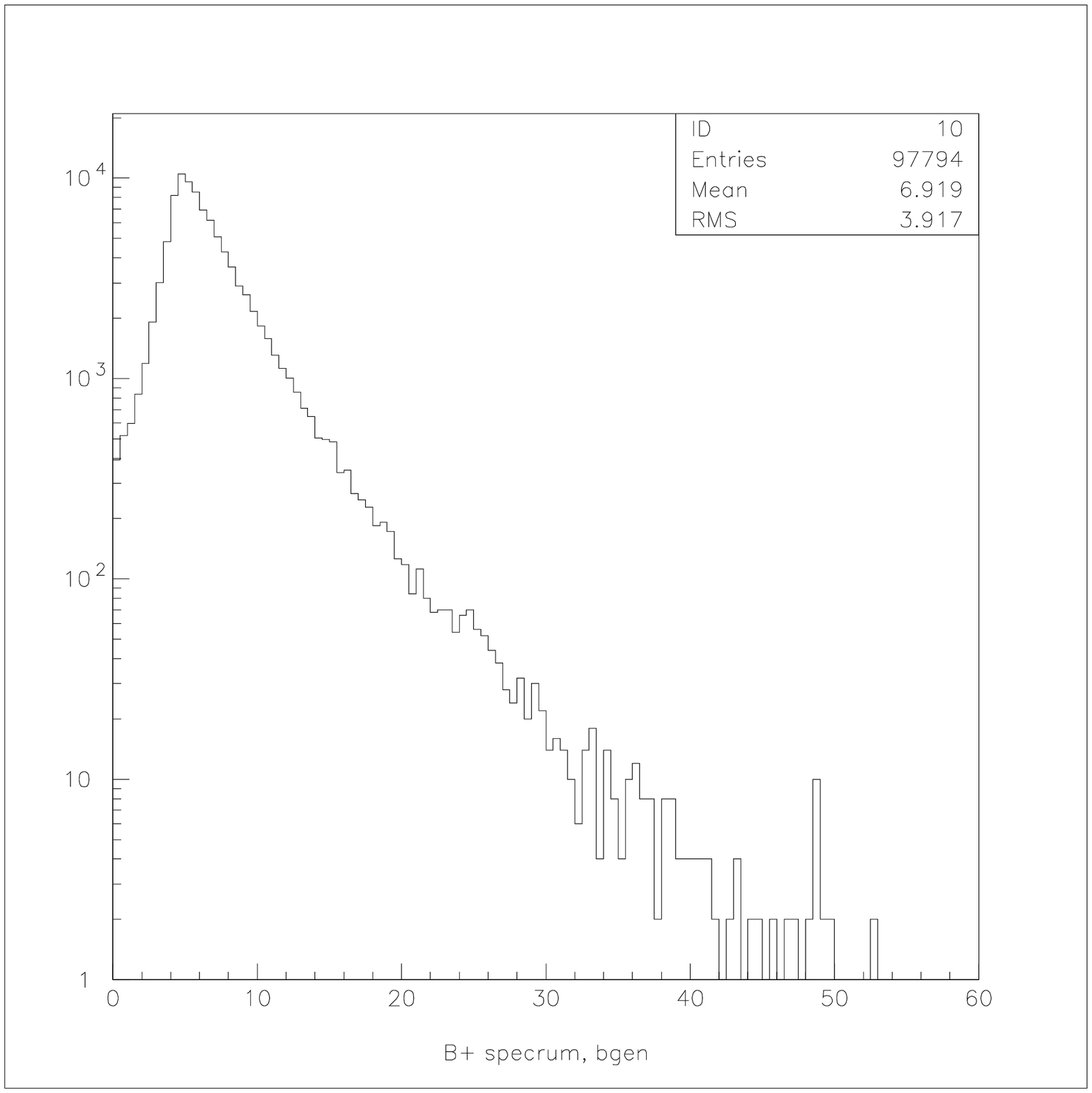}}
\vspace{0.1cm}
\caption[$P_t$ distribution of $B^+$ mesons]
{$P_t$ distribution of $B^+$ mesons in GeV/c, at the generation level.
}
\label{pt_b}
\end{center}
\end{figure}

The theoretical cross sections used for the Monte Carlo generations were
calculated for $\sqrt{s} = 1.8$ TeV.

\subsubsection{Selection Criteria for the $B_c^+  \rightarrow J/\psi \pi^+$ 
decay channel \label{glob}}
\index{decay!$B_c \to J/\psi \pi^+$}

To select $J/\psi \rightarrow \mu^+\mu^-$ candidates, we require that
the transverse momentum, $P_t$, of each muon is greater than
2.0 GeV/c. The CMU muon chambers, at a radius of 3.5 m from the beam axis, 
cover
the pseudorapidity region $|\eta|<$ 0.6. These chambers are complemented
by the central muon upgrade system (CMP) which consists of four layers of 
drift tubes behind two feet of steel. In addition we have the CMX muon
chambers covering the pseudorapidity region 0.6$<|\eta|<$1.0.   
The position matchings between muon chamber track and 
the track as measured in the Central Tracking Chamber (CTC) are required
to have $\chi^2<9.0$ in r-$\phi$ and $\chi^2<12$ in r-z.
There is no requirement on a specific trigger path.
We require that at least one
muon is reconstructed in the Silicon VerteX detector (SVX)
({\it i.e.} at least 3 associated SVX hits were found).
We then keep the $J/\psi$ candidates so that the muon pair passes a vertex
constrained fit and the mass of the pair is within 50 MeV of the known
value \cite{pdg} of the $J/\psi$ mass.

After the $J/\psi$ candidates are found, events are scanned for other
daughter particle candidates, $\pi^+$'s in this case, 
from $B_c^+$ decays. We reconstruct the $B_c^+$ by performing a
vertex-constrained fit. The tracks with more than 3
associated SVX hits are considered to be SVX tracks, and all others are
considered to be CTC tracks. We apply no further quality cuts to the SVX
tracks.
In the vertex constrained fit the
invariant mass of the $\mu^+ \mu^-$ pair is
constrained to the known $J/\psi$ mass \cite{pdg}. For each $B_c^+$ candidate,
we require that the $\chi^2$ probability of the fit be
greater than $1\%$.

After the $B_c^+$ is reconstructed, the following cuts are applied:
all the $B_c^+$ candidates are required to have
transverse momentum $P_t(B_c^+)$ greater than 6.0 GeV/c;
proper lifetime $ct(B_c^+)$ greater than 80 $\mu$m;
impact parameter with respect to the beam line $|I_{xy}(B_c^+)|$ less than
80 $\mu$m. 
We also require that the $\pi^+$ is
reconstructed in the SVX and that
$P_t(\pi^+) > 2.5$ GeV/c.

\subsubsection{Acceptance Calculation using QFL$^{\prime}$}

Using the QFL$^{\prime}$ Monte Carlo we find that the geometric/kinematic
acceptance for $B_c^{+} \rightarrow J/\psi \pi^{+}$ is equal to $\sim$0.018
while the same acceptance, and with the same selection criteria, for 
$B^{+} \rightarrow J/\psi K^{+}$ is equal to 
$\sim$0.04 \cite{4884}. These acceptances are calculated in the region 
$6.0< P_t < 30.0$ and
$|y|<0.9$ where $P_t$ and $y$ are the transverse momentum and rapidity,
respectively, of the $B_c^+$ or $B^+$ mesons. 
The acceptances are calculated with the default kinematic cuts of
$P_t(B_c^+,B^+) > 6.0 GeV/c$, $P_t(\pi^+,K^+) > 2.5 GeV/c$, 
$|I_{xy}(B_c^+,B^+)| < 80
\mu m$ and $ct(B_c^+,B^+) > 80 \mu m$. At least one of the two muons forming 
the 
$J/\psi$ has to be reconstructed in the SVX.
These acceptances do not include any trigger
efficiency corrections.

\subsubsection{Yield estimate for the $B_c^+  \rightarrow J/\psi \pi^+$
decay channel  \label{svx12}} 
\index{decay!$B_c \to J/\psi \pi^+$}

We know that 
$\sigma(B_c^{+}) \cdot BR(B_c^{+} \rightarrow J/\psi \pi^{+})$ is equal 
to: 
$$ R_{l} \cdot \sigma(B^{+}) \cdot 
BR(B^{+} \rightarrow J/\psi K^{+}) \cdot \frac{BR(B_c^{+} 
\rightarrow J/\psi \pi^{+})}
{BR(B_c^{+} \rightarrow J/\psi l^{+} \nu)},$$
where 
\beq
R_{l} = \frac{\sigma(B_c^{+}) 
\cdot BR(B_c^{+} \rightarrow J/\psi l^{+} \nu)}{\sigma(B^{+})
 \cdot BR(B^{+} \rightarrow J/\psi K^{+})}
\eeq 
and $\sigma$ stands for production cross section and $BR$ stands for
branching ratio. 
We get the ratio 
$R_{l}$ from \cite{cdfbc},  
$\sigma(B^{+})$ from \cite{todd} and
$BR(B^{+} \rightarrow J/\psi K^{+})$ from \cite{pdg}.
We tabulate the theoretical expectations for the value of the ratio 
\beq
R=\frac{BR(B_c^{+} \rightarrow J/\psi \pi^{+})}
{BR(B_c^{+} \rightarrow J/\psi l^{+} \nu)}
\eeq
in Table 1. These values cover 
a wide spectrum from 0.06 to 0.32 according to References 
\cite{changchen}-\cite{BJ}.
For this study we use as a default value the one from Reference 
\cite{changchen}. 
$\sigma(B_c^{+}) \cdot BR(B_c^{+} \rightarrow J/\psi \pi^{+})$ is equal
to 
$(0.132^{+0.061}_{-0.052}) \cdot (3.52 \pm 0.61 \mu b) \cdot (9.9 \pm 1.0) 
\times 10^{-4} \cdot (0.091)=$
(4.6$\pm$2.3)$\cdot$ 10$^{-4} \cdot$ 0.091 $\mu$b. Then we multiply the 
calculated value of
$\sigma(B_c^{+}) \cdot BR(B_c^{+} \rightarrow J/\psi \pi^{+})$, which is 
the cross section times the branching ratio of either positively charged or 
negatively charged $B_c$'s, by 
the branching ratio of the decay $J/\psi \rightarrow \mu^+ \mu^-$ \cite{pdg},
 by 
the total integrated luminosity (100 $pb^{-1}$) and by the kinematic/geometric
acceptance of the decay $B_c^{+} \rightarrow J/\psi \pi^{+}$ (0.018).
This way we get the number, S, of $B_c^{+} \rightarrow J/\psi \pi^{+}$
events expected in our Run~I sample. Then we multiply that number by 2 to
account for both positively and negatively charged particles. S is 
approximately equal to 9 events.
If we consider the variations in $R$ discussed above, then the 
expected number of $B_c^{+} \rightarrow J/\psi \pi^{+}$ events in Run~I varies
between 6 and 32.

One could think that we could possibly exclude the possibility of an $R$
in the range of 0.3 based on the number of events we currently observe
or based on the limit we set on
\beq
R_{\pi} = \frac{\sigma(B_c^{+}) \cdot BR(B_c^{+} \rightarrow J/\psi \pi^+)}
{\sigma(B^{+}) \cdot
 BR(B^{+} \rightarrow J/\psi K^{+})}
\eeq
in \cite{bcoldprl} using Run~I 
data. As can be seen from Fig.~\ref{limit}, which is taken from Reference 
 \cite{bcoldprl}, 
$R_{\pi}<$ 0.10 for a $B_c^+$ lifetime of 0.33 ps
and $R_{\pi}<$ 0.07 for a $B_c^+$ lifetime of 0.50 ps. 
We can write
$R_{\pi}$ as equal to:
$$ R_{l} \cdot 
\frac{BR(B_c^{+} \rightarrow J/\psi \pi^+)}
{BR(B_c^{+} \rightarrow J/\psi l^+ \nu)}.$$
We get the ratio $R_{l}$ from ref.\cite{cdfbc} to be equal to
0.132 and the ratio $\frac{BR(B_c^{+} \rightarrow J/\psi \pi^+)}
{BR(B_c^{+} \rightarrow J/\psi l^+ \nu)}$ to be in the range 0.06-0.32.
This results in 
$R_{\pi}$ in the range 0.008-0.04.
We see that we cannot really exclude the value 0.32 for $R$ based on our 
published limit for $R_{\pi}$.

\begin{figure}[thb]
\begin{center}
\mbox{\epsfxsize=0.72\textwidth\epsffile {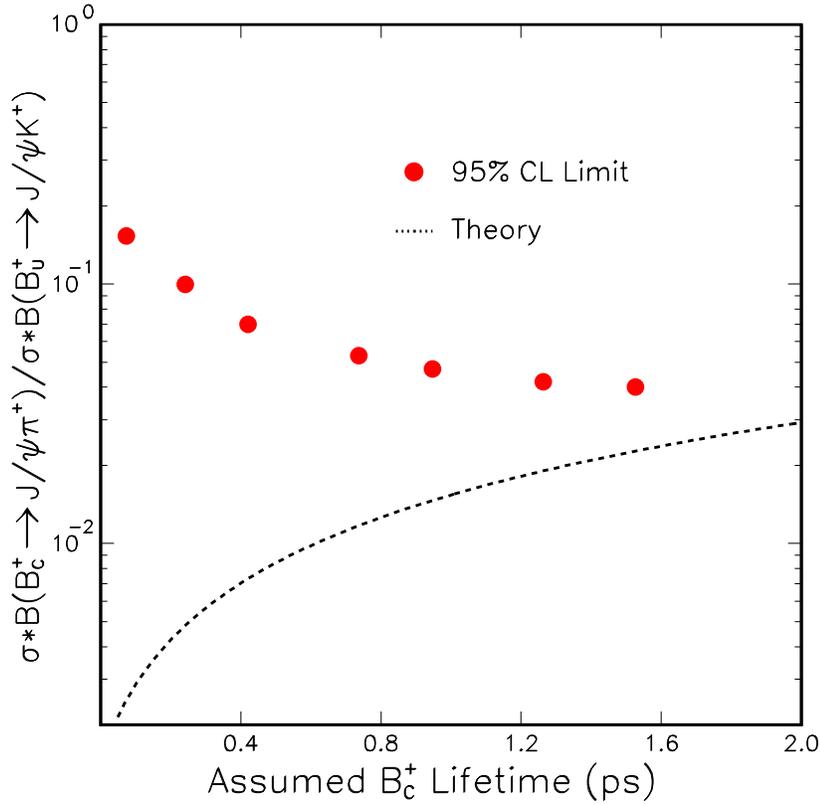}}
\vspace{-0.2cm}
\caption[The ratio 
$\sigma \times BR(B_c^+ \rightarrow J/\psi \pi^+)$/
$\sigma \times BR(B^+ \rightarrow J/\psi K^+$]
{The circular points show the different 95\% CL limits on 
the ratio of cross section times branching fraction for 
$B_c^+ \rightarrow J/\psi \pi^+$ relative to 
$B^+ \rightarrow J/\psi K^+$ as a function of the 
$B_c^+$ lifetime. The dotted curve represents a calculation of this ratio
based on the assumption that the $B_c^+$ is produced 1.5 $\times$ 10$^{-3}$
times as often as all other $B$ mesons and that $\Gamma(B^+_c \rightarrow 
J/\psi \pi^+)=4.2 \times 10^9$ s$^{-1}$. 
}
\label{limit}
\end{center}
\end{figure}

For Run~IIa we expect to have an integrated luminosity of 2 fb$^{-1}$,
that is an increase by a factor of 20 in comparison with Run~I.
In Run~II we will take data with an upgraded silicon detector that has extended
coverage in comparison to Run~I by a factor of 1.4. 
We also plan to run with extended muon coverage for muons of type CMP (x1.2) 
and CMX (x1.3) and with lower muon $P_t$ thresholds. For CMU type muons we 
plan to 
lower the thresholds from 2.0 GeV/c to 1.5 GeV/c. If we lower our $P_t$ muon
thresholds from 2.0 GeV/c to 1.5 GeV/c in our current Run~I analysis we get
an increase of acceptance by a factor of 1.43.
According to reference \cite{PAC99} there will be a factor of 2 increase in the
$B^0 \rightarrow J/\psi K_s^0$ yield due to the extended muon coverage and due
to the lowering of the muon $P_t$ threshold from 2.0 GeV/c to 1.5 GeV/c.
Assuming conservatively an increase in yield in Run~IIa by a factor
of 40 in comparison to Run~I we expect to have approximately 360 events 
in the decay channel $B_c^+ \rightarrow J/\psi \pi^+$.

\begin{table}[bhtp]
\begin{center}
\begin{tabular}{|c|c|c|c|}
\hline  Reference & $B_c^+ \rightarrow J/\psi \pi^+$ 
& $B_c^+ \rightarrow J/\psi l^+ \nu$ & R \\
\hline Chang, Chen (1994) \cite{changchen}
&  $\Gamma=$ 3.14 $\cdot$ 10$^{-6}$ eV
&  $\Gamma=$ 34.4 $\cdot$ 10$^{-6}$ eV   &  0.091     \\
\hline Gershtein et al (1998) \cite{gers2}
&  BR $=$ 0.2\%  & BR $=$ 2.5\%  &   0.08   \\
\hline Gershtein et al (1995) \cite{gers}
&  $\Gamma=$ 3.14 $\cdot$ 10$^{-6}$ eV
 & $\Gamma=$ 38.5 $\cdot$ 10$^{-6}$ eV (ISGW1) &   0.08\\
& $\Gamma=$ 3.14 $\cdot$ 10$^{-6}$ eV
 & $\Gamma=$ 53.1 $\cdot$ 10$^{-6}$ eV (ISGW2) &   0.06\\
\hline Kiselev et al. (1999) \cite{ellis}  & BR $=$ 0.67\% &  BR $=$ 2.1\%  &  0.32 \\ 
\hline Bjorken (1986)  \cite{BJ}&    BR $=$ 0.6\% &   BR $=$ 2.0\%&       0.29\\
\hline
\end{tabular}\vspace*{4pt}
\caption{Theoretical estimations of the branching ratios of two 
$B_c^+$ decay modes and of their ratio R.}
\label{list}
\end{center}
\end{table}

\begin{figure}[thb]
\begin{center}
\mbox{\epsfxsize=0.72\textwidth\epsffile {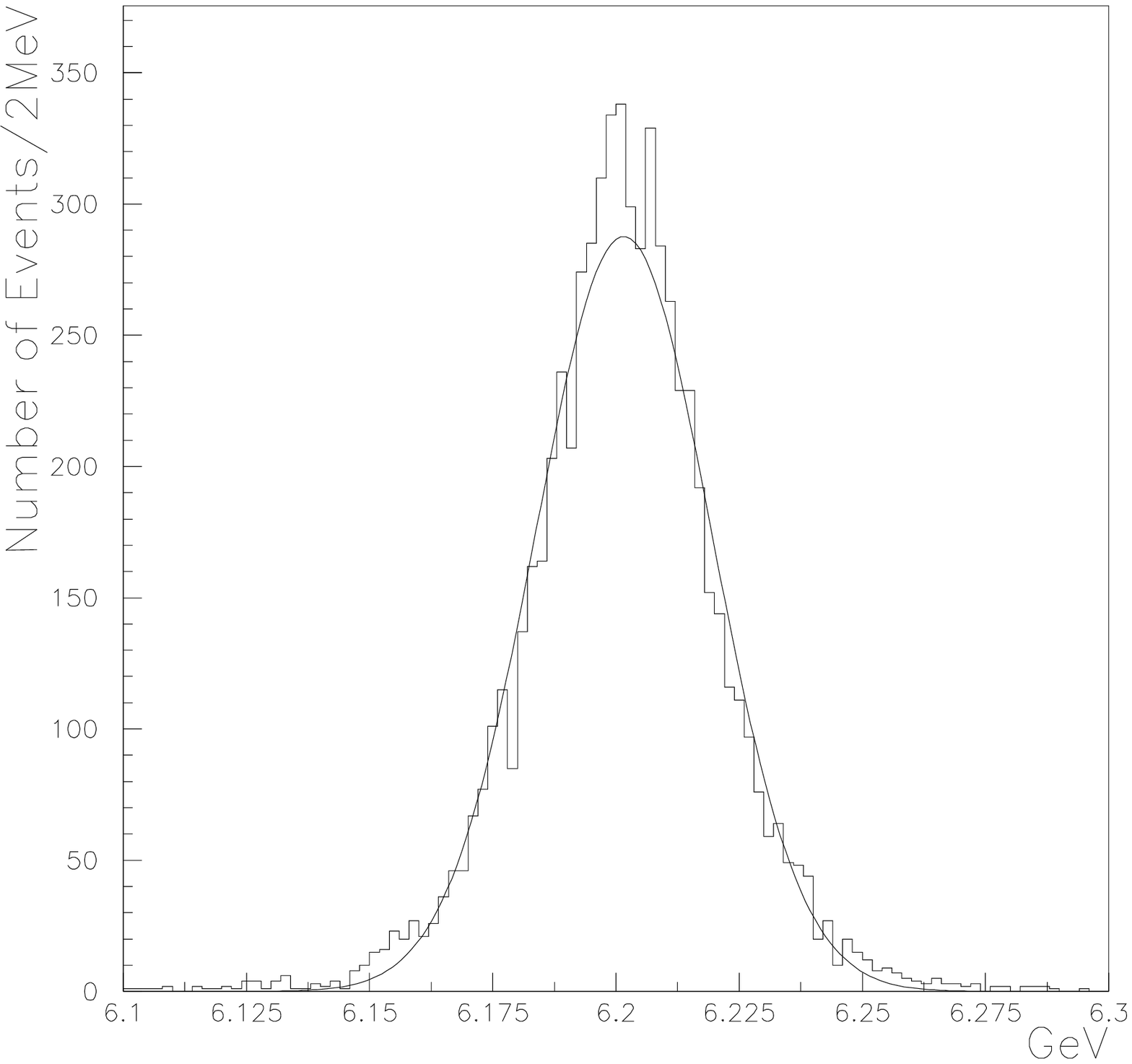}}
\vspace{-0.4cm}
\caption[Mass distribution of $B_c^+$ mesons in GeV/c$^2$]
{Mass distribution of $B_c^+$ mesons in GeV/c$^2$.}
\label{mass_bc}
\end{center}
\end{figure}

In Fig.~\ref{mass_bc} we show the distribution of the $B_c^+$ mass from 
Monte Carlo after 
the selection requirements described above. Here we have lowered the 
muon $P_t$ threshold to 1.5 GeV/c. The mass resolution is 17.9$\pm$0.3 MeV.

\subsubsection{Yield estimate for the $B_c^+ \rightarrow B_s^0 \pi^+$ 
decay channel \label{impt}}
\index{decay!$B_c \to B_s^0 \pi^+$}

From Reference \cite{changchen} we see that 
$\Gamma=$ 73.3 $\cdot$ 10$^{-6}$ eV for the decay channel 
$B_c^+ \rightarrow B_s^0 \pi^+$ and that
$\Gamma=$ 34.4 $\cdot$ 10$^{-6}$ eV
for the decay channel $B_c^+ \rightarrow J/\psi l^+ \nu$. 
Using the fact that 
from References \cite{gers2,ellis,BJ} the branching ratio of 
$B_c^+ \rightarrow J/\psi l^+ \nu$ is expected to be $\sim$2.2\%, we derive
the branching ratio for $B_c^+ \rightarrow B_s^0 \pi^+$ to be equal to 
4.7\%. 

According to Reference \cite{bspsiphi} we had 58$\pm$12 $B_s^0 \rightarrow
J/\psi \phi$ events in Run~I and if we multiply the yield by a factor of 40 
for Run~IIa we will have $\sim$ 2,400 events.

From Reference \cite{PAC99} we see that we expect 23,400 fully reconstructed
$B_s^0$ events in the decay channels $B_s^0 \rightarrow D_s^- \pi^+$ and 
$B_s^0 \rightarrow D_s^- \pi^+ \pi^- \pi^+$ for scenario A and 15,300 
events for scenario C. Scenario A corresponds to a Tevatron running 
with bunch spacing of 396 ns and instantaneous luminosity of 
0.7 $\times$ 10$^{32}$ cm$^{-2}$s$^{-1}$ and scenario C 
corresponds to a Tevatron running
with bunch spacing of 396 ns and instantaneous luminosity of
1.7 $\times$ 10$^{32}$ cm$^{-2}$s$^{-1}$. 
Based on these we conservatively assume a total of 
25,000 fully reconstructed $B_s^0$ events in Run~IIa.

From the CDF measurement of  
$R_{l}$ \cite{cdfbc}, from the 
branching ratios of $B_c^+ \rightarrow J/\psi l^+ \nu = 2.2\%$ and 
$B^+ \rightarrow J/\psi K^+ = 0.099\%$ \cite{pdg} and from the fact that 39.7\%
of $b$ quarks fragment into $B^+$ mesons \cite{pdg} we get that 
$\sigma(B_c^+)/\sigma(b)$ is equal to 2.3 $\cdot$ 10$^{-3}$.
On the other hand we know that $\sigma(B_s^0)/\sigma(b)$ is equal to
10.5\% \cite{pdg}, that is, $\sigma(B_s^0)/\sigma(B_c^+)$ is equal to
45.6. Therefore the observed number of $B_c^+ \rightarrow B_s^0 \pi^+$
events in Run~IIa will be equal to $25,000/45.6 \cdot 4.7\% \cdot A_{\pi}$
where $A_{\pi}$ is the acceptance for finding the pion in 
$B_c^+ \rightarrow B_s^0 \pi^+$ after having found the $B_s^0$. For 
$A_{\pi} = 100\%$, we would observe 26 such decays.  

If we wanted to estimate how many of the expected 25,000 fully reconstructed
$B_s^0$ events in Run~IIa will originate from $B_c^{+}$ decays, we would
have to multiply $25,000/45.6 = 548.2$ by the branching ratio
$BR(B_c^+ \rightarrow B_s^0/B_s^{0*} X)$ and the acceptance of reconstructing 
$X$ in the presence of the $B_s^0/B_s^{0*}$.

\begin{table}[bhtp]
\begin{center}
\begin{tabular}{|c|c|c|c|}
\hline  Decay Mode & Width in 10$^{-6}$ eV  \cite{changchen}\\
\hline $B_c^+ \rightarrow B_s^0 e^+ \bar{\nu_e}$ 
&  26.6\\
\hline $B_c^+ \rightarrow B_s^{0*} e^+ \bar{\nu_e}$
& 44.0\\
\hline $B_c^+ \rightarrow B_s^0 \pi^+$
&  73.3\\
\hline $B_c^+ \rightarrow B_s^0 \rho^+$
& 56.1\\ 
\hline $B_c^+ \rightarrow B_s^{0*} \pi^+$ & 64.7\\
\hline $B_c^+ \rightarrow B_s^{0*} \rho^+$ & 188.0\\
\hline $B_c^+ \rightarrow B_s^{0} K^+$ & 5.27\\
\hline $B_c^+ \rightarrow B_s^{0*} K^+$ & 3.72\\
\hline Total& 461.69\\
\hline
\end{tabular}\vspace*{4pt}
\caption{Theoretical estimates of the width of
$B_c^+$ decay modes involving $B_s^{0}$ or $B_s^{0*}$ in the
final states.\label{list2}}
\end{center}
\end{table}

In Table \ref{list2} we show the widths of various $B_c^+$ decays involving
a $B_s^{0}$ or a $B_s^{0*}$ meson from Reference \cite{changchen}. If this is a
complete list of the $B_c^+$ decays involving
a $B_s^{0}$ or a $B_s^{0*}$ and if we use the fact (as in the beginning of 
this section)
that 
$\Gamma=$ 34.4 $\cdot$ 10$^{-6}$ eV
for the decay $B_c^+ \rightarrow J/\psi l^+ \nu$ and
the corresponding branching ratio 
is expected to be $\sim$2.2\%, then $BR(B_c^+ \rightarrow B^0_s/B_s^{0*} X)$
is expected to be equal to 29.5\%. It will be very useful to measure the
above branching ratios. If the expectations are correct though, the 
above discussion indicates that we will not have many $B_c^+$ candidates
decaying to a $B_s^0$ or a $B_s^{0*}$. We have to also take into account
that it will not be easy to detect the photon from the $B_s^{0*}$ decay or
to reconstruct the $\rho$ meson in some of the above decays.
 
\subsubsection{Yield estimate for the $B_c^+ \rightarrow J/\psi l^+ \nu$ 
decay channel \label{imptq}}
\index{decay!$B_c \to J/\psi \ell^+\nu$}

For the decay channel $B_c^+ \rightarrow J/\psi l^+ \nu$ we expect to observe
$\sim$ 800 events in Run~IIa based on the 20.4 candidates of Run~I 
\cite{cdfbc} and the factor of 40 expected increase in the yield.

\subsubsection{Conclusion}

In Run~IIa we expect to see $\sim$ 800 $B_c^+ \rightarrow J/\psi l^+ \nu$
events, $\sim$ 360 $B_c^+ \rightarrow J/\psi \pi^+ $ events and maybe 
a small number of
events from other exclusive decay channels. We can use these events to
measure accurately the $B_c^+$ mass and lifetime as well as ratios of
various $B_c^+$ branching ratios.



\def\bentarrow{\:\raisebox{1.3ex}{\rlap{$\vert$}}\!\rightarrow}
\def\dk#1#2#3{
\begin{equation}
\begin{array}{r c l}
#1 & \rightarrow & #2 \\
 & & \bentarrow #3
\end{array}
\end{equation}
}

\def\mydek#1#2#3{
$\begin{array}{l}
#1 \rightarrow #2 \\
\bentarrow #3
\end{array}$
}

\def\shmydek#1#2{
$\begin{array}{l}
#1  \\
\bentarrow #2
\end{array}$
}


\def\threehalf{{3 \over 2}}
\def\half{{1 \over 2}}
\def\dkp#1#2#3#4{
\begin{equation}
\begin{array}{r c l}
#1 & \rightarrow & #2#3 \\
 & & \phantom{\; #2}\bentarrow #4
\end{array}
\end{equation}
}

\def\longbent{\:\raisebox{3.5ex}{\rlap{$\vert$}}\raisebox{1.3ex}%
	{\rlap{$\vert$}}\!\rightarrow}

\def\onedk#1#2{
\begin{equation}
\begin{array}{l}
 #1 \\
 \bentarrow #2
\end{array}
\end{equation}
}

\def\bthdk#1#2#3#4{
\begin{displaymath}
\begin{array}{c l}
  & #1#2 \\
  & \:\raisebox{1.3ex}{\rlap{$\vert$}}\raisebox{-0.5ex}{$\vert$}%
\phantom{#1}\!\bentarrow #3 \\
  & \bentarrow #4
\end{array}
\end{displaymath}
}


\section{Doubly-heavy Baryons}

\index{doubly-heavy baryons}
\index{baryons!doubly-heavy}
\index{baryons!triply-heavy}
Doubly-heavy baryons are baryons that contain two heavy quarks,
either $c c$, $b c$ or $b b$.  These hadrons provide yet another window 
onto the dynamics of heavy quarks. 
Doubly-heavy baryons are expected to be produced 
at respectable rates at the Tevatron; the basic production cross
sections are in the nanobarn range. Also of interest are the states
formed from three heavy quarks: $bbb$, $bbc$, $bcc$ or $ccc$. We do 
not provide cross section estimates for these triply-heavy baryons.


\subsection[Spectroscopy]
{Spectroscopy\authorfootnote{Author: R.K. Ellis}}
\index{doubly-heavy baryons!spectrum}
The spectroscopy of baryons containing two heavy quarks
$QQ$ is of interest because of similarities both to 
a quarkonium state, $Q\bar{Q}$ and to a heavy-light meson, $\bar{Q}q$.
On the one hand, the slow 
relative motion of the two heavy quarks is similar to quarkonium.
On the other hand, the lighter degree of freedom moves relativistically
around the slowly moving $QQ$. Since the $QQ$ is in a color antitriplet
state, in the heavy quark limit the system is very similar to a 
$\bar{Q}q$ system. 

A rich spectrum of excitations is expected, both excitations of the 
$QQ$ system as well as the light degrees of freedom. However from the 
experimental point of view, a detailed discussion of the excited states
is probably premature. We will limit our discussion here to 
the estimates of the masses of the lowest lying states,
$\Xi_{QQ^\prime}=(QQ^{\prime}q)$, $\Omega_{QQ^\prime}=(QQ^{\prime}s)$
containing two heavy quarks 
and the states $\Xi_{QQQ^\prime}=(QQQ^\prime)$, $\Omega_{QQQ}=(QQQ)$
containing three heavy quarks. Where the corresponding 
$J=\half$ state exists, the  $J=\threehalf$ states are 
unstable decaying to the ground state by photon emission.
The hyperfine splittings with the spin-$\half $ states are calculated 
using the procedure of De Rujula et al.~\cite{DeRujula:1975ge}. Ignoring 
electromagnetism we have that
\begin{equation}
H = H_0 + \alpha \sum_{i<j} \frac{s_i \cdot s_j}{m_i m_j} \; ,
\end{equation}
where the sum runs over the three pairs of quarks and
$\alpha$ is a constant fixed using the normal decuplet-octet splitting.

Estimates for the masses and spectra of the baryons containing two or 
more heavy quarks have been considered by many 
authors~\cite{Bjorkenccc}-\cite{Gershmass}.  
We are not aware of a reference which gives the masses of all the states
we are interested in. This is of some importance since the 
separation between states may be more reliable than the absolute energy scale. 
We have therefore created a complete list, using the 
simple procedure suggested by Bjorken\cite{Bjorkenccc}. 
The mass of the spin-$\threehalf$, 
$QQQ$ baryons are calculated by scaling from the $Q\bar{Q}$ state,
\begin{equation}
\frac{M_{QQQ}}{M_{Q\bar{Q}}} = \frac{3}{2}
 + \frac{k}{m^{\frac{4}{3}}}
 + \frac{k^\prime}{\ln^2 m/m_0} \; .
\end{equation}
Experience from the $\Delta,\rho$ and $\Omega, \phi$ systems suggest the values
\begin{equation}
M_{ccc} = (1.59 \pm 0.03) M_{c \bar{c}},\;\;\;
M_{bbb} = (1.56 \pm 0.02) M_{b \bar{b}}\; .
\end{equation}
The other spin-$\threehalf$ baryons are estimated using an equal 
spacing rule to interpolate between the $QQQ$ state and normal 
$J=\threehalf$ baryons. This results in the rule 
that replacing a $b$-quark by a $c$-quark costs $3.280$~GeV,
a $c$-quark by an $s$-quark costs $1.085$~GeV, and an $s$-quark
by a $u$-quark costs $0.145$~GeV.
Finally the masses of the spin-$\half$ baryons are calculated using
the procedure and mass values of DGG~\cite{DeRujula:1975ge}.
The results for the hyperfine splitting are
\begin{eqnarray}
E(abc,J=\threehalf) &=& E_0 + \frac{\alpha}{4} 
\Big( \frac{1}{m_a m_b}+\frac{1}{m_b m_c}+\frac{1}{m_c m_a} \Big),\nonumber \\
E(bbc,J=\half) &=& E_0 + \frac{\alpha}{4} \Big( \frac{1}{m_b^2}-\frac{3}{m_b m_c}\Big), \nonumber \\
E(abc,J=\half) &=& E_0 + \frac{\alpha}{4 m_a m_b m_c} 
\Big(-\Sigma-2 \sqrt{\Delta}\Big),\nonumber \\
E^\prime(abc,J=\half) &=& E_0 + \frac{\alpha}{4 m_a m_b m_c} 
\Big(-\Sigma+2 \sqrt{\Delta}\Big) \; ,
\end{eqnarray}
where $\Sigma=m_a+m_b+m_c,\Delta=\Sigma^2-3(m_a m_b+m_b m_c+m_c m_a)$.
The values found using this procedure 
are compared with the results of other authors in Table~\ref{tab:dhb0}. 
We quote only the central values
and refer the reader to the original publications for the estimated errors.
There is substantial agreement between all estimates if these 
errors are taken  into account. 

\begin{table}[htb]
\begin{center}
\begin{tabular}{|l|c|c|c|c|}
\hline
State                          & Mass~ref.~\cite{Richard:1994ae}& Mass~ref.~\cite{Ebert:1997ec} & Mass~ref.~\cite{Lichtenberg:1996kg}  & Mass \\ 
\hline
$\Xi^{++}_{cc}(ccu,J=\threehalf)      $     &     3.735                    &        3.81            &                3.746             &      3.711 \\  
\hline
$\Omega^{+}_{cc}(ccs,J=\threehalf)$         &     3.84                     &        3.89            &                3.851             &      3.848 \\
\hline
$\Omega^{++}_{ccc}(ccc,J=\threehalf)$       &                           &                        &                                  &      4.925\\
\hline
\hline
$\Xi^{++}_{cc}(ccu,J=\half)$                &     3.70                     &        3.66            &                 3.676            &      3.651 \\  
\hline
$\Omega^{+}_{cc}(ccs,J=\half)$              &    3.72                      &        3.76            &                3.787             &      3.811 \\
\hline
\hline
$\Xi^{+}_{bc}(bcu,J=\threehalf)$            &    7.02                      &        7.02            &                7.083             &      7.000 \\
\hline
$\Omega^{0}_{bc}(bcs,J=\threehalf)$         &    7.105                     &        7.11            &                7.165             &      7.128 \\
\hline
$\Xi^{+}_{bcc}(bcc,J=\threehalf)$           &                              &                        &                                  &      8.202 \\
\hline 
\hline
$\Xi^{+}_{bc}(bcu,J=\half)$                 &    6.93                      &        6.95            &                7.029             &      6.938 \\
\hline
$\Xi^{\prime \; +}_{bc}(bcu,J=\half)$       &                              &        7.00            &                7.053             &      6.971 \\
\hline
$\Omega^{0}_{bc}(bcs,J=\half)$              &    7.00                      &        7.05            &                7.126             &      7.095 \\
\hline
$\Omega^{\prime \; 0}_{bc}(bcs,J=\half)$    &                              &        7.09            &                7.148             &      7.115 \\
\hline
$\Xi^{+}_{bcc}(bcc,J=\half)$                &                              &                        &                              &      8.198 \\
\hline
\hline
$\Xi^{0}_{bb}(bbu,J=\threehalf)$            &   10.255                     &        10.28           &                10.398            &      10.257\\
\hline
$\Omega^{-}_{bb}(bbs,J=\threehalf)$         &   10.315                     &        10.36           &                10.483            &      10.399\\
\hline
$\Xi^{0}_{bbc}(bbc,J=\threehalf)$           &                              &                        &                                  &      11.481\\
\hline
$\Omega^{-}_{bbb}(bbb,J=\threehalf)$        &                              &                        &                                  &      14.760\\
\hline
\hline
$\Xi^{0}_{bb}(bbu,J=\half)$                 &   10.21                      &        10.23          &                                  &      10.235\\
\hline
$\Omega^{-}_{bb}(bbs,J=\half)$              &   10.27                      &        10.32          &                                  &      10.385\\
\hline
$\Xi^{0}_{bbc}(bbc,J=\half)$                &                              &                        &                                  &      11.476\\
\hline
\end{tabular}
\end{center}
\caption[Mass estimates for baryons with two or more heavy quarks.]
{Mass estimates in GeV for low-lying baryons with two or more 
heavy quarks. 
The last column shows our values derived using the methods
of Ref~\cite{Bjorkenccc}.}
\label{tab:dhb0}
\end{table}

\subsection[Production]
{Production\authorfootnote{Authors: E. Braaten, A. Likhoded}}
\index{doubly-heavy baryons!production}
The direct production of a doubly-heavy baryon
can be treated within the NRQCD factorization framework described in
Section~\ref{Quarkonium_production}.
The cross section for the direct production of a $b c$ baryon $H$ 
can be expressed in the form (\ref{chnine:ppbarcross}) and 
(\ref{chnine:partoncross}), except that $b \bar b(n)$ is replaced by $b c(n)$.
The short-distance cross section $d\hat{\sigma}[ij\to  b c(n)+X]$
for creating the $b c$ in the color and angular-momentum state $n$
can be calculated as a perturbative expansion in $\alpha_s$ 
at scales of order $m_c$ or larger.  
The nonperturbative matrix element
$\langle O^H(n)\rangle$ encodes the probability for a $b c$ 
in the state $n$ to bind to form the baryon $H$. 
The matrix element scales as a definite power 
of the relative velocity $v$ of the charm quark.

The production mechanisms for $b c$ are similar to
those for $\bar b c$, because two heavy quark-antiquark pairs 
must be created in the collision.  
The lowest order mechanisms for creating $b c$ are the 
order-$\alpha_s^4$ processes $q \bar q, g g \to (b c) + \bar b \bar c$.
The color state of the quark-quark system can be 
either color-antitriplet or color-sextet.
From counting the color states, we expect the color-sextet cross section 
to be about a factor of 2 larger than the color-antitriplet cross section.
Thus the relative importance of the two color states should be
determined primarily by the probability for the $b$ and $c$ to bind with a 
light quark to produce a doubly-heavy baryon.

In the ground state $\Xi_{bc}$ of the $b c$ baryon system,
the $b c$ is in a color-antitriplet S-wave state that is relatively 
compact compared to the size of the baryon.  
The $b c$ diquark behaves very much like the heavy antiquark
in a heavy-light meson.
The probability that $b$ and $c$ quarks that are created with small 
relative momentum in a parton collision will hadronize into the $\Xi_{bc}$ 
is therefore expected to be 
greatest if the $b c$ is in a color-antitriplet S-wave state.
All other NRQCD matrix elements are suppressed by powers of $v$.
If we keep only the color-antitriplet S-wave contribution, 
the formula for the cross section of $\Xi_{bc}$ is very similar to 
that for the $B_c$ in the color-singlet model.  
The short-distance cross section for creating a color-singlet $\bar b c$ 
is replaced by the cross section for creating a color-antitriplet $b c$.
The long-distance factor for $B_c$, which is proportional to the square
of the radial wavefunction at the origin $R_{1S}(0)$, is replaced by
a color-antitriplet matrix element for the $\Xi_{bc}$.
This matrix element can be treated as a phenomenological parameter
analogous to the color-octet matrix elements for quarkonium production.
It can also be estimated using a quark potential model 
for the doubly-heavy baryon,
in which case it is proportional to the square of an effective 
radial wavefunction at the origin $R^{bc}_{1S}(0)$ for the $b c$ diquark.
The quark model estimate is probably much less reliable than the 
potential model estimate of $R_{1S}(0)$ for $B_c$.
We will refer to the cross sections obtained by using the 
leading order $b c$ cross section and a
quark model estimate for the S-wave color-antitriplet matrix element 
as the {\it diquark model} for doubly-heavy baryon production.

The production of doubly-heavy baryons in the diquark model 
at order $\alpha_s^4$ was
studied in detail in the series of papers~\cite{prod}. 
The resulting differential cross sections  ${d\sigma}/{dp_T}$ 
for $\Xi_{cc}$, $\Xi_{bc}$ and $\Xi_{bb}$ baryons
are presented in Fig.~\ref{tevpic} and compared to that of the $B_c$ meson.
The cross sections integrated over $p_T$ greater than ${p_T}^{min}$ 
are shown in Fig.~\ref{tevpic1}.
The cross sections are evaluated at 2.0~TeV and integrated over $|y|<1$.
The quark masses were set to $m_c=1.5$ GeV and $m_b=4.8$ GeV.  
The QCD coupling constant was fixed at $\alpha_s=0.23$.  
The radial wave functions at the origin for the
heavy diquarks were taken to be $R^{cc}_{1S}(0)=0.601$ GeV$^{3/2}$, 
$R^{bc}_{1S}(0)=0.714$ GeV$^{3/2}$, 
and $R^{bb}_{1S}(0)=1.35$ GeV$^{3/2}$ \cite{Spectr}.  
The largest uncertainties come from the choice of scale in the
overall factor $\alpha_s^4(\mu)$ and from the radial wave functions 
at the origin for the heavy diquarks.
The production rates for $\Xi_{cc}$ and $\Xi_{bc}$ are predicted 
to be as large as 50\% of the
production rate for $(B_c + B^*_c)$ for ${p_T}^{min} = 10$~GeV.

The uncertainties in the prediction of the $\Xi_{bc}$ cross section 
can be greatly reduced by normalizing the cross section to that of 
the $B_c$.  The reason is that the short-distance cross sections 
come from exactly the same Feynman diagrams, except that the color-singlet
$\bar b c$ is replaced by a color-antitriplet $b c$. 
Regions of phase space with various gluon virtualities $\mu$ 
are weighted in almost the same way, so the factor of $\alpha_s^4(\mu)$
that gives the largest uncertainty in the $B_c$ cross section 
cancels in the ratio $\sigma (\Xi_{bc})/\sigma(B_c)$
given that $\mu_{bc} = \mu_{b\bar{c}}$.  
The largest remaining uncertainty comes from the radial wave function 
at the origin $R^{bc}_{1S}(0)$ for the heavy diquark in the $\Xi_{bc}$.  
Using the value $R^{bc}_{1S}(0)=0.714~{\rm GeV}^{3/2}$,
one gets the estimate
\begin{equation}
\sigma (\Xi_{bc})/\sigma (B_c) \simeq 0.5\,.
\end{equation}
Taking into account the cascade decays of excited $b c$ baryon
states, the inclusive $\Xi_{bc}$ 
cross-section integrated over $p_T>6$ GeV and $|y|<1$ is predicted to be 
\begin{equation}
\sigma (\Xi_{bc})=1.5~nb\,.
\end{equation}
Alternatively, taking the experimental value for the $B_c$ cross 
section~\cite{cdfbc}, 
we get 
\begin{equation}
\sigma (\Xi_{bc})=5\pm 3~nb\,.
\end{equation}
With an integrated luminosity of about 100 pb$^{-1}$ in Run I,
this corresponds to more than $10^5$ events with $\Xi_{bc}$ baryons 
in the considered kinematical region. 

The predicted cross sections for $bb$ and $cc$ baryons have larger 
uncertainties than those for $bc$ baryons.
The large uncertainties from the $\alpha_s^4(\mu)$ factor
can not be removed by normalizing to $b \bar b$ or $c \bar c$
quarkonium cross sections, since they arise from very different 
short-distance parton processes.
However, from the predictions in Fig.~\ref{tevpic},
one can expect roughly the same number of $\Xi_{cc}$ and $\Xi_{bc}$ 
events in the kinematical region of $p_T>10$~GeV and $|y|<1$. 
The total number of $\Xi_{bb}$ events will be about a 
factor of 10 smaller.


\subsection[Decays]{Decays\authorfootnote{Authors: A. Likhoded, R. Van Kooten}}
\label{DHB:decay}
\index{doubly-heavy baryons!decays}
\index{doubly-heavy baryons!lifetimes}

Since the major thrust of this report is $B$-physics,
we may be tempted to limit discussion to baryons containing $b$ quarks.
However, such baryons often decay by cascades into baryons containing
$c$ quarks and identification of the latter may be an important
first step in reconstructing doubly-heavy baryons containing $b$ quarks.

The lifetimes for the ground states of the doubly-heavy baryons
have been calculated in
the  framework of the operator product expansion,
which involves an expansion in inverse powers of the 
heavy quark mass~\cite{lifetime}.  In leading order of $1/m_Q$,
the inclusive widths are determined by the spectator approximation 
with QCD corrections.  The next order in $1/m_Q$ 
takes into account corrections connected with quark motion in the 
doubly-heavy baryon and with the chromomagnetic quark interaction.
In doubly-heavy baryons, there is Pauli interference (PI) of
quark decay products with identical quarks from the initial state,
as well as weak scattering (WS, or weak exchange) where exchange
of a $W^{\pm}$ occurs between quarks. Both effects play important
roles in the mechanism of doubly-heavy baryon decay. 
For example, Pauli interference leads to the increase of the 
$b$-quark decay contribution to $bc$-baryon by a factor of two. 
In this case, the sign is basically determined from the interference
of the charm quark of the initial state with the charm quark from
the $b$-quark decay. The antisymmetric color structure of the
baryon wave function leads to the positive sign for the Pauli interference. 
The overall effect of the corrections to the spectator mechanism can reach
40--50\%. In Tables~\ref{tab:one} and \ref{tab:two}, we show the 
contributions of the different
modes to the total decay widths of $cc$ and $bc$ baryons.
One can see from these tables that the weak scattering contributions 
are comparable with the spectator contributions. 
The estimates of the lifetimes of the $\Xi_{cc}$ and $\Xi_{bc}$ baryons 
from the operator product expansion 
are\cite{Likhoded:1999yv,Kiselev:2000kh,Kiselev:1999sy}:
\begin{eqnarray}
\tau_{\Xi_{cc}^{++}}&=&0.43\pm0.1~{\rm ps}, \nonumber\\*
\tau_{\Xi_{cc}^{+}}&=&0.12\pm0.1~{\rm ps}, \nonumber\\*
\tau_{\Xi_{cc}^{0}}&=&0.28\pm0.07~{\rm ps}, \nonumber\\*
\tau_{\Xi_{bc}^{+}}&=&0.33\pm0.08~{\rm ps}. 
\end{eqnarray}

\begin{table}[thp]
\begin{center}
\begin{tabular}{|c|c|c|c|}
\hline
Mode & width, ps$^{-1}$ & Fraction of $\Xi_{cc}^{++}$ width &
Fraction of $\Xi_{cc}^{+}$ width \\
\hline
$c\to sdu$ & 2.894 & 1.24 & 0.32 \\
\hline
$c \to s {\ell}^{+} \nu$ & 0.760 & 0.32 & 0.09 \\
\hline 
PI & $-1.317$ & $-0.56 $& $-$\\
\hline 
WS & 5.254 & $-$ & 0.59 \\
\hline \hline
$\Gamma_{\Xi_{cc}^{++}}$ & 2.337 & 1& $-$ \\
\hline
$\Gamma_{\Xi_{cc}^{+}}$ & 8.909 & $-$ & 1 \\
\hline
\end{tabular}
\end{center}
\caption[Fractional contributions of different modes to the total
decay width of doubly-charmed baryons, $\Xi_{cc}$.]
{Fractional contributions of different modes to the total
decay width of doubly-charmed baryons, $\Xi_{cc}$.
PI and WS are Pauli interference and weak scattering effects,
respectively.}
\label{tab:one}
\end{table}

\begin{table}[thb]
\begin{center}
\begin{tabular}{|c|c|c|}
\hline
Mode & Fraction of $\Xi_{bc}^{+}$ width & Fraction of $\Xi_{bc}^{0}$ width  \\
\hline
$b\to c+X$ & 0.120 & 0.17  \\
\hline
$c \to s+X$ & 0.37 & 0.31 \\
\hline 
PI & 0.23 & 0.20 \\
\hline 
WS & 0.20 & 0.31 \\
\hline
\end{tabular}
\end{center}
\caption[Fractional 
contributions of different modes to the total
decay width of $\Xi_{bc}$ baryons.]
{Fractional contributions of different modes to the total
decay width of $\Xi_{bc}$ baryons.
PI and WS are Pauli interference and weak scattering effects,
respectively.}
\label{tab:two}
\end{table}

\begin{table}[thb]
\begin{center}
\begin{tabular}{|c|c|c||c|c|c|}
\hline
Baryon                & Mode                                   & Br (\%) &  
Baryon                & Mode                                   & Br (\%) \\ 
\hline
$\Xi_{bb}^{\diamond}$ & $\Xi_{bc}^{\diamond} \; l^- \bar \nu_l$ & 11.2 &
                      &                                        &       \\
                      & $\Xi_{bc}^{\diamond} \; \pi^{-}$       &  0.3  & 
                      &                                        &       \\
                      & $\Xi_{bc}^{\diamond} \; \rho^{-}$      &  3.4  &
                      &                                        &       \\ 
\hline 
$\Xi_{bc}^{0}$        & $\Xi_{cc}^{+}  \; l^- \bar \nu_l$      &  3.3  & 
$\Xi_{bc}^{+}$        & $\Xi_{cc}^{++} \; l^- \bar \nu_l$      &  3.5  \\
                      & $\Xi_{bs}^{-}  \; l^+ \nu_l$           &  2.8  & 
                      & $\Xi_{bs}^{0}  \; l^+ \nu_l$           &  3.0  \\
                      & $\Xi_{cc}^{+}  \; \pi^{-}$             &  0.55 & 
                      & $\Xi_{cc}^{++} \; \pi^{-}$             &  0.6  \\
                      & $\Xi_{cc}^{+}  \; \rho^{-}$            &  1.4  & 
                      & $\Xi_{cc}^{++} \; \rho^{-}$            &  1.5  \\
                      & $\Xi_{bs}^{-}  \; \pi^{+}$             & 12.3  & 
                      & $\Xi_{bs}^{0}  \; \pi^{+}$             & 13.1  \\
                      & $\Xi_{bs}^{-}  \; \rho^{+}$            &  5.2  & 
                      & $\Xi_{bs}^{0}  \; \rho^{+}$            &  5.6  \\
\hline 
$\Xi_{cc}^{+}$        & $\Xi_{cs}^{0}  \; l^+ \nu_l$           &  6.9  & 
$\Xi_{cc}^{++}$       & $\Xi_{cs}^{+}  \; l^+ \nu_l$           & 14.9  \\
                      & $\Xi_{cs}^{0}  \; \pi^{+}$             &  4.2  & 
                      & $\Xi_{cs}^{+}  \; \pi^{+}$             &  8.1  \\
                      & $\Xi_{cs}^{0}  \; \rho^{+}$            & 24.8  &  
                      & $\Xi_{cs}^{+}  \; \rho^{+}$            & 45.1  \\
\hline
\end{tabular}
\end{center}
\caption[Exclusive (spectator) decay modes of doubly-heavy baryon 
calculated in the framework of QCD sum rules.]
{Exclusive (spectator) decay modes of doubly-heavy baryon 
calculated in the framework of QCD sum rules. The symbol 
$^{\diamond}$ represents electric charge, i.e., $^{\diamond = \pm , 0}$.}
\label{tab:three}
\end{table}

We proceed to discuss exclusive decay modes of the 
doubly-heavy baryons that may be observable.
Let us first consider spectator decay modes, 
in which the doubly-heavy baryon decays into either
a lighter doubly-heavy baryon or a baryon containing a single heavy quark. 
In Table~\ref{tab:three},
we give branching fractions for exclusive spectator 
decay modes that were calculated in the
framework of QCD sum rules \cite{SR}. Some of the decay modes have
surprisingly large branching fractions, particularly 
$\Xi_{cc}^{+}\to\Xi_{cs}^0 \pi^{+} (\rho^+)$,
$\Xi_{cc}^{++}\to\Xi_{cs}^{+}\pi^{+}(\rho^+)$ and semileptonic decays.
A recent work from Onishchenko calculating these branching ratios
including results from three-point NRQCD sum rules~\cite{Onish} give
even larger values.

We also consider another class of exclusive decay modes for
doubly-heavy baryons that may be observable.
As pointed out above, the contribution from weak scattering to the 
$\Xi_{bc}^{+}$ and $\Xi_{bc}^{0}$ decay width is about $20\%$. 
This type of decay is characterized by specific kinematics. 
In the rest frame of the $\Xi_{bc}$ baryon, 
the $c$ and $s$ quarks from the scattering process $b c \to c s$
move in the opposite directions with a high momentum of about 3.2 GeV. 
Both the $c$ and $s$ quarks then fragment, 
producing multiparticle final states.
These states include the 3-particle states $ D^{(*)}K^{(*)}N$. 
The $D^{(*)}$ and $K^{(*)}$ can be produced by fragmentation
processes: $c \to D^{(*)}+X_q$ and $s \to K^{(*)}+X_q$.
However, in order for only one additional particle $N$
to be produced in the decay of $\Xi_{bc}$,
the $D^{(*)}$ and $K^{(*)}$ must be in the hard part of
fragmentation spectrum with $z>0.8$. 
We can use a well-known parametrization of the fragmentation functions
to estimate the probability for such a decay:
\begin{eqnarray}
Br(\Xi_{bc} \to D^{(*)}K^{(*)}N)
&\approx&Br(WS) \times W(z_D>0.8)\times
	W(z_K>0.8)\nonumber\\
&=& 0.2 \times 0.2 \times 0.04\; .
\end{eqnarray}
The resulting rough estimate of the branching fraction for  
$\Xi_{bc} \to D^{(*)}K^{(*)}N$ is 0.2\%.
We conclude that the branching fractions for these modes
may be large enough to be observed.



\subsection[Experimental observability]
{Experimental Observability\authorfootnote{Authors: R.K. Ellis, R. Van Kooten}}
\index{doubly-heavy baryons!observability}

To set the scale for observability of doubly-heavy baryons, 
we note that CDF has
observed~\cite{cdfbc} about 20 events of the type
$B_c \rightarrow J/\psi \ell^{\pm} \nu$ in 
0.11~fb$^{-1}$ using the $\mu^+ \mu^-$ decay of the $J/\psi$. 
We take this process to have a total branching ratio of 0.3\%.
We assume comparable trigger and reconstruction efficiencies
for the $B_c$ and typical doubly-heavy baryon decay modes.
In Section~\ref{DHB:decay}, the cross sections for baryons 
containing $cc$ or $bc$ were predicted to be within about a factor
of 2 of the cross section for $B_c$ for $p_T > 10$~GeV.
With data samples of 3, 10, and 30~fb$^{-1}$,
we are therefore initially restricted to doubly-heavy baryon decay modes 
with branching ratios greater than
of order $10^{-4}$, $5 \times 10^{-5}$, and $1.5 \times 10^{-5}$
respectively. Special purpose detectors with better triggering abilities
(such as BTeV) will be able to investigate rarer modes.

Historically, larger samples of rarer $b$-quark states are available 
for measuring properties such as lifetimes and polarization through 
``semi-exclusive'' decays where not all the decay products are
reconstructed and/or there is an escaping neutrino from
semileptonic decay.  As an example, the $\Lambda_b$ state was
first observed at LEP~\cite{Akers} through an excess of 
``correct-sign'' $\Lambda$-$\ell^-$ and $\Lambda$-$\ell^+$ correlations
over the ``wrong-sign'' correlations.  A similar situation will exist for
doubly-heavy baryons, but now there can be two leptons in the decay chain
exhibiting charge correlations.

If a doubly-heavy baryon decays semileptonically into either
a lighter doubly-heavy baryon or a baryon containing a single heavy quark,
and if the second baryon also decays semileptonically,
they will give rise to two leptons associated with the same jet. 
Due to the large masses of the heavy baryons, both of these leptons
tend to be at large values of $p_T$ relative to the jet axis. 
It is also interesting to note that cascading semileptonic decays 
in the case of 
$\Xi_{bb}$ or $\Xi_{cc}$ can result in same-sign leptons in the same jet, 
a process with very little background.
The largest irreducible background comes from a gluon splitting into
$b\bar{b}$ to produce a jet containing two $b$-hadrons,
one of which decays semileptonically and the other
mixes before decaying semileptonically. 
Another background is the decay of one $b$ quark semileptonically
along with a same-sign lepton from the cascade decay through charm
of the other $\bar{b}$ quark. However, this
background level can be reduced by appropriate kinematic cuts on 
the lepton from the cascade decay.
More problematic in this case would be lepton misidentification faking the
same-sign lepton signal.

Other doubly-heavy baryon decays resulting in two opposite-sign leptons
may still be distinctive due to the kinematics of the decay, but can suffer
from a large physics background due to generic 
$b \rightarrow c \rightarrow s$ decays.  Typical branching ratios
for both types of decays are collected in 
Table~\ref{tab:semicasc} using the
semileptonic rates predicted in ref.~\cite{Onish}.

\vspace*{0.5cm}
\begin{table}[htb]
\begin{center}
\begin{tabular}{|lll|c|c|c|}
\hline Mode & & & Br (\%) &  Lepton charge correlation \\ \hline
$\Xi_{bb}^-$ & $\rightarrow \Xi_{bc}^0 \ell^- \nu$ 
             & $\rightarrow \Xi_{cc}^+ \ell^- \nu$ & 0.69\% & same-sign \\
             &  
             & $\rightarrow \Xi_{bs}^- \ell^+ \nu$ & 0.61\% & opp-sign \\ \hline

$\Xi_{bb}^0$ & $\rightarrow \Xi_{bc}^+ \ell^- \nu$ 
             & $\rightarrow \Xi_{cc}^{++} \ell^- \nu$ &  0.73\% & same-sign \\
             &
             & $\rightarrow \Xi_{bs}^- \ell^+ \nu$ & 0.61\% & opp-sign \\ \hline

$\Xi_{bc}^0$ & $\rightarrow \Xi_{cc}^+ \ell^- \nu$
             & $\rightarrow \Xi_{cs}^0 \ell^+ \nu$ & 0.35\% & opp-sign \\
             & $\rightarrow \Xi_{bs}^- \ell^+ \nu$ 
             & $\rightarrow \Xi_{cs}^0 \ell^- \nu$ & 0.37\% & opp-sign \\ \hline

$\Xi_{bc}^+$ & $\rightarrow \Xi_{cc}^{++} \ell^- \nu$
             & $\rightarrow \Xi_{cs}^+ \ell^+ \nu$ & 0.82\% & opp-sign \\
             & $\rightarrow \Xi_{bs}^0 \ell^+ \nu$ 
             & $\rightarrow \Xi_{cs}^+ \ell^- \nu$ & 0.40\% & opp-sign \\ \hline

$\Xi_{cc}^{++}$ & $\rightarrow \Xi_{cs}^+ \ell^+ \nu$
               & $\rightarrow \Xi \ell^+ X$ &        1.5\% & same-sign \\ \hline
$\Xi_{cc}^+$   & $\rightarrow \Xi_{cs}^0 \ell^+ \nu$
             & $\rightarrow \Xi \ell^+ X$   &       0.68\% & same-sign \\ \hline
\end{tabular}
\end{center}
\caption{ Decay rates of cascading semileptonic decays from
doubly-heavy baryons.}
\label{tab:semicasc}
\end{table}

For fully exclusive decays,
decay diagrams involving either spectator
decays or W-exchange (including Cabibbo-suppressed decays) were
considered. For a doubly-heavy baryon of the form $bbq$ or $bcq$,
the decays into $J/\psi$'s are the favored decay modes, 
because of the ease of
triggering on subsequent decays into $\mu^+ \mu^-$ and $e^+e^-$ 
and also because they absorb $Q$-value which would otherwise
produce pion multiplicity contributing to combinatorial
background. It is predicted that ``golden" exclusive decay modes such
as $\Xi_{bc}^+ \rightarrow \Lambda_c^+ J/\psi$ or 
$\Xi_{bb}^0 \rightarrow \Lambda_b^0 J/\psi$ into triggerable modes 
have small total branching ratios between $10^{-5}$
and $10^{-7}$, although the inclusive rate for
all decays containing a $J/\psi$ may be as high as 
$10^{-3}$~\cite{Anatoli_comm}.

Decays without a distinctive $J/\psi$ may still be accessible due to 
the abundance of cascading decays. 
These cascade decays can then result in triggerable decays either 
from jets containing multiple leptons
due to sequences of semi-leptonic decays as discussed above or from
many tracks with relatively large impact parameter significance.

\begin{figure}[htbp]
\begin{center}
\mbox{\epsfig{file=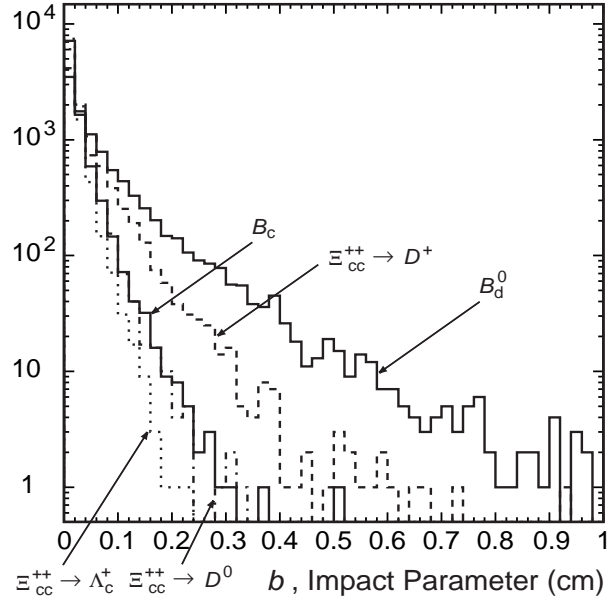,height=8cm,width=8cm}}
\end{center}
\caption{ 
Impact distribution of all charged decay products in the cascade 
decays of $\Xi_{cc}^{++}$ via $\Lambda_c$, $D^0$, and $D^+$ 
compared to the same distribution of the decay products from
$B_c$ and $B_d^0$.}
\protect\label{fig:dhbimpact}
\end{figure}

A doubly-heavy baryon decaying into either a $b$-baryon or another
doubly-heavy baryon will have a decay length that
gives rise to a large number of high-impact parameter tracks that
could allow the use of a vertex/silicon track trigger. A doubly-heavy
baryon with a short lifetime (e.g., 
$\tau(\Xi_{cc}^{++}) \approx 0.4$~ps, 
$\tau(\Xi_{bc}^0) \approx 0.3$~ps, and 
$\tau(\Xi_{bb}^0) \approx 0.7$~ps) can
decay into a charm meson with a reasonably long lifetime
(e.g., $\tau(D^+) = 1.06$ ps). 
The large mass of the doubly-heavy baryon parent can give
substantial transverse momentum kick with respect to the original
flight direction. Including a fairly long charm decay length, the
final decay products can have impact parameters comparable to those
of $b$ hadrons with typical lifetimes of $1.5$~ps. These decays 
can be found with reasonable efficiency using a vertex/silicon track trigger.
To test this idea, a Monte Carlo study was made at the four-vector level.
$\Xi_{cc}^{++}$ baryons were generated with a $p_T$ distribution 
according to Ref.~\cite{anatoliprod} with a lifetime of 0.43~ps.
This doubly-heavy baryon was allowed to decay to a singly-charmed
meson or baryon in a two-body decay for simplicity. The impact 
parameters of the resultant decay products of the $\Lambda_c$, 
$D^0$, or $D^{\pm}$ were found. Distributions of these impact parameters
are shown in Fig.~\ref{fig:dhbimpact} which, for the case of 
$\Xi_{cc}^{++} \rightarrow D^+$, is intermediate in extent between
that of the products from $B_c$ decay and from $B^0_d$ decay. 
Considering the earlier description of the efficiency of a typical
vertex/silicon track trigger (see \ref{sec:results}), there is potential 
promise in triggering on doubly-heavy baryons in cascade decays.

Finally, the decay mode $\Xi_{bc} \rightarrow D^{(*)} K^{(*)} N$
discussed in Section~\ref{DHB:decay}
shows promise due to its potentially relatively large rate and 
the nature of its decay products which allows a fairly clean reconstruction.
If $N = p$, the proton would allow for the use of particle identification. 
The $D^*$ can be identified by the standard procedure of cutting on the
small value of $\Delta_m = m(D^*) - m(D)$, where $m(D^*)$ is the reconstructed
mass of the $D^*$ using the soft pion in $D^* \rightarrow D \pi$.
Triggering would rely on a silicon track trigger as described above.

Decay modes containing a $\Lambda_c$ in the cascade chain also 
have potential due to the possibility of the clean reconstruction
of this charm baryon in the decay $\Lambda_c \rightarrow p K \pi$.
Of note is that CDF cleanly reconstructed 197 signal events with
$\Lambda_c$ for a $\Lambda_b$ lifetime measurement\cite{cdflblife}.
They used the lepton from the semileptonic decay 
$\Lambda_b \rightarrow \Lambda_c \ell \nu X$ as part of their trigger.
Triggering certainly is a serious issue for any all-hadronic decay mode.

\begin{table}[phtb]
\begin{center}
\begin{tabular}{|l|c|c|l|c|}
\hline
State                    & Lifetime~[ps]               & Mass~[GeV] & Interesting Exclusive  & Estimated\\ 
                         &                             &            & Decay Modes            &   Br     \\
\hline
$\Xi^{++}_{cc}(ccu)$     &  0.43\cite{Kiselev:1999sy}          &   3.651     & 
\mydek{D^{*+}  \pi^+ \Lambda^0}{p \pi^-}{K^- \pi^+ \pi^+}
& $9.4 \times 10^{-4}$ \\  
                         &                                     &                            &          
\mydek{D^{*+} p \bar{K}^0}{\pi^+ \pi^-}{K^- \pi^+ \pi^+}
& $4.7 \times 10^{-4}$ \\ 
                         &                                     &                            &          
\mydek{\Lambda_c^+(cdu) \; \pi^+ \; \bar{K}^0}{\pi^+ \pi^-}{p K^- \pi^+}
& $4 \times 10^{-4}$ \\          
\hline
$\Xi^{+}_{cc}(ccd)$      &  0.11\cite{Kiselev:1999sy}          &   3.651     & 
\mydek{D^{*+} \; \Lambda^0}{ p \pi^-}{K^- \pi^+ \pi^+}
& $4 \times 10^{-4}$ \\ 
                         &                                     &                            &          
\mydek{D^0 p \bar{K}^0}{ \pi^+ \pi^-}{K^- \pi^+}
& $2 \times 10^{-4}$ \\  
                         &                                     &                            &          
\mydek{\Lambda_c^+(cdu) \;  \bar{K}^0}{ \pi^+ \pi^-}{p K^- \pi^+}
& $1.6 \times 10^{-4}$ \\      
\hline
$\Omega^{+}_{cc}(ccs)$   &  0.5 \cite{BJ}                &   3.811     &$\Lambda^0 K^0 D^+$  &  \\
\hline
$\Omega^{++}_{ccc}(ccc)$ &  0.3\cite{BJ}                 &   4.925     & Cascades to $\Xi^{+}_{cc}$   & -- \\
\hline
\end{tabular}
\end{center}
\caption{Properties and interesting exclusive hadronic decay modes of
multiple-charm baryons.}
\label{tab:dhb1}
\end{table}
\index{decay!$\Xi^{++}_{cc}(ccu)$}
\index{decay!$\Xi^{+}_{cc}(ccd) $}
\index{decay!$\Omega^{+}_{cc}(ccs)$}
\index{decay!$\Omega^{++}_{ccc}(ccc)$}

\begin{table}[htbp]
\begin{center}
\begin{tabular}{|l|c|c|l|c|}
\hline
State                    & Lifetime~[ps]               & Mass~[GeV] & Interesting Exclusive  & Estimated\\ 
                         &                             &            & Decay Modes            &   Br     \\
\hline
$\Xi^{+}_{bc}(bcu)$      &  $0.33\pm0.08$ \cite{Kiselev:2000kh}  &     6.971                  &
\mydek{\Lambda_c^+ \; J/\psi}{\mu^+ \mu^-}{p K^- \pi^+}
& $< 2 \times 10^{-7}$ \\ 
                         &                                     &                            &          
\mydek{D^{(*)0} \; p  \bar{K}^0}{\pi^+ \pi^-}{K^- \pi^+}
& $2.1 \times 10^{-5}$ \\  
                         &                                     &                            &          
\shmydek{D^{(*)+} \; p  K^{(*)-}}{K^- \pi^+ \pi^+}
& $4 \times 10^{-5}$  \\  
\hline
$\Xi^{0}_{bc}(bcd)$      &  $0.28 \pm 0.07$\cite{Kiselev:2000kh} &     6.971                  & 
\mydek{\Xi_c^0(csd) \; J/\psi}{\mu^+ \mu^-}{\Lambda^0 \bar{K}^0}
& $< 2 \times 10^{-7}$ \\ 
                         &                                     &                            &          
\shmydek{D^{(*)0} \; p  K^{(*)-}}{K^- \pi^+}
& $6.2 \times 10^{-5}$ \\  
                         &                                     &                            &          
\shmydek{D^{(*)+} \; p \pi^- K^{(*)-}}{K^- \pi^+ \pi^+}
& $4 \times 10^{-5}$ \\  
\hline
$\Omega^{0}_{bc}(bcs)$   &  --                      &     7.095          & 
\mydek{\Omega_c^0(ssc) \; J/\psi}{\mu^+ \mu^-}{\Omega^- \pi^+}
& $< 1.44 \times 10^{-6}$ \\
\hline
$\Xi^{+}_{bcc}(bcc)$     &  --                         &     8.198           & Cascades to above states
 & \\
\hline
\end{tabular}
\end{center}
\caption[Baryons containing a $b$ and a $c$ quark]
{Properties and interesting exclusive hadronic decay modes of
	heavy baryons containing both a $b$ and a $c$ quark.}
\label{tab:dhb2}
\end{table}
\index{decay!$\Xi^{+}_{bc}(bcu)$}
\index{decay!$\Xi^{0}_{bc}(bcd)$}
\index{decay!$\Omega^{0}_{bc}(bcs)$}
\index{decay!$\Xi^{+}_{bcc}(bcc)$}

With these considerations, a list of potentially interesting 
exclusive hadronic decay modes
for doubly-heavy baryons are collected in Tables~\ref{tab:dhb1} 
and~\ref{tab:dhb2}.
These decay modes, in addition to the semi-exclusive decays 
from semileptonic decays, deserve further study as possible discovery 
modes for doubly-heavy baryons.

\begin{table}[htbp]
\begin{center}
\begin{tabular}{|l|c|c|l|c|}
\hline
State                    & Lifetime~[ps]               & Mass~[GeV] & Interesting Exclusive  & Estimated\\ 
                         &                             &            & Decay Modes            &   Br     \\
\hline
$\Xi^{0}_{bb}(bbu)$      & $0.79$\cite{Likhoded:1999yv}      &    10.235         &     
\shmydek{\Xi_{bc}^+(bcu) \;  \pi^-}{\mathrm{as~in~Table~\ref{tab:dhb2}}}
&     \\
                         &                                     &                            &          
\mydek{\Lambda_b^0(bdu) \; J/\psi}{\mu^+ \mu^-}{\Lambda_c^+ \pi^-}
& $3 \times 10^{-6}$ \\ 
                         &                                     &                            &    
& $2.4 \times 10^{-7}$  if \\ 
                         &                                     &                            &    
& $\Lambda^0_b \rightarrow \Lambda_c \ell \nu X$ \\ 
\hline
$\Xi^{-}_{bb}(bbd)$      & $0.8$\cite{Likhoded:1999yv}      &     10.235         &    
\shmydek{\Xi_{bc}^0(bcd) \;  \pi^-}{\mathrm{as~in~Table~\ref{tab:dhb2}}}
&     \\
                         &                                     &                            &          
\mydek{\Xi_b^-(bsd) \; J/\psi}{\mu^+ \mu^-}{\Xi_c^0 \pi^-}
& $6 \times 10^{-5}$ \\
\hline
$\Omega^{-}_{bb}(bbs)$      & $0.8$\cite{Likhoded:1999yv}      &     10.385         & Cascades to above states & \\
\hline
$\Xi^{0}_{bbc}(bbc)$     & --      &     11.476   & Cascades to above states & \\
\hline
$\Omega^{-}_{bbb}(bbb)$  & --
          &     14.76    & Cascades to above states& \\ 
\hline
\end{tabular}
\end{center}
\caption[Baryons containing two $b$ quarks]
{Properties and interesting exclusive hadronic decay modes of
	heavy baryons containing two $b$ quarks.}
\label{tab:dhb3}
\end{table}
\index{decay!$\Xi^{0}_{bb}(bbu)$}
\index{decay!$\Xi^{-}_{bb}(bbd)$}
\index{decay!$\Omega^{-}_{bb}(bbs)$}
\index{decay!$\Xi^{0}_{bbc}(bbc)$}

\newcommand{\g}{{g}}
\renewcommand{\p}{{p}}
\newcommand{\q}{{q}}
\renewcommand{\u}{{u}}
\renewcommand{\B}{{B}}
\renewcommand{\H}{{H}}
\newcommand{\Q}{{Q}}
\newcommand{\cbar}{\bar{{c}}}
\newcommand{\dbar}{\bar{{d}}}
\newcommand{\qbar}{\bar{{q}}}
\newcommand{\sbar}{\bar{{s}}}
\newcommand{\ubar}{\bar{{u}}}
\newcommand{\bbbbar}{\bar{{b}}}
\newcommand{\Qbar}{\bar{{Q}}}

\relax
\def\pl#1#2#3{{\it Phys.\ Lett.\ }{\bf #1}\ (#2)\ #3}
\def\zp#1#2#3{{\it Z.\ Phys.\ }{\bf #1}\ (#2)\ #3}
\def\prl#1#2#3{{\it Phys.\ Rev.\ Lett.\ }{\bf #1}\ (#2)\ #3}
\def\rmp#1#2#3{{\it Rev.\ Mod.\ Phys.\ }{\bf#1}\ (#2)\ #3}
\def\prep#1#2#3{{\it Phys.\ Rep.\ }{\bf #1}\ (#2)\ #3}
\def\pr#1#2#3{{\it Phys.\ Rev.\ }{\bf #1}\ (#2)\ #3}
\def\np#1#2#3{{\it Nucl.\ Phys.\ }{\bf #1}\ (#2)\ #3}
\def\sjnp#1#2#3{{\it Sov.\ J.\ Nucl.\ Phys.\ }{\bf #1}\ (#2)\ #3}
\def\app#1#2#3{{\it Acta Phys.\ Polon.\ }{\bf #1}\ (#2)\ #3}
\def\jmp#1#2#3{{\it J.\ Math.\ Phys.\ }{\bf #1}\ (#2)\ #3}
\def\jp#1#2#3{{\it J.\ Phys.\ }{\bf #1}\ (#2)\ #3}
\def\nc#1#2#3{{\it Nuovo Cim.\ }{\bf #1}\ (#2)\ #3}
\def\lnc#1#2#3{{\it Lett.\ Nuovo Cim.\ }{\bf #1}\ (#2)\ #3}
\def\ptp#1#2#3{{\it Progr. Theor. Phys.\ }{\bf #1}\ (#2)\ #3}
\def\tmf#1#2#3{{\it Teor.\ Mat.\ Fiz.\ }{\bf #1}\ (#2)\ #3}
\def\tmp#1#2#3{{\it Theor.\ Math.\ Phys.\ }{\bf #1}\ (#2)\ #3}
\def\jhep#1#2#3{{\it J.\ High\ Energy\ Phys.\ }{\bf #1}\ (#2)\ #3}
\def\epj#1#2#3{{\it Eur.\ Phys. J.\ }{\bf #1}\ (#2)\ #3}

\relax
\newcommand\hepph[1]{{\tt hep-ph/#1}}
\newcommand\hepex[1]{{\tt hep-ex/#1}}

\def\s{\sigma}
\def\Dh{\hat{D}}
\def\({\left(}
\def\){\right)}
\newcommand\LambdaQCD{\Lambda_{\scriptscriptstyle \rm QCD}}

\section[Fragmentation]{Fragmentation}
\label{sec:fragmentation}
\index{fragmentation}
\index{fragmentation function}

The fragmentation of quarks and gluons into hadrons involves confinement
dynamics, and occurs at time scales that are long compared to
those of the hard scattering that produced the quarks and gluons.
Accordingly, in single-particle inclusive hard-scattering processes,
the fragmentation process is factorized in perturbative QCD
(see \cite{Collins:1989gx} and references therein) from the hard interaction 
and summarized in a nonperturbative fragmentation function (FF) $D_i^H(z,\mu)$ 
\cite{Collins:1982uw}. 
$D_i^H(z,\mu)$ is the probability density of a hadron $H$ to form from 
parton $i$ with momentum fraction $z$ at factorization scale $\mu$.
Just as initial-state parton distribution functions
(PDFs), fragmentation functions are not completely calculable 
in perturbation 
theory, although their evolution with $\mu$ is. 
The evolution equations for FFs are
identical in form to those for PDFs, although
the evolution kernels differ from second order onwards
\cite{Furmanski:1980cm,Curci:1980uw,Ellis:1996qj}.  
Though nonperturbative, these
FFs are universal and so, they may be determined for each hadron $H$
in a few calibration experiments at some fixed scale $\mu_0$, for 
subsequent use
in other experiments and at other values of $\mu$ (see for example 
\cite{Binnewies:1995ju,Greco:1995bn,Kniehl:2000fe,Kretzer:2000yf,Bourhis:2000gs}).  

The fragmentation of heavy quarks\footnote{By heavy quarks we mean charm and
bottom quarks. The top quark  decays by the weak interaction before it has time
to hadronize.}  is somewhat different.
\index{fragmentation function!heavy quarks}
When the heavy quark is produced with an energy $E$ not much 
larger than its mass $m$, the 
fragmentation process consists
mainly of the nonperturbative transition of the heavy quark to the
hadron H, which one assumes can be described by a nonperturbative
FF.  One may make a general ansatz for the functional form of this FF, 
the parameters of which are to be fixed by fitting to data.
One may also be inspired by physical considerations in motivating a 
functional form.  
Within this philosophy, the best known form is that of Peterson
et al.\ \cite{Peterson:1983ak}.  More recent forms are based on
heavy-quark effective field theory.  Both are described in section
\ref{sec:non-pert-fragm}.
When the heavy quark is produced with an energy $E$ much 
larger than its mass $m$, 
large logarithms of $E/m$ occur in the
perturbative expression for the heavy-quark inclusive
cross section, which must be resummed. 
These logarithms may be traced to the fragmentation stage
of the reaction, and the resummation may be 
achieved using the formalism of perturbative FFs (PFFs).  
They describe the fragmentation process from scale $E$ 
down to scale $m$.  These PFFs are perturbative because, first,
the coupling constant is small enough in the range from $m$ to
$E$ and second, the heavy-quark mass regulates the
collinear divergences, which would otherwise have to be absorbed into 
nonperturbative FFs.  
These perturbative and
nonperturbative FFs must be properly matched together
to avoid miscounting contributions and to enable a comparison with data.

Outside the context of factorization-based perturbative QCD, there is
the successful string model of the Lund group.  Yet other
approaches to heavy-quark fragmentation exist
\cite{Berezhnoy:1999rd,Berezhnoy:1999yj}, but for lack of space
we shall not discuss them here.  

We begin with a discussion of perturbative fragmentation
and describe an attempt to learn more about the nonperturbative FF
from the very precise SLD data. Next we examine what 
heavy-quark production data in $p\bar{p}$ collisions tell us about
the nonperturbative FF.
Then in section \ref{sec:non-pert-fragm} we review and clarify the theory of the
nonperturbative FFs.  Next we discuss the
phenomenon of beam drag in the context of the Lund string model, as
(perhaps) observed at HERA in charm DIS and photoproduction, and
present a study of the possible impact of this effect on CDF/D0 and
BTeV studies.  In section \ref{sec:heavy-quark-fragm}, we discuss 
various ``systematic errors'' in heavy quark FFs, and we conclude
with observations on the experimental impact of the
various issues in heavy-quark fragmentation. 

We refer to the recent LHC
Workshop report on bottom production \cite{Nason:1999ta} for
additional studies involving heavy-quark fragmentation.\footnote{In
particular,  in the context of section \ref{sec:heavy-quark-fragm}, one can
find there a study by Frixione and Mangano on effective $z>1$ support for
fragmentation functions in certain event generators.}

\subsection{Perturbative fragmentation}
\label{sec:pert-fragm-1}

\subsubsection[Heavy-quark fragmentation in $e^+e^-$ collisions]
{Heavy-quark fragmentation in $e^+e^-$ collisions
{\authorfootnote{Author: C. Oleari}}}
\label{sec:heavy-quark-fragm-1}

A well-defined fragmentation function is a universal function,
\index{fragmentation!perturbative}
\index{fragmentation!heavy quark!$e^+ e^-$ collisions}
so that its study can be undertaken in the context of $e^+e^-$
collisions without the complication due to initial state hadrons.
We are interested in the situation where the scale $E$
of the process is much larger than the heavy-quark mass $m$,
a typical situation at the Tevatron.
This requires the resummation of 
large logarithms $\ln(E/m)$
for a reliable computation of a differential cross section.
This is achieved via the formalism of the perturbative FF (PFF),
which we briefly review.
The PFF satisfies the DGLAP \cite{Altarelli:1977zs,Gribov:1972ri,Dokshitzer:1977sg}
evolution equation:
\begin{equation}
  \label{eq:8}
\frac{d D_{i,{\rm pert}}(x,\mu)}{d\ln\mu}
 = \sum_j \int_x^1 \frac{dz}{z} P_{ij}\left(\frac{x}{z},\alpha_s(\mu)\right)
D_{j,{\rm pert}}(z,\mu) \,,
\end{equation}
where $i,j$ label parton flavors. With  initial condition at $\mu_0\simeq m$, 
Eq.~(\ref{eq:8}) determines the PFF at $x,\mu$, and resums logarithms $\ln(\mu/\mu_0)$ to
all orders. This initial condition for the PFF  was first computed to NLO in
\cite{Mele:1991cw} and is, for $i=Q$, given by the distribution
\begin{eqnarray}
\label{eq:1}
 D_{Q,{\rm pert}}(z,\mu_0) = \delta(1-z)+\frac{\alpha_s(\mu_0)C_F}{2\pi} 
\left[\Bigg(\frac{1+z^2}{1-z}\Bigg)\Bigg(\ln\left(\frac{\mu_0^2}{m^2}\right)-1-2\ln(1-z)\Bigg)\right]_+\,.
\end{eqnarray}
For heavy quark production at $E \gg m$, the choice $\mu=E$ 
in the solution of (\ref{eq:8})
then resums terms containing $\ln^n(E/m)$ to all
orders in $\alpha_s$, to next-to-leading logarithmic (NLL) accuracy.

To study the PFF in the context of $e^+e^-$ collisions at center-of-mass
energy $E$
\begin{equation}
  \label{eq:4}
  e^+\,e^- \rightarrow Z/\gamma(q) \rightarrow Q(p)+X, \qquad
\end{equation}
one defines
\begin{equation}
  \label{eq:5}
  x_E = x = \frac{2p\cdot q}{q^2}\,,
\end{equation}
and factorizes the single heavy \textit{quark} inclusive
cross section as 
\begin{equation}
\label{eq:factorization}
   \frac{d\s}{dx}(x,E,m) =  \sum_i \int_x^1 \frac{dz}{z}\, 
 \frac{d\hat\s_i}{dz}\(z,E,\mu_F\) \,
  D_{i,{\rm pert}}\(\frac{x}{z},\mu_F,m\) \;,
\end{equation}
where $d\hat\s_i(x,E,\mu_F)/dx$ are the (scheme-dependent)
partonic cross sections for producing the parton $i$, and $D_{i,{\rm
pert}}(x,\mu_F,m)$ are the (scheme- but not process-dependent) perturbative
fragmentation functions (PFFs) for parton $i$ to evolve into the heavy quark
$Q$. 
The factorization scale $\mu_F$ must be chosen of order $E$,  to
avoid the appearance of large logarithms  $\ln(E/\mu_F)$ in the partonic cross
sections.

The single-\textit{hadron} inclusive cross section, including
nonperturbative corrections, is then usually written as multiple convolution
\begin{eqnarray}
\label{eq:3}
   \frac{d\s^H}{dx}(x,E,m) &=& 
\sum_i  \frac{d\hat\s_i}{dx}\(x,E,\mu_F\) \,
 \otimes D_{i,{\rm pert}}\(x,\mu_F,m\) \otimes D_{\rm NP}^H (x) \;.
\end{eqnarray}
Therefore the full fragmentation function requires the combination of 
the perturbative FF with a nonperturbative part which models the final 
hadronization.
Note that for heavy-quark production in hadronic collisions at 
hadron transverse momentum $p_T \gg m$, one must 
replace the CM energy $E$ with the boost-invariant $p_T$.

In Eq.~(\ref{eq:1}) the 
initial condition is expressed as an expansion in terms of $\as(m)$.
This is a beneficial property of heavy-quark fragmentation, but one
should keep in mind that higher-order terms in this condition may
be important.
Moreover, irreducible, nonperturbative uncertainties of
order $\LambdaQCD/m$ are present.
We assume that all these
effects are described by a nonperturbative fragmentation function
$D_{\rm NP}^H$, that takes into account all low-energy effects, including
the final nonperturbative hadronization of the heavy quark.
There are theoretical approaches to the fragmentation-function calculation
that employ heavy-quark effective theory in order to study
nonperturbative effects~\cite{Braaten:1995bz,Jaffe:1994ie} 
more systematically. These will be discussed in the next subsection.
Here  we want to establish a connection with the most
commonly used parameterizations, and thus we will use the Peterson form
and the ``Euler'' form $x^\alpha(1-x)^\beta$.

\subsubsection{Impact of SLD data}
\label{sec:impact-sld-data}

Let us first briefly review the present situation.
Heavy-flavor production in  $e^+e^-$ collisions
has been thoroughly studied both at fixed order 
\cite{Nason:1997nw,Rodrigo:1999qg,Bernreuther:1997jn}
and in a combined fixed order plus next-to-leading logarithmic 
(here $\ln(E/m)$ with $E$ the center of mass energy) resummed approach
\cite{Nason:1999zj}, using the Peterson function
\index{fragmentation function!Peterson}
for the nonperturbative transition\footnote{This combined approach is actually
a variable flavor number scheme (VFNS) for heavy-quark fragmentation.}.
The results of this study \cite{Nason:1999zj}, based
on LEP \cite{Buskulic:1995gp,Akers:1995jc}
and ARGUS \cite{Albrecht:1991ss} data
can be summarized as follows:
\begin{itemize}
\item
Either increasing the order of the finite order perturbative expansion or
including next-to-leading log (NLL) effects in a resummed approach reduces the
Peterson parameter $\epsilon$ obtained from fitting to the data, 
corresponding to a harder nonperturbative fragmentation function.
\item The differential cross section obtained by matching the
$\as^2$ fixed order and the NLL resummed expressions, the ``NLL improved'', is
harder than pure NLL, so that the $\epsilon$ parameter needed in the
nonperturbative part is, in general, larger, but only slightly so, than in the NLL case.
\item At LEP energies, the importance of ${\cal O}(m/E)$ terms is found to be
minor.
\end{itemize}
\begin{figure}[htbp]
\centerline{\epsfig{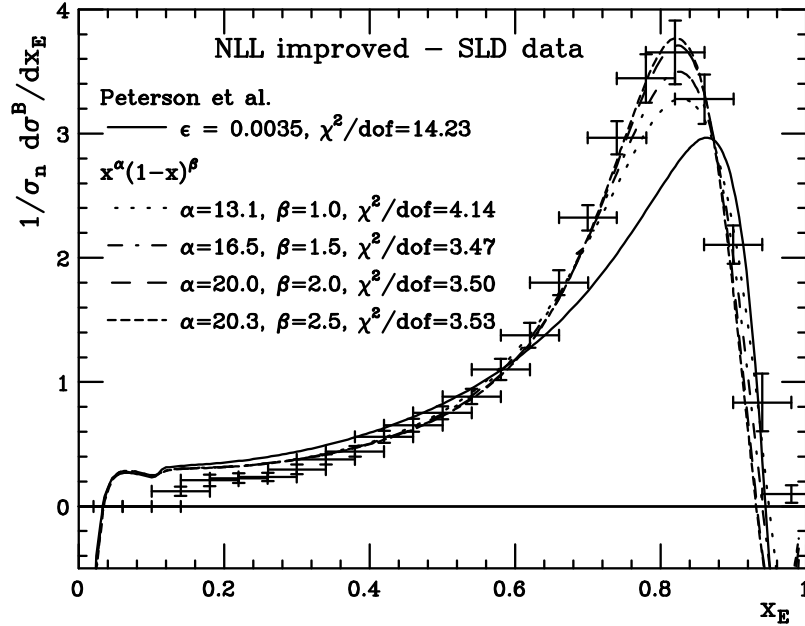}}
\caption[Fit to SLD data of Peterson and Euler nonperturbative
fragmentation functions]{ \label{fig:fitsld}
Fit to SLD data of Peterson and Euler nonperturbative
fragmentation function via NLL improved calculation.
The $\chi^2$ are per degree of freedom.}
\end{figure}
Let us now show some new results obtained from a fit to 
SLD~\cite{Abe:1999ki} $B$-production data in
$e^+e^-$ collision at $E=91.2$~GeV, in Fig.~\ref{fig:fitsld}.
The theoretical curves have been determined from Eq.~(\ref{eq:3}),
at the same level of accuracy as used in \cite{Nason:1999zj}.
We have used two families of nonperturbative FF: the Peterson form and the
Euler form $x^\alpha(1-x)^\beta$.  We have fitted the data by
$\chi^2$ minimization, keeping $\Lambda_{\rm QCD}^{(5)}$ fixed to $200$~MeV.
With this procedure we have fitted both the value of
$\epsilon$ and the normalization (which was allowed to float) for the Peterson FF,
and for the Euler form the value of $\alpha$ and the normalization, varying
$\beta$ in the range between~1.0 and~2.5, with an increment of~0.5.

We should caution at this point that the values of $\epsilon$ for the Peterson
FF fitted above \textit{cannot} be used for LEP studies, as the SLD and LEP collaborations
use different values for some key input parameters such as the fraction of $b$ quarks
producing $B^{**}$ mesons. 

We find the Peterson form to have a very poor $\chi^2$/d.o.f.
The Euler form can accommodate the data better, at the
values $\alpha=16.5, \beta=1.5$. To compare the fit results,
we plot in Fig.~\ref{fig:npfrag} the 
nonperturbative FF at their best $\chi^2$/d.o.f. values.
\begin{figure}[htbp]
\centerline{\epsfig{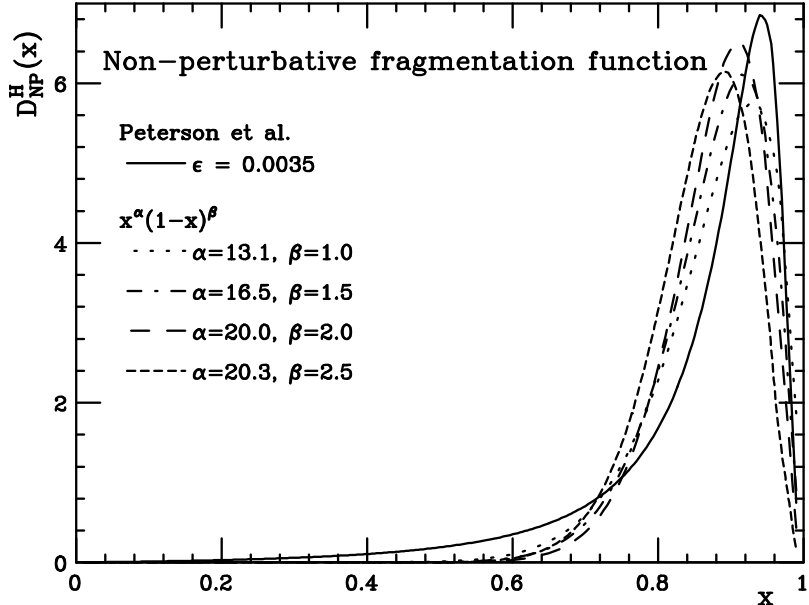}}
\caption[Functional forms of the nonperturbative fragmentation functions]{
\label{fig:npfrag}
Functional forms of the nonperturbative fragmentation functions that
produced the best fit to SLD data.}
\end{figure}
We see that all curves are strongly peaked near $x=1$, with the
best fit value corresponding to a fairly hard fragmentation.

We note that in Ref.~\cite{Cacciari:1994mq}, and more recently in~\cite{Cacciari:1998it},
the formalism used in the previous analysis of $e^+e^-$ collisions has
been applied to the Tevatron $b$-quark $p_T$ cross section, for which
the data exceed the central NLO-theory estimate by a factor of
two\footnote{This is somewhat less the case for the $b$-jet cross section,  see
section 1.3 in this chapter.}.
At large $p_T$ the theoretical uncertainty due to scale variations was found to
be reduced with respect to the fixed-order approach, but the cross section
decreased. At moderate $p_T$, the cross section gets somewhat enhanced, but
not enough to explain the data-theory discrepancy.  
Another study involving FFs in heavy-quark hadroproduction, in the context
of the so-called ACOT VFNS, was performed in~\cite{Olness:1997yc}. 
The PFF formalism was also
applied to $\gamma p$~\cite{Cacciari:1996fs} and $\gamma
\gamma$~\cite{Cacciari:1996ej} charm production. Similar conclusions were
reached as for $p\bar{p}$ $b$-quark production.

\subsubsection[Heavy-quark fragmentation in $p\bar{p}$ collisions]
{Heavy-quark fragmentation in $p\bar{p}$ collisions
{\authorfootnote{Authors: P. Nason, G. Ridolfi}}}
\label{sec:heavy-quark-fragm-2}

It is well known that Tevatron data for the integrated transverse
momentum spectrum in $b$ production are systematically larger than QCD
predictions. This problem has been around for a long time,
although it has become less severe with time.
\index{fragmentation!heavy quark!$p\bar{p}$ collisions}
The present status of this issue has been previously presented in
Fig.~\ref{allbxs}.
A similar discrepancy is also observed in UA1 data
(see ref.~\cite{Frixione:1997ma}
for details).

The theoretical prediction has a considerable uncertainty, which is
mainly due to neglected higher-order terms in the perturbative
expansion. In our opinion, it is not unlikely that we may have to live
with this discrepancy, which is certainly disturbing, but not strong
enough to question the validity of perturbative QCD calculations. In
other words, the QCD ${\cal O}(\as^3)$ corrections for this process
are above $100\%$ of the Born term, and thus it is not impossible that
higher order terms may give contributions of the same
size. Nevertheless, it is conceivable that also nonperturbative effects
contribute to enhance the cross section for this observable. 

In this note, we present a study of the effects of $b$-quark
fragmentation on the predicted single-inclusive $p_T$ spectrum.  In
analogy with the case of charm production, the agreement between
theory and data improves if one does not include any fragmentation
effects. It is then natural to ask whether the fragmentation functions
commonly used in these calculations are appropriate. The
LEP~\cite{Abreu:1993jn,Buskulic:1995gp,Alexander:1995aj} and SLD~\cite{Abe:1999ki} measurements have shown
that fragmentation functions are harder than previously thought.

The effect of a nonperturbative fragmentation function on the $p_T$ spectrum
is easily quantified if one assumes a steeply-falling transverse momentum
distribution for the produced $b$ quark
\begin{equation}
\frac{d\sigma}{d p_T} = A p_T^{-M}\, .
\end{equation}
The corresponding distribution for the hadron is
\begin{equation}
\frac{d\sigma_{\rm had}}{d p_T} = A \int {\hat p}_T^{-M}\,
       \delta(p_T-z{\hat p}_T) D(z)\, dz \,d{\hat p}_T
=A  p_T^{-M} \int_0^1 dz z^{M-1} D(z)\,.
\end{equation}
We can see that the hadron spectrum is proportional to the quark
spectrum times the $M^{\rm th}$ moment of the fragmentation function $D(z)$.
Thus, the larger the moment,
the larger the enhancement of the spectrum.

In practice, the value of $M$ will be slightly dependent upon $p_T$.
We thus define a $p_T$ dependent $M$ value
\begin{equation}\label{apprfragmeff}
\frac{d\log\sigma(p_T>p_T^{\rm cut})}{d\log p_T^{\rm cut}}=
-M(p_T^{\rm cut})+1
\end{equation}
and
\begin{equation}
\sigma_{\rm had}(p_T>p_T^{\rm cut})
=\sigma(p_T>p_T^{\rm cut})\times
\int_0^1 dz z^{M(p_T^{\rm cut})-1} D(z)\,.
\end{equation}
This gives an excellent approximation to the effect of the
fragmentation function, as can be seen from fig.~\ref{fig:frageff}.
\begin{figure}
\begin{center}
    \includegraphics[width=0.6\textwidth,clip]{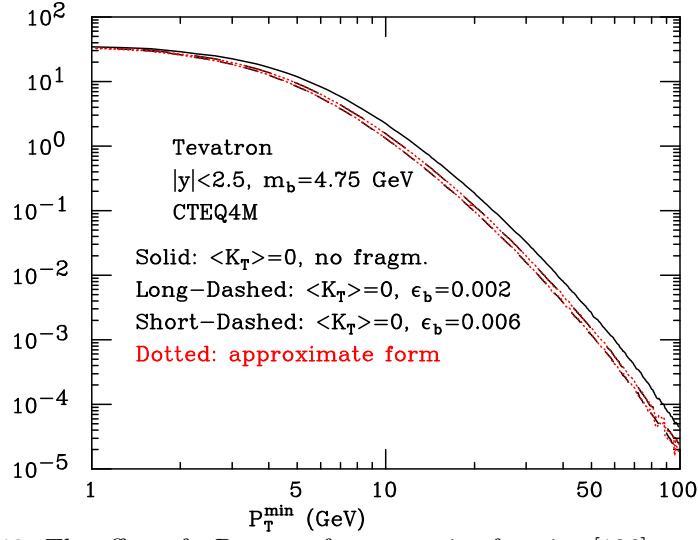}
    \caption[The effect of a Peterson fragmentation function
on the inclusive $b$ cross section.]{
The effect of a Peterson fragmentation function~\cite{Peterson:1983ak}
on the inclusive $b$
cross section. The (red) dotted lines correspond to the approximation
Eq.~(\ref{apprfragmeff}), and are almost indistinguishable from
the exact results.}
    \label{fig:frageff}
\end{center}
\end{figure}

Since the second moment of the fragmentation function is well
constrained by $e^+e^-$ data, it is sensible to ask for what shapes of
the fragmentation function, for fixed $\langle z \rangle$, one gets the
highest value for $\langle z^{M-1} \rangle$. We convinced ourselves that
the maximum is achieved by the functional form
\begin{equation}
D(z) = A\delta(z)+B\delta(1-z) \; ,
\end{equation}
which gives
\begin{equation}
\langle z\rangle = \frac{B}{A+B}\,;\quad
\langle z^{M-1}\rangle = \frac{B}{A+B}\,.
\end{equation}
This is however not very realistic: somehow, we expect a
fragmentation function which is concentrated at high values of $z$,
and has a tail at small $z$. We convinced ourselves that, if we impose
the further constraint that $D(z)$ should be monotonically increasing,
one gets instead the functional form
\begin{equation}
\label{maxform}
D(z) = A+B\delta(1-z)\,,
\end{equation}
which gives
\begin{equation}
\langle z\rangle = \frac{A/2 + B}{A+B}\,;\quad
\langle z^{M-1}\rangle = \frac{A/M + B}{A+B}\,.
\end{equation}
We computed numerically the $M^{th}$ moments of the Peterson form:
\begin{equation}
D(z)\propto \frac{1}{z\left(1-\frac{1}{z}-\frac{\epsilon}{1-z}\right)^2} \; ,
\end{equation}
of the form
\begin{equation}
\label{powerform}
D(z) \propto z^\alpha (1-z)^\beta \; ,
\end{equation}
for $\beta=1$ (Kartvelishvili~\cite{Kartvelishvili:1978pi}), for which
\begin{equation}
\langle z^{M-1} \rangle =\frac{\Gamma(\alpha+M)\Gamma(\alpha+\beta+2)}
{\Gamma(\alpha+1)\Gamma(\alpha+\beta+M+1)}\,,
\end{equation}
of the form of Collins and Spiller~\cite{Collins:1985ms}
\begin{equation}
D(z)\propto\frac{
\left(\frac{1-z}{z}+\frac{(2-z)\epsilon}{1-z}\right)\left(1+z^2\right)}
{\left(1-\frac{1}{z}-\frac{\epsilon}{1-z}\right)^2}\; ,
\end{equation}
and of the form in Eq.~(\ref{maxform}),
at fixed values of $\langle z \rangle$ corresponding to the
choices $\epsilon_b=0.002$ and $0.006$ in the Peterson form.
We found that the $p_T$ distribution at the Tevatron, for $p_T$ 
in the range $10$ to $100$ GeV, behaves like $p_T^{-M}$, with $M$ around 5.
Therefore, we present in Tables~\ref{tabfrag1} and \ref{tabfrag2}
values of the $4^{\rm th}$,
$5^{\rm th}$ and $6^{\rm th}$ moments of the above-mentioned
fragmentation functions.
\begin{table}[htbp]
\begin{center}
\leavevmode
\begin{tabular}{|l|r|r|r|}
 \hline
 $\langle z \rangle=0.879$ & $M=4$ & $M=5$ & $M=6$ \\
\hline
Peterson  & 0.711  &  0.649 & 0.595 \\
\hline
Kartvelishvili  & 0.694  &  0.622 & 0.562 \\
\hline
Collins-Spiller & 0.729  &  0.677 & 0.633 \\
\hline
Maximal (Eq. (\ref{maxform})) & 0.818  &  0.806 & 0.798 \\
\hline
\end{tabular}
\end{center}
\caption[Moments of the fragmentation functions 
at fixed $\langle z \rangle$ for $\epsilon_b=0.002$]{
Values of the $4^{\rm th}$, $5^{\rm th}$ and $6^{\rm th}$ moment,
at fixed $\langle z \rangle$ (corresponding to $\epsilon_b=0.002$
in the Peterson form), for different forms of the fragmentation function.}
\label{tabfrag1}
\end{table}
\begin{table}[htbp]
 \begin{center}
 \leavevmode
\begin{tabular}{|l|r|r|r|}
 \hline
 $\langle z \rangle=0.828$ & $M=4$ & $M=5$ & $M=6$ \\
\hline
Peterson & 0.611  &  0.535 & 0.474 \\
\hline
Kartvelishvili & 0.594  &  0.513 & 0.447 \\
\hline
Collins-Spiller & 0.626  &  0.559 & 0.505 \\
\hline
Maximal (Eq. (\ref{maxform})) & 0.742  &  0.724 & 0.713 \\
\hline
\end{tabular}
\end{center}
\caption[Moments of the fragmentation functions 
at fixed $\langle z \rangle$ for $\epsilon_b=0.006$]{
Values of the $4^{\rm th}$, $5^{\rm th}$ and $6^{\rm th}$ moment,
at fixed $\langle z \rangle$ (corresponding to $\epsilon_b=0.006$
in the Peterson form), for different forms of the fragmentation function.}
\label{tabfrag2}
\end{table}
We thus find that keeping the second moment fixed the variation of the
hadronic $p_T$ distribution obtained by varying the shape of the
fragmentation function among commonly used models is between 5\% and
13\% for both values of $\epsilon_b$. Therefore, it seems difficult to enhance
the transverse momentum distribution by suitable choices of the form
of the fragmentation function. With the extreme choice of
Eq.~(\ref{maxform}), one gets at most a variation of 50\% for the
largest values of $\epsilon_b$ and $M$. It would be interesting to see
if such an extreme choice is compatible with $e^+e^-$ fragmentation function
measurements.

\subsection[Fragmentation in the nonperturbative regime]
{Fragmentation in the nonperturbative regime
\authorfootnote{Authors: G. Bodwin, B. Harris}}
\label{sec:non-pert-fragm}
\index{fragmentation!non-perturbative}
\index{fragmentation function!Peterson}

\def\slash#1{\rlap{\hbox{$\mskip 1 mu /$}}#1}   
\def\bra#1{\left\langle #1\right|}              
\def\ket#1{\left| #1\right\rangle}              

In this subsection we review the theory of the nonperturbative 
transition of a heavy quark into a heavy meson.

First, we examine the model of Peterson {\em et al}.\ \cite{Peterson:1983ak}
for the fragmentation of a fast-moving heavy quark $Q$ with mass $m_Q$ into 
a heavy hadron $H$ (consisting of $Q\overline{q}$) with mass $m_H$ and a
light quark $q$ with mass $m_q$. The basic assumption in this model is
that the amplitude for the fragmentation is proportional to $1/(\Delta
E)$, where $\Delta E = E_H + E_q - E_Q$ is the energy denominator for
the process in old-fashioned perturbation theory. It follows that the
probability for the transition $Q \rightarrow H + q$ is proportional to
$1/(\Delta E)^2$. Taking the momentum of the heavy quark to define the
longitudinal axis, one can express $\Delta E$ in terms of the
magnitude of $P_Q$, the heavy-quark momentum, the fraction $z$ of the
heavy-quark momentum that is carried by the heavy hadron, and the
transverse momentum $p_\bot$ of the heavy hadron or the light quark:
\begin{eqnarray}
\Delta E &=& \sqrt{m_H^2+p_\bot^2+z^2P_Q^2} + \sqrt{m_q^2+p_\bot^2
+(1-z)^2P_Q^2} - \sqrt{m_Q^2+P_Q^2} \nonumber \\
&\approx& \frac{m_H^2+p_\bot^2}{2zP_Q} + 
\frac{m_q^2+p_\bot^2}{2(1-z)P_Q}
- \frac{m_Q^2}{2P_Q} + \cdots \nonumber \\ 
&\approx & -{m_Q^2\over 2P_Q}[1 - 1/z - \epsilon/(1-z)].
\label{energyd}
\end{eqnarray}
In the last line, we have set $m_H\approx m_Q$ and neglected $p_\bot^2$
relative to $m_Q^2$ and used the definition  
$\epsilon \equiv (m_q^2+p_\bot^2)/m_Q^2$. 
Multiplying $1/(\Delta E)^2$ by a
factor $1/z$ for the longitudinal phase space, one arrives at the
following {\it ansatz} for the fragmentation function \cite{Peterson:1983ak}:
\begin{equation}
D_Q^H(z) = \frac{N}{z[1 - 1/z - \epsilon/(1-z)]^2},
\end{equation}
where the normalization $N$ is fixed by the condition
\begin{equation}
\sum_H \int dz D_Q^H(z) = 1,
\end{equation}
and the sum extends over all hadrons that contain $Q$. Contrary to the
claims in Ref.\ \cite{Peterson:1983ak}, we find that $D_Q^H(z)$ has a
maximum at $z \approx 1 - \sqrt\epsilon$ and a 
width of order $\sqrt\epsilon$.  Previously it was  believed that 
the shape of the Peterson et al.\ form is incompatible with 
results obtained from heavy-quark effective theory (HQET).  However, 
as we shall discuss below, our new results for the maximum point and width 
are compatible with the HQET analysis.

Next we discuss the work of Jaffe and Randall \cite{Jaffe:1994ie}, which
provides a QCD-based interpretation of heavy-quark fragmentation in
terms of the heavy-quark mass expansion. One begins with the standard
Collins-Soper \cite{Collins:1982uw} definition of the fragmentation function
for a heavy quark into a heavy hadron:
\index{fragmentation function!in HQET}
\begin{equation}
\hat f(z,\mu^2)=\frac{z}{4\pi} \int d\lambda e^{i \lambda/z} 
{1\over 2N_c}{\rm Tr}\, \slash{n} \bra{0} 
h(\lambda n)\ket{H'(P)}\bra{H'(P)} \overline h(0) \ket{0},
\label{np_op}
\end{equation}
where the trace is over color and Dirac indices, $N_c$ is the number of
colors, $h(x)$ is the heavy-quark field at space-time position $x$, $P$
is the four-momentum of the heavy hadron, and $n$ is defined by $n^2=0$
and $n\cdot p=1$. The state $\ket{H'(P)}$ consists of the heavy hadron
plus any number of additional hadrons. The matrix element is understood
to be evaluated in the light-cone gauge $n\cdot A=0$. We note that the
definition (\ref{np_op}) contains a factor $z$ relative to the
definition of the fragmentation function used in Ref.~\cite{Jaffe:1994ie}.
This factor $z$ will be important in comparing with the work of Braaten
{\em et al}.\ below.

Following the standard method for obtaining the heavy-quark mass
expansion, one decomposes the field $h(x)$ into the sum of large
$h_v(x)$ and small $\underline{h}_v(x)$ components:
\begin{eqnarray}
h_v(x) &=& e^{-im_Q v \cdot x} P_+ h(x) \\
\underline{h}_v(x) &=& e^{-im_Q v \cdot x} P_- h(x),
\end{eqnarray}
with $P_\pm = (1\pm \slash{v})/2$ and $v$ the hadron's four velocity.
The leading term in the mass expansion of $f(x,\mu^2)$ is contained in
the large-large combination of fields:
\begin{eqnarray}
\hat f(z,\mu^2)&=&\frac{z}{4\pi} \int d\lambda e^{i \lambda/z} 
{1\over 2N_c} {\rm Tr}\, \slash{n}\nonumber\\
&&\times\,\bra{0} P_+h(\lambda n)\ket{H'(P)}\bra{H'(P)} 
\overline{P_+h(0)} \ket{0} e^{-im_Q\lambda n \cdot v}+\ldots\; .
\label{np_expan}
\end{eqnarray}
In Ref.\ \cite{Jaffe:1994ie}, it is argued that the matrix element in
Eq.~(\ref{np_expan}) is a dimensionless function ${\cal F}(\lambda
\delta)$. This function may be written in terms of its Fourier
transform:
\begin{equation}
{\cal F}(\lambda \delta) = 2 \int_{-\infty}^{\infty} d\alpha 
e^{-i\alpha \lambda \delta} \, a(\alpha) \, ,
\label{np_fourier}
\end{equation}
where $\delta=1-m_Q/m_H$ which in terms of the parameter 
$\epsilon$ appearing in the Peterson et al.\ form 
is $\delta\approx\sqrt\epsilon$.
Inserting Eq.\ (\ref{np_fourier}) into Eq.\ (\ref{np_expan}), one can 
evaluate the integral, with the result
\begin{equation}
\hat f(z,\mu^2)=\frac{z}{\delta} \hat a \left( \frac{1/z-m_Q/m_H}
{\delta} \right)+\ldots \; .
\end{equation}
A more complete analysis in Ref.~\cite{Jaffe:1994ie} also yields the
next-to-leading term in the hadron mass expansion:
\begin{equation}
\hat{f}(z,\mu^2) = z\left[\frac{1}{\delta} \hat{a}(y) 
+ \hat{b}(y) + \cdots \right] \, ,
\label{np_rj1}
\end{equation}
where $y=(1/z-m_Q/m_H) / \delta$. The analysis in Ref.~\cite{Jaffe:1994ie}
does not yield a precise prediction for the functional form of $a$ and
$b$, but some general properties may be deduced. The function $a$
describes, in the limit of infinite heavy-quark mass, the effects of
binding in the heavy hadron on the heavy-quark momentum distribution. For
a free heavy quark, $a(y)$ would be a $\delta$-function at $y=1$. In a
heavy hadron, the binding smears the heavy-quark momentum distribution.
It can be shown \cite{Jaffe:1994ie} that the distribution has a maximum at
$z \approx 1 - \delta$ and a width of order $\delta$. 
Using the fact the $\delta\approx\sqrt\epsilon$, we find that the maximum 
is at $z \approx 1 - \sqrt\epsilon$ and the width is of order 
$\sqrt\epsilon$ in agreement with the Peterson {\em et al}.\ model, 
as described above.

Braaten {\em et al}.\ \cite{Braaten:1995bz} present a QCD-inspired model for
the fragmentation of a heavy quark into an $S-$wave light-heavy meson.
In this model, the fragmentation function is computed in perturbative
QCD (Born level) in an expansion in inverse powers of the heavy-quark
mass. For the projection of the $Qq$ state onto the meson, Braaten
{\em et al}.\ take the standard nonrelativistic-bound-state expression.
For example, in the case of a $^1S_0$ meson, they assume the Feynman
rule for the $QqH$ vertex to be
\begin{equation}
\frac{\delta_{ij}}{\sqrt{3}} \frac{R(0)\sqrt{m_H}}{\sqrt{4\pi}}
\gamma_5 (1+\slash{v})/2,
\end{equation}
where $R(0)$ is the radial wave function at the origin.  Braaten {\em et
al}.\ make use of a definition of the fragmentation function that is
equivalent to the Collins-Soper definition (\ref{np_op}). From the
terms of leading order in the heavy-quark mass expansion, they obtain, 
in the case of a $^1S_0$ meson,
\begin{equation}
\hat f\approx N\left[{1\over \delta}\frac{(1-y)^2}{y^6} (3y^2+4y+8)
-\frac{(1-y)^3}{y^6} (3y^2+4y+8)\right],
\label{np_b1}
\end{equation}
where $N=2\alpha_s^2|R(0)|^2/(81\pi m_q^3)$. At first glance, this
result may seem to contradict the Jaffe-Randall analysis, which shows
that the terms of leading order in the heavy-quark mass expansion give a
contribution that is contained entirely in the function $a(y)$ in
Eq.~(\ref{np_rj1}). However, the factor $z$ in the definition of the
fragmentation function (\ref{np_op}) is crucial here. From the
definitions of $y$ and $\delta$, we have $z=1/[1-\delta(1-y)]$, and,
so, we can re-write Eq.~(\ref{np_b1}) as
\begin{equation}
\hat f/z\approx {N\over \delta}\frac{(1-y)^2}{y^6} (3y^2+4y+8),
\end{equation}
which is of the form of $a(y)$ in Eq.~(\ref{np_rj1}). 

There are several additional studies of the  perturbative-QCD
fragmentation function in the limit of a large heavy-quark mass. See
the paper of Braaten {\em et al}.\ \cite{Braaten:1995bz} for references.

\subsection[Beam drag]{Beam drag\,\authorfootnote{Author: E. Norrbin}}
\index{beam drag}
\label{sec:beam-drag}

A puzzling situation has arisen in charm electroproduction and 
photoproduction at HERA.
The data are well-described by NLO QCD \cite{Harris:1998zq,Frixione:1994dg} 
plus Peterson fragmentation \cite{Peterson:1983ak}, except at
low $p_t$ and large rapidity. To show this effect the
charm electroproduction data of the ZEUS collaboration \cite{Breitweg:1999ad}
is shown in Fig.~\ref{fig:herabeamdrag}.
\begin{figure}[htbp]
\centerline{\epsfig{figure=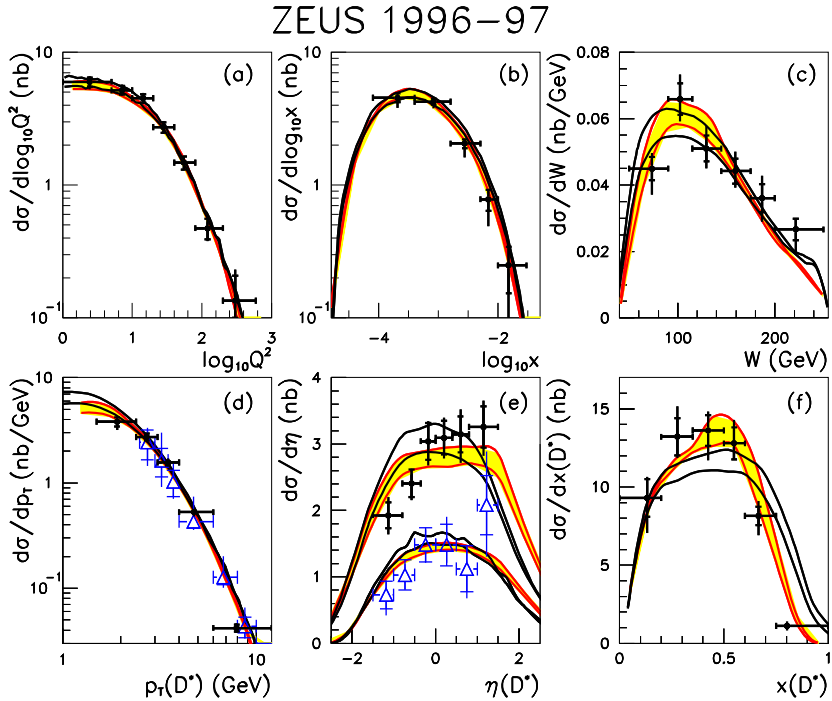,width=0.75\textwidth,clip=}}
\caption[Effects of beam drag at HERA vs NLO theory and data]
{\label{fig:herabeamdrag} Effects of beam drag at HERA vs NLO 
theory and data from \protect{\cite{Breitweg:1999ad}}.}
\end{figure}
The figure shows the $D^*$ meson cross sections
differential in momentum transfer $Q^2$, Bjorken $x$, hadronic energy $W$,
transverse momentum $p_t$, pseudo-rapidity $\eta$, and $D^*$ momentum
fraction $x(D^*)=2 | \vec{p}_{\gamma P{\rm cms}} | / W$, compared
with theory \cite{Harris:1998zq}.  The boundaries of the bands correspond to
varying the charm quark mass of $1.4$ GeV by $\pm 0.1$ GeV. The overall
agreement is good, but the theory underestimates the data in the forward
region and overestimates it in the backward region.  Additionally, the
$D^*$ momentum fraction data, which is particularly sensitive to the charm
hadronization process, is poorly described.  Similar effects are seen in
the $D^*$ photoproduction data \cite{Breitweg:1998yt} at HERA.   
Variations of the parton distribution set, renormalization/factorization
scale, charm mass, or fragmentation parameter $\epsilon$ are unable to
account for the differences between data and theory. It also appears
unlikely that an evolving fragmentation function would help since the $p_t$ 
range covered is so small.
A similar effect has been observed by the H1 collaboration in their
charm electroproduction data \cite{Adloff:1996xq}.

One explanation \cite{Norrbin:1999mz} proposed for the photoproduction data 
\cite{Breitweg:1998yt} appears to
work for the DIS data as well. One imagines
a color string connecting the hadronizing charm quark and the proton
remnant which pulls (drags) the charmed meson to the forward region. This
beam-remnant drag effect is made quantitative in the Lund String model 
modified for heavy-flavor production \cite{Bowler:1981sb},
as implemented in Pythia \cite{Sjostrand:1994yb}. 
The shaded band in Fig.~\ref{fig:herabeamdrag} shows that agreement
is better when the Peterson {\it et al.}
model is replaced by an effective fragmentation model extracted from the
Pythia-based Monte Carlo RAPGAP \cite{Jung:1995gf}.

Another way to improve the agreement between data and theory is to simply
raise the minimum $p_t$ of the events that are selected.  Data from a
slightly different decay chain, but higher minimum $p_t$ cut are shown as
open triangles.  Here the Peterson and RAPGAP improved NLO predictions
give essentially the same results, as expected from a power-suppressed
effect.  Note however that raising $p_t^{\rm min}$ may not be an option 
in some situations.

Let us adopt the view that beam remnant drag effects, which appear naturally
in the string fragmentation model \cite{Andersson:1983ia}, exert influence 
here.
This effect is closely related to the collapse of small strings,
which is the most extreme case of string drag when all the energy and
momentum of the remnant is taken up by the produced hadron. This latter effect
has been used to describe the large charm asymmetries observed
at several fixed target experiments \cite{Charmasym,Norrbin:1998bw}.
It is important to understand this effect,
as it affects $B-\bar{B}$ asymmetries at e.g. HERA-B
\cite{Norrbin:2000zc}, and thus CP violation measurements.
An open problem is how to match the string
model for beam drag to factorized QCD.
In the rest of this section the Lund string fragmentation model is summarized and its influence on the
distribution of final state hadrons in $p {\bar p}$ collisions is reviewed.

The Lund string fragmentation model provides a different approach to the problem
\index{fragmentation!color string model}
\index{Lund string model}
of hadronization than the methods discussed so far. The Lund model in its basic
form is simply a prescription for turning a partonic state (no matter how it was produced)
into a hadronic final state. A string is the tube-like QCD force field stretched between a
triplet and an anti-triplet color charge. In a high-energy process the string contains a
lot of energy and decays into hadrons by the production of new $\q\qbar$ pairs along
the force field. The decay dynamics is constrained by a few physical assumptions
such as Lorentz invariance, confinement and independence of the final state on the
order in which string breaks are considered in the fragmentation process 
(`left-right symmetry'). This results in a fragmentation function with two free parameters,
called the Lund symmetric FF \cite{Andersson:1983ia}
\begin{equation}
f(z) \propto \frac{1}{z}(1-z)^a e^{-bm_\perp^2/z}
\end{equation}
with a modification needed for heavy flavors \cite{Bowler:1981sb}
\begin{equation}
f(z) \propto \frac{1}{z^{1+bm^2_{Q}}}(1-z)^a e^{-bm_\perp^2/z},
\end{equation}
where $m_\perp$ is the transverse mass of the produced hadron, $m_{Q}$
is the heavy-quark mass and $a$ and $b$ are the two free parameters, which are
common for all flavors and hadrons. The $z$ variable represents the light cone
fraction along the string direction and unlike `stand alone' fragmentation functions no ambiguity is involved (cf. section \ref{sec:heavy-quark-fragm}).

To be practically useful, the Lund model must be hooked on to a
perturbative description of the underlying hard process (such as
$e^+e^- \rightarrow \q\qbar$ or $\g\g \rightarrow \q\qbar$). Normally
a parton shower approximation is used, where the final state of a LO
process is evolved down in virtuality by parton emission to some
cut-off scale $Q_0$.  Using a fixed lower scale for the perturbative
treatment it has been possible in $e^+e^-$ experiments to fit the two
parameters of the Lund model to data and get excellent results. If the
$Q_0$ scale is changed, the nonperturbative parameters of the Lund
model will have to be re-tuned, so $Q_0$ can be seen as the scale at
which perturbation theory is abandoned and replaced by the
nonperturbative model. In the ideal case the value of $Q_0$ should not
matter. In practice this is true only as long as it is at the order of
1 GeV. Any small change in $Q_0$ around this value is absorbed by a
change of the fragmentation parameters.  In a Monte Carlo program some
additional parameters are needed to fully describe the process $e^+e^-
\rightarrow$ hadrons, such as the transverse smearing along the string
direction and the flavor composition of the new $\q\qbar$ pairs
\cite{Sjostrand:1994yb}.

When the model is carried over to hadron collisions several new
aspects have to be considered. First of all, the particles entering the
hard subprocess are not color singlets but quarks or gluons confined
to hadrons.  This problem is well known and solved by introducing PDFs
which have been measured to good accuracy at e.g. HERA. Less well
known is the consideration of color flow, beam remnants, small strings
in the hadronization and beam drag.  The remainder of this section
will cover these aspects including a discussion of their possible
implications in $p {\bar p}$ physics.

\begin{figure}[t]
\begin{picture}(210,130)(0,-10)
\Text(100,-5)[]{(a)}
\Text(10,30)[r]{$\p$}
\Text(10,100)[r]{${\bar p}$}
\Line(15,30)(25,30)
\Line(15,100)(25,100)
\GOval(30,30)(20,5)(0){0.5}
\GOval(30,100)(20,5)(0){0.5}
\Gluon(33,45)(75,65){4}{4}
\Gluon(33,85)(75,65){4}{4}
\Gluon(75,65)(120,65){4}{4}
\Line(120,65)(160,45)
\Line(120,65)(160,85)
\Text(165,45)[l]{$\bbbbar$}
\Text(165,85)[l]{${b}$}
\Line(33,15)(160,15)
\Line(33,115)(160,115)
\Text(165,15)[l]{$\u{d}$}
\Text(165,115)[l]{$\ubar{\dbar}$}
\Line(35,30)(160,30)
\Line(35,100)(160,100)
\Text(165,30)[l]{$\u$}
\Text(165,100)[l]{$\ubar$}
\put(178,37.5){\oval(7,16)[r]}
\put(182,65){\oval(15,100)[r]}
\put(178,92.5){\oval(7,16)[r]}
\end{picture}
\hfill
\begin{picture}(210,130)(5,-10)
\Text(110,-5)[]{(b)}
\LongArrow(0,60)(50,60)
\Text(0,70)[l]{${\bar p}$}
\LongArrow(210,60)(160,60)
\Text(210,70)[r]{$\p$}
\LongArrow(100,60)(170,65)
\Text(175,70)[l]{$\ubar$}
\LongArrow(100,60)(160,90)
\Text(165,94)[l]{${b}$}
\LongArrow(100,60)(30,50)
\Text(26,47)[t]{$\u$}
\LongArrow(100,60)(80,10)
\Text(90,10)[r]{$\bbbbar$}
\LongArrow(100,60)(35,75)
\Text(35,87)[]{$\u{d}$}
\LongArrow(100,60)(170,55)
\Text(175,49)[l]{$\ubar\dbar$}
\DashLine(170,65)(160,90){4}
\DashLine(30,50)(80,10){4}
\DashLine(35,75)(170,55){4}
\end{picture}\\[4pt]
\caption[Example of a string configuration in a $p {\bar p}$ collision.]
{Example of a string configuration in a $p {\bar p}$ collision. 
(a) Graph of the process, with brackets denoting the final color singlet
subsystems. (b) Corresponding momentum space picture, with dashed lines
denoting the strings. 
\label{fragstrings}}
\end{figure}
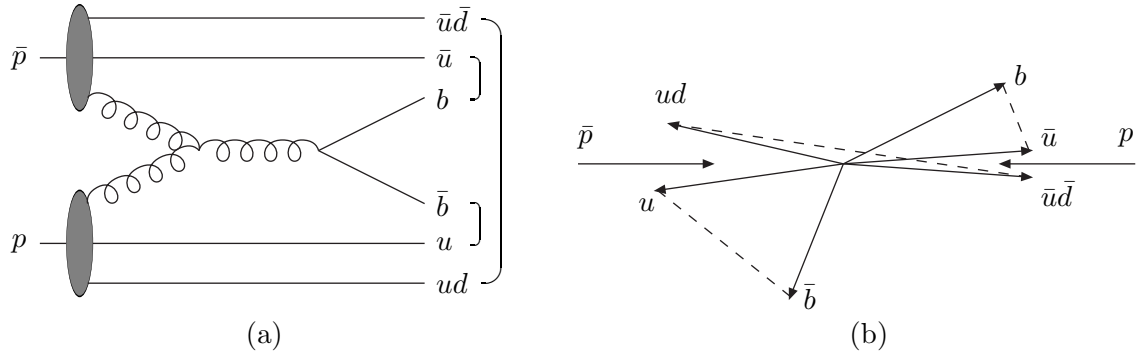

To be able to use the Lund string fragmentation model for
hadronization, the strings in the event must be constructed. The
string topology can be derived from the color flow of the hard
process. For example, consider the LO process $u {\bar u} \to
{b}\bbbbar$ in a $p {\bar p}$ collision. The color of the
incoming $\u$ is inherited by the outgoing ${b}$, so the
${b}$ will form a color-singlet together with the proton
remnant, here represented by a color anti-triplet $\u{d}$
diquark.  In total, the event will thus contain two strings, one
${b}$--$\u{d}$ and one $\bbbbar$--$\ubar\dbar$. In $\g\g
\to {b}\bbbbar$ a similar inspection shows that two distinct
color topologies are possible. Representing the proton remnant by a
$\u$ quark and a $\u{d}$ diquark (alternatively ${d}$
plus $\u\u$), one possibility is to have three strings
${b}$--$\ubar$, $\bbbbar$--$\u$ and
$\u{d}$--$\ubar\dbar$, Fig.~\ref{fragstrings}, and the other is
the three strings ${b}$--$\u{d}$,
$\bbbbar$--$\ubar\dbar$ and $\u$--$\ubar$.  When the remnant energy is
to be shared between two objects, as e.g. in Fig.~\ref{fragstrings},
further nonperturbative parameters are introduced, with a limited but
not always negligible impact on the uncertainty of the results
\cite{Norrbin:2000zc}.

Once the string topology has been determined, the Lund string 
fragmentation model \cite{Andersson:1983ia} can be applied to describe the
nonperturbative hadronization. Presupposing that the 
fragmentation mechanism is universal, i.e. process-independent,
the good description of $e^+e^-$ annihilation data should carry over.
The main difference between $e^+e^-$ and hadron--hadron events is that 
the latter contain beam remnants which are color-connected with the
hard-scattering partons.

Depending on the invariant mass of a string, practical considerations
lead to the need to distinguish three hadronization prescriptions:
\begin{enumerate}
\item {\em Normal string fragmentation}.
In the ideal situation, each string has a large invariant mass. Then the 
standard iterative fragmentation scheme, for which the assumption of a 
continuum of phase-space states is essential, works well. The average 
multiplicity increases linearly with the string `length', which means 
logarithmically with the string mass. In practice, this approach can be 
used for all strings above some cut-off mass of a few GeV. 
\item {\em Cluster decay}.
If a string is produced with a small invariant mass, maybe only two-body
final states are kinematically accessible. The continuum assumption
above then is not valid, and the traditional iterative Lund scheme is 
not applicable. We call such a low-mass string a cluster, and consider
it separately from above. When kinematically possible, a $\Q$--$\qbar$ 
cluster will decay into one heavy and one light hadron by the production 
of a light quark--antiquark pair in the color force field between the 
two cluster endpoints, with the new quark flavor selected according to 
the same rules as in normal string fragmentation. Close to the two-body
threshold the decay is isotropic, while it should smoothly attach to the
string picture of a preferential longitudinal direction for heavier
clusters.
\item {\em Cluster collapse}.
This is the extreme case of the above situation, where the string 
mass is so small that the cluster cannot decay into two hadrons.
It is then assumed to collapse directly into a single hadron, which
inherits the flavor content of the string endpoints. The original 
continuum of string/cluster masses is replaced by a discrete set
of hadron masses, mainly $D/B$ and $D^*/B^*$ (or corresponding 
baryon states). In order to preserve overall energy and momentum,
nearby string pieces have to absorb a recoil of the collapse,
according to a procedure intended to minimize disturbances to the event.
This mechanism plays a special r\^ole, in that it allows 
large flavor asymmetries in favor of hadron species that can inherit 
some of the beam-remnant flavor content. 
\end{enumerate}

Thus the cluster collapse mechanism tends to enhance the production of
heavy hadrons which share its light flavor with the hadron beam.
For charm this has been observed at several fixed-target experiments, where the
effect is very large for large $x_{F}$ \cite{Charmasym}.
This effect is yet to be studied for $B$ mesons. It is expected
to be fairly large at HERA-B \cite{Norrbin:2000zc} but small for
the Tevatron, see \cite{Norrbin:2000zc} and below.

The color connection between the produced heavy quarks and the beam
remnants in the string model gives rise to an effect called
beam remnant drag. \index{beam drag}
In an independent fragmentation scenario, a quark jet
fragments symmetrically around the quark direction. The light cone
(along the quark axis) energy-momentum of the quark
is then simply scaled by some factor, picked from a fragmentation function,
in order to give the momentum of the hadron.
Thus, on the average, the rapidity would be conserved in the fragmentation process.
This is not necessarily so in string fragmentation where
both string ends contribute to the four-momentum of the produced
heavy hadron. If the other end of the string is a beam remnant,
the hadron will typically be shifted in rapidity in the direction of the beam remnant,
often resulting in an increase in $|y|$. This beam-drag is shown
qualitatively in Fig.~\ref{drag}, where the rapidity shift for bottom
hadrons in a 2 TeV $p {\bar p}$ collision is shown as a
function of rapidity and transverse momentum. We use two different measures of
the rapidity shift. The first is the average rapidity shift
$\Delta y = \langle y_{B} - y_{b} \rangle$. Here the heavy quark
can be connected to a beam remnant on either side of the event, giving rise
to shifts in both directions which tend to cancel in inclusive
measures. A better definition is therefore
\begin{equation}
\Delta y_{sign} = \langle (y_{B}-y_{b})
\cdot {sign}(y_{other~end}) \rangle,
\end{equation}
which measures the rapidity shift in the direction of the other end of the string.
This shift should almost always be positive. The rapidity shift is not directly accessible
experimentally, only indirectly as a discrepancy between the shape of perturbatively
calculated quark distributions and data. As can be seen from the figures this effect
is large only at large rapidities and small $p_{T}$.

\begin{figure}[t]
\begin{center}
\epsfig{file=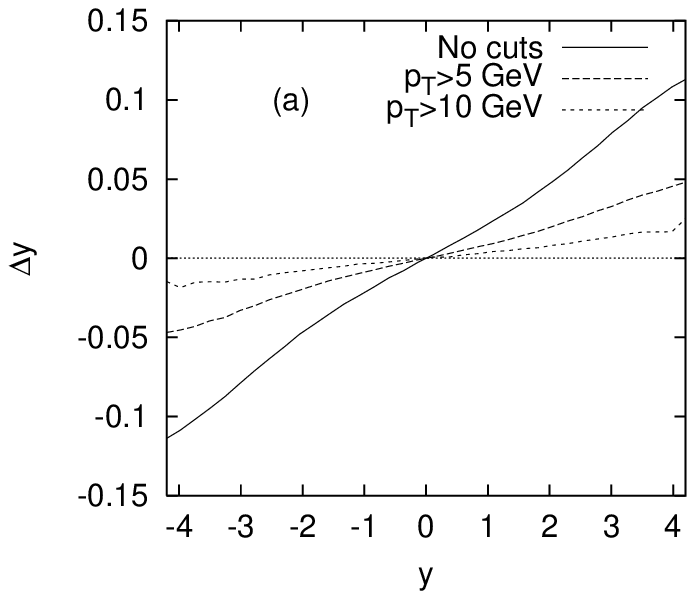,width=75mm}
\epsfig{file=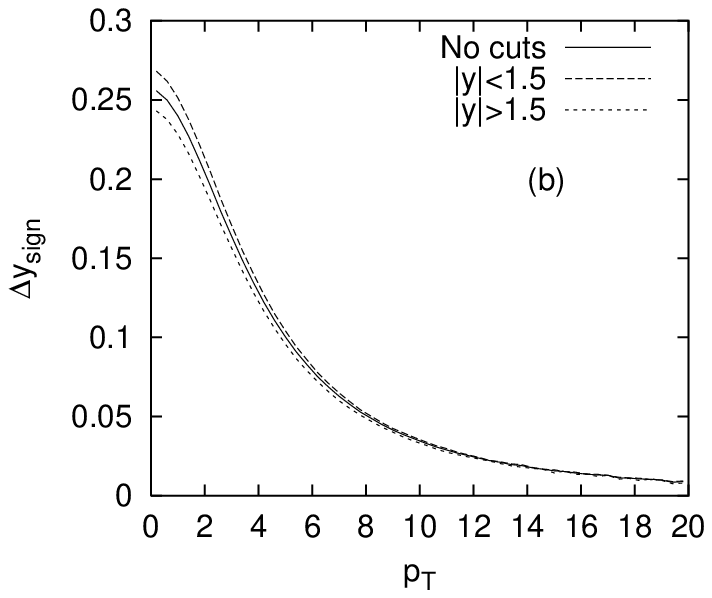,width=75mm}
\end{center}
\caption{(a) Average rapidity shift $\Delta y$ as a function of $y$ for some 
different $p_{T}$ cuts for a $p {\bar p}$ collider at 2 TeV.
(b) Average rapidity shift $\Delta y_{sign}$ as a function of $p_{T}$
for some different rapidity cuts.
\label{drag}}
\end{figure}

\begin{figure}[thb]
\begin{center}
\epsfig{file=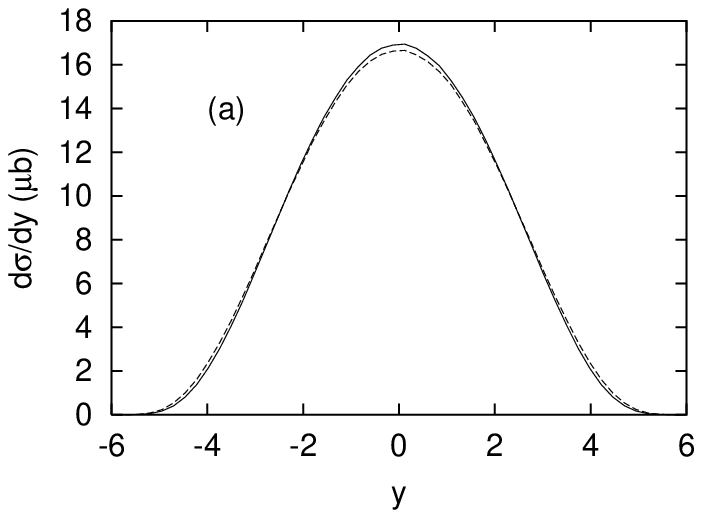,width=75mm}
\epsfig{file=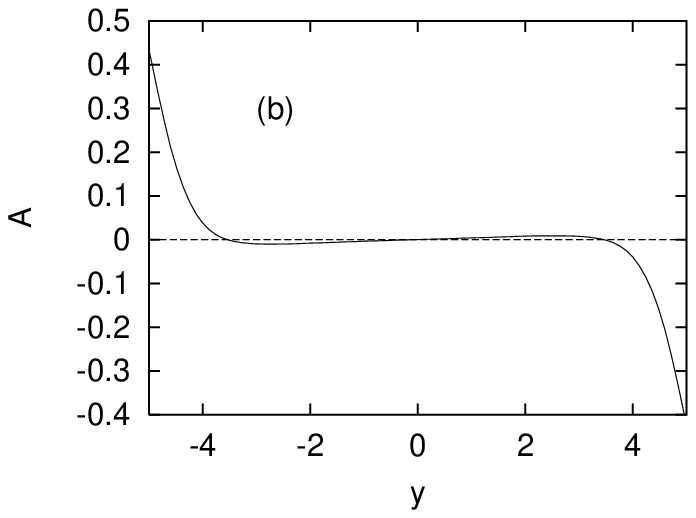,width=75mm}
\end{center}
\caption[Bottom production at the Tevatron]
{Bottom production at the Tevatron.
(a) Rapidity distribution of bottom quarks (full)
and the B hadrons produced from them (dashed).
(b) The asymmetry $A=\frac{\sigma({B}^0)-\sigma(\Bbar^0)}
{\sigma({B}^0)+\sigma(\Bbar^0)}$
as a function of rapidity. For simplicity, only pair production is included.}
\label{tev_generic}
\end{figure}

A possible observable consequence of the beam drag effect is the
asymmetry between $B^0$ and $\bar{B}^0$ for large
rapidities.  Fig.~\ref{tev_generic} shows the distribution of bottom
quarks and the hadrons produced from them, as well as the asymmetry
between $\B^0$ and $\Bbar^0$ without any kinematic cuts, using pair
production only. The asymmetry is antisymmetric because of the
asymmetry of the initial state.  Therefore the asymmetry is zero at
$y=0$ and increasing in different directions for increasing/decreasing
rapidities.  Consider the situation when the kinematic limit at large
positive rapidities (the proton fragmentation region) is
approached. Here the asymmetry changes sign for large rapidities
because of the drag-effect; ${b}$ quarks connected to diquarks
from the proton beam remnant which carry most of the remnant energy
often produce $\Bbar^0$ hadrons which are shifted more in rapidity
than the ${B}^0$'s are.  The $\Bbar^0$ rapidity distribution is
thus harder than for ${B}^0$.  Cluster collapse, on the other
hand, tends to enhance the production of `leading' particles (in this
case ${B}^0$) so the two mechanisms give rise to asymmetries
with different signs. Collapse is the main effect at central
rapidities while eventually at very large $y$, the drag effect
dominates. The situation is reversed in the ${\bar p}$ fragmentation
region. Despite the superficial differences between cluster collapse
and beam drag it should be realized that they are simply two
consequences of the same thing, namely the color connection between
the products of the hard process and the beam remnants and the
subsequent hadronization of the resulting string.

\begin{figure}[thb]
\begin{center}
\epsfig{file=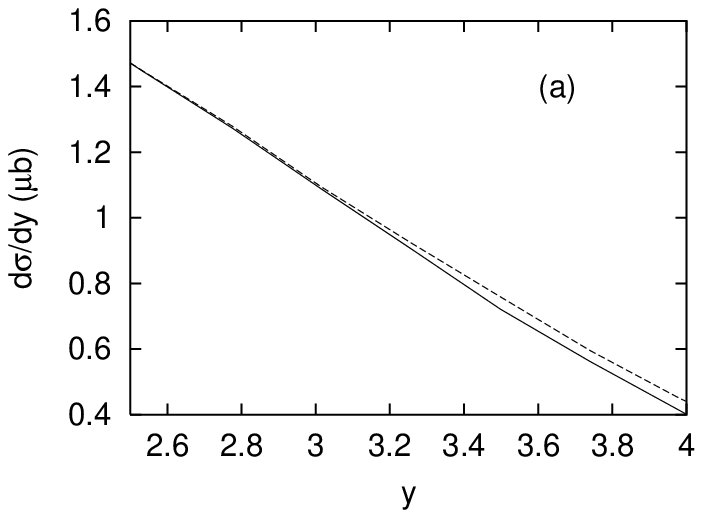,width=75mm}
\epsfig{file=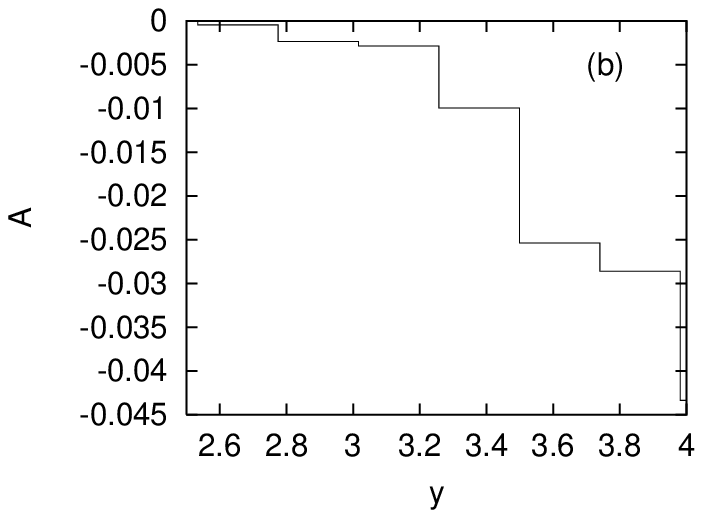,width=75mm}
\end{center}
\caption[Bottom production at the Tevatron]
{Bottom production at the Tevatron for $2.5<|y|<4$ and $p_T<5$ GeV.
(a) $\B^0$ (full) and $\Bbar^0$ (dashed) rapidity spectra.
(b) The asymmetry $A=\frac{\sigma(\B^0)-\sigma(\Bbar^0)}{\sigma(\B^0)+\sigma(\Bbar^0)}$
as a function of rapidity. For simplicity, only pair production is included.
\label{tev_btev}}
\end{figure}

In Fig.~\ref{tev_btev} we introduce cuts in order to study the region of 
large (positive) rapidities and small $p_{T}$. The $\Bbar^0$ spectrum is 
slightly harder than the $\B^0$ one, but the size of the effect is quite 
small, approaching $4\%$ at very large rapidities. Still, if large precision 
is desired in CP violation studies, this effect could be non-negligible.  The 
effect is much larger at HERA-B because of the much smaller CM-energy, so 
the effect should be studied there first to assess its size.  Further details 
and applications are given in \cite{Norrbin:2000zc}.

\subsection[Heavy-quark fragmentation ambiguities]
{Heavy-quark fragmentation ambiguities\authorfootnote{Authors: B. Harris, E. Laenen}}
\label{sec:heavy-quark-fragm}
\index{fragmentation!heavy quark!ambiguities}

Unlike massless quarks, heavy quarks cannot have all of their four-momentum 
components degraded and stay on their mass shell simultaneously.
Hence, only some components get scaled by $z$; which ones is to some 
degree a matter of choice, and a source of ambiguity.  
Moreover, the action of boosting between the parton-parton center-of-mass 
and the laboratory frame, and fragmenting do not commute for heavy quarks, 
generating another ambiguity.  We have studied these ambiguities
numerically at leading order, in the context of $B$-meson 
production at the Tevatron.

When fragmenting a heavy quark to a heavy hadron one must
choose the momentum component to be scaled. Some common choices are 
\begin{enumerate}
\item scaling the space component $\vec{p}_H = z\vec{p}_Q$ and adjusting 
the energy via the mass-shell condition $E_H = \sqrt{\vec{p}_H+m_H^2}$ ;
\item scaling the plus component $(E+p_{||})_H = z(E+p_{||})_Q$ and adjusting 
the minus component via the mass-shell condition 
$(E-p_{||})_H = m_H^2/(E+p_{||})_H$ ( here $||$ refers 
to the $Q$ direction, not the beam );
\item scaling the full four-momentum, ignoring the mass-shell requirement.
\end{enumerate}
In Fig.\ \ref{fig:harris4} we show the effects of these choices on the
$P_T$ distribution of a generic $B$-meson in $p\bar{p}$ collisions
at 1.8 TeV.  We have taken $m_B=m_b=5$ GeV, a Peterson et al.\ 
\cite{Peterson:1983ak} fragmentation function with $\epsilon=0.006$, 
and performed the fragmentation in the parton-parton center-of-mass.
Choices 1 and 3 handle the $p_x$ and $p_y$ components in the same fashion so 
no effect is observed.  The second method mixes the $p_x$ and $p_y$ components 
with the $E$ and $p_z$ components producing a pronounced difference at 
small $p_T$.  All methods are identical in the limit of large $p_T$ 
indicating that this is a source of power suppressed corrections.
\begin{figure}[htbp]
\centerline{\epsfig{figure=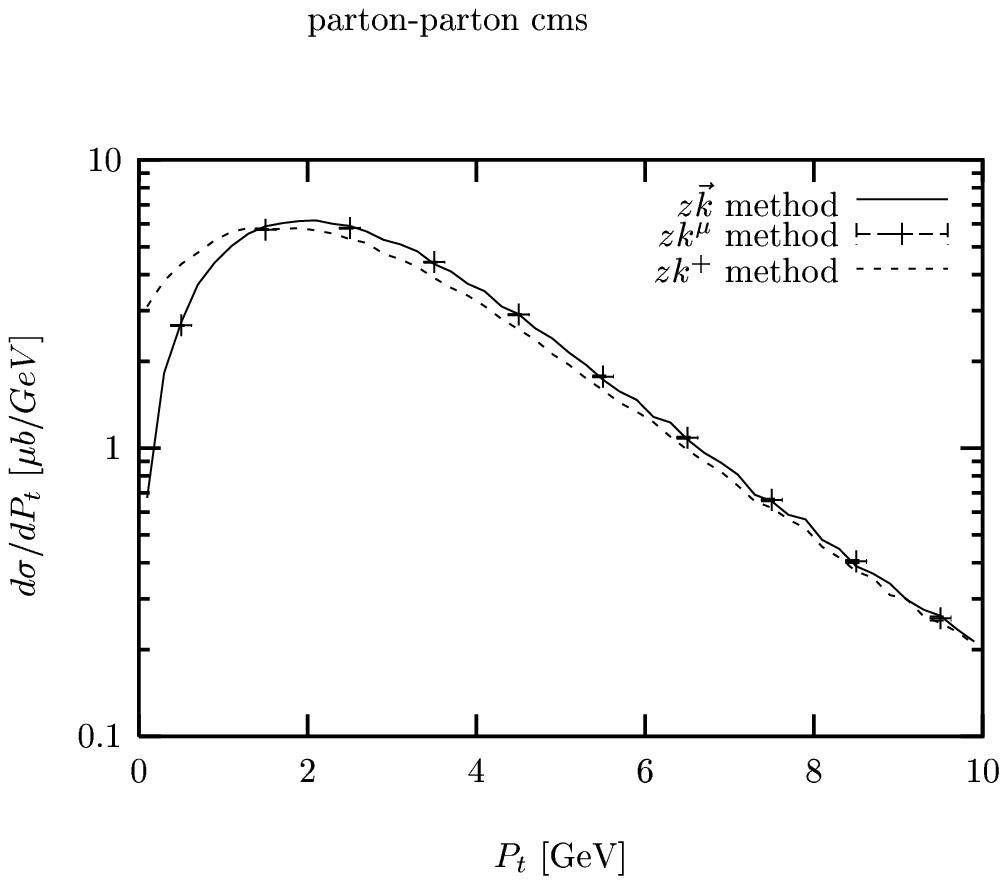,width=0.6\textwidth,angle=0,clip=}}
\caption{\label{fig:harris4} Effect of the different four-vector smearing 
methods on the $P_T$ distribution of the $B$-meson.}
\end{figure}

There is also an ambiguity in the order in which the
fragmentation of the heavy quark, and the boosting from the parton-parton 
center-of-mass to the lab frame is implemented.  
To see why this is so, one may compute, e.g.\ for the fragmentation choice 
$\vec{p}_H = z\vec{p}_Q$ with $m_H\simeq m_Q$, the difference in energy of a 
massive particle arising from the order of boosting 
(along $z$-axis with rapidity $\eta$) and fragmenting
\begin{equation}
  \label{eq:2}
  \Delta E =\frac{m_Q^2}{2|\vec{p}_Q|}\sinh(\eta) \frac{1-z^2}{z}
\end{equation}
for small $m_Q/|\vec{p}_Q|$. 

The result of this non-commutativity (for the $\vec{p}_H = z\vec{p}_Q$ method) 
is shown in Figs.\ \ref{commbf} and \ref{fig:harris3} for the pseudo-rapidity 
and transverse momentum distribution, respectively, of the $B-$meson.  
The interplay of such ambiguities with acceptance cuts is given 
in Table \ref{tab:ffone}.  For a transverse momentum above 5 GeV the
effect is negligible.

\begin{figure}[htbp]
\begin{center}
\epsfig{figure=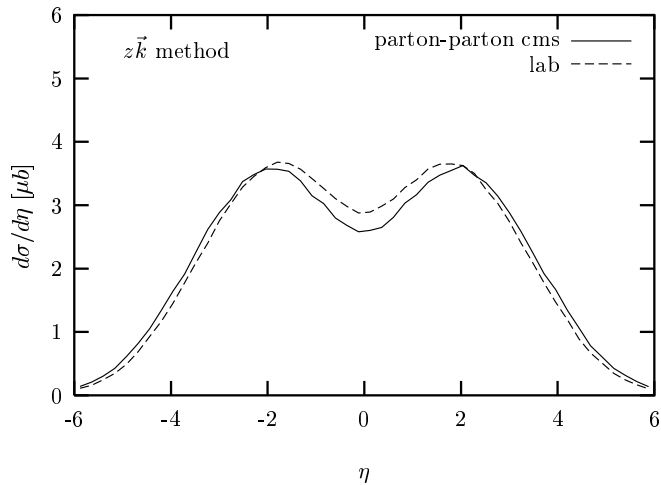,width=0.6\textwidth,angle=0,clip=}
\end{center}
\caption{\label{commbf} Commutativity of boosting and fragmenting
for pseudo-rapidity distribution of the $B$-meson.}
\end{figure}

\begin{figure}[htbp]
\centerline{\epsfig{figure=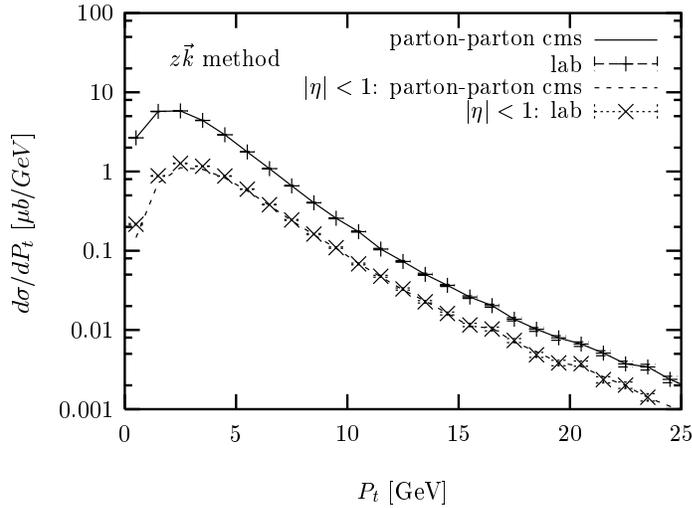,width=0.6\textwidth,angle=0,clip=}}
\caption{\label{fig:harris3} Commutativity of boosting and fragmenting for 
$P_T$ distribution of the $B$-meson.}
\end{figure}

\begin{table}
 \begin{center}
 \leavevmode
\begin{tabular}{c||c|c|c}
 \hline\hline
acceptance & $\sigma[\mu b] $ lab.& parton cms $\sigma[\mu b]$ & difference \\
 \hline\hline
no cuts & 26.24 & 26.24 & - \\
 $p_T > 0.5,|\eta| < 1.5$ & 9.66 & 8.87 & 8\% \\
 $p_T > 5,|\eta| < 1.5$ & 2.58 & 2.53 & 2\% \\
      \hline\hline
    \end{tabular}
  \end{center}
\caption{\label{tab:ffone}Interplay of boosting ambiguity with acceptance cuts.}
\end{table}

\subsection[Experimental Impact]
{Experimental impact\authorfootnote{Author: S. Menary}}
\label{sec:experimental-impact}
\index{fragmentation!experimental impact}

In this section we shall try to address the extent to which fragmentation
issues are relevant for heavy-quark studies at the Tevatron, specifically
with respect to detector design and optimization.

The results from LEP and SLD on $b$ fragmentation in $e^+e^-$ collisions
clearly indicate that the spectrum is harder than can be described by
the standard Peterson {\it et al.} fragmentation function. This has little
experimental impact other than in analyses which attempt to extract the
$b$ quark cross-section from the observed $b$ hadron momentum spectrum.
This could potentially have a small effect on the efficiency of `same-side'
tagging using the $\pi$ from $B^{**}$ decays. That is, a harder or
softer fragmentation function will directly feed into the relation of
this pion to the $B$ from the decay.

There are some interesting correlations, both in $b$ production and
fragmentation, which feed, perhaps only weakly, into predictions of
tagging efficiencies and strategies. The correlation
of $b$ and $\bar{b}$ directions in the forward and backward regions
increases the "away-side" tagging power of
a forward detector like BTeV since it implies that if one $b$ is
produced going forward (or backward) then most likely the other is
also to be found in the same fiducial volume. On the other hand,
`same-side' tagging will be affected by the distribution of fragmentation
particles produced along with the $b$ hadron.
This has been studied by CDF who rely on data
to extract efficiencies and tune the Monte Carlos.

The beam drag effect discussed in section 1.5.3 is potentially an issue
for measurements in the very forward direction. The fact that BTeV is
a two arm spectrometer can minimize this considerably since they can
measure asymmetries in both the $p$ and $\bar{p}$ arms and essentially
subtract out the beam drag effect. In particular this can be done using
very high statistics non-CP modes where an asymmetry is dominantly
produced by this effect. This could be more problematic at LHCb which is
a single arm spectrometer at a $pp$ collider.

\clearpage{\pagestyle{empty}\cleardoublepage}
\chapter{Summary and Highlights}

During the coming decade a wide range of measurements in the production and
decay of $b$ flavored hadrons will be made at the Tevatron. In this chapter, we
summarize the most important ones. Particularly in the decays, some observables
are key to pinning down the flavor structure of the Standard Model and,
possibly, uncovering new physics. Some are unique to the Tevatron, and some are
expected to be competitive with measurements at the $e^+e^-$ $B$ factories. The
Tevatron enjoys a few advantages that compensate for the clean environment in
$e^+e^-$ collisions.  First, the $b\bar{b}$ production cross section is much
higher at the Tevatron than at $e^+e^-$ machines. Thus, despite the higher
background rates at the Tevatron, it is possible to use selective triggers to
obtain high statistics samples with good signal-to-noise ratio.  Second, all
$b$ flavored hadrons are created ($B_s$, $\Lambda_b$, etc.), whereas an
$e^+e^-$ machine operating at the $\Upsilon(4S)$ resonance produces only
$B^\pm$, $\Bbar^0_d$ and $B^0_d$. Consequently, the $B$ decay program at hadron
colliders complements that at the $e^+e^-$ $B$ factories.

The CDF, D\O, and BTeV detectors and simulation tools used for this report are
described in Chapters 2--5. Since Run~I, both CDF and D\O\ have gone through
major upgrades, and both will be much more powerful for doing $B$ physics in
Run~II. Both detectors feature excellent charged particle tracking using
solenoidal magnetic fields, and both have silicon (Si) vertex detectors capable
of tracking in three dimensions. The magnetic field is 2\,T at D\O\ and 1.4\,T
at CDF. D\O\ has a smaller radius tracking volume ($50$\,cm), which allows for
Si and scintillating fiber tracking out to $\eta = 1.6$, and Si disk tracking
to $\eta=3$. CDF has a larger tracking radius ($140$\,cm), with full Si and
drift chamber tracking out to $\eta = 1$, and Si-only tracking (to $30$\,cm
radius) out to $\eta = 2$. D\O\ muon coverage extends to $\eta=2$; CDF to $\eta
= 1.5$. Both detectors have track based triggers at level-1 and Si vertex
impact triggers at level-2. Because of the larger level-1 bandwidth ($40$\,kHz)
and the use of deadtimeless SVX3 chip readout, CDF can deploy a Si vertex
hadronic decay trigger at level-2, which is directly sensitive to decays like
$B\to\pi^+\pi^-$ and $B_s\to D_s^-\pi^+$. With a smaller level-1 bandwidth
($10$\,kHz), D\O\ does not plan to implement such a strategy. CDF has two
methods for particle ID --- time-of-flight at low momentum ($p<2\,{\rm GeV}/c$)
and relativistic rise $dE/dx$ for high momentum ($p>2\,{\rm GeV}/c$); the
latter is crucial for channels like $B\to\pi^+\pi^-$, where a statistical
separation of $\pi\pi$ and $K\pi$ signals can be used. In summary, both CDF and
D\O\ have comparable $B$ physics reach; CDF has an advantage in particle ID and
in Si vertex triggering.

BTeV is slated to run in earnest after Run~II, following construction and some
running during Run~IIb. It will have competition from LHC-$b$ at CERN, and from
experiments at the $e^+e^-$ $B$ factories, which by then are expected deliver
an order of magnitude higher luminosity than at present. BTeV is not a central
detector (as are CDF and D\O), but a two-arm forward spectrometer. The key
features of its design are a silicon pixel detector, a flexible trigger based
on detached vertices, particle ID with excellent $K/\pi$ separation, and an
electromagnetic calorimeter capable of identifying $\pi^0$'s and photons.

In this workshop, simulations of the CDF, D\O\ and BTeV detectors have made
common assumptions about production rates, branching ratios, and flavor tagging
efficiencies. The common starting point is needed for comparing the reaches of
the three detectors, but it is nevertheless difficult to make meaningful
comparisons without Run~II data. At hadron colliders all signal channels, as
well as issues like flavor tagging, have backgrounds that will be detector
dependent and that, in many cases, cannot be reliably predicted from a heavy
flavor production Monte Carlo. These backgrounds affect both
signal-to-background statistics and also the strategies used to reject
background, which in turn affect signal yields. For example, for
$B\to\pi^+\pi^-$, to reject the combinatorics from all sources of backgrounds,
making harder cuts on the detachment of the secondary vertex and other
quantities will affect both the statistical accuracy and the ability to
separate the direct and mixing induced $CP$ asymmetries. While CDF has Run~I
data on many channels and flavor tags, D\O\ and BTeV will need real data to
understand fully the effects of backgrounds.

Most simulations carried out during this workshop considered an integrated
luminosity of 2\,fb$^{-1}$, corresponding to Run~IIa for D\O\ and CDF and to the
first year of running for BTeV. Other workshops in this series, which focused
on high $p_T$ physics, considered also the potential for 10\,fb$^{-1}$ and
30~fb$^{-1}$. In the case of $B$ physics, the role of real data in optimizing
event selection makes it difficult to make sensible estimates for such high
luminosity until more experience is at hand.

The most important measurements at the Tevatron can be divided roughly into two
categories. Some modes have a relatively simple theoretical interpretation, in
the context of the Standard Model, either because hadronic uncertainties are
under good control, or because loops, small CKM angles, or GIM effects suppress
the Standard Model rate. These modes test the CKM mechanism and probe for
non-CKM sources of \CP\ and flavor violation. Other measurements are more
sensitive to QCD in ways that are theoretically challenging. They do not (yet)
probe flavor dynamics, but, clearly, better understanding of QCD in $B$ physics
can be reinvested in understanding the flavor sector. We therefore present two
semi-prioritized lists labeled ``tests of flavor and \CP\ violation'' and
``test of QCD'', but both labels should be construed loosely. Moreover, the
composition of these lists is colored by our  understanding during the course
of the workshop. In the coming decade theoretical and experimental developments
are likely to spur changes in the lists. Many other interesting observables,
not discussed in this summary, are covered throughout Chapters 6--9.

\vspace*{8pt}
\noindent{\bf\em Tests of flavor and \boldmath\CP\ violation}
\begin{itemize} \vspace*{-8pt}

\item \bbms: Both CDF and D\O\ can measure $x_s \lesssim 30$ from events with a
semileptonic $B_s$ decay. This covers some of the expected range $x_s \lesssim
45$. With nonleptonic modes, CDF can measure $x_s \lesssim 59$--74, depending
on assumptions, and BTeV $x_s \lesssim 75$. As soon as a 5$\sigma$ observation
of mixing is made, the statistical error of $x_s$ is very small, about
$\pm0.14$. In the Standard Model, $x_s/x_d$ can be used to determine
$|V_{ts}/V_{td}|$, relying on input from lattice QCD for the size of $SU(3)$
breaking.

\item $CP$ asymmetries in $B_s \to\psi\phi$, $\psi\eta^{(\prime)}$: These
measure the relative phase between the amplitudes for \bbms\ and $b\to
c\bar{c}s$ decay, $\beta_s$.  In the Standard Model $\sin2\beta_s$ is a few
percent, so observation of a large $CP$ asymmetry would be a clear sign of new
physics.  The expected error at CDF is about $1.6$ times that of $\sin2\beta$,
further diluted by the $CP$-odd contribution to the $\psi\phi$ final state. 
Although this $CP$-odd contribution is expected to be small, it can be avoided
by using the decay modes $B_s\to\psi\eta^{(\prime)}$, which are pure $CP$-even.
With its excellent photon detection, BTeV is well optimized to measure the
asymmetries in these neutral modes.

\item $CP$ asymmetries in $B_s\to D_s^\pm K^\mp$: Combining the
time dependent asymmetries in these four modes allows the cleanest
determination of $\gamma-2\beta_s$. These measurements must be carried out in
the presence of the large Cabibbo allowed $B_s \to D_s \pi$ background.
Combined with the measurement (or bound) on $\beta_s$, one obtains $\gamma$,
one of the angles of the unitarity triangle. At the Tevatron, this measurement
will probably be possible only at BTeV, with an expected precision of
$\sigma(\gamma) \simeq 10^\circ$.

\item $CP$ asymmetries in $B_d\to\rho\pi$: These asymmetries seem to be the
cleanest way to measure $\alpha$, the angle at the apex of the unitarity
triangle. Once enough events are available to isolate the $\Delta I=3/2$
channel, it is possible to measure $\alpha$ without uncertainties from penguin
amplitudes, which contribute only to $\Delta I=1/2$. With its excellent
electromagnetic calorimetry, BTeV should compete well with BaBar and Belle, on
a similar time scale.

\item $CP$ asymmetry in $B_d \to\psi K_S$: These asymmetries measure the
relative phase between the amplitudes for \bbmd\ and $b\to c\bar{c}s$ decay. In
the Standard Model, this is the angle $\beta$ of the unitarity triangle, and it
is obtained with very small theoretical uncertainty. By the end of Run~IIa, the
CDF and D\O\ measurements may become competitive with BaBar and Belle. Based on
$B\to J/\psi K_S$ only, D\O\ and CDF anticipate a precision on $\sin2\beta$ in
the range 0.04--0.05, with 2\,fb$^{-1}$.

\item $CP$ asymmetries in $B_{d,s}\to h^+_1h^-_2$, $h_i=K$, $\pi$: The utility
of these modes depends on how well the uncertainty from flavor $SU(3)$ breaking
can be controlled. Data for these and other processes will tell us the range of
such effects; the resulting Standard Model constraints could be quite
stringent.  A study by CDF shows that 20\% effects from $SU(3)$ breaking lead
to an uncertainty of only $\sim 3^\circ$ on $\gamma$, which is much smaller
than CDF's expected statistical error with 2\,fb$^{-1}$ of $\sim 10^\circ$.

\item Rare semileptonic and radiative decays, such as $B\to K^*\ell^+\ell^-$ or
$B\to K^*\gamma$: Although the rate for $B\to K^*\gamma$ is higher, the
Tevatron detectors will probably not compete with BaBar and Belle. In the case
of $B\to K^*\ell^+\ell^-$, the high rate of a hadron machine is needed, and the
lepton pair provides a good trigger. In $B\to K^*\ell^+\ell^-$ the
forward-backward asymmetry is especially intriguing, because it is sensitive to
short distance physics.

\item Search for $B_{d,s}\to \ell^+\ell^-$: These flavor-changing neutral
current processes are highly suppressed in the Standard Model. Limits on the
rate constrain non-standard models, which often have other ways to mediate such
decays. Late in Run~II or at BTeV, it may be possible to observe a handful of
events at the standard model rate.

\item \CP\ asymmetries in flavor specific final states: These can be used to
measure the relative phase between the mass and width mixing amplitudes,
analogous to $\Re(\epsilon_K)$ in the kaon sector. The most often cited example
is semileptonic decays, but if this asymmetry is sufficiently large, it may be
possible to detect it using fully reconstructed modes such as $B_s\to
D_s^-\pi^+$.  In the $B$ system, the SM predicts the relative phase to be very
small, but new physics could make these asymmetries measurable without pushing
the measured value of $\sin2\beta$ outside its expected range.

\end{itemize} \vspace*{-6pt}

The most striking thing about this list is that the whole program, taken
together, is much more interesting than any single measurement. Indeed, if one
adds in measurements from the $e^+e^-$ machines, the list becomes even more
compelling. In the near future, BaBar, Belle, CDF, and D\O\ will measure
$\sin2\beta$ at the few percent level; the Tevatron measurement of $x_s$
combined with the known value of $x_d$ will determine $|V_{td}/V_{ts}|$;
BaBar's and Belle's measurements of semileptonic decays will determine
$|V_{ub}/V_{cb}|$. Apart from $\sin2\beta$, some input from hadronic physics is
needed, but the combination of these results will still test the unitarity of
the CKM matrix below the $10\%$ level.

\vspace*{8pt}
\noindent{\bf\em Tests of QCD}
\begin{itemize} \vspace*{-8pt}

\item $B_s$ width difference $\Delta\Gamma_s$: The theoretical prediction
of $\Delta\Gamma_s$ requires hadronic matrix elements of two four-quark
operators, so it relies on lattice QCD calculations to a greater extent
than $x_s$. It is still an interesting measurement, especially if it is
smaller than expected in the Standard Model.

\item $\Lambda_b$ lifetime: The measured value of the $\Lambda_b$ lifetime does
not agree with theoretical expectations. It is important to improve the
measurements with fully reconstructed hadronic decays. If the discrepancy
remains, it would presumably imply a failure of the operator product expansion
employed in inclusive decays and lifetimes, and one would have to reconcile the
failure here with success for other lifetimes and for inclusive semileptonic
decay distributions.

\item Semileptonic form factors: The $q^2$ dependence of the decay
distribution for \mbox{$\Lambda_b \to \Lambda_c \ell \bar\nu$}, $B\to
K^{(*)}\ell^+\ell^-$, etc., can be combined with theoretical information,
potentially reducing the (theoretical) error on $|V_{cb}|$, $|V_{ub}|$, and
$|V_{ts}|$.

\item Quarkonium production and polarization: Run~I data for charmonium
production disagrees at large $p_T$ with predictions from NRQCD. If the
discrepancies persist, and are confirmed through bottomonium production, they
would pose an important riddle. To make meaningful measurements, experiments
must collect higher statistics, extend to higher $p_T$, cleanly separate states
with different quantum numbers, and distinguish feed-down from direct
production.

\item $b\,\bar b$ production cross section: The cross section
$\sigma(p_T>p_T^{\rm min})$ measured in Run~I data is about twice the
prediction from perturbative QCD. With all scales ($m_b$, $\sqrt{\hat{s}}$,
$p_T$) much larger than $\lqcd$, perturbative QCD should be reliable.  A
variety of QCD effects have been studied, but do not seem to account for the
excess. If the excess is confirmed (with much higher statistics) in Run~II, one
must ask whether something in the theory has gone awry, or whether a
non-standard mechanism produces $b\,\bar b$ events.

\item Spectra of $B_c$ and other doubly heavy hadrons: CDF's $B_c$ mass and
lifetime measurements from Run~I will be improved. Fully reconstructed
nonleptonic decays are needed to obtain a good measurement of the mass, which
can be compared to calculations based on potential models or from lattice QCD.
Because there are now two heavy quarks, the lifetimes and decay widths of $B_c$
and other doubly heavy hadrons provide novel tests of the theory.

\end{itemize} \vspace*{-6pt}

Each of these is interesting for its interaction with QCD theory,
and, at least for the next several years,
each
can be studied only at the Tevatron.

In conclusion, one sees that $B$ physics at the Tevatron will produce a
program of many interesting and, in some cases, essential measurements.
In the case of $B$ decay measurements, the information gained is
competitive with and complementary to that gained from experiments
in $e^+e^-$ accelerators at the $\Upsilon$(4S).
Indeed, the full suite of measurements from $B_u$, $B_d$, and $B_s$
decays is much more interesting than any one or two measurements
taken in isolation.


\clearpage{\pagestyle{empty}\cleardoublepage}

\backmatter

\addcontentsline{toc}{chapter}{Index}
\printindex
 

\begin{thebibliography}{99}
\bibitem{btev_auth} A list of the BTeV Collaboration
                    can be found in Ref.~\cite{btev_prop}.
\bibitem{execsum} The executive summary is part of the BTeV Proposal,
                 Ref.~\cite{btev_prop} and can be found separately on
                 the BTeV Proposal web page.
\end{thebibliography}
\end{document}